# Perturbation of CO $A^1\Pi$ by $a^3\Pi$ should have been undetectable: detection, mechanism, and surprises

**Stanisław Ryzner,[1,2] Robert W. Field,[3] Wim Ubachs,[4] Alan N. Heays,[5,6] Asen Pashov,[7] Patryk Jasik,[8,9] Józef E. Sienkiewicz,[8,10] Nelson de Oliveira,[11] Kin F. Lai,[12] Marzena I. Malicka,[13] Aleksandra Stasik,[1,2] Wojciech Szajna,[1] Izabela Piotrowska,[1] and Rafał Hakalla[1,a)]**

[1] Materials Spectroscopy Laboratory, Institute of Physics, University of Rzeszów, 1 Pigonia Street, 35-310 Rzeszów, Poland.
[2] Doctoral School of the University of Rzeszów, University of Rzeszów, 16C Rejtana Street, 35-959 Rzeszów, Poland
[3] Department of Chemistry, Massachusetts Institute of Technology, Cambridge, MA02139 USA.
[4] Department of Physics and Astronomy, and LaserLaB, Vrije Universiteit, De Boelelaan 1100, 1081 HZ Amsterdam, Netherlands.
[5] J. Heyrovsky Institute of Physical Chemistry, Dolejskova 2155/3, 182 23 Prague 8, Czech Republic.
[6] Manaaki Whenua Landcare Research, Bioeconomy Science Institute, PO Box 69040, Lincoln 7640, New Zealand.
[7] Faculty of Physics, Sofia University St. Kliment Ohridski, 5 James Bourchier Boulevard, 1164 Sofia, Bulgaria.
[8] Faculty of Applied Physics and Mathematics, Gdańsk University of Technology, Gdańsk 80-233, Poland.
[9] BioTechMed Center, Gdańsk University of Technology, Gdańsk 80-233, Poland.
[10] Advanced Materials Center, Gdańsk University of Technology, Gdańsk 80-233, Poland.
[11] Synchrotron SOLEIL, Orme de Merisiers, St. Aubin, BP 48, F-91192 Gif sur Yvette Cedex, France.
[12] Department of Physics, The University of Hong Kong, Hong Kong SAR, China.
[13] The Faculty of Mathematics and Applied Physics, Rzeszów University of Technology, 8 Powstańców Warszawy Street, 35-959 Rzeszów, Poland.

[a)] **Author to whom correspondence should be addressed:** rhakalla@ur.edu.pl

## ABSTRACT

A direct perturbation between the $a^3\Pi$ and $A^1\Pi$ vibrational levels of two electronic states that belong to the same electronic configuration of the carbon monoxide molecule should be unobservably weak because the vibrational overlap integral $\langle a^3\Pi, v' | A^1\Pi, v'' \rangle$ for their near-degenerate levels is extremely small. Yet level-shift and transition-intensity signatures of the $a^3\Pi(v = 13) \sim A^1\Pi(v = 2)$ perturbation are unmistakably present in the $^{13}C^{18}O$ isotopologue. A detailed study of, among others, the $A^1\Pi - X^1\Sigma^+(2, 0)$, $B^1\Sigma^+ - A^1\Pi(0, 2)$, and $C^1\Sigma^+ - A^1\Pi(0, 2)$ spectra, obtained by the VUV- and Vis-FT complementary techniques in the energy regions of the six crossing points (*e*- and *f*-symmetry, $|\Omega| = 0, 1, 2$) between the $a^3\Pi(v = 13)$ and $A^1\Pi(v = 2)$ states, reveal that the perturbation pathways are indirect, primarily facilitated by the $e^3\Sigma^-(v = 4)$ state, and that quantum mechanical interferences via its $F_{1e}$ and $F_{3e}$ components occur.

# I. INTRODUCTION

Perturbations, as the word suggests, are rare events. They are a disruption of the expected regular pattern in a spectrum. They arise from inter-electronic state interaction terms in the molecular Hamiltonian that, guided by the Born-Oppenheimer (BO) approximation, are left out in the first approximation of the model for the spectrum. These terms include: spin-orbit, spin-electronic, **L**-uncoupling, **S**-uncoupling and electrostatic interactions between different electronic states. When two same-*J*, same-parity levels of two electronic states are accidentally near-degenerate, the regular level structures of both states are disrupted: this is a perturbation! The off-diagonal matrix elements responsible for the inter-electronic state interactions depend on calculable electronic, vibrational, and rotational factors and the energy differences between the perturbed and perturber electronic states. A complete deperturbation model consists of factors calculable from the vibration-rotation wavefunctions of the BO potential energy curves and an electronic factor for the non-adiabatic interaction.

One of us (RWF) created a complete deperturbation model for the interactions of the CO $a^3\Pi$ and $A^1\Pi$ states of the $\pi^4\sigma\pi^*$ electronic configuration and the $a'^3\Sigma^+$, $e^3\Sigma^-$ $d^3\Delta$, $I^1\Sigma^-$ and $D^1\Delta$ electronic states of the $\pi^3\sigma^2\pi^*$ electronic configuration.[1–3] All electronic factors of these $5 \times 2 = 10$ inter-configuration perturbation matrix elements depend on a single π to σ orbital promotion. The vibration-rotation factor consists of (*J*-dependent) vibrational $\langle v'|v''\rangle$ overlap and rotational $\langle v'|\mathbf{B}(\mathbf{R})|v''\rangle$ integrals, which are accurately calculable from the BO potential energy curves. The only electronic matrix element excluded from this deperturbation model is the isoconfigurational $\langle a^3\Pi|\mathbf{H^{SO}}|A^1\Pi\rangle$ factor. It is missing because it is known *a priori* to be *negligibly* small.

The off-diagonal interaction between same-*J*, same-parity perturbed and perturber states gives rise to observable level-shifts, intensity-borrowing, and quantum mechanical interference effects. These effects are modelled by nondegenerate and quasi-degenerate perturbation theory. The electronic state interaction factors are determined by least-squares adjustment of the parameters in a multi-state deperturbation model. Over several decades many such deperturbation studies have been performed, in which the CO molecule and its isotopologues have served as a benchmark system.[4–6] In particular, the $A^1\Pi$ state and its perturbers have in recent years been investigated at high resolution for various carbon monoxide isotopologues, using the techniques of Fourier-transform (FT) emission in the visible domain and synchrotron-based FT absorption spectroscopy in the vacuum ultraviolet domain,[7–12] in some cases supported by laser spectroscopy.[13–15]

Why is the $\langle a^3\Pi|\mathbf{H^{SO}}|A^1\Pi\rangle$ matrix element extremely small for the CO $a^3\Pi$ and $A^1\Pi$ states? The reason is that the isoconfigurational electronic states have similar rotation-vibration constants. There is a very strong $\Delta v = 0$ propensity rule for vibrational overlap integrals between isoconfigurational electronic states. The energy difference between the $a^3\Pi$ and $A^1\Pi$ states is about eleven times the vibrational frequency, $\omega_e$. Therefore the $\langle v_{a^3\Pi}|v_{A^1\Pi}\rangle$ overlap integral between near degenerate vibrational levels is expected to be extremely small ($< 10^{-6}$) and the electronic factor of the spin-orbit matrix element in CO, is expected to be too small in this case (< 100 $cm^{-1}$), so that the *direct* $A^1\Pi \sim a^3\Pi$ spin-orbit matrix element will be undetectable at the current precision limit of modern spectroscopy in the visible and ultraviolet regions, ca. 0.001 – 0.01 $cm^{-1}$.

However, *indirect* perturbation interactions between the $\pi^4\sigma\pi^*$ configuration $a^3\Pi$ and $A^1\Pi$ states, mediated by the $\pi^3\sigma^2\pi^*$ configuration $e^3\Sigma^-$, $d^3\Delta$, $a'^3\Sigma^+$, $I^1\Sigma^-$, and $D^1\Delta$ electronic states, are in principle observable. Several stringent conditions must be met in order to observe these indirect perturbations:

- A near coincidence must be identified among the $A^1\Pi(v)$, $a^3\Pi(v + 11)$ and one vibrational member of the $e^3\Sigma^-$, $d^3\Delta$, $a'^3\Sigma^+$, $I^1\Sigma^-$ and $D^1\Delta$ group of states. The vibration-rotation constants of all seven electronic states are known for $^{12}C^{16}O$, and isotope shifts for $^{12}C^{18}O$, $^{13}C^{16}O$, and $^{13}C^{18}O$ are calculated. It turns out that the $^{13}C^{18}O$ $A^1\Pi(v = 2)$, $a^3\Pi(v = 13)$ pair offers the best three-state vibronic near-coincidence.
- This vibronic near-coincidence must occur for the same angular momentum state, *J,* and for levels of the same *e/f* symmetry in each of the three interacting states.
- The electronic factors of each of the 5 × 2 pairs of inter-configuration pairs of perturbing states must be known. These factors are indeed available.[11,12]
- The vibrational overlap and **B(R)** integrals must be evaluated. Although these quantities are not expected to be strongly *J*-dependent, they have been calculated for a complete set of 1 < *J* < 100 rotational levels.
- A CO source capable of populating moderately high-*J* levels and not consuming prohibitive amounts of $^{13}C$ and $^{18}O$ rare isotopes must be available. Such a source has been implemented in the present study.

Thus, indirect $A^1\Pi(v = 2) \sim a^3\Pi(v = 13)$ perturbations mediated by $e^3\Sigma^-(v = 4)$, $d^3\Delta(v = 7)$, and $I^1\Sigma^-(v = 3)$ vibronic levels, are observed. The existence of each of these interactions is verified by the effect on the quality of fit by turning off each of the interaction matrix elements in the deperturbation model for which the PGOPHER software is used.[16,17] The indirect perturbation with the $a^3\Pi(v = 13)$ state was observed through a comprehensive spectroscopic analysis of the $A^1\Pi(v = 2)$ and all of its perturbers in $^{13}C^{18}O$, based on experimental VUV-FT and Vis-FT data obtained in this work.

# II. EXPERIMENTAL DETAILS

## A. Vis-FT emission spectroscopy

The Bruker IFS 125HR Fourier-transform spectrometer, working under vacuum conditions, fed by $^{13}C^{18}O$ fluorescence radiation generated in an air-cooled hollow-cathode (HC) discharge lamp, was used. The construction of the HC lamp is described in our previous works.[13,18,19] All details and the procedure of producing the radiation from the $^{13}C^{18}O$ molecules was described in literature.[7] Only the most important information is provided below. The instrumental resolution of the FT spectrometer was 0.018 $cm^{-1}$ during acquisition of 128 scans. Experimental spectra for the $^{13}C^{18}O$ $B^1\Sigma^+ - A^1\Pi(0, 2)$ and $C^1\Sigma^+ - A^1\Pi(0, 2)$ bands, with the maximum signal-to-noise ratio ($SNR_{max}$) 80:1 and 15:1, respectively, are shown in Figures 1 and 2 along with simulations generated with the PGOPHER program[16] after the final deperturbation fit.

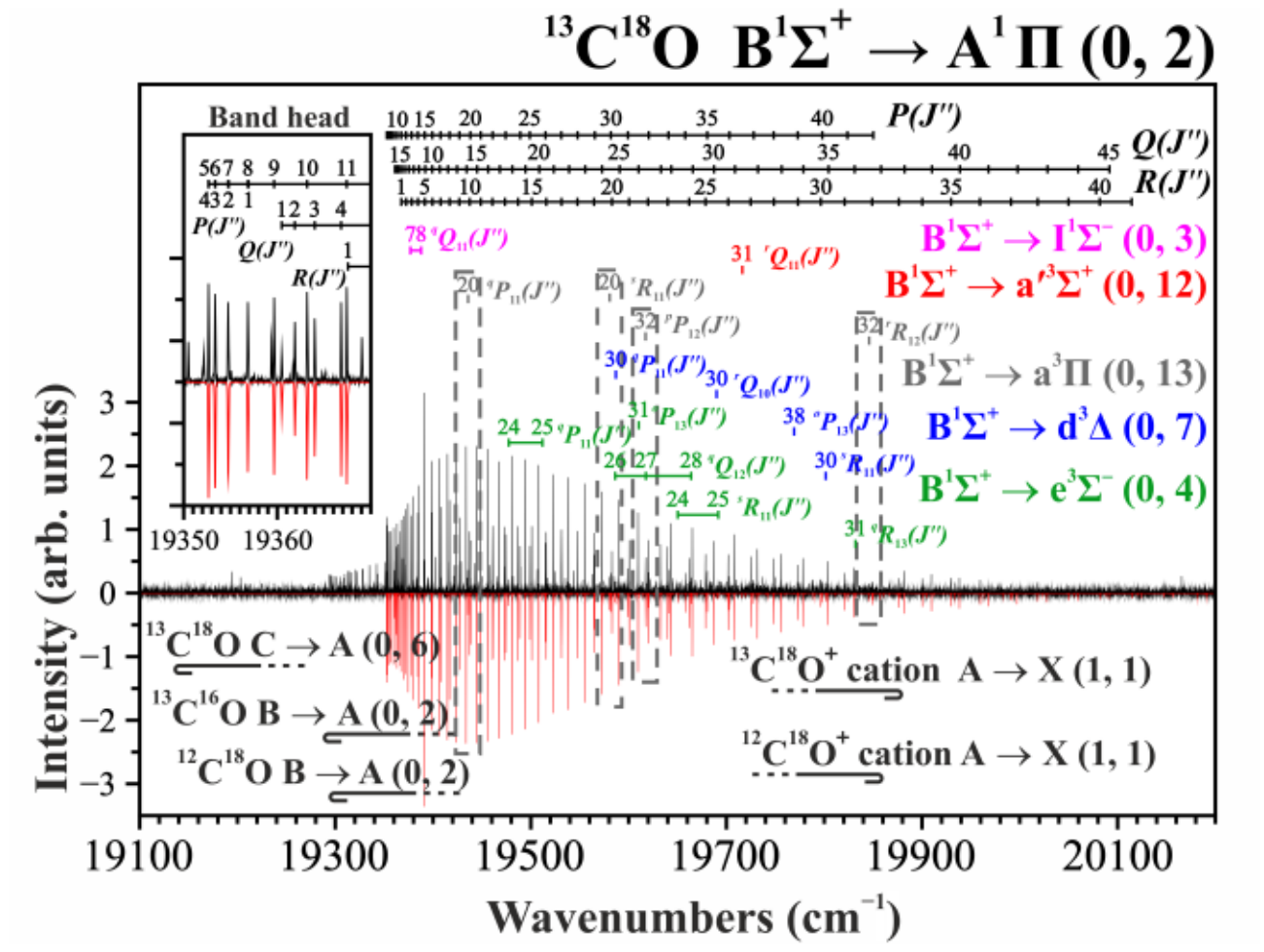

**FIG. 1.** The $^{13}C^{18}O$ $B^1\Sigma^+ - A^1\Pi(0, 2)$ emission spectrum obtained by means of Vis-FT spectroscopy technique (upper, black trace) as well as the simulated one (lower, red trace) prepared after the final deperturbation analysis using the PGOPHER program.[16,17] The $^{13}C^{18}O$ $C^1\Sigma^+ - A^1\Pi(0, 6)$, $^{13}C^{16}O$ $B^1\Sigma^+ - A^1\Pi(0, 2)$, $^{12}C^{18}O$ $B^1\Sigma^+ - A^1\Pi(0, 2)$, $^{12}C^{18}O^+$ $A^2\Pi - X^2\Sigma^+$ (1, 1) and $^{13}C^{18}O^+$ $A^2\Pi - X^2\Sigma^+$ (1, 1) bands coexisting in the spectrum, have been taken into account in the analysis as contaminations of the band under consideration. An inset shows the band head region of the $^{13}C^{18}O$ $B^1\Sigma^+ - A^1\Pi(0, 2)$ band. It should be noted that, from the point of view of observing a novel class of indirect perturbation in a diatomic molecule, the key region (marked by gray dashed-line rectangles) is for $^{13}C^{18}O$ $B^1\Sigma^+ - a^3\Pi(0, 13)$ $^sR_{11ee}$ (20), $^qP_{11ee}$ (20), $^rR_{12ee}$ (32), $^pP_{12ee}$ (32), where these unexpected extra-lines appeared. This phenomenon is interpreted here as an extremely weak perturbation that becomes observable via an indirect mechanisms

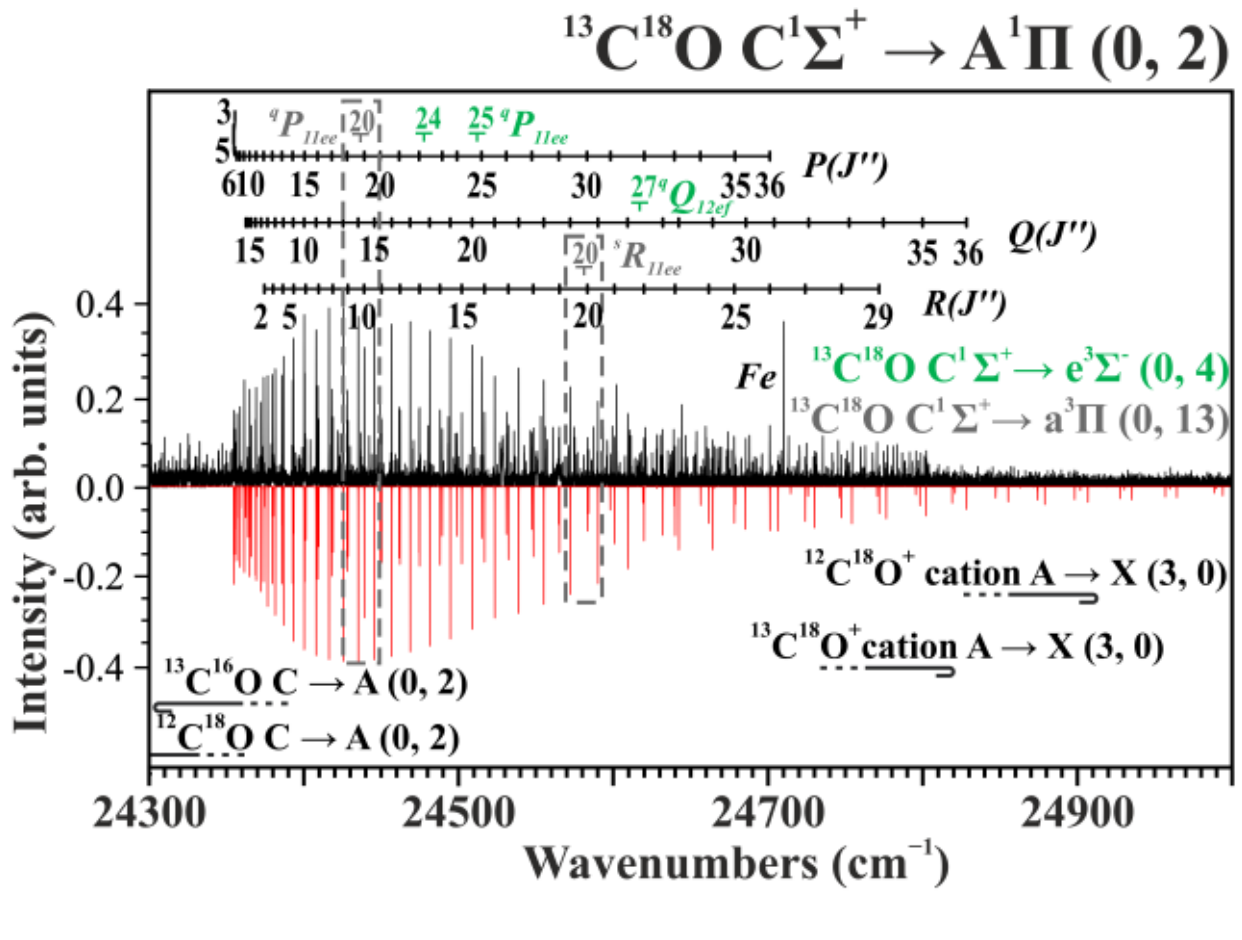

**FIG. 2.** The $^{13}C^{18}O$ $C^1\Sigma^+ - A^1\Pi(0, 2)$ emission spectrum obtained by means of Vis-FT spectroscopy technique (upper, black trace) as well as the simulated one (lower, red trace) prepared after the final deperturbation analysis using the PGOPHER program.[16,17] The $^{13}C^{16}O$ $C^1\Sigma^+ - A^1\Pi(0, 2)$, $^{12}C^{18}O$ $C^1\Sigma^+ - A^1\Pi(0, 2)$, $^{12}C^{18}O^+$ $A^2\Pi - X^2\Sigma^+$ (3, 0) and $^{13}C^{18}O^+$ $A^2\Pi - X^2\Sigma^+$ (3, 0) bands coexisting in the spectrum, have been taken into account in the analysis as contaminations of the band under consideration. It can be seen that, from the point of view of observing a novel class of indirect perturbation in a diatomic molecule, the key region (marked by gray dashed-line rectangles) is for $^{13}C^{18}O$ $C^1\Sigma^+ - a^3\Pi(0, 13)$ $^sR_{11ee}$ (20) and $^qP_{11ee}$ (20), where these unexpected extra-lines appeared. This phenomenon is interpreted here as an extremely weak perturbation that becomes observable via an indirect mechanisms.

The He-Ne line (633 nm), generated by a stabilized interferometric laser (±1.5 MHz / 2 h) was used for calibration of the frequency axis, resulting in a standard deviation of 0.004 $cm^{-1}$. Positions of the molecular transitions were determined by commercial Bruker software, OPUS 8.5,[20] which fits Voigt profiles to the measured spectral line contours. Each single line contour was analyzed individually. The absolute accuracy of the wavenumber measurements was computed by Gaussian error propagation, including calibration and line profile fitting uncertainties using the empirical Brault formula.[21] The formula does not consider line-correlation effects within the complex (multi-line) contours, therefore for blended lines, the uncertainties were arbitrarily doubled. As a result, the wavenumbers of $B^1\Sigma^+ - A^1\Pi(0, 2)$ and $C^1\Sigma^+ - A^1\Pi(0, 2)$ were obtained at accuracies for single lines and blended lines of about 0.005 $cm^{-1}$ and 0.02 $cm^{-1}$, respectively. Results are collected in Table S1 for $B^1\Sigma^+ - A^1\Pi(0, 2)$ and in Table S2 for $C^1\Sigma^+ - A^1\Pi(0, 2)$ of the supplementary material. In the recorded spectra there appear 25 extra-lines, which originate from forbidden transitions. The extra-lines are listed in Table S3.

The recorded $^{13}C^{18}O$ spectra are contaminated by the 1.9% presence of $^{16}O_2$ in the isotopically-enriched oxygen $^{18}O_2$ sample (Sigma Aldrich) in the lamp mixture and from the 0.02% $^{12}C$ residue after deposition of the $^{13}C$ (Cambridge Isotopes) into the hollow cathode.[15,22] In the $^{13}C^{18}O$ $B^1\Sigma^+ - A^1\Pi(0, 2)$ region, the unwanted bands include: $^{12}C^{18}O$ $B^1\Sigma^+ - A^1\Pi(0, 2)$;[9] $^{13}C^{16}O$ $B^1\Sigma^+ - A^1\Pi(0, 2)$;[23] as well as $^{13}C^{18}O$ $C^1\Sigma^+ - A^1\Pi(0, 6)$, $^{13}C^{18}O^+$ $A^2\Pi - X^2\Sigma^+(1, 1)$ and $^{12}C^{18}O^+$ $A^2\Pi - X^2\Sigma^+(1, 1)$, which were simulated by PGOPHER based on the molecular constants from literature.[7,15,24,25] In the $^{13}C^{18}O$ $C^1\Sigma^+ - A^1\Pi(0, 2)$ region, the contaminants include $^{12}C^{18}O$ $C^1\Sigma^+ - A^1\Pi(0, 2)$,[9] $^{13}C^{16}O$ $C^1\Sigma^+ - A^1\Pi(0, 2)$,[26] $^{12}C^{18}O^+$ $A^2\Pi - X^2\Sigma^+(3, 0)$[27] and $^{13}C^{18}O^+$ $A^2\Pi - X^2\Sigma^+(3, 0)$, simulated by PGOPHER based on the molecular constants given in the literature.[7,15,24]

### B. VUV-FT absorption spectroscopy

High-resolution absorption spectra of the $A^1\Pi - X^1\Sigma^+(2, 0)$, $a^3\Pi - X^1\Sigma^+(13, 0)$, $e^3\Sigma^- - X^1\Sigma^+(4, 0)$, $a'^3\Sigma^+ - X^1\Sigma^+(12, 0)$, $D^1\Delta - X^1\Sigma^+(3, 0)$, $d^3\Delta - X^1\Sigma^+(7, 0)$, and $I^1\Sigma^- - X^1\Sigma^+(3, 0)$ bands were obtained using the VUV-FT spectrometer on the DESIRS beamline at the SOLEIL synchrotron facility. A detailed description of this layout can be found in the literature.[28–31] In the framework of the current study, two room-temperature spectra of $^{13}C^{18}O$ have been recorded in the 59,800 - 73,100 $cm^{-1}$ range at a gas pressure at the entrance of the cell of about 0.04 mbar, in a windowless cell set-up and 73 mb, inside a 1.8 cm long $MgF_2$ windowed cell. The SNR of the low- and high-pressure spectra reached maxima of 530:1 and 80:1, respectively. The instrumental resolution was set to 0.15 $cm^{-1}$. The complete VUV-FT spectra with assignments are shown in Figure 3, while the pure $A^1\Pi - X^1\Sigma^+(2, 0)$ spectrum is presented in Figure 4. Additionally, the band-heads of $A^1\Pi - X^1\Sigma^+(2, 0)$, $a'^3\Sigma^+ - X^1\Sigma^+(12, 0)$ and $d^3\Delta - X^1\Sigma^+(7, 0)$ are shown in Figure 5. Figure 3 shows the observed transitions of the $^{13}C^{18}O$ $a^3\Pi - X^1\Sigma^+(13, 0)$ band. Many additional bands, which can be seen in Figures 3 and 4, come from a variety of isotopologues (especially in the higher-pressure spectrum) and have been taken into account in the final analysis.

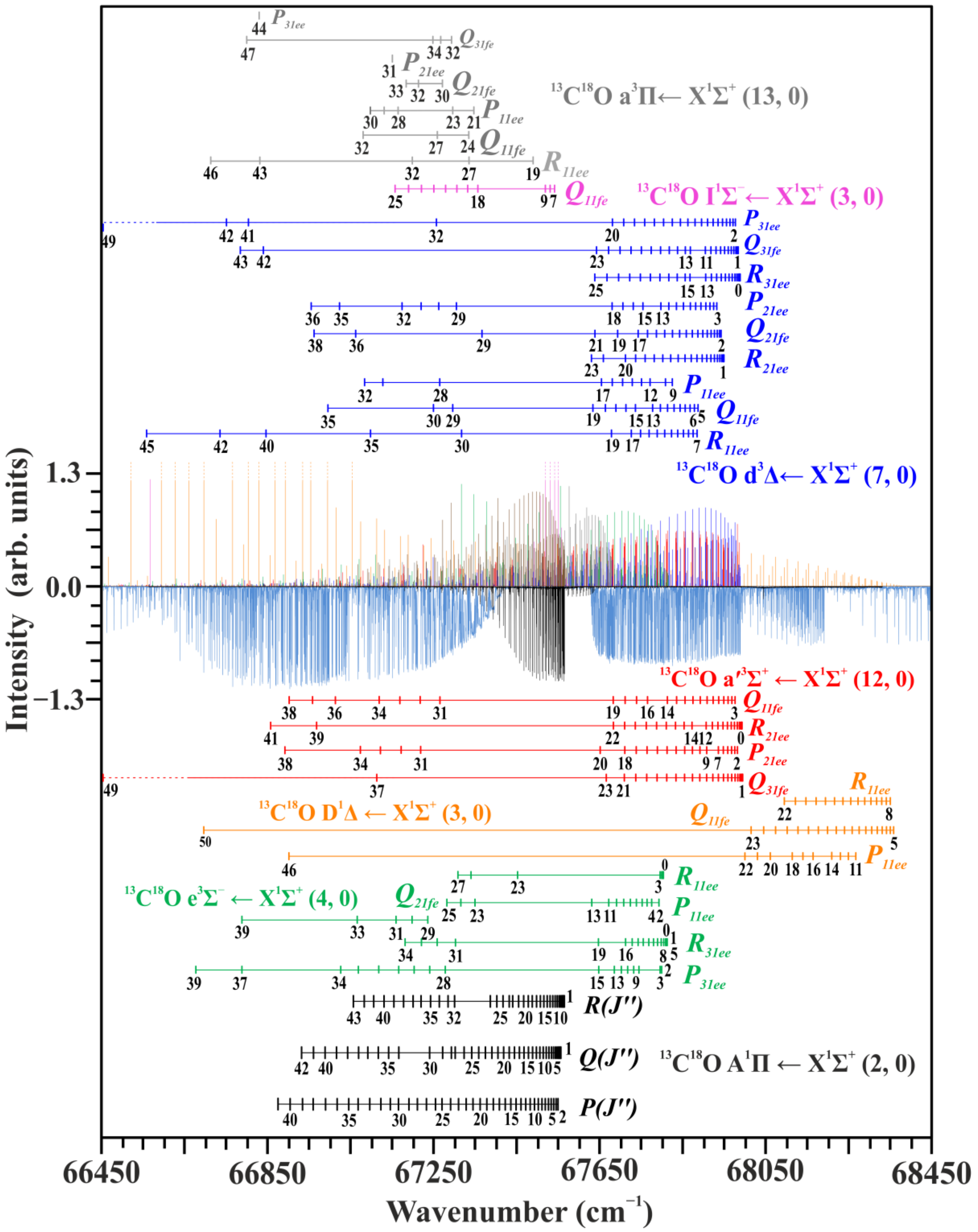

**FIG. 3.** The $^{13}C^{18}O$ $A^1\Pi - X^1\Sigma^+(2, 0)$, $D^1\Delta - X^1\Sigma^+(3, 0)$, $I^1\Sigma^- - X^1\Sigma^+(3, 0)$, $a'^3\Sigma^+ - X^1\Sigma^+(12, 0)$, $d^3\Delta - X^1\Sigma^+(7, 0)$, $e^3\Sigma^- - X^1\Sigma^+(4, 0)$, and $a^3\Pi - X^1\Sigma^+(13, 0)$ absorption bands together with their extra-lines, recorded using VUV-FT spectroscopy at 0.0378 mb and 300 K (lower, black trace) and at 73 mb and 300 K (lower, blue trace; with the latter spectrum showing saturation of the most intense A–X lines). At the top of the figure there are tie lines, refer to the $J''$ rotational levels, marking observed transitions in the $^{13}C^{18}O$ $a^3\Pi - X^1\Sigma^+(13, 0)$ band. At the present measurement sensitivity and resolution, these transitions would not have been observable through direct interaction between the $A^1\Pi(v = 2)$ and $a^3\Pi(v = 13)$ states. Their presence in the spectrum results from indirect interactions, for which the $e^3\Sigma^-(v = 4)$ perturber is primarily responsible.

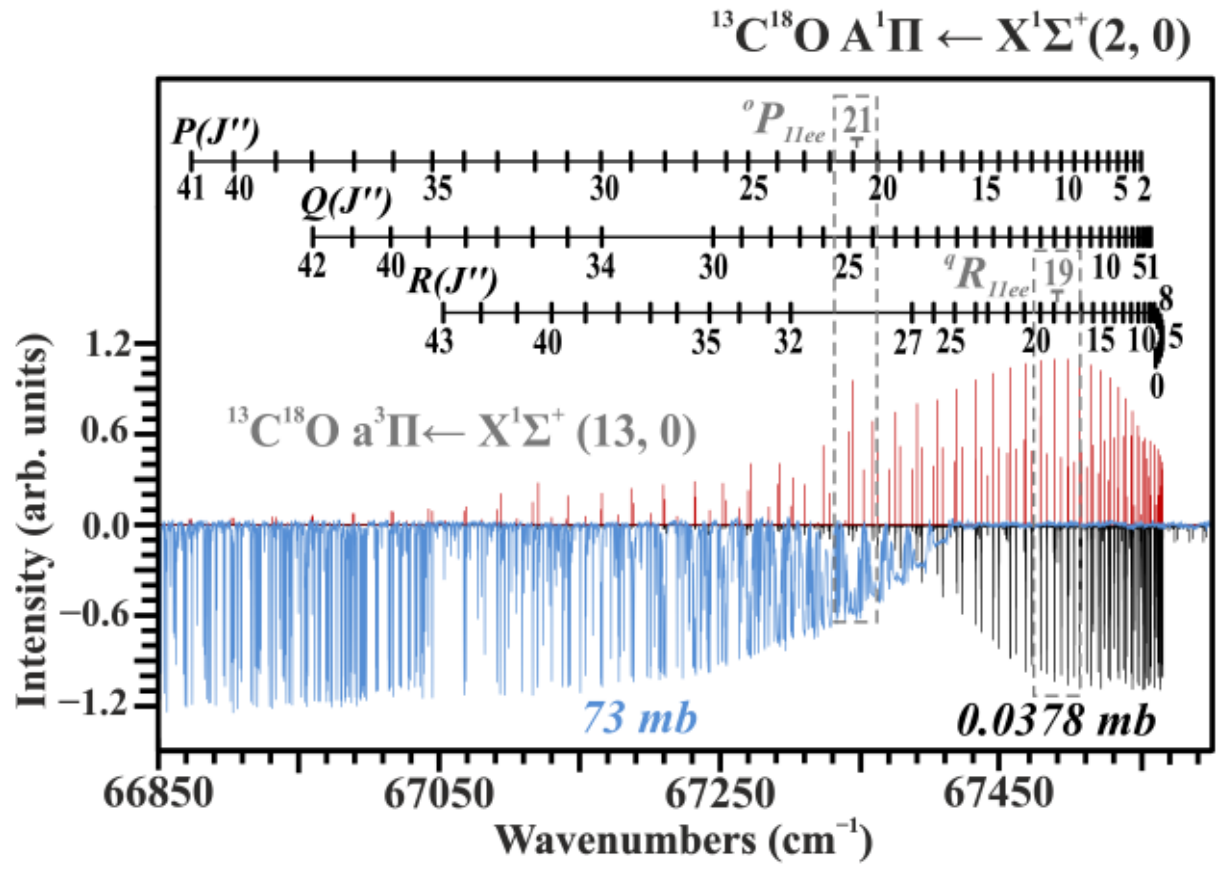

**FIG. 4.** The pure spectrum of the $^{13}C^{18}O$ $A^1\Pi - X^1\Sigma^+(2, 0)$ absorption band together with its extra lines, obtained by means of the VUV-FT spectroscopy technique at a pressure of 0.0378 mb and temperature 300 K (lower, black trace) and 73 mb and temperature 300 K (lower, blue trace; with saturation for the most intense lines of the A–X band). The simulation of the $A^1\Pi - X^1\Sigma^+(2, 0)$ spectrum (upper, red trace) was prepared after the final deperturbation analysis using the PGOPHER program.[16,17]

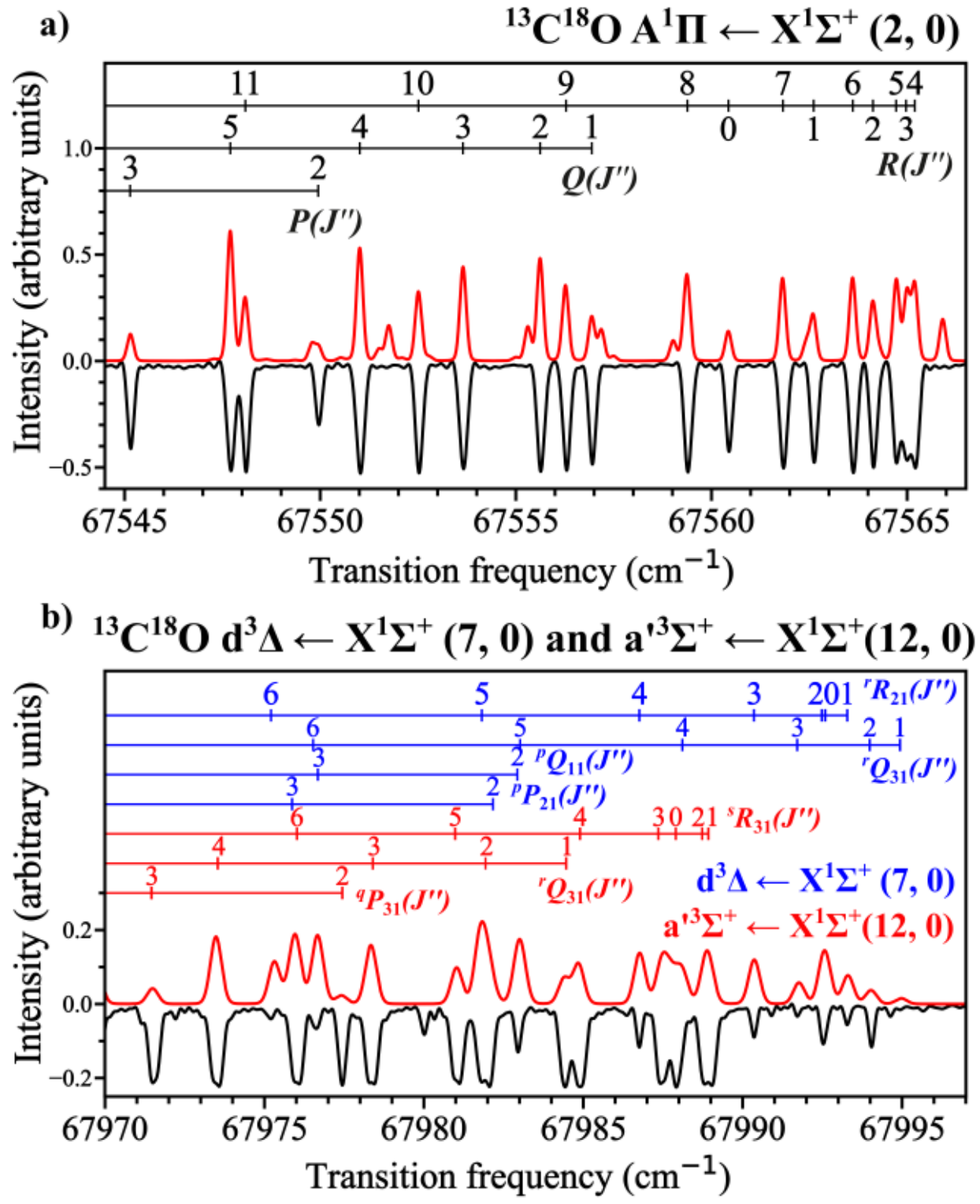

**FIG. 5.** (a) The isolated band-head region of the $A^1\Pi \leftarrow X^1\Sigma^+(2, 0)$ band. (b) The isolated band region of the $d^3\Delta \leftarrow X^1\Sigma^+(7, 0)$ and $a'^3\Sigma^+ \leftarrow X^1\Sigma^+(12, 0)$ bands. The lower black traces represent the experimental spectra, whereas the upper red traces show the simulated spectra. The simulations of the spectra were prepared after the final deperturbation analysis using the PGOPHER program.[16,17] Many additional bands from a variety of isotopologues (especially in the higher-pressure spectrum) have been taken into account in the final analysis. One can see that the line shapes of many partially blended features were reproduced in the post-analysis calculated spectrum.

Details about the analysis: *(i)* the absorption spectra were converted to transmission spectra; *(ii)* the baseline was corrected to the horizontal one using the *rubber band correction* method,[32] which calculates the baseline as a frequency polygon, consisting of *n* baseline points. The number of baseline points can be defined in the corresponding entry field of the OPUS software;[20] *(iii)* the post zero-filling procedure was implemented using a factor of 8 in the FT algorithm; *(iv)* absolute frequency calibrations of the spectra were performed with reference to the known frequencies of the overlapping $^{13}C^{18}O$ $A^1\Pi - X^1\Sigma^+(1, 0)$.[7] Finally, the uncertainties of the absolute wavenumber calibrations were 0.01 and 0.025 $cm^{-1}$, for the lower-pressure and higher-pressure spectra, respectively; *(v)* the PGOPHER program[16] and the literature data[1–4,7,11–14,33–36] were used to optimize the spectrum assignment process; and *(vi)* the transition frequencies were measured for each line separately with OPUS 8.5 software,[20] enabling Voigt profile fitting to the line contours. The absolute accuracy of the wavenumber determinations was assessed by incorporating uncertainties arising from calibration and line-shape fitting, following the empirical Brault approach.[21] For complex multi-component contours, the uncertainties were increased by a factor of 2. Under these assumptions, the resulting uncertainty in the line positions was about 0.04 $cm^{-1}$. Wavenumbers of the $A^1\Pi - X^1\Sigma^+(2, 0)$, $a^3\Pi - X^1\Sigma^+(13, 0)$, $a'^3\Sigma^+ - X^1\Sigma^+(12, 0)$, $I^1\Sigma^- - X^1\Sigma^+(3, 0)$, $D^1\Delta - X^1\Sigma^+(3, 0)$, $d^3\Delta - X^1\Sigma^+(7, 0)$ and $e^3\Sigma^- - X^1\Sigma^+(4, 0)$ bands, together with their uncertainties, are listed in Tables S4 – S9 of the supplementary material.

Based on the above-described complementary experimental Vis-FT and VUV-FT methods, a high quality analysis of the $B^1\Sigma^+ - A^1\Pi$, $C^1\Sigma^+ - A^1\Pi$, and $A^1\Pi - X^1\Sigma^+$, $a^3\Pi - X^1\Sigma^+$, $e^3\Sigma^- - X^1\Sigma^+$, $a'^3\Sigma^+ - X^1\Sigma^+$, $d^3\Delta - X^1\Sigma^+$, $D^1\Delta - X^1\Sigma^+$, and $I^1\Sigma^- - X^1\Sigma^+$ spectra has been obtained and the $^{13}C^{18}O$ $A^1\Pi(v = 2)$ vibrational level has been fully deperturbed.

## III. DEPERTURBATION ANALYSIS

The properties of $A^1\Pi(v = 2)$ and the levels suspected of perturbing it in the $^{13}C^{18}O$ isotopologue are expressed in the PGOPHER model[17] via the matrix elements of the effective Hamiltonian.[16] The perturbers were selected based on the rovibronic term diagram (Figure S1) in the neighborhood of the level under consideration in the region of 67,000 – 73,000 $cm^{-1}$.The perturbations of the $^1\Pi$ and $^3\Pi$ states by all five of the $\pi^3\sigma^2\pi^*$ configuration states can be quantitatively predicted. They have been extensively tested for statistically justified direct and indirect interactions with the $A^1\Pi(v = 2)$ and the $a^3\Pi(v = 13)$ levels. Detailed information is included in Table S17. The $B^1\Sigma^+(v = 0)$ and $C^1\Sigma^+(v = 0)$ states contain some previously-unidentified, very weak ($\leq 0.04$ $cm^{-1}$) perturbations.[15] For that reason, the energy levels are represented in the model by their term energies. This approach eliminates the influence of these B and C state perturbations on the description of the $A^1\Pi(v = 2)$ level and its perturbers. The model was iteratively fitted to 1008 experimental wavenumbers derived from the $A^1\Pi - X^1\Sigma^+(2, 0)$, $a^3\Pi - X^1\Sigma^+(13, 0)$, $a'^3\Sigma^+ - X^1\Sigma^+(12, 0)$, $d^3\Delta - X^1\Sigma^+(7, 0)$, $e^3\Sigma^- - X^1\Sigma^+(4, 0)$, $D^1\Delta - X^1\Sigma^+(3, 0)$, $I^1\Sigma^- - X^1\Sigma^+(3, 0)$, $B^1\Sigma^+ - A^1\Pi(0, 2)$, $C^1\Sigma^+ - A^1\Pi(0, 2)$ bands, together with their extra-lines, comprising 16 bands obtained in present work, as well as from the $B^1\Sigma^+ - X^1\Sigma^+(0, 0)$ and $C^1\Sigma^+ - X^1\Sigma^+(0, 0)$ bands obtained previously by our group.[15]

An initial model has been defined, based on molecular parameter values taken from the literature[1,2,4,13,14,34] and isotopically scaled. Their values have been recalculated for use with the total angular momentum operator excluding spin ($N$). Molecular constants for the $^{13}C^{18}O$ $X^1\Sigma^+(v = 0)$ ground state were taken from the work of Coxon and Hajigeorgiou[33] and held fixed in all fits. The input values of the perturbation parameters were calculated as follows:

*(i)* for the $A^1\Pi \sim (e^3\Sigma^-, a'^3\Sigma^+, d^3\Delta)$ spin-orbit interactions, parametrized by $\eta$, and the $A^1\Pi \sim (I^1\Sigma^-, D^1\Delta)$ rotation-electronic (**L**-uncoupling) perturbations, parametrized by $\xi$, were calculated on the basis of isotopologue-independent $\boldsymbol{a}_{\mathbf{A\sim d,e,a'}}$ and $\boldsymbol{b}_{\mathbf{A\sim D,I}}$ electronic parameters taken from the literature,[4,11] based on equations (1) and (2) - (4) from the work of Le Floch *et al.*[4] and Ryzner *et al.*,[7] respectively;

*(ii)* for the $a^3\Pi \sim (I^1\Sigma^-, D^1\Delta)$ spin-orbit perturbations as well as the $a^3\Pi \sim (a'^3\Sigma^+, e^3\Sigma^-, d^3\Delta)$ spin-orbit/spin-electronic/rotation-electronic(**L**-uncoupling) combined interactions, perturbation parameters were calculated on the basis of the isotopologue-independent $\boldsymbol{a}_{\mathbf{a\sim D,I}}$ as well as $\boldsymbol{a}_{\mathbf{a\sim e,d,a'}}$ and $\boldsymbol{b}_{\mathbf{a\sim e,d,a'}}$ parameters determined in the literature.[3,37] Note that the spin-orbit and spin-electronic interactions follow the same $\Omega$-dependence, so they cannot be determined independently, thus the $\eta_{a\sim e,d,a'}$ parameters represent linear combinations of the two parameters. The $\alpha_{a\sim e,d,a'}(\boldsymbol{a}_{\mathbf{a\sim e,d,a'}})$ and $\beta_{\mathbf{a\sim e,d,a'}}(\boldsymbol{b}_{\mathbf{a\sim e,d,a'}})$ dependences,[3,36] were used in the calculations. The $\eta_{a\sim e,d,a'}(\alpha_{a\sim e,d,a'})$ and $\xi_{a\sim e,d,a'}(\beta_{a\sim e,d,a'})$ relations, resulting from *e*/*f*-symmetrized matrix elements of the $a^3\Pi \sim e^3\Sigma^-$, $d^3\Delta$, $a'^3\Sigma^+$ interactions, were taken from the work of Ryzner *et al.*;[7]

*(iii)* The vibrational overlap integrals $\langle v_A | v_{a',e,d} \rangle$ and $\langle v_a | v_{a',e,d} \rangle$ as well as rotational operator integrals $\langle v_A | \mathbf{B(R)} | v_{I,D} \rangle$and $\langle v_a | \mathbf{B(R)} | v_{e,d,a'} \rangle$ were calculated as described in the literature.[11,12]

Some of these perturbation parameters were kept fixed at their previously determined values throughout the entire deperturbation analysis. This included those, which were numerically significant, but could not be determined in the fits for the reasons specified in Table S17. During the deperturbation analysis, 35 intra-molecular interactions between the $A^1\Pi(v = 2)$ level and its expected perturbers have been tested for their significance, and, as a result, 15 of them were used in the final model, of which 10 were fitted. The significance of a given interaction was tested by verifying the influence of its fit parameters on the residuals of the deperturbation fit by comparing the results to two cases: (*i*) when the interaction parameters were fixed in the fit at its initial, theoretical value; (*ii*) when the interaction was turned off by setting its parameters to zero.

To prevent overfitting of the multiparameter model, the correlation matrix was continuously monitored throughout the analysis to detect significant correlations between parameters. The root-mean-square error (RMSE) of the unweighted residuals of all transitions included in the final model was 0.030 $cm^{-1}$. This value is within the arithmetic mean of the medians of line position uncertainties in the individual bands used in this analysis, thus the model is statistically satisfactory. The final values of all molecular constants and interaction parameters are presented in Table I. The PGOPHER output (*.log) file, containing the explicitly-formulated effective Hamiltonian and matrix elements, is provided as supplementary material.

**TABLE I. Deperturbed molecular constants of the $A^1\Pi(v = 2)$ level and its perturbers as well as interaction parameters in the $^{13}C^{18}O$ isotopologue.**[a,b]

| Constant | $A^1\Pi(v = 2)$ | $D^1\Delta(v = 3)$ | $I^1\Sigma^-(v = 3)$ | $I^1\Sigma^-(v = 4)$ |
|---|---|---|---|---|
| $T_v$ | 67557.92638(88) | 68377.912(50) | 67579.238(15) | 68544.29[g] |
| $B$ | 1.4177051(35) | 1.089219(188) | 1.09919(23) | 1.08405[h] |
| $D \times 10^6$ | 6.2440(25) | 5.855(80) | 5.78(40) | 5.71[d] |
| $H \times 10^{12}$ | −12.81[c] | −2.25[d] | 2.25[d] | 2.25[d] |
| $q \times 10^5$ | −1.19[c] | | | |
| $\xi(A^1\Pi_{v'=2}\sim Y^*_{v''}) \times 10^2$ | | 2.60[e] | −5.042(48) | 6.50[e] |
| $\eta(a^3\Pi_{v'=13}\sim Y^*_{v''})$ | | −22.86[f] | −11.53[f] | −16.44[f] |
| Constant | $a^3\Pi(v = 13)$ | $a'^3\Sigma^+(v = 12)$ | $d^3\Delta(v = 7)$ | $e^3\Sigma^-(v = 4)$ |
| $T_v$ | 67670.6779(51) | 67991.2957(59) | 67946.2487(36) | 67802.7849(56) |
| $B$ | 1.30813765(56) | 1.037593(22) | 1.079739(14) | 1.097154(30) |
| $D \times 10^6$ | 6.0616(46) | 5.186(12) | 5.1278(80) | 5.343(25) |
| $H \times 10^{12}$ | | −0.30[d] | −0.60[d] | −1.50[d] |
| $A$ | 38.115(13) | | -16.9899(33) | |
| $A_D \times 10^5$ | −19.41[i] | | −9.17[h] | |
| $\lambda$ | −0.018[i] | −1.1457(91) | 1.2316(54) | 0.425(12) |
| $\gamma \times 10^2$ | 0.32[h] | −0.527(31) | −0.291(26) | |
| $o$ | 0.65[j] | | | |
| $p \times 10^3$ | 8.83[k] | | | |
| $q \times 10^5$ | 2.79[l] | | | |
| $\eta(A^1\Pi_{v'=2}\sim Y^*_{v''})$ | | −5.902(11) | −9.519(10) | −13.1136(46) |
| $\eta(a^3\Pi_{v'=13}\sim Y^*_{v''})$ | | 2.39(20) | 23.219(44) | 24.729(20) |
| $\xi(a^3\Pi_{v'=13}\sim Y^*_{v''}) \times 10^2$ | | −1.12(13) | −6.279(45) | −5.982(26) |

[a] In $cm^{-1}$. The $\eta$ and $\xi$ symbols indicate a spin-orbit and **L**-uncoupling interaction parameter, respectively. We adopt the definitions given by Niu *et al.*,[14] which differ from those used elsewhere in the literature.[36,38] Note that the spin-orbit and spin-electronic interactions follow the same $\Omega$-dependence, so they cannot be determined independently, thus the $\eta_{a\sim e,d,a'}$ parameters represent a linear combination of both effects.
[b] Numbers in parentheses are $1\sigma$ standard uncertainties of parameters varied in the analysis and given in units of the last-significant digit. When expressing the results with their uncertainties, the approach described by Watson[39] was used to avoid a non-physical precision when recreating energy levels using data obtained in this work, so some uncertainty values were rounded to more than two significant digits.
[c] Calculated based on published data[13,14] and isotopic scaling procedure.
[d] Calculated based on published data[4] and isotopic scaling procedure.
[e] The **L**-uncoupling interaction parameter was calculated based on the isotopologue-independent $\boldsymbol{b}_{A\sim I}$ electronic parameter[4] based on the Eq. (4) from the work of Hakalla *et al.*[11] and Eq. (3) given in the work of Ryzner *et al.*[7] The rotational operator integral $\langle v_A|\mathbf{B(R)}|v_I\rangle$ was calculated as it was described in the literature.[11,12]
[f] The spin-orbit interaction parameters were calculated based on the $\boldsymbol{a}_{a\sim I}$ or $\boldsymbol{a}_{a\sim D}$ isotopologue-independent parameters given by Field *et al.*[3] or Garetz *et al.*,[37] respectively. The $\alpha_{a\sim I}(\boldsymbol{a}_{a\sim I})$ and $\alpha_{a\sim D}(\boldsymbol{a}_{a\sim D})$ dependences of the perturbation parameters used in the calculations were taken from published data.[3,36] The $\alpha_{a\sim I}(\boldsymbol{a}_{a\sim I})$ relationships resulted from symmetrized matrix elements of the $a^3\Pi \sim I^1\Sigma^-$ and $a^3\Pi \sim D^1\Delta$ interactions and they are given as equation (5) in the work of Ryzner *et al.*[7] The vibrational overlap integrals $\langle v_a|v_{I,D}\rangle$ were calculated as described in the literature.[11,12]
[g] Calculated based on published data[1,33] and isotopic scaling procedure.
[h] Calculated based on published data,[1] if needed, converted from MHz to $cm^{-1}$, and isotopically scaled.
[i] Calculated based on published data[34] converted from MHz to $cm^{-1}$ and isotopically scaled. For the λ constant, the relation $\lambda = \frac{3}{2} \times \varepsilon$ was used.
[j] Calculated based on published data[2] ($o = C^{\delta}$ or $q = 2 \times B_0^{+}$) and isotopic scaling procedure.
[k] Calculated based on published data[34,35] data and unpublished data on Cameron system in $^{12}C^{16}O$ as well as isotopic scaling procedure.
[l] Calculated based on published data[2] ($q = 2 \times B_{0+}$) and isotopic scaling procedure.

* $Y^{*}_{v''}$ denotes the perturber given in the column header where the value of the corresponding perturbation parameter is given.

High-quality term values of all relevant levels, with their corresponding $^1\Pi$, $^3\Pi$ and $^3\Sigma^-$ fractional characters, are presented in Tables S10 – S16.

## IV. DIRECT $A^1\Pi(v = 2) \sim a^3\Pi(v = 13)$ SPIN-ORBIT INTERACTION IN $^{13}C^{18}O$?

### A. Circumstantial evidence

Analysis of the resulting experimental data using:

*(i)* the $^1\Pi$ fractional characters of the investigated levels, presented in Figure 6, and calculated as $C_{ik}^2 \cdot 100\%$, where $C_{ik} = \langle\psi_i|\psi_k\rangle$, is the mixing coefficient extracted from the eigenvectors of the diagonalized energy matrix in the final fit of the deperturbation analysis;

*(ii)* the reduced term values of $A^1\Pi(v = 2)$ and its perturbers, defined as $T_{red}(J) = T(J) - B_{A(2)}J(J+1) + D_{A(2)}J^2(J+1)^2 - H_{A(2)}J^3(J+1)^3$, displayed in Figure 7 to highlight deviations associated with level interactions;

*(iii)* the intensity irregularities and local weakenings of spectral lines, as well as the presence of detectable extra-lines in the region where the diabatic intersections of the $A^1\Pi(v = 2)$ and $a^3\Pi(v = 13)$ levels in $^{13}C^{18}O$ occur, all illustrated in Figure 8, suggest the possible presence of a direct spin-orbit perturbation between the $A^1\Pi(v = 2)$ and $a^3\Pi(v = 13)$ levels in $^{13}C^{18}O$. This impression may result from the following facts:

- Figure 6 shows that:

(*i*) there is a distinctly observed unexpected, substantial mixing of the $A^1\Pi(v = 2)$ and $a^3\Pi(v = 13)$ level characters, which is particularly pronounced for $J = 20$, 32, and 46;

- Figure 7 shows that:

(*ii*) only a single anticrossing, highlighted with the red arrows in the figure, exists for $J = 20$, namely between the $A^1\Pi(v = 2)$ and the $a^3\Pi(v = 13)$ levels, and the nearby $J$-neighborhood is free of other perturbers;

(*iii*) in this region ($J = 20$) there is a very clear shift of about 1 cm$^{-1}$ of the $A^1\Pi(v = 2, F_{1e}, J = 20)$ component, which suggests a direct $A^1\Pi(v = 2) \sim a^3\Pi(v = 13)$ interaction;

- Figure 8 shows:

(*iv*) the significant weakening of the $B^1\Sigma^+ - A^1\Pi(0, 2)$ $^rR_{11ee}(20)$ line. The recovery of the branch intensities for $J > 20$ is clearly visible in Figure 8(d). The $^rR_{11ee}(20)$ line is weakened due to intensity borrowing by the $B^1\Sigma^+ - a^3\Pi(0, 13)$ $^sR_{11ee}(20)$ line;

($v$) the observation of 6 extra-lines and the accompanying weakening of the "main" lines, which might otherwise have been interpreted as evidence for the impossible to observe $A^1\Pi(v = 2) \sim a^3\Pi(v = 13)$ direct interaction. They are observed in three different band systems obtained using two independent FT techniques, both in emission and in absorption. Such intensity depletion is rarely measured with sufficient precision to serve as decisive element evidence for the mechanism of an indirect perturbation;

- Removal of the $a^3\Pi$ state from the fit model leaves statistically unacceptable level-shifts. These are shown in Figure 9(b) as the unweighted residuals of the wavenumbers for all spectral lines included in the final deperturbation fit when intensity borrowing by $a^3\Pi(v = 13)$ is neglected, in contrast to Figure 9(a), where the $a^3\Pi(v = 13)$ level is included. It can be seen that the Figure 9 demonstrates the quality of the fit, with unweighted residuals confined to within $\pm 0.1$ cm$^{-1}$, in contrast to the deviations of the calculated (unrefined) spectrum, which range from +8 to −13 cm$^{-1}$, in the case where the influence of the $a^3\Pi$ state is excluded.

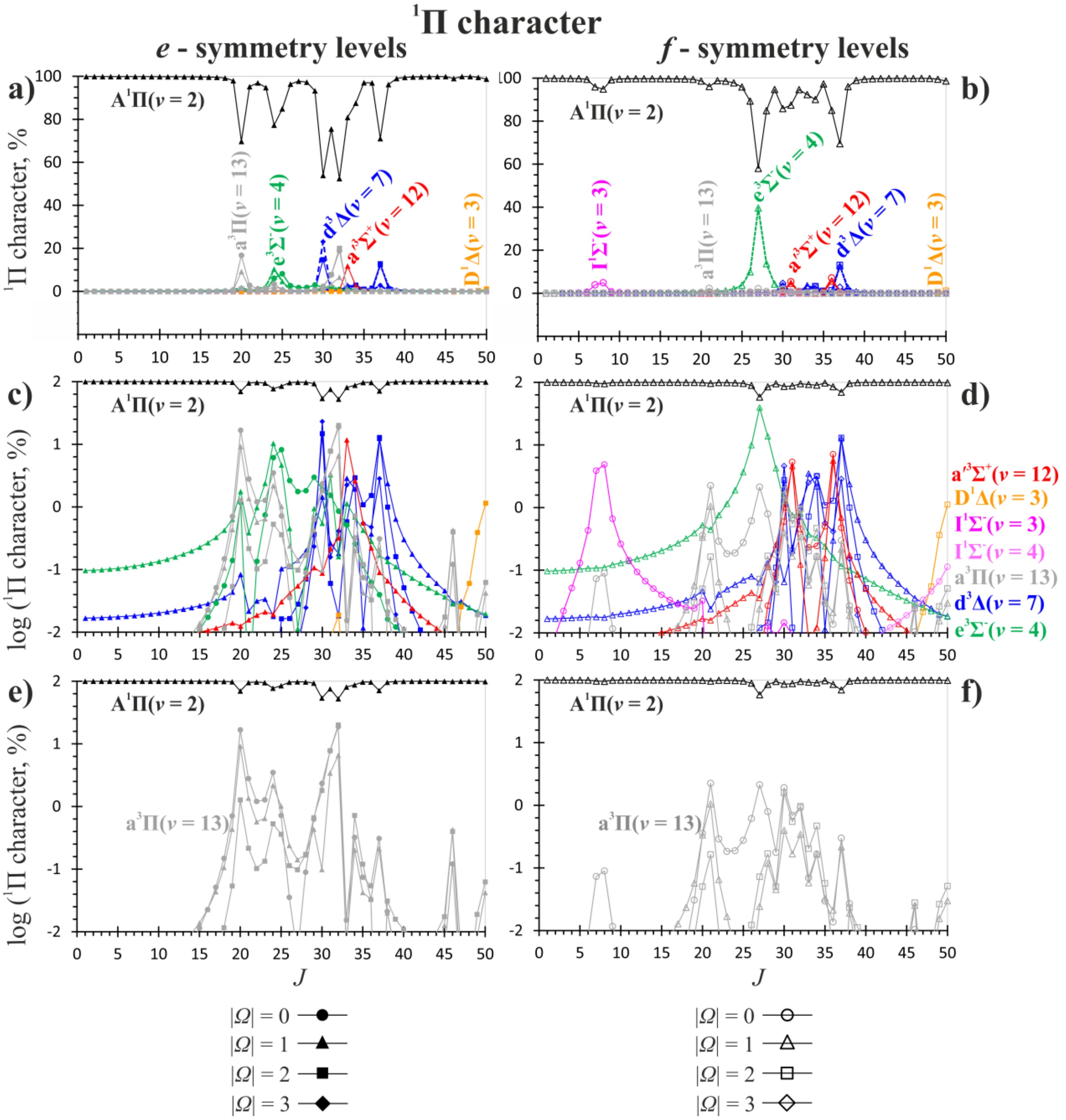

**FIG. 6.** Fractional ¹Π character, (a) and (b), as well as logarithm of the fractional ¹Π character, (c) and (d), of the *e*- and *f*-symmetry components of the $^{13}C^{18}O$ $A^1\Pi(v = 2)$ level, respectively, and all adjacent rovibronic levels that have a noticeable influence on it, within the accuracy of the current measurements and observed spectral range. Figures (e) and (f) are copies of the (c) and (d) with the characteristics of only the $a^3\Pi(v = 13)$ and $A^1\Pi(v = 2)$ states remaining, so that the effects of the $a^3\Pi$ state borrowing the $^1\Pi$ character are clearly displayed. The $|\Omega|$ designations are given for the hypothetical case of no interactions between the spin components of a given electronic state, although for the $^3\Sigma^{\pm}$ states they are only pictorial labels, since the $F_1$, $F_2$, and $F_3$ components are more properly Hund's-case-(b) spin-rotational components rather than pure-$\Omega$ components.

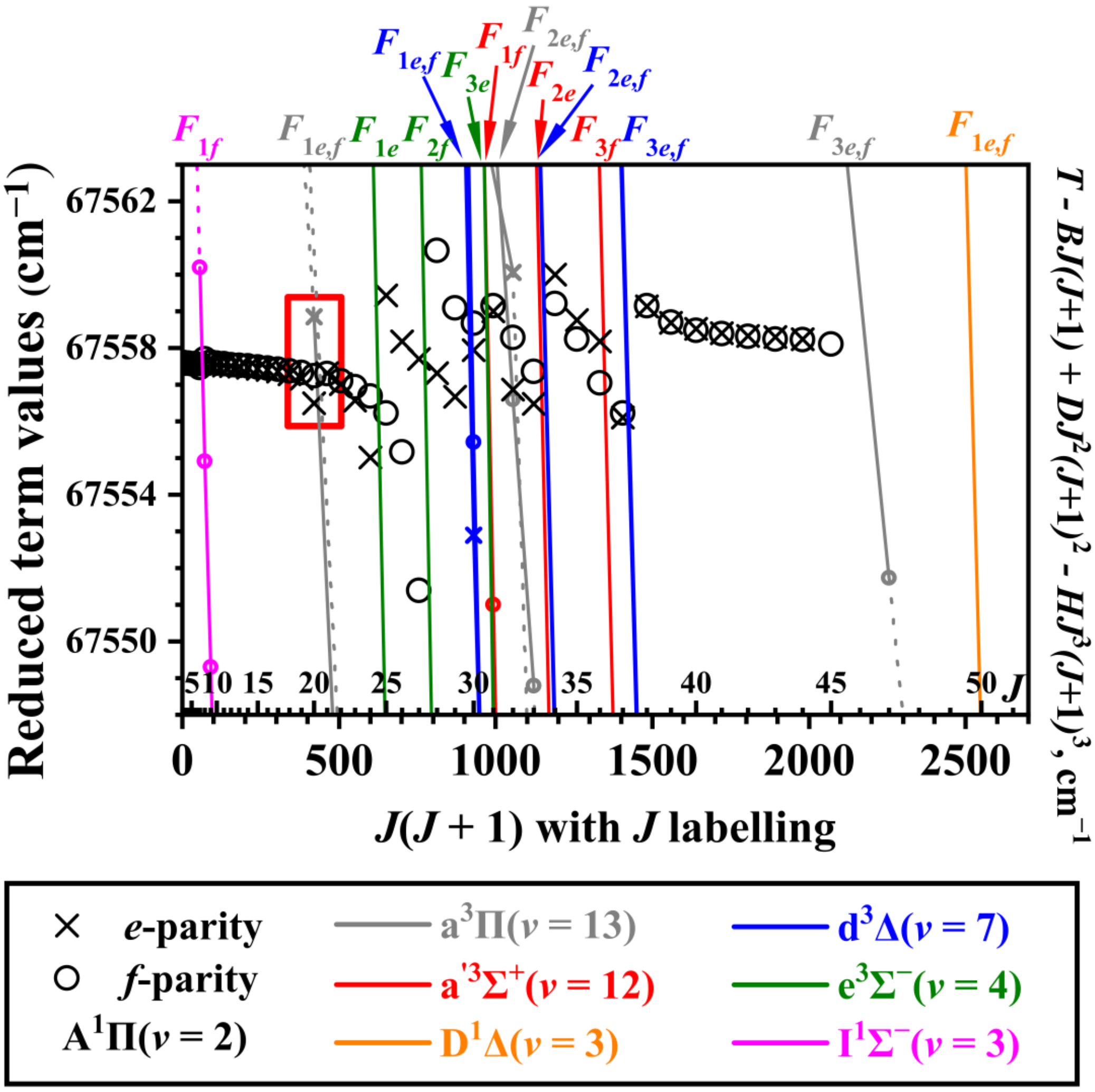

**FIG. 7.** Reduced rovibronic term values of the $A^1\Pi(v = 2)$ level and its perturbers in the $^{13}C^{18}O$ isotopologue. Solid lines represent term values derived from the experimental extra-line wavenumbers, whereas the dotted lines were plotted using isotopically scaled equilibrium constants. The $T$, $B$, $D$, and $H$ values of $A^1\Pi(v = 2)$ were determined in a final deperturbation fit (see Table I). The red rectangle highlight the only anticrossing between the $A^1\Pi(v = 2)$ and the $a^3\Pi(v = 13)$ levels in a nearby $J$-neighborhood free of other perturbers.

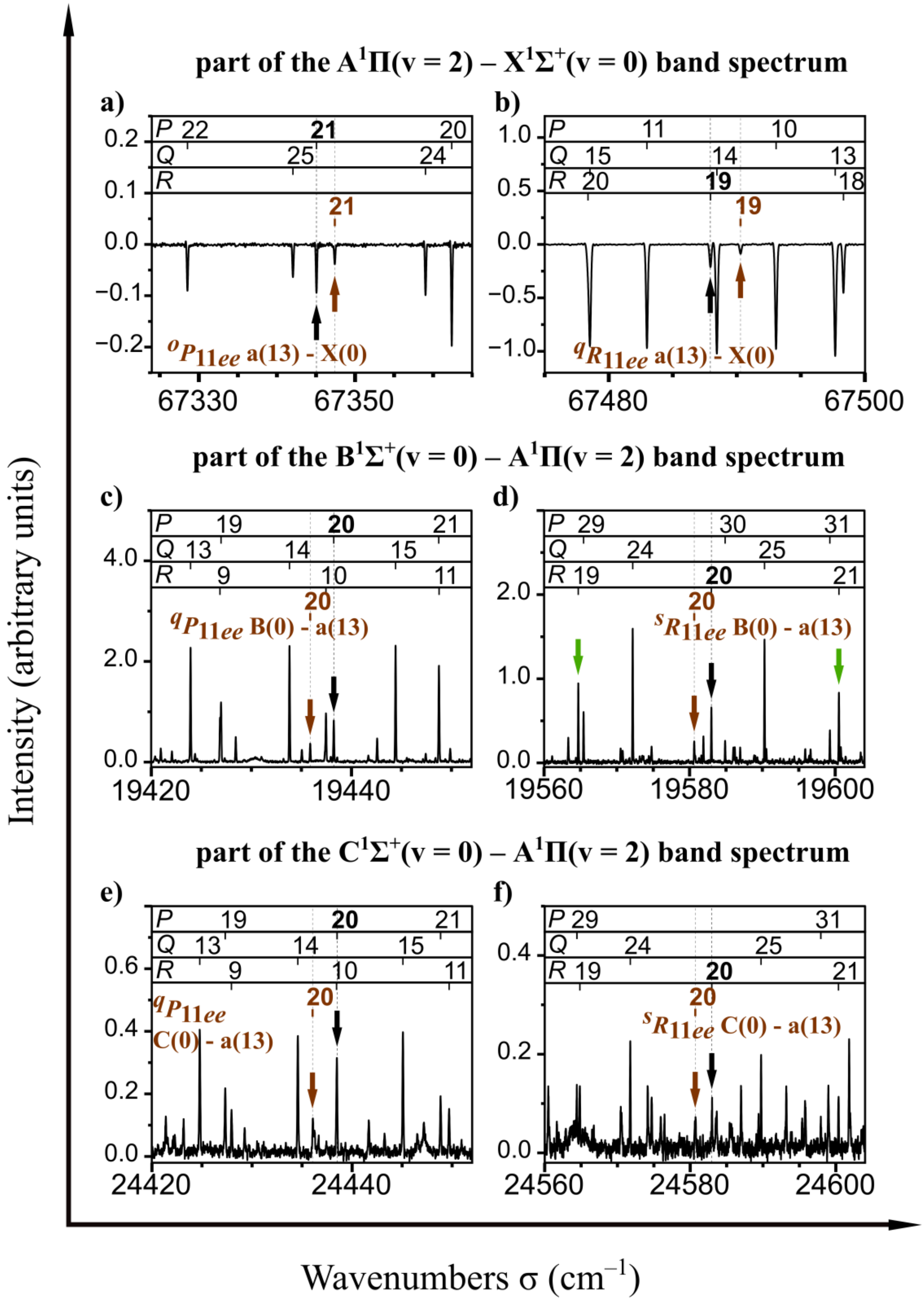

**FIG. 8.** (a)-(b), (c)-(d), and (e)-(f) regions of the $^{13}C^{18}O$ $A^1\Pi - X^1\Sigma^+(2, 0)$, $B^1\Sigma^+ - A^1\Pi(0, 2)$, and $C^1\Sigma^+ - A^1\Pi(0, 2)$ band spectra, respectively, displaying features attributed to the unobservable $A^1\Pi(v = 2) \sim a^3\Pi(v = 13)$ direct interaction between the *e*-symmetry components at $J = 20$. Panels (a), (c) and (e) correspond to the *P*-branch, while (b), (d) and (f) refer to the *R*-branch. The extra-lines related to the $a^3\Pi(v = 13)$ level are marked with grey fonts and arrows. The lines associated with the $A^1\Pi(v = 2, F_{1e}, J = 20)$ level, weakened as a result of intensity-borrowing by the lines related to the $a^3\Pi(v = 13, F_{1e}, J = 20)$ level, are marked with the black downward-pointing arrows. The green arrows in part (d) indicate reappearance of the $B^1\Sigma^+ - A^1\Pi(0, 2)$ $^rR_{11ee}$ branch in the region of $J = 19$-$21$.

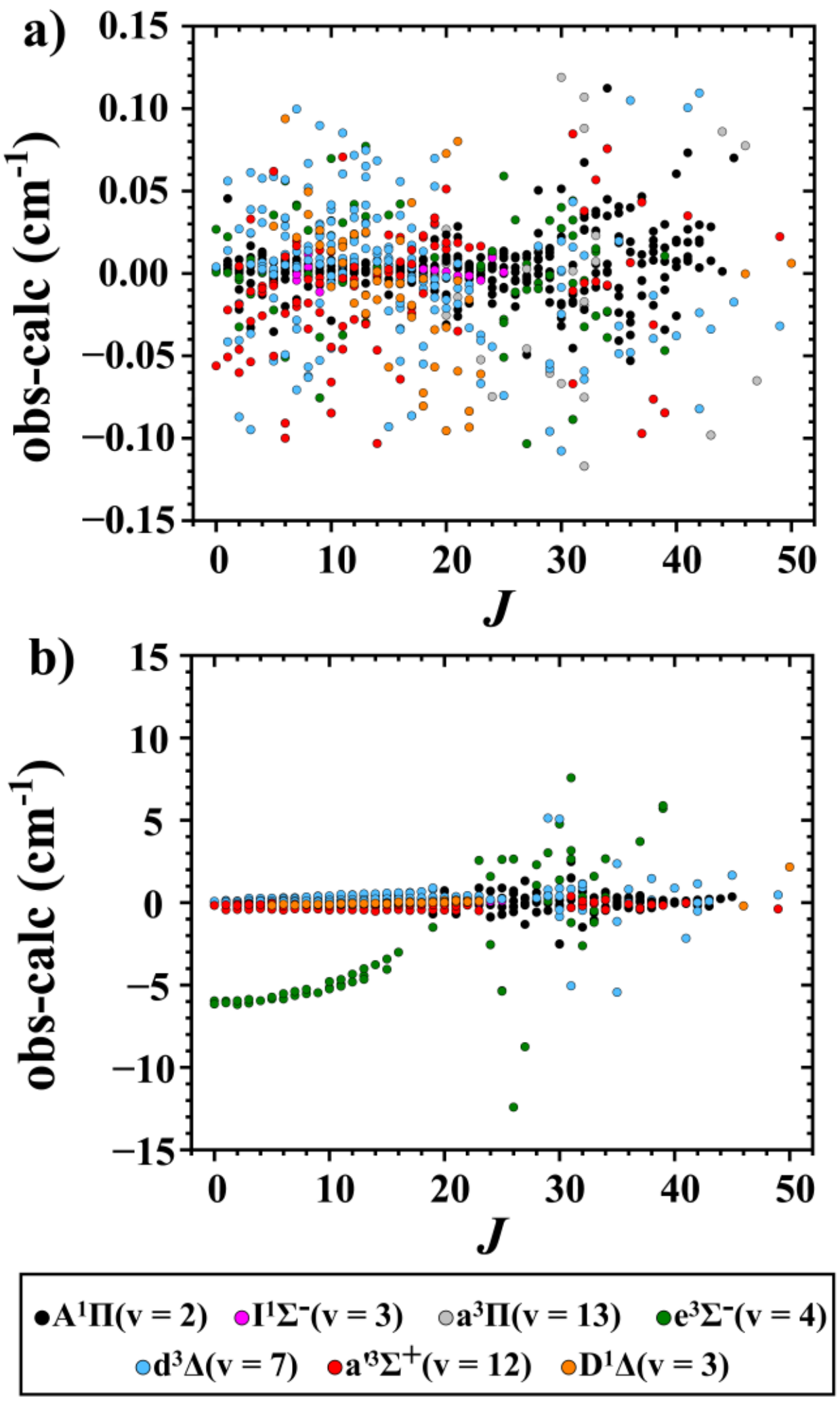

**FIG. 9.** Unweighted residuals of the wavenumbers for all spectral lines included in the final deperturbation fit: (a) with the $a^3\Pi(v = 13)$ level included; (b) if intensity-borrowing by $a^3\Pi(v = 13)$ is not included. The fit was performed using the PGOPHER program.[16,17]

Based on these findings, we decided to examine the possibility that a direct $A^1\Pi(v = 2) \sim a^3\Pi(v = 13)$ perturbation in $^{13}C^{18}O$ could have been observed.

### B. Assessment of possible anomalies in the *J*-dependence of $\langle a^3\Pi, v = 13 | A^1\Pi, v = 2\rangle$

The $A^1\Pi - X^1\Sigma^+$ transition is electric-dipole allowed, but the $a^3\Pi - X^1\Sigma^+$ transition is spin-forbidden. This forbidden-ness is absolute. However, if an excited electronic state exists that has an electric-dipole allowed transition from the $X^1\Sigma^+$ state and can perturb the $a^3\Pi$ state, then simple intensity borrowing, with predictable $\Omega$, $J$, and $e,f$-symmetry intensity dependences, could occur. The mechanism of $A^1\Pi - X^1\Sigma^+$ intensity borrowing by the $a^3\Pi - X^1\Sigma^+$ Cameron bands is explained in detail in Appendix A. Let us call such an intermediate state by 'Y', the $\Omega$, $J$, and $e,f$-symmetry-dependent fractional intensity borrowing is then given by the square of the mixing-amplitude:

$$\textit{Mixing-amplitude} = \frac{\langle a_v | \mathbf{H}^{\mathbf{SO}} | Y_{v'} \rangle}{[E_{a_v} - E_{Y_{v'}}]}. \qquad (1)$$

The $\mathbf{H}^{\mathbf{SO}}$ matrix element is a product of a vibrational and an electronic factor. Since the iso-configurational $A^1\Pi$ and $a^3\Pi$ states have similar shapes of potential energy curves, only the $\Delta v = 0$ vibrational factor is ~1, but the fractional mixing is extremely small owing to the enormous $v_a = v_A = 0$ energy denominator (16,275 cm$^{-1}$). This would lead to a predictable pattern of $a^3\Pi - X^1\Sigma^+$ rotation-fine structure transition strengths. When $v_A \neq v_a$ ($\Delta v$ = 11 in the present case), the vibrational factor of the $\mathbf{H}^{\mathbf{SO}}$ matrix element is extremely small, but could still lead to observable effects in regions where the $\Delta v$ = 11 vibrational overlap integrals become locally enhanced due to a possible strong $J$-dependence. Thus, at a first stage of the verification, $J$-dependent vibrational overlap integrals $\langle a^3\Pi, v = 13 | A^1\Pi, v = 2\rangle$ in $^{13}C^{18}O$ were calculated using two independent methods, described in detail in Appendix B and presented in Figure 11. Crucially, the $J$-dependence highlighted there defines the possible range, from $+1.2\times10^{-5}$ to $-5\times10^{-6}$, of the vibrational overlap between the $A^1\Pi$ and $a^3\Pi$ states that make the direct perturbation unobservable within the accuracies of the current measurements. It should be emphasized that the $\langle v|\Delta v\rangle$ overlap integrals for large $\Delta v$ are very small because the integrand is a small rapidly oscillating function. The $\Delta v$ overlap integrals exhibit small fluctuations with respect to $\Delta v$. However, the $J$-dependence of these large $\Delta v$ overlap integrals exhibits no fluctuations. Thus, the vibrational overlap integral $\langle a^3\Pi, v = 13 | A^1\Pi, v = 2\rangle$ in $^{13}C^{18}O$ for $J$ = 20 is equal to $7.8\times10^{-6}$ or $10.0\times10^{-6}$, depending on the calculation method and rotational branch considered. This indicates a certain degree of model dependence of the overlap integral. The two methods differ by 13%, which we therefore adopt as the uncertainty of the value.

## C. The calculated spin-orbit perturbation parameter of the $A^1\Pi \sim a^3\Pi$ interaction in $^{13}C^{18}O$

Following the book of Lefebvre-Brion and Field[36] page 188, the spin-orbit perturbation parameter of the $A^1\Pi \sim a^3\Pi$ interaction is given as

$$\left\langle A^1\Pi, v' \middle| \mathbf{H}^{\mathbf{SO}} \middle| a^3\Pi, v'' \right\rangle = \frac{1}{2} a_{A^1\Pi \sim a^3\Pi} \left\langle A^1\Pi, v' \middle| a^3\Pi, v'' \right\rangle \quad (2)$$

Since both states belong to the same configuration, the following hypothesis can be made: if the electronic wavefunctions for the $^1\Pi$ and $^3\Pi$ states are built from identical orbitals, both states would have identical potential curves, thus

$$a_{A^1\Pi \sim a^3\Pi} = 2A_{a^3\Pi}\,, \quad (3)$$

where

$a_{A^1\Pi \sim a^3\Pi}$ - is the isotopologue-independent perturbation parameter of the spin-orbit $A^1\Pi \sim a^3\Pi$ interaction,

and $A_{a^3\Pi}$ - is the diagonal spin-orbit molecular constant of the $a^3\Pi$ state.

The fitted spin-orbit constant of the $a^3\Pi(v = 13)$ in $^{13}C^{18}O$ (from Table I) is

$$A_{a^3\Pi(v=13)} = 38.115(13)\ \text{cm}^{-1}. \quad (4)$$

Gilijamse *et al.* [40] reported an $A^1\Pi \sim a^3\Pi$ electronic spin-orbit coupling of image cm$^{-1}$ at $R = 2.3\ a_0$, obtained at the CASSCF+MRCI level. Within the notation adopted here, combining the approximation image $\boldsymbol{a}_{\mathrm{A}^1\Pi\sim \mathrm{a}^3\Pi} = 2A_{\mathrm{a}^3\Pi}$ with the value image $A_{\mathrm{a}^3\Pi(v=13)}$ = 38.115(13) $\mathrm{cm}^{-1}$ yields an electronic prefactor of essentially comparable magnitude. The two quantities should not, however, be regarded as strictly identical: the *ab initio* result refers to the an electronic matrix element evaluated at a fixed internuclear distance, whereas the spectroscopic value employed here is inferred from a vibrationally dependent spin-orbit constant. Nevertheless, the close correspondence between them is physically meaningful. Both the present spectroscopic estimate and the independent *ab initio* result[40] indicated that the relevant electronic couplingis of the order of only a few tens of $\mathrm{cm}^{-1}$.

Ultimately,

$$\langle \mathrm{A}^1\Pi, v = 2|\mathbf{H}^{\mathbf{SO}}|\mathrm{a}^3\Pi, v = 13\rangle_{\mathrm{calc}} = (3.40 \pm 0.44) \times 10^{-4}\ \mathrm{cm}^{-1} \text{ for } J = 20 \text{ in } {}^{13}\mathrm{C}^{18}\mathrm{O}. \quad (5)$$

The observed phenomenon that the $\mathrm{A}^1\Pi \sim \mathrm{a}^3\Pi$ perturbation could be modelled using a single parameter to describe the direct $\mathrm{A}^1\Pi(v = 2) \sim \mathrm{a}^3\Pi(v = 13)$ interaction; however, the required matrix element would then need to be several orders of magnitude larger than predicted by the following experimental spectroscopy calculation as well as electronic-structure theory prediction. Such a large electronic factor is not plausible. Thus, the observed perturbations are interpreted here not as a direct perturbation, but as an effect that arises via an *indirect* perturbation mechanism.

## V. THE INDIRECT PERTURBATIONS PATHWAYS IN $^{13}\mathrm{C}^{18}\mathrm{O}$

Since observation of a *direct* $\mathrm{A}^1\Pi(v = 2) \sim \mathrm{a}^3\Pi(v = 13)$ perturbation is impossible at the precision of the current experiment, we examined possible *indirect* paths via all electronic states of the $\pi^3\sigma^2\pi^*$ configuration. Each of these $\pi^3\sigma^2\pi^*$ electronic states is sufficiently well characterized experimentally that the locations and strengths of all perturbations of both the $^1\Pi$ state and the $^3\Pi$ state by $\pi^3\sigma^2\pi^*$ states can be predicted and modelled. The locations and predicted perturbation parameters of 3 observable indirect paths are identified. The $\mathrm{A}^1\Pi(v = 2) \sim \mathrm{a}^3\Pi(v = 13)$ interaction is facilitated via:

*(i)* $\mathrm{e}^3\Sigma^-(v = 4)$ for $J$ = 20, 24 and 32 *e*-components (see Figure 6(c)) and the $J$ = 27 *f*-component (see Figure 6(d)) via spin-orbit/spin-electronic/rotation-electronic (**L**-uncoupling) interactions;

*(ii)* $\mathrm{I}^1\Sigma^-(v = 3)$ for $J$ = 7-8 (see Figure 6(d)) via the rotation-electronic (**L**-uncoupling)/spin-orbit interactions;

*(iii)* $\mathrm{d}^3\Delta(v = 7)$ for $J$ = 37 (see Figures 6(c) and 6(d)) via spin-orbit/spin-electronic/rotation-electronic (**L**-uncoupling) interactions.

Identification and discussion of all components of the indirect interaction mechanism are complete. However, the observed energy level patterns and systematic tests of the fit-model suggest that the existence of the extra-lines involving the *e*-components of the $\mathrm{a}^3\Pi(v = 13)$ level in the region of $J$ = 20 cannot be explained by the isolated contribution of the indirect $\mathrm{A}^1\Pi(v = 2) \sim \mathrm{e}^3\Sigma^-(v = 4) \sim \mathrm{a}^3\Pi(v = 13)$ perturbation in $^{13}\mathrm{C}^{18}\mathrm{O}$. This indicates that the crossing of the $\mathrm{A}^1\Pi(v = 2)$ and $\mathrm{e}^3\Sigma^-(v = 4)$ states occurs at $J$ values quite far from the $J$-level region where the $\mathrm{A}^1\Pi(v = 2)$ and $\mathrm{a}^3\Pi(v = 13)$ diabatic intersections occur, i.e. at $J$ = 24 - 25 (see Figure 7). Furthermore, a spin-orbit interaction is allowed only between components with $\Delta|\Omega| = 0$ and the closest $\mathrm{a}^3\Pi(v = 13, F_{2e}, |\Omega| = 1) \sim \mathrm{e}^3\Sigma^-(v = 4)$ intersections occur as much as about 67 cm$^{-1}$ above the a$^3\Pi$($v$ = 13, $F_{1e}$, |$\Omega$| = 0) ~ A$^1\Pi$($v$ = 2, $F_{1e}$, |$\Omega$| = 1) anti-crossing at $J$ = 20.

## VI. INTERFERENCES BETWEEN MULTIPLE INDIRECT PERTURBATION PATHWAYS

As neither the direct A$^1\Pi$($v$ = 2) ~ a$^3\Pi$($v$ = 13) perturbation nor the isolated-path indirect A$^1\Pi$($v$ = 2) ~ e$^3\Sigma^-$($v$ = 4) ~ a$^3\Pi$($v$ = 13) pathway sufficiently explains the detection of the extra-lines shown in Figure 8, we examined a model involving *interferences of multiple indirect perturbation paths between* the A$^1\Pi$($v$ = 2) and a$^3\Pi$($v$ = 13) basis states in $^{13}$C$^{18}$O. These multiple paths are mediated by the a$^3\Pi$($v$ = 13), $^3\Pi_1$ ~ $^3\Pi_0$, $^3\Pi_1$ ~ $^3\Pi_2$, and e$^3\Sigma^-$($v$ = 4), $F_{1e}$ ~ $F_{3e}$ intra-state **S**-uncoupling interactions, described in detail in section III.1 of supplementary material.

### A. The special advantages of $^{13}$C$^{18}$O

Quantum mechanical interference occurs when there are multiple intermediate basis-state pathways between initial and final basis states. It turned out that the A$^1\Pi$($v$ = 2), a$^3\Pi$($v$ = 13) pair in $^{13}$C$^{18}$O offers the most favorable conditions for observation of this phenomenon, because the maximum of the e$^3\Sigma^-$($v$ = 4, $F_1$) ~ a$^3\Pi$($v$ = 13, $F_2$) interaction in $^{13}$C$^{18}$O occurs almost exactly at $J$ = 20, where also, but at much lower energy, there is a diabatic crossing of the A$^1\Pi$($v$ = 2, $F_1$) level with the a$^3\Pi$($v$ = 13, $F_1$) level. In addition, all three spin components of the e$^3\Sigma^-$($v$ = 4) level perturb A$^1\Pi$($v$ = 2): $F_1$ and $F_3$ perturb the $e$-symmetry component and $F_2$ perturbs the $f$-symmetry component. Due to the large value of the A$^1\Pi$ ~ e$^3\Sigma^-$ perturbation matrix elements, there is some small, but sufficient A$^1\Pi$($v$ = 2) character in rovibronic levels of the e$^3\Sigma^-$($v$ = 4) state even at $J$ values far from the A$^1\Pi$($v$ = 2) – e$^3\Sigma^-$($v$ = 4) level crossings, i.e. $J$ = 24-25, 27-28, and 30-31 ( compare Figures 6 and 7). All three spin components of the e$^3\Sigma^-$($v$ = 4) state perturb spin components of the a$^3\Pi$($v$ = 13) state via $e \leftrightarrow e$ and $f \leftrightarrow f$ selection rules. In addition, there are significant, intra-state **S**-uncoupling interactions of the $\Omega$ = 1 and 0 characters of both the e$^3\Sigma^-$($v$ = 4) and a$^3\Pi$($v$ = 13) levels. These interactions are responsible for the existence of a small admixture of A$^1\Pi$($v$ = 2) character in the e$^3\Sigma^-$($v$ = 4) and a$^3\Pi$($v$ = 13) levels and, in consequence, are crucial for a description of the observed relative intensities of the extra-line transitions into levels of dominant a$^3\Pi$($v$ = 13) character (see Figure 8).

### B. Interferences in CO

The pathways between $^3\Pi_{\Omega e}$ and $^1\Pi_{1e}$ basis states exist via $^3\Sigma^-_{F_1 e}$ and $^3\Sigma^-_{F_3 e}$ basis states. There are, in principle, four pairs of interfering intermediate basis-state pathways. They are listed in Table II.

**TABLE II. The interferences among the multiple intermediate basis-state pathways between the $A^1\Pi(v = 2)$ and $a^3\Pi(v = 13)$ levels in $^{13}C^{18}O$ via specified intermediate states**[a]

| Interference number | Indirect perturbation number | Indirect interaction path | Interaction nature | Interference intermediaries |
|---|---|---|---|---|
| I | 1 | $a^3\Pi_1 \sim e^3\Sigma^-_{F_{1e}} \sim A^1\Pi_1$ | $(H^{SO},H^{L\text{-}U})$, $H^{SO}$ | $^3\Sigma^-_{F_{1e}} \sim {}^3\Sigma^-_{F_{3e}}$ |
| | 2 | $a^3\Pi_1 \sim e^3\Sigma^-_{F_{3e}} \sim A^1\Pi_1$ | $(H^{SO},H^{L\text{-}U})$, $H^{SO}$ | |
| II | 3 | $a^3\Pi_0 \sim e^3\Sigma^-_{F_{1e}} \sim A^1\Pi_1$ | $(H^{SO},H^{L\text{-}U})$, $H^{SO}$ | $^3\Sigma^-_{F_{1e}} \sim {}^3\Sigma^-_{F_{3e}}$ |
| | 4 | $a^3\Pi_0 \sim e^3\Sigma^-_{F_{3e}} \sim A^1\Pi_1$ | $(H^{SO},H^{L\text{-}U})$, $H^{SO}$ | |
| III | 5 | $a^3\Pi_2 \sim a^3\Pi_1 \sim e^3\Sigma^-_{F_{1e}} \sim A^1\Pi_1$ | $H^{S\text{-}U}$, $(H^{SO},H^{L\text{-}U})$, $H^{SO}$ | $^3\Sigma^-_{F_{1e}} \sim {}^3\Sigma^-_{F_{3e}}$ |
| | 6 | $a^3\Pi_2 \sim a^3\Pi_1 \sim e^3\Sigma^-_{F_{3e}} \sim A^1\Pi_1$ | $H^{S\text{-}U}$, $(H^{SO},H^{L\text{-}U})$, $H^{SO}$ | and $^3\Pi_2 \sim {}^3\Pi_1$ |
| IV | 7 | $a^3\Pi_0 \sim a^3\Pi_1 \sim e^3\Sigma^-_{F_{1e}} \sim A^1\Pi_1$ | $H^{S\text{-}U}$, $(H^{SO},H^{L\text{-}U})$, $H^{SO}$ | $^3\Sigma^-_{F_{1e}} \sim {}^3\Sigma^-_{F_{3e}}$ |
| | 8 | $a^3\Pi_0 \sim a^3\Pi_1 \sim e^3\Sigma^-_{F_{3e}} \sim A^1\Pi_1$ | $H^{S\text{-}U}$, $(H^{SO},H^{L\text{-}U})$, $H^{SO}$ | and $^3\Pi_0 \sim {}^3\Pi_1$ |

[a] Here, rectangles mark the pairs of interacting components responsible for interference between specified indirect perturbation paths.

Paths 1, 2 are the strongest, particularly at low-$J$. Paths 3, 4 are weaker than paths 1, 2, and are critically dependent on near-degeneracy of $^3\Pi_{1e}$ and $^3\Sigma^-_{\Omega e}$ states. Paths 5, 6 and 7, 8 populate the $\Omega = 0$ and $\Omega = 2$ components of $^3\Pi_{\Omega e}$ equally, and their intensities are proportional to $J^2$. To assess the interferences connecting the initial $A^1\Pi(v = 2)$ and final $a^3\Pi(v = 13)$ basis states in $^{13}C^{18}O$, the final model was formulated in an explicitly component-resolved basis. The $e^3\Sigma^-(v = 4)$ and $a^3\Pi(v = 13)$ levels were separated into their $F_1$, $F_2$, and $F_3$ *e/f*-parity components, and the corresponding off-diagonal intra-state interaction matrix elements were introduced explicitly. These terms were implemented using Western's[16,17] perturbation-based rotational-Hamiltonian formalism. Details of the procedure used here are provided in section III.2 of supplementary material. Then, the intra-manifold couplings between the interfering *e*-symmetry components of the $e^3\Sigma^-(v = 4)$ and the $a^3\Pi(v = 13)$ levels were tested using an on/off benchmark. The effects of their removal from the fit-model on the $A^1\Pi(v = 2, F_{1e})$ character were monitored against the final model via the squared eigenvector coefficient $\left(C^2_{ik} \cdot 100\%\right)$. The results, summarized in Figure 10 and displaying the sensitivity of the mechanism to each of several competing effects, are based on the final deperturbation model as a reference (black line with triangular markers), and on specific tests within this model:

***(i)*** with the turned-off $e^3\Sigma^-(v = 4)$, $F_{1e} \sim F_{3e}$ intra-level interaction (green line with circular markers), for which I-IV interferences are turned off, whilst the 1-8 indirect perturbation pathways are active (see Table II). Relative to the reference trace (all interactions and interferences turned on), the comparison reveals a predominantly constructive interference (positive) net contribution of the four interference terms I–IV across essentially the entire rotational manifold, with a diagnostically revealing exception at $J$ = 26-28, where the combined effect of interferences I-IV becomes destructive (negative);

*(ii)* with the turned-off $a^3\Pi(v = 13)$, $F_{1e} \sim F_{2e}$ and $F_{3e} \sim F_{2e}$ interactions (grey line with square markers), for which the III and IV interferences and the 5-8 perturbation paths are suppressed. A pronounced intensity reduction, compared to the reference (black) trace, occurs for most rotational levels in the transfer of the $A^1\Pi(v = 2, F_{1e})$ character into other interacting states. This behavior reveals the net constructive effect of interferences III and IV, in combination with the contributions from the isolated indirect perturbation paths 5-8. Exceptions occur for $J$ = 22-23, 26-28, 30-31, 34-36, and 38 where the cumulative effects of these processes are destructive;

*(iii)* and with the $e^3\Sigma^-(v = 4)$ and $a^3\Pi(v = 13)$ intra-state interactions both turned off (orange line with cross-shaped markers), for which the I – IV interferences and 5-8 perturbation paths are turned off. In this case, over almost the entire range (excluding $J$ = 26-28, 34-36, and 38), the constructive results are revealed arising from the sum of the I – IV interferences in combination with the isolated 5 - 8 indirect perturbation paths.

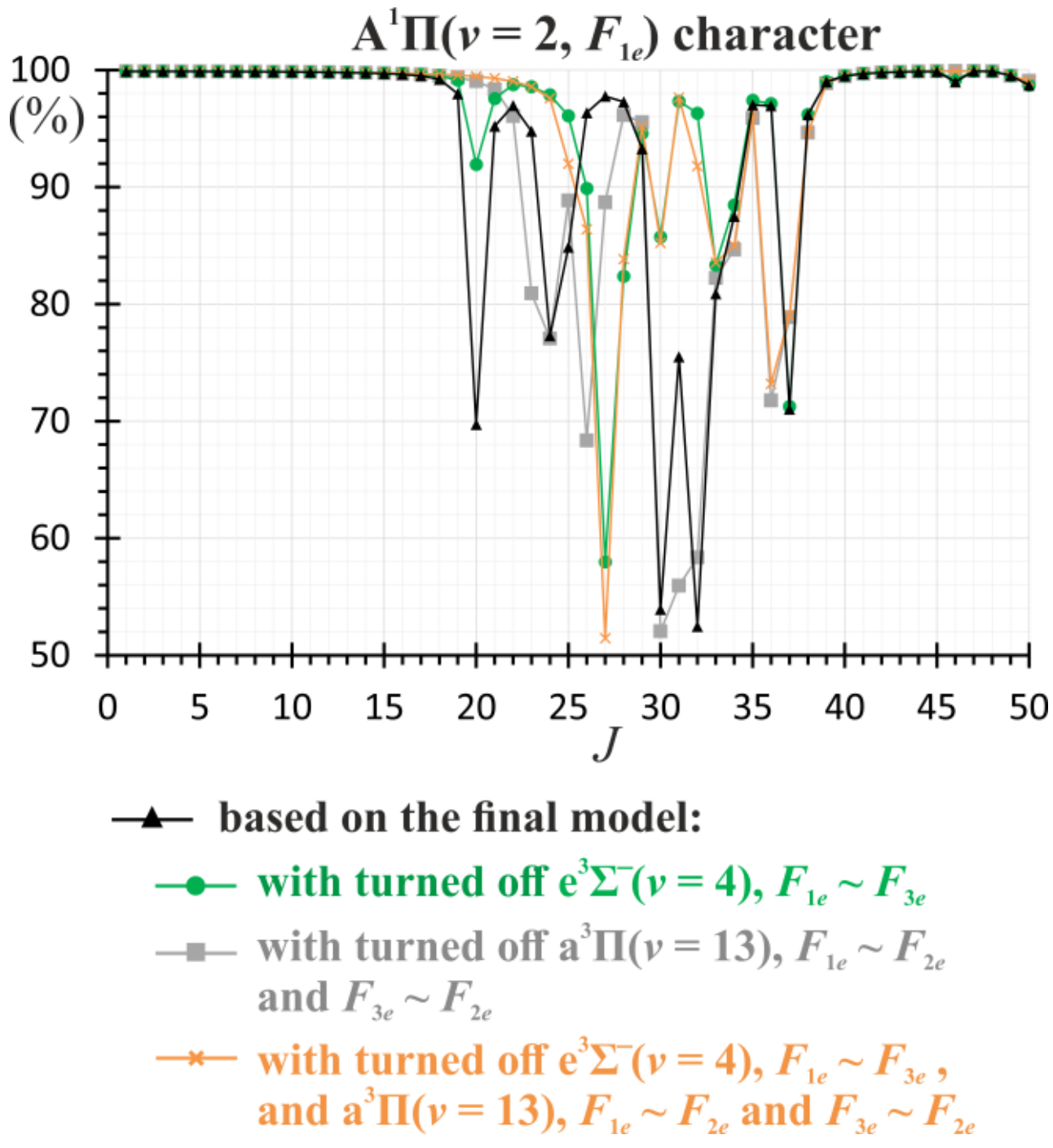

**FIG. 10.** The inherent fractional $^{13}C^{18}O$ $A^1\Pi(v = 2, F_{1e})$ character based on the final deperturbation model as a reference (black line with triangular markers), as well as specific tests described in a legend of this figure. For details, see the Table II and section III.2 of supplementary material.

From the perspective of our analysis of the perturbation between the $a^3\Pi(v = 13)$ and $A^1\Pi(v = 2)$ levels in $^{13}C^{18}O$, the region around $J = 20$ for the $e$-symmetry components is of primary interest with the addition of significant effects at $J = 24$ and $J = 32$. In these regions, the $e$-symmetry levels of $a^3\Pi$ show the strongest borrowing of $^1\Pi$ character (see Figure 6(e)). The total interference I-IV is strongly constructive there, with the only exception being the very weakly destructive combined contribution of interferences III and IV, mediated by the $a^3\Pi$, $F_{1e} \sim F_{2e}$ and $F_{2e} \sim F_{3e}$ interactions at $J = 24$. Overall, constructive interference enhances the effect on the multiple-step coupling chain that mediates the $A^1\Pi(v = 2) \sim a^3\Pi(v = 13)$ interaction.

## VII. CONCLUSIONS

This work describes a complete picture of well-studied perturbations in CO, which consists of all (5 × 2 = 10) perturbation mechanisms between the electronic states of the $\pi^3\sigma^2\pi^*$ and $\pi^4\sigma\pi^*$ configurations. The model exploits rigorous rotation-, vibration-, and isotopologue-dependences, most based on the existence of experimentally-refined potential energy curves of all seven electronic states. Among all CO isotopologues and vibrational levels, the best possibility for $A^1\Pi \sim a^3\Pi$ near-degeneracy was found for $^{13}C^{18}O$ $A^1\Pi(v = 2) \sim a^3\Pi(v = 13)$. The spectroscopic data required was harvested from state-of-the-art spectra of $A^1\Pi - X^1\Sigma^+$, $B^1\Sigma^+ - A^1\Pi$, $C^1\Sigma^+ - A^1\Pi$, $e^3\Sigma^- - X^1\Sigma^+$, $a'^3\Sigma^+ - X^1\Sigma^+$, $D^1\Delta - X^1\Sigma^+$, $d^3\Delta - X^1\Sigma^+$, and $I^1\Sigma^- - X^1\Sigma^+$ electronic transitions, providing 16 bands comprising 787 spectral lines and 428 rotational term values up to $J_{max} = 50$ of $^{13}C^{18}O$. This paper is notable in that it extends the study of spectroscopic perturbations into a previously unexplored regime. The $A^1\Pi$ and $a^3\Pi$ electronic states belong to the same electronic configuration, but their energy separation is approximately eleven times their $\omega_e$ vibrational frequency. The vibrational overlap integral between near degenerate ($\Delta v = 11$) vibrational levels of these two electronic states is so small that local spin-orbit perturbation level-shifts will be much smaller than the 0.001 - 0.01 $cm^{-1}$ precision of modern electronic spectroscopy. But in this paper we report observation and full characterization of this "impossible perturbation." It is a *perturbation facilitated by perturbations*, in $^{13}C^{18}O$, of *both* the $A^1\Pi(v = 2)$ and $a^3\Pi(v = 13)$ states by the $e^3\Sigma^-(v = 4)$ state. Moreover, since the model involves simultaneous interactions among three or more rovibronic states, the $e^3\Sigma^-(v = 4)$ and $a^3\Pi(v = 13)$ intermediate levels were decomposed into its interacting triplet components and each analysis conclusion was verified by turning-off one fit-parameter at a time and observation of whether and how the fit was degraded. As a result, quantum mechanical constructive interference is observed between the same pair of nominally $a^3\Pi$ and $A^1\Pi$ rovibronic levels by the $F_{1e}$ and $F_{3e}$ spin-components of the $e^3\Sigma^-$ state.

## SUPPLEMENTARY MATERIAL

The supplementary material provides: (*i*) the wavenumbers of the lines recorded in this work using Vis-FT and VUV-FT spectroscopy; (*ii*) the experimental term values; and (*iii*) details of the present deperturbation analysis ("Supplementary material (1).docx" file).

## ACKNOWLEDGMENTS

RH thanks LASERLAB-EUROPE for support of this research [grant numbers EUH2020-RIP-654148 and EC's-SPF-284464]. RH, SR, WSz, AS, and IP thank the European Regional Development Fund and the Polish state budget within the framework of the Carpathian Regional Operational Program [grant number RPPK.01.03.0 0-18-0 01/10] through the funding of the Centre for Innovation and Transfer of Natural Sciences and Engineering Knowledge of the University of Rzeszów. RWF thanks the US National Science Foundation (grant number CHE-1800410) for support of his research, which included substantial collaborations. AP acknowledges partial support by the Bulgarian national plan for recovery and resilience, contract BG-RRP-2.004-0008-C01 SUMMIT: Sofia University Marking Momentum for Innovation and Technological Transfer, project number 3.1.4. The authors are grateful to the general and technical staff of SOLEIL synchrotron for running the facility and providing beam time under project numbers 20120653 and 20160118.

## AUTHOR DECLARATIONS

### Conflict of Interest

The authors have no conflicts to disclose.

### Author Contributions

**Stanisław Ryzner**: Conceptualization, Methodology, Software, Validation, Formal analysis, Investigation, Resources, Data curation, Writing – original draft, Writing – review & editing, Visualization. **Robert W. Field**: Conceptualization, Methodology, Writing – original draft, Writing – review & editing, Supervision, Funding acquisition. **Wim Ubachs**: Investigation, Resources, Writing – original draft, Writing – review & editing, Supervision. **Alan N. Heays**: Software, Validation, Formal analysis, Investigation, Data curation. **Asen Pashov**: Methodology, Software, Formal analysis, Data curation, Writing – original draft, Funding acquisition. **Patryk Jasik**: Methodology, Software, Formal analysis, Investigation, Data curation, Writing – original draft. **Józef E. Sienkiewicz**: Software, Validation, Supervision. **Nelson de Oliveira**: Validation, Investigation, Writing – review & editing. **Kin F. Lai**: Methodology, Software, Formal analysis, Data curation. **Marzena I. Malicka**: Software, Formal analysis, Data curation, Visualization. **Aleksandra Stasik**: Formal analysis, Investigation, Data curation, Visualization. **Wojciech Szajna**: Writing – review & editing, Visualization. **Izabela Piotrowska**: Writing – review & editing, Visualization. **Rafał Hakalla**: Conceptualization, Methodology, Software, Validation, Formal analysis, Investigation, Resources, Data curation, Writing – original draft, Writing – review & editing, Visualization, Supervision, Project administration, Funding acquisition.

## DATA AVAILABILITY

The data that support the findings of this study are available within the article and its supplementary material as the output LOG of the final fit obtained with the PGOPHER program ("Supplementary material (2).txt" file).

## APPENDIX A: INTENSITY OF THE $a^3\Pi - X^1\Sigma^+$ CAMERON BANDS IN CO

In the above it is argued that the spin-orbit interaction matrix element $\langle A^1\Pi|\mathbf{H^{SO}}|a^3\Pi\rangle$ is too small to cause observable line shifts in the spectra, at least at the current state-of-the-art accuracies achieved. However, it has been established in the literature that the intensity of the $a^3\Pi - X^1\Sigma^+$ system, first observed by Cameron,[41] is produced by intensity borrowing from the fourth positive system in CO, the $A^1\Pi - X^1\Sigma^+$ system, via the same extremely small spin-orbit interaction matrix element. Though this may be regarded as counterintuitive, an explanation for this is based on the strength of the Cameron bands indeed being very small. While the band oscillator strength of the $A^1\Pi - X^1\Sigma^+(0, 0)$ band has a value of $f_{0,0} = 1.65 \times 10^{-2}$,[42] for the (0, 0) band in the Cameron system it reaches only $f_{0,0} = 1.63 \times 10^{-7}$, so it is five orders of magnitude weaker.[43–45] Band intensity measurements and comparisons with lifetime measurements of the $a^3\Pi$ state revealed that 99% of the intensity of the Cameron bands is produced by the spin-orbit interaction between $a^3\Pi$ and $A^1\Pi$.[35,46] This is also proven by the observation that for low-$J$ levels the $a^3\Pi_1$ components receive their intensity, via borrowing from $A^1\Pi_1$. For higher $J$ levels, transitions to the $a^3\Pi_0$ and $a^3\Pi_2$ components obtain intensity, caused by the mixing of $\Omega$-components for increased rotational energy in the triplet manifold (see also Figure S2). In a highly sensitive laser excitation study the $P(1)$ transition in excitation to $a^3\Pi_0(v = 0, J = 0)$ was observed, although extremely weakly.[35] This provides indication that the Cameron bands also draw their intensity from interaction with a $^1\Sigma^+$ state, albeit for a minute fraction.

## APPENDIX B: $J$-DEPENDENT VIBRATIONAL OVERLAP INTEGRALS $\langle a^3\Pi, v = 13|A^1\Pi, v = 2\rangle$ IN $^{13}C^{18}O$

$J$-dependent vibrational overlap integrals $\langle a^3\Pi, v = 13|A^1\Pi, v = 2\rangle$ in $^{13}C^{18}O$ were calculated in two ways:

**Method 1:** The vibrational wavefunctions $|A^1\Pi, v = 2\rangle$ and $|a^3\Pi, v = 13\rangle$ were obtained by numerically solving the radial Schrödinger equation for the rotational quantum number $J\epsilon\langle 1, 100\rangle$ using an equidistant grid of 5000 potential energy curve points in the interval $\langle 0.99\ \text{Å}, 2.4\ \text{Å}\rangle$ for the $A^1\Pi$ state and in the interval $\langle 0.8\ \text{Å}, 2.6\ \text{Å}\rangle$ for the $a^3\Pi$ state. The $\langle a^3\Pi, v = 13|A^1\Pi, v = 2\rangle$ vibrational overlap integrals were computed using the RKR potential curves for both states. The RKR curves were calculated based on isotopically adjusted equilibrium vibration-rotation constants of Field,[1] Field *et al.*,[3],

Lefebvre-Brion and Field,[36] Le Floch *et al.*,[4] and Kittrell *et al.*[47] using the FRACONB program developed by Jung[48] and Jakubek[49]. The results are shown as circles in Figure 11.

**Method 2:** Vibrational overlap integrals $\langle a^3\Pi, v = 13 | A^1\Pi, v = 2\rangle$ were calculated using vibrational wavefunctions obtained by numerically solving the radial Schrödinger equation. The *J*-dependent potential energy curves for the $a^3\Pi$ and $A^1\Pi$ electronic states of $^{13}C^{18}O$ were constructed using the RKR method, which is based on molecular constants. The Python code was developed to compute the two Klein integrals in order to derive the classical turning points on which the RKR method is based.[50] The vibrational and rotational integrals (*f* and *g* integrals) of the original RKR method were slightly modified to prevent the occurrence of a singularity at the upper integration limit $v' \rightarrow v$ by adding a correction term. These equations are:

$$f(v) = \frac{h}{\pi\sqrt{2\mu}}\left(\int_{v_{min}}^{v-\delta} \frac{dv\prime}{\sqrt{G(v)-G(v\prime)}} + 2\sqrt{\frac{\delta}{Q(v)}}\right), \tag{B1}$$

$$g(v) = \frac{4\pi\sqrt{2\mu}}{h}\left(\int_{v_{min}}^{v-\delta} \frac{B(v\prime)}{\sqrt{G(v)-G(v\prime)}} dv' + 2B(v)\sqrt{\frac{\delta}{Q(v)}}\right), \tag{B2}$$

where $G(v)$ and $B(v)$ are standard *v*-dependent functions based on a power series in $(v + 1/2)$ associated with the Dunham series. $Q(v) = \frac{dG(v)}{d\left(v+\frac{1}{2}\right)}$ is the derivative of the function $G(v)$ with respect to $(v + 1/2)$, while $\delta$ is a small parameter, usually set to $10^{-5}$, and adjusted to ensure convergence. In the convention used here, the functions $f(v)$ and $g(v)$ are related to the classical inner $r_-(v)$ and outer $r_+(v)$ turning points by $f(v) = r_+(v) - r_-(v)$ and $g(v) = \frac{1}{r_-(v)} - \frac{1}{r_+(v)}$. Thus, the functions $f(v)$ and $g(v)$ are twice the corresponding functions defined by Le Roy.[49] With this convention, the turning points are given by

$$r_\pm(v) = \frac{f(v)}{2}\left[\sqrt{1 + \frac{4}{f(v)g(v)}} \pm 1\right], \tag{B3}$$

which is algebraically equivalent to Le Roy's turning-point expressions.

Vibrational and rotational equilibrium constants for the $a^3\Pi$ and $A^1\Pi$ electronic states of the CO molecule needed to obtain $f(v)$ and $g(v)$ functions were taken from the dissertation of Field[1] (Appendix 2b) and recalculated for the $^{13}C^{18}O$ molecular isotopologue in the way described previously.

Using the RKR potential energy curves of the $a^3\Pi$ and $A^1\Pi$ electronic states, the vibrational wavefunctions $v = 13$ for $a^3\Pi$ and $v = 2$ for $A^1\Pi$ were calculated for *J* ranging from 1 to 100. All wavefunctions were obtained using the sPYtroscopy program[51] which implements the discrete variable representation method in the Fourier sine basis. The resulting wavefunctions were used to calculate *J*-dependent vibrational overlap integrals $\langle a^3\Pi, v = 13, J \mid A^1\Pi, v = 2, J\rangle$. The results are shown as crosses in Figure 11.

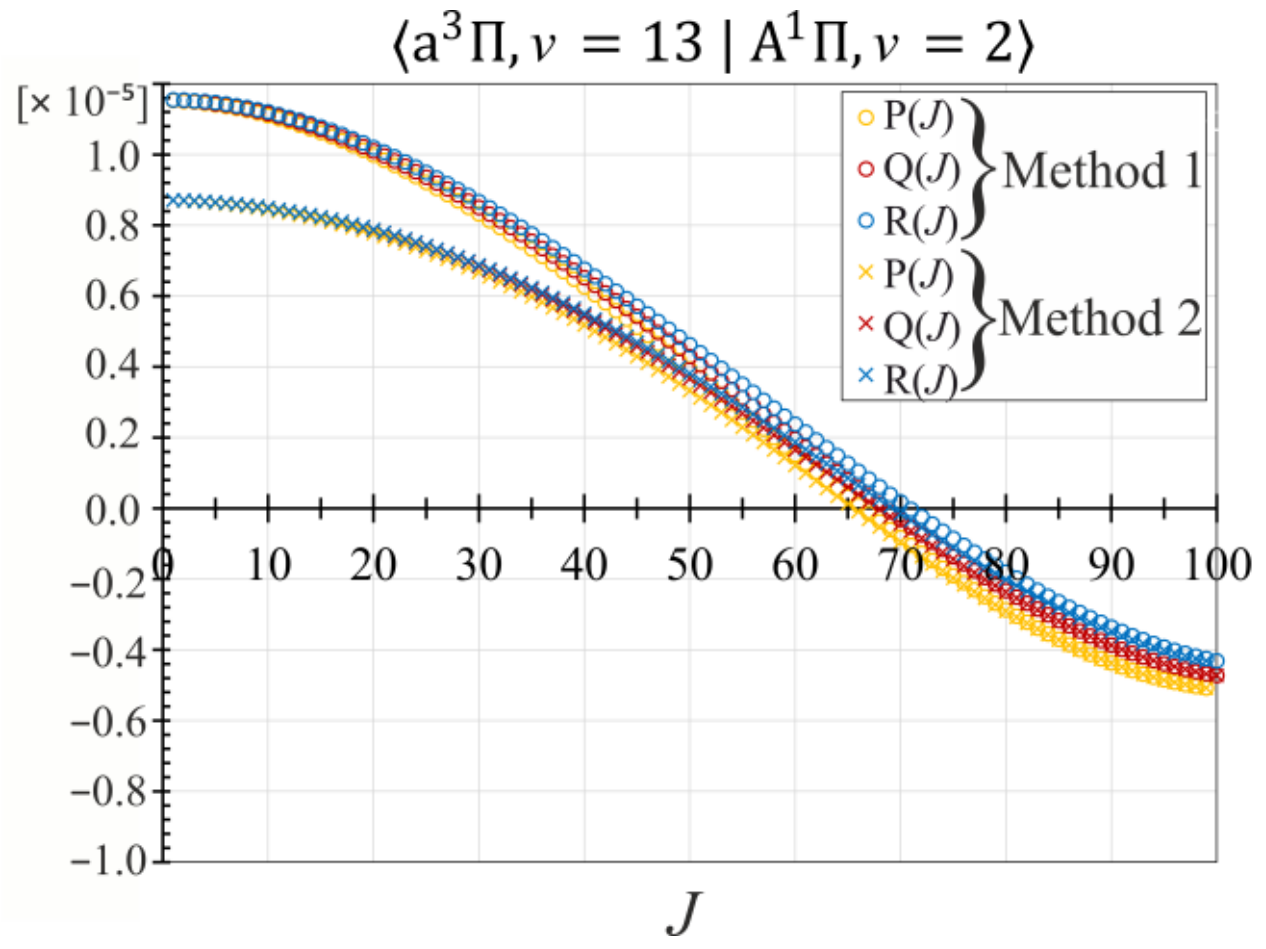

**FIG. 11.** The *J*-dependence of vibrational overlap integrals between $A^1\Pi(v = 2)$ and $a^3\Pi(v = 13)$ in $^{13}C^{18}O$.

# Supplementary Material for "Perturbation of CO $A^1\Pi$ by $a^3\Pi$ should have been undetectable: detection, mechanism, and surprises"

The supplementary material provides: (I) the wavenumbers of the lines recorded and analysed in this work using two complementary, Vis-FT and VUV-FT, spectroscopy techniques; (II) the experimental term values; and (III) details of the present deperturbation analysis.

# Table of Contents

# I. WAVENUMBERS OF THE LINES RECORDED IN THIS WORK USING VIS-FT AND VUV-FT SPECTROSCOPY.

**TABLE S1.** Wavenumbers (in $cm^{-1}$) of the $B^1\Sigma^+ \rightarrow A^1\Pi$ (0, 2), Vis-FT emission band of $^{13}C^{18}O$.[a]

| | $^rR_{11ee}(J'')$ | | $^qQ_{11ef}$ | | $^pP_{11ee}(J'')$ | |
|---|---|---|---|---|---|---|
| $J''$ | $\sigma$ | u($\sigma$) | $\sigma$ | u($\sigma$) | $\sigma$ | u($\sigma$) |
| 0 | | | | | | |
| 1 | 19367.55[b] | 0.02 | 19360.477[b] | 0.007 | 19356.94[b] | 0.02 |
| 2 | 19372.50[b] | 0.01 | 19361.875 | 0.005 | 19354.810[b] | 0.009 |
| 3 | 19378.158[b] | 0.007 | 19364.002 | 0.005 | 19353.374[b] | 0.009 |
| 4 | 19384.512[b] | 0.006 | 19366.825 | 0.005 | 19352.674[b] | 0.005 |
| 5 | 19391.578[b] | 0.006 | 19370.357 | 0.005 | 19352.642[b] | 0.005 |
| 6 | 19399.343[b] | 0.006 | 19374.604 | 0.005 | 19353.348[b] | 0.005 |
| 7 | 19407.821[b] | 0.005 | 19379.616 | 0.005 | 19354.744[b] | 0.005 |
| 8 | 19417.005[b] | 0.006 | 19385.013 | 0.005 | 19356.850[b] | 0.005 |
| 9 | 19426.875[b] | 0.005 | 19391.442[b] | 0.005 | 19359.662 | 0.005 |
| 10 | 19437.463[b] | 0.005 | 19398.515 | 0.005 | 19363.184 | 0.005 |
| 11 | 19448.762[b] | 0.005 | 19406.283 | 0.005 | 19367.415[b] | 0.005 |
| 12 | 19460.755 | 0.005 | 19414.757 | 0.005 | 19372.359[b] | 0.005 |
| 13 | 19473.461 | 0.005 | 19423.933 | 0.005 | 19378.005[b] | 0.005 |
| 14 | 19486.875 | 0.005 | 19433.819 | 0.005 | 19384.364[b] | 0.005 |
| 15 | 19500.997 | 0.005 | 19444.414 | 0.005 | 19391.439[b] | 0.005 |
| 16 | 19515.790 | 0.005 | 19455.720 | 0.005 | 19399.227[b] | 0.005 |
| 17 | 19531.381 | 0.005 | 19467.695 | 0.005 | 19407.740[b] | 0.005 |
| 18 | 19547.663[b] | 0.005 | 19480.467 | 0.005 | 19416.944[b] | 0.005 |
| 19 | 19564.700 | 0.005 | 19493.926 | 0.005 | 19426.980[b] | 0.005 |
| 20 | 19582.978 | 0.005 | 19508.112 | 0.005 | 19438.239 | 0.005 |
| 21 | 19600.493 | 0.005 | 19522.867[b] | 0.005 | 19448.731[b] | 0.005 |
| 22 | 19619.814 | 0.005 | 19538.577 | 0.005 | 19461.040 | 0.005 |
| 23 | 19639.992[b] | 0.005 | 19554.975 | 0.005 | 19474.210 | 0.005 |
| 24 | 19661.974 | 0.005 | 19572.168 | 0.005 | 19489.192 | 0.005 |
| 25 | 19678.712[b] | 0.006 | 19590.271 | 0.005 | 19498.926[b] | 0.005 |
| 26 | 19701.822 | 0.005 | 19609.697[b] | 0.005 | 19515.046 | 0.005 |
| 27 | 19724.847 | 0.005 | 19632.545 | 0.005 | 19531.094 | 0.005 |
| 28 | 19748.500 | 0.005 | 19643.051 | 0.005 | 19547.772[b] | 0.005 |
| 29 | 19773.112 | 0.005 | 19665.116 | 0.005 | 19565.419 | 0.005 |
| 30 | 19796.52[b] | 0.02 | 19686.738 | 0.005 | 19581.884[b] | 0.008 |
| 31 | 19820.848 | 0.006 | 19708.164 | 0.005 | 19599.253 | 0.005 |
| 32 | 19849.09[b] | 0.02 | 19731.655 | 0.005 | 19620.538 | 0.006 |
| 33 | 19876.28[b] | 0.01 | 19755.933 | 0.005 | 19640.794[b] | 0.008 |
| 34 | 19900.300 | 0.006 | 19778.141 | 0.005 | 19657.868[b] | 0.008 |
| 35 | 19929.75[b] | 0.01 | 19803.895 | 0.005 | 19680.416 | 0.006 |
| 36 | 19959.263 | 0.007 | 19830.546 | 0.005 | 19703.047 | 0.006 |
| 37 | 19990.98[b] | 0.02 | 19857.570[b] | 0.008 | 19727.852 | 0.008 |
| 38 | 20018.317 | 0.007 | 19881.564 | 0.005 | 19748.299 | 0.007 |
| 39 | 20049.83[b] | 0.02 | 19909.667[b] | 0.008 | 19772.943 | 0.007 |
| 40 | 20081.81[bw] | 0.06 | 19938.204 | 0.006 | 19798.026 | 0.007 |
| 41 | 20114.4[bw] | 0.2 | 19967.40[bw] | 0.03 | 19823.771 | 0.009 |
| 42 | | | 19997.25[bw] | 0.03 | 19850.22[w] | 0.02 |
| 43 | | | 20027.83[bw] | 0.04 | | |
| 44 | | | 20059.10[bw] | 004 | | |
| 45 | | | 20091.16[bw] | 005 | | |

[a] The u($\sigma$) values indicate one standard deviation, representing combinations of the calibration and fitting errors of the wavenumber measurements. The instrumental resolution was 0.0179 $cm^{-1}$. The absolute calibration uncertainty was 0.004 $cm^{-1}$. The absolute accuracy of the wavenumber measurements was estimated to be 0.005 $cm^{-1}$. The superscripts *p, q,* and *r* denote change in the total angular momentum excluding spin. Blended and/or weak lines (SNR < 3) are indicated by the superscripts “b” and/or “w,” respectively.

**Alt Text for Table S1:** The table lists *J*-dependent wavenumbers with uncertainties for assigned rotational branches of the $^{13}C^{18}O$ $B^1\Sigma^+ \rightarrow A^1\Pi(0,\ 2)$ Vis-FT emission band. Superscripts identify branch type, blended features, and weak lines.

**TABLE S2.** Wavenumbers (in $cm^{-1}$) of the $C^1\Sigma^+ \rightarrow A^1\Pi$ (0, 2), Vis-FT emission band of $^{13}C^{18}O$.[a]

| | ${}^{r}R_{11ee}(J'')$ | | ${}^{q}Q_{11ef}(J'')$ | | ${}^{p}P_{11ee}(J'')$ | |
|---|---|---|---|---|---|---|
| $J''$ | $\sigma$ | u($\sigma$) | $\sigma$ | u($\sigma$) | $\sigma$ | u($\sigma$) |
| 0 | | | | | | |
| 1 | | | 24361.97[bw] | 0.08 | | |
| 2 | 24373.89[bw] | 0.04 | 24363.32[b] | 0.03 | | |
| 3 | 24379.51[bw] | 0.05 | 24365.42[b] | 0.03 | 24354.83[bw] | 0.05 |
| 4 | 24385.85[bw] | 0.04 | 24368.22[b] | 0.02 | 24354.08[b] | 0.02 |
| 5 | 24392.89 | 0.02 | 24371.709 | 0.008 | 24354.00[b] | 0.03 |
| 6 | 24400.59[bw] | 0.05 | 24375.922 | 0.007 | 24354.69[b] | 0.03 |
| 7 | 24409.03[b] | 0.02 | 24380.89[b] | 0.01 | 24356.04[b] | 0.02 |
| 8 | 24418.15 | 0.01 | 24386.227 | 0.007 | 24358.11[b] | 0.02 |
| 9 | 24427.96[b] | 0.02 | 24392.604[b] | 0.009 | 24360.90[b] | 0.02 |
| 10 | 24438.46[b] | 0.02 | 24399.603[b] | 0.007 | 24364.339 | 0.008 |
| 11 | 24449.70[b] | 0.02 | 24407.303 | 0.006 | 24368.504 | 0.008 |
| 12 | 24461.614 | 0.009 | 24415.703[b] | 0.008 | 24373.37[b] | 0.02 |
| 13 | 24474.23[b] | 0.02 | 24424.792[b] | 0.008 | 24378.945 | 0.007 |
| 14 | 24487.55[b] | 0.02 | 24434.589 | 0.006 | 24385.220 | 0.007 |
| 15 | 24501.561 | 0.009 | 24445.087[b] | 0.008 | 24392.209 | 0.007 |
| 16 | 24516.30[b] | 0.02 | 24456.292 | 0.006 | 24399.90[b] | 0.02 |
| 17 | 24531.75[b] | 0.02 | 24468.201 | 0.006 | 24408.31[b] | 0.02 |
| 18 | 24547.91[b] | 0.02 | 24480.825 | 0.006 | 24417.45[b] | 0.02 |
| 19 | 24564.82[b] | 0.03 | 24494.158 | 0.006 | 24427.334 | 0.008 |
| 20 | 24582.95[bw] | 0.03 | 24508.24[b] | 0.02 | 24438.50[b] | 0.02 |
| 21 | 24600.35 | 0.02 | 24522.854 | 0.007 | 24448.84[b] | 0.02 |
| 22 | 2461952[bw] | 0.04 | 24538.437 | 0.007 | 24461.03[b] | 0.02 |
| 23 | 2463958 | 0.02 | 24554.70[b] | 0.02 | 24474.07[b] | 0.02 |
| 24 | 24661.43[bw] | 0.03 | 24571.749 | 0.007 | 24488.93[b] | 0.02 |
| 25 | 24678.00[bw] | 0.04 | 24589.716 | 0.008 | 24498.51[b] | 0.02 |
| 26 | 24700.97[bw] | 0.09 | 24608.991 | 0.009 | 24514.47[b] | 0.02 |
| 27 | 24723.79[bw] | 0.04 | 24631.70[b] | 0.02 | 24530.39[b] | 0.03 |
| 28 | 24747.34[bw] | 0.04 | 24642.05[b] | 0.03 | 24546.97[bw] | 0.04 |
| 29 | 24771.81[bw] | 0.06 | 24663.96 | 0.02 | 24564.42[b] | 0.02 |
| 30 | | | 24685.42[b] | 0.02 | 24580.73 | 0.03 |
| 31 | | | 24706.68[bw] | 0.04 | 24597.92[bw] | 0.04 |
| 32 | | | 24730.02[w] | 0.02 | 24619.06[bw] | 0.6 |
| 33 | | | 24754.15[bw] | 0.06 | 24639.16[bw] | 0.06 |
| 34 | | | 24776.18[bw] | 0.04 | 24656.05[bw] | 0.07 |
| 35 | | | 24801.78[bw] | 0.07 | 24678.44[bw] | 0.07 |
| 36 | | | 24828.22[w] | 0.04 | 24700.88[bw] | 0.05 |

[a] The u($\sigma$) values indicate one standard deviation, representing combinations of the calibration and fitting errors of the wavenumber measurements. The instrumental resolution was 0.0179 $cm^{-1}$. The absolute calibration uncertainty was 0.004 $cm^{-1}$. The absolute accuracy of the wavenumber measurements was estimated to be 0.007 $cm^{-1}$. Blended and/or weak lines (SNR < 3) are indicated by the superscripts "b" and/or "w," respectively. The superscripts *p, q,* and *r* denote change in the total angular momentum excluding spin.

**Alt Text for Table S2:** The table reports *J*-dependent wavenumbers with uncertainties for assigned rotational branches of the $^{13}C^{18}O$ $C^1\Sigma^+ \rightarrow A^1\Pi$ (0, 2) Vis-FT emission band. Line annotations indicate branch type, blending, and weak signal-to-noise features.

**TABLE S3.** Wavenumbers (in cm⁻¹) of the interaction-induced lines observed in the $B^1\Sigma^+ \rightarrow A^1\Pi(0, 2)$ and $C^1\Sigma^+ \rightarrow A^1\Pi(0, 2)$, Vis-FT emission bands of $^{13}C^{18}O$.[a]

| | $^sR_{11ee}$ | | $^rR_{12ee}$ | | $^qR_{13ee}$ | | $^rQ_{11ef}$ | | $^qQ_{12ef}$ | | $^qP_{11ee}$ | | $^pP_{12ee}$ | | $^oP_{13ee}$ | |
|---|---|---|---|---|---|---|---|---|---|---|---|---|---|---|---|---|
| $J''$ | $\sigma$ | u($\sigma$) | $\sigma$ | u($\sigma$) | $\sigma$ | u($\sigma$) | $\sigma$ | u($\sigma$) | $\sigma$ | u($\sigma$) | $\sigma$ | u($\sigma$) | $\sigma$ | u($\sigma$) | $\sigma$ | u($\sigma$) |
| | | | | | | | | | **$B^1\Sigma^+$ - $I^1\Sigma^-(0, 3)$** | | | | | | | |
| 7 | | | | | | | 19376.87[bw] | 0.03 | | | | | | | | |
| 8 | | | | | | | 19387.81[b] | 0.03 | | | | | | | | |
| | | | | | | | | | **$B^1\Sigma^+$ - $a^3\Pi(0, 13)$** | | | | | | | |
| 20 | 19580.621 | 0.006 | | | | | | | | | 19435.887 | 0.005 | | | | |
| … | | | | | | | | | | | | | | | | |
| 32 | | | 19845.89[b] | 0.02 | | | | | | | | | 19617.340[b] | 0.008 | | |
| | | | | | | | | | **$B^1\Sigma^+$ - $e^3\Sigma^-(0, 4)$** | | | | | | | |
| 24 | 19650.21[b] | 0.02 | | | | | | | | | 19477.424[b] | 0.009 | | | | |
| 25 | 19691.71[b] | 0.03 | | | | | | | | | 19511.93[b] | 0.02 | | | | |
| 26 | | | | | | | | | 19586.20[b] | 0.02 | | | | | | |
| 27 | | | | | | | | | 19617.640[b] | 0.006 | | | | | | |
| 28 | | | | | | | | | 19663.96[b] | 0.02 | | | | | | |
| … | | | | | | | | | | | | | | | | |
| 31 | | | | | 19832.03[bw] | 0.04 | | | | | | | | | 19610.45[b] | 0.03 |
| | | | | | | | | | **$B^1\Sigma^+$ - $d^3\Delta(0, 7)$** | | | | | | | |
| 30 | 19801.58[b] | 0.02 | | | | | 19689.97 | 0.01 | | | 19586.93[b] | 0.06 | | | | |
| … | | | | | | | | | | | | | | | | |
| 38 | | | | | | | | | | | | | | | 19769.20[b] | 0.02 |
| | | | | | | | | | **$B^1\Sigma^+$ - $a'^3\Sigma^+(0, 12)$** | | | | | | | |
| 31 | | | | | | | 19716.32[b] | 0.02 | | | | | | | | |
| | | | | | | | | | **$C^1\Sigma^+$ - $a^3\Pi(0, 13)$** | | | | | | | |
| 20 | 24580.68[bw] | 0.04 | | | | | | | | | 24436.09[b] | 0.03 | | | | |
| | | | | | | | | | **$C^1\Sigma^+$ - $e^3\Sigma^-(0, 4)$** | | | | | | | |
| 24 | | | | | | | | | | | 24477.15[w] | 0.03 | | | | |
| 25 | | | | | | | | | | | 24511.50[w] | 0.03 | | | | |
| … | | | | | | | | | | | | | | | | |
| 27 | | | | | | | | | 24616.78[w] | 0.02 | | | | | | |

[a] The u($\sigma$) values indicate one standard deviation, representing combinations of the calibration and fitting errors of the wavenumber measurements. The instrumental resolution was 0.0179 cm⁻¹. The absolute calibration uncertainty was 0.004 cm⁻¹. The absolute accuracy of the wavenumber measurements was estimated to be 0.02 cm⁻¹. Blended and/or weak lines (SNR < 3) are indicated by the superscripts "b" and/or "w," respectively. The superscripts *o, p, q*, and *s* denote change in the total angular momentum excluding spin.

**Alt Text for Table S3:** The table lists wavenumbers with uncertainties of interaction-induced extra-lines observed in the $^{13}C^{18}O$ $B^1\Sigma^+ \rightarrow A^1\Pi$ (0, 2) and $^{13}C^{18}O$ $C^1\Sigma^+ \rightarrow A^1\Pi$ (0, 2) Vis-FT emission bands. It organizes the lines by rotational branch and upper-state assignment.

**TABLE S4.** Wavenumbers (in cm$^{-1}$) of the $A^1\Pi \leftarrow X^1\Sigma^+(2, 0)$, VUV-FT absorption band of $^{13}C^{18}O$.[a]

| | $^rR_{11ee}(J'')$ | | $^qQ_{11fe}$ | | $^pP_{11ee}(J'')$ | |
|---|---|---|---|---|---|---|
| $J''$ | $\sigma$ | u($\sigma$) | $\sigma$ | u($\sigma$) | $\sigma$ | u($\sigma$) |
| 0 | 67560.43 | 0.02 | | | | |
| 1 | 67562.61 | 0.02 | 67556.94 | 0.02 | | |
| 2 | 67564.12 | 0.02 | 67555.62 | 0.02 | 67549.95 | 0.02 |
| 3 | 67564.97[b] | 0.02 | 67553.65 | 0.02 | 67545.15 | 0.02 |
| 4 | 67565.19[b] | 0.02 | 67551.01 | 0.02 | 67539.68 | 0.02 |
| 5 | 67564.72[b] | 0.02 | 67547.71[b] | 0.02 | 67533.56 | 0.02 |
| 6 | 67563.61 | 0.02 | 67543.75 | 0.02 | 67526.77 | 0.02 |
| 7 | 67561.82 | 0.02 | 67539.06 | 0.02 | 67519.33 | 0.02 |
| 8 | 67559.38 | 0.02 | 67534.03 | 0.02 | 67511.23 | 0.02 |
| 9 | 67556.28 | 0.02 | 67528.02 | 0.02 | 67502.47 | 0.02 |
| 10 | 67552.51 | 0.02 | 67521.42 | 0.02 | 67493.05 | 0.02 |
| 11 | 67548.09[b] | 0.02 | 67514.16 | 0.02 | 67482.98 | 0.02 |
| 12 | 67543.00 | 0.02 | 67506.24 | 0.02 | 67472.24 | 0.02 |
| 13 | 67537.24 | 0.02 | 67497.67 | 0.02 | 67460.84 | 0.02 |
| 14 | 67530.81 | 0.02 | 67488.43 | 0.02 | 67448.79 | 0.02 |
| 15 | 67523.71 | 0.02 | 67478.53[b] | 0.02 | 67436.07 | 0.02 |
| 16 | 67515.94 | 0.02 | 67467.96 | 0.02 | 67422.68 | 0.02 |
| 17 | 67507.49 | 0.02 | 67456.73 | 0.02 | 67408.63 | 0.02 |
| 18 | 67498.31 | 0.02 | 67444.82 | 0.02 | 67393.91 | 0.02 |
| 19 | 67487.93 | 0.02 | 67432.25[b] | 0.02 | 67378.51 | 0.02 |
| 20 | 67478.38[b] | 0.02 | 67418.98 | 0.02 | 67362.39 | 0.02 |
| 21 | 67467.01 | 0.02 | 67405.19 | 0.02 | 67345.08 | 0.02 |
| 22 | 67454.85 | 0.02 | 67390.48 | 0.02 | 67328.57 | 0.02 |
| 23 | 67440.92 | 0.02 | 67375.13 | 0.02 | 67310.29 | 0.02 |
| 24 | 67432.3[bw] | 0.2 | 67359.04 | 0.02 | 67291.21 | 0.02 |
| 25 | 67417.31 | 0.02 | 67342.06 | 0.02 | 67270.37 | 0.02 |
| 26 | 67402.43 | 0.02 | 67323.82 | 0.02 | 67254.82 | 0.02 |
| 27 | 67386.97 | 0.03 | 67302.20[b] | 0.04 | 67232.93[b] | 0.06 |
| 28 | | | 67292.98[b] | 0.04 | 67211.18 | 0.05 |
| 29 | | | 67272.18[b] | 0.05 | 67188.8[bw] | 0.2 |
| 30 | | | 67251.93[b] | 0.09 | 67165.61 | 0.05 |
| 31 | | | | | 67143.48[b] | 0.04 |
| 32 | 67300.75[b] | 0.04 | | | 67120.64[b] | 0.04 |
| 33 | 67285.10[b] | 0.04 | | | 67093.93[b] | 0.03 |
| 34 | 67264.07[b] | 0.04 | 67166.44[b] | 0.03 | 67068.22[b] | 0.03 |
| 35 | 67243.00[b] | 0.03 | 67142.17[b] | 0.03 | 67045.76 | 0.03 |
| 36 | 67219.77 | 0.03 | 67117.08 | 0.03 | 67017.88 | 0.03 |
| 37 | 67200.95[b] | 0.03 | 67091.66[b] | 0.03 | 66989.99 | 0.03 |
| 38 | 67177.97 | 0.03 | 67069.34[b] | 0.03 | 66959.93 | 0.03 |
| 39 | 67154.53[b] | 0.03 | 67042.95 | 0.03 | 66934.34[b] | 0.04 |
| 40 | 67130.62 | 0.03 | 67016.14[b] | 0.04 | 66904.52[b] | 0.04 |
| 41 | 67105.97[b] | 0.06 | 66988.77[b] | 0.05 | 66874.33[b] | 0.06 |
| 42 | 67080.59 | 0.03 | 66960.71 | 0.04 | | |
| 43 | 67054.6[w] | 0.2 | | | | |

[a] The u($\sigma$) values indicate one standard deviation, representing combinations of the calibration and fitting errors of the wavenumber measurements. The instrumental resolution was 0.15 cm$^{-1}$, absolute calibration uncertainty was 0.01 cm$^{-1}$ and the absolute accuracy of the line frequency measurements were estimated to be 0.03 cm$^{-1}$ for the medium-strong and isolated lines. The superscripts $p$, $q$, and $r$ denote change in the total angular momentum excluding spin. Blended and/or weak lines (SNR < 3) are indicated by the superscripts "b" and/or "w," respectively.

**Alt Text for Table S4:** The table reports *J*-dependent wavenumbers with uncertainties for assigned rotational branches of the $^{13}C^{18}O$ $A^1\Pi \leftarrow X^1\Sigma^+(2, 0)$ VUV-FT absorption band. Superscripts mark branch type, blended lines, and weak spectral features.

**TABLE S5.** Wavenumbers (in $cm^{-1}$) of the $a^3\Pi \leftarrow X^1\Sigma^+(13, 0)$, VUV-FT absorption band of $^{13}C^{18}O$.[a]

| | $^qR_{11ee}$ | | $^pQ_{11fe}$ | | $^qQ_{21fe}$ | | $^rQ_{31fe}$ | | $^oP_{11ee}$ | | $^pP_{21ee}$ | | $^qP_{31ee}$ | |
|---|---|---|---|---|---|---|---|---|---|---|---|---|---|---|
| $J''$ | $\sigma$ | u($\sigma$) | $\sigma$ | u($\sigma$) | $\sigma$ | u($\sigma$) | $\sigma$ | u($\sigma$) | $\sigma$ | u($\sigma$) | $\sigma$ | u($\sigma$) | $\sigma$ | u($\sigma$) |
| 19 | 67490.29 | 0.02 | | | | | | | | | | | | |
| … | | | | | | | | | | | | | | |
| 21 | | | | | | | | | 67347.42 | 0.02 | | | | |
| … | | | | | | | | | | | | | | |
| 23 | | | | | | | | | 67296.47[b] | 0.04 | | | | |
| 24 | | | 67335.15[b] | 0.05 | | | | | | | | | | |
| … | | | | | | | | | | | | | | |
| 27 | 67337.30[b] | 0.04 | 67259.45[b] | 0.07 | | | | | | | | | | |
| 28 | | | | | | | | | 67172.87[b] | 0.03 | | | | |
| 29 | | | | | | | | | 67139.13 | 0.03 | | | | |
| 30 | | | | | | 0.03 | | | 67105.82[b] | 0.09 | | | | |
| 31 | | | | | | | | | | | 67151.05 | 0.03 | | |
| 32 | 67200.59[b] | 0.03 | 67110.18 | 0.04 | 67208.13 | 0.03 | 67297.15[b] | 0.03 | | | | | | |
| 33 | | | | | 67178.46[b] | 0.03 | 67268.5[bw] | 0.2 | | | | | | |
| 34 | | | | | | | 67248.82[b] | 0.03 | | | | | | |
| … | | | | | | | | | | | | | | |
| 43 | 66834.20 | 0.03 | | | | | | | | | | | | |
| 44 | | | | | | | | | | | | | 66803.81[b] | 0.04 |
| … | | | | | | | | | | | | | | |
| 46 | 66716.15 | 0.03 | | | | | | | | | | | | |
| 47 | | | | | | | 66803.90[b] | 0.06 | | | | | | |

[a] The u($\sigma$) values indicate one standard deviation, representing combinations of the calibration and fitting errors of the wavenumber measurements. The instrumental resolution was 0.15 $cm^{-1}$. The absolute calibration uncertainty was 0.025 $cm^{-1}$. The absolute accuracy of the wavenumber measurements was estimated to be 0.04 $cm^{-1}$. Blended and/or weak lines (SNR < 3) are indicated by the superscripts "b" and/or "w," respectively. The superscripts *o, p, q, r,* and s denote change in the total angular momentum excluding spin.

**Alt Text for Table S5:** The table lists *J*-dependent wavenumbers with uncertainties for assigned rotational branches of the $^{13}C^{18}O$ $a^3\Pi \leftarrow X^1\Sigma^+(13, 0)$ VUV-FT absorption band. Line annotations identify branch type, blending, and weak signal-to-noise features.

**TABLE S6.** Wavenumbers (in $cm^{-1}$) of the $a'^3\Sigma^+ \leftarrow X^1\Sigma^+$ (12, 0) and $I^1\Sigma^- \leftarrow X^1\Sigma^+(3, 0)$, VUV-FT absorption band of $^{13}C^{18}O$.[a]

| | $a'^3\Sigma^+ \leftarrow X^1\Sigma^+$ (12, 0) | | | | | | | | $I^1\Sigma^- \leftarrow X^1\Sigma^+(3, 0)$ | |
|---|---|---|---|---|---|---|---|---|---|---|
| | $^{r}R_{21ee}(J'')$ | | $^{p}Q_{11fe}(J'')$ | | $^{r}Q_{31fe}(J'')$ | | $^{p}P_{21ee}(J'')$ | | $^{q}Q_{11fe}(J'')$ | |
| $J''$ | $\sigma$ | u($\sigma$) | $\sigma$ | u($\sigma$) | $\sigma$ | u($\sigma$) | $\sigma$ | u($\sigma$) | $\sigma$ | u($\sigma$) |
| 0 | 67992.59[b] | 0.05 | | | | | | | | |
| 1 | 67993.28 | 0.04 | | | 67994.9[bw] | 0.3 | | | | |
| 2 | 67992.48[b] | 0.05 | 67982.94[b] | 0.04 | 67993.98[b] | 0.06 | 67982.17[b] | 0.05 | | |
| 3 | 67990.35 | 0.03 | 67976.68[b] | 0.07 | 67991.71[b] | 0.08 | 67975.87[b] | 0.04 | | |
| 4 | 67986.76[b] | 0.04 | 67969.00[b] | 0.07 | 67988.11[b] | 0.06 | 67968.07[b] | 0.08 | | |
| 5 | 67981.82[b] | 0.05 | 67959.87[b] | 0.05 | 67983.01[b] | 0.07 | 67958.93[b] | 0.05 | | |
| 6 | 67975.19[b] | 0.06 | 67949.29[b] | 0.06 | 67976.53[b] | 0.08 | 67948.36[b] | 0.04 | | |
| 7 | 67967.48[b] | 0.07 | 67937.32 | 0.03 | 67968.74[b] | 0.06 | 67936.38[b] | 0.05 | 67541.79 | 0.02 |
| 8 | 67958.19[b] | 0.04 | 67923.90[b] | 0.04 | 67959.46[b] | 0.07 | | | 67531.24 | 0.02 |
| 9 | 67947.48[b] | 0.04 | 67909.11[b] | 0.04 | 67948.78[b] | 0.06 | 67908.13[b] | 0.06 | 67519.71[b] | 0.03 |
| 10 | 67935.31[b] | 0.05 | 67892.85[b] | 0.04 | 67936.6[b] | 0.1 | 67891.88[b] | 0.05 | | |
| 11 | 67921.87 | 0.03 | 67875.16[b] | 0.04 | 67923.11[b] | 0.08 | 67874.27[b] | 0.04 | | |
| 12 | 67906.92[b] | 0.04 | 67856.13[b] | 0.05 | 67908.12[b] | 0.06 | 67855.12[b] | 0.06 | | |
| 13 | | | 67835.61[b] | 0.04 | 67891.80[b] | 0.05 | 67834.62[b] | 0.05 | | |
| 14 | 67872.78 | 0.03 | 67813.67[b] | 0.04 | 67873.95[b] | 0.04 | 67812.72[b] | 0.03 | | |
| 15 | 67853.57 | 0.03 | | | 67854.84[b] | 0.05 | 67789.43[b] | 0.08 | | |
| 16 | 67832.95 | 0.03 | 67765.63[b] | 0.04 | 67834.24[b] | 0.06 | 67764.59[b] | 0.05 | | |
| 17 | 67810.91 | 0.03 | 67739.46[b] | 0.04 | 67812.15[b] | 0.07 | 67738.50[b] | 0.05 | | |
| 18 | 67787.46[b] | 0.04 | 67711.90[b] | 0.04 | 67788.73[b] | 0.09 | 67710.91 | 0.04 | 67357.06[b] | 0.05 |
| 19 | 67762.57[b] | 0.05 | 67682.94[b] | 0.04 | 67763.86[b] | 0.03 | | | 67332.48[b] | 0.04 |
| 20 | 67736.30[b] | 0.06 | | | 67737.54 | 0.05 | 67651.54[b] | 0.03 | 67306.61 | 0.03 |
| 21 | 67708.49[b] | 0.04 | | | 67709.81[b] | 0.08 | | | 67279.43 | 0.03 |
| 22 | 67679.39[b] | 0.06 | | | | | | | 67250.96[b] | 0.04 |
| 23 | | | | | 67650.10[b] | 0.06 | | | 67221.18[b] | 0.05 |
| 24 | | | | | | | | | 67190.12[b] | 0.07 |
| 25 | | | | | | | | | 67157.73 | 0.06 |
| … | | | | | | | | | | |
| 31 | | | 67223.80[b] | 0.03 | | | 67223.63[b] | 0.04 | | |
| 32 | | | 67176.99 | 0.03 | | | 67176.42[b] | 0.06 | | |
| 33 | | | 67128.24[b] | 0.08 | | | 67127.65 | 0.03 | | |
| 34 | | | 67078.17[b] | 0.04 | | | 67078.08[b] | 0.04 | | |
| 35 | | | | | | | | | | |
| 36 | | | 66973.22[b] | 0.03 | | | | | | |
| 37 | | | 66918.53 | 0.03 | 67072.54 | 0.04 | | | | |
| 38 | | | 66862.79[w] | 0.06 | | | 66861.89[b] | 0.04 | | |
| 39 | 66965.94 | 0.03 | | | | | | | | |
| 40 | | | | | | | | | | |
| 41 | 66855.13[b] | 0.03 | | | | | | | | |

[a] The u($\sigma$) values indicate one standard deviation, representing combinations of the calibration and fitting errors of the wavenumber measurements. The instrumental resolution was 0.15 $cm^{-1}$, absolute calibration uncertainty was 0.025 $cm^{-1}$ as well as absolute accuracy of the wavenumber measurements was estimated to be 0.05 and 0.04 $cm^{-1}$ for $a'^3\Sigma^+ \leftarrow X^1\Sigma^+(12, 0)$ and $I^1\Sigma^- \leftarrow X^1\Sigma^+(3, 0)$, respectively. Blended and/or weak lines (SNR < 3) are indicated by the superscripts "b" and/or "w," respectively. The superscripts *o, p, q, r,* and *s* denote change in the total angular momentum excluding spin.

**Alt Text for Table S6:** The table reports wavenumbers with uncertainties for assigned branches of the $^{13}C^{18}O$ $a'^3\Sigma^+ \leftarrow X^1\Sigma^+(12, 0)$ and $I^1\Sigma^- \leftarrow X^1\Sigma^+(3, 0)$ VUV-FT absorption bands. It includes annotations for branch type, blended lines, and weak transitions.

**TABLE S7.** Wavenumbers (in cm$^{-1}$) of the $D^1\Delta \leftarrow X^1\Sigma^+(3, 0)$, VUV-FT absorption band of $^{13}C^{18}O$.[a]

| | $^rR_{11ee}(J'')$ | | $^qQ_{11fe}(J'')$ | | $^pP_{11ee}(J'')$ | |
|---|---|---|---|---|---|---|
| $J''$ | $\sigma$ | u($\sigma$) | $\sigma$ | u($\sigma$) | $\sigma$ | u($\sigma$) |
| … | | | | | | |
| 5 | | | 68358.5[bw] | 0.3 | | |
| 6 | | | 68350.7[bw] | 0.5 | | |
| 7 | | | 68341.4[bw] | 0.2 | | |
| 8 | 68350.5[bw] | 0.2 | 68330.9[bw] | 0.3 | | |
| 9 | 68340.8[bw] | 0.2 | 68319.0[bw] | 0.2 | | |
| 10 | 68329.8[bw] | 0.3 | 68305.9[bw] | 0.2 | | |
| 11 | 68317.5[bw] | 0.2 | 68291.4[bw] | 0.2 | 68267.5[bw] | 0.4 |
| 12 | 68303.94[b] | 0.05 | 68275.6[bw] | 0.2 | 68249.5[bw] | 0.3 |
| 13 | 68289.0[bw] | 0.2 | 68258.53[w] | 0.07 | 68230.3[bw] | 0.2 |
| 14 | 68272.7[bw] | 0.2 | 68240.1[bw] | 0.2 | 68209.7[bw] | 0.4 |
| 15 | 68255.1[bw] | 0.2 | 68220.4[bw] | 0.2 | | |
| 16 | 68236.3[b] | 0.2 | 68199.4[bw] | 0.2 | 68164.61[b] | 0.05 |
| 17 | 68216.08 | 0.06 | 68177.07[b] | 0.09 | 68140.08 | 0.04 |
| 18 | 68194.5[bw] | 0.3 | 68153.28[b] | 0.09 | 68114.27[b] | 0.04 |
| 19 | 68171.69[b] | 0.07 | 68128.3[bw] | 0.2 | | |
| 20 | 68147.5[bw] | 0.3 | 68102.0[bw] | 0.2 | 68058.7[b] | 0.2 |
| 21 | 68122.1[bw] | 0.3 | 68074.4[bw] | 0.3 | 68028.97 | 0.04 |
| 22 | 68095.2[w] | 0.1 | 68045.5[bw] | 0.3 | 67997.7[bw] | 0.2 |
| 23 | | | 68015.2[bw] | 0.3 | | |
| … | | | | | | |
| 46 | | | | | 66856.25[b] | 0.05 |
| … | | | | | | |
| 50 | | | 66700.8[bw] | 0.2 | | |

[a] The u($\sigma$) values indicate one standard deviation, representing combinations of the calibration and fitting errors of the wavenumber measurements. The instrumental resolution was 0.15 cm$^{-1}$, absolute calibration uncertainty was 0.025 cm$^{-1}$ as well as the best absolute accuracy of the line frequency measurements were estimated to be 0.2 cm$^{-1}$ for strong single lines. The superscripts *p, q,* and *r* denote change in the total angular momentum excluding spin. Blended and/or weak lines (SNR < 3) are indicated by the superscripts "b" and/or "w," respectively.

**Alt Text for Table S7:** The table lists *J*-dependent wavenumbers with uncertainties for assigned rotational branches of the $^{13}C^{18}O$ $D^1\Delta \leftarrow X^1\Sigma^+(3, 0)$ VUV-FT absorption band. Superscripts identify branch type, blended features, and weak spectral lines.

**TABLE S8.** Wavenumbers (in cm$^{-1}$) of the d$^3\Delta \leftarrow$ X$^1\Sigma^+$ (7, 0), VUV-FT absorption band of $^{13}$C$^{18}$O.[a]

| | ${}^{q}R_{11ee}(J'')$ | | ${}^{r}R_{21ee}(J'')$ | | ${}^{s}R_{31ee}(J'')$ | | ${}^{p}Q_{11fe}(J'')$ | | ${}^{q}Q_{21fe}(J'')$ | | ${}^{r}Q_{31fe}(J'')$ | | ${}^{o}P_{11ee}(J'')$ | | ${}^{p}P_{21ee}(J'')$ | | ${}^{q}P_{31ee}(J'')$ | |
|---|---|---|---|---|---|---|---|---|---|---|---|---|---|---|---|---|---|---|
| $J''$ | $\sigma$ | u($\sigma$) | $\sigma$ | u($\sigma$) | $\sigma$ | u($\sigma$) | $\sigma$ | u($\sigma$) | $\sigma$ | u($\sigma$)) | $\sigma$ | u($\sigma$) | $\sigma$ | u($\sigma$) | $\sigma$ | u($\sigma$) | $\sigma$ | u($\sigma$) |
| 0 | | | | | 67987.90[b] | 0.04 | | | | | | | | | | | | |
| 1 | | | 67949.9[b] | 0.1 | 67988.92[b] | 0.04 | | | | | 67984.46[b] | 0.03 | | | | | | |
| 2 | | | 67949.37[b] | 0.06 | 67988.73[b] | 0.04 | | | 67942.91[b] | 0.06 | 67981.94[b] | 0.05 | | | | | 67977.44[b] | 0.03 |
| 3 | 67908.59[b] | 0.04 | 67947.58[b] | 0.04 | 67987.35[b] | 0.04 | | | 67938.92 | 0.03 | 67978.40[b] | 0.05 | | | 67932.47[b] | 0.08 | 67971.46[b] | 0.03 |
| 4 | 67904.80[b] | 0.05 | 67944.40[b] | 0.03 | 67984.90[b] | 0.04 | | | 67933.60[b] | 0.03 | 67973.54[b] | 0.04 | | | 67924.94[b] | 0.06 | 67964.40[b] | 0.03 |
| 5 | 67899.59[b] | 0.05 | 67939.91 | 0.03 | 67980.98[b] | 0.04 | | | 67926.99[b] | 0.04 | 67967.39[b] | 0.04 | | | 67916.13[b] | 0.04 | 67956.08[b] | 0.03 |
| 6 | 67892.90[b] | 0.06 | 67934.07[b] | 0.03 | 67976.02[b] | 0.04 | 67878.7[bw] | 0.3 | 67918.96[b] | 0.03 | 67960.13[b] | 0.04 | | | 67906.01[b] | 0.04 | 67946.50[b] | 0.03 |
| 7 | 67885.01[b] | 0.09 | 67926.88[b] | 0.03 | 67969.72[b] | 0.04 | 67868.52[w] | 0.07 | 67909.63[b] | 0.03 | 67951.52[b] | 0.04 | | | 67894.51 | 0.03 | 67935.74[b] | 0.03 |
| 8 | 67875.41[b] | 0.09 | 67918.35[b] | 0.03 | 67962.24[b] | 0.04 | 67856.89[b] | 0.03 | 67898.98[b] | 0.03 | 67941.82[b] | 0.04 | | | 67881.71[b] | 0.04 | 67923.56[b] | 0.03 |
| 9 | 67864.81 | 0.06 | 67908.53[b] | 0.04 | 67953.41[b] | 0.04 | 67844.00[b] | 0.09 | 67887.00[b] | 0.03 | 67930.73[b] | 0.04 | 67825.62[b] | 0.06 | 67867.56[b] | 0.03 | 67910.38[b] | 0.03 |
| 10 | 67852.20[b] | 0.07 | 67897.42 | 0.03 | 67943.36[b] | 0.04 | 67829.62[b] | 0.07 | 67873.62[b] | 0.03 | 67918.47[b] | 0.04 | 67809.10[b] | 0.07 | 67852.08[b] | 0.03 | 67895.86[b] | 0.03 |
| 11 | 67838.53[b] | 0.06 | 67884.92[b] | 0.03 | 67932.03[b] | 0.04 | 67813.79[b] | 0.04 | 67859.03[b] | 0.03 | 67904.94[b] | 0.04 | | | 67835.29[b] | 0.03 | 67880.14[b] | 0.03 |
| 12 | 67823.35 | 0.04 | 67871.10[b] | 0.03 | 67919.46[b] | 0.04 | 67796.64[b] | 0.03 | 67843.05 | 0.03 | | | 67771.92 | 0.05 | 67817.18[b] | 0.03 | 67863.10[b] | 0.03 |
| 13 | 67806.88[b] | 0.05 | 67855.90[b] | 0.03 | 67905.67[b] | 0.04 | 67777.98[b] | 0.04 | 67825.75[b] | 0.03 | 67874.17[b] | 0.04 | 67751.25[b] | 0.05 | 67797.71[b] | 0.03 | 67844.82[b] | 0.03 |
| 14 | 67788.93[b] | 0.05 | 67839.42[b] | 0.03 | | | | | 67807.11[b] | 0.03 | 67856.83[b] | 0.04 | 67729.13 | 0.04 | | | 67825.25[b] | 0.03 |
| 15 | 67769.65[b] | 0.04 | 67821.57[b] | 0.03 | 67874.04 | 0.04 | 67736.60[b] | 0.04 | 67787.10[b] | 0.03 | 67838.07[b] | 0.03 | 67705.65[b] | 0.07 | 67754.77[b] | 0.03 | 67804.45[b] | 0.04 |
| 16 | 67748.96[b] | 0.05 | 67802.35[b] | 0.03 | 67856.36[b] | 0.03 | 67713.82[b] | 0.04 | 67765.82[b] | 0.03 | 67818.32[b] | 0.03 | 67680.76[b] | 0.05 | 67731.30[b] | 0.03 | 67782.32[b] | 0.03 |
| 17 | 67726.87 | 0.03 | 67781.89[b] | 0.03 | 67837.38[b] | 0.03 | 67689.70 | 0.03 | 67743.13 | 0.03 | 67797.12[b] | 0.04 | 67654.57[b] | 0.06 | 67706.50 | 0.03 | 67759.00 | 0.03 |
| 18 | | | 67760.05 | 0.03 | 67817.04[b] | 0.04 | 67664.19[b] | 0.04 | | | 67774.62[b] | 0.03 | | | 67680.32[b] | 0.03 | 67734.35[b] | 0.03 |
| 19 | 67678.87[b] | 0.03 | 67736.87[b] | 0.03 | 67795.53[b] | 0.04 | 67637.41[b] | 0.03 | 67693.82[b] | 0.03 | 67750.84[b] | 0.03 | | | | | 67708.38[b] | 0.04 |
| 20 | | | 67712.34[b] | 0.03 | 67772.60[b] | 0.03 | | | | | 67725.80[b] | 0.03 | | | | | 67681.18[b] | 0.04 |
| 21 | | | | | 67748.40 | 0.03 | | | 67639.20[b] | 0.04 | 67699.44[b] | 0.03 | | | | | | |
| 22 | | | 67659.32[b] | 0.03 | 67722.89[b] | 0.03 | | | | | 67671.79[b] | 0.03 | | | | | | |
| 23 | | | 67630.75[b] | 0.06 | 67696.08[b] | 0.03 | | | | | 67642.80[b] | 0.04 | | | | | | |
| 24 | | | | | 67667.97[b] | 0.03 | | | | | | | | | | | | |
| 25 | | | | | 67638.52 | 0.03 | | | | | | | | | | | | |
| … | | | | | | | | | | | | | | | | | | |
| 28 | | | | | | | | | | | | | 67274.95[b] | 0.04 | | | | |
| 29 | | | | | | | 67297.51[b] | 0.04 | 67367.09[b] | 0.06 | | | | | 67305.35[b] | 0.09 | | |
| 30 | 67314.38 | 0.03 | | | | | 67248.70[b] | 0.05 | | | | | | | 67262.83 | 0.03 | | |
| 31 | | | | | | | | | | | 67364.17 | 0.05 | 67138.49[b] | 0.03 | 67219.81[b] | 0.03 | | |
| 32 | | | | | | | | | | | | | 67095.56 | 0.04 | 67175.12[b] | 0.05 | 67253.39[b] | 0.06 |
| 33 | | | | | | | | | | | | | | | 67129.34 | 0.03 | | |
| 34 | | | | | | | | | | | | | | | 67082.91 | 0.03 | | |
| 35 | 67094.52[b] | 0.05 | | | | | 67015.38 | 0.03 | | | | | | | 67033.37[b] | 0.03 | | |

**TABLE S8.** – Continuation

| | $^{q}R_{11ee}(J'')$ | | $^{r}R_{21ee}(J'')$ | | $^{s}R_{31ee}(J'')$ | | $^{p}Q_{11fe}(J'')$ | | $^{q}Q_{21fe}(J'')$ | | $^{r}Q_{31fe}(J'')$ | | $^{o}P_{11ee}(J'')$ | | $^{p}P_{21ee}(J'')$ | | $^{q}P_{31ee}(J'')$ | |
|---|---|---|---|---|---|---|---|---|---|---|---|---|---|---|---|---|---|---|
| $J''$ | $\sigma$ | u($\sigma$) | $\sigma$ | u($\sigma$) | $\sigma$ | u($\sigma$) | $\sigma$ | u($\sigma$) | $\sigma$ | u($\sigma$)) | $\sigma$ | u($\sigma$) | $\sigma$ | u($\sigma$) | $\sigma$ | u($\sigma$) | $\sigma$ | u($\sigma$) |
| 36 | | | | | | | | | 67058.82 | 0.04 | | | | | 66981.19[b] | 0.03 | | |
| 37 | | | | | | | | | | | | | | | | | | |
| 38 | | | | | | | | | 66959.39 | 0.05 | | | | | | | | |
| 39 | | | | | | | | | | | | | | | | | | |
| 40 | 66843.61 | 0.04 | | | | | | | | | | | | | | | | |
| 41 | | | | | | | | | | | | | | | | | 66803.29 | 0.05 |
| 42 | | | | | | | | | | | 66840.25 | 0.03 | | | | | 66748.72 | 0.03 |
| 43 | | | | | | | | | | | 66784.80[b] | 0.03 | | | | | | |
| 44 | | | | | | | | | | | | | | | | | | |
| 45 | 66557.75 | 0.03 | | | | | | | | | | | | | | | | |
| ... | | | | | | | | | | | | | | | | | | |
| 49 | | | | | | | | | | | | | | | | | 66318.53 | 0.04 |

[a] The u($\sigma$) values indicate one standard deviation, representing combinations of the calibration and fitting errors of the wavenumber measurements. The instrumental resolution was 0.15 cm$^{-1}$. The absolute calibration uncertainty was 0.025 cm$^{-1}$. The absolute accuracy of the wavenumber measurements was estimated to be 0.04 cm$^{-1}$. Blended and/or weak lines (SNR < 3) are indicated by the superscripts "b" and/or "w," respectively. The superscripts *o, p, q, r,* and s denote change in the total angular momentum excluding spin.

**Alt Text for Table S8:** The table reports *J*-dependent wavenumbers with uncertainties for assigned rotational branches of the $^{13}C^{18}O$ $d^3\Delta \leftarrow X^1\Sigma^+(7, 0)$ VUV-FT absorption band. Line annotations indicate branch type, blending, and weak signal-to-noise features.

**TABLE S9.** Wavenumbers (in $cm^{-1}$) of the $e^3\Sigma^- \leftarrow X^1\Sigma^+(4, 0)$, VUV-FT absorption band of $^{13}C^{18}O$.[a]

| | $^{q}R_{11ee}$ | | $^{s}R_{31ee}$ | | $^{q}Q_{21fe}$ | | $^{o}P_{11ee}$ | | $^{q}P_{31ee}$ | |
|---|---|---|---|---|---|---|---|---|---|---|
| $J''$ | $\sigma$ | u($\sigma$) | $\sigma$ | u($\sigma$) | $\sigma$ | u($\sigma$) | $\sigma$ | u($\sigma$) | $\sigma$ | u($\sigma$) |
| 0 | 67803.89[b] | 0.03 | 67810.21[b] | 0.04 | | | | | | |
| 1 | 67802.57[b] | 0.03 | 67813.33[b] | 0.03 | | | | | | |
| 2 | 67799.96[b] | 0.03 | | | | | 67793.38 | 0.03 | 67799.70[b] | 0.04 |
| 3 | 67796.09[b] | 0.03 | | | | | | | 67795.86[b] | 0.04 |
| 4 | | | | | | | 67775.51[b] | 0.03 | | |
| 5 | | | 67812.70[b] | 0.04 | | | 67764.69[b] | 0.03 | | |
| 6 | | | 67809.38[b] | 0.03 | | | 67752.45[b] | 0.04 | | |
| 7 | | | 67804.67[b] | 0.03 | | | 67739.11[b] | 0.03 | | |
| 8 | | | 67798.64[b] | 0.03 | | | 67724.37[b] | 0.03 | | |
| 9 | | | | | | | 67708.25[b] | 0.07 | 67745.24[b] | 0.03 |
| 10 | | | 67782.83[b] | 0.03 | | | 67691.05[b] | 0.03 | 67732.33[b] | 0.03 |
| 11 | | | 67772.84[b] | 0.04 | | | 67672.46[b] | 0.03 | 67718.04[b] | 0.04 |
| 12 | | | 67761.72[b] | 0.03 | | | | | 67702.53[b] | 0.03 |
| 13 | | | 67749.30[b] | 0.03 | | | 67631.47[b] | 0.05 | 67685.60[b] | 0.04 |
| 14 | | | 67735.46[b] | 0.03 | | | | | | |
| 15 | | | 67720.42[b] | 0.03 | | | | | 67648.09[b] | 0.03 |
| 16 | | | 67704.12[b] | 0.04 | | | | | | |
| … | | | | | | | | | | |
| 19 | | | 67647.56[b] | 0.03 | | | | | | |
| … | | | | | | | | | | |
| 23 | 67452.69 | 0.03 | | | | | 67350.26[b] | 0.04 | | |
| 24 | | | | | | | 67316.07[b] | 0.04 | | |
| 25 | | | | | | | 67282.2[b] | 0.1 | | |
| 26 | 67335.34[b] | 0.08 | | | | | | | | |
| 27 | 67303.58[b] | 0.05 | | | | | | | | |
| 28 | | | | | | | | | 67278.39 | 0.03 |
| 29 | | | | | 67236.34[b] | 0.03 | | | 67241.04[b] | 0.03 |
| 30 | | | | | 67198.57 | 0.03 | | | 67203.51[b] | 0.03 |
| 31 | | | 67295.13[b] | 0.03 | 67159.75 | 0.03 | | | 67166.20[b] | 0.03 |
| 32 | | | 67259.15[b] | 0.04 | | | | | 67109.46 | 0.03 |
| 33 | | | 67221.28[b] | 0.08 | 67070.68 | 0.04 | | | 67069.50[b] | 0.04 |
| 34 | | | 67181.81[b] | 0.03 | | | | | 67026.64[b] | 0.07 |
| … | | | | | | | | | | |
| 37 | | | | | | | | | 66887.93[b] | 0.03 |
| … | | | | | | | | | | |
| 39 | | | | | 66788.53[b] | 0.05 | | | 66788.45[b] | 0.06 |

[a] The u($\sigma$) values indicate one standard deviation, representing combinations of the calibration and fitting errors of the wavenumber measurements. The instrumental resolution was 0.15 $cm^{-1}$. The absolute calibration uncertainty was 0.025 $cm^{-1}$. The absolute accuracy of the wavenumber measurements was estimated to be 0.04 $cm^{-1}$. Blended and/or weak lines (SNR < 3) are indicated by the superscripts "b" and/or "w," respectively. The superscripts *o*, *q*, and *s* denote change in the total angular momentum excluding spin.

**Alt Text for Table S9:** The table lists *J*-dependent wavenumbers with uncertainties for assigned rotational branches of the $^{13}C^{18}O$ $e^3\Sigma^- \leftarrow X^1\Sigma^+(4, 0)$ VUV-FT absorption band. Superscripts mark branch type, blended features, and weak spectral lines.

Based on the above-described complementary experimental Vis-FT and VUV-FT methods, a high quality analysis of the $B^1\Sigma^+ - A^1\Pi$, $C^1\Sigma^+ - A^1\Pi$, and $A^1\Pi - X^1\Sigma^+$, $a^3\Pi - X^1\Sigma^+$, $e^3\Sigma^- - X^1\Sigma^+$, $a'^3\Sigma^+ - X^1\Sigma^+$, $d^3\Delta - X^1\Sigma^+$, $D^1\Delta - X^1\Sigma^+$, and $I^1\Sigma^- - X^1\Sigma^+$ spectra has been obtained and the $^{13}C^{18}O$ $A^1\Pi(v = 2)$ vibrational level has been fully deperturbed.

## II. EXPERIMENTAL TERM VALUES

**TABLE S10.** Term values (in $cm^{-1}$) of the $A^1\Pi(v = 2)$ level in $^{13}C^{18}O$.[a,b,c]

| $J$ | $A^1\Pi(v = 2)$ | | | | | | | |
|---|---|---|---|---|---|---|---|---|
| | $F_{1e}$ ($cm^{-1}$) | $^1\Pi$ (%) | $^3\Pi$ (%) | $^3\Sigma^-$ (%) | $F_{1f}$ ($cm^{-1}$) | $^1\Pi$ (%) | $^3\Pi$ (%) | $^3\Sigma^-$ (%) |
| 1 | 67560.427(6) | 99.9 | 4.4E-03 | 0.1 | 67560.426(6) | 99.9 | 4.4E-03 | 9.4E-02 |
| 2 | 67566.096(5) | 99.9 | 4.6E-03 | 0.1 | 67566.104(4) | 99.9 | 4.5E-03 | 9.5E-02 |
| 3 | 67574.601(5) | 99.9 | 4.8E-03 | 0.1 | 67574.597(4) | 99.9 | 4.6E-03 | 9.6E-02 |
| 4 | 67585.930(4) | 99.9 | 5.1E-03 | 0.1 | 67585.929(4) | 99.8 | 4.8E-03 | 9.8E-02 |
| 5 | 67600.109(3) | 99.9 | 5.5E-03 | 0.1 | 67600.093(4) | 99.7 | 5.0E-03 | 0.1 |
| 6 | 67617.105(3) | 99.9 | 6.0E-03 | 0.1 | 67617.078(4) | 99.4 | 5.4E-03 | 0.1 |
| 7 | 67636.938(3) | 99.9 | 6.8E-03 | 0.1 | 67636.835(4) | 95.9 | 5.8E-03 | 0.1 |
| 8 | 67659.599(3) | 99.9 | 7.8E-03 | 0.1 | 67659.744(4) | 94.9 | 6.3E-03 | 0.1 |
| 9 | 67685.092(3) | 99.8 | 9.2E-03 | 0.1 | 67685.153(4) | 99.2 | 7.1E-03 | 0.1 |
| 10 | 67713.412(3) | 99.8 | 1.1E-02 | 0.1 | 67713.454(4) | 99.6 | 8.0E-03 | 0.1 |
| 11 | 67744.552(3) | 99.8 | 1.4E-02 | 0.1 | 67744.591(3) | 99.7 | 9.3E-03 | 0.1 |
| 12 | 67778.518(3) | 99.8 | 1.8E-02 | 0.2 | 67778.554(4) | 99.7 | 1.1E-02 | 0.1 |
| 13 | 67815.307(3) | 99.8 | 2.6E-02 | 0.2 | 67815.342(4) | 99.7 | 1.4E-02 | 0.2 |
| 14 | 67854.911(3) | 99.7 | 4.6E-02 | 0.2 | 67854.948(3) | 99.7 | 1.8E-02 | 0.2 |
| 15 | 67897.329(3) | 99.7 | 6.6E-02 | 0.2 | 67897.372(4) | 99.7 | 2.5E-02 | 0.2 |
| 16 | 67942.557(3) | 99.6 | 7.7E-02 | 0.3 | 67942.607(3) | 99.7 | 4.0E-02 | 0.2 |
| 17 | 67990.587(3) | 99.5 | 0.2 | 0.3 | 67990.653(3) | 99.6 | 7.7E-02 | 0.2 |
| 18 | 68041.407(3) | 99.2 | 0.4 | 0.4 | 68041.502(3) | 99.6 | 0.2 | 0.2 |
| 19 | 68094.988(3) | 98.0 | 1.5 | 0.5 | 68095.150(3) | 99.4 | 0.2 | 0.3 |
| 20 | 68150.832(3) | 69.7 | 29.6 | 0.7 | 68151.573(4) | 98.6 | 0.9 | 0.4 |
| 21 | 68210.955(3) | 95.2 | 3.8 | 1.0 | 68210.944(4) | 96.1 | 3.3 | 0.5 |
| 22 | 68272.771(3) | 97.0 | 1.2 | 1.8 | 68272.867(4) | 98.8 | 0.4 | 0.7 |
| 23 | 68337.235(3) | 94.8 | 0.8 | 4.4 | 68337.609(4) | 98.6 | 0.4 | 0.9 |
| 24 | 68403.392(3) | 77.3 | 0.6 | 22.0 | 68405.062(4) | 97.8 | 0.9 | 1.1 |
| 25 | 68478.299(4) | 84.9 | 0.6 | 14.4 | 68475.097(4) | 95.9 | 0.9 | 3.1 |
| 26 | 68550.322(3) | 96.3 | 0.6 | 2.8 | 68547.313(4) | 89.3 | 0.4 | 10.1 |
| 27 | 68625.915(3) | 97.8 | 1.1 | 0.9 | 68619.598(4) | 58.0 | 0.2 | 41.5 |
| 28 | 68704.370(4) | 97.3 | 1.2 | 1.3 | 68707.712(4) | 84.9 | 0.1 | 14.6 |
| 29 | 68785.345(3) | 93.3 | 4.5 | 1.9 | 68787.757(4) | 94.7 | 0.1 | 4.4 |
| 30 | 68870.988(6) | 53.9 | 10.3 | 3.2 | 68871.720(4) | 85.8 | 0.5 | 1.9 |
| 31 | 68959.205(4) | 75.5 | 6.3 | 16.6 | 68959.354(4) | 87.4 | 1.2 | 0.9 |
| 32 | 69046.983(5) | 52.5 | 41.4 | 3.8 | 69048.400(4) | 94.6 | 3.2 | 0.3 |
| 33 | 69139.264(6) | 80.9 | 0.9 | 1.5 | 69140.141(5) | 90.0 | 0.1 | 0.4 |
| 34 | 69238.203(5) | 87.5 | 0.3 | 0.8 | 69237.410(5) | 97.3 | 0.5 | 0.3 |
| 35 | 69335.130(5) | 97.0 | 0.7 | 0.5 | 69334.607(5) | 85.0 | 0.4 | 0.2 |
| 36 | 69435.453(4) | 96.9 | 0.2 | 0.3 | 69434.322(5) | 69.5 | 0.1 | 0.2 |
| 37 | 69537.019(7) | 71.0 | 8.9E-02 | 0.2 | 69537.143(8) | 96.0 | 4.3E-02 | 0.1 |
| 38 | 69646.420(5) | 96.2 | 6.4E-02 | 0.2 | 69646.440(5) | 98.9 | 1.1E-02 | 0.1 |
| 39 | 69755.060(6) | 99.0 | 9.2E-02 | 0.1 | 69755.075(7) | 99.5 | 5.6E-02 | 9.0E-02 |
| 40 | 69866.712(7) | 99.5 | 9.0E-02 | 0.1 | 69866.670(6) | 99.7 | 8.8E-02 | 7.5E-02 |
| 41 | 69981.108(8) | 99.7 | 2.1E-02 | 7.8E-02 | 69981.10(3) | 99.8 | 2.6E-02 | 6.4E-02 |
| 42 | 70098.27(2) | 99.8 | 1.0E-02 | 6.5E-02 | 70098.24(3) | 99.8 | 9.8E-03 | 5.2E-02 |
| 43 | 70218.16(3) | 99.8 | 9.8E-03 | 5.5E-02 | 70218.10(4) | 90.0 | 6.7E-03 | 4.5E-02 |
| 44 | 70340.7(2) | 99.9 | 1.3E-02 | 4.6E-02 | 70340.70(4) | 99.9 | 5.7E-03 | 3.9E-02 |
| 45 | | | | | 70465.89(4) | 99.9 | 8.0E-03 | 3.4E-02 |

[a] All term values are given in relation to $X^1\Sigma^+(v = 0, J = 0)$ level in $^{13}C^{18}O$.
[b] Numbers in parentheses are the random (fitting) uncertainties in units of the last-significant digit;
[c] "$^1\Pi$ (%)", "$^3\Pi$ (%)", and "$^3\Sigma^-$ (%)" denotes fractional $^{13}C^{18}O$ $A^1\Pi(v = 2)$, $a^3\Pi(v = 13)$, and $e^3\Sigma^-(v = 4)$ characters of the rovibronic levels involved in the perturbations of $A^1\Pi(v = 2)$, respectively.

**Alt Text for Table S10:** The table lists $J$-dependent term values of the $^{13}C^{18}O$ $A^1\Pi(v = 2)$ level, referenced to $X^1\Sigma^+(v = 0, J = 0)$. It includes fitting uncertainties and fractional $^1\Pi$, $^3\Pi$, and $^3\Sigma^-$ characters describing perturbation-induced mixing.

**TABLE S11.** Term values (in cm$^{-1}$) of the $d^3\Delta(v = 7)$ level in $^{13}C^{18}O$.[a,b,c]

| $J$ | d³Δ(v = 7) $F_{1e}$ (cm$^{-1}$) | $^1\Pi$ (%) | $^3\Pi$ (%) | $^3\Sigma^-$ (%) | $F_{1f}$ (cm$^{-1}$) | $^1\Pi$ (%) | $^3\Pi$ (%) | $^3\Sigma^-$ (%) |
|---|---|---|---|---|---|---|---|---|
| 4 | 67929.54(4) | 1.6E-09 | 5.5E-04 | 2.0E-09 | | | | |
| 5 | 67939.73(6) | 4.7E-09 | 9.7E-04 | 9.2E-09 | | | | |
| 6 | 67951.98(6) | 1.1E-08 | 1.5E-03 | 3.3E-08 | 67952.0(4) | 1.1E-08 | 1.5E-03 | 1.2E-08 |
| 7 | 67966.24(8) | 2.2E-08 | 2.2E-03 | 1.0E-07 | 67966.3(1) | 2.4E-08 | 2.2E-03 | 4.3E-08 |
| 8 | 67982.67(6) | 4.1E-08 | 3.1E-03 | 3.1E-07 | 67982.60(4) | 3.3E-08 | 3.1E-03 | 1.3E-07 |
| 9 | 68001.14(7) | 7.4E-08 | 4.2E-03 | 8.6E-07 | 68001.1(2) | 6.8E-08 | 4.2E-03 | 3.9E-07 |
| 10 | 68021.65(5) | 1.3E-07 | 5.7E-03 | 2.4E-06 | 68021.7(1) | 1.2E-07 | 5.7E-03 | 1.1E-06 |
| 11 | 68044.24(5) | 2.2E-07 | 7.5E-03 | 7.1E-06 | 68044.23(6) | 2.0E-07 | 7.5E-03 | 2.9E-06 |
| 12 | 68068.95(5) | 3.6E-07 | 9.9E-03 | 2.3E-05 | 68068.96(4) | 3.4E-07 | 1.0E-02 | 7.9E-06 |
| 13 | 68095.66(4) | 6.0E-07 | 1.3E-02 | 9.6E-05 | 68095.66(5) | 5.5E-07 | 1.3E-02 | 2.2E-05 |
| 14 | 68124.54(5) | 1.0E-06 | 1.7E-02 | 7.6E-04 | | | | |
| 15 | 68155.43(5) | 1.7E-06 | 1.9E-02 | 4.2E-03 | 68155.45(5) | 1.5E-06 | 2.3E-02 | 1.9E-04 |
| 16 | 68188.50(4) | 3.0E-06 | 3.1E-02 | 7.0E-04 | 68188.46(5) | 2.5E-06 | 3.1E-02 | 6.9E-04 |
| 17 | 68223.61(6) | 5.6E-06 | 4.4E-02 | 3.8E-04 | 68223.63(4) | 4.3E-06 | 4.1E-02 | 3.2E-03 |
| 18 | 68260.80(4) | 1.2E-05 | 6.3E-02 | 3.6E-04 | 68260.88(5) | 7.9E-06 | 4.5E-02 | 1.8E-02 |
| 19 | | | | | | | | |
| 20 | 68341.78(4) | 4.5E-04 | 0.2 | 2.9E-03 | | | | |
| … | | | | | | | | |
| 27 | 68689.71(5) | 6.1E-03 | 0.9 | 20.2 | | | | |
| … | | | | | | | | |
| 29 | | | | | 68813.06(5) | 6.5E-02 | 21.4 | 1.1E-03 |
| 30 | 68865.94(2) | 23.1 | 13.6 | 2.7 | 68868.49(2) | 4.6 | 12.1 | 2.5E-03 |
| 31 | 68934.16(3) | 0.2 | 3.7 | 4.2 | 68934.83(7) | 0.2 | 3.7 | 2.2E-02 |
| … | | | | | | | | |
| 35 | | | | | 69207.84(4) | 0.6 | 22.8 | 2.4E-02 |
| 36 | 69286.99(7) | 0.5 | 7.7 | 0.1 | | | | |
| … | | | | | | | | |
| 41 | 69694.16(5) | 1.7E-03 | 12.5 | 1.6E-03 | | | | |
| … | | | | | | | | |
| 43 | 69870.84(7) | 3.5E-04 | 0.8 | 3.7E-04 | | | | |
| … | | | | | | | | |
| 46 | 70151.22(4) | 9.2E-05 | 0.6 | 3.9E-03 | | | | |
| $J$ | $F_{2e}$ (cm$^{-1}$) | $^1\Pi$ (%) | $^3\Pi$ (%) | $^3\Sigma^-$ (%) | $F_{2f}$ (cm$^{-1}$) | $^1\Pi$ (%) | $^3\Pi$ (%) | $^3\Sigma^-$ (%) |
| 2 | 67953.41(8) | 1.1E-06 | 0.2 | 3.7E-07 | 67953.39(8) | 1.1E-06 | 0.2 | 3.6E-07 |
| 3 | 67959.86(6) | 2.9E-06 | 0.2 | 9.1E-07 | 67959.87(4) | 2.9E-06 | 0.2 | 8.5E-07 |
| 4 | 67968.53(4) | 5.4E-06 | 0.2 | 1.6E-06 | 67968.52(4) | 5.4E-06 | 0.2 | 1.4E-06 |
| 5 | 67979.33(3) | 8.7E-06 | 0.2 | 2.4E-06 | 67979.38(5) | 8.8E-06 | 0.2 | 1.9E-06 |
| 6 | 67992.29(3) | 1.3E-05 | 0.2 | 3.3E-06 | 67992.30(4) | 1.3E-05 | 0.2 | 2.2E-06 |
| 7 | 68007.42(3) | 1.9E-05 | 0.2 | 4.7E-06 | 68007.41(4) | 1.9E-05 | 0.2 | 2.2E-06 |
| 8 | 68024.68(3) | 2.6E-05 | 0.2 | 7.0E-06 | 68024.70(4) | 2.2E-05 | 0.2 | 1.8E-06 |
| 9 | 68044.10(3) | 3.5E-05 | 0.2 | 1.2E-05 | 68044.13(4) | 3.3E-05 | 0.2 | 8.2E-07 |
| 10 | 68065.70(3) | 4.6E-05 | 0.2 | 2.7E-05 | 68065.67(4) | 4.4E-05 | 0.2 | 1.1E-08 |
| 11 | 68089.48(3) | 6.1E-05 | 0.2 | 7.2E-05 | 68089.47(4) | 5.9E-05 | 0.2 | 1.7E-06 |
| 12 | | | | | 68115.37(4) | 7.7E-05 | 0.2 | 1.3E-05 |
| 13 | 68143.41(4) | 1.1E-04 | 0.2 | 9.6E-04 | 68143.43(4) | 1.0E-04 | 0.2 | 5.9E-05 |
| 14 | 68173.60(3) | 1.4E-04 | 0.2 | 7.7E-03 | 68173.63(4) | 1.3E-04 | 0.3 | 2.1E-04 |
| 15 | 68205.94(3) | 1.9E-04 | 0.2 | 4.4E-02 | 68205.95(4) | 1.8E-04 | 0.3 | 7.6E-04 |
| 16 | 68240.42(3) | 2.7E-04 | 0.3 | 7.1E-03 | 68240.47(4) | 2.4E-04 | 0.3 | 2.9E-03 |
| 17 | 68277.00(3) | 3.9E-04 | 0.3 | 3.4E-03 | 68277.06(4) | 3.2E-04 | 0.3 | 1.3E-02 |
| 18 | 68315.82(4) | 6.1E-04 | 0.4 | 2.2E-03 | | | | |
| 19 | 68356.73(4) | 1.3E-03 | 0.5 | 1.5E-03 | 68356.73(4) | 7.2E-04 | 0.4 | 9.0E-02 |
| 20 | 68399.77(4) | 9.5E-03 | 0.5 | 4.7E-03 | | | | |
| 21 | 68444.94(4) | 3.1E-05 | 0.7 | 3.5E-02 | 68444.97(4) | 1.4E-03 | 0.7 | 5.2E-02 |
| … | | | | | | | | |
| 23 | 68541.71(4) | 8.2E-05 | 0.9 | 0.8 | | | | |
| 24 | 68593.23(8) | 6.1E-04 | 16.7 | 0.7 | | | | |
| … | | | | | | | | |
| 28 | 68820.9(2) | 7.05E-02 | 2.8 | 2.0 | | | | |
| 29 | 68882.62(4) | 0.4 | 9.4 | 0.9 | 68882.64(7) | 0.1 | 17.8 | 4.5E-07 |
| 30 | 68947.28(4) | 14.8 | 7.3 | 0.5 | | | | |
| 31 | 69013.71(6) | 0.1 | 1.9 | 1.6 | | | | |
| … | | | | | | | | |
| 34 | 69225.84(4) | 2.9 | 17.4 | 2.6E-02 | | | | |
| 35 | 69298.46(4) | 1.1 | 24.3 | 1.1E-02 | | | | |
| 36 | | | | | 69376.09(5) | 1.3 | 4.4 | 3.8E-02 |
| … | | | | | | | | |
| 38 | | | | | 69536.49(6) | 1.0 | 0.4 | 2.9E-02 |

**Table S11.** – Continuation

| $J$ | d³Δ(v = 7) | | | | | | | |
|---|---|---|---|---|---|---|---|---|
| | $F_{3e}$ (cm⁻¹) | ¹Π (%) | ³Π (%) | ³Σ⁻ (%) | $F_{3f}$ (cm⁻¹) | ¹Π (%) | ³Π (%) | ³Σ⁻ (%) |
| 1 | 67987.91(3) | 1.7E-02 | 9.1E-02 | 2.4E-03 | 67987.95(4) | 1.7E-02 | 9.1E-02 | 2.4E-03 |
| 2 | 67992.42(3) | 1.7E-02 | 9.1E-02 | 2.5E-03 | 67992.41(6) | 1.7E-02 | 9.1E-02 | 2.5E-03 |
| 3 | 67999.27(3) | 1.7E-02 | 9.2E-02 | 2.5E-03 | 67999.36(6) | 1.7E-02 | 9.2E-02 | 2.5E-03 |
| 4 | 68008.40(3) | 1.7E-02 | 9.3E-02 | 2.6E-03 | 68008.47(5) | 1.7E-02 | 9.3E-02 | 2.6E-03 |
| 5 | 68019.83(3) | 1.8E-02 | 9.4E-02 | 2.7E-03 | 68019.77(5) | 1.8E-02 | 9.4E-02 | 2.7E-03 |
| 6 | 68033.47(3) | 1.8E-02 | 9.6E-02 | 2.9E-03 | 68033.47(5) | 1.8E-02 | 9.6E-02 | 2.9E-03 |
| 7 | 68049.32(3) | 1.8E-02 | 9.8E-02 | 3.0E-03 | 68049.30(5) | 1.8E-02 | 9.8E-02 | 3.0E-03 |
| 8 | 68067.51(3) | 1.9E-02 | 0.1 | 3.3E-03 | 68067.54(6) | 1.8E-02 | 0.1 | 3.3E-03 |
| 9 | 68087.93(3) | 2.0E-02 | 0.1 | 3.5E-03 | 68087.87(6) | 1.9E-02 | 0.1 | 3.5E-03 |
| 10 | 68110.56(3) | 2.0E-02 | 0.1 | 3.9E-03 | 68110.52(5) | 2.0E-02 | 0.1 | 3.9E-03 |
| 11 | 68135.41(3) | 2.1E-02 | 0.1 | 4.4E-03 | 68135.38(5) | 2.1E-02 | 0.1 | 4.4E-03 |
| 12 | 68162.48(3) | 2.2E-02 | 0.1 | 5.2E-03 | | | | |
| 13 | 68191.77(3) | 2.3E-02 | 0.1 | 6.6E-03 | 68191.85(5) | 2.3E-02 | 0.1 | 5.9E-03 |
| 14 | 68223.33(3) | 2.5E-02 | 0.1 | 1.1E-02 | 68223.35(5) | 2.4E-02 | 0.1 | 7.1E-03 |
| 15 | 68256.97(4) | 2.6E-02 | 0.1 | 4.8E-03 | 68256.91(4) | 2.6E-02 | 0.1 | 9.2E-03 |
| 16 | 68292.92(3) | 2.9E-02 | 0.1 | 4.2E-03 | 68292.97(4) | 2.8E-02 | 0.1 | 1.3E-02 |
| 17 | 68331.02(3) | 3.1E-02 | 0.1 | 6.0E-03 | 68331.04(5) | 3.0E-02 | 0.1 | 2.1E-02 |
| 18 | 68371.30(3) | 3.6E-02 | 0.2 | 7.9E-03 | 68371.30(4) | 3.3E-02 | 0.1 | 4.1E-02 |
| 19 | 68413.75(4) | 4.4E-02 | 0.2 | 9.8E-03 | 68413.75(4) | 3.8E-02 | 0.2 | 4.0E-04 |
| 20 | 68458.44(5) | 8.3E-02 | 0.1 | 1.0E-02 | 68458.40(4) | 4.7E-02 | 0.2 | 1.4E-03 |
| 21 | 68505.20(4) | 2.2E-02 | 0.2 | 3.4E-02 | 68505.20(4) | 2.4E-02 | 0.2 | 5.5E-03 |
| 22 | 68554.17(4) | 3.1E-02 | 0.2 | 3.9E-02 | 68554.18(4) | 4.1E-02 | 0.2 | 1.2E-02 |
| 23 | 68605.28(4) | 3.4E-02 | 0.2 | 3.1E-02 | 68605.28(4) | 5.0E-02 | 0.2 | 2.6E-02 |
| 24 | 68658.56(4) | 1.8E-02 | 1.3 | 8.3E-02 | | | | |
| 25 | 68713.99(4) | 0.1 | 1.1 | 9.2E-02 | | | | |
| 26 | 68771.55(4) | 0.1 | 1.0 | 0.5 | | | | |
| ... | | | | | | | | |
| 31 | 69091.97(8) | 0.4 | 1.5 | 0.1 | | | | |
| ... | | | | | | | | |
| 38 | 69625.51(2) | 2.4 | 2.3 | 9.1E-02 | | | | |
| ... | | | | | | | | |
| 40 | 69795.65(6) | 0.3 | 21.0 | 1.1E-02 | | | | |
| 41 | 69886.29(4) | 0.2 | 14.9 | 3.0E-03 | | | | |
| 42 | | | | | 69977.82(4) | 0.1 | 3.5 | 1.6E-02 |
| 43 | | | | | 70070.96(4) | 8.3E-02 | 1.1 | 2.0E-02 |
| ... | | | | | | | | |
| 48 | 70566.98(4) | 2.5E-02 | 1.5 | 0.6 | | | | |

[a] All term values are given in relation to X¹Σ⁺($v$ = 0, $J$ = 0) level in ¹³C¹⁸O.
[b] Numbers in parentheses are the random (fitting) uncertainties in units of the last-significant digit.
[c] "¹Π (%)", "³Π (%)", and "³Σ⁻ (%)" denotes fractional ¹³C¹⁸O A¹Π($v$ = 2), a³Π($v$ = 13), and e³Σ⁻($v$ = 4) characters of the rovibronic levels involved in the perturbations of d³Δ($v$ = 7), respectively.

**Alt Text for Table S11:** The table reports $J$-dependent term values of the ¹³C¹⁸O d³Δ($v$ = 7) level, referenced to X¹Σ⁺($v$ = 0, $J$ = 0). It includes fitting uncertainties and fractional electronic-state characters relevant to perturbations of A¹Π($v$ = 2).

**TABLE S12.** Term values (in cm$^{-1}$) of the a′ $^3\Sigma^+(v = 12)$ levels in $^{13}C^{18}O$.[a,b,c]

| $J$ | a′$^3\Sigma^+(v = 12)$ | | | | | | | | | | | |
|---|---|---|---|---|---|---|---|---|---|---|---|---|
| | $F_{1f}$ (cm$^{-1}$) | $^1\Pi$ (%) | $^3\Pi$ (%) | $^3\Sigma^-$ (%) | $F_{2e}$ (cm$^{-1}$) | $^1\Pi$ (%) | $^3\Pi$ (%) | $^3\Sigma^-$ (%) | $F_{3f}$ (cm$^{-1}$) | $^1\Pi$ (%) | $^3\Pi$ (%) | $^3\Sigma^-$ (%) |
| 1 | | | | | 67992.62(4) | 6.2E-03 | 8.9E-04 | 9.3E-05 | 67998.4(4) | 3.1E-07 | 1.5E-03 | 1.3E-08 |
| 2 | 67993.41(5) | 6.3E-03 | 9.0E-04 | 9.4E-05 | 67996.80(3) | 6.3E-03 | 9.0E-04 | 9.4E-05 | 68004.46(7) | 9.5E-07 | 1.5E-03 | 4.1E-08 |
| 3 | 67997.63(8) | 6.3E-03 | 9.1E-04 | 9.6E-05 | 68002.97(5) | 6.3E-03 | 9.1E-04 | 9.6E-05 | 68012.7(1) | 2.0E-06 | 1.5E-03 | 8.3E-08 |
| 4 | 68003.93(8) | 6.4E-03 | 9.2E-04 | 9.9E-05 | 68011.31(4) | 6.4E-03 | 9.3E-04 | 9.9E-05 | 68023.03(7) | 3.4E-06 | 1.5E-03 | 1.4E-07 |
| 5 | 68012.26(6) | 6.5E-03 | 9.4E-04 | 1.0E-04 | 68021.69(4) | 6.5E-03 | 9.4E-04 | 1.0E-04 | 68035.41(9) | 5.5E-06 | 1.5E-03 | 2.2E-07 |
| 6 | 68022.64(7) | 6.7E-03 | 9.6E-04 | 1.1E-04 | 68034.19(4) | 6.7E-03 | 9.7E-04 | 1.1E-04 | 68049.87(9) | 8.5E-06 | 1.6E-03 | 3.2E-07 |
| 7 | 68035.10(4) | 6.6E-03 | 9.8E-04 | 1.1E-04 | 68048.5(2) | 6.9E-03 | 9.9E-04 | 1.1E-04 | 68066.52(7) | 1.6E-05 | 1.6E-03 | 4.4E-07 |
| 8 | 68049.61(5) | 6.7E-03 | 1.0E-03 | 1.2E-04 | 68065.26(5) | 7.1E-03 | 1.0E-03 | 1.2E-04 | 68085.17(8) | 6.3E-06 | 1.6E-03 | 5.9E-07 |
| 9 | 68066.24(5) | 7.3E-03 | 1.0E-03 | 1.3E-04 | 68083.91(4) | 7.3E-03 | 1.1E-03 | 1.3E-04 | 68105.91(7) | 1.5E-05 | 1.7E-03 | 7.8E-07 |
| 10 | 68084.87(5) | 7.6E-03 | 1.1E-03 | 1.4E-04 | 68104.67(4) | 7.6E-03 | 1.1E-03 | 1.5E-04 | 68128.6(2) | 2.1E-05 | 1.8E-03 | 1.0E-06 |
| 11 | 68105.60(4) | 8.0E-03 | 1.1E-03 | 1.6E-04 | 68127.39(4) | 7.9E-03 | 1.2E-03 | 1.6E-04 | 68153.5(1) | 2.9E-05 | 1.8E-03 | 1.3E-06 |
| 12 | 68128.45(6) | 8.4E-03 | 1.2E-03 | 1.8E-04 | 68152.30(6) | 8.3E-03 | 1.2E-03 | 1.9E-04 | 68180.48(5) | 3.8E-05 | 1.9E-03 | 1.6E-06 |
| 13 | 68153.29(4) | 8.9E-03 | 1.2E-03 | 2.0E-04 | 68179.24(3) | 8.8E-03 | 1.3E-03 | 2.2E-04 | 68209.47(6) | 5.1E-05 | 2.0E-03 | 2.0E-06 |
| 14 | 68180.20(5) | 9.4E-03 | 1.3E-03 | 2.4E-04 | 68208.26(9) | 9.3E-03 | 1.3E-03 | 3.1E-04 | 68240.46(5) | 6.7E-05 | 2.1E-03 | 2.4E-06 |
| 15 | | | | | 68239.28(3) | 1.0E-02 | 1.6E-03 | 1.6E-04 | 68273.68(6) | 8.8E-05 | 2.3E-03 | 2.9E-06 |
| 16 | 68240.28(5) | 1.1E-02 | 1.4E-03 | 3.7E-04 | 68272.42(3) | 1.1E-02 | 1.7E-03 | 2.3E-04 | 68308.88(7) | 1.2E-04 | 2.4E-03 | 3.2E-06 |
| 17 | 68273.39(4) | 1.2E-02 | 1.4E-03 | 5.2E-04 | 68307.60(3) | 1.2E-02 | 1.8E-03 | 3.1E-04 | 68346.07(9) | 1.6E-04 | 2.6E-03 | 3.0E-06 |
| 18 | 68308.58(5) | 1.3E-02 | 1.3E-03 | 7.6E-04 | 68344.84(4) | 1.3E-02 | 1.9E-03 | 3.9E-04 | 68385.4(2) | 2.2E-04 | 2.8E-03 | 1.1E-06 |
| 19 | 68345.85(5) | 1.5E-02 | 2.1E-03 | 7.3E-05 | 68384.14(3) | 1.4E-02 | 2.1E-03 | 5.0E-04 | 68426.76(4) | 3.2E-04 | 3.0E-03 | 1.4E-05 |
| 20 | | | | | 68425.48(5) | 1.3E-02 | 5.5E-03 | 5.6E-04 | 68470.14(6) | 5.5E-04 | 1.5E-03 | 1.5E-05 |
| 21 | | | | | 68468.90(7) | 1.6E-02 | 3.8E-03 | 1.5E-03 | 68515.6(1) | 1.5E-04 | 1.5E-03 | 1.5E-05 |
| 22 | | | | | 68514.25(4) | 1.9E-02 | 3.3E-03 | 2.3E-03 | | | | |
| 23 | | | | | 68561.78(7) | 2.1E-02 | 5.4E-03 | 2.1E-03 | 68612.58(7) | 9.3E-04 | 5.1E-03 | 4.4E-06 |
| … | | | | | | | | | | | | |
| 30 | | | | | 68951.10(4) | 0.2 | 0.1 | 3.6E-02 | | | | |
| 31 | 68951.21(2) | 4.6 | 2.1E-02 | 1.0E-03 | 69015.00(7) | 0.3 | 0.1 | 7.0E-02 | | | | |
| 32 | 69015.57(4) | 0.1 | 5.8E-02 | 1.4E-02 | 69080.77(4) | 11.8 | 0.8 | 2.9E-02 | | | | |
| 33 | 69081.36(9) | 8.0E-03 | 0.4 | 1.0E-02 | 69149.17(5) | 2.6 | 0.1 | 1.7E-02 | | | | |
| 34 | 69149.26(4) | 1.2E-02 | 0.2 | 6.0E-03 | | | | | | | | |
| … | | | | | | | | | | | | |
| 36 | 69290.49(4) | 5.6 | 0.6 | 3.8E-03 | | | | | | | | |
| 37 | 69364.01(4) | 0.7 | 0.2 | 3.9E-03 | 69438.99(4) | 9.0E-02 | 3.0E-02 | 1.7E-02 | 69518.02(5) | 0.5 | 0.2 | 2.8E-03 |
| 38 | 69439.90(7) | 0.2 | 9.5 | 4.5E-03 | | | | | | | | |
| … | | | | | | | | | | | | |
| 40 | | | | | 69678.07(4) | 3.2E-02 | 0.2 | 0.1 | | | | |
| … | | | | | | | | | | | | |
| 42 | | | | | 69847.49(4) | 1.8E-02 | 1.0E-02 | 1.0E-02 | | | | |
| … | | | | | | | | | | | | |
| 49 | | | | | | | | | 70604.11(4) | 2.2E-04 | 3.5E-03 | 1.3E-04 |

[a] All term values are given in relation to $X^1\Sigma^+(v = 0, J = 0)$ level in $^{13}C^{18}O$.

[b] Numbers in parentheses are the random (fitting) uncertainties in units of the last-significant digit.

[c] "$^1\Pi$ (%)", "$^3\Pi$ (%)", and "$^3\Sigma^-$ (%)" denotes fractional $^{13}C^{18}O$ $A^1\Pi(v = 2)$, $a^3\Pi(v = 13)$, and $e^3\Sigma^-(v = 4)$ characters of the rovibronic levels involved in the perturbations of a′$^3\Sigma^+(v = 12)$, respectively.

**Alt Text for Table S12:** The table lists $J$-dependent term values of the $^{13}C^{18}O$ a′$^3\Sigma^+$ level, referenced to $X^1\Sigma^+(v = 0, J = 0)$. It gives fitting uncertainties and fractional $^1\Pi$, $^3\Pi$, and $^3\Sigma^-$ character contributions.

**TABLE S13.** Term values (in $cm^{-1}$) of the $e^3\Sigma^-(v = 4)$ levels in $^{13}C^{18}O$.[a,b,c]

| $J$ | $e^3\Sigma^-(v = 4)$ | | | | | | | | | | | |
|---|---|---|---|---|---|---|---|---|---|---|---|---|
| | $F_{1e}$ ($cm^{-1}$) | $^1\Pi$ (%) | $^3\Pi$ (%) | $^3\Sigma^-$ (%) | $F_{2f}$ ($cm^{-1}$) | $^1\Pi$ (%) | $^3\Pi$ (%) | $^3\Sigma^-$ (%) | $F_{3e}$ ($cm^{-1}$) | $^1\Pi$ (%) | $^3\Pi$ (%) | $^3\Sigma^-$ (%) |
| 1 | 67803.87(3) | 9.7E-02 | 0.6 | 99.3 | | | | | 67810.20(3) | 1.7E-05 | 0.7 | 99.3 |
| 2 | 67806.06(4) | 9.8E-02 | 0.6 | 99.3 | | | | | 67816.82(3) | 5.1E-05 | 0.7 | 99.3 |
| 3 | 67810.44(3) | 9.9E-02 | 0.6 | 99.3 | | | | | | | | |
| 4 | 67817.06(3) | 0.1 | 0.6 | 99.3 | | | | | | | | |
| 5 | 67825.79(4) | 0.1 | 0.7 | 99.2 | | | | | | | | |
| 6 | 67836.89(4) | 0.1 | 0.7 | 99.2 | | | | | 67865.09(4) | 4.4E-04 | 0.8 | 99.2 |
| 7 | 67850.08(4) | 0.1 | 0.7 | 99.1 | | | | | 67882.72(4) | 6.4E-04 | 0.8 | 99.2 |
| 8 | 67865.38(9) | 0.1 | 0.8 | 99.1 | | | | | 67902.41(3) | 9.1E-04 | 0.8 | 99.2 |
| 9 | 67883.10(4) | 0.1 | 0.9 | 99.0 | | | | | 67924.36(3) | 1.3E-03 | 0.9 | 99.1 |
| 10 | 67902.90(4) | 0.1 | 1.0 | 98.9 | | | | | 67948.48(4) | 1.8E-03 | 0.9 | 99.1 |
| 11 | | | | | | | | | 67974.86(3) | 2.5E-03 | 1.0 | 99.0 |
| 12 | 67949.15(6) | 0.2 | 1.5 | 98.4 | | | | | 68003.28(3) | 3.5E-03 | 1.1 | 98.9 |
| 13 | | | | | | | | | 68034.03(4) | 5.0E-03 | 1.4 | 98.6 |
| 14 | | | | | | | | | 68066.96(3) | 7.3E-03 | 3.1 | 96.9 |
| 15 | | | | | | | | | 68101.99(4) | 1.1E-02 | 14.3 | 85.7 |
| 16 | | | | | | | | | 68139.26(4) | 1.7E-02 | 4.1 | 95.9 |
| 17 | | | | | | | | | 68178.77(4) | 2.9E-02 | 3.4 | 96.6 |
| ... | | | | | | | | | | | | |
| 20 | | | | | | | | | 68310.46(4) | 1.2 | 9.1 | 89.7 |
| ... | | | | | | | | | | | | |
| 22 | 68312.74(5) | 1.0 | 12.9 | 86.1 | | | | | | | | |
| 23 | 68362.10(5) | 2.4 | 29.8 | 67.8 | | | | | | | | |
| 24 | 68415.159(8) | 10.3 | 15.5 | 72.4 | | | | | | | | |
| 25 | 68465.30(2) | 4.7 | 16.8 | 76.8 | | | | | | | | |
| 26 | | | | | 68570.81(2) | 9.9 | 6.9 | 53.4 | | | | |
| 27 | 68558.8(1) | 1.1E-02 | 24.1 | 60.0 | 68634.504(7) | 39.6 | 4.6 | 81.4 | 68693.15(4) | 1.8 | 19.1 | 62.3 |
| 28 | 68620.98(6) | 0.1 | 17.4 | 77.8 | 68686.80(2) | 13.7 | 4.2 | 91.0 | 68756.60(4) | 1.8 | 12.9 | 80.1 |
| 29 | | | | | 68751.90(4) | 4.2 | 7.1 | 90.9 | 68823.31(4) | 3.0 | 9.3 | 81.6 |
| 30 | | | | | 68818.37(4) | 1.2 | 16.7 | 82.4 | 68893.68(4) | 1.9 | 17.0 | 74.7 |
| 31 | | | | | 68887.22(4) | 0.8 | 49.8 | 49.4 | 68948.02(2) | 1.5 | 18.5 | 79.3 |
| 32 | | | | | | | | | 69022.61(3) | 0.9 | 8.3 | 90.7 |
| 33 | | | | | 69023.80(5) | 0.4 | 15.9 | 83.6 | 69097.73(4) | 0.5 | 4.2 | 95.1 |
| 34 | | | | | | | | | 69174.4(1) | 0.3 | 1.9 | 97.1 |
| 35 | | | | | | | | | 69252.89(4) | 0.1 | 1.4 | 98.2 |
| 36 | | | | | | | | | 69333.41(4) | 6.3E-02 | 0.9 | 98.8 |
| 37 | | | | | | | | | | | | |
| 38 | | | | | | | | | 69500.57(7) | 1.6E-02 | 1.7 | 98.2 |
| 39 | | | | | 69500.66(6) | 8.7E-02 | 0.6 | 99.2 | | | | |

[a] All term values are given in relation to $X^1\Sigma^+(v = 0, J = 0)$ level in $^{13}C^{18}O$.

[b] Numbers in parentheses are the random (fitting) uncertainties in units of the last-significant digit.

[c] "$^1\Pi$ (%)", "$^3\Pi$ (%)", and "$^3\Sigma^-$ (%)" denotes fractional $^{13}C^{18}O$ $A^1\Pi(v = 2)$, $a^3\Pi(v = 13)$, and $e^3\Sigma^-(v = 4)$ characters of the rovibronic levels involved in the perturbations of $e^3\Sigma^-(v = 4)$, respectively.

**Alt Text for Table S13:** The table reports $J$-dependent term values of the $^{13}C^{18}O$ $e^3\Sigma^-(v = 4)$ level, referenced to $X^1\Sigma^+(v = 0, J = 0)$. It includes fitting uncertainties and fractional characters involved in $A^1\Pi(v = 2)$ perturbations.

**TABLE S14.** Term values (in $cm^{-1}$) of the $D^1\Delta(v = 3)$ levels in $^{13}C^{18}O$.[a,b,c]

| *J* | $D^1\Delta(v = 3)$ | | | | | | | |
|---|---|---|---|---|---|---|---|---|
| | $F_{1e}$ ($cm^{-1}$) | $^1\Pi$ (%) | $^3\Pi$ (%) | $^3\Sigma^-$ (%) | $F_{1f}$ ($cm^{-1}$) | $^1\Pi$ (%) | $^3\Pi$ (%) | $^3\Sigma^-$ (%) |
| 5 | | | | | 68410.9(4) | 2.7E-06 | 3.9E-02 | 1.6E-06 |
| 6 | | | | | 68424.0(6) | 3.8E-06 | 4.0E-02 | 2.7E-06 |
| 7 | | | | | 68439.2(2) | 4.8E-06 | 4.0E-02 | 4.2E-06 |
| 8 | | | | | 68456.6(3) | 6.9E-06 | 4.0E-02 | 6.6E-06 |
| 9 | 68476.2(3) | 8.6E-06 | 4.1E-02 | 1.4E-05 | 68476.2(3) | 8.9E-06 | 4.1E-02 | 1.0E-05 |
| 10 | 68497.9(2) | 1.1E-05 | 4.1E-02 | 2.4E-05 | 68497.9(2) | 1.1E-05 | 4.1E-02 | 1.7E-05 |
| 11 | 68521.9(2) | 1.3E-05 | 4.2E-02 | 4.7E-05 | 68521.9(2) | 1.3E-05 | 4.2E-02 | 2.7E-05 |
| 12 | 68548.0(2) | 1.5E-05 | 4.3E-02 | 1.0E-04 | 68547.9(3) | 1.6E-05 | 4.3E-02 | 4.5E-05 |
| 13 | 68576.26(6) | 1.8E-05 | 4.3E-02 | 3.1E-04 | 68576.21(8) | 1.9E-05 | 4.3E-02 | 8.1E-05 |
| 14 | 68606.7(3) | 2.1E-05 | 4.3E-02 | 1.8E-03 | 68606.7(2) | 2.2E-05 | 4.4E-02 | 1.5E-04 |
| 15 | 68639.26(6) | 2.3E-05 | 3.8E-02 | 7.4E-03 | 68639.3(2) | 2.5E-05 | 4.5E-02 | 3.3E-04 |
| 16 | 68674.01(4) | 2.5E-05 | 4.6E-02 | 9.5E-04 | 68674.0(2) | 2.8E-05 | 4.6E-02 | 8.3E-04 |
| 17 | 68710.95(5) | 2.6E-05 | 4.7E-02 | 4.0E-04 | 68711.0(2) | 3.1E-05 | 4.5E-02 | 2.7E-03 |
| 18 | 68750.01(7) | 2.2E-05 | 4.9E-02 | 2.7E-04 | 68750.0(1) | 3.3E-05 | 3.8E-02 | 1.1E-02 |
| 19 | 68791.3(2) | 7.7E-06 | 5.0E-02 | 2.5E-04 | 68791.2(2) | 3.1E-05 | 4.1E-02 | 9.3E-03 |
| 20 | 68834.70(4) | 2.3E-04 | 5.1E-02 | 7.7E-04 | 68834.6(2) | 1.2E-05 | 4.7E-02 | 4.7E-03 |
| 21 | 68880.1(2) | 3.0E-04 | 5.1E-02 | 2.6E-03 | 68880.1(3) | 2.6E-04 | 5.0E-02 | 3.4E-03 |
| 22 | 68927.9(3) | 2.3E-04 | 5.0E-02 | 4.9E-03 | 68927.9(3) | 1.2E-04 | 5.2E-02 | 3.4E-03 |
| 23 | 68977.6(2) | 2.8E-04 | 3.6E-02 | 2.0E-02 | 68977.6(4) | 1.1E-04 | 5.1E-02 | 4.6E-03 |
| … | | | | | | | | |
| 45 | 70608.43(6) | 5.1E-03 | 0.5 | 1.1E-05 | | | | |
| … | | | | | | | | |
| 50 | | | | | 71121.3(3) | 1.1 | 15.6 | 6.7E-06 |

[a] All term values are given in relation to $X^1\Sigma^+(v = 0, J = 0)$ level in $^{13}C^{18}O$.
[b] Numbers in parentheses are the random (fitting) uncertainties in units of the last-significant digit.
[c] "$^1\Pi$ (%)", "$^3\Pi$ (%)", and "$^3\Sigma^-$ (%)" denotes fractional $^{13}C^{18}O$ $A^1\Pi(v = 2)$, $a^3\Pi(v = 13)$, and $e^3\Sigma^-(v = 4)$ characters of the rovibronic levels involved in the perturbations of $D^1\Delta(v = 3)$, respectively.

**Alt Text for Table S14:** The table lists *J*-dependent term values of the $^{13}C^{18}O$ $D^1\Delta(v = 3)$ level, referenced to $X^1\Sigma^+(v = 0, J = 0)$. It includes fitting uncertainties and fractional state-character information for levels interacting with $A^1\Pi(v = 2)$.

**TABLE S15.** Term values (in cm$^{-1}$) of the $a^3\Pi(v = 13)$ level in $^{13}C^{18}O$.[a,b,c]

| $J$ | $a^3\Pi(v = 13)$ | | | | | | | |
|---|---|---|---|---|---|---|---|---|
| | $F_{1e}$ (cm$^{-1}$) | $^1\Pi$ (%) | $^3\Pi$ (%) | $^3\Sigma^-$ (%) | $F_{1f}$ (cm$^{-1}$) | $^1\Pi$ (%) | $^3\Pi$ (%) | $^3\Sigma^-$ (%) |
| 20 | 68153.188(5) | 16.8 | 73.0 | 10.2 | | | | |
| 21 | | | | | | | | |
| 22 | 68258.95(5) | 1.2 | 93.1 | 5.7 | | | | |
| … | | | | | | | | |
| 24 | | | | | 68381.18(6) | 0.2 | 90.8 | 6.5 |
| … | | | | | | | | |
| 27 | 68587.63(4) | 4.3E-03 | 76.8 | 23.1 | 68576.8(1) | 2.1 | 93.0 | 3.7 |
| 28 | 68654.70(3) | 8.9E-02 | 83.0 | 16.3 | | | | |
| 29 | 68725.6(2) | 0.6 | 78.3 | 14.8 | | | | |
| … | | | | | | | | |
| 32 | | | | | 68948.76(5) | 0.9 | 68.4 | 5.0 |
| 33 | 69039.17(4) | 1.5E-02 | 91.2 | 3.5 | | | | |
| … | | | | | | | | |
| 44 | 70120.37(4) | 2.0E-03 | 99.1 | 0.4 | | | | |
| … | | | | | | | | |
| 47 | 70468.33(4) | 1.2E-03 | 89.1 | 0.3 | | | | |
| $J$ | $F_{2e}$ (cm$^{-1}$) | $^1\Pi$ (%) | $^3\Pi$ (%) | $^3\Sigma^-$ (%) | $F_{2f}$ (cm$^{-1}$) | $^1\Pi$ (%) | $^3\Pi$ (%) | $^3\Sigma^-$ (%) |
| 30 | 68878.52(4) | 9.9E-02 | 88.5 | 7.7 | 68885.56(5) | 0.4 | 97.2 | 0.9 |
| … | | | | | | | | |
| 32 | 69050.178(9) | 6.6 | 85.2 | 4.8 | 69046.72(4) | 0.3 | 78.3 | 17.7 |
| 33 | | | | | 69131.58(4) | 5.8E-02 | 78.6 | 8.6 |
| $J$ | $F_{3e}$ (cm$^{-1}$) | $^1\Pi$ (%) | $^3\Pi$ (%) | $^3\Sigma^-$ (%) | $F_{3f}$ (cm$^{-1}$) | $^1\Pi$ (%) | $^3\Pi$ (%) | $^3\Sigma^-$ (%) |
| 32 | | | | | 69135.74(4) | 1.0 | 89.5 | 4.1 |
| 33 | | | | | 69221.6(2) | 0.2 | 82.9 | 2.0 |
| 34 | | | | | 69319.90(4) | 0.5 | 75.3 | 0.8 |
| … | | | | | | | | |
| 43 | 70241.94(5) | 7.7E-03 | 98.5 | 2.7E-02 | | | | |
| … | | | | | | | | |
| 47 | | | | | 70718.15(8) | 1.2E-03 | 96.4 | 1.1E-02 |

[a] All term values are given in relation to $X^1\Sigma^+(v = 0, J = 0)$ level in $^{13}C^{18}O$.
[b] Numbers in parentheses are the random (fitting) uncertainties in units of the last-significant digit.
[c] "$^1\Pi$ (%)", "$^3\Pi$ (%)", and "$^3\Sigma^-$ (%)" denotes fractional $^{13}C^{18}O$ $A^1\Pi(v = 2)$, $a^3\Pi(v = 13)$, and $e^3\Sigma^-(v = 4)$ characters of the rovibronic levels involved in the perturbations of $a^3\Pi(v = 13)$, respectively.

**Alt Text for Table S15:** The table reports $J$-dependent term values of the $^{13}C^{18}O$ $a^3\Pi(v = 13)$ level, referenced to $X^1\Sigma^+(v = 0, J = 0)$. It includes fitting uncertainties and fractional $^1\Pi$, $^3\Pi$, and $^3\Sigma^-$ characters quantifying perturbation-induced mixing.

**TABLE S16.** Term values (in cm$^{-1}$) of the $I^1\Sigma^-(v = 3)$ levels in $^{13}C^{18}O$.[a,b,c]

| *J* | $I^1\Sigma^-$ ($v$ = 3) | | | |
|---|---|---|---|---|
| | $F_{1f}$ (cm$^{-1}$) | $^1\Pi$ (%) | $^3\Pi$ (%) | $^3\Sigma^-$ (%) |
| 7 | 67639.57(2) | 3.9 | 1.9 | 1.3E-05 |
| 8 | 67656.95(2) | 4.9 | 1.8 | 1.9E-05 |
| 9 | 67676.85(3) | 0.7 | 1.6 | 2.7E-05 |
| ... | | | | |
| 18 | 67953.74(6) | 2.5E-02 | 0.6 | 4.0E-03 |
| 19 | 67995.39(4) | 2.5E-02 | 0.6 | 8.2E-04 |
| 20 | 68039.21(4) | 3.2E-02 | 0.5 | 6.0E-05 |
| 21 | 68085.20(4) | 5.5E-05 | 0.5 | 4.4E-05 |
| 22 | 68133.35(5) | 6.1E-03 | 0.4 | 5.0E-04 |
| 23 | 68183.66(5) | 6.8E-03 | 0.4 | 2.3E-03 |
| 24 | 68236.15(9) | 6.1E-03 | 0.3 | 9.5E-03 |
| 25 | 68290.77(7) | 4.9E-03 | 0.2 | 4.1E-02 |

[a] All term values are given in relation to $X^1\Sigma^+(v = 0, J = 0)$ level in $^{13}C^{18}O$.

[b] Numbers in parentheses are the random (fitting) uncertainties in units of the last-significant digit

[c] "$^1\Pi$ (%)", "$^3\Pi$ (%)", and "$^3\Sigma^-$ (%)" denotes fractional $^{13}C^{18}O$ $A^1\Pi(v = 2)$, $a^3\Pi(v = 13)$, and $e^3\Sigma^-(v = 4)$ characters of the rovibronic levels involved in the perturbations of $I^1\Sigma^-(v = 3)$, respectively

**Alt Text for Table S16:** The table lists *J*-dependent term values of the $^{13}C^{18}O$ $I^1\Sigma^-(v = 3)$ level, referenced to $X^1\Sigma^+(v = 0, J = 0)$. It includes fitting uncertainties and fractional electronic-state characters relevant to indirect perturbation pathways involving $A^1\Pi(v = 2)$.

# III. DETAILS OF THE PRESENT DEPERTURBATION ANALYSIS

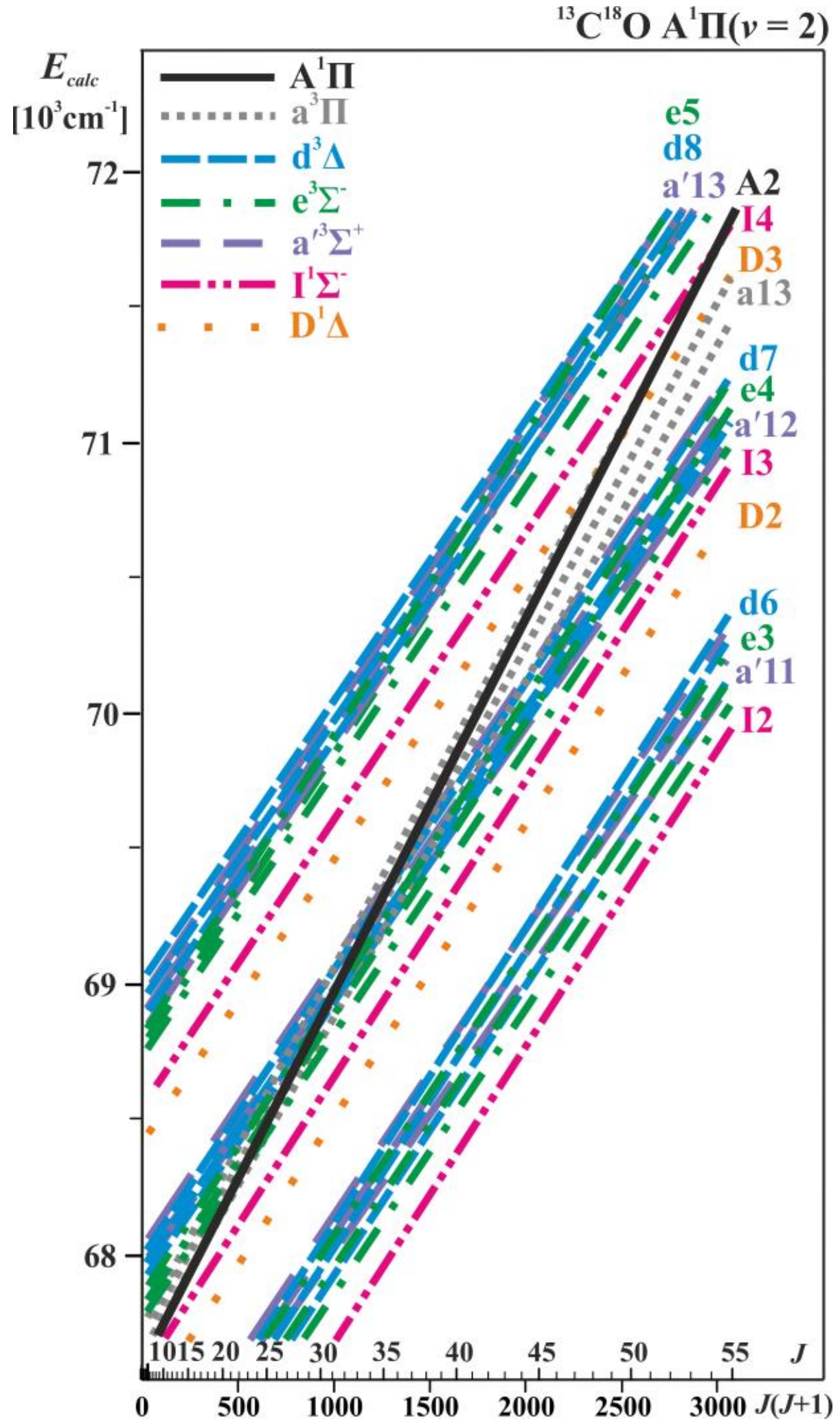

**FIG. S1.** Rovibronic term series in the neighborhood of the $^{13}C^{18}O\ A^1\Pi(v = 2)$ level in the region of 67,500 – 72,000 $cm^{-1}$. Symbols used at the upper end of each term series denote electronic state and vibrational quantum number of that electronic state. It is evident that transitions into every electronic state predicted to lie near $A^1\Pi(v = 2)$ are capable of appearing, via perturbation, as detectable features in the spectrum.

**Alt Text for Figure S1:** Rovibronic term series near the $^{13}C^{18}O\ A^1\Pi(v = 2)$ level are shown over the 67,500–72,000 $cm^{-1}$ region. The figure indicates that several neighboring electronic states can acquire sufficient $A^1\Pi(v = 2)$ character through perturbation to appear as detectable spectral features.

**TABLE S17.** Description of all examined interstate interactions within the deperturbation analysis of the $^{13}C^{18}O$ $A^1\Pi(v = 2)$ level.

| Nº | Interaction | | | Nature of the interaction | $\langle v_i \| v_j \rangle$[a] or $\langle v_i \| \mathbf{B}(\mathbf{R}) \| v_j \rangle$[b] | Taken into account [c] | Status | Comment[d] |
|---|---|---|---|---|---|---|---|---|
| 1 | $A^1\Pi(v = 2)$ | ~ | $e^3\Sigma^-(v = 3)$ | Spin-orbit | 0.282[a] | No | - | Undetected. |
| 2 | -\|\|- | ~ | $e^3\Sigma^-(v = 4)$ | -\|\|- | −0.304[a] | Yes | Fitted | - |
| 3 | -\|\|- | ~ | $e^3\Sigma^-(v = 5)$ | -\|\|- | 0.216[a] | No | - | Undetected. |
| 4 | -\|\|- | ~ | $d^3\Delta(v = 6)$ | -\|\|- | 0.050[a] | No | - | Undetected. |
| 5 | -\|\|- | ~ | $d^3\Delta(v = 7)$ | -\|\|- | −0.165[a] | Yes | Fitted | - |
| 6 | -\|\|- | ~ | $d^3\Delta(v = 8)$ | -\|\|- | 0.233[a] | No | - | Undetected. |
| 7 | -\|\|- | ~ | $a'^3\Sigma^+(v = 11)$ | -\|\|- | −0.200[a] | No | - | Undetected. |
| 8 | -\|\|- | ~ | $a'^3\Sigma^+(v = 12)$ | -\|\|- | 0.163[a] | Yes | Fitted | - |
| 9 | -\|\|- | ~ | $a'^3\Sigma^+(v = 13)$ | -\|\|- | −0.130[a] | No | - | Undetected. |
| 10 | -\|\|- | ~ | $I^1\Sigma^-(v = 2)$ | Rotation-electronic (**L**-uncoupling) | −0.014[b] | No | - | Undetected. |
| 11 | -\|\|- | ~ | $I^1\Sigma^-(v = 3)$ | -\|\|- | 0.303[b] | Yes | Fitted | - |
| 12 | -\|\|- | ~ | $I^1\Sigma^-(v = 4)$ | -\|\|- | −0.405[b] | Yes | Fixed | Detected. Statistically justified. The anti-crossing is outside the experimental range. No extra-lines originated from I(4). Thus, the fitted value is far from a predicted or reliable one. |
| 13 | -\|\|- | ~ | $I^1\Sigma^-(v = 5)$ | -\|\|- | 0.349[b] | No | - | Undetected. |
| 14 | -\|\|- | ~ | $D^1\Delta(v = 2)$ | -\|\|- | 0.067[b] | No | - | Undetected. |
| 15 | -\|\|- | ~ | $D^1\Delta(v = 3)$ | -\|\|- | 0.235[b] | Yes | Fixed | Detected. Fitting results in divergence. |
| 16 | -\|\|- | ~ | $D^1\Delta(v = 4)$ | -\|\|- | −0.381[b] | No | - | Undetected. |
| 17 | $a^3\Pi(v = 13)$ | ~ | $e^3\Sigma^-(v = 3)$ | Spin-orbit / spin-electronic | 0.228[a] | No | - | Undetected. |
| 18 | -\|\|- | ~ | -\|\|- | Rotation-electronic (**L**-uncoupling) | 0.197[b] | No | - | Undetected. |
| 19 | -\|\|- | ~ | $e^3\Sigma^-(v = 4)$ | Spin-orbit / spin-electronic | 0.395[a] | Yes | Fitted | - |
| 20 | -\|\|- | ~ | -\|\|- | Rotation-electronic (**L**-uncoupling) | 0.353[b] | Yes | Fitted | - |
| 21 | -\|\|- | ~ | $e^3\Sigma^-(v = 5)$ | Spin-orbit / spin-electronic | 0.294[a] | No | - | Undetected. |
| 22 | -\|\|- | ~ | -\|\|- | Rotation-electronic (**L**-uncoupling) | 0.271[b] | No | - | Undetected. |
| 23 | -\|\|- | ~ | $d^3\Delta(v = 7)$ | Spin-orbit / spin-electronic | −0.257[a] | Yes | Fitted | - |
| 24 | -\|\|- | ~ | -\|\|- | Rotation-electronic (**L**-uncoupling) | −0.253[b] | Yes | Fitted | - |
| 25 | -\|\|- | ~ | $a'^3\Sigma^+(v = 11)$ | Spin-orbit / spin-electronic | −0.167[a] | No | - | Undetected. |
| 26 | -\|\|- | ~ | -\|\|- | Rotation-electronic (**L**-uncoupling) | −0.187[b] | No | - | Undetected. |
| 27 | -\|\|- | ~ | $a'^3\Sigma^+(v = 12)$ | Spin-orbit / spin-electronic | −0.057[a] | Yes | Fitted | - |
| 28 | -\|\|- | ~ | -\|\|- | Rotation-electronic (**L**-uncoupling) | −0.065[b] | Yes | Fitted | - |
| 29 | -\|\|- | ~ | $a'^3\Sigma^+(v = 13)$ | Spin-orbit / spin-electronic | 0.196[a] | No | - | Undetected. |
| 30 | -\|\|- | ~ | -\|\|- | Rotation-electronic (**L**-uncoupling) | 0.232[b] | No | - | Undetected. |
| 31 | -\|\|- | ~ | $I^1\Sigma^-(v = 2)$ | Spin-orbit | 0.108[a] | No | - | Undetected. |
| 32 | -\|\|- | ~ | $I^1\Sigma^-(v = 3)$ | -\|\|- | 0.281[a] | Yes | Fixed | Detected. Fitting results in divergence. |
| 33 | -\|\|- | ~ | $I^1\Sigma^-(v = 4)$ | -\|\|- | 0.401[a] | Yes | Fixed | Detected. Fitting results in divergence. |
| 34 | -\|\|- | ~ | $D^1\Delta(v = 2)$ | -\|\|- | 0.143[a] | No | - | Undetected. |
| 35 | -\|\|- | ~ | $D^1\Delta(v = 3)$ | -\|\|- | 0.325[a] | Yes | Fixed | Detected. Fitting results in divergence. |

[a] The vibrational overlap integrals $\langle v_i|v_j\rangle$, calculated as described in the literature.[11,12]

[b] The rotational operator integrals $\langle v_i|\mathbf{B(R)}|v_j\rangle$, in cm$^{-1}$, calculated as described in the literature.[11,12]

[c] In the final fit.

[d] The significance of the parameters was tested as described in detail in the footnotes to Table 8 of our previous work.[7]

**Alt Text for Table S17:** The table summarizes all interstate interactions tested in the deperturbation analysis of the $^{13}C^{18}O$ $A^1\Pi(v = 2)$ level, including vibrational overlap and rotational-operator integrals, fitted parameters, and their statistical significance in the final model.

## III.1. THE INTRA-STATE $^3\Pi$ AND $^3\Sigma^-$ INTERACTIONS

The magnitudes of the intra-state a($^3\Pi_1 \sim {}^3\Pi_0$, $^3\Pi_1 \sim {}^3\Pi_2$) and e$^3\Sigma^-$($F_{1e} \sim F_{3e}$) **S**-uncoupling interactions, quantified by the degree of mutual mixing of the characters of the states participating in the indirect perturbations, are illustrated in Figures S2 and S3.

Figure S2 shows that both the a$^3\Pi$($v$ = 13, $F_{1e}$), $|\Omega|$ = 0 and 2 spin components borrow a substantial amount of a$^3\Pi$($v$ = 13, $F_2$, $|\Omega|$ = 1) character at $J$ = 20 (ca. 35%) and even more at higher $J$'s (up to 47%). Near $J$ = 19, all three components of the $^3\Pi$ state have similar $^3\Pi_1$ content, which is a consequence of the transition from Hund's case (a) towards Hund's case (b). As $J$ increases, the $^3\Pi_1$ component increasingly becomes a linear combination of $^3\Pi_2$ and $^3\Pi_0$, and the character of the nominal $^3\Pi_1$ state decreases further, falling below 10% above $J$ = 38. The **S**-uncoupling interactions among the $\Omega$-components of the a$^3\Pi$ state in CO molecule are well known[52]. Mosyagin *et al.*[53] report the following formula:

$$\langle {}^3\Pi_0|\mathbf{H}|\,{}^3\Pi_1\rangle = -B\sqrt{2X}, \tag{1}$$

$$\langle {}^3\Pi_1|\mathbf{H}|\,{}^3\Pi_2\rangle = -B\sqrt{2(X-2)}, \tag{2}$$

where $X = J'(J'+1)$ and $B(R) = \frac{\hbar^2}{2\mu R^2}$.

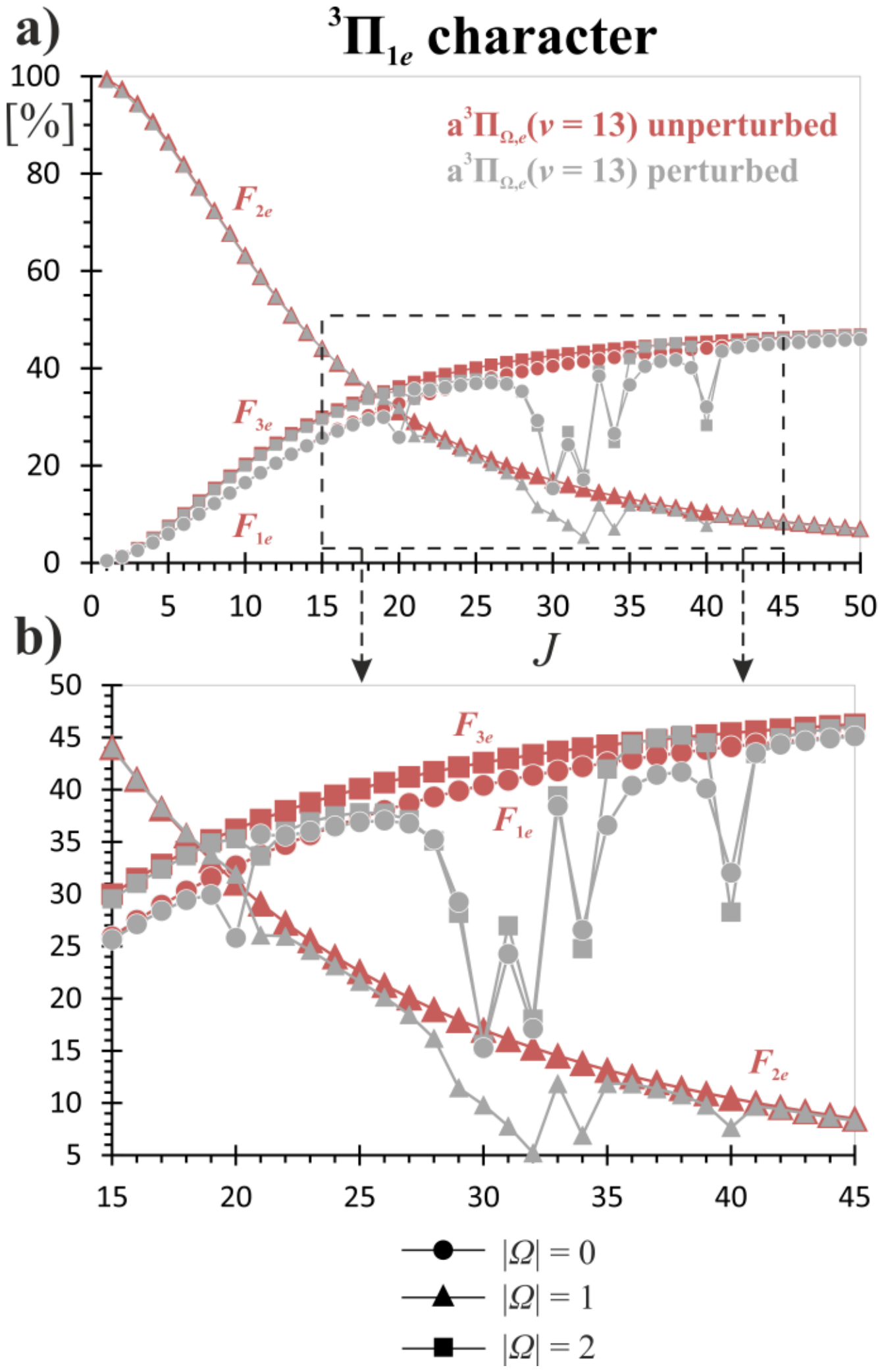

**FIG. S2.** (a) The fractional $^3\Pi_{1e}$ character of the three *e*-parity spin components of the $^{13}C^{18}O$ $a^3\Pi(v = 13)$ level. (b) An enlarged view of the most relevant region, *J* = 15-45, of the figure (a). The |*Ω*| designations are given for the hypothetical case of no interactions between the spin components of the $a^3\Pi$ state.

**Alt Text for Figure S2:** The fractional $^3\Pi_{1e}$ character is shown for the three *e*-parity spin components of the $^{13}C^{18}O$ $a^3\Pi(v = 13)$ level. For *J* > 2, all the components have $^3\Pi_{1e}$ character, which determines the strength of the indirect $a^3\Pi(v = 13) \sim A^1\Pi(v = 2)$ interaction and the interferences between indirect perturbation paths in $^{13}C^{18}O$.

In turn, in Figure S3 something even more important is revealed, namely, there is nearly 50:50 **S**-uncoupling mixing of |*Ω*| = 0 and 1 characters for the two *e*-symmetry components, $F_{1e}$ and $F_{3e}$, of the $e^3\Sigma^-$ state. It means that the two *e*-symmetry basis states of $e^3\Sigma^-$ are a mixture of 25% (*Ω* = +1), 25% (*Ω* = −1), and 50% (*Ω* = 0).

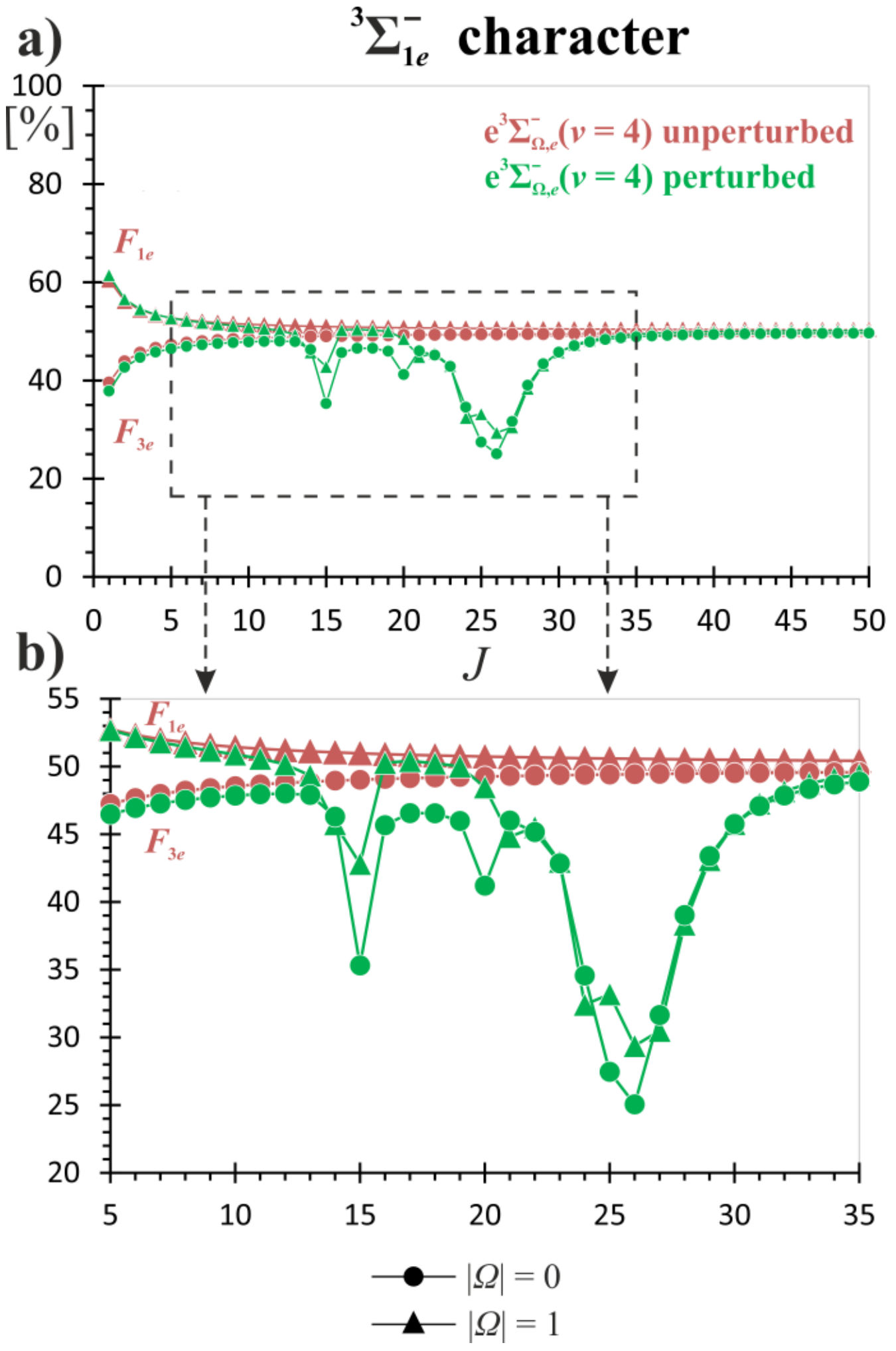

**FIG. S3.** (a) The fractional $^3\Sigma^-_{1e}$ character of the $e$-parity spin components $F_{1e}$ and $F_{3e}$ of the $^{13}C^{18}O$ $e^3\Sigma^-(v = 4)$ level. (b) An enlarged view of the most relevant region, $J$ = 5-35, of the figure (a). The $|\Omega|$ designations are only pictorial labels, since the $F_1$, $F_2$, and $F_3$ components of $e^3\Sigma^-$ are more properly Hund's-case-(b) spin-rotational components rather than pure-$\Omega$ components.

**Alt Text for Figure S3:** The fractional $^3\Sigma^-_{1e}$ character is shown for the $e$-parity spin-rotational components of $^{13}C^{18}O$ $e^3\Sigma^-(v = 4)$ . The strong $^3\Sigma^-_{1e}$ admixture in $F_{3e}$ determines the strength of the indirect $a^3\Pi(v = 13) \sim A^1\Pi(v = 2)$ interaction and the interferences between indirect perturbation paths in $^{13}C^{18}O$.

Figure S2 shows that all $F_{1e}$, $F_{2e}$ and $F_{3e}$ components of the $^3\Pi$ state for $J > 2$ possess $^3\Pi_{1e}$ character, while Figure S3 shows that the $e^3\Sigma^-(F_{3e})$ component exhibits a very strong admixture of $^3\Sigma^-_{1e}$ character over the entire range of rotational levels $J$. These features determine the strength of the indirect $a^3\Pi(v = 13) \sim A^1\Pi(v = 2)$ interaction and the interferences between indirect perturbation pathways in $^{13}C^{18}O$ (see Table II for details).

## III.2. METHODOLOGY FOR THE ANALYSIS OF INTERFERENCE BETWEEN INDIRECT PERTURBATION PATHWAYS

The presence and nature of the interferences between the multiple intermediate basis-state pathways (see Table II) linking the initial $A^1\Pi(v = 2)$ and the final $a^3\Pi(v = 13)$ basis states in $^{13}C^{18}O$ were investigated in detail by resolving the $e^3\Sigma^-(v = 4)$ and $a^3\Pi(v = 13)$ levels into their multiplet components ($F_1$, $F_2$, and $F_3$) and incorporating explicit forms representing the intercomponent interactions into the modified model. The standard rotational Hamiltonian can be expressed entirely in terms of diagonal perturbations, as shown by Western within the PGOPHER program.[16,17] The specific factors defined in it facilitate this task. The method is useful not only for generating an alternative representation of the molecular constants, thereby enabling complete interaction mechanism between the multiplet components of a given electronic state, but also for introducing additional terms, such as higher-order centrifugal distortion or lambda doubling terms.

The $e^3\Sigma^-(v = 4)$ level was partitioned into the $F_{1e}$, $F_{2f}$, and $F_{3e}$ components. Its effective rotational Hamiltonian

$$\mathbf{H}_{\mathrm{rot}} = T + BN^2 - DN^4 + HN^6 + \gamma\,\mathbf{N}\cdot\mathbf{S} + \tfrac{2}{3}\lambda(3S_z^2 - S^2), \tag{3}$$

with matrix elements listed in Table S18, was then reconstructed by introducing the corresponding interactions explicitly between the separated components. In the PGOPHER implementation, the $F_{1e} \sim F_{3e}$ intra-state interaction within $e^3\Sigma^-(v = 4)$ was represented through the perturbation terms

$$B\,\langle e^3\Sigma^-(v = 4, F_{1e})|N^2|e^3\Sigma^-(v = 4, F_{3e})\rangle, \tag{4}$$

$$-D\,\langle e^3\Sigma^-(v = 4, F_{1e})|N^4|e^3\Sigma^-(v = 4, F_{3e})\rangle, \tag{5}$$

$$H\,\langle e^3\Sigma^-(v = 4, F_{1e})|N^6|e^3\Sigma^-(v = 4, F_{3e})\rangle, \tag{6}$$

and

$$\gamma\,\langle e^3\Sigma^-(v = 4, F_{1e})|\mathbf{N}\cdot\mathbf{S}|e^3\Sigma^-(v = 4, F_{3e})\rangle. \tag{7}$$

In the final calculations, however, the spin-rotation interaction was set to zero, because there is no literature value of $\gamma$ for $e^3\Sigma^-$ available and attempts to determine it in the present deperturbation analysis led either to statistically insignificant values or to strong correlation with the molecular constants of $a^3\Pi(v = 13)$, in particular $A$ and $D$. In turn, the spin-spin term contributes only to the diagonal structure of the separated $e^3\Sigma^-(v = 4)$ components and does not contribute to the $F_{1e} \sim F_{3e}$ interaction. In addition, the inter-state interactions between $A^1\Pi(v = 2)$ and the separate $e^3\Sigma^-(v = 4)$ components were introduced as

$$\langle A^1\Pi(v = 2)|\mathbf{H}^{\mathrm{SO}}|e^3\Sigma^-(v = 4, F_{1e})\rangle, \tag{8}$$

$$\langle A^1\Pi(v = 2)|\mathbf{H}^{\mathrm{SO}}|e^3\Sigma^-(v = 4, F_{2f})\rangle, \tag{9}$$

and

$$\langle A^1\Pi(v = 2)|\mathbf{H}^{\mathrm{SO}}|e^3\Sigma^-(v = 4, F_{3e})\rangle. \tag{10}$$

**TABLE S18.** Matrix elements of the effective Hamiltonian for the $e^3\Sigma^-$ state.[a]

| $e^3\Sigma^-$ | $F_{1e}$ | $F_{2f}$ | $F_{3e}$ |
|---|---|---|---|
| $F_{1e}$ | $T_v + Bx - D(x^2 + 4x)$ $+ H(x^3 + 12x^2 + 8x)$ $+ \frac{2}{3}\lambda - \gamma$ | | $-2B\sqrt{x} + 4D(x + 1)\sqrt{x}$ $-2H(3x^2 + 10x + 4)\sqrt{x}$ $+\gamma\sqrt{x}$ |
| $F_{2f}$ | | $T_v + Bx - Dx^2 + Hx^3$ $+ \frac{2}{3}\lambda - \gamma$ | |
| $F_{3e}$ | | | $T_v + B(x + 2) - D(x^2 + 8x + 4)$ $+ H(x^3+18x^2+28x+8)$ $-\frac{4}{3}\lambda - 2\gamma$ |

[a] $x = J(J + 1)$; the matrix is symmetric, thus the left-sided, non-diagonal elements, not shown in the Hamiltonian, are equivalent to the right-sided, non-diagonal ones; $T_v$ – means rotation-less energies in relation to the $X^1\Sigma^+(v = 0)$ level; the basis functions have been abbreviated as $F_{1e}$, $F_{2f}$, $F_{3e}$.

**Alt Text for Table S18:** Table of matrix elements for the effective Hamiltonian of the $e^3\Sigma^-$ state, arranged for the $F_{1e}$, $F_{2f}$, and $F_{3e}$ components. Diagonal elements contain term energy and rotational contributions, while off-diagonal elements represent couplings between spin components.

A more explicit component-resolved construction was adopted for the $a^3\Pi(v = 13)$ level. Here the state was separated into the $F_{1e,f}$, $F_{2e,f}$, and $F_{3e,f}$ components, corresponding to $\Omega = 0$, 1, and 2, respectively. The effective Hamiltonian

$$T_v + BN^2 - DN^4 + HN^6 + \tfrac{2}{3}\lambda(3S_z^2 - S^2) + \gamma(\mathbf{N}\cdot\mathbf{S}) + A(L_zS_z) + \tfrac{1}{2}A_D(N^2L_zS_z + L_zS_zN^2) + o(S_+^2L_-^2 + S_-^2L_+^2) - p(N_+S_+L_-^2 + N_-S_-L_+^2) + q(N_+^2L_-^2 + N_-^2L_+^2), \quad (11)$$

with matrix elements gathered in Table S19, was reconstructed by introducing the intra-state interactions explicitly between the separated multiplet components. In this form, the $F_{1e,f} \sim F_{2e,f}$, $F_{2e,f} \sim F_{3e,f}$, and $F_{1e,f} \sim F_{3e,f}$ interactions were defined separately, rather than being embedded implicitly in the effective $a^3\Pi$ Hamiltonian. In PGOPHER notation, the $F_1 \sim F_2$ and $F_2 \sim F_3$ interactions were represented by the perturbation terms

$$B\,\langle a^3\Pi(v = 13, F_{1e,f})|N^2|a^3\Pi(v = 13, F_{2e,f})\rangle, \quad (12)$$

$$-D\,\langle a^3\Pi(v = 13, F_{1e,f})|N^4|a^3\Pi(v = 13, F_{2e,f})\rangle, \quad (13)$$

$$H\,\langle a^3\Pi(v = 13, F_{1e,f})|N^6|a^3\Pi(v = 13, F_{2e,f})\rangle, \quad (14)$$

$$\gamma\,\langle a^3\Pi(v = 13, F_{1e,f})|\mathbf{N}\cdot\mathbf{S}|a^3\Pi(v = 13, F_{2e,f})\rangle, \quad (15)$$

and

$$A_D\,\left\langle a^3\Pi(v = 13, F_{1e,f})\middle|\tfrac{1}{2}\,[N^2, L_zS_z]_+\middle|a^3\Pi(v = 13, F_{2e,f})\right\rangle, \quad (16)$$

together with the corresponding $p$- and $q$-type interactions,

$$p\,\langle a^3\Pi(v = 13, F_{1e,f})|N_+S_+L_-^2 + N_-S_-L_+^2|a^3\Pi(v = 13, F_{2e,f})\rangle, \quad (17)$$

and

$$q\,\langle a^3\Pi(v = 13, F_{1e,f})|N_+^2L_-^2 + N_-^2L_+^2|a^3\Pi(v = 13, F_{2e,f})\rangle, \quad (18)$$

and, analogously, between $F_2 \sim F_3$, except $p$- and $q$-type perturbations. In turn, the $F_1 \sim F_3$ interaction within the $a^3\Pi$ state was introduced through the terms responsible for the $\Delta\Omega = 2$ structure, namely

$$-D \left\langle a^3\Pi(v = 13, F_{1e,f}) \middle| N^4 \middle| a^3\Pi(v = 13, F_{3e,f}) \right\rangle, \quad (19)$$

$$q \left\langle a^3\Pi(v = 13, F_{1e,f}) \middle| N_+^2 L_-^2 + N_-^2 L_+^2 \middle| a^3\Pi(v = 13, F_{3e,f}) \right\rangle. \quad (20)$$

**TABLE S19.** Matrix elements of the effective Hamiltonian for the $a^3\Pi$ state.[a]

| $a^3\Pi$ | $F_{1e,f}$ | $F_{2e,f}$ | $F_{3e,f}$ |
|---|---|---|---|
| $F_{1e,f}$ | $T_v - A + (B - A_D)(x + 2) - D(x^2 + 6x + 4) + \frac{2}{3}\lambda - 2\gamma \pm (o + p + q)$ | $-\sqrt{2x}\,[B - 2D(x + 2) - \frac{1}{2}\gamma - \frac{1}{2}A_D \pm (\frac{1}{2}p + q)\,]$ | $-\sqrt{x(x - 2)}\,(2D \mp \frac{1}{2}q)$ |
| $F_{2e,f}$ | | $T_v + B(x + 2) - D(x^2 + 8x) - \frac{4}{3}\lambda - 2\gamma \pm \frac{1}{2}qx$ | $-\sqrt{2(x - 2)}[B - 2Dx - \frac{1}{2}\gamma + \frac{1}{2}A_D\,]$ |
| $F_{3e,f}$ | | | $T_v + A + (B + A_D)(x - 2) - D(x^2 - 4x) + \frac{2}{3}\lambda$ |

[a] $x = J(J + 1)$; the matrix is symmetric, thus the left-sided, non-diagonal elements, not shown in the Hamiltonian, are equivalent to the right-sided, non-diagonal ones; $T_v$ – means rotation-less energies in relation to the $X^1\Sigma^+(v = 0)$ level; top and bottom signs ($\pm$ or $\mp$) refer to the $e$ and $f$ components, respectively.

**Alt Text for Table S19:** Table of matrix elements for the effective Hamiltonian of the $a^3\Pi$ state, arranged for the $F_{1ef}$, $F_{2ef}$, and $F_{3ef}$ components. Diagonal elements contain term-energy and rotational contributions, while off-diagonal elements represent couplings between spin components, including parity-dependent terms.

These component-resolved representations are algebraically equivalent to the compact effective-state representations when all corresponding interactions are retained. Their advantage is methodological rather than formal, because each basis-state channel is represented explicitly, selected interactions can be switched on or off individually, allowing the role of specific intermediate pathways in the observed indirect perturbation and interference patterns to be tested directly.

## SUPPLEMENTARY MATERIAL (no 2)

## Perturbation of CO A1Π by a3Π should have been undetectable: detection, mechanism, and surprises

```
PGOPHER 11.0.142 1 Apr 2021 8:30 64 bit Unicode (Delphi 33/33)
   pgopher 44b8c50a90149a6ce53e669f4a0e44afc86f1abd
    lutils 65f951dfa553c7f9bc501632d3360a85cd1a0165
     utils ec0cac7863199e65355ae4c6ab78dbe3037fae3b
    lforms d4dfa0031586b0cf5ffd76ff4d8f7aee2a9f8932
    wutils 9bb7abdf4436dd4d7dfed61ce7fb7c6bc914386c
 GraphicEx bfb0df9494191c3a10d6089c21fe439fcd763090
Mixture: 13C18O Deperturbation of the A(v=2) level.pgo
  Units            cm1
  PrintLevel       LineList
  Precision        4
  BasisOrder       boDefault
  QuantumNumberFormat J
  FitCycles        1
  FitDetails       False
  AutoReplot       True
  AllowComplex     True
  NoLineList       False
  ShowEstUnc       False
  ShowDerivatives  False
  SVDlimit         5
  ShowSubBasis     False
  AllowZeroI       False
  ScaledParameters   True
  FixTransitionPhase  False
  OldHyperfineOrder   False
  OldColourOrder      False
  SingleResidualPlot  False
  UpperStateColour    False
  SVDThresh    1E-6          Discard Singular Value Decomposition eigenvalues below this fraction of the largest eigenvalue; <0 to use QR factorisation.
  ScaleChanges 1             Scale changes produced by fitting by this - normally 1
  SmallE      2E-18          Matrix elements smaller than this are discarded
  Species: CO
    Jmax 63
    Concentration 1          Relative concentration of species
    LinearMolecule: 13C18O
      JAdjustSym  True
      BlockMatrix True
      AbsoluteE   False
      Symmetric   False
      SymWt       1
      AsymWt      1
      RSquaredH   False
      ShowJ       True
      ShowOmega   False
      ShowN       True
      ShowFNumber True
      Showef      True
      Abundance     1          Abundance of this isotope; default is 1
      AssumedOrigin 1E99          Energy origin for Boltzmann calculations; 1e99 implies automatic estimate. Currently 0
      LinearManifold: X

        Initial         True
        EigenSearch     True
        LimitSearch     True
        AutoQConverge   True
        UsePopParams    False
        OverrideEnergies True
        Linear: X Xv=0
          Status    Active
          Lambda    Sigma+
          OmegaOrder Auto => Regular
          B  1.746408199          Rotational constant.
          D  5.0488146E-6          Quartic centrifugal distortion.
          H  4.35471E-12          Sextic centrifugal distortion.
          L  -2.4821E-17          J^8  centrifugal distortion.
          M  2.99E-23          J^10  centrifugal distortion.
          PP -4.5E-28          J^12 centrifugal distortion. (Not P to avoid confusion with p.)
          <J,Lambda=0,Omega=0|B|J,Lambda=0,Omega=0> = J*(J+1)
            Symmetrized:
          <J,Lambda=0,Omega=0|B|J,Lambda=0,Omega=0> = J*(J+1)

          <J,Lambda=0,Omega=0|D|J,Lambda=0,Omega=0> = -J^2*(J*(J+2)+1)
            Symmetrized:
          <J,Lambda=0,Omega=0|D|J,Lambda=0,Omega=0> = -J^2*(J*(J+2)+1)

          <J,Lambda=0,Omega=0|H|J,Lambda=0,Omega=0> = J^3*(J*(J*(J+3)+3)+1)
            Symmetrized:
          <J,Lambda=0,Omega=0|H|J,Lambda=0,Omega=0> = J^3*(J*(J*(J+3)+3)+1)

          <J,Lambda=0,Omega=0|L|J,Lambda=0,Omega=0> = J^4*(J*(J*(J*(J+4)+6)+4)+1)
```

Symmetrized:
<J,Lambda=0,Omega=0|L|J,Lambda=0,Omega=0> = J^4*(J*(J*(J*(J+4)+6)+4)+1)

<J,Lambda=0,Omega=0|M|J,Lambda=0,Omega=0> = J^5*(J*(J*(J*(J*(J+5)+10)+10)+5)+1)
Symmetrized:
<J,Lambda=0,Omega=0|M|J,Lambda=0,Omega=0> = J^5*(J*(J*(J*(J*(J+5)+10)+10)+5)+1)

<J,Lambda=0,Omega=0|PP|J,Lambda=0,Omega=0> = J^6*(J*(J*(J*(J*(J*(J+6)+15)+20)+15)+6)+1)
Symmetrized:
<J,Lambda=0,Omega=0|PP|J,Lambda=0,Omega=0> = J^6*(J*(J*(J*(J*(J*(J+6)+15)+20)+15)+6)+1)

LinearPerturbation: X <Xv=0|N^14|Xv=0>
O (rotational constant)
Value -2.2E-33 Size of perturbation.
<J,Lambda=0,Omega=0|N^14|J,Lambda=0,Omega=0> = J^7*(J*(J*(J*(J*(J*(J*(J+7)+21)+35)+35)+21)+7)+1)

Symmetrized Matrix Elements of <Xv=0|N^14|Xv=0>
<J,Lambda=0,Omega=0|N^14|J,Lambda=0,Omega=0> = J^7*(J*(J*(J*(J*(J*(J*(J+7)+21)+35)+35)+21)+7)+1)

LinearManifold: C
Initial True
EigenSearch True
LimitSearch False
AutoQConverge True
UsePopParams False
OverrideEnergies True
E 0 e 1 91918.81461769 Energy for J=E 0 e 1 level:C Cv=0 0 0 F1e
E 1 e 1 91922.3501979315 Energy for J=E 1 e 1 level:C Cv=0 1 1 F1e
E 2 e 1 91929.41460931 Energy for J=E 2 e 1 level:C Cv=0 2 2 F1e
E 3 e 1 91940.0087911758 Energy for J=E 3 e 1 level:C Cv=0 3 3 F1e
E 4 e 1 91954.1391754158 Energy for J=E 4 e 1 level:C Cv=0 4 4 F1e
E 5 e 1 91971.8047483651 Energy for J=E 5 e 1 level:C Cv=0 5 5 F1e
E 6 e 1 91992.9963415625 Energy for J=E 6 e 1 level:C Cv=0 6 6 F1e
E 7 e 1 92017.7210158785 Energy for J=E 7 e 1 level:C Cv=0 7 7 F1e
E 8 e 1 92045.9729147246 Energy for J=E 8 e 1 level:C Cv=0 8 8 F1e
E 9 e 1 92077.7548931101 Energy for J=E 9 e 1 level:C Cv=0 9 9 F1e
E 10 e 1 92113.0587701987 Energy for J=E 10 e 1 level:C Cv=0 10 10 F1e
E 11 e 1 92151.8966952636 Energy for J=E 11 e 1 level:C Cv=0 11 11 F1e
E 12 e 1 92194.2576663392 Energy for J=E 12 e 1 level:C Cv=0 12 12 F1e
E 13 e 1 92240.1328568783 Energy for J=E 13 e 1 level:C Cv=0 13 13 F1e
E 14 e 1 92289.5374084555 Energy for J=E 14 e 1 level:C Cv=0 14 14 F1e
E 15 e 1 92342.4615531966 Energy for J=E 15 e 1 level:C Cv=0 15 15 F1e
E 16 e 1 92398.9025159008 Energy for J=E 16 e 1 level:C Cv=0 16 16 F1e
E 17 e 1 92458.8587353746 Energy for J=E 17 e 1 level:C Cv=0 17 17 F1e
E 18 e 1 92522.3290731028 Energy for J=E 18 e 1 level:C Cv=0 18 18 F1e
E 19 e 1 92589.3126987891 Energy for J=E 19 e 1 level:C Cv=0 19 19 F1e
E 20 e 1 92659.8038563447 Energy for J=E 20 e 1 level:C Cv=0 20 20 F1e
E 21 e 1 92733.8006230299 Energy for J=E 21 e 1 level:C Cv=0 21 21 F1e
E 22 e 1 92811.3031429715 Energy for J=E 22 e 1 level:C Cv=0 22 22 F1e
E 23 e 1 92892.3079897717 Energy for J=E 23 e 1 level:C Cv=0 23 23 F1e
E 24 e 1 92976.822314948 Energy for J=E 24 e 1 level:C Cv=0 24 24 F1e
E 25 e 1 93064.8231465903 Energy for J=E 25 e 1 level:C Cv=0 25 25 F1e
E 26 e 1 93156.3176170672 Energy for J=E 26 e 1 level:C Cv=0 26 26 F1e
E 27 e 1 93251.2988822767 Energy for J=E 27 e 1 level:C Cv=0 27 27 F1e
E 28 e 1 93349.7667497678 Energy for J=E 28 e 1 level:C Cv=0 28 28 F1e
E 29 e 1 93451.7189577959 Energy for J=E 29 e 1 level:C Cv=0 29 29 F1e
E 30 e 1 93557.148271242 Energy for J=E 30 e 1 level:C Cv=0 30 30 F1e
E 31 e 1 93666.0549149757 Energy for J=E 31 e 1 level:C Cv=0 31 31 F1e
E 32 e 1 93778.4343213518 Energy for J=E 32 e 1 level:C Cv=0 32 32 F1e
E 33 e 1 93894.2815297818 Energy for J=E 33 e 1 level:C Cv=0 33 33 F1e
E 34 e 1 94013.591760117 Energy for J=E 34 e 1 level:C Cv=0 34 34 F1e
E 35 e 1 94136.3569127458 Energy for J=E 35 e 1 level:C Cv=0 35 35 F1e
E 36 e 1 94262.5809686937 Energy for J=E 36 e 1 level:C Cv=0 36 36 F1e
E 37 e 1 94392.253689726 Energy for J=E 37 e 1 level:C Cv=0 37 37 F1e
E 38 e 1 94525.3757684531 Energy for J=E 38 e 1 level:C Cv=0 38 38 F1e
E 39 e 1 94661.9366784378 Energy for J=E 39 e 1 level:C Cv=0 39 39 F1e
E 40 e 1 94801.9378743053 Energy for J=E 40 e 1 level:C Cv=0 40 40 F1e
E 41 e 1 94945.3716418562 Energy for J=E 41 e 1 level:C Cv=0 41 41 F1e
E 42 e 1 95092.2322481804 Energy for J=E 42 e 1 level:C Cv=0 42 42 F1e
E 43 e 1 95242.507641775 Energy for J=E 43 e 1 level:C Cv=0 43 43 F1e
E 44 e 1 95396.2059026634 Energy for J=E 44 e 1 level:C Cv=0 44 44 F1e
E 45 e 1 95553.3147425169 Energy for J=E 45 e 1 level:C Cv=0 45 45 F1e
E 46 e 1 95713.8389547786 Energy for J=E 46 e 1 level:C Cv=0 46 46 F1e
E 47 e 1 95877.7430647896 Energy for J=E 47 e 1 level:C Cv=0 47 47 F1e
E 48 e 1 96045.0535799167 Energy for J=E 48 e 1 level:C Cv=0 48 48 F1e
E 49 e 1 96215.7478896832 Energy for J=E 49 e 1 level:C Cv=0 49 49 F1e
E 50 e 1 96389.8321659005 Energy for J=E 50 e 1 level:C Cv=0 50 50 F1e
E 51 e 1 96567.2766128028 Energy for J=E 51 e 1 level:C Cv=0 51 51 F1e
E 52 e 1 96748.0881671828 Energy for J=E 52 e 1 level:C Cv=0 52 52 F1e
E 53 e 1 96932.2837485302 Energy for J=E 53 e 1 level:C Cv=0 53 53 F1e
E 54 e 1 97119.8371091712 Energy for J=E 54 e 1 level:C Cv=0 54 54 F1e
E 55 e 1 97310.7208844101 Energy for J=E 55 e 1 level:C Cv=0 55 55 F1e
E 56 e 1 97504.930642673 Energy for J=E 56 e 1 level:C Cv=0 56 56 F1e
E 57 e 1 97702.5267856525 Energy for J=E 57 e 1 level:C Cv=0 57 57 F1e
E 58 e 1 97903.4396484547 Energy for J=E 58 e 1 level:C Cv=0 58 58 F1e
E 59 e 1 98107.6243997471 Energy for J=E 59 e 1 level:C Cv=0 59 59 F1e
E 60 e 1 98315.1990419085 Energy for J=E 60 e 1 level:C Cv=0 60 60 F1e
E 61 e 1 98526.0359657192 Energy for J=E 61 e 1 level
E 62 e 1 98740.1821395715 Energy for J=E 62 e 1 level
E 63 e 1 98957.6207414145 Energy for J=E 63 e 1 level
E 64 e 1 99178.3440172057 Energy for J=E 64 e 1 level
E 65 e 1 99402.3440898222 Energy for J=E 65 e 1 level

Linear: C Cv=0
Status Active
Lambda Sigma+
OmegaOrder Auto => Regular
Origin 91918.79 State Origin. Estimated lowest level = 91918.79
B 1.76641191315359 Rotational constant.
D 5.12833508732762E-6 Quartic centrifugal distortion.
<J,Lambda=0,Omega=0|B|J,Lambda=0,Omega=0> = J*(J+1)
Symmetrized:
<J,Lambda=0,Omega=0|B|J,Lambda=0,Omega=0> = J*(J+1)

<J,Lambda=0,Omega=0|D|J,Lambda=0,Omega=0> = -J^2*(J*(J+2)+1)
Symmetrized:
<J,Lambda=0,Omega=0|D|J,Lambda=0,Omega=0> = -J^2*(J*(J+2)+1)

LinearManifold: B
Initial True
EigenSearch True
LimitSearch False
AutoQConverge True
UsePopParams False
OverrideEnergies True
E 0 e 1 86917.3573438145 Energy for J=E 0 e 1 level:B Bv=0 0 0 F1e
E 1 e 1 86920.8990798371 Energy for J=E 1 e 1 level:B Bv=0 1 1 F1e
E 2 e 1 86927.9802287834 Energy for J=E 2 e 1 level:B Bv=0 2 2 F1e
E 3 e 1 86938.5975955474 Energy for J=E 3 e 1 level:B Bv=0 3 3 F1e
E 4 e 1 86952.7535165513 Energy for J=E 4 e 1 level:B Bv=0 4 4 F1e
E 5 e 1 86970.4485900942 Energy for J=E 5 e 1 level:B Bv=0 5 5 F1e
E 6 e 1 86991.6808705965 Energy for J=E 6 e 1 level:B Bv=0 6 6 F1e
E 7 e 1 87016.4501212347 Energy for J=E 7 e 1 level:B Bv=0 7 7 F1e
E 8 e 1 87044.7550139408 Energy for J=E 8 e 1 level:B Bv=0 8 8 F1e
E 9 e 1 87076.5945102899 Energy for J=E 9 e 1 level:B Bv=0 9 9 F1e
E 10 e 1 87111.9673665656 Energy for J=E 10 e 1 level:B Bv=0 10 10 F1e
E 11 e 1 87150.8722462898 Energy for J=E 11 e 1 level:B Bv=0 11 11 F1e
E 12 e 1 87193.3103335359 Energy for J=E 12 e 1 level:B Bv=0 12 12 F1e
E 13 e 1 87239.2738059414 Energy for J=E 13 e 1 level:B Bv=0 13 13 F1e
E 14 e 1 87288.7659451284 Energy for J=E 14 e 1 level:B Bv=0 14 14 F1e
E 15 e 1 87341.7838423505 Energy for J=E 15 e 1 level:B Bv=0 15 15 F1e
E 16 e 1 87398.3239892327 Energy for J=E 16 e 1 level:B Bv=0 16 16 F1e
E 17 e 1 87458.3457386672 Energy for J=E 17 e 1 level:B Bv=0 17 17 F1e
E 18 e 1 87521.9669176787 Energy for J=E 18 e 1 level:B Bv=0 18 18 F1e
E 19 e 1 87589.0729980635 Energy for J=E 19 e 1 level:B Bv=0 19 19 F1e
E 20 e 1 87659.6848539994 Energy for J=E 20 e 1 level:B Bv=0 20 20 F1e
E 21 e 1 87733.8093501407 Energy for J=E 21 e 1 level:B Bv=0 21 21 F1e
E 22 e 1 87811.4434750138 Energy for J=E 22 e 1 level:B Bv=0 22 22 F1e
E 23 e 1 87892.5833524621 Energy for J=E 23 e 1 level:B Bv=0 23 23 F1e
E 24 e 1 87977.2254359612 Energy for J=E 24 e 1 level:B Bv=0 24 24 F1e
E 25 e 1 88065.3655819688 Energy for J=E 25 e 1 level:B Bv=0 25 25 F1e
E 26 e 1 88157.0074638094 Energy for J=E 26 e 1 level:B Bv=0 26 26 F1e
E 27 e 1 88252.1428727514 Energy for J=E 27 e 1 level:B Bv=0 27 27 F1e
E 28 e 1 88350.7622038786 Energy for J=E 28 e 1 level:B Bv=0 28 28 F1e
E 29 e 1 88452.8714734024 Energy for J=E 29 e 1 level:B Bv=0 29 29 F1e
E 30 e 1 88558.4573652084 Energy for J=E 30 e 1 level:B Bv=0 30 30 F1e
E 31 e 1 88667.5170675484 Energy for J=E 31 e 1 level:B Bv=0 31 31 F1e
E 32 e 1 88780.0530157549 Energy for J=E 32 e 1 level:B Bv=0 32 32 F1e
E 33 e 1 88896.0732697678 Energy for J=E 33 e 1 level:B Bv=0 33 33 F1e
E 34 e 1 89015.5479117331 Energy for J=E 34 e 1 level:B Bv=0 34 34 F1e
E 35 e 1 89138.5007753021 Energy for J=E 35 e 1 level:B Bv=0 35 35 F1e
E 36 e 1 89264.8671434128 Energy for J=E 36 e 1 level:B Bv=0 36 36 F1e
E 37 e 1 89394.712613204 Energy for J=E 37 e 1 level:B Bv=0 37 37 F1e
E 38 e 1 89528.0033885644 Energy for J=E 38 e 1 level:B Bv=0 38 38 F1e
E 39 e 1 89664.7427100465 Energy for J=E 39 e 1 level:B Bv=0 39 39 F1e
E 40 e 1 89804.8732821819 Energy for J=E 40 e 1 level:B Bv=0 40 40 F1e
E 41 e 1 89948.4826520769 Energy for J=E 41 e 1 level:B Bv=0 41 41 F1e
E 42 e 1 90095.4817872327 Energy for J=E 42 e 1 level:B Bv=0 42 42 F1e
E 43 e 1 90245.9288501256 Energy for J=E 43 e 1 level:B Bv=0 43 43 F1e
E 44 e 1 90399.7945833577 Energy for J=E 44 e 1 level:B Bv=0 44 44 F1e
E 45 e 1 90557.048472636 Energy for J=E 45 e 1 level:B Bv=0 45 45 F1e
E 46 e 1 90717.7471747185 Energy for J=E 46 e 1 level:B Bv=0 46 46 F1e
E 47 e 1 90881.7572647896 Energy for J=E 47 e 1 level:B Bv=0 47 47 F1e
E 48 e 1 91049.1774976252 Energy for J=E 48 e 1 level:B Bv=0 48 48 F1e
E 49 e 1 91220.0533896832 Energy for J=E 49 e 1 level:B Bv=0 49 49 F1e
E 50 e 1 91394.1969159005 Energy for J=E 50 e 1 level:B Bv=0 50 50 F1e
E 51 e 1 91571.8211856604 Energy for J=E 51 e 1 level
E 52 e 1 91752.7615313151 Energy for J=E 52 e 1 level
E 53 e 1 91937.0582069911 Energy for J=E 53 e 1 level
E 54 e 1 92124.7041674308 Energy for J=E 54 e 1 level
E 55 e 1 92315.6922344469 Energy for J=E 55 e 1 level
Linear: B Bv=0
Status Active
Lambda Sigma+
OmegaOrder Auto => Regular
Origin 86917.360320496 State Origin. Estimated lowest level = 86917.360320496
B 1.76976443875216 Rotational constant.
D 5.53872463797865E-6 Quartic centrifugal distortion.
Matrix elements as for C Cv=0
LinearManifold: Excited
Initial False
EigenSearch True
LimitSearch True
AutoQConverge True
UsePopParams False

```
OverrideEnergies True
Linear: Excited Av=2
  Status     Active
  Lambda     Pi
  OmegaOrder Auto => Regular
  Origin 67557.9263807482           Float, Std Dev = 0.000878566892884828  State Origin. Estimated lowest level = 67557.9263807482
  B      1.41770507032897           Float, Std Dev = 3.51295994438537E-6  Rotational constant.
  q      -1.187706E-5               Lambda doubling constant.
  D      6.24399485577261E-6        Float, Std Dev = 2.53609150054505E-9  Quartic centrifugal distortion.
  H      -1.280902E-11              Sextic centrifugal distortion.
  <J,Lambda=-1,Omega=-1|B|J,Lambda=-1,Omega=-1> = J*(J+1)
  <J,Lambda=1,Omega=1|B|J,Lambda=1,Omega=1> = J*(J+1)
    Symmetrized:
  <J,Lambda=1,Omega=1|B|J,Lambda=1,Omega=1> = J*(J+1)
  <J,Lambda=1,Omega=1|B|J,Lambda=1,Omega=1> = J*(J+1)

  <J,Lambda=-1,Omega=-1|q|J,Lambda=1,Omega=1> = (J*(J+1))/2
  <J,Lambda=1,Omega=1|q|J,Lambda=-1,Omega=-1> = (J*(J+1))/2
    Symmetrized:
  <J,Lambda=1,Omega=1|q|J,Lambda=1,Omega=1> = (J*(J+1))/2
  <J,Lambda=1,Omega=1|q|J,Lambda=1,Omega=1> = (J*(J+1))/2

  <J,Lambda=-1,Omega=-1|D|J,Lambda=-1,Omega=-1> = -J^2*(J*(J+2)+1)
  <J,Lambda=1,Omega=1|D|J,Lambda=1,Omega=1> = -J^2*(J*(J+2)+1)
    Symmetrized:
  <J,Lambda=1,Omega=1|D|J,Lambda=1,Omega=1> = -J^2*(J*(J+2)+1)
  <J,Lambda=1,Omega=1|D|J,Lambda=1,Omega=1> = -J^2*(J*(J+2)+1)

  <J,Lambda=-1,Omega=-1|H|J,Lambda=-1,Omega=-1> = J^3*(J*(J*(J+3)+3)+1)
  <J,Lambda=1,Omega=1|H|J,Lambda=1,Omega=1> = J^3*(J*(J*(J+3)+3)+1)
    Symmetrized:
  <J,Lambda=1,Omega=1|H|J,Lambda=1,Omega=1> = J^3*(J*(J*(J+3)+3)+1)
  <J,Lambda=1,Omega=1|H|J,Lambda=1,Omega=1> = J^3*(J*(J*(J+3)+3)+1)

Linear: Excited Iv=3
  Status     Active
  Colour     Fuchsia
  Lambda     Sigma-
  OmegaOrder Auto => Regular
  Origin 67579.2383904013           Float, Std Dev = 0.0154787716797877  State Origin. Estimated lowest level = 67579.2383904013
  B      1.09918701526773           Float, Std Dev = 0.000226636002436598  Rotational constant.
  D      5.77973239037241E-6        Float, Std Dev = 4.04347248069572E-7  Quartic centrifugal distortion.
  H      2.247197E-12               Sextic centrifugal distortion.
  <J,Lambda=0,Omega=0|B|J,Lambda=0,Omega=0> = J*(J+1)
    Symmetrized:
  <J,Lambda=0,Omega=0|B|J,Lambda=0,Omega=0> = J*(J+1)

  <J,Lambda=0,Omega=0|D|J,Lambda=0,Omega=0> = -J^2*(J*(J+2)+1)
    Symmetrized:
  <J,Lambda=0,Omega=0|D|J,Lambda=0,Omega=0> = -J^2*(J*(J+2)+1)

  <J,Lambda=0,Omega=0|H|J,Lambda=0,Omega=0> = J^3*(J*(J*(J+3)+3)+1)
    Symmetrized:
  <J,Lambda=0,Omega=0|H|J,Lambda=0,Omega=0> = J^3*(J*(J*(J+3)+3)+1)

Linear: Excited Iv=4
  Status     Active
  Colour     Fuchsia
  Lambda     Sigma-
  OmegaOrder Auto => Regular
  Origin 68544.2939028              State Origin. Estimated lowest level = 68544.2939028
  B      1.0840499034               Rotational constant.
  D      5.712347E-6                Quartic centrifugal distortion.
  H      2.247197E-12               Sextic centrifugal distortion.
  <J,Lambda=0,Omega=0|B|J,Lambda=0,Omega=0> = J*(J+1)
    Symmetrized:
  <J,Lambda=0,Omega=0|B|J,Lambda=0,Omega=0> = J*(J+1)

  <J,Lambda=0,Omega=0|D|J,Lambda=0,Omega=0> = -J^2*(J*(J+2)+1)
    Symmetrized:
  <J,Lambda=0,Omega=0|D|J,Lambda=0,Omega=0> = -J^2*(J*(J+2)+1)

  <J,Lambda=0,Omega=0|H|J,Lambda=0,Omega=0> = J^3*(J*(J*(J+3)+3)+1)
    Symmetrized:
  <J,Lambda=0,Omega=0|H|J,Lambda=0,Omega=0> = J^3*(J*(J*(J+3)+3)+1)

LinearPerturbation: Excited <Av=2|J+L-|Iv=3>
  Value -0.0504224240425148        Float, Std Dev = 0.000477162968858135  Size of perturbation.
<J,Lambda=-1,Omega=-1|J+L-|J,Lambda=0,Omega=0> = -sqrt(J*(J+1))
<J,Lambda=1,Omega=1|J+L-|J,Lambda=0,Omega=0> = sqrt(J*(J+1))

Symmetrized Matrix Elements of <Av=2|J+L-|Iv=3>
<J,Lambda=1,Omega=1|J+L-|J,Lambda=0,Omega=0> = -sqrt(J*(J+1))

LinearPerturbation: Excited <Av=2|J+L-|Iv=4>
  Value 0.06495                    Size of perturbation.
<J,Lambda=-1,Omega=-1|J+L-|J,Lambda=0,Omega=0> = -sqrt(J*(J+1))
<J,Lambda=1,Omega=1|J+L-|J,Lambda=0,Omega=0> = sqrt(J*(J+1))

Symmetrized Matrix Elements of <Av=2|J+L-|Iv=4>
<J,Lambda=1,Omega=1|J+L-|J,Lambda=0,Omega=0> = -sqrt(J*(J+1))
```

Linear: Excited ev=4
Status Active
Colour Lime
S 1
Lambda Sigma-
OmegaOrder Auto => Inverted
Origin 67802.7849165451 Float, Std Dev = 0.00556026375325388 State Origin. Estimated lowest level = 67802.2187789372
B 1.09715336874161 Float, Std Dev = 3.00716570600556E-5 Rotational constant.
LambdaSS 0.424603205929654 Float, Std Dev = 0.0115162602834212 Spin - spin coupling constant (=1.5*epsilon).
D 5.34314012526997E-6 Float, Std Dev = 2.51620227317152E-8 Quartic centrifugal distortion.
H -1.498131E-12 Sextic centrifugal distortion.
<J,Lambda=0,Omega=-1|B|J,Lambda=0,Omega=-1> = J*(J+1)
<J,Lambda=0,Omega=-1|B|J,Lambda=0,Omega=0> = -sqrt(2*J*(J+1))
<J,Lambda=0,Omega=0|B|J,Lambda=0,Omega=-1> = -sqrt(2*J*(J+1))
<J,Lambda=0,Omega=0|B|J,Lambda=0,Omega=0> = J*(J+1)+2
<J,Lambda=0,Omega=0|B|J,Lambda=0,Omega=1> = -sqrt(2*J*(J+1))
<J,Lambda=0,Omega=1|B|J,Lambda=0,Omega=0> = -sqrt(2*J*(J+1))
<J,Lambda=0,Omega=1|B|J,Lambda=0,Omega=1> = J*(J+1)
Symmetrized:
<J,Lambda=0,Omega=1|B|J,Lambda=0,Omega=1> = J*(J+1)
<J,Lambda=0,Omega=1|B|J,Lambda=0,Omega=1> = -sqrt(2*J*(J+1))
<J,Lambda=0,Omega=1|B|J,Lambda=0,Omega=0> = -sqrt(2*J*(J+1))
<J,Lambda=0,Omega=0|B|J,Lambda=0,Omega=1> = J*(J+1)+2
<J,Lambda=0,Omega=0|B|J,Lambda=0,Omega=0> = -sqrt(2*J*(J+1))

<J,Lambda=0,Omega=-1|LambdaSS|J,Lambda=0,Omega=-1> = 2/3
<J,Lambda=0,Omega=0|LambdaSS|J,Lambda=0,Omega=0> = -4/3
<J,Lambda=0,Omega=1|LambdaSS|J,Lambda=0,Omega=1> = 2/3
Symmetrized:
<J,Lambda=0,Omega=1|LambdaSS|J,Lambda=0,Omega=1> = 2/3
<J,Lambda=0,Omega=1|LambdaSS|J,Lambda=0,Omega=1> = -4/3
<J,Lambda=0,Omega=0|LambdaSS|J,Lambda=0,Omega=0> = 2/3

<J,Lambda=0,Omega=-1|D|J,Lambda=0,Omega=-1> = -J*(J*(J*(J+2)+3)+2)
<J,Lambda=0,Omega=-1|D|J,Lambda=0,Omega=0> = 2*(J*(J+1)+1)*sqrt(2*J*(J+1))
<J,Lambda=0,Omega=-1|D|J,Lambda=0,Omega=1> = -2*J*(J+1)
<J,Lambda=0,Omega=0|D|J,Lambda=0,Omega=-1> = 2*(J*(J+1)+1)*sqrt(2*J*(J+1))
<J,Lambda=0,Omega=0|D|J,Lambda=0,Omega=0> = -J*(J*(J*(J+2)+9)+8)-4
<J,Lambda=0,Omega=0|D|J,Lambda=0,Omega=1> = 2*(J*(J+1)+1)*sqrt(2*J*(J+1))
<J,Lambda=0,Omega=1|D|J,Lambda=0,Omega=-1> = -2*J*(J+1)
<J,Lambda=0,Omega=1|D|J,Lambda=0,Omega=0> = 2*(J*(J+1)+1)*sqrt(2*J*(J+1))
<J,Lambda=0,Omega=1|D|J,Lambda=0,Omega=1> = -J*(J*(J*(J+2)+3)+2)
Symmetrized:
<J,Lambda=0,Omega=1|D|J,Lambda=0,Omega=1> = -J*(J*(J*(J+2)+3)+2)
<J,Lambda=0,Omega=1|D|J,Lambda=0,Omega=1> = 2*(J*(J+1)+1)*sqrt(2*J*(J+1))
<J,Lambda=0,Omega=1|D|J,Lambda=0,Omega=0> = -2*J*(J+1)
<J,Lambda=0,Omega=0|D|J,Lambda=0,Omega=1> = 2*(J*(J+1)+1)*sqrt(2*J*(J+1))
<J,Lambda=0,Omega=0|D|J,Lambda=0,Omega=0> = -J*(J*(J*(J+2)+9)+8)-4

<J,Lambda=0,Omega=-1|H|J,Lambda=0,Omega=-1> = J*(J*(J*(J*(J*(J+3)+9)+13)+10)+4)
<J,Lambda=0,Omega=-1|H|J,Lambda=0,Omega=0> = (J*(J*(-3*J*(J+2)-13)-10)-4)*sqrt(2*J*(J+1))
<J,Lambda=0,Omega=-1|H|J,Lambda=0,Omega=1> = 2*J*(J*(3*J*(J+2)+5)+2)
<J,Lambda=0,Omega=0|H|J,Lambda=0,Omega=-1> = (J*(J*(-3*J*(J+2)-13)-10)-4)*sqrt(2*J*(J+1))
<J,Lambda=0,Omega=0|H|J,Lambda=0,Omega=0> = J*(J*(J*(J*(J*(J+3)+21)+37)+46)+28)+8
<J,Lambda=0,Omega=0|H|J,Lambda=0,Omega=1> = (J*(J*(-3*J*(J+2)-13)-10)-4)*sqrt(2*J*(J+1))
<J,Lambda=0,Omega=1|H|J,Lambda=0,Omega=-1> = 2*J*(J*(3*J*(J+2)+5)+2)
<J,Lambda=0,Omega=1|H|J,Lambda=0,Omega=0> = (J*(J*(-3*J*(J+2)-13)-10)-4)*sqrt(2*J*(J+1))
<J,Lambda=0,Omega=1|H|J,Lambda=0,Omega=1> = J*(J*(J*(J*(J*(J+3)+9)+13)+10)+4)
Symmetrized:
<J,Lambda=0,Omega=1|H|J,Lambda=0,Omega=1> = J*(J*(J*(J*(J*(J+3)+9)+13)+10)+4)
<J,Lambda=0,Omega=1|H|J,Lambda=0,Omega=1> = (J*(J*(-3*J*(J+2)-13)-10)-4)*sqrt(2*J*(J+1))
<J,Lambda=0,Omega=1|H|J,Lambda=0,Omega=0> = 2*J*(J*(3*J*(J+2)+5)+2)
<J,Lambda=0,Omega=0|H|J,Lambda=0,Omega=1> = (J*(J*(-3*J*(J+2)-13)-10)-4)*sqrt(2*J*(J+1))
<J,Lambda=0,Omega=0|H|J,Lambda=0,Omega=0> = J*(J*(J*(J*(J*(J+3)+21)+37)+46)+28)+8

LinearPerturbation: Excited <Av=2|LS|ev=4>
Value -13.1135785220975 Float, Std Dev = 0.00455861064639965 Size of perturbation.
<J,Lambda=-1,Omega=-1|LS|J,Lambda=0,Omega=-1> = -1/sqrt(3)
<J,Lambda=1,Omega=1|LS|J,Lambda=0,Omega=1> = -1/sqrt(3)

Symmetrized Matrix Elements of <Av=2|LS|ev=4>
<J,Lambda=1,Omega=1|LS|J,Lambda=0,Omega=1> = -1/sqrt(3)
<J,Lambda=1,Omega=1|LS|J,Lambda=0,Omega=1> = -1/sqrt(3)

Linear: Excited dv=7
Status Active
Colour Blue
S 1
Lambda Delta
OmegaOrder Auto => Inverted
Origin 67946.2486612564 Float, Std Dev = 0.00364331290227751 State Origin. Estimated lowest level = 67913.089899836
B 1.07973878060191 Float, Std Dev = 1.39032711556845E-5 Rotational constant.
A -16.9899246280871 Float, Std Dev = 0.00331391148225554 Spin-orbit coupling constant.
LambdaSS 1.23163175363889 Float, Std Dev = 0.00538387977440938 Spin - spin coupling constant (=1.5*epsilon).
gamma -0.00290463321309688 Float, Std Dev = 0.000257537521412191 Spin rotation coupling constant.
D 5.12786503880413E-6 Float, Std Dev = 8.04194689239351E-9 Quartic centrifugal distortion.
H -5.992526E-13 Sextic centrifugal distortion.
AD -9.172647E-5 Centrifugal distortion of A.
<J,Lambda=-2,Omega=-3|B|J,Lambda=-2,Omega=-3> = J*(J+1)-4
<J,Lambda=-2,Omega=-3|B|J,Lambda=-2,Omega=-2> = -sqrt(2*(J*(J+1)-6))
<J,Lambda=-2,Omega=-2|B|J,Lambda=-2,Omega=-3> = -sqrt(2*(J*(J+1)-6))
<J,Lambda=-2,Omega=-2|B|J,Lambda=-2,Omega=-2> = J*(J+1)+2

<J,Lambda=-2,Omega=-2|B|J,Lambda=-2,Omega=-1> = -sqrt(2*(J*(J+1)-2))
<J,Lambda=-2,Omega=-1|B|J,Lambda=-2,Omega=-2> = -sqrt(2*(J*(J+1)-2))
<J,Lambda=-2,Omega=-1|B|J,Lambda=-2,Omega=-1> = J*(J+1)+4
<J,Lambda=2,Omega=1|B|J,Lambda=2,Omega=1> = J*(J+1)+4
<J,Lambda=2,Omega=1|B|J,Lambda=2,Omega=2> = -sqrt(2*(J*(J+1)-2))
<J,Lambda=2,Omega=2|B|J,Lambda=2,Omega=1> = -sqrt(2*(J*(J+1)-2))
<J,Lambda=2,Omega=2|B|J,Lambda=2,Omega=2> = J*(J+1)+2
<J,Lambda=2,Omega=2|B|J,Lambda=2,Omega=3> = -sqrt(2*(J*(J+1)-6))
<J,Lambda=2,Omega=3|B|J,Lambda=2,Omega=2> = -sqrt(2*(J*(J+1)-6))
<J,Lambda=2,Omega=3|B|J,Lambda=2,Omega=3> = J*(J+1)-4
Symmetrized:
<J,Lambda=2,Omega=3|B|J,Lambda=2,Omega=3> = J*(J+1)-4
<J,Lambda=2,Omega=3|B|J,Lambda=2,Omega=3> = -sqrt(2*(J*(J+1)-6))
<J,Lambda=2,Omega=3|B|J,Lambda=2,Omega=2> = -sqrt(2*(J*(J+1)-6))
<J,Lambda=2,Omega=3|B|J,Lambda=2,Omega=2> = J*(J+1)+2
<J,Lambda=2,Omega=2|B|J,Lambda=2,Omega=3> = -sqrt(2*(J*(J+1)-2))
<J,Lambda=2,Omega=2|B|J,Lambda=2,Omega=3> = -sqrt(2*(J*(J+1)-2))
<J,Lambda=2,Omega=2|B|J,Lambda=2,Omega=2> = J*(J+1)+4
<J,Lambda=2,Omega=2|B|J,Lambda=2,Omega=2> = J*(J+1)+4
<J,Lambda=2,Omega=2|B|J,Lambda=2,Omega=1> = -sqrt(2*(J*(J+1)-2))
<J,Lambda=2,Omega=2|B|J,Lambda=2,Omega=1> = -sqrt(2*(J*(J+1)-2))
<J,Lambda=2,Omega=1|B|J,Lambda=2,Omega=2> = J*(J+1)+2
<J,Lambda=2,Omega=1|B|J,Lambda=2,Omega=2> = -sqrt(2*(J*(J+1)-6))
<J,Lambda=2,Omega=1|B|J,Lambda=2,Omega=1> = -sqrt(2*(J*(J+1)-6))
<J,Lambda=2,Omega=1|B|J,Lambda=2,Omega=1> = J*(J+1)-4

<J,Lambda=-2,Omega=-3|A|J,Lambda=-2,Omega=-3> = 2
<J,Lambda=-2,Omega=-1|A|J,Lambda=-2,Omega=-1> = -2
<J,Lambda=2,Omega=1|A|J,Lambda=2,Omega=1> = -2
<J,Lambda=2,Omega=3|A|J,Lambda=2,Omega=3> = 2
Symmetrized:
<J,Lambda=2,Omega=3|A|J,Lambda=2,Omega=3> = 2
<J,Lambda=2,Omega=3|A|J,Lambda=2,Omega=3> = -2
<J,Lambda=2,Omega=1|A|J,Lambda=2,Omega=1> = -2
<J,Lambda=2,Omega=1|A|J,Lambda=2,Omega=1> = 2

<J,Lambda=-2,Omega=-3|LambdaSS|J,Lambda=-2,Omega=-3> = 2/3
<J,Lambda=-2,Omega=-2|LambdaSS|J,Lambda=-2,Omega=-2> = -4/3
<J,Lambda=-2,Omega=-1|LambdaSS|J,Lambda=-2,Omega=-1> = 2/3
<J,Lambda=2,Omega=1|LambdaSS|J,Lambda=2,Omega=1> = 2/3
<J,Lambda=2,Omega=2|LambdaSS|J,Lambda=2,Omega=2> = -4/3
<J,Lambda=2,Omega=3|LambdaSS|J,Lambda=2,Omega=3> = 2/3
Symmetrized:
<J,Lambda=2,Omega=3|LambdaSS|J,Lambda=2,Omega=3> = 2/3
<J,Lambda=2,Omega=3|LambdaSS|J,Lambda=2,Omega=3> = -4/3
<J,Lambda=2,Omega=2|LambdaSS|J,Lambda=2,Omega=2> = 2/3
<J,Lambda=2,Omega=2|LambdaSS|J,Lambda=2,Omega=2> = 2/3
<J,Lambda=2,Omega=1|LambdaSS|J,Lambda=2,Omega=1> = -4/3
<J,Lambda=2,Omega=1|LambdaSS|J,Lambda=2,Omega=1> = 2/3

<J,Lambda=-2,Omega=-3|gamma|J,Lambda=-2,Omega=-3> = 1
<J,Lambda=-2,Omega=-3|gamma|J,Lambda=-2,Omega=-2> = sqrt(2*(J*(J+1)-6))/2
<J,Lambda=-2,Omega=-2|gamma|J,Lambda=-2,Omega=-3> = sqrt(2*(J*(J+1)-6))/2
<J,Lambda=-2,Omega=-2|gamma|J,Lambda=-2,Omega=-2> = -2
<J,Lambda=-2,Omega=-2|gamma|J,Lambda=-2,Omega=-1> = sqrt(2*(J*(J+1)-2))/2
<J,Lambda=-2,Omega=-1|gamma|J,Lambda=-2,Omega=-2> = sqrt(2*(J*(J+1)-2))/2
<J,Lambda=-2,Omega=-1|gamma|J,Lambda=-2,Omega=-1> = -3
<J,Lambda=2,Omega=1|gamma|J,Lambda=2,Omega=1> = -3
<J,Lambda=2,Omega=1|gamma|J,Lambda=2,Omega=2> = sqrt(2*(J*(J+1)-2))/2
<J,Lambda=2,Omega=2|gamma|J,Lambda=2,Omega=1> = sqrt(2*(J*(J+1)-2))/2
<J,Lambda=2,Omega=2|gamma|J,Lambda=2,Omega=2> = -2
<J,Lambda=2,Omega=2|gamma|J,Lambda=2,Omega=3> = sqrt(2*(J*(J+1)-6))/2
<J,Lambda=2,Omega=3|gamma|J,Lambda=2,Omega=2> = sqrt(2*(J*(J+1)-6))/2
<J,Lambda=2,Omega=3|gamma|J,Lambda=2,Omega=3> = 1
Symmetrized:
<J,Lambda=2,Omega=3|gamma|J,Lambda=2,Omega=3> = 1
<J,Lambda=2,Omega=3|gamma|J,Lambda=2,Omega=3> = sqrt(2*(J*(J+1)-6))/2
<J,Lambda=2,Omega=3|gamma|J,Lambda=2,Omega=2> = sqrt(2*(J*(J+1)-6))/2
<J,Lambda=2,Omega=3|gamma|J,Lambda=2,Omega=2> = -2
<J,Lambda=2,Omega=2|gamma|J,Lambda=2,Omega=3> = sqrt(2*(J*(J+1)-2))/2
<J,Lambda=2,Omega=2|gamma|J,Lambda=2,Omega=3> = sqrt(2*(J*(J+1)-2))/2
<J,Lambda=2,Omega=2|gamma|J,Lambda=2,Omega=2> = -3
<J,Lambda=2,Omega=2|gamma|J,Lambda=2,Omega=2> = -3
<J,Lambda=2,Omega=2|gamma|J,Lambda=2,Omega=1> = sqrt(2*(J*(J+1)-2))/2
<J,Lambda=2,Omega=2|gamma|J,Lambda=2,Omega=1> = sqrt(2*(J*(J+1)-2))/2
<J,Lambda=2,Omega=1|gamma|J,Lambda=2,Omega=2> = -2
<J,Lambda=2,Omega=1|gamma|J,Lambda=2,Omega=2> = sqrt(2*(J*(J+1)-6))/2
<J,Lambda=2,Omega=1|gamma|J,Lambda=2,Omega=1> = sqrt(2*(J*(J+1)-6))/2
<J,Lambda=2,Omega=1|gamma|J,Lambda=2,Omega=1> = 1

<J,Lambda=-2,Omega=-3|D|J,Lambda=-2,Omega=-3> = -J*(J*(J*(J+2)-5)-6)-4
<J,Lambda=-2,Omega=-3|D|J,Lambda=-2,Omega=-2> = -2*(-J*(J+1)+1)*sqrt(2*(J*(J+1)-6))
<J,Lambda=-2,Omega=-3|D|J,Lambda=-2,Omega=-1> = -2*sqrt(J*(J*(J*(J+2)-7)-8)+12)
<J,Lambda=-2,Omega=-2|D|J,Lambda=-2,Omega=-3> = -2*(-J*(J+1)+1)*sqrt(2*(J*(J+1)-6))
<J,Lambda=-2,Omega=-2|D|J,Lambda=-2,Omega=-2> = -J*(J*(J*(J+2)+9)+8)+12
<J,Lambda=-2,Omega=-2|D|J,Lambda=-2,Omega=-1> = 2*(J*(J+1)+3)*sqrt(2*(J*(J+1)-2))
<J,Lambda=-2,Omega=-1|D|J,Lambda=-2,Omega=-3> = -2*sqrt(J*(J*(J*(J+2)-7)-8)+12)
<J,Lambda=-2,Omega=-1|D|J,Lambda=-2,Omega=-2> = 2*(J*(J+1)+3)*sqrt(2*(J*(J+1)-2))
<J,Lambda=-2,Omega=-1|D|J,Lambda=-2,Omega=-1> = -J*(J*(J*(J+2)+11)+10)-12
<J,Lambda=2,Omega=1|D|J,Lambda=2,Omega=1> = -J*(J*(J*(J+2)+11)+10)-12
<J,Lambda=2,Omega=1|D|J,Lambda=2,Omega=2> = 2*(J*(J+1)+3)*sqrt(2*(J*(J+1)-2))
<J,Lambda=2,Omega=1|D|J,Lambda=2,Omega=3> = -2*sqrt(J*(J*(J*(J+2)-7)-8)+12)

<J,Lambda=2,Omega=2|D|J,Lambda=2,Omega=1> = 2*(J*(J+1)+3)*sqrt(2*(J*(J+1)-2))
<J,Lambda=2,Omega=2|D|J,Lambda=2,Omega=2> = -J*(J*(J*(J+2)+9)+8)+12
<J,Lambda=2,Omega=2|D|J,Lambda=2,Omega=3> = -2*(-J*(J+1)+1)*sqrt(2*(J*(J+1)-6))
<J,Lambda=2,Omega=3|D|J,Lambda=2,Omega=1> = -2*sqrt(J*(J*(J*(J+2)-7)-8)+12)
<J,Lambda=2,Omega=3|D|J,Lambda=2,Omega=2> = -2*(-J*(J+1)+1)*sqrt(2*(J*(J+1)-6))
<J,Lambda=2,Omega=3|D|J,Lambda=2,Omega=3> = -J*(J*(J*(J+2)-5)-6)-4
Symmetrized:
<J,Lambda=2,Omega=3|D|J,Lambda=2,Omega=3> = -J*(J*(J*(J+2)-5)-6)-4
<J,Lambda=2,Omega=3|D|J,Lambda=2,Omega=3> = -2*(-J*(J+1)+1)*sqrt(2*(J*(J+1)-6))
<J,Lambda=2,Omega=3|D|J,Lambda=2,Omega=2> = -2*sqrt(J*(J*(J*(J+2)-7)-8)+12)
<J,Lambda=2,Omega=3|D|J,Lambda=2,Omega=2> = -2*(-J*(J+1)+1)*sqrt(2*(J*(J+1)-6))
<J,Lambda=2,Omega=3|D|J,Lambda=2,Omega=1> = -J*(J*(J*(J+2)+9)+8)+12
<J,Lambda=2,Omega=3|D|J,Lambda=2,Omega=1> = 2*(J*(J+1)+3)*sqrt(2*(J*(J+1)-2))
<J,Lambda=2,Omega=2|D|J,Lambda=2,Omega=3> = -2*sqrt(J*(J*(J*(J+2)-7)-8)+12)
<J,Lambda=2,Omega=2|D|J,Lambda=2,Omega=3> = 2*(J*(J+1)+3)*sqrt(2*(J*(J+1)-2))
<J,Lambda=2,Omega=2|D|J,Lambda=2,Omega=2> = -J*(J*(J*(J+2)+11)+10)-12
<J,Lambda=2,Omega=2|D|J,Lambda=2,Omega=2> = -J*(J*(J*(J+2)+11)+10)-12
<J,Lambda=2,Omega=2|D|J,Lambda=2,Omega=1> = 2*(J*(J+1)+3)*sqrt(2*(J*(J+1)-2))
<J,Lambda=2,Omega=2|D|J,Lambda=2,Omega=1> = -2*sqrt(J*(J*(J*(J+2)-7)-8)+12)
<J,Lambda=2,Omega=1|D|J,Lambda=2,Omega=3> = 2*(J*(J+1)+3)*sqrt(2*(J*(J+1)-2))
<J,Lambda=2,Omega=1|D|J,Lambda=2,Omega=3> = -J*(J*(J*(J+2)+9)+8)+12
<J,Lambda=2,Omega=1|D|J,Lambda=2,Omega=2> = -2*(-J*(J+1)+1)*sqrt(2*(J*(J+1)-6))
<J,Lambda=2,Omega=1|D|J,Lambda=2,Omega=2> = -2*sqrt(J*(J*(J*(J+2)-7)-8)+12)
<J,Lambda=2,Omega=1|D|J,Lambda=2,Omega=1> = -2*(-J*(J+1)+1)*sqrt(2*(J*(J+1)-6))
<J,Lambda=2,Omega=1|D|J,Lambda=2,Omega=1> = -J*(J*(J*(J+2)-5)-6)-4

<J,Lambda=-2,Omega=-3|H|J,Lambda=-2,Omega=-3> = J^2*(J*(J*(J*(J+3)-3)-11)-6)+8
<J,Lambda=-2,Omega=-3|H|J,Lambda=-2,Omega=-2> = (J*(J*(-3*J*(J+2)-1)+2)+4)*sqrt(2*(J*(J+1)-6))
<J,Lambda=-2,Omega=-3|H|J,Lambda=-2,Omega=-1> = 2*(3*J*(J+1)+2)*sqrt(J*(J*(J*(J+2)-7)-8)+12)
<J,Lambda=-2,Omega=-2|H|J,Lambda=-2,Omega=-3> = J*(J*(-3*J*(J+2)+1)+4)*sqrt(2*(J*(J+1)-6))-2*sqrt(2*(J*(J*(J*(J*(J*(J+3)-7)-19)+18)+28)-24))
<J,Lambda=-2,Omega=-2|H|J,Lambda=-2,Omega=-2> = J*(J*(J*(J*(J*(J+3)+21)+37)-2)-20)-24
<J,Lambda=-2,Omega=-2|H|J,Lambda=-2,Omega=-1> = J*(J*(-3*J*(J+2)-23)-20)*sqrt(2*(J*(J+1)-2))-2*sqrt(2*(J*(J*(J*(J*(J*(J+3)-11)-27)+46)+60)-72))-24*sqrt(2*(J*(J+1)-2))
<J,Lambda=-2,Omega=-1|H|J,Lambda=-2,Omega=-3> = 2*(3*J*(J+1)+2)*sqrt(J*(J*(J*(J+2)-7)-8)+12)
<J,Lambda=-2,Omega=-1|H|J,Lambda=-2,Omega=-2> = (J*(J*(-3*J*(J+2)-25)-22)-12)*sqrt(2*(J*(J+1)-2))
<J,Lambda=-2,Omega=-1|H|J,Lambda=-2,Omega=-1> = J*(J*(J*(J*(J*(J+3)+21)+37)+74)+56)+24
<J,Lambda=2,Omega=1|H|J,Lambda=2,Omega=1> = J*(J*(J*(J*(J*(J+3)+21)+37)+74)+56)+24
<J,Lambda=2,Omega=1|H|J,Lambda=2,Omega=2> = (J*(J*(-3*J*(J+2)-25)-22)-12)*sqrt(2*(J*(J+1)-2))
<J,Lambda=2,Omega=1|H|J,Lambda=2,Omega=3> = 2*(3*J*(J+1)+2)*sqrt(J*(J*(J*(J+2)-7)-8)+12)
<J,Lambda=2,Omega=2|H|J,Lambda=2,Omega=1> = J*(J*(-3*J*(J+2)-23)-20)*sqrt(2*(J*(J+1)-2))-2*sqrt(2*(J*(J*(J*(J*(J*(J+3)-11)-27)+46)+60)-72))-24*sqrt(2*(J*(J+1)-2))
<J,Lambda=2,Omega=2|H|J,Lambda=2,Omega=2> = J*(J*(J*(J*(J*(J+3)+21)+37)-2)-20)-24
<J,Lambda=2,Omega=2|H|J,Lambda=2,Omega=3> = J*(J*(-3*J*(J+2)+1)+4)*sqrt(2*(J*(J+1)-6))-2*sqrt(2*(J*(J*(J*(J*(J*(J+3)-7)-19)+18)+28)-24))
<J,Lambda=2,Omega=3|H|J,Lambda=2,Omega=1> = 2*(3*J*(J+1)+2)*sqrt(J*(J*(J*(J+2)-7)-8)+12)
<J,Lambda=2,Omega=3|H|J,Lambda=2,Omega=2> = (J*(J*(-3*J*(J+2)-1)+2)+4)*sqrt(2*(J*(J+1)-6))
<J,Lambda=2,Omega=3|H|J,Lambda=2,Omega=3> = J^2*(J*(J*(J*(J+3)-3)-11)-6)+8
Symmetrized:
<J,Lambda=2,Omega=3|H|J,Lambda=2,Omega=3> = J^2*(J*(J*(J*(J+3)-3)-11)-6)+8
<J,Lambda=2,Omega=3|H|J,Lambda=2,Omega=3> = (J*(J*(-3*J*(J+2)-1)+2)+4)*sqrt(2*(J*(J+1)-6))
<J,Lambda=2,Omega=3|H|J,Lambda=2,Omega=2> = 2*(3*J*(J+1)+2)*sqrt(J*(J*(J*(J+2)-7)-8)+12)
<J,Lambda=2,Omega=3|H|J,Lambda=2,Omega=2> = J*(J*(-3*J*(J+2)+1)+4)*sqrt(2*(J*(J+1)-6))-2*sqrt(2*(J*(J*(J*(J*(J*(J+3)-7)-19)+18)+28)-24))
<J,Lambda=2,Omega=3|H|J,Lambda=2,Omega=1> = J*(J*(J*(J*(J*(J+3)+21)+37)-2)-20)-24
<J,Lambda=2,Omega=3|H|J,Lambda=2,Omega=1> = J*(J*(-3*J*(J+2)-23)-20)*sqrt(2*(J*(J+1)-2))-2*sqrt(2*(J*(J*(J*(J*(J*(J+3)-11)-27)+46)+60)-72))-24*sqrt(2*(J*(J+1)-2))
<J,Lambda=2,Omega=2|H|J,Lambda=2,Omega=3> = 2*(3*J*(J+1)+2)*sqrt(J*(J*(J*(J+2)-7)-8)+12)
<J,Lambda=2,Omega=2|H|J,Lambda=2,Omega=3> = (J*(J*(-3*J*(J+2)-25)-22)-12)*sqrt(2*(J*(J+1)-2))
<J,Lambda=2,Omega=2|H|J,Lambda=2,Omega=2> = J*(J*(J*(J*(J*(J+3)+21)+37)+74)+56)+24
<J,Lambda=2,Omega=2|H|J,Lambda=2,Omega=2> = J*(J*(J*(J*(J*(J+3)+21)+37)+74)+56)+24
<J,Lambda=2,Omega=2|H|J,Lambda=2,Omega=1> = (J*(J*(-3*J*(J+2)-25)-22)-12)*sqrt(2*(J*(J+1)-2))
<J,Lambda=2,Omega=2|H|J,Lambda=2,Omega=1> = 2*(3*J*(J+1)+2)*sqrt(J*(J*(J*(J+2)-7)-8)+12)
<J,Lambda=2,Omega=1|H|J,Lambda=2,Omega=3> = J*(J*(-3*J*(J+2)-23)-20)*sqrt(2*(J*(J+1)-2))-2*sqrt(2*(J*(J*(J*(J*(J*(J+3)-11)-27)+46)+60)-72))-24*sqrt(2*(J*(J+1)-2))
<J,Lambda=2,Omega=1|H|J,Lambda=2,Omega=3> = J*(J*(J*(J*(J*(J+3)+21)+37)-2)-20)-24
<J,Lambda=2,Omega=1|H|J,Lambda=2,Omega=2> = J*(J*(-3*J*(J+2)+1)+4)*sqrt(2*(J*(J+1)-6))-2*sqrt(2*(J*(J*(J*(J*(J*(J+3)-7)-19)+18)+28)-24))
<J,Lambda=2,Omega=1|H|J,Lambda=2,Omega=2> = 2*(3*J*(J+1)+2)*sqrt(J*(J*(J*(J+2)-7)-8)+12)
<J,Lambda=2,Omega=1|H|J,Lambda=2,Omega=1> = (J*(J*(-3*J*(J+2)-1)+2)+4)*sqrt(2*(J*(J+1)-6))
<J,Lambda=2,Omega=1|H|J,Lambda=2,Omega=1> = J^2*(J*(J*(J*(J+3)-3)-11)-6)+8

<J,Lambda=-2,Omega=-3|AD|J,Lambda=-2,Omega=-3> = 2*(J*(J+1)-4)
<J,Lambda=-2,Omega=-3|AD|J,Lambda=-2,Omega=-2> = -sqrt(2*(J*(J+1)-6))
<J,Lambda=-2,Omega=-2|AD|J,Lambda=-2,Omega=-3> = -sqrt(2*(J*(J+1)-6))
<J,Lambda=-2,Omega=-2|AD|J,Lambda=-2,Omega=-1> = sqrt(2*(J*(J+1)-2))
<J,Lambda=-2,Omega=-1|AD|J,Lambda=-2,Omega=-2> = sqrt(2*(J*(J+1)-2))
<J,Lambda=-2,Omega=-1|AD|J,Lambda=-2,Omega=-1> = 2*(-J*(J+1)-4)
<J,Lambda=2,Omega=1|AD|J,Lambda=2,Omega=1> = 2*(-J*(J+1)-4)
<J,Lambda=2,Omega=1|AD|J,Lambda=2,Omega=2> = sqrt(2*(J*(J+1)-2))
<J,Lambda=2,Omega=2|AD|J,Lambda=2,Omega=1> = sqrt(2*(J*(J+1)-2))
<J,Lambda=2,Omega=2|AD|J,Lambda=2,Omega=3> = -sqrt(2*(J*(J+1)-6))
<J,Lambda=2,Omega=3|AD|J,Lambda=2,Omega=2> = -sqrt(2*(J*(J+1)-6))
<J,Lambda=2,Omega=3|AD|J,Lambda=2,Omega=3> = 2*(J*(J+1)-4)
Symmetrized:
<J,Lambda=2,Omega=3|AD|J,Lambda=2,Omega=3> = 2*(J*(J+1)-4)
<J,Lambda=2,Omega=3|AD|J,Lambda=2,Omega=3> = -sqrt(2*(J*(J+1)-6))
<J,Lambda=2,Omega=3|AD|J,Lambda=2,Omega=2> = -sqrt(2*(J*(J+1)-6))
<J,Lambda=2,Omega=3|AD|J,Lambda=2,Omega=2> = sqrt(2*(J*(J+1)-2))
<J,Lambda=2,Omega=2|AD|J,Lambda=2,Omega=3> = sqrt(2*(J*(J+1)-2))
<J,Lambda=2,Omega=2|AD|J,Lambda=2,Omega=3> = 2*(-J*(J+1)-4)
<J,Lambda=2,Omega=2|AD|J,Lambda=2,Omega=1> = 2*(-J*(J+1)-4)
<J,Lambda=2,Omega=2|AD|J,Lambda=2,Omega=1> = sqrt(2*(J*(J+1)-2))
<J,Lambda=2,Omega=1|AD|J,Lambda=2,Omega=2> = sqrt(2*(J*(J+1)-2))
<J,Lambda=2,Omega=1|AD|J,Lambda=2,Omega=2> = -sqrt(2*(J*(J+1)-6))
<J,Lambda=2,Omega=1|AD|J,Lambda=2,Omega=1> = -sqrt(2*(J*(J+1)-6))
<J,Lambda=2,Omega=1|AD|J,Lambda=2,Omega=1> = 2*(J*(J+1)-4)

LinearPerturbation: Excited <Av=2|LS|dv=7>
Value -9.51850056737858 Float, Std Dev = 0.0104778117014898 Size of perturbation.

<J,Lambda=-1,Omega=-1|LS|J,Lambda=-2,Omega=-1> = 1/sqrt(3)
<J,Lambda=1,Omega=1|LS|J,Lambda=2,Omega=1> = -1/sqrt(3)

Symmetrized Matrix Elements of <Av=2|LS|dv=7>
<J,Lambda=1,Omega=1|LS|J,Lambda=2,Omega=1> = 1/sqrt(3)
<J,Lambda=1,Omega=1|LS|J,Lambda=2,Omega=1> = -1/sqrt(3)

Linear: Excited a'v=12
Status Active
Colour Red
S 1
Lambda Sigma+
OmegaOrder Auto => Inverted
Origin 67991.2956897487 Float, Std Dev = 0.00589179423262914 State Origin. Estimated lowest level = 67990.5318805452
B 1.03759347733243 Float, Std Dev = 2.24926190351211E-5 Rotational constant.
LambdaSS -1.14571380522651 Float, Std Dev = 0.00908941623482264 Spin - spin coupling constant (=1.5*epsilon).
gamma -0.00527072942354211 Float, Std Dev = 0.000311699571001926 Spin rotation coupling constant.
D 5.18562812216526E-6 Float, Std Dev = 1.22271031507727E-8 Quartic centrifugal distortion.
H -2.996263E-13 Sextic centrifugal distortion.
<J,Lambda=0,Omega=-1|B|J,Lambda=0,Omega=-1> = J*(J+1)
<J,Lambda=0,Omega=-1|B|J,Lambda=0,Omega=0> = -sqrt(2*J*(J+1))
<J,Lambda=0,Omega=0|B|J,Lambda=0,Omega=-1> = -sqrt(2*J*(J+1))
<J,Lambda=0,Omega=0|B|J,Lambda=0,Omega=0> = J*(J+1)+2
<J,Lambda=0,Omega=0|B|J,Lambda=0,Omega=1> = -sqrt(2*J*(J+1))
<J,Lambda=0,Omega=1|B|J,Lambda=0,Omega=0> = -sqrt(2*J*(J+1))
<J,Lambda=0,Omega=1|B|J,Lambda=0,Omega=1> = J*(J+1)
Symmetrized:
<J,Lambda=0,Omega=1|B|J,Lambda=0,Omega=1> = J*(J+1)
<J,Lambda=0,Omega=1|B|J,Lambda=0,Omega=1> = -sqrt(2*J*(J+1))
<J,Lambda=0,Omega=1|B|J,Lambda=0,Omega=0> = -sqrt(2*J*(J+1))
<J,Lambda=0,Omega=0|B|J,Lambda=0,Omega=1> = J*(J+1)+2
<J,Lambda=0,Omega=0|B|J,Lambda=0,Omega=0> = -sqrt(2*J*(J+1))

<J,Lambda=0,Omega=-1|LambdaSS|J,Lambda=0,Omega=-1> = 2/3
<J,Lambda=0,Omega=0|LambdaSS|J,Lambda=0,Omega=0> = -4/3
<J,Lambda=0,Omega=1|LambdaSS|J,Lambda=0,Omega=1> = 2/3
Symmetrized:
<J,Lambda=0,Omega=1|LambdaSS|J,Lambda=0,Omega=1> = 2/3
<J,Lambda=0,Omega=1|LambdaSS|J,Lambda=0,Omega=1> = -4/3
<J,Lambda=0,Omega=0|LambdaSS|J,Lambda=0,Omega=0> = 2/3

<J,Lambda=0,Omega=-1|gamma|J,Lambda=0,Omega=-1> = -1
<J,Lambda=0,Omega=-1|gamma|J,Lambda=0,Omega=0> = sqrt(2*J*(J+1))/2
<J,Lambda=0,Omega=0|gamma|J,Lambda=0,Omega=-1> = sqrt(2*J*(J+1))/2
<J,Lambda=0,Omega=0|gamma|J,Lambda=0,Omega=0> = -2
<J,Lambda=0,Omega=0|gamma|J,Lambda=0,Omega=1> = sqrt(2*J*(J+1))/2
<J,Lambda=0,Omega=1|gamma|J,Lambda=0,Omega=0> = sqrt(2*J*(J+1))/2
<J,Lambda=0,Omega=1|gamma|J,Lambda=0,Omega=1> = -1
Symmetrized:
<J,Lambda=0,Omega=1|gamma|J,Lambda=0,Omega=1> = -1
<J,Lambda=0,Omega=1|gamma|J,Lambda=0,Omega=1> = sqrt(2*J*(J+1))/2
<J,Lambda=0,Omega=1|gamma|J,Lambda=0,Omega=0> = sqrt(2*J*(J+1))/2
<J,Lambda=0,Omega=0|gamma|J,Lambda=0,Omega=1> = -2
<J,Lambda=0,Omega=0|gamma|J,Lambda=0,Omega=0> = sqrt(2*J*(J+1))/2

<J,Lambda=0,Omega=-1|D|J,Lambda=0,Omega=-1> = -J*(J*(J+2)+3)+2)
<J,Lambda=0,Omega=-1|D|J,Lambda=0,Omega=0> = 2*(J*(J+1)+1)*sqrt(2*J*(J+1))
<J,Lambda=0,Omega=-1|D|J,Lambda=0,Omega=1> = -2*J*(J+1)
<J,Lambda=0,Omega=0|D|J,Lambda=0,Omega=-1> = 2*(J*(J+1)+1)*sqrt(2*J*(J+1))
<J,Lambda=0,Omega=0|D|J,Lambda=0,Omega=0> = -J*(J*(J*(J+2)+9)+8)-4
<J,Lambda=0,Omega=0|D|J,Lambda=0,Omega=1> = 2*(J*(J+1)+1)*sqrt(2*J*(J+1))
<J,Lambda=0,Omega=1|D|J,Lambda=0,Omega=-1> = -2*J*(J+1)
<J,Lambda=0,Omega=1|D|J,Lambda=0,Omega=0> = 2*(J*(J+1)+1)*sqrt(2*J*(J+1))
<J,Lambda=0,Omega=1|D|J,Lambda=0,Omega=1> = -J*(J*(J*(J+2)+3)+2)
Symmetrized:
<J,Lambda=0,Omega=1|D|J,Lambda=0,Omega=1> = -J*(J*(J*(J+2)+3)+2)
<J,Lambda=0,Omega=1|D|J,Lambda=0,Omega=1> = 2*(J*(J+1)+1)*sqrt(2*J*(J+1))
<J,Lambda=0,Omega=1|D|J,Lambda=0,Omega=0> = -2*J*(J+1)
<J,Lambda=0,Omega=0|D|J,Lambda=0,Omega=1> = 2*(J*(J+1)+1)*sqrt(2*J*(J+1))
<J,Lambda=0,Omega=0|D|J,Lambda=0,Omega=0> = -J*(J*(J*(J+2)+9)+8)-4

<J,Lambda=0,Omega=-1|H|J,Lambda=0,Omega=-1> = J*(J*(J*(J*(J*(J+3)+9)+13)+10)+4)
<J,Lambda=0,Omega=-1|H|J,Lambda=0,Omega=0> = (J*(J*(-3*J*(J+2)-13)-10)-4)*sqrt(2*J*(J+1))
<J,Lambda=0,Omega=-1|H|J,Lambda=0,Omega=1> = 2*J*(J*(3*J*(J+2)+5)+2)
<J,Lambda=0,Omega=0|H|J,Lambda=0,Omega=-1> = (J*(J*(-3*J*(J+2)-13)-10)-4)*sqrt(2*J*(J+1))
<J,Lambda=0,Omega=0|H|J,Lambda=0,Omega=0> = J*(J*(J*(J*(J*(J+3)+21)+37)+46)+28)+8
<J,Lambda=0,Omega=0|H|J,Lambda=0,Omega=1> = (J*(J*(-3*J*(J+2)-13)-10)-4)*sqrt(2*J*(J+1))
<J,Lambda=0,Omega=1|H|J,Lambda=0,Omega=-1> = 2*J*(J*(3*J*(J+2)+5)+2)
<J,Lambda=0,Omega=1|H|J,Lambda=0,Omega=0> = (J*(J*(-3*J*(J+2)-13)-10)-4)*sqrt(2*J*(J+1))
<J,Lambda=0,Omega=1|H|J,Lambda=0,Omega=1> = J*(J*(J*(J*(J*(J+3)+9)+13)+10)+4)
Symmetrized:
<J,Lambda=0,Omega=1|H|J,Lambda=0,Omega=1> = J*(J*(J*(J*(J*(J+3)+9)+13)+10)+4)
<J,Lambda=0,Omega=1|H|J,Lambda=0,Omega=1> = (J*(J*(-3*J*(J+2)-13)-10)-4)*sqrt(2*J*(J+1))
<J,Lambda=0,Omega=1|H|J,Lambda=0,Omega=0> = 2*J*(J*(3*J*(J+2)+5)+2)
<J,Lambda=0,Omega=0|H|J,Lambda=0,Omega=1> = (J*(J*(-3*J*(J+2)-13)-10)-4)*sqrt(2*J*(J+1))
<J,Lambda=0,Omega=0|H|J,Lambda=0,Omega=0> = J*(J*(J*(J*(J*(J+3)+21)+37)+46)+28)+8

LinearPerturbation: Excited <Av=2|LS|a'v=12>
Value -5.90222571840932 Float, Std Dev = 0.0114841933161004 Size of perturbation.
<J,Lambda=-1,Omega=-1|LS|J,Lambda=0,Omega=-1> = 1/sqrt(3)
<J,Lambda=1,Omega=1|LS|J,Lambda=0,Omega=1> = -1/sqrt(3)

Symmetrized Matrix Elements of <Av=2|LS|a'v=12>
<J,Lambda=1,Omega=1|LS|J,Lambda=0,Omega=1> = 1/sqrt(3)
<J,Lambda=1,Omega=1|LS|J,Lambda=0,Omega=1> = -1/sqrt(3)

Linear: Excited Dsv=3
Status Active
Colour Orange
Lambda Delta
OmegaOrder Auto => Regular
Origin 68377.9123376615 Float, Std Dev = 0.0503654794741688 State Origin. Estimated lowest level = 68377.9123376615
B 1.08921917955507 Float, Std Dev = 0.000187517727013344 Rotational constant.
D 5.85540454591073E-6 Float, Std Dev = 7.96670714380823E-8 Quartic centrifugal distortion.
H -2.247197E-13 Sextic centrifugal distortion.
<J,Lambda=-2,Omega=-2|B|J,Lambda=-2,Omega=-2> = J*(J+1)
<J,Lambda=2,Omega=2|B|J,Lambda=2,Omega=2> = J*(J+1)
Symmetrized:
<J,Lambda=2,Omega=2|B|J,Lambda=2,Omega=2> = J*(J+1)
<J,Lambda=2,Omega=2|B|J,Lambda=2,Omega=2> = J*(J+1)

<J,Lambda=-2,Omega=-2|D|J,Lambda=-2,Omega=-2> = -J^2*(J*(J+2)+1)
<J,Lambda=2,Omega=2|D|J,Lambda=2,Omega=2> = -J^2*(J*(J+2)+1)
Symmetrized:
<J,Lambda=2,Omega=2|D|J,Lambda=2,Omega=2> = -J^2*(J*(J+2)+1)
<J,Lambda=2,Omega=2|D|J,Lambda=2,Omega=2> = -J^2*(J*(J+2)+1)

<J,Lambda=-2,Omega=-2|H|J,Lambda=-2,Omega=-2> = J^3*(J*(J*(J+3)+3)+1)
<J,Lambda=2,Omega=2|H|J,Lambda=2,Omega=2> = J^3*(J*(J*(J+3)+3)+1)
Symmetrized:
<J,Lambda=2,Omega=2|H|J,Lambda=2,Omega=2> = J^3*(J*(J*(J+3)+3)+1)
<J,Lambda=2,Omega=2|H|J,Lambda=2,Omega=2> = J^3*(J*(J*(J+3)+3)+1)

LinearPerturbation: Excited <Av=2|J+L-|Dsv=3>
Value 0.02597 Size of perturbation.
<J,Lambda=-1,Omega=-1|J+L-|J,Lambda=-2,Omega=-2> = sqrt(J*(J+1)-2)
<J,Lambda=1,Omega=1|J+L-|J,Lambda=2,Omega=2> = sqrt(J*(J+1)-2)

Symmetrized Matrix Elements of <Av=2|J+L-|Dsv=3>
<J,Lambda=1,Omega=1|J+L-|J,Lambda=2,Omega=2> = sqrt(J*(J+1)-2)
<J,Lambda=1,Omega=1|J+L-|J,Lambda=2,Omega=2> = sqrt(J*(J+1)-2)

Linear: Excited a3Pv=13
Status Active
Colour Silver
S 1
Lambda Pi
OmegaOrder Auto => Regular
Origin 67670.6782977856 Float, Std Dev = 0.00506807720509925 State Origin. Estimated lowest level = 67632.5507030479
B 1.30813665955102 Float, Std Dev = 5.56682832240352E-7 Rotational constant.
A 38.1154482644111 Float, Std Dev = 0.0129516607117093 Spin-orbit coupling constant.
LambdaSS -0.01821971 Spin - spin coupling constant (=1.5*epsilon).
o 0.6497647 Spin-spin coupling constant giving lambda doubling for Omega = 0 states.
gamma 0.003153713 Spin rotation coupling constant.
p 0.008831129 Lambda doubling constant.
q 2.786443E-5 Lambda doubling constant.
D 6.06110427245085E-6 Float, Std Dev = 4.64500829284345E-9 Quartic centrifugal distortion.
AD -0.0001940953 Centrifugal distortion of A.
<J,Lambda=-1,Omega=-2|B|J,Lambda=-1,Omega=-2> = J*(J+1)-2
<J,Lambda=-1,Omega=-2|B|J,Lambda=-1,Omega=-1> = -sqrt(2*(J*(J+1)-2))
<J,Lambda=-1,Omega=-1|B|J,Lambda=-1,Omega=-2> = -sqrt(2*(J*(J+1)-2))
<J,Lambda=-1,Omega=-1|B|J,Lambda=-1,Omega=-1> = J*(J+1)+2
<J,Lambda=-1,Omega=-1|B|J,Lambda=-1,Omega=0> = -sqrt(2*J*(J+1))
<J,Lambda=-1,Omega=0|B|J,Lambda=-1,Omega=-1> = -sqrt(2*J*(J+1))
<J,Lambda=-1,Omega=0|B|J,Lambda=-1,Omega=0> = J*(J+1)+2
<J,Lambda=1,Omega=0|B|J,Lambda=1,Omega=0> = J*(J+1)+2
<J,Lambda=1,Omega=0|B|J,Lambda=1,Omega=1> = -sqrt(2*J*(J+1))
<J,Lambda=1,Omega=1|B|J,Lambda=1,Omega=0> = -sqrt(2*J*(J+1))
<J,Lambda=1,Omega=1|B|J,Lambda=1,Omega=1> = J*(J+1)+2
<J,Lambda=1,Omega=1|B|J,Lambda=1,Omega=2> = -sqrt(2*(J*(J+1)-2))
<J,Lambda=1,Omega=2|B|J,Lambda=1,Omega=1> = -sqrt(2*(J*(J+1)-2))
<J,Lambda=1,Omega=2|B|J,Lambda=1,Omega=2> = J*(J+1)-2
Symmetrized:
<J,Lambda=1,Omega=2|B|J,Lambda=1,Omega=2> = J*(J+1)-2
<J,Lambda=1,Omega=2|B|J,Lambda=1,Omega=2> = -sqrt(2*(J*(J+1)-2))
<J,Lambda=1,Omega=2|B|J,Lambda=1,Omega=1> = -sqrt(2*(J*(J+1)-2))
<J,Lambda=1,Omega=2|B|J,Lambda=1,Omega=1> = J*(J+1)+2
<J,Lambda=1,Omega=1|B|J,Lambda=1,Omega=2> = -sqrt(2*J*(J+1))
<J,Lambda=1,Omega=1|B|J,Lambda=1,Omega=2> = -sqrt(2*J*(J+1))
<J,Lambda=1,Omega=1|B|J,Lambda=1,Omega=1> = J*(J+1)+2
<J,Lambda=1,Omega=1|B|J,Lambda=1,Omega=1> = J*(J+1)+2
<J,Lambda=1,Omega=1|B|J,Lambda=1,Omega=0> = -sqrt(2*J*(J+1))
<J,Lambda=1,Omega=1|B|J,Lambda=1,Omega=0> = -sqrt(2*J*(J+1))
<J,Lambda=1,Omega=0|B|J,Lambda=1,Omega=1> = J*(J+1)+2
<J,Lambda=1,Omega=0|B|J,Lambda=1,Omega=1> = -sqrt(2*(J*(J+1)-2))
<J,Lambda=1,Omega=0|B|J,Lambda=1,Omega=0> = -sqrt(2*(J*(J+1)-2))
<J,Lambda=1,Omega=0|B|J,Lambda=1,Omega=0> = J*(J+1)-2

<J,Lambda=-1,Omega=-2|A|J,Lambda=-1,Omega=-2> = 1
<J,Lambda=-1,Omega=0|A|J,Lambda=-1,Omega=0> = -1
<J,Lambda=1,Omega=0|A|J,Lambda=1,Omega=0> = -1
<J,Lambda=1,Omega=2|A|J,Lambda=1,Omega=2> = 1
Symmetrized:
<J,Lambda=1,Omega=2|A|J,Lambda=1,Omega=2> = 1

<J,Lambda=1,Omega=2|A|J,Lambda=1,Omega=2> = -1
<J,Lambda=1,Omega=0|A|J,Lambda=1,Omega=0> = -1
<J,Lambda=1,Omega=0|A|J,Lambda=1,Omega=0> = 1

<J,Lambda=-1,Omega=-2|LambdaSS|J,Lambda=-1,Omega=-2> = 2/3
<J,Lambda=-1,Omega=-1|LambdaSS|J,Lambda=-1,Omega=-1> = -4/3
<J,Lambda=-1,Omega=0|LambdaSS|J,Lambda=-1,Omega=0> = 2/3
<J,Lambda=1,Omega=0|LambdaSS|J,Lambda=1,Omega=0> = 2/3
<J,Lambda=1,Omega=1|LambdaSS|J,Lambda=1,Omega=1> = -4/3
<J,Lambda=1,Omega=2|LambdaSS|J,Lambda=1,Omega=2> = 2/3
Symmetrized:
<J,Lambda=1,Omega=2|LambdaSS|J,Lambda=1,Omega=2> = 2/3
<J,Lambda=1,Omega=2|LambdaSS|J,Lambda=1,Omega=2> = -4/3
<J,Lambda=1,Omega=1|LambdaSS|J,Lambda=1,Omega=1> = 2/3
<J,Lambda=1,Omega=1|LambdaSS|J,Lambda=1,Omega=1> = 2/3
<J,Lambda=1,Omega=0|LambdaSS|J,Lambda=1,Omega=0> = -4/3
<J,Lambda=1,Omega=0|LambdaSS|J,Lambda=1,Omega=0> = 2/3

<J,Lambda=-1,Omega=0|o|J,Lambda=1,Omega=0> = 1
<J,Lambda=1,Omega=0|o|J,Lambda=-1,Omega=0> = 1
Symmetrized:
<J,Lambda=1,Omega=0|o|J,Lambda=1,Omega=0> = 1
<J,Lambda=1,Omega=0|o|J,Lambda=1,Omega=0> = 1

<J,Lambda=-1,Omega=-2|gamma|J,Lambda=-1,Omega=-1> = sqrt(2*(J*(J+1)-2))/2
<J,Lambda=-1,Omega=-1|gamma|J,Lambda=-1,Omega=-2> = sqrt(2*(J*(J+1)-2))/2
<J,Lambda=-1,Omega=-1|gamma|J,Lambda=-1,Omega=-1> = -2
<J,Lambda=-1,Omega=-1|gamma|J,Lambda=-1,Omega=0> = sqrt(2*J*(J+1))/2
<J,Lambda=-1,Omega=0|gamma|J,Lambda=-1,Omega=-1> = sqrt(2*J*(J+1))/2
<J,Lambda=-1,Omega=0|gamma|J,Lambda=-1,Omega=0> = -2
<J,Lambda=1,Omega=0|gamma|J,Lambda=1,Omega=0> = -2
<J,Lambda=1,Omega=0|gamma|J,Lambda=1,Omega=1> = sqrt(2*J*(J+1))/2
<J,Lambda=1,Omega=1|gamma|J,Lambda=1,Omega=0> = sqrt(2*J*(J+1))/2
<J,Lambda=1,Omega=1|gamma|J,Lambda=1,Omega=1> = -2
<J,Lambda=1,Omega=1|gamma|J,Lambda=1,Omega=2> = sqrt(2*(J*(J+1)-2))/2
<J,Lambda=1,Omega=2|gamma|J,Lambda=1,Omega=1> = sqrt(2*(J*(J+1)-2))/2
Symmetrized:
<J,Lambda=1,Omega=2|gamma|J,Lambda=1,Omega=1> = sqrt(2*(J*(J+1)-2))/2
<J,Lambda=1,Omega=2|gamma|J,Lambda=1,Omega=1> = sqrt(2*(J*(J+1)-2))/2
<J,Lambda=1,Omega=1|gamma|J,Lambda=1,Omega=2> = -2
<J,Lambda=1,Omega=1|gamma|J,Lambda=1,Omega=2> = sqrt(2*J*(J+1))/2
<J,Lambda=1,Omega=1|gamma|J,Lambda=1,Omega=1> = sqrt(2*J*(J+1))/2
<J,Lambda=1,Omega=1|gamma|J,Lambda=1,Omega=1> = -2
<J,Lambda=1,Omega=1|gamma|J,Lambda=1,Omega=0> = -2
<J,Lambda=1,Omega=1|gamma|J,Lambda=1,Omega=0> = sqrt(2*J*(J+1))/2
<J,Lambda=1,Omega=0|gamma|J,Lambda=1,Omega=1> = sqrt(2*J*(J+1))/2
<J,Lambda=1,Omega=0|gamma|J,Lambda=1,Omega=1> = -2
<J,Lambda=1,Omega=0|gamma|J,Lambda=1,Omega=0> = sqrt(2*(J*(J+1)-2))/2
<J,Lambda=1,Omega=0|gamma|J,Lambda=1,Omega=0> = sqrt(2*(J*(J+1)-2))/2

<J,Lambda=-1,Omega=-1|p|J,Lambda=1,Omega=0> = (-sqrt(2*J*(J+1)))/2
<J,Lambda=-1,Omega=0|p|J,Lambda=1,Omega=0> = 1
<J,Lambda=-1,Omega=0|p|J,Lambda=1,Omega=1> = (-sqrt(2*J*(J+1)))/2
<J,Lambda=1,Omega=0|p|J,Lambda=-1,Omega=-1> = (-sqrt(2*J*(J+1)))/2
<J,Lambda=1,Omega=0|p|J,Lambda=-1,Omega=0> = 1
<J,Lambda=1,Omega=1|p|J,Lambda=-1,Omega=0> = (-sqrt(2*J*(J+1)))/2
Symmetrized:
<J,Lambda=1,Omega=1|p|J,Lambda=1,Omega=0> = (-sqrt(2*J*(J+1)))/2
<J,Lambda=1,Omega=1|p|J,Lambda=1,Omega=0> = 1
<J,Lambda=1,Omega=0|p|J,Lambda=1,Omega=0> = (-sqrt(2*J*(J+1)))/2
<J,Lambda=1,Omega=0|p|J,Lambda=1,Omega=0> = (-sqrt(2*J*(J+1)))/2
<J,Lambda=1,Omega=0|p|J,Lambda=1,Omega=1> = 1
<J,Lambda=1,Omega=0|p|J,Lambda=1,Omega=1> = (-sqrt(2*J*(J+1)))/2

<J,Lambda=-1,Omega=-2|q|J,Lambda=1,Omega=0> = sqrt(J*(J*(J+1)-2)*(J+1))/2
<J,Lambda=-1,Omega=-1|q|J,Lambda=1,Omega=0> = -sqrt(2*J*(J+1))
<J,Lambda=-1,Omega=-1|q|J,Lambda=1,Omega=1> = (J*(J+1))/2
<J,Lambda=-1,Omega=0|q|J,Lambda=1,Omega=0> = 1
<J,Lambda=-1,Omega=0|q|J,Lambda=1,Omega=1> = -sqrt(2*J*(J+1))
<J,Lambda=-1,Omega=0|q|J,Lambda=1,Omega=2> = sqrt(J*(J*(J+1)-2)*(J+1))/2
<J,Lambda=1,Omega=0|q|J,Lambda=-1,Omega=-2> = sqrt(J*(J*(J+1)-2)*(J+1))/2
<J,Lambda=1,Omega=0|q|J,Lambda=-1,Omega=-1> = -sqrt(2*J*(J+1))
<J,Lambda=1,Omega=0|q|J,Lambda=-1,Omega=0> = 1
<J,Lambda=1,Omega=1|q|J,Lambda=-1,Omega=-1> = (J*(J+1))/2
<J,Lambda=1,Omega=1|q|J,Lambda=-1,Omega=0> = -sqrt(2*J*(J+1))
<J,Lambda=1,Omega=2|q|J,Lambda=-1,Omega=0> = sqrt(J*(J*(J+1)-2)*(J+1))/2
Symmetrized:
<J,Lambda=1,Omega=2|q|J,Lambda=1,Omega=0> = sqrt(J*(J*(J+1)-2)*(J+1))/2
<J,Lambda=1,Omega=2|q|J,Lambda=1,Omega=0> = -sqrt(2*J*(J+1))
<J,Lambda=1,Omega=1|q|J,Lambda=1,Omega=0> = (J*(J+1))/2
<J,Lambda=1,Omega=1|q|J,Lambda=1,Omega=0> = 1
<J,Lambda=1,Omega=1|q|J,Lambda=1,Omega=1> = -sqrt(2*J*(J+1))
<J,Lambda=1,Omega=1|q|J,Lambda=1,Omega=1> = sqrt(J*(J*(J+1)-2)*(J+1))/2
<J,Lambda=1,Omega=0|q|J,Lambda=1,Omega=0> = sqrt(J*(J*(J+1)-2)*(J+1))/2
<J,Lambda=1,Omega=0|q|J,Lambda=1,Omega=0> = -sqrt(2*J*(J+1))
<J,Lambda=1,Omega=0|q|J,Lambda=1,Omega=1> = 1
<J,Lambda=1,Omega=0|q|J,Lambda=1,Omega=1> = (J*(J+1))/2
<J,Lambda=1,Omega=0|q|J,Lambda=1,Omega=2> = -sqrt(2*J*(J+1))
<J,Lambda=1,Omega=0|q|J,Lambda=1,Omega=2> = sqrt(J*(J*(J+1)-2)*(J+1))/2

<J,Lambda=-1,Omega=-2|D|J,Lambda=-1,Omega=-2> = -J*(J*(J*(J+2)-1)-2)
<J,Lambda=-1,Omega=-2|D|J,Lambda=-1,Omega=-1> = 2*J*(J+1)*sqrt(2*(J*(J+1)-2))

<J,Lambda=-1,Omega=-2|D|J,Lambda=-1,Omega=0> = -2*sqrt(J*(J*(J*(J+2)-1)-2))
<J,Lambda=-1,Omega=-1|D|J,Lambda=-1,Omega=-2> = 2*J*(J+1)*sqrt(2*(J*(J+1)-2))
<J,Lambda=-1,Omega=-1|D|J,Lambda=-1,Omega=-1> = -J*(J*(J*(J+2)+9)+8)
<J,Lambda=-1,Omega=-1|D|J,Lambda=-1,Omega=0> = 2*(J*(J+1)+2)*sqrt(2*J*(J+1))
<J,Lambda=-1,Omega=0|D|J,Lambda=-1,Omega=-2> = -2*sqrt(J*(J*(J*(J+2)-1)-2))
<J,Lambda=-1,Omega=0|D|J,Lambda=-1,Omega=-1> = 2*(J*(J+1)+2)*sqrt(2*J*(J+1))
<J,Lambda=-1,Omega=0|D|J,Lambda=-1,Omega=0> = -J*(J*(J*(J+2)+7)+6)-4
<J,Lambda=1,Omega=0|D|J,Lambda=1,Omega=0> = -J*(J*(J*(J+2)+7)+6)-4
<J,Lambda=1,Omega=0|D|J,Lambda=1,Omega=1> = 2*(J*(J+1)+2)*sqrt(2*J*(J+1))
<J,Lambda=1,Omega=0|D|J,Lambda=1,Omega=2> = -2*sqrt(J*(J*(J*(J+2)-1)-2))
<J,Lambda=1,Omega=1|D|J,Lambda=1,Omega=0> = 2*(J*(J+1)+2)*sqrt(2*J*(J+1))
<J,Lambda=1,Omega=1|D|J,Lambda=1,Omega=1> = -J*(J*(J*(J+2)+9)+8)
<J,Lambda=1,Omega=1|D|J,Lambda=1,Omega=2> = 2*J*(J+1)*sqrt(2*(J*(J+1)-2))
<J,Lambda=1,Omega=2|D|J,Lambda=1,Omega=0> = -2*sqrt(J*(J*(J*(J+2)-1)-2))
<J,Lambda=1,Omega=2|D|J,Lambda=1,Omega=1> = 2*J*(J+1)*sqrt(2*(J*(J+1)-2))
<J,Lambda=1,Omega=2|D|J,Lambda=1,Omega=2> = -J*(J*(J*(J+2)-1)-2)
Symmetrized:
<J,Lambda=1,Omega=2|D|J,Lambda=1,Omega=2> = -J*(J*(J*(J+2)-1)-2)
<J,Lambda=1,Omega=2|D|J,Lambda=1,Omega=2> = 2*J*(J+1)*sqrt(2*(J*(J+1)-2))
<J,Lambda=1,Omega=2|D|J,Lambda=1,Omega=1> = -2*sqrt(J*(J*(J*(J+2)-1)-2))
<J,Lambda=1,Omega=2|D|J,Lambda=1,Omega=1> = 2*J*(J+1)*sqrt(2*(J*(J+1)-2))
<J,Lambda=1,Omega=2|D|J,Lambda=1,Omega=0> = -J*(J*(J*(J+2)+9)+8)
<J,Lambda=1,Omega=2|D|J,Lambda=1,Omega=0> = 2*(J*(J+1)+2)*sqrt(2*J*(J+1))
<J,Lambda=1,Omega=1|D|J,Lambda=1,Omega=2> = -2*sqrt(J*(J*(J*(J+2)-1)-2))
<J,Lambda=1,Omega=1|D|J,Lambda=1,Omega=2> = 2*(J*(J+1)+2)*sqrt(2*J*(J+1))
<J,Lambda=1,Omega=1|D|J,Lambda=1,Omega=1> = -J*(J*(J*(J+2)+7)+6)-4
<J,Lambda=1,Omega=1|D|J,Lambda=1,Omega=1> = -J*(J*(J*(J+2)+7)+6)-4
<J,Lambda=1,Omega=1|D|J,Lambda=1,Omega=0> = 2*(J*(J+1)+2)*sqrt(2*J*(J+1))
<J,Lambda=1,Omega=1|D|J,Lambda=1,Omega=0> = -2*sqrt(J*(J*(J*(J+2)-1)-2))
<J,Lambda=1,Omega=0|D|J,Lambda=1,Omega=2> = 2*(J*(J+1)+2)*sqrt(2*J*(J+1))
<J,Lambda=1,Omega=0|D|J,Lambda=1,Omega=2> = -J*(J*(J*(J+2)+9)+8)
<J,Lambda=1,Omega=0|D|J,Lambda=1,Omega=1> = 2*J*(J+1)*sqrt(2*(J*(J+1)-2))
<J,Lambda=1,Omega=0|D|J,Lambda=1,Omega=1> = -2*sqrt(J*(J*(J*(J+2)-1)-2))
<J,Lambda=1,Omega=0|D|J,Lambda=1,Omega=0> = 2*J*(J+1)*sqrt(2*(J*(J+1)-2))
<J,Lambda=1,Omega=0|D|J,Lambda=1,Omega=0> = -J*(J*(J*(J+2)-1)-2)

<J,Lambda=-1,Omega=-2|AD|J,Lambda=-1,Omega=-2> = J*(J+1)-2
<J,Lambda=-1,Omega=-2|AD|J,Lambda=-1,Omega=-1> = (-sqrt(2*(J*(J+1)-2)))/2
<J,Lambda=-1,Omega=-1|AD|J,Lambda=-1,Omega=-2> = (-sqrt(2*(J*(J+1)-2)))/2
<J,Lambda=-1,Omega=-1|AD|J,Lambda=-1,Omega=0> = sqrt(2*J*(J+1))/2
<J,Lambda=-1,Omega=0|AD|J,Lambda=-1,Omega=-1> = sqrt(2*J*(J+1))/2
<J,Lambda=-1,Omega=0|AD|J,Lambda=-1,Omega=0> = -J*(J+1)-2
<J,Lambda=1,Omega=0|AD|J,Lambda=1,Omega=0> = -J*(J+1)-2
<J,Lambda=1,Omega=0|AD|J,Lambda=1,Omega=1> = sqrt(2*J*(J+1))/2
<J,Lambda=1,Omega=1|AD|J,Lambda=1,Omega=0> = sqrt(2*J*(J+1))/2
<J,Lambda=1,Omega=1|AD|J,Lambda=1,Omega=2> = (-sqrt(2*(J*(J+1)-2)))/2
<J,Lambda=1,Omega=2|AD|J,Lambda=1,Omega=1> = (-sqrt(2*(J*(J+1)-2)))/2
<J,Lambda=1,Omega=2|AD|J,Lambda=1,Omega=2> = J*(J+1)-2
Symmetrized:
<J,Lambda=1,Omega=2|AD|J,Lambda=1,Omega=2> = J*(J+1)-2
<J,Lambda=1,Omega=2|AD|J,Lambda=1,Omega=2> = (-sqrt(2*(J*(J+1)-2)))/2
<J,Lambda=1,Omega=2|AD|J,Lambda=1,Omega=1> = (-sqrt(2*(J*(J+1)-2)))/2
<J,Lambda=1,Omega=2|AD|J,Lambda=1,Omega=1> = sqrt(2*J*(J+1))/2
<J,Lambda=1,Omega=1|AD|J,Lambda=1,Omega=2> = sqrt(2*J*(J+1))/2
<J,Lambda=1,Omega=1|AD|J,Lambda=1,Omega=2> = -J*(J+1)-2
<J,Lambda=1,Omega=1|AD|J,Lambda=1,Omega=0> = -J*(J+1)-2
<J,Lambda=1,Omega=1|AD|J,Lambda=1,Omega=0> = sqrt(2*J*(J+1))/2
<J,Lambda=1,Omega=0|AD|J,Lambda=1,Omega=1> = sqrt(2*J*(J+1))/2
<J,Lambda=1,Omega=0|AD|J,Lambda=1,Omega=1> = (-sqrt(2*(J*(J+1)-2)))/2
<J,Lambda=1,Omega=0|AD|J,Lambda=1,Omega=0> = (-sqrt(2*(J*(J+1)-2)))/2
<J,Lambda=1,Omega=0|AD|J,Lambda=1,Omega=0> = J*(J+1)-2

LinearPerturbation: Excited <a3Pv=13|LS|Iv=3>
Value -11.5278 Size of perturbation.
<J,Lambda=-1,Omega=0|LS|J,Lambda=0,Omega=0> = -1/sqrt(3)
<J,Lambda=1,Omega=0|LS|J,Lambda=0,Omega=0> = -1/sqrt(3)

Symmetrized Matrix Elements of <a3Pv=13|LS|Iv=3>
<J,Lambda=1,Omega=0|LS|J,Lambda=0,Omega=0> = -1/sqrt(3)

LinearPerturbation: Excited <a3Pv=13|LS|Iv=4>
Value -16.442 Size of perturbation.
<J,Lambda=-1,Omega=0|LS|J,Lambda=0,Omega=0> = -1/sqrt(3)
<J,Lambda=1,Omega=0|LS|J,Lambda=0,Omega=0> = -1/sqrt(3)

Symmetrized Matrix Elements of <a3Pv=13|LS|Iv=4>
<J,Lambda=1,Omega=0|LS|J,Lambda=0,Omega=0> = -1/sqrt(3)

LinearPerturbation: Excited <a3Pv=13|LS|ev=4>
Value 24.7288685468983 Float, Std Dev = 0.0197211631393428 Size of perturbation.
<J,Lambda=-1,Omega=-1|LS|J,Lambda=0,Omega=-1> = 1/sqrt(6)
<J,Lambda=-1,Omega=0|LS|J,Lambda=0,Omega=0> = 1/sqrt(6)
<J,Lambda=1,Omega=0|LS|J,Lambda=0,Omega=0> = -1/sqrt(6)
<J,Lambda=1,Omega=1|LS|J,Lambda=0,Omega=1> = -1/sqrt(6)

Symmetrized Matrix Elements of <a3Pv=13|LS|ev=4>
<J,Lambda=1,Omega=1|LS|J,Lambda=0,Omega=1> = 1/sqrt(6)
<J,Lambda=1,Omega=1|LS|J,Lambda=0,Omega=1> = 1/sqrt(6)
<J,Lambda=1,Omega=0|LS|J,Lambda=0,Omega=0> = -1/sqrt(6)

LinearPerturbation: Excited <a3Pv=13|J+L-|ev=4>
Value -0.0598221863162786 Float, Std Dev = 0.000255625514810117 Size of perturbation.

<J,Lambda=-1,Omega=-2|J+L-|J,Lambda=0,Omega=-1> = -sqrt(J*(J+1)-2)
<J,Lambda=-1,Omega=-1|J+L-|J,Lambda=0,Omega=0> = -sqrt(J*(J+1))
<J,Lambda=-1,Omega=0|J+L-|J,Lambda=0,Omega=1> = -sqrt(J*(J+1))
<J,Lambda=1,Omega=0|J+L-|J,Lambda=0,Omega=-1> = sqrt(J*(J+1))
<J,Lambda=1,Omega=1|J+L-|J,Lambda=0,Omega=0> = sqrt(J*(J+1))
<J,Lambda=1,Omega=2|J+L-|J,Lambda=0,Omega=1> = sqrt(J*(J+1)-2)

Symmetrized Matrix Elements of <a3Pv=13|J+L-|ev=4>
<J,Lambda=1,Omega=2|J+L-|J,Lambda=0,Omega=1> = -sqrt(J*(J+1)-2)
<J,Lambda=1,Omega=2|J+L-|J,Lambda=0,Omega=1> = -sqrt(J*(J+1))
<J,Lambda=1,Omega=1|J+L-|J,Lambda=0,Omega=0> = -sqrt(J*(J+1))
<J,Lambda=1,Omega=0|J+L-|J,Lambda=0,Omega=1> = sqrt(J*(J+1))
<J,Lambda=1,Omega=0|J+L-|J,Lambda=0,Omega=1> = sqrt(J*(J+1))

LinearPerturbation: Excited <a3Pv=13|LS|dv=7>
Value 23.2188654276278 Float, Std Dev = 0.0441226740988454 Size of perturbation.
<J,Lambda=-1,Omega=-2|LS|J,Lambda=-2,Omega=-2> = -1/sqrt(6)
<J,Lambda=-1,Omega=-1|LS|J,Lambda=-2,Omega=-1> = -1/sqrt(6)
<J,Lambda=1,Omega=1|LS|J,Lambda=2,Omega=1> = -1/sqrt(6)
<J,Lambda=1,Omega=2|LS|J,Lambda=2,Omega=2> = -1/sqrt(6)

Symmetrized Matrix Elements of <a3Pv=13|LS|dv=7>
<J,Lambda=1,Omega=2|LS|J,Lambda=2,Omega=2> = -1/sqrt(6)
<J,Lambda=1,Omega=2|LS|J,Lambda=2,Omega=2> = -1/sqrt(6)
<J,Lambda=1,Omega=1|LS|J,Lambda=2,Omega=1> = -1/sqrt(6)
<J,Lambda=1,Omega=1|LS|J,Lambda=2,Omega=1> = -1/sqrt(6)

LinearPerturbation: Excited <a3Pv=13|J+L-|dv=7>
Value -0.062795149954351 Float, Std Dev = 0.000453447699962488 Size of perturbation.
<J,Lambda=-1,Omega=-2|J+L-|J,Lambda=-2,Omega=-3> = sqrt(J*(J+1)-6)
<J,Lambda=-1,Omega=-1|J+L-|J,Lambda=-2,Omega=-2> = sqrt(J*(J+1)-2)
<J,Lambda=-1,Omega=0|J+L-|J,Lambda=-2,Omega=-1> = sqrt(J*(J+1))
<J,Lambda=1,Omega=0|J+L-|J,Lambda=2,Omega=1> = sqrt(J*(J+1))
<J,Lambda=1,Omega=1|J+L-|J,Lambda=2,Omega=2> = sqrt(J*(J+1)-2)
<J,Lambda=1,Omega=2|J+L-|J,Lambda=2,Omega=3> = sqrt(J*(J+1)-6)

Symmetrized Matrix Elements of <a3Pv=13|J+L-|dv=7>
<J,Lambda=1,Omega=2|J+L-|J,Lambda=2,Omega=3> = sqrt(J*(J+1)-6)
<J,Lambda=1,Omega=2|J+L-|J,Lambda=2,Omega=3> = sqrt(J*(J+1)-2)
<J,Lambda=1,Omega=1|J+L-|J,Lambda=2,Omega=2> = sqrt(J*(J+1))
<J,Lambda=1,Omega=1|J+L-|J,Lambda=2,Omega=2> = sqrt(J*(J+1))
<J,Lambda=1,Omega=0|J+L-|J,Lambda=2,Omega=1> = sqrt(J*(J+1)-2)
<J,Lambda=1,Omega=0|J+L-|J,Lambda=2,Omega=1> = sqrt(J*(J+1)-6)

LinearPerturbation: Excited <a3Pv=13|LS|a'v=12>
Value 2.39113881014602 Float, Std Dev = 0.203999893082939 Size of perturbation.
<J,Lambda=-1,Omega=-1|LS|J,Lambda=0,Omega=-1> = -1/sqrt(6)
<J,Lambda=-1,Omega=0|LS|J,Lambda=0,Omega=0> = -1/sqrt(6)
<J,Lambda=1,Omega=0|LS|J,Lambda=0,Omega=0> = -1/sqrt(6)
<J,Lambda=1,Omega=1|LS|J,Lambda=0,Omega=1> = -1/sqrt(6)

Symmetrized Matrix Elements of <a3Pv=13|LS|a'v=12>
<J,Lambda=1,Omega=1|LS|J,Lambda=0,Omega=1> = -1/sqrt(6)
<J,Lambda=1,Omega=1|LS|J,Lambda=0,Omega=1> = -1/sqrt(6)
<J,Lambda=1,Omega=0|LS|J,Lambda=0,Omega=0> = -1/sqrt(6)

LinearPerturbation: Excited <a3Pv=13|J+L-|a'v=12>
Value -0.0112468031812311 Float, Std Dev = 0.00125606908675765 Size of perturbation.
<J,Lambda=-1,Omega=-2|J+L-|J,Lambda=0,Omega=-1> = sqrt(J*(J+1)-2)
<J,Lambda=-1,Omega=-1|J+L-|J,Lambda=0,Omega=0> = sqrt(J*(J+1))
<J,Lambda=-1,Omega=0|J+L-|J,Lambda=0,Omega=1> = sqrt(J*(J+1))
<J,Lambda=1,Omega=0|J+L-|J,Lambda=0,Omega=-1> = sqrt(J*(J+1))
<J,Lambda=1,Omega=1|J+L-|J,Lambda=0,Omega=0> = sqrt(J*(J+1))
<J,Lambda=1,Omega=2|J+L-|J,Lambda=0,Omega=1> = sqrt(J*(J+1)-2)

Symmetrized Matrix Elements of <a3Pv=13|J+L-|a'v=12>
<J,Lambda=1,Omega=2|J+L-|J,Lambda=0,Omega=1> = sqrt(J*(J+1)-2)
<J,Lambda=1,Omega=2|J+L-|J,Lambda=0,Omega=1> = sqrt(J*(J+1))
<J,Lambda=1,Omega=1|J+L-|J,Lambda=0,Omega=0> = sqrt(J*(J+1))
<J,Lambda=1,Omega=0|J+L-|J,Lambda=0,Omega=1> = sqrt(J*(J+1))
<J,Lambda=1,Omega=0|J+L-|J,Lambda=0,Omega=1> = sqrt(J*(J+1))

LinearPerturbation: Excited <a3Pv=13|LS|Dsv=3>
Value -22.8568 Size of perturbation.
<J,Lambda=-1,Omega=-2|LS|J,Lambda=-2,Omega=-2> = 1/sqrt(3)
<J,Lambda=1,Omega=2|LS|J,Lambda=2,Omega=2> = -1/sqrt(3)

Symmetrized Matrix Elements of <a3Pv=13|LS|Dsv=3>
<J,Lambda=1,Omega=2|LS|J,Lambda=2,Omega=2> = 1/sqrt(3)
<J,Lambda=1,Omega=2|LS|J,Lambda=2,Omega=2> = -1/sqrt(3)

TransitionMoments: <Excited|mu|X>
SphericalTransitionMoment: <Excited|mu|X> <Av=2|T(1)|Xv=0>
Rank 1
Component Auto => 1
Strength 1 Transition (dipole) moment / Debye - note that relative intensities are proportional to the square of this value.
Reduced Matrix Elements of <Av=2|T(1)|Xv=0>
<J-1,Lambda=-1,Omega=-1|T(1)|J,Lambda=0,Omega=0> = sqrt(2*(J-1))/2
<J-1,Lambda=1,Omega=1|T(1)|J,Lambda=0,Omega=0> = sqrt(2*(J-1))/2
<J,Lambda=-1,Omega=-1|T(1)|J,Lambda=0,Omega=0> = (2*J+1)/sqrt(2*(2*J+1))
<J,Lambda=1,Omega=1|T(1)|J,Lambda=0,Omega=0> = (-2*J-1)/sqrt(2*(2*J+1))
<J+1,Lambda=-1,Omega=-1|T(1)|J,Lambda=0,Omega=0> = sqrt(2*(J+2))/2

<J+1,Lambda=1,Omega=1|T(1)|J,Lambda=0,Omega=0> = sqrt(2*(J+2))/2
Line Strengths from <Av=2|T(1)|Xv=0>
S<J-1,Lambda=-1,Omega=-1|T(1)|J,Lambda=0,Omega=0> = (J-1)/2
S<J-1,Lambda=1,Omega=1|T(1)|J,Lambda=0,Omega=0> = (J-1)/2
S<J,Lambda=-1,Omega=-1|T(1)|J,Lambda=0,Omega=0> = (2*J+1)/2
S<J,Lambda=1,Omega=1|T(1)|J,Lambda=0,Omega=0> = (-2*J-1)/-2
S<J+1,Lambda=-1,Omega=-1|T(1)|J,Lambda=0,Omega=0> = (J+2)/2
S<J+1,Lambda=1,Omega=1|T(1)|J,Lambda=0,Omega=0> = (J+2)/2
Symmetrized Reduced Matrix Elements of <Av=2|T(1)|Xv=0>
<J-1,Lambda=1,Omega=1 + |T(1)|J,Lambda=0,Omega=0> = sqrt(J-1)
<J,Lambda=1,Omega=1 - |T(1)|J,Lambda=0,Omega=0> = (-2*J-1)/sqrt(2*J+1)
<J+1,Lambda=1,Omega=1 + |T(1)|J,Lambda=0,Omega=0> = sqrt(J+2)
Symmetrized Line Strengths from <Av=2|T(1)|Xv=0>
S<J-1,Lambda=1,Omega=1 + |T(1)|J,Lambda=0,Omega=0> = J-1
S<J,Lambda=1,Omega=1 - |T(1)|J,Lambda=0,Omega=0> = (-2*J-1)/-1
S<J+1,Lambda=1,Omega=1 + |T(1)|J,Lambda=0,Omega=0> = J+2

ForbiddenTransitionMoment: <Excited|mu|X> <a3Pv=13|fT(1,1)|Xv=0>
Rank 1
Component 1
Magnetic False
Strength 1 Transition (dipole) moment / Debye - note that relative intensities are proportional to the square of this value.
Reduced Matrix Elements of <a3Pv=13|fT(1,1)|Xv=0>
<J-1,Lambda=-1,Omega=-1|fT(1,1)|J,Lambda=0,Omega=0> = (-sqrt(2*(J-1)))/2
<J-1,Lambda=1,Omega=1|fT(1,1)|J,Lambda=0,Omega=0> = sqrt(2*(J-1))/2
<J,Lambda=-1,Omega=-1|fT(1,1)|J,Lambda=0,Omega=0> = (-2*J-1)/sqrt(2*(2*J+1))
<J,Lambda=1,Omega=1|fT(1,1)|J,Lambda=0,Omega=0> = (-2*J-1)/sqrt(2*(2*J+1))
<J+1,Lambda=-1,Omega=-1|fT(1,1)|J,Lambda=0,Omega=0> = (-sqrt(2*(J+2)))/2
<J+1,Lambda=1,Omega=1|fT(1,1)|J,Lambda=0,Omega=0> = sqrt(2*(J+2))/2
Line Strengths from <a3Pv=13|fT(1,1)|Xv=0>
S<J-1,Lambda=-1,Omega=-1|fT(1,1)|J,Lambda=0,Omega=0> = (J-1)/2
S<J-1,Lambda=1,Omega=1|fT(1,1)|J,Lambda=0,Omega=0> = (J-1)/2
S<J,Lambda=-1,Omega=-1|fT(1,1)|J,Lambda=0,Omega=0> = (-2*J-1)/-2
S<J,Lambda=1,Omega=1|fT(1,1)|J,Lambda=0,Omega=0> = (-2*J-1)/-2
S<J+1,Lambda=-1,Omega=-1|fT(1,1)|J,Lambda=0,Omega=0> = (J+2)/2
S<J+1,Lambda=1,Omega=1|fT(1,1)|J,Lambda=0,Omega=0> = (J+2)/2
Symmetrized Reduced Matrix Elements of <a3Pv=13|fT(1,1)|Xv=0>
<J-1,Lambda=1,Omega=1 - |fT(1,1)|J,Lambda=0,Omega=0> = sqrt(J-1)
<J,Lambda=1,Omega=1 + |fT(1,1)|J,Lambda=0,Omega=0> = (-2*J-1)/sqrt(2*J+1)
<J+1,Lambda=1,Omega=1 - |fT(1,1)|J,Lambda=0,Omega=0> = sqrt(J+2)
Symmetrized Line Strengths from <a3Pv=13|fT(1,1)|Xv=0>
S<J-1,Lambda=1,Omega=1 - |fT(1,1)|J,Lambda=0,Omega=0> = J-1
S<J,Lambda=1,Omega=1 + |fT(1,1)|J,Lambda=0,Omega=0> = (-2*J-1)/-1
S<J+1,Lambda=1,Omega=1 - |fT(1,1)|J,Lambda=0,Omega=0> = J+2

ForbiddenTransitionMoment: <Excited|mu|X> <dv=7|fT(1,1)|Xv=0>
Rank 1
Component 1
Magnetic False
Strength 1 Transition (dipole) moment / Debye - note that relative intensities are proportional to the square of this value.
Reduced Matrix Elements of <dv=7|fT(1,1)|Xv=0>
<J-1,Lambda=-2,Omega=-1|fT(1,1)|J,Lambda=0,Omega=0> = (-sqrt(2*(J-1)))/2
<J-1,Lambda=2,Omega=1|fT(1,1)|J,Lambda=0,Omega=0> = sqrt(2*(J-1))/2
<J,Lambda=-2,Omega=-1|fT(1,1)|J,Lambda=0,Omega=0> = (-2*J-1)/sqrt(2*(2*J+1))
<J,Lambda=2,Omega=1|fT(1,1)|J,Lambda=0,Omega=0> = (-2*J-1)/sqrt(2*(2*J+1))
<J+1,Lambda=-2,Omega=-1|fT(1,1)|J,Lambda=0,Omega=0> = (-sqrt(2*(J+2)))/2
<J+1,Lambda=2,Omega=1|fT(1,1)|J,Lambda=0,Omega=0> = sqrt(2*(J+2))/2
Line Strengths from <dv=7|fT(1,1)|Xv=0>
S<J-1,Lambda=-2,Omega=-1|fT(1,1)|J,Lambda=0,Omega=0> = (J-1)/2
S<J-1,Lambda=2,Omega=1|fT(1,1)|J,Lambda=0,Omega=0> = (J-1)/2
S<J,Lambda=-2,Omega=-1|fT(1,1)|J,Lambda=0,Omega=0> = (-2*J-1)/-2
S<J,Lambda=2,Omega=1|fT(1,1)|J,Lambda=0,Omega=0> = (-2*J-1)/-2
S<J+1,Lambda=-2,Omega=-1|fT(1,1)|J,Lambda=0,Omega=0> = (J+2)/2
S<J+1,Lambda=2,Omega=1|fT(1,1)|J,Lambda=0,Omega=0> = (J+2)/2
Symmetrized Reduced Matrix Elements of <dv=7|fT(1,1)|Xv=0>
<J-1,Lambda=2,Omega=1 - |fT(1,1)|J,Lambda=0,Omega=0> = sqrt(J-1)
<J,Lambda=2,Omega=1 + |fT(1,1)|J,Lambda=0,Omega=0> = (-2*J-1)/sqrt(2*J+1)
<J+1,Lambda=2,Omega=1 - |fT(1,1)|J,Lambda=0,Omega=0> = sqrt(J+2)
Symmetrized Line Strengths from <dv=7|fT(1,1)|Xv=0>
S<J-1,Lambda=2,Omega=1 - |fT(1,1)|J,Lambda=0,Omega=0> = J-1
S<J,Lambda=2,Omega=1 + |fT(1,1)|J,Lambda=0,Omega=0> = (-2*J-1)/-1
S<J+1,Lambda=2,Omega=1 - |fT(1,1)|J,Lambda=0,Omega=0> = J+2

ForbiddenTransitionMoment: <Excited|mu|X> <a'v=12|fT(1,1)|Xv=0>
Rank 1
Component 1
Magnetic False
Strength 1 Transition (dipole) moment / Debye - note that relative intensities are proportional to the square of this value.
Reduced Matrix Elements of <a'v=12|fT(1,1)|Xv=0>
<J-1,Lambda=0,Omega=-1|fT(1,1)|J,Lambda=0,Omega=0> = (-sqrt(2*(J-1)))/2
<J-1,Lambda=0,Omega=1|fT(1,1)|J,Lambda=0,Omega=0> = sqrt(2*(J-1))/2
<J,Lambda=0,Omega=-1|fT(1,1)|J,Lambda=0,Omega=0> = (-2*J-1)/sqrt(2*(2*J+1))
<J,Lambda=0,Omega=1|fT(1,1)|J,Lambda=0,Omega=0> = (-2*J-1)/sqrt(2*(2*J+1))
<J+1,Lambda=0,Omega=-1|fT(1,1)|J,Lambda=0,Omega=0> = (-sqrt(2*(J+2)))/2
<J+1,Lambda=0,Omega=1|fT(1,1)|J,Lambda=0,Omega=0> = sqrt(2*(J+2))/2
Line Strengths from <a'v=12|fT(1,1)|Xv=0>
S<J-1,Lambda=0,Omega=-1|fT(1,1)|J,Lambda=0,Omega=0> = (J-1)/2
S<J-1,Lambda=0,Omega=1|fT(1,1)|J,Lambda=0,Omega=0> = (J-1)/2
S<J,Lambda=0,Omega=-1|fT(1,1)|J,Lambda=0,Omega=0> = (-2*J-1)/-2
S<J,Lambda=0,Omega=1|fT(1,1)|J,Lambda=0,Omega=0> = (-2*J-1)/-2
S<J+1,Lambda=0,Omega=-1|fT(1,1)|J,Lambda=0,Omega=0> = (J+2)/2
S<J+1,Lambda=0,Omega=1|fT(1,1)|J,Lambda=0,Omega=0> = (J+2)/2

Symmetrized Reduced Matrix Elements of <a'v=12|fT(1,1)|Xv=0>
<J-1,Lambda=0,Omega=1 - |fT(1,1)|J,Lambda=0,Omega=0> = sqrt(J-1)
<J,Lambda=0,Omega=1 + |fT(1,1)|J,Lambda=0,Omega=0> = (-2*J-1)/sqrt(2*J+1)
<J+1,Lambda=0,Omega=1 - |fT(1,1)|J,Lambda=0,Omega=0> = sqrt(J+2)
Symmetrized Line Strengths from <a'v=12|fT(1,1)|Xv=0>
S<J-1,Lambda=0,Omega=1 - |fT(1,1)|J,Lambda=0,Omega=0> = J-1
S<J,Lambda=0,Omega=1 + |fT(1,1)|J,Lambda=0,Omega=0> = (-2*J-1)/-1
S<J+1,Lambda=0,Omega=1 - |fT(1,1)|J,Lambda=0,Omega=0> = J+2

ForbiddenTransitionMoment: <Excited|mu|X> <ev=4|fT(1,1)|Xv=0>
Rank 1
Component 1
Magnetic False
Strength 1 Transition (dipole) moment / Debye - note that relative intensities are proportional to the square of this value.
Reduced Matrix Elements of <ev=4|fT(1,1)|Xv=0>
<J-1,Lambda=0,Omega=-1|fT(1,1)|J,Lambda=0,Omega=0> = sqrt(2*(J-1))/2
<J-1,Lambda=0,Omega=1|fT(1,1)|J,Lambda=0,Omega=0> = sqrt(2*(J-1))/2
<J,Lambda=0,Omega=-1|fT(1,1)|J,Lambda=0,Omega=0> = (2*J+1)/sqrt(2*(2*J+1))
<J,Lambda=0,Omega=1|fT(1,1)|J,Lambda=0,Omega=0> = (-2*J-1)/sqrt(2*(2*J+1))
<J+1,Lambda=0,Omega=-1|fT(1,1)|J,Lambda=0,Omega=0> = sqrt(2*(J+2))/2
<J+1,Lambda=0,Omega=1|fT(1,1)|J,Lambda=0,Omega=0> = sqrt(2*(J+2))/2
Line Strengths from <ev=4|fT(1,1)|Xv=0>
S<J-1,Lambda=0,Omega=-1|fT(1,1)|J,Lambda=0,Omega=0> = (J-1)/2
S<J-1,Lambda=0,Omega=1|fT(1,1)|J,Lambda=0,Omega=0> = (J-1)/2
S<J,Lambda=0,Omega=-1|fT(1,1)|J,Lambda=0,Omega=0> = (2*J+1)/2
S<J,Lambda=0,Omega=1|fT(1,1)|J,Lambda=0,Omega=0> = (-2*J-1)/-2
S<J+1,Lambda=0,Omega=-1|fT(1,1)|J,Lambda=0,Omega=0> = (J+2)/2
S<J+1,Lambda=0,Omega=1|fT(1,1)|J,Lambda=0,Omega=0> = (J+2)/2
Symmetrized Reduced Matrix Elements of <ev=4|fT(1,1)|Xv=0>
<J-1,Lambda=0,Omega=1 + |fT(1,1)|J,Lambda=0,Omega=0> = sqrt(J-1)
<J,Lambda=0,Omega=1 - |fT(1,1)|J,Lambda=0,Omega=0> = (-2*J-1)/sqrt(2*J+1)
<J+1,Lambda=0,Omega=1 + |fT(1,1)|J,Lambda=0,Omega=0> = sqrt(J+2)
Symmetrized Line Strengths from <ev=4|fT(1,1)|Xv=0>
S<J-1,Lambda=0,Omega=1 + |fT(1,1)|J,Lambda=0,Omega=0> = J-1
S<J,Lambda=0,Omega=1 - |fT(1,1)|J,Lambda=0,Omega=0> = (-2*J-1)/-1
S<J+1,Lambda=0,Omega=1 + |fT(1,1)|J,Lambda=0,Omega=0> = J+2

ForbiddenTransitionMoment: <Excited|mu|X> <Iv=3|fT(1,1)|Xv=0>
Rank 1
Component 1
Magnetic False
Strength 1 Transition (dipole) moment / Debye - note that relative intensities are proportional to the square of this value.
ForbiddenTransitionMoment: <Excited|mu|X> <Dsv=3|fT(1,1)|Xv=0>
Rank 1
Component 1
Magnetic False
Strength 1 Transition (dipole) moment / Debye - note that relative intensities are proportional to the square of this value.
TransitionMoments: <B|mu|X>
SphericalTransitionMoment: <B|mu|X> <Bv=0|T(1)|Xv=0>
Rank 1
Component Auto => 0
Strength 1 Transition (dipole) moment / Debye - note that relative intensities are proportional to the square of this value.
Reduced Matrix Elements of <Bv=0|T(1)|Xv=0>
<J-1,Lambda=0,Omega=0|T(1)|J,Lambda=0,Omega=0> = -sqrt(J)
<J+1,Lambda=0,Omega=0|T(1)|J,Lambda=0,Omega=0> = (J+1)/sqrt(J+1)
Line Strengths from <Bv=0|T(1)|Xv=0>
S<J-1,Lambda=0,Omega=0|T(1)|J,Lambda=0,Omega=0> = J
S<J+1,Lambda=0,Omega=0|T(1)|J,Lambda=0,Omega=0> = J+1
Symmetrized Reduced Matrix Elements of <Bv=0|T(1)|Xv=0>
<J-1,Lambda=0,Omega=0|T(1)|J,Lambda=0,Omega=0> = -sqrt(J)
<J+1,Lambda=0,Omega=0|T(1)|J,Lambda=0,Omega=0> = (J+1)/sqrt(J+1)
Symmetrized Line Strengths from <Bv=0|T(1)|Xv=0>
S<J-1,Lambda=0,Omega=0|T(1)|J,Lambda=0,Omega=0> = J
S<J+1,Lambda=0,Omega=0|T(1)|J,Lambda=0,Omega=0> = J+1

TransitionMoments: <B|mu|Excited>
SphericalTransitionMoment: <B|mu|Excited> <Bv=0|T(1)|Av=2>
Rank 1
Component Auto => 1
Strength 1 Transition (dipole) moment / Debye - note that relative intensities are proportional to the square of this value.
Reduced Matrix Elements of <Bv=0|T(1)|Av=2>
<J-1,Lambda=0,Omega=0|T(1)|J,Lambda=-1,Omega=-1> = sqrt(2*(J+1))/2
<J-1,Lambda=0,Omega=0|T(1)|J,Lambda=1,Omega=1> = sqrt(2*(J+1))/2
<J,Lambda=0,Omega=0|T(1)|J,Lambda=-1,Omega=-1> = (-2*J-1)/sqrt(2*(2*J+1))
<J,Lambda=0,Omega=0|T(1)|J,Lambda=1,Omega=1> = (2*J+1)/sqrt(2*(2*J+1))
<J+1,Lambda=0,Omega=0|T(1)|J,Lambda=-1,Omega=-1> = sqrt(2*J)/2
<J+1,Lambda=0,Omega=0|T(1)|J,Lambda=1,Omega=1> = sqrt(2*J)/2
Line Strengths from <Bv=0|T(1)|Av=2>
S<J-1,Lambda=0,Omega=0|T(1)|J,Lambda=-1,Omega=-1> = (J+1)/2
S<J-1,Lambda=0,Omega=0|T(1)|J,Lambda=1,Omega=1> = (J+1)/2
S<J,Lambda=0,Omega=0|T(1)|J,Lambda=-1,Omega=-1> = (-2*J-1)/-2
S<J,Lambda=0,Omega=0|T(1)|J,Lambda=1,Omega=1> = (2*J+1)/2
S<J+1,Lambda=0,Omega=0|T(1)|J,Lambda=-1,Omega=-1> = J/2
S<J+1,Lambda=0,Omega=0|T(1)|J,Lambda=1,Omega=1> = J/2
Symmetrized Reduced Matrix Elements of <Bv=0|T(1)|Av=2>
<J-1,Lambda=0,Omega=0|T(1)| + J,Lambda=1,Omega=1> = sqrt(J+1)
<J,Lambda=0,Omega=0|T(1)| - J,Lambda=1,Omega=1> = (2*J+1)/sqrt(2*J+1)
<J+1,Lambda=0,Omega=0|T(1)| + J,Lambda=1,Omega=1> = sqrt(J)
Symmetrized Line Strengths from <Bv=0|T(1)|Av=2>
S<J-1,Lambda=0,Omega=0|T(1)| + J,Lambda=1,Omega=1> = J+1
S<J,Lambda=0,Omega=0|T(1)| - J,Lambda=1,Omega=1> = 2*J+1
S<J+1,Lambda=0,Omega=0|T(1)| + J,Lambda=1,Omega=1> = J

TransitionMoments: <C|mu|X>
SphericalTransitionMoment: <C|mu|X> <Cv=0|T(1)|Xv=0>
Rank 1
Component Auto => 0
Strength 1 Transition (dipole) moment / Debye - note that relative intensities are proportional to the square of this value.
Matrix elements as for <B|mu|X> <Bv=0|T(1)|Xv=0>
TransitionMoments: <C|mu|Excited>
SphericalTransitionMoment: <C|mu|Excited> <Cv=0|T(1)|Av=2>
Rank 1
Component Auto => 1
Strength 1 Transition (dipole) moment / Debye - note that relative intensities are proportional to the square of this value.
Matrix elements as for <B|mu|Excited> <Bv=0|T(1)|Av=2>
Simulation: Simulation
Intensity factor:0.333333333333333 (1 from Spol)
IntensityUnits Arbitrary
LifeModel lmNone
PlotUnits cm1
Fluorescence False
nDF 50000
WidthMult 6
AutoMin True
AutoMax True
ShowSum False
ShowParts True
ShowFortrat True
UseUpper False
ShowSymmetry False
ShowDeltaJ True
ScaleMarkSize True
UseSymmetry False
UseStateNumber False
FortratQno J
PreserveArea False
SeparateScales False
LogScale False
Show2DPlot False
Auto2DY True
Fmin 19713.4264274956 Left edge of plot range.
Fmax 19725.7352925044 Right edge of plot range.
Ymax 7.90431110974814E-183 Top of plot range; updated automatically if AutoMax set.
Temperature 300 Rotational temeperature (Kelvin).
Gaussian 0.08 Gaussian contribution to linewidth (full width half maximum). If both this and a Lorentzian width (below) is set the result is a convolution of the two, a Voigt profile.
Lorentzian 0.02 Lorentzian contribution to linewidth (full width half maximum). This and the Gaussian width can also be set from the main window.
SMargin -1 Extra points to calculate off each end of required spectrum.
OThreshold 0.0001 Ignore peaks smaller than this fraction of the maximum peak intensity.
Tvib -1 Vibrational Temperature (Kelvin); set to -1 (default) to use rotational temperature for all Boltzmann factors.
MinI 1E-15 Weakest transitions to plot in Fortrat plots, as a fraction of strongest.
Tspin -1 Spin Temperature; set to -1 (default) to equilibrate different spinstates
IScale 1 Scale factor applied to all line intensities
PlotSplit 0.5 Fraction of plot area to use for simulation where plots separated (-1..1)
Fundamental Constants from CODATA 2018: h=6.62607015E-34 k=1.380649E-23 NA=6.02214076E23 e=1.602176634E-19 me=9.1093837015E-31

| Molecu | Manifol' | J' | S' | #' | Manifol" | J" | S" | #" | Position | I | Eupper | Elower | Spol | A | Width | |
|---|---|---|---|---|---|---|---|---|---|---|---|---|---|---|---|---|
| 13C18O | C | 1 | e | 1 | X | 0 | e | 1 | 91922.3502 | .3333333 | 91922.3502 | 0.0000 | .3333333 | 81197804 | 0.000000 | : rR(0) : C Cv=0 1 1 F1e - X Xv=0 0 0 F1e |
| 13C18O | C | 0 | e | 1 | X | 1 | e | 1 | 91915.3218 | .3277961 | 91918.8146 | 3.4928 | .3333333 | 2.4354e8 | 0.000000 | : pP(1) : C Cv=0 0 0 F1e - X Xv=0 1 1 F1e |
| 13C18O | C | 2 | e | 1 | X | 1 | e | 1 | 91925.9218 | .6555922 | 91929.4146 | 3.4928 | .6666667 | 97448723 | 0.000000 | : rR(1) : C Cv=0 2 2 F1e - X Xv=0 1 1 F1e |
| 13C18O | C | 1 | e | 1 | X | 2 | e | 1 | 91911.8719 | .6339926 | 91922.3502 | 10.4783 | .6666667 | 1.6234e8 | 0.000000 | : pP(2) : C Cv=0 1 1 F1e - X Xv=0 2 2 F1e |
| 13C18O | C | 3 | e | 1 | X | 2 | e | 1 | 91929.5305 | .9509888 | 91940.0088 | 10.4783 | 1.000000 | 1.0442e8 | 0.000000 | : rR(2) : C Cv=0 3 3 F1e - X Xv=0 2 2 F1e |
| 13C18O | C | 2 | e | 1 | X | 3 | e | 1 | 91908.4584 | .9043813 | 91929.4146 | 20.9562 | 1.000000 | 1.4609e8 | 0.000000 | : pP(3) : C Cv=0 2 2 F1e - X Xv=0 3 3 F1e |
| 13C18O | C | 4 | e | 1 | X | 3 | e | 1 | 91933.1830 | 1.205842 | 91954.1392 | 20.9562 | 1.333333 | 1.0830e8 | 0.000000 | : rR(3) : C Cv=0 4 4 F1e - X Xv=0 3 3 F1e |
| 13C18O | C | 3 | e | 1 | X | 4 | e | 1 | 91905.0826 | 1.127699 | 91940.0088 | 34.9261 | 1.333333 | 1.3912e8 | 0.000000 | : pP(4) : C Cv=0 3 3 F1e - X Xv=0 4 4 F1e |
| 13C18O | C | 5 | e | 1 | X | 4 | e | 1 | 91936.8786 | 1.409623 | 91971.8047 | 34.9261 | 1.666667 | 1.1078e8 | 0.000000 | : rR(4) : C Cv=0 5 5 F1e - X Xv=0 4 4 F1e |
| 13C18O | C | 4 | e | 1 | X | 5 | e | 1 | 91901.7515 | 1.296383 | 91954.1392 | 52.3877 | 1.666667 | 1.3524e8 | 0.000000 | : pP(5) : C Cv=0 4 4 F1e - X Xv=0 5 5 F1e |
| 13C18O | C | 6 | e | 1 | X | 5 | e | 1 | 91940.6086 | 1.555660 | 91992.9963 | 52.3877 | 2.000000 | 1.1249e8 | 0.000000 | : rR(5) : C Cv=0 6 6 F1e - X Xv=0 5 5 F1e |
| 13C18O | C | 5 | e | 1 | X | 6 | e | 1 | 91898.4645 | 1.406934 | 91971.8047 | 73.3402 | 2.000000 | 1.3277e8 | 0.000000 | : pP(6) : C Cv=0 5 5 F1e - X Xv=0 6 6 F1e |
| 13C18O | C | 7 | e | 1 | X | 6 | e | 1 | 91944.3808 | 1.641423 | 92017.7210 | 73.3402 | 2.333333 | 1.1376e8 | 0.000000 | : rR(6) : C Cv=0 7 7 F1e - X Xv=0 6 6 F1e |
| 13C18O | C | 6 | e | 1 | X | 7 | e | 1 | 91895.2133 | 1.459856 | 91992.9963 | 97.7830 | 2.333333 | 1.3105e8 | 0.000000 | : pP(7) : C Cv=0 6 6 F1e - X Xv=0 7 7 F1e |
| 13C18O | C | 8 | e | 1 | X | 7 | e | 1 | 91948.1899 | 1.668407 | 92045.9729 | 97.7830 | 2.666667 | 1.1473e8 | 0.000000 | : rR(7) : C Cv=0 8 8 F1e - X Xv=0 7 7 F1e |
| 13C18O | C | 7 | e | 1 | X | 8 | e | 1 | 91892.0058 | 1.459230 | 92017.7210 | 125.7152 | 2.666667 | 1.2979e8 | 0.000000 | : pP(8) : C Cv=0 7 7 F1e - X Xv=0 8 8 F1e |
| 13C18O | C | 9 | e | 1 | X | 8 | e | 1 | 91952.0397 | 1.641634 | 92077.7549 | 125.7152 | 3.000000 | 1.1550e8 | 0.000000 | : rR(8) : C Cv=0 9 9 F1e - X Xv=0 8 8 F1e |
| 13C18O | C | 8 | e | 1 | X | 9 | e | 1 | 91888.8371 | 1.411991 | 92045.9729 | 157.1358 | 3.000000 | 1.2882e8 | 0.000000 | : pP(9) : C Cv=0 8 8 F1e - X Xv=0 9 9 F1e |
| 13C18O | C | 10 | e | 1 | X | 9 | e | 1 | 91955.9229 | 1.568879 | 92113.0588 | 157.1358 | 3.333333 | 1.1612e8 | 0.000000 | : rR(9) : C Cv=0 10 10 F1e - X Xv=0 9 9 F1e |
| 13C18O | C | 9 | e | 1 | X | 10 | e | 1 | 91885.7111 | 1.327033 | 92077.7549 | 192.0438 | 3.333333 | 1.2805e8 | 0.000000 | : pP(10) : C Cv=0 9 9 F1e - X Xv=0 10 10 F1e |
| 13C18O | C | 11 | e | 1 | X | 10 | e | 1 | 91959.8529 | 1.459736 | 92151.8967 | 192.0438 | 3.666667 | 1.1664e8 | 0.000000 | : rR(10) : C Cv=0 11 11 F1e - X Xv=0 10 10 F1e |
| 13C18O | C | 10 | e | 1 | X | 11 | e | 1 | 91882.6208 | 1.214243 | 92113.0588 | 230.4379 | 3.666667 | 1.2743e8 | 0.000000 | : pP(11) : C Cv=0 10 10 F1e - X Xv=0 11 11 F1e |
| 13C18O | C | 12 | e | 1 | X | 11 | e | 1 | 91963.8197 | 1.324628 | 92194.2577 | 230.4379 | 4.000000 | 1.1708e8 | 0.000000 | : rR(11) : C Cv=0 12 12 F1e - X Xv=0 11 11 F1e |
| 13C18O | C | 11 | e | 1 | X | 12 | e | 1 | 91879.5799 | 1.083595 | 92151.8967 | 272.3168 | 4.000000 | 1.2691e8 | 0.000000 | : pP(12) : C Cv=0 11 11 F1e - X Xv=0 12 12 F1e |
| 13C18O | C | 13 | e | 1 | X | 12 | e | 1 | 91967.8160 | 1.173894 | 92240.1329 | 272.3168 | 4.333333 | 1.1746e8 | 0.000000 | : rR(12) : C Cv=0 13 13 F1e - X Xv=0 12 12 F1e |
| 13C18O | C | 12 | e | 1 | X | 13 | e | 1 | 91876.5786 | .9443794 | 92194.2577 | 317.6791 | 4.333333 | 1.2648e8 | 0.000000 | : pP(13) : C Cv=0 12 12 F1e - X Xv=0 13 13 F1e |
| 13C18O | C | 14 | e | 1 | X | 13 | e | 1 | 91971.8583 | 1.017024 | 92289.5374 | 317.6791 | 4.666667 | 1.1779e8 | 0.000000 | : rR(13) : C Cv=0 14 14 F1e - X Xv=0 13 13 F1e |
| 13C18O | C | 13 | e | 1 | X | 14 | e | 1 | 91873.6097 | .8046310 | 92240.1329 | 366.5231 | 4.666667 | 1.2611e8 | 0.000000 | : pP(14) : C Cv=0 13 13 F1e - X Xv=0 14 14 F1e |
| 13C18O | C | 15 | e | 1 | X | 14 | e | 1 | 91975.9384 | .8621046 | 92342.4616 | 366.5231 | 5.000000 | 1.1807e8 | 0.000000 | : rR(14) : C Cv=0 15 15 F1e - X Xv=0 14 14 F1e |
| 13C18O | C | 14 | e | 1 | X | 15 | e | 1 | 91870.6902 | .6707753 | 92289.5374 | 418.8472 | 5.000000 | 1.2578e8 | 0.000000 | : pP(15) : C Cv=0 14 14 F1e - X Xv=0 15 15 F1e |
| 13C18O | C | 16 | e | 1 | X | 15 | e | 1 | 91980.0553 | .7154937 | 92398.9025 | 418.8472 | 5.333333 | 1.1833e8 | 0.000000 | : rR(15) : C Cv=0 16 16 F1e - X Xv=0 15 15 F1e |
| 13C18O | C | 15 | e | 1 | X | 16 | e | 1 | 91867.8120 | .5474926 | 92342.4616 | 474.6496 | 5.333333 | 1.2550e8 | 0.000000 | : pP(16) : C Cv=0 15 15 F1e - X Xv=0 16 16 F1e |
| 13C18O | C | 17 | e | 1 | X | 16 | e | 1 | 91984.2091 | .5817108 | 92458.8587 | 474.6496 | 5.666667 | 1.1856e8 | 0.000000 | : rR(16) : C Cv=0 17 17 F1e - X Xv=0 16 16 F1e |
| 13C18O | C | 16 | e | 1 | X | 17 | e | 1 | 91864.9742 | .4377629 | 92398.9025 | 533.9283 | 5.666667 | 1.2525e8 | 0.000000 | : pP(17) : C Cv=0 16 16 F1e - X Xv=0 17 17 F1e |
| 13C18O | C | 18 | e | 1 | X | 17 | e | 1 | 91988.4008 | .4635137 | 92522.3291 | 533.9283 | 6.000000 | 1.1876e8 | 0.000000 | : rR(17) : C Cv=0 18 18 F1e - X Xv=0 17 17 F1e |
| 13C18O | C | 17 | e | 1 | X | 18 | e | 1 | 91862.1775 | .3430504 | 92458.8587 | 596.6812 | 6.000000 | 1.2503e8 | 0.000000 | : pP(18) : C Cv=0 17 17 F1e - X Xv=0 18 18 F1e |
| 13C18O | C | 19 | e | 1 | X | 18 | e | 1 | 91992.6315 | .3621088 | 92589.3127 | 596.6812 | 6.333333 | 1.1895e8 | 0.000000 | : rR(18) : C Cv=0 19 19 F1e - X Xv=0 18 18 F1e |

| | | | | | | | | | | |
|---|---|---|---|---|---|---|---|---|---|---|
| 13C18O | C 18 e 1 | X 19 e 1 | 91859.4228 | .2635741 | 92522.3291 | 662.9063 | 6.333333 | 1.2483e8 | 0.000000 : | pP(19) : C Cv=0 18 18 F1e - X Xv=0 19 19 F1e |
| 13C18O | C 20 e 1 | X 19 e 1 | 91996.8976 | .2774464 | 92659.8039 | 662.9063 | 6.666667 | 1.1912e8 | 0.000000 : | rR(19) : C Cv=0 20 20 F1e - X Xv=0 19 19 F1e |
| 13C18O | C 19 e 1 | X 20 e 1 | 91856.7115 | .1986167 | 92589.3127 | 732.6012 | 6.666667 | 1.2465e8 | 0.000000 : | pP(20) : C Cv=0 19 19 F1e - X Xv=0 20 20 F1e |
| 13C18O | C 21 e 1 | X 20 e 1 | 92001.1995 | .2085475 | 92733.8006 | 732.6012 | 7.000000 | 1.1927e8 | 0.000000 : | rR(20) : C Cv=0 21 21 F1e - X Xv=0 20 20 F1e |
| 13C18O | C 20 e 1 | X 21 e 1 | 91854.0405 | .1468316 | 92659.8039 | 805.7634 | 7.000000 | 1.2449e8 | 0.000000 : | pP(21) : C Cv=0 20 20 F1e - X Xv=0 21 21 F1e |
| 13C18O | C 22 e 1 | X 21 e 1 | 92005.5398 | .1538236 | 92811.3031 | 805.7634 | 7.333333 | 1.1941e8 | 0.000000 : | rR(21) : C Cv=0 22 22 F1e - X Xv=0 21 21 F1e |
| 13C18O | C 21 e 1 | X 22 e 1 | 91851.4102 | .1065175 | 92733.8006 | 882.3904 | 7.333333 | 1.2434e8 | 0.000000 : | pP(22) : C Cv=0 21 21 F1e - X Xv=0 22 22 F1e |
| 13C18O | C 23 e 1 | X 22 e 1 | 92009.9176 | .1113592 | 92892.3080 | 882.3904 | 7.666667 | 1.1955e8 | 0.000000 : | rR(22) : C Cv=0 23 23 F1e - X Xv=0 22 22 F1e |
| 13C18O | C 22 e 1 | X 23 e 1 | 91848.8235 | .0758425 | 92811.3031 | 962.4797 | 7.666667 | 1.2420e8 | 0.000000 : | pP(23) : C Cv=0 22 22 F1e - X Xv=0 23 23 F1e |
| 13C18O | C 24 e 1 | X 23 e 1 | 92014.3427 | .0791400 | 92976.8223 | 962.4797 | 8.000000 | 1.1967e8 | 0.000000 : | rR(23) : C Cv=0 24 24 F1e - X Xv=0 23 23 F1e |
| 13C18O | C 23 e 1 | X 24 e 1 | 91846.2797 | .0530124 | 92892.3080 | 1046.0283 | 8.000000 | 1.2408e8 | 0.000000 : | pP(24) : C Cv=0 23 23 F1e - X Xv=0 24 24 F1e |
| 13C18O | C 25 e 1 | X 24 e 1 | 92018.7949 | .0552212 | 93064.8231 | 1046.0283 | 8.333333 | 1.1978e8 | 0.000000 : | rR(24) : C Cv=0 25 25 F1e - X Xv=0 24 24 F1e |
| 13C18O | C 24 e 1 | X 25 e 1 | 91843.7889 | .0363821 | 92976.8223 | 1133.0334 | 8.333333 | 1.2396e8 | 0.000000 : | pP(25) : C Cv=0 24 24 F1e - X Xv=0 25 25 F1e |
| 13C18O | C 26 e 1 | X 25 e 1 | 92023.2842 | .0378374 | 93156.3176 | 1133.0334 | 8.666667 | 1.1989e8 | 0.000000 : | rR(25) : C Cv=0 26 26 F1e - X Xv=0 25 25 F1e |
| 13C18O | C 25 e 1 | X 26 e 1 | 91841.3312 | .0245194 | 93064.8231 | 1223.4920 | 8.666667 | 1.2386e8 | 0.000000 : | pP(26) : C Cv=0 25 25 F1e - X Xv=0 26 26 F1e |
| 13C18O | C 27 e 1 | X 26 e 1 | 92027.8069 | .0254625 | 93251.2989 | 1223.4920 | 9.000000 | 1.1999e8 | 0.000000 : | rR(26) : C Cv=0 27 27 F1e - X Xv=0 26 26 F1e |
| 13C18O | C 26 e 1 | X 27 e 1 | 91838.9167 | .0162294 | 93156.3176 | 1317.4009 | 9.000000 | 1.2376e8 | 0.000000 : | pP(27) : C Cv=0 26 26 F1e - X Xv=0 27 27 F1e |
| 13C18O | C 28 e 1 | X 27 e 1 | 92032.3659 | .0168305 | 93349.7667 | 1317.4009 | 9.333333 | 1.2009e8 | 0.000000 : | rR(27) : C Cv=0 28 28 F1e - X Xv=0 27 27 F1e |
| 13C18O | C 27 e 1 | X 28 e 1 | 91836.5420 | .0105517 | 93251.2989 | 1414.7569 | 9.333333 | 1.2366e8 | 0.000000 : | pP(28) : C Cv=0 27 27 F1e - X Xv=0 28 28 F1e |
| 13C18O | C 29 e 1 | X 28 e 1 | 92036.9621 | .0109285 | 93451.7190 | 1414.7569 | 9.666667 | 1.2018e8 | 0.000000 : | rR(28) : C Cv=0 29 29 F1e - X Xv=0 28 28 F1e |
| 13C18O | C 28 e 1 | X 29 e 1 | 91834.2102 | .0067392 | 93349.7667 | 1515.5565 | 9.666667 | 1.2358e8 | 0.000000 : | pP(29) : C Cv=0 28 28 F1e - X Xv=0 29 29 F1e |
| 13C18O | C 30 e 1 | X 29 e 1 | 92041.5917 | .0069716 | 93557.1483 | 1515.5565 | 10.00000 | 1.2027e8 | 0.000000 : | rR(29) : C Cv=0 30 30 F1e - X Xv=0 29 29 F1e |
| 13C18O | C 29 e 1 | X 30 e 1 | 91831.9226 | .0042288 | 93451.7190 | 1619.7964 | 10.00000 | 1.2350e8 | 0.000000 : | pP(30) : C Cv=0 29 29 F1e - X Xv=0 30 30 F1e |
| 13C18O | C 31 e 1 | X 30 e 1 | 92046.2585 | .0043698 | 93666.0549 | 1619.7964 | 10.33333 | 1.2035e8 | 0.000000 : | rR(30) : C Cv=0 31 31 F1e - X Xv=0 30 30 F1e |
| 13C18O | C 30 e 1 | X 31 e 1 | 91829.6755 | .0026073 | 93557.1483 | 1727.4728 | 10.33333 | 1.2342e8 | 0.000000 : | pP(31) : C Cv=0 30 30 F1e - X Xv=0 31 31 F1e |
| 13C18O | C 32 e 1 | X 31 e 1 | 92050.9615 | .0026914 | 93778.4343 | 1727.4728 | 10.66667 | 1.2043e8 | 0.000000 : | rR(31) : C Cv=0 32 32 F1e - X Xv=0 31 31 F1e |
| 13C18O | C 31 e 1 | X 32 e 1 | 91827.4729 | .0015796 | 93666.0549 | 1838.5820 | 10.66667 | 1.2335e8 | 0.000000 : | pP(32) : C Cv=0 31 31 F1e - X Xv=0 32 32 F1e |
| 13C18O | C 33 e 1 | X 32 e 1 | 92055.6995 | .0016290 | 93894.2815 | 1838.5820 | 11.00000 | 1.2050e8 | 0.000000 : | rR(32) : C Cv=0 33 33 F1e - X Xv=0 32 32 F1e |
| 13C18O | C 32 e 1 | X 33 e 1 | 91825.3141 | .0009405 | 93778.4343 | 1953.1202 | 11.00000 | 1.2328e8 | 0.000000 : | pP(33) : C Cv=0 32 32 F1e - X Xv=0 33 33 F1e |
| 13C18O | C 34 e 1 | X 33 e 1 | 92060.4715 | .0009690 | 94013.5918 | 1953.1202 | 11.33333 | 1.2057e8 | 0.000000 : | rR(33) : C Cv=0 34 34 F1e - X Xv=0 33 33 F1e |
| 13C18O | C 33 e 1 | X 34 e 1 | 91823.1981 | .0005503 | 93894.2815 | 2071.0834 | 11.33333 | 1.2321e8 | 0.000000 : | pP(34) : C Cv=0 33 33 F1e - X Xv=0 34 34 F1e |
| 13C18O | C 35 e 1 | X 34 e 1 | 92065.2735 | .0005665 | 94136.3569 | 2071.0834 | 11.66667 | 1.2064e8 | 0.000000 : | rR(34) : C Cv=0 35 35 F1e - X Xv=0 34 34 F1e |
| 13C18O | C 34 e 1 | X 35 e 1 | 91821.1243 | .0003165 | 94013.5918 | 2192.4675 | 11.66667 | 1.2315e8 | 0.000000 : | pP(35) : C Cv=0 34 34 F1e - X Xv=0 35 35 F1e |
| 13C18O | C 36 e 1 | X 35 e 1 | 92070.1135 | .0003255 | 94262.5810 | 2192.4675 | 12.00000 | 1.2071e8 | 0.000000 : | rR(35) : C Cv=0 36 36 F1e - X Xv=0 35 35 F1e |
| 13C18O | C 35 e 1 | X 36 e 1 | 91819.0887 | .0001789 | 94136.3569 | 2317.2682 | 12.00000 | 1.2310e8 | 0.000000 : | pP(36) : C Cv=0 35 35 F1e - X Xv=0 36 36 F1e |
| 13C18O | C 37 e 1 | X 36 e 1 | 92074.9855 | .0001839 | 94392.2537 | 2317.2682 | 12.33333 | 1.2077e8 | 0.000000 : | rR(36) : C Cv=0 37 37 F1e - X Xv=0 36 36 F1e |
| 13C18O | C 38 e 1 | X 37 e 1 | 92079.8945 | .0001021 | 94525.3758 | 2445.4813 | 12.66667 | 1.2083e8 | 0.000000 : | rR(37) : C Cv=0 38 38 F1e - X Xv=0 37 37 F1e |
| 13C18O | B 1 e 1 | X 0 e 1 | 86920.8991 | .3333333 | 86920.8991 | 0.0000 | .3333333 | 68652057 | 0.000000 : | rR(0) : B Bv=0 1 1 F1e - X Xv=0 0 0 F1e |
| 13C18O | B 0 e 1 | X 1 e 1 | 86913.8645 | .3277961 | 86917.3573 | 3.4928 | .3333333 | 2.0591e8 | 0.000000 : | pP(1) : B Bv=0 0 0 F1e - X Xv=0 1 1 F1e |
| 13C18O | B 2 e 1 | X 1 e 1 | 86924.4874 | .6555922 | 86927.9802 | 3.4928 | .6666667 | 82392672 | 0.000000 : | rR(1) : B Bv=0 2 2 F1e - X Xv=0 1 1 F1e |
| 13C18O | B 1 e 1 | X 2 e 1 | 86910.4208 | .6339926 | 86920.8991 | 10.4783 | .6666667 | 1.3725e8 | 0.000000 : | pP(2) : B Bv=0 1 1 F1e - X Xv=0 2 2 F1e |
| 13C18O | B 3 e 1 | X 2 e 1 | 86928.1193 | .9509888 | 86938.5976 | 10.4783 | 1.000000 | 88288929 | 0.000000 : | rR(2) : B Bv=0 3 3 F1e - X Xv=0 2 2 F1e |
| 13C18O | B 2 e 1 | X 3 e 1 | 86907.0241 | .9043813 | 86927.9802 | 20.9562 | 1.000000 | 1.2351e8 | 0.000000 : | pP(3) : B Bv=0 2 2 F1e - X Xv=0 3 3 F1e |
| 13C18O | B 4 e 1 | X 3 e 1 | 86931.7973 | 1.205842 | 86952.7535 | 20.9562 | 1.333333 | 91570511 | 0.000000 : | rR(3) : B Bv=0 4 4 F1e - X Xv=0 3 3 F1e |
| 13C18O | B 3 e 1 | X 4 e 1 | 86903.6715 | 1.127699 | 86938.5976 | 34.9261 | 1.333333 | 1.1762e8 | 0.000000 : | pP(4) : B Bv=0 3 3 F1e - X Xv=0 4 4 F1e |
| 13C18O | B 5 e 1 | X 4 e 1 | 86935.5224 | 1.409623 | 86970.4486 | 34.9261 | 1.666667 | 93663699 | 0.000000 : | rR(4) : B Bv=0 5 5 F1e - X Xv=0 4 4 F1e |
| 13C18O | B 4 e 1 | X 5 e 1 | 86900.3658 | 1.296383 | 86952.7535 | 52.3877 | 1.666667 | 1.1434e8 | 0.000000 : | pP(5) : B Bv=0 4 4 F1e - X Xv=0 5 5 F1e |
| 13C18O | B 6 e 1 | X 5 e 1 | 86939.2932 | 1.555660 | 86991.6809 | 52.3877 | 2.000000 | 95117055 | 0.000000 : | rR(5) : B Bv=0 6 6 F1e - X Xv=0 5 5 F1e |
| 13C18O | B 5 e 1 | X 6 e 1 | 86897.1084 | 1.406934 | 86970.4486 | 73.3402 | 2.000000 | 1.1225e8 | 0.000000 : | pP(6) : B Bv=0 5 5 F1e - X Xv=0 6 6 F1e |
| 13C18O | B 7 e 1 | X 6 e 1 | 86943.1099 | 1.641423 | 87016.4501 | 73.3402 | 2.333333 | 96186578 | 0.000000 : | rR(6) : B Bv=0 7 7 F1e - X Xv=0 6 6 F1e |
| 13C18O | B 6 e 1 | X 7 e 1 | 86893.8978 | 1.459856 | 86991.6809 | 97.7830 | 2.333333 | 1.1080e8 | 0.000000 : | pP(7) : B Bv=0 6 6 F1e - X Xv=0 7 7 F1e |
| 13C18O | B 8 e 1 | X 7 e 1 | 86946.9720 | 1.668407 | 87044.7550 | 97.7830 | 2.666667 | 97007795 | 0.000000 : | rR(7) : B Bv=0 8 8 F1e - X Xv=0 7 7 F1e |
| 13C18O | B 7 e 1 | X 8 e 1 | 86890.7349 | 1.459230 | 87016.4501 | 125.7152 | 2.666667 | 1.0973e8 | 0.000000 : | pP(8) : B Bv=0 7 7 F1e - X Xv=0 8 8 F1e |
| 13C18O | B 9 e 1 | X 8 e 1 | 86950.8793 | 1.641634 | 87076.5945 | 125.7152 | 3.000000 | 97659169 | 0.000000 : | rR(8) : B Bv=0 9 9 F1e - X Xv=0 8 8 F1e |
| 13C18O | B 8 e 1 | X 9 e 1 | 86887.6192 | 1.411991 | 87044.7550 | 157.1358 | 3.000000 | 1.0891e8 | 0.000000 : | pP(9) : B Bv=0 8 8 F1e - X Xv=0 9 9 F1e |
| 13C18O | B 10 e 1 | X 9 e 1 | 86954.8315 | 1.568879 | 87111.9674 | 157.1358 | 3.333333 | 98189272 | 0.000000 : | rR(9) : B Bv=0 10 10 F1e - X Xv=0 9 9 F1e |
| 13C18O | B 9 e 1 | X 10 e 1 | 86884.5507 | 1.327033 | 87076.5945 | 192.0438 | 3.333333 | 1.0826e8 | 0.000000 : | pP(10) : B Bv=0 9 9 F1e - X Xv=0 10 10 F1e |
| 13C18O | B 11 e 1 | X 10 e 1 | 86958.8284 | 1.459736 | 87150.8722 | 192.0438 | 3.666667 | 98629781 | 0.000000 : | rR(10) : B Bv=0 11 11 F1e - X Xv=0 10 10 F1e |
| 13C18O | B 10 e 1 | X 11 e 1 | 86881.5294 | 1.214243 | 87111.9674 | 230.4379 | 3.666667 | 1.0774e8 | 0.000000 : | pP(11) : B Bv=0 10 10 F1e - X Xv=0 11 11 F1e |
| 13C18O | B 12 e 1 | X 11 e 1 | 86962.8724 | 1.324628 | 87193.3103 | 230.4379 | 4.000000 | 99002246 | 0.000000 : | rR(11) : B Bv=0 12 12 F1e - X Xv=0 11 11 F1e |
| 13C18O | B 11 e 1 | X 12 e 1 | 86878.5554 | 1.083595 | 87150.8722 | 272.3168 | 4.000000 | 1.0730e8 | 0.000000 : | pP(12) : B Bv=0 11 11 F1e - X Xv=0 12 12 F1e |
| 13C18O | B 13 e 1 | X 12 e 1 | 86966.9570 | 1.173894 | 87239.2738 | 272.3168 | 4.333333 | 99321802 | 0.000000 : | rR(12) : B Bv=0 13 13 F1e - X Xv=0 12 12 F1e |
| 13C18O | B 12 e 1 | X 13 e 1 | 86875.6313 | .9443794 | 87193.3103 | 317.6791 | 4.333333 | 1.0693e8 | 0.000000 : | pP(13) : B Bv=0 12 12 F1e - X Xv=0 13 13 F1e |
| 13C18O | B 14 e 1 | X 13 e 1 | 86971.0869 | 1.017024 | 87288.7659 | 317.6791 | 4.666667 | 99599443 | 0.000000 : | rR(13) : B Bv=0 14 14 F1e - X Xv=0 13 13 F1e |
| 13C18O | B 13 e 1 | X 14 e 1 | 86872.7507 | .8046310 | 87239.2738 | 366.5231 | 4.666667 | 1.0661e8 | 0.000000 : | pP(14) : B Bv=0 13 13 F1e - X Xv=0 14 14 F1e |
| 13C18O | B 15 e 1 | X 14 e 1 | 86975.2607 | .8621046 | 87341.7838 | 366.5231 | 5.000000 | 99843309 | 0.000000 : | rR(14) : B Bv=0 15 15 F1e - X Xv=0 14 14 F1e |
| 13C18O | B 14 e 1 | X 15 e 1 | 86869.9187 | .6707753 | 87288.7659 | 418.8472 | 5.000000 | 1.0634e8 | 0.000000 : | pP(15) : B Bv=0 14 14 F1e - X Xv=0 15 15 F1e |
| 13C18O | B 16 e 1 | X 15 e 1 | 86979.4768 | .7154937 | 87398.3240 | 418.8472 | 5.333333 | 1.0006e8 | 0.000000 : | rR(15) : B Bv=0 16 16 F1e - X Xv=0 15 15 F1e |
| 13C18O | B 15 e 1 | X 16 e 1 | 86867.1343 | .5474926 | 87341.7838 | 474.6496 | 5.333333 | 1.0610e8 | 0.000000 : | pP(16) : B Bv=0 15 15 F1e - X Xv=0 16 16 F1e |
| 13C18O | B 17 e 1 | X 16 e 1 | 86983.6962 | .5817108 | 87458.3457 | 474.6496 | 5.666667 | 1.0025e8 | 0.000000 : | rR(16) : B Bv=0 17 17 F1e - X Xv=0 16 16 F1e |
| 13C18O | B 16 e 1 | X 17 e 1 | 86864.3957 | .4377629 | 87398.3240 | 533.9283 | 5.666667 | 1.0589e8 | 0.000000 : | pP(17) : B Bv=0 16 16 F1e - X Xv=0 17 17 F1e |
| 13C18O | B 18 e 1 | X 17 e 1 | 86988.0386 | .4635137 | 87521.9669 | 533.9283 | 6.000000 | 1.0043e8 | 0.000000 : | rR(17) : B Bv=0 18 18 F1e - X Xv=0 17 17 F1e |
| 13C18O | B 17 e 1 | X 18 e 1 | 86861.6645 | .3430504 | 87458.3457 | 596.6812 | 6.000000 | 1.0570e8 | 0.000000 : | pP(18) : B Bv=0 17 17 F1e - X Xv=0 18 18 F1e |
| 13C18O | B 19 e 1 | X 18 e 1 | 86992.3917 | .3621088 | 87589.0730 | 596.6812 | 6.333333 | 1.0059e8 | 0.000000 : | rR(18) : B Bv=0 19 19 F1e - X Xv=0 18 18 F1e |
| 13C18O | B 18 e 1 | X 19 e 1 | 86859.0606 | .2635741 | 87521.9669 | 662.9063 | 6.333333 | 1.0554e8 | 0.000000 : | pP(19) : B Bv=0 18 18 F1e - X Xv=0 19 19 F1e |
| 13C18O | B 20 e 1 | X 19 e 1 | 86996.7785 | .2774464 | 87659.6849 | 662.9063 | 6.666667 | 1.0073e8 | 0.000000 : | rR(19) : B Bv=0 20 20 F1e - X Xv=0 19 19 F1e |
| 13C18O | B 19 e 1 | X 20 e 1 | 86856.4718 | .1986167 | 87589.0730 | 732.6012 | 6.666667 | 1.0538e8 | 0.000000 : | pP(20) : B Bv=0 19 19 F1e - X Xv=0 20 20 F1e |
| 13C18O | B 21 e 1 | X 20 e 1 | 87001.2082 | .2085475 | 87733.8094 | 732.6012 | 7.000000 | 1.0086e8 | 0.000000 : | rR(20) : B Bv=0 21 21 F1e - X Xv=0 20 20 F1e |
| 13C18O | B 20 e 1 | X 21 e 1 | 86853.9215 | .1468316 | 87659.6849 | 805.7634 | 7.000000 | 1.0525e8 | 0.000000 : | pP(21) : B Bv=0 20 20 F1e - X Xv=0 21 21 F1e |
| 13C18O | B 22 e 1 | X 21 e 1 | 87005.6801 | .1538236 | 87811.4435 | 805.7634 | 7.333333 | 1.0098e8 | 0.000000 : | rR(21) : B Bv=0 22 22 F1e - X Xv=0 21 21 F1e |
| 13C18O | B 21 e 1 | X 22 e 1 | 86851.4189 | .1065175 | 87733.8094 | 882.3904 | 7.333333 | 1.0512e8 | 0.000000 : | pP(22) : B Bv=0 21 21 F1e - X Xv=0 22 22 F1e |
| 13C18O | B 23 e 1 | X 22 e 1 | 87010.1929 | .1113592 | 87892.5834 | 882.3904 | 7.666667 | 1.0110e8 | 0.000000 : | rR(22) : B Bv=0 23 23 F1e - X Xv=0 22 22 F1e |
| 13C18O | B 22 e 1 | X 23 e 1 | 86848.9638 | .0758425 | 87811.4435 | 962.4797 | 7.666667 | 1.0501e8 | 0.000000 : | pP(23) : B Bv=0 22 22 F1e - X Xv=0 23 23 F1e |
| 13C18O | B 24 e 1 | X 23 e 1 | 87014.7458 | .0791400 | 87977.2254 | 962.4797 | 8.000000 | 1.0120e8 | 0.000000 : | rR(23) : B Bv=0 24 24 F1e - X Xv=0 23 23 F1e |
| 13C18O | B 23 e 1 | X 24 e 1 | 86846.5551 | .0530124 | 87892.5834 | 1046.0283 | 8.000000 | 1.0490e8 | 0.000000 : | pP(24) : B Bv=0 23 23 F1e - X Xv=0 24 24 F1e |
| 13C18O | B 25 e 1 | X 24 e 1 | 87019.3373 | .0552212 | 88065.3656 | 1046.0283 | 8.333333 | 1.0130e8 | 0.000000 : | rR(24) : B Bv=0 25 25 F1e - X Xv=0 24 24 F1e |
| 13C18O | B 24 e 1 | X 25 e 1 | 86844.1920 | .0363821 | 87977.2254 | 1133.0334 | 8.333333 | 1.0480e8 | 0.000000 : | pP(25) : B Bv=0 24 24 F1e - X Xv=0 25 25 F1e |
| 13C18O | B 26 e 1 | X 25 e 1 | 87023.9741 | .0378374 | 88157.0075 | 1133.0334 | 8.666667 | 1.0139e8 | 0.000000 : | rR(25) : B Bv=0 26 26 F1e - X Xv=0 25 25 F1e |
| 13C18O | B 25 e 1 | X 26 e 1 | 86841.8736 | .0245194 | 88065.3656 | 1223.4920 | 8.666667 | 1.0471e8 | 0.000000 : | pP(26) : B Bv=0 25 25 F1e - X Xv=0 26 26 F1e |
| 13C18O | B 27 e 1 | X 26 e 1 | 87028.6509 | .0254625 | 88252.1429 | 1223.4920 | 9.000000 | 1.0148e8 | 0.000000 : | rR(26) : B Bv=0 27 27 F1e - X Xv=0 26 26 F1e |
| 13C18O | B 26 e 1 | X 27 e 1 | 86839.6066 | .0162294 | 88157.0075 | 1317.4009 | 9.000000 | 1.0463e8 | 0.000000 : | pP(27) : B Bv=0 26 26 F1e - X Xv=0 27 27 F1e |
| 13C18O | B 28 e 1 | X 27 e 1 | 87033.3613 | .0168305 | 88350.7622 | 1317.4009 | 9.333333 | 1.0156e8 | 0.000000 : | rR(27) : B Bv=0 28 28 F1e - X Xv=0 27 27 F1e |
| 13C18O | B 27 e 1 | X 28 e 1 | 86837.3860 | .0105517 | 88252.1429 | 1414.7569 | 9.333333 | 1.0455e8 | 0.000000 : | pP(28) : B Bv=0 27 27 F1e - X Xv=0 28 28 F1e |
| 13C18O | B 29 e 1 | X 28 e 1 | 87038.1146 | .0109285 | 88452.8715 | 1414.7569 | 9.666667 | 1.0164e8 | 0.000000 : | rR(28) : B Bv=0 29 29 F1e - X Xv=0 28 28 F1e |
| 13C18O | B 28 e 1 | X 29 e 1 | 86835.2057 | .0067392 | 88350.7622 | 1515.5565 | 9.666667 | 1.0448e8 | 0.000000 : | pP(29) : B Bv=0 28 28 F1e - X Xv=0 29 29 F1e |
| 13C18O | B 30 e 1 | X 29 e 1 | 87042.9008 | .0069716 | 88558.4574 | 1515.5565 | 10.00000 | 1.0172e8 | 0.000000 : | rR(29) : B Bv=0 30 30 F1e - X Xv=0 29 29 F1e |
| 13C18O | B 29 e 1 | X 30 e 1 | 86833.0751 | .0042288 | 88452.8715 | 1619.7964 | 10.00000 | 1.0441e8 | 0.000000 : | pP(30) : B Bv=0 29 29 F1e - X Xv=0 30 30 F1e |
| 13C18O | B 31 e 1 | X 30 e 1 | 87047.7207 | .0043698 | 88667.5171 | 1619.7964 | 10.33333 | 1.0179e8 | 0.000000 : | rR(30) : B Bv=0 31 31 F1e - X Xv=0 30 30 F1e |
| 13C18O | B 30 e 1 | X 31 e 1 | 86830.9846 | .0026073 | 88558.4574 | 1727.4728 | 10.33333 | 1.0434e8 | 0.000000 : | pP(31) : B Bv=0 30 30 F1e - X Xv=0 31 31 F1e |
| 13C18O | B 32 e 1 | X 31 e 1 | 87052.5802 | .0026914 | 88780.0530 | 1727.4728 | 10.66667 | 1.0186e8 | 0.000000 : | rR(31) : B Bv=0 32 32 F1e - X Xv=0 31 31 F1e |
| 13C18O | B 31 e 1 | X 32 e 1 | 86828.9350 | .0015796 | 88667.5171 | 1838.5820 | 10.66667 | 1.0428e8 | 0.000000 : | pP(32) : B Bv=0 31 31 F1e - X Xv=0 32 32 F1e |
| 13C18O | B 33 e 1 | X 32 e 1 | 87057.4912 | .0016290 | 88896.0733 | 1838.5820 | 11.00000 | 1.0192e8 | 0.000000 : | rR(32) : B Bv=0 33 33 F1e - X Xv=0 32 32 F1e |
| 13C18O | B 32 e 1 | X 33 e 1 | 86826.9328 | .0009405 | 88780.0530 | 1953.1202 | 11.00000 | 1.0422e8 | 0.000000 : | pP(33) : B Bv=0 32 32 F1e - X Xv=0 33 33 F1e |
| 13C18O | B 34 e 1 | X 33 e 1 | 87062.4277 | .0009690 | 89015.5479 | 1953.1202 | 11.33333 | 1.0198e8 | 0.000000 : | rR(33) : B Bv=0 34 34 F1e - X Xv=0 33 33 F1e |
| 13C18O | B 33 e 1 | X 34 e 1 | 86824.9899 | .0005503 | 88896.0733 | 2071.0834 | 11.33333 | 1.0417e8 | 0.000000 : | pP(34) : B Bv=0 33 33 F1e - X Xv=0 34 34 F1e |
| 13C18O | B 35 e 1 | X 34 e 1 | 87067.4174 | .0005665 | 89138.5008 | 2071.0834 | 11.66667 | 1.0204e8 | 0.000000 : | rR(34) : B Bv=0 35 35 F1e - X Xv=0 34 34 F1e |
| 13C18O | B 34 e 1 | X 35 e 1 | 86823.0804 | .0003165 | 89015.5479 | 2192.4675 | 11.66667 | 1.0412e8 | 0.000000 : | pP(35) : B Bv=0 34 34 F1e - X Xv=0 35 35 F1e |
| 13C18O | B 36 e 1 | X 35 e 1 | 87072.3997 | .0003255 | 89264.8671 | 2192.4675 | 12.00000 | 1.0210e8 | 0.000000 : | rR(35) : B Bv=0 36 36 F1e - X Xv=0 35 35 F1e |
| 13C18O | B 35 e 1 | X 36 e 1 | 86821.2326 | .0001789 | 89138.5008 | 2317.2682 | 12.00000 | 1.0407e8 | 0.000000 : | pP(36) : B Bv=0 35 35 F1e - X Xv=0 36 36 F1e |
| 13C18O | B 37 e 1 | X 36 e 1 | 87077.4444 | .0001839 | 89394.7126 | 2317.2682 | 12.33333 | 1.0216e8 | 0.000000 : | rR(36) : B Bv=0 37 37 F1e - X Xv=0 36 36 F1e |
| 13C18O | B 38 e 1 | X 37 e 1 | 87082.5221 | .0001021 | 89528.0034 | 2445.4813 | 12.66667 | 1.0221e8 | 0.000000 : | rR(37) : B Bv=0 38 38 F1e - X Xv=0 37 37 F1e |
| 13C18O Excited | 1 e 1 | X 0 e 1 | 67560.4283 | .5936401 | 67560.4283 | 0.0000 | .5936401 | 57411990 | 0.000000 : | rR(0) : Excited Av=2 1 1 F1e - X Xv=0 0 0 F1e |
| 13C18O Excited | 1 e 2 | X 0 e 1 | 67635.7484 | .0038055 | 67635.7484 | 0.0000 | .0038055 | 369271.7 | 0.000000 : | rR1(0) : Excited a3Pv=13 1 1 F1e - X Xv=0 0 0 F1e |
| 13C18O Excited | 1 e 3 | X 0 e 1 | 67675.0379 | .8221444 | 67675.0379 | 0.0000 | .8221444 | 79916381 | 0.000000 : | sR21(0) : Excited a3Pv=13 1 2 F2e - X Xv=0 0 0 F1e |
| 13C18O Excited | 1 e 4 | X 0 e 1 | 67803.8618 | .3684105 | 67803.8618 | 0.0000 | .3684105 | 36016169 | 0.000000 : | qR1(0) : Excited ev=4 1 0 F1e - X Xv=0 0 0 F1e |
| 13C18O Excited | 1 e 5 | X 0 e 1 | 67810.2130 | .2262324 | 67810.2130 | 0.0000 | .2262324 | 22122914 | 0.000000 : | sR31(0) : Excited ev=4 1 2 F3e - X Xv=0 0 0 F1e |
| 13C18O Excited | 1 e 6 | X 0 e 1 | 67987.8961 | .6252122 | 67987.8961 | 0.0000 | .6252122 | 61620389 | 0.000000 : | sR31(0) : Excited dv=7 1 2 F3e - X Xv=0 0 0 F1e |
| 13C18O Excited | 1 e 7 | X 0 e 1 | 67992.6434 | .6938881 | 67992.6434 | 0.0000 | .6938881 | 68403357 | 0.000000 : | rR21(0) : Excited a'v=12 1 1 F2e - X Xv=0 0 0 F1e |
| 13C18O Excited | 1 f 1 | X 1 e 1 | 67556.9350 | .8756358 | 67560.4278 | 3.4928 | .8904273 | 86101447 | 0.000000 : | qQ(1) : Excited Av=2 1 1 F1f - X Xv=0 1 1 F1e |
| 13C18O Excited | 1 f 3 | X 1 e 1 | 67636.0701 | .0061962 | 67639.5629 | 3.4928 | .0063009 | 611420.3 | 0.000000 : | qQ1(1) : Excited a3Pv=13 1 1 F1f - X Xv=0 1 1 F1e |
| 13C18O Excited | 1 f 4 | X 1 e 1 | 67671.5588 | 1.212154 | 67675.0516 | 3.4928 | 1.232630 | 1.1980e8 | 0.000000 : | rQ21(1) : Excited a3Pv=13 1 2 F2f - X Xv=0 1 1 F1e |
| 13C18O Excited | 1 f 5 | X 1 e 1 | 67802.7807 | .8770838 | 67806.2735 | 3.4928 | .8918997 | 87188805 | 0.000000 : | qQ21(1) : Excited ev=4 1 1 F2f - X Xv=0 1 1 F1e |
| 13C18O Excited | 1 f 6 | X 1 e 1 | 67984.4030 | .9145238 | 67987.8958 | 3.4928 | .9299722 | 91643150 | 0.000000 : | rQ31(1) : Excited dv=7 1 2 F3f - X Xv=0 1 1 F1e |
| 13C18O Excited | 1 f 7 | X 1 e 1 | 67987.6666 | .8269967 | 67991.1594 | 3.4928 | .8409665 | 82884114 | 0.000000 : | pQ1(1) : Excited a'v=12 1 0 F1f - X Xv=0 1 1 F1e |
| 13C18O Excited | 1 f 8 | X 1 e 1 | 67994.9809 | .2042622 | 67998.4737 | 3.4928 | .2077126 | 20478385 | 0.000000 : | rQ31(1) : Excited a'v=12 1 2 F3f - X Xv=0 1 1 F1e |
| 13C18O Excited | 2 e 1 | X 1 e 1 | 67562.6044 | .8751036 | 67566.0972 | 3.4928 | .8898860 | 51642466 | 0.000000 : | rR(1) : Excited Av=2 2 2 F1e - X Xv=0 1 1 F1e |
| 13C18O Excited | 2 e 2 | X 1 e 1 | 67637.1322 | .0165777 | 67640.6250 | 3.4928 | .0168578 | 981541.8 | 0.000000 : | qR1(1) : Excited a3Pv=13 2 1 F1e - X Xv=0 1 1 F1e |
| 13C18O Excited | 2 e 3 | X 1 e 1 | 67676.7124 | 1.188186 | 67680.2051 | 3.4928 | 1.208257 | 70474278 | 0.000000 : | rR21(1) : Excited a3Pv=13 2 2 F2e - X Xv=0 1 1 F1e |
| 13C18O Excited | 2 e 4 | X 1 e 1 | 67710.2950 | .0150915 | 67713.7878 | 3.4928 | .0153464 | 896446.3 | 0.000000 : | sR31(1) : Excited a3Pv=13 2 3 F3e - X Xv=0 1 1 F1e |
| 13C18O Excited | 2 e 5 | X 1 e 1 | 67802.5469 | .4989530 | 67806.0397 | 3.4928 | .5073815 | 29759530 | 0.000000 : | qR1(1) : Excited ev=4 2 1 F1e - X Xv=0 1 1 F1e |
| 13C18O Excited | 2 e 6 | X 1 e 1 | 67813.3319 | .3772959 | 67816.8247 | 3.4928 | .3836693 | 22514157 | 0.000000 : | sR31(1) : Excited ev=4 2 3 F3e - X Xv=0 1 1 F1e |
| 13C18O Excited | 2 e 7 | X 1 e 1 | 67949.8912 | .0058195 | 67953.3840 | 3.4928 | .0059178 | 349367.4 | 0.000000 : | rR21(1) : Excited dv=7 2 2 F2e - X Xv=0 1 1 F1e |
| 13C18O Excited | 2 e 8 | X 1 e 1 | 67988.9639 | .9136136 | 67992.4567 | 3.4928 | .9290466 | 54942218 | 0.000000 : | sR31(1) : Excited dv=7 2 3 F3e - X Xv=0 1 1 F1e |
| 13C18O Excited | 2 e 9 | X 1 e 1 | 67993.3010 | 1.026301 | 67996.7938 | 3.4928 | 1.043637 | 61730728 | 0.000000 : | rR21(1) : Excited a'v=12 2 2 F2e - X Xv=0 1 1 F1e |
| 13C18O Excited | 1 e 1 | X 2 e 1 | 67549.9500 | .2822726 | 67560.4283 | 10.4783 | .2968201 | 28692641 | 0.000000 : | pP(2) : Excited Av=2 1 1 F1e - X Xv=0 2 2 F1e |
| 13C18O Excited | 1 e 2 | X 2 e 1 | 67625.2701 | .0018095 | 67635.7484 | 10.4783 | .0019028 | 184550.1 | 0.000000 : | pP1(2) : Excited a3Pv=13 1 1 F1e - X Xv=0 2 2 F1e |
| 13C18O Excited | 1 e 3 | X 2 e 1 | 67664.5596 | .3909251 | 67675.0379 | 10.4783 | .4110722 | 39939633 | 0.000000 : | qP21(2) : Excited a3Pv=13 1 2 F2e - X Xv=0 2 2 F1e |
| 13C18O Excited | 1 e 4 | X 2 e 1 | 67793.3836 | .1751772 | 67803.8618 | 10.4783 | .1842053 | 17999737 | 0.000000 : | oP1(2) : Excited ev=4 1 0 F1e - X Xv=0 2 2 F1e |
| 13C18O Excited | 1 e 5 | X 2 e 1 | 67799.7347 | .1075722 | 67810.2130 | 10.4783 | .1131162 | 11056330 | 0.000000 : | qP31(2) : Excited ev=4 1 2 F3e - X Xv=0 2 2 F1e |
| 13C18O Excited | 1 e 6 | X 2 e 1 | 67977.4179 | .2972849 | 67987.8961 | 10.4783 | .3126061 | 30795951 | 0.000000 : | qP31(2) : Excited dv=7 1 2 F3e - X Xv=0 2 2 F1e |
| 13C18O Excited | 1 e 7 | X 2 e 1 | 67982.1651 | .3299399 | 67992.6434 | 10.4783 | .3469441 | 34185869 | 0.000000 : | pP21(2) : Excited a'v=12 1 1 F2e - X Xv=0 2 2 F1e |
| 13C18O Excited | 2 f 1 | X 2 e 1 | 67555.6174 | 1.410271 | 67566.0957 | 10.4783 | 1.482952 | 86032992 | 0.000000 : | qQ(2) : Excited Av=2 2 2 F1f - X Xv=0 2 2 F1e |
| 13C18O Excited | 2 f 2 | X 2 e 1 | 67573.8534 | .0004560 | 67584.3317 | 10.4783 | .0004795 | 27838.58 | 0.000000 : | qQ(2) : Excited Iv=3 2 2 F1f - X Xv=0 2 2 F1e |
| 13C18O Excited | 2 f 3 | X 2 e 1 | 67633.9224 | .0293554 | 67644.4007 | 10.4783 | .0308683 | 1797051 | 0.000000 : | pQ1(2) : Excited a3Pv=13 2 1 F1f - X Xv=0 2 2 F1e |
| 13C18O Excited | 2 f 4 | X 2 e 1 | 67669.7673 | 1.912436 | 67680.2455 | 10.4783 | 2.010998 | 1.1726e8 | 0.000000 : | qQ21(2) : Excited a3Pv=13 2 2 F2f - X Xv=0 2 2 F1e |
| 13C18O Excited | 2 f 5 | X 2 e 1 | 67703.3096 | .0243487 | 67713.7879 | 10.4783 | .0256035 | 1495143 | 0.000000 : | rQ31(2) : Excited a3Pv=13 2 3 F3f - X Xv=0 2 2 F1e |
| 13C18O Excited | 2 f 6 | X 2 e 1 | 67800.1912 | 1.411988 | 67810.6694 | 10.4783 | 1.484758 | 87076685 | 0.000000 : | qQ21(2) : Excited ev=4 2 2 F2f - X Xv=0 2 2 F1e |
| 13C18O Excited | 2 f 7 | X 2 e 1 | 67942.9057 | .0093790 | 67953.3840 | 10.4783 | .0098624 | 582059.4 | 0.000000 : | qQ21(2) : Excited dv=7 2 2 F2f - X Xv=0 2 2 F1e |
| 13C18O Excited | 2 f 8 | X 2 e 1 | 67981.9756 | 1.352777 | 67992.4539 | 10.4783 | 1.422495 | 84097958 | 0.000000 : | rQ31(2) : Excited dv=7 2 3 F3f - X Xv=0 2 2 F1e |
| 13C18O Excited | 2 f 9 | X 2 e 1 | 67982.9521 | 1.281734 | 67993.4304 | 10.4783 | 1.347791 | 79684911 | 0.000000 : | pQ1(2) : Excited a'v=12 2 1 F1f - X Xv=0 2 2 F1e |
| 13C18O Excited | 2 f 10 | X 2 e 1 | 67994.0273 | .4921611 | 68004.5056 | 10.4783 | .5175256 | 30612414 | 0.000000 : | rQ31(2) : Excited a'v=12 2 3 F3f - X Xv=0 2 2 F1e |
| 13C18O Excited | 3 e 1 | X 2 e 1 | 67564.1219 | 1.127253 | 67574.6002 | 10.4783 | 1.185348 | 49138264 | 0.000000 : | rR(2) : Excited Av=2 3 3 F1e - X Xv=0 2 2 F1e |
| 13C18O Excited | 3 e 2 | X 2 e 1 | 67637.4665 | .0417760 | 67647.9448 | 10.4783 | .0439290 | 1827001 | 0.000000 : | qR1(2) : Excited a3Pv=13 3 2 F1e - X Xv=0 2 2 F1e |
| 13C18O Excited | 3 e 3 | X 2 e 1 | 67677.4821 | 1.487100 | 67687.9604 | 10.4783 | 1.563741 | 65151259 | 0.000000 : | rR21(2) : Excited a3Pv=13 3 3 F2e - X Xv=0 2 2 F1e |
| 13C18O Excited | 3 e 4 | X 2 e 1 | 67711.7562 | .0469985 | 67722.2345 | 10.4783 | .0494207 | 2062179 | 0.000000 : | sR31(2) : Excited a3Pv=13 3 4 F3e - X Xv=0 2 2 F1e |
| 13C18O Excited | 3 e 5 | X 2 e 1 | 67799.9593 | .6184552 | 67810.4376 | 10.4783 | .6503286 | 27242475 | 0.000000 : | qR1(2) : Excited ev=4 3 2 F1e - X Xv=0 2 2 F1e |
| 13C18O Excited | 3 e 6 | X 2 e 1 | 67815.1327 | .5095982 | 67825.6110 | 10.4783 | .5358614 | 22462479 | 0.000000 : | sR31(2) : Excited ev=4 3 4 F3e - X Xv=0 2 2 F1e |
| 13C18O Excited | 3 e 8 | X 2 e 1 | 67949.3909 | .0184130 | 67959.8692 | 10.4783 | .0193619 | 816453.0 | 0.000000 : | rR21(2) : Excited dv=7 3 3 F2e - X Xv=0 2 2 F1e |
| 13C18O Excited | 3 e 9 | X 2 e 1 | 67988.8161 | 1.160224 | 67999.2944 | 10.4783 | 1.220019 | 51535247 | 0.000000 : | sR31(2) : Excited dv=7 3 4 F3e - X Xv=0 2 2 F1e |
| 13C18O Excited | 3 e 10 | X 2 e 1 | 67992.5409 | 1.330062 | 68003.0192 | 10.4783 | 1.398609 | 59088866 | 0.000000 : | rR21(2) : Excited a'v=12 3 3 F2e - X Xv=0 2 2 F1e |
| 13C18O Excited | 2 e 1 | X 3 e 1 | 67545.1411 | .5365309 | 67566.0972 | 20.9562 | .5932574 | 34401621 | 0.000000 : | pP(3) : Excited Av=2 2 2 F1e - X Xv=0 3 3 F1e |
| 13C18O Excited | 2 e 2 | X 3 e 1 | 67619.6688 | .0101639 | 67640.6250 | 20.9562 | .0112385 | 653854.5 | 0.000000 : | oP1(3) : Excited a3Pv=13 2 1 F1e - X Xv=0 3 3 F1e |
| 13C18O Excited | 2 e 3 | X 3 e 1 | 67659.2490 | .7284836 | 67680.2051 | 20.9562 | .8055049 | 46946491 | 0.000000 : | pP21(3) : Excited a3Pv=13 2 2 F2e - X Xv=0 3 3 F1e |
| 13C18O Excited | 2 e 4 | X 3 e 1 | 67692.8316 | .0092527 | 67713.7878 | 20.9562 | .0102309 | 597168.6 | 0.000000 : | qP31(3) : Excited a3Pv=13 2 3 F3e - X Xv=0 3 3 F1e |
| 13C18O Excited | 2 e 5 | X 3 e 1 | 67785.0836 | .3059109 | 67806.0397 | 20.9562 | .3382543 | 19824361 | 0.000000 : | oP1(3) : Excited ev=4 2 1 F1e - X Xv=0 3 3 F1e |
| 13C18O Excited | 2 e 6 | X 3 e 1 | 67795.8685 | .2313222 | 67816.8247 | 20.9562 | .2557795 | 14997845 | 0.000000 : | qP31(3) : Excited ev=4 2 3 F3e - X Xv=0 3 3 F1e |
| 13C18O Excited | 2 e 7 | X 3 e 1 | 67932.4278 | .0035680 | 67953.3840 | 20.9562 | .0039452 | 232732.1 | 0.000000 : | pP21(3) : Excited dv=7 2 2 F2e - X Xv=0 3 3 F1e |
| 13C18O Excited | 2 e 8 | X 3 e 1 | 67971.5005 | .5601416 | 67992.4567 | 20.9562 | .6193644 | 36599928 | 0.000000 : | qP31(3) : Excited dv=7 2 3 F3e - X Xv=0 3 3 F1e |
| 13C18O Excited | 2 e 9 | X 3 e 1 | 67975.8376 | .6292307 | 67996.7938 | 20.9562 | .6957582 | 41122117 | 0.000000 : | pP21(3) : Excited a'v=12 2 2 F2e - X Xv=0 3 3 F1e |
| 13C18O Excited | 3 f 1 | X 3 e 1 | 67553.6406 | 1.875400 | 67574.5968 | 20.9562 | 2.073682 | 85923883 | 0.000000 : | qQ(3) : Excited Av=2 3 3 F1f - X Xv=0 3 3 F1e |
| 13C18O Excited | 3 f 2 | X 3 e 1 | 67569.9977 | .0013338 | 67590.9538 | 20.9562 | .0014749 | 61155.68 | 0.000000 : | qQ(3) : Excited Iv=3 3 3 F1f - X Xv=0 3 3 F1e |
| 13C18O Excited | 3 f 3 | X 3 e 1 | 67630.7086 | .0758726 | 67651.6648 | 20.9562 | .0838944 | 3488112 | 0.000000 : | pQ1(3) : Excited a3Pv=13 3 2 F1f - X Xv=0 3 3 F1e |
| 13C18O Excited | 3 f 4 | X 3 e 1 | 67667.0824 | 2.468487 | 67688.0385 | 20.9562 | 2.729476 | 1.1367e8 | 0.000000 : | qQ21(3) : Excited a3Pv=13 3 3 F2f - X Xv=0 3 3 F1e |
| 13C18O Excited | 3 f 5 | X 3 e 1 | 67701.2788 | .0783690 | 67722.2350 | 20.9562 | .0866548 | 3614170 | 0.000000 : | rQ31(3) : Excited a3Pv=13 3 4 F3f - X Xv=0 3 3 F1e |
| 13C18O Excited | 3 f 6 | X 3 e 1 | 67796.3071 | 1.876507 | 67817.2633 | 20.9562 | 2.074906 | 86904451 | 0.000000 : | qQ21(3) : Excited ev=4 3 3 F2f - X Xv=0 3 3 F1e |
| 13C18O Excited | 3 f 8 | X 3 e 1 | 67938.9130 | .0306392 | 67959.8692 | 20.9562 | .0338786 | 1427929 | 0.000000 : | qQ21(3) : Excited dv=7 3 3 F2f - X Xv=0 3 3 F1e |
| 13C18O Excited | 3 f 9 | X 3 e 1 | 67976.7064 | 1.252944 | 67997.6626 | 20.9562 | 1.385415 | 58490562 | 0.000000 : | pQ1(3) : Excited a'v=12 3 2 F1f - X Xv=0 3 3 F1e |
| 13C18O Excited | 3 f 10 | X 3 e 1 | 67978.3417 | 2.125488 | 67999.2979 | 20.9562 | 2.350212 | 99230264 | 0.000000 : | rQ31(3) : Excited dv=7 3 4 F3f - X Xv=0 3 3 F1e |
| 13C18O Excited | 3 f 11 | X 3 e 1 | 67991.7609 | .7660011 | 68012.7171 | 20.9562 | .8469891 | 35782615 | 0.000000 : | rQ31(3) : Excited a'v=12 3 4 F3f - X Xv=0 3 3 F1e |
| 13C18O Excited | 4 e 1 | X 3 e 1 | 67564.9806 | 1.338211 | 67585.9368 | 20.9562 | 1.479698 | 47711043 | 0.000000 : | rR(3) : Excited Av=2 4 4 F1e - X Xv=0 3 3 F1e |
| 13C18O Excited | 4 e 2 | X 3 e 1 | 67636.7569 | .0803359 | 67657.7131 | 20.9562 | .0888297 | 2873343 | 0.000000 : | qR1(3) : Excited a3Pv=13 4 3 F1e - X Xv=0 3 3 F1e |
| 13C18O Excited | 4 e 3 | X 3 e 1 | 67677.3519 | 1.701584 | 67698.3080 | 20.9562 | 1.881489 | 60969519 | 0.000000 : | rR21(3) : Excited a3Pv=13 4 4 F2e - X Xv=0 3 3 F1e |
| 13C18O Excited | 4 e 4 | X 3 e 1 | 67712.5220 | .0962723 | 67733.4782 | 20.9562 | .1064510 | 3454916 | 0.000000 : | sR31(3) : Excited a3Pv=13 4 5 F3e - X Xv=0 3 3 F1e |
| 13C18O Excited | 4 e 5 | X 3 e 1 | 67796.0846 | .7175174 | 67817.0407 | 20.9562 | .7933793 | 25844947 | 0.000000 : | qR1(3) : Excited ev=4 4 3 F1e - X Xv=0 3 3 F1e |
| 13C18O Excited | 4 e 6 | X 3 e 1 | 67815.6299 | .6204992 | 67836.5861 | 20.9562 | .6861035 | 22369688 | 0.000000 : | sR31(3) : Excited ev=4 4 5 F3e - X Xv=0 3 3 F1e |
| 13C18O Excited | 4 e 7 | X 3 e 1 | 67908.5618 | .0002169 | 67929.5180 | 20.9562 | .0002399 | 7852.716 | 0.000000 : | qR1(3) : Excited dv=7 4 3 F1e - X Xv=0 3 3 F1e |
| 13C18O Excited | 4 e 8 | X 3 e 1 | 67947.5591 | .0384518 | 67968.5152 | 20.9562 | .0425173 | 1394338 | 0.000000 : | rR21(3) : Excited dv=7 4 4 F2e - X Xv=0 3 3 F1e |
| 13C18O Excited | 4 e 9 | X 3 e 1 | 67987.4493 | 1.346585 | 68008.4055 | 20.9562 | 1.488957 | 48915810 | 0.000000 : | sR31(3) : Excited dv=7 4 5 F3e - X Xv=0 3 3 F1e |
| 13C18O Excited | 4 e 10 | X 3 e 1 | 67990.3630 | 1.596838 | 68011.3192 | 20.9562 | 1.765669 | 58013926 | 0.000000 : | rR21(3) : Excited a'v=12 4 4 F2e - X Xv=0 3 3 F1e |
| 13C18O Excited | 3 e 1 | X 4 e 1 | 67539.6741 | .7519026 | 67574.6002 | 34.9261 | .8890113 | 36813706 | 0.000000 : | pP(4) : Excited Av=2 3 3 F1e - X Xv=0 4 4 F1e |
| 13C18O Excited | 3 e 2 | X 4 e 1 | 67613.0186 | .0278655 | 67647.9448 | 34.9261 | .0329468 | 1368766 | 0.000000 : | oP1(4) : Excited a3Pv=13 3 2 F1e - X Xv=0 4 4 F1e |
| 13C18O Excited | 3 e 3 | X 4 e 1 | 67653.0343 | .9919286 | 67687.9604 | 34.9261 | 1.172806 | 48810509 | 0.000000 : | pP21(4) : Excited a3Pv=13 3 3 F2e - X Xv=0 4 4 F1e |
| 13C18O Excited | 3 e 4 | X 4 e 1 | 67687.3083 | .0313490 | 67722.2345 | 34.9261 | .0370655 | 1544959 | 0.000000 : | qP31(4) : Excited a3Pv=13 3 4 F3e - X Xv=0 4 4 F1e |
| 13C18O Excited | 3 e 5 | X 4 e 1 | 67775.5114 | .4125233 | 67810.4376 | 34.9261 | .4877465 | 20409762 | 0.000000 : | oP1(4) : Excited ev=4 3 2 F1e - X Xv=0 4 4 F1e |
| 13C18O Excited | 3 e 6 | X 4 e 1 | 67790.6848 | .3399132 | 67825.6110 | 34.9261 | .4018960 | 16828646 | 0.000000 : | qP31(4) : Excited ev=4 3 4 F3e - X Xv=0 4 4 F1e |
| 13C18O Excited | 3 e 8 | X 4 e 1 | 67924.9430 | .0122819 | 67959.8692 | 34.9261 | .0145215 | 611679.0 | 0.000000 : | pP21(4) : Excited dv=7 3 3 F2e - X Xv=0 4 4 F1e |
| 13C18O Excited | 3 e 9 | X 4 e 1 | 67964.3682 | .7738951 | 67999.2944 | 34.9261 | .9150141 | 38609754 | 0.000000 : | qP31(4) : Excited dv=7 3 4 F3e - X Xv=0 4 4 F1e |
| 13C18O Excited | 3 e 10 | X 4 e 1 | 67968.0930 | .8871806 | 68003.0192 | 34.9261 | 1.048957 | 44268863 | 0.000000 : | pP21(4) : Excited a'v=12 3 3 F2e - X Xv=0 4 4 F1e |
| 13C18O Excited | 4 f 1 | X 4 e 1 | 67551.0038 | 2.250952 | 67585.9299 | 34.9261 | 2.661411 | 85760682 | 0.000000 : | qQ(4) : Excited Av=2 4 4 F1f - X Xv=0 4 4 F1e |
| 13C18O Excited | 4 f 2 | X 4 e 1 | 67564.8563 | .0031554 | 67599.7824 | 34.9261 | .0037308 | 120293.1 | 0.000000 : | qQ(4) : Excited Iv=3 4 4 F1f - X Xv=0 4 4 F1e |
| 13C18O Excited | 4 f 3 | X 4 e 1 | 67626.4374 | .1463351 | 67661.3636 | 34.9261 | .1730191 | 5594027 | 0.000000 : | pQ1(4) : Excited a3Pv=13 4 3 F1f - X Xv=0 4 4 F1e |
| 13C18O Excited | 4 f 4 | X 4 e 1 | 67663.5066 | 2.853027 | 67698.4328 | 34.9261 | 3.373274 | 1.0924e8 | 0.000000 : | qQ21(4) : Excited a3Pv=13 4 4 F2f - X Xv=0 4 4 F1e |
| 13C18O Excited | 4 f 5 | X 4 e 1 | 67698.5533 | .1625550 | 67733.4795 | 34.9261 | .1921967 | 6233976 | 0.000000 : | rQ31(4) : Excited a3Pv=13 4 5 F3f - X Xv=0 4 4 F1e |
| 13C18O Excited | 4 f 6 | X 4 e 1 | 67791.1288 | 2.250649 | 67826.0549 | 34.9261 | 2.661053 | 86666848 | 0.000000 : | qQ21(4) : Excited ev=4 4 4 F2f - X Xv=0 4 4 F1e |
| 13C18O Excited | 4 f 7 | X 4 e 1 | 67894.5919 | .0003651 | 67929.5180 | 34.9261 | .0004317 | 14124.37 | 0.000000 : | pQ1(4) : Excited dv=7 4 3 F1f - X Xv=0 4 4 F1e |
| 13C18O Excited | 4 f 8 | X 4 e 1 | 67933.5891 | .0647120 | 67968.5152 | 34.9261 | .0765122 | 2507640 | 0.000000 : | qQ21(4) : Excited dv=7 4 4 F2f - X Xv=0 4 4 F1e |
| 13C18O Excited | 4 f 9 | X 4 e 1 | 67969.0113 | 1.522801 | 68003.9375 | 34.9261 | 1.800483 | 59102071 | 0.000000 : | pQ1(4) : Excited a'v=12 4 3 F1f - X Xv=0 4 4 F1e |
| 13C18O Excited | 4 f 10 | X 4 e 1 | 67973.4818 | 2.432568 | 68008.4080 | 34.9261 | 2.876144 | 94430017 | 0.000000 : | rQ31(4) : Excited dv=7 4 5 F3f - X Xv=0 4 4 F1e |
| 13C18O Excited | 4 f 11 | X 4 e 1 | 67988.1161 | .9994898 | 68023.0422 | 34.9261 | 1.181746 | 38824330 | 0.000000 : | rQ31(4) : Excited a'v=12 4 5 F3f - X Xv=0 4 4 F1e |
| 13C18O Excited | 5 e 1 | X 4 e 1 | 67565.1800 | 1.499192 | 67600.1062 | 34.9261 | 1.772568 | 46763002 | 0.000000 : | rR(4) : Excited Av=2 5 5 F1e - X Xv=0 4 4 F1e |
| 13C18O Excited | 5 e 2 | X 4 e 1 | 67635.0106 | .1304710 | 67669.9367 | 34.9261 | .1542623 | 4082303 | 0.000000 : | qR1(4) : Excited a3Pv=13 5 4 F1e - X Xv=0 4 4 F1e |
| 13C18O Excited | 5 e 3 | X 4 e 1 | 67676.3267 | 1.824761 | 67711.2528 | 34.9261 | 2.157505 | 57199579 | 0.000000 : | rR21(4) : Excited a3Pv=13 5 5 F2e - X Xv=0 4 4 F1e |
| 13C18O Excited | 5 e 4 | X 4 e 1 | 67712.5788 | .1597001 | 67747.5049 | 34.9261 | .1888212 | 5014061 | 0.000000 : | sR31(4) : Excited a3Pv=13 5 6 F3e - X Xv=0 4 4 F1e |
| 13C18O Excited | 5 e 5 | X 4 e 1 | 67790.9193 | .7909811 | 67825.8454 | 34.9261 | .9352157 | 24920519 | 0.000000 : | qR1(4) : Excited ev=4 5 4 F1e - X Xv=0 4 4 F1e |
| 13C18O Excited | 5 e 6 | X 4 e 1 | 67814.8273 | .7062360 | 67849.7534 | 34.9261 | .8350174 | 22274105 | 0.000000 : | sR31(4) : Excited ev=4 5 6 F3e - X Xv=0 4 4 F1e |
| 13C18O Excited | 5 e 7 | X 4 e 1 | 67904.7964 | .0006204 | 67939.7225 | 34.9261 | .0007336 | 19645.50 | 0.000000 : | qR1(4) : Excited dv=7 5 4 F1e - X Xv=0 4 4 F1e |
| 13C18O Excited | 5 e 8 | X 4 e 1 | 67944.3952 | .0651661 | 67979.3213 | 34.9261 | .0770491 | 2067089 | 0.000000 : | rR21(4) : Excited dv=7 5 5 F2e - X Xv=0 4 4 F1e |
| 13C18O Excited | 5 e 9 | X 4 e 1 | 67984.8590 | 1.446077 | 68019.7851 | 34.9261 | 1.709768 | 45952010 | 0.000000 : | sR31(4) : Excited dv=7 5 6 F3e - X Xv=0 4 4 F1e |
| 13C18O Excited | 5 e 10 | X 4 e 1 | 67986.7676 | 1.834535 | 68021.6938 | 34.9261 | 2.169060 | 58300945 | 0.000000 : | rR21(4) : Excited a'v=12 5 5 F2e - X Xv=0 4 4 F1e |
| 13C18O Excited | 4 e 1 | X 5 e 1 | 67533.5491 | .9207626 | 67585.9368 | 52.3877 | 1.183758 | 38115591 | 0.000000 : | pP(5) : Excited Av=2 4 4 F1e - X Xv=0 5 5 F1e |
| 13C18O Excited | 4 e 2 | X 5 e 1 | 67605.3254 | .0552755 | 67657.7131 | 52.3877 | .0710638 | 2295471 | 0.000000 : | oP1(5) : Excited a3Pv=13 4 3 F1e - X Xv=0 5 5 F1e |
| 13C18O Excited | 4 e 3 | X 5 e 1 | 67645.9203 | 1.170783 | 67698.3080 | 52.3877 | 1.505191 | 48707688 | 0.000000 : | pP21(5) : Excited a3Pv=13 4 4 F2e - X Xv=0 5 5 F1e |
| 13C18O Excited | 4 e 4 | X 5 e 1 | 67681.0905 | .0662406 | 67733.4782 | 52.3877 | .0851608 | 2760086 | 0.000000 : | qP31(5) : Excited a3Pv=13 4 5 F3e - X Xv=0 5 5 F1e |
| 13C18O Excited | 4 e 5 | X 5 e 1 | 67764.6530 | .4936913 | 67817.0407 | 52.3877 | .6347035 | 20647213 | 0.000000 : | oP1(5) : Excited ev=4 4 3 F1e - X Xv=0 5 5 F1e |
| 13C18O Excited | 4 e 6 | X 5 e 1 | 67784.1984 | .4269374 | 67836.5861 | 52.3877 | .5488828 | 17870879 | 0.000000 : | qP31(5) : Excited ev=4 4 5 F3e - X Xv=0 5 5 F1e |
| 13C18O Excited | 4 e 7 | X 5 e 1 | 67877.1303 | .0001493 | 67929.5180 | 52.3877 | .0001919 | 6273.453 | 0.000000 : | oP1(5) : Excited dv=7 4 3 F1e - X Xv=0 5 5 F1e |
| 13C18O Excited | 4 e 8 | X 5 e 1 | 67916.1275 | .0264570 | 67968.5152 | 52.3877 | .0340138 | 1113923 | 0.000000 : | pP21(5) : Excited dv=7 4 4 F2e - X Xv=0 5 5 F1e |
| 13C18O Excited | 4 e 9 | X 5 e 1 | 67956.0178 | .9265241 | 68008.4055 | 52.3877 | 1.191165 | 39078398 | 0.000000 : | qP31(5) : Excited dv=7 4 5 F3e - X Xv=0 5 5 F1e |
| 13C18O Excited | 4 e 10 | X 5 e 1 | 67958.9315 | 1.098712 | 68011.3192 | 52.3877 | 1.412535 | 46346804 | 0.000000 : | pP21(5) : Excited a'v=12 4 4 F2e - X Xv=0 5 5 F1e |
| 13C18O Excited | 5 f 1 | X 5 e 1 | 67547.7051 | 2.522906 | 67600.0928 | 52.3877 | 3.243518 | 85502513 | 0.000000 : | qQ(5) : Excited Av=2 5 5 F1f - X Xv=0 5 5 F1e |
| 13C18O Excited | 5 f 2 | X 5 e 1 | 67558.4301 | .0071856 | 67610.8178 | 52.3877 | .0092380 | 243638.0 | 0.000000 : | qQ(5) : Excited Iv=3 5 5 F1f - X Xv=0 5 5 F1e |
| 13C18O Excited | 5 f 3 | X 5 e 1 | 67621.1196 | .2356965 | 67673.5073 | 52.3877 | .3030180 | 8013943 | 0.000000 : | pQ1(5) : Excited a3Pv=13 5 4 F1f - X Xv=0 5 5 F1e |
| 13C18O Excited | 5 f 4 | X 5 e 1 | 67659.0428 | 3.060305 | 67711.4305 | 52.3877 | 3.934414 | 1.0423e8 | 0.000000 : | qQ21(5) : Excited a3Pv=13 5 5 F2f - X Xv=0 5 5 F1e |
| 13C18O Excited | 5 f 5 | X 5 e 1 | 67695.1199 | .2704059 | 67747.5076 | 52.3877 | .3476414 | 9224320 | 0.000000 : | rQ31(5) : Excited a3Pv=13 5 6 F3f - X Xv=0 5 5 F1e |
| 13C18O Excited | 5 f 6 | X 5 e 1 | 67784.6566 | 2.521465 | 67837.0443 | 52.3877 | 3.241666 | 86356152 | 0.000000 : | qQ21(5) : Excited ev=4 5 5 F2f - X Xv=0 5 5 F1e |
| 13C18O Excited | 5 f 7 | X 5 e 1 | 67887.3348 | .0010459 | 67939.7225 | 52.3877 | .0013446 | 35981.90 | 0.000000 : | pQ1(5) : Excited dv=7 5 4 F1f - X Xv=0 5 5 F1e |
| 13C18O Excited | 5 f 8 | X 5 e 1 | 67926.9336 | .1098305 | 67979.3213 | 52.3877 | .1412012 | 3785255 | 0.000000 : | qQ21(5) : Excited dv=7 5 5 F2f - X Xv=0 5 5 F1e |
| 13C18O Excited | 5 f 9 | X 5 e 1 | 67959.8791 | 1.678517 | 68012.2668 | 52.3877 | 2.157949 | 57933484 | 0.000000 : | pQ1(5) : Excited a'v=12 5 4 F1f - X Xv=0 5 5 F1e |
| 13C18O Excited | 5 f 10 | X 5 e 1 | 67967.4005 | 2.671773 | 68019.7882 | 52.3877 | 3.434906 | 92246010 | 0.000000 : | rQ31(5) : Excited dv=7 5 6 F3f - X Xv=0 5 5 F1e |
| 13C18O Excited | 5 f 11 | X 5 e 1 | 67983.0719 | 1.181086 | 68035.4596 | 52.3877 | 1.518437 | 40806545 | 0.000000 : | rQ31(5) : Excited a'v=12 5 6 F3f - X Xv=0 5 5 F1e |
| 13C18O Excited | 6 e 1 | X 5 e 1 | 67564.7200 | 1.605082 | 67617.1077 | 52.3877 | 2.063538 | 46063031 | 0.000000 : | rR(5) : Excited Av=2 6 6 F1e - X Xv=0 5 5 F1e |
| 13C18O Excited | 6 e 2 | X 5 e 1 | 67632.2358 | .1881011 | 67684.6235 | 52.3877 | .2418280 | 5414370 | 0.000000 : | qR1(5) : Excited a3Pv=13 6 5 F1e - X Xv=0 5 5 F1e |
| 13C18O Excited | 6 e 3 | X 5 e 1 | 67674.4117 | 1.859534 | 67726.7994 | 52.3877 | 2.390669 | 53625705 | 0.000000 : | rR21(5) : Excited a3Pv=13 6 6 F2e - X Xv=0 5 5 F1e |
| 13C18O Excited | 6 e 4 | X 5 e 1 | 67711.9117 | .2313094 | 67764.2994 | 52.3877 | .2973779 | 6681654 | 0.000000 : | sR31(5) : Excited a3Pv=13 6 7 F3e - X Xv=0 5 5 F1e |
| 13C18O Excited | 6 e 5 | X 5 e 1 | 67784.4626 | .8360968 | 67836.8503 | 52.3877 | 1.074910 | 24229390 | 0.000000 : | qR1(5) : Excited ev=4 6 5 F1e - X Xv=0 5 5 F1e |
| 13C18O Excited | 6 e 6 | X 5 e 1 | 67812.7260 | .7643332 | 67865.1137 | 52.3877 | .9826482 | 22177457 | 0.000000 : | sR31(5) : Excited ev=4 6 7 F3e - X Xv=0 5 5 F1e |
| 13C18O Excited | 6 e 7 | X 5 e 1 | 67899.5909 | .0013532 | 67951.9786 | 52.3877 | .0017398 | 39415.78 | 0.000000 : | qR1(5) : Excited dv=7 6 5 F1e - X Xv=0 5 5 F1e |
| 13C18O Excited | 6 e 8 | X 5 e 1 | 67939.8989 | .0965021 | 67992.2866 | 52.3877 | .1240658 | 2815834 | 0.000000 : | rR21(5) : Excited dv=7 6 6 F2e - X Xv=0 5 5 F1e |
| 13C18O Excited | 6 e 9 | X 5 e 1 | 67981.0370 | 1.307750 | 68033.4247 | 52.3877 | 1.681280 | 38228163 | 0.000000 : | sR31(5) : Excited dv=7 6 7 F3e - X Xv=0 5 5 F1e |
| 13C18O Excited | 6 e 10 | X 5 e 1 | 67981.7577 | 2.184621 | 68034.1454 | 52.3877 | 2.808610 | 63862916 | 0.000000 : | rR21(5) : Excited a'v=12 6 6 F2e - X Xv=0 5 5 F1e |
| 13C18O Excited | 5 e 1 | X 6 e 1 | 67526.7660 | 1.039119 | 67600.1062 | 73.3402 | 1.477140 | 38902738 | 0.000000 : | pP(6) : Excited Av=2 5 5 F1e - X Xv=0 6 6 F1e |
| 13C18O Excited | 5 e 2 | X 6 e 1 | 67596.5965 | .0904320 | 67669.9367 | 73.3402 | .1285519 | 3396126 | 0.000000 : | oP1(6) : Excited a3Pv=13 5 4 F1e - X Xv=0 6 6 F1e |
| 13C18O Excited | 5 e 3 | X 6 e 1 | 67637.9126 | 1.264778 | 67711.2528 | 73.3402 | 1.797921 | 47585194 | 0.000000 : | pP21(6) : Excited a3Pv=13 5 5 F2e - X Xv=0 6 6 F1e |
| 13C18O Excited | 5 e 4 | X 6 e 1 | 67674.1647 | .1106913 | 67747.5049 | 73.3402 | .1573510 | 4171277 | 0.000000 : | qP31(6) : Excited a3Pv=13 5 6 F3e - X Xv=0 6 6 F1e |
| 13C18O Excited | 5 e 5 | X 6 e 1 | 67752.5052 | .5482446 | 67825.8454 | 73.3402 | .7793464 | 20731816 | 0.000000 : | oP1(6) : Excited ev=4 5 4 F1e - X Xv=0 6 6 F1e |
| 13C18O Excited | 5 e 6 | X 6 e 1 | 67776.4132 | .4895061 | 67849.7534 | 73.3402 | .6958478 | 18530229 | 0.000000 : | qP31(6) : Excited ev=4 5 6 F3e - X Xv=0 6 6 F1e |
| 13C18O Excited | 5 e 7 | X 6 e 1 | 67866.3823 | .0004300 | 67939.7225 | 73.3402 | .0006113 | 16343.48 | 0.000000 : | oP1(6) : Excited dv=7 5 4 F1e - X Xv=0 6 6 F1e |
| 13C18O Excited | 5 e 8 | X 6 e 1 | 67905.9811 | .0451679 | 67979.3213 | 73.3402 | .0642076 | 1719654 | 0.000000 : | pP21(6) : Excited dv=7 5 5 F2e - X Xv=0 6 6 F1e |
| 13C18O Excited | 5 e 9 | X 6 e 1 | 67946.4449 | 1.002305 | 68019.7851 | 73.3402 | 1.424807 | 38228467 | 0.000000 : | qP31(6) : Excited dv=7 5 6 F3e - X Xv=0 6 6 F1e |
| 13C18O Excited | 5 e 10 | X 6 e 1 | 67948.3535 | 1.271552 | 68021.6938 | 73.3402 | 1.807550 | 48501814 | 0.000000 : | pP21(6) : Excited a'v=12 5 5 F2e - X Xv=0 6 6 F1e |
| 13C18O Excited | 6 f 1 | X 6 e 1 | 67543.7379 | 2.679802 | 67617.0781 | 73.3402 | 3.809421 | 84956051 | 0.000000 : | qQ(6) : Excited Av=2 6 6 F1f - X Xv=0 6 6 F1e |
| 13C18O Excited | 6 f 2 | X 6 e 1 | 67550.7243 | .0195970 | 67624.0645 | 73.3402 | .0278577 | 621463.5 | 0.000000 : | qQ(6) : Excited Iv=3 6 6 F1f - X Xv=0 6 6 F1e |
| 13C18O Excited | 6 f 3 | X 6 e 1 | 67614.7670 | .3348028 | 67688.1073 | 73.3402 | .4759324 | 10647559 | 0.000000 : | pQ1(6) : Excited a3Pv=13 6 5 F1f - X Xv=0 6 6 F1e |
| 13C18O Excited | 6 f 4 | X 6 e 1 | 67653.6934 | 3.103161 | 67727.0337 | 73.3402 | 4.411238 | 98858747 | 0.000000 : | qQ21(6) : Excited a3Pv=13 6 6 F2f - X Xv=0 6 6 F1e |
| 13C18O Excited | 6 f 5 | X 6 e 1 | 67690.9639 | .3906124 | 67764.3041 | 73.3402 | .5552674 | 12464486 | 0.000000 : | rQ31(6) : Excited a3Pv=13 6 7 F3f - X Xv=0 6 6 F1e |
| 13C18O Excited | 6 f 6 | X 6 e 1 | 67776.8912 | 2.683627 | 67850.2314 | 73.3402 | 3.814858 | 85961385 | 0.000000 : | qQ21(6) : Excited ev=4 6 6 F2f - X Xv=0 6 6 F1e |
| 13C18O Excited | 6 f 7 | X 6 e 1 | 67878.6384 | .0022722 | 67951.9786 | 73.3402 | .0032301 | 73112.17 | 0.000000 : | pQ1(6) : Excited dv=7 6 5 F1f - X Xv=0 6 6 F1e |
| 13C18O Excited | 6 f 8 | X 6 e 1 | 67918.9464 | .1619889 | 67992.2866 | 73.3402 | .2302722 | 5221489 | 0.000000 : | qQ21(6) : Excited dv=7 6 6 F2f - X Xv=0 6 6 F1e |
| 13C18O Excited | 6 f 9 | X 6 e 1 | 67949.3192 | 1.762262 | 68022.6595 | 73.3402 | 2.505110 | 56880337 | 0.000000 : | pQ1(6) : Excited a'v=12 6 5 F1f - X Xv=0 6 6 F1e |
| 13C18O Excited | 6 f 10 | X 6 e 1 | 67960.0920 | 2.797999 | 68033.4323 | 73.3402 | 3.977442 | 90353666 | 0.000000 : | rQ31(6) : Excited dv=7 6 7 F3f - X Xv=0 6 6 F1e |
| 13C18O Excited | 6 f 11 | X 6 e 1 | 67976.6198 | 1.305662 | 68049.9600 | 73.3402 | 1.856038 | 42193503 | 0.000000 : | rQ31(6) : Excited a'v=12 6 7 F3f - X Xv=0 6 6 F1e |
| 13C18O Excited | 7 e 1 | X 6 e 1 | 67563.6000 | 1.654636 | 67636.9402 | 73.3402 | 2.352115 | 45501854 | 0.000000 : | rR(6) : Excited Av=2 7 7 F1e - X Xv=0 6 6 F1e |
| 13C18O Excited | 7 e 2 | X 6 e 1 | 67628.4414 | .2477355 | 67701.7816 | 73.3402 | .3521635 | 6832265 | 0.000000 : | qR1(6) : Excited a3Pv=13 7 6 F1e - X Xv=0 6 6 F1e |
| 13C18O Excited | 7 e 3 | X 6 e 1 | 67671.6113 | 1.816499 | 67744.9515 | 73.3402 | 2.582209 | 50192996 | 0.000000 : | rR21(6) : Excited a3Pv=13 7 7 F2e - X Xv=0 6 6 F1e |
| 13C18O Excited | 7 e 4 | X 6 e 1 | 67710.5054 | .3038720 | 67783.8456 | 73.3402 | .4319633 | 8410991 | 0.000000 : | sR31(6) : Excited a3Pv=13 7 8 F3e - X Xv=0 6 6 F1e |
| 13C18O Excited | 7 e 5 | X 6 e 1 | 67776.7146 | .8522155 | 67850.0549 | 73.3402 | 1.211450 | 23658070 | 0.000000 : | qR1(6) : Excited ev=4 7 6 F1e - X Xv=0 6 6 F1e |

| | | | | | | | | | | | |
|---|---|---|---|---|---|---|---|---|---|---|---|
| 13C18O Excited | 7 e 6 | X | 6 e 1 | 67809.3268 | .7940898 | 67882.6671 | 73.3402 | 1.128823 | 22076301 | 0.000000 : | sR31(6) : Excited ev=4 7 8 F3e - X Xv=0 6 6 F1e |
| 13C18O Excited | 7 e 7 | X | 6 e 1 | 67892.9510 | .0024716 | 67966.2912 | 73.3402 | .0035135 | 68968.18 | 0.000000 : | qR1(6) : Excited dv=7 7 6 F1e - X Xv=0 6 6 F1e |
| 13C18O Excited | 7 e 8 | X | 6 e 1 | 67934.0698 | .1295309 | 68007.4101 | 73.3402 | .1841322 | 3620968 | 0.000000 : | rR21(6) : Excited dv=7 7 7 F2e - X Xv=0 6 6 F1e |
| 13C18O Excited | 7 e 9 | X | 6 e 1 | 67975.3097 | 1.519712 | 68048.6499 | 73.3402 | 2.160317 | 42560159 | 0.000000 : | rR21(6) : Excited a'v=12 7 7 F2e - X Xv=0 6 6 F1e |
| 13C18O Excited | 7 e 10 | X | 6 e 1 | 67976.0013 | 2.058800 | 68049.3416 | 73.3402 | 2.926646 | 57659289 | 0.000000 : | sR31(6) : Excited dv=7 7 8 F3e - X Xv=0 6 6 F1e |
| 13C18O Excited | 6 e 1 | X | 7 e 1 | 67519.3247 | 1.106621 | 67617.1077 | 97.7830 | 1.768747 | 39403069 | 0.000000 : | pP(7) : Excited Av=2 6 6 F1e - X Xv=0 7 7 F1e |
| 13C18O Excited | 6 e 2 | X | 7 e 1 | 67586.8405 | .1296860 | 67684.6235 | 97.7830 | .2072812 | 4631550 | 0.000000 : | oP1(7) : Excited a3Pv=13 6 5 F1e - X Xv=0 7 7 F1e |
| 13C18O Excited | 6 e 3 | X | 7 e 1 | 67629.0163 | 1.282053 | 67726.7994 | 97.7830 | 2.049145 | 45872454 | 0.000000 : | pP21(7) : Excited a3Pv=13 6 6 F2e - X Xv=0 7 7 F1e |
| 13C18O Excited | 6 e 4 | X | 7 e 1 | 67666.5163 | .1594760 | 67764.2994 | 97.7830 | .2548953 | 5715621 | 0.000000 : | qP31(7) : Excited a3Pv=13 6 7 F3e - X Xv=0 7 7 F1e |
| 13C18O Excited | 6 e 5 | X | 7 e 1 | 67739.0672 | .5764458 | 67836.8503 | 97.7830 | .9213511 | 20726351 | 0.000000 : | oP1(7) : Excited ev=4 6 5 F1e - X Xv=0 7 7 F1e |
| 13C18O Excited | 6 e 6 | X | 7 e 1 | 67767.3307 | .5269684 | 67865.1137 | 97.7830 | .8422699 | 18971099 | 0.000000 : | qP31(7) : Excited ev=4 6 7 F3e - X Xv=0 7 7 F1e |
| 13C18O Excited | 6 e 7 | X | 7 e 1 | 67854.1956 | .0009330 | 67951.9786 | 97.7830 | .0014912 | 33717.23 | 0.000000 : | oP1(7) : Excited dv=7 6 5 F1e - X Xv=0 7 7 F1e |
| 13C18O Excited | 6 e 8 | X | 7 e 1 | 67894.5036 | .0665333 | 67992.2866 | 97.7830 | .1063422 | 2408737 | 0.000000 : | pP21(7) : Excited dv=7 6 6 F2e - X Xv=0 7 7 F1e |
| 13C18O Excited | 6 e 9 | X | 7 e 1 | 67935.6416 | .9016262 | 68033.4247 | 97.7830 | 1.441097 | 32701399 | 0.000000 : | qP31(7) : Excited dv=7 6 7 F3e - X Xv=0 7 7 F1e |
| 13C18O Excited | 6 e 10 | X | 7 e 1 | 67936.3624 | 1.506184 | 68034.1454 | 97.7830 | 2.407380 | 54630057 | 0.000000 : | pP21(7) : Excited a'v=12 6 6 F2e - X Xv=0 7 7 F1e |
| 13C18O Excited | 7 f 1 | X | 7 e 1 | 67539.0504 | 2.631313 | 67636.8335 | 97.7830 | 4.205709 | 81271120 | 0.000000 : | qQ(7) : Excited Av=2 7 7 F1f - X Xv=0 7 7 F1e |
| 13C18O Excited | 7 f 2 | X | 7 e 1 | 67541.7893 | .1327384 | 67639.5724 | 97.7830 | .2121598 | 4100276 | 0.000000 : | qQ(7) : Excited Iv=3 7 7 F1f - X Xv=0 7 7 F1e |
| 13C18O Excited | 7 f 3 | X | 7 e 1 | 67607.3925 | .4326530 | 67705.1755 | 97.7830 | .6915227 | 13403591 | 0.000000 : | pQ1(7) : Excited a3Pv=13 7 6 F1f - X Xv=0 7 7 F1e |
| 13C18O Excited | 7 f 4 | X | 7 e 1 | 67647.4609 | 3.007651 | 67745.2439 | 97.7830 | 4.807221 | 93342779 | 0.000000 : | qQ21(7) : Excited a3Pv=13 7 7 F2f - X Xv=0 7 7 F1e |
| 13C18O Excited | 7 f 5 | X | 7 e 1 | 67686.0700 | .5100054 | 67783.8530 | 97.7830 | .8151574 | 15855193 | 0.000000 : | rQ31(7) : Excited a3Pv=13 7 8 F3f - X Xv=0 7 7 F1e |
| 13C18O Excited | 7 f 6 | X | 7 e 1 | 67767.8333 | 2.739235 | 67865.6163 | 97.7830 | 4.378203 | 85467087 | 0.000000 : | qQ21(7) : Excited ev=4 7 7 F2f - X Xv=0 7 7 F1e |
| 13C18O Excited | 7 f 7 | X | 7 e 1 | 67868.5082 | .0041201 | 67966.2912 | 97.7830 | .0065852 | 129124.6 | 0.000000 : | pQ1(7) : Excited dv=7 7 6 F1f - X Xv=0 7 7 F1e |
| 13C18O Excited | 7 f 8 | X | 7 e 1 | 67909.6270 | .2158203 | 68007.4101 | 97.7830 | .3449523 | 6776185 | 0.000000 : | qQ21(7) : Excited dv=7 7 7 F2f - X Xv=0 7 7 F1e |
| 13C18O Excited | 7 f 9 | X | 7 e 1 | 67937.3361 | 1.782314 | 68035.1191 | 97.7830 | 2.848728 | 56028463 | 0.000000 : | pQ1(7) : Excited a'v=12 7 6 F1f - X Xv=0 7 7 F1e |
| 13C18O Excited | 7 f 10 | X | 7 e 1 | 67951.5510 | 2.812667 | 68049.3340 | 97.7830 | 4.495572 | 88473921 | 0.000000 : | rQ31(7) : Excited dv=7 7 8 F3f - X Xv=0 7 7 F1e |
| 13C18O Excited | 7 f 11 | X | 7 e 1 | 67968.7555 | 1.372800 | 68066.5385 | 97.7830 | 2.194188 | 43214948 | 0.000000 : | rQ31(7) : Excited a'v=12 7 8 F3f - X Xv=0 7 7 F1e |
| 13C18O Excited | 8 e 1 | X | 7 e 1 | 67561.8196 | 1.650285 | 67659.6027 | 97.7830 | 2.637702 | 45019864 | 0.000000 : | rR(7) : Excited Av=2 8 8 F1e - X Xv=0 7 7 F1e |
| 13C18O Excited | 8 e 2 | X | 7 e 1 | 67623.6365 | .3035469 | 67721.4195 | 97.7830 | .4851684 | 8303527 | 0.000000 : | qR1(7) : Excited a3Pv=13 8 7 F1e - X Xv=0 7 7 F1e |
| 13C18O Excited | 8 e 3 | X | 7 e 1 | 67667.9293 | 1.711181 | 67765.7124 | 97.7830 | 2.735034 | 46901407 | 0.000000 : | rR21(7) : Excited a3Pv=13 8 8 F2e - X Xv=0 7 7 F1e |
| 13C18O Excited | 8 e 4 | X | 7 e 1 | 67708.3446 | .3704383 | 67806.1276 | 97.7830 | .5920829 | 10171468 | 0.000000 : | sR31(7) : Excited a3Pv=13 8 9 F3e - X Xv=0 7 7 F1e |
| 13C18O Excited | 8 e 5 | X | 7 e 1 | 67767.6764 | .8405361 | 67865.4594 | 97.7830 | 1.343455 | 23140107 | 0.000000 : | qR1(7) : Excited ev=4 8 7 F1e - X Xv=0 7 7 F1e |
| 13C18O Excited | 8 e 6 | X | 7 e 1 | 67804.6300 | .7966035 | 67902.4131 | 97.7830 | 1.273236 | 21966530 | 0.000000 : | sR31(7) : Excited ev=4 8 9 F3e - X Xv=0 7 7 F1e |
| 13C18O Excited | 8 e 7 | X | 7 e 1 | 67884.8825 | .0039652 | 67982.6656 | 97.7830 | .0063377 | 109730.7 | 0.000000 : | qR1(7) : Excited dv=7 8 7 F1e - X Xv=0 7 7 F1e |
| 13C18O Excited | 8 e 8 | X | 7 e 1 | 67926.9075 | .1609936 | 68024.6905 | 97.7830 | .2573210 | 4463497 | 0.000000 : | rR21(7) : Excited dv=7 8 8 F2e - X Xv=0 7 7 F1e |
| 13C18O Excited | 8 e 9 | X | 7 e 1 | 67967.4636 | 1.783716 | 68065.2466 | 97.7830 | 2.850969 | 49541614 | 0.000000 : | rR21(7) : Excited a'v=12 8 8 F2e - X Xv=0 7 7 F1e |
| 13C18O Excited | 8 e 10 | X | 7 e 1 | 67969.7061 | 1.763524 | 68067.4891 | 97.7830 | 2.818695 | 48985640 | 0.000000 : | sR31(7) : Excited dv=7 8 9 F3e - X Xv=0 7 7 F1e |
| 13C18O Excited | 7 e 1 | X | 8 e 1 | 67511.2250 | 1.126216 | 67636.9402 | 125.7152 | 2.058101 | 39721603 | 0.000000 : | pP(8) : Excited Av=2 7 7 F1e - X Xv=0 8 8 F1e |
| 13C18O Excited | 7 e 2 | X | 8 e 1 | 67576.0664 | .1686193 | 67701.7816 | 125.7152 | .3081431 | 5964353 | 0.000000 : | oP1(8) : Excited a3Pv=13 7 6 F1e - X Xv=0 8 8 F1e |
| 13C18O Excited | 7 e 3 | X | 8 e 1 | 67619.2363 | 1.236387 | 67744.9515 | 125.7152 | 2.259433 | 43816976 | 0.000000 : | pP21(8) : Excited a3Pv=13 7 7 F2e - X Xv=0 8 8 F1e |
| 13C18O Excited | 7 e 4 | X | 8 e 1 | 67658.1304 | .2068282 | 67783.8456 | 125.7152 | .3779679 | 7342552 | 0.000000 : | qP31(8) : Excited a3Pv=13 7 8 F3e - X Xv=0 8 8 F1e |
| 13C18O Excited | 7 e 5 | X | 8 e 1 | 67724.3397 | .5800542 | 67850.0549 | 125.7152 | 1.060019 | 20652858 | 0.000000 : | oP1(8) : Excited ev=4 7 6 F1e - X Xv=0 8 8 F1e |
| 13C18O Excited | 7 e 6 | X | 8 e 1 | 67756.9519 | .5404915 | 67882.6671 | 125.7152 | .9877201 | 19272038 | 0.000000 : | qP31(8) : Excited ev=4 7 8 F3e - X Xv=0 8 8 F1e |
| 13C18O Excited | 7 e 7 | X | 8 e 1 | 67840.5760 | .0016823 | 67966.2912 | 125.7152 | .0030743 | 60207.61 | 0.000000 : | oP1(8) : Excited dv=7 7 6 F1e - X Xv=0 8 8 F1e |
| 13C18O Excited | 7 e 8 | X | 8 e 1 | 67881.6949 | .0881643 | 68007.4101 | 125.7152 | .1611156 | 3161025 | 0.000000 : | pP21(8) : Excited dv=7 7 7 F2e - X Xv=0 8 8 F1e |
| 13C18O Excited | 7 e 9 | X | 8 e 1 | 67922.9347 | 1.034381 | 68048.6499 | 125.7152 | 1.890277 | 37154125 | 0.000000 : | pP21(8) : Excited a'v=12 7 7 F2e - X Xv=0 8 8 F1e |
| 13C18O Excited | 7 e 10 | X | 8 e 1 | 67923.6264 | 1.401307 | 68049.3416 | 125.7152 | 2.560816 | 50335350 | 0.000000 : | qP31(8) : Excited dv=7 7 8 F3e - X Xv=0 8 8 F1e |
| 13C18O Excited | 8 f 1 | X | 8 e 1 | 67531.2381 | .1109338 | 67656.9533 | 125.7152 | .2027257 | 3455394 | 0.000000 : | qQ(8) : Excited Iv=3 8 8 F1f - X Xv=0 8 8 F1e |
| 13C18O Excited | 8 f 2 | X | 8 e 1 | 67534.0283 | 2.621588 | 67659.7435 | 125.7152 | 4.790815 | 81667996 | 0.000000 : | qQ(8) : Excited Av=2 8 8 F1f - X Xv=0 8 8 F1e |
| 13C18O Excited | 8 f 3 | X | 8 e 1 | 67599.0088 | .5186946 | 67724.7240 | 125.7152 | .9478874 | 16205120 | 0.000000 : | pQ1(8) : Excited a3Pv=13 8 7 F1f - X Xv=0 8 8 F1e |
| 13C18O Excited | 8 f 4 | X | 8 e 1 | 67640.3470 | 2.806919 | 67766.0623 | 125.7152 | 5.129498 | 87855078 | 0.000000 : | qQ21(8) : Excited a3Pv=13 8 8 F2f - X Xv=0 8 8 F1e |
| 13C18O Excited | 8 f 5 | X | 8 e 1 | 67680.4228 | .6163484 | 67806.1380 | 125.7152 | 1.126345 | 19325690 | 0.000000 : | rQ31(8) : Excited a3Pv=13 8 9 F3f - X Xv=0 8 8 F1e |
| 13C18O Excited | 8 f 6 | X | 8 e 1 | 67757.4841 | 2.696916 | 67883.1993 | 125.7152 | 4.928473 | 84851343 | 0.000000 : | qQ21(8) : Excited ev=4 8 8 F2f - X Xv=0 8 8 F1e |
| 13C18O Excited | 8 f 7 | X | 8 e 1 | 67856.9503 | .0065473 | 67982.6656 | 125.7152 | .0119649 | 206903.0 | 0.000000 : | pQ1(8) : Excited dv=7 8 7 F1f - X Xv=0 8 8 F1e |
| 13C18O Excited | 8 f 8 | X | 8 e 1 | 67898.9753 | .2656480 | 68024.6905 | 125.7152 | .4854578 | 8410382 | 0.000000 : | qQ21(8) : Excited dv=7 8 8 F2f - X Xv=0 8 8 F1e |
| 13C18O Excited | 8 f 9 | X | 8 e 1 | 67923.9320 | 1.746109 | 68049.6472 | 125.7152 | 3.190923 | 55342561 | 0.000000 : | pQ1(8) : Excited a'v=12 8 7 F1f - X Xv=0 8 8 F1e |
| 13C18O Excited | 8 f 10 | X | 8 e 1 | 67941.7716 | 2.728658 | 68067.4868 | 125.7152 | 4.986481 | 86552435 | 0.000000 : | rQ31(8) : Excited dv=7 8 9 F3f - X Xv=0 8 8 F1e |
| 13C18O Excited | 8 f 11 | X | 8 e 1 | 67959.4766 | 1.385955 | 68085.1918 | 125.7152 | 2.532761 | 43996564 | 0.000000 : | rQ31(8) : Excited a'v=12 8 9 F3f - X Xv=0 8 8 F1e |
| 13C18O Excited | 9 e 1 | X | 8 e 1 | 67559.3783 | 1.597613 | 67685.0936 | 125.7152 | 2.919555 | 44580347 | 0.000000 : | rR(8) : Excited Av=2 9 9 F1e - X Xv=0 8 8 F1e |
| 13C18O Excited | 9 e 2 | X | 8 e 1 | 67617.8304 | .3503696 | 67743.5456 | 125.7152 | .6402822 | 9802232 | 0.000000 : | qR1(8) : Excited a3Pv=13 9 8 F1e - X Xv=0 8 8 F1e |
| 13C18O Excited | 9 e 3 | X | 8 e 1 | 67663.3686 | 1.561241 | 67789.0838 | 125.7152 | 2.853088 | 43766899 | 0.000000 : | rR21(8) : Excited a3Pv=13 9 9 F2e - X Xv=0 8 8 F1e |
| 13C18O Excited | 9 e 4 | X | 8 e 1 | 67705.4141 | .4255777 | 67831.1293 | 125.7152 | .7777211 | 11952642 | 0.000000 : | sR31(8) : Excited a3Pv=13 9 10 F3e - X Xv=0 8 8 F1e |
| 13C18O Excited | 9 e 5 | X | 8 e 1 | 67757.3493 | .8037185 | 67883.0645 | 125.7152 | 1.468754 | 22624974 | 0.000000 : | qR1(8) : Excited ev=4 9 8 F1e - X Xv=0 8 8 F1e |
| 13C18O Excited | 9 e 6 | X | 8 e 1 | 67798.6359 | .7745556 | 67924.3511 | 125.7152 | 1.415460 | 21843908 | 0.000000 : | sR31(8) : Excited ev=4 9 10 F3e - X Xv=0 8 8 F1e |
| 13C18O Excited | 9 e 7 | X | 8 e 1 | 67875.3917 | .0057490 | 68001.1070 | 125.7152 | .0105060 | 162683.4 | 0.000000 : | qR1(8) : Excited dv=7 9 8 F1e - X Xv=0 8 8 F1e |
| 13C18O Excited | 9 e 8 | X | 8 e 1 | 67918.4116 | .1878530 | 68044.1268 | 125.7152 | .3432916 | 5325931 | 0.000000 : | rR21(8) : Excited dv=7 9 9 F2e - X Xv=0 8 8 F1e |
| 13C18O Excited | 9 e 9 | X | 8 e 1 | 67958.1946 | 1.783243 | 68083.9098 | 125.7152 | 3.258784 | 50646675 | 0.000000 : | rR21(8) : Excited a'v=12 9 9 F2e - X Xv=0 8 8 F1e |
| 13C18O Excited | 9 e 10 | X | 8 e 1 | 67962.1699 | 1.630265 | 68087.8851 | 125.7152 | 2.979225 | 46310002 | 0.000000 : | sR31(8) : Excited dv=7 9 10 F3e - X Xv=0 8 8 F1e |
| 13C18O Excited | 8 e 1 | X | 9 e 1 | 67502.4668 | 1.103529 | 67659.6027 | 157.1358 | 2.344624 | 39912284 | 0.000000 : | pP(9) : Excited Av=2 8 8 F1e - X Xv=0 9 9 F1e |
| 13C18O Excited | 8 e 2 | X | 9 e 1 | 67564.2837 | .2029788 | 67721.4195 | 157.1358 | .4312608 | 7361495 | 0.000000 : | oP1(9) : Excited a3Pv=13 8 7 F1e - X Xv=0 9 9 F1e |
| 13C18O Excited | 8 e 3 | X | 9 e 1 | 67608.5765 | 1.144250 | 67765.7124 | 157.1358 | 2.431141 | 41580534 | 0.000000 : | pP21(9) : Excited a3Pv=13 8 8 F2e - X Xv=0 9 9 F1e |
| 13C18O Excited | 8 e 4 | X | 9 e 1 | 67648.9918 | .2477084 | 67806.1276 | 157.1358 | .5262959 | 9017549 | 0.000000 : | qP31(9) : Excited a3Pv=13 8 9 F3e - X Xv=0 9 9 F1e |
| 13C18O Excited | 8 e 5 | X | 9 e 1 | 67708.3235 | .5620582 | 67865.4594 | 157.1358 | 1.194182 | 20514987 | 0.000000 : | oP1(9) : Excited ev=4 8 7 F1e - X Xv=0 9 9 F1e |
| 13C18O Excited | 8 e 6 | X | 9 e 1 | 67745.2772 | .5326809 | 67902.4131 | 157.1358 | 1.131765 | 19474573 | 0.000000 : | qP31(9) : Excited ev=4 8 9 F3e - X Xv=0 9 9 F1e |
| 13C18O Excited | 8 e 7 | X | 9 e 1 | 67825.5297 | .0026515 | 67982.6656 | 157.1358 | .0056335 | 97282.75 | 0.000000 : | oP1(9) : Excited dv=7 8 7 F1e - X Xv=0 9 9 F1e |
| 13C18O Excited | 8 e 8 | X | 9 e 1 | 67867.5547 | .1076548 | 68024.6905 | 157.1358 | .2287298 | 3957162 | 0.000000 : | pP21(9) : Excited dv=7 8 8 F2e - X Xv=0 9 9 F1e |
| 13C18O Excited | 8 e 9 | X | 9 e 1 | 67908.1108 | 1.192753 | 68065.2466 | 157.1358 | 2.534194 | 43921724 | 0.000000 : | pP21(9) : Excited a'v=12 8 8 F2e - X Xv=0 9 9 F1e |
| 13C18O Excited | 8 e 10 | X | 9 e 1 | 67910.3532 | 1.179251 | 68067.4891 | 157.1358 | 2.505507 | 43428823 | 0.000000 : | qP31(9) : Excited dv=7 8 9 F3e - X Xv=0 9 9 F1e |
| 13C18O Excited | 9 f 1 | X | 9 e 1 | 67519.7238 | .0092461 | 67676.8596 | 157.1358 | .0196449 | 299441.2 | 0.000000 : | qQ(9) : Excited Iv=3 9 9 F1f - X Xv=0 9 9 F1e |
| 13C18O Excited | 9 f 2 | X | 9 e 1 | 67528.0167 | 2.609114 | 67685.1525 | 157.1358 | 5.543478 | 84528685 | 0.000000 : | qQ(9) : Excited Av=2 9 9 F1f - X Xv=0 9 9 F1e |
| 13C18O Excited | 9 f 3 | X | 9 e 1 | 67589.6287 | .5846166 | 67746.7645 | 157.1358 | 1.242111 | 18991990 | 0.000000 : | pQ1(9) : Excited a3Pv=13 9 8 F1f - X Xv=0 9 9 F1e |
| 13C18O Excited | 9 f 4 | X | 9 e 1 | 67632.3531 | 2.535687 | 67789.4890 | 157.1358 | 5.387470 | 82531205 | 0.000000 : | qQ21(9) : Excited a3Pv=13 9 9 F2f - X Xv=0 9 9 F1e |
| 13C18O Excited | 9 f 5 | X | 9 e 1 | 67674.0072 | .7003669 | 67831.1431 | 157.1358 | 1.488041 | 22837595 | 0.000000 : | rQ31(9) : Excited a3Pv=13 9 10 F3f - X Xv=0 9 9 F1e |
| 13C18O Excited | 9 f 6 | X | 9 e 1 | 67745.8452 | 2.570388 | 67902.9811 | 157.1358 | 5.461198 | 84082533 | 0.000000 : | qQ21(9) : Excited ev=4 9 9 F2f - X Xv=0 9 9 F1e |
| 13C18O Excited | 9 f 7 | X | 9 e 1 | 67843.9711 | .0093885 | 68001.1070 | 157.1358 | .0199473 | 308452.5 | 0.000000 : | pQ1(9) : Excited dv=7 9 8 F1f - X Xv=0 9 9 F1e |
| 13C18O Excited | 9 f 8 | X | 9 e 1 | 67886.9910 | .3064623 | 68044.1268 | 157.1358 | .6511279 | 10087785 | 0.000000 : | qQ21(9) : Excited dv=7 9 9 F2f - X Xv=0 9 9 F1e |
| 13C18O Excited | 9 f 9 | X | 9 e 1 | 67909.1083 | 1.662700 | 68066.2441 | 157.1358 | 3.532672 | 54784434 | 0.000000 : | pQ1(9) : Excited a'v=12 9 8 F1f - X Xv=0 9 9 F1e |
| 13C18O Excited | 9 f 10 | X | 9 e 1 | 67930.7480 | 2.564764 | 68087.8839 | 157.1358 | 5.449248 | 84587392 | 0.000000 : | rQ31(9) : Excited dv=7 9 10 F3f - X Xv=0 9 9 F1e |
| 13C18O Excited | 9 f 11 | X | 9 e 1 | 67948.7817 | 1.351617 | 68105.9175 | 157.1358 | 2.871726 | 44612632 | 0.000000 : | rQ31(9) : Excited a'v=12 9 10 F3f - X Xv=0 9 9 F1e |
| 13C18O Excited | 10 e 1 | X | 9 e 1 | 67556.2755 | 1.504583 | 67713.4114 | 157.1358 | 3.196725 | 44157712 | 0.000000 : | rR(9) : Excited Av=2 10 10 F1e - X Xv=0 9 9 F1e |
| 13C18O Excited | 10 e 2 | X | 9 e 1 | 67611.0318 | .3844293 | 67768.1677 | 157.1358 | .8167812 | 11310000 | 0.000000 : | qR1(9) : Excited a3Pv=13 10 9 F1e - X Xv=0 9 9 F1e |
| 13C18O Excited | 10 e 3 | X | 9 e 1 | 67657.9309 | 1.384126 | 67815.0667 | 157.1358 | 2.940796 | 40806105 | 0.000000 : | rR21(9) : Excited a3Pv=13 10 10 F2e - X Xv=0 9 9 F1e |
| 13C18O Excited | 10 e 4 | X | 9 e 1 | 67701.6988 | .4662146 | 67858.8347 | 157.1358 | .9905470 | 13771395 | 0.000000 : | sR31(9) : Excited a3Pv=13 10 11 F3e - X Xv=0 9 9 F1e |

13C18O Excited 10 e 5 X 9 e 1 67745.7362 .7453082 67902.8720 157.1358 1.583526 22058461 0.000000 : qR1(9) : Excited ev=4 10 9 F1e - X Xv=0 9 9 F1e
13C18O Excited 10 e 6 X 9 e 1 67791.3447 .7318474 67948.4806 157.1358 1.554926 21703846 0.000000 : sR31(9) : Excited ev=4 10 11 F3e - X Xv=0 9 9 F1e
13C18O Excited 10 e 7 X 9 e 1 67864.4848 .0076751 68021.6207 157.1358 .0163070 228353.2 0.000000 : qR1(9) : Excited dv=7 10 9 F1e - X Xv=0 9 9 F1e
13C18O Excited 10 e 8 X 9 e 1 67908.5818 .2077412 68065.7176 157.1358 .4413792 6192842 0.000000 : rR21(9) : Excited dv=7 10 10 F2e - X Xv=0 9 9 F1e
13C18O Excited 10 e 9 X 9 e 1 67947.5062 1.707175 68104.6420 157.1358 3.627165 50979089 0.000000 : rR21(9) : Excited a'v=12 10 10 F2e - X Xv=0 9 9 F1e
13C18O Excited 10 e 10 X 9 e 1 67953.3831 1.489735 68110.5190 157.1358 3.165180 44497518 0.000000 : sR31(9) : Excited dv=7 10 11 F3e - X Xv=0 9 9 F1e
13C18O Excited 9 e 1 X 10 e 1 67493.0498 1.046073 67685.0936 192.0438 2.627600 40004254 0.000000 : pP(10) : Excited Av=2 9 9 F1e - X Xv=0 10 10 F1e
13C18O Excited 9 e 2 X 10 e 1 67551.5018 .2294124 67743.5456 192.0438 .5762540 8796073 0.000000 : oP1(10) : Excited a3Pv=13 9 8 F1e - X Xv=0 10 10 F1e
13C18O Excited 9 e 3 X 10 e 1 67597.0400 1.022258 67789.0838 192.0438 2.567779 39274483 0.000000 : pP21(10) : Excited a3Pv=13 9 9 F2e - X Xv=0 10 10 F1e
13C18O Excited 9 e 4 X 10 e 1 67639.0855 .2786566 67831.1293 192.0438 .6999490 10725793 0.000000 : qP31(10) : Excited a3Pv=13 9 10 F3e - X Xv=0 10 10 F1e
13C18O Excited 9 e 5 X 10 e 1 67691.0207 .5262528 67883.0645 192.0438 1.321879 20302736 0.000000 : oP1(10) : Excited ev=4 9 8 F1e - X Xv=0 10 10 F1e
13C18O Excited 9 e 6 X 10 e 1 67732.3073 .5071577 67924.3511 192.0438 1.273914 19601874 0.000000 : qP31(10) : Excited ev=4 9 10 F3e - X Xv=0 10 10 F1e
13C18O Excited 9 e 7 X 10 e 1 67809.0631 .0037643 68001.1070 192.0438 .0094554 145986.2 0.000000 : oP1(10) : Excited dv=7 9 8 F1e - X Xv=0 10 10 F1e
13C18O Excited 9 e 8 X 10 e 1 67852.0830 .1230010 68044.1268 192.0438 .3089624 4779308 0.000000 : pP21(10) : Excited dv=7 9 9 F2e - X Xv=0 10 10 F1e
13C18O Excited 9 e 9 X 10 e 1 67891.8660 1.167619 68083.9098 192.0438 2.932906 45448671 0.000000 : pP21(10) : Excited a'v=12 9 9 F2e - X Xv=0 10 10 F1e
13C18O Excited 9 e 10 X 10 e 1 67895.8413 1.067453 68087.8851 192.0438 2.681302 41557089 0.000000 : qP31(10) : Excited dv=7 9 10 F3e - X Xv=0 10 10 F1e
13C18O Excited 10 f 1 X 10 e 1 67506.8499 .0019734 67698.8937 192.0438 .0049570 68322.55 0.000000 : qQ(10) : Excited Iv=3 10 10 F1f - X Xv=0 10 10 F1e
13C18O Excited 10 f 2 X 10 e 1 67521.4105 2.436527 67713.4543 192.0438 6.120237 84410587 0.000000 : qQ(10) : Excited Av=2 10 10 F1f - X Xv=0 10 10 F1e
13C18O Excited 10 f 3 X 10 e 1 67579.2640 .6253703 67771.3078 192.0438 1.570849 21720951 0.000000 : pQ1(10) : Excited a3Pv=13 10 9 F1f - X Xv=0 10 10 F1e
13C18O Excited 10 f 4 X 10 e 1 67623.4799 2.226099 67815.5238 192.0438 5.591671 77470841 0.000000 : qQ21(10) : Excited a3Pv=13 10 10 F2f - X Xv=0 10 10 F1e
13C18O Excited 10 f 5 X 10 e 1 67666.8084 .7567975 67858.8522 192.0438 1.900977 26388093 0.000000 : rQ31(10) : Excited a3Pv=13 10 11 F3f - X Xv=0 10 10 F1e
13C18O Excited 10 f 6 X 10 e 1 67732.9190 2.376689 67924.9628 192.0438 5.969932 83113761 0.000000 : qQ21(10) : Excited ev=4 10 10 F2f - X Xv=0 10 10 F1e
13C18O Excited 10 f 7 X 10 e 1 67829.5768 .0123825 68021.6207 192.0438 .0311033 434878.0 0.000000 : pQ1(10) : Excited dv=7 10 9 F1f - X Xv=0 10 10 F1e
13C18O Excited 10 f 8 X 10 e 1 67873.6738 .3346376 68065.7176 192.0438 .8405660 11775518 0.000000 : qQ21(10) : Excited dv=7 10 10 F2f - X Xv=0 10 10 F1e
13C18O Excited 10 f 9 X 10 e 1 67892.8660 1.542522 68084.9098 192.0438 3.874615 54325671 0.000000 : pQ1(10) : Excited a'v=12 10 9 F1f - X Xv=0 10 10 F1e
13C18O Excited 10 f 10 X 10 e 1 67918.4743 2.342473 68110.5181 192.0438 5.883988 82592320 0.000000 : rQ31(10) : Excited dv=7 10 11 F3f - X Xv=0 10 10 F1e
13C18O Excited 10 f 11 X 10 e 1 67936.6699 1.278371 68128.7137 192.0438 3.211102 45109811 0.000000 : rQ31(10) : Excited a'v=12 10 11 F3f - X Xv=0 10 10 F1e
13C18O Excited 11 e 1 X 10 e 1 67552.5104 1.380627 67744.5542 192.0438 3.467954 43731407 0.000000 : rR(10) : Excited Av=2 11 11 F1e - X Xv=0 10 10 F1e
13C18O Excited 11 e 2 X 10 e 1 67603.2490 .4037161 67795.2928 192.0438 1.014082 12816561 0.000000 : qR1(10) : Excited a3Pv=13 11 10 F1e - X Xv=0 10 10 F1e
13C18O Excited 11 e 3 X 10 e 1 67651.6173 1.195378 67843.6612 192.0438 3.002634 38030551 0.000000 : rR21(10) : Excited a3Pv=13 11 11 F2e - X Xv=0 10 10 F1e
13C18O Excited 11 e 4 X 10 e 1 67697.1828 .4922209 67889.2267 192.0438 1.236395 15691503 0.000000 : sR31(10) : Excited a3Pv=13 11 12 F3e - X Xv=0 10 10 F1e
13C18O Excited 11 e 5 X 10 e 1 67732.8418 .6688594 67924.8856 192.0438 1.680088 21356271 0.000000 : qR1(10) : Excited ev=4 11 10 F1e - X Xv=0 10 10 F1e
13C18O Excited 11 e 6 X 10 e 1 67782.7570 .6731548 67974.8008 192.0438 1.690877 21540974 0.000000 : sR31(10) : Excited ev=4 11 12 F3e - X Xv=0 10 10 F1e
13C18O Excited 11 e 7 X 10 e 1 67852.1679 .0095601 68044.2117 192.0438 .0240136 306863.2 0.000000 : qR1(10) : Excited dv=7 11 10 F1e - X Xv=0 10 10 F1e
13C18O Excited 11 e 8 X 10 e 1 67897.4177 .2192332 68089.4615 192.0438 .5506852 7051131 0.000000 : rR21(10) : Excited dv=7 11 11 F2e - X Xv=0 10 10 F1e
13C18O Excited 11 e 9 X 10 e 1 67935.3986 1.585564 68127.4424 192.0438 3.982729 51081635 0.000000 : rR21(10) : Excited a'v=12 11 11 F2e - X Xv=0 10 10 F1e
13C18O Excited 11 e 10 X 10 e 1 67943.3393 1.333883 68135.3831 192.0438 3.350541 42988390 0.000000 : sR31(10) : Excited dv=7 11 12 F3e - X Xv=0 10 10 F1e
13C18O Excited 10 e 1 X 11 e 1 67482.9734 .9623802 67713.4114 230.4379 2.906114 40012843 0.000000 : pP(11) : Excited Av=2 10 10 F1e - X Xv=0 11 11 F1e
13C18O Excited 10 e 2 X 11 e 1 67537.7298 .2458935 67768.1677 230.4379 .7425284 10248412 0.000000 : oP1(11) : Excited a3Pv=13 10 9 F1e - X Xv=0 11 11 F1e
13C18O Excited 10 e 3 X 11 e 1 67584.6288 .8853321 67815.0667 230.4379 2.673451 36976016 0.000000 : pP21(11) : Excited a3Pv=13 10 10 F2e - X Xv=0 11 11 F1e
13C18O Excited 10 e 4 X 11 e 1 67628.3968 .2982060 67858.8347 230.4379 .9004973 12478829 0.000000 : qP31(11) : Excited a3Pv=13 10 11 F3e - X Xv=0 11 11 F1e
13C18O Excited 10 e 5 X 11 e 1 67672.4341 .4767234 67902.8720 230.4379 1.439569 19988124 0.000000 : oP1(11) : Excited ev=4 10 9 F1e - X Xv=0 11 11 F1e
13C18O Excited 10 e 6 X 11 e 1 67718.0426 .4681135 67948.4806 230.4379 1.413569 19666834 0.000000 : qP31(11) : Excited ev=4 10 11 F3e - X Xv=0 11 11 F1e
13C18O Excited 10 e 7 X 11 e 1 67791.1827 .0049093 68021.6207 230.4379 .0148246 206921.8 0.000000 : oP1(11) : Excited dv=7 10 9 F1e - X Xv=0 11 11 F1e
13C18O Excited 10 e 8 X 11 e 1 67835.2797 .1328780 68065.7176 230.4379 .4012538 5611645 0.000000 : pP21(11) : Excited dv=7 10 10 F2e - X Xv=0 11 11 F1e
13C18O Excited 10 e 9 X 11 e 1 67874.2041 1.091965 68104.6420 230.4379 3.297423 46194798 0.000000 : pP21(11) : Excited a'v=12 10 10 F2e - X Xv=0 11 11 F1e
13C18O Excited 10 e 10 X 11 e 1 67880.0811 .9528833 68110.5190 230.4379 2.877436 40321521 0.000000 : qP31(11) : Excited dv=7 10 11 F3e - X Xv=0 11 11 F1e
13C18O Excited 11 f 1 X 11 e 1 67492.6711 .0004973 67723.1091 230.4379 .0015016 18884.54 0.000000 : qQ(11) : Excited Iv=3 11 11 F1f - X Xv=0 11 11 F1e
13C18O Excited 11 f 2 X 11 e 1 67514.1534 2.210969 67744.5914 230.4379 6.676497 84048265 0.000000 : qQ(11) : Excited Av=2 11 11 F1f - X Xv=0 11 11 F1e
13C18O Excited 11 f 3 X 11 e 1 67567.9259 .6394001 67798.3638 230.4379 1.930806 24364417 0.000000 : pQ1(11) : Excited a3Pv=13 11 10 F1f - X Xv=0 11 11 F1e
13C18O Excited 11 f 4 X 11 e 1 67613.7277 1.905145 67844.1656 230.4379 5.752994 72743493 0.000000 : qQ21(11) : Excited a3Pv=13 11 11 F2f - X Xv=0 11 11 F1e
13C18O Excited 11 f 5 X 11 e 1 67658.8113 .7845464 67889.2493 230.4379 2.369107 30016033 0.000000 : rQ31(11) : Excited a3Pv=13 11 12 F3f - X Xv=0 11 11 F1e
13C18O Excited 11 f 6 X 11 e 1 67718.7090 2.134284 67949.1469 230.4379 6.444929 81872819 0.000000 : qQ21(11) : Excited ev=4 11 11 F2f - X Xv=0 11 11 F1e
13C18O Excited 11 f 7 X 11 e 1 67813.7738 .0152241 68044.2117 230.4379 .0459724 586471.3 0.000000 : pQ1(11) : Excited dv=7 11 10 F1f - X Xv=0 11 11 F1e
13C18O Excited 11 f 8 X 11 e 1 67859.0235 .3482919 68089.4615 230.4379 1.051742 13443977 0.000000 : qQ21(11) : Excited dv=7 11 11 F2f - X Xv=0 11 11 F1e
13C18O Excited 11 f 9 X 11 e 1 67875.2057 1.396613 68105.6436 230.4379 4.217374 53947488 0.000000 : pQ1(11) : Excited a'v=12 11 10 F1f - X Xv=0 11 11 F1e
13C18O Excited 11 f 10 X 11 e 1 67904.9446 2.083464 68135.3825 230.4379 6.291468 80584551 0.000000 : rQ31(11) : Excited dv=7 11 12 F3f - X Xv=0 11 11 F1e
13C18O Excited 11 f 11 X 11 e 1 67923.1404 1.175918 68153.5783 230.4379 3.550937 45518908 0.000000 : rQ31(11) : Excited a'v=12 11 12 F3f - X Xv=0 11 11 F1e
13C18O Excited 12 e 1 X 11 e 1 67548.0821 1.235717 67778.5200 230.4379 3.731512 43282007 0.000000 : rR(11) : Excited Av=2 12 12 F1e - X Xv=0 11 11 F1e
13C18O Excited 12 e 2 X 11 e 1 67594.4893 .4080089 67824.9272 230.4379 1.232071 14320328 0.000000 : qR1(11) : Excited a3Pv=13 12 11 F1e - X Xv=0 11 11 F1e
13C18O Excited 12 e 3 X 11 e 1 67644.4282 1.007648 67874.8661 230.4379 3.042811 35444962 0.000000 : rR21(11) : Excited a3Pv=13 12 12 F2e - X Xv=0 11 11 F1e
13C18O Excited 12 e 4 X 11 e 1 67691.8457 .5074739 67922.2837 230.4379 1.532427 17888429 0.000000 : sR31(11) : Excited a3Pv=13 12 13 F3e - X Xv=0 11 11 F1e
13C18O Excited 12 e 5 X 11 e 1 67718.6768 .5762005 67949.1147 230.4379 1.739961 20335198 0.000000 : qR1(11) : Excited ev=4 12 11 F1e - X Xv=0 11 11 F1e
13C18O Excited 12 e 6 X 11 e 1 67772.8733 .6034644 68003.3112 230.4379 1.822290 21348567 0.000000 : sR31(11) : Excited ev=4 12 13 F3e - X Xv=0 11 11 F1e
13C18O Excited 12 e 7 X 11 e 1 67838.4471 .0112185 68068.8850 230.4379 .0338765 398024.7 0.000000 : qR1(11) : Excited dv=7 12 11 F1e - X Xv=0 11 11 F1e
13C18O Excited 12 e 8 X 11 e 1 67884.9191 .2219277 68115.3570 230.4379 .6701584 7890069 0.000000 : rR21(11) : Excited dv=7 12 12 F2e - X Xv=0 11 11 F1e
13C18O Excited 12 e 9 X 11 e 1 67921.8719 1.434811 68152.3098 230.4379 4.332719 51094352 0.000000 : rR21(11) : Excited a'v=12 12 12 F2e - X Xv=0 11 11 F1e
13C18O Excited 12 e 10 X 11 e 1 67932.0325 1.168600 68162.4704 230.4379 3.528840 41633140 0.000000 : sR31(11) : Excited dv=7 12 13 F3e - X Xv=0 11 11 F1e
13C18O Excited 11 e 1 X 12 e 1 67472.2374 .8611755 67744.5542 272.3168 3.178958 39944385 0.000000 : pP(12) : Excited Av=2 11 11 F1e - X Xv=0 12 12 F1e
13C18O Excited 11 e 2 X 12 e 1 67522.9760 .2518207 67795.2928 272.3168 .9295754 11706713 0.000000 : oP1(12) : Excited a3Pv=13 11 10 F1e - X Xv=0 12 12 F1e
13C18O Excited 11 e 3 X 12 e 1 67571.3443 .7456255 67843.6612 272.3168 2.752415 34737390 0.000000 : pP21(12) : Excited a3Pv=13 11 11 F2e - X Xv=0 12 12 F1e
13C18O Excited 11 e 4 X 12 e 1 67616.9098 .3070262 67889.2267 272.3168 1.133362 14332771 0.000000 : qP31(12) : Excited a3Pv=13 11 12 F3e - X Xv=0 12 12 F1e
13C18O Excited 11 e 5 X 12 e 1 67652.5688 .4172057 67924.8856 272.3168 1.540080 19507062 0.000000 : oP1(12) : Excited ev=4 11 10 F1e - X Xv=0 12 12 F1e
13C18O Excited 11 e 6 X 12 e 1 67702.4840 .4198850 67974.8008 272.3168 1.549971 19675823 0.000000 : qP31(12) : Excited ev=4 11 12 F3e - X Xv=0 12 12 F1e
13C18O Excited 11 e 7 X 12 e 1 67771.8949 .0059632 68044.2117 272.3168 .0220125 280294.1 0.000000 : oP1(12) : Excited dv=7 11 10 F1e - X Xv=0 12 12 F1e
13C18O Excited 11 e 8 X 12 e 1 67817.1447 .1367482 68089.4615 272.3168 .5047947 6440639 0.000000 : pP21(12) : Excited dv=7 11 11 F2e - X Xv=0 12 12 F1e
13C18O Excited 11 e 9 X 12 e 1 67855.1256 .9890064 68127.4424 272.3168 3.650835 46659043 0.000000 : pP21(12) : Excited a'v=12 11 11 F2e - X Xv=0 12 12 F1e
13C18O Excited 11 e 10 X 12 e 1 67863.0663 .8320189 68135.3831 272.3168 3.071329 39266518 0.000000 : qP31(12) : Excited dv=7 11 12 F3e - X Xv=0 12 12 F1e
13C18O Excited 12 f 1 X 12 e 1 67477.1932 .0001134 67749.5100 272.3168 .0004186 4840.348 0.000000 : qQ(12) : Excited Iv=3 12 12 F1f - X Xv=0 12 12 F1e
13C18O Excited 12 f 2 X 12 e 1 67506.2383 1.955585 67778.5552 272.3168 7.218880 83576650 0.000000 : qQ(12) : Excited Av=2 12 12 F1f - X Xv=0 12 12 F1e
13C18O Excited 12 f 3 X 12 e 1 67555.6246 .6282485 67827.9414 272.3168 2.319127 26908692 0.000000 : pQ1(12) : Excited a3Pv=13 12 11 F1f - X Xv=0 12 12 F1e
13C18O Excited 12 f 4 X 12 e 1 67603.0960 1.593499 67875.4128 272.3168 5.882271 68395605 0.000000 : qQ21(12) : Excited a3Pv=13 12 12 F2f - X Xv=0 12 12 F1e
13C18O Excited 12 f 5 X 12 e 1 67650.0005 .7862336 67922.3173 272.3168 2.902316 33816728 0.000000 : rQ31(12) : Excited a3Pv=13 12 13 F3f - X Xv=0 12 12 F1e
13C18O Excited 12 f 6 X 12 e 1 67703.2207 1.861239 67975.5375 272.3168 6.870609 80242904 0.000000 : qQ21(12) : Excited ev=4 12 12 F2f - X Xv=0 12 12 F1e
13C18O Excited 12 f 7 X 12 e 1 67796.5682 .0176221 68068.8850 272.3168 .0650507 762883.9 0.000000 : pQ1(12) : Excited dv=7 12 11 F1f - X Xv=0 12 12 F1e
13C18O Excited 12 f 8 X 12 e 1 67843.0402 .3472822 68115.3570 272.3168 1.281964 15065205 0.000000 : qQ21(12) : Excited dv=7 12 12 F2f - X Xv=0 12 12 F1e
13C18O Excited 12 f 9 X 12 e 1 67856.1279 1.235797 68128.4447 272.3168 4.561844 53640286 0.000000 : pQ1(12) : Excited a'v=12 12 11 F1f - X Xv=0 12 12 F1e
13C18O Excited 12 f 10 X 12 e 1 67890.1531 1.807674 68162.4699 272.3168 6.672878 78580893 0.000000 : rQ31(12) : Excited dv=7 12 13 F3f - X Xv=0 12 12 F1e
13C18O Excited 12 f 11 X 12 e 1 67908.1928 1.054149 68180.5096 272.3168 3.891302 45861147 0.000000 : rQ31(12) : Excited a'v=12 12 13 F3f - X Xv=0 12 12 F1e
13C18O Excited 13 e 1 X 12 e 1 67542.9896 1.079506 67815.3064 272.3168 3.984909 42787696 0.000000 : rR(12) : Excited Av=2 13 13 F1e - X Xv=0 12 12 F1e
13C18O Excited 13 e 2 X 12 e 1 67584.7589 .3986298 67857.0758 272.3168 1.471509 15829563 0.000000 : qR1(12) : Excited a3Pv=13 13 12 F1e - X Xv=0 12 12 F1e
13C18O Excited 13 e 3 X 12 e 1 67636.3630 .8303092 67908.6799 272.3168 3.065018 33047105 0.000000 : rR21(12) : Excited a3Pv=13 13 13 F2e - X Xv=0 12 12 F1e

| | | | | | | | | | | |
|---|---|---|---|---|---|---|---|---|---|---|
| 13C18O Excited | 13 e 4 | X 12 e 1 | 67685.6482 | .5247111 | 67957.9650 | 272.3168 | 1.936928 | 20929691 | 0.000000 | : sR31(12) : Excited a3Pv=13 13 14 F3e - X Xv=0 12 12 F1e |
| 13C18O Excited | 13 e 5 | X 12 e 1 | 67703.2719 | .4617005 | 67975.5888 | 272.3168 | 1.704329 | 18430715 | 0.000000 | : qR1(12) : Excited ev=4 13 12 F1e - X Xv=0 12 12 F1e |
| 13C18O Excited | 13 e 6 | X 12 e 1 | 67761.6946 | .5276441 | 68034.0115 | 272.3168 | 1.947755 | 21117702 | 0.000000 | : sR31(12) : Excited ev=4 13 14 F3e - X Xv=0 12 12 F1e |
| 13C18O Excited | 13 e 7 | X 12 e 1 | 67823.3282 | .0124953 | 68095.6450 | 272.3168 | .0461255 | 501462.3 | 0.000000 | : qR1(12) : Excited dv=7 13 12 F1e - X Xv=0 12 12 F1e |
| 13C18O Excited | 13 e 8 | X 12 e 1 | 67871.0859 | .2163570 | 68143.4027 | 272.3168 | .7986639 | 8701179 | 0.000000 | : rR21(12) : Excited dv=7 13 13 F2e - X Xv=0 12 12 F1e |
| 13C18O Excited | 13 e 9 | X 12 e 1 | 67906.9259 | 1.267795 | 68179.2428 | 272.3168 | 4.679960 | 51067429 | 0.000000 | : rR21(12) : Excited a'v=12 13 13 F2e - X Xv=0 12 12 F1e |
| 13C18O Excited | 13 e 10 | X 12 e 1 | 67919.4570 | 1.001820 | 68191.7738 | 272.3168 | 3.698137 | 40376181 | 0.000000 | : sR31(12) : Excited dv=7 13 14 F3e - X Xv=0 12 12 F1e |
| 13C18O Excited | 12 e 1 | X 13 e 1 | 67460.8409 | .7506667 | 67778.5200 | 317.6791 | 3.444473 | 39798020 | 0.000000 | : pP(13) : Excited Av=2 12 12 F1e - X Xv=0 13 13 F1e |
| 13C18O Excited | 12 e 2 | X 13 e 1 | 67507.2481 | .2478552 | 67824.9272 | 317.6791 | 1.137296 | 13167648 | 0.000000 | : oP1(13) : Excited a3Pv=13 12 11 F1e - X Xv=0 13 13 F1e |
| 13C18O Excited | 12 e 3 | X 13 e 1 | 67557.1870 | .6121210 | 67874.8661 | 317.6791 | 2.808749 | 32591998 | 0.000000 | : pP21(13) : Excited a3Pv=13 12 12 F2e - X Xv=0 13 13 F1e |
| 13C18O Excited | 12 e 4 | X 13 e 1 | 67604.6046 | .3082776 | 67922.2837 | 317.6791 | 1.414548 | 16448635 | 0.000000 | : qP31(13) : Excited a3Pv=13 12 13 F3e - X Xv=0 13 13 F1e |
| 13C18O Excited | 12 e 5 | X 13 e 1 | 67631.4356 | .3500273 | 67949.1147 | 317.6791 | 1.606118 | 18698498 | 0.000000 | : oP1(13) : Excited ev=4 12 11 F1e - X Xv=0 13 13 F1e |
| 13C18O Excited | 12 e 6 | X 13 e 1 | 67685.6321 | .3665894 | 68003.3112 | 317.6791 | 1.682114 | 19630366 | 0.000000 | : qP31(13) : Excited ev=4 12 13 F3e - X Xv=0 13 13 F1e |
| 13C18O Excited | 12 e 7 | X 13 e 1 | 67751.2059 | .0068149 | 68068.8850 | 317.6791 | .0312706 | 365991.8 | 0.000000 | : oP1(13) : Excited dv=7 12 11 F1e - X Xv=0 13 13 F1e |
| 13C18O Excited | 12 e 8 | X 13 e 1 | 67797.6780 | .1348155 | 68115.3570 | 317.6791 | .6186078 | 7255098 | 0.000000 | : pP21(13) : Excited dv=7 12 12 F2e - X Xv=0 13 13 F1e |
| 13C18O Excited | 12 e 9 | X 13 e 1 | 67834.6307 | .8716112 | 68152.3098 | 317.6791 | 3.999433 | 46982513 | 0.000000 | : pP21(13) : Excited a'v=12 12 12 F2e - X Xv=0 13 13 F1e |
| 13C18O Excited | 12 e 10 | X 13 e 1 | 67844.7913 | .7098951 | 68162.4704 | 317.6791 | 3.257390 | 38282718 | 0.000000 | : qP31(13) : Excited dv=7 12 13 F3e - X Xv=0 13 13 F1e |
| 13C18O Excited | 13 f 2 | X 13 e 1 | 67497.6626 | 1.688130 | 67815.3417 | 317.6791 | 7.746072 | 83005603 | 0.000000 | : qQ(13) : Excited Av=2 13 13 F1f - X Xv=0 13 13 F1e |
| 13C18O Excited | 13 f 3 | X 13 e 1 | 67542.3692 | .5957712 | 67860.0483 | 317.6791 | 2.733727 | 29352402 | 0.000000 | : pQ1(13) : Excited a3Pv=13 13 12 F1f - X Xv=0 13 13 F1e |
| 13C18O Excited | 13 f 4 | X 13 e 1 | 67591.5841 | 1.305460 | 67909.2632 | 317.6791 | 5.990172 | 64458002 | 0.000000 | : qQ21(13) : Excited a3Pv=13 13 13 F2f - X Xv=0 13 13 F1e |
| 13C18O Excited | 13 f 5 | X 13 e 1 | 67640.3576 | .7674910 | 67958.0367 | 317.6791 | 3.521672 | 37977488 | 0.000000 | : rQ31(13) : Excited a3Pv=13 13 14 F3f - X Xv=0 13 13 F1e |
| 13C18O Excited | 13 f 6 | X 13 e 1 | 67686.4636 | 1.573546 | 68004.1427 | 317.6791 | 7.220296 | 78022546 | 0.000000 | : qQ21(13) : Excited ev=4 13 13 F2f - X Xv=0 13 13 F1e |
| 13C18O Excited | 13 f 7 | X 13 e 1 | 67777.9659 | .0193504 | 68095.6450 | 317.6791 | .0887905 | 963366.6 | 0.000000 | : pQ1(13) : Excited dv=7 13 12 F1f - X Xv=0 13 13 F1e |
| 13C18O Excited | 13 f 8 | X 13 e 1 | 67825.7235 | .3328774 | 68143.4026 | 317.6791 | 1.527425 | 16607449 | 0.000000 | : qQ21(13) : Excited dv=7 13 13 F2f - X Xv=0 13 13 F1e |
| 13C18O Excited | 13 f 9 | X 13 e 1 | 67835.6329 | 1.070016 | 68153.3120 | 317.6791 | 4.909822 | 53407107 | 0.000000 | : pQ1(13) : Excited a'v=12 13 12 F1f - X Xv=0 13 13 F1e |
| 13C18O Excited | 13 f 10 | X 13 e 1 | 67874.0943 | 1.531997 | 68191.7734 | 317.6791 | 7.029648 | 76595876 | 0.000000 | : rQ31(13) : Excited dv=7 13 14 F3f - X Xv=0 13 13 F1e |
| 13C18O Excited | 13 f 11 | X 13 e 1 | 67891.8266 | .9223592 | 68209.5057 | 317.6791 | 4.232293 | 46151717 | 0.000000 | : rQ31(13) : Excited a'v=12 13 14 F3f - X Xv=0 13 13 F1e |
| 13C18O Excited | 14 e 1 | X 13 e 1 | 67537.2315 | .9206267 | 67854.9105 | 317.6791 | 4.224343 | 42219624 | 0.000000 | : rR(13) : Excited Av=2 14 14 F1e - X Xv=0 13 13 F1e |
| 13C18O Excited | 14 e 2 | X 13 e 1 | 67574.0632 | .3780415 | 67891.7423 | 317.6791 | 1.734663 | 17365232 | 0.000000 | : qR1(13) : Excited a3Pv=13 14 13 F1e - X Xv=0 13 13 F1e |
| 13C18O Excited | 14 e 3 | X 13 e 1 | 67627.4209 | .6695327 | 67945.1000 | 317.6791 | 3.072185 | 30827712 | 0.000000 | : rR21(13) : Excited a3Pv=13 14 14 F2e - X Xv=0 13 13 F1e |
| 13C18O Excited | 14 e 4 | X 13 e 1 | 67678.4126 | .6024906 | 67996.0917 | 317.6791 | 2.764559 | 27803650 | 0.000000 | : sR31(13) : Excited a3Pv=13 14 15 F3e - X Xv=0 13 13 F1e |
| 13C18O Excited | 14 e 5 | X 13 e 1 | 67686.7978 | .2744106 | 68004.4768 | 317.6791 | 1.259147 | 12668168 | 0.000000 | : qR1(13) : Excited ev=4 14 13 F1e - X Xv=0 13 13 F1e |
| 13C18O Excited | 14 e 6 | X 13 e 1 | 67749.2225 | .4500903 | 68066.9016 | 317.6791 | 2.065262 | 20835964 | 0.000000 | : sR31(13) : Excited ev=4 14 15 F3e - X Xv=0 13 13 F1e |
| 13C18O Excited | 14 e 7 | X 13 e 1 | 67806.8168 | .0132893 | 68124.4959 | 317.6791 | .0609787 | 616770.3 | 0.000000 | : qR1(13) : Excited dv=7 14 13 F1e - X Xv=0 13 13 F1e |
| 13C18O Excited | 14 e 8 | X 13 e 1 | 67855.9178 | .2037756 | 68173.5968 | 317.6791 | .9350350 | 9477995 | 0.000000 | : rR21(13) : Excited dv=7 14 14 F2e - X Xv=0 13 13 F1e |
| 13C18O Excited | 14 e 9 | X 13 e 1 | 67890.5606 | 1.095296 | 68208.2397 | 317.6791 | 5.025824 | 51022398 | 0.000000 | : rR21(13) : Excited a'v=12 14 14 F2e - X Xv=0 13 13 F1e |
| 13C18O Excited | 14 e 10 | X 13 e 1 | 67905.6075 | .8407888 | 68223.2866 | 317.6791 | 3.858003 | 39192674 | 0.000000 | : sR31(13) : Excited dv=7 14 15 F3e - X Xv=0 13 13 F1e |
| 13C18O Excited | 13 e 1 | X 14 e 1 | 67448.7833 | .6380044 | 67815.3064 | 366.5231 | 3.700273 | 39565416 | 0.000000 | : pP(14) : Excited Av=2 13 13 F1e - X Xv=0 14 14 F1e |
| 13C18O Excited | 13 e 2 | X 14 e 1 | 67490.5527 | .2355962 | 67857.0758 | 366.5231 | 1.366401 | 14637499 | 0.000000 | : oP1(14) : Excited a3Pv=13 13 12 F1e - X Xv=0 14 14 F1e |
| 13C18O Excited | 13 e 3 | X 14 e 1 | 67542.1567 | .4907251 | 67908.6799 | 366.5231 | 2.846088 | 30558552 | 0.000000 | : pP21(14) : Excited a3Pv=13 13 13 F2e - X Xv=0 14 14 F1e |
| 13C18O Excited | 13 e 4 | X 14 e 1 | 67591.4419 | .3101121 | 67957.9650 | 366.5231 | 1.798576 | 19353677 | 0.000000 | : qP31(14) : Excited a3Pv=13 13 14 F3e - X Xv=0 14 14 F1e |
| 13C18O Excited | 13 e 5 | X 14 e 1 | 67609.0657 | .2728719 | 67975.5888 | 366.5231 | 1.582592 | 17042893 | 0.000000 | : oP1(14) : Excited ev=4 13 12 F1e - X Xv=0 14 14 F1e |
| 13C18O Excited | 13 e 6 | X 14 e 1 | 67667.4884 | .3118455 | 68034.0115 | 366.5231 | 1.808629 | 19527623 | 0.000000 | : qP31(14) : Excited ev=4 13 14 F3e - X Xv=0 14 14 F1e |
| 13C18O Excited | 13 e 7 | X 14 e 1 | 67729.1219 | .0073849 | 68095.6450 | 366.5231 | .0428308 | 463706.0 | 0.000000 | : oP1(14) : Excited dv=7 13 12 F1e - X Xv=0 14 14 F1e |
| 13C18O Excited | 13 e 8 | X 14 e 1 | 67776.8796 | .1278702 | 68143.4027 | 366.5231 | .7416165 | 8046069 | 0.000000 | : pP21(14) : Excited dv=7 13 13 F2e - X Xv=0 14 14 F1e |
| 13C18O Excited | 13 e 9 | X 14 e 1 | 67812.7197 | .7492856 | 68179.2428 | 366.5231 | 4.345677 | 47222675 | 0.000000 | : pP21(14) : Excited a'v=12 13 13 F2e - X Xv=0 14 14 F1e |
| 13C18O Excited | 13 e 10 | X 14 e 1 | 67825.2507 | .5920907 | 68191.7738 | 366.5231 | 3.433984 | 37336376 | 0.000000 | : qP31(14) : Excited dv=7 13 14 F3e - X Xv=0 14 14 F1e |
| 13C18O Excited | 14 f 2 | X 14 e 1 | 67488.4248 | 1.423249 | 67854.9479 | 366.5231 | 8.254503 | 82319791 | 0.000000 | : qQ(14) : Excited Av=2 14 14 F1f - X Xv=0 14 14 F1e |
| 13C18O Excited | 14 f 3 | X 14 e 1 | 67528.1678 | .5472049 | 67894.6909 | 366.5231 | 3.173657 | 31705918 | 0.000000 | : pQ1(14) : Excited a3Pv=13 14 13 F1f - X Xv=0 14 14 F1e |
| 13C18O Excited | 14 f 4 | X 14 e 1 | 67579.1907 | 1.049598 | 67945.7138 | 366.5231 | 6.087414 | 60953302 | 0.000000 | : qQ21(14) : Excited a3Pv=13 14 14 F2f - X Xv=0 14 14 F1e |
| 13C18O Excited | 14 f 5 | X 14 e 1 | 67629.8580 | .7364653 | 67996.3811 | 366.5231 | 4.271322 | 42865037 | 0.000000 | : rQ31(14) : Excited a3Pv=13 14 15 F3f - X Xv=0 14 14 F1e |
| 13C18O Excited | 14 f 6 | X 14 e 1 | 67668.4550 | 1.283520 | 68034.9782 | 366.5231 | 7.444111 | 74833672 | 0.000000 | : qQ21(14) : Excited ev=4 14 14 F2f - X Xv=0 14 14 F1e |
| 13C18O Excited | 14 f 7 | X 14 e 1 | 67757.9728 | .0202796 | 68124.4959 | 366.5231 | .1176171 | 1187072 | 0.000000 | : pQ1(14) : Excited dv=7 14 13 F1f - X Xv=0 14 14 F1e |
| 13C18O Excited | 14 f 8 | X 14 e 1 | 67807.0736 | .3070716 | 68173.5967 | 366.5231 | 1.780942 | 18013583 | 0.000000 | : qQ21(14) : Excited dv=7 14 14 F2f - X Xv=0 14 14 F1e |
| 13C18O Excited | 14 f 9 | X 14 e 1 | 67813.7210 | .9080398 | 68180.2441 | 366.5231 | 5.266413 | 53283543 | 0.000000 | : pQ1(14) : Excited a'v=12 14 13 F1f - X Xv=0 14 14 F1e |
| 13C18O Excited | 14 f 10 | X 14 e 1 | 67856.7631 | 1.269589 | 68223.2862 | 366.5231 | 7.363312 | 74641104 | 0.000000 | : rQ31(14) : Excited dv=7 14 15 F3f - X Xv=0 14 14 F1e |
| 13C18O Excited | 14 f 11 | X 14 e 1 | 67874.0415 | .7886586 | 68240.5646 | 366.5231 | 4.574031 | 46401891 | 0.000000 | : rQ31(14) : Excited a'v=12 14 15 F3f - X Xv=0 14 14 F1e |
| 13C18O Excited | 15 e 1 | X 14 e 1 | 67530.8059 | .7661624 | 67897.3290 | 366.5231 | 4.443558 | 41533489 | 0.000000 | : rR(14) : Excited Av=2 15 15 F1e - X Xv=0 14 14 F1e |
| 13C18O Excited | 15 e 2 | X 14 e 1 | 67562.4059 | .3494087 | 67928.9290 | 366.5231 | 2.026487 | 18967967 | 0.000000 | : qR1(14) : Excited a3Pv=13 15 14 F1e - X Xv=0 14 14 F1e |
| 13C18O Excited | 15 e 3 | X 14 e 1 | 67617.6002 | .5286589 | 67984.1233 | 366.5231 | 3.066095 | 28769123 | 0.000000 | : rR21(14) : Excited a3Pv=13 15 15 F2e - X Xv=0 14 14 F1e |
| 13C18O Excited | 15 e 4 | X 14 e 1 | 67667.1477 | .7098679 | 68033.6708 | 366.5231 | 4.117064 | 38715330 | 0.000000 | : qR1(14) : Excited ev=4 15 14 F1e - X Xv=0 14 14 F1e |
| 13C18O Excited | 15 e 5 | X 14 e 1 | 67672.2397 | .0526100 | 68038.7628 | 366.5231 | .3051255 | 2869934 | 0.000000 | : sR31(14) : Excited a3Pv=13 15 16 F3e - X Xv=0 14 14 F1e |
| 13C18O Excited | 15 e 6 | X 14 e 1 | 67735.4591 | .3744738 | 68101.9822 | 366.5231 | 2.171859 | 20485261 | 0.000000 | : sR31(14) : Excited ev=4 15 16 F3e - X Xv=0 14 14 F1e |
| 13C18O Excited | 15 e 7 | X 14 e 1 | 67788.9186 | .0135630 | 68155.4417 | 366.5231 | .0786623 | 743711.5 | 0.000000 | : qR1(14) : Excited dv=7 15 14 F1e - X Xv=0 14 14 F1e |
| 13C18O Excited | 15 e 8 | X 14 e 1 | 67839.4146 | .1858887 | 68205.9378 | 366.5231 | 1.078110 | 10215768 | 0.000000 | : rR21(14) : Excited dv=7 15 15 F2e - X Xv=0 14 14 F1e |
| 13C18O Excited | 15 e 9 | X 14 e 1 | 67872.7757 | .9260901 | 68239.2988 | 366.5231 | 5.371100 | 50969672 | 0.000000 | : rR21(14) : Excited a'v=12 15 15 F2e - X Xv=0 14 14 F1e |
| 13C18O Excited | 15 e 10 | X 14 e 1 | 67890.4790 | .6911675 | 68257.0021 | 366.5231 | 4.008606 | 38069896 | 0.000000 | : sR31(14) : Excited dv=7 15 16 F3e - X Xv=0 14 14 F1e |
| 13C18O Excited | 14 e 1 | X 15 e 1 | 67436.0633 | .5289359 | 67854.9105 | 418.8472 | 3.942720 | 39228166 | 0.000000 | : pP(15) : Excited Av=2 14 14 F1e - X Xv=0 15 15 F1e |
| 13C18O Excited | 14 e 2 | X 15 e 1 | 67472.8951 | .2171995 | 67891.7423 | 418.8472 | 1.619019 | 16134864 | 0.000000 | : oP1(15) : Excited a3Pv=13 14 13 F1e - X Xv=0 15 15 F1e |
| 13C18O Excited | 14 e 3 | X 15 e 1 | 67526.2528 | .3846726 | 67945.1000 | 418.8472 | 2.867373 | 28643596 | 0.000000 | : pP21(15) : Excited a3Pv=13 14 14 F2e - X Xv=0 15 15 F1e |
| 13C18O Excited | 14 e 4 | X 15 e 1 | 67577.2445 | .3461543 | 67996.0917 | 418.8472 | 2.580255 | 25833874 | 0.000000 | : qP31(15) : Excited a3Pv=13 14 15 F3e - X Xv=0 15 15 F1e |
| 13C18O Excited | 14 e 5 | X 15 e 1 | 67585.6296 | .1576596 | 68004.4768 | 418.8472 | 1.175204 | 11770686 | 0.000000 | : oP1(15) : Excited ev=4 14 13 F1e - X Xv=0 15 15 F1e |
| 13C18O Excited | 14 e 6 | X 15 e 1 | 67648.0544 | .2585944 | 68066.9016 | 418.8472 | 1.927578 | 19359911 | 0.000000 | : qP31(15) : Excited ev=4 14 15 F3e - X Xv=0 15 15 F1e |
| 13C18O Excited | 14 e 7 | X 15 e 1 | 67705.6487 | .0076352 | 68124.4959 | 418.8472 | .0569134 | 573079.5 | 0.000000 | : oP1(15) : Excited dv=7 14 13 F1e - X Xv=0 15 15 F1e |
| 13C18O Excited | 14 e 8 | X 15 e 1 | 67754.7496 | .1170770 | 68173.5968 | 418.8472 | .8726993 | 8806621 | 0.000000 | : pP21(15) : Excited dv=7 14 14 F2e - X Xv=0 15 15 F1e |
| 13C18O Excited | 14 e 9 | X 15 e 1 | 67789.3924 | .6292904 | 68208.2397 | 418.8472 | 4.690769 | 47408333 | 0.000000 | : pP21(15) : Excited a'v=12 14 14 F2e - X Xv=0 15 15 F1e |
| 13C18O Excited | 14 e 10 | X 15 e 1 | 67804.4394 | .4830659 | 68223.2866 | 418.8472 | 3.600803 | 36416579 | 0.000000 | : qP31(15) : Excited dv=7 14 15 F3e - X Xv=0 15 15 F1e |
| 13C18O Excited | 15 f 2 | X 15 e 1 | 67478.5235 | 1.172313 | 67897.3707 | 418.8472 | 8.738493 | 81488250 | 0.000000 | : qQ(15) : Excited Av=2 15 15 F1f - X Xv=0 15 15 F1e |
| 13C18O Excited | 15 f 3 | X 15 e 1 | 67513.0276 | .4882821 | 67931.8748 | 418.8472 | 3.639685 | 33992905 | 0.000000 | : pQ1(15) : Excited a3Pv=13 15 14 F1f - X Xv=0 15 15 F1e |
| 13C18O Excited | 15 f 4 | X 15 e 1 | 67565.9136 | .8297885 | 67984.7608 | 418.8472 | 6.185294 | 57903528 | 0.000000 | : qQ21(15) : Excited a3Pv=13 15 15 F2f - X Xv=0 15 15 F1e |
| 13C18O Excited | 15 f 5 | X 15 e 1 | 67618.4594 | .7042664 | 68037.3066 | 418.8472 | 5.249645 | 49259211 | 0.000000 | : rQ31(15) : Excited a3Pv=13 15 16 F3f - X Xv=0 15 15 F1e |
| 13C18O Excited | 15 f 6 | X 15 e 1 | 67649.2312 | .9977659 | 68068.0784 | 418.8472 | 7.437408 | 69883063 | 0.000000 | : qQ21(15) : Excited ev=4 15 15 F2f - X Xv=0 15 15 F1e |
| 13C18O Excited | 15 f 7 | X 15 e 1 | 67736.5944 | .0203871 | 68155.4416 | 418.8472 | .1519667 | 1433443 | 0.000000 | : pQ1(15) : Excited dv=7 15 14 F1f - X Xv=0 15 15 F1e |
| 13C18O Excited | 15 f 8 | X 15 e 1 | 67787.0901 | .2703142 | 68205.9373 | 418.8472 | 2.014938 | 19048663 | 0.000000 | : qQ21(15) : Excited dv=7 15 15 F2f - X Xv=0 15 15 F1e |
| 13C18O Excited | 15 f 9 | X 15 e 1 | 67790.3927 | .7589333 | 68209.2399 | 418.8472 | 5.657135 | 53488784 | 0.000000 | : pQ1(15) : Excited a'v=12 15 14 F1f - X Xv=0 15 15 F1e |
| 13C18O Excited | 15 f 10 | X 15 e 1 | 67838.1546 | 1.029693 | 68257.0018 | 418.8472 | 7.675392 | 72725108 | 0.000000 | : rQ31(15) : Excited dv=7 15 16 F3f - X Xv=0 15 15 F1e |
| 13C18O Excited | 15 f 11 | X 15 e 1 | 67854.8372 | .6595969 | 68273.6844 | 418.8472 | 4.916675 | 46620367 | 0.000000 | : rQ31(15) : Excited a'v=12 15 16 F3f - X Xv=0 15 15 F1e |
| 13C18O Excited | 16 e 1 | X 15 e 1 | 67523.7099 | .6212896 | 67942.5572 | 418.8472 | 4.631130 | 40650451 | 0.000000 | : rR(15) : Excited Av=2 16 16 F1e - X Xv=0 15 15 F1e |
| 13C18O Excited | 16 e 2 | X 15 e 1 | 67549.7894 | .3162486 | 67968.6367 | 418.8472 | 2.357336 | 20715864 | 0.000000 | : qR1(15) : Excited a3Pv=13 16 15 F1e - X Xv=0 15 15 F1e |
| 13C18O Excited | 16 e 3 | X 15 e 1 | 67606.8987 | .4087042 | 68025.7459 | 418.8472 | 3.046506 | 26840124 | 0.000000 | : rR21(15) : Excited a3Pv=13 16 16 F2e - X Xv=0 15 15 F1e |
| 13C18O Excited | 16 e 4 | X 15 e 1 | 67648.5771 | .4365768 | 68067.4243 | 418.8472 | 3.254270 | 28723613 | 0.000000 | : qR1(15) : Excited ev=4 16 15 F1e - X Xv=0 15 15 F1e |
| 13C18O Excited | 16 e 5 | X 15 e 1 | 67662.8699 | .2130673 | 68081.7171 | 418.8472 | 1.588217 | 14027186 | 0.000000 | : sR31(15) : Excited a3Pv=13 16 17 F3e - X Xv=0 15 15 F1e |
| 13C18O Excited | 16 e 6 | X 15 e 1 | 67720.4079 | .3035931 | 68139.2551 | 418.8472 | 2.263001 | 20037936 | 0.000000 | : sR31(15) : Excited ev=4 16 17 F3e - X Xv=0 15 15 F1e |
| 13C18O Excited | 16 e 7 | X 15 e 1 | 67769.6387 | .0133415 | 68188.4859 | 418.8472 | .0994484 | 882495.8 | 0.000000 | : qR1(15) : Excited dv=7 16 15 F1e - X Xv=0 15 15 F1e |
| 13C18O Excited | 16 e 8 | X 15 e 1 | 67821.5765 | .1645750 | 68240.4237 | 418.8472 | 1.226752 | 10911134 | 0.000000 | : rR21(15) : Excited dv=7 16 16 F2e - X Xv=0 15 15 F1e |
| 13C18O Excited | 16 e 9 | X 15 e 1 | 67853.5711 | .7668732 | 68272.4183 | 418.8472 | 5.716319 | 50914806 | 0.000000 | : rR21(15) : Excited a'v=12 16 16 F2e - X Xv=0 15 15 F1e |
| 13C18O Excited | 16 e 10 | X 15 e 1 | 67874.0668 | .5567908 | 68292.9141 | 418.8472 | 4.150353 | 37000375 | 0.000000 | : sR31(15) : Excited dv=7 16 17 F3e - X Xv=0 15 15 F1e |
| 13C18O Excited | 15 e 1 | X 16 e 1 | 67422.6794 | .4276433 | 67897.3290 | 474.6496 | 4.165836 | 38750911 | 0.000000 | : pP(16) : Excited Av=2 15 15 F1e - X Xv=0 16 16 F1e |
| 13C18O Excited | 15 e 2 | X 16 e 1 | 67454.2794 | .1950269 | 67928.9290 | 474.6496 | 1.899831 | 17697229 | 0.000000 | : oP1(16) : Excited a3Pv=13 15 14 F1e - X Xv=0 16 16 F1e |
| 13C18O Excited | 15 e 3 | X 16 e 1 | 67509.4737 | .2950777 | 67984.1233 | 474.6496 | 2.874464 | 26841872 | 0.000000 | : pP21(16) : Excited a3Pv=13 15 15 F2e - X Xv=0 16 16 F1e |
| 13C18O Excited | 15 e 4 | X 16 e 1 | 67559.0213 | .3962218 | 68033.6708 | 474.6496 | 3.859747 | 36121907 | 0.000000 | : oP1(16) : Excited ev=4 15 14 F1e - X Xv=0 16 16 F1e |
| 13C18O Excited | 15 e 5 | X 16 e 1 | 67564.1132 | .0293650 | 68038.7628 | 474.6496 | .2860552 | 2677687 | 0.000000 | : qP31(16) : Excited a3Pv=13 15 16 F3e - X Xv=0 16 16 F1e |
| 13C18O Excited | 15 e 6 | X 16 e 1 | 67627.3326 | .2090173 | 68101.9822 | 474.6496 | 2.036117 | 19113108 | 0.000000 | : qP31(16) : Excited ev=4 15 16 F3e - X Xv=0 16 16 F1e |
| 13C18O Excited | 15 e 7 | X 16 e 1 | 67680.7921 | .0075704 | 68155.4417 | 474.6496 | .0737459 | 693898.5 | 0.000000 | : oP1(16) : Excited dv=7 15 14 F1e - X Xv=0 16 16 F1e |
| 13C18O Excited | 15 e 8 | X 16 e 1 | 67731.2882 | .1037561 | 68205.9378 | 474.6496 | 1.010728 | 9531561 | 0.000000 | : pP21(16) : Excited dv=7 15 15 F2e - X Xv=0 16 16 F1e |
| 13C18O Excited | 15 e 9 | X 16 e 1 | 67764.6492 | .5169089 | 68239.2988 | 474.6496 | 5.035407 | 47556060 | 0.000000 | : pP21(16) : Excited a'v=12 15 15 F2e - X Xv=0 16 16 F1e |
| 13C18O Excited | 15 e 10 | X 16 e 1 | 67782.3525 | .3857839 | 68257.0021 | 474.6496 | 3.758068 | 35520271 | 0.000000 | : qP31(16) : Excited dv=7 15 16 F3e - X Xv=0 16 16 F1e |
| 13C18O Excited | 16 f 2 | X 16 e 1 | 67467.9570 | .9432934 | 67942.6066 | 474.6496 | 9.188980 | 80458042 | 0.000000 | : qQ(16) : Excited Av=2 16 16 F1f - X Xv=0 16 16 F1e |
| 13C18O Excited | 16 f 3 | X 16 e 1 | 67496.9548 | .4245281 | 67971.6044 | 474.6496 | 4.135489 | 36256755 | 0.000000 | : pQ1(16) : Excited a3Pv=13 16 15 F1f - X Xv=0 16 16 F1e |
| 13C18O Excited | 16 f 4 | X 16 e 1 | 67551.7500 | .6463837 | 68026.3995 | 474.6496 | 6.296669 | 55338858 | 0.000000 | : qQ21(16) : Excited a3Pv=13 16 16 F2f - X Xv=0 16 16 F1e |
| 13C18O Excited | 16 f 5 | X 16 e 1 | 67606.0675 | .6881784 | 68080.7171 | 474.6496 | 6.703807 | 59059265 | 0.000000 | : rQ31(16) : Excited a3Pv=13 16 17 F3f - X Xv=0 16 16 F1e |
| 13C18O Excited | 16 f 6 | X 16 e 1 | 67628.8821 | .7129834 | 68103.5316 | 474.6496 | 6.945443 | 61249994 | 0.000000 | : qQ21(16) : Excited ev=4 16 16 F2f - X Xv=0 16 16 F1e |
| 13C18O Excited | 16 f 7 | X 16 e 1 | 67713.8363 | .0197466 | 68188.4859 | 474.6496 | .1923591 | 1702764 | 0.000000 | : pQ1(16) : Excited dv=7 16 15 F1f - X Xv=0 16 16 F1e |
| 13C18O Excited | 16 f 8 | X 16 e 1 | 67765.6333 | .3549437 | 68240.2828 | 474.6496 | 3.457641 | 30677356 | 0.000000 | : pQ1(16) : Excited a'v=12 16 15 F1f - X Xv=0 16 16 F1e |
| 13C18O Excited | 16 f 9 | X 16 e 1 | 67765.7879 | .4980597 | 68240.4375 | 474.6496 | 4.851789 | 43047001 | 0.000000 | : qQ21(16) : Excited dv=7 16 16 F2f - X Xv=0 16 16 F1e |
| 13C18O Excited | 16 f 10 | X 16 e 1 | 67818.2642 | .8178821 | 68292.9137 | 474.6496 | 7.967301 | 70853406 | 0.000000 | : rQ31(16) : Excited dv=7 16 17 F3f - X Xv=0 16 16 F1e |
| 13C18O Excited | 16 f 11 | X 16 e 1 | 67834.2134 | .5400089 | 68308.8629 | 474.6496 | 5.260433 | 46814171 | 0.000000 | : rQ31(16) : Excited a'v=12 16 17 F3f - X Xv=0 16 16 F1e |
| 13C18O Excited | 17 e 1 | X 16 e 1 | 67515.9390 | .4889252 | 67990.5885 | 474.6496 | 4.762807 | 39403726 | 0.000000 | : rR(16) : Excited Av=2 17 17 F1e - X Xv=0 16 16 F1e |
| 13C18O Excited | 17 e 2 | X 16 e 1 | 67536.2151 | .2823588 | 68010.8647 | 474.6496 | 2.750564 | 22776516 | 0.000000 | : qR1(16) : Excited a3Pv=13 17 16 F1e - X Xv=0 16 16 F1e |
| 13C18O Excited | 17 e 3 | X 16 e 1 | 67595.3128 | .3088524 | 68069.9624 | 474.6496 | 3.008648 | 24979087 | 0.000000 | : rR21(16) : Excited a3Pv=13 17 17 F2e - X Xv=0 16 16 F1e |
| 13C18O Excited | 17 e 4 | X 16 e 1 | 67628.1266 | .3232774 | 68102.7761 | 474.6496 | 3.149168 | 26183838 | 0.000000 | : qR1(16) : Excited ev=4 17 16 F1e - X Xv=0 16 16 F1e |
| 13C18O Excited | 17 e 5 | X 16 e 1 | 67653.2593 | .2205316 | 68127.9089 | 474.6496 | 2.148282 | 17881868 | 0.000000 | : sR31(16) : Excited a3Pv=13 17 18 F3e - X Xv=0 16 16 F1e |
| 13C18O Excited | 17 e 6 | X 16 e 1 | 67704.0748 | .2393226 | 68178.7244 | 474.6496 | 2.331333 | 19449303 | 0.000000 | : sR31(16) : Excited ev=4 17 18 F3e - X Xv=0 16 16 F1e |
| 13C18O Excited | 17 e 7 | X 16 e 1 | 67748.9826 | .0127007 | 68223.6322 | 474.6496 | .1237228 | 1034221 | 0.000000 | : qR1(16) : Excited dv=7 17 16 F1e - X Xv=0 16 16 F1e |
| 13C18O Excited | 17 e 8 | X 16 e 1 | 67802.4031 | .1416488 | 68277.0527 | 474.6496 | 1.379855 | 11561766 | 0.000000 | : rR21(16) : Excited dv=7 17 17 F2e - X Xv=0 16 16 F1e |
| 13C18O Excited | 17 e 9 | X 16 e 1 | 67832.9467 | .6222829 | 68307.5963 | 474.6496 | 6.061894 | 50861105 | 0.000000 | : rR21(16) : Excited a'v=12 17 17 F2e - X Xv=0 16 16 F1e |
| 13C18O Excited | 17 e 10 | X 16 e 1 | 67856.3667 | .4397452 | 68331.0163 | 474.6496 | 4.283725 | 35978973 | 0.000000 | : sR31(16) : Excited dv=7 17 18 F3e - X Xv=0 16 16 F1e |
| 13C18O Excited | 16 e 1 | X 17 e 1 | 67408.6289 | .3367203 | 67942.5572 | 533.9283 | 4.358711 | 38063965 | 0.000000 | : pP(17) : Excited Av=2 16 16 F1e - X Xv=0 17 17 F1e |
| 13C18O Excited | 16 e 2 | X 17 e 1 | 67434.7084 | .1713973 | 67968.6367 | 533.9283 | 2.218669 | 19397804 | 0.000000 | : oP1(17) : Excited a3Pv=13 16 15 F1e - X Xv=0 17 17 F1e |
| 13C18O Excited | 16 e 3 | X 17 e 1 | 67491.8176 | .2215054 | 68025.7459 | 533.9283 | 2.867299 | 25132513 | 0.000000 | : pP21(17) : Excited a3Pv=13 16 16 F2e - X Xv=0 17 17 F1e |
| 13C18O Excited | 16 e 4 | X 17 e 1 | 67533.4960 | .2366115 | 68067.4243 | 533.9283 | 3.062842 | 26896256 | 0.000000 | : oP1(17) : Excited ev=4 16 15 F1e - X Xv=0 17 17 F1e |
| 13C18O Excited | 16 e 5 | X 17 e 1 | 67547.7888 | .1154761 | 68081.7171 | 533.9283 | 1.494792 | 13134809 | 0.000000 | : qP31(17) : Excited a3Pv=13 16 17 F3e - X Xv=0 17 17 F1e |
| 13C18O Excited | 16 e 6 | X 17 e 1 | 67605.3268 | .1645384 | 68139.2551 | 533.9283 | 2.129884 | 18763252 | 0.000000 | : qP31(17) : Excited ev=4 16 17 F3e - X Xv=0 17 17 F1e |
| 13C18O Excited | 16 e 7 | X 17 e 1 | 67654.5576 | .0072307 | 68188.4859 | 533.9283 | .0935985 | 826360.1 | 0.000000 | : oP1(17) : Excited dv=7 16 15 F1e - X Xv=0 17 17 F1e |
| 13C18O Excited | 16 e 8 | X 17 e 1 | 67706.4954 | .0891947 | 68240.4237 | 533.9283 | 1.154590 | 10217116 | 0.000000 | : pP21(17) : Excited dv=7 16 16 F2e - X Xv=0 17 17 F1e |
| 13C18O Excited | 16 e 9 | X 17 e 1 | 67738.4900 | .4156223 | 68272.4183 | 533.9283 | 5.380065 | 47676412 | 0.000000 | : pP21(17) : Excited a'v=12 16 16 F2e - X Xv=0 17 17 F1e |
| 13C18O Excited | 16 e 10 | X 17 e 1 | 67758.9858 | .3017640 | 68292.9141 | 533.9283 | 3.906214 | 34647050 | 0.000000 | : qP31(17) : Excited dv=7 16 17 F3e - X Xv=0 17 17 F1e |
| 13C18O Excited | 17 f 2 | X 17 e 1 | 67456.7236 | .7408527 | 67990.6518 | 533.9283 | 9.590043 | 79131914 | 0.000000 | : qQ(17) : Excited Av=2 17 17 F1f - X Xv=0 17 17 F1e |
| 13C18O Excited | 17 f 3 | X 17 e 1 | 67479.9549 | .3608263 | 68013.8832 | 533.9283 | 4.670753 | 38580396 | 0.000000 | : pQ1(17) : Excited a3Pv=13 17 16 F1f - X Xv=0 17 17 F1e |
| 13C18O Excited | 17 f 4 | X 17 e 1 | 67536.6955 | .4973260 | 68070.6238 | 533.9283 | 6.437687 | 53309506 | 0.000000 | : qQ21(17) : Excited a3Pv=13 17 17 F2f - X Xv=0 17 17 F1e |
| 13C18O Excited | 17 f 5 | X 17 e 1 | 67592.3900 | .7210755 | 68126.3183 | 533.9283 | 9.334036 | 77485115 | 0.000000 | : rQ31(17) : Excited a3Pv=13 17 18 F3f - X Xv=0 17 17 F1e |
| 13C18O Excited | 17 f 6 | X 17 e 1 | 67607.6974 | .4061711 | 68141.6257 | 533.9283 | 5.257724 | 43675874 | 0.000000 | : qQ21(17) : Excited ev=4 17 17 F2f - X Xv=0 17 17 F1e |
| 13C18O Excited | 17 f 7 | X 17 e 1 | 67689.7039 | .0185046 | 68223.6322 | 533.9283 | .2395345 | 1997061 | 0.000000 | : pQ1(17) : Excited dv=7 17 16 F1f - X Xv=0 17 17 F1e |
| 13C18O Excited | 17 f 8 | X 17 e 1 | 67739.4850 | .4751861 | 68273.4133 | 533.9283 | 6.151095 | 51396491 | 0.000000 | : pQ1(17) : Excited a'v=12 17 16 F1f - X Xv=0 17 17 F1e |
| 13C18O Excited | 17 f 9 | X 17 e 1 | 67743.1251 | .2168071 | 68277.0534 | 533.9283 | 2.806481 | 23453800 | 0.000000 | : qQ21(17) : Excited dv=7 17 17 F2f - X Xv=0 17 17 F1e |
| 13C18O Excited | 17 f 10 | X 17 e 1 | 67797.0877 | .6365786 | 68331.0160 | 533.9283 | 8.240257 | 69028630 | 0.000000 | : rQ31(17) : Excited dv=7 17 18 F3f - X Xv=0 17 17 F1e |
| 13C18O Excited | 17 f 11 | X 17 e 1 | 67812.1698 | .4330440 | 68346.0981 | 533.9283 | 5.605582 | 46989307 | 0.000000 | : rQ31(17) : Excited a'v=12 17 18 F3f - X Xv=0 17 17 F1e |
| 13C18O Excited | 18 e 1 | X 17 e 1 | 67507.4823 | .3687470 | 68041.4106 | 533.9283 | 4.773283 | 37341745 | 0.000000 | : rR(17) : Excited Av=2 18 18 F1e - X Xv=0 17 17 F1e |
| 13C18O Excited | 18 e 2 | X 17 e 1 | 67521.6854 | .2526565 | 68055.6136 | 533.9283 | 3.270537 | 25601808 | 0.000000 | : qR1(17) : Excited a3Pv=13 18 17 F1e - X Xv=0 17 17 F1e |
| 13C18O Excited | 18 e 3 | X 17 e 1 | 67582.8362 | .2266751 | 68116.7645 | 533.9283 | 2.934219 | 23031570 | 0.000000 | : rR21(17) : Excited a3Pv=13 18 18 F2e - X Xv=0 17 17 F1e |
| 13C18O Excited | 18 e 4 | X 17 e 1 | 67606.3488 | .2491568 | 68140.2771 | 533.9283 | 3.225236 | 25342276 | 0.000000 | : qR1(17) : Excited ev=4 18 17 F1e - X Xv=0 17 17 F1e |
| 13C18O Excited | 18 e 5 | X 17 e 1 | 67642.8545 | .2003492 | 68176.7828 | 533.9283 | 2.593441 | 20410979 | 0.000000 | : sR31(17) : Excited a3Pv=13 18 19 F3e - X Xv=0 17 17 F1e |
| 13C18O Excited | 18 e 6 | X 17 e 1 | 67686.4700 | .1826333 | 68220.3983 | 533.9283 | 2.364115 | 18642150 | 0.000000 | : sR31(17) : Excited ev=4 18 19 F3e - X Xv=0 17 17 F1e |
| 13C18O Excited | 18 e 7 | X 17 e 1 | 67726.9561 | .0117515 | 68260.8844 | 533.9283 | .1521185 | 1201679 | 0.000000 | : qR1(17) : Excited dv=7 18 17 F1e - X Xv=0 17 17 F1e |
| 13C18O Excited | 18 e 8 | X 17 e 1 | 67781.8947 | .1186853 | 68315.8230 | 533.9283 | 1.536333 | 12166014 | 0.000000 | : rR21(17) : Excited dv=7 18 18 F2e - X Xv=0 17 17 F1e |
| 13C18O Excited | 18 e 9 | X 17 e 1 | 67810.9024 | .4950480 | 68344.8306 | 533.9283 | 6.408199 | 50810828 | 0.000000 | : rR21(17) : Excited a'v=12 18 18 F2e - X Xv=0 17 17 F1e |
| 13C18O Excited | 18 e 10 | X 17 e 1 | 67837.3746 | .3406194 | 68371.3029 | 533.9283 | 4.409183 | 35001521 | 0.000000 | : sR31(17) : Excited dv=7 18 19 F3e - X Xv=0 17 17 F1e |
| 13C18O Excited | 17 e 1 | X 18 e 1 | 67393.9073 | .2571853 | 67990.5885 | 596.6812 | 4.498206 | 37013203 | 0.000000 | : pP(18) : Excited Av=2 17 17 F1e - X Xv=0 18 18 F1e |
| 13C18O Excited | 17 e 2 | X 18 e 1 | 67414.1834 | .1485268 | 68010.8647 | 596.6812 | 2.597755 | 21394759 | 0.000000 | : oP1(18) : Excited a3Pv=13 17 16 F1e - X Xv=0 18 18 F1e |
| 13C18O Excited | 17 e 3 | X 18 e 1 | 67473.2811 | .1624630 | 68069.9624 | 596.6812 | 2.841501 | 23463820 | 0.000000 | : pP21(18) : Excited a3Pv=13 17 17 F2e - X Xv=0 18 18 F1e |
| 13C18O Excited | 17 e 4 | X 18 e 1 | 67506.0949 | .1700509 | 68102.7761 | 596.6812 | 2.974214 | 24595554 | 0.000000 | : oP1(18) : Excited ev=4 17 16 F1e - X Xv=0 18 18 F1e |
| 13C18O Excited | 17 e 5 | X 18 e 1 | 67531.2276 | .1160044 | 68127.9089 | 596.6812 | 2.028933 | 16797207 | 0.000000 | : qP31(18) : Excited a3Pv=13 17 18 F3e - X Xv=0 18 18 F1e |
| 13C18O Excited | 17 e 6 | X 18 e 1 | 67582.0431 | .1258889 | 68178.7244 | 596.6812 | 2.201814 | 18269640 | 0.000000 | : qP31(18) : Excited ev=4 17 18 F3e - X Xv=0 18 18 F1e |
| 13C18O Excited | 17 e 7 | X 18 e 1 | 67626.9510 | .0066809 | 68223.6322 | 596.6812 | .1168493 | 971495.6 | 0.000000 | : oP1(18) : Excited dv=7 17 16 F1e - X Xv=0 18 18 F1e |
| 13C18O Excited | 17 e 8 | X 18 e 1 | 67680.3715 | .0745104 | 68277.0527 | 596.6812 | 1.303197 | 10860593 | 0.000000 | : pP21(18) : Excited dv=7 17 17 F2e - X Xv=0 18 18 F1e |
| 13C18O Excited | 17 e 9 | X 18 e 1 | 67710.9150 | .3273343 | 68307.5963 | 596.6812 | 5.725122 | 47776706 | 0.000000 | : pP21(18) : Excited a'v=12 17 17 F2e - X Xv=0 18 18 F1e |
| 13C18O Excited | 17 e 10 | X 18 e 1 | 67734.3350 | .2313155 | 68331.0163 | 596.6812 | 4.045740 | 33797143 | 0.000000 | : qP31(18) : Excited dv=7 17 18 F3e - X Xv=0 18 18 F1e |
| 13C18O Excited | 18 f 2 | X 18 e 1 | 67444.8202 | .5665017 | 68041.5015 | 596.6812 | 9.908194 | 77296886 | 0.000000 | : qQ(18) : Excited Av=2 18 18 F1f - X Xv=0 18 18 F1e |
| 13C18O Excited | 18 f 3 | X 18 e 1 | 67462.0328 | .3013795 | 68058.7141 | 596.6812 | 5.271169 | 41153514 | 0.000000 | : pQ1(18) : Excited a3Pv=13 18 17 F1f - X Xv=0 18 18 F1e |
| 13C18O Excited | 18 f 4 | X 18 e 1 | 67520.7439 | .3791249 | 68117.4251 | 596.6812 | 6.630947 | 51904968 | 0.000000 | : qQ21(18) : Excited a3Pv=13 18 18 F2f - X Xv=0 18 18 F1e |
| 13C18O Excited | 18 f 5 | X 18 e 1 | 67576.2495 | .8083237 | 68172.9307 | 596.6812 | 14.13770 | 1.1094e8 | 0.000000 | : rQ31(18) : Excited a3Pv=13 18 19 F3f - X Xv=0 18 18 F1e |
| 13C18O Excited | 18 f 6 | X 18 e 1 | 67586.8547 | .0778302 | 68183.5359 | 596.6812 | 1.361262 | 10686860 | 0.000000 | : qQ21(18) : Excited ev=4 18 18 F2f - X Xv=0 18 18 F1e |
| 13C18O Excited | 18 f 7 | X 18 e 1 | 67664.2031 | .0168512 | 68260.8844 | 596.6812 | .2947295 | 2321786 | 0.000000 | : pQ1(18) : Excited dv=7 18 17 F1f - X Xv=0 18 18 F1e |
| 13C18O Excited | 18 f 8 | X 18 e 1 | 67711.9078 | .3739517 | 68308.5890 | 596.6812 | 6.540467 | 51632796 | 0.000000 | : pQ1(18) : Excited a'v=12 18 17 F1f - X Xv=0 18 18 F1e |
| 13C18O Excited | 18 f 9 | X 18 e 1 | 67719.1422 | .1757747 | 68315.8234 | 596.6812 | 3.074324 | 24277595 | 0.000000 | : qQ21(18) : Excited dv=7 18 18 F2f - X Xv=0 18 18 F1e |
| 13C18O Excited | 18 f 10 | X 18 e 1 | 67774.6214 | .4857130 | 68371.3026 | 596.6812 | 8.495189 | 67250580 | 0.000000 | : rQ31(18) : Excited dv=7 18 19 F3f - X Xv=0 18 18 F1e |
| 13C18O Excited | 18 f 11 | X 18 e 1 | 67788.7063 | .3403352 | 68385.3875 | 596.6812 | 5.952511 | 47151328 | 0.000000 | : rQ31(18) : Excited a'v=12 18 19 F3f - X Xv=0 18 18 F1e |
| 13C18O Excited | 19 e 1 | X 18 e 1 | 67498.3030 | .2510098 | 68094.9843 | 596.6812 | 4.390196 | 32570265 | 0.000000 | : rR(18) : Excited Av=2 19 19 F1e - X Xv=0 18 18 F1e |
| 13C18O Excited | 19 e 2 | X 18 e 1 | 67506.2223 | .2394519 | 68102.9036 | 596.6812 | 4.188048 | 31081492 | 0.000000 | : qR1(18) : Excited a3Pv=13 19 18 F1e - X Xv=0 18 18 F1e |
| 13C18O Excited | 19 e 3 | X 18 e 1 | 67569.4510 | .1568490 | 68166.1322 | 596.6812 | 2.743311 | 20416676 | 0.000000 | : rR21(18) : Excited a3Pv=13 19 19 F2e - X Xv=0 18 18 F1e |
| 13C18O Excited | 19 e 4 | X 18 e 1 | 67583.3318 | .1980064 | 68180.0131 | 596.6812 | 3.463159 | 25789923 | 0.000000 | : qR1(18) : Excited ev=4 19 18 F1e - X Xv=0 18 18 F1e |
| 13C18O Excited | 19 e 5 | X 18 e 1 | 67631.5745 | .1742496 | 68228.2557 | 596.6812 | 3.047650 | 22744294 | 0.000000 | : sR31(18) : Excited a3Pv=13 19 20 F3e - X Xv=0 18 18 F1e |
| 13C18O Excited | 19 e 6 | X 18 e 1 | 67667.6132 | .1336384 | 68264.2945 | 596.6812 | 2.337354 | 17471325 | 0.000000 | : sR31(18) : Excited ev=4 19 20 F3e - X Xv=0 18 18 F1e |
| 13C18O Excited | 19 e 7 | X 18 e 1 | 67703.5655 | .0106231 | 68300.2468 | 596.6812 | .1857996 | 1391035 | 0.000000 | : qR1(18) : Excited dv=7 19 18 F1e - X Xv=0 18 18 F1e |

13C18O Excited 19 e 8 X 18 e 1 67760.0512 .0969169 68356.7324 596.6812 1.695091 12722516 0.000000 : rR21(18) : Excited dv=7 19 19 F2e - X Xv=0 18 18 F1e
13C18O Excited 19 e 9 X 18 e 1 67787.4379 .3862528 68384.1192 596.6812 6.755616 50765826 0.000000 : rR21(18) : Excited a'v=12 19 19 F2e - X Xv=0 18 18 F1e
13C18O Excited 19 e 10 X 18 e 1 67817.0871 .2588377 68413.7683 596.6812 4.527107 34064108 0.000000 : sR31(18) : Excited dv=7 19 20 F3e - X Xv=0 18 18 F1e
13C18O Excited 18 e 1 X 19 e 1 67378.5043 .1881943 68041.4106 662.9063 4.522057 35174010 0.000000 : pP(19) : Excited Av=2 18 18 F1e - X Xv=0 19 19 F1e
13C18O Excited 18 e 2 X 19 e 1 67392.7073 .1289461 68055.6136 662.9063 3.098404 24115620 0.000000 : oP1(19) : Excited a3Pv=13 18 17 F1e - X Xv=0 19 19 F1e
13C18O Excited 18 e 3 X 19 e 1 67453.8582 .1156863 68116.7645 662.9063 2.779787 21694697 0.000000 : pP21(19) : Excited a3Pv=13 18 18 F2e - X Xv=0 19 19 F1e
13C18O Excited 18 e 4 X 19 e 1 67477.3708 .1271601 68140.2771 662.9063 3.055486 23871325 0.000000 : oP1(19) : Excited ev=4 18 17 F1e - X Xv=0 19 19 F1e
13C18O Excited 18 e 5 X 19 e 1 67513.8765 .1022505 68176.7828 662.9063 2.456944 19226317 0.000000 : qP31(19) : Excited a3Pv=13 18 19 F3e - X Xv=0 19 19 F1e
13C18O Excited 18 e 6 X 19 e 1 67557.4920 .0932090 68220.3983 662.9063 2.239688 17560216 0.000000 : qP31(19) : Excited ev=4 18 19 F3e - X Xv=0 19 19 F1e
13C18O Excited 18 e 7 X 19 e 1 67597.9781 .0059975 68260.8844 662.9063 .1441122 1131941 0.000000 : oP1(19) : Excited dv=7 18 17 F1e - X Xv=0 19 19 F1e
13C18O Excited 18 e 8 X 19 e 1 67652.9167 .0605724 68315.8230 662.9063 1.455474 11460028 0.000000 : pP21(19) : Excited dv=7 18 18 F2e - X Xv=0 19 19 F1e
13C18O Excited 18 e 9 X 19 e 1 67681.9243 .2526534 68344.8306 662.9063 6.070925 47862425 0.000000 : pP21(19) : Excited a'v=12 18 18 F2e - X Xv=0 19 19 F1e
13C18O Excited 18 e 10 X 19 e 1 67708.3966 .1738391 68371.3029 662.9063 4.177121 32970559 0.000000 : qP31(19) : Excited dv=7 18 19 F3e - X Xv=0 19 19 F1e
13C18O Excited 19 f 2 X 19 e 1 67432.2419 .4180452 68095.1482 662.9063 10.04507 74304400 0.000000 : qQ(19) : Excited Av=2 19 19 F1f - X Xv=0 19 19 F1e
13C18O Excited 19 f 3 X 19 e 1 67443.1944 .2507221 68106.1007 662.9063 6.024519 44585698 0.000000 : pQ1(19) : Excited a3Pv=13 19 18 F1f - X Xv=0 19 19 F1e
13C18O Excited 19 f 4 X 19 e 1 67503.8849 .2876470 68166.7912 662.9063 6.911774 51290228 0.000000 : qQ21(19) : Excited a3Pv=13 19 19 F2f - X Xv=0 19 19 F1e
13C18O Excited 19 f 5 X 19 e 1 67555.3109 .6700099 68218.2172 662.9063 16.09944 1.1974e8 0.000000 : qQ21(19) : Excited ev=4 19 19 F2f - X Xv=0 19 19 F1e
13C18O Excited 19 f 6 X 19 e 1 67568.6928 .0101891 68231.5991 662.9063 .2448311 1822057 0.000000 : rQ31(19) : Excited a3Pv=13 19 20 F3f - X Xv=0 19 19 F1e
13C18O Excited 19 f 7 X 19 e 1 67637.3405 .0149942 68300.2468 662.9063 .3602898 2689493 0.000000 : pQ1(19) : Excited dv=7 19 18 F1f - X Xv=0 19 19 F1e
13C18O Excited 19 f 8 X 19 e 1 67682.9144 .2872193 68345.8207 662.9063 6.901496 51622520 0.000000 : pQ1(19) : Excited a'v=12 19 18 F1f - X Xv=0 19 19 F1e
13C18O Excited 19 f 9 X 19 e 1 67693.8265 .1405811 68356.7328 662.9063 3.377977 25279162 0.000000 : qQ21(19) : Excited dv=7 19 19 F2f - X Xv=0 19 19 F1e
13C18O Excited 19 f 10 X 19 e 1 67750.8617 .3634251 68413.7680 662.9063 8.732622 65516086 0.000000 : rQ31(19) : Excited dv=7 19 20 F3f - X Xv=0 19 19 F1e
13C18O Excited 19 f 11 X 19 e 1 67763.8225 .2622607 68426.7288 662.9063 6.301775 47305913 0.000000 : rQ31(19) : Excited a'v=12 19 20 F3f - X Xv=0 19 19 F1e
13C18O Excited 20 e 1 X 19 e 1 67487.9125 .0375326 68150.8188 662.9063 .9018578 6361443 0.000000 : rR(19) : Excited Av=2 20 20 F1e - X Xv=0 19 19 F1e
13C18O Excited 20 e 2 X 19 e 1 67490.2877 .3417081 68153.1940 662.9063 8.210789 57922635 0.000000 : qR1(19) : Excited a3Pv=13 20 19 F1e - X Xv=0 19 19 F1e
13C18O Excited 20 e 3 X 19 e 1 67554.9933 .0678009 68217.8996 662.9063 1.629166 11525962 0.000000 : rR21(19) : Excited a3Pv=13 20 20 F2e - X Xv=0 19 19 F1e
13C18O Excited 20 e 4 X 19 e 1 67559.2784 .1934064 68222.1847 662.9063 4.647298 32884791 0.000000 : qR1(19) : Excited ev=4 20 19 F1e - X Xv=0 19 19 F1e
13C18O Excited 20 e 5 X 19 e 1 67619.3549 .1492279 68282.2612 662.9063 3.585747 25440893 0.000000 : sR31(19) : Excited a3Pv=13 20 21 F3e - X Xv=0 19 19 F1e
13C18O Excited 20 e 6 X 19 e 1 67647.5475 .0915863 68310.4538 662.9063 2.200698 15633498 0.000000 : sR31(19) : Excited ev=4 20 21 F3e - X Xv=0 19 19 F1e
13C18O Excited 20 e 7 X 19 e 1 67678.8196 .0094533 68341.7259 662.9063 .2271500 1615885 0.000000 : qR1(19) : Excited dv=7 20 19 F1e - X Xv=0 19 19 F1e
13C18O Excited 20 e 8 X 19 e 1 67736.8727 .0771981 68399.7790 662.9063 1.854966 13229732 0.000000 : rR21(19) : Excited dv=7 20 20 F2e - X Xv=0 19 19 F1e
13C18O Excited 20 e 9 X 19 e 1 67762.5533 .2956709 68425.4596 662.9063 7.104577 50727923 0.000000 : rR21(19) : Excited a'v=12 20 20 F2e - X Xv=0 19 19 F1e
13C18O Excited 20 e 10 X 19 e 1 67795.5008 .1930090 68458.4071 662.9063 4.637749 33162667 0.000000 : sR31(19) : Excited dv=7 20 21 F3e - X Xv=0 19 19 F1e
13C18O Excited 19 e 1 X 20 e 1 67362.3831 .1242552 68094.9843 732.6012 4.170686 30755208 0.000000 : pP(20) : Excited Av=2 19 19 F1e - X Xv=0 20 20 F1e
13C18O Excited 19 e 2 X 20 e 1 67370.3024 .1185338 68102.9036 732.6012 3.978645 29349421 0.000000 : oP1(20) : Excited a3Pv=13 19 18 F1e - X Xv=0 20 20 F1e
13C18O Excited 19 e 3 X 20 e 1 67433.5311 .0776436 68166.1322 732.6012 2.606145 19279030 0.000000 : pP21(20) : Excited a3Pv=13 19 19 F2e - X Xv=0 20 20 F1e
13C18O Excited 19 e 4 X 20 e 1 67447.4119 .0980173 68180.0131 732.6012 3.290001 24352902 0.000000 : oP1(20) : Excited ev=4 19 18 F1e - X Xv=0 20 20 F1e
13C18O Excited 19 e 5 X 20 e 1 67495.6546 .0862573 68228.2557 732.6012 2.895268 21477069 0.000000 : qP31(20) : Excited a3Pv=13 19 20 F3e - X Xv=0 20 20 F1e
13C18O Excited 19 e 6 X 20 e 1 67531.6933 .0661538 68264.2945 732.6012 2.220486 16497943 0.000000 : qP31(20) : Excited ev=4 19 20 F3e - X Xv=0 20 20 F1e
13C18O Excited 19 e 7 X 20 e 1 67567.6456 .0052587 68300.2468 732.6012 .1765096 1313541 0.000000 : oP1(20) : Excited dv=7 19 18 F1e - X Xv=0 20 20 F1e
13C18O Excited 19 e 8 X 20 e 1 67624.1313 .0479759 68356.7324 732.6012 1.610336 12013804 0.000000 : pP21(20) : Excited dv=7 19 19 F2e - X Xv=0 20 20 F1e
13C18O Excited 19 e 9 X 20 e 1 67651.5180 .1912033 68384.1192 732.6012 6.417835 47938014 0.000000 : pP21(20) : Excited a'v=12 19 19 F2e - X Xv=0 20 20 F1e
13C18O Excited 19 e 10 X 20 e 1 67681.1671 .1281301 68413.7683 732.6012 4.300752 32166718 0.000000 : qP31(20) : Excited dv=7 19 20 F3e - X Xv=0 20 20 F1e
13C18O Excited 20 f 2 X 20 e 1 67418.9693 .2792958 68151.5704 732.6012 9.374703 65923982 0.000000 : qQ(20) : Excited Av=2 20 20 F1f - X Xv=0 20 20 F1e
13C18O Excited 20 f 3 X 20 e 1 67423.4576 .2246714 68156.0588 732.6012 7.541207 53041223 0.000000 : pQ1(20) : Excited a3Pv=13 20 19 F1f - X Xv=0 20 20 F1e
13C18O Excited 20 f 4 X 20 e 1 67486.1012 .2187315 68218.7024 732.6012 7.341830 51782965 0.000000 : qQ21(20) : Excited a3Pv=13 20 20 F2f - X Xv=0 20 20 F1e
13C18O Excited 20 f 5 X 20 e 1 67530.7047 .4558734 68263.3059 732.6012 15.30162 1.0814e8 0.000000 : qQ21(20) : Excited ev=4 20 20 F2f - X Xv=0 20 20 F1e
13C18O Excited 20 f 6 X 20 e 1 67552.0920 .0529038 68284.6932 732.6012 1.775741 12561333 0.000000 : rQ31(20) : Excited a3Pv=13 20 21 F3f - X Xv=0 20 20 F1e
13C18O Excited 20 f 7 X 20 e 1 67609.1247 .0131452 68341.7259 732.6012 .4412257 3129078 0.000000 : pQ1(20) : Excited dv=7 20 19 F1f - X Xv=0 20 20 F1e
13C18O Excited 20 f 8 X 20 e 1 67652.5052 .2162016 68385.1063 732.6012 7.256913 51563592 0.000000 : pQ1(20) : Excited a'v=12 20 19 F1f - X Xv=0 20 20 F1e
13C18O Excited 20 f 9 X 20 e 1 67667.1782 .1100283 68399.7794 732.6012 3.693156 26258591 0.000000 : qQ21(20) : Excited dv=7 20 20 F2f - X Xv=0 20 20 F1e
13C18O Excited 20 f 10 X 20 e 1 67725.8057 .2667174 68458.4069 732.6012 8.952502 63818494 0.000000 : rQ31(20) : Excited dv=7 20 21 F3f - X Xv=0 20 20 F1e
13C18O Excited 20 f 11 X 20 e 1 67737.5184 .1982459 68470.1195 732.6012 6.654221 47459662 0.000000 : rQ31(20) : Excited a'v=12 20 21 F3f - X Xv=0 20 20 F1e
13C18O Excited 21 e 1 X 20 e 1 67471.7850 .0672120 68204.3861 732.6012 2.256004 15162175 0.000000 : qR1(20) : Excited a3Pv=13 21 20 F1e - X Xv=0 20 20 F1e
13C18O Excited 21 e 2 X 20 e 1 67478.3580 .2198295 68210.9591 732.6012 7.378685 49605231 0.000000 : rR(20) : Excited Av=2 21 21 F1e - X Xv=0 20 20 F1e
13C18O Excited 21 e 3 X 20 e 1 67533.3597 .0616999 68265.9608 732.6012 2.070989 13956865 0.000000 : qR1(20) : Excited ev=4 21 20 F1e - X Xv=0 20 20 F1e
13C18O Excited 21 e 4 X 20 e 1 67540.3427 .1281603 68272.9438 732.6012 4.301765 28999564 0.000000 : rR21(20) : Excited a3Pv=13 21 21 F2e - X Xv=0 20 20 F1e
13C18O Excited 21 e 5 X 20 e 1 67606.0647 .1290312 68338.6658 732.6012 4.330996 29281930 0.000000 : sR31(20) : Excited a3Pv=13 21 22 F3e - X Xv=0 20 20 F1e
13C18O Excited 21 e 6 X 20 e 1 67626.3872 .0546707 68358.9884 732.6012 1.835050 12417992 0.000000 : sR31(20) : Excited ev=4 21 22 F3e - X Xv=0 20 20 F1e
13C18O Excited 21 e 7 X 20 e 1 67652.7323 .0083946 68385.3335 732.6012 .2817682 1908987 0.000000 : qR1(20) : Excited dv=7 21 20 F1e - X Xv=0 20 20 F1e
13C18O Excited 21 e 8 X 20 e 1 67712.3594 .0600208 68444.9606 732.6012 2.014630 13685291 0.000000 : rR21(20) : Excited dv=7 21 21 F2e - X Xv=0 20 20 F1e
13C18O Excited 21 e 9 X 20 e 1 67736.2484 .2221210 68468.8496 732.6012 7.455601 50699194 0.000000 : rR21(20) : Excited a'v=12 21 21 F2e - X Xv=0 20 20 F1e
13C18O Excited 21 e 10 X 20 e 1 67772.6132 .1412515 68505.2143 732.6012 4.741176 32292654 0.000000 : sR31(20) : Excited dv=7 21 22 F3e - X Xv=0 20 20 F1e
13C18O Excited 20 e 1 X 21 e 1 67345.0554 .0180165 68150.8188 805.7634 .8589122 6020124 0.000000 : pP(21) : Excited Av=2 20 20 F1e - X Xv=0 21 21 F1e
13C18O Excited 20 e 2 X 21 e 1 67347.4307 .1640277 68153.1940 805.7634 7.819799 54814855 0.000000 : oP1(21) : Excited a3Pv=13 20 19 F1e - X Xv=0 21 21 F1e
13C18O Excited 20 e 3 X 21 e 1 67412.1363 .0325460 68217.8996 805.7634 1.551586 10907615 0.000000 : pP21(21) : Excited a3Pv=13 20 20 F2e - X Xv=0 21 21 F1e
13C18O Excited 20 e 4 X 21 e 1 67416.4213 .0928395 68222.1847 805.7634 4.425998 31120593 0.000000 : oP1(21) : Excited ev=4 20 19 F1e - X Xv=0 21 21 F1e
13C18O Excited 20 e 5 X 21 e 1 67476.4978 .0716328 68282.2612 805.7634 3.414998 24076180 0.000000 : qP31(21) : Excited a3Pv=13 20 21 F3e - X Xv=0 21 21 F1e
13C18O Excited 20 e 6 X 21 e 1 67504.6904 .0439635 68310.4538 805.7634 2.095903 14794917 0.000000 : qP31(21) : Excited ev=4 20 21 F3e - X Xv=0 21 21 F1e
13C18O Excited 20 e 7 X 21 e 1 67535.9625 .0045378 68341.7259 805.7634 .2163333 1529214 0.000000 : oP1(21) : Excited dv=7 20 19 F1e - X Xv=0 21 21 F1e
13C18O Excited 20 e 8 X 21 e 1 67594.0156 .0370568 68399.7790 805.7634 1.766635 12520194 0.000000 : pP21(21) : Excited dv=7 20 20 F2e - X Xv=0 21 21 F1e
13C18O Excited 20 e 9 X 21 e 1 67619.6962 .1419288 68425.4596 805.7634 6.766264 48007395 0.000000 : pP21(21) : Excited a'v=12 20 20 F2e - X Xv=0 21 21 F1e
13C18O Excited 20 e 10 X 21 e 1 67652.6438 .0926487 68458.4071 805.7634 4.416904 31384257 0.000000 : qP31(21) : Excited dv=7 20 21 F3e - X Xv=0 21 21 F1e
13C18O Excited 21 f 2 X 21 e 1 67402.6636 .0777400 68208.4270 805.7634 3.706152 24831866 0.000000 : pQ1(21) : Excited a3Pv=13 21 20 F1f - X Xv=0 21 21 F1e
13C18O Excited 21 f 3 X 21 e 1 67405.1567 .2933993 68210.9201 805.7634 13.98742 93728550 0.000000 : qQ(21) : Excited Av=2 21 21 F1f - X Xv=0 21 21 F1e
13C18O Excited 21 f 4 X 21 e 1 67467.3595 .1686923 68273.1228 805.7634 8.042180 54039324 0.000000 : qQ21(21) : Excited a3Pv=13 21 21 F2f - X Xv=0 21 21 F1e
13C18O Excited 21 f 5 X 21 e 1 67504.1970 .3059312 68309.9604 805.7634 14.58486 98163405 0.000000 : qQ21(21) : Excited ev=4 21 21 F2f - X Xv=0 21 21 F1e
13C18O Excited 21 f 6 X 21 e 1 67535.3094 .0630504 68341.0728 805.7634 3.005841 20258799 0.000000 : rQ31(21) : Excited a3Pv=13 21 22 F3f - X Xv=0 21 21 F1e
13C18O Excited 21 f 7 X 21 e 1 67579.5701 .0115359 68385.3335 805.7634 .5499608 3713924 0.000000 : pQ1(21) : Excited dv=7 21 20 F1f - X Xv=0 21 21 F1e
13C18O Excited 21 f 8 X 21 e 1 67620.6804 .1596648 68426.4437 805.7634 7.611805 51496918 0.000000 : pQ1(21) : Excited a'v=12 21 20 F1f - X Xv=0 21 21 F1e
13C18O Excited 21 f 9 X 21 e 1 67639.1976 .0841792 68444.9610 805.7634 4.013130 27172750 0.000000 : qQ21(21) : Excited dv=7 21 21 F2f - X Xv=0 21 21 F1e
13C18O Excited 21 f 10 X 21 e 1 67699.4507 .1920114 68505.2140 805.7634 9.153886 62146393 0.000000 : rQ31(21) : Excited dv=7 21 22 F3f - X Xv=0 21 21 F1e
13C18O Excited 21 f 11 X 21 e 1 67709.7936 .1470668 68515.5570 805.7634 7.011214 47621455 0.000000 : rQ31(21) : Excited a'v=12 21 22 F3f - X Xv=0 21 21 F1e
13C18O Excited 22 e 1 X 21 e 1 67453.2433 .0727832 68259.0067 805.7634 3.469843 22265335 0.000000 : qR1(21) : Excited a3Pv=13 22 21 F1e - X Xv=0 21 21 F1e
13C18O Excited 22 e 2 X 21 e 1 67467.0114 .1390851 68272.7747 805.7634 6.630696 42574001 0.000000 : rR(21) : Excited Av=2 22 22 F1e - X Xv=0 21 21 F1e
13C18O Excited 22 e 3 X 21 e 1 67506.9737 .0578729 68312.7371 805.7634 2.759011 17746400 0.000000 : qR1(21) : Excited ev=4 22 21 F1e - X Xv=0 21 21 F1e
13C18O Excited 22 e 4 X 21 e 1 67524.1954 .0792320 68329.9588 805.7634 3.777279 24314661 0.000000 : rR21(21) : Excited a3Pv=13 22 22 F2e - X Xv=0 21 21 F1e
13C18O Excited 22 e 5 X 21 e 1 67591.2757 .1161970 68397.0391 805.7634 5.539534 35764827 0.000000 : sR31(21) : Excited a3Pv=13 22 23 F3e - X Xv=0 21 21 F1e
13C18O Excited 22 e 6 X 21 e 1 67604.5308 .0203874 68410.2941 805.7634 .9719428 6278834 0.000000 : sR31(21) : Excited ev=4 22 23 F3e - X Xv=0 21 21 F1e
13C18O Excited 22 e 7 X 21 e 1 67625.3362 .0076790 68431.0995 805.7634 .3660844 2367121 0.000000 : qR1(21) : Excited dv=7 22 21 F1e - X Xv=0 21 21 F1e
13C18O Excited 22 e 8 X 21 e 1 67686.5118 .0455680 68492.2752 805.7634 2.172392 14084956 0.000000 : rR21(21) : Excited dv=7 22 22 F2e - X Xv=0 21 21 F1e
13C18O Excited 22 e 9 X 21 e 1 67708.5232 .1638083 68514.2866 805.7634 7.809343 50682197 0.000000 : rR21(21) : Excited a'v=12 22 22 F2e - X Xv=0 21 21 F1e

13C18O Excited 22 e 10 X 21 e 1 67748.4218 .1014649 68554.1851 805.7634 4.837205 31448716 0.000000 : sR31(21) : Excited dv=7 22 23 F3e - X Xv=0 21 21 F1e
13C18O Excited 21 e 1 X 22 e 1 67321.9957 .0312792 68204.3861 882.3904 2.153459 14376808 0.000000 : oP1(22) : Excited a3Pv=13 21 20 F1e - X Xv=0 22 22 F1e
13C18O Excited 21 e 2 X 22 e 1 67328.5687 .1023046 68210.9591 882.3904 7.043290 47035820 0.000000 : pP(22) : Excited Av=2 21 21 F1e - X Xv=0 22 22 F1e
13C18O Excited 21 e 3 X 22 e 1 67383.5704 .0287140 68265.9608 882.3904 1.976853 13234010 0.000000 : oP1(22) : Excited ev=4 21 20 F1e - X Xv=0 22 22 F1e
13C18O Excited 21 e 4 X 22 e 1 67390.5534 .0596434 68272.9438 882.3904 4.106230 27497637 0.000000 : pP21(22) : Excited a3Pv=13 21 21 F2e - X Xv=0 22 22 F1e
13C18O Excited 21 e 5 X 22 e 1 67456.2754 .0600487 68338.6658 882.3904 4.134132 27765558 0.000000 : qP31(22) : Excited a3Pv=13 21 22 F3e - X Xv=0 22 22 F1e
13C18O Excited 21 e 6 X 22 e 1 67476.5979 .0254427 68358.9884 882.3904 1.751638 11774947 0.000000 : qP31(22) : Excited ev=4 21 22 F3e - X Xv=0 22 22 F1e
13C18O Excited 21 e 7 X 22 e 1 67502.9430 .0039067 68385.3335 882.3904 .2689606 1810138 0.000000 : oP1(22) : Excited dv=7 21 20 F1e - X Xv=0 22 22 F1e
13C18O Excited 21 e 8 X 22 e 1 67562.5702 .0279326 68444.9606 882.3904 1.923056 12976731 0.000000 : pP21(22) : Excited dv=7 21 21 F2e - X Xv=0 22 22 F1e
13C18O Excited 21 e 9 X 22 e 1 67586.4591 .1033710 68468.8496 882.3904 7.116710 48074340 0.000000 : pP21(22) : Excited a'v=12 21 21 F2e - X Xv=0 22 22 F1e
13C18O Excited 21 e 10 X 22 e 1 67622.8239 .0657358 68505.2143 882.3904 4.525668 30620873 0.000000 : qP31(22) : Excited dv=7 21 22 F3e - X Xv=0 22 22 F1e
13C18O Excited 22 f 2 X 22 e 1 67381.1303 .0911147 68263.5208 882.3904 6.272912 40123149 0.000000 : pQ1(22) : Excited a3Pv=13 22 21 F1f - X Xv=0 22 22 F1e
13C18O Excited 22 f 3 X 22 e 1 67390.4764 .1755632 68272.8668 882.3904 12.08687 77342910 0.000000 : qQ(22) : Excited Av=2 22 22 F1f - X Xv=0 22 22 F1e
13C18O Excited 22 f 4 X 22 e 1 67447.5871 .1348555 68329.9775 882.3904 9.284305 59560668 0.000000 : qQ21(22) : Excited a3Pv=13 22 22 F2f - X Xv=0 22 22 F1e
13C18O Excited 22 f 5 X 22 e 1 67476.3382 .1996277 68358.7287 882.3904 13.74363 88280927 0.000000 : qQ21(22) : Excited ev=4 22 22 F2f - X Xv=0 22 22 F1e
13C18O Excited 22 f 6 X 22 e 1 67517.8434 .0562870 68400.2338 882.3904 3.875152 24937640 0.000000 : rQ31(22) : Excited a3Pv=13 22 23 F3f - X Xv=0 22 22 F1e
13C18O Excited 22 f 7 X 22 e 1 67548.7091 .0105380 68431.0995 882.3904 .7255018 4675204 0.000000 : pQ1(22) : Excited dv=7 22 21 F1f - X Xv=0 22 22 F1e
13C18O Excited 22 f 8 X 22 e 1 67587.4402 .1157389 68469.8306 882.3904 7.968198 51436220 0.000000 : pQ1(22) : Excited a'v=12 22 21 F1f - X Xv=0 22 22 F1e
13C18O Excited 22 f 9 X 22 e 1 67609.8852 .0629482 68492.2756 882.3904 4.333749 28003043 0.000000 : qQ21(22) : Excited dv=7 22 22 F2f - X Xv=0 22 22 F1e
13C18O Excited 22 f 10 X 22 e 1 67671.7944 .1355818 68554.1848 882.3904 9.334308 60480602 0.000000 : rQ31(22) : Excited dv=7 22 23 F3f - X Xv=0 22 22 F1e
13C18O Excited 22 f 11 X 22 e 1 67680.6480 .1071250 68563.0384 882.3904 7.375158 47805272 0.000000 : rQ31(22) : Excited a'v=12 22 23 F3f - X Xv=0 22 22 F1e
13C18O Excited 23 e 1 X 22 e 1 67433.4417 .0625313 68315.8322 882.3904 4.305049 26425890 0.000000 : qR1(22) : Excited a3Pv=13 23 22 F1e - X Xv=0 22 22 F1e
13C18O Excited 23 e 2 X 22 e 1 67454.8458 .0872866 68337.2362 882.3904 6.009356 36922660 0.000000 : rR(22) : Excited Av=2 23 23 F1e - X Xv=0 22 22 F1e
13C18O Excited 23 e 3 X 22 e 1 67479.6941 .0470949 68362.0846 882.3904 3.242312 19943423 0.000000 : qR1(22) : Excited ev=4 23 22 F1e - X Xv=0 22 22 F1e
13C18O Excited 23 e 4 X 22 e 1 67507.2161 .0540475 68389.6065 882.3904 3.720972 22915673 0.000000 : rR21(22) : Excited a3Pv=13 23 23 F2e - X Xv=0 22 22 F1e
13C18O Excited 23 e 5 X 22 e 1 67573.3538 .0980024 68455.7442 882.3904 6.747103 41674401 0.000000 : sR31(22) : Excited ev=4 23 24 F3e - X Xv=0 22 22 F1e
13C18O Excited 23 e 7 X 22 e 1 67596.7607 .0080694 68479.1512 882.3904 .5555471 3434979 0.000000 : qR1(22) : Excited dv=7 23 22 F1e - X Xv=0 22 22 F1e
13C18O Excited 23 e 8 X 22 e 1 67659.3303 .0337828 68541.7207 882.3904 2.325821 14420654 0.000000 : rR21(22) : Excited dv=7 23 23 F2e - X Xv=0 22 22 F1e
13C18O Excited 23 e 9 X 22 e 1 67679.3776 .1186214 68561.7680 882.3904 8.166642 50680184 0.000000 : rR21(22) : Excited a'v=12 23 23 F2e - X Xv=0 22 22 F1e
13C18O Excited 23 e 10 X 22 e 1 67722.9247 .0715405 68605.3152 882.3904 4.925298 30624229 0.000000 : sR31(22) : Excited dv=7 23 24 F3e - X Xv=0 22 22 F1e
13C18O Excited 22 e 1 X 23 e 1 67296.5270 .0328330 68259.0067 962.4797 3.318980 21149179 0.000000 : oP1(23) : Excited a3Pv=13 22 21 F1e - X Xv=0 23 23 F1e
13C18O Excited 22 e 2 X 23 e 1 67310.2951 .0627422 68272.7747 962.4797 6.342405 40439836 0.000000 : pP(23) : Excited Av=2 22 22 F1e - X Xv=0 23 23 F1e
13C18O Excited 22 e 3 X 23 e 1 67350.2575 .0261068 68312.7371 962.4797 2.639054 16856871 0.000000 : oP1(23) : Excited ev=4 22 21 F1e - X Xv=0 23 23 F1e
13C18O Excited 22 e 4 X 23 e 1 67367.4792 .0357421 68329.9588 962.4797 3.613049 23095943 0.000000 : pP21(23) : Excited a3Pv=13 22 22 F2e - X Xv=0 23 23 F1e
13C18O Excited 22 e 5 X 23 e 1 67434.5595 .0524172 68397.0391 962.4797 5.298685 33972430 0.000000 : qP31(23) : Excited a3Pv=13 22 23 F3e - X Xv=0 23 23 F1e
13C18O Excited 22 e 6 X 23 e 1 67447.8145 .0091969 68410.2941 962.4797 .9296844 5964171 0.000000 : qP31(23) : Excited ev=4 22 23 F3e - X Xv=0 23 23 F1e
13C18O Excited 22 e 7 X 23 e 1 67468.6199 .0034640 68431.0995 962.4797 .3501677 2248498 0.000000 : oP1(23) : Excited dv=7 22 21 F1e - X Xv=0 23 23 F1e
13C18O Excited 22 e 8 X 23 e 1 67529.7955 .0205560 68492.2752 962.4797 2.077940 13379203 0.000000 : pP21(23) : Excited dv=7 22 22 F2e - X Xv=0 23 23 F1e
13C18O Excited 22 e 9 X 23 e 1 67551.8069 .0738951 68514.2866 962.4797 7.469806 48142780 0.000000 : pP21(23) : Excited a'v=12 22 22 F2e - X Xv=0 23 23 F1e
13C18O Excited 22 e 10 X 23 e 1 67591.7055 .0457715 68554.1851 962.4797 4.626891 29873110 0.000000 : qP31(23) : Excited dv=7 22 23 F3e - X Xv=0 23 23 F1e
13C18O Excited 23 f 2 X 23 e 1 67358.6358 .0715027 68321.1155 962.4797 7.227966 44220283 0.000000 : pQ1(23) : Excited a3Pv=13 23 22 F1f - X Xv=0 23 23 F1e
13C18O Excited 23 f 3 X 23 e 1 67375.1303 .1148074 68337.6100 962.4797 11.60550 71053961 0.000000 : qQ(23) : Excited Av=2 23 23 F1f - X Xv=0 23 23 F1e
13C18O Excited 23 f 4 X 23 e 1 67426.5936 .1164052 68389.0733 962.4797 11.76702 72208046 0.000000 : qQ21(23) : Excited a3Pv=13 23 23 F2f - X Xv=0 23 23 F1e
13C18O Excited 23 f 5 X 23 e 1 67447.4643 .1189014 68409.9440 962.4797 12.01935 73825008 0.000000 : qQ21(23) : Excited ev=4 23 23 F2f - X Xv=0 23 23 F1e
13C18O Excited 23 f 6 X 23 e 1 67499.4421 .0428933 68461.9218 962.4797 4.335939 26693730 0.000000 : rQ31(23) : Excited a3Pv=13 23 24 F3f - X Xv=0 23 23 F1e
13C18O Excited 23 f 7 X 23 e 1 67516.6661 .0114841 68479.1457 962.4797 1.160887 7152348 0.000000 : pQ1(23) : Excited dv=7 23 22 F1f - X Xv=0 23 23 F1e
13C18O Excited 23 f 8 X 23 e 1 67552.7850 .0823717 68515.2647 962.4797 8.326676 51383895 0.000000 : pQ1(23) : Excited a'v=12 23 22 F1f - X Xv=0 23 23 F1e
13C18O Excited 23 f 9 X 23 e 1 67579.2416 .0460071 68541.7212 962.4797 4.650705 28733223 0.000000 : qQ21(23) : Excited dv=7 23 23 F2f - X Xv=0 23 23 F1e
13C18O Excited 23 f 10 X 23 e 1 67642.8351 .0938628 68605.3148 962.4797 9.488273 58786576 0.000000 : rQ31(23) : Excited dv=7 23 24 F3f - X Xv=0 23 23 F1e
13C18O Excited 23 f 11 X 23 e 1 67650.0814 .0766749 68612.5611 962.4797 7.750815 48037227 0.000000 : rQ31(23) : Excited a'v=12 23 24 F3f - X Xv=0 23 23 F1e
13C18O Excited 24 e 1 X 23 e 1 67412.2343 .0509455 68374.7139 962.4797 5.149907 30293043 0.000000 : qR1(23) : Excited a3Pv=13 24 23 F1e - X Xv=0 23 23 F1e
13C18O Excited 24 e 2 X 23 e 1 67440.9076 .0378965 68403.3873 962.4797 3.830830 22562667 0.000000 : rR(23) : Excited Av=2 24 24 F1e - X Xv=0 23 23 F1e
13C18O Excited 24 e 3 X 23 e 1 67452.6850 .0509944 68415.1647 962.4797 5.154855 30376764 0.000000 : qR1(23) : Excited ev=4 24 23 F1e - X Xv=0 23 23 F1e
13C18O Excited 24 e 4 X 23 e 1 67489.3227 .0375711 68451.8024 962.4797 3.797937 22417144 0.000000 : rR21(23) : Excited a3Pv=13 24 24 F2e - X Xv=0 23 23 F1e
13C18O Excited 24 e 5 X 23 e 1 67550.5200 .0598207 68512.9996 962.4797 6.047075 35789758 0.000000 : sR31(23) : Excited ev=4 24 25 F3e - X Xv=0 23 23 F1e
13C18O Excited 24 e 6 X 23 e 1 67563.2998 .0006354 68525.7795 962.4797 .0642274 380347.0 0.000000 : sR31(23) : Excited a3Pv=13 24 25 F3e - X Xv=0 23 23 F1e
13C18O Excited 24 e 7 X 23 e 1 67568.7627 .0160060 68531.2424 962.4797 1.617993 9583892 0.000000 : qR1(23) : Excited dv=7 24 23 F1e - X Xv=0 23 23 F1e
13C18O Excited 24 e 8 X 23 e 1 67630.8159 .0244432 68593.2956 962.4797 2.470885 14676206 0.000000 : rR21(23) : Excited dv=7 24 24 F2e - X Xv=0 23 23 F1e
13C18O Excited 24 e 9 X 23 e 1 67648.8117 .0843688 68611.2914 962.4797 8.528557 50697139 0.000000 : rR21(23) : Excited a'v=12 24 24 F2e - X Xv=0 23 23 F1e
13C18O Excited 24 e 10 X 23 e 1 67696.1208 .0495060 68658.6005 962.4797 5.004395 29810572 0.000000 : sR31(23) : Excited dv=7 24 25 F3e - X Xv=0 23 23 F1e
13C18O Excited 23 e 1 X 24 e 1 67269.8039 .0273390 68315.8322 1046.0283 4.125672 25140894 0.000000 : oP1(24) : Excited a3Pv=13 23 22 F1e - X Xv=0 24 24 F1e
13C18O Excited 23 e 2 X 24 e 1 67291.2080 .0381621 68337.2362 1046.0283 5.758966 35127326 0.000000 : pP(24) : Excited Av=2 23 23 F1e - X Xv=0 24 24 F1e
13C18O Excited 23 e 3 X 24 e 1 67316.0563 .0205901 68362.0846 1046.0283 3.107216 18973742 0.000000 : oP1(24) : Excited ev=4 23 22 F1e - X Xv=0 24 24 F1e
13C18O Excited 23 e 4 X 24 e 1 67343.5782 .0236298 68389.6065 1046.0283 3.565932 21801540 0.000000 : pP21(24) : Excited a3Pv=13 23 23 F2e - X Xv=0 24 24 F1e
13C18O Excited 23 e 5 X 24 e 1 67409.7160 .0428471 68455.7442 1046.0283 6.465974 39648525 0.000000 : qP31(24) : Excited ev=4 23 24 F3e - X Xv=0 24 24 F1e
13C18O Excited 23 e 7 X 24 e 1 67433.1229 .0035280 68479.1512 1046.0283 .5323993 3268006 0.000000 : oP1(24) : Excited dv=7 23 22 F1e - X Xv=0 24 24 F1e
13C18O Excited 23 e 8 X 24 e 1 67495.6925 .0147700 68541.7207 1046.0283 2.228912 13719764 0.000000 : pP21(24) : Excited dv=7 23 23 F2e - X Xv=0 24 24 F1e
13C18O Excited 23 e 9 X 24 e 1 67515.7397 .0518618 68561.7680 1046.0283 7.826366 48217068 0.000000 : pP21(24) : Excited a'v=12 23 23 F2e - X Xv=0 24 24 F1e
13C18O Excited 23 e 10 X 24 e 1 67559.2869 .0312778 68605.3152 1046.0283 4.720077 29135993 0.000000 : qP31(24) : Excited dv=7 23 24 F3e - X Xv=0 24 24 F1e
13C18O Excited 24 f 2 X 24 e 1 67335.2219 .0532922 68381.2502 1046.0283 8.042228 47144458 0.000000 : pQ1(24) : Excited a3Pv=13 24 23 F1f - X Xv=0 24 24 F1e
13C18O Excited 24 f 3 X 24 e 1 67359.0321 .0722299 68405.0604 1046.0283 10.90008 63965306 0.000000 : qQ(24) : Excited Av=2 24 24 F1f - X Xv=0 24 24 F1e
13C18O Excited 24 f 4 X 24 e 1 67403.7725 .1141176 68449.8008 1046.0283 17.22127 1.0126e8 0.000000 : qQ21(24) : Excited a3Pv=13 24 24 F2f - X Xv=0 24 24 F1e
13C18O Excited 24 f 5 X 24 e 1 67418.2650 .0515218 68464.2933 1046.0283 7.775063 45747147 0.000000 : qQ21(24) : Excited ev=4 24 24 F2f - X Xv=0 24 24 F1e
13C18O Excited 24 f 6 X 24 e 1 67478.5930 .0138756 68524.6213 1046.0283 2.093935 12353462 0.000000 : rQ31(24) : Excited a3Pv=13 24 25 F3f - X Xv=0 24 24 F1e
13C18O Excited 24 f 7 X 24 e 1 67484.9118 .0280965 68530.9400 1046.0283 4.239989 25021428 0.000000 : pQ1(24) : Excited dv=7 24 23 F1f - X Xv=0 24 24 F1e
13C18O Excited 24 f 8 X 24 e 1 67516.7153 .0575450 68562.7436 1046.0283 8.684005 51319374 0.000000 : pQ1(24) : Excited a'v=12 24 23 F1f - X Xv=0 24 24 F1e
13C18O Excited 24 f 9 X 24 e 1 67547.2679 .0328521 68593.2962 1046.0283 4.957655 29337767 0.000000 : qQ21(24) : Excited dv=7 24 24 F2f - X Xv=0 24 24 F1e
13C18O Excited 24 f 10 X 24 e 1 67612.5717 .0636338 68658.6000 1046.0283 9.602861 56991537 0.000000 : rQ31(24) : Excited dv=7 24 25 F3f - X Xv=0 24 24 F1e
13C18O Excited 24 f 11 X 24 e 1 67618.0937 .0540023 68664.1220 1046.0283 8.149390 48377254 0.000000 : rQ31(24) : Excited a'v=12 24 25 F3f - X Xv=0 24 24 F1e
13C18O Excited 25 e 1 X 24 e 1 67389.2493 .0402626 68435.2776 1046.0283 6.075950 34303567 0.000000 : qR1(24) : Excited a3Pv=13 25 24 F1e - X Xv=0 24 24 F1e
13C18O Excited 25 e 2 X 24 e 1 67419.2386 .0007528 68465.2669 1046.0283 .1136035 642238.5 0.000000 : qR1(24) : Excited ev=4 25 24 F1e - X Xv=0 24 24 F1e
13C18O Excited 25 e 3 X 24 e 1 67432.2809 .0559221 68478.3092 1046.0283 8.439098 47736746 0.000000 : rR(24) : Excited Av=2 25 25 F1e - X Xv=0 24 24 F1e
13C18O Excited 25 e 4 X 24 e 1 67470.4802 .0263985 68516.5085 1046.0283 3.983755 22572894 0.000000 : rR21(24) : Excited a3Pv=13 25 25 F2e - X Xv=0 24 24 F1e
13C18O Excited 25 e 5 X 24 e 1 67524.6437 .0347277 68570.6720 1046.0283 5.240692 29766572 0.000000 : sR31(24) : Excited ev=4 25 26 F3e - X Xv=0 24 24 F1e
13C18O Excited 25 e 6 X 24 e 1 67534.1085 .0017492 68580.1367 1046.0283 .2639665 1499932 0.000000 : qR1(24) : Excited dv=7 25 24 F1e - X Xv=0 24 24 F1e
13C18O Excited 25 e 7 X 24 e 1 67549.3497 .0175446 68595.3780 1046.0283 2.647628 15054754 0.000000 : sR31(24) : Excited a3Pv=13 25 26 F3e - X Xv=0 24 24 F1e
13C18O Excited 25 e 8 X 24 e 1 67600.9703 .0172274 68646.9986 1046.0283 2.599752 14816441 0.000000 : rR21(24) : Excited dv=7 25 25 F2e - X Xv=0 24 24 F1e
13C18O Excited 25 e 9 X 24 e 1 67616.8258 .0589515 68662.8541 1046.0283 8.896258 50737022 0.000000 : rR21(24) : Excited a'v=12 25 25 F2e - X Xv=0 24 24 F1e
13C18O Excited 25 e 10 X 24 e 1 67668.0096 .0336140 68714.0379 1046.0283 5.072625 28995875 0.000000 : sR31(24) : Excited dv=7 25 26 F3e - X Xv=0 24 24 F1e
13C18O Excited 24 e 1 X 25 e 1 67241.6805 .0215844 68374.7139 1133.0334 4.943911 28861151 0.000000 : oP1(25) : Excited a3Pv=13 24 23 F1e - X Xv=0 25 25 F1e
13C18O Excited 24 e 2 X 25 e 1 67270.3539 .0160559 68403.3873 1133.0334 3.677597 21496244 0.000000 : pP(25) : Excited Av=2 24 24 F1e - X Xv=0 25 25 F1e
13C18O Excited 24 e 3 X 25 e 1 67282.1313 .0216051 68415.1647 1133.0334 4.948660 28941046 0.000000 : oP1(25) : Excited ev=4 24 23 F1e - X Xv=0 25 25 F1e

| | | | | | | | | | | |
|---|---|---|---|---|---|---|---|---|---|---|
| 13C18O Excited | 24 e 4 | X 25 e 1 | 67318.7690 | .0159180 | 68451.8024 | 1133.0334 | 3.646019 | 21357716 | 0.000000 | : pP21(25) : Excited a3Pv=13 24 24 F2e - X Xv=0 25 25 F1e |
| 13C18O Excited | 24 e 5 | X 25 e 1 | 67379.9662 | .0253446 | 68512.9996 | 1133.0334 | 5.805192 | 34098579 | 0.000000 | : qP31(25) : Excited ev=4 24 25 F3e - X Xv=0 25 25 F1e |
| 13C18O Excited | 24 e 6 | X 25 e 1 | 67392.7461 | .0002692 | 68525.7795 | 1133.0334 | .0616583 | 362374.9 | 0.000000 | : qP31(25) : Excited a3Pv=13 24 25 F3e - X Xv=0 25 25 F1e |
| 13C18O Excited | 24 e 7 | X 25 e 1 | 67398.2090 | .0067814 | 68531.2424 | 1133.0334 | 1.553273 | 9131041 | 0.000000 | : oP1(25) : Excited dv=7 24 23 F1e - X Xv=0 25 25 F1e |
| 13C18O Excited | 24 e 8 | X 25 e 1 | 67460.2622 | .0103560 | 68593.2956 | 1133.0334 | 2.372050 | 13982835 | 0.000000 | : pP21(25) : Excited dv=7 24 24 F2e - X Xv=0 25 25 F1e |
| 13C18O Excited | 24 e 9 | X 25 e 1 | 67478.2580 | .0357451 | 68611.2914 | 1133.0334 | 8.187415 | 48302071 | 0.000000 | : pP21(25) : Excited a'v=12 24 24 F2e - X Xv=0 25 25 F1e |
| 13C18O Excited | 24 e 10 | X 25 e 1 | 67525.5671 | .0209745 | 68658.6005 | 1133.0334 | 4.804219 | 28402392 | 0.000000 | : qP31(25) : Excited dv=7 24 25 F3e - X Xv=0 25 25 F1e |
| 13C18O Excited | 25 f 2 | X 25 e 1 | 67310.8893 | .0387247 | 68443.9227 | 1133.0334 | 8.869906 | 49903186 | 0.000000 | : pQ1(25) : Excited a3Pv=13 25 24 F1f - X Xv=0 25 25 F1e |
| 13C18O Excited | 25 f 3 | X 25 e 1 | 67342.0644 | .0410139 | 68475.0978 | 1133.0334 | 9.394237 | 52926609 | 0.000000 | : qQ(25) : Excited Av=2 25 25 F1f - X Xv=0 25 25 F1e |
| 13C18O Excited | 25 f 4 | X 25 e 1 | 67377.4995 | .1112783 | 68510.5329 | 1133.0334 | 25.48831 | 1.4383e8 | 0.000000 | : qQ21(25) : Excited ev=4 25 25 F2f - X Xv=0 25 25 F1e |
| 13C18O Excited | 25 f 5 | X 25 e 1 | 67390.4863 | .0067549 | 68523.5197 | 1133.0334 | 1.547203 | 8735674 | 0.000000 | : qQ21(25) : Excited a3Pv=13 25 25 F2f - X Xv=0 25 25 F1e |
| 13C18O Excited | 25 f 6 | X 25 e 1 | 67446.8410 | .0001338 | 68579.8744 | 1133.0334 | .0306578 | 173531.6 | 0.000000 | : pQ1(25) : Excited dv=7 25 24 F1f - X Xv=0 25 25 F1e |
| 13C18O Excited | 25 f 7 | X 25 e 1 | 67461.8905 | .0313133 | 68594.9239 | 1133.0334 | 7.172319 | 40624534 | 0.000000 | : rQ31(25) : Excited a3Pv=13 25 26 F3f - X Xv=0 25 25 F1e |
| 13C18O Excited | 25 f 8 | X 25 e 1 | 67479.2320 | .0393323 | 68612.2654 | 1133.0334 | 9.009076 | 51067420 | 0.000000 | : pQ1(25) : Excited a'v=12 25 24 F1f - X Xv=0 25 25 F1e |
| 13C18O Excited | 25 f 9 | X 25 e 1 | 67513.9660 | .0228876 | 68646.9994 | 1133.0334 | 5.242414 | 29762226 | 0.000000 | : qQ21(25) : Excited dv=7 25 25 F2f - X Xv=0 25 25 F1e |
| 13C18O Excited | 25 f 10 | X 25 e 1 | 67581.0037 | .0420857 | 68714.0371 | 1133.0334 | 9.639724 | 54889818 | 0.000000 | : rQ31(25) : Excited dv=7 25 26 F3f - X Xv=0 25 25 F1e |
| 13C18O Excited | 25 f 11 | X 25 e 1 | 67584.6847 | .0375724 | 68717.7181 | 1133.0334 | 8.605968 | 49011487 | 0.000000 | : rQ31(25) : Excited a'v=12 25 26 F3f - X Xv=0 25 25 F1e |
| 13C18O Excited | 26 e 1 | X 25 e 1 | 67363.8466 | .0303019 | 68496.8800 | 1133.0334 | 6.940654 | 37664185 | 0.000000 | : qR1(25) : Excited a3Pv=13 26 25 F1e - X Xv=0 25 25 F1e |
| 13C18O Excited | 26 e 3 | X 25 e 1 | 67417.2963 | .0351661 | 68550.3297 | 1133.0334 | 8.054807 | 43814386 | 0.000000 | : rR(25) : Excited Av=2 26 26 F1e - X Xv=0 25 25 F1e |
| 13C18O Excited | 26 e 4 | X 25 e 1 | 67450.6397 | .0188471 | 68583.6731 | 1133.0334 | 4.316936 | 23516971 | 0.000000 | : rR21(25) : Excited a3Pv=13 26 26 F2e - X Xv=0 25 25 F1e |
| 13C18O Excited | 26 e 5 | X 25 e 1 | 67496.8704 | .0188316 | 68629.9038 | 1133.0334 | 4.313391 | 23546007 | 0.000000 | : sR31(25) : Excited ev=4 26 27 F3e - X Xv=0 25 25 F1e |
| 13C18O Excited | 26 e 6 | X 25 e 1 | 67501.8146 | .0040085 | 68634.8480 | 1133.0334 | .9181459 | 5013091 | 0.000000 | : qR1(25) : Excited dv=7 26 25 F1e - X Xv=0 25 25 F1e |
| 13C18O Excited | 26 e 7 | X 25 e 1 | 67531.5243 | .0146118 | 68664.5577 | 1133.0334 | 3.346832 | 18297897 | 0.000000 | : sR31(25) : Excited a3Pv=13 26 27 F3e - X Xv=0 25 25 F1e |
| 13C18O Excited | 26 e 8 | X 25 e 1 | 67569.7972 | .0117609 | 68702.8305 | 1133.0334 | 2.693840 | 14752894 | 0.000000 | : rR21(25) : Excited dv=7 26 26 F2e - X Xv=0 25 25 F1e |
| 13C18O Excited | 26 e 9 | X 25 e 1 | 67583.4205 | .0404722 | 68716.4539 | 1133.0334 | 9.270165 | 50799042 | 0.000000 | : rR21(25) : Excited a'v=12 26 26 F2e - X Xv=0 25 25 F1e |
| 13C18O Excited | 26 e 10 | X 25 e 1 | 67638.5918 | .0223827 | 68771.6252 | 1133.0334 | 5.126769 | 28162747 | 0.000000 | : sR31(25) : Excited dv=7 26 27 F3e - X Xv=0 25 25 F1e |
| 13C18O Excited | 25 e 1 | X 26 e 1 | 67211.7856 | .0165287 | 68435.2776 | 1223.4920 | 5.842259 | 32724301 | 0.000000 | : oP1(26) : Excited a3Pv=13 25 24 F1e - X Xv=0 26 26 F1e |
| 13C18O Excited | 25 e 2 | X 26 e 1 | 67241.7749 | .0003090 | 68465.2669 | 1223.4920 | .1092341 | 612673.3 | 0.000000 | : oP1(26) : Excited ev=4 25 24 F1e - X Xv=0 26 26 F1e |
| 13C18O Excited | 25 e 3 | X 26 e 1 | 67254.8172 | .0229573 | 68478.3092 | 1223.4920 | 8.114517 | 45539275 | 0.000000 | : pP(26) : Excited Av=2 25 25 F1e - X Xv=0 26 26 F1e |
| 13C18O Excited | 25 e 4 | X 26 e 1 | 67293.0165 | .0108372 | 68516.5085 | 1223.4920 | 3.830533 | 21533890 | 0.000000 | : pP21(26) : Excited a3Pv=13 25 25 F2e - X Xv=0 26 26 F1e |
| 13C18O Excited | 25 e 5 | X 26 e 1 | 67347.1800 | .0142565 | 68570.6720 | 1223.4920 | 5.039127 | 28396631 | 0.000000 | : qP31(26) : Excited ev=4 25 26 F3e - X Xv=0 26 26 F1e |
| 13C18O Excited | 25 e 6 | X 26 e 1 | 67356.6448 | .0007181 | 68580.1367 | 1223.4920 | .2538140 | 1430903 | 0.000000 | : oP1(26) : Excited dv=7 25 24 F1e - X Xv=0 26 26 F1e |
| 13C18O Excited | 25 e 7 | X 26 e 1 | 67371.8861 | .0072025 | 68595.3780 | 1223.4920 | 2.545796 | 14361934 | 0.000000 | : qP31(26) : Excited a3Pv=13 25 26 F3e - X Xv=0 26 26 F1e |
| 13C18O Excited | 25 e 8 | X 26 e 1 | 67423.5066 | .0070722 | 68646.9986 | 1223.4920 | 2.499761 | 14134673 | 0.000000 | : pP21(26) : Excited dv=7 25 25 F2e - X Xv=0 26 26 F1e |
| 13C18O Excited | 25 e 9 | X 26 e 1 | 67439.3621 | .0242010 | 68662.8541 | 1223.4920 | 8.554094 | 48402485 | 0.000000 | : pP21(26) : Excited a'v=12 25 25 F2e - X Xv=0 26 26 F1e |
| 13C18O Excited | 25 e 10 | X 26 e 1 | 67490.5459 | .0137993 | 68714.0379 | 1223.4920 | 4.877524 | 27661867 | 0.000000 | : qP31(26) : Excited dv=7 25 26 F3e - X Xv=0 26 26 F1e |
| 13C18O Excited | 26 f 2 | X 26 e 1 | 67285.6334 | .0276674 | 68509.1253 | 1223.4920 | 9.779359 | 52884082 | 0.000000 | : pQ1(26) : Excited a3Pv=13 26 25 F1f - X Xv=0 26 26 F1e |
| 13C18O Excited | 26 f 3 | X 26 e 1 | 67323.8216 | .0165860 | 68547.3136 | 1223.4920 | 5.862490 | 31756741 | 0.000000 | : qQ(26) : Excited Av=2 26 26 F1f - X Xv=0 26 26 F1e |
| 13C18O Excited | 26 f 4 | X 26 e 1 | 67347.3457 | .0877366 | 68570.8376 | 1223.4920 | 31.01145 | 1.6816e8 | 0.000000 | : qQ21(26) : Excited ev=4 26 26 F2f - X Xv=0 26 26 F1e |
| 13C18O Excited | 26 f 6 | X 26 e 1 | 67410.9328 | .0005797 | 68634.4248 | 1223.4920 | .2049166 | 1114335 | 0.000000 | : pQ1(26) : Excited dv=7 26 25 F1f - X Xv=0 26 26 F1e |
| 13C18O Excited | 26 f 7 | X 26 e 1 | 67440.2938 | .0387086 | 68663.7857 | 1223.4920 | 13.68199 | 74499804 | 0.000000 | : pQ1(26) : Excited a'v=12 26 25 F1f - X Xv=0 26 26 F1e |
| 13C18O Excited | 26 f 8 | X 26 e 1 | 67440.9115 | .0101441 | 68664.4035 | 1223.4920 | 3.585554 | 19524239 | 0.000000 | : rQ31(26) : Excited a3Pv=13 26 27 F3f - X Xv=0 26 26 F1e |
| 13C18O Excited | 26 f 9 | X 26 e 1 | 67479.3398 | .0154892 | 68702.8317 | 1223.4920 | 5.474834 | 29862834 | 0.000000 | : qQ21(26) : Excited dv=7 26 26 F2f - X Xv=0 26 26 F1e |
| 13C18O Excited | 26 f 10 | X 26 e 1 | 67548.1314 | .0265694 | 68771.6234 | 1223.4920 | 9.391257 | 51382035 | 0.000000 | : rQ31(26) : Excited dv=7 26 27 F3f - X Xv=0 26 26 F1e |
| 13C18O Excited | 26 f 11 | X 26 e 1 | 67549.8551 | .0263744 | 68773.3471 | 1223.4920 | 9.322310 | 51008714 | 0.000000 | : rQ31(26) : Excited a'v=12 26 27 F3f - X Xv=0 26 26 F1e |
| 13C18O Excited | 27 e 1 | X 26 e 1 | 67335.3497 | .0209350 | 68558.8417 | 1223.4920 | 7.399706 | 38646000 | 0.000000 | : qR1(26) : Excited ev=4 27 26 F1e - X Xv=0 26 26 F1e |
| 13C18O Excited | 27 e 2 | X 26 e 1 | 67364.1174 | .0006905 | 68587.6093 | 1223.4920 | .2440739 | 1276345 | 0.000000 | : qR1(26) : Excited a3Pv=13 27 26 F1e - X Xv=0 26 26 F1e |
| 13C18O Excited | 27 e 3 | X 26 e 1 | 67402.4329 | .0214302 | 68625.9249 | 1223.4920 | 7.574730 | 39678442 | 0.000000 | : rR(26) : Excited Av=2 27 27 F1e - X Xv=0 26 26 F1e |
| 13C18O Excited | 27 e 4 | X 26 e 1 | 67429.7071 | .0138713 | 68653.1991 | 1223.4920 | 4.902963 | 25714209 | 0.000000 | : rR21(26) : Excited a3Pv=13 27 27 F2e - X Xv=0 26 26 F1e |
| 13C18O Excited | 27 e 5 | X 26 e 1 | 67466.1991 | .0032286 | 68689.6911 | 1223.4920 | 1.141180 | 5994784 | 0.000000 | : qR1(26) : Excited dv=7 27 26 F1e - X Xv=0 26 26 F1e |
| 13C18O Excited | 27 e 6 | X 26 e 1 | 67469.6665 | .0108632 | 68693.1584 | 1223.4920 | 3.839731 | 20173780 | 0.000000 | : sR31(26) : Excited ev=4 27 28 F3e - X Xv=0 26 26 F1e |
| 13C18O Excited | 27 e 7 | X 26 e 1 | 67513.1350 | .0115022 | 68736.6269 | 1223.4920 | 4.065575 | 21401662 | 0.000000 | : sR31(26) : Excited a3Pv=13 27 28 F3e - X Xv=0 26 26 F1e |
| 13C18O Excited | 27 e 8 | X 26 e 1 | 67537.3067 | .0076189 | 68760.7987 | 1223.4920 | 2.692969 | 14191337 | 0.000000 | : rR21(26) : Excited dv=7 27 27 F2e - X Xv=0 26 26 F1e |
| 13C18O Excited | 27 e 9 | X 26 e 1 | 67548.5973 | .0272859 | 68772.0893 | 1223.4920 | 9.644494 | 50849795 | 0.000000 | : rR21(26) : Excited a'v=12 27 27 F2e - X Xv=0 26 26 F1e |
| 13C18O Excited | 27 e 10 | X 26 e 1 | 67607.8701 | .0146020 | 68831.3620 | 1223.4920 | 5.161234 | 27283876 | 0.000000 | : sR31(26) : Excited dv=7 27 28 F3e - X Xv=0 26 26 F1e |
| 13C18O Excited | 26 e 1 | X 27 e 1 | 67179.4791 | .0120523 | 68496.8800 | 1317.4009 | 6.683592 | 35972235 | 0.000000 | : oP1(27) : Excited a3Pv=13 26 25 F1e - X Xv=0 27 27 F1e |
| 13C18O Excited | 26 e 3 | X 27 e 1 | 67232.9288 | .0139870 | 68550.3297 | 1317.4009 | 7.756481 | 41846429 | 0.000000 | : pP(27) : Excited Av=2 26 26 F1e - X Xv=0 27 27 F1e |
| 13C18O Excited | 26 e 4 | X 27 e 1 | 67266.2722 | .0074963 | 68583.6731 | 1317.4009 | 4.157049 | 22460780 | 0.000000 | : pP21(27) : Excited a3Pv=13 26 26 F2e - X Xv=0 27 27 F1e |
| 13C18O Excited | 26 e 5 | X 27 e 1 | 67312.5029 | .0074901 | 68629.9038 | 1317.4009 | 4.153635 | 22488638 | 0.000000 | : qP31(27) : Excited ev=4 26 27 F3e - X Xv=0 27 27 F1e |
| 13C18O Excited | 26 e 6 | X 27 e 1 | 67317.4471 | .0015943 | 68634.8480 | 1317.4009 | .8841405 | 4787974 | 0.000000 | : oP1(27) : Excited dv=7 26 25 F1e - X Xv=0 27 27 F1e |
| 13C18O Excited | 26 e 7 | X 27 e 1 | 67347.1568 | .0058117 | 68664.5577 | 1317.4009 | 3.222876 | 17476276 | 0.000000 | : qP31(27) : Excited a3Pv=13 26 27 F3e - X Xv=0 27 27 F1e |
| 13C18O Excited | 26 e 8 | X 27 e 1 | 67385.4297 | .0046778 | 68702.8305 | 1317.4009 | 2.594068 | 14090518 | 0.000000 | : pP21(27) : Excited dv=7 26 26 F2e - X Xv=0 27 27 F1e |
| 13C18O Excited | 26 e 9 | X 27 e 1 | 67399.0530 | .0160975 | 68716.4539 | 1317.4009 | 8.926825 | 48518346 | 0.000000 | : pP21(27) : Excited a'v=12 26 26 F2e - X Xv=0 27 27 F1e |
| 13C18O Excited | 26 e 10 | X 27 e 1 | 67454.2243 | .0089025 | 68771.6252 | 1317.4009 | 4.936889 | 26898519 | 0.000000 | : qP31(27) : Excited dv=7 26 27 F3e - X Xv=0 27 27 F1e |
| 13C18O Excited | 27 f 2 | X 27 e 1 | 67259.4435 | .0195526 | 68576.8444 | 1317.4009 | 10.84288 | 56437170 | 0.000000 | : pQ1(27) : Excited a3Pv=13 27 26 F1f - X Xv=0 27 27 F1e |
| 13C18O Excited | 27 f 3 | X 27 e 1 | 67302.1995 | .0001728 | 68619.6004 | 1317.4009 | .0958071 | 499627.3 | 0.000000 | : qQ(27) : Excited Av=2 27 27 F1f - X Xv=0 27 27 F1e |
| 13C18O Excited | 27 f 4 | X 27 e 1 | 67317.1076 | .0656349 | 68634.5085 | 1317.4009 | 36.39772 | 1.8994e8 | 0.000000 | : qQ21(27) : Excited ev=4 27 27 F2f - X Xv=0 27 27 F1e |
| 13C18O Excited | 27 f 5 | X 27 e 1 | 67339.8820 | .0012495 | 68657.2829 | 1317.4009 | .6929178 | 3619591 | 0.000000 | : qQ21(27) : Excited a3Pv=13 27 27 F2f - X Xv=0 27 27 F1e |
| 13C18O Excited | 27 f 6 | X 27 e 1 | 67373.5880 | .0004942 | 68690.9888 | 1317.4009 | .2740371 | 1433637 | 0.000000 | : pQ1(27) : Excited dv=7 27 26 F1f - X Xv=0 27 27 F1e |
| 13C18O Excited | 27 f 7 | X 27 e 1 | 67400.0250 | .0184140 | 68717.4259 | 1317.4009 | 10.21146 | 53484592 | 0.000000 | : pQ1(27) : Excited a'v=12 27 26 F1f - X Xv=0 27 27 F1e |
| 13C18O Excited | 27 f 8 | X 27 e 1 | 67419.1908 | .0152489 | 68736.5917 | 1317.4009 | 8.456238 | 44329061 | 0.000000 | : rQ31(27) : Excited a3Pv=13 27 28 F3f - X Xv=0 27 27 F1e |
| 13C18O Excited | 27 f 9 | X 27 e 1 | 67443.3998 | .0100083 | 68760.8007 | 1317.4009 | 5.550060 | 29125726 | 0.000000 | : qQ21(27) : Excited dv=7 27 27 F2f - X Xv=0 27 27 F1e |
| 13C18O Excited | 27 f 10 | X 27 e 1 | 67513.5902 | .0107240 | 68830.9911 | 1317.4009 | 5.946965 | 31306153 | 0.000000 | : rQ31(27) : Excited a'v=12 27 28 F3f - X Xv=0 27 27 F1e |
| 13C18O Excited | 27 f 11 | X 27 e 1 | 67513.9719 | .0238003 | 68831.3728 | 1317.4009 | 13.19841 | 69480575 | 0.000000 | : rQ31(27) : Excited dv=7 27 28 F3f - X Xv=0 27 27 F1e |
| 13C18O Excited | 28 e 1 | X 27 e 1 | 67303.6838 | .0133308 | 68621.0847 | 1317.4009 | 7.392560 | 37201457 | 0.000000 | : qR1(27) : Excited ev=4 28 27 F1e - X Xv=0 27 27 F1e |
| 13C18O Excited | 28 e 2 | X 27 e 1 | 67337.3499 | .0014798 | 68654.7508 | 1317.4009 | .8206154 | 4135769 | 0.000000 | : qR1(27) : Excited a3Pv=13 28 27 F1e - X Xv=0 27 27 F1e |
| 13C18O Excited | 28 e 3 | X 27 e 1 | 67386.9785 | .0123126 | 68704.3794 | 1317.4009 | 6.827904 | 34487677 | 0.000000 | : rR(27) : Excited Av=2 28 28 F1e - X Xv=0 27 27 F1e |
| 13C18O Excited | 28 e 4 | X 27 e 1 | 67407.4606 | .0107994 | 68724.8615 | 1317.4009 | 5.988790 | 30276912 | 0.000000 | : rR21(27) : Excited a3Pv=13 28 28 F2e - X Xv=0 27 27 F1e |
| 13C18O Excited | 28 e 5 | X 27 e 1 | 67431.6230 | .0004109 | 68749.0239 | 1317.4009 | .2278394 | 1153103 | 0.000000 | : qR1(27) : Excited dv=7 28 27 F1e - X Xv=0 27 27 F1e |
| 13C18O Excited | 28 e 6 | X 27 e 1 | 67439.1628 | .0079650 | 68756.5637 | 1317.4009 | 4.416973 | 22361957 | 0.000000 | : sR31(27) : Excited ev=4 28 29 F3e - X Xv=0 27 27 F1e |
| 13C18O Excited | 28 e 7 | X 27 e 1 | 67493.8871 | .0096095 | 68811.2880 | 1317.4009 | 5.328950 | 27044785 | 0.000000 | : sR31(27) : Excited a3Pv=13 28 29 F3e - X Xv=0 27 27 F1e |
| 13C18O Excited | 28 e 8 | X 27 e 1 | 67503.5638 | .0039782 | 68820.9647 | 1317.4009 | 2.206116 | 11201006 | 0.000000 | : rR21(27) : Excited dv=7 28 28 F2e - X Xv=0 27 27 F1e |
| 13C18O Excited | 28 e 9 | X 27 e 1 | 67512.3613 | .0179564 | 68829.7621 | 1317.4009 | 9.957682 | 50577424 | 0.000000 | : rR21(27) : Excited a'v=12 28 28 F2e - X Xv=0 27 27 F1e |
| 13C18O Excited | 28 e 10 | X 27 e 1 | 67575.8503 | .0093155 | 68893.2512 | 1317.4009 | 5.165889 | 26312870 | 0.000000 | : sR31(27) : Excited dv=7 28 29 F3e - X Xv=0 27 27 F1e |
| 13C18O Excited | 27 e 1 | X 28 e 1 | 67144.0849 | .0080668 | 68558.8417 | 1414.7569 | 7.135430 | 36949129 | 0.000000 | : oP1(28) : Excited ev=4 27 26 F1e - X Xv=0 28 28 F1e |
| 13C18O Excited | 27 e 2 | X 28 e 1 | 67172.8525 | .0002661 | 68587.6093 | 1414.7569 | .2353570 | 1220307 | 0.000000 | : oP1(28) : Excited a3Pv=13 27 26 F1e - X Xv=0 28 28 F1e |
| 13C18O Excited | 27 e 3 | X 28 e 1 | 67211.1680 | .0082577 | 68625.9249 | 1414.7569 | 7.304204 | 37936560 | 0.000000 | : pP(28) : Excited Av=2 27 27 F1e - X Xv=0 28 28 F1e |
| 13C18O Excited | 27 e 4 | X 28 e 1 | 67238.4422 | .0053450 | 68653.1991 | 1414.7569 | 4.727857 | 24585441 | 0.000000 | : pP21(28) : Excited a3Pv=13 27 27 F2e - X Xv=0 28 28 F1e |
| 13C18O Excited | 27 e 5 | X 28 e 1 | 67274.9342 | .0012441 | 68689.6911 | 1414.7569 | 1.100423 | 5731660 | 0.000000 | : oP1(28) : Excited dv=7 27 26 F1e - X Xv=0 28 28 F1e |
| 13C18O Excited | 27 e 6 | X 28 e 1 | 67278.4016 | .0041859 | 68693.1584 | 1414.7569 | 3.702598 | 19288317 | 0.000000 | : qP31(28) : Excited ev=4 27 28 F3e - X Xv=0 28 28 F1e |
| 13C18O Excited | 27 e 7 | X 28 e 1 | 67321.8701 | .0044321 | 68736.6269 | 1414.7569 | 3.920375 | 20462416 | 0.000000 | : qP31(28) : Excited a3Pv=13 27 28 F3e - X Xv=0 28 28 F1e |
| 13C18O Excited | 27 e 8 | X 28 e 1 | 67346.0418 | .0029358 | 68760.7987 | 1414.7569 | 2.596792 | 13568569 | 0.000000 | : pP21(28) : Excited dv=7 27 27 F2e - X Xv=0 28 28 F1e |

13C18O Excited 27 e 9 X 28 e 1 67357.3324 .0105140 68772.0893 1414.7569 9.300048 48618390 0.000000 : pP21(28) : Excited a'v=12 27 27 F2e - X Xv=0 28 28 F1e
13C18O Excited 27 e 10 X 28 e 1 67416.6052 .0056266 68831.3620 1414.7569 4.976904 26086792 0.000000 : qP31(28) : Excited dv=7 27 28 F3e - X Xv=0 28 28 F1e
13C18O Excited 28 f 2 X 28 e 1 67232.2985 .0137691 68647.0553 1414.7569 12.17929 61094821 0.000000 : pQ1(28) : Excited a3Pv=13 28 27 F1f - X Xv=0 28 28 F1e
13C18O Excited 28 f 3 X 28 e 1 67272.0381 .0063375 68686.7950 1414.7569 5.605796 28170181 0.000000 : qQ21(28) : Excited ev=4 28 28 F2f - X Xv=0 28 28 F1e
13C18O Excited 28 f 4 X 28 e 1 67292.9542 .0339443 68707.7111 1414.7569 30.02502 1.5102e8 0.000000 : qQ(28) : Excited Av=2 28 28 F1f - X Xv=0 28 28 F1e
13C18O Excited 28 f 5 X 28 e 1 67314.0268 .0017307 68728.7837 1414.7569 1.530850 7707224 0.000000 : qQ21(28) : Excited a3Pv=13 28 28 F2f - X Xv=0 28 28 F1e
13C18O Excited 28 f 6 X 28 e 1 67335.2557 .0002441 68750.0126 1414.7569 .2159083 1088041 0.000000 : pQ1(28) : Excited dv=7 28 27 F1f - X Xv=0 28 28 F1e
13C18O Excited 28 f 7 X 28 e 1 67358.3126 .0121571 68773.0695 1414.7569 10.75342 54246139 0.000000 : pQ1(28) : Excited a'v=12 28 27 F1f - X Xv=0 28 28 F1e
13C18O Excited 28 f 8 X 28 e 1 67396.5985 .0118113 68811.3554 1414.7569 10.44758 52793215 0.000000 : rQ31(28) : Excited a3Pv=13 28 29 F3f - X Xv=0 28 28 F1e
13C18O Excited 28 f 9 X 28 e 1 67406.2137 .0053366 68820.9706 1414.7569 4.720385 23863040 0.000000 : qQ21(28) : Excited dv=7 28 28 F2f - X Xv=0 28 28 F1e
13C18O Excited 28 f 10 X 28 e 1 67475.9248 .0100583 68890.6816 1414.7569 8.896982 45116751 0.000000 : rQ31(28) : Excited a'v=12 28 29 F3f - X Xv=0 28 28 F1e
13C18O Excited 28 f 11 X 28 e 1 67478.4966 .0120115 68893.2535 1414.7569 10.62459 53883647 0.000000 : rQ31(28) : Excited dv=7 28 29 F3f - X Xv=0 28 28 F1e
13C18O Excited 29 e 1 X 28 e 1 67269.3875 .0081558 68684.1444 1414.7569 7.214132 35019338 0.000000 : qR1(28) : Excited ev=4 29 28 F1e - X Xv=0 28 28 F1e
13C18O Excited 29 e 2 X 28 e 1 67310.9277 .0017155 68725.6846 1414.7569 1.517459 7379811 0.000000 : qR1(28) : Excited a3Pv=13 29 28 F1e - X Xv=0 28 28 F1e
13C18O Excited 29 e 3 X 28 e 1 67370.6013 .0056583 68785.3582 1414.7569 5.004961 24405256 0.000000 : rR(28) : Excited Av=2 29 29 F1e - X Xv=0 28 28 F1e
13C18O Excited 29 e 4 X 28 e 1 67383.2023 .0090234 68797.9592 1414.7569 7.981520 38941434 0.000000 : rR21(28) : Excited a3Pv=13 29 29 F2e - X Xv=0 28 28 F1e
13C18O Excited 29 e 5 X 28 e 1 67395.8910 .0008832 68810.6478 1414.7569 .7812264 3813718 0.000000 : qR1(28) : Excited dv=7 29 28 F1e - X Xv=0 28 28 F1e
13C18O Excited 29 e 6 X 28 e 1 67408.5256 .0042113 68823.2825 1414.7569 3.725054 18194850 0.000000 : sR31(28) : Excited ev=4 29 30 F3e - X Xv=0 28 28 F1e
13C18O Excited 29 e 7 X 28 e 1 67467.9757 .0057310 68882.7326 1414.7569 5.069270 24826179 0.000000 : rR21(28) : Excited dv=7 29 29 F2e - X Xv=0 28 28 F1e
13C18O Excited 29 e 8 X 28 e 1 67473.9579 .0121910 68888.7148 1414.7569 10.78335 52824312 0.000000 : sR31(28) : Excited a3Pv=13 29 30 F3e - X Xv=0 28 28 F1e
13C18O Excited 29 e 9 X 28 e 1 67475.0575 .0031676 68889.8144 1414.7569 2.801895 13726286 0.000000 : rR21(28) : Excited a'v=12 29 29 F2e - X Xv=0 28 28 F1e
13C18O Excited 29 e 10 X 28 e 1 67542.5451 .0057896 68957.3019 1414.7569 5.121110 25163313 0.000000 : sR31(28) : Excited dv=7 29 30 F3e - X Xv=0 28 28 F1e
13C18O Excited 28 e 1 X 29 e 1 67105.5282 .0049761 68621.0847 1515.5565 7.137644 35602326 0.000000 : oP1(29) : Excited ev=4 28 27 F1e - X Xv=0 29 29 F1e
13C18O Excited 28 e 2 X 29 e 1 67139.1942 .0005524 68654.7508 1515.5565 .7923183 3958008 0.000000 : oP1(29) : Excited a3Pv=13 28 27 F1e - X Xv=0 29 29 F1e
13C18O Excited 28 e 3 X 29 e 1 67188.8228 .0045960 68704.3794 1515.5565 6.592459 33005562 0.000000 : pP(29) : Excited Av=2 28 28 F1e - X Xv=0 29 29 F1e
13C18O Excited 28 e 4 X 29 e 1 67209.3049 .0040312 68724.8615 1515.5565 5.782280 28975832 0.000000 : pP21(29) : Excited a3Pv=13 28 28 F2e - X Xv=0 29 29 F1e
13C18O Excited 28 e 5 X 29 e 1 67233.4674 .0001534 68749.0239 1515.5565 .2199829 1103555 0.000000 : oP1(29) : Excited dv=7 28 27 F1e - X Xv=0 29 29 F1e
13C18O Excited 28 e 6 X 29 e 1 67241.0071 .0029732 68756.5637 1515.5565 4.264663 21401093 0.000000 : qP31(29) : Excited ev=4 28 29 F3e - X Xv=0 29 29 F1e
13C18O Excited 28 e 7 X 29 e 1 67295.7315 .0035870 68811.2880 1515.5565 5.145193 25882892 0.000000 : qP31(29) : Excited a3Pv=13 28 29 F3e - X Xv=0 29 29 F1e
13C18O Excited 28 e 8 X 29 e 1 67305.4082 .0014850 68820.9647 1515.5565 2.130043 10719804 0.000000 : pP21(29) : Excited dv=7 28 28 F2e - X Xv=0 29 29 F1e
13C18O Excited 28 e 9 X 29 e 1 67314.2056 .0067027 68829.7621 1515.5565 9.614313 48404643 0.000000 : pP21(29) : Excited a'v=12 28 28 F2e - X Xv=0 29 29 F1e
13C18O Excited 28 e 10 X 29 e 1 67377.6946 .0034773 68893.2512 1515.5565 4.987755 25182691 0.000000 : qP31(29) : Excited dv=7 28 29 F3e - X Xv=0 29 29 F1e
13C18O Excited 29 f 2 X 29 e 1 67204.1511 .0097824 68719.7076 1515.5565 14.03167 67915527 0.000000 : pQ1(29) : Excited a3Pv=13 29 28 F1f - X Xv=0 29 29 F1e
13C18O Excited 29 f 3 X 29 e 1 67236.3402 .0069684 68751.8967 1515.5565 9.995313 48448463 0.000000 : qQ21(29) : Excited ev=4 29 29 F2f - X Xv=0 29 29 F1e
13C18O Excited 29 f 4 X 29 e 1 67272.1964 .0165891 68787.7530 1515.5565 23.79518 1.1552e8 0.000000 : qQ(29) : Excited Av=2 29 29 F1f - X Xv=0 29 29 F1e
13C18O Excited 29 f 5 X 29 e 1 67285.5008 .0018033 68801.0574 1515.5565 2.586700 12565562 0.000000 : qQ21(29) : Excited a3Pv=13 29 29 F2f - X Xv=0 29 29 F1e
13C18O Excited 29 f 7 X 29 e 1 67315.2112 .0081454 68830.7678 1515.5565 11.68362 56831411 0.000000 : pQ1(29) : Excited a'v=12 29 28 F1f - X Xv=0 29 29 F1e
13C18O Excited 29 f 8 X 29 e 1 67367.1819 .0072118 68882.7385 1515.5565 10.34451 50434373 0.000000 : qQ21(29) : Excited dv=7 29 29 F2f - X Xv=0 29 29 F1e
13C18O Excited 29 f 9 X 29 e 1 67373.6495 .0042476 68889.2060 1515.5565 6.092629 29712990 0.000000 : rQ31(29) : Excited a3Pv=13 29 30 F3f - X Xv=0 29 29 F1e
13C18O Excited 29 f 10 X 29 e 1 67436.8296 .0066075 68952.3862 1515.5565 9.477706 46351744 0.000000 : rQ31(29) : Excited a'v=12 29 30 F3f - X Xv=0 29 29 F1e
13C18O Excited 29 f 11 X 29 e 1 67441.7475 .0071986 68957.3040 1515.5565 10.32550 50509038 0.000000 : rQ31(29) : Excited dv=7 29 30 F3f - X Xv=0 29 29 F1e
13C18O Excited 30 e 1 X 29 e 1 67233.0110 .0049203 68748.5675 1515.5565 7.057593 33082465 0.000000 : qR1(29) : Excited ev=4 30 29 F1e - X Xv=0 29 29 F1e
13C18O Excited 30 e 2 X 29 e 1 67284.3287 .0015148 68799.8852 1515.5565 2.172751 10208109 0.000000 : qR1(29) : Excited a3Pv=13 30 29 F1e - X Xv=0 29 29 F1e
13C18O Excited 30 e 4 X 29 e 1 67355.4123 .0044065 68870.9688 1515.5565 6.320610 29789972 0.000000 : rR(29) : Excited Av=2 30 30 F1e - X Xv=0 29 29 F1e
13C18O Excited 30 e 5 X 29 e 1 67362.9643 .0061076 68878.5208 1515.5565 8.760708 41304406 0.000000 : rR21(29) : Excited a3Pv=13 30 30 F2e - X Xv=0 29 29 F1e
13C18O Excited 30 e 6 X 29 e 1 67378.2078 .0016436 68893.7644 1515.5565 2.357583 11122924 0.000000 : sR31(29) : Excited ev=4 30 31 F3e - X Xv=0 29 29 F1e
13C18O Excited 30 e 7 X 29 e 1 67431.6803 .0025783 68947.2368 1515.5565 3.698324 17490124 0.000000 : rR21(29) : Excited dv=7 30 30 F2e - X Xv=0 29 29 F1e
13C18O Excited 30 e 8 X 29 e 1 67435.6091 .0084726 68951.1656 1515.5565 12.15300 57483713 0.000000 : rR21(29) : Excited a'v=12 30 30 F2e - X Xv=0 29 29 F1e
13C18O Excited 30 e 9 X 29 e 1 67453.2469 .0028939 68968.8034 1515.5565 4.150989 19649602 0.000000 : sR31(29) : Excited a3Pv=13 30 31 F3e - X Xv=0 29 29 F1e
13C18O Excited 30 e 10 X 29 e 1 67507.9821 .0034753 69023.5387 1515.5565 4.984930 23654734 0.000000 : sR31(29) : Excited dv=7 30 31 F3e - X Xv=0 29 29 F1e
13C18O Excited 29 e 1 X 30 e 1 67064.3480 .0029490 68684.1444 1619.7964 6.973661 33543423 0.000000 : oP1(30) : Excited ev=4 29 28 F1e - X Xv=0 30 30 F1e
13C18O Excited 29 e 2 X 30 e 1 67105.8882 .0006203 68725.6846 1619.7964 1.466877 7068823 0.000000 : oP1(30) : Excited a3Pv=13 29 28 F1e - X Xv=0 30 30 F1e
13C18O Excited 29 e 3 X 30 e 1 67165.5618 .0020460 68785.3582 1619.7964 4.838129 23377001 0.000000 : pP(30) : Excited Av=2 29 29 F1e - X Xv=0 30 30 F1e
13C18O Excited 29 e 4 X 30 e 1 67178.1628 .0032627 68797.9592 1619.7964 7.715469 37300797 0.000000 : pP21(30) : Excited a3Pv=13 29 29 F2e - X Xv=0 30 30 F1e
13C18O Excited 29 e 5 X 30 e 1 67190.8515 .0003194 68810.6478 1619.7964 .7551855 3653049 0.000000 : oP1(30) : Excited dv=7 29 28 F1e - X Xv=0 30 30 F1e
13C18O Excited 29 e 6 X 30 e 1 67203.4861 .0015227 68823.2825 1619.7964 3.600885 17428344 0.000000 : qP31(30) : Excited ev=4 29 30 F3e - X Xv=0 30 30 F1e
13C18O Excited 29 e 7 X 30 e 1 67262.9362 .0020722 68882.7326 1619.7964 4.900294 23780504 0.000000 : pP21(30) : Excited dv=7 29 29 F2e - X Xv=0 30 30 F1e
13C18O Excited 29 e 8 X 30 e 1 67268.9184 .0044081 68888.7148 1619.7964 10.42391 50599401 0.000000 : qP31(30) : Excited a3Pv=13 29 30 F3e - X Xv=0 30 30 F1e
13C18O Excited 29 e 9 X 30 e 1 67270.0180 .0011454 68889.8144 1619.7964 2.708498 13148149 0.000000 : pP21(30) : Excited a'v=12 29 29 F2e - X Xv=0 30 30 F1e
13C18O Excited 29 e 10 X 30 e 1 67337.5055 .0020934 68957.3019 1619.7964 4.950407 24103681 0.000000 : qP31(30) : Excited dv=7 29 30 F3e - X Xv=0 30 30 F1e
13C18O Excited 30 f 2 X 30 e 1 67174.8790 .0071843 68794.6754 1619.7964 16.98885 79428843 0.000000 : pQ1(30) : Excited a3Pv=13 30 29 F1f - X Xv=0 30 30 F1e
13C18O Excited 30 f 3 X 30 e 1 67198.5348 .0043917 68818.3311 1619.7964 10.38509 48605242 0.000000 : qQ21(30) : Excited ev=4 30 30 F2f - X Xv=0 30 30 F1e
13C18O Excited 30 f 5 X 30 e 1 67251.9040 .0090526 68871.7004 1619.7964 21.40705 1.0043e8 0.000000 : qQ(30) : Excited Av=2 30 30 F1f - X Xv=0 30 30 F1e
13C18O Excited 30 f 6 X 30 e 1 67265.6476 .0002502 68885.4440 1619.7964 .5917392 2777819 0.000000 : qQ21(30) : Excited a3Pv=13 30 30 F2f - X Xv=0 30 30 F1e
13C18O Excited 30 f 7 X 30 e 1 67270.8246 .0060716 68890.6210 1619.7964 14.35766 67415156 0.000000 : pQ1(30) : Excited a'v=12 30 29 F1f - X Xv=0 30 30 F1e
13C18O Excited 30 f 8 X 30 e 1 67327.4496 .0036325 68947.2460 1619.7964 8.589966 40435382 0.000000 : qQ21(30) : Excited dv=7 30 30 F2f - X Xv=0 30 30 F1e
13C18O Excited 30 f 9 X 30 e 1 67349.1655 .0039681 68968.9619 1619.7964 9.383508 44213562 0.000000 : rQ31(30) : Excited a3Pv=13 30 31 F3f - X Xv=0 30 30 F1e
13C18O Excited 30 f 10 X 30 e 1 67396.3126 .0041966 69016.1090 1619.7964 9.923710 46857172 0.000000 : rQ31(30) : Excited a'v=12 30 31 F3f - X Xv=0 30 30 F1e
13C18O Excited 30 f 11 X 30 e 1 67403.7451 .0042206 69023.5415 1619.7964 9.980616 47141459 0.000000 : rQ31(30) : Excited dv=7 30 31 F3f - X Xv=0 30 30 F1e
13C18O Excited 31 e 1 X 30 e 1 67194.8908 .0029461 68814.6872 1619.7964 6.966627 31565610 0.000000 : qR1(30) : Excited ev=4 31 30 F1e - X Xv=0 30 30 F1e
13C18O Excited 31 e 2 X 30 e 1 67257.2166 .0011686 68877.0130 1619.7964 2.763401 12555775 0.000000 : qR1(30) : Excited a3Pv=13 31 30 F1e - X Xv=0 30 30 F1e
13C18O Excited 31 e 3 X 30 e 1 67314.4096 .0003937 68934.2060 1619.7964 .9310754 4241230 0.000000 : qR1(30) : Excited dv=7 31 30 F1e - X Xv=0 30 30 F1e
13C18O Excited 31 e 4 X 30 e 1 67328.2461 .0005767 68948.0425 1619.7964 1.363700 6215753 0.000000 : sR31(30) : Excited ev=4 31 32 F3e - X Xv=0 30 30 F1e
13C18O Excited 31 e 5 X 30 e 1 67339.3907 .0058303 68959.1871 1619.7964 13.78720 62873315 0.000000 : rR(30) : Excited Av=2 31 31 F1e - X Xv=0 30 30 F1e
13C18O Excited 31 e 6 X 30 e 1 67349.3961 .0004534 68969.1925 1619.7964 1.072281 4892070 0.000000 : rR21(30) : Excited a3Pv=13 31 31 F2e - X Xv=0 30 30 F1e
13C18O Excited 31 e 7 X 30 e 1 67393.8988 .0007260 69013.6952 1619.7964 1.716737 7847808 0.000000 : rR21(30) : Excited dv=7 31 31 F2e - X Xv=0 30 30 F1e
13C18O Excited 31 e 8 X 30 e 1 67395.2069 .0062519 69015.0033 1619.7964 14.78402 67586876 0.000000 : rR21(30) : Excited a'v=12 31 31 F2e - X Xv=0 30 30 F1e
13C18O Excited 31 e 9 X 30 e 1 67431.2998 .0022388 69051.0962 1619.7964 5.294177 24241861 0.000000 : sR31(30) : Excited a3Pv=13 31 32 F3e - X Xv=0 30 30 F1e
13C18O Excited 31 e 10 X 30 e 1 67472.2309 .0019681 69092.0273 1619.7964 4.654091 21349758 0.000000 : sR31(30) : Excited dv=7 31 32 F3e - X Xv=0 30 30 F1e
13C18O Excited 30 e 1 X 31 e 1 67021.0947 .0017233 68748.5675 1727.4728 6.829928 31713509 0.000000 : oP1(31) : Excited ev=4 30 29 F1e - X Xv=0 31 31 F1e
13C18O Excited 30 e 2 X 31 e 1 67072.4124 .0005305 68799.8852 1727.4728 2.102662 9785767 0.000000 : oP1(31) : Excited a3Pv=13 30 29 F1e - X Xv=0 31 31 F1e
13C18O Excited 30 e 4 X 31 e 1 67143.4960 .0015433 68870.9688 1727.4728 6.116719 28557752 0.000000 : pP(31) : Excited Av=2 30 30 F1e - X Xv=0 31 31 F1e
13C18O Excited 30 e 5 X 31 e 1 67151.0480 .0021392 68878.5208 1727.4728 8.478105 39595949 0.000000 : pP21(31) : Excited a3Pv=13 30 30 F2e - X Xv=0 31 31 F1e
13C18O Excited 30 e 6 X 31 e 1 67166.2916 .0005757 68893.7644 1727.4728 2.281532 10662874 0.000000 : qP31(31) : Excited ev=4 30 31 F3e - X Xv=0 31 31 F1e
13C18O Excited 30 e 7 X 31 e 1 67219.7640 .0009031 68947.2368 1727.4728 3.579043 16766849 0.000000 : pP21(31) : Excited dv=7 30 30 F2e - X Xv=0 31 31 F1e
13C18O Excited 30 e 8 X 31 e 1 67223.6928 .0029675 68951.1656 1727.4728 11.76096 55106600 0.000000 : pP21(31) : Excited a'v=12 30 30 F2e - X Xv=0 31 31 F1e
13C18O Excited 30 e 9 X 31 e 1 67241.3306 .0010136 68968.8034 1727.4728 4.017086 18837083 0.000000 : qP31(31) : Excited a3Pv=13 30 31 F3e - X Xv=0 31 31 F1e
13C18O Excited 30 e 10 X 31 e 1 67296.0659 .0012172 69023.5387 1727.4728 4.824126 22676774 0.000000 : qP31(31) : Excited dv=7 30 31 F3e - X Xv=0 31 31 F1e
13C18O Excited 31 f 2 X 31 e 1 67144.0705 .0057010 68871.5433 1727.4728 22.59480 1.0214e8 0.000000 : pQ1(31) : Excited a3Pv=13 31 30 F1f - X Xv=0 31 31 F1e
13C18O Excited 31 f 3 X 31 e 1 67159.7064 .0017960 68887.1792 1727.4728 7.118003 32200855 0.000000 : qQ21(31) : Excited ev=4 31 31 F2f - X Xv=0 31 31 F1e
13C18O Excited 31 f 4 X 31 e 1 67207.0598 .0001809 68934.5326 1727.4728 .7170919 3250891 0.000000 : pQ1(31) : Excited dv=7 31 30 F1f - X Xv=0 31 31 F1e
13C18O Excited 31 f 5 X 31 e 1 67223.7141 .0005821 68951.1869 1727.4728 2.306898 10465954 0.000000 : pQ1(31) : Excited a'v=12 31 30 F1f - X Xv=0 31 31 F1e

13C18O Excited 31 f 6 X 31 e 1 67231.8857 .0080157 68959.3585 1727.4728 31.76824 1.4418e8 0.000000 : qQ(31) : Excited Av=2 31 31 F1f - X Xv=0 31 31 F1e
13C18O Excited 31 f 7 X 31 e 1 67236.9592 .0001964 68964.4320 1727.4728 .7783627 3533370 0.000000 : qQ21(31) : Excited a3Pv=13 31 31 F2f - X Xv=0 31 31 F1e
13C18O Excited 31 f 8 X 31 e 1 67286.2835 .0022258 69013.7563 1727.4728 8.821300 40132396 0.000000 : qQ21(31) : Excited dv=7 31 31 F2f - X Xv=0 31 31 F1e
13C18O Excited 31 f 9 X 31 e 1 67323.8267 .0028489 69051.2996 1727.4728 11.29114 51454950 0.000000 : rQ31(31) : Excited a3Pv=13 31 32 F3f - X Xv=0 31 31 F1e
13C18O Excited 31 f 10 X 31 e 1 67354.3746 .0026041 69081.8474 1727.4728 10.32084 47097246 0.000000 : rQ31(31) : Excited a'v=12 31 32 F3f - X Xv=0 31 31 F1e
13C18O Excited 31 f 11 X 31 e 1 67364.5598 .0023423 69092.0326 1727.4728 9.283081 42380831 0.000000 : rQ31(31) : Excited dv=7 31 32 F3f - X Xv=0 31 31 F1e
13C18O Excited 32 e 1 X 31 e 1 67155.2079 .0017497 68882.6807 1727.4728 6.934649 30400003 0.000000 : qR1(31) : Excited ev=4 32 31 F1e - X Xv=0 31 31 F1e
13C18O Excited 32 e 2 X 31 e 1 67229.3865 .0008373 68956.8594 1727.4728 3.318328 14595092 0.000000 : qR1(31) : Excited a3Pv=13 32 31 F1e - X Xv=0 31 31 F1e
13C18O Excited 32 e 3 X 31 e 1 67274.1489 .0001691 69001.6217 1727.4728 .6702509 2953875 0.000000 : qR1(31) : Excited dv=7 32 31 F1e - X Xv=0 31 31 F1e
13C18O Excited 32 e 4 X 31 e 1 67295.1392 .0011516 69022.6120 1727.4728 4.564262 20134076 0.000000 : sR31(31) : Excited ev=4 32 33 F3e - X Xv=0 31 31 F1e
13C18O Excited 32 e 5 X 31 e 1 67319.5422 .0019666 69047.0151 1727.4728 7.794225 34419651 0.000000 : rR(31) : Excited Av=2 32 32 F1e - X Xv=0 31 31 F1e
13C18O Excited 32 e 6 X 31 e 1 67322.6876 .0007629 69050.1604 1727.4728 3.023729 13354795 0.000000 : rR21(31) : Excited a3Pv=13 32 32 F2e - X Xv=0 31 31 F1e
13C18O Excited 32 e 7 X 31 e 1 67353.2371 .0022134 69080.7099 1727.4728 8.772164 38796473 0.000000 : rR21(31) : Excited a'v=12 32 32 F2e - X Xv=0 31 31 F1e
13C18O Excited 32 e 8 X 31 e 1 67355.1270 .0023667 69082.5998 1727.4728 9.379756 41487148 0.000000 : rR21(31) : Excited dv=7 32 32 F2e - X Xv=0 31 31 F1e
13C18O Excited 32 e 9 X 31 e 1 67408.1522 .0016955 69135.6250 1727.4728 6.719713 29791893 0.000000 : sR31(31) : Excited a3Pv=13 32 33 F3e - X Xv=0 31 31 F1e
13C18O Excited 32 e 10 X 31 e 1 67435.5109 .0009646 69162.9837 1727.4728 3.822884 16969427 0.000000 : sR31(31) : Excited dv=7 32 33 F3e - X Xv=0 31 31 F1e
13C18O Excited 31 e 1 X 32 e 1 66976.1051 .0009994 68814.6872 1838.5820 6.748920 30281460 0.000000 : oP1(32) : Excited ev=4 31 30 F1e - X Xv=0 32 32 F1e
13C18O Excited 31 e 2 X 32 e 1 67038.4309 .0003964 68877.0130 1838.5820 2.677045 12045091 0.000000 : oP1(32) : Excited a3Pv=13 31 30 F1e - X Xv=0 32 32 F1e
13C18O Excited 31 e 3 X 32 e 1 67095.6239 .0001336 68934.2060 1838.5820 .9019793 4068760 0.000000 : oP1(32) : Excited dv=7 31 30 F1e - X Xv=0 32 32 F1e
13C18O Excited 31 e 4 X 32 e 1 67109.4604 .0001956 68948.0425 1838.5820 1.321085 5963000 0.000000 : qP31(32) : Excited ev=4 31 32 F3e - X Xv=0 32 32 F1e
13C18O Excited 31 e 5 X 32 e 1 67120.6051 .0019779 68959.1871 1838.5820 13.35635 60316775 0.000000 : pP(32) : Excited Av=2 31 31 F1e - X Xv=0 32 32 F1e
13C18O Excited 31 e 6 X 32 e 1 67130.6104 .0001538 68969.1925 1838.5820 1.038772 4693157 0.000000 : pP21(32) : Excited a3Pv=13 31 31 F2e - X Xv=0 32 32 F1e
13C18O Excited 31 e 7 X 32 e 1 67175.1132 .0002463 69013.6952 1838.5820 1.663089 7528762 0.000000 : pP21(32) : Excited dv=7 31 31 F2e - X Xv=0 32 32 F1e
13C18O Excited 31 e 8 X 32 e 1 67176.4213 .0021209 69015.0033 1838.5820 14.32202 64839200 0.000000 : pP21(32) : Excited a'v=12 31 31 F2e - X Xv=0 32 32 F1e
13C18O Excited 31 e 9 X 32 e 1 67212.5142 .0007595 69051.0962 1838.5820 5.128734 23256455 0.000000 : qP31(32) : Excited a3Pv=13 31 32 F3e - X Xv=0 32 32 F1e
13C18O Excited 31 e 10 X 32 e 1 67253.4452 .0006677 69092.0273 1838.5820 4.508651 20482034 0.000000 : qP31(32) : Excited dv=7 31 32 F3e - X Xv=0 32 32 F1e
13C18O Excited 32 f 2 X 32 e 1 67110.0910 .0046108 68948.6730 1838.5820 31.13514 1.3621e8 0.000000 : pQ1(32) : Excited a3Pv=13 32 31 F1f - X Xv=0 32 32 F1e
13C18O Excited 32 f 4 X 32 e 1 67163.1357 .0001013 69001.7177 1838.5820 .6843133 3000944 0.000000 : pQ1(32) : Excited dv=7 32 31 F1f - X Xv=0 32 32 F1e
13C18O Excited 32 f 5 X 32 e 1 67176.9535 .0012352 69015.5355 1838.5820 8.340675 36599251 0.000000 : pQ1(32) : Excited a'v=12 32 31 F1f - X Xv=0 32 32 F1e
13C18O Excited 32 f 6 X 32 e 1 67208.0269 .0008439 69046.6090 1838.5820 5.698567 25040279 0.000000 : qQ21(32) : Excited a3Pv=13 32 32 F2f - X Xv=0 32 32 F1e
13C18O Excited 32 f 7 X 32 e 1 67209.8059 .0029254 69048.3880 1838.5820 19.75433 86810098 0.000000 : qQ(32) : Excited Av=2 32 32 F1f - X Xv=0 32 32 F1e
13C18O Excited 32 f 8 X 32 e 1 67243.8956 .0014449 69082.4776 1838.5820 9.756851 42941613 0.000000 : qQ21(32) : Excited dv=7 32 32 F2f - X Xv=0 32 32 F1e
13C18O Excited 32 f 9 X 32 e 1 67297.2714 .0021029 69135.8535 1838.5820 14.20013 62646186 0.000000 : rQ31(32) : Excited a3Pv=13 32 33 F3f - X Xv=0 32 32 F1e
13C18O Excited 32 f 10 X 32 e 1 67311.0177 .0015676 69149.5997 1838.5820 10.58526 46727235 0.000000 : rQ31(32) : Excited a'v=12 32 33 F3f - X Xv=0 32 32 F1e
13C18O Excited 32 f 11 X 32 e 1 67324.4157 .0011239 69162.9977 1838.5820 7.589656 33523547 0.000000 : rQ31(32) : Excited dv=7 32 33 F3f - X Xv=0 32 32 F1e
13C18O Excited 33 e 1 X 32 e 1 67114.0605 .0010288 68952.6425 1838.5820 6.947394 29492468 0.000000 : qR1(32) : Excited ev=4 33 32 F1e - X Xv=0 32 32 F1e
13C18O Excited 33 e 2 X 32 e 1 67200.6606 .0005762 69039.2426 1838.5820 3.890917 16581403 0.000000 : qR1(32) : Excited a3Pv=13 33 32 F1e - X Xv=0 32 32 F1e
13C18O Excited 33 e 4 X 32 e 1 67259.1811 .0008552 69097.7631 1838.5820 5.774647 24673390 0.000000 : sR31(32) : Excited ev=4 33 34 F3e - X Xv=0 32 32 F1e
13C18O Excited 33 e 5 X 32 e 1 67294.0244 .0002352 69132.6065 1838.5820 1.588349 6797106 0.000000 : rR21(32) : Excited a3Pv=13 33 33 F2e - X Xv=0 32 32 F1e
13C18O Excited 33 e 6 X 32 e 1 67300.6823 .0004217 69139.2644 1838.5820 2.847381 12188568 0.000000 : rR(32) : Excited Av=2 33 33 F1e - X Xv=0 32 32 F1e
13C18O Excited 33 e 7 X 32 e 1 67310.5903 .0017688 69149.1723 1838.5820 11.94450 51152498 0.000000 : rR21(32) : Excited a'v=12 33 33 F2e - X Xv=0 32 32 F1e
13C18O Excited 33 e 8 X 32 e 1 67315.5901 .0017928 69154.1722 1838.5820 12.10640 51857392 0.000000 : rR21(32) : Excited dv=7 33 33 F2e - X Xv=0 32 32 F1e
13C18O Excited 33 e 9 X 32 e 1 67382.7737 .0014109 69221.3558 1838.5820 9.527190 40931732 0.000000 : sR31(32) : Excited a3Pv=13 33 34 F3e - X Xv=0 32 32 F1e
13C18O Excited 33 e 10 X 32 e 1 67398.8332 .0002355 69237.4152 1838.5820 1.590360 6837563 0.000000 : sR31(32) : Excited dv=7 33 34 F3e - X Xv=0 32 32 F1e
13C18O Excited 32 e 1 X 33 e 1 66929.5605 .0005749 68882.6807 1953.1202 6.724508 29182635 0.000000 : oP1(33) : Excited ev=4 32 31 F1e - X Xv=0 33 33 F1e
13C18O Excited 32 e 2 X 33 e 1 67003.7391 .0002751 68956.8594 1953.1202 3.217773 14010787 0.000000 : oP1(33) : Excited a3Pv=13 32 31 F1e - X Xv=0 33 33 F1e
13C18O Excited 32 e 4 X 33 e 1 67069.4918 .0003784 69022.6120 1953.1202 4.425951 19328213 0.000000 : qP31(33) : Excited ev=4 32 33 F3e - X Xv=0 33 33 F1e
13C18O Excited 32 e 5 X 33 e 1 67093.8948 .0006462 69047.0151 1953.1202 7.558036 33042131 0.000000 : pP(33) : Excited Av=2 32 32 F1e - X Xv=0 33 33 F1e
13C18O Excited 32 e 6 X 33 e 1 67097.0402 .0002507 69050.1604 1953.1202 2.932100 12820324 0.000000 : pP21(33) : Excited a3Pv=13 32 32 F2e - X Xv=0 33 33 F1e
13C18O Excited 32 e 7 X 33 e 1 67127.5897 .0007273 69080.7099 1953.1202 8.506341 37243975 0.000000 : pP21(33) : Excited a'v=12 32 32 F2e - X Xv=0 33 33 F1e
13C18O Excited 32 e 8 X 33 e 1 67129.4796 .0007777 69082.5998 1953.1202 9.095521 39826990 0.000000 : pP21(33) : Excited dv=7 32 32 F2e - X Xv=0 33 33 F1e
13C18O Excited 32 e 9 X 33 e 1 67182.5048 .0005571 69135.6250 1953.1202 6.516086 28599961 0.000000 : qP31(33) : Excited a3Pv=13 32 33 F3e - X Xv=0 33 33 F1e
13C18O Excited 32 e 10 X 33 e 1 67209.8634 .0003169 69162.9837 1953.1202 3.707039 16290570 0.000000 : qP31(33) : Excited dv=7 32 33 F3e - X Xv=0 33 33 F1e
13C18O Excited 33 f 2 X 33 e 1 67070.6644 .0027050 69023.7846 1953.1202 31.63795 1.3405e8 0.000000 : qQ21(33) : Excited ev=4 33 33 F2f - X Xv=0 33 33 F1e
13C18O Excited 33 f 3 X 33 e 1 67086.5939 .0001810 69039.7142 1953.1202 2.117095 8976276 0.000000 : pQ1(33) : Excited a3Pv=13 33 32 F1f - X Xv=0 33 33 F1e
13C18O Excited 33 f 5 X 33 e 1 67128.2400 .0008112 69081.3603 1953.1202 9.488144 40303772 0.000000 : pQ1(33) : Excited a'v=12 33 32 F1f - X Xv=0 33 33 F1e
13C18O Excited 33 f 6 X 33 e 1 67178.4412 .0002734 69131.5615 1953.1202 3.197939 13614715 0.000000 : qQ21(33) : Excited a3Pv=13 33 33 F2f - X Xv=0 33 33 F1e
13C18O Excited 33 f 7 X 33 e 1 67187.0463 .0014392 69140.1665 1953.1202 16.83310 71691800 0.000000 : qQ(33) : Excited Av=2 33 33 F1f - X Xv=0 33 33 F1e
13C18O Excited 33 f 8 X 33 e 1 67200.7746 .0011919 69153.8948 1953.1202 13.94045 59408466 0.000000 : qQ21(33) : Excited dv=7 33 33 F2f - X Xv=0 33 33 F1e
13C18O Excited 33 f 9 X 33 e 1 67266.2238 .0011971 69219.3441 1953.1202 14.00169 59843953 0.000000 : rQ31(33) : Excited a'v=12 33 34 F3f - X Xv=0 33 33 F1e
13C18O Excited 33 f 10 X 33 e 1 67268.4657 .0014455 69221.5859 1953.1202 16.90726 72269726 0.000000 : rQ31(33) : Excited a3Pv=13 33 34 F3f - X Xv=0 33 33 F1e
13C18O Excited 33 f 11 X 33 e 1 67284.3548 .0002624 69237.4751 1953.1202 3.069627 13130355 0.000000 : rQ31(33) : Excited dv=7 33 34 F3f - X Xv=0 33 33 F1e
13C18O Excited 34 e 1 X 33 e 1 67071.5040 .0005979 69024.6243 1953.1202 6.992856 28770212 0.000000 : qR1(33) : Excited ev=4 34 33 F1e - X Xv=0 33 33 F1e
13C18O Excited 34 e 2 X 33 e 1 67170.6533 .0003905 69123.7736 1953.1202 4.567164 18873815 0.000000 : qR1(33) : Excited a3Pv=13 34 33 F1e - X Xv=0 33 33 F1e
13C18O Excited 34 e 4 X 33 e 1 67221.3017 .0005456 69174.4220 1953.1202 6.381762 26432346 0.000000 : sR31(33) : Excited ev=4 34 35 F3e - X Xv=0 33 33 F1e
13C18O Excited 34 e 5 X 33 e 1 67263.8143 .0008132 69216.9346 1953.1202 9.511418 39469723 0.000000 : rR21(33) : Excited a'v=12 34 34 F2e - X Xv=0 33 33 F1e
13C18O Excited 34 e 8 X 33 e 1 67285.0860 .0016211 69238.2062 1953.1202 18.96099 78757459 0.000000 : rR(33) : Excited Av=2 34 34 F1e - X Xv=0 33 33 F1e
13C18O Excited 34 e 9 X 33 e 1 67350.8680 .0009879 69303.9882 1953.1202 11.55482 48135693 0.000000 : sR31(33) : Excited dv=7 34 35 F3e - X Xv=0 33 33 F1e
13C18O Excited 33 e 1 X 34 e 1 66881.5591 .0003274 68952.6425 2071.0834 6.743059 28328577 0.000000 : oP1(34) : Excited ev=4 33 32 F1e - X Xv=0 34 34 F1e
13C18O Excited 33 e 2 X 34 e 1 66968.1592 .0001834 69039.2426 2071.0834 3.776479 15927248 0.000000 : oP1(34) : Excited a3Pv=13 33 32 F1e - X Xv=0 34 34 F1e
13C18O Excited 33 e 4 X 34 e 1 67026.6797 .0002722 69097.7631 2071.0834 5.604805 23700212 0.000000 : qP31(34) : Excited ev=4 33 34 F3e - X Xv=0 34 34 F1e
13C18O Excited 33 e 6 X 34 e 1 67068.1810 .0001342 69139.2644 2071.0834 2.763635 11707896 0.000000 : pP(34) : Excited Av=2 33 33 F1e - X Xv=0 34 34 F1e
13C18O Excited 33 e 7 X 34 e 1 67078.0889 .0005629 69149.1723 2071.0834 11.59319 49135312 0.000000 : pP21(34) : Excited a'v=12 33 33 F2e - X Xv=0 34 34 F1e
13C18O Excited 33 e 8 X 34 e 1 67083.0887 .0005706 69154.1722 2071.0834 11.75033 49812447 0.000000 : pP21(34) : Excited dv=7 33 33 F2e - X Xv=0 34 34 F1e
13C18O Excited 33 e 9 X 34 e 1 67150.2723 .0004490 69221.3558 2071.0834 9.246979 39318037 0.000000 : qP31(34) : Excited a3Pv=13 33 34 F3e - X Xv=0 34 34 F1e
13C18O Excited 34 f 2 X 34 e 1 67027.6741 .0014376 69098.7575 2071.0834 29.60679 1.2157e8 0.000000 : qQ21(34) : Excited ev=4 34 34 F2f - X Xv=0 34 34 F1e
13C18O Excited 34 f 3 X 34 e 1 67052.3972 .0003207 69123.4807 2071.0834 6.604021 27147244 0.000000 : pQ1(34) : Excited a3Pv=13 34 33 F1f - X Xv=0 34 34 F1e
13C18O Excited 34 f 5 X 34 e 1 67078.0983 .0004730 69149.1817 2071.0834 9.741191 40089313 0.000000 : pQ1(34) : Excited a'v=12 34 33 F1f - X Xv=0 34 34 F1e
13C18O Excited 34 f 6 X 34 e 1 67146.9245 .0002427 69218.0079 2071.0834 4.999009 20636528 0.000000 : qQ21(34) : Excited a3Pv=13 34 34 F2f - X Xv=0 34 34 F1e
13C18O Excited 34 f 8 X 34 e 1 67166.3243 .0013572 69237.4077 2071.0834 27.95002 1.1548e8 0.000000 : qQ(34) : Excited Av=2 34 34 F1f - X Xv=0 34 34 F1e
13C18O Excited 34 f 9 X 34 e 1 67220.0532 .0006143 69291.1366 2071.0834 12.65186 52399261 0.000000 : rQ31(34) : Excited a'v=12 34 35 F3f - X Xv=0 34 34 F1e
13C18O Excited 34 f 10 X 34 e 1 67232.9949 .0011091 69304.0784 2071.0834 22.84168 94656354 0.000000 : rQ31(34) : Excited dv=7 34 35 F3f - X Xv=0 34 34 F1e
13C18O Excited 35 e 1 X 34 e 1 67027.5717 .0003429 69098.6551 2071.0834 7.062311 28182036 0.000000 : qR1(34) : Excited ev=4 35 34 F1e - X Xv=0 34 34 F1e
13C18O Excited 35 e 2 X 34 e 1 67137.0763 .0002242 69208.1597 2071.0834 4.617071 18514794 0.000000 : qR1(34) : Excited a3Pv=13 35 34 F1e - X Xv=0 34 34 F1e
13C18O Excited 35 e 4 X 34 e 1 67181.8286 .0003318 69252.9120 2071.0834 6.832094 27452024 0.000000 : sR31(34) : Excited ev=4 35 36 F3e - X Xv=0 34 34 F1e
13C18O Excited 35 e 5 X 34 e 1 67218.2972 .0005117 69289.3806 2071.0834 10.53721 42408547 0.000000 : rR21(34) : Excited a'v=12 35 35 F2e - X Xv=0 34 34 F1e
13C18O Excited 35 e 6 X 34 e 1 67227.4253 .0001342 69298.5087 2071.0834 2.762793 11123791 0.000000 : rR21(34) : Excited dv=7 35 35 F2e - X Xv=0 34 34 F1e
13C18O Excited 35 e 8 X 34 e 1 67264.0243 .0007247 69335.1077 2071.0834 14.92419 60187254 0.000000 : rR(34) : Excited Av=2 35 35 F1e - X Xv=0 34 34 F1e
13C18O Excited 35 e 9 X 34 e 1 67311.8189 .0005202 69382.9024 2071.0834 10.71226 43293255 0.000000 : sR31(34) : Excited dv=7 35 36 F3e - X Xv=0 34 34 F1e
13C18O Excited 34 e 1 X 35 e 1 66832.1568 .0001843 69024.6243 2192.4675 6.793060 27650070 0.000000 : oP1(35) : Excited ev=4 34 33 F1e - X Xv=0 35 35 F1e
13C18O Excited 34 e 2 X 35 e 1 66931.3061 .0001204 69123.7736 2192.4675 4.436674 18139268 0.000000 : oP1(35) : Excited a3Pv=13 34 33 F1e - X Xv=0 35 35 F1e
13C18O Excited 34 e 4 X 35 e 1 66981.9545 .0001682 69174.4220 2192.4675 6.199426 25403835 0.000000 : qP31(35) : Excited ev=4 34 35 F3e - X Xv=0 35 35 F1e

```
13C18O Excited  34  e  5      X  35  e  1    67024.4671 .0002507    69216.9346     2192.4675 9.239663 37934170 0.000000 : pP21(35) : Excited a'v=12 34 34 F2e - X Xv=0 35 35 F1e
13C18O Excited  34  e  8      X  35  e  1    67045.7387 .0004997    69238.2062     2192.4675 18.41924 75693689 0.000000 :  pP(35) : Excited Av=2 34 34 F1e - X Xv=0 35 35 F1e
13C18O Excited  34  e  9      X  35  e  1    67111.5208 .0003045    69303.9882     2192.4675 11.22469 46263635 0.000000 : qP31(35) : Excited dv=7 34 35 F3e - X Xv=0 35 35 F1e
13C18O Excited  35  f  2      X  35  e  1    66982.7329 .0007780    69175.2003     2192.4675 28.67870 1.1421e8 0.000000 : qQ21(35) : Excited ev=4 35 35 F2f - X Xv=0 35 35 F1e
13C18O Excited  35  f  3      X  35  e  1    67015.3572 .0002598    69207.8247     2192.4675 9.576183 38192716 0.000000 :  pQ1(35) : Excited dv=7 35 34 F1f - X Xv=0 35 35 F1e
13C18O Excited  35  f  4      X  35  e  1    67025.2788 .0001904    69217.7463     2192.4675 7.018812 28005579 0.000000 :  pQ1(35) : Excited a3Pv=13 35 34 F1f - X Xv=0 35 35 F1e
13C18O Excited  35  f  6      X  35  e  1    67105.9123 .0001977    69298.3798     2192.4675 7.287411 29182380 0.000000 : qQ21(35) : Excited dv=7 35 35 F2f - X Xv=0 35 35 F1e
13C18O Excited  35  f  8      X  35  e  1    67142.1297 .0006275    69334.5972     2192.4675 23.12933 92771254 0.000000 :   qQ(35) : Excited Av=2 35 35 F1f - X Xv=0 35 35 F1e
13C18O Excited  35  f  9      X  35  e  1    67172.4976 .0003787    69364.9651     2192.4675 13.95820 56062049 0.000000 : rQ31(35) : Excited a'v=12 35 36 F3f - X Xv=0 35 35 F1e
13C18O Excited  35  f 10      X  35  e  1    67190.4674 .0005773    69382.9348     2192.4675 21.28081 85541381 0.000000 : rQ31(35) : Excited dv=7 35 36 F3f - X Xv=0 35 35 F1e
13C18O Excited  35  f 11      X  35  e  1    67218.0413 .0001075    69410.5088     2192.4675 3.962613 15947929 0.000000 : rQ31(35) : Excited a3Pv=13 35 36 F3f - X Xv=0 35 35 F1e
13C18O Excited  36  e  1      X  35  e  1    66982.2842 .0001940    69174.7517     2192.4675 7.149623 27692589 0.000000 :  qR1(35) : Excited ev=4 36 35 F1e - X Xv=0 35 35 F1e
13C18O Excited  36  e  4      X  35  e  1    67140.9034 .0001983    69333.3709     2192.4675 7.309154 28512103 0.000000 : sR31(35) : Excited ev=4 36 37 F3e - X Xv=0 35 35 F1e
13C18O Excited  36  e  5      X  35  e  1    67170.7687 .0003033    69363.2362     2192.4675 11.17831 43663425 0.000000 : rR21(35) : Excited a'v=12 36 36 F2e - X Xv=0 35 35 F1e
13C18O Excited  36  e  8      X  35  e  1    67242.9566 .0003385    69435.4241     2192.4675 12.47861 48899799 0.000000 :   rR(35) : Excited Av=2 36 36 F1e - X Xv=0 35 35 F1e
13C18O Excited  36  e  9      X  35  e  1    67270.0769 .0003099    69462.5444     2192.4675 11.42270 44816227 0.000000 : sR31(35) : Excited dv=7 36 37 F3e - X Xv=0 35 35 F1e
13C18O Excited  35  e  1      X  36  e  1    66781.3869 .0001024    69098.6551     2317.2682 6.866135 27098407 0.000000 :  oP1(36) : Excited ev=4 35 34 F1e - X Xv=0 36 36 F1e
13C18O Excited  35  e  5      X  36  e  1    66972.1124 .0001528    69289.3806     2317.2682 10.24451 40779173 0.000000 : pP21(36) : Excited a'v=12 35 35 F2e - X Xv=0 36 36 F1e
13C18O Excited  35  e  8      X  36  e  1    67017.8395 .0002163    69335.1077     2317.2682 14.50963 57875240 0.000000 :   pP(36) : Excited Av=2 35 35 F1e - X Xv=0 36 36 F1e
13C18O Excited  35  e  9      X  36  e  1    67065.6342 .0001553    69382.9024     2317.2682 10.41470 41630527 0.000000 : qP31(36) : Excited dv=7 35 36 F3e - X Xv=0 36 36 F1e
13C18O Excited  36  f  2      X  36  e  1    66936.2565 .0004238    69253.5247     2317.2682 28.42543 1.0987e8 0.000000 : qQ21(36) : Excited ev=4 36 36 F2f - X Xv=0 36 36 F1e
13C18O Excited  36  f  3      X  36  e  1    66969.6516 .0001022    69286.9198     2317.2682 6.853421 26530296 0.000000 :  pQ1(36) : Excited dv=7 36 35 F1f - X Xv=0 36 36 F1e
13C18O Excited  36  f  4      X  36  e  1    66973.2115 .0001606    69290.4797     2317.2682 10.76856 41692845 0.000000 :  pQ1(36) : Excited a'v=12 36 35 F1f - X Xv=0 36 36 F1e
13C18O Excited  36  f  6      X  36  e  1    67058.7190 .0001125    69375.9872     2317.2682 7.542085 29312832 0.000000 : qQ21(36) : Excited dv=7 36 36 F2f - X Xv=0 36 36 F1e
13C18O Excited  36  f  8      X  36  e  1    67117.0591 .0001560    69434.3273     2317.2682 10.46567 40781780 0.000000 :   qQ(36) : Excited Av=2 36 36 F1f - X Xv=0 36 36 F1e
13C18O Excited  36  f  9      X  36  e  1    67124.1501 .0003479    69441.4183     2317.2682 23.33222 90947969 0.000000 : rQ31(36) : Excited a'v=12 36 37 F3f - X Xv=0 36 36 F1e
13C18O Excited  36  f 10      X  36  e  1    67145.2986 .0003406    69462.5668     2317.2682 22.84518 89133714 0.000000 : rQ31(36) : Excited dv=7 36 37 F3f - X Xv=0 36 36 F1e
13C18O Excited  37  e  1      X  36  e  1    66935.6550 .0001081    69252.9232     2317.2682 7.250469 27277264 0.000000 :  qR1(36) : Excited ev=4 37 36 F1e - X Xv=0 36 36 F1e
13C18O Excited  37  e  4      X  36  e  1    67098.6131 .0001198    69415.8813     2317.2682 8.035617 30452434 0.000000 : sR31(36) : Excited ev=4 37 38 F3e - X Xv=0 36 36 F1e
13C18O Excited  37  e  5      X  36  e  1    67121.7522 .0001729    69439.0204     2317.2682 11.59530 43987973 0.000000 : rR21(36) : Excited a'v=12 37 37 F2e - X Xv=0 36 36 F1e
13C18O Excited  37  e  9      X  36  e  1    67228.2648 .0002966    69545.5330     2317.2682 19.89328 75827078 0.000000 : sR31(36) : Excited dv=7 37 38 F3e - X Xv=0 36 36 F1e
13C18O Excited  37  f  2      X  37  e  1    66888.3683 .0002300    69333.8496     2445.4813 28.52933 1.0710e8 0.000000 : qQ21(37) : Excited ev=4 37 37 F2f - X Xv=0 37 37 F1e
13C18O Excited  37  f  3      X  37  e  1    66918.6252 .0001299    69364.1064     2445.4813 16.11062 60564137 0.000000 :  pQ1(37) : Excited a'v=12 37 36 F1f - X Xv=0 37 37 F1e
13C18O Excited  37  f 10      X  37  e  1    67100.1015 .0003257    69545.5828     2445.4813 40.40129 1.5312e8 0.000000 : rQ31(37) : Excited dv=7 37 38 F3f - X Xv=0 37 37 F1e
13C18O Excited  38  e  9      X  37  e  1    67200.9529 .0001538    69646.4342     2445.4813 19.07568 70735758 0.000000 :   rR(37) : Excited Av=2 38 38 F1e - X Xv=0 37 37 F1e
13C18O Excited  38  f  2      X  38  e  1    66839.1163 .0001237    69416.2185     2577.1022 28.84764 1.0525e8 0.000000 : qQ21(38) : Excited ev=4 38 38 F2f - X Xv=0 38 38 F1e
13C18O Excited  38  f 10      X  38  e  1    67069.3632 .0001642    69646.4654     2577.1022 38.29435 1.4117e8 0.000000 :   qQ(38) : Excited Av=2 38 38 F1f - X Xv=0 38 38 F1e
1178 Lines above OThreshold, Total intensity = 479.2413324369
PGOPHER 11.0.142 1 Apr 2021 8:30 64 bit Unicode (Delphi 33/33)
   pgopher 44b8c50a90149a6ce53e669f4a0e44afc86f1abd
    lutils 65f951dfa553c7f9bc501632d3360a85cd1a0165
     utils ec0cac7863199e65355ae4c6ab78dbe3037fae3b
    lforms d4dfa0031586b0cf5ffd76ff4d8f7aee2a9f8932
    wutils 9bb7abdf4436dd4d7dfed61ce7fb7c6bc914386c
 GraphicEx bfb0df9494191c3a10d6089c21fe439fcd763090
Mixture: 13C18O Deperturbation of the A(v=2) level.pgo
  Units            cm1
  PrintLevel       LineList
  Precision        4
  BasisOrder       boDefault
  QuantumNumberFormat J
  FitCycles        1
  FitDetails       False
  AutoReplot       True
  AllowComplex     True
  NoLineList       False
  ShowEstUnc       False
  ShowDerivatives  False
  SVDlimit         5
  ShowSubBasis     False
  AllowZeroI       False
  ScaledParameters True
  FixTransitionPhase False
  OldHyperfineOrder False
  OldColourOrder   False
  SingleResidualPlot False
  UpperStateColour False
  SVDThresh  1E-6                      Discard Singular Value Decomposition eigenvalues below this fraction of the largest eigenvalue; <0 to use QR factorisation.
  ScaleChanges 1                       Scale changes produced by fitting by this - normally 1
  SmallE     2E-18                     Matrix elements smaller than this are discarded
  Species: CO
    Jmax 63
    Concentration 1                    Relative concentration of species
    LinearMolecule: 13C18O
      JAdjustSym  True
      BlockMatrix True
      AbsoluteE   False
      Symmetric   False
      SymWt       1
      AsymWt      1
      RSquaredH   False
      ShowJ       True
      ShowOmega   False
      ShowN       True
      ShowFNumber True
      Showef      True
      Abundance       1                Abundance of this isotope; default is 1
      AssumedOrigin 1E99                   Energy origin for Boltzmann calculations; 1e99 implies automatic estimate. Currently 0
      LinearManifold: X

        Initial        True
        EigenSearch    True
```

```
  LimitSearch      True
  AutoQConverge    True
  UsePopParams     False
  OverrideEnergies True
  Linear: X Xv=0
    Status     Active
    Lambda     Sigma+
    OmegaOrder Auto => Regular
    B  1.746408199                    Rotational constant.
    D  5.0488146E-6                   Quartic centrifugal distortion.
    H  4.35471E-12                    Sextic centrifugal distortion.
    L  -2.4821E-17                    J^8  centrifugal distortion.
    M  2.99E-23                       J^10  centrifugal distortion.
    PP -4.5E-28                       J^12 centrifugal distortion. (Not P to avoid confusion with p.)
    <J,Lambda=0,Omega=0|B|J,Lambda=0,Omega=0> = J*(J+1)
      Symmetrized:
    <J,Lambda=0,Omega=0|B|J,Lambda=0,Omega=0> = J*(J+1)

    <J,Lambda=0,Omega=0|D|J,Lambda=0,Omega=0> = -J^2*(J*(J+2)+1)
      Symmetrized:
    <J,Lambda=0,Omega=0|D|J,Lambda=0,Omega=0> = -J^2*(J*(J+2)+1)

    <J,Lambda=0,Omega=0|H|J,Lambda=0,Omega=0> = J^3*(J*(J*(J+3)+3)+1)
      Symmetrized:
    <J,Lambda=0,Omega=0|H|J,Lambda=0,Omega=0> = J^3*(J*(J*(J+3)+3)+1)

    <J,Lambda=0,Omega=0|L|J,Lambda=0,Omega=0> = J^4*(J*(J*(J*(J+4)+6)+4)+1)
      Symmetrized:
    <J,Lambda=0,Omega=0|L|J,Lambda=0,Omega=0> = J^4*(J*(J*(J*(J+4)+6)+4)+1)

    <J,Lambda=0,Omega=0|M|J,Lambda=0,Omega=0> = J^5*(J*(J*(J*(J*(J+5)+10)+10)+5)+1)
      Symmetrized:
    <J,Lambda=0,Omega=0|M|J,Lambda=0,Omega=0> = J^5*(J*(J*(J*(J*(J+5)+10)+10)+5)+1)

    <J,Lambda=0,Omega=0|PP|J,Lambda=0,Omega=0> = J^6*(J*(J*(J*(J*(J*(J+6)+15)+20)+15)+6)+1)
      Symmetrized:
    <J,Lambda=0,Omega=0|PP|J,Lambda=0,Omega=0> = J^6*(J*(J*(J*(J*(J*(J+6)+15)+20)+15)+6)+1)

  LinearPerturbation: X <Xv=0|N^14|Xv=0>
    O (rotational constant)
    Value -2.2E-33                    Size of perturbation.
  <J,Lambda=0,Omega=0|N^14|J,Lambda=0,Omega=0> = J^7*(J*(J*(J*(J*(J*(J*(J+7)+21)+35)+35)+21)+7)+1)

  Symmetrized Matrix Elements of <Xv=0|N^14|Xv=0>
  <J,Lambda=0,Omega=0|N^14|J,Lambda=0,Omega=0> = J^7*(J*(J*(J*(J*(J*(J*(J+7)+21)+35)+35)+21)+7)+1)

LinearManifold: C
  Initial          True
  EigenSearch      True
  LimitSearch      False
  AutoQConverge    True
  UsePopParams     False
  OverrideEnergies True
  E 0 e 1  91918.81461769          Energy for J=E 0 e 1 level:C Cv=0  0  0 F1e
  E 1 e 1  91922.3501979315        Energy for J=E 1 e 1 level:C Cv=0  1  1 F1e
  E 2 e 1  91929.41460931          Energy for J=E 2 e 1 level:C Cv=0  2  2 F1e
  E 3 e 1  91940.0087911758        Energy for J=E 3 e 1 level:C Cv=0  3  3 F1e
  E 4 e 1  91954.1391754158        Energy for J=E 4 e 1 level:C Cv=0  4  4 F1e
  E 5 e 1  91971.8047483651        Energy for J=E 5 e 1 level:C Cv=0  5  5 F1e
  E 6 e 1  91992.9963415625        Energy for J=E 6 e 1 level:C Cv=0  6  6 F1e
  E 7 e 1  92017.7210158785        Energy for J=E 7 e 1 level:C Cv=0  7  7 F1e
  E 8 e 1  92045.9729147246        Energy for J=E 8 e 1 level:C Cv=0  8  8 F1e
  E 9 e 1  92077.7548931101        Energy for J=E 9 e 1 level:C Cv=0  9  9 F1e
  E 10 e 1 92113.0587701987        Energy for J=E 10 e 1 level:C Cv=0 10 10 F1e
  E 11 e 1 92151.8966952636        Energy for J=E 11 e 1 level:C Cv=0 11 11 F1e
  E 12 e 1 92194.2576663392        Energy for J=E 12 e 1 level:C Cv=0 12 12 F1e
  E 13 e 1 92240.1328568783        Energy for J=E 13 e 1 level:C Cv=0 13 13 F1e
  E 14 e 1 92289.5374084555        Energy for J=E 14 e 1 level:C Cv=0 14 14 F1e
  E 15 e 1 92342.4615531966        Energy for J=E 15 e 1 level:C Cv=0 15 15 F1e
  E 16 e 1 92398.9025159008        Energy for J=E 16 e 1 level:C Cv=0 16 16 F1e
  E 17 e 1 92458.8587353746        Energy for J=E 17 e 1 level:C Cv=0 17 17 F1e
  E 18 e 1 92522.3290731028        Energy for J=E 18 e 1 level:C Cv=0 18 18 F1e
  E 19 e 1 92589.3126987891        Energy for J=E 19 e 1 level:C Cv=0 19 19 F1e
  E 20 e 1 92659.8038563447        Energy for J=E 20 e 1 level:C Cv=0 20 20 F1e
  E 21 e 1 92733.8006230299        Energy for J=E 21 e 1 level:C Cv=0 21 21 F1e
  E 22 e 1 92811.3031429715        Energy for J=E 22 e 1 level:C Cv=0 22 22 F1e
  E 23 e 1 92892.3079897717        Energy for J=E 23 e 1 level:C Cv=0 23 23 F1e
  E 24 e 1 92976.822314948         Energy for J=E 24 e 1 level:C Cv=0 24 24 F1e
  E 25 e 1 93064.8231465903        Energy for J=E 25 e 1 level:C Cv=0 25 25 F1e
  E 26 e 1 93156.3176170672        Energy for J=E 26 e 1 level:C Cv=0 26 26 F1e
  E 27 e 1 93251.2988822767        Energy for J=E 27 e 1 level:C Cv=0 27 27 F1e
  E 28 e 1 93349.7667497678        Energy for J=E 28 e 1 level:C Cv=0 28 28 F1e
  E 29 e 1 93451.7189577959        Energy for J=E 29 e 1 level:C Cv=0 29 29 F1e
  E 30 e 1 93557.148271242         Energy for J=E 30 e 1 level:C Cv=0 30 30 F1e
  E 31 e 1 93666.0549149757        Energy for J=E 31 e 1 level:C Cv=0 31 31 F1e
  E 32 e 1 93778.4343213518        Energy for J=E 32 e 1 level:C Cv=0 32 32 F1e
  E 33 e 1 93894.2815297818        Energy for J=E 33 e 1 level:C Cv=0 33 33 F1e
  E 34 e 1 94013.591760117         Energy for J=E 34 e 1 level:C Cv=0 34 34 F1e
  E 35 e 1 94136.3569127458        Energy for J=E 35 e 1 level:C Cv=0 35 35 F1e
  E 36 e 1 94262.5809686937        Energy for J=E 36 e 1 level:C Cv=0 36 36 F1e
  E 37 e 1 94392.253689726         Energy for J=E 37 e 1 level:C Cv=0 37 37 F1e
  E 38 e 1 94525.3757684531        Energy for J=E 38 e 1 level:C Cv=0 38 38 F1e
```

```
E 39 e 1 94661.9366784378        Energy for J=E 39 e 1 level:C Cv=0 39 39 F1e
E 40 e 1 94801.9378743053        Energy for J=E 40 e 1 level:C Cv=0 40 40 F1e
E 41 e 1 94945.3716418562        Energy for J=E 41 e 1 level:C Cv=0 41 41 F1e
E 42 e 1 95092.2322481804        Energy for J=E 42 e 1 level:C Cv=0 42 42 F1e
E 43 e 1 95242.507641775         Energy for J=E 43 e 1 level:C Cv=0 43 43 F1e
E 44 e 1 95396.2059026634        Energy for J=E 44 e 1 level:C Cv=0 44 44 F1e
E 45 e 1 95553.3147425169        Energy for J=E 45 e 1 level:C Cv=0 45 45 F1e
E 46 e 1 95713.8389547786        Energy for J=E 46 e 1 level:C Cv=0 46 46 F1e
E 47 e 1 95877.7430647896        Energy for J=E 47 e 1 level:C Cv=0 47 47 F1e
E 48 e 1 96045.0535799167        Energy for J=E 48 e 1 level:C Cv=0 48 48 F1e
E 49 e 1 96215.7478896832        Energy for J=E 49 e 1 level:C Cv=0 49 49 F1e
E 50 e 1 96389.8321659005        Energy for J=E 50 e 1 level:C Cv=0 50 50 F1e
E 51 e 1 96567.2766128028        Energy for J=E 51 e 1 level:C Cv=0 51 51 F1e
E 52 e 1 96748.0881671828        Energy for J=E 52 e 1 level:C Cv=0 52 52 F1e
E 53 e 1 96932.2837485302        Energy for J=E 53 e 1 level:C Cv=0 53 53 F1e
E 54 e 1 97119.8371091712        Energy for J=E 54 e 1 level:C Cv=0 54 54 F1e
E 55 e 1 97310.7208844101        Energy for J=E 55 e 1 level:C Cv=0 55 55 F1e
E 56 e 1 97504.930642673         Energy for J=E 56 e 1 level:C Cv=0 56 56 F1e
E 57 e 1 97702.5267856525        Energy for J=E 57 e 1 level:C Cv=0 57 57 F1e
E 58 e 1 97903.4396484547        Energy for J=E 58 e 1 level:C Cv=0 58 58 F1e
E 59 e 1 98107.6243997471        Energy for J=E 59 e 1 level:C Cv=0 59 59 F1e
E 60 e 1 98315.1990419085        Energy for J=E 60 e 1 level:C Cv=0 60 60 F1e
E 61 e 1 98526.0359657192        Energy for J=E 61 e 1 level:C Cv=0 61 61 F1e
E 62 e 1 98740.1821395715        Energy for J=E 62 e 1 level:C Cv=0 62 62 F1e
E 63 e 1 98957.6207414145        Energy for J=E 63 e 1 level:C Cv=0 63 63 F1e
E 64 e 1 99178.3440172057        Energy for J=E 64 e 1 level
E 65 e 1 99402.3440898222        Energy for J=E 65 e 1 level
Linear: C Cv=0
  Status     Active
  Lambda     Sigma+
  OmegaOrder Auto => Regular
  Origin 91918.79             State Origin. Estimated lowest level = 91918.79
  B      1.76641191315359          Rotational constant.
  D      5.12833508732762E-6       Quartic centrifugal distortion.
  <J,Lambda=0,Omega=0|B|J,Lambda=0,Omega=0> = J*(J+1)
   Symmetrized:
  <J,Lambda=0,Omega=0|B|J,Lambda=0,Omega=0> = J*(J+1)

  <J,Lambda=0,Omega=0|D|J,Lambda=0,Omega=0> = -J^2*(J*(J+2)+1)
   Symmetrized:
  <J,Lambda=0,Omega=0|D|J,Lambda=0,Omega=0> = -J^2*(J*(J+2)+1)

LinearManifold: B
 Initial          True
 EigenSearch      True
 LimitSearch      False
 AutoQConverge    True
 UsePopParams     False
 OverrideEnergies True
 E 0 e 1  86917.3573438145       Energy for J=E 0 e 1 level:B Bv=0  0  0 F1e
 E 1 e 1  86920.8990798371       Energy for J=E 1 e 1 level:B Bv=0  1  1 F1e
 E 2 e 1  86927.9802287834       Energy for J=E 2 e 1 level:B Bv=0  2  2 F1e
 E 3 e 1  86938.5975955474       Energy for J=E 3 e 1 level:B Bv=0  3  3 F1e
 E 4 e 1  86952.7535165513       Energy for J=E 4 e 1 level:B Bv=0  4  4 F1e
 E 5 e 1  86970.4485900942       Energy for J=E 5 e 1 level:B Bv=0  5  5 F1e
 E 6 e 1  86991.6808705965       Energy for J=E 6 e 1 level:B Bv=0  6  6 F1e
 E 7 e 1  87016.4501212347       Energy for J=E 7 e 1 level:B Bv=0  7  7 F1e
 E 8 e 1  87044.7550139408       Energy for J=E 8 e 1 level:B Bv=0  8  8 F1e
 E 9 e 1  87076.5945102899       Energy for J=E 9 e 1 level:B Bv=0  9  9 F1e
 E 10 e 1 87111.9673665656       Energy for J=E 10 e 1 level:B Bv=0 10 10 F1e
 E 11 e 1 87150.8722462898       Energy for J=E 11 e 1 level:B Bv=0 11 11 F1e
 E 12 e 1 87193.3103335359       Energy for J=E 12 e 1 level:B Bv=0 12 12 F1e
 E 13 e 1 87239.2738059414       Energy for J=E 13 e 1 level:B Bv=0 13 13 F1e
 E 14 e 1 87288.7659451284       Energy for J=E 14 e 1 level:B Bv=0 14 14 F1e
 E 15 e 1 87341.7838423505       Energy for J=E 15 e 1 level:B Bv=0 15 15 F1e
 E 16 e 1 87398.3239892327       Energy for J=E 16 e 1 level:B Bv=0 16 16 F1e
 E 17 e 1 87458.3457386672       Energy for J=E 17 e 1 level:B Bv=0 17 17 F1e
 E 18 e 1 87521.9669176787       Energy for J=E 18 e 1 level:B Bv=0 18 18 F1e
 E 19 e 1 87589.0729980635       Energy for J=E 19 e 1 level:B Bv=0 19 19 F1e
 E 20 e 1 87659.6848539994       Energy for J=E 20 e 1 level:B Bv=0 20 20 F1e
 E 21 e 1 87733.8093501407       Energy for J=E 21 e 1 level:B Bv=0 21 21 F1e
 E 22 e 1 87811.4434750138       Energy for J=E 22 e 1 level:B Bv=0 22 22 F1e
 E 23 e 1 87892.5833524621       Energy for J=E 23 e 1 level:B Bv=0 23 23 F1e
 E 24 e 1 87977.2254359612       Energy for J=E 24 e 1 level:B Bv=0 24 24 F1e
 E 25 e 1 88065.3655819688       Energy for J=E 25 e 1 level:B Bv=0 25 25 F1e
 E 26 e 1 88157.0074638094       Energy for J=E 26 e 1 level:B Bv=0 26 26 F1e
 E 27 e 1 88252.1428727514       Energy for J=E 27 e 1 level:B Bv=0 27 27 F1e
 E 28 e 1 88350.7622038786       Energy for J=E 28 e 1 level:B Bv=0 28 28 F1e
 E 29 e 1 88452.8714734024       Energy for J=E 29 e 1 level:B Bv=0 29 29 F1e
 E 30 e 1 88558.4573652084       Energy for J=E 30 e 1 level:B Bv=0 30 30 F1e
 E 31 e 1 88667.5170675484       Energy for J=E 31 e 1 level:B Bv=0 31 31 F1e
 E 32 e 1 88780.0530157549       Energy for J=E 32 e 1 level:B Bv=0 32 32 F1e
 E 33 e 1 88896.0732697678       Energy for J=E 33 e 1 level:B Bv=0 33 33 F1e
 E 34 e 1 89015.5479117331       Energy for J=E 34 e 1 level:B Bv=0 34 34 F1e
 E 35 e 1 89138.5007753021       Energy for J=E 35 e 1 level:B Bv=0 35 35 F1e
 E 36 e 1 89264.8671434128       Energy for J=E 36 e 1 level:B Bv=0 36 36 F1e
 E 37 e 1 89394.712613204        Energy for J=E 37 e 1 level:B Bv=0 37 37 F1e
 E 38 e 1 89528.0033885644       Energy for J=E 38 e 1 level:B Bv=0 38 38 F1e
 E 39 e 1 89664.7427100465       Energy for J=E 39 e 1 level:B Bv=0 39 39 F1e
 E 40 e 1 89804.8732821819       Energy for J=E 40 e 1 level:B Bv=0 40 40 F1e
 E 41 e 1 89948.4826520769       Energy for J=E 41 e 1 level:B Bv=0 41 41 F1e
 E 42 e 1 90095.4817872327       Energy for J=E 42 e 1 level:B Bv=0 42 42 F1e
```

```
 E 43 e 1 90245.9288501256            Energy for J=E 43 e 1 level:B Bv=0 43 43 F1e
 E 44 e 1 90399.7945833577            Energy for J=E 44 e 1 level:B Bv=0 44 44 F1e
 E 45 e 1 90557.048472636            Energy for J=E 45 e 1 level:B Bv=0 45 45 F1e
 E 46 e 1 90717.7471747185            Energy for J=E 46 e 1 level:B Bv=0 46 46 F1e
 E 47 e 1 90881.7572647896            Energy for J=E 47 e 1 level:B Bv=0 47 47 F1e
 E 48 e 1 91049.1774976252            Energy for J=E 48 e 1 level:B Bv=0 48 48 F1e
 E 49 e 1 91220.0533896832            Energy for J=E 49 e 1 level:B Bv=0 49 49 F1e
 E 50 e 1 91394.1969159005            Energy for J=E 50 e 1 level:B Bv=0 50 50 F1e
 E 51 e 1 91571.8211856604            Energy for J=E 51 e 1 level:B Bv=0 51 51 F1e
 E 52 e 1 91752.7615313151            Energy for J=E 52 e 1 level:B Bv=0 52 52 F1e
 E 53 e 1 91937.0582069911            Energy for J=E 53 e 1 level:B Bv=0 53 53 F1e
 E 54 e 1 92124.7041674308            Energy for J=E 54 e 1 level:B Bv=0 54 54 F1e
 E 55 e 1 92315.6922344469            Energy for J=E 55 e 1 level:B Bv=0 55 55 F1e
 Linear: B Bv=0
  Status     Active
  Lambda     Sigma+
  OmegaOrder Auto => Regular
  Origin 86917.360320496              State Origin. Estimated lowest level = 86917.360320496
  B      1.76976443875216             Rotational constant.
  D      5.53872463797865E-6          Quartic centrifugal distortion.
  Matrix elements as for C Cv=0
LinearManifold: Excited
 Initial          False
 EigenSearch      True
 LimitSearch      True
 AutoQConverge    True
 UsePopParams     False
 OverrideEnergies True
 Linear: Excited Av=2
  Status     Active
  Lambda     Pi
  OmegaOrder Auto => Regular
  Origin 67557.9263807482          Float, Std Dev = 0.000878566892884828  State Origin. Estimated lowest level = 67557.9263807482
  B      1.41770507032897          Float, Std Dev = 3.51295994438537E-6  Rotational constant.
  q      -1.187706E-5              Lambda doubling constant.
  D      6.24399485577261E-6       Float, Std Dev = 2.53609150054505E-9  Quartic centrifugal distortion.
  H      -1.280902E-11             Sextic centrifugal distortion.
  <J,Lambda=-1,Omega=-1|B|J,Lambda=-1,Omega=-1> = J*(J+1)
  <J,Lambda=1,Omega=1|B|J,Lambda=1,Omega=1> = J*(J+1)
   Symmetrized:
  <J,Lambda=1,Omega=1|B|J,Lambda=1,Omega=1> = J*(J+1)
  <J,Lambda=1,Omega=1|B|J,Lambda=1,Omega=1> = J*(J+1)

  <J,Lambda=-1,Omega=-1|q|J,Lambda=1,Omega=1> = (J*(J+1))/2
  <J,Lambda=1,Omega=1|q|J,Lambda=-1,Omega=-1> = (J*(J+1))/2
   Symmetrized:
  <J,Lambda=1,Omega=1|q|J,Lambda=1,Omega=1> = (J*(J+1))/2
  <J,Lambda=1,Omega=1|q|J,Lambda=1,Omega=1> = (J*(J+1))/2

  <J,Lambda=-1,Omega=-1|D|J,Lambda=-1,Omega=-1> = -J^2*(J*(J+2)+1)
  <J,Lambda=1,Omega=1|D|J,Lambda=1,Omega=1> = -J^2*(J*(J+2)+1)
   Symmetrized:
  <J,Lambda=1,Omega=1|D|J,Lambda=1,Omega=1> = -J^2*(J*(J+2)+1)
  <J,Lambda=1,Omega=1|D|J,Lambda=1,Omega=1> = -J^2*(J*(J+2)+1)

  <J,Lambda=-1,Omega=-1|H|J,Lambda=-1,Omega=-1> = J^3*(J*(J*(J+3)+3)+1)
  <J,Lambda=1,Omega=1|H|J,Lambda=1,Omega=1> = J^3*(J*(J*(J+3)+3)+1)
   Symmetrized:
  <J,Lambda=1,Omega=1|H|J,Lambda=1,Omega=1> = J^3*(J*(J*(J+3)+3)+1)
  <J,Lambda=1,Omega=1|H|J,Lambda=1,Omega=1> = J^3*(J*(J*(J+3)+3)+1)

 Linear: Excited Iv=3
  Status     Active
  Colour     Fuchsia
  Lambda     Sigma-
  OmegaOrder Auto => Regular
  Origin 67579.2383904013          Float, Std Dev = 0.0154787716797877  State Origin. Estimated lowest level = 67579.2383904013
  B      1.09918701526773          Float, Std Dev = 0.000226636002436598  Rotational constant.
  D      5.77973239037241E-6       Float, Std Dev = 4.04347248069572E-7  Quartic centrifugal distortion.
  H      2.247197E-12              Sextic centrifugal distortion.
  <J,Lambda=0,Omega=0|B|J,Lambda=0,Omega=0> = J*(J+1)
   Symmetrized:
  <J,Lambda=0,Omega=0|B|J,Lambda=0,Omega=0> = J*(J+1)

  <J,Lambda=0,Omega=0|D|J,Lambda=0,Omega=0> = -J^2*(J*(J+2)+1)
   Symmetrized:
  <J,Lambda=0,Omega=0|D|J,Lambda=0,Omega=0> = -J^2*(J*(J+2)+1)

  <J,Lambda=0,Omega=0|H|J,Lambda=0,Omega=0> = J^3*(J*(J*(J+3)+3)+1)
   Symmetrized:
  <J,Lambda=0,Omega=0|H|J,Lambda=0,Omega=0> = J^3*(J*(J*(J+3)+3)+1)

 Linear: Excited Iv=4
  Status     Active
  Colour     Fuchsia
  Lambda     Sigma-
  OmegaOrder Auto => Regular
  Origin 68544.2939028             State Origin. Estimated lowest level = 68544.2939028
  B      1.0840499034              Rotational constant.
  D      5.712347E-6               Quartic centrifugal distortion.
  H      2.247197E-12              Sextic centrifugal distortion.
  <J,Lambda=0,Omega=0|B|J,Lambda=0,Omega=0> = J*(J+1)
```

Symmetrized:
<J,Lambda=0,Omega=0|B|J,Lambda=0,Omega=0> = J*(J+1)

<J,Lambda=0,Omega=0|D|J,Lambda=0,Omega=0> = -J^2*(J*(J+2)+1)
Symmetrized:
<J,Lambda=0,Omega=0|D|J,Lambda=0,Omega=0> = -J^2*(J*(J+2)+1)

<J,Lambda=0,Omega=0|H|J,Lambda=0,Omega=0> = J^3*(J*(J*(J+3)+3)+1)
Symmetrized:
<J,Lambda=0,Omega=0|H|J,Lambda=0,Omega=0> = J^3*(J*(J*(J+3)+3)+1)

LinearPerturbation: Excited <Av=2|J+L-|Iv=3>
Value -0.0504224240425148 Float, Std Dev = 0.000477162968858135 Size of perturbation.
<J,Lambda=-1,Omega=-1|J+L-|J,Lambda=0,Omega=0> = -sqrt(J*(J+1))
<J,Lambda=1,Omega=1|J+L-|J,Lambda=0,Omega=0> = sqrt(J*(J+1))

Symmetrized Matrix Elements of <Av=2|J+L-|Iv=3>
<J,Lambda=1,Omega=1|J+L-|J,Lambda=0,Omega=0> = -sqrt(J*(J+1))

LinearPerturbation: Excited <Av=2|J+L-|Iv=4>
Value 0.06495 Size of perturbation.
<J,Lambda=-1,Omega=-1|J+L-|J,Lambda=0,Omega=0> = -sqrt(J*(J+1))
<J,Lambda=1,Omega=1|J+L-|J,Lambda=0,Omega=0> = sqrt(J*(J+1))

Symmetrized Matrix Elements of <Av=2|J+L-|Iv=4>
<J,Lambda=1,Omega=1|J+L-|J,Lambda=0,Omega=0> = -sqrt(J*(J+1))

Linear: Excited ev=4
Status Active
Colour Lime
S 1
Lambda Sigma-
OmegaOrder Auto => Inverted
Origin 67802.7849165451 Float, Std Dev = 0.00556026375325388 State Origin. Estimated lowest level = 67802.2187789372
B 1.09715336874161 Float, Std Dev = 3.00716570600556E-5 Rotational constant.
LambdaSS 0.424603205929654 Float, Std Dev = 0.0115162602834212 Spin - spin coupling constant (=1.5*epsilon).
D 5.34314012526997E-6 Float, Std Dev = 2.51620227317152E-8 Quartic centrifugal distortion.
H -1.498131E-12 Sextic centrifugal distortion.
<J,Lambda=0,Omega=-1|B|J,Lambda=0,Omega=-1> = J*(J+1)
<J,Lambda=0,Omega=-1|B|J,Lambda=0,Omega=0> = -sqrt(2*J*(J+1))
<J,Lambda=0,Omega=0|B|J,Lambda=0,Omega=-1> = -sqrt(2*J*(J+1))
<J,Lambda=0,Omega=0|B|J,Lambda=0,Omega=0> = J*(J+1)+2
<J,Lambda=0,Omega=0|B|J,Lambda=0,Omega=1> = -sqrt(2*J*(J+1))
<J,Lambda=0,Omega=1|B|J,Lambda=0,Omega=0> = -sqrt(2*J*(J+1))
<J,Lambda=0,Omega=1|B|J,Lambda=0,Omega=1> = J*(J+1)
Symmetrized:
<J,Lambda=0,Omega=1|B|J,Lambda=0,Omega=1> = J*(J+1)
<J,Lambda=0,Omega=1|B|J,Lambda=0,Omega=1> = -sqrt(2*J*(J+1))
<J,Lambda=0,Omega=1|B|J,Lambda=0,Omega=0> = -sqrt(2*J*(J+1))
<J,Lambda=0,Omega=0|B|J,Lambda=0,Omega=1> = J*(J+1)+2
<J,Lambda=0,Omega=0|B|J,Lambda=0,Omega=0> = -sqrt(2*J*(J+1))

<J,Lambda=0,Omega=-1|LambdaSS|J,Lambda=0,Omega=-1> = 2/3
<J,Lambda=0,Omega=0|LambdaSS|J,Lambda=0,Omega=0> = -4/3
<J,Lambda=0,Omega=1|LambdaSS|J,Lambda=0,Omega=1> = 2/3
Symmetrized:
<J,Lambda=0,Omega=1|LambdaSS|J,Lambda=0,Omega=1> = 2/3
<J,Lambda=0,Omega=1|LambdaSS|J,Lambda=0,Omega=1> = -4/3
<J,Lambda=0,Omega=0|LambdaSS|J,Lambda=0,Omega=0> = 2/3

<J,Lambda=0,Omega=-1|D|J,Lambda=0,Omega=-1> = -J*(J*(J*(J+2)+3)+2)
<J,Lambda=0,Omega=-1|D|J,Lambda=0,Omega=0> = 2*(J*(J+1)+1)*sqrt(2*J*(J+1))
<J,Lambda=0,Omega=-1|D|J,Lambda=0,Omega=1> = -2*J*(J+1)
<J,Lambda=0,Omega=0|D|J,Lambda=0,Omega=-1> = 2*(J*(J+1)+1)*sqrt(2*J*(J+1))
<J,Lambda=0,Omega=0|D|J,Lambda=0,Omega=0> = -J*(J*(J*(J+2)+9)+8)-4
<J,Lambda=0,Omega=0|D|J,Lambda=0,Omega=1> = 2*(J*(J+1)+1)*sqrt(2*J*(J+1))
<J,Lambda=0,Omega=1|D|J,Lambda=0,Omega=-1> = -2*J*(J+1)
<J,Lambda=0,Omega=1|D|J,Lambda=0,Omega=0> = 2*(J*(J+1)+1)*sqrt(2*J*(J+1))
<J,Lambda=0,Omega=1|D|J,Lambda=0,Omega=1> = -J*(J*(J*(J+2)+3)+2)
Symmetrized:
<J,Lambda=0,Omega=1|D|J,Lambda=0,Omega=1> = -J*(J*(J*(J+2)+3)+2)
<J,Lambda=0,Omega=1|D|J,Lambda=0,Omega=1> = 2*(J*(J+1)+1)*sqrt(2*J*(J+1))
<J,Lambda=0,Omega=1|D|J,Lambda=0,Omega=0> = -2*J*(J+1)
<J,Lambda=0,Omega=0|D|J,Lambda=0,Omega=1> = 2*(J*(J+1)+1)*sqrt(2*J*(J+1))
<J,Lambda=0,Omega=0|D|J,Lambda=0,Omega=0> = -J*(J*(J*(J+2)+9)+8)-4

<J,Lambda=0,Omega=-1|H|J,Lambda=0,Omega=-1> = J*(J*(J*(J*(J*(J+3)+9)+13)+10)+4)
<J,Lambda=0,Omega=-1|H|J,Lambda=0,Omega=0> = (J*(J*(-3*J*(J+2)-13)-10)-4)*sqrt(2*J*(J+1))
<J,Lambda=0,Omega=-1|H|J,Lambda=0,Omega=1> = 2*J*(J*(3*J*(J+2)+5)+2)
<J,Lambda=0,Omega=0|H|J,Lambda=0,Omega=-1> = (J*(J*(-3*J*(J+2)-13)-10)-4)*sqrt(2*J*(J+1))
<J,Lambda=0,Omega=0|H|J,Lambda=0,Omega=0> = J*(J*(J*(J*(J*(J+3)+21)+37)+46)+28)+8
<J,Lambda=0,Omega=0|H|J,Lambda=0,Omega=1> = (J*(J*(-3*J*(J+2)-13)-10)-4)*sqrt(2*J*(J+1))
<J,Lambda=0,Omega=1|H|J,Lambda=0,Omega=-1> = 2*J*(J*(3*J*(J+2)+5)+2)
<J,Lambda=0,Omega=1|H|J,Lambda=0,Omega=0> = (J*(J*(-3*J*(J+2)-13)-10)-4)*sqrt(2*J*(J+1))
<J,Lambda=0,Omega=1|H|J,Lambda=0,Omega=1> = J*(J*(J*(J*(J*(J+3)+9)+13)+10)+4)
Symmetrized:
<J,Lambda=0,Omega=1|H|J,Lambda=0,Omega=1> = J*(J*(J*(J*(J*(J+3)+9)+13)+10)+4)
<J,Lambda=0,Omega=1|H|J,Lambda=0,Omega=1> = (J*(J*(-3*J*(J+2)-13)-10)-4)*sqrt(2*J*(J+1))
<J,Lambda=0,Omega=1|H|J,Lambda=0,Omega=0> = 2*J*(J*(3*J*(J+2)+5)+2)
<J,Lambda=0,Omega=0|H|J,Lambda=0,Omega=1> = (J*(J*(-3*J*(J+2)-13)-10)-4)*sqrt(2*J*(J+1))
<J,Lambda=0,Omega=0|H|J,Lambda=0,Omega=0> = J*(J*(J*(J*(J*(J+3)+21)+37)+46)+28)+8

LinearPerturbation: Excited <Av=2|LS|ev=4>
Value -13.1135785220975 Float, Std Dev = 0.00455861064639965 Size of perturbation.
<J,Lambda=-1,Omega=-1|LS|J,Lambda=0,Omega=-1> = -1/sqrt(3)
<J,Lambda=1,Omega=1|LS|J,Lambda=0,Omega=1> = -1/sqrt(3)

Symmetrized Matrix Elements of <Av=2|LS|ev=4>
<J,Lambda=1,Omega=1|LS|J,Lambda=0,Omega=1> = -1/sqrt(3)
<J,Lambda=1,Omega=1|LS|J,Lambda=0,Omega=1> = -1/sqrt(3)

Linear: Excited dv=7
Status Active
Colour Blue
S 1
Lambda Delta
OmegaOrder Auto => Inverted
Origin 67946.2486612564 Float, Std Dev = 0.00364331290227751 State Origin. Estimated lowest level = 67913.089899836
B 1.07973878060191 Float, Std Dev = 1.39032711556845E-5 Rotational constant.
A -16.9899246280871 Float, Std Dev = 0.00331391148225554 Spin-orbit coupling constant.
LambdaSS 1.23163175363889 Float, Std Dev = 0.00538387977440938 Spin - spin coupling constant (=1.5*epsilon).
gamma -0.00290463321309688 Float, Std Dev = 0.000257537521412191 Spin rotation coupling constant.
D 5.12786503880413E-6 Float, Std Dev = 8.04194689239351E-9 Quartic centrifugal distortion.
H -5.992526E-13 Sextic centrifugal distortion.
AD -9.172647E-5 Centrifugal distortion of A.
<J,Lambda=-2,Omega=-3|B|J,Lambda=-2,Omega=-3> = J*(J+1)-4
<J,Lambda=-2,Omega=-3|B|J,Lambda=-2,Omega=-2> = -sqrt(2*(J*(J+1)-6))
<J,Lambda=-2,Omega=-2|B|J,Lambda=-2,Omega=-3> = -sqrt(2*(J*(J+1)-6))
<J,Lambda=-2,Omega=-2|B|J,Lambda=-2,Omega=-2> = J*(J+1)+2
<J,Lambda=-2,Omega=-2|B|J,Lambda=-2,Omega=-1> = -sqrt(2*(J*(J+1)-2))
<J,Lambda=-2,Omega=-1|B|J,Lambda=-2,Omega=-2> = -sqrt(2*(J*(J+1)-2))
<J,Lambda=-2,Omega=-1|B|J,Lambda=-2,Omega=-1> = J*(J+1)+4
<J,Lambda=2,Omega=1|B|J,Lambda=2,Omega=1> = J*(J+1)+4
<J,Lambda=2,Omega=1|B|J,Lambda=2,Omega=2> = -sqrt(2*(J*(J+1)-2))
<J,Lambda=2,Omega=2|B|J,Lambda=2,Omega=1> = -sqrt(2*(J*(J+1)-2))
<J,Lambda=2,Omega=2|B|J,Lambda=2,Omega=2> = J*(J+1)+2
<J,Lambda=2,Omega=2|B|J,Lambda=2,Omega=3> = -sqrt(2*(J*(J+1)-6))
<J,Lambda=2,Omega=3|B|J,Lambda=2,Omega=2> = -sqrt(2*(J*(J+1)-6))
<J,Lambda=2,Omega=3|B|J,Lambda=2,Omega=3> = J*(J+1)-4
Symmetrized:
<J,Lambda=2,Omega=3|B|J,Lambda=2,Omega=3> = J*(J+1)-4
<J,Lambda=2,Omega=3|B|J,Lambda=2,Omega=3> = -sqrt(2*(J*(J+1)-6))
<J,Lambda=2,Omega=3|B|J,Lambda=2,Omega=2> = -sqrt(2*(J*(J+1)-6))
<J,Lambda=2,Omega=3|B|J,Lambda=2,Omega=2> = J*(J+1)+2
<J,Lambda=2,Omega=2|B|J,Lambda=2,Omega=3> = -sqrt(2*(J*(J+1)-2))
<J,Lambda=2,Omega=2|B|J,Lambda=2,Omega=3> = -sqrt(2*(J*(J+1)-2))
<J,Lambda=2,Omega=2|B|J,Lambda=2,Omega=2> = J*(J+1)+4
<J,Lambda=2,Omega=2|B|J,Lambda=2,Omega=2> = J*(J+1)+4
<J,Lambda=2,Omega=2|B|J,Lambda=2,Omega=1> = -sqrt(2*(J*(J+1)-2))
<J,Lambda=2,Omega=2|B|J,Lambda=2,Omega=1> = -sqrt(2*(J*(J+1)-2))
<J,Lambda=2,Omega=1|B|J,Lambda=2,Omega=2> = J*(J+1)+2
<J,Lambda=2,Omega=1|B|J,Lambda=2,Omega=2> = -sqrt(2*(J*(J+1)-6))
<J,Lambda=2,Omega=1|B|J,Lambda=2,Omega=1> = -sqrt(2*(J*(J+1)-6))
<J,Lambda=2,Omega=1|B|J,Lambda=2,Omega=1> = J*(J+1)-4

<J,Lambda=-2,Omega=-3|A|J,Lambda=-2,Omega=-3> = 2
<J,Lambda=-2,Omega=-1|A|J,Lambda=-2,Omega=-1> = -2
<J,Lambda=2,Omega=1|A|J,Lambda=2,Omega=1> = -2
<J,Lambda=2,Omega=3|A|J,Lambda=2,Omega=3> = 2
Symmetrized:
<J,Lambda=2,Omega=3|A|J,Lambda=2,Omega=3> = 2
<J,Lambda=2,Omega=3|A|J,Lambda=2,Omega=3> = -2
<J,Lambda=2,Omega=1|A|J,Lambda=2,Omega=1> = -2
<J,Lambda=2,Omega=1|A|J,Lambda=2,Omega=1> = 2

<J,Lambda=-2,Omega=-3|LambdaSS|J,Lambda=-2,Omega=-3> = 2/3
<J,Lambda=-2,Omega=-2|LambdaSS|J,Lambda=-2,Omega=-2> = -4/3
<J,Lambda=-2,Omega=-1|LambdaSS|J,Lambda=-2,Omega=-1> = 2/3
<J,Lambda=2,Omega=1|LambdaSS|J,Lambda=2,Omega=1> = 2/3
<J,Lambda=2,Omega=2|LambdaSS|J,Lambda=2,Omega=2> = -4/3
<J,Lambda=2,Omega=3|LambdaSS|J,Lambda=2,Omega=3> = 2/3
Symmetrized:
<J,Lambda=2,Omega=3|LambdaSS|J,Lambda=2,Omega=3> = 2/3
<J,Lambda=2,Omega=3|LambdaSS|J,Lambda=2,Omega=3> = -4/3
<J,Lambda=2,Omega=2|LambdaSS|J,Lambda=2,Omega=2> = 2/3
<J,Lambda=2,Omega=2|LambdaSS|J,Lambda=2,Omega=2> = 2/3
<J,Lambda=2,Omega=1|LambdaSS|J,Lambda=2,Omega=1> = -4/3
<J,Lambda=2,Omega=1|LambdaSS|J,Lambda=2,Omega=1> = 2/3

<J,Lambda=-2,Omega=-3|gamma|J,Lambda=-2,Omega=-3> = 1
<J,Lambda=-2,Omega=-3|gamma|J,Lambda=-2,Omega=-2> = sqrt(2*(J*(J+1)-6))/2
<J,Lambda=-2,Omega=-2|gamma|J,Lambda=-2,Omega=-3> = sqrt(2*(J*(J+1)-6))/2
<J,Lambda=-2,Omega=-2|gamma|J,Lambda=-2,Omega=-2> = -2
<J,Lambda=-2,Omega=-2|gamma|J,Lambda=-2,Omega=-1> = sqrt(2*(J*(J+1)-2))/2
<J,Lambda=-2,Omega=-1|gamma|J,Lambda=-2,Omega=-2> = sqrt(2*(J*(J+1)-2))/2
<J,Lambda=-2,Omega=-1|gamma|J,Lambda=-2,Omega=-1> = -3
<J,Lambda=2,Omega=1|gamma|J,Lambda=2,Omega=1> = -3
<J,Lambda=2,Omega=1|gamma|J,Lambda=2,Omega=2> = sqrt(2*(J*(J+1)-2))/2
<J,Lambda=2,Omega=2|gamma|J,Lambda=2,Omega=1> = sqrt(2*(J*(J+1)-2))/2
<J,Lambda=2,Omega=2|gamma|J,Lambda=2,Omega=2> = -2
<J,Lambda=2,Omega=2|gamma|J,Lambda=2,Omega=3> = sqrt(2*(J*(J+1)-6))/2
<J,Lambda=2,Omega=3|gamma|J,Lambda=2,Omega=2> = sqrt(2*(J*(J+1)-6))/2
<J,Lambda=2,Omega=3|gamma|J,Lambda=2,Omega=3> = 1
Symmetrized:

<J,Lambda=2,Omega=3|gamma|J,Lambda=2,Omega=3> = 1
<J,Lambda=2,Omega=3|gamma|J,Lambda=2,Omega=3> = sqrt(2*(J*(J+1)-6))/2
<J,Lambda=2,Omega=3|gamma|J,Lambda=2,Omega=2> = sqrt(2*(J*(J+1)-6))/2
<J,Lambda=2,Omega=3|gamma|J,Lambda=2,Omega=2> = -2
<J,Lambda=2,Omega=2|gamma|J,Lambda=2,Omega=3> = sqrt(2*(J*(J+1)-2))/2
<J,Lambda=2,Omega=2|gamma|J,Lambda=2,Omega=3> = sqrt(2*(J*(J+1)-2))/2
<J,Lambda=2,Omega=2|gamma|J,Lambda=2,Omega=2> = -3
<J,Lambda=2,Omega=2|gamma|J,Lambda=2,Omega=2> = -3
<J,Lambda=2,Omega=2|gamma|J,Lambda=2,Omega=1> = sqrt(2*(J*(J+1)-2))/2
<J,Lambda=2,Omega=2|gamma|J,Lambda=2,Omega=1> = sqrt(2*(J*(J+1)-2))/2
<J,Lambda=2,Omega=1|gamma|J,Lambda=2,Omega=2> = -2
<J,Lambda=2,Omega=1|gamma|J,Lambda=2,Omega=2> = sqrt(2*(J*(J+1)-6))/2
<J,Lambda=2,Omega=1|gamma|J,Lambda=2,Omega=1> = sqrt(2*(J*(J+1)-6))/2
<J,Lambda=2,Omega=1|gamma|J,Lambda=2,Omega=1> = 1

<J,Lambda=-2,Omega=-3|D|J,Lambda=-2,Omega=-3> = -J*(J*(J*(J+2)-5)-6)-4
<J,Lambda=-2,Omega=-3|D|J,Lambda=-2,Omega=-2> = -2*(-J*(J+1)+1)*sqrt(2*(J*(J+1)-6))
<J,Lambda=-2,Omega=-3|D|J,Lambda=-2,Omega=-1> = -2*sqrt(J*(J*(J*(J+2)-7)-8)+12)
<J,Lambda=-2,Omega=-2|D|J,Lambda=-2,Omega=-3> = -2*(-J*(J+1)+1)*sqrt(2*(J*(J+1)-6))
<J,Lambda=-2,Omega=-2|D|J,Lambda=-2,Omega=-2> = -J*(J*(J*(J+2)+9)+8)+12
<J,Lambda=-2,Omega=-2|D|J,Lambda=-2,Omega=-1> = 2*(J*(J+1)+3)*sqrt(2*(J*(J+1)-2))
<J,Lambda=-2,Omega=-1|D|J,Lambda=-2,Omega=-3> = -2*sqrt(J*(J*(J*(J+2)-7)-8)+12)
<J,Lambda=-2,Omega=-1|D|J,Lambda=-2,Omega=-2> = 2*(J*(J+1)+3)*sqrt(2*(J*(J+1)-2))
<J,Lambda=-2,Omega=-1|D|J,Lambda=-2,Omega=-1> = -J*(J*(J*(J+2)+11)+10)-12
<J,Lambda=2,Omega=1|D|J,Lambda=2,Omega=1> = -J*(J*(J*(J+2)+11)+10)-12
<J,Lambda=2,Omega=1|D|J,Lambda=2,Omega=2> = 2*(J*(J+1)+3)*sqrt(2*(J*(J+1)-2))
<J,Lambda=2,Omega=1|D|J,Lambda=2,Omega=3> = -2*sqrt(J*(J*(J*(J+2)-7)-8)+12)
<J,Lambda=2,Omega=2|D|J,Lambda=2,Omega=1> = 2*(J*(J+1)+3)*sqrt(2*(J*(J+1)-2))
<J,Lambda=2,Omega=2|D|J,Lambda=2,Omega=2> = -J*(J*(J*(J+2)+9)+8)+12
<J,Lambda=2,Omega=2|D|J,Lambda=2,Omega=3> = -2*(-J*(J+1)+1)*sqrt(2*(J*(J+1)-6))
<J,Lambda=2,Omega=3|D|J,Lambda=2,Omega=1> = -2*sqrt(J*(J*(J*(J+2)-7)-8)+12)
<J,Lambda=2,Omega=3|D|J,Lambda=2,Omega=2> = -2*(-J*(J+1)+1)*sqrt(2*(J*(J+1)-6))
<J,Lambda=2,Omega=3|D|J,Lambda=2,Omega=3> = -J*(J*(J*(J+2)-5)-6)-4
Symmetrized:
<J,Lambda=2,Omega=3|D|J,Lambda=2,Omega=3> = -J*(J*(J*(J+2)-5)-6)-4
<J,Lambda=2,Omega=3|D|J,Lambda=2,Omega=3> = -2*(-J*(J+1)+1)*sqrt(2*(J*(J+1)-6))
<J,Lambda=2,Omega=3|D|J,Lambda=2,Omega=2> = -2*sqrt(J*(J*(J*(J+2)-7)-8)+12)
<J,Lambda=2,Omega=3|D|J,Lambda=2,Omega=2> = -2*(-J*(J+1)+1)*sqrt(2*(J*(J+1)-6))
<J,Lambda=2,Omega=3|D|J,Lambda=2,Omega=1> = -J*(J*(J*(J+2)+9)+8)+12
<J,Lambda=2,Omega=3|D|J,Lambda=2,Omega=1> = 2*(J*(J+1)+3)*sqrt(2*(J*(J+1)-2))
<J,Lambda=2,Omega=2|D|J,Lambda=2,Omega=3> = -2*sqrt(J*(J*(J*(J+2)-7)-8)+12)
<J,Lambda=2,Omega=2|D|J,Lambda=2,Omega=3> = 2*(J*(J+1)+3)*sqrt(2*(J*(J+1)-2))
<J,Lambda=2,Omega=2|D|J,Lambda=2,Omega=2> = -J*(J*(J*(J+2)+11)+10)-12
<J,Lambda=2,Omega=2|D|J,Lambda=2,Omega=2> = -J*(J*(J*(J+2)+11)+10)-12
<J,Lambda=2,Omega=2|D|J,Lambda=2,Omega=1> = 2*(J*(J+1)+3)*sqrt(2*(J*(J+1)-2))
<J,Lambda=2,Omega=2|D|J,Lambda=2,Omega=1> = -2*sqrt(J*(J*(J*(J+2)-7)-8)+12)
<J,Lambda=2,Omega=1|D|J,Lambda=2,Omega=3> = 2*(J*(J+1)+3)*sqrt(2*(J*(J+1)-2))
<J,Lambda=2,Omega=1|D|J,Lambda=2,Omega=3> = -J*(J*(J*(J+2)+9)+8)+12
<J,Lambda=2,Omega=1|D|J,Lambda=2,Omega=2> = -2*(-J*(J+1)+1)*sqrt(2*(J*(J+1)-6))
<J,Lambda=2,Omega=1|D|J,Lambda=2,Omega=2> = -2*sqrt(J*(J*(J*(J+2)-7)-8)+12)
<J,Lambda=2,Omega=1|D|J,Lambda=2,Omega=1> = -2*(-J*(J+1)+1)*sqrt(2*(J*(J+1)-6))
<J,Lambda=2,Omega=1|D|J,Lambda=2,Omega=1> = -J*(J*(J*(J+2)-5)-6)-4

<J,Lambda=-2,Omega=-3|H|J,Lambda=-2,Omega=-3> = J^2*(J*(J*(J*(J+3)-3)-11)-6)+8
<J,Lambda=-2,Omega=-3|H|J,Lambda=-2,Omega=-2> = (J*(J*(-3*J*(J+2)-1)+2)+4)*sqrt(2*(J*(J+1)-6))
<J,Lambda=-2,Omega=-3|H|J,Lambda=-2,Omega=-1> = 2*(3*J*(J+1)+2)*sqrt(J*(J*(J*(J+2)-7)-8)+12)
<J,Lambda=-2,Omega=-2|H|J,Lambda=-2,Omega=-3> = J*(J*(-3*J*(J+2)+1)+4)*sqrt(2*(J*(J+1)-6))-2*sqrt(2*(J*(J*(J*(J*(J*(J+3)-7)-19)+18)+28)-24))
<J,Lambda=-2,Omega=-2|H|J,Lambda=-2,Omega=-2> = J*(J*(J*(J*(J*(J+3)+21)+37)-2)-20)-24
<J,Lambda=-2,Omega=-2|H|J,Lambda=-2,Omega=-1> = J*(J*(-3*J*(J+2)-23)-20)*sqrt(2*(J*(J+1)-2))-2*sqrt(2*(J*(J*(J*(J*(J*(J+3)-11)-27)+46)+60)-72))-24*sqrt(2*(J*(J+1)-2))
<J,Lambda=-2,Omega=-1|H|J,Lambda=-2,Omega=-3> = 2*(3*J*(J+1)+2)*sqrt(J*(J*(J*(J+2)-7)-8)+12)
<J,Lambda=-2,Omega=-1|H|J,Lambda=-2,Omega=-2> = (J*(J*(-3*J*(J+2)-25)-22)-12)*sqrt(2*(J*(J+1)-2))
<J,Lambda=-2,Omega=-1|H|J,Lambda=-2,Omega=-1> = J*(J*(J*(J*(J*(J+3)+21)+37)+74)+56)+24
<J,Lambda=2,Omega=1|H|J,Lambda=2,Omega=1> = J*(J*(J*(J*(J*(J+3)+21)+37)+74)+56)+24
<J,Lambda=2,Omega=1|H|J,Lambda=2,Omega=2> = (J*(J*(-3*J*(J+2)-25)-22)-12)*sqrt(2*(J*(J+1)-2))
<J,Lambda=2,Omega=1|H|J,Lambda=2,Omega=3> = 2*(3*J*(J+1)+2)*sqrt(J*(J*(J*(J+2)-7)-8)+12)
<J,Lambda=2,Omega=2|H|J,Lambda=2,Omega=1> = J*(J*(-3*J*(J+2)-23)-20)*sqrt(2*(J*(J+1)-2))-2*sqrt(2*(J*(J*(J*(J*(J*(J+3)-11)-27)+46)+60)-72))-24*sqrt(2*(J*(J+1)-2))
<J,Lambda=2,Omega=2|H|J,Lambda=2,Omega=2> = J*(J*(J*(J*(J*(J+3)+21)+37)-2)-20)-24
<J,Lambda=2,Omega=2|H|J,Lambda=2,Omega=3> = J*(J*(-3*J*(J+2)+1)+4)*sqrt(2*(J*(J+1)-6))-2*sqrt(2*(J*(J*(J*(J*(J*(J+3)-7)-19)+18)+28)-24))
<J,Lambda=2,Omega=3|H|J,Lambda=2,Omega=1> = 2*(3*J*(J+1)+2)*sqrt(J*(J*(J*(J+2)-7)-8)+12)
<J,Lambda=2,Omega=3|H|J,Lambda=2,Omega=2> = (J*(J*(-3*J*(J+2)-1)+2)+4)*sqrt(2*(J*(J+1)-6))
<J,Lambda=2,Omega=3|H|J,Lambda=2,Omega=3> = J^2*(J*(J*(J*(J+3)-3)-11)-6)+8
Symmetrized:
<J,Lambda=2,Omega=3|H|J,Lambda=2,Omega=3> = J^2*(J*(J*(J*(J+3)-3)-11)-6)+8
<J,Lambda=2,Omega=3|H|J,Lambda=2,Omega=3> = (J*(J*(-3*J*(J+2)-1)+2)+4)*sqrt(2*(J*(J+1)-6))
<J,Lambda=2,Omega=3|H|J,Lambda=2,Omega=2> = 2*(3*J*(J+1)+2)*sqrt(J*(J*(J*(J+2)-7)-8)+12)
<J,Lambda=2,Omega=3|H|J,Lambda=2,Omega=2> = J*(J*(-3*J*(J+2)+1)+4)*sqrt(2*(J*(J+1)-6))-2*sqrt(2*(J*(J*(J*(J*(J*(J+3)-7)-19)+18)+28)-24))
<J,Lambda=2,Omega=3|H|J,Lambda=2,Omega=1> = J*(J*(J*(J*(J*(J+3)+21)+37)-2)-20)-24
<J,Lambda=2,Omega=3|H|J,Lambda=2,Omega=1> = J*(J*(-3*J*(J+2)-23)-20)*sqrt(2*(J*(J+1)-2))-2*sqrt(2*(J*(J*(J*(J*(J*(J+3)-11)-27)+46)+60)-72))-24*sqrt(2*(J*(J+1)-2))
<J,Lambda=2,Omega=2|H|J,Lambda=2,Omega=3> = 2*(3*J*(J+1)+2)*sqrt(J*(J*(J*(J+2)-7)-8)+12)
<J,Lambda=2,Omega=2|H|J,Lambda=2,Omega=3> = (J*(J*(-3*J*(J+2)-25)-22)-12)*sqrt(2*(J*(J+1)-2))
<J,Lambda=2,Omega=2|H|J,Lambda=2,Omega=2> = J*(J*(J*(J*(J*(J+3)+21)+37)+74)+56)+24
<J,Lambda=2,Omega=2|H|J,Lambda=2,Omega=2> = J*(J*(J*(J*(J*(J+3)+21)+37)+74)+56)+24
<J,Lambda=2,Omega=2|H|J,Lambda=2,Omega=1> = (J*(J*(-3*J*(J+2)-25)-22)-12)*sqrt(2*(J*(J+1)-2))
<J,Lambda=2,Omega=2|H|J,Lambda=2,Omega=1> = 2*(3*J*(J+1)+2)*sqrt(J*(J*(J*(J+2)-7)-8)+12)
<J,Lambda=2,Omega=1|H|J,Lambda=2,Omega=3> = J*(J*(-3*J*(J+2)-23)-20)*sqrt(2*(J*(J+1)-2))-2*sqrt(2*(J*(J*(J*(J*(J*(J+3)-11)-27)+46)+60)-72))-24*sqrt(2*(J*(J+1)-2))
<J,Lambda=2,Omega=1|H|J,Lambda=2,Omega=3> = J*(J*(J*(J*(J*(J+3)+21)+37)-2)-20)-24
<J,Lambda=2,Omega=1|H|J,Lambda=2,Omega=2> = J*(J*(-3*J*(J+2)+1)+4)*sqrt(2*(J*(J+1)-6))-2*sqrt(2*(J*(J*(J*(J*(J*(J+3)-7)-19)+18)+28)-24))
<J,Lambda=2,Omega=1|H|J,Lambda=2,Omega=2> = 2*(3*J*(J+1)+2)*sqrt(J*(J*(J*(J+2)-7)-8)+12)
<J,Lambda=2,Omega=1|H|J,Lambda=2,Omega=1> = (J*(J*(-3*J*(J+2)-1)+2)+4)*sqrt(2*(J*(J+1)-6))
<J,Lambda=2,Omega=1|H|J,Lambda=2,Omega=1> = J^2*(J*(J*(J*(J+3)-3)-11)-6)+8

<J,Lambda=-2,Omega=-3|AD|J,Lambda=-2,Omega=-3> = 2*(J*(J+1)-4)

```
  <J,Lambda=-2,Omega=-3|AD|J,Lambda=-2,Omega=-2> = -sqrt(2*(J*(J+1)-6))
  <J,Lambda=-2,Omega=-2|AD|J,Lambda=-2,Omega=-3> = -sqrt(2*(J*(J+1)-6))
  <J,Lambda=-2,Omega=-2|AD|J,Lambda=-2,Omega=-1> = sqrt(2*(J*(J+1)-2))
  <J,Lambda=-2,Omega=-1|AD|J,Lambda=-2,Omega=-2> = sqrt(2*(J*(J+1)-2))
  <J,Lambda=-2,Omega=-1|AD|J,Lambda=-2,Omega=-1> = 2*(-J*(J+1)-4)
  <J,Lambda=2,Omega=1|AD|J,Lambda=2,Omega=1> = 2*(-J*(J+1)-4)
  <J,Lambda=2,Omega=1|AD|J,Lambda=2,Omega=2> = sqrt(2*(J*(J+1)-2))
  <J,Lambda=2,Omega=2|AD|J,Lambda=2,Omega=1> = sqrt(2*(J*(J+1)-2))
  <J,Lambda=2,Omega=2|AD|J,Lambda=2,Omega=3> = -sqrt(2*(J*(J+1)-6))
  <J,Lambda=2,Omega=3|AD|J,Lambda=2,Omega=2> = -sqrt(2*(J*(J+1)-6))
  <J,Lambda=2,Omega=3|AD|J,Lambda=2,Omega=3> = 2*(J*(J+1)-4)
   Symmetrized:
  <J,Lambda=2,Omega=3|AD|J,Lambda=2,Omega=3> = 2*(J*(J+1)-4)
  <J,Lambda=2,Omega=3|AD|J,Lambda=2,Omega=3> = -sqrt(2*(J*(J+1)-6))
  <J,Lambda=2,Omega=3|AD|J,Lambda=2,Omega=2> = -sqrt(2*(J*(J+1)-6))
  <J,Lambda=2,Omega=3|AD|J,Lambda=2,Omega=2> = sqrt(2*(J*(J+1)-2))
  <J,Lambda=2,Omega=2|AD|J,Lambda=2,Omega=3> = sqrt(2*(J*(J+1)-2))
  <J,Lambda=2,Omega=2|AD|J,Lambda=2,Omega=3> = 2*(-J*(J+1)-4)
  <J,Lambda=2,Omega=2|AD|J,Lambda=2,Omega=1> = 2*(-J*(J+1)-4)
  <J,Lambda=2,Omega=2|AD|J,Lambda=2,Omega=1> = sqrt(2*(J*(J+1)-2))
  <J,Lambda=2,Omega=1|AD|J,Lambda=2,Omega=2> = sqrt(2*(J*(J+1)-2))
  <J,Lambda=2,Omega=1|AD|J,Lambda=2,Omega=2> = -sqrt(2*(J*(J+1)-6))
  <J,Lambda=2,Omega=1|AD|J,Lambda=2,Omega=1> = -sqrt(2*(J*(J+1)-6))
  <J,Lambda=2,Omega=1|AD|J,Lambda=2,Omega=1> = 2*(J*(J+1)-4)

LinearPerturbation: Excited <Av=2|LS|dv=7>
  Value -9.51850056737858          Float, Std Dev = 0.0104778117014898  Size of perturbation.
<J,Lambda=-1,Omega=-1|LS|J,Lambda=-2,Omega=-1> = 1/sqrt(3)
<J,Lambda=1,Omega=1|LS|J,Lambda=2,Omega=1> = -1/sqrt(3)

Symmetrized Matrix Elements of <Av=2|LS|dv=7>
<J,Lambda=1,Omega=1|LS|J,Lambda=2,Omega=1> = 1/sqrt(3)
<J,Lambda=1,Omega=1|LS|J,Lambda=2,Omega=1> = -1/sqrt(3)

Linear: Excited a'v=12
  Status     Active
  Colour     Red
  S          1
  Lambda     Sigma+
  OmegaOrder Auto => Inverted
  Origin   67991.2956897487          Float, Std Dev = 0.00589179423262914  State Origin. Estimated lowest level = 67990.5318805452
  B        1.03759347733243          Float, Std Dev = 2.24926190351211E-5  Rotational constant.
  LambdaSS -1.14571380522651         Float, Std Dev = 0.00908941623482264  Spin - spin coupling constant (=1.5*epsilon).
  gamma    -0.00527072942354211      Float, Std Dev = 0.000311699571001926  Spin rotation coupling constant.
  D        5.18562812216526E-6       Float, Std Dev = 1.22271031507727E-8  Quartic centrifugal distortion.
  H        -2.996263E-13             Sextic centrifugal distortion.
  <J,Lambda=0,Omega=-1|B|J,Lambda=0,Omega=-1> = J*(J+1)
  <J,Lambda=0,Omega=-1|B|J,Lambda=0,Omega=0> = -sqrt(2*J*(J+1))
  <J,Lambda=0,Omega=0|B|J,Lambda=0,Omega=-1> = -sqrt(2*J*(J+1))
  <J,Lambda=0,Omega=0|B|J,Lambda=0,Omega=0> = J*(J+1)+2
  <J,Lambda=0,Omega=0|B|J,Lambda=0,Omega=1> = -sqrt(2*J*(J+1))
  <J,Lambda=0,Omega=1|B|J,Lambda=0,Omega=0> = -sqrt(2*J*(J+1))
  <J,Lambda=0,Omega=1|B|J,Lambda=0,Omega=1> = J*(J+1)
   Symmetrized:
  <J,Lambda=0,Omega=1|B|J,Lambda=0,Omega=1> = J*(J+1)
  <J,Lambda=0,Omega=1|B|J,Lambda=0,Omega=1> = -sqrt(2*J*(J+1))
  <J,Lambda=0,Omega=1|B|J,Lambda=0,Omega=0> = -sqrt(2*J*(J+1))
  <J,Lambda=0,Omega=0|B|J,Lambda=0,Omega=1> = J*(J+1)+2
  <J,Lambda=0,Omega=0|B|J,Lambda=0,Omega=0> = -sqrt(2*J*(J+1))

  <J,Lambda=0,Omega=-1|LambdaSS|J,Lambda=0,Omega=-1> = 2/3
  <J,Lambda=0,Omega=0|LambdaSS|J,Lambda=0,Omega=0> = -4/3
  <J,Lambda=0,Omega=1|LambdaSS|J,Lambda=0,Omega=1> = 2/3
   Symmetrized:
  <J,Lambda=0,Omega=1|LambdaSS|J,Lambda=0,Omega=1> = 2/3
  <J,Lambda=0,Omega=1|LambdaSS|J,Lambda=0,Omega=1> = -4/3
  <J,Lambda=0,Omega=0|LambdaSS|J,Lambda=0,Omega=0> = 2/3

  <J,Lambda=0,Omega=-1|gamma|J,Lambda=0,Omega=-1> = -1
  <J,Lambda=0,Omega=-1|gamma|J,Lambda=0,Omega=0> = sqrt(2*J*(J+1))/2
  <J,Lambda=0,Omega=0|gamma|J,Lambda=0,Omega=-1> = sqrt(2*J*(J+1))/2
  <J,Lambda=0,Omega=0|gamma|J,Lambda=0,Omega=0> = -2
  <J,Lambda=0,Omega=0|gamma|J,Lambda=0,Omega=1> = sqrt(2*J*(J+1))/2
  <J,Lambda=0,Omega=1|gamma|J,Lambda=0,Omega=0> = sqrt(2*J*(J+1))/2
  <J,Lambda=0,Omega=1|gamma|J,Lambda=0,Omega=1> = -1
   Symmetrized:
  <J,Lambda=0,Omega=1|gamma|J,Lambda=0,Omega=1> = -1
  <J,Lambda=0,Omega=1|gamma|J,Lambda=0,Omega=1> = sqrt(2*J*(J+1))/2
  <J,Lambda=0,Omega=1|gamma|J,Lambda=0,Omega=0> = sqrt(2*J*(J+1))/2
  <J,Lambda=0,Omega=0|gamma|J,Lambda=0,Omega=1> = -2
  <J,Lambda=0,Omega=0|gamma|J,Lambda=0,Omega=0> = sqrt(2*J*(J+1))/2

  <J,Lambda=0,Omega=-1|D|J,Lambda=0,Omega=-1> = -J*(J*(J*(J+2)+3)+2)
  <J,Lambda=0,Omega=-1|D|J,Lambda=0,Omega=0> = 2*(J*(J+1)+1)*sqrt(2*J*(J+1))
  <J,Lambda=0,Omega=-1|D|J,Lambda=0,Omega=1> = -2*J*(J+1)
  <J,Lambda=0,Omega=0|D|J,Lambda=0,Omega=-1> = 2*(J*(J+1)+1)*sqrt(2*J*(J+1))
  <J,Lambda=0,Omega=0|D|J,Lambda=0,Omega=0> = -J*(J*(J*(J+2)+9)+8)-4
  <J,Lambda=0,Omega=0|D|J,Lambda=0,Omega=1> = 2*(J*(J+1)+1)*sqrt(2*J*(J+1))
  <J,Lambda=0,Omega=1|D|J,Lambda=0,Omega=-1> = -2*J*(J+1)
  <J,Lambda=0,Omega=1|D|J,Lambda=0,Omega=0> = 2*(J*(J+1)+1)*sqrt(2*J*(J+1))
  <J,Lambda=0,Omega=1|D|J,Lambda=0,Omega=1> = -J*(J*(J*(J+2)+3)+2)
   Symmetrized:
```

<J,Lambda=0,Omega=1|D|J,Lambda=0,Omega=1> = -J*(J*(J*(J+2)+3)+2)
<J,Lambda=0,Omega=1|D|J,Lambda=0,Omega=1> = 2*(J*(J+1)+1)*sqrt(2*J*(J+1))
<J,Lambda=0,Omega=1|D|J,Lambda=0,Omega=0> = -2*J*(J+1)
<J,Lambda=0,Omega=0|D|J,Lambda=0,Omega=1> = 2*(J*(J+1)+1)*sqrt(2*J*(J+1))
<J,Lambda=0,Omega=0|D|J,Lambda=0,Omega=0> = -J*(J*(J*(J+2)+9)+8)-4

<J,Lambda=0,Omega=-1|H|J,Lambda=0,Omega=-1> = J*(J*(J*(J*(J*(J+3)+9)+13)+10)+4)
<J,Lambda=0,Omega=-1|H|J,Lambda=0,Omega=0> = (J*(J*(-3*J*(J+2)-13)-10)-4)*sqrt(2*J*(J+1))
<J,Lambda=0,Omega=-1|H|J,Lambda=0,Omega=1> = 2*J*(J*(3*J*(J+2)+5)+2)
<J,Lambda=0,Omega=0|H|J,Lambda=0,Omega=-1> = (J*(J*(-3*J*(J+2)-13)-10)-4)*sqrt(2*J*(J+1))
<J,Lambda=0,Omega=0|H|J,Lambda=0,Omega=0> = J*(J*(J*(J*(J*(J+3)+21)+37)+46)+28)+8
<J,Lambda=0,Omega=0|H|J,Lambda=0,Omega=1> = (J*(J*(-3*J*(J+2)-13)-10)-4)*sqrt(2*J*(J+1))
<J,Lambda=0,Omega=1|H|J,Lambda=0,Omega=-1> = 2*J*(J*(3*J*(J+2)+5)+2)
<J,Lambda=0,Omega=1|H|J,Lambda=0,Omega=0> = (J*(J*(-3*J*(J+2)-13)-10)-4)*sqrt(2*J*(J+1))
<J,Lambda=0,Omega=1|H|J,Lambda=0,Omega=1> = J*(J*(J*(J*(J*(J+3)+9)+13)+10)+4)
Symmetrized:
<J,Lambda=0,Omega=1|H|J,Lambda=0,Omega=1> = J*(J*(J*(J*(J*(J+3)+9)+13)+10)+4)
<J,Lambda=0,Omega=1|H|J,Lambda=0,Omega=1> = (J*(J*(-3*J*(J+2)-13)-10)-4)*sqrt(2*J*(J+1))
<J,Lambda=0,Omega=1|H|J,Lambda=0,Omega=0> = 2*J*(J*(3*J*(J+2)+5)+2)
<J,Lambda=0,Omega=0|H|J,Lambda=0,Omega=1> = (J*(J*(-3*J*(J+2)-13)-10)-4)*sqrt(2*J*(J+1))
<J,Lambda=0,Omega=0|H|J,Lambda=0,Omega=0> = J*(J*(J*(J*(J*(J+3)+21)+37)+46)+28)+8

LinearPerturbation: Excited <Av=2|LS|a'v=12>
Value -5.90222571840932 Float, Std Dev = 0.0114841933161004 Size of perturbation.
<J,Lambda=-1,Omega=-1|LS|J,Lambda=0,Omega=-1> = 1/sqrt(3)
<J,Lambda=1,Omega=1|LS|J,Lambda=0,Omega=1> = -1/sqrt(3)

Symmetrized Matrix Elements of <Av=2|LS|a'v=12>
<J,Lambda=1,Omega=1|LS|J,Lambda=0,Omega=1> = 1/sqrt(3)
<J,Lambda=1,Omega=1|LS|J,Lambda=0,Omega=1> = -1/sqrt(3)

Linear: Excited Dsv=3
Status Active
Colour Orange
Lambda Delta
OmegaOrder Auto => Regular
Origin 68377.9123376615 Float, Std Dev = 0.0503654794741688 State Origin. Estimated lowest level = 68377.9123376615
B 1.08921917955507 Float, Std Dev = 0.000187517727013344 Rotational constant.
D 5.85540454591073E-6 Float, Std Dev = 7.96670714380823E-8 Quartic centrifugal distortion.
H -2.247197E-13 Sextic centrifugal distortion.
<J,Lambda=-2,Omega=-2|B|J,Lambda=-2,Omega=-2> = J*(J+1)
<J,Lambda=2,Omega=2|B|J,Lambda=2,Omega=2> = J*(J+1)
Symmetrized:
<J,Lambda=2,Omega=2|B|J,Lambda=2,Omega=2> = J*(J+1)
<J,Lambda=2,Omega=2|B|J,Lambda=2,Omega=2> = J*(J+1)

<J,Lambda=-2,Omega=-2|D|J,Lambda=-2,Omega=-2> = -J^2*(J*(J+2)+1)
<J,Lambda=2,Omega=2|D|J,Lambda=2,Omega=2> = -J^2*(J*(J+2)+1)
Symmetrized:
<J,Lambda=2,Omega=2|D|J,Lambda=2,Omega=2> = -J^2*(J*(J+2)+1)
<J,Lambda=2,Omega=2|D|J,Lambda=2,Omega=2> = -J^2*(J*(J+2)+1)

<J,Lambda=-2,Omega=-2|H|J,Lambda=-2,Omega=-2> = J^3*(J*(J*(J+3)+3)+1)
<J,Lambda=2,Omega=2|H|J,Lambda=2,Omega=2> = J^3*(J*(J*(J+3)+3)+1)
Symmetrized:
<J,Lambda=2,Omega=2|H|J,Lambda=2,Omega=2> = J^3*(J*(J*(J+3)+3)+1)
<J,Lambda=2,Omega=2|H|J,Lambda=2,Omega=2> = J^3*(J*(J*(J+3)+3)+1)

LinearPerturbation: Excited <Av=2|J+L-|Dsv=3>
Value 0.02597 Size of perturbation.
<J,Lambda=-1,Omega=-1|J+L-|J,Lambda=-2,Omega=-2> = sqrt(J*(J+1)-2)
<J,Lambda=1,Omega=1|J+L-|J,Lambda=2,Omega=2> = sqrt(J*(J+1)-2)

Symmetrized Matrix Elements of <Av=2|J+L-|Dsv=3>
<J,Lambda=1,Omega=1|J+L-|J,Lambda=2,Omega=2> = sqrt(J*(J+1)-2)
<J,Lambda=1,Omega=1|J+L-|J,Lambda=2,Omega=2> = sqrt(J*(J+1)-2)

Linear: Excited a3Pv=13
Status Active
Colour Silver
S 1
Lambda Pi
OmegaOrder Auto => Regular
Origin 67670.6782977856 Float, Std Dev = 0.00506807720509925 State Origin. Estimated lowest level = 67632.5507030479
B 1.30813665955102 Float, Std Dev = 5.56682832240352E-7 Rotational constant.
A 38.1154482644111 Float, Std Dev = 0.0129516607117093 Spin-orbit coupling constant.
LambdaSS -0.01821971 Spin - spin coupling constant (=1.5*epsilon).
o 0.6497647 Spin-spin coupling constant giving lambda doubling for Omega = 0 states.
gamma 0.003153713 Spin rotation coupling constant.
p 0.008831129 Lambda doubling constant.
q 2.786443E-5 Lambda doubling constant.
D 6.06110427245085E-6 Float, Std Dev = 4.64500829284345E-9 Quartic centrifugal distortion.
AD -0.0001940953 Centrifugal distortion of A.
<J,Lambda=-1,Omega=-2|B|J,Lambda=-1,Omega=-2> = J*(J+1)-2
<J,Lambda=-1,Omega=-2|B|J,Lambda=-1,Omega=-1> = -sqrt(2*(J*(J+1)-2))
<J,Lambda=-1,Omega=-1|B|J,Lambda=-1,Omega=-2> = -sqrt(2*(J*(J+1)-2))
<J,Lambda=-1,Omega=-1|B|J,Lambda=-1,Omega=-1> = J*(J+1)+2
<J,Lambda=-1,Omega=-1|B|J,Lambda=-1,Omega=0> = -sqrt(2*J*(J+1))
<J,Lambda=-1,Omega=0|B|J,Lambda=-1,Omega=-1> = -sqrt(2*J*(J+1))
<J,Lambda=-1,Omega=0|B|J,Lambda=-1,Omega=0> = J*(J+1)+2
<J,Lambda=1,Omega=0|B|J,Lambda=1,Omega=0> = J*(J+1)+2
<J,Lambda=1,Omega=0|B|J,Lambda=1,Omega=1> = -sqrt(2*J*(J+1))

<J,Lambda=1,Omega=1|B|J,Lambda=1,Omega=0> = -sqrt(2*J*(J+1))
<J,Lambda=1,Omega=1|B|J,Lambda=1,Omega=1> = J*(J+1)+2
<J,Lambda=1,Omega=1|B|J,Lambda=1,Omega=2> = -sqrt(2*(J*(J+1)-2))
<J,Lambda=1,Omega=2|B|J,Lambda=1,Omega=1> = -sqrt(2*(J*(J+1)-2))
<J,Lambda=1,Omega=2|B|J,Lambda=1,Omega=2> = J*(J+1)-2
Symmetrized:
<J,Lambda=1,Omega=2|B|J,Lambda=1,Omega=2> = J*(J+1)-2
<J,Lambda=1,Omega=2|B|J,Lambda=1,Omega=2> = -sqrt(2*(J*(J+1)-2))
<J,Lambda=1,Omega=2|B|J,Lambda=1,Omega=1> = -sqrt(2*(J*(J+1)-2))
<J,Lambda=1,Omega=2|B|J,Lambda=1,Omega=1> = J*(J+1)+2
<J,Lambda=1,Omega=1|B|J,Lambda=1,Omega=2> = -sqrt(2*J*(J+1))
<J,Lambda=1,Omega=1|B|J,Lambda=1,Omega=2> = -sqrt(2*J*(J+1))
<J,Lambda=1,Omega=1|B|J,Lambda=1,Omega=1> = J*(J+1)+2
<J,Lambda=1,Omega=1|B|J,Lambda=1,Omega=1> = J*(J+1)+2
<J,Lambda=1,Omega=1|B|J,Lambda=1,Omega=0> = -sqrt(2*J*(J+1))
<J,Lambda=1,Omega=1|B|J,Lambda=1,Omega=0> = -sqrt(2*J*(J+1))
<J,Lambda=1,Omega=0|B|J,Lambda=1,Omega=1> = J*(J+1)+2
<J,Lambda=1,Omega=0|B|J,Lambda=1,Omega=1> = -sqrt(2*(J*(J+1)-2))
<J,Lambda=1,Omega=0|B|J,Lambda=1,Omega=0> = -sqrt(2*(J*(J+1)-2))
<J,Lambda=1,Omega=0|B|J,Lambda=1,Omega=0> = J*(J+1)-2

<J,Lambda=-1,Omega=-2|A|J,Lambda=-1,Omega=-2> = 1
<J,Lambda=-1,Omega=0|A|J,Lambda=-1,Omega=0> = -1
<J,Lambda=1,Omega=0|A|J,Lambda=1,Omega=0> = -1
<J,Lambda=1,Omega=2|A|J,Lambda=1,Omega=2> = 1
Symmetrized:
<J,Lambda=1,Omega=2|A|J,Lambda=1,Omega=2> = 1
<J,Lambda=1,Omega=2|A|J,Lambda=1,Omega=2> = -1
<J,Lambda=1,Omega=0|A|J,Lambda=1,Omega=0> = -1
<J,Lambda=1,Omega=0|A|J,Lambda=1,Omega=0> = 1

<J,Lambda=-1,Omega=-2|LambdaSS|J,Lambda=-1,Omega=-2> = 2/3
<J,Lambda=-1,Omega=-1|LambdaSS|J,Lambda=-1,Omega=-1> = -4/3
<J,Lambda=-1,Omega=0|LambdaSS|J,Lambda=-1,Omega=0> = 2/3
<J,Lambda=1,Omega=0|LambdaSS|J,Lambda=1,Omega=0> = 2/3
<J,Lambda=1,Omega=1|LambdaSS|J,Lambda=1,Omega=1> = -4/3
<J,Lambda=1,Omega=2|LambdaSS|J,Lambda=1,Omega=2> = 2/3
Symmetrized:
<J,Lambda=1,Omega=2|LambdaSS|J,Lambda=1,Omega=2> = 2/3
<J,Lambda=1,Omega=2|LambdaSS|J,Lambda=1,Omega=2> = -4/3
<J,Lambda=1,Omega=1|LambdaSS|J,Lambda=1,Omega=1> = 2/3
<J,Lambda=1,Omega=1|LambdaSS|J,Lambda=1,Omega=1> = 2/3
<J,Lambda=1,Omega=0|LambdaSS|J,Lambda=1,Omega=0> = -4/3
<J,Lambda=1,Omega=0|LambdaSS|J,Lambda=1,Omega=0> = 2/3

<J,Lambda=-1,Omega=0|o|J,Lambda=1,Omega=0> = 1
<J,Lambda=1,Omega=0|o|J,Lambda=-1,Omega=0> = 1
Symmetrized:
<J,Lambda=1,Omega=0|o|J,Lambda=1,Omega=0> = 1
<J,Lambda=1,Omega=0|o|J,Lambda=1,Omega=0> = 1

<J,Lambda=-1,Omega=-2|gamma|J,Lambda=-1,Omega=-1> = sqrt(2*(J*(J+1)-2))/2
<J,Lambda=-1,Omega=-1|gamma|J,Lambda=-1,Omega=-2> = sqrt(2*(J*(J+1)-2))/2
<J,Lambda=-1,Omega=-1|gamma|J,Lambda=-1,Omega=-1> = -2
<J,Lambda=-1,Omega=-1|gamma|J,Lambda=-1,Omega=0> = sqrt(2*J*(J+1))/2
<J,Lambda=-1,Omega=0|gamma|J,Lambda=-1,Omega=-1> = sqrt(2*J*(J+1))/2
<J,Lambda=-1,Omega=0|gamma|J,Lambda=-1,Omega=0> = -2
<J,Lambda=1,Omega=0|gamma|J,Lambda=1,Omega=0> = -2
<J,Lambda=1,Omega=0|gamma|J,Lambda=1,Omega=1> = sqrt(2*J*(J+1))/2
<J,Lambda=1,Omega=1|gamma|J,Lambda=1,Omega=0> = sqrt(2*J*(J+1))/2
<J,Lambda=1,Omega=1|gamma|J,Lambda=1,Omega=1> = -2
<J,Lambda=1,Omega=1|gamma|J,Lambda=1,Omega=2> = sqrt(2*(J*(J+1)-2))/2
<J,Lambda=1,Omega=2|gamma|J,Lambda=1,Omega=1> = sqrt(2*(J*(J+1)-2))/2
Symmetrized:
<J,Lambda=1,Omega=2|gamma|J,Lambda=1,Omega=1> = sqrt(2*(J*(J+1)-2))/2
<J,Lambda=1,Omega=2|gamma|J,Lambda=1,Omega=1> = sqrt(2*(J*(J+1)-2))/2
<J,Lambda=1,Omega=1|gamma|J,Lambda=1,Omega=2> = -2
<J,Lambda=1,Omega=1|gamma|J,Lambda=1,Omega=2> = sqrt(2*J*(J+1))/2
<J,Lambda=1,Omega=1|gamma|J,Lambda=1,Omega=1> = sqrt(2*J*(J+1))/2
<J,Lambda=1,Omega=1|gamma|J,Lambda=1,Omega=1> = -2
<J,Lambda=1,Omega=1|gamma|J,Lambda=1,Omega=0> = -2
<J,Lambda=1,Omega=1|gamma|J,Lambda=1,Omega=0> = sqrt(2*J*(J+1))/2
<J,Lambda=1,Omega=0|gamma|J,Lambda=1,Omega=1> = sqrt(2*J*(J+1))/2
<J,Lambda=1,Omega=0|gamma|J,Lambda=1,Omega=1> = -2
<J,Lambda=1,Omega=0|gamma|J,Lambda=1,Omega=0> = sqrt(2*(J*(J+1)-2))/2
<J,Lambda=1,Omega=0|gamma|J,Lambda=1,Omega=0> = sqrt(2*(J*(J+1)-2))/2

<J,Lambda=-1,Omega=-1|p|J,Lambda=1,Omega=0> = (-sqrt(2*J*(J+1)))/2
<J,Lambda=-1,Omega=0|p|J,Lambda=1,Omega=0> = 1
<J,Lambda=-1,Omega=0|p|J,Lambda=1,Omega=1> = (-sqrt(2*J*(J+1)))/2
<J,Lambda=1,Omega=0|p|J,Lambda=-1,Omega=-1> = (-sqrt(2*J*(J+1)))/2
<J,Lambda=1,Omega=0|p|J,Lambda=-1,Omega=0> = 1
<J,Lambda=1,Omega=1|p|J,Lambda=-1,Omega=0> = (-sqrt(2*J*(J+1)))/2
Symmetrized:
<J,Lambda=1,Omega=1|p|J,Lambda=1,Omega=0> = (-sqrt(2*J*(J+1)))/2
<J,Lambda=1,Omega=1|p|J,Lambda=1,Omega=0> = 1
<J,Lambda=1,Omega=0|p|J,Lambda=1,Omega=0> = (-sqrt(2*J*(J+1)))/2
<J,Lambda=1,Omega=0|p|J,Lambda=1,Omega=0> = (-sqrt(2*J*(J+1)))/2
<J,Lambda=1,Omega=0|p|J,Lambda=1,Omega=1> = 1
<J,Lambda=1,Omega=0|p|J,Lambda=1,Omega=1> = (-sqrt(2*J*(J+1)))/2

<J,Lambda=-1,Omega=-2|q|J,Lambda=1,Omega=0> = sqrt(J*(J*(J+1)-2)*(J+1))/2

<J,Lambda=-1,Omega=-1|q|J,Lambda=1,Omega=0> = -sqrt(2*J*(J+1))
<J,Lambda=-1,Omega=-1|q|J,Lambda=1,Omega=1> = (J*(J+1))/2
<J,Lambda=-1,Omega=0|q|J,Lambda=1,Omega=0> = 1
<J,Lambda=-1,Omega=0|q|J,Lambda=1,Omega=1> = -sqrt(2*J*(J+1))
<J,Lambda=-1,Omega=0|q|J,Lambda=1,Omega=2> = sqrt(J*(J*(J+1)-2)*(J+1))/2
<J,Lambda=1,Omega=0|q|J,Lambda=-1,Omega=-2> = sqrt(J*(J*(J+1)-2)*(J+1))/2
<J,Lambda=1,Omega=0|q|J,Lambda=-1,Omega=-1> = -sqrt(2*J*(J+1))
<J,Lambda=1,Omega=0|q|J,Lambda=-1,Omega=0> = 1
<J,Lambda=1,Omega=1|q|J,Lambda=-1,Omega=-1> = (J*(J+1))/2
<J,Lambda=1,Omega=1|q|J,Lambda=-1,Omega=0> = -sqrt(2*J*(J+1))
<J,Lambda=1,Omega=2|q|J,Lambda=-1,Omega=0> = sqrt(J*(J*(J+1)-2)*(J+1))/2
Symmetrized:
<J,Lambda=1,Omega=2|q|J,Lambda=1,Omega=0> = sqrt(J*(J*(J+1)-2)*(J+1))/2
<J,Lambda=1,Omega=2|q|J,Lambda=1,Omega=0> = -sqrt(2*J*(J+1))
<J,Lambda=1,Omega=1|q|J,Lambda=1,Omega=0> = (J*(J+1))/2
<J,Lambda=1,Omega=1|q|J,Lambda=1,Omega=0> = 1
<J,Lambda=1,Omega=1|q|J,Lambda=1,Omega=1> = -sqrt(2*J*(J+1))
<J,Lambda=1,Omega=1|q|J,Lambda=1,Omega=1> = sqrt(J*(J*(J+1)-2)*(J+1))/2
<J,Lambda=1,Omega=0|q|J,Lambda=1,Omega=0> = sqrt(J*(J*(J+1)-2)*(J+1))/2
<J,Lambda=1,Omega=0|q|J,Lambda=1,Omega=0> = -sqrt(2*J*(J+1))
<J,Lambda=1,Omega=0|q|J,Lambda=1,Omega=1> = 1
<J,Lambda=1,Omega=0|q|J,Lambda=1,Omega=1> = (J*(J+1))/2
<J,Lambda=1,Omega=0|q|J,Lambda=1,Omega=2> = -sqrt(2*J*(J+1))
<J,Lambda=1,Omega=0|q|J,Lambda=1,Omega=2> = sqrt(J*(J*(J+1)-2)*(J+1))/2

<J,Lambda=-1,Omega=-2|D|J,Lambda=-1,Omega=-2> = -J*(J*(J*(J+2)-1)-2)
<J,Lambda=-1,Omega=-2|D|J,Lambda=-1,Omega=-1> = 2*J*(J+1)*sqrt(2*(J*(J+1)-2))
<J,Lambda=-1,Omega=-2|D|J,Lambda=-1,Omega=0> = -2*sqrt(J*(J*(J*(J+2)-1)-2))
<J,Lambda=-1,Omega=-1|D|J,Lambda=-1,Omega=-2> = 2*J*(J+1)*sqrt(2*(J*(J+1)-2))
<J,Lambda=-1,Omega=-1|D|J,Lambda=-1,Omega=-1> = -J*(J*(J*(J+2)+9)+8)
<J,Lambda=-1,Omega=-1|D|J,Lambda=-1,Omega=0> = 2*(J*(J+1)+2)*sqrt(2*J*(J+1))
<J,Lambda=-1,Omega=0|D|J,Lambda=-1,Omega=-2> = -2*sqrt(J*(J*(J*(J+2)-1)-2))
<J,Lambda=-1,Omega=0|D|J,Lambda=-1,Omega=-1> = 2*(J*(J+1)+2)*sqrt(2*J*(J+1))
<J,Lambda=-1,Omega=0|D|J,Lambda=-1,Omega=0> = -J*(J*(J*(J+2)+7)+6)-4
<J,Lambda=1,Omega=0|D|J,Lambda=1,Omega=0> = -J*(J*(J*(J+2)+7)+6)-4
<J,Lambda=1,Omega=0|D|J,Lambda=1,Omega=1> = 2*(J*(J+1)+2)*sqrt(2*J*(J+1))
<J,Lambda=1,Omega=0|D|J,Lambda=1,Omega=2> = -2*sqrt(J*(J*(J*(J+2)-1)-2))
<J,Lambda=1,Omega=1|D|J,Lambda=1,Omega=0> = 2*(J*(J+1)+2)*sqrt(2*J*(J+1))
<J,Lambda=1,Omega=1|D|J,Lambda=1,Omega=1> = -J*(J*(J*(J+2)+9)+8)
<J,Lambda=1,Omega=1|D|J,Lambda=1,Omega=2> = 2*J*(J+1)*sqrt(2*(J*(J+1)-2))
<J,Lambda=1,Omega=2|D|J,Lambda=1,Omega=0> = -2*sqrt(J*(J*(J*(J+2)-1)-2))
<J,Lambda=1,Omega=2|D|J,Lambda=1,Omega=1> = 2*J*(J+1)*sqrt(2*(J*(J+1)-2))
<J,Lambda=1,Omega=2|D|J,Lambda=1,Omega=2> = -J*(J*(J*(J+2)-1)-2)
Symmetrized:
<J,Lambda=1,Omega=2|D|J,Lambda=1,Omega=2> = -J*(J*(J*(J+2)-1)-2)
<J,Lambda=1,Omega=2|D|J,Lambda=1,Omega=2> = 2*J*(J+1)*sqrt(2*(J*(J+1)-2))
<J,Lambda=1,Omega=2|D|J,Lambda=1,Omega=1> = -2*sqrt(J*(J*(J*(J+2)-1)-2))
<J,Lambda=1,Omega=2|D|J,Lambda=1,Omega=1> = 2*J*(J+1)*sqrt(2*(J*(J+1)-2))
<J,Lambda=1,Omega=2|D|J,Lambda=1,Omega=0> = -J*(J*(J*(J+2)+9)+8)
<J,Lambda=1,Omega=2|D|J,Lambda=1,Omega=0> = 2*(J*(J+1)+2)*sqrt(2*J*(J+1))
<J,Lambda=1,Omega=1|D|J,Lambda=1,Omega=2> = -2*sqrt(J*(J*(J*(J+2)-1)-2))
<J,Lambda=1,Omega=1|D|J,Lambda=1,Omega=2> = 2*(J*(J+1)+2)*sqrt(2*J*(J+1))
<J,Lambda=1,Omega=1|D|J,Lambda=1,Omega=1> = -J*(J*(J*(J+2)+7)+6)-4
<J,Lambda=1,Omega=1|D|J,Lambda=1,Omega=1> = -J*(J*(J*(J+2)+7)+6)-4
<J,Lambda=1,Omega=1|D|J,Lambda=1,Omega=0> = 2*(J*(J+1)+2)*sqrt(2*J*(J+1))
<J,Lambda=1,Omega=1|D|J,Lambda=1,Omega=0> = -2*sqrt(J*(J*(J*(J+2)-1)-2))
<J,Lambda=1,Omega=0|D|J,Lambda=1,Omega=2> = 2*(J*(J+1)+2)*sqrt(2*J*(J+1))
<J,Lambda=1,Omega=0|D|J,Lambda=1,Omega=2> = -J*(J*(J*(J+2)+9)+8)
<J,Lambda=1,Omega=0|D|J,Lambda=1,Omega=1> = 2*J*(J+1)*sqrt(2*(J*(J+1)-2))
<J,Lambda=1,Omega=0|D|J,Lambda=1,Omega=1> = -2*sqrt(J*(J*(J*(J+2)-1)-2))
<J,Lambda=1,Omega=0|D|J,Lambda=1,Omega=0> = 2*J*(J+1)*sqrt(2*(J*(J+1)-2))
<J,Lambda=1,Omega=0|D|J,Lambda=1,Omega=0> = -J*(J*(J*(J+2)-1)-2)

<J,Lambda=-1,Omega=-2|AD|J,Lambda=-1,Omega=-2> = J*(J+1)-2
<J,Lambda=-1,Omega=-2|AD|J,Lambda=-1,Omega=-1> = (-sqrt(2*(J*(J+1)-2)))/2
<J,Lambda=-1,Omega=-1|AD|J,Lambda=-1,Omega=-2> = (-sqrt(2*(J*(J+1)-2)))/2
<J,Lambda=-1,Omega=-1|AD|J,Lambda=-1,Omega=0> = sqrt(2*J*(J+1))/2
<J,Lambda=-1,Omega=0|AD|J,Lambda=-1,Omega=-1> = sqrt(2*J*(J+1))/2
<J,Lambda=-1,Omega=0|AD|J,Lambda=-1,Omega=0> = -J*(J+1)-2
<J,Lambda=1,Omega=0|AD|J,Lambda=1,Omega=0> = -J*(J+1)-2
<J,Lambda=1,Omega=0|AD|J,Lambda=1,Omega=1> = sqrt(2*J*(J+1))/2
<J,Lambda=1,Omega=1|AD|J,Lambda=1,Omega=0> = sqrt(2*J*(J+1))/2
<J,Lambda=1,Omega=1|AD|J,Lambda=1,Omega=2> = (-sqrt(2*(J*(J+1)-2)))/2
<J,Lambda=1,Omega=2|AD|J,Lambda=1,Omega=1> = (-sqrt(2*(J*(J+1)-2)))/2
<J,Lambda=1,Omega=2|AD|J,Lambda=1,Omega=2> = J*(J+1)-2
Symmetrized:
<J,Lambda=1,Omega=2|AD|J,Lambda=1,Omega=2> = J*(J+1)-2
<J,Lambda=1,Omega=2|AD|J,Lambda=1,Omega=2> = (-sqrt(2*(J*(J+1)-2)))/2
<J,Lambda=1,Omega=2|AD|J,Lambda=1,Omega=1> = (-sqrt(2*(J*(J+1)-2)))/2
<J,Lambda=1,Omega=2|AD|J,Lambda=1,Omega=1> = sqrt(2*J*(J+1))/2
<J,Lambda=1,Omega=1|AD|J,Lambda=1,Omega=2> = sqrt(2*J*(J+1))/2
<J,Lambda=1,Omega=1|AD|J,Lambda=1,Omega=2> = -J*(J+1)-2
<J,Lambda=1,Omega=1|AD|J,Lambda=1,Omega=0> = -J*(J+1)-2
<J,Lambda=1,Omega=1|AD|J,Lambda=1,Omega=0> = sqrt(2*J*(J+1))/2
<J,Lambda=1,Omega=0|AD|J,Lambda=1,Omega=1> = sqrt(2*J*(J+1))/2
<J,Lambda=1,Omega=0|AD|J,Lambda=1,Omega=1> = (-sqrt(2*(J*(J+1)-2)))/2
<J,Lambda=1,Omega=0|AD|J,Lambda=1,Omega=0> = (-sqrt(2*(J*(J+1)-2)))/2
<J,Lambda=1,Omega=0|AD|J,Lambda=1,Omega=0> = J*(J+1)-2

LinearPerturbation: Excited <a3Pv=13|LS|Iv=3>
Value -11.5278 Size of perturbation.
<J,Lambda=-1,Omega=0|LS|J,Lambda=0,Omega=0> = -1/sqrt(3)

<J,Lambda=1,Omega=0|LS|J,Lambda=0,Omega=0> = -1/sqrt(3)

Symmetrized Matrix Elements of <a3Pv=13|LS|Iv=3>
<J,Lambda=1,Omega=0|LS|J,Lambda=0,Omega=0> = -1/sqrt(3)

LinearPerturbation: Excited <a3Pv=13|LS|Iv=4>
Value -16.442 Size of perturbation.
<J,Lambda=-1,Omega=0|LS|J,Lambda=0,Omega=0> = -1/sqrt(3)
<J,Lambda=1,Omega=0|LS|J,Lambda=0,Omega=0> = -1/sqrt(3)

Symmetrized Matrix Elements of <a3Pv=13|LS|Iv=4>
<J,Lambda=1,Omega=0|LS|J,Lambda=0,Omega=0> = -1/sqrt(3)

LinearPerturbation: Excited <a3Pv=13|LS|ev=4>
Value 24.7288685468983 Float, Std Dev = 0.0197211631393428 Size of perturbation.
<J,Lambda=-1,Omega=-1|LS|J,Lambda=0,Omega=-1> = 1/sqrt(6)
<J,Lambda=-1,Omega=0|LS|J,Lambda=0,Omega=0> = 1/sqrt(6)
<J,Lambda=1,Omega=0|LS|J,Lambda=0,Omega=0> = -1/sqrt(6)
<J,Lambda=1,Omega=1|LS|J,Lambda=0,Omega=1> = -1/sqrt(6)

Symmetrized Matrix Elements of <a3Pv=13|LS|ev=4>
<J,Lambda=1,Omega=1|LS|J,Lambda=0,Omega=1> = 1/sqrt(6)
<J,Lambda=1,Omega=1|LS|J,Lambda=0,Omega=1> = 1/sqrt(6)
<J,Lambda=1,Omega=0|LS|J,Lambda=0,Omega=0> = -1/sqrt(6)

LinearPerturbation: Excited <a3Pv=13|J+L-|ev=4>
Value -0.0598221863162786 Float, Std Dev = 0.000255625514810117 Size of perturbation.
<J,Lambda=-1,Omega=-2|J+L-|J,Lambda=0,Omega=-1> = -sqrt(J*(J+1)-2)
<J,Lambda=-1,Omega=-1|J+L-|J,Lambda=0,Omega=0> = -sqrt(J*(J+1))
<J,Lambda=-1,Omega=0|J+L-|J,Lambda=0,Omega=1> = -sqrt(J*(J+1))
<J,Lambda=1,Omega=0|J+L-|J,Lambda=0,Omega=-1> = sqrt(J*(J+1))
<J,Lambda=1,Omega=1|J+L-|J,Lambda=0,Omega=0> = sqrt(J*(J+1))
<J,Lambda=1,Omega=2|J+L-|J,Lambda=0,Omega=1> = sqrt(J*(J+1)-2)

Symmetrized Matrix Elements of <a3Pv=13|J+L-|ev=4>
<J,Lambda=1,Omega=2|J+L-|J,Lambda=0,Omega=1> = -sqrt(J*(J+1)-2)
<J,Lambda=1,Omega=2|J+L-|J,Lambda=0,Omega=1> = -sqrt(J*(J+1))
<J,Lambda=1,Omega=1|J+L-|J,Lambda=0,Omega=0> = -sqrt(J*(J+1))
<J,Lambda=1,Omega=0|J+L-|J,Lambda=0,Omega=1> = sqrt(J*(J+1))
<J,Lambda=1,Omega=0|J+L-|J,Lambda=0,Omega=1> = sqrt(J*(J+1))

LinearPerturbation: Excited <a3Pv=13|LS|dv=7>
Value 23.2188654276278 Float, Std Dev = 0.0441226740988454 Size of perturbation.
<J,Lambda=-1,Omega=-2|LS|J,Lambda=-2,Omega=-2> = -1/sqrt(6)
<J,Lambda=-1,Omega=-1|LS|J,Lambda=-2,Omega=-1> = -1/sqrt(6)
<J,Lambda=1,Omega=1|LS|J,Lambda=2,Omega=1> = -1/sqrt(6)
<J,Lambda=1,Omega=2|LS|J,Lambda=2,Omega=2> = -1/sqrt(6)

Symmetrized Matrix Elements of <a3Pv=13|LS|dv=7>
<J,Lambda=1,Omega=2|LS|J,Lambda=2,Omega=2> = -1/sqrt(6)
<J,Lambda=1,Omega=2|LS|J,Lambda=2,Omega=2> = -1/sqrt(6)
<J,Lambda=1,Omega=1|LS|J,Lambda=2,Omega=1> = -1/sqrt(6)
<J,Lambda=1,Omega=1|LS|J,Lambda=2,Omega=1> = -1/sqrt(6)

LinearPerturbation: Excited <a3Pv=13|J+L-|dv=7>
Value -0.062795149954351 Float, Std Dev = 0.000453447699962488 Size of perturbation.
<J,Lambda=-1,Omega=-2|J+L-|J,Lambda=-2,Omega=-3> = sqrt(J*(J+1)-6)
<J,Lambda=-1,Omega=-1|J+L-|J,Lambda=-2,Omega=-2> = sqrt(J*(J+1)-2)
<J,Lambda=-1,Omega=0|J+L-|J,Lambda=-2,Omega=-1> = sqrt(J*(J+1))
<J,Lambda=1,Omega=0|J+L-|J,Lambda=2,Omega=1> = sqrt(J*(J+1))
<J,Lambda=1,Omega=1|J+L-|J,Lambda=2,Omega=2> = sqrt(J*(J+1)-2)
<J,Lambda=1,Omega=2|J+L-|J,Lambda=2,Omega=3> = sqrt(J*(J+1)-6)

Symmetrized Matrix Elements of <a3Pv=13|J+L-|dv=7>
<J,Lambda=1,Omega=2|J+L-|J,Lambda=2,Omega=3> = sqrt(J*(J+1)-6)
<J,Lambda=1,Omega=2|J+L-|J,Lambda=2,Omega=3> = sqrt(J*(J+1)-2)
<J,Lambda=1,Omega=1|J+L-|J,Lambda=2,Omega=2> = sqrt(J*(J+1))
<J,Lambda=1,Omega=1|J+L-|J,Lambda=2,Omega=2> = sqrt(J*(J+1))
<J,Lambda=1,Omega=0|J+L-|J,Lambda=2,Omega=1> = sqrt(J*(J+1)-2)
<J,Lambda=1,Omega=0|J+L-|J,Lambda=2,Omega=1> = sqrt(J*(J+1)-6)

LinearPerturbation: Excited <a3Pv=13|LS|a'v=12>
Value 2.39113881014602 Float, Std Dev = 0.203999893082939 Size of perturbation.
<J,Lambda=-1,Omega=-1|LS|J,Lambda=0,Omega=-1> = -1/sqrt(6)
<J,Lambda=-1,Omega=0|LS|J,Lambda=0,Omega=0> = -1/sqrt(6)
<J,Lambda=1,Omega=0|LS|J,Lambda=0,Omega=0> = -1/sqrt(6)
<J,Lambda=1,Omega=1|LS|J,Lambda=0,Omega=1> = -1/sqrt(6)

Symmetrized Matrix Elements of <a3Pv=13|LS|a'v=12>
<J,Lambda=1,Omega=1|LS|J,Lambda=0,Omega=1> = -1/sqrt(6)
<J,Lambda=1,Omega=1|LS|J,Lambda=0,Omega=1> = -1/sqrt(6)
<J,Lambda=1,Omega=0|LS|J,Lambda=0,Omega=0> = -1/sqrt(6)

LinearPerturbation: Excited <a3Pv=13|J+L-|a'v=12>
Value -0.0112468031812311 Float, Std Dev = 0.00125606908675765 Size of perturbation.
<J,Lambda=-1,Omega=-2|J+L-|J,Lambda=0,Omega=-1> = sqrt(J*(J+1)-2)
<J,Lambda=-1,Omega=-1|J+L-|J,Lambda=0,Omega=0> = sqrt(J*(J+1))
<J,Lambda=-1,Omega=0|J+L-|J,Lambda=0,Omega=1> = sqrt(J*(J+1))
<J,Lambda=1,Omega=0|J+L-|J,Lambda=0,Omega=-1> = sqrt(J*(J+1))
<J,Lambda=1,Omega=1|J+L-|J,Lambda=0,Omega=0> = sqrt(J*(J+1))
<J,Lambda=1,Omega=2|J+L-|J,Lambda=0,Omega=1> = sqrt(J*(J+1)-2)

```
  Symmetrized Matrix Elements of <a3Pv=13|J+L-|a'v=12>
  <J,Lambda=1,Omega=2|J+L-|J,Lambda=0,Omega=1> = sqrt(J*(J+1)-2)
  <J,Lambda=1,Omega=2|J+L-|J,Lambda=0,Omega=1> = sqrt(J*(J+1))
  <J,Lambda=1,Omega=1|J+L-|J,Lambda=0,Omega=0> = sqrt(J*(J+1))
  <J,Lambda=1,Omega=0|J+L-|J,Lambda=0,Omega=1> = sqrt(J*(J+1))
  <J,Lambda=1,Omega=0|J+L-|J,Lambda=0,Omega=1> = sqrt(J*(J+1))

  LinearPerturbation: Excited <a3Pv=13|LS|Dsv=3>
    Value -22.8568                    Size of perturbation.
  <J,Lambda=-1,Omega=-2|LS|J,Lambda=-2,Omega=-2> = 1/sqrt(3)
  <J,Lambda=1,Omega=2|LS|J,Lambda=2,Omega=2> = -1/sqrt(3)

  Symmetrized Matrix Elements of <a3Pv=13|LS|Dsv=3>
  <J,Lambda=1,Omega=2|LS|J,Lambda=2,Omega=2> = 1/sqrt(3)
  <J,Lambda=1,Omega=2|LS|J,Lambda=2,Omega=2> = -1/sqrt(3)

TransitionMoments: <Excited|mu|X>
  SphericalTransitionMoment: <Excited|mu|X> <Av=2|T(1)|Xv=0>
    Rank      1
    Component Auto => 1
    Strength 1                        Transition (dipole) moment / Debye - note that relative intensities are proportional to the square of this value.
  Reduced Matrix Elements of <Av=2|T(1)|Xv=0>
  <J-1,Lambda=-1,Omega=-1|T(1)|J,Lambda=0,Omega=0> = sqrt(2*(J-1))/2
  <J-1,Lambda=1,Omega=1|T(1)|J,Lambda=0,Omega=0> = sqrt(2*(J-1))/2
  <J,Lambda=-1,Omega=-1|T(1)|J,Lambda=0,Omega=0> = (2*J+1)/sqrt(2*(2*J+1))
  <J,Lambda=1,Omega=1|T(1)|J,Lambda=0,Omega=0> = (-2*J-1)/sqrt(2*(2*J+1))
  <J+1,Lambda=-1,Omega=-1|T(1)|J,Lambda=0,Omega=0> = sqrt(2*(J+2))/2
  <J+1,Lambda=1,Omega=1|T(1)|J,Lambda=0,Omega=0> = sqrt(2*(J+2))/2
  Line Strengths from <Av=2|T(1)|Xv=0>
  S<J-1,Lambda=-1,Omega=-1|T(1)|J,Lambda=0,Omega=0> = (J-1)/2
  S<J-1,Lambda=1,Omega=1|T(1)|J,Lambda=0,Omega=0> = (J-1)/2
  S<J,Lambda=-1,Omega=-1|T(1)|J,Lambda=0,Omega=0> = (2*J+1)/2
  S<J,Lambda=1,Omega=1|T(1)|J,Lambda=0,Omega=0> = (-2*J-1)/-2
  S<J+1,Lambda=-1,Omega=-1|T(1)|J,Lambda=0,Omega=0> = (J+2)/2
  S<J+1,Lambda=1,Omega=1|T(1)|J,Lambda=0,Omega=0> = (J+2)/2
  Symmetrized Reduced Matrix Elements of <Av=2|T(1)|Xv=0>
  <J-1,Lambda=1,Omega=1 + |T(1)|J,Lambda=0,Omega=0> = sqrt(J-1)
  <J,Lambda=1,Omega=1 - |T(1)|J,Lambda=0,Omega=0> = (-2*J-1)/sqrt(2*J+1)
  <J+1,Lambda=1,Omega=1 + |T(1)|J,Lambda=0,Omega=0> = sqrt(J+2)
  Symmetrized Line Strengths from <Av=2|T(1)|Xv=0>
  S<J-1,Lambda=1,Omega=1 + |T(1)|J,Lambda=0,Omega=0> = J-1
  S<J,Lambda=1,Omega=1 - |T(1)|J,Lambda=0,Omega=0> = (-2*J-1)/-1
  S<J+1,Lambda=1,Omega=1 + |T(1)|J,Lambda=0,Omega=0> = J+2

  ForbiddenTransitionMoment: <Excited|mu|X> <a3Pv=13|fT(1,1)|Xv=0>
    Rank      1
    Component 1
    Magnetic  False
    Strength 1                        Transition (dipole) moment / Debye - note that relative intensities are proportional to the square of this value.
  Reduced Matrix Elements of <a3Pv=13|fT(1,1)|Xv=0>
  <J-1,Lambda=-1,Omega=-1|fT(1,1)|J,Lambda=0,Omega=0> = (-sqrt(2*(J-1)))/2
  <J-1,Lambda=1,Omega=1|fT(1,1)|J,Lambda=0,Omega=0> = sqrt(2*(J-1))/2
  <J,Lambda=-1,Omega=-1|fT(1,1)|J,Lambda=0,Omega=0> = (-2*J-1)/sqrt(2*(2*J+1))
  <J,Lambda=1,Omega=1|fT(1,1)|J,Lambda=0,Omega=0> = (-2*J-1)/sqrt(2*(2*J+1))
  <J+1,Lambda=-1,Omega=-1|fT(1,1)|J,Lambda=0,Omega=0> = (-sqrt(2*(J+2)))/2
  <J+1,Lambda=1,Omega=1|fT(1,1)|J,Lambda=0,Omega=0> = sqrt(2*(J+2))/2
  Line Strengths from <a3Pv=13|fT(1,1)|Xv=0>
  S<J-1,Lambda=-1,Omega=-1|fT(1,1)|J,Lambda=0,Omega=0> = (J-1)/2
  S<J-1,Lambda=1,Omega=1|fT(1,1)|J,Lambda=0,Omega=0> = (J-1)/2
  S<J,Lambda=-1,Omega=-1|fT(1,1)|J,Lambda=0,Omega=0> = (-2*J-1)/-2
  S<J,Lambda=1,Omega=1|fT(1,1)|J,Lambda=0,Omega=0> = (-2*J-1)/-2
  S<J+1,Lambda=-1,Omega=-1|fT(1,1)|J,Lambda=0,Omega=0> = (J+2)/2
  S<J+1,Lambda=1,Omega=1|fT(1,1)|J,Lambda=0,Omega=0> = (J+2)/2
  Symmetrized Reduced Matrix Elements of <a3Pv=13|fT(1,1)|Xv=0>
  <J-1,Lambda=1,Omega=1 - |fT(1,1)|J,Lambda=0,Omega=0> = sqrt(J-1)
  <J,Lambda=1,Omega=1 + |fT(1,1)|J,Lambda=0,Omega=0> = (-2*J-1)/sqrt(2*J+1)
  <J+1,Lambda=1,Omega=1 - |fT(1,1)|J,Lambda=0,Omega=0> = sqrt(J+2)
  Symmetrized Line Strengths from <a3Pv=13|fT(1,1)|Xv=0>
  S<J-1,Lambda=1,Omega=1 - |fT(1,1)|J,Lambda=0,Omega=0> = J-1
  S<J,Lambda=1,Omega=1 + |fT(1,1)|J,Lambda=0,Omega=0> = (-2*J-1)/-1
  S<J+1,Lambda=1,Omega=1 - |fT(1,1)|J,Lambda=0,Omega=0> = J+2

  ForbiddenTransitionMoment: <Excited|mu|X> <dv=7|fT(1,1)|Xv=0>
    Rank      1
    Component 1
    Magnetic  False
    Strength 1                        Transition (dipole) moment / Debye - note that relative intensities are proportional to the square of this value.
  Reduced Matrix Elements of <dv=7|fT(1,1)|Xv=0>
  <J-1,Lambda=-2,Omega=-1|fT(1,1)|J,Lambda=0,Omega=0> = (-sqrt(2*(J-1)))/2
  <J-1,Lambda=2,Omega=1|fT(1,1)|J,Lambda=0,Omega=0> = sqrt(2*(J-1))/2
  <J,Lambda=-2,Omega=-1|fT(1,1)|J,Lambda=0,Omega=0> = (-2*J-1)/sqrt(2*(2*J+1))
  <J,Lambda=2,Omega=1|fT(1,1)|J,Lambda=0,Omega=0> = (-2*J-1)/sqrt(2*(2*J+1))
  <J+1,Lambda=-2,Omega=-1|fT(1,1)|J,Lambda=0,Omega=0> = (-sqrt(2*(J+2)))/2
  <J+1,Lambda=2,Omega=1|fT(1,1)|J,Lambda=0,Omega=0> = sqrt(2*(J+2))/2
  Line Strengths from <dv=7|fT(1,1)|Xv=0>
  S<J-1,Lambda=-2,Omega=-1|fT(1,1)|J,Lambda=0,Omega=0> = (J-1)/2
  S<J-1,Lambda=2,Omega=1|fT(1,1)|J,Lambda=0,Omega=0> = (J-1)/2
  S<J,Lambda=-2,Omega=-1|fT(1,1)|J,Lambda=0,Omega=0> = (-2*J-1)/-2
  S<J,Lambda=2,Omega=1|fT(1,1)|J,Lambda=0,Omega=0> = (-2*J-1)/-2
  S<J+1,Lambda=-2,Omega=-1|fT(1,1)|J,Lambda=0,Omega=0> = (J+2)/2
  S<J+1,Lambda=2,Omega=1|fT(1,1)|J,Lambda=0,Omega=0> = (J+2)/2
  Symmetrized Reduced Matrix Elements of <dv=7|fT(1,1)|Xv=0>
```

<J-1,Lambda=2,Omega=1 - |fT(1,1)|J,Lambda=0,Omega=0> = sqrt(J-1)
<J,Lambda=2,Omega=1 + |fT(1,1)|J,Lambda=0,Omega=0> = (-2*J-1)/sqrt(2*J+1)
<J+1,Lambda=2,Omega=1 - |fT(1,1)|J,Lambda=0,Omega=0> = sqrt(J+2)
Symmetrized Line Strengths from <dv=7|fT(1,1)|Xv=0>
S<J-1,Lambda=2,Omega=1 - |fT(1,1)|J,Lambda=0,Omega=0> = J-1
S<J,Lambda=2,Omega=1 + |fT(1,1)|J,Lambda=0,Omega=0> = (-2*J-1)/-1
S<J+1,Lambda=2,Omega=1 - |fT(1,1)|J,Lambda=0,Omega=0> = J+2

ForbiddenTransitionMoment: <Excited|mu|X> <a'v=12|fT(1,1)|Xv=0>
Rank 1
Component 1
Magnetic False
Strength 1 Transition (dipole) moment / Debye - note that relative intensities are proportional to the square of this value.
Reduced Matrix Elements of <a'v=12|fT(1,1)|Xv=0>
<J-1,Lambda=0,Omega=-1|fT(1,1)|J,Lambda=0,Omega=0> = (-sqrt(2*(J-1)))/2
<J-1,Lambda=0,Omega=1|fT(1,1)|J,Lambda=0,Omega=0> = sqrt(2*(J-1))/2
<J,Lambda=0,Omega=-1|fT(1,1)|J,Lambda=0,Omega=0> = (-2*J-1)/sqrt(2*(2*J+1))
<J,Lambda=0,Omega=1|fT(1,1)|J,Lambda=0,Omega=0> = (-2*J-1)/sqrt(2*(2*J+1))
<J+1,Lambda=0,Omega=-1|fT(1,1)|J,Lambda=0,Omega=0> = (-sqrt(2*(J+2)))/2
<J+1,Lambda=0,Omega=1|fT(1,1)|J,Lambda=0,Omega=0> = sqrt(2*(J+2))/2
Line Strengths from <a'v=12|fT(1,1)|Xv=0>
S<J-1,Lambda=0,Omega=-1|fT(1,1)|J,Lambda=0,Omega=0> = (J-1)/2
S<J-1,Lambda=0,Omega=1|fT(1,1)|J,Lambda=0,Omega=0> = (J-1)/2
S<J,Lambda=0,Omega=-1|fT(1,1)|J,Lambda=0,Omega=0> = (-2*J-1)/-2
S<J,Lambda=0,Omega=1|fT(1,1)|J,Lambda=0,Omega=0> = (-2*J-1)/-2
S<J+1,Lambda=0,Omega=-1|fT(1,1)|J,Lambda=0,Omega=0> = (J+2)/2
S<J+1,Lambda=0,Omega=1|fT(1,1)|J,Lambda=0,Omega=0> = (J+2)/2
Symmetrized Reduced Matrix Elements of <a'v=12|fT(1,1)|Xv=0>
<J-1,Lambda=0,Omega=1 - |fT(1,1)|J,Lambda=0,Omega=0> = sqrt(J-1)
<J,Lambda=0,Omega=1 + |fT(1,1)|J,Lambda=0,Omega=0> = (-2*J-1)/sqrt(2*J+1)
<J+1,Lambda=0,Omega=1 - |fT(1,1)|J,Lambda=0,Omega=0> = sqrt(J+2)
Symmetrized Line Strengths from <a'v=12|fT(1,1)|Xv=0>
S<J-1,Lambda=0,Omega=1 - |fT(1,1)|J,Lambda=0,Omega=0> = J-1
S<J,Lambda=0,Omega=1 + |fT(1,1)|J,Lambda=0,Omega=0> = (-2*J-1)/-1
S<J+1,Lambda=0,Omega=1 - |fT(1,1)|J,Lambda=0,Omega=0> = J+2

ForbiddenTransitionMoment: <Excited|mu|X> <ev=4|fT(1,1)|Xv=0>
Rank 1
Component 1
Magnetic False
Strength 1 Transition (dipole) moment / Debye - note that relative intensities are proportional to the square of this value.
Reduced Matrix Elements of <ev=4|fT(1,1)|Xv=0>
<J-1,Lambda=0,Omega=-1|fT(1,1)|J,Lambda=0,Omega=0> = sqrt(2*(J-1))/2
<J-1,Lambda=0,Omega=1|fT(1,1)|J,Lambda=0,Omega=0> = sqrt(2*(J-1))/2
<J,Lambda=0,Omega=-1|fT(1,1)|J,Lambda=0,Omega=0> = (2*J+1)/sqrt(2*(2*J+1))
<J,Lambda=0,Omega=1|fT(1,1)|J,Lambda=0,Omega=0> = (-2*J-1)/sqrt(2*(2*J+1))
<J+1,Lambda=0,Omega=-1|fT(1,1)|J,Lambda=0,Omega=0> = sqrt(2*(J+2))/2
<J+1,Lambda=0,Omega=1|fT(1,1)|J,Lambda=0,Omega=0> = sqrt(2*(J+2))/2
Line Strengths from <ev=4|fT(1,1)|Xv=0>
S<J-1,Lambda=0,Omega=-1|fT(1,1)|J,Lambda=0,Omega=0> = (J-1)/2
S<J-1,Lambda=0,Omega=1|fT(1,1)|J,Lambda=0,Omega=0> = (J-1)/2
S<J,Lambda=0,Omega=-1|fT(1,1)|J,Lambda=0,Omega=0> = (2*J+1)/2
S<J,Lambda=0,Omega=1|fT(1,1)|J,Lambda=0,Omega=0> = (-2*J-1)/-2
S<J+1,Lambda=0,Omega=-1|fT(1,1)|J,Lambda=0,Omega=0> = (J+2)/2
S<J+1,Lambda=0,Omega=1|fT(1,1)|J,Lambda=0,Omega=0> = (J+2)/2
Symmetrized Reduced Matrix Elements of <ev=4|fT(1,1)|Xv=0>
<J-1,Lambda=0,Omega=1 + |fT(1,1)|J,Lambda=0,Omega=0> = sqrt(J-1)
<J,Lambda=0,Omega=1 - |fT(1,1)|J,Lambda=0,Omega=0> = (-2*J-1)/sqrt(2*J+1)
<J+1,Lambda=0,Omega=1 + |fT(1,1)|J,Lambda=0,Omega=0> = sqrt(J+2)
Symmetrized Line Strengths from <ev=4|fT(1,1)|Xv=0>
S<J-1,Lambda=0,Omega=1 + |fT(1,1)|J,Lambda=0,Omega=0> = J-1
S<J,Lambda=0,Omega=1 - |fT(1,1)|J,Lambda=0,Omega=0> = (-2*J-1)/-1
S<J+1,Lambda=0,Omega=1 + |fT(1,1)|J,Lambda=0,Omega=0> = J+2

ForbiddenTransitionMoment: <Excited|mu|X> <Iv=3|fT(1,1)|Xv=0>
Rank 1
Component 1
Magnetic False
Strength 1 Transition (dipole) moment / Debye - note that relative intensities are proportional to the square of this value.
ForbiddenTransitionMoment: <Excited|mu|X> <Dsv=3|fT(1,1)|Xv=0>
Rank 1
Component 1
Magnetic False
Strength 1 Transition (dipole) moment / Debye - note that relative intensities are proportional to the square of this value.
TransitionMoments: <B|mu|X>
SphericalTransitionMoment: <B|mu|X> <Bv=0|T(1)|Xv=0>
Rank 1
Component Auto => 0
Strength 1 Transition (dipole) moment / Debye - note that relative intensities are proportional to the square of this value.
Reduced Matrix Elements of <Bv=0|T(1)|Xv=0>
<J-1,Lambda=0,Omega=0|T(1)|J,Lambda=0,Omega=0> = -sqrt(J)
<J+1,Lambda=0,Omega=0|T(1)|J,Lambda=0,Omega=0> = (J+1)/sqrt(J+1)
Line Strengths from <Bv=0|T(1)|Xv=0>
S<J-1,Lambda=0,Omega=0|T(1)|J,Lambda=0,Omega=0> = J
S<J+1,Lambda=0,Omega=0|T(1)|J,Lambda=0,Omega=0> = J+1
Symmetrized Reduced Matrix Elements of <Bv=0|T(1)|Xv=0>
<J-1,Lambda=0,Omega=0|T(1)|J,Lambda=0,Omega=0> = -sqrt(J)
<J+1,Lambda=0,Omega=0|T(1)|J,Lambda=0,Omega=0> = (J+1)/sqrt(J+1)
Symmetrized Line Strengths from <Bv=0|T(1)|Xv=0>
S<J-1,Lambda=0,Omega=0|T(1)|J,Lambda=0,Omega=0> = J
S<J+1,Lambda=0,Omega=0|T(1)|J,Lambda=0,Omega=0> = J+1

```
    TransitionMoments: <B|mu|Excited>
      SphericalTransitionMoment: <B|mu|Excited> <Bv=0|T(1)|Av=2>
        Rank     1
        Component Auto => 1
        Strength 1                    Transition (dipole) moment / Debye - note that relative intensities are proportional to the square of this value.
      Reduced Matrix Elements of <Bv=0|T(1)|Av=2>
      <J-1,Lambda=0,Omega=0|T(1)|J,Lambda=-1,Omega=-1> = sqrt(2*(J+1))/2
      <J-1,Lambda=0,Omega=0|T(1)|J,Lambda=1,Omega=1> = sqrt(2*(J+1))/2
      <J,Lambda=0,Omega=0|T(1)|J,Lambda=-1,Omega=-1> = (-2*J-1)/sqrt(2*(2*J+1))
      <J,Lambda=0,Omega=0|T(1)|J,Lambda=1,Omega=1> = (2*J+1)/sqrt(2*(2*J+1))
      <J+1,Lambda=0,Omega=0|T(1)|J,Lambda=-1,Omega=-1> = sqrt(2*J)/2
      <J+1,Lambda=0,Omega=0|T(1)|J,Lambda=1,Omega=1> = sqrt(2*J)/2
      Line Strengths from <Bv=0|T(1)|Av=2>
      S<J-1,Lambda=0,Omega=0|T(1)|J,Lambda=-1,Omega=-1> = (J+1)/2
      S<J-1,Lambda=0,Omega=0|T(1)|J,Lambda=1,Omega=1> = (J+1)/2
      S<J,Lambda=0,Omega=0|T(1)|J,Lambda=-1,Omega=-1> = (-2*J-1)/-2
      S<J,Lambda=0,Omega=0|T(1)|J,Lambda=1,Omega=1> = (2*J+1)/2
      S<J+1,Lambda=0,Omega=0|T(1)|J,Lambda=-1,Omega=-1> = J/2
      S<J+1,Lambda=0,Omega=0|T(1)|J,Lambda=1,Omega=1> = J/2
      Symmetrized Reduced Matrix Elements of <Bv=0|T(1)|Av=2>
      <J-1,Lambda=0,Omega=0|T(1)| + J,Lambda=1,Omega=1> = sqrt(J+1)
      <J,Lambda=0,Omega=0|T(1)| - J,Lambda=1,Omega=1> = (2*J+1)/sqrt(2*J+1)
      <J+1,Lambda=0,Omega=0|T(1)| + J,Lambda=1,Omega=1> = sqrt(J)
      Symmetrized Line Strengths from <Bv=0|T(1)|Av=2>
      S<J-1,Lambda=0,Omega=0|T(1)| + J,Lambda=1,Omega=1> = J+1
      S<J,Lambda=0,Omega=0|T(1)| - J,Lambda=1,Omega=1> = 2*J+1
      S<J+1,Lambda=0,Omega=0|T(1)| + J,Lambda=1,Omega=1> = J

    TransitionMoments: <C|mu|X>
      SphericalTransitionMoment: <C|mu|X> <Cv=0|T(1)|Xv=0>
        Rank     1
        Component Auto => 0
        Strength 1                    Transition (dipole) moment / Debye - note that relative intensities are proportional to the square of this value.
      Matrix elements as for <B|mu|X> <Bv=0|T(1)|Xv=0>
    TransitionMoments: <C|mu|Excited>
      SphericalTransitionMoment: <C|mu|Excited> <Cv=0|T(1)|Av=2>
        Rank     1
        Component Auto => 1
        Strength 1                    Transition (dipole) moment / Debye - note that relative intensities are proportional to the square of this value.
      Matrix elements as for <B|mu|Excited> <Bv=0|T(1)|Av=2>
  Simulation: Simulation
    Intensity factor:0.333333333333333 (1 from Spol)
    IntensityUnits Arbitrary
    LifeModel     lmNone
    PlotUnits     cm1
    Fluorescence  False
    nDF           50000
    WidthMult     6
    AutoMin       True
    AutoMax       True
    ShowSum       False
    ShowParts     True
    ShowFortrat   True
    UseUpper      False
    ShowSymmetry  False
    ShowDeltaJ    True
    ScaleMarkSize True
    UseSymmetry   False
    UseStateNumber False
    FortratQno    J
    PreserveArea  False
    SeparateScales False
    LogScale      False
    Show2DPlot    False
    Auto2DY       True
    Fmin       19713.4264274956        Left edge of plot range.
    Fmax       19725.7352925044        Right edge of plot range.
    Ymax       7.90431110974814E-183        Top of plot range; updated automatically if AutoMax set.
    Temperature 300                    Rotational temeperature (Kelvin).
    Gaussian    0.08                   Gaussian contribution to linewidth (full width half maximum). If both this and a Lorentzian width (below) is set the result is a convolution of the two, a Voigt
profile.
    Lorentzian  0.02                   Lorentzian contribution to linewidth (full width half maximum). This and the Gaussian width can also be set from the main window.
    SMargin     -1                     Extra points to calculate off each end of required spectrum.
    OThreshold  0.0001                 Ignore peaks smaller than this fraction of the maximum peak intensity.
    Tvib        -1                     Vibrational Temperature (Kelvin); set to -1 (default) to use rotational temperature for all Boltzmann factors.
    MinI        1E-15                  Weakest transitions to plot in Fortrat plots, as a fraction of strongest.
    Tspin       -1                     Spin Temperature; set to -1 (default) to equilibrate different spinstates
    IScale      1                      Scale factor applied to all line intensities
    PlotSplit   0.5                    Fraction of plot area to use for simulation where plots separated (-1..1)
  Fundamental Constants from CODATA 2018: h=6.62607015E-34 k=1.380649E-23 NA=6.02214076E23 e=1.602176634E-19 me=9.1093837015E-31

PGOPHER 11.0.142 1 Apr 2021 8:30 64 bit Unicode (Delphi 33/33)
   pgopher 44b8c50a90149a6ce53e669f4a0e44afc86f1abd
    lutils 65f951dfa553c7f9bc501632d3360a85cd1a0165
     utils ec0cac7863199e65355ae4c6ab78dbe3037fae3b
    lforms d4dfa0031586b0cf5ffd76ff4d8f7aee2a9f8932
    wutils 9bb7abdf4436dd4d7dfed61ce7fb7c6bc914386c
 GraphicEx bfb0df9494191c3a10d6089c21fe439fcd763090
Mixture: 13C18O Deperturbation of the A(v=2) level.pgo
  Units           cm1
  PrintLevel      LineList
  Precision       4
```

```
BasisOrder          boDefault
QuantumNumberFormat J
FitCycles           1
FitDetails          False
AutoReplot          True
AllowComplex        True
NoLineList          False
ShowEstUnc          False
ShowDerivatives     False
SVDlimit            5
ShowSubBasis        False
AllowZeroI          False
ScaledParameters    True
FixTransitionPhase  False
OldHyperfineOrder   False
OldColourOrder      False
SingleResidualPlot  False
UpperStateColour    False
SVDThresh    1E-6                      Discard Singular Value Decomposition eigenvalues below this fraction of the largest eigenvalue; <0 to use QR factorisation.
ScaleChanges 1                        Scale changes produced by fitting by this - normally 1
SmallE       2E-18                    Matrix elements smaller than this are discarded
Species: CO
  Jmax 63
  Concentration 1                      Relative concentration of species
  LinearMolecule: 13C18O
    JAdjustSym  True
    BlockMatrix True
    AbsoluteE   False
    Symmetric   False
    SymWt       1
    AsymWt      1
    RSquaredH   False
    ShowJ       True
    ShowOmega   False
    ShowN       True
    ShowFNumber True
    Showef      True
    Abundance     1                     Abundance of this isotope; default is 1
    AssumedOrigin 1E99                       Energy origin for Boltzmann calculations; 1e99 implies automatic estimate. Currently 0
    LinearManifold: X

      Initial          True
      EigenSearch      True
      LimitSearch      True
      AutoQConverge    True
      UsePopParams     False
      OverrideEnergies True
      Linear: X Xv=0
        Status     Active
        Lambda     Sigma+
        OmegaOrder Auto => Regular
        B  1.746408199                Rotational constant.
        D  5.0488146E-6                Quartic centrifugal distortion.
        H  4.35471E-12                Sextic centrifugal distortion.
        L  -2.4821E-17               J^8  centrifugal distortion.
        M  2.99E-23                  J^10  centrifugal distortion.
        PP -4.5E-28                  J^12 centrifugal distortion. (Not P to avoid confusion with p.)
        <J,Lambda=0,Omega=0|B|J,Lambda=0,Omega=0> = J*(J+1)
          Symmetrized:
        <J,Lambda=0,Omega=0|B|J,Lambda=0,Omega=0> = J*(J+1)

        <J,Lambda=0,Omega=0|D|J,Lambda=0,Omega=0> = -J^2*(J*(J+2)+1)
          Symmetrized:
        <J,Lambda=0,Omega=0|D|J,Lambda=0,Omega=0> = -J^2*(J*(J+2)+1)

        <J,Lambda=0,Omega=0|H|J,Lambda=0,Omega=0> = J^3*(J*(J*(J+3)+3)+1)
          Symmetrized:
        <J,Lambda=0,Omega=0|H|J,Lambda=0,Omega=0> = J^3*(J*(J*(J+3)+3)+1)

        <J,Lambda=0,Omega=0|L|J,Lambda=0,Omega=0> = J^4*(J*(J*(J*(J+4)+6)+4)+1)
          Symmetrized:
        <J,Lambda=0,Omega=0|L|J,Lambda=0,Omega=0> = J^4*(J*(J*(J*(J+4)+6)+4)+1)

        <J,Lambda=0,Omega=0|M|J,Lambda=0,Omega=0> = J^5*(J*(J*(J*(J*(J+5)+10)+10)+5)+1)
          Symmetrized:
        <J,Lambda=0,Omega=0|M|J,Lambda=0,Omega=0> = J^5*(J*(J*(J*(J*(J+5)+10)+10)+5)+1)

        <J,Lambda=0,Omega=0|PP|J,Lambda=0,Omega=0> = J^6*(J*(J*(J*(J*(J*(J+6)+15)+20)+15)+6)+1)
          Symmetrized:
        <J,Lambda=0,Omega=0|PP|J,Lambda=0,Omega=0> = J^6*(J*(J*(J*(J*(J*(J+6)+15)+20)+15)+6)+1)

      LinearPerturbation: X <Xv=0|N^14|Xv=0>
        O (rotational constant)
        Value -2.2E-33                 Size of perturbation.
      <J,Lambda=0,Omega=0|N^14|J,Lambda=0,Omega=0> = J^7*(J*(J*(J*(J*(J*(J*(J+7)+21)+35)+35)+21)+7)+1)

      Symmetrized Matrix Elements of <Xv=0|N^14|Xv=0>
      <J,Lambda=0,Omega=0|N^14|J,Lambda=0,Omega=0> = J^7*(J*(J*(J*(J*(J*(J*(J+7)+21)+35)+35)+21)+7)+1)

    LinearManifold: C
      Initial          True
      EigenSearch      True
```

```
  LimitSearch      False
  AutoQConverge    True
  UsePopParams     False
  OverrideEnergies True
  E 0 e 1  91918.81461769            Energy for J=E 0 e 1 level:C Cv=0  0  0 F1e
  E 1 e 1  91922.3501979315          Energy for J=E 1 e 1 level:C Cv=0  1  1 F1e
  E 2 e 1  91929.41460931            Energy for J=E 2 e 1 level:C Cv=0  2  2 F1e
  E 3 e 1  91940.0087911758          Energy for J=E 3 e 1 level:C Cv=0  3  3 F1e
  E 4 e 1  91954.1391754158          Energy for J=E 4 e 1 level:C Cv=0  4  4 F1e
  E 5 e 1  91971.8047483651          Energy for J=E 5 e 1 level:C Cv=0  5  5 F1e
  E 6 e 1  91992.9963415625          Energy for J=E 6 e 1 level:C Cv=0  6  6 F1e
  E 7 e 1  92017.7210158785          Energy for J=E 7 e 1 level:C Cv=0  7  7 F1e
  E 8 e 1  92045.9729147246          Energy for J=E 8 e 1 level:C Cv=0  8  8 F1e
  E 9 e 1  92077.7548931101          Energy for J=E 9 e 1 level:C Cv=0  9  9 F1e
  E 10 e 1 92113.0587701987          Energy for J=E 10 e 1 level:C Cv=0 10 10 F1e
  E 11 e 1 92151.8966952636          Energy for J=E 11 e 1 level:C Cv=0 11 11 F1e
  E 12 e 1 92194.2576663392          Energy for J=E 12 e 1 level:C Cv=0 12 12 F1e
  E 13 e 1 92240.1328568783          Energy for J=E 13 e 1 level:C Cv=0 13 13 F1e
  E 14 e 1 92289.5374084555          Energy for J=E 14 e 1 level:C Cv=0 14 14 F1e
  E 15 e 1 92342.4615531966          Energy for J=E 15 e 1 level:C Cv=0 15 15 F1e
  E 16 e 1 92398.9025159008          Energy for J=E 16 e 1 level:C Cv=0 16 16 F1e
  E 17 e 1 92458.8587353746          Energy for J=E 17 e 1 level:C Cv=0 17 17 F1e
  E 18 e 1 92522.3290731028          Energy for J=E 18 e 1 level:C Cv=0 18 18 F1e
  E 19 e 1 92589.3126987891          Energy for J=E 19 e 1 level:C Cv=0 19 19 F1e
  E 20 e 1 92659.8038563447          Energy for J=E 20 e 1 level:C Cv=0 20 20 F1e
  E 21 e 1 92733.8006230299          Energy for J=E 21 e 1 level:C Cv=0 21 21 F1e
  E 22 e 1 92811.3031429715          Energy for J=E 22 e 1 level:C Cv=0 22 22 F1e
  E 23 e 1 92892.3079897717          Energy for J=E 23 e 1 level:C Cv=0 23 23 F1e
  E 24 e 1 92976.822314948           Energy for J=E 24 e 1 level:C Cv=0 24 24 F1e
  E 25 e 1 93064.8231465903          Energy for J=E 25 e 1 level:C Cv=0 25 25 F1e
  E 26 e 1 93156.3176170672          Energy for J=E 26 e 1 level:C Cv=0 26 26 F1e
  E 27 e 1 93251.2988822767          Energy for J=E 27 e 1 level:C Cv=0 27 27 F1e
  E 28 e 1 93349.7667497678          Energy for J=E 28 e 1 level:C Cv=0 28 28 F1e
  E 29 e 1 93451.7189577959          Energy for J=E 29 e 1 level:C Cv=0 29 29 F1e
  E 30 e 1 93557.148271242           Energy for J=E 30 e 1 level:C Cv=0 30 30 F1e
  E 31 e 1 93666.0549149757          Energy for J=E 31 e 1 level:C Cv=0 31 31 F1e
  E 32 e 1 93778.4343213518          Energy for J=E 32 e 1 level:C Cv=0 32 32 F1e
  E 33 e 1 93894.2815297818          Energy for J=E 33 e 1 level:C Cv=0 33 33 F1e
  E 34 e 1 94013.591760117           Energy for J=E 34 e 1 level:C Cv=0 34 34 F1e
  E 35 e 1 94136.3569127458          Energy for J=E 35 e 1 level:C Cv=0 35 35 F1e
  E 36 e 1 94262.5809686937          Energy for J=E 36 e 1 level:C Cv=0 36 36 F1e
  E 37 e 1 94392.253689726           Energy for J=E 37 e 1 level:C Cv=0 37 37 F1e
  E 38 e 1 94525.3757684531          Energy for J=E 38 e 1 level:C Cv=0 38 38 F1e
  E 39 e 1 94661.9366784378          Energy for J=E 39 e 1 level:C Cv=0 39 39 F1e
  E 40 e 1 94801.9378743053          Energy for J=E 40 e 1 level:C Cv=0 40 40 F1e
  E 41 e 1 94945.3716418562          Energy for J=E 41 e 1 level:C Cv=0 41 41 F1e
  E 42 e 1 95092.2322481804          Energy for J=E 42 e 1 level:C Cv=0 42 42 F1e
  E 43 e 1 95242.507641775           Energy for J=E 43 e 1 level:C Cv=0 43 43 F1e
  E 44 e 1 95396.2059026634          Energy for J=E 44 e 1 level:C Cv=0 44 44 F1e
  E 45 e 1 95553.3147425169          Energy for J=E 45 e 1 level:C Cv=0 45 45 F1e
  E 46 e 1 95713.8389547786          Energy for J=E 46 e 1 level:C Cv=0 46 46 F1e
  E 47 e 1 95877.7430647896          Energy for J=E 47 e 1 level:C Cv=0 47 47 F1e
  E 48 e 1 96045.0535799167          Energy for J=E 48 e 1 level:C Cv=0 48 48 F1e
  E 49 e 1 96215.7478896832          Energy for J=E 49 e 1 level:C Cv=0 49 49 F1e
  E 50 e 1 96389.8321659005          Energy for J=E 50 e 1 level:C Cv=0 50 50 F1e
  E 51 e 1 96567.2766128028          Energy for J=E 51 e 1 level:C Cv=0 51 51 F1e
  E 52 e 1 96748.0881671828          Energy for J=E 52 e 1 level:C Cv=0 52 52 F1e
  E 53 e 1 96932.2837485302          Energy for J=E 53 e 1 level:C Cv=0 53 53 F1e
  E 54 e 1 97119.8371091712          Energy for J=E 54 e 1 level:C Cv=0 54 54 F1e
  E 55 e 1 97310.7208844101          Energy for J=E 55 e 1 level:C Cv=0 55 55 F1e
  E 56 e 1 97504.930642673           Energy for J=E 56 e 1 level:C Cv=0 56 56 F1e
  E 57 e 1 97702.5267856525          Energy for J=E 57 e 1 level:C Cv=0 57 57 F1e
  E 58 e 1 97903.4396484547          Energy for J=E 58 e 1 level:C Cv=0 58 58 F1e
  E 59 e 1 98107.6243997471          Energy for J=E 59 e 1 level:C Cv=0 59 59 F1e
  E 60 e 1 98315.1990419085          Energy for J=E 60 e 1 level:C Cv=0 60 60 F1e
  E 61 e 1 98526.0359657192          Energy for J=E 61 e 1 level:C Cv=0 61 61 F1e
  E 62 e 1 98740.1821395715          Energy for J=E 62 e 1 level:C Cv=0 62 62 F1e
  E 63 e 1 98957.6207414145          Energy for J=E 63 e 1 level:C Cv=0 63 63 F1e
  E 64 e 1 99178.3440172057          Energy for J=E 64 e 1 level
  E 65 e 1 99402.3440898222          Energy for J=E 65 e 1 level
  Linear: C Cv=0
    Status     Active
    Lambda     Sigma+
    OmegaOrder Auto => Regular
    Origin 91918.79                  State Origin. Estimated lowest level = 91918.79
    B     1.76641191315359           Rotational constant.
    D     5.12833508732762E-6        Quartic centrifugal distortion.
    <J,Lambda=0,Omega=0|B|J,Lambda=0,Omega=0> = J*(J+1)
      Symmetrized:
    <J,Lambda=0,Omega=0|B|J,Lambda=0,Omega=0> = J*(J+1)

    <J,Lambda=0,Omega=0|D|J,Lambda=0,Omega=0> = -J^2*(J*(J+2)+1)
      Symmetrized:
    <J,Lambda=0,Omega=0|D|J,Lambda=0,Omega=0> = -J^2*(J*(J+2)+1)

LinearManifold: B
  Initial          True
  EigenSearch      True
  LimitSearch      False
  AutoQConverge    True
  UsePopParams     False
  OverrideEnergies True
```

```
E 0 e 1  86917.3573438145           Energy for J=E 0 e 1 level:B Bv=0  0  0 F1e
E 1 e 1  86920.8990798371           Energy for J=E 1 e 1 level:B Bv=0  1  1 F1e
E 2 e 1  86927.9802287834           Energy for J=E 2 e 1 level:B Bv=0  2  2 F1e
E 3 e 1  86938.5975955474           Energy for J=E 3 e 1 level:B Bv=0  3  3 F1e
E 4 e 1  86952.7535165513           Energy for J=E 4 e 1 level:B Bv=0  4  4 F1e
E 5 e 1  86970.4485900942           Energy for J=E 5 e 1 level:B Bv=0  5  5 F1e
E 6 e 1  86991.6808705965           Energy for J=E 6 e 1 level:B Bv=0  6  6 F1e
E 7 e 1  87016.4501212347           Energy for J=E 7 e 1 level:B Bv=0  7  7 F1e
E 8 e 1  87044.7550139408           Energy for J=E 8 e 1 level:B Bv=0  8  8 F1e
E 9 e 1  87076.5945102899           Energy for J=E 9 e 1 level:B Bv=0  9  9 F1e
E 10 e 1 87111.9673665656           Energy for J=E 10 e 1 level:B Bv=0 10 10 F1e
E 11 e 1 87150.8722462898           Energy for J=E 11 e 1 level:B Bv=0 11 11 F1e
E 12 e 1 87193.3103335359           Energy for J=E 12 e 1 level:B Bv=0 12 12 F1e
E 13 e 1 87239.2738059414           Energy for J=E 13 e 1 level:B Bv=0 13 13 F1e
E 14 e 1 87288.7659451284           Energy for J=E 14 e 1 level:B Bv=0 14 14 F1e
E 15 e 1 87341.7838423505           Energy for J=E 15 e 1 level:B Bv=0 15 15 F1e
E 16 e 1 87398.3239892327           Energy for J=E 16 e 1 level:B Bv=0 16 16 F1e
E 17 e 1 87458.3457386672           Energy for J=E 17 e 1 level:B Bv=0 17 17 F1e
E 18 e 1 87521.9669176787           Energy for J=E 18 e 1 level:B Bv=0 18 18 F1e
E 19 e 1 87589.0729980635           Energy for J=E 19 e 1 level:B Bv=0 19 19 F1e
E 20 e 1 87659.6848539994           Energy for J=E 20 e 1 level:B Bv=0 20 20 F1e
E 21 e 1 87733.8093501407           Energy for J=E 21 e 1 level:B Bv=0 21 21 F1e
E 22 e 1 87811.4434750138           Energy for J=E 22 e 1 level:B Bv=0 22 22 F1e
E 23 e 1 87892.5833524621           Energy for J=E 23 e 1 level:B Bv=0 23 23 F1e
E 24 e 1 87977.2254359612           Energy for J=E 24 e 1 level:B Bv=0 24 24 F1e
E 25 e 1 88065.3655819688           Energy for J=E 25 e 1 level:B Bv=0 25 25 F1e
E 26 e 1 88157.0074638094           Energy for J=E 26 e 1 level:B Bv=0 26 26 F1e
E 27 e 1 88252.1428727514           Energy for J=E 27 e 1 level:B Bv=0 27 27 F1e
E 28 e 1 88350.7622038786           Energy for J=E 28 e 1 level:B Bv=0 28 28 F1e
E 29 e 1 88452.8714734024           Energy for J=E 29 e 1 level:B Bv=0 29 29 F1e
E 30 e 1 88558.4573652084           Energy for J=E 30 e 1 level:B Bv=0 30 30 F1e
E 31 e 1 88667.5170675484           Energy for J=E 31 e 1 level:B Bv=0 31 31 F1e
E 32 e 1 88780.0530157549           Energy for J=E 32 e 1 level:B Bv=0 32 32 F1e
E 33 e 1 88896.0732697678           Energy for J=E 33 e 1 level:B Bv=0 33 33 F1e
E 34 e 1 89015.5479117331           Energy for J=E 34 e 1 level:B Bv=0 34 34 F1e
E 35 e 1 89138.5007753021           Energy for J=E 35 e 1 level:B Bv=0 35 35 F1e
E 36 e 1 89264.8671434128           Energy for J=E 36 e 1 level:B Bv=0 36 36 F1e
E 37 e 1 89394.712613204           Energy for J=E 37 e 1 level:B Bv=0 37 37 F1e
E 38 e 1 89528.0033885644           Energy for J=E 38 e 1 level:B Bv=0 38 38 F1e
E 39 e 1 89664.7427100465           Energy for J=E 39 e 1 level:B Bv=0 39 39 F1e
E 40 e 1 89804.8732821819           Energy for J=E 40 e 1 level:B Bv=0 40 40 F1e
E 41 e 1 89948.4826520769           Energy for J=E 41 e 1 level:B Bv=0 41 41 F1e
E 42 e 1 90095.4817872327           Energy for J=E 42 e 1 level:B Bv=0 42 42 F1e
E 43 e 1 90245.9288501256           Energy for J=E 43 e 1 level:B Bv=0 43 43 F1e
E 44 e 1 90399.7945833577           Energy for J=E 44 e 1 level:B Bv=0 44 44 F1e
E 45 e 1 90557.048472636           Energy for J=E 45 e 1 level:B Bv=0 45 45 F1e
E 46 e 1 90717.7471747185           Energy for J=E 46 e 1 level:B Bv=0 46 46 F1e
E 47 e 1 90881.7572647896           Energy for J=E 47 e 1 level:B Bv=0 47 47 F1e
E 48 e 1 91049.1774976252           Energy for J=E 48 e 1 level:B Bv=0 48 48 F1e
E 49 e 1 91220.0533896832           Energy for J=E 49 e 1 level:B Bv=0 49 49 F1e
E 50 e 1 91394.1969159005           Energy for J=E 50 e 1 level:B Bv=0 50 50 F1e
E 51 e 1 91571.8211856604           Energy for J=E 51 e 1 level:B Bv=0 51 51 F1e
E 52 e 1 91752.7615313151           Energy for J=E 52 e 1 level:B Bv=0 52 52 F1e
E 53 e 1 91937.0582069911           Energy for J=E 53 e 1 level:B Bv=0 53 53 F1e
E 54 e 1 92124.7041674308           Energy for J=E 54 e 1 level:B Bv=0 54 54 F1e
E 55 e 1 92315.6922344469           Energy for J=E 55 e 1 level:B Bv=0 55 55 F1e
Linear: B Bv=0
  Status     Active
  Lambda     Sigma+
  OmegaOrder Auto => Regular
  Origin 86917.360320496          State Origin. Estimated lowest level = 86917.360320496
  B      1.76976443875216         Rotational constant.
  D      5.53872463797865E-6      Quartic centrifugal distortion.
  Matrix elements as for C Cv=0
LinearManifold: Excited
 Initial         False
 EigenSearch     True
 LimitSearch     True
 AutoQConverge   True
 UsePopParams    False
 OverrideEnergies True
 Linear: Excited Av=2
  Status     Active
  Lambda     Pi
  OmegaOrder Auto => Regular
  Origin 67557.9263807482         Float  State Origin. Estimated lowest level = 67557.9263807482
  B      1.41770507032897         Float  Rotational constant.
  q      -1.187706E-5             Lambda doubling constant.
  D      6.24399485577261E-6      Float  Quartic centrifugal distortion.
  H      -1.280902E-11            Sextic centrifugal distortion.
  <J,Lambda=-1,Omega=-1|B|J,Lambda=-1,Omega=-1> = J*(J+1)
  <J,Lambda=1,Omega=1|B|J,Lambda=1,Omega=1> = J*(J+1)
    Symmetrized:
  <J,Lambda=1,Omega=1|B|J,Lambda=1,Omega=1> = J*(J+1)
  <J,Lambda=1,Omega=1|B|J,Lambda=1,Omega=1> = J*(J+1)

  <J,Lambda=-1,Omega=-1|q|J,Lambda=1,Omega=1> = (J*(J+1))/2
  <J,Lambda=1,Omega=1|q|J,Lambda=-1,Omega=-1> = (J*(J+1))/2
    Symmetrized:
  <J,Lambda=1,Omega=1|q|J,Lambda=1,Omega=1> = (J*(J+1))/2
  <J,Lambda=1,Omega=1|q|J,Lambda=1,Omega=1> = (J*(J+1))/2
```

<J,Lambda=-1,Omega=-1|D|J,Lambda=-1,Omega=-1> = -J^2*(J*(J+2)+1)
<J,Lambda=1,Omega=1|D|J,Lambda=1,Omega=1> = -J^2*(J*(J+2)+1)
Symmetrized:
<J,Lambda=1,Omega=1|D|J,Lambda=1,Omega=1> = -J^2*(J*(J+2)+1)
<J,Lambda=1,Omega=1|D|J,Lambda=1,Omega=1> = -J^2*(J*(J+2)+1)

<J,Lambda=-1,Omega=-1|H|J,Lambda=-1,Omega=-1> = J^3*(J*(J*(J+3)+3)+1)
<J,Lambda=1,Omega=1|H|J,Lambda=1,Omega=1> = J^3*(J*(J*(J+3)+3)+1)
Symmetrized:
<J,Lambda=1,Omega=1|H|J,Lambda=1,Omega=1> = J^3*(J*(J*(J+3)+3)+1)
<J,Lambda=1,Omega=1|H|J,Lambda=1,Omega=1> = J^3*(J*(J*(J+3)+3)+1)

Linear: Excited Iv=3
Status Active
Colour Fuchsia
Lambda Sigma-
OmegaOrder Auto => Regular
Origin 67579.2383904013 Float State Origin. Estimated lowest level = 67579.2383904013
B 1.09918701526773 Float Rotational constant.
D 5.77973239037241E-6 Float Quartic centrifugal distortion.
H 2.247197E-12 Sextic centrifugal distortion.
<J,Lambda=0,Omega=0|B|J,Lambda=0,Omega=0> = J*(J+1)
Symmetrized:
<J,Lambda=0,Omega=0|B|J,Lambda=0,Omega=0> = J*(J+1)

<J,Lambda=0,Omega=0|D|J,Lambda=0,Omega=0> = -J^2*(J*(J+2)+1)
Symmetrized:
<J,Lambda=0,Omega=0|D|J,Lambda=0,Omega=0> = -J^2*(J*(J+2)+1)

<J,Lambda=0,Omega=0|H|J,Lambda=0,Omega=0> = J^3*(J*(J*(J+3)+3)+1)
Symmetrized:
<J,Lambda=0,Omega=0|H|J,Lambda=0,Omega=0> = J^3*(J*(J*(J+3)+3)+1)

Linear: Excited Iv=4
Status Active
Colour Fuchsia
Lambda Sigma-
OmegaOrder Auto => Regular
Origin 68544.2939028 State Origin. Estimated lowest level = 68544.2939028
B 1.0840499034 Rotational constant.
D 5.712347E-6 Quartic centrifugal distortion.
H 2.247197E-12 Sextic centrifugal distortion.
<J,Lambda=0,Omega=0|B|J,Lambda=0,Omega=0> = J*(J+1)
Symmetrized:
<J,Lambda=0,Omega=0|B|J,Lambda=0,Omega=0> = J*(J+1)

<J,Lambda=0,Omega=0|D|J,Lambda=0,Omega=0> = -J^2*(J*(J+2)+1)
Symmetrized:
<J,Lambda=0,Omega=0|D|J,Lambda=0,Omega=0> = -J^2*(J*(J+2)+1)

<J,Lambda=0,Omega=0|H|J,Lambda=0,Omega=0> = J^3*(J*(J*(J+3)+3)+1)
Symmetrized:
<J,Lambda=0,Omega=0|H|J,Lambda=0,Omega=0> = J^3*(J*(J*(J+3)+3)+1)

LinearPerturbation: Excited <Av=2|J+L-|Iv=3>
Value -0.0504224240425148 Float Size of perturbation.
<J,Lambda=-1,Omega=-1|J+L-|J,Lambda=0,Omega=0> = -sqrt(J*(J+1))
<J,Lambda=1,Omega=1|J+L-|J,Lambda=0,Omega=0> = sqrt(J*(J+1))

Symmetrized Matrix Elements of <Av=2|J+L-|Iv=3>
<J,Lambda=1,Omega=1|J+L-|J,Lambda=0,Omega=0> = -sqrt(J*(J+1))

LinearPerturbation: Excited <Av=2|J+L-|Iv=4>
Value 0.06495 Size of perturbation.
<J,Lambda=-1,Omega=-1|J+L-|J,Lambda=0,Omega=0> = -sqrt(J*(J+1))
<J,Lambda=1,Omega=1|J+L-|J,Lambda=0,Omega=0> = sqrt(J*(J+1))

Symmetrized Matrix Elements of <Av=2|J+L-|Iv=4>
<J,Lambda=1,Omega=1|J+L-|J,Lambda=0,Omega=0> = -sqrt(J*(J+1))

Linear: Excited ev=4
Status Active
Colour Lime
S 1
Lambda Sigma-
OmegaOrder Auto => Inverted
Origin 67802.7849165451 Float State Origin. Estimated lowest level = 67802.2187789372
B 1.09715336874161 Float Rotational constant.
LambdaSS 0.424603205929654 Float Spin - spin coupling constant (=1.5*epsilon).
D 5.34314012526997E-6 Float Quartic centrifugal distortion.
H -1.498131E-12 Sextic centrifugal distortion.
<J,Lambda=0,Omega=-1|B|J,Lambda=0,Omega=-1> = J*(J+1)
<J,Lambda=0,Omega=-1|B|J,Lambda=0,Omega=0> = -sqrt(2*J*(J+1))
<J,Lambda=0,Omega=0|B|J,Lambda=0,Omega=-1> = -sqrt(2*J*(J+1))
<J,Lambda=0,Omega=0|B|J,Lambda=0,Omega=0> = J*(J+1)+2
<J,Lambda=0,Omega=0|B|J,Lambda=0,Omega=1> = -sqrt(2*J*(J+1))
<J,Lambda=0,Omega=1|B|J,Lambda=0,Omega=0> = -sqrt(2*J*(J+1))
<J,Lambda=0,Omega=1|B|J,Lambda=0,Omega=1> = J*(J+1)
Symmetrized:
<J,Lambda=0,Omega=1|B|J,Lambda=0,Omega=1> = J*(J+1)
<J,Lambda=0,Omega=1|B|J,Lambda=0,Omega=1> = -sqrt(2*J*(J+1))
<J,Lambda=0,Omega=1|B|J,Lambda=0,Omega=0> = -sqrt(2*J*(J+1))

```
<J,Lambda=0,Omega=0|B|J,Lambda=0,Omega=1> = J*(J+1)+2
<J,Lambda=0,Omega=0|B|J,Lambda=0,Omega=0> = -sqrt(2*J*(J+1))

<J,Lambda=0,Omega=-1|LambdaSS|J,Lambda=0,Omega=-1> = 2/3
<J,Lambda=0,Omega=0|LambdaSS|J,Lambda=0,Omega=0> = -4/3
<J,Lambda=0,Omega=1|LambdaSS|J,Lambda=0,Omega=1> = 2/3
  Symmetrized:
<J,Lambda=0,Omega=1|LambdaSS|J,Lambda=0,Omega=1> = 2/3
<J,Lambda=0,Omega=1|LambdaSS|J,Lambda=0,Omega=1> = -4/3
<J,Lambda=0,Omega=0|LambdaSS|J,Lambda=0,Omega=0> = 2/3

<J,Lambda=0,Omega=-1|D|J,Lambda=0,Omega=-1> = -J*(J*(J*(J+2)+3)+2)
<J,Lambda=0,Omega=-1|D|J,Lambda=0,Omega=0> = 2*(J*(J+1)+1)*sqrt(2*J*(J+1))
<J,Lambda=0,Omega=-1|D|J,Lambda=0,Omega=1> = -2*J*(J+1)
<J,Lambda=0,Omega=0|D|J,Lambda=0,Omega=-1> = 2*(J*(J+1)+1)*sqrt(2*J*(J+1))
<J,Lambda=0,Omega=0|D|J,Lambda=0,Omega=0> = -J*(J*(J*(J+2)+9)+8)-4
<J,Lambda=0,Omega=0|D|J,Lambda=0,Omega=1> = 2*(J*(J+1)+1)*sqrt(2*J*(J+1))
<J,Lambda=0,Omega=1|D|J,Lambda=0,Omega=-1> = -2*J*(J+1)
<J,Lambda=0,Omega=1|D|J,Lambda=0,Omega=0> = 2*(J*(J+1)+1)*sqrt(2*J*(J+1))
<J,Lambda=0,Omega=1|D|J,Lambda=0,Omega=1> = -J*(J*(J*(J+2)+3)+2)
  Symmetrized:
<J,Lambda=0,Omega=1|D|J,Lambda=0,Omega=1> = -J*(J*(J*(J+2)+3)+2)
<J,Lambda=0,Omega=1|D|J,Lambda=0,Omega=1> = 2*(J*(J+1)+1)*sqrt(2*J*(J+1))
<J,Lambda=0,Omega=1|D|J,Lambda=0,Omega=0> = -2*J*(J+1)
<J,Lambda=0,Omega=0|D|J,Lambda=0,Omega=1> = 2*(J*(J+1)+1)*sqrt(2*J*(J+1))
<J,Lambda=0,Omega=0|D|J,Lambda=0,Omega=0> = -J*(J*(J*(J+2)+9)+8)-4

<J,Lambda=0,Omega=-1|H|J,Lambda=0,Omega=-1> = J*(J*(J*(J*(J*(J+3)+9)+13)+10)+4)
<J,Lambda=0,Omega=-1|H|J,Lambda=0,Omega=0> = (J*(J*(-3*J*(J+2)-13)-10)-4)*sqrt(2*J*(J+1))
<J,Lambda=0,Omega=-1|H|J,Lambda=0,Omega=1> = 2*J*(J*(3*J*(J+2)+5)+2)
<J,Lambda=0,Omega=0|H|J,Lambda=0,Omega=-1> = (J*(J*(-3*J*(J+2)-13)-10)-4)*sqrt(2*J*(J+1))
<J,Lambda=0,Omega=0|H|J,Lambda=0,Omega=0> = J*(J*(J*(J*(J*(J+3)+21)+37)+46)+28)+8
<J,Lambda=0,Omega=0|H|J,Lambda=0,Omega=1> = (J*(J*(-3*J*(J+2)-13)-10)-4)*sqrt(2*J*(J+1))
<J,Lambda=0,Omega=1|H|J,Lambda=0,Omega=-1> = 2*J*(J*(3*J*(J+2)+5)+2)
<J,Lambda=0,Omega=1|H|J,Lambda=0,Omega=0> = (J*(J*(-3*J*(J+2)-13)-10)-4)*sqrt(2*J*(J+1))
<J,Lambda=0,Omega=1|H|J,Lambda=0,Omega=1> = J*(J*(J*(J*(J*(J+3)+9)+13)+10)+4)
  Symmetrized:
<J,Lambda=0,Omega=1|H|J,Lambda=0,Omega=1> = J*(J*(J*(J*(J*(J+3)+9)+13)+10)+4)
<J,Lambda=0,Omega=1|H|J,Lambda=0,Omega=1> = (J*(J*(-3*J*(J+2)-13)-10)-4)*sqrt(2*J*(J+1))
<J,Lambda=0,Omega=1|H|J,Lambda=0,Omega=0> = 2*J*(J*(3*J*(J+2)+5)+2)
<J,Lambda=0,Omega=0|H|J,Lambda=0,Omega=1> = (J*(J*(-3*J*(J+2)-13)-10)-4)*sqrt(2*J*(J+1))
<J,Lambda=0,Omega=0|H|J,Lambda=0,Omega=0> = J*(J*(J*(J*(J*(J+3)+21)+37)+46)+28)+8

LinearPerturbation: Excited <Av=2|LS|ev=4>
  Value -13.1135785220975          Float  Size of perturbation.
<J,Lambda=-1,Omega=-1|LS|J,Lambda=0,Omega=-1> = -1/sqrt(3)
<J,Lambda=1,Omega=1|LS|J,Lambda=0,Omega=1> = -1/sqrt(3)

Symmetrized Matrix Elements of <Av=2|LS|ev=4>
<J,Lambda=1,Omega=1|LS|J,Lambda=0,Omega=1> = -1/sqrt(3)
<J,Lambda=1,Omega=1|LS|J,Lambda=0,Omega=1> = -1/sqrt(3)

Linear: Excited dv=7
  Status     Active
  Colour     Blue
  S          1
  Lambda     Delta
  OmegaOrder Auto => Inverted
  Origin   67946.2486612564          Float  State Origin. Estimated lowest level = 67913.089899836
  B        1.07973878060191          Float  Rotational constant.
  A        -16.9899246280871         Float  Spin-orbit coupling constant.
  LambdaSS 1.23163175363889          Float  Spin - spin coupling constant (=1.5*epsilon).
  gamma    -0.00290463321309688      Float  Spin rotation coupling constant.
  D        5.12786503880413E-6       Float  Quartic centrifugal distortion.
  H        -5.992526E-13             Sextic centrifugal distortion.
  AD       -9.172647E-5              Centrifugal distortion of A.
<J,Lambda=-2,Omega=-3|B|J,Lambda=-2,Omega=-3> = J*(J+1)-4
<J,Lambda=-2,Omega=-3|B|J,Lambda=-2,Omega=-2> = -sqrt(2*(J*(J+1)-6))
<J,Lambda=-2,Omega=-2|B|J,Lambda=-2,Omega=-3> = -sqrt(2*(J*(J+1)-6))
<J,Lambda=-2,Omega=-2|B|J,Lambda=-2,Omega=-2> = J*(J+1)+2
<J,Lambda=-2,Omega=-2|B|J,Lambda=-2,Omega=-1> = -sqrt(2*(J*(J+1)-2))
<J,Lambda=-2,Omega=-1|B|J,Lambda=-2,Omega=-2> = -sqrt(2*(J*(J+1)-2))
<J,Lambda=-2,Omega=-1|B|J,Lambda=-2,Omega=-1> = J*(J+1)+4
<J,Lambda=2,Omega=1|B|J,Lambda=2,Omega=1> = J*(J+1)+4
<J,Lambda=2,Omega=1|B|J,Lambda=2,Omega=2> = -sqrt(2*(J*(J+1)-2))
<J,Lambda=2,Omega=2|B|J,Lambda=2,Omega=1> = -sqrt(2*(J*(J+1)-2))
<J,Lambda=2,Omega=2|B|J,Lambda=2,Omega=2> = J*(J+1)+2
<J,Lambda=2,Omega=2|B|J,Lambda=2,Omega=3> = -sqrt(2*(J*(J+1)-6))
<J,Lambda=2,Omega=3|B|J,Lambda=2,Omega=2> = -sqrt(2*(J*(J+1)-6))
<J,Lambda=2,Omega=3|B|J,Lambda=2,Omega=3> = J*(J+1)-4
  Symmetrized:
<J,Lambda=2,Omega=3|B|J,Lambda=2,Omega=3> = J*(J+1)-4
<J,Lambda=2,Omega=3|B|J,Lambda=2,Omega=3> = -sqrt(2*(J*(J+1)-6))
<J,Lambda=2,Omega=3|B|J,Lambda=2,Omega=2> = -sqrt(2*(J*(J+1)-6))
<J,Lambda=2,Omega=3|B|J,Lambda=2,Omega=2> = J*(J+1)+2
<J,Lambda=2,Omega=2|B|J,Lambda=2,Omega=3> = -sqrt(2*(J*(J+1)-2))
<J,Lambda=2,Omega=2|B|J,Lambda=2,Omega=3> = -sqrt(2*(J*(J+1)-2))
<J,Lambda=2,Omega=2|B|J,Lambda=2,Omega=2> = J*(J+1)+4
<J,Lambda=2,Omega=2|B|J,Lambda=2,Omega=2> = J*(J+1)+4
<J,Lambda=2,Omega=2|B|J,Lambda=2,Omega=1> = -sqrt(2*(J*(J+1)-2))
<J,Lambda=2,Omega=2|B|J,Lambda=2,Omega=1> = -sqrt(2*(J*(J+1)-2))
<J,Lambda=2,Omega=1|B|J,Lambda=2,Omega=2> = J*(J+1)+2
```

<J,Lambda=2,Omega=1|B|J,Lambda=2,Omega=2> = -sqrt(2*(J*(J+1)-6))
<J,Lambda=2,Omega=1|B|J,Lambda=2,Omega=1> = -sqrt(2*(J*(J+1)-6))
<J,Lambda=2,Omega=1|B|J,Lambda=2,Omega=1> = J*(J+1)-4

<J,Lambda=-2,Omega=-3|A|J,Lambda=-2,Omega=-3> = 2
<J,Lambda=-2,Omega=-1|A|J,Lambda=-2,Omega=-1> = -2
<J,Lambda=2,Omega=1|A|J,Lambda=2,Omega=1> = -2
<J,Lambda=2,Omega=3|A|J,Lambda=2,Omega=3> = 2
Symmetrized:
<J,Lambda=2,Omega=3|A|J,Lambda=2,Omega=3> = 2
<J,Lambda=2,Omega=3|A|J,Lambda=2,Omega=3> = -2
<J,Lambda=2,Omega=1|A|J,Lambda=2,Omega=1> = -2
<J,Lambda=2,Omega=1|A|J,Lambda=2,Omega=1> = 2

<J,Lambda=-2,Omega=-3|LambdaSS|J,Lambda=-2,Omega=-3> = 2/3
<J,Lambda=-2,Omega=-2|LambdaSS|J,Lambda=-2,Omega=-2> = -4/3
<J,Lambda=-2,Omega=-1|LambdaSS|J,Lambda=-2,Omega=-1> = 2/3
<J,Lambda=2,Omega=1|LambdaSS|J,Lambda=2,Omega=1> = 2/3
<J,Lambda=2,Omega=2|LambdaSS|J,Lambda=2,Omega=2> = -4/3
<J,Lambda=2,Omega=3|LambdaSS|J,Lambda=2,Omega=3> = 2/3
Symmetrized:
<J,Lambda=2,Omega=3|LambdaSS|J,Lambda=2,Omega=3> = 2/3
<J,Lambda=2,Omega=3|LambdaSS|J,Lambda=2,Omega=3> = -4/3
<J,Lambda=2,Omega=2|LambdaSS|J,Lambda=2,Omega=2> = 2/3
<J,Lambda=2,Omega=2|LambdaSS|J,Lambda=2,Omega=2> = 2/3
<J,Lambda=2,Omega=1|LambdaSS|J,Lambda=2,Omega=1> = -4/3
<J,Lambda=2,Omega=1|LambdaSS|J,Lambda=2,Omega=1> = 2/3

<J,Lambda=-2,Omega=-3|gamma|J,Lambda=-2,Omega=-3> = 1
<J,Lambda=-2,Omega=-3|gamma|J,Lambda=-2,Omega=-2> = sqrt(2*(J*(J+1)-6))/2
<J,Lambda=-2,Omega=-2|gamma|J,Lambda=-2,Omega=-3> = sqrt(2*(J*(J+1)-6))/2
<J,Lambda=-2,Omega=-2|gamma|J,Lambda=-2,Omega=-2> = -2
<J,Lambda=-2,Omega=-2|gamma|J,Lambda=-2,Omega=-1> = sqrt(2*(J*(J+1)-2))/2
<J,Lambda=-2,Omega=-1|gamma|J,Lambda=-2,Omega=-2> = sqrt(2*(J*(J+1)-2))/2
<J,Lambda=-2,Omega=-1|gamma|J,Lambda=-2,Omega=-1> = -3
<J,Lambda=2,Omega=1|gamma|J,Lambda=2,Omega=1> = -3
<J,Lambda=2,Omega=1|gamma|J,Lambda=2,Omega=2> = sqrt(2*(J*(J+1)-2))/2
<J,Lambda=2,Omega=2|gamma|J,Lambda=2,Omega=1> = sqrt(2*(J*(J+1)-2))/2
<J,Lambda=2,Omega=2|gamma|J,Lambda=2,Omega=2> = -2
<J,Lambda=2,Omega=2|gamma|J,Lambda=2,Omega=3> = sqrt(2*(J*(J+1)-6))/2
<J,Lambda=2,Omega=3|gamma|J,Lambda=2,Omega=2> = sqrt(2*(J*(J+1)-6))/2
<J,Lambda=2,Omega=3|gamma|J,Lambda=2,Omega=3> = 1
Symmetrized:
<J,Lambda=2,Omega=3|gamma|J,Lambda=2,Omega=3> = 1
<J,Lambda=2,Omega=3|gamma|J,Lambda=2,Omega=3> = sqrt(2*(J*(J+1)-6))/2
<J,Lambda=2,Omega=3|gamma|J,Lambda=2,Omega=2> = sqrt(2*(J*(J+1)-6))/2
<J,Lambda=2,Omega=3|gamma|J,Lambda=2,Omega=2> = -2
<J,Lambda=2,Omega=2|gamma|J,Lambda=2,Omega=3> = sqrt(2*(J*(J+1)-2))/2
<J,Lambda=2,Omega=2|gamma|J,Lambda=2,Omega=3> = sqrt(2*(J*(J+1)-2))/2
<J,Lambda=2,Omega=2|gamma|J,Lambda=2,Omega=2> = -3
<J,Lambda=2,Omega=2|gamma|J,Lambda=2,Omega=2> = -3
<J,Lambda=2,Omega=2|gamma|J,Lambda=2,Omega=1> = sqrt(2*(J*(J+1)-2))/2
<J,Lambda=2,Omega=2|gamma|J,Lambda=2,Omega=1> = sqrt(2*(J*(J+1)-2))/2
<J,Lambda=2,Omega=1|gamma|J,Lambda=2,Omega=2> = -2
<J,Lambda=2,Omega=1|gamma|J,Lambda=2,Omega=2> = sqrt(2*(J*(J+1)-6))/2
<J,Lambda=2,Omega=1|gamma|J,Lambda=2,Omega=1> = sqrt(2*(J*(J+1)-6))/2
<J,Lambda=2,Omega=1|gamma|J,Lambda=2,Omega=1> = 1

<J,Lambda=-2,Omega=-3|D|J,Lambda=-2,Omega=-3> = -J*(J*(J*(J+2)-5)-6)-4
<J,Lambda=-2,Omega=-3|D|J,Lambda=-2,Omega=-2> = -2*(-J*(J+1)+1)*sqrt(2*(J*(J+1)-6))
<J,Lambda=-2,Omega=-3|D|J,Lambda=-2,Omega=-1> = -2*sqrt(J*(J*(J*(J+2)-7)-8)+12)
<J,Lambda=-2,Omega=-2|D|J,Lambda=-2,Omega=-3> = -2*(-J*(J+1)+1)*sqrt(2*(J*(J+1)-6))
<J,Lambda=-2,Omega=-2|D|J,Lambda=-2,Omega=-2> = -J*(J*(J*(J+2)+9)+8)+12
<J,Lambda=-2,Omega=-2|D|J,Lambda=-2,Omega=-1> = 2*(J*(J+1)+3)*sqrt(2*(J*(J+1)-2))
<J,Lambda=-2,Omega=-1|D|J,Lambda=-2,Omega=-3> = -2*sqrt(J*(J*(J*(J+2)-7)-8)+12)
<J,Lambda=-2,Omega=-1|D|J,Lambda=-2,Omega=-2> = 2*(J*(J+1)+3)*sqrt(2*(J*(J+1)-2))
<J,Lambda=-2,Omega=-1|D|J,Lambda=-2,Omega=-1> = -J*(J*(J*(J+2)+11)+10)-12
<J,Lambda=2,Omega=1|D|J,Lambda=2,Omega=1> = -J*(J*(J*(J+2)+11)+10)-12
<J,Lambda=2,Omega=1|D|J,Lambda=2,Omega=2> = 2*(J*(J+1)+3)*sqrt(2*(J*(J+1)-2))
<J,Lambda=2,Omega=1|D|J,Lambda=2,Omega=3> = -2*sqrt(J*(J*(J*(J+2)-7)-8)+12)
<J,Lambda=2,Omega=2|D|J,Lambda=2,Omega=1> = 2*(J*(J+1)+3)*sqrt(2*(J*(J+1)-2))
<J,Lambda=2,Omega=2|D|J,Lambda=2,Omega=2> = -J*(J*(J*(J+2)+9)+8)+12
<J,Lambda=2,Omega=2|D|J,Lambda=2,Omega=3> = -2*(-J*(J+1)+1)*sqrt(2*(J*(J+1)-6))
<J,Lambda=2,Omega=3|D|J,Lambda=2,Omega=1> = -2*sqrt(J*(J*(J*(J+2)-7)-8)+12)
<J,Lambda=2,Omega=3|D|J,Lambda=2,Omega=2> = -2*(-J*(J+1)+1)*sqrt(2*(J*(J+1)-6))
<J,Lambda=2,Omega=3|D|J,Lambda=2,Omega=3> = -J*(J*(J*(J+2)-5)-6)-4
Symmetrized:
<J,Lambda=2,Omega=3|D|J,Lambda=2,Omega=3> = -J*(J*(J*(J+2)-5)-6)-4
<J,Lambda=2,Omega=3|D|J,Lambda=2,Omega=3> = -2*(-J*(J+1)+1)*sqrt(2*(J*(J+1)-6))
<J,Lambda=2,Omega=3|D|J,Lambda=2,Omega=2> = -2*sqrt(J*(J*(J*(J+2)-7)-8)+12)
<J,Lambda=2,Omega=3|D|J,Lambda=2,Omega=2> = -2*(-J*(J+1)+1)*sqrt(2*(J*(J+1)-6))
<J,Lambda=2,Omega=3|D|J,Lambda=2,Omega=1> = -J*(J*(J*(J+2)+9)+8)+12
<J,Lambda=2,Omega=3|D|J,Lambda=2,Omega=1> = 2*(J*(J+1)+3)*sqrt(2*(J*(J+1)-2))
<J,Lambda=2,Omega=2|D|J,Lambda=2,Omega=3> = -2*sqrt(J*(J*(J*(J+2)-7)-8)+12)
<J,Lambda=2,Omega=2|D|J,Lambda=2,Omega=3> = 2*(J*(J+1)+3)*sqrt(2*(J*(J+1)-2))
<J,Lambda=2,Omega=2|D|J,Lambda=2,Omega=2> = -J*(J*(J*(J+2)+11)+10)-12
<J,Lambda=2,Omega=2|D|J,Lambda=2,Omega=2> = -J*(J*(J*(J+2)+11)+10)-12
<J,Lambda=2,Omega=2|D|J,Lambda=2,Omega=1> = 2*(J*(J+1)+3)*sqrt(2*(J*(J+1)-2))
<J,Lambda=2,Omega=2|D|J,Lambda=2,Omega=1> = -2*sqrt(J*(J*(J*(J+2)-7)-8)+12)
<J,Lambda=2,Omega=1|D|J,Lambda=2,Omega=3> = 2*(J*(J+1)+3)*sqrt(2*(J*(J+1)-2))
<J,Lambda=2,Omega=1|D|J,Lambda=2,Omega=3> = -J*(J*(J*(J+2)+9)+8)+12
<J,Lambda=2,Omega=1|D|J,Lambda=2,Omega=2> = -2*(-J*(J+1)+1)*sqrt(2*(J*(J+1)-6))

```
  <J,Lambda=2,Omega=1|D|J,Lambda=2,Omega=2> = -2*sqrt(J*(J*(J*(J+2)-7)-8)+12)
  <J,Lambda=2,Omega=1|D|J,Lambda=2,Omega=1> = -2*(-J*(J+1)+1)*sqrt(2*(J*(J+1)-6))
  <J,Lambda=2,Omega=1|D|J,Lambda=2,Omega=1> = -J*(J*(J*(J+2)-5)-6)-4

  <J,Lambda=-2,Omega=-3|H|J,Lambda=-2,Omega=-3> = J^2*(J*(J*(J*(J+3)-3)-11)-6)+8
  <J,Lambda=-2,Omega=-3|H|J,Lambda=-2,Omega=-2> = (J*(J*(-3*J*(J+2)-1)+2)+4)*sqrt(2*(J*(J+1)-6))
  <J,Lambda=-2,Omega=-3|H|J,Lambda=-2,Omega=-1> = 2*(3*J*(J+1)+2)*sqrt(J*(J*(J*(J+2)-7)-8)+12)
  <J,Lambda=-2,Omega=-2|H|J,Lambda=-2,Omega=-3> = J*(J*(-3*J*(J+2)+1)+4)*sqrt(2*(J*(J+1)-6))-2*sqrt(2*(J*(J*(J*(J*(J+3)-7)-19)+18)+28)-24))
  <J,Lambda=-2,Omega=-2|H|J,Lambda=-2,Omega=-2> = J*(J*(J*(J*(J*(J+3)+21)+37)-2)-20)-24
  <J,Lambda=-2,Omega=-2|H|J,Lambda=-2,Omega=-1> = J*(J*(-3*J*(J+2)-23)-20)*sqrt(2*(J*(J+1)-2))-2*sqrt(2*(J*(J*(J*(J*(J+3)-11)-27)+46)+60)-72))-24*sqrt(2*(J*(J+1)-2))
  <J,Lambda=-2,Omega=-1|H|J,Lambda=-2,Omega=-3> = 2*(3*J*(J+1)+2)*sqrt(J*(J*(J*(J+2)-7)-8)+12)
  <J,Lambda=-2,Omega=-1|H|J,Lambda=-2,Omega=-2> = (J*(J*(-3*J*(J+2)-25)-22)-12)*sqrt(2*(J*(J+1)-2))
  <J,Lambda=-2,Omega=-1|H|J,Lambda=-2,Omega=-1> = J*(J*(J*(J*(J*(J+3)+21)+37)+74)+56)+24
  <J,Lambda=2,Omega=1|H|J,Lambda=2,Omega=1> = J*(J*(J*(J*(J*(J+3)+21)+37)+74)+56)+24
  <J,Lambda=2,Omega=1|H|J,Lambda=2,Omega=2> = (J*(J*(-3*J*(J+2)-25)-22)-12)*sqrt(2*(J*(J+1)-2))
  <J,Lambda=2,Omega=1|H|J,Lambda=2,Omega=3> = 2*(3*J*(J+1)+2)*sqrt(J*(J*(J*(J+2)-7)-8)+12)
  <J,Lambda=2,Omega=2|H|J,Lambda=2,Omega=1> = J*(J*(-3*J*(J+2)-23)-20)*sqrt(2*(J*(J+1)-2))-2*sqrt(2*(J*(J*(J*(J*(J+3)-11)-27)+46)+60)-72))-24*sqrt(2*(J*(J+1)-2))
  <J,Lambda=2,Omega=2|H|J,Lambda=2,Omega=2> = J*(J*(J*(J*(J*(J+3)+21)+37)-2)-20)-24
  <J,Lambda=2,Omega=2|H|J,Lambda=2,Omega=3> = J*(J*(-3*J*(J+2)+1)+4)*sqrt(2*(J*(J+1)-6))-2*sqrt(2*(J*(J*(J*(J*(J+3)-7)-19)+18)+28)-24))
  <J,Lambda=2,Omega=3|H|J,Lambda=2,Omega=1> = 2*(3*J*(J+1)+2)*sqrt(J*(J*(J*(J+2)-7)-8)+12)
  <J,Lambda=2,Omega=3|H|J,Lambda=2,Omega=2> = (J*(J*(-3*J*(J+2)-1)+2)+4)*sqrt(2*(J*(J+1)-6))
  <J,Lambda=2,Omega=3|H|J,Lambda=2,Omega=3> = J^2*(J*(J*(J*(J+3)-3)-11)-6)+8
   Symmetrized:
  <J,Lambda=2,Omega=3|H|J,Lambda=2,Omega=3> = J^2*(J*(J*(J*(J+3)-3)-11)-6)+8
  <J,Lambda=2,Omega=3|H|J,Lambda=2,Omega=3> = (J*(J*(-3*J*(J+2)-1)+2)+4)*sqrt(2*(J*(J+1)-6))
  <J,Lambda=2,Omega=3|H|J,Lambda=2,Omega=2> = 2*(3*J*(J+1)+2)*sqrt(J*(J*(J*(J+2)-7)-8)+12)
  <J,Lambda=2,Omega=3|H|J,Lambda=2,Omega=2> = J*(J*(-3*J*(J+2)+1)+4)*sqrt(2*(J*(J+1)-6))-2*sqrt(2*(J*(J*(J*(J*(J+3)-7)-19)+18)+28)-24))
  <J,Lambda=2,Omega=3|H|J,Lambda=2,Omega=1> = J*(J*(J*(J*(J*(J+3)+21)+37)-2)-20)-24
  <J,Lambda=2,Omega=3|H|J,Lambda=2,Omega=1> = J*(J*(-3*J*(J+2)-23)-20)*sqrt(2*(J*(J+1)-2))-2*sqrt(2*(J*(J*(J*(J*(J+3)-11)-27)+46)+60)-72))-24*sqrt(2*(J*(J+1)-2))
  <J,Lambda=2,Omega=2|H|J,Lambda=2,Omega=3> = 2*(3*J*(J+1)+2)*sqrt(J*(J*(J*(J+2)-7)-8)+12)
  <J,Lambda=2,Omega=2|H|J,Lambda=2,Omega=3> = (J*(J*(-3*J*(J+2)-25)-22)-12)*sqrt(2*(J*(J+1)-2))
  <J,Lambda=2,Omega=2|H|J,Lambda=2,Omega=2> = J*(J*(J*(J*(J*(J+3)+21)+37)+74)+56)+24
  <J,Lambda=2,Omega=2|H|J,Lambda=2,Omega=2> = J*(J*(J*(J*(J*(J+3)+21)+37)+74)+56)+24
  <J,Lambda=2,Omega=2|H|J,Lambda=2,Omega=1> = (J*(J*(-3*J*(J+2)-25)-22)-12)*sqrt(2*(J*(J+1)-2))
  <J,Lambda=2,Omega=2|H|J,Lambda=2,Omega=1> = 2*(3*J*(J+1)+2)*sqrt(J*(J*(J*(J+2)-7)-8)+12)
  <J,Lambda=2,Omega=1|H|J,Lambda=2,Omega=3> = J*(J*(-3*J*(J+2)-23)-20)*sqrt(2*(J*(J+1)-2))-2*sqrt(2*(J*(J*(J*(J*(J+3)-11)-27)+46)+60)-72))-24*sqrt(2*(J*(J+1)-2))
  <J,Lambda=2,Omega=1|H|J,Lambda=2,Omega=3> = J*(J*(J*(J*(J*(J+3)+21)+37)-2)-20)-24
  <J,Lambda=2,Omega=1|H|J,Lambda=2,Omega=2> = J*(J*(-3*J*(J+2)+1)+4)*sqrt(2*(J*(J+1)-6))-2*sqrt(2*(J*(J*(J*(J*(J+3)-7)-19)+18)+28)-24))
  <J,Lambda=2,Omega=1|H|J,Lambda=2,Omega=2> = 2*(3*J*(J+1)+2)*sqrt(J*(J*(J*(J+2)-7)-8)+12)
  <J,Lambda=2,Omega=1|H|J,Lambda=2,Omega=1> = (J*(J*(-3*J*(J+2)-1)+2)+4)*sqrt(2*(J*(J+1)-6))
  <J,Lambda=2,Omega=1|H|J,Lambda=2,Omega=1> = J^2*(J*(J*(J*(J+3)-3)-11)-6)+8

  <J,Lambda=-2,Omega=-3|AD|J,Lambda=-2,Omega=-3> = 2*(J*(J+1)-4)
  <J,Lambda=-2,Omega=-3|AD|J,Lambda=-2,Omega=-2> = -sqrt(2*(J*(J+1)-6))
  <J,Lambda=-2,Omega=-2|AD|J,Lambda=-2,Omega=-3> = -sqrt(2*(J*(J+1)-6))
  <J,Lambda=-2,Omega=-2|AD|J,Lambda=-2,Omega=-1> = sqrt(2*(J*(J+1)-2))
  <J,Lambda=-2,Omega=-1|AD|J,Lambda=-2,Omega=-2> = sqrt(2*(J*(J+1)-2))
  <J,Lambda=-2,Omega=-1|AD|J,Lambda=-2,Omega=-1> = 2*(-J*(J+1)-4)
  <J,Lambda=2,Omega=1|AD|J,Lambda=2,Omega=1> = 2*(-J*(J+1)-4)
  <J,Lambda=2,Omega=1|AD|J,Lambda=2,Omega=2> = sqrt(2*(J*(J+1)-2))
  <J,Lambda=2,Omega=2|AD|J,Lambda=2,Omega=1> = sqrt(2*(J*(J+1)-2))
  <J,Lambda=2,Omega=2|AD|J,Lambda=2,Omega=3> = -sqrt(2*(J*(J+1)-6))
  <J,Lambda=2,Omega=3|AD|J,Lambda=2,Omega=2> = -sqrt(2*(J*(J+1)-6))
  <J,Lambda=2,Omega=3|AD|J,Lambda=2,Omega=3> = 2*(J*(J+1)-4)
   Symmetrized:
  <J,Lambda=2,Omega=3|AD|J,Lambda=2,Omega=3> = 2*(J*(J+1)-4)
  <J,Lambda=2,Omega=3|AD|J,Lambda=2,Omega=3> = -sqrt(2*(J*(J+1)-6))
  <J,Lambda=2,Omega=3|AD|J,Lambda=2,Omega=2> = -sqrt(2*(J*(J+1)-6))
  <J,Lambda=2,Omega=3|AD|J,Lambda=2,Omega=2> = sqrt(2*(J*(J+1)-2))
  <J,Lambda=2,Omega=2|AD|J,Lambda=2,Omega=3> = sqrt(2*(J*(J+1)-2))
  <J,Lambda=2,Omega=2|AD|J,Lambda=2,Omega=3> = 2*(-J*(J+1)-4)
  <J,Lambda=2,Omega=2|AD|J,Lambda=2,Omega=1> = 2*(-J*(J+1)-4)
  <J,Lambda=2,Omega=2|AD|J,Lambda=2,Omega=1> = sqrt(2*(J*(J+1)-2))
  <J,Lambda=2,Omega=1|AD|J,Lambda=2,Omega=2> = sqrt(2*(J*(J+1)-2))
  <J,Lambda=2,Omega=1|AD|J,Lambda=2,Omega=2> = -sqrt(2*(J*(J+1)-6))
  <J,Lambda=2,Omega=1|AD|J,Lambda=2,Omega=1> = -sqrt(2*(J*(J+1)-6))
  <J,Lambda=2,Omega=1|AD|J,Lambda=2,Omega=1> = 2*(J*(J+1)-4)

LinearPerturbation: Excited <Av=2|LS|dv=7>
  Value -9.51850056737858          Float  Size of perturbation.
<J,Lambda=-1,Omega=-1|LS|J,Lambda=-2,Omega=-1> = 1/sqrt(3)
<J,Lambda=1,Omega=1|LS|J,Lambda=2,Omega=1> = -1/sqrt(3)

Symmetrized Matrix Elements of <Av=2|LS|dv=7>
<J,Lambda=1,Omega=1|LS|J,Lambda=2,Omega=1> = 1/sqrt(3)
<J,Lambda=1,Omega=1|LS|J,Lambda=2,Omega=1> = -1/sqrt(3)

Linear: Excited a'v=12
  Status     Active
  Colour     Red
  S          1
  Lambda     Sigma+
  OmegaOrder Auto => Inverted
  Origin   67991.2956897487           Float  State Origin. Estimated lowest level = 67990.5318805452
  B        1.03759347733243           Float  Rotational constant.
  LambdaSS -1.14571380522651          Float  Spin - spin coupling constant (=1.5*epsilon).
  gamma    -0.00527072942354211       Float  Spin rotation coupling constant.
  D        5.18562812216526E-6        Float  Quartic centrifugal distortion.
  H        -2.996263E-13              Sextic centrifugal distortion.
  <J,Lambda=0,Omega=-1|B|J,Lambda=0,Omega=-1> = J*(J+1)
  <J,Lambda=0,Omega=-1|B|J,Lambda=0,Omega=0> = -sqrt(2*J*(J+1))
  <J,Lambda=0,Omega=0|B|J,Lambda=0,Omega=-1> = -sqrt(2*J*(J+1))
```

```
  <J,Lambda=0,Omega=0|B|J,Lambda=0,Omega=0> = J*(J+1)+2
  <J,Lambda=0,Omega=0|B|J,Lambda=0,Omega=1> = -sqrt(2*J*(J+1))
  <J,Lambda=0,Omega=1|B|J,Lambda=0,Omega=0> = -sqrt(2*J*(J+1))
  <J,Lambda=0,Omega=1|B|J,Lambda=0,Omega=1> = J*(J+1)
    Symmetrized:
  <J,Lambda=0,Omega=1|B|J,Lambda=0,Omega=1> = J*(J+1)
  <J,Lambda=0,Omega=1|B|J,Lambda=0,Omega=1> = -sqrt(2*J*(J+1))
  <J,Lambda=0,Omega=1|B|J,Lambda=0,Omega=0> = -sqrt(2*J*(J+1))
  <J,Lambda=0,Omega=0|B|J,Lambda=0,Omega=1> = J*(J+1)+2
  <J,Lambda=0,Omega=0|B|J,Lambda=0,Omega=0> = -sqrt(2*J*(J+1))

  <J,Lambda=0,Omega=-1|LambdaSS|J,Lambda=0,Omega=-1> = 2/3
  <J,Lambda=0,Omega=0|LambdaSS|J,Lambda=0,Omega=0> = -4/3
  <J,Lambda=0,Omega=1|LambdaSS|J,Lambda=0,Omega=1> = 2/3
    Symmetrized:
  <J,Lambda=0,Omega=1|LambdaSS|J,Lambda=0,Omega=1> = 2/3
  <J,Lambda=0,Omega=1|LambdaSS|J,Lambda=0,Omega=1> = -4/3
  <J,Lambda=0,Omega=0|LambdaSS|J,Lambda=0,Omega=0> = 2/3

  <J,Lambda=0,Omega=-1|gamma|J,Lambda=0,Omega=-1> = -1
  <J,Lambda=0,Omega=-1|gamma|J,Lambda=0,Omega=0> = sqrt(2*J*(J+1))/2
  <J,Lambda=0,Omega=0|gamma|J,Lambda=0,Omega=-1> = sqrt(2*J*(J+1))/2
  <J,Lambda=0,Omega=0|gamma|J,Lambda=0,Omega=0> = -2
  <J,Lambda=0,Omega=0|gamma|J,Lambda=0,Omega=1> = sqrt(2*J*(J+1))/2
  <J,Lambda=0,Omega=1|gamma|J,Lambda=0,Omega=0> = sqrt(2*J*(J+1))/2
  <J,Lambda=0,Omega=1|gamma|J,Lambda=0,Omega=1> = -1
    Symmetrized:
  <J,Lambda=0,Omega=1|gamma|J,Lambda=0,Omega=1> = -1
  <J,Lambda=0,Omega=1|gamma|J,Lambda=0,Omega=1> = sqrt(2*J*(J+1))/2
  <J,Lambda=0,Omega=1|gamma|J,Lambda=0,Omega=0> = sqrt(2*J*(J+1))/2
  <J,Lambda=0,Omega=0|gamma|J,Lambda=0,Omega=1> = -2
  <J,Lambda=0,Omega=0|gamma|J,Lambda=0,Omega=0> = sqrt(2*J*(J+1))/2

  <J,Lambda=0,Omega=-1|D|J,Lambda=0,Omega=-1> = -J*(J*(J*(J+2)+3)+2)
  <J,Lambda=0,Omega=-1|D|J,Lambda=0,Omega=0> = 2*(J*(J+1)+1)*sqrt(2*J*(J+1))
  <J,Lambda=0,Omega=-1|D|J,Lambda=0,Omega=1> = -2*J*(J+1)
  <J,Lambda=0,Omega=0|D|J,Lambda=0,Omega=-1> = 2*(J*(J+1)+1)*sqrt(2*J*(J+1))
  <J,Lambda=0,Omega=0|D|J,Lambda=0,Omega=0> = -J*(J*(J*(J+2)+9)+8)-4
  <J,Lambda=0,Omega=0|D|J,Lambda=0,Omega=1> = 2*(J*(J+1)+1)*sqrt(2*J*(J+1))
  <J,Lambda=0,Omega=1|D|J,Lambda=0,Omega=-1> = -2*J*(J+1)
  <J,Lambda=0,Omega=1|D|J,Lambda=0,Omega=0> = 2*(J*(J+1)+1)*sqrt(2*J*(J+1))
  <J,Lambda=0,Omega=1|D|J,Lambda=0,Omega=1> = -J*(J*(J*(J+2)+3)+2)
    Symmetrized:
  <J,Lambda=0,Omega=1|D|J,Lambda=0,Omega=1> = -J*(J*(J*(J+2)+3)+2)
  <J,Lambda=0,Omega=1|D|J,Lambda=0,Omega=1> = 2*(J*(J+1)+1)*sqrt(2*J*(J+1))
  <J,Lambda=0,Omega=1|D|J,Lambda=0,Omega=0> = -2*J*(J+1)
  <J,Lambda=0,Omega=0|D|J,Lambda=0,Omega=1> = 2*(J*(J+1)+1)*sqrt(2*J*(J+1))
  <J,Lambda=0,Omega=0|D|J,Lambda=0,Omega=0> = -J*(J*(J*(J+2)+9)+8)-4

  <J,Lambda=0,Omega=-1|H|J,Lambda=0,Omega=-1> = J*(J*(J*(J*(J*(J+3)+9)+13)+10)+4)
  <J,Lambda=0,Omega=-1|H|J,Lambda=0,Omega=0> = (J*(J*(-3*J*(J+2)-13)-10)-4)*sqrt(2*J*(J+1))
  <J,Lambda=0,Omega=-1|H|J,Lambda=0,Omega=1> = 2*J*(J*(3*J*(J+2)+5)+2)
  <J,Lambda=0,Omega=0|H|J,Lambda=0,Omega=-1> = (J*(J*(-3*J*(J+2)-13)-10)-4)*sqrt(2*J*(J+1))
  <J,Lambda=0,Omega=0|H|J,Lambda=0,Omega=0> = J*(J*(J*(J*(J*(J+3)+21)+37)+46)+28)+8
  <J,Lambda=0,Omega=0|H|J,Lambda=0,Omega=1> = (J*(J*(-3*J*(J+2)-13)-10)-4)*sqrt(2*J*(J+1))
  <J,Lambda=0,Omega=1|H|J,Lambda=0,Omega=-1> = 2*J*(J*(3*J*(J+2)+5)+2)
  <J,Lambda=0,Omega=1|H|J,Lambda=0,Omega=0> = (J*(J*(-3*J*(J+2)-13)-10)-4)*sqrt(2*J*(J+1))
  <J,Lambda=0,Omega=1|H|J,Lambda=0,Omega=1> = J*(J*(J*(J*(J*(J+3)+9)+13)+10)+4)
    Symmetrized:
  <J,Lambda=0,Omega=1|H|J,Lambda=0,Omega=1> = J*(J*(J*(J*(J*(J+3)+9)+13)+10)+4)
  <J,Lambda=0,Omega=1|H|J,Lambda=0,Omega=1> = (J*(J*(-3*J*(J+2)-13)-10)-4)*sqrt(2*J*(J+1))
  <J,Lambda=0,Omega=1|H|J,Lambda=0,Omega=0> = 2*J*(J*(3*J*(J+2)+5)+2)
  <J,Lambda=0,Omega=0|H|J,Lambda=0,Omega=1> = (J*(J*(-3*J*(J+2)-13)-10)-4)*sqrt(2*J*(J+1))
  <J,Lambda=0,Omega=0|H|J,Lambda=0,Omega=0> = J*(J*(J*(J*(J*(J+3)+21)+37)+46)+28)+8

LinearPerturbation: Excited <Av=2|LS|a'v=12>
  Value -5.90222571840932          Float  Size of perturbation.
<J,Lambda=-1,Omega=-1|LS|J,Lambda=0,Omega=-1> = 1/sqrt(3)
<J,Lambda=1,Omega=1|LS|J,Lambda=0,Omega=1> = -1/sqrt(3)

Symmetrized Matrix Elements of <Av=2|LS|a'v=12>
<J,Lambda=1,Omega=1|LS|J,Lambda=0,Omega=1> = 1/sqrt(3)
<J,Lambda=1,Omega=1|LS|J,Lambda=0,Omega=1> = -1/sqrt(3)

Linear: Excited Dsv=3
  Status     Active
  Colour     Orange
  Lambda     Delta
  OmegaOrder Auto => Regular
  Origin 68377.9123376615          Float  State Origin. Estimated lowest level = 68377.9123376615
  B      1.08921917955507          Float  Rotational constant.
  D      5.85540454591073E-6       Float  Quartic centrifugal distortion.
  H      -2.247197E-13                    Sextic centrifugal distortion.
  <J,Lambda=-2,Omega=-2|B|J,Lambda=-2,Omega=-2> = J*(J+1)
  <J,Lambda=2,Omega=2|B|J,Lambda=2,Omega=2> = J*(J+1)
    Symmetrized:
  <J,Lambda=2,Omega=2|B|J,Lambda=2,Omega=2> = J*(J+1)
  <J,Lambda=2,Omega=2|B|J,Lambda=2,Omega=2> = J*(J+1)

  <J,Lambda=-2,Omega=-2|D|J,Lambda=-2,Omega=-2> = -J^2*(J*(J+2)+1)
  <J,Lambda=2,Omega=2|D|J,Lambda=2,Omega=2> = -J^2*(J*(J+2)+1)
    Symmetrized:
```

<J,Lambda=2,Omega=2|D|J,Lambda=2,Omega=2> = -J^2*(J*(J+2)+1)
<J,Lambda=2,Omega=2|D|J,Lambda=2,Omega=2> = -J^2*(J*(J+2)+1)

<J,Lambda=-2,Omega=-2|H|J,Lambda=-2,Omega=-2> = J^3*(J*(J*(J+3)+3)+1)
<J,Lambda=2,Omega=2|H|J,Lambda=2,Omega=2> = J^3*(J*(J*(J+3)+3)+1)
Symmetrized:
<J,Lambda=2,Omega=2|H|J,Lambda=2,Omega=2> = J^3*(J*(J*(J+3)+3)+1)
<J,Lambda=2,Omega=2|H|J,Lambda=2,Omega=2> = J^3*(J*(J*(J+3)+3)+1)

LinearPerturbation: Excited <Av=2|J+L-|Dsv=3>
Value 0.02597 Size of perturbation.
<J,Lambda=-1,Omega=-1|J+L-|J,Lambda=-2,Omega=-2> = sqrt(J*(J+1)-2)
<J,Lambda=1,Omega=1|J+L-|J,Lambda=2,Omega=2> = sqrt(J*(J+1)-2)

Symmetrized Matrix Elements of <Av=2|J+L-|Dsv=3>
<J,Lambda=1,Omega=1|J+L-|J,Lambda=2,Omega=2> = sqrt(J*(J+1)-2)
<J,Lambda=1,Omega=1|J+L-|J,Lambda=2,Omega=2> = sqrt(J*(J+1)-2)

Linear: Excited a3Pv=13
Status Active
Colour Silver
S 1
Lambda Pi
OmegaOrder Auto => Regular
Origin 67670.6782977856 Float State Origin. Estimated lowest level = 67632.5507030479
B 1.30813665955102 Float Rotational constant.
A 38.1154482644111 Float Spin-orbit coupling constant.
LambdaSS -0.01821971 Spin - spin coupling constant (=1.5*epsilon).
o 0.6497647 Spin-spin coupling constant giving lambda doubling for Omega = 0 states.
gamma 0.003153713 Spin rotation coupling constant.
p 0.008831129 Lambda doubling constant.
q 2.786443E-5 Lambda doubling constant.
D 6.06110427245085E-6 Float Quartic centrifugal distortion.
AD -0.0001940953 Centrifugal distortion of A.
<J,Lambda=-1,Omega=-2|B|J,Lambda=-1,Omega=-2> = J*(J+1)-2
<J,Lambda=-1,Omega=-2|B|J,Lambda=-1,Omega=-1> = -sqrt(2*(J*(J+1)-2))
<J,Lambda=-1,Omega=-1|B|J,Lambda=-1,Omega=-2> = -sqrt(2*(J*(J+1)-2))
<J,Lambda=-1,Omega=-1|B|J,Lambda=-1,Omega=-1> = J*(J+1)+2
<J,Lambda=-1,Omega=-1|B|J,Lambda=-1,Omega=0> = -sqrt(2*J*(J+1))
<J,Lambda=-1,Omega=0|B|J,Lambda=-1,Omega=-1> = -sqrt(2*J*(J+1))
<J,Lambda=-1,Omega=0|B|J,Lambda=-1,Omega=0> = J*(J+1)+2
<J,Lambda=1,Omega=0|B|J,Lambda=1,Omega=0> = J*(J+1)+2
<J,Lambda=1,Omega=0|B|J,Lambda=1,Omega=1> = -sqrt(2*J*(J+1))
<J,Lambda=1,Omega=1|B|J,Lambda=1,Omega=0> = -sqrt(2*J*(J+1))
<J,Lambda=1,Omega=1|B|J,Lambda=1,Omega=1> = J*(J+1)+2
<J,Lambda=1,Omega=1|B|J,Lambda=1,Omega=2> = -sqrt(2*(J*(J+1)-2))
<J,Lambda=1,Omega=2|B|J,Lambda=1,Omega=1> = -sqrt(2*(J*(J+1)-2))
<J,Lambda=1,Omega=2|B|J,Lambda=1,Omega=2> = J*(J+1)-2
Symmetrized:
<J,Lambda=1,Omega=2|B|J,Lambda=1,Omega=2> = J*(J+1)-2
<J,Lambda=1,Omega=2|B|J,Lambda=1,Omega=2> = -sqrt(2*(J*(J+1)-2))
<J,Lambda=1,Omega=2|B|J,Lambda=1,Omega=1> = -sqrt(2*(J*(J+1)-2))
<J,Lambda=1,Omega=2|B|J,Lambda=1,Omega=1> = J*(J+1)+2
<J,Lambda=1,Omega=1|B|J,Lambda=1,Omega=2> = -sqrt(2*J*(J+1))
<J,Lambda=1,Omega=1|B|J,Lambda=1,Omega=2> = -sqrt(2*J*(J+1))
<J,Lambda=1,Omega=1|B|J,Lambda=1,Omega=1> = J*(J+1)+2
<J,Lambda=1,Omega=1|B|J,Lambda=1,Omega=1> = J*(J+1)+2
<J,Lambda=1,Omega=1|B|J,Lambda=1,Omega=0> = -sqrt(2*J*(J+1))
<J,Lambda=1,Omega=1|B|J,Lambda=1,Omega=0> = -sqrt(2*J*(J+1))
<J,Lambda=1,Omega=0|B|J,Lambda=1,Omega=1> = J*(J+1)+2
<J,Lambda=1,Omega=0|B|J,Lambda=1,Omega=1> = -sqrt(2*(J*(J+1)-2))
<J,Lambda=1,Omega=0|B|J,Lambda=1,Omega=0> = -sqrt(2*(J*(J+1)-2))
<J,Lambda=1,Omega=0|B|J,Lambda=1,Omega=0> = J*(J+1)-2

<J,Lambda=-1,Omega=-2|A|J,Lambda=-1,Omega=-2> = 1
<J,Lambda=-1,Omega=0|A|J,Lambda=-1,Omega=0> = -1
<J,Lambda=1,Omega=0|A|J,Lambda=1,Omega=0> = -1
<J,Lambda=1,Omega=2|A|J,Lambda=1,Omega=2> = 1
Symmetrized:
<J,Lambda=1,Omega=2|A|J,Lambda=1,Omega=2> = 1
<J,Lambda=1,Omega=2|A|J,Lambda=1,Omega=2> = -1
<J,Lambda=1,Omega=0|A|J,Lambda=1,Omega=0> = -1
<J,Lambda=1,Omega=0|A|J,Lambda=1,Omega=0> = 1

<J,Lambda=-1,Omega=-2|LambdaSS|J,Lambda=-1,Omega=-2> = 2/3
<J,Lambda=-1,Omega=-1|LambdaSS|J,Lambda=-1,Omega=-1> = -4/3
<J,Lambda=-1,Omega=0|LambdaSS|J,Lambda=-1,Omega=0> = 2/3
<J,Lambda=1,Omega=0|LambdaSS|J,Lambda=1,Omega=0> = 2/3
<J,Lambda=1,Omega=1|LambdaSS|J,Lambda=1,Omega=1> = -4/3
<J,Lambda=1,Omega=2|LambdaSS|J,Lambda=1,Omega=2> = 2/3
Symmetrized:
<J,Lambda=1,Omega=2|LambdaSS|J,Lambda=1,Omega=2> = 2/3
<J,Lambda=1,Omega=2|LambdaSS|J,Lambda=1,Omega=2> = -4/3
<J,Lambda=1,Omega=1|LambdaSS|J,Lambda=1,Omega=1> = 2/3
<J,Lambda=1,Omega=1|LambdaSS|J,Lambda=1,Omega=1> = 2/3
<J,Lambda=1,Omega=0|LambdaSS|J,Lambda=1,Omega=0> = -4/3
<J,Lambda=1,Omega=0|LambdaSS|J,Lambda=1,Omega=0> = 2/3

<J,Lambda=-1,Omega=0|o|J,Lambda=1,Omega=0> = 1
<J,Lambda=1,Omega=0|o|J,Lambda=-1,Omega=0> = 1
Symmetrized:
<J,Lambda=1,Omega=0|o|J,Lambda=1,Omega=0> = 1

<J,Lambda=1,Omega=0|o|J,Lambda=1,Omega=0> = 1

<J,Lambda=-1,Omega=-2|gamma|J,Lambda=-1,Omega=-1> = sqrt(2*(J*(J+1)-2))/2
<J,Lambda=-1,Omega=-1|gamma|J,Lambda=-1,Omega=-2> = sqrt(2*(J*(J+1)-2))/2
<J,Lambda=-1,Omega=-1|gamma|J,Lambda=-1,Omega=-1> = -2
<J,Lambda=-1,Omega=-1|gamma|J,Lambda=-1,Omega=0> = sqrt(2*J*(J+1))/2
<J,Lambda=-1,Omega=0|gamma|J,Lambda=-1,Omega=-1> = sqrt(2*J*(J+1))/2
<J,Lambda=-1,Omega=0|gamma|J,Lambda=-1,Omega=0> = -2
<J,Lambda=1,Omega=0|gamma|J,Lambda=1,Omega=0> = -2
<J,Lambda=1,Omega=0|gamma|J,Lambda=1,Omega=1> = sqrt(2*J*(J+1))/2
<J,Lambda=1,Omega=1|gamma|J,Lambda=1,Omega=0> = sqrt(2*J*(J+1))/2
<J,Lambda=1,Omega=1|gamma|J,Lambda=1,Omega=1> = -2
<J,Lambda=1,Omega=1|gamma|J,Lambda=1,Omega=2> = sqrt(2*(J*(J+1)-2))/2
<J,Lambda=1,Omega=2|gamma|J,Lambda=1,Omega=1> = sqrt(2*(J*(J+1)-2))/2
Symmetrized:
<J,Lambda=1,Omega=2|gamma|J,Lambda=1,Omega=1> = sqrt(2*(J*(J+1)-2))/2
<J,Lambda=1,Omega=2|gamma|J,Lambda=1,Omega=1> = sqrt(2*(J*(J+1)-2))/2
<J,Lambda=1,Omega=1|gamma|J,Lambda=1,Omega=2> = -2
<J,Lambda=1,Omega=1|gamma|J,Lambda=1,Omega=2> = sqrt(2*J*(J+1))/2
<J,Lambda=1,Omega=1|gamma|J,Lambda=1,Omega=1> = sqrt(2*J*(J+1))/2
<J,Lambda=1,Omega=1|gamma|J,Lambda=1,Omega=1> = -2
<J,Lambda=1,Omega=1|gamma|J,Lambda=1,Omega=0> = -2
<J,Lambda=1,Omega=1|gamma|J,Lambda=1,Omega=0> = sqrt(2*J*(J+1))/2
<J,Lambda=1,Omega=0|gamma|J,Lambda=1,Omega=1> = sqrt(2*J*(J+1))/2
<J,Lambda=1,Omega=0|gamma|J,Lambda=1,Omega=1> = -2
<J,Lambda=1,Omega=0|gamma|J,Lambda=1,Omega=0> = sqrt(2*(J*(J+1)-2))/2
<J,Lambda=1,Omega=0|gamma|J,Lambda=1,Omega=0> = sqrt(2*(J*(J+1)-2))/2

<J,Lambda=-1,Omega=-1|p|J,Lambda=1,Omega=0> = (-sqrt(2*J*(J+1)))/2
<J,Lambda=-1,Omega=0|p|J,Lambda=1,Omega=0> = 1
<J,Lambda=-1,Omega=0|p|J,Lambda=1,Omega=1> = (-sqrt(2*J*(J+1)))/2
<J,Lambda=1,Omega=0|p|J,Lambda=-1,Omega=-1> = (-sqrt(2*J*(J+1)))/2
<J,Lambda=1,Omega=0|p|J,Lambda=-1,Omega=0> = 1
<J,Lambda=1,Omega=1|p|J,Lambda=-1,Omega=0> = (-sqrt(2*J*(J+1)))/2
Symmetrized:
<J,Lambda=1,Omega=1|p|J,Lambda=1,Omega=0> = (-sqrt(2*J*(J+1)))/2
<J,Lambda=1,Omega=1|p|J,Lambda=1,Omega=0> = 1
<J,Lambda=1,Omega=0|p|J,Lambda=1,Omega=0> = (-sqrt(2*J*(J+1)))/2
<J,Lambda=1,Omega=0|p|J,Lambda=1,Omega=0> = (-sqrt(2*J*(J+1)))/2
<J,Lambda=1,Omega=0|p|J,Lambda=1,Omega=1> = 1
<J,Lambda=1,Omega=0|p|J,Lambda=1,Omega=1> = (-sqrt(2*J*(J+1)))/2

<J,Lambda=-1,Omega=-2|q|J,Lambda=1,Omega=0> = sqrt(J*(J*(J+1)-2)*(J+1))/2
<J,Lambda=-1,Omega=-1|q|J,Lambda=1,Omega=0> = -sqrt(2*J*(J+1))
<J,Lambda=-1,Omega=-1|q|J,Lambda=1,Omega=1> = (J*(J+1))/2
<J,Lambda=-1,Omega=0|q|J,Lambda=1,Omega=0> = 1
<J,Lambda=-1,Omega=0|q|J,Lambda=1,Omega=1> = -sqrt(2*J*(J+1))
<J,Lambda=-1,Omega=0|q|J,Lambda=1,Omega=2> = sqrt(J*(J*(J+1)-2)*(J+1))/2
<J,Lambda=1,Omega=0|q|J,Lambda=-1,Omega=-2> = sqrt(J*(J*(J+1)-2)*(J+1))/2
<J,Lambda=1,Omega=0|q|J,Lambda=-1,Omega=-1> = -sqrt(2*J*(J+1))
<J,Lambda=1,Omega=0|q|J,Lambda=-1,Omega=0> = 1
<J,Lambda=1,Omega=1|q|J,Lambda=-1,Omega=-1> = (J*(J+1))/2
<J,Lambda=1,Omega=1|q|J,Lambda=-1,Omega=0> = -sqrt(2*J*(J+1))
<J,Lambda=1,Omega=2|q|J,Lambda=-1,Omega=0> = sqrt(J*(J*(J+1)-2)*(J+1))/2
Symmetrized:
<J,Lambda=1,Omega=2|q|J,Lambda=1,Omega=0> = sqrt(J*(J*(J+1)-2)*(J+1))/2
<J,Lambda=1,Omega=2|q|J,Lambda=1,Omega=0> = -sqrt(2*J*(J+1))
<J,Lambda=1,Omega=1|q|J,Lambda=1,Omega=0> = (J*(J+1))/2
<J,Lambda=1,Omega=1|q|J,Lambda=1,Omega=0> = 1
<J,Lambda=1,Omega=1|q|J,Lambda=1,Omega=1> = -sqrt(2*J*(J+1))
<J,Lambda=1,Omega=1|q|J,Lambda=1,Omega=1> = sqrt(J*(J*(J+1)-2)*(J+1))/2
<J,Lambda=1,Omega=0|q|J,Lambda=1,Omega=0> = sqrt(J*(J*(J+1)-2)*(J+1))/2
<J,Lambda=1,Omega=0|q|J,Lambda=1,Omega=0> = -sqrt(2*J*(J+1))
<J,Lambda=1,Omega=0|q|J,Lambda=1,Omega=1> = 1
<J,Lambda=1,Omega=0|q|J,Lambda=1,Omega=1> = (J*(J+1))/2
<J,Lambda=1,Omega=0|q|J,Lambda=1,Omega=2> = -sqrt(2*J*(J+1))
<J,Lambda=1,Omega=0|q|J,Lambda=1,Omega=2> = sqrt(J*(J*(J+1)-2)*(J+1))/2

<J,Lambda=-1,Omega=-2|D|J,Lambda=-1,Omega=-2> = -J*(J*(J*(J+2)-1)-2)
<J,Lambda=-1,Omega=-2|D|J,Lambda=-1,Omega=-1> = 2*J*(J+1)*sqrt(2*(J*(J+1)-2))
<J,Lambda=-1,Omega=-2|D|J,Lambda=-1,Omega=0> = -2*sqrt(J*(J*(J*(J+2)-1)-2))
<J,Lambda=-1,Omega=-1|D|J,Lambda=-1,Omega=-2> = 2*J*(J+1)*sqrt(2*(J*(J+1)-2))
<J,Lambda=-1,Omega=-1|D|J,Lambda=-1,Omega=-1> = -J*(J*(J*(J+2)+9)+8)
<J,Lambda=-1,Omega=-1|D|J,Lambda=-1,Omega=0> = 2*(J*(J+1)+2)*sqrt(2*J*(J+1))
<J,Lambda=-1,Omega=0|D|J,Lambda=-1,Omega=-2> = -2*sqrt(J*(J*(J*(J+2)-1)-2))
<J,Lambda=-1,Omega=0|D|J,Lambda=-1,Omega=-1> = 2*(J*(J+1)+2)*sqrt(2*J*(J+1))
<J,Lambda=-1,Omega=0|D|J,Lambda=-1,Omega=0> = -J*(J*(J*(J+2)+7)+6)-4
<J,Lambda=1,Omega=0|D|J,Lambda=1,Omega=0> = -J*(J*(J*(J+2)+7)+6)-4
<J,Lambda=1,Omega=0|D|J,Lambda=1,Omega=1> = 2*(J*(J+1)+2)*sqrt(2*J*(J+1))
<J,Lambda=1,Omega=0|D|J,Lambda=1,Omega=2> = -2*sqrt(J*(J*(J*(J+2)-1)-2))
<J,Lambda=1,Omega=1|D|J,Lambda=1,Omega=0> = 2*(J*(J+1)+2)*sqrt(2*J*(J+1))
<J,Lambda=1,Omega=1|D|J,Lambda=1,Omega=1> = -J*(J*(J*(J+2)+9)+8)
<J,Lambda=1,Omega=1|D|J,Lambda=1,Omega=2> = 2*J*(J+1)*sqrt(2*(J*(J+1)-2))
<J,Lambda=1,Omega=2|D|J,Lambda=1,Omega=0> = -2*sqrt(J*(J*(J*(J+2)-1)-2))
<J,Lambda=1,Omega=2|D|J,Lambda=1,Omega=1> = 2*J*(J+1)*sqrt(2*(J*(J+1)-2))
<J,Lambda=1,Omega=2|D|J,Lambda=1,Omega=2> = -J*(J*(J*(J+2)-1)-2)
Symmetrized:
<J,Lambda=1,Omega=2|D|J,Lambda=1,Omega=2> = -J*(J*(J*(J+2)-1)-2)
<J,Lambda=1,Omega=2|D|J,Lambda=1,Omega=2> = 2*J*(J+1)*sqrt(2*(J*(J+1)-2))
<J,Lambda=1,Omega=2|D|J,Lambda=1,Omega=1> = -2*sqrt(J*(J*(J*(J+2)-1)-2))
<J,Lambda=1,Omega=2|D|J,Lambda=1,Omega=1> = 2*J*(J+1)*sqrt(2*(J*(J+1)-2))
<J,Lambda=1,Omega=2|D|J,Lambda=1,Omega=0> = -J*(J*(J*(J+2)+9)+8)

<J,Lambda=1,Omega=2|D|J,Lambda=1,Omega=0> = 2*(J*(J+1)+2)*sqrt(2*J*(J+1))
<J,Lambda=1,Omega=1|D|J,Lambda=1,Omega=2> = -2*sqrt(J*(J*(J*(J+2)-1)-2))
<J,Lambda=1,Omega=1|D|J,Lambda=1,Omega=2> = 2*(J*(J+1)+2)*sqrt(2*J*(J+1))
<J,Lambda=1,Omega=1|D|J,Lambda=1,Omega=1> = -J*(J*(J*(J+2)+7)+6)-4
<J,Lambda=1,Omega=1|D|J,Lambda=1,Omega=1> = -J*(J*(J*(J+2)+7)+6)-4
<J,Lambda=1,Omega=1|D|J,Lambda=1,Omega=0> = 2*(J*(J+1)+2)*sqrt(2*J*(J+1))
<J,Lambda=1,Omega=1|D|J,Lambda=1,Omega=0> = -2*sqrt(J*(J*(J*(J+2)-1)-2))
<J,Lambda=1,Omega=0|D|J,Lambda=1,Omega=2> = 2*(J*(J+1)+2)*sqrt(2*J*(J+1))
<J,Lambda=1,Omega=0|D|J,Lambda=1,Omega=2> = -J*(J*(J*(J+2)+9)+8)
<J,Lambda=1,Omega=0|D|J,Lambda=1,Omega=1> = 2*J*(J+1)*sqrt(2*(J*(J+1)-2))
<J,Lambda=1,Omega=0|D|J,Lambda=1,Omega=1> = -2*sqrt(J*(J*(J*(J+2)-1)-2))
<J,Lambda=1,Omega=0|D|J,Lambda=1,Omega=0> = 2*J*(J+1)*sqrt(2*(J*(J+1)-2))
<J,Lambda=1,Omega=0|D|J,Lambda=1,Omega=0> = -J*(J*(J*(J+2)-1)-2)

<J,Lambda=-1,Omega=-2|AD|J,Lambda=-1,Omega=-2> = J*(J+1)-2
<J,Lambda=-1,Omega=-2|AD|J,Lambda=-1,Omega=-1> = (-sqrt(2*(J*(J+1)-2)))/2
<J,Lambda=-1,Omega=-1|AD|J,Lambda=-1,Omega=-2> = (-sqrt(2*(J*(J+1)-2)))/2
<J,Lambda=-1,Omega=-1|AD|J,Lambda=-1,Omega=0> = sqrt(2*J*(J+1))/2
<J,Lambda=-1,Omega=0|AD|J,Lambda=-1,Omega=-1> = sqrt(2*J*(J+1))/2
<J,Lambda=-1,Omega=0|AD|J,Lambda=-1,Omega=0> = -J*(J+1)-2
<J,Lambda=1,Omega=0|AD|J,Lambda=1,Omega=0> = -J*(J+1)-2
<J,Lambda=1,Omega=0|AD|J,Lambda=1,Omega=1> = sqrt(2*J*(J+1))/2
<J,Lambda=1,Omega=1|AD|J,Lambda=1,Omega=0> = sqrt(2*J*(J+1))/2
<J,Lambda=1,Omega=1|AD|J,Lambda=1,Omega=2> = (-sqrt(2*(J*(J+1)-2)))/2
<J,Lambda=1,Omega=2|AD|J,Lambda=1,Omega=1> = (-sqrt(2*(J*(J+1)-2)))/2
<J,Lambda=1,Omega=2|AD|J,Lambda=1,Omega=2> = J*(J+1)-2
Symmetrized:
<J,Lambda=1,Omega=2|AD|J,Lambda=1,Omega=2> = J*(J+1)-2
<J,Lambda=1,Omega=2|AD|J,Lambda=1,Omega=2> = (-sqrt(2*(J*(J+1)-2)))/2
<J,Lambda=1,Omega=2|AD|J,Lambda=1,Omega=1> = (-sqrt(2*(J*(J+1)-2)))/2
<J,Lambda=1,Omega=2|AD|J,Lambda=1,Omega=1> = sqrt(2*J*(J+1))/2
<J,Lambda=1,Omega=1|AD|J,Lambda=1,Omega=2> = sqrt(2*J*(J+1))/2
<J,Lambda=1,Omega=1|AD|J,Lambda=1,Omega=2> = -J*(J+1)-2
<J,Lambda=1,Omega=1|AD|J,Lambda=1,Omega=0> = -J*(J+1)-2
<J,Lambda=1,Omega=1|AD|J,Lambda=1,Omega=0> = sqrt(2*J*(J+1))/2
<J,Lambda=1,Omega=0|AD|J,Lambda=1,Omega=1> = sqrt(2*J*(J+1))/2
<J,Lambda=1,Omega=0|AD|J,Lambda=1,Omega=1> = (-sqrt(2*(J*(J+1)-2)))/2
<J,Lambda=1,Omega=0|AD|J,Lambda=1,Omega=0> = (-sqrt(2*(J*(J+1)-2)))/2
<J,Lambda=1,Omega=0|AD|J,Lambda=1,Omega=0> = J*(J+1)-2

LinearPerturbation: Excited <a3Pv=13|LS|Iv=3>
Value -11.5278 Size of perturbation.
<J,Lambda=-1,Omega=0|LS|J,Lambda=0,Omega=0> = -1/sqrt(3)
<J,Lambda=1,Omega=0|LS|J,Lambda=0,Omega=0> = -1/sqrt(3)

Symmetrized Matrix Elements of <a3Pv=13|LS|Iv=3>
<J,Lambda=1,Omega=0|LS|J,Lambda=0,Omega=0> = -1/sqrt(3)

LinearPerturbation: Excited <a3Pv=13|LS|Iv=4>
Value -16.442 Size of perturbation.
<J,Lambda=-1,Omega=0|LS|J,Lambda=0,Omega=0> = -1/sqrt(3)
<J,Lambda=1,Omega=0|LS|J,Lambda=0,Omega=0> = -1/sqrt(3)

Symmetrized Matrix Elements of <a3Pv=13|LS|Iv=4>
<J,Lambda=1,Omega=0|LS|J,Lambda=0,Omega=0> = -1/sqrt(3)

LinearPerturbation: Excited <a3Pv=13|LS|ev=4>
Value 24.7288685468983 Float Size of perturbation.
<J,Lambda=-1,Omega=-1|LS|J,Lambda=0,Omega=-1> = 1/sqrt(6)
<J,Lambda=-1,Omega=0|LS|J,Lambda=0,Omega=0> = 1/sqrt(6)
<J,Lambda=1,Omega=0|LS|J,Lambda=0,Omega=0> = -1/sqrt(6)
<J,Lambda=1,Omega=1|LS|J,Lambda=0,Omega=1> = -1/sqrt(6)

Symmetrized Matrix Elements of <a3Pv=13|LS|ev=4>
<J,Lambda=1,Omega=1|LS|J,Lambda=0,Omega=1> = 1/sqrt(6)
<J,Lambda=1,Omega=1|LS|J,Lambda=0,Omega=1> = 1/sqrt(6)
<J,Lambda=1,Omega=0|LS|J,Lambda=0,Omega=0> = -1/sqrt(6)

LinearPerturbation: Excited <a3Pv=13|J+L-|ev=4>
Value -0.0598221863162786 Float Size of perturbation.
<J,Lambda=-1,Omega=-2|J+L-|J,Lambda=0,Omega=-1> = -sqrt(J*(J+1)-2)
<J,Lambda=-1,Omega=-1|J+L-|J,Lambda=0,Omega=0> = -sqrt(J*(J+1))
<J,Lambda=-1,Omega=0|J+L-|J,Lambda=0,Omega=1> = -sqrt(J*(J+1))
<J,Lambda=1,Omega=0|J+L-|J,Lambda=0,Omega=-1> = sqrt(J*(J+1))
<J,Lambda=1,Omega=1|J+L-|J,Lambda=0,Omega=0> = sqrt(J*(J+1))
<J,Lambda=1,Omega=2|J+L-|J,Lambda=0,Omega=1> = sqrt(J*(J+1)-2)

Symmetrized Matrix Elements of <a3Pv=13|J+L-|ev=4>
<J,Lambda=1,Omega=2|J+L-|J,Lambda=0,Omega=1> = -sqrt(J*(J+1)-2)
<J,Lambda=1,Omega=2|J+L-|J,Lambda=0,Omega=1> = -sqrt(J*(J+1))
<J,Lambda=1,Omega=1|J+L-|J,Lambda=0,Omega=0> = -sqrt(J*(J+1))
<J,Lambda=1,Omega=0|J+L-|J,Lambda=0,Omega=1> = sqrt(J*(J+1))
<J,Lambda=1,Omega=0|J+L-|J,Lambda=0,Omega=1> = sqrt(J*(J+1))

LinearPerturbation: Excited <a3Pv=13|LS|dv=7>
Value 23.2188654276278 Float Size of perturbation.
<J,Lambda=-1,Omega=-2|LS|J,Lambda=-2,Omega=-2> = -1/sqrt(6)
<J,Lambda=-1,Omega=-1|LS|J,Lambda=-2,Omega=-1> = -1/sqrt(6)
<J,Lambda=1,Omega=1|LS|J,Lambda=2,Omega=1> = -1/sqrt(6)
<J,Lambda=1,Omega=2|LS|J,Lambda=2,Omega=2> = -1/sqrt(6)

Symmetrized Matrix Elements of <a3Pv=13|LS|dv=7>

<J,Lambda=1,Omega=2|LS|J,Lambda=2,Omega=2> = -1/sqrt(6)
<J,Lambda=1,Omega=2|LS|J,Lambda=2,Omega=2> = -1/sqrt(6)
<J,Lambda=1,Omega=1|LS|J,Lambda=2,Omega=1> = -1/sqrt(6)
<J,Lambda=1,Omega=1|LS|J,Lambda=2,Omega=1> = -1/sqrt(6)

LinearPerturbation: Excited <a3Pv=13|J+L-|dv=7>
Value -0.062795149954351 Float Size of perturbation.
<J,Lambda=-1,Omega=-2|J+L-|J,Lambda=-2,Omega=-3> = sqrt(J*(J+1)-6)
<J,Lambda=-1,Omega=-1|J+L-|J,Lambda=-2,Omega=-2> = sqrt(J*(J+1)-2)
<J,Lambda=-1,Omega=0|J+L-|J,Lambda=-2,Omega=-1> = sqrt(J*(J+1))
<J,Lambda=1,Omega=0|J+L-|J,Lambda=2,Omega=1> = sqrt(J*(J+1))
<J,Lambda=1,Omega=1|J+L-|J,Lambda=2,Omega=2> = sqrt(J*(J+1)-2)
<J,Lambda=1,Omega=2|J+L-|J,Lambda=2,Omega=3> = sqrt(J*(J+1)-6)

Symmetrized Matrix Elements of <a3Pv=13|J+L-|dv=7>
<J,Lambda=1,Omega=2|J+L-|J,Lambda=2,Omega=3> = sqrt(J*(J+1)-6)
<J,Lambda=1,Omega=2|J+L-|J,Lambda=2,Omega=3> = sqrt(J*(J+1)-2)
<J,Lambda=1,Omega=1|J+L-|J,Lambda=2,Omega=2> = sqrt(J*(J+1))
<J,Lambda=1,Omega=1|J+L-|J,Lambda=2,Omega=2> = sqrt(J*(J+1))
<J,Lambda=1,Omega=0|J+L-|J,Lambda=2,Omega=1> = sqrt(J*(J+1)-2)
<J,Lambda=1,Omega=0|J+L-|J,Lambda=2,Omega=1> = sqrt(J*(J+1)-6)

LinearPerturbation: Excited <a3Pv=13|LS|a'v=12>
Value 2.39113881014602 Float Size of perturbation.
<J,Lambda=-1,Omega=-1|LS|J,Lambda=0,Omega=-1> = -1/sqrt(6)
<J,Lambda=-1,Omega=0|LS|J,Lambda=0,Omega=0> = -1/sqrt(6)
<J,Lambda=1,Omega=0|LS|J,Lambda=0,Omega=0> = -1/sqrt(6)
<J,Lambda=1,Omega=1|LS|J,Lambda=0,Omega=1> = -1/sqrt(6)

Symmetrized Matrix Elements of <a3Pv=13|LS|a'v=12>
<J,Lambda=1,Omega=1|LS|J,Lambda=0,Omega=1> = -1/sqrt(6)
<J,Lambda=1,Omega=1|LS|J,Lambda=0,Omega=1> = -1/sqrt(6)
<J,Lambda=1,Omega=0|LS|J,Lambda=0,Omega=0> = -1/sqrt(6)

LinearPerturbation: Excited <a3Pv=13|J+L-|a'v=12>
Value -0.0112468031812311 Float Size of perturbation.
<J,Lambda=-1,Omega=-2|J+L-|J,Lambda=0,Omega=-1> = sqrt(J*(J+1)-2)
<J,Lambda=-1,Omega=-1|J+L-|J,Lambda=0,Omega=0> = sqrt(J*(J+1))
<J,Lambda=-1,Omega=0|J+L-|J,Lambda=0,Omega=1> = sqrt(J*(J+1))
<J,Lambda=1,Omega=0|J+L-|J,Lambda=0,Omega=-1> = sqrt(J*(J+1))
<J,Lambda=1,Omega=1|J+L-|J,Lambda=0,Omega=0> = sqrt(J*(J+1))
<J,Lambda=1,Omega=2|J+L-|J,Lambda=0,Omega=1> = sqrt(J*(J+1)-2)

Symmetrized Matrix Elements of <a3Pv=13|J+L-|a'v=12>
<J,Lambda=1,Omega=2|J+L-|J,Lambda=0,Omega=1> = sqrt(J*(J+1)-2)
<J,Lambda=1,Omega=2|J+L-|J,Lambda=0,Omega=1> = sqrt(J*(J+1))
<J,Lambda=1,Omega=1|J+L-|J,Lambda=0,Omega=0> = sqrt(J*(J+1))
<J,Lambda=1,Omega=0|J+L-|J,Lambda=0,Omega=1> = sqrt(J*(J+1))
<J,Lambda=1,Omega=0|J+L-|J,Lambda=0,Omega=1> = sqrt(J*(J+1))

LinearPerturbation: Excited <a3Pv=13|LS|Dsv=3>
Value -22.8568 Size of perturbation.
<J,Lambda=-1,Omega=-2|LS|J,Lambda=-2,Omega=-2> = 1/sqrt(3)
<J,Lambda=1,Omega=2|LS|J,Lambda=2,Omega=2> = -1/sqrt(3)

Symmetrized Matrix Elements of <a3Pv=13|LS|Dsv=3>
<J,Lambda=1,Omega=2|LS|J,Lambda=2,Omega=2> = 1/sqrt(3)
<J,Lambda=1,Omega=2|LS|J,Lambda=2,Omega=2> = -1/sqrt(3)

TransitionMoments: <Excited|mu|X>
SphericalTransitionMoment: <Excited|mu|X> <Av=2|T(1)|Xv=0>
Rank 1
Component Auto => 1
Strength 1 Transition (dipole) moment / Debye - note that relative intensities are proportional to the square of this value.
Reduced Matrix Elements of <Av=2|T(1)|Xv=0>
<J-1,Lambda=-1,Omega=-1|T(1)|J,Lambda=0,Omega=0> = sqrt(2*(J-1))/2
<J-1,Lambda=1,Omega=1|T(1)|J,Lambda=0,Omega=0> = sqrt(2*(J-1))/2
<J,Lambda=-1,Omega=-1|T(1)|J,Lambda=0,Omega=0> = (2*J+1)/sqrt(2*(2*J+1))
<J,Lambda=1,Omega=1|T(1)|J,Lambda=0,Omega=0> = (-2*J-1)/sqrt(2*(2*J+1))
<J+1,Lambda=-1,Omega=-1|T(1)|J,Lambda=0,Omega=0> = sqrt(2*(J+2))/2
<J+1,Lambda=1,Omega=1|T(1)|J,Lambda=0,Omega=0> = sqrt(2*(J+2))/2
Line Strengths from <Av=2|T(1)|Xv=0>
S<J-1,Lambda=-1,Omega=-1|T(1)|J,Lambda=0,Omega=0> = (J-1)/2
S<J-1,Lambda=1,Omega=1|T(1)|J,Lambda=0,Omega=0> = (J-1)/2
S<J,Lambda=-1,Omega=-1|T(1)|J,Lambda=0,Omega=0> = (2*J+1)/2
S<J,Lambda=1,Omega=1|T(1)|J,Lambda=0,Omega=0> = (-2*J-1)/-2
S<J+1,Lambda=-1,Omega=-1|T(1)|J,Lambda=0,Omega=0> = (J+2)/2
S<J+1,Lambda=1,Omega=1|T(1)|J,Lambda=0,Omega=0> = (J+2)/2
Symmetrized Reduced Matrix Elements of <Av=2|T(1)|Xv=0>
<J-1,Lambda=1,Omega=1 + |T(1)|J,Lambda=0,Omega=0> = sqrt(J-1)
<J,Lambda=1,Omega=1 - |T(1)|J,Lambda=0,Omega=0> = (-2*J-1)/sqrt(2*J+1)
<J+1,Lambda=1,Omega=1 + |T(1)|J,Lambda=0,Omega=0> = sqrt(J+2)
Symmetrized Line Strengths from <Av=2|T(1)|Xv=0>
S<J-1,Lambda=1,Omega=1 + |T(1)|J,Lambda=0,Omega=0> = J-1
S<J,Lambda=1,Omega=1 - |T(1)|J,Lambda=0,Omega=0> = (-2*J-1)/-1
S<J+1,Lambda=1,Omega=1 + |T(1)|J,Lambda=0,Omega=0> = J+2

ForbiddenTransitionMoment: <Excited|mu|X> <a3Pv=13|fT(1,1)|Xv=0>
Rank 1
Component 1
Magnetic False
Strength 1 Transition (dipole) moment / Debye - note that relative intensities are proportional to the square of this value.

Reduced Matrix Elements of <a3Pv=13|fT(1,1)|Xv=0>
<J-1,Lambda=-1,Omega=-1|fT(1,1)|J,Lambda=0,Omega=0> = (-sqrt(2*(J-1)))/2
<J-1,Lambda=1,Omega=1|fT(1,1)|J,Lambda=0,Omega=0> = sqrt(2*(J-1))/2
<J,Lambda=-1,Omega=-1|fT(1,1)|J,Lambda=0,Omega=0> = (-2*J-1)/sqrt(2*(2*J+1))
<J,Lambda=1,Omega=1|fT(1,1)|J,Lambda=0,Omega=0> = (-2*J-1)/sqrt(2*(2*J+1))
<J+1,Lambda=-1,Omega=-1|fT(1,1)|J,Lambda=0,Omega=0> = (-sqrt(2*(J+2)))/2
<J+1,Lambda=1,Omega=1|fT(1,1)|J,Lambda=0,Omega=0> = sqrt(2*(J+2))/2
Line Strengths from <a3Pv=13|fT(1,1)|Xv=0>
S<J-1,Lambda=-1,Omega=-1|fT(1,1)|J,Lambda=0,Omega=0> = (J-1)/2
S<J-1,Lambda=1,Omega=1|fT(1,1)|J,Lambda=0,Omega=0> = (J-1)/2
S<J,Lambda=-1,Omega=-1|fT(1,1)|J,Lambda=0,Omega=0> = (-2*J-1)/-2
S<J,Lambda=1,Omega=1|fT(1,1)|J,Lambda=0,Omega=0> = (-2*J-1)/-2
S<J+1,Lambda=-1,Omega=-1|fT(1,1)|J,Lambda=0,Omega=0> = (J+2)/2
S<J+1,Lambda=1,Omega=1|fT(1,1)|J,Lambda=0,Omega=0> = (J+2)/2
Symmetrized Reduced Matrix Elements of <a3Pv=13|fT(1,1)|Xv=0>
<J-1,Lambda=1,Omega=1 - |fT(1,1)|J,Lambda=0,Omega=0> = sqrt(J-1)
<J,Lambda=1,Omega=1 + |fT(1,1)|J,Lambda=0,Omega=0> = (-2*J-1)/sqrt(2*J+1)
<J+1,Lambda=1,Omega=1 - |fT(1,1)|J,Lambda=0,Omega=0> = sqrt(J+2)
Symmetrized Line Strengths from <a3Pv=13|fT(1,1)|Xv=0>
S<J-1,Lambda=1,Omega=1 - |fT(1,1)|J,Lambda=0,Omega=0> = J-1
S<J,Lambda=1,Omega=1 + |fT(1,1)|J,Lambda=0,Omega=0> = (-2*J-1)/-1
S<J+1,Lambda=1,Omega=1 - |fT(1,1)|J,Lambda=0,Omega=0> = J+2

ForbiddenTransitionMoment: <Excited|mu|X> <dv=7|fT(1,1)|Xv=0>
Rank 1
Component 1
Magnetic False
Strength 1 Transition (dipole) moment / Debye - note that relative intensities are proportional to the square of this value.
Reduced Matrix Elements of <dv=7|fT(1,1)|Xv=0>
<J-1,Lambda=-2,Omega=-1|fT(1,1)|J,Lambda=0,Omega=0> = (-sqrt(2*(J-1)))/2
<J-1,Lambda=2,Omega=1|fT(1,1)|J,Lambda=0,Omega=0> = sqrt(2*(J-1))/2
<J,Lambda=-2,Omega=-1|fT(1,1)|J,Lambda=0,Omega=0> = (-2*J-1)/sqrt(2*(2*J+1))
<J,Lambda=2,Omega=1|fT(1,1)|J,Lambda=0,Omega=0> = (-2*J-1)/sqrt(2*(2*J+1))
<J+1,Lambda=-2,Omega=-1|fT(1,1)|J,Lambda=0,Omega=0> = (-sqrt(2*(J+2)))/2
<J+1,Lambda=2,Omega=1|fT(1,1)|J,Lambda=0,Omega=0> = sqrt(2*(J+2))/2
Line Strengths from <dv=7|fT(1,1)|Xv=0>
S<J-1,Lambda=-2,Omega=-1|fT(1,1)|J,Lambda=0,Omega=0> = (J-1)/2
S<J-1,Lambda=2,Omega=1|fT(1,1)|J,Lambda=0,Omega=0> = (J-1)/2
S<J,Lambda=-2,Omega=-1|fT(1,1)|J,Lambda=0,Omega=0> = (-2*J-1)/-2
S<J,Lambda=2,Omega=1|fT(1,1)|J,Lambda=0,Omega=0> = (-2*J-1)/-2
S<J+1,Lambda=-2,Omega=-1|fT(1,1)|J,Lambda=0,Omega=0> = (J+2)/2
S<J+1,Lambda=2,Omega=1|fT(1,1)|J,Lambda=0,Omega=0> = (J+2)/2
Symmetrized Reduced Matrix Elements of <dv=7|fT(1,1)|Xv=0>
<J-1,Lambda=2,Omega=1 - |fT(1,1)|J,Lambda=0,Omega=0> = sqrt(J-1)
<J,Lambda=2,Omega=1 + |fT(1,1)|J,Lambda=0,Omega=0> = (-2*J-1)/sqrt(2*J+1)
<J+1,Lambda=2,Omega=1 - |fT(1,1)|J,Lambda=0,Omega=0> = sqrt(J+2)
Symmetrized Line Strengths from <dv=7|fT(1,1)|Xv=0>
S<J-1,Lambda=2,Omega=1 - |fT(1,1)|J,Lambda=0,Omega=0> = J-1
S<J,Lambda=2,Omega=1 + |fT(1,1)|J,Lambda=0,Omega=0> = (-2*J-1)/-1
S<J+1,Lambda=2,Omega=1 - |fT(1,1)|J,Lambda=0,Omega=0> = J+2

ForbiddenTransitionMoment: <Excited|mu|X> <a'v=12|fT(1,1)|Xv=0>
Rank 1
Component 1
Magnetic False
Strength 1 Transition (dipole) moment / Debye - note that relative intensities are proportional to the square of this value.
Reduced Matrix Elements of <a'v=12|fT(1,1)|Xv=0>
<J-1,Lambda=0,Omega=-1|fT(1,1)|J,Lambda=0,Omega=0> = (-sqrt(2*(J-1)))/2
<J-1,Lambda=0,Omega=1|fT(1,1)|J,Lambda=0,Omega=0> = sqrt(2*(J-1))/2
<J,Lambda=0,Omega=-1|fT(1,1)|J,Lambda=0,Omega=0> = (-2*J-1)/sqrt(2*(2*J+1))
<J,Lambda=0,Omega=1|fT(1,1)|J,Lambda=0,Omega=0> = (-2*J-1)/sqrt(2*(2*J+1))
<J+1,Lambda=0,Omega=-1|fT(1,1)|J,Lambda=0,Omega=0> = (-sqrt(2*(J+2)))/2
<J+1,Lambda=0,Omega=1|fT(1,1)|J,Lambda=0,Omega=0> = sqrt(2*(J+2))/2
Line Strengths from <a'v=12|fT(1,1)|Xv=0>
S<J-1,Lambda=0,Omega=-1|fT(1,1)|J,Lambda=0,Omega=0> = (J-1)/2
S<J-1,Lambda=0,Omega=1|fT(1,1)|J,Lambda=0,Omega=0> = (J-1)/2
S<J,Lambda=0,Omega=-1|fT(1,1)|J,Lambda=0,Omega=0> = (-2*J-1)/-2
S<J,Lambda=0,Omega=1|fT(1,1)|J,Lambda=0,Omega=0> = (-2*J-1)/-2
S<J+1,Lambda=0,Omega=-1|fT(1,1)|J,Lambda=0,Omega=0> = (J+2)/2
S<J+1,Lambda=0,Omega=1|fT(1,1)|J,Lambda=0,Omega=0> = (J+2)/2
Symmetrized Reduced Matrix Elements of <a'v=12|fT(1,1)|Xv=0>
<J-1,Lambda=0,Omega=1 - |fT(1,1)|J,Lambda=0,Omega=0> = sqrt(J-1)
<J,Lambda=0,Omega=1 + |fT(1,1)|J,Lambda=0,Omega=0> = (-2*J-1)/sqrt(2*J+1)
<J+1,Lambda=0,Omega=1 - |fT(1,1)|J,Lambda=0,Omega=0> = sqrt(J+2)
Symmetrized Line Strengths from <a'v=12|fT(1,1)|Xv=0>
S<J-1,Lambda=0,Omega=1 - |fT(1,1)|J,Lambda=0,Omega=0> = J-1
S<J,Lambda=0,Omega=1 + |fT(1,1)|J,Lambda=0,Omega=0> = (-2*J-1)/-1
S<J+1,Lambda=0,Omega=1 - |fT(1,1)|J,Lambda=0,Omega=0> = J+2

ForbiddenTransitionMoment: <Excited|mu|X> <ev=4|fT(1,1)|Xv=0>
Rank 1
Component 1
Magnetic False
Strength 1 Transition (dipole) moment / Debye - note that relative intensities are proportional to the square of this value.
Reduced Matrix Elements of <ev=4|fT(1,1)|Xv=0>
<J-1,Lambda=0,Omega=-1|fT(1,1)|J,Lambda=0,Omega=0> = sqrt(2*(J-1))/2
<J-1,Lambda=0,Omega=1|fT(1,1)|J,Lambda=0,Omega=0> = sqrt(2*(J-1))/2
<J,Lambda=0,Omega=-1|fT(1,1)|J,Lambda=0,Omega=0> = (2*J+1)/sqrt(2*(2*J+1))
<J,Lambda=0,Omega=1|fT(1,1)|J,Lambda=0,Omega=0> = (-2*J-1)/sqrt(2*(2*J+1))
<J+1,Lambda=0,Omega=-1|fT(1,1)|J,Lambda=0,Omega=0> = sqrt(2*(J+2))/2
<J+1,Lambda=0,Omega=1|fT(1,1)|J,Lambda=0,Omega=0> = sqrt(2*(J+2))/2
Line Strengths from <ev=4|fT(1,1)|Xv=0>

```
    S<J-1,Lambda=0,Omega=-1|fT(1,1)|J,Lambda=0,Omega=0> = (J-1)/2
    S<J-1,Lambda=0,Omega=1|fT(1,1)|J,Lambda=0,Omega=0> = (J-1)/2
    S<J,Lambda=0,Omega=-1|fT(1,1)|J,Lambda=0,Omega=0> = (2*J+1)/2
    S<J,Lambda=0,Omega=1|fT(1,1)|J,Lambda=0,Omega=0> = (-2*J-1)/-2
    S<J+1,Lambda=0,Omega=-1|fT(1,1)|J,Lambda=0,Omega=0> = (J+2)/2
    S<J+1,Lambda=0,Omega=1|fT(1,1)|J,Lambda=0,Omega=0> = (J+2)/2
    Symmetrized Reduced Matrix Elements of <ev=4|fT(1,1)|Xv=0>
    <J-1,Lambda=0,Omega=1 + |fT(1,1)|J,Lambda=0,Omega=0> = sqrt(J-1)
    <J,Lambda=0,Omega=1 - |fT(1,1)|J,Lambda=0,Omega=0> = (-2*J-1)/sqrt(2*J+1)
    <J+1,Lambda=0,Omega=1 + |fT(1,1)|J,Lambda=0,Omega=0> = sqrt(J+2)
    Symmetrized Line Strengths from <ev=4|fT(1,1)|Xv=0>
    S<J-1,Lambda=0,Omega=1 + |fT(1,1)|J,Lambda=0,Omega=0> = J-1
    S<J,Lambda=0,Omega=1 - |fT(1,1)|J,Lambda=0,Omega=0> = (-2*J-1)/-1
    S<J+1,Lambda=0,Omega=1 + |fT(1,1)|J,Lambda=0,Omega=0> = J+2

    ForbiddenTransitionMoment: <Excited|mu|X> <Iv=3|fT(1,1)|Xv=0>
      Rank      1
      Component 1
      Magnetic  False
      Strength 1                Transition (dipole) moment / Debye - note that relative intensities are proportional to the square of this value.
    ForbiddenTransitionMoment: <Excited|mu|X> <Dsv=3|fT(1,1)|Xv=0>
      Rank      1
      Component 1
      Magnetic  False
      Strength 1                Transition (dipole) moment / Debye - note that relative intensities are proportional to the square of this value.
  TransitionMoments: <B|mu|X>
    SphericalTransitionMoment: <B|mu|X> <Bv=0|T(1)|Xv=0>
      Rank      1
      Component Auto => 0
      Strength 1                Transition (dipole) moment / Debye - note that relative intensities are proportional to the square of this value.
    Reduced Matrix Elements of <Bv=0|T(1)|Xv=0>
    <J-1,Lambda=0,Omega=0|T(1)|J,Lambda=0,Omega=0> = -sqrt(J)
    <J+1,Lambda=0,Omega=0|T(1)|J,Lambda=0,Omega=0> = (J+1)/sqrt(J+1)
    Line Strengths from <Bv=0|T(1)|Xv=0>
    S<J-1,Lambda=0,Omega=0|T(1)|J,Lambda=0,Omega=0> = J
    S<J+1,Lambda=0,Omega=0|T(1)|J,Lambda=0,Omega=0> = J+1
    Symmetrized Reduced Matrix Elements of <Bv=0|T(1)|Xv=0>
    <J-1,Lambda=0,Omega=0|T(1)|J,Lambda=0,Omega=0> = -sqrt(J)
    <J+1,Lambda=0,Omega=0|T(1)|J,Lambda=0,Omega=0> = (J+1)/sqrt(J+1)
    Symmetrized Line Strengths from <Bv=0|T(1)|Xv=0>
    S<J-1,Lambda=0,Omega=0|T(1)|J,Lambda=0,Omega=0> = J
    S<J+1,Lambda=0,Omega=0|T(1)|J,Lambda=0,Omega=0> = J+1

  TransitionMoments: <B|mu|Excited>
    SphericalTransitionMoment: <B|mu|Excited> <Bv=0|T(1)|Av=2>
      Rank      1
      Component Auto => 1
      Strength 1                Transition (dipole) moment / Debye - note that relative intensities are proportional to the square of this value.
    Reduced Matrix Elements of <Bv=0|T(1)|Av=2>
    <J-1,Lambda=0,Omega=0|T(1)|J,Lambda=-1,Omega=-1> = sqrt(2*(J+1))/2
    <J-1,Lambda=0,Omega=0|T(1)|J,Lambda=1,Omega=1> = sqrt(2*(J+1))/2
    <J,Lambda=0,Omega=0|T(1)|J,Lambda=-1,Omega=-1> = (-2*J-1)/sqrt(2*(2*J+1))
    <J,Lambda=0,Omega=0|T(1)|J,Lambda=1,Omega=1> = (2*J+1)/sqrt(2*(2*J+1))
    <J+1,Lambda=0,Omega=0|T(1)|J,Lambda=-1,Omega=-1> = sqrt(2*J)/2
    <J+1,Lambda=0,Omega=0|T(1)|J,Lambda=1,Omega=1> = sqrt(2*J)/2
    Line Strengths from <Bv=0|T(1)|Av=2>
    S<J-1,Lambda=0,Omega=0|T(1)|J,Lambda=-1,Omega=-1> = (J+1)/2
    S<J-1,Lambda=0,Omega=0|T(1)|J,Lambda=1,Omega=1> = (J+1)/2
    S<J,Lambda=0,Omega=0|T(1)|J,Lambda=-1,Omega=-1> = (-2*J-1)/-2
    S<J,Lambda=0,Omega=0|T(1)|J,Lambda=1,Omega=1> = (2*J+1)/2
    S<J+1,Lambda=0,Omega=0|T(1)|J,Lambda=-1,Omega=-1> = J/2
    S<J+1,Lambda=0,Omega=0|T(1)|J,Lambda=1,Omega=1> = J/2
    Symmetrized Reduced Matrix Elements of <Bv=0|T(1)|Av=2>
    <J-1,Lambda=0,Omega=0|T(1)| + J,Lambda=1,Omega=1> = sqrt(J+1)
    <J,Lambda=0,Omega=0|T(1)| - J,Lambda=1,Omega=1> = (2*J+1)/sqrt(2*J+1)
    <J+1,Lambda=0,Omega=0|T(1)| + J,Lambda=1,Omega=1> = sqrt(J)
    Symmetrized Line Strengths from <Bv=0|T(1)|Av=2>
    S<J-1,Lambda=0,Omega=0|T(1)| + J,Lambda=1,Omega=1> = J+1
    S<J,Lambda=0,Omega=0|T(1)| - J,Lambda=1,Omega=1> = 2*J+1
    S<J+1,Lambda=0,Omega=0|T(1)| + J,Lambda=1,Omega=1> = J

  TransitionMoments: <C|mu|X>
    SphericalTransitionMoment: <C|mu|X> <Cv=0|T(1)|Xv=0>
      Rank      1
      Component Auto => 0
      Strength 1                Transition (dipole) moment / Debye - note that relative intensities are proportional to the square of this value.
    Matrix elements as for <B|mu|X> <Bv=0|T(1)|Xv=0>
  TransitionMoments: <C|mu|Excited>
    SphericalTransitionMoment: <C|mu|Excited> <Cv=0|T(1)|Av=2>
      Rank      1
      Component Auto => 1
      Strength 1                Transition (dipole) moment / Debye - note that relative intensities are proportional to the square of this value.
    Matrix elements as for <B|mu|Excited> <Bv=0|T(1)|Av=2>
Simulation: Simulation
  Intensity factor:0.333333333333333 (1 from Spol)
  IntensityUnits Arbitrary
  LifeModel      lmNone
  PlotUnits      cm1
  Fluorescence   False
  nDF            50000
  WidthMult      6
  AutoMin        True
```

AutoMax True
ShowSum False
ShowParts True
ShowFortrat True
UseUpper False
ShowSymmetry False
ShowDeltaJ True
ScaleMarkSize True
UseSymmetry False
UseStateNumber False
FortratQno J
PreserveArea False
SeparateScales False
LogScale False
Show2DPlot False
Auto2DY True
Fmin 19713.4264274956 Left edge of plot range.
Fmax 19725.7352925044 Right edge of plot range.
Ymax 7.90431110974814E-183 Top of plot range; updated automatically if AutoMax set.
Temperature 300 Rotational temeperature (Kelvin).
Gaussian 0.08 Gaussian contribution to linewidth (full width half maximum). If both this and a Lorentzian width (below) is set the result is a convolution of the two, a Voigt profile.
Lorentzian 0.02 Lorentzian contribution to linewidth (full width half maximum). This and the Gaussian width can also be set from the main window.
SMargin -1 Extra points to calculate off each end of required spectrum.
OThreshold 0.0001 Ignore peaks smaller than this fraction of the maximum peak intensity.
Tvib -1 Vibrational Temperature (Kelvin); set to -1 (default) to use rotational temperature for all Boltzmann factors.
MinI 1E-15 Weakest transitions to plot in Fortrat plots, as a fraction of strongest.
Tspin -1 Spin Temperature; set to -1 (default) to equilibrate different spinstates
IScale 1 Scale factor applied to all line intensities
PlotSplit 0.5 Fraction of plot area to use for simulation where plots separated (-1..1)
Fundamental Constants from CODATA 2018: h=6.62607015E-34 k=1.380649E-23 NA=6.02214076E23 e=1.602176634E-19 me=9.1093837015E-31

Residuals before fit for 13C18O

| | J'S' | #' | J"S" | #" | Observed | Calculated | Obs-Calc | StdDev | |
|---|---|---|---|---|---|---|---|---|---|
| B | 0 e | 1 Excited | 1 e | 1 | 19356.9364 | 19356.9290 | 0.0074 | .010400 | 0.0000 : pP(1) : B Bv=0 0 0 F1e - Excited Av=2 1 1 F1e : 1:<Linelist> |
| B | 1 e | 1 Excited | 2 e | 1 | 19354.8102 | 19354.8018 | 0.0084 | .008800 | 0.0000 : pP(2) : B Bv=0 1 1 F1e - Excited Av=2 2 2 F1e : 2:<Linelist> |
| B | 2 e | 1 Excited | 3 e | 1 | 19353.3740 | 19353.3800 | -0.0060 | .008200 | 0.0000 : pP(3) : B Bv=0 2 2 F1e - Excited Av=2 3 3 F1e : 3:<Linelist> |
| B | 3 e | 1 Excited | 4 e | 1 | 19352.6739 | 19352.6608 | 0.0131 | .005000 | 0.0000 : pP(4) : B Bv=0 3 3 F1e - Excited Av=2 4 4 F1e : 4:<Linelist> |
| B | 4 e | 1 Excited | 5 e | 1 | 19352.6421 | 19352.6473 | -0.0052 | .004600 | 0.0000 : pP(5) : B Bv=0 4 4 F1e - Excited Av=2 5 5 F1e : 5:<Linelist> |
| B | 5 e | 1 Excited | 6 e | 1 | 19353.3482 | 19353.3409 | 0.0073 | .004800 | 0.0000 : pP(6) : B Bv=0 5 5 F1e - Excited Av=2 6 6 F1e : 6:<Linelist> |
| B | 6 e | 1 Excited | 7 e | 1 | 19354.7444 | 19354.7406 | 0.0038 | .004700 | 0.0000 : pP(7) : B Bv=0 6 6 F1e - Excited Av=2 7 7 F1e : 7:<Linelist> |
| B | 7 e | 1 Excited | 8 e | 1 | 19356.8501 | 19356.8475 | 0.0026 | .004500 | 0.0000 : pP(8) : B Bv=0 7 7 F1e - Excited Av=2 8 8 F1e : 8:<Linelist> |
| B | 8 e | 1 Excited | 9 e | 1 | 19359.6624 | 19359.6614 | 0.0010 | .004200 | 0.0000 : pP(9) : B Bv=0 8 8 F1e - Excited Av=2 9 9 F1e : 9:<Linelist> |
| B | 9 e | 1 Excited | 10 e | 1 | 19363.1843 | 19363.1832 | 0.0011 | .004100 | 0.0000 : pP(10) : B Bv=0 9 9 F1e - Excited Av=2 10 10 F1e : 10:<Linelist> |
| B | 10 e | 1 Excited | 11 e | 1 | 19367.4148 | 19367.4132 | 0.0016 | .004400 | 0.0000 : pP(11) : B Bv=0 10 10 F1e - Excited Av=2 11 11 F1e : 11:<Linelist> |
| B | 11 e | 1 Excited | 12 e | 1 | 19372.3592 | 19372.3522 | 0.0070 | .004300 | 0.0000 : pP(12) : B Bv=0 11 11 F1e - Excited Av=2 12 12 F1e : 12:<Linelist> |
| B | 12 e | 1 Excited | 13 e | 1 | 19378.0054 | 19378.0039 | 0.0015 | .004300 | 0.0000 : pP(13) : B Bv=0 12 12 F1e - Excited Av=2 13 13 F1e : 13:<Linelist> |
| B | 13 e | 1 Excited | 14 e | 1 | 19384.3636 | 19384.3633 | 0.0003 | .004300 | 0.0000 : pP(14) : B Bv=0 13 13 F1e - Excited Av=2 14 14 F1e : 14:<Linelist> |
| B | 14 e | 1 Excited | 15 e | 1 | 19391.4392 | 19391.4370 | 0.0022 | .004400 | 0.0000 : pP(15) : B Bv=0 14 14 F1e - Excited Av=2 15 15 F1e : 15:<Linelist> |
| B | 15 e | 1 Excited | 16 e | 1 | 19399.2271 | 19399.2267 | 0.0004 | .004300 | 0.0000 : pP(16) : B Bv=0 15 15 F1e - Excited Av=2 16 16 F1e : 16:<Linelist> |
| B | 16 e | 1 Excited | 17 e | 1 | 19407.7396 | 19407.7355 | 0.0041 | .004300 | 0.0000 : pP(17) : B Bv=0 16 16 F1e - Excited Av=2 17 17 F1e : 17:<Linelist> |
| B | 17 e | 1 Excited | 18 e | 1 | 19416.9437 | 19416.9351 | 0.0086 | .004400 | 0.0000 : pP(18) : B Bv=0 17 17 F1e - Excited Av=2 18 18 F1e : 18:<Linelist> |
| B | 18 e | 1 Excited | 19 e | 1 | 19426.9802 | 19426.9826 | -0.0024 | .004400 | 0.0000 : pP(19) : B Bv=0 18 18 F1e - Excited Av=2 19 19 F1e : 19:<Linelist> |
| B | 19 e | 1 Excited | 20 e | 1 | 19438.2394 | 19438.2542 | -0.0148 | .004200 | 0.0000 : pP(20) : B Bv=0 19 19 F1e - Excited Av=2 20 20 F1e : 20:<Linelist> |
| B | 20 e | 1 Excited | 21 e | 2 | 19448.7305 | 19448.7257 | 0.0048 | .004400 | 0.0000 : pP(21) : B Bv=0 20 20 F1e - Excited Av=2 21 21 F1e : 21:<Linelist> |
| B | 21 e | 1 Excited | 22 e | 2 | 19461.0398 | 19461.0346 | 0.0052 | .004200 | 0.0000 : pP(22) : B Bv=0 21 21 F1e - Excited Av=2 22 22 F1e : 22:<Linelist> |
| B | 22 e | 1 Excited | 23 e | 2 | 19474.2101 | 19474.2072 | 0.0029 | .004200 | 0.0000 : pP(23) : B Bv=0 22 22 F1e - Excited Av=2 23 23 F1e : 23:<Linelist> |
| B | 23 e | 1 Excited | 24 e | 2 | 19489.1916 | 19489.1960 | -0.0044 | .004300 | 0.0000 : pP(24) : B Bv=0 23 23 F1e - Excited Av=2 24 24 F1e : 24:<Linelist> |
| B | 24 e | 1 Excited | 25 e | 3 | 19498.9255 | 19498.9162 | 0.0093 | .004900 | 0.0000 : pP(25) : B Bv=0 24 24 F1e - Excited Av=2 25 25 F1e : 25:<Linelist> |
| B | 25 e | 1 Excited | 26 e | 3 | 19515.0461 | 19515.0358 | 0.0103 | .004200 | 0.0000 : pP(26) : B Bv=0 25 25 F1e - Excited Av=2 26 26 F1e : 26:<Linelist> |
| B | 26 e | 1 Excited | 27 e | 3 | 19531.0940 | 19531.0825 | 0.0115 | .004300 | 0.0000 : pP(27) : B Bv=0 26 26 F1e - Excited Av=2 27 27 F1e : 27:<Linelist> |
| B | 27 e | 1 Excited | 28 e | 3 | 19547.7725 | 19547.7635 | 0.0090 | .004900 | 0.0000 : pP(28) : B Bv=0 27 27 F1e - Excited Av=2 28 28 F1e : 28:<Linelist> |
| B | 28 e | 1 Excited | 29 e | 3 | 19565.4193 | 19565.4040 | 0.0153 | .004400 | 0.0000 : pP(29) : B Bv=0 28 28 F1e - Excited Av=2 29 29 F1e : 29:<Linelist> |
| B | 29 e | 1 Excited | 30 e | 4 | 19581.8845 | 19581.9027 | -0.0182 | .007200 | 0.0000 : pP(30) : B Bv=0 29 29 F1e - Excited Av=2 30 30 F1e : 30:<Linelist> |
| B | 30 e | 1 Excited | 31 e | 5 | 19599.2533 | 19599.2703 | -0.0170 | .004700 | 0.0000 : pP(31) : B Bv=0 30 30 F1e - Excited Av=2 31 31 F1e : 31:<Linelist> |
| B | 31 e | 1 Excited | 32 e | 5 | 19620.5380 | 19620.5020 | 0.0360 | .005100 | 0.0000 : pP(32) : B Bv=0 31 31 F1e - Excited Av=2 32 32 F1e : 32:<Linelist> |
| B | 32 e | 1 Excited | 33 e | 6 | 19640.7943 | 19640.7886 | 0.0057 | .007100 | 0.0000 : pP(33) : B Bv=0 32 32 F1e - Excited Av=2 33 33 F1e : 33:<Linelist> |
| B | 33 e | 1 Excited | 34 e | 8 | 19657.8684 | 19657.8671 | 0.0013 | .007600 | 0.0000 : pP(34) : B Bv=0 33 33 F1e - Excited Av=2 34 34 F1e : 34:<Linelist> |
| B | 34 e | 1 Excited | 35 e | 8 | 19680.4164 | 19680.4402 | -0.0238 | .005100 | 0.0000 : pP(35) : B Bv=0 34 34 F1e - Excited Av=2 35 35 F1e : 35:<Linelist> |
| B | 35 e | 1 Excited | 36 e | 8 | 19703.0470 | 19703.0767 | -0.0297 | .005600 | 0.0000 : pP(36) : B Bv=0 35 35 F1e - Excited Av=2 36 36 F1e : 36:<Linelist> |
| B | 36 e | 1 Excited | 37 e | 8 | 19727.8522 | 19727.8463 | 0.0059 | .007300 | 0.0000 : pP(37) : B Bv=0 36 36 F1e - Excited Av=2 37 37 F1e : 37:<Linelist> |
| B | 37 e | 1 Excited | 38 e | 9 | 19748.2986 | 19748.2785 | 0.0201 | .006100 | 0.0000 : pP(38) : B Bv=0 37 37 F1e - Excited Av=2 38 38 F1e : 38:<Linelist> |
| B | 38 e | 1 Excited | 39 e | 9 | 19772.9425 | 19772.9449 | -0.0024 | .006600 | 0.0000 : pP(39) : B Bv=0 38 38 F1e - Excited Av=2 39 39 F1e : 39:<Linelist> |
| B | 39 e | 1 Excited | 40 e | 9 | 19798.0262 | 19798.0520 | -0.0258 | .006700 | 0.0000 : pP(40) : B Bv=0 39 39 F1e - Excited Av=2 40 40 F1e : 40:<Linelist> |
| B | 40 e | 1 Excited | 41 e | 9 | 19823.7713 | 19823.7639 | 0.0074 | .008200 | 0.0000 : pP(41) : B Bv=0 40 40 F1e - Excited Av=2 41 41 F1e : 41:<Linelist> |
| B | 41 e | 1 Excited | 42 e | 9 | 19850.2203 | 19850.2169 | 0.0034 | .015900 | 0.0000 : pP(42) : B Bv=0 41 41 F1e - Excited Av=2 42 42 F1e : 42:<Linelist> |
| B | 1 e | 1 Excited | 1 f | 1 | 19360.4765 | 19360.4712 | 0.0053 | .006900 | 0.0000 : qQ(1) : B Bv=0 1 1 F1e - Excited Av=2 1 1 F1f : 51:<Linelist> |
| B | 2 e | 1 Excited | 2 f | 1 | 19361.8747 | 19361.8845 | -0.0098 | .004300 | 0.0000 : qQ(2) : B Bv=0 2 2 F1e - Excited Av=2 2 2 F1f : 52:<Linelist> |
| B | 3 e | 1 Excited | 3 f | 1 | 19364.0018 | 19364.0008 | 0.0010 | .004200 | 0.0000 : qQ(3) : B Bv=0 3 3 F1e - Excited Av=2 3 3 F1f : 53:<Linelist> |
| B | 4 e | 1 Excited | 4 f | 1 | 19366.8254 | 19366.8236 | 0.0018 | .004200 | 0.0000 : qQ(4) : B Bv=0 4 4 F1e - Excited Av=2 4 4 F1f : 54:<Linelist> |
| B | 5 e | 1 Excited | 5 f | 1 | 19370.3568 | 19370.3558 | 0.0010 | .004100 | 0.0000 : qQ(5) : B Bv=0 5 5 F1e - Excited Av=2 5 5 F1f : 55:<Linelist> |
| B | 6 e | 1 Excited | 6 f | 1 | 19374.6037 | 19374.6027 | 0.0010 | .004100 | 0.0000 : qQ(6) : B Bv=0 6 6 F1e - Excited Av=2 6 6 F1f : 56:<Linelist> |
| B | 7 e | 1 Excited | 7 f | 1 | 19379.6160 | 19379.6167 | -0.0007 | .004100 | 0.0000 : qQ(7) : B Bv=0 7 7 F1e - Excited Av=2 7 7 F1f : 57:<Linelist> |
| B | 8 e | 1 Excited | 8 f | 2 | 19385.0128 | 19385.0115 | 0.0013 | .004100 | 0.0000 : qQ(8) : B Bv=0 8 8 F1e - Excited Av=2 8 8 F1f : 58:<Linelist> |
| B | 9 e | 1 Excited | 9 f | 2 | 19391.4421 | 19391.4420 | 0.0001 | .004200 | 0.0000 : qQ(9) : B Bv=0 9 9 F1e - Excited Av=2 9 9 F1f : 59:<Linelist> |
| B | 10 e | 1 Excited | 10 f | 2 | 19398.5151 | 19398.5131 | 0.0020 | .004100 | 0.0000 : qQ(10) : B Bv=0 10 10 F1e - Excited Av=2 10 10 F1f : 60:<Linelist> |
| B | 11 e | 1 Excited | 11 f | 2 | 19406.2832 | 19406.2809 | 0.0023 | .004100 | 0.0000 : qQ(11) : B Bv=0 11 11 F1e - Excited Av=2 11 11 F1f : 61:<Linelist> |
| B | 12 e | 1 Excited | 12 f | 2 | 19414.7569 | 19414.7552 | 0.0017 | .004100 | 0.0000 : qQ(12) : B Bv=0 12 12 F1e - Excited Av=2 12 12 F1f : 62:<Linelist> |
| B | 13 e | 1 Excited | 13 f | 2 | 19423.9331 | 19423.9321 | 0.0010 | .004100 | 0.0000 : qQ(13) : B Bv=0 13 13 F1e - Excited Av=2 13 13 F1f : 63:<Linelist> |
| B | 14 e | 1 Excited | 14 f | 2 | 19433.8190 | 19433.8180 | 0.0010 | .004100 | 0.0000 : qQ(14) : B Bv=0 14 14 F1e - Excited Av=2 14 14 F1f : 64:<Linelist> |
| B | 15 e | 1 Excited | 15 f | 2 | 19444.4144 | 19444.4131 | 0.0013 | .004100 | 0.0000 : qQ(15) : B Bv=0 15 15 F1e - Excited Av=2 15 15 F1f : 65:<Linelist> |

| | | | | | | | |
|---|---|---|---|---|---|---|---|
| B | 16 e | 1 Excited | 16 f | 2 | 19455.7195 | 19455.7174 | 0.0021 .004100 0.0000 : qQ(16) : B Bv=0 16 16 F1e - Excited Av=2 16 16 F1f : 66:<Linelist> |
| B | 17 e | 1 Excited | 17 f | 2 | 19467.6954 | 19467.6939 | 0.0015 .004100 0.0000 : qQ(17) : B Bv=0 17 17 F1e - Excited Av=2 17 17 F1f : 67:<Linelist> |
| B | 18 e | 1 Excited | 18 f | 2 | 19480.4670 | 19480.4655 | 0.0015 .004100 0.0000 : qQ(18) : B Bv=0 18 18 F1e - Excited Av=2 18 18 F1f : 68:<Linelist> |
| B | 19 e | 1 Excited | 19 f | 2 | 19493.9260 | 19493.9248 | 0.0012 .004100 0.0000 : qQ(19) : B Bv=0 19 19 F1e - Excited Av=2 19 19 F1f : 69:<Linelist> |
| B | 20 e | 1 Excited | 20 f | 2 | 19508.1115 | 19508.1144 | -0.0029 .004100 0.0000 : qQ(20) : B Bv=0 20 20 F1e - Excited Av=2 20 20 F1f : 70:<Linelist> |
| B | 21 e | 1 Excited | 21 f | 3 | 19522.8673 | 19522.8893 | -0.0220 .004200 0.0000 : qQ(21) : B Bv=0 21 21 F1e - Excited Av=2 21 21 F1f : 71:<Linelist> |
| B | 22 e | 1 Excited | 22 f | 3 | 19538.5774 | 19538.5766 | 0.0008 .004100 0.0000 : qQ(22) : B Bv=0 22 22 F1e - Excited Av=2 22 22 F1f : 72:<Linelist> |
| B | 23 e | 1 Excited | 23 f | 3 | 19554.9752 | 19554.9734 | 0.0018 .004100 0.0000 : qQ(23) : B Bv=0 23 23 F1e - Excited Av=2 23 23 F1f : 73:<Linelist> |
| B | 24 e | 1 Excited | 24 f | 3 | 19572.1678 | 19572.1651 | 0.0027 .004100 0.0000 : qQ(24) : B Bv=0 24 24 F1e - Excited Av=2 24 24 F1f : 74:<Linelist> |
| B | 25 e | 1 Excited | 25 f | 3 | 19590.2714 | 19590.2678 | 0.0036 .004100 0.0000 : qQ(25) : B Bv=0 25 25 F1e - Excited Av=2 25 25 F1f : 75:<Linelist> |
| B | 26 e | 1 Excited | 26 f | 3 | 19609.6974 | 19609.6939 | 0.0035 .004300 0.0000 : qQ(26) : B Bv=0 26 26 F1e - Excited Av=2 26 26 F1f : 76:<Linelist> |
| B | 27 e | 1 Excited | 27 f | 3 | 19632.5452 | 19632.5425 | 0.0027 .004200 0.0000 : qQ(27) : B Bv=0 27 27 F1e - Excited Av=2 27 27 F1f : 77:<Linelist> |
| B | 28 e | 1 Excited | 28 f | 4 | 19643.0506 | 19643.0511 | -0.0005 .004200 0.0000 : qQ(28) : B Bv=0 28 28 F1e - Excited Av=2 28 28 F1f : 78:<Linelist> |
| B | 29 e | 1 Excited | 29 f | 4 | 19665.1158 | 19665.1185 | -0.0027 .004200 0.0000 : qQ(29) : B Bv=0 29 29 F1e - Excited Av=2 29 29 F1f : 79:<Linelist> |
| B | 30 e | 1 Excited | 30 f | 5 | 19686.7378 | 19686.7570 | -0.0192 .004200 0.0000 : qQ(30) : B Bv=0 30 30 F1e - Excited Av=2 30 30 F1f : 80:<Linelist> |
| B | 31 e | 1 Excited | 31 f | 6 | 19708.1635 | 19708.1585 | 0.0050 .004200 0.0000 : qQ(31) : B Bv=0 31 31 F1e - Excited Av=2 31 31 F1f : 81:<Linelist> |
| B | 32 e | 1 Excited | 32 f | 7 | 19731.6545 | 19731.6650 | -0.0105 .004300 0.0000 : qQ(32) : B Bv=0 32 32 F1e - Excited Av=2 32 32 F1f : 82:<Linelist> |
| B | 33 e | 1 Excited | 33 f | 7 | 19755.9326 | 19755.9068 | 0.0258 .004300 0.0000 : qQ(33) : B Bv=0 33 33 F1e - Excited Av=2 33 33 F1f : 83:<Linelist> |
| B | 34 e | 1 Excited | 34 f | 8 | 19778.1407 | 19778.1402 | 0.0005 .004400 0.0000 : qQ(34) : B Bv=0 34 34 F1e - Excited Av=2 34 34 F1f : 84:<Linelist> |
| B | 35 e | 1 Excited | 35 f | 8 | 19803.8946 | 19803.9036 | -0.0090 .004500 0.0000 : qQ(35) : B Bv=0 35 35 F1e - Excited Av=2 35 35 F1f : 85:<Linelist> |
| B | 36 e | 1 Excited | 36 f | 8 | 19830.5464 | 19830.5398 | 0.0066 .004700 0.0000 : qQ(36) : B Bv=0 36 36 F1e - Excited Av=2 36 36 F1f : 86:<Linelist> |
| B | 37 e | 1 Excited | 37 f | 9 | 19857.5698 | 19857.5834 | -0.0136 .007600 0.0000 : qQ(37) : B Bv=0 37 37 F1e - Excited Av=2 37 37 F1f : 87:<Linelist> |
| B | 38 e | 1 Excited | 38 f | 10 | 19881.5635 | 19881.5380 | 0.0255 .004900 0.0000 : qQ(38) : B Bv=0 38 38 F1e - Excited Av=2 38 38 F1f : 88:<Linelist> |
| B | 39 e | 1 Excited | 39 f | 10 | 19909.6670 | 19909.6770 | -0.0100 .007500 0.0000 : qQ(39) : B Bv=0 39 39 F1e - Excited Av=2 39 39 F1f : 89:<Linelist> |
| B | 40 e | 1 Excited | 40 f | 10 | 19938.2037 | 19938.1835 | 0.0202 .005800 0.0000 : qQ(40) : B Bv=0 40 40 F1e - Excited Av=2 40 40 F1f : 90:<Linelist> |
| B | 41 e | 1 Excited | 41 f | 10 | 19967.3972 | 19967.3783 | 0.0189 .029500 0.0000 : qQ(41) : B Bv=0 41 41 F1e - Excited Av=2 41 41 F1f : 91:<Linelist> |
| B | 42 e | 1 Excited | 42 f | 10 | 19997.2535 | 19997.2241 | 0.0294 .026200 0.0000 : qQ(42) : B Bv=0 42 42 F1e - Excited Av=2 42 42 F1f : 92:<Linelist> |
| B | 43 e | 1 Excited | 43 f | 10 | 20027.8299 | 20027.8017 | 0.0282 .033900 0.0000 : qQ(43) : B Bv=0 43 43 F1e - Excited Av=2 43 43 F1f : 93:<Linelist> |
| B | 44 e | 1 Excited | 44 f | 10 | 20059.0969 | 20059.0957 | 0.0012 .039200 0.0000 : qQ(44) : B Bv=0 44 44 F1e - Excited Av=2 44 44 F1f : 94:<Linelist> |
| B | 45 e | 1 Excited | 45 f | 10 | 20091.1567 | 20091.0867 | 0.0700 .040800 0.0000 : qQ(45) : B Bv=0 45 45 F1e - Excited Av=2 45 45 F1f : 95:<Linelist> |
| B | 2 e | 1 Excited | 1 e | 1 | 19367.5524 | 19367.5519 | 0.0005 .018800 0.0000 : rR(1) : B Bv=0 2 2 F1e - Excited Av=2 1 1 F1e : 101:<Linelist> |
| B | 3 e | 1 Excited | 2 e | 1 | 19372.5035 | 19372.5004 | 0.0031 .011000 0.0000 : rR(2) : B Bv=0 3 3 F1e - Excited Av=2 2 2 F1e : 102:<Linelist> |
| B | 4 e | 1 Excited | 3 e | 1 | 19378.1584 | 19378.1533 | 0.0051 .006900 0.0000 : rR(3) : B Bv=0 4 4 F1e - Excited Av=2 3 3 F1e : 103:<Linelist> |
| B | 5 e | 1 Excited | 4 e | 1 | 19384.5121 | 19384.5118 | 0.0003 .006000 0.0000 : rR(4) : B Bv=0 5 5 F1e - Excited Av=2 4 4 F1e : 104:<Linelist> |
| B | 6 e | 1 Excited | 5 e | 1 | 19391.5775 | 19391.5747 | 0.0028 .005300 0.0000 : rR(5) : B Bv=0 6 6 F1e - Excited Av=2 5 5 F1e : 105:<Linelist> |
| B | 7 e | 1 Excited | 6 e | 1 | 19399.3426 | 19399.3424 | 0.0002 .005100 0.0000 : rR(6) : B Bv=0 7 7 F1e - Excited Av=2 6 6 F1e : 106:<Linelist> |
| B | 8 e | 1 Excited | 7 e | 1 | 19407.8211 | 19407.8148 | 0.0063 .004900 0.0000 : rR(7) : B Bv=0 8 8 F1e - Excited Av=2 7 7 F1e : 107:<Linelist> |
| B | 9 e | 1 Excited | 8 e | 1 | 19417.0053 | 19416.9918 | 0.0135 .005100 0.0000 : rR(8) : B Bv=0 9 9 F1e - Excited Av=2 8 8 F1e : 108:<Linelist> |
| B | 10 e | 1 Excited | 9 e | 1 | 19426.8748 | 19426.8738 | 0.0010 .004600 0.0000 : rR(9) : B Bv=0 10 10 F1e - Excited Av=2 9 9 F1e : 109:<Linelist> |
| B | 11 e | 1 Excited | 10 e | 1 | 19437.4625 | 19437.4609 | 0.0016 .004500 0.0000 : rR(10) : B Bv=0 11 11 F1e - Excited Av=2 10 10 F1e : 110:<Linelist> |
| B | 12 e | 1 Excited | 11 e | 1 | 19448.7620 | 19448.7561 | 0.0059 .004400 0.0000 : rR(11) : B Bv=0 12 12 F1e - Excited Av=2 11 11 F1e : 111:<Linelist> |
| B | 13 e | 1 Excited | 12 e | 1 | 19460.7552 | 19460.7538 | 0.0014 .004200 0.0000 : rR(12) : B Bv=0 13 13 F1e - Excited Av=2 12 12 F1e : 112:<Linelist> |
| B | 14 e | 1 Excited | 13 e | 1 | 19473.4610 | 19473.4595 | 0.0015 .004200 0.0000 : rR(13) : B Bv=0 14 14 F1e - Excited Av=2 13 13 F1e : 113:<Linelist> |
| B | 15 e | 1 Excited | 14 e | 1 | 19486.8753 | 19486.8733 | 0.0020 .004200 0.0000 : rR(14) : B Bv=0 15 15 F1e - Excited Av=2 14 14 F1e : 114:<Linelist> |
| B | 16 e | 1 Excited | 15 e | 1 | 19500.9974 | 19500.9950 | 0.0024 .004200 0.0000 : rR(15) : B Bv=0 16 16 F1e - Excited Av=2 15 15 F1e : 115:<Linelist> |
| B | 17 e | 1 Excited | 16 e | 1 | 19515.7901 | 19515.7886 | 0.0015 .004200 0.0000 : rR(16) : B Bv=0 17 17 F1e - Excited Av=2 16 16 F1e : 116:<Linelist> |
| B | 18 e | 1 Excited | 17 e | 1 | 19531.3806 | 19531.3784 | 0.0022 .004200 0.0000 : rR(17) : B Bv=0 18 18 F1e - Excited Av=2 17 17 F1e : 117:<Linelist> |
| B | 19 e | 1 Excited | 18 e | 1 | 19547.6628 | 19547.6624 | 0.0004 .004500 0.0000 : rR(18) : B Bv=0 19 19 F1e - Excited Av=2 18 18 F1e : 118:<Linelist> |
| B | 20 e | 1 Excited | 19 e | 1 | 19564.6997 | 19564.7006 | -0.0009 .004200 0.0000 : rR(19) : B Bv=0 20 20 F1e - Excited Av=2 19 19 F1e : 119:<Linelist> |
| B | 21 e | 1 Excited | 20 e | 1 | 19582.9777 | 19582.9906 | -0.0129 .004300 0.0000 : rR(20) : B Bv=0 21 21 F1e - Excited Av=2 20 20 F1e : 120:<Linelist> |
| B | 22 e | 1 Excited | 21 e | 2 | 19600.4930 | 19600.4843 | 0.0087 .004200 0.0000 : rR(21) : B Bv=0 22 22 F1e - Excited Av=2 21 21 F1e : 121:<Linelist> |
| B | 23 e | 1 Excited | 22 e | 2 | 19619.8141 | 19619.8086 | 0.0055 .004200 0.0000 : rR(22) : B Bv=0 23 23 F1e - Excited Av=2 22 22 F1e : 122:<Linelist> |
| B | 24 e | 1 Excited | 23 e | 2 | 19639.9923 | 19639.9892 | 0.0031 .004800 0.0000 : rR(23) : B Bv=0 24 24 F1e - Excited Av=2 23 23 F1e : 123:<Linelist> |
| B | 25 e | 1 Excited | 24 e | 2 | 19661.9741 | 19661.9783 | -0.0042 .004400 0.0000 : rR(24) : B Bv=0 25 25 F1e - Excited Av=2 24 24 F1e : 124:<Linelist> |
| B | 26 e | 1 Excited | 25 e | 3 | 19678.7118 | 19678.6982 | 0.0136 .005200 0.0000 : rR(25) : B Bv=0 26 26 F1e - Excited Av=2 25 25 F1e : 125:<Linelist> |
| B | 27 e | 1 Excited | 26 e | 3 | 19701.8223 | 19701.8131 | 0.0092 .004400 0.0000 : rR(26) : B Bv=0 27 27 F1e - Excited Av=2 26 26 F1e : 126:<Linelist> |
| B | 28 e | 1 Excited | 27 e | 3 | 19724.8468 | 19724.8373 | 0.0095 .004400 0.0000 : rR(27) : B Bv=0 28 28 F1e - Excited Av=2 27 27 F1e : 127:<Linelist> |
| B | 29 e | 1 Excited | 28 e | 3 | 19748.5004 | 19748.4921 | 0.0083 .004400 0.0000 : rR(28) : B Bv=0 29 29 F1e - Excited Av=2 28 28 F1e : 128:<Linelist> |
| B | 30 e | 1 Excited | 29 e | 3 | 19773.1118 | 19773.0992 | 0.0126 .004600 0.0000 : rR(29) : B Bv=0 30 30 F1e - Excited Av=2 29 29 F1e : 129:<Linelist> |
| B | 31 e | 1 Excited | 30 e | 4 | 19796.5213 | 19796.5483 | -0.0270 .011600 0.0000 : rR(30) : B Bv=0 31 31 F1e - Excited Av=2 30 30 F1e : 130:<Linelist> |
| B | 32 e | 1 Excited | 31 e | 5 | 19820.8478 | 19820.8659 | -0.0181 .005100 0.0000 : rR(31) : B Bv=0 32 32 F1e - Excited Av=2 31 31 F1e : 131:<Linelist> |
| B | 33 e | 1 Excited | 32 e | 5 | 19849.0850 | 19849.0582 | 0.0268 .012100 0.0000 : rR(32) : B Bv=0 33 33 F1e - Excited Av=2 32 32 F1e : 132:<Linelist> |
| B | 34 e | 1 Excited | 33 e | 6 | 19876.2837 | 19876.2835 | 0.0002 .009900 0.0000 : rR(33) : B Bv=0 34 34 F1e - Excited Av=2 33 33 F1e : 133:<Linelist> |
| B | 35 e | 1 Excited | 34 e | 8 | 19900.3002 | 19900.2946 | 0.0056 .006000 0.0000 : rR(34) : B Bv=0 35 35 F1e - Excited Av=2 34 34 F1e : 134:<Linelist> |
| B | 36 e | 1 Excited | 35 e | 8 | 19929.7457 | 19929.7595 | -0.0138 .009500 0.0000 : rR(35) : B Bv=0 36 36 F1e - Excited Av=2 35 35 F1e : 135:<Linelist> |
| B | 37 e | 1 Excited | 36 e | 8 | 19959.2632 | 19959.2885 | -0.0253 .006100 0.0000 : rR(36) : B Bv=0 37 37 F1e - Excited Av=2 36 36 F1e : 136:<Linelist> |
| B | 38 e | 1 Excited | 37 e | 8 | 19990.9782 | 19990.9825 | -0.0043 .016400 0.0000 : rR(37) : B Bv=0 38 38 F1e - Excited Av=2 37 37 F1e : 137:<Linelist> |
| B | 39 e | 1 Excited | 38 e | 9 | 20018.3165 | 20018.3086 | 0.0079 .006700 0.0000 : rR(38) : B Bv=0 39 39 F1e - Excited Av=2 38 38 F1e : 138:<Linelist> |
| B | 40 e | 1 Excited | 39 e | 9 | 20049.8304 | 20049.8148 | 0.0156 .018400 0.0000 : rR(39) : B Bv=0 40 40 F1e - Excited Av=2 39 39 F1e : 139:<Linelist> |
| B | 41 e | 1 Excited | 40 e | 9 | 20081.8108 | 20081.7920 | 0.0188 .058500 0.0000 : rR(40) : B Bv=0 41 41 F1e - Excited Av=2 40 40 F1e : 140:<Linelist> |
| B | 42 e | 1 Excited | 41 e | 9 | 20114.3826 | 20114.3724 | 0.0102 .103900 0.0000 : rR(41) : B Bv=0 42 42 F1e - Excited Av=2 41 41 F1e : 141:<Linelist> |
| B | 7 e | 1 Excited | 7 f | 2 | 19376.8735 | 19376.8777 | -0.0042 .023800 4.0000 : qQ(7) : B Bv=0 7 7 F1e - Excited Iv=3 7 7 F1f : 151:<Linelist> |
| B | 8 e | 1 Excited | 8 f | 1 | 19387.8133 | 19387.8017 | 0.0116 .022600 4.0000 : qQ(8) : B Bv=0 8 8 F1e - Excited Iv=3 8 8 F1f : 152:<Linelist> |
| B | 19 e | 1 Excited | 20 e | 2 | 19435.8867 | 19435.8790 | 0.0077 .004900 3.0000 : qP1(20) : B Bv=0 19 19 F1e - Excited a3Pv=13 20 19 F1e : 153:<Linelist> |
| B | 21 e | 1 Excited | 20 e | 2 | 19580.6212 | 19580.6153 | 0.0059 .005400 3.0000 : sR1(20) : B Bv=0 21 21 F1e - Excited a3Pv=13 20 19 F1e : 154:<Linelist> |
| B | 33 e | 1 Excited | 32 e | 6 | 19845.8934 | 19845.9129 | -0.0195 .012700 3.0000 : rR12(32) : B Bv=0 33 33 F1e - Excited a3Pv=13 32 32 F2e : 155:<Linelist> |
| B | 23 e | 1 Excited | 24 e | 3 | 19477.4238 | 19477.4187 | 0.0051 .008600 2.0000 : qP1(24) : B Bv=0 23 23 F1e - Excited ev=4 24 23 F1e : 156:<Linelist> |
| B | 24 e | 1 Excited | 25 e | 2 | 19511.9306 | 19511.9585 | -0.0279 .014200 2.0000 : qP1(25) : B Bv=0 24 24 F1e - Excited ev=4 25 24 F1e : 157:<Linelist> |
| B | 25 e | 1 Excited | 24 e | 3 | 19650.2118 | 19650.2009 | 0.0109 .013000 2.0000 : sR1(24) : B Bv=0 25 25 F1e - Excited ev=4 24 23 F1e : 158:<Linelist> |
| B | 26 e | 1 Excited | 25 e | 2 | 19691.7106 | 19691.7405 | -0.0299 .022200 2.0000 : sR1(25) : B Bv=0 26 26 F1e - Excited ev=4 25 24 F1e : 159:<Linelist> |
| B | 26 e | 1 Excited | 26 f | 4 | 19586.2022 | 19586.1698 | 0.0324 .012600 2.0000 : qQ12(26) : B Bv=0 26 26 F1e - Excited ev=4 26 26 F2f : 160:<Linelist> |
| B | 27 e | 1 Excited | 27 f | 4 | 19617.6400 | 19617.6344 | 0.0056 .005300 2.0000 : qQ12(27) : B Bv=0 27 27 F1e - Excited ev=4 27 27 F2f : 161:<Linelist> |
| B | 28 e | 1 Excited | 28 f | 3 | 19663.9614 | 19663.9672 | -0.0058 .011100 2.0000 : qQ12(28) : B Bv=0 28 28 F1e - Excited ev=4 28 28 F2f : 162:<Linelist> |
| B | 30 e | 1 Excited | 30 f | 4 | 19689.9707 | 19689.9792 | -0.0085 .009500 1.0000 : rQ1(30) : B Bv=0 30 30 F1e - Excited dv=7 30 29 F1f : 163:<Linelist> |
| B | 37 e | 1 Excited | 38 e | 8 | 19769.2016 | 19769.2150 | -0.0134 .015100 1.0000 : oP13(38) : B Bv=0 37 37 F1e - Excited dv=7 38 39 F3e : 164:<Linelist> |
| C | 2 e | 1 Excited | 3 e | 1 | 24354.8311 | 24354.8144 | 0.0167 .042100 0.0000 : pP(3) : C Cv=0 2 2 F1e - Excited Av=2 3 3 F1e : 169:<Linelist> |
| C | 3 e | 1 Excited | 4 e | 1 | 24354.0840 | 24354.0720 | 0.0120 .018700 0.0000 : pP(4) : C Cv=0 3 3 F1e - Excited Av=2 4 4 F1e : 170:<Linelist> |
| C | 4 e | 1 Excited | 5 e | 1 | 24353.9976 | 24354.0330 | -0.0354 .025800 0.0000 : pP(5) : C Cv=0 4 4 F1e - Excited Av=2 5 5 F1e : 171:<Linelist> |
| C | 5 e | 1 Excited | 6 e | 1 | 24354.6935 | 24354.6971 | -0.0036 .022200 0.0000 : pP(6) : C Cv=0 5 5 F1e - Excited Av=2 6 6 F1e : 172:<Linelist> |
| C | 6 e | 1 Excited | 7 e | 1 | 24356.0403 | 24356.0561 | -0.0158 .016300 0.0000 : pP(7) : C Cv=0 6 6 F1e - Excited Av=2 7 7 F1e : 173:<Linelist> |
| C | 7 e | 1 Excited | 8 e | 1 | 24358.1104 | 24358.1183 | -0.0079 .012600 0.0000 : pP(8) : C Cv=0 7 7 F1e - Excited Av=2 8 8 F1e : 174:<Linelist> |
| C | 8 e | 1 Excited | 9 e | 1 | 24360.9032 | 24360.8793 | 0.0239 .010100 0.0000 : pP(9) : C Cv=0 8 8 F1e - Excited Av=2 9 9 F1e : 175:<Linelist> |

C 9 e 1 Excited 10 e 1 24364.3392 24364.3435 -0.0043 .007200 0.0000 : pP(10) : C Cv=0 9 9 F1e - Excited Av=2 10 10 F1e : 176:<Linelist>
C 10 e 1 Excited 11 e 1 24368.5041 24368.5046 -0.0005 .007100 0.0000 : pP(11) : C Cv=0 10 10 F1e - Excited Av=2 11 11 F1e : 177:<Linelist>
C 11 e 1 Excited 12 e 1 24373.3699 24373.3767 -0.0068 .011400 0.0000 : pP(12) : C Cv=0 11 11 F1e - Excited Av=2 12 12 F1e : 178:<Linelist>
C 12 e 1 Excited 13 e 1 24378.9452 24378.9512 -0.0060 .006600 0.0000 : pP(13) : C Cv=0 12 12 F1e - Excited Av=2 13 13 F1e : 179:<Linelist>
C 13 e 1 Excited 14 e 1 24385.2201 24385.2223 -0.0022 .006300 0.0000 : pP(14) : C Cv=0 13 13 F1e - Excited Av=2 14 14 F1e : 180:<Linelist>
C 14 e 1 Excited 15 e 1 24392.2089 24392.2084 0.0005 .006600 0.0000 : pP(15) : C Cv=0 14 14 F1e - Excited Av=2 15 15 F1e : 181:<Linelist>
C 15 e 1 Excited 16 e 1 24399.8980 24399.9044 -0.0064 .011400 0.0000 : pP(16) : C Cv=0 15 15 F1e - Excited Av=2 16 16 F1e : 182:<Linelist>
C 16 e 1 Excited 17 e 1 24408.3124 24408.3140 -0.0016 .010700 0.0000 : pP(17) : C Cv=0 16 16 F1e - Excited Av=2 17 17 F1e : 183:<Linelist>
C 17 e 1 Excited 18 e 1 24417.4487 24417.4481 0.0006 .012700 0.0000 : pP(18) : C Cv=0 17 17 F1e - Excited Av=2 18 18 F1e : 184:<Linelist>
C 18 e 1 Excited 19 e 1 24427.3343 24427.3448 -0.0105 .007100 0.0000 : pP(19) : C Cv=0 18 18 F1e - Excited Av=2 19 19 F1e : 185:<Linelist>
C 19 e 1 Excited 20 e 1 24438.5042 24438.4939 0.0103 .011300 0.0000 : pP(20) : C Cv=0 19 19 F1e - Excited Av=2 20 20 F1e : 186:<Linelist>
C 20 e 1 Excited 21 e 2 24448.8413 24448.8447 -0.0034 .013200 0.0000 : pP(21) : C Cv=0 20 20 F1e - Excited Av=2 21 21 F1e : 187:<Linelist>
C 21 e 1 Excited 22 e 2 24461.0288 24461.0259 0.0029 .015300 0.0000 : pP(22) : C Cv=0 21 21 F1e - Excited Av=2 22 22 F1e : 188:<Linelist>
C 22 e 1 Excited 23 e 2 24474.0657 24474.0669 -0.0012 .015200 0.0000 : pP(23) : C Cv=0 22 22 F1e - Excited Av=2 23 23 F1e : 189:<Linelist>
C 23 e 1 Excited 24 e 2 24488.9330 24488.9207 0.0123 .014600 0.0000 : pP(24) : C Cv=0 23 23 F1e - Excited Av=2 24 24 F1e : 190:<Linelist>
C 24 e 1 Excited 25 e 3 24498.5130 24498.5131 -0.0001 .016600 0.0000 : pP(25) : C Cv=0 24 24 F1e - Excited Av=2 25 25 F1e : 191:<Linelist>
C 25 e 1 Excited 26 e 3 24514.4732 24514.4934 -0.0202 .017300 0.0000 : pP(26) : C Cv=0 25 25 F1e - Excited Av=2 26 26 F1e : 192:<Linelist>
C 26 e 1 Excited 27 e 3 24530.3903 24530.3927 -0.0024 .024100 0.0000 : pP(27) : C Cv=0 26 26 F1e - Excited Av=2 27 27 F1e : 193:<Linelist>
C 27 e 1 Excited 28 e 3 24546.9699 24546.9195 0.0504 .031800 0.0000 : pP(28) : C Cv=0 27 27 F1e - Excited Av=2 28 28 F1e : 194:<Linelist>
C 28 e 1 Excited 29 e 3 24564.4189 24564.4086 0.0103 .019200 0.0000 : pP(29) : C Cv=0 28 28 F1e - Excited Av=2 29 29 F1e : 195:<Linelist>
C 29 e 1 Excited 30 e 4 24580.7313 24580.7502 -0.0189 .030600 0.0000 : pP(30) : C Cv=0 29 29 F1e - Excited Av=2 30 30 F1e : 196:<Linelist>
C 30 e 1 Excited 31 e 5 24597.9158 24597.9612 -0.0454 .036300 0.0000 : pP(31) : C Cv=0 30 30 F1e - Excited Av=2 31 31 F1e : 197:<Linelist>
C 31 e 1 Excited 32 e 5 24619.0661 24619.0399 0.0262 .051700 0.0000 : pP(32) : C Cv=0 31 31 F1e - Excited Av=2 32 32 F1e : 198:<Linelist>
C 32 e 1 Excited 33 e 6 24639.1615 24639.1699 -0.0084 .052300 0.0000 : pP(33) : C Cv=0 32 32 F1e - Excited Av=2 33 33 F1e : 199:<Linelist>
C 33 e 1 Excited 34 e 8 24656.0518 24656.0753 -0.0235 .061300 0.0000 : pP(34) : C Cv=0 33 33 F1e - Excited Av=2 34 34 F1e : 200:<Linelist>
C 34 e 1 Excited 35 e 8 24678.4432 24678.4841 -0.0409 .067800 0.0000 : pP(35) : C Cv=0 34 34 F1e - Excited Av=2 35 35 F1e : 201:<Linelist>
C 35 e 1 Excited 36 e 8 24700.8798 24700.9328 -0.0530 .046500 0.0000 : pP(36) : C Cv=0 35 35 F1e - Excited Av=2 36 36 F1e : 202:<Linelist>
C 1 e 1 Excited 1 f 1 24361.9678 24361.9224 0.0454 .072300 0.0000 : qQ(1) : C Cv=0 1 1 F1e - Excited Av=2 1 1 F1f : 208:<Linelist>
C 2 e 1 Excited 2 f 1 24363.3207 24363.3189 0.0018 .029700 0.0000 : qQ(2) : C Cv=0 2 2 F1e - Excited Av=2 2 2 F1f : 209:<Linelist>
C 3 e 1 Excited 3 f 1 24365.4217 24365.4120 0.0097 .029000 0.0000 : qQ(3) : C Cv=0 3 3 F1e - Excited Av=2 3 3 F1f : 210:<Linelist>
C 4 e 1 Excited 4 f 1 24368.2151 24368.2092 0.0059 .013700 0.0000 : qQ(4) : C Cv=0 4 4 F1e - Excited Av=2 4 4 F1f : 211:<Linelist>
C 5 e 1 Excited 5 f 1 24371.7092 24371.7120 -0.0028 .008000 0.0000 : qQ(5) : C Cv=0 5 5 F1e - Excited Av=2 5 5 F1f : 212:<Linelist>
C 6 e 1 Excited 6 f 1 24375.9216 24375.9182 0.0034 .006500 0.0000 : qQ(6) : C Cv=0 6 6 F1e - Excited Av=2 6 6 F1f : 213:<Linelist>
C 7 e 1 Excited 7 f 1 24380.8890 24380.8876 0.0014 .009900 0.0000 : qQ(7) : C Cv=0 7 7 F1e - Excited Av=2 7 7 F1f : 214:<Linelist>
C 8 e 1 Excited 8 f 2 24386.2273 24386.2294 -0.0021 .006100 0.0000 : qQ(8) : C Cv=0 8 8 F1e - Excited Av=2 8 8 F1f : 215:<Linelist>
C 9 e 1 Excited 9 f 2 24392.6039 24392.6024 0.0015 .008100 0.0000 : qQ(9) : C Cv=0 9 9 F1e - Excited Av=2 9 9 F1f : 216:<Linelist>
C 10 e 1 Excited 10 f 2 24399.6030 24399.6045 -0.0015 .007000 0.0000 : qQ(10) : C Cv=0 10 10 F1e - Excited Av=2 10 10 F1f : 217:<Linelist>
C 11 e 1 Excited 11 f 2 24407.3028 24407.3053 -0.0025 .005500 0.0000 : qQ(11) : C Cv=0 11 11 F1e - Excited Av=2 11 11 F1f : 218:<Linelist>
C 12 e 1 Excited 12 f 2 24415.7027 24415.7025 0.0002 .007900 0.0000 : qQ(12) : C Cv=0 12 12 F1e - Excited Av=2 12 12 F1f : 219:<Linelist>
C 13 e 1 Excited 13 f 2 24424.7920 24424.7911 0.0009 .007500 0.0000 : qQ(13) : C Cv=0 13 13 F1e - Excited Av=2 13 13 F1f : 220:<Linelist>
C 14 e 1 Excited 14 f 2 24434.5887 24434.5895 -0.0008 .005200 0.0000 : qQ(14) : C Cv=0 14 14 F1e - Excited Av=2 14 14 F1f : 221:<Linelist>
C 15 e 1 Excited 15 f 2 24445.0872 24445.0909 -0.0037 .007300 0.0000 : qQ(15) : C Cv=0 15 15 F1e - Excited Av=2 15 15 F1f : 222:<Linelist>
C 16 e 1 Excited 16 f 2 24456.2918 24456.2959 -0.0041 .005100 0.0000 : qQ(16) : C Cv=0 16 16 F1e - Excited Av=2 16 16 F1f : 223:<Linelist>
C 17 e 1 Excited 17 f 2 24468.2014 24468.2069 -0.0055 .005300 0.0000 : qQ(17) : C Cv=0 17 17 F1e - Excited Av=2 17 17 F1f : 224:<Linelist>
C 18 e 1 Excited 18 f 2 24480.8249 24480.8276 -0.0027 .005300 0.0000 : qQ(18) : C Cv=0 18 18 F1e - Excited Av=2 18 18 F1f : 225:<Linelist>
C 19 e 1 Excited 19 f 2 24494.1579 24494.1645 -0.0066 .005500 0.0000 : qQ(19) : C Cv=0 19 19 F1e - Excited Av=2 19 19 F1f : 226:<Linelist>
C 20 e 1 Excited 20 f 2 24508.2408 24508.2334 0.0074 .010200 0.0000 : qQ(20) : C Cv=0 20 20 F1e - Excited Av=2 20 20 F1f : 227:<Linelist>
C 21 e 1 Excited 21 f 3 24522.8544 24522.8805 -0.0261 .006500 0.0000 : qQ(21) : C Cv=0 21 21 F1e - Excited Av=2 21 21 F1f : 228:<Linelist>
C 22 e 1 Excited 22 f 3 24538.4372 24538.4363 0.0009 .006100 0.0000 : qQ(22) : C Cv=0 22 22 F1e - Excited Av=2 22 22 F1f : 229:<Linelist>
C 23 e 1 Excited 23 f 3 24554.6975 24554.6980 -0.0005 .011100 0.0000 : qQ(23) : C Cv=0 23 23 F1e - Excited Av=2 23 23 F1f : 230:<Linelist>
C 24 e 1 Excited 24 f 3 24571.7492 24571.7619 -0.0127 .006900 0.0000 : qQ(24) : C Cv=0 24 24 F1e - Excited Av=2 24 24 F1f : 231:<Linelist>
C 25 e 1 Excited 25 f 3 24589.7160 24589.7253 -0.0093 .007300 0.0000 : qQ(25) : C Cv=0 25 25 F1e - Excited Av=2 25 25 F1f : 232:<Linelist>
C 26 e 1 Excited 26 f 3 24608.9912 24609.0041 -0.0129 .008600 0.0000 : qQ(26) : C Cv=0 26 26 F1e - Excited Av=2 26 26 F1f : 233:<Linelist>
C 27 e 1 Excited 27 f 3 24631.6970 24631.6985 -0.0015 .015200 0.0000 : qQ(27) : C Cv=0 27 27 F1e - Excited Av=2 27 27 F1f : 234:<Linelist>
C 28 e 1 Excited 28 f 4 24642.0465 24642.0557 -0.0092 .022800 0.0000 : qQ(28) : C Cv=0 28 28 F1e - Excited Av=2 28 28 F1f : 235:<Linelist>
C 29 e 1 Excited 29 f 4 24663.9563 24663.9660 -0.0097 .010500 0.0000 : qQ(29) : C Cv=0 29 29 F1e - Excited Av=2 29 29 F1f : 236:<Linelist>
C 30 e 1 Excited 30 f 5 24685.4160 24685.4479 -0.0319 .017200 0.0000 : qQ(30) : C Cv=0 30 30 F1e - Excited Av=2 30 30 F1f : 237:<Linelist>
C 31 e 1 Excited 31 f 6 24706.6836 24706.6964 -0.0128 .032500 0.0000 : qQ(31) : C Cv=0 31 31 F1e - Excited Av=2 31 31 F1f : 238:<Linelist>
C 32 e 1 Excited 32 f 7 24730.0243 24730.0463 -0.0220 .014000 0.0000 : qQ(32) : C Cv=0 32 32 F1e - Excited Av=2 32 32 F1f : 239:<Linelist>
C 33 e 1 Excited 33 f 7 24754.1507 24754.1150 0.0357 .056300 0.0000 : qQ(33) : C Cv=0 33 33 F1e - Excited Av=2 33 33 F1f : 240:<Linelist>
C 34 e 1 Excited 34 f 8 24776.1762 24776.1840 -0.0078 .041000 0.0000 : qQ(34) : C Cv=0 34 34 F1e - Excited Av=2 34 34 F1f : 241:<Linelist>
C 35 e 1 Excited 35 f 8 24801.7820 24801.7597 0.0223 .067400 0.0000 : qQ(35) : C Cv=0 35 35 F1e - Excited Av=2 35 35 F1f : 242:<Linelist>
C 36 e 1 Excited 36 f 8 24828.2163 24828.2537 -0.0374 .038000 0.0000 : qQ(36) : C Cv=0 36 36 F1e - Excited Av=2 36 36 F1f : 243:<Linelist>
C 3 e 1 Excited 2 e 1 24373.8897 24373.9116 -0.0219 .038600 0.0000 : rR(2) : C Cv=0 3 3 F1e - Excited Av=2 2 2 F1e : 250:<Linelist>
C 4 e 1 Excited 3 e 1 24379.5061 24379.5390 -0.0329 .044900 0.0000 : rR(3) : C Cv=0 4 4 F1e - Excited Av=2 3 3 F1e : 251:<Linelist>
C 5 e 1 Excited 4 e 1 24385.8550 24385.8680 -0.0130 .031800 0.0000 : rR(4) : C Cv=0 5 5 F1e - Excited Av=2 4 4 F1e : 252:<Linelist>
C 6 e 1 Excited 5 e 1 24392.8915 24392.8902 0.0013 .012900 0.0000 : rR(5) : C Cv=0 6 6 F1e - Excited Av=2 5 5 F1e : 253:<Linelist>
C 7 e 1 Excited 6 e 1 24400.5933 24400.6133 -0.0200 .041400 0.0000 : rR(6) : C Cv=0 7 7 F1e - Excited Av=2 6 6 F1e : 254:<Linelist>
C 8 e 1 Excited 7 e 1 24409.0340 24409.0327 0.0013 .015900 0.0000 : rR(7) : C Cv=0 8 8 F1e - Excited Av=2 7 7 F1e : 255:<Linelist>
C 9 e 1 Excited 8 e 1 24418.1499 24418.1522 -0.0023 .009300 0.0000 : rR(8) : C Cv=0 9 9 F1e - Excited Av=2 8 8 F1e : 256:<Linelist>
C 10 e 1 Excited 9 e 1 24427.9582 24427.9652 -0.0070 .017000 0.0000 : rR(9) : C Cv=0 10 10 F1e - Excited Av=2 9 9 F1e : 257:<Linelist>
C 11 e 1 Excited 10 e 1 24438.4566 24438.4853 -0.0287 .018600 0.0000 : rR(10) : C Cv=0 11 11 F1e - Excited Av=2 10 10 F1e : 258:<Linelist>
C 12 e 1 Excited 11 e 1 24449.6990 24449.7035 -0.0045 .018000 0.0000 : rR(11) : C Cv=0 12 12 F1e - Excited Av=2 11 11 F1e : 259:<Linelist>
C 13 e 1 Excited 12 e 1 24461.6142 24461.6128 0.0014 .008600 0.0000 : rR(12) : C Cv=0 13 13 F1e - Excited Av=2 12 12 F1e : 260:<Linelist>
C 14 e 1 Excited 13 e 1 24474.2315 24474.2310 0.0005 .014000 0.0000 : rR(13) : C Cv=0 14 14 F1e - Excited Av=2 13 13 F1e : 261:<Linelist>
C 15 e 1 Excited 14 e 1 24487.5500 24487.5510 -0.0010 .014200 0.0000 : rR(14) : C Cv=0 15 15 F1e - Excited Av=2 14 14 F1e : 262:<Linelist>
C 16 e 1 Excited 15 e 1 24501.5606 24501.5736 -0.0130 .008400 0.0000 : rR(15) : C Cv=0 16 16 F1e - Excited Av=2 15 15 F1e : 263:<Linelist>
C 17 e 1 Excited 16 e 1 24516.3027 24516.3016 0.0011 .014900 0.0000 : rR(16) : C Cv=0 17 17 F1e - Excited Av=2 16 16 F1e : 264:<Linelist>
C 18 e 1 Excited 17 e 1 24531.7453 24531.7405 0.0048 .019000 0.0000 : rR(17) : C Cv=0 18 18 F1e - Excited Av=2 17 17 F1e : 265:<Linelist>
C 19 e 1 Excited 18 e 1 24547.9100 24547.9021 0.0079 .019300 0.0000 : rR(18) : C Cv=0 19 19 F1e - Excited Av=2 18 18 F1e : 266:<Linelist>
C 20 e 1 Excited 19 e 1 24564.8237 24564.8196 0.0041 .020200 0.0000 : rR(19) : C Cv=0 20 20 F1e - Excited Av=2 19 19 F1e : 267:<Linelist>
C 21 e 1 Excited 20 e 1 24582.9508 24582.9819 -0.0311 .027600 0.0000 : rR(20) : C Cv=0 21 21 F1e - Excited Av=2 20 20 F1e : 268:<Linelist>
C 22 e 1 Excited 21 e 2 24600.3543 24600.3440 0.0103 .010900 0.0000 : rR(21) : C Cv=0 22 22 F1e - Excited Av=2 21 21 F1e : 269:<Linelist>
C 23 e 1 Excited 22 e 2 24619.5151 24619.5333 -0.0182 .032800 0.0000 : rR(22) : C Cv=0 23 23 F1e - Excited Av=2 22 22 F1e : 270:<Linelist>
C 24 e 1 Excited 23 e 2 24639.5830 24639.5861 -0.0031 .014500 0.0000 : rR(23) : C Cv=0 24 24 F1e - Excited Av=2 23 23 F1e : 271:<Linelist>
C 25 e 1 Excited 24 e 2 24661.4271 24661.4358 -0.0087 .026100 0.0000 : rR(24) : C Cv=0 25 25 F1e - Excited Av=2 24 24 F1e : 272:<Linelist>
C 26 e 1 Excited 25 e 3 24678.0000 24678.0084 -0.0084 .036200 0.0000 : rR(25) : C Cv=0 26 26 F1e - Excited Av=2 25 25 F1e : 273:<Linelist>
C 27 e 1 Excited 26 e 3 24700.9698 24700.9691 0.0007 .082500 0.0000 : rR(26) : C Cv=0 27 27 F1e - Excited Av=2 26 26 F1e : 274:<Linelist>
C 28 e 1 Excited 27 e 3 24723.7925 24723.8418 -0.0493 .035400 0.0000 : rR(27) : C Cv=0 28 28 F1e - Excited Av=2 27 27 F1e : 275:<Linelist>
C 29 e 1 Excited 28 e 3 24747.3426 24747.3396 0.0030 .035000 0.0000 : rR(28) : C Cv=0 29 29 F1e - Excited Av=2 28 28 F1e : 276:<Linelist>
C 30 e 1 Excited 29 e 3 24771.8045 24771.7901 0.0144 .058100 0.0000 : rR(29) : C Cv=0 30 30 F1e - Excited Av=2 29 29 F1e : 277:<Linelist>
C 19 e 1 Excited 20 e 2 24436.0932 24436.1187 -0.0255 .021800 0.0000 : qP1(20) : C Cv=0 19 19 F1e - Excited a3Pv=13 20 19 F1e : 289:<Linelist>

C 21 e 1 Excited 20 e 2 24580.6333 24580.6066 0.0267 .032200 0.0000 : sR1(20) : C Cv=0 21 21 F1e - Excited a3Pv=13 20 19 F1e : 290:<Linelist>
C 23 e 1 Excited 24 e 3 24477.1546 24477.1433 0.0113 .025700 0.0000 : qP1(24) : C Cv=0 23 23 F1e - Excited ev=4 24 23 F1e : 291:<Linelist>
C 24 e 1 Excited 25 e 2 24511.5045 24511.5554 -0.0509 .028200 0.0000 : qP1(25) : C Cv=0 24 24 F1e - Excited ev=4 25 24 F1e : 292:<Linelist>
C 27 e 1 Excited 27 f 4 24616.7806 24616.7904 -0.0098 .015900 0.0000 : qQ12(27) : C Cv=0 27 27 F1e - Excited ev=4 27 27 F2f : 293:<Linelist>
Excited 1 e 1 X 2 e 1 67549.9502 67549.9500 0.0002 .010100 0.0000 : pP(2) : Excited Av=2 1 1 F1e - X Xv=0 2 2 F1e : 294:<Linelist>
Excited 2 e 1 X 3 e 1 67545.1455 67545.1411 0.0044 .010100 0.0000 : pP(3) : Excited Av=2 2 2 F1e - X Xv=0 3 3 F1e : 295:<Linelist>
Excited 3 e 1 X 4 e 1 67539.6797 67539.6741 0.0056 .010100 0.0000 : pP(4) : Excited Av=2 3 3 F1e - X Xv=0 4 4 F1e : 296:<Linelist>
Excited 4 e 1 X 5 e 1 67533.5561 67533.5491 0.0070 .010100 0.0000 : pP(5) : Excited Av=2 4 4 F1e - X Xv=0 5 5 F1e : 297:<Linelist>
Excited 5 e 1 X 6 e 1 67526.7715 67526.7660 0.0055 .010100 0.0000 : pP(6) : Excited Av=2 5 5 F1e - X Xv=0 6 6 F1e : 298:<Linelist>
Excited 6 e 1 X 7 e 1 67519.3292 67519.3247 0.0045 .010100 0.0000 : pP(7) : Excited Av=2 6 6 F1e - X Xv=0 7 7 F1e : 299:<Linelist>
Excited 7 e 1 X 8 e 1 67511.2293 67511.2250 0.0043 .010100 0.0000 : pP(8) : Excited Av=2 7 7 F1e - X Xv=0 8 8 F1e : 300:<Linelist>
Excited 8 e 1 X 9 e 1 67502.4707 67502.4668 0.0039 .010100 0.0000 : pP(9) : Excited Av=2 8 8 F1e - X Xv=0 9 9 F1e : 301:<Linelist>
Excited 9 e 1 X 10 e 1 67493.0543 67493.0498 0.0045 .010100 0.0000 : pP(10) : Excited Av=2 9 9 F1e - X Xv=0 10 10 F1e : 302:<Linelist>
Excited 10 e 1 X 11 e 1 67482.9777 67482.9734 0.0043 .010100 0.0000 : pP(11) : Excited Av=2 10 10 F1e - X Xv=0 11 11 F1e : 303:<Linelist>
Excited 11 e 1 X 12 e 1 67472.2405 67472.2374 0.0031 .010100 0.0000 : pP(12) : Excited Av=2 11 11 F1e - X Xv=0 12 12 F1e : 304:<Linelist>
Excited 12 e 1 X 13 e 1 67460.8434 67460.8409 0.0025 .010100 0.0000 : pP(13) : Excited Av=2 12 12 F1e - X Xv=0 13 13 F1e : 305:<Linelist>
Excited 13 e 1 X 14 e 1 67448.7861 67448.7833 0.0028 .010100 0.0000 : pP(14) : Excited Av=2 13 13 F1e - X Xv=0 14 14 F1e : 306:<Linelist>
Excited 14 e 1 X 15 e 1 67436.0669 67436.0633 0.0036 .010100 0.0000 : pP(15) : Excited Av=2 14 14 F1e - X Xv=0 15 15 F1e : 307:<Linelist>
Excited 15 e 1 X 16 e 1 67422.6827 67422.6794 0.0033 .010100 0.0000 : pP(16) : Excited Av=2 15 15 F1e - X Xv=0 16 16 F1e : 308:<Linelist>
Excited 16 e 1 X 17 e 1 67408.6317 67408.6289 0.0028 .010100 0.0000 : pP(17) : Excited Av=2 16 16 F1e - X Xv=0 17 17 F1e : 309:<Linelist>
Excited 17 e 1 X 18 e 1 67393.9116 67393.9073 0.0043 .010100 0.0000 : pP(18) : Excited Av=2 17 17 F1e - X Xv=0 18 18 F1e : 310:<Linelist>
Excited 18 e 1 X 19 e 1 67378.5076 67378.5043 0.0033 .010100 0.0000 : pP(19) : Excited Av=2 18 18 F1e - X Xv=0 19 19 F1e : 311:<Linelist>
Excited 19 e 1 X 20 e 1 67362.3915 67362.3831 0.0084 .010200 0.0000 : pP(20) : Excited Av=2 19 19 F1e - X Xv=0 20 20 F1e : 312:<Linelist>
Excited 20 e 1 X 21 e 1 67345.0774 67345.0554 0.0220 .010500 0.0000 : pP(21) : Excited Av=2 20 20 F1e - X Xv=0 21 21 F1e : 313:<Linelist>
Excited 21 e 2 X 22 e 1 67328.5654 67328.5687 -0.0033 .010600 0.0000 : pP(22) : Excited Av=2 21 21 F1e - X Xv=0 22 22 F1e : 314:<Linelist>
Excited 22 e 2 X 23 e 1 67310.2943 67310.2951 -0.0008 .011100 0.0000 : pP(23) : Excited Av=2 22 22 F1e - X Xv=0 23 23 F1e : 315:<Linelist>
Excited 23 e 2 X 24 e 1 67291.2127 67291.2080 0.0047 .012500 0.0000 : pP(24) : Excited Av=2 23 23 F1e - X Xv=0 24 24 F1e : 316:<Linelist>
Excited 24 e 2 X 25 e 1 67270.3661 67270.3539 0.0122 .016000 0.0000 : pP(25) : Excited Av=2 24 24 F1e - X Xv=0 25 25 F1e : 317:<Linelist>
Excited 25 e 3 X 26 e 1 67254.8180 67254.8172 0.0008 .017900 0.0000 : pP(26) : Excited Av=2 25 25 F1e - X Xv=0 26 26 F1e : 318:<Linelist>
Excited 26 e 3 X 27 e 1 67232.9269 67232.9288 -0.0019 .059700 0.0000 : pP(27) : Excited Av=2 26 26 F1e - X Xv=0 27 27 F1e : 319:<Linelist>
Excited 27 e 3 X 28 e 1 67211.1808 67211.1680 0.0128 .048200 0.0000 : pP(28) : Excited Av=2 27 27 F1e - X Xv=0 28 28 F1e : 320:<Linelist>
Excited 28 e 3 X 29 e 1 67188.8238 67188.8228 0.0010 .193300 0.0000 : pP(29) : Excited Av=2 28 28 F1e - X Xv=0 29 29 F1e : 321:<Linelist>
Excited 29 e 3 X 30 e 1 67165.6131 67165.5618 0.0513 .048200 0.0000 : pP(30) : Excited Av=2 29 29 F1e - X Xv=0 30 30 F1e : 322:<Linelist>
Excited 30 e 4 X 31 e 1 67143.4840 67143.4960 -0.0120 .031500 0.0000 : pP(31) : Excited Av=2 30 30 F1e - X Xv=0 31 31 F1e : 323:<Linelist>
Excited 31 e 5 X 32 e 1 67120.6390 67120.6051 0.0339 .030200 0.0000 : pP(32) : Excited Av=2 31 31 F1e - X Xv=0 32 32 F1e : 324:<Linelist>
Excited 32 e 5 X 33 e 1 67093.9326 67093.8948 0.0378 .027500 0.0000 : pP(33) : Excited Av=2 32 32 F1e - X Xv=0 33 33 F1e : 325:<Linelist>
Excited 33 e 6 X 34 e 1 67068.2158 67068.1810 0.0348 .027600 0.0000 : pP(34) : Excited Av=2 33 33 F1e - X Xv=0 34 34 F1e : 326:<Linelist>
Excited 34 e 8 X 35 e 1 67045.7573 67045.7387 0.0186 .025600 0.0000 : pP(35) : Excited Av=2 34 34 F1e - X Xv=0 35 35 F1e : 327:<Linelist>
Excited 35 e 8 X 36 e 1 67017.8793 67017.8395 0.0398 .025500 0.0000 : pP(36) : Excited Av=2 35 35 F1e - X Xv=0 36 36 F1e : 328:<Linelist>
Excited 36 e 8 X 37 e 1 66989.9894 66989.9428 0.0466 .025600 0.0000 : pP(37) : Excited Av=2 36 36 F1e - X Xv=0 37 37 F1e : 329:<Linelist>
Excited 37 e 8 X 38 e 1 66959.9316 66959.9187 0.0129 .026500 0.0000 : pP(38) : Excited Av=2 37 37 F1e - X Xv=0 38 38 F1e : 330:<Linelist>
Excited 38 e 9 X 39 e 1 66934.3375 66934.3078 0.0297 .038800 0.0000 : pP(39) : Excited Av=2 38 38 F1e - X Xv=0 39 39 F1e : 331:<Linelist>
Excited 39 e 9 X 40 e 1 66904.5166 66904.5093 0.0073 .036700 0.0000 : pP(40) : Excited Av=2 39 39 F1e - X Xv=0 40 40 F1e : 332:<Linelist>
Excited 40 e 9 X 41 e 1 66874.3333 66874.3249 0.0084 .051400 0.0000 : pP(41) : Excited Av=2 40 40 F1e - X Xv=0 41 41 F1e : 333:<Linelist>
Excited 1 f 1 X 1 e 1 67556.9399 67556.9350 0.0049 .010100 0.0000 : qQ(1) : Excited Av=2 1 1 F1f - X Xv=0 1 1 F1e : 349:<Linelist>
Excited 2 f 1 X 2 e 1 67555.6207 67555.6174 0.0033 .010100 0.0000 : qQ(2) : Excited Av=2 2 2 F1f - X Xv=0 2 2 F1e : 350:<Linelist>
Excited 3 f 1 X 3 e 1 67553.6473 67553.6406 0.0067 .010100 0.0000 : qQ(3) : Excited Av=2 3 3 F1f - X Xv=0 3 3 F1e : 351:<Linelist>
Excited 4 f 1 X 4 e 1 67551.0122 67551.0038 0.0084 .010100 0.0000 : qQ(4) : Excited Av=2 4 4 F1f - X Xv=0 4 4 F1e : 352:<Linelist>
Excited 5 f 1 X 5 e 1 67547.7061 67547.7051 0.0010 .010100 0.0000 : qQ(5) : Excited Av=2 5 5 F1f - X Xv=0 5 5 F1e : 353:<Linelist>
Excited 6 f 1 X 6 e 1 67543.7463 67543.7379 0.0084 .010100 0.0000 : qQ(6) : Excited Av=2 6 6 F1f - X Xv=0 6 6 F1e : 354:<Linelist>
Excited 7 f 1 X 7 e 1 67539.0589 67539.0504 0.0085 .010100 0.0000 : qQ(7) : Excited Av=2 7 7 F1f - X Xv=0 7 7 F1e : 355:<Linelist>
Excited 8 f 2 X 8 e 1 67534.0337 67534.0283 0.0054 .010100 0.0000 : qQ(8) : Excited Av=2 8 8 F1f - X Xv=0 8 8 F1e : 356:<Linelist>
Excited 9 f 2 X 9 e 1 67528.0212 67528.0167 0.0045 .010100 0.0000 : qQ(9) : Excited Av=2 9 9 F1f - X Xv=0 9 9 F1e : 357:<Linelist>
Excited 10 f 2 X 10 e 1 67521.4185 67521.4105 0.0080 .010100 0.0000 : qQ(10) : Excited Av=2 10 10 F1f - X Xv=0 10 10 F1e : 358:<Linelist>
Excited 11 f 2 X 11 e 1 67514.1596 67514.1534 0.0062 .010100 0.0000 : qQ(11) : Excited Av=2 11 11 F1f - X Xv=0 11 11 F1e : 359:<Linelist>
Excited 12 f 2 X 12 e 1 67506.2432 67506.2383 0.0049 .010100 0.0000 : qQ(12) : Excited Av=2 12 12 F1f - X Xv=0 12 12 F1e : 360:<Linelist>
Excited 13 f 2 X 13 e 1 67497.6691 67497.6626 0.0065 .010100 0.0000 : qQ(13) : Excited Av=2 13 13 F1f - X Xv=0 13 13 F1e : 361:<Linelist>
Excited 14 f 2 X 14 e 1 67488.4325 67488.4248 0.0077 .010100 0.0000 : qQ(14) : Excited Av=2 14 14 F1f - X Xv=0 14 14 F1e : 362:<Linelist>
Excited 15 f 2 X 15 e 1 67478.5322 67478.5235 0.0087 .010100 0.0000 : qQ(15) : Excited Av=2 15 15 F1f - X Xv=0 15 15 F1e : 363:<Linelist>
Excited 16 f 2 X 16 e 1 67467.9609 67467.9570 0.0039 .010100 0.0000 : qQ(16) : Excited Av=2 16 16 F1f - X Xv=0 16 16 F1e : 364:<Linelist>
Excited 17 f 2 X 17 e 1 67456.7272 67456.7236 0.0036 .010100 0.0000 : qQ(17) : Excited Av=2 17 17 F1f - X Xv=0 17 17 F1e : 365:<Linelist>
Excited 18 f 2 X 18 e 1 67444.8243 67444.8202 0.0041 .010100 0.0000 : qQ(18) : Excited Av=2 18 18 F1f - X Xv=0 18 18 F1e : 366:<Linelist>
Excited 19 f 2 X 19 e 1 67432.2464 67432.2419 0.0045 .010100 0.0000 : qQ(19) : Excited Av=2 19 19 F1f - X Xv=0 19 19 F1e : 367:<Linelist>
Excited 20 f 2 X 20 e 1 67418.9794 67418.9693 0.0101 .010100 0.0000 : qQ(20) : Excited Av=2 20 20 F1f - X Xv=0 20 20 F1e : 368:<Linelist>
Excited 21 f 3 X 21 e 1 67405.1852 67405.1567 0.0285 .010100 0.0000 : qQ(21) : Excited Av=2 21 21 F1f - X Xv=0 21 21 F1e : 369:<Linelist>
Excited 22 f 3 X 22 e 1 67390.4821 67390.4764 0.0057 .010100 0.0000 : qQ(22) : Excited Av=2 22 22 F1f - X Xv=0 22 22 F1e : 370:<Linelist>
Excited 23 f 3 X 23 e 1 67375.1345 67375.1303 0.0042 .010200 0.0000 : qQ(23) : Excited Av=2 23 23 F1f - X Xv=0 23 23 F1e : 371:<Linelist>
Excited 24 f 3 X 24 e 1 67359.0361 67359.0321 0.0040 .010500 0.0000 : qQ(24) : Excited Av=2 24 24 F1f - X Xv=0 24 24 F1e : 372:<Linelist>
Excited 25 f 3 X 25 e 1 67342.0646 67342.0644 0.0002 .011100 0.0000 : qQ(25) : Excited Av=2 25 25 F1f - X Xv=0 25 25 F1e : 373:<Linelist>
Excited 26 f 3 X 26 e 1 67323.8176 67323.8216 -0.0040 .012200 0.0000 : qQ(26) : Excited Av=2 26 26 F1f - X Xv=0 26 26 F1e : 374:<Linelist>
Excited 27 f 3 X 27 e 1 67302.1959 67302.1995 -0.0036 .031900 0.0000 : qQ(27) : Excited Av=2 27 27 F1f - X Xv=0 27 27 F1e : 375:<Linelist>
Excited 28 f 4 X 28 e 1 67292.9786 67292.9542 0.0244 .036400 0.0000 : qQ(28) : Excited Av=2 28 28 F1f - X Xv=0 28 28 F1e : 376:<Linelist>
Excited 29 f 4 X 29 e 1 67272.1787 67272.1964 -0.0177 .049900 0.0000 : qQ(29) : Excited Av=2 29 29 F1f - X Xv=0 29 29 F1e : 377:<Linelist>
Excited 30 f 5 X 30 e 1 67251.9261 67251.9040 0.0221 .083300 0.0000 : qQ(30) : Excited Av=2 30 30 F1f - X Xv=0 30 30 F1e : 378:<Linelist>
Excited 34 f 8 X 34 e 1 67166.4366 67166.3243 0.1123 .028200 0.0000 : qQ(34) : Excited Av=2 34 34 F1f - X Xv=0 34 34 F1e : 382:<Linelist>
Excited 35 f 8 X 35 e 1 67142.1663 67142.1297 0.0366 .028100 0.0000 : qQ(35) : Excited Av=2 35 35 F1f - X Xv=0 35 35 F1e : 383:<Linelist>
Excited 36 f 8 X 36 e 1 67117.0783 67117.0591 0.0192 .025600 0.0000 : qQ(36) : Excited Av=2 36 36 F1f - X Xv=0 36 36 F1e : 384:<Linelist>
Excited 37 f 9 X 37 e 1 67091.6594 67091.6480 0.0114 .027200 0.0000 : qQ(37) : Excited Av=2 37 37 F1f - X Xv=0 37 37 F1e : 385:<Linelist>
Excited 38 f 10 X 38 e 1 67069.3438 67069.3632 -0.0194 .027600 0.0000 : qQ(38) : Excited Av=2 38 38 F1f - X Xv=0 38 38 F1e : 386:<Linelist>
Excited 39 f 10 X 39 e 1 67042.9465 67042.9393 0.0072 .025800 0.0000 : qQ(39) : Excited Av=2 39 39 F1f - X Xv=0 39 39 F1e : 387:<Linelist>
Excited 40 f 10 X 40 e 1 67016.1445 67016.1406 0.0039 .032300 0.0000 : qQ(40) : Excited Av=2 40 40 F1f - X Xv=0 40 40 F1e : 388:<Linelist>
Excited 41 f 10 X 41 e 1 66988.7653 66988.7386 0.0267 .040700 0.0000 : qQ(41) : Excited Av=2 41 41 F1f - X Xv=0 41 41 F1e : 389:<Linelist>
Excited 42 f 10 X 42 e 1 66960.7067 66960.6864 0.0203 .039500 0.0000 : qQ(42) : Excited Av=2 42 42 F1f - X Xv=0 42 42 F1e : 390:<Linelist>
Excited 1 e 1 X 0 e 1 67560.4306 67560.4283 0.0023 .010100 0.0000 : rR(0) : Excited Av=2 1 1 F1e - X Xv=0 0 0 F1e : 405:<Linelist>
Excited 2 e 1 X 1 e 1 67562.6070 67562.6044 0.0026 .010100 0.0000 : rR(1) : Excited Av=2 2 2 F1e - X Xv=0 1 1 F1e : 406:<Linelist>
Excited 3 e 1 X 2 e 1 67564.1235 67564.1219 0.0016 .010100 0.0000 : rR(2) : Excited Av=2 3 3 F1e - X Xv=0 2 2 F1e : 407:<Linelist>
Excited 4 e 1 X 3 e 1 67564.9682 67564.9806 -0.0124 .010100 0.0000 : rR(3) : Excited Av=2 4 4 F1e - X Xv=0 3 3 F1e : 408:<Linelist>
Excited 5 e 1 X 4 e 1 67565.1911 67565.1800 0.0111 .010100 0.0000 : rR(4) : Excited Av=2 5 5 F1e - X Xv=0 4 4 F1e : 409:<Linelist>
Excited 6 e 1 X 5 e 1 67564.7158 67564.7200 -0.0042 .010100 0.0000 : rR(5) : Excited Av=2 6 6 F1e - X Xv=0 5 5 F1e : 410:<Linelist>
Excited 7 e 1 X 6 e 1 67563.6074 67563.6000 0.0074 .010100 0.0000 : rR(6) : Excited Av=2 7 7 F1e - X Xv=0 6 6 F1e : 411:<Linelist>
Excited 8 e 1 X 7 e 1 67561.8234 67561.8196 0.0038 .010100 0.0000 : rR(7) : Excited Av=2 8 8 F1e - X Xv=0 7 7 F1e : 412:<Linelist>
Excited 9 e 1 X 8 e 1 67559.3812 67559.3783 0.0029 .010100 0.0000 : rR(8) : Excited Av=2 9 9 F1e - X Xv=0 8 8 F1e : 413:<Linelist>

| | | | | | | | | | | |
|---|---|---|---|---|---|---|---|---|---|---|
| Excited | 10 | e | 1 | X | 9 e | 1 | 67556.2814 | 67556.2755 | 0.0059 .010100 0.0000 : rR(9) : Excited Av=2 10 10 F1e - X Xv=0 9 9 F1e | : 414:<Linelist> |
| Excited | 11 | e | 1 | X | 10 e | 1 | 67552.5123 | 67552.5104 | 0.0019 .010100 0.0000 : rR(10) : Excited Av=2 11 11 F1e - X Xv=0 10 10 F1e | : 415:<Linelist> |
| Excited | 12 | e | 1 | X | 11 e | 1 | 67548.0899 | 67548.0821 | 0.0078 .010100 0.0000 : rR(11) : Excited Av=2 12 12 F1e - X Xv=0 11 11 F1e | : 416:<Linelist> |
| Excited | 13 | e | 1 | X | 12 e | 1 | 67542.9953 | 67542.9896 | 0.0057 .010100 0.0000 : rR(12) : Excited Av=2 13 13 F1e - X Xv=0 12 12 F1e | : 417:<Linelist> |
| Excited | 14 | e | 1 | X | 13 e | 1 | 67537.2370 | 67537.2315 | 0.0055 .010100 0.0000 : rR(13) : Excited Av=2 14 14 F1e - X Xv=0 13 13 F1e | : 418:<Linelist> |
| Excited | 15 | e | 1 | X | 14 e | 1 | 67530.8100 | 67530.8059 | 0.0041 .010100 0.0000 : rR(14) : Excited Av=2 15 15 F1e - X Xv=0 14 14 F1e | : 419:<Linelist> |
| Excited | 16 | e | 1 | X | 15 e | 1 | 67523.7148 | 67523.7099 | 0.0049 .010100 0.0000 : rR(15) : Excited Av=2 16 16 F1e - X Xv=0 15 15 F1e | : 420:<Linelist> |
| Excited | 17 | e | 1 | X | 16 e | 1 | 67515.9440 | 67515.9390 | 0.0050 .010100 0.0000 : rR(16) : Excited Av=2 17 17 F1e - X Xv=0 16 16 F1e | : 421:<Linelist> |
| Excited | 18 | e | 1 | X | 17 e | 1 | 67507.4864 | 67507.4823 | 0.0041 .010100 0.0000 : rR(17) : Excited Av=2 18 18 F1e - X Xv=0 17 17 F1e | : 422:<Linelist> |
| Excited | 19 | e | 1 | X | 18 e | 1 | 67498.3111 | 67498.3030 | 0.0081 .010100 0.0000 : rR(18) : Excited Av=2 19 19 F1e - X Xv=0 18 18 F1e | : 423:<Linelist> |
| Excited | 20 | e | 1 | X | 19 e | 1 | 67487.9331 | 67487.9125 | 0.0206 .010100 0.0000 : rR(19) : Excited Av=2 20 20 F1e - X Xv=0 19 19 F1e | : 424:<Linelist> |
| Excited | 21 | e | 2 | X | 20 e | 1 | 67478.3814 | 67478.3580 | 0.0234 .010400 0.0000 : rR(20) : Excited Av=2 21 21 F1e - X Xv=0 20 20 F1e | : 425:<Linelist> |
| Excited | 22 | e | 2 | X | 21 e | 1 | 67467.0147 | 67467.0114 | 0.0033 .010200 0.0000 : rR(21) : Excited Av=2 22 22 F1e - X Xv=0 21 21 F1e | : 426:<Linelist> |
| Excited | 23 | e | 2 | X | 22 e | 1 | 67454.8501 | 67454.8458 | 0.0043 .010400 0.0000 : rR(22) : Excited Av=2 23 23 F1e - X Xv=0 22 22 F1e | : 427:<Linelist> |
| Excited | 24 | e | 2 | X | 23 e | 1 | 67440.9173 | 67440.9076 | 0.0097 .011100 0.0000 : rR(23) : Excited Av=2 24 24 F1e - X Xv=0 23 23 F1e | : 428:<Linelist> |
| Excited | 25 | e | 3 | X | 24 e | 1 | 67432.2626 | 67432.2809 | -0.0183 .199000 0.0000 : rR(24) : Excited Av=2 25 25 F1e - X Xv=0 24 24 F1e | : 429:<Linelist> |
| Excited | 26 | e | 3 | X | 25 e | 1 | 67417.3064 | 67417.2963 | 0.0101 .013800 0.0000 : rR(25) : Excited Av=2 26 26 F1e - X Xv=0 25 25 F1e | : 430:<Linelist> |
| Excited | 27 | e | 3 | X | 26 e | 1 | 67402.4283 | 67402.4329 | -0.0046 .017900 0.0000 : rR(26) : Excited Av=2 27 27 F1e - X Xv=0 26 26 F1e | : 431:<Linelist> |
| Excited | 28 | e | 3 | X | 27 e | 1 | 67386.9732 | 67386.9785 | -0.0053 .025700 0.0000 : rR(27) : Excited Av=2 28 28 F1e - X Xv=0 27 27 F1e | : 432:<Linelist> |
| Excited | 33 | e | 6 | X | 32 e | 1 | 67300.7496 | 67300.6823 | 0.0673 .032600 0.0000 : rR(32) : Excited Av=2 33 33 F1e - X Xv=0 32 32 F1e | : 437:<Linelist> |
| Excited | 34 | e | 8 | X | 33 e | 1 | 67285.1008 | 67285.0860 | 0.0148 .032900 0.0000 : rR(33) : Excited Av=2 34 34 F1e - X Xv=0 33 33 F1e | : 438:<Linelist> |
| Excited | 35 | e | 8 | X | 34 e | 1 | 67264.0692 | 67264.0243 | 0.0449 .032500 0.0000 : rR(34) : Excited Av=2 35 35 F1e - X Xv=0 34 34 F1e | : 439:<Linelist> |
| Excited | 36 | e | 8 | X | 35 e | 1 | 67242.9971 | 67242.9566 | 0.0405 .028700 0.0000 : rR(35) : Excited Av=2 36 36 F1e - X Xv=0 35 35 F1e | : 440:<Linelist> |
| Excited | 37 | e | 8 | X | 36 e | 1 | 67219.7656 | 67219.7527 | 0.0129 .025800 0.0000 : rR(36) : Excited Av=2 37 37 F1e - X Xv=0 36 36 F1e | : 441:<Linelist> |
| Excited | 38 | e | 9 | X | 37 e | 1 | 67200.9484 | 67200.9529 | -0.0045 .027600 0.0000 : rR(37) : Excited Av=2 38 38 F1e - X Xv=0 37 37 F1e | : 442:<Linelist> |
| Excited | 39 | e | 9 | X | 38 e | 1 | 67177.9730 | 67177.9563 | 0.0167 .025900 0.0000 : rR(38) : Excited Av=2 39 39 F1e - X Xv=0 38 38 F1e | : 443:<Linelist> |
| Excited | 40 | e | 9 | X | 39 e | 1 | 67154.5301 | 67154.5643 | -0.0342 .027500 0.0000 : rR(39) : Excited Av=2 40 40 F1e - X Xv=0 39 39 F1e | : 444:<Linelist> |
| Excited | 41 | e | 9 | X | 40 e | 1 | 67130.6206 | 67130.5602 | 0.0604 .025600 0.0000 : rR(40) : Excited Av=2 41 41 F1e - X Xv=0 40 40 F1e | : 445:<Linelist> |
| Excited | 42 | e | 9 | X | 41 e | 1 | 67105.9732 | 67105.9000 | 0.0732 .050500 0.0000 : rR(41) : Excited Av=2 42 42 F1e - X Xv=0 41 41 F1e | : 446:<Linelist> |
| Excited | 43 | e | 9 | X | 42 e | 1 | 67080.5860 | 67080.5674 | 0.0186 .027100 0.0000 : rR(42) : Excited Av=2 43 43 F1e - X Xv=0 42 42 F1e | : 447:<Linelist> |
| Excited | 44 | e | 9 | X | 43 e | 1 | 67054.5617 | 67054.5541 | 0.0076 .173500 0.0000 : rR(43) : Excited Av=2 44 44 F1e - X Xv=0 43 43 F1e | : 448:<Linelist> |
| Excited | 7 | f | 2 | X | 7 e | 1 | 67541.7886 | 67541.7893 | -0.0007 .010300 0.0000 : qQ(7) : Excited Iv=3 7 7 F1f - X Xv=0 7 7 F1e | : 467:<Linelist> |
| Excited | 8 | f | 1 | X | 8 e | 1 | 67531.2419 | 67531.2381 | 0.0038 .010200 0.0000 : qQ(8) : Excited Iv=3 8 8 F1f - X Xv=0 8 8 F1e | : 468:<Linelist> |
| Excited | 9 | f | 1 | X | 9 e | 1 | 67519.7125 | 67519.7238 | -0.0113 .024600 0.0000 : qQ(9) : Excited Iv=3 9 9 F1f - X Xv=0 9 9 F1e | : 469:<Linelist> |
| Excited | 18 | f | 1 | X | 18 e | 1 | 67357.0554 | 67357.0529 | 0.0025 .047300 0.0000 : qQ(18) : Excited Iv=3 18 18 F1f - X Xv=0 18 18 F1e | : 478:<Linelist> |
| Excited | 19 | f | 1 | X | 19 e | 1 | 67332.4832 | 67332.4812 | 0.0020 .031000 0.0000 : qQ(19) : Excited Iv=3 19 19 F1f - X Xv=0 19 19 F1e | : 479:<Linelist> |
| Excited | 20 | f | 1 | X | 20 e | 1 | 67306.6099 | 67306.6090 | 0.0009 .026700 0.0000 : qQ(20) : Excited Iv=3 20 20 F1f - X Xv=0 20 20 F1e | : 480:<Linelist> |
| Excited | 21 | f | 1 | X | 21 e | 1 | 67279.4347 | 67279.4362 | -0.0015 .027100 0.0000 : qQ(21) : Excited Iv=3 21 21 F1f - X Xv=0 21 21 F1e | : 481:<Linelist> |
| Excited | 22 | f | 1 | X | 22 e | 1 | 67250.9609 | 67250.9624 | -0.0015 .037400 0.0000 : qQ(22) : Excited Iv=3 22 22 F1f - X Xv=0 22 22 F1e | : 482:<Linelist> |
| Excited | 23 | f | 1 | X | 23 e | 1 | 67221.1832 | 67221.1874 | -0.0042 .041500 0.0000 : qQ(23) : Excited Iv=3 23 23 F1f - X Xv=0 23 23 F1e | : 483:<Linelist> |
| Excited | 24 | f | 1 | X | 24 e | 1 | 67190.1202 | 67190.1108 | 0.0094 .069000 0.0000 : qQ(24) : Excited Iv=3 24 24 F1f - X Xv=0 24 24 F1e | : 484:<Linelist> |
| Excited | 25 | f | 1 | X | 25 e | 1 | 67157.7332 | 67157.7325 | 0.0007 .057300 0.0000 : qQ(25) : Excited Iv=3 25 25 F1f - X Xv=0 25 25 F1e | : 485:<Linelist> |
| Excited | 20 | e | 2 | X | 21 e | 1 | 67347.4163 | 67347.4307 | -0.0144 .012700 0.0000 : oP1(21) : Excited a3Pv=13 20 19 F1e - X Xv=0 21 21 F1e | : 529:<Linelist> |
| Excited | 22 | e | 1 | X | 23 e | 1 | 67296.4746 | 67296.5270 | -0.0524 .030400 0.0000 : oP1(23) : Excited a3Pv=13 22 21 F1e - X Xv=0 23 23 F1e | : 531:<Linelist> |
| Excited | 27 | e | 2 | X | 28 e | 1 | 67172.8684 | 67172.8525 | 0.0159 .027800 0.0000 : oP1(28) : Excited a3Pv=13 27 26 F1e - X Xv=0 28 28 F1e | : 536:<Linelist> |
| Excited | 28 | e | 2 | X | 29 e | 1 | 67139.1336 | 67139.1942 | -0.0606 .026900 0.0000 : oP1(29) : Excited a3Pv=13 28 27 F1e - X Xv=0 29 29 F1e | : 537:<Linelist> |
| Excited | 29 | e | 2 | X | 30 e | 1 | 67105.8214 | 67105.8882 | -0.0668 .082900 0.0000 : oP1(30) : Excited a3Pv=13 29 28 F1e - X Xv=0 30 30 F1e | : 538:<Linelist> |
| Excited | 30 | e | 5 | X | 31 e | 1 | 67151.0494 | 67151.0480 | 0.0014 .025600 0.0000 : pP21(31) : Excited a3Pv=13 30 30 F2e - X Xv=0 31 31 F1e | : 588:<Linelist> |
| Excited | 24 | f | 2 | X | 24 e | 1 | 67335.1471 | 67335.2219 | -0.0748 .040700 0.0000 : pQ1(24) : Excited a3Pv=13 24 23 F1f - X Xv=0 24 24 F1e | : 630:<Linelist> |
| Excited | 27 | f | 2 | X | 27 e | 1 | 67259.4462 | 67259.4435 | 0.0027 .067100 0.0000 : pQ1(27) : Excited a3Pv=13 27 26 F1f - X Xv=0 27 27 F1e | : 633:<Linelist> |
| Excited | 32 | f | 2 | X | 32 e | 1 | 67110.1790 | 67110.0910 | 0.0880 .033500 0.0000 : pQ1(32) : Excited a3Pv=13 32 31 F1f - X Xv=0 32 32 F1e | : 638:<Linelist> |
| Excited | 43 | e | 10 | X | 44 e | 1 | 66803.8107 | 66803.7247 | 0.0860 .034300 0.0000 : qP31(44) : Excited a3Pv=13 43 44 F3e - X Xv=0 44 44 F1e | : 700:<Linelist> |
| Excited | 30 | f | 6 | X | 30 e | 1 | 67265.7665 | 67265.6476 | 0.1189 .029900 0.0000 : qQ21(30) : Excited a3Pv=13 30 30 F2f - X Xv=0 30 30 F1e | : 736:<Linelist> |
| Excited | 32 | f | 6 | X | 32 e | 1 | 67208.1338 | 67208.0269 | 0.1069 .025700 0.0000 : qQ21(32) : Excited a3Pv=13 32 32 F2f - X Xv=0 32 32 F1e | : 738:<Linelist> |
| Excited | 33 | f | 6 | X | 33 e | 1 | 67178.4641 | 67178.4412 | 0.0229 .029500 0.0000 : qQ21(33) : Excited a3Pv=13 33 33 F2f - X Xv=0 33 33 F1e | : 739:<Linelist> |
| Excited | 20 | e | 2 | X | 19 e | 1 | 67490.2850 | 67490.2877 | -0.0027 .010600 0.0000 : qR1(19) : Excited a3Pv=13 20 19 F1e - X Xv=0 19 19 F1e | : 775:<Linelist> |
| Excited | 28 | e | 2 | X | 27 e | 1 | 67337.3041 | 67337.3499 | -0.0458 .035900 0.0000 : qR1(27) : Excited a3Pv=13 28 27 F1e - X Xv=0 27 27 F1e | : 783:<Linelist> |
| Excited | 33 | e | 2 | X | 32 e | 1 | 67200.5854 | 67200.6606 | -0.0752 .027300 0.0000 : qR1(32) : Excited a3Pv=13 33 32 F1e - X Xv=0 32 32 F1e | : 788:<Linelist> |
| Excited | 44 | e | 6 | X | 43 e | 1 | 66834.2048 | 66834.3029 | -0.0981 .029000 0.0000 : qR1(43) : Excited a3Pv=13 44 43 F1e - X Xv=0 43 43 F1e | : 799:<Linelist> |
| Excited | 47 | e | 7 | X | 46 e | 1 | 66716.1535 | 66716.0761 | 0.0774 .025600 0.0000 : qR1(46) : Excited a3Pv=13 47 46 F1e - X Xv=0 46 46 F1e | : 802:<Linelist> |
| Excited | 32 | f | 9 | X | 32 e | 1 | 67297.1544 | 67297.2714 | -0.1170 .028800 0.0000 : rQ31(32) : Excited a3Pv=13 32 33 F3f - X Xv=0 32 32 F1e | : 837:<Linelist> |
| Excited | 33 | f | 10 | X | 33 e | 1 | 67268.4728 | 67268.4657 | 0.0071 .130700 0.0000 : rQ31(33) : Excited a3Pv=13 33 34 F3f - X Xv=0 33 33 F1e | : 838:<Linelist> |
| Excited | 34 | f | 11 | X | 34 e | 1 | 67248.8191 | 67248.7434 | 0.0757 .028700 0.0000 : rQ31(34) : Excited a3Pv=13 34 35 F3f - X Xv=0 34 34 F1e | : 839:<Linelist> |
| Excited | 47 | f | 10 | X | 47 e | 1 | 66803.9009 | 66803.9660 | -0.0651 .054400 0.0000 : rQ31(47) : Excited a3Pv=13 47 48 F3f - X Xv=0 47 47 F1e | : 852:<Linelist> |
| Excited | 1 | e | 4 | X | 2 e | 1 | 67793.3796 | 67793.3836 | -0.0040 .025800 0.0000 : oP1(2) : Excited ev=4 1 0 F1e - X Xv=0 2 2 F1e | : 956:<Linelist> |
| Excited | 3 | e | 5 | X | 4 e | 1 | 67775.5119 | 67775.5114 | 0.0005 .029200 0.0000 : oP1(4) : Excited ev=4 3 2 F1e - X Xv=0 4 4 F1e | : 958:<Linelist> |
| Excited | 4 | e | 5 | X | 5 e | 1 | 67764.6884 | 67764.6530 | 0.0354 .028800 0.0000 : oP1(5) : Excited ev=4 4 3 F1e - X Xv=0 5 5 F1e | : 959:<Linelist> |
| Excited | 5 | e | 5 | X | 6 e | 1 | 67752.4543 | 67752.5052 | -0.0509 .030300 0.0000 : oP1(6) : Excited ev=4 5 4 F1e - X Xv=0 6 6 F1e | : 960:<Linelist> |
| Excited | 6 | e | 5 | X | 7 e | 1 | 67739.1087 | 67739.0672 | 0.0415 .028400 0.0000 : oP1(7) : Excited ev=4 6 5 F1e - X Xv=0 7 7 F1e | : 961:<Linelist> |
| Excited | 7 | e | 5 | X | 8 e | 1 | 67724.3658 | 67724.3397 | 0.0261 .028600 0.0000 : oP1(8) : Excited ev=4 7 6 F1e - X Xv=0 8 8 F1e | : 962:<Linelist> |
| Excited | 8 | e | 5 | X | 9 e | 1 | 67708.2480 | 67708.3235 | -0.0755 .069500 0.0000 : oP1(9) : Excited ev=4 8 7 F1e - X Xv=0 9 9 F1e | : 963:<Linelist> |
| Excited | 9 | e | 5 | X | 10 e | 1 | 67691.0526 | 67691.0207 | 0.0319 .028700 0.0000 : oP1(10) : Excited ev=4 9 8 F1e - X Xv=0 10 10 F1e | : 964:<Linelist> |
| Excited | 10 | e | 5 | X | 11 e | 1 | 67672.4648 | 67672.4341 | 0.0307 .028700 0.0000 : oP1(11) : Excited ev=4 10 9 F1e - X Xv=0 11 11 F1e | : 965:<Linelist> |
| Excited | 12 | e | 5 | X | 13 e | 1 | 67631.4701 | 67631.4356 | 0.0345 .042800 0.0000 : oP1(13) : Excited ev=4 12 11 F1e - X Xv=0 13 13 F1e | : 967:<Linelist> |
| Excited | 22 | e | 3 | X | 23 e | 1 | 67350.2608 | 67350.2575 | 0.0033 .038700 0.0000 : oP1(23) : Excited ev=4 22 21 F1e - X Xv=0 23 23 F1e | : 977:<Linelist> |
| Excited | 23 | e | 3 | X | 24 e | 1 | 67316.0698 | 67316.0563 | 0.0135 .039400 0.0000 : oP1(24) : Excited ev=4 23 22 F1e - X Xv=0 24 24 F1e | : 978:<Linelist> |
| Excited | 24 | e | 3 | X | 25 e | 1 | 67282.1903 | 67282.1313 | 0.0590 .094300 0.0000 : oP1(25) : Excited ev=4 24 23 F1e - X Xv=0 25 25 F1e | : 979:<Linelist> |
| Excited | 1 | e | 5 | X | 2 e | 1 | 67799.7022 | 67799.7347 | -0.0325 .030500 0.0000 : qP31(2) : Excited ev=4 1 2 F3e - X Xv=0 2 2 F1e | :1006:<Linelist> |
| Excited | 2 | e | 6 | X | 3 e | 1 | 67795.8558 | 67795.8685 | -0.0127 .032100 0.0000 : qP31(3) : Excited ev=4 2 3 F3e - X Xv=0 3 3 F1e | :1007:<Linelist> |
| Excited | 8 | e | 6 | X | 9 e | 1 | 67745.2384 | 67745.2772 | -0.0388 .029500 0.0000 : qP31(9) : Excited ev=4 8 9 F3e - X Xv=0 9 9 F1e | :1013:<Linelist> |
| Excited | 9 | e | 6 | X | 10 e | 1 | 67732.3303 | 67732.3073 | 0.0230 .029600 0.0000 : qP31(10) : Excited ev=4 9 10 F3e - X Xv=0 10 10 F1e | :1014:<Linelist> |
| Excited | 10 | e | 6 | X | 11 e | 1 | 67718.0372 | 67718.0426 | -0.0054 .030100 0.0000 : qP31(11) : Excited ev=4 10 11 F3e - X Xv=0 11 11 F1e | :1015:<Linelist> |
| Excited | 11 | e | 6 | X | 12 e | 1 | 67702.5258 | 67702.4840 | 0.0418 .029300 0.0000 : qP31(12) : Excited ev=4 11 12 F3e - X Xv=0 12 12 F1e | :1016:<Linelist> |
| Excited | 12 | e | 6 | X | 13 e | 1 | 67685.6022 | 67685.6321 | -0.0299 .035200 0.0000 : qP31(13) : Excited ev=4 12 13 F3e - X Xv=0 13 13 F1e | :1017:<Linelist> |
| Excited | 14 | e | 6 | X | 15 e | 1 | 67648.0898 | 67648.0544 | 0.0354 .030000 0.0000 : qP31(15) : Excited ev=4 14 15 F3e - X Xv=0 15 15 F1e | :1019:<Linelist> |
| Excited | 27 | e | 6 | X | 28 e | 1 | 67278.3922 | 67278.4016 | -0.0094 .026200 0.0000 : qP31(28) : Excited ev=4 27 28 F3e - X Xv=0 28 28 F1e | :1032:<Linelist> |
| Excited | 28 | e | 6 | X | 29 e | 1 | 67241.0392 | 67241.0071 | 0.0321 .029800 0.0000 : qP31(29) : Excited ev=4 28 29 F3e - X Xv=0 29 29 F1e | :1033:<Linelist> |
| Excited | 29 | e | 6 | X | 30 e | 1 | 67203.5134 | 67203.4861 | 0.0273 .028200 0.0000 : qP31(30) : Excited ev=4 29 30 F3e - X Xv=0 30 30 F1e | :1034:<Linelist> |
| Excited | 30 | e | 6 | X | 31 e | 1 | 67166.2030 | 67166.2916 | -0.0886 .028400 0.0000 : qP31(31) : Excited ev=4 30 31 F3e - X Xv=0 31 31 F1e | :1035:<Linelist> |
| Excited | 31 | e | 4 | X | 32 e | 1 | 67109.4561 | 67109.4604 | -0.0043 .025600 0.0000 : qP31(32) : Excited ev=4 31 32 F3e - X Xv=0 32 32 F1e | :1036:<Linelist> |
| Excited | 32 | e | 4 | X | 33 e | 1 | 67069.5048 | 67069.4918 | 0.0130 .034600 0.0000 : qP31(33) : Excited ev=4 32 33 F3e - X Xv=0 33 33 F1e | :1037:<Linelist> |
| Excited | 33 | e | 4 | X | 34 e | 1 | 67026.6408 | 67026.6797 | -0.0389 .061500 0.0000 : qP31(34) : Excited ev=4 33 34 F3e - X Xv=0 34 34 F1e | :1038:<Linelist> |

| | | | | | | | | | | | | | | |
|---|---|---|---|---|---|---|---|---|---|---|---|---|---|---|
| Excited | 36 | e | 4 | X | 37 | e | 1 | 66887.9332 | 66887.8896 | 0.0436 | .026900 | 0.0000 | : qP31(37) : Excited ev=4 36 37 F3e - X Xv=0 37 37 F1e | :1041:<Linelist> |
| Excited | 38 | e | 4 | X | 39 | e | 1 | 66788.4476 | 66788.4944 | -0.0468 | .052200 | 0.0000 | : qP31(39) : Excited ev=4 38 39 F3e - X Xv=0 39 39 F1e | :1043:<Linelist> |
| Excited | 29 | f | 3 | X | 29 | e | 1 | 67236.3390 | 67236.3402 | -0.0012 | .028600 | 0.0000 | : qQ21(29) : Excited ev=4 29 29 F2f - X Xv=0 29 29 F1e | :1083:<Linelist> |
| Excited | 30 | f | 3 | X | 30 | e | 1 | 67198.5748 | 67198.5348 | 0.0400 | .025700 | 0.0000 | : qQ21(30) : Excited ev=4 30 30 F2f - X Xv=0 30 30 F1e | :1084:<Linelist> |
| Excited | 31 | f | 3 | X | 31 | e | 1 | 67159.7505 | 67159.7064 | 0.0441 | .026000 | 0.0000 | : qQ21(31) : Excited ev=4 31 31 F2f - X Xv=0 31 31 F1e | :1085:<Linelist> |
| Excited | 33 | f | 2 | X | 33 | e | 1 | 67070.6794 | 67070.6644 | 0.0150 | .036000 | 0.0000 | : qQ21(33) : Excited ev=4 33 33 F2f - X Xv=0 33 33 F1e | :1087:<Linelist> |
| Excited | 39 | f | 2 | X | 39 | e | 1 | 66788.5333 | 66788.5227 | 0.0106 | .046100 | 0.0000 | : qQ21(39) : Excited ev=4 39 39 F2f - X Xv=0 39 39 F1e | :1093:<Linelist> |
| Excited | 1 | e | 4 | X | 0 | e | 1 | 67803.8885 | 67803.8618 | 0.0267 | .028800 | 0.0000 | : qR1(0) : Excited ev=4 1 0 F1e - X Xv=0 0 0 F1e | :1105:<Linelist> |
| Excited | 2 | e | 5 | X | 1 | e | 1 | 67802.5691 | 67802.5469 | 0.0222 | .029300 | 0.0000 | : qR1(1) : Excited ev=4 2 1 F1e - X Xv=0 1 1 F1e | :1106:<Linelist> |
| Excited | 3 | e | 5 | X | 2 | e | 1 | 67799.9569 | 67799.9593 | -0.0024 | .029200 | 0.0000 | : qR1(2) : Excited ev=4 3 2 F1e - X Xv=0 2 2 F1e | :1107:<Linelist> |
| Excited | 4 | e | 5 | X | 3 | e | 1 | 67796.0929 | 67796.0846 | 0.0083 | .029100 | 0.0000 | : qR1(3) : Excited ev=4 4 3 F1e - X Xv=0 3 3 F1e | :1108:<Linelist> |
| Excited | 24 | e | 3 | X | 23 | e | 1 | 67452.6898 | 67452.6850 | 0.0048 | .020400 | 0.0000 | : qR1(23) : Excited ev=4 24 23 F1e - X Xv=0 23 23 F1e | :1128:<Linelist> |
| Excited | 27 | e | 1 | X | 26 | e | 1 | 67335.3384 | 67335.3497 | -0.0113 | .080000 | 0.0000 | : qR1(26) : Excited ev=4 27 26 F1e - X Xv=0 26 26 F1e | :1131:<Linelist> |
| Excited | 28 | e | 1 | X | 27 | e | 1 | 67303.5804 | 67303.6838 | -0.1034 | .045500 | 0.0000 | : qR1(27) : Excited ev=4 28 27 F1e - X Xv=0 27 27 F1e | :1132:<Linelist> |
| Excited | 1 | e | 5 | X | 0 | e | 1 | 67810.2137 | 67810.2130 | 0.0007 | .031900 | 0.0000 | : sR31(0) : Excited ev=4 1 2 F3e - X Xv=0 0 0 F1e | :1155:<Linelist> |
| Excited | 2 | e | 6 | X | 1 | e | 1 | 67813.3323 | 67813.3319 | 0.0004 | .028400 | 0.0000 | : sR31(1) : Excited ev=4 2 3 F3e - X Xv=0 1 1 F1e | :1156:<Linelist> |
| Excited | 6 | e | 6 | X | 5 | e | 1 | 67812.7037 | 67812.7260 | -0.0223 | .031500 | 0.0000 | : sR31(5) : Excited ev=4 6 7 F3e - X Xv=0 5 5 F1e | :1160:<Linelist> |
| Excited | 7 | e | 6 | X | 6 | e | 1 | 67809.3828 | 67809.3268 | 0.0560 | .029000 | 0.0000 | : sR31(6) : Excited ev=4 7 8 F3e - X Xv=0 6 6 F1e | :1161:<Linelist> |
| Excited | 8 | e | 6 | X | 7 | e | 1 | 67804.6707 | 67804.6300 | 0.0407 | .029900 | 0.0000 | : sR31(7) : Excited ev=4 8 9 F3e - X Xv=0 7 7 F1e | :1162:<Linelist> |
| Excited | 9 | e | 6 | X | 8 | e | 1 | 67798.6351 | 67798.6359 | -0.0008 | .029900 | 0.0000 | : sR31(8) : Excited ev=4 9 10 F3e - X Xv=0 8 8 F1e | :1163:<Linelist> |
| Excited | 11 | e | 6 | X | 10 | e | 1 | 67782.8266 | 67782.7570 | 0.0696 | .029500 | 0.0000 | : sR31(10) : Excited ev=4 11 12 F3e - X Xv=0 10 10 F1e | :1165:<Linelist> |
| Excited | 12 | e | 6 | X | 11 | e | 1 | 67772.8358 | 67772.8733 | -0.0375 | .030300 | 0.0000 | : sR31(11) : Excited ev=4 12 13 F3e - X Xv=0 11 11 F1e | :1166:<Linelist> |
| Excited | 13 | e | 6 | X | 12 | e | 1 | 67761.7169 | 67761.6946 | 0.0223 | .029700 | 0.0000 | : sR31(12) : Excited ev=4 13 14 F3e - X Xv=0 12 12 F1e | :1167:<Linelist> |
| Excited | 14 | e | 6 | X | 13 | e | 1 | 67749.2996 | 67749.2225 | 0.0771 | .029800 | 0.0000 | : sR31(13) : Excited ev=4 14 15 F3e - X Xv=0 13 13 F1e | :1168:<Linelist> |
| Excited | 15 | e | 6 | X | 14 | e | 1 | 67735.4627 | 67735.4591 | 0.0036 | .029200 | 0.0000 | : sR31(14) : Excited ev=4 15 16 F3e - X Xv=0 14 14 F1e | :1169:<Linelist> |
| Excited | 16 | e | 6 | X | 15 | e | 1 | 67720.4165 | 67720.4079 | 0.0086 | .029400 | 0.0000 | : sR31(15) : Excited ev=4 16 17 F3e - X Xv=0 15 15 F1e | :1170:<Linelist> |
| Excited | 17 | e | 6 | X | 16 | e | 1 | 67704.1169 | 67704.0748 | 0.0421 | .032600 | 0.0000 | : sR31(16) : Excited ev=4 17 18 F3e - X Xv=0 16 16 F1e | :1171:<Linelist> |
| Excited | 20 | e | 6 | X | 19 | e | 1 | 67647.5571 | 67647.5475 | 0.0096 | .029100 | 0.0000 | : sR31(19) : Excited ev=4 20 21 F3e - X Xv=0 19 19 F1e | :1174:<Linelist> |
| Excited | 32 | e | 4 | X | 31 | e | 1 | 67295.1332 | 67295.1392 | -0.0060 | .029500 | 0.0000 | : sR31(31) : Excited ev=4 32 33 F3e - X Xv=0 31 31 F1e | :1186:<Linelist> |
| Excited | 33 | e | 4 | X | 32 | e | 1 | 67259.1487 | 67259.1811 | -0.0324 | .036000 | 0.0000 | : sR31(32) : Excited ev=4 33 34 F3e - X Xv=0 32 32 F1e | :1187:<Linelist> |
| Excited | 34 | e | 4 | X | 33 | e | 1 | 67221.2759 | 67221.3017 | -0.0258 | .079800 | 0.0000 | : sR31(33) : Excited ev=4 34 35 F3e - X Xv=0 33 33 F1e | :1188:<Linelist> |
| Excited | 35 | e | 4 | X | 34 | e | 1 | 67181.8056 | 67181.8286 | -0.0230 | .028100 | 0.0000 | : sR31(34) : Excited ev=4 35 36 F3e - X Xv=0 34 34 F1e | :1189:<Linelist> |
| Excited | 8 | e | 7 | X | 9 | e | 1 | 67825.6194 | 67825.5297 | 0.0897 | .058700 | 0.0000 | : oP1(9) : Excited dv=7 8 7 F1e - X Xv=0 9 9 F1e | :1210:<Linelist> |
| Excited | 9 | e | 7 | X | 10 | e | 1 | 67809.1018 | 67809.0631 | 0.0387 | .062600 | 0.0000 | : oP1(10) : Excited dv=7 9 8 F1e - X Xv=0 10 10 F1e | :1211:<Linelist> |
| Excited | 11 | e | 7 | X | 12 | e | 1 | 67771.9188 | 67771.8949 | 0.0239 | .042600 | 0.0000 | : oP1(12) : Excited dv=7 11 10 F1e - X Xv=0 12 12 F1e | :1213:<Linelist> |
| Excited | 12 | e | 7 | X | 13 | e | 1 | 67751.2473 | 67751.2059 | 0.0414 | .049500 | 0.0000 | : oP1(13) : Excited dv=7 12 11 F1e - X Xv=0 13 13 F1e | :1214:<Linelist> |
| Excited | 13 | e | 7 | X | 14 | e | 1 | 67729.1337 | 67729.1219 | 0.0118 | .036100 | 0.0000 | : oP1(14) : Excited dv=7 13 12 F1e - X Xv=0 14 14 F1e | :1215:<Linelist> |
| Excited | 14 | e | 7 | X | 15 | e | 1 | 67705.6540 | 67705.6487 | 0.0053 | .060400 | 0.0000 | : oP1(15) : Excited dv=7 14 13 F1e - X Xv=0 15 15 F1e | :1216:<Linelist> |
| Excited | 15 | e | 7 | X | 16 | e | 1 | 67680.7586 | 67680.7921 | -0.0335 | .046800 | 0.0000 | : oP1(16) : Excited dv=7 15 14 F1e - X Xv=0 16 16 F1e | :1217:<Linelist> |
| Excited | 16 | e | 7 | X | 17 | e | 1 | 67654.5721 | 67654.5576 | 0.0145 | .051500 | 0.0000 | : oP1(17) : Excited dv=7 16 15 F1e - X Xv=0 17 17 F1e | :1218:<Linelist> |
| Excited | 27 | e | 5 | X | 28 | e | 1 | 67274.9507 | 67274.9342 | 0.0165 | .034900 | 0.0000 | : oP1(28) : Excited dv=7 27 26 F1e - X Xv=0 28 28 F1e | :1229:<Linelist> |
| Excited | 30 | e | 3 | X | 31 | e | 1 | 67138.4857 | 67138.4698 | 0.0159 | .029200 | 0.0000 | : oP1(31) : Excited dv=7 30 29 F1e - X Xv=0 31 31 F1e | :1232:<Linelist> |
| Excited | 31 | e | 3 | X | 32 | e | 1 | 67095.5596 | 67095.6239 | -0.0643 | .032200 | 0.0000 | : oP1(32) : Excited dv=7 31 30 F1e - X Xv=0 32 32 F1e | :1233:<Linelist> |
| Excited | 2 | e | 7 | X | 3 | e | 1 | 67932.4665 | 67932.4278 | 0.0387 | .075100 | 0.0000 | : pP21(3) : Excited dv=7 2 2 F2e - X Xv=0 3 3 F1e | :1252:<Linelist> |
| Excited | 3 | e | 8 | X | 4 | e | 1 | 67924.9449 | 67924.9430 | 0.0019 | .056200 | 0.0000 | : pP21(4) : Excited dv=7 3 3 F2e - X Xv=0 4 4 F1e | :1253:<Linelist> |
| Excited | 4 | e | 8 | X | 5 | e | 1 | 67916.1301 | 67916.1275 | 0.0026 | .037200 | 0.0000 | : pP21(5) : Excited dv=7 4 4 F2e - X Xv=0 5 5 F1e | :1254:<Linelist> |
| Excited | 5 | e | 8 | X | 6 | e | 1 | 67906.0098 | 67905.9811 | 0.0287 | .034800 | 0.0000 | : pP21(6) : Excited dv=7 5 5 F2e - X Xv=0 6 6 F1e | :1255:<Linelist> |
| Excited | 6 | e | 8 | X | 7 | e | 1 | 67894.5083 | 67894.5036 | 0.0047 | .026200 | 0.0000 | : pP21(7) : Excited dv=7 6 6 F2e - X Xv=0 7 7 F1e | :1256:<Linelist> |
| Excited | 7 | e | 8 | X | 8 | e | 1 | 67881.7113 | 67881.6949 | 0.0164 | .030500 | 0.0000 | : pP21(8) : Excited dv=7 7 7 F2e - X Xv=0 8 8 F1e | :1257:<Linelist> |
| Excited | 8 | e | 8 | X | 9 | e | 1 | 67867.5636 | 67867.5547 | 0.0089 | .028200 | 0.0000 | : pP21(9) : Excited dv=7 8 8 F2e - X Xv=0 9 9 F1e | :1258:<Linelist> |
| Excited | 9 | e | 8 | X | 10 | e | 1 | 67852.0771 | 67852.0830 | -0.0059 | .028900 | 0.0000 | : pP21(10) : Excited dv=7 9 9 F2e - X Xv=0 10 10 F1e | :1259:<Linelist> |
| Excited | 10 | e | 8 | X | 11 | e | 1 | 67835.2880 | 67835.2797 | 0.0083 | .028200 | 0.0000 | : pP21(11) : Excited dv=7 10 10 F2e - X Xv=0 11 11 F1e | :1260:<Linelist> |
| Excited | 11 | e | 8 | X | 12 | e | 1 | 67817.1833 | 67817.1447 | 0.0386 | .029800 | 0.0000 | : pP21(12) : Excited dv=7 11 11 F2e - X Xv=0 12 12 F1e | :1261:<Linelist> |
| Excited | 12 | e | 8 | X | 13 | e | 1 | 67797.7085 | 67797.6780 | 0.0305 | .028800 | 0.0000 | : pP21(13) : Excited dv=7 12 12 F2e - X Xv=0 13 13 F1e | :1262:<Linelist> |
| Excited | 14 | e | 8 | X | 15 | e | 1 | 67754.7654 | 67754.7496 | 0.0158 | .028400 | 0.0000 | : pP21(15) : Excited dv=7 14 14 F2e - X Xv=0 15 15 F1e | :1264:<Linelist> |
| Excited | 15 | e | 8 | X | 16 | e | 1 | 67731.2950 | 67731.2882 | 0.0068 | .028100 | 0.0000 | : pP21(16) : Excited dv=7 15 15 F2e - X Xv=0 16 16 F1e | :1265:<Linelist> |
| Excited | 16 | e | 8 | X | 17 | e | 1 | 67706.5009 | 67706.4954 | 0.0055 | .025800 | 0.0000 | : pP21(17) : Excited dv=7 16 16 F2e - X Xv=0 17 17 F1e | :1266:<Linelist> |
| Excited | 17 | e | 8 | X | 18 | e | 1 | 67680.3165 | 67680.3715 | -0.0550 | .029100 | 0.0000 | : pP21(18) : Excited dv=7 17 17 F2e - X Xv=0 18 18 F1e | :1267:<Linelist> |
| Excited | 28 | e | 8 | X | 29 | e | 1 | 67305.3504 | 67305.4082 | -0.0578 | .084700 | 0.0000 | : pP21(29) : Excited dv=7 28 28 F2e - X Xv=0 29 29 F1e | :1278:<Linelist> |
| Excited | 29 | e | 7 | X | 30 | e | 1 | 67262.8285 | 67262.9362 | -0.1077 | .028900 | 0.0000 | : pP21(30) : Excited dv=7 29 29 F2e - X Xv=0 30 30 F1e | :1279:<Linelist> |
| Excited | 30 | e | 7 | X | 31 | e | 1 | 67219.8074 | 67219.7640 | 0.0434 | .029600 | 0.0000 | : pP21(31) : Excited dv=7 30 30 F2e - X Xv=0 31 31 F1e | :1280:<Linelist> |
| Excited | 31 | e | 7 | X | 32 | e | 1 | 67175.1240 | 67175.1132 | 0.0108 | .042000 | 0.0000 | : pP21(32) : Excited dv=7 31 31 F2e - X Xv=0 32 32 F1e | :1281:<Linelist> |
| Excited | 34 | e | 7 | X | 35 | e | 1 | 67033.3721 | 67033.4208 | -0.0487 | .027100 | 0.0000 | : pP21(35) : Excited dv=7 34 34 F2e - X Xv=0 35 35 F1e | :1284:<Linelist> |
| Excited | 35 | e | 6 | X | 36 | e | 1 | 66981.1925 | 66981.2405 | -0.0480 | .026700 | 0.0000 | : pP21(36) : Excited dv=7 35 35 F2e - X Xv=0 36 36 F1e | :1285:<Linelist> |
| Excited | 6 | f | 7 | X | 6 | e | 1 | 67878.6718 | 67878.6384 | 0.0334 | .232600 | 0.0000 | : pQ1(6) : Excited dv=7 6 5 F1f - X Xv=0 6 6 F1e | :1303:<Linelist> |
| Excited | 7 | f | 7 | X | 7 | e | 1 | 67868.5212 | 67868.5082 | 0.0130 | .073500 | 0.0000 | : pQ1(7) : Excited dv=7 7 6 F1f - X Xv=0 7 7 F1e | :1304:<Linelist> |
| Excited | 8 | f | 7 | X | 8 | e | 1 | 67856.8879 | 67856.9503 | -0.0624 | .030000 | 0.0000 | : pQ1(8) : Excited dv=7 8 7 F1f - X Xv=0 8 8 F1e | :1305:<Linelist> |
| Excited | 9 | f | 7 | X | 9 | e | 1 | 67844.0010 | 67843.9711 | 0.0299 | .086800 | 0.0000 | : pQ1(9) : Excited dv=7 9 8 F1f - X Xv=0 9 9 F1e | :1306:<Linelist> |
| Excited | 10 | f | 7 | X | 10 | e | 1 | 67829.6184 | 67829.5768 | 0.0416 | .070000 | 0.0000 | : pQ1(10) : Excited dv=7 10 9 F1f - X Xv=0 10 10 F1e | :1307:<Linelist> |
| Excited | 11 | f | 7 | X | 11 | e | 1 | 67813.7915 | 67813.7738 | 0.0177 | .039700 | 0.0000 | : pQ1(11) : Excited dv=7 11 10 F1f - X Xv=0 11 11 F1e | :1308:<Linelist> |
| Excited | 12 | f | 7 | X | 12 | e | 1 | 67796.6398 | 67796.5682 | 0.0716 | .029400 | 0.0000 | : pQ1(12) : Excited dv=7 12 11 F1f - X Xv=0 12 12 F1e | :1309:<Linelist> |
| Excited | 13 | f | 7 | X | 13 | e | 1 | 67777.9809 | 67777.9659 | 0.0150 | .036300 | 0.0000 | : pQ1(13) : Excited dv=7 13 12 F1f - X Xv=0 13 13 F1e | :1310:<Linelist> |
| Excited | 15 | f | 7 | X | 15 | e | 1 | 67736.5992 | 67736.5944 | 0.0048 | .034200 | 0.0000 | : pQ1(15) : Excited dv=7 15 14 F1f - X Xv=0 15 15 F1e | :1312:<Linelist> |
| Excited | 16 | f | 7 | X | 16 | e | 1 | 67713.8154 | 67713.8363 | -0.0209 | .034600 | 0.0000 | : pQ1(16) : Excited dv=7 16 15 F1f - X Xv=0 16 16 F1e | :1313:<Linelist> |
| Excited | 17 | f | 7 | X | 17 | e | 1 | 67689.7018 | 67689.7039 | -0.0021 | .028000 | 0.0000 | : pQ1(17) : Excited dv=7 17 16 F1f - X Xv=0 17 17 F1e | :1314:<Linelist> |
| Excited | 18 | f | 7 | X | 18 | e | 1 | 67664.1939 | 67664.2031 | -0.0092 | .034800 | 0.0000 | : pQ1(18) : Excited dv=7 18 17 F1f - X Xv=0 18 18 F1e | :1315:<Linelist> |
| Excited | 19 | f | 7 | X | 19 | e | 1 | 67637.4103 | 67637.3405 | 0.0698 | .031500 | 0.0000 | : pQ1(19) : Excited dv=7 19 18 F1f - X Xv=0 19 19 F1e | :1316:<Linelist> |
| Excited | 29 | f | 6 | X | 29 | e | 1 | 67297.5066 | 67297.5614 | -0.0548 | .032900 | 0.0000 | : pQ1(29) : Excited dv=7 29 28 F1f - X Xv=0 29 29 F1e | :1326:<Linelist> |
| Excited | 30 | f | 4 | X | 30 | e | 1 | 67248.7013 | 67248.6818 | 0.0195 | .041200 | 0.0000 | : pQ1(30) : Excited dv=7 30 29 F1f - X Xv=0 30 30 F1e | :1327:<Linelist> |
| Excited | 35 | f | 3 | X | 35 | e | 1 | 67015.3768 | 67015.3572 | 0.0196 | .025600 | 0.0000 | : pQ1(35) : Excited dv=7 35 34 F1f - X Xv=0 35 35 F1e | :1332:<Linelist> |
| Excited | 1 | e | 6 | X | 2 | e | 1 | 67977.4448 | 67977.4179 | 0.0269 | .029000 | 0.0000 | : qP31(2) : Excited dv=7 1 2 F3e - X Xv=0 2 2 F1e | :1348:<Linelist> |
| Excited | 2 | e | 8 | X | 3 | e | 1 | 67971.4639 | 67971.5005 | -0.0366 | .029600 | 0.0000 | : qP31(3) : Excited dv=7 2 3 F3e - X Xv=0 3 3 F1e | :1349:<Linelist> |
| Excited | 3 | e | 9 | X | 4 | e | 1 | 67964.3989 | 67964.3682 | 0.0307 | .029500 | 0.0000 | : qP31(4) : Excited dv=7 3 4 F3e - X Xv=0 4 4 F1e | :1350:<Linelist> |
| Excited | 4 | e | 9 | X | 5 | e | 1 | 67956.0797 | 67956.0178 | 0.0619 | .029300 | 0.0000 | : qP31(5) : Excited dv=7 4 5 F3e - X Xv=0 5 5 F1e | :1351:<Linelist> |
| Excited | 5 | e | 9 | X | 6 | e | 1 | 67946.5016 | 67946.4449 | 0.0567 | .029800 | 0.0000 | : qP31(6) : Excited dv=7 5 6 F3e - X Xv=0 6 6 F1e | :1352:<Linelist> |
| Excited | 6 | e | 9 | X | 7 | e | 1 | 67935.7413 | 67935.6416 | 0.0997 | .028900 | 0.0000 | : qP31(7) : Excited dv=7 6 7 F3e - X Xv=0 7 7 F1e | :1353:<Linelist> |
| Excited | 7 | e | 10 | X | 8 | e | 1 | 67923.5632 | 67923.6264 | -0.0632 | .029700 | 0.0000 | : qP31(8) : Excited dv=7 7 8 F3e - X Xv=0 8 8 F1e | :1354:<Linelist> |
| Excited | 8 | e | 10 | X | 9 | e | 1 | 67910.3798 | 67910.3532 | 0.0266 | .029400 | 0.0000 | : qP31(9) : Excited dv=7 8 9 F3e - X Xv=0 9 9 F1e | :1355:<Linelist> |
| Excited | 9 | e | 10 | X | 10 | e | 1 | 67895.8638 | 67895.8413 | 0.0225 | .029100 | 0.0000 | : qP31(10) : Excited dv=7 9 10 F3e - X Xv=0 10 10 F1e | :1356:<Linelist> |
| Excited | 10 | e | 10 | X | 11 | e | 1 | 67880.1415 | 67880.0811 | 0.0604 | .029500 | 0.0000 | : qP31(11) : Excited dv=7 10 11 F3e - X Xv=0 11 11 F1e | :1357:<Linelist> |
| Excited | 11 | e | 10 | X | 12 | e | 1 | 67863.1001 | 67863.0663 | 0.0338 | .029800 | 0.0000 | : qP31(12) : Excited dv=7 11 12 F3e - X Xv=0 12 12 F1e | :1358:<Linelist> |
| Excited | 12 | e | 10 | X | 13 | e | 1 | 67844.8157 | 67844.7913 | 0.0244 | .028400 | 0.0000 | : qP31(13) : Excited dv=7 12 13 F3e - X Xv=0 13 13 F1e | :1359:<Linelist> |

Excited 13 e 10 X 14 e 1 67825.2462 67825.2507 -0.0045 .028800 0.0000 : qP31(14) : Excited dv=7 13 14 F3e - X Xv=0 14 14 F1e :1360:<Linelist>
Excited 14 e 10 X 15 e 1 67804.4537 67804.4394 0.0143 .030400 0.0000 : qP31(15) : Excited dv=7 14 15 F3e - X Xv=0 15 15 F1e :1361:<Linelist>
Excited 15 e 10 X 16 e 1 67782.3184 67782.3525 -0.0341 .028800 0.0000 : qP31(16) : Excited dv=7 15 16 F3e - X Xv=0 16 16 F1e :1362:<Linelist>
Excited 16 e 10 X 17 e 1 67759.0029 67758.9858 0.0171 .025900 0.0000 : qP31(17) : Excited dv=7 16 17 F3e - X Xv=0 17 17 F1e :1363:<Linelist>
Excited 17 e 10 X 18 e 1 67734.3483 67734.3350 0.0133 .028100 0.0000 : qP31(18) : Excited dv=7 17 18 F3e - X Xv=0 18 18 F1e :1364:<Linelist>
Excited 18 e 10 X 19 e 1 67708.3820 67708.3966 -0.0146 .039000 0.0000 : qP31(19) : Excited dv=7 18 19 F3e - X Xv=0 19 19 F1e :1365:<Linelist>
Excited 19 e 10 X 20 e 1 67681.1843 67681.1671 0.0172 .036500 0.0000 : qP31(20) : Excited dv=7 19 20 F3e - X Xv=0 20 20 F1e :1366:<Linelist>
Excited 31 e 10 X 32 e 1 67253.3859 67253.4452 -0.0593 .055000 0.0000 : qP31(32) : Excited dv=7 31 32 F3e - X Xv=0 32 32 F1e :1378:<Linelist>
Excited 40 e 7 X 41 e 1 66803.2851 66803.1846 0.1005 .043800 0.0000 : qP31(41) : Excited dv=7 40 41 F3e - X Xv=0 41 41 F1e :1387:<Linelist>
Excited 41 e 7 X 42 e 1 66748.7177 66748.7416 -0.0239 .025600 0.0000 : qP31(42) : Excited dv=7 41 42 F3e - X Xv=0 42 42 F1e :1388:<Linelist>
Excited 48 e 6 X 49 e 1 66318.5261 66318.5580 -0.0319 .030800 0.0000 : qP31(49) : Excited dv=7 48 49 F3e - X Xv=0 49 49 F1e :1395:<Linelist>
Excited 2 f 7 X 2 e 1 67942.9100 67942.9057 0.0043 .057900 0.0000 : qQ21(2) : Excited dv=7 2 2 F2f - X Xv=0 2 2 F1e :1397:<Linelist>
Excited 3 f 8 X 3 e 1 67938.9185 67938.9130 0.0055 .027900 0.0000 : qQ21(3) : Excited dv=7 3 3 F2f - X Xv=0 3 3 F1e :1398:<Linelist>
Excited 4 f 8 X 4 e 1 67933.5956 67933.5891 0.0065 .029900 0.0000 : qQ21(4) : Excited dv=7 4 4 F2f - X Xv=0 4 4 F1e :1399:<Linelist>
Excited 5 f 8 X 5 e 1 67926.9923 67926.9336 0.0587 .031300 0.0000 : qQ21(5) : Excited dv=7 5 5 F2f - X Xv=0 5 5 F1e :1400:<Linelist>
Excited 6 f 8 X 6 e 1 67918.9579 67918.9464 0.0115 .028500 0.0000 : qQ21(6) : Excited dv=7 6 6 F2f - X Xv=0 6 6 F1e :1401:<Linelist>
Excited 7 f 8 X 7 e 1 67909.6306 67909.6270 0.0036 .028300 0.0000 : qQ21(7) : Excited dv=7 7 7 F2f - X Xv=0 7 7 F1e :1402:<Linelist>
Excited 8 f 8 X 8 e 1 67898.9833 67898.9753 0.0080 .028400 0.0000 : qQ21(8) : Excited dv=7 8 8 F2f - X Xv=0 8 8 F1e :1403:<Linelist>
Excited 9 f 8 X 9 e 1 67886.9965 67886.9910 0.0055 .028300 0.0000 : qQ21(9) : Excited dv=7 9 9 F2f - X Xv=0 9 9 F1e :1404:<Linelist>
Excited 10 f 8 X 10 e 1 67873.6283 67873.6738 -0.0455 .028300 0.0000 : qQ21(10) : Excited dv=7 10 10 F2f - X Xv=0 10 10 F1e :1405:<Linelist>
Excited 11 f 8 X 11 e 1 67859.0326 67859.0235 0.0091 .028400 0.0000 : qQ21(11) : Excited dv=7 11 11 F2f - X Xv=0 11 11 F1e :1406:<Linelist>
Excited 12 f 8 X 12 e 1 67843.0498 67843.0402 0.0096 .025900 0.0000 : qQ21(12) : Excited dv=7 12 12 F2f - X Xv=0 12 12 F1e :1407:<Linelist>
Excited 13 f 8 X 13 e 1 67825.7535 67825.7235 0.0300 .028600 0.0000 : qQ21(13) : Excited dv=7 13 13 F2f - X Xv=0 13 13 F1e :1408:<Linelist>
Excited 14 f 8 X 14 e 1 67807.1071 67807.0736 0.0335 .028300 0.0000 : qQ21(14) : Excited dv=7 14 14 F2f - X Xv=0 14 14 F1e :1409:<Linelist>
Excited 15 f 8 X 15 e 1 67787.0993 67787.0901 0.0092 .028200 0.0000 : qQ21(15) : Excited dv=7 15 15 F2f - X Xv=0 15 15 F1e :1410:<Linelist>
Excited 16 f 9 X 16 e 1 67765.8231 67765.7879 0.0352 .028300 0.0000 : qQ21(16) : Excited dv=7 16 16 F2f - X Xv=0 16 16 F1e :1411:<Linelist>
Excited 17 f 9 X 17 e 1 67743.1287 67743.1251 0.0036 .025800 0.0000 : qQ21(17) : Excited dv=7 17 17 F2f - X Xv=0 17 17 F1e :1412:<Linelist>
Excited 19 f 9 X 19 e 1 67693.8230 67693.8265 -0.0035 .028200 0.0000 : qQ21(19) : Excited dv=7 19 19 F2f - X Xv=0 19 19 F1e :1414:<Linelist>
Excited 21 f 9 X 21 e 1 67639.2035 67639.1976 0.0059 .030100 0.0000 : qQ21(21) : Excited dv=7 21 21 F2f - X Xv=0 21 21 F1e :1416:<Linelist>
Excited 29 f 8 X 29 e 1 67367.0860 67367.1819 -0.0959 .052500 0.0000 : qQ21(29) : Excited dv=7 29 29 F2f - X Xv=0 29 29 F1e :1424:<Linelist>
Excited 36 f 6 X 36 e 1 67058.8239 67058.7190 0.1049 .034900 0.0000 : qQ21(36) : Excited dv=7 36 36 F2f - X Xv=0 36 36 F1e :1431:<Linelist>
Excited 38 f 6 X 38 e 1 66959.3872 66959.4267 -0.0395 .046100 0.0000 : qQ21(38) : Excited dv=7 38 38 F2f - X Xv=0 38 38 F1e :1433:<Linelist>
Excited 4 e 7 X 3 e 1 67908.5861 67908.5618 0.0243 .030500 0.0000 : qR1(3) : Excited dv=7 4 3 F1e - X Xv=0 3 3 F1e :1447:<Linelist>
Excited 5 e 7 X 4 e 1 67904.8006 67904.7964 0.0042 .043600 0.0000 : qR1(4) : Excited dv=7 5 4 F1e - X Xv=0 4 4 F1e :1448:<Linelist>
Excited 6 e 7 X 5 e 1 67899.5901 67899.5909 -0.0008 .041800 0.0000 : qR1(5) : Excited dv=7 6 5 F1e - X Xv=0 5 5 F1e :1449:<Linelist>
Excited 7 e 7 X 6 e 1 67892.9017 67892.9510 -0.0493 .055500 0.0000 : qR1(6) : Excited dv=7 7 6 F1e - X Xv=0 6 6 F1e :1450:<Linelist>
Excited 8 e 7 X 7 e 1 67884.8118 67884.8825 -0.0707 .054600 0.0000 : qR1(7) : Excited dv=7 8 7 F1e - X Xv=0 7 7 F1e :1451:<Linelist>
Excited 9 e 7 X 8 e 1 67875.4055 67875.3917 0.0138 .087700 0.0000 : qR1(8) : Excited dv=7 9 8 F1e - X Xv=0 8 8 F1e :1452:<Linelist>
Excited 10 e 7 X 9 e 1 67864.5139 67864.4848 0.0291 .036500 0.0000 : qR1(9) : Excited dv=7 10 9 F1e - X Xv=0 9 9 F1e :1453:<Linelist>
Excited 11 e 7 X 10 e 1 67852.1994 67852.1679 0.0315 .060200 0.0000 : qR1(10) : Excited dv=7 11 10 F1e - X Xv=0 10 10 F1e :1454:<Linelist>
Excited 12 e 7 X 11 e 1 67838.5323 67838.4471 0.0852 .056700 0.0000 : qR1(11) : Excited dv=7 12 11 F1e - X Xv=0 11 11 F1e :1455:<Linelist>
Excited 13 e 7 X 12 e 1 67823.3475 67823.3282 0.0193 .031700 0.0000 : qR1(12) : Excited dv=7 13 12 F1e - X Xv=0 12 12 F1e :1456:<Linelist>
Excited 14 e 7 X 13 e 1 67806.8753 67806.8168 0.0585 .042100 0.0000 : qR1(13) : Excited dv=7 14 13 F1e - X Xv=0 13 13 F1e :1457:<Linelist>
Excited 15 e 7 X 14 e 1 67788.9325 67788.9186 0.0139 .045300 0.0000 : qR1(14) : Excited dv=7 15 14 F1e - X Xv=0 14 14 F1e :1458:<Linelist>
Excited 16 e 7 X 15 e 1 67769.6464 67769.6387 0.0077 .038400 0.0000 : qR1(15) : Excited dv=7 16 15 F1e - X Xv=0 15 15 F1e :1459:<Linelist>
Excited 17 e 7 X 16 e 1 67748.9646 67748.9826 -0.0180 .044500 0.0000 : qR1(16) : Excited dv=7 17 16 F1e - X Xv=0 16 16 F1e :1460:<Linelist>
Excited 18 e 7 X 17 e 1 67726.8697 67726.9561 -0.0864 .026000 0.0000 : qR1(17) : Excited dv=7 18 17 F1e - X Xv=0 17 17 F1e :1461:<Linelist>
Excited 20 e 7 X 19 e 1 67678.8724 67678.8196 0.0528 .029400 0.0000 : qR1(19) : Excited dv=7 20 19 F1e - X Xv=0 19 19 F1e :1463:<Linelist>
Excited 31 e 3 X 30 e 1 67314.3848 67314.4096 -0.0248 .028500 0.0000 : qR1(30) : Excited dv=7 31 30 F1e - X Xv=0 30 30 F1e :1474:<Linelist>
Excited 36 e 2 X 35 e 1 67094.5189 67094.5508 -0.0319 .049200 0.0000 : qR1(35) : Excited dv=7 36 35 F1e - X Xv=0 35 35 F1e :1479:<Linelist>
Excited 41 e 2 X 40 e 1 66843.6069 66843.6448 -0.0379 .033700 0.0000 : qR1(40) : Excited dv=7 41 40 F1e - X Xv=0 40 40 F1e :1484:<Linelist>
Excited 43 e 2 X 42 e 1 66733.2659 66733.3481 -0.0822 .053000 0.0000 : qR1(42) : Excited dv=7 43 42 F1e - X Xv=0 42 42 F1e :1486:<Linelist>
Excited 46 e 2 X 45 e 1 66557.7520 66557.7694 -0.0174 .025100 0.0000 : qR1(45) : Excited dv=7 46 45 F1e - X Xv=0 45 45 F1e :1489:<Linelist>
Excited 1 f 6 X 1 e 1 67984.4590 67984.4030 0.0560 .029600 0.0000 : rQ31(1) : Excited dv=7 1 2 F3f - X Xv=0 1 1 F1e :1494:<Linelist>
Excited 2 f 8 X 2 e 1 67981.9350 67981.9756 -0.0406 .043100 0.0000 : rQ31(2) : Excited dv=7 2 3 F3f - X Xv=0 2 2 F1e :1495:<Linelist>
Excited 3 f 10 X 3 e 1 67978.4029 67978.3417 0.0612 .042400 0.0000 : rQ31(3) : Excited dv=7 3 4 F3f - X Xv=0 3 3 F1e :1496:<Linelist>
Excited 4 f 10 X 4 e 1 67973.5394 67973.4818 0.0576 .033300 0.0000 : rQ31(4) : Excited dv=7 4 5 F3f - X Xv=0 4 4 F1e :1497:<Linelist>
Excited 5 f 10 X 5 e 1 67967.3855 67967.4005 -0.0150 .034700 0.0000 : rQ31(5) : Excited dv=7 5 6 F3f - X Xv=0 5 5 F1e :1498:<Linelist>
Excited 6 f 10 X 6 e 1 67960.1257 67960.0920 0.0337 .031400 0.0000 : rQ31(6) : Excited dv=7 6 7 F3f - X Xv=0 6 6 F1e :1499:<Linelist>
Excited 7 f 10 X 7 e 1 67951.5173 67951.5510 -0.0337 .034300 0.0000 : rQ31(7) : Excited dv=7 7 8 F3f - X Xv=0 7 7 F1e :1500:<Linelist>
Excited 8 f 10 X 8 e 1 67941.8235 67941.7716 0.0519 .039200 0.0000 : rQ31(8) : Excited dv=7 8 9 F3f - X Xv=0 8 8 F1e :1501:<Linelist>
Excited 9 f 10 X 9 e 1 67930.7299 67930.7480 -0.0181 .038900 0.0000 : rQ31(9) : Excited dv=7 9 10 F3f - X Xv=0 9 9 F1e :1502:<Linelist>
Excited 10 f 10 X 10 e 1 67918.4727 67918.4743 -0.0016 .037800 0.0000 : rQ31(10) : Excited dv=7 10 11 F3f - X Xv=0 10 10 F1e :1503:<Linelist>
Excited 11 f 10 X 11 e 1 67904.9422 67904.9446 -0.0024 .032200 0.0000 : rQ31(11) : Excited dv=7 11 12 F3f - X Xv=0 11 11 F1e :1504:<Linelist>
Excited 13 f 10 X 13 e 1 67874.1690 67874.0943 0.0747 .038100 0.0000 : rQ31(13) : Excited dv=7 13 14 F3f - X Xv=0 13 13 F1e :1506:<Linelist>
Excited 14 f 10 X 14 e 1 67856.8314 67856.7631 0.0683 .035200 0.0000 : rQ31(14) : Excited dv=7 14 15 F3f - X Xv=0 14 14 F1e :1507:<Linelist>
Excited 15 f 10 X 15 e 1 67838.0615 67838.1546 -0.0931 .029700 0.0000 : rQ31(15) : Excited dv=7 15 16 F3f - X Xv=0 15 15 F1e :1508:<Linelist>
Excited 16 f 10 X 16 e 1 67818.3199 67818.2642 0.0557 .029900 0.0000 : rQ31(16) : Excited dv=7 16 17 F3f - X Xv=0 16 16 F1e :1509:<Linelist>
Excited 17 f 10 X 17 e 1 67797.1151 67797.0877 0.0274 .034100 0.0000 : rQ31(17) : Excited dv=7 17 18 F3f - X Xv=0 17 17 F1e :1510:<Linelist>
Excited 18 f 10 X 18 e 1 67774.6192 67774.6214 -0.0022 .029100 0.0000 : rQ31(18) : Excited dv=7 18 19 F3f - X Xv=0 18 18 F1e :1511:<Linelist>
Excited 19 f 10 X 19 e 1 67750.8395 67750.8617 -0.0222 .028700 0.0000 : rQ31(19) : Excited dv=7 19 20 F3f - X Xv=0 19 19 F1e :1512:<Linelist>
Excited 20 f 10 X 20 e 1 67725.8035 67725.8057 -0.0022 .028500 0.0000 : rQ31(20) : Excited dv=7 20 21 F3f - X Xv=0 20 20 F1e :1513:<Linelist>
Excited 21 f 10 X 21 e 1 67699.4416 67699.4507 -0.0091 .028200 0.0000 : rQ31(21) : Excited dv=7 21 22 F3f - X Xv=0 21 21 F1e :1514:<Linelist>
Excited 22 f 10 X 22 e 1 67671.7875 67671.7944 -0.0069 .028400 0.0000 : rQ31(22) : Excited dv=7 22 23 F3f - X Xv=0 22 22 F1e :1515:<Linelist>
Excited 23 f 10 X 23 e 1 67642.7977 67642.8351 -0.0374 .030500 0.0000 : rQ31(23) : Excited dv=7 23 24 F3f - X Xv=0 23 23 F1e :1516:<Linelist>
Excited 42 f 8 X 42 e 1 66840.2515 66840.1421 0.1094 .028300 0.0000 : rQ31(42) : Excited dv=7 42 43 F3f - X Xv=0 42 42 F1e :1535:<Linelist>
Excited 43 f 8 X 43 e 1 66784.8024 66784.8363 -0.0339 .028800 0.0000 : rQ31(43) : Excited dv=7 43 44 F3f - X Xv=0 43 43 F1e :1536:<Linelist>
Excited 2 e 7 X 1 e 1 67949.9051 67949.8912 0.0139 .095400 0.0000 : rR21(1) : Excited dv=7 2 2 F2e - X Xv=0 1 1 F1e :1544:<Linelist>
Excited 3 e 8 X 2 e 1 67949.3726 67949.3909 -0.0183 .056600 0.0000 : rR21(2) : Excited dv=7 3 3 F2e - X Xv=0 2 2 F1e :1545:<Linelist>
Excited 4 e 8 X 3 e 1 67947.5836 67947.5591 0.0245 .034800 0.0000 : rR21(3) : Excited dv=7 4 4 F2e - X Xv=0 3 3 F1e :1546:<Linelist>
Excited 5 e 8 X 4 e 1 67944.3977 67944.3952 0.0025 .029500 0.0000 : rR21(4) : Excited dv=7 5 5 F2e - X Xv=0 4 4 F1e :1547:<Linelist>
Excited 6 e 8 X 5 e 1 67939.9065 67939.8989 0.0076 .026100 0.0000 : rR21(5) : Excited dv=7 6 6 F2e - X Xv=0 5 5 F1e :1548:<Linelist>
Excited 7 e 8 X 6 e 1 67934.0749 67934.0698 0.0051 .028400 0.0000 : rR21(6) : Excited dv=7 7 7 F2e - X Xv=0 6 6 F1e :1549:<Linelist>
Excited 8 e 8 X 7 e 1 67926.8791 67926.9075 -0.0284 .029000 0.0000 : rR21(7) : Excited dv=7 8 8 F2e - X Xv=0 7 7 F1e :1550:<Linelist>
Excited 9 e 8 X 8 e 1 67918.3549 67918.4116 -0.0567 .029600 0.0000 : rR21(8) : Excited dv=7 9 9 F2e - X Xv=0 8 8 F1e :1551:<Linelist>
Excited 10 e 8 X 9 e 1 67908.5290 67908.5818 -0.0528 .030500 0.0000 : rR21(9) : Excited dv=7 10 10 F2e - X Xv=0 9 9 F1e :1552:<Linelist>
Excited 11 e 8 X 10 e 1 67897.4250 67897.4177 0.0073 .025900 0.0000 : rR21(10) : Excited dv=7 11 11 F2e - X Xv=0 10 10 F1e :1553:<Linelist>
Excited 12 e 8 X 11 e 1 67884.9248 67884.9191 0.0057 .030100 0.0000 : rR21(11) : Excited dv=7 12 12 F2e - X Xv=0 11 11 F1e :1554:<Linelist>
Excited 13 e 8 X 12 e 1 67871.0957 67871.0859 0.0098 .028000 0.0000 : rR21(12) : Excited dv=7 13 13 F2e - X Xv=0 12 12 F1e :1555:<Linelist>
Excited 14 e 8 X 13 e 1 67855.9047 67855.9178 -0.0131 .028900 0.0000 : rR21(13) : Excited dv=7 14 14 F2e - X Xv=0 13 13 F1e :1556:<Linelist>
Excited 15 e 8 X 14 e 1 67839.4217 67839.4146 0.0071 .028200 0.0000 : rR21(14) : Excited dv=7 15 15 F2e - X Xv=0 14 14 F1e :1557:<Linelist>
Excited 16 e 8 X 15 e 1 67821.5702 67821.5765 -0.0063 .028600 0.0000 : rR21(15) : Excited dv=7 16 16 F2e - X Xv=0 15 15 F1e :1558:<Linelist>

Excited 17 e 8 X 16 e 1 67802.3498 67802.4031 -0.0533 .030000 0.0000 : rR21(16) : Excited dv=7 17 17 F2e - X Xv=0 16 16 F1e :1559:<Linelist>
Excited 18 e 8 X 17 e 1 67781.8912 67781.8947 -0.0035 .028000 0.0000 : rR21(17) : Excited dv=7 18 18 F2e - X Xv=0 17 17 F1e :1560:<Linelist>
Excited 19 e 8 X 18 e 1 67760.0463 67760.0512 -0.0049 .025900 0.0000 : rR21(18) : Excited dv=7 19 19 F2e - X Xv=0 18 18 F1e :1561:<Linelist>
Excited 20 e 8 X 19 e 1 67736.8650 67736.8727 -0.0077 .028000 0.0000 : rR21(19) : Excited dv=7 20 20 F2e - X Xv=0 19 19 F1e :1562:<Linelist>
Excited 21 e 8 X 20 e 1 67712.3382 67712.3594 -0.0212 .028500 0.0000 : rR21(20) : Excited dv=7 21 21 F2e - X Xv=0 20 20 F1e :1563:<Linelist>
Excited 23 e 8 X 22 e 1 67659.3201 67659.3303 -0.0102 .029200 0.0000 : rR21(22) : Excited dv=7 23 23 F2e - X Xv=0 22 22 F1e :1565:<Linelist>
Excited 24 e 8 X 23 e 1 67630.7490 67630.8159 -0.0669 .055300 0.0000 : rR21(23) : Excited dv=7 24 24 F2e - X Xv=0 23 23 F1e :1566:<Linelist>
Excited 1 e 6 X 0 e 1 67987.9000 67987.8961 0.0039 .030300 0.0000 : sR31(0) : Excited dv=7 1 2 F3e - X Xv=0 0 0 F1e :1593:<Linelist>
Excited 2 e 8 X 1 e 1 67988.9224 67988.9639 -0.0415 .033200 0.0000 : sR31(1) : Excited dv=7 2 3 F3e - X Xv=0 1 1 F1e :1594:<Linelist>
Excited 3 e 9 X 2 e 1 67988.7290 67988.8161 -0.0871 .031300 0.0000 : sR31(2) : Excited dv=7 3 4 F3e - X Xv=0 2 2 F1e :1595:<Linelist>
Excited 4 e 9 X 3 e 1 67987.3545 67987.4493 -0.0948 .033100 0.0000 : sR31(3) : Excited dv=7 4 5 F3e - X Xv=0 3 3 F1e :1596:<Linelist>
Excited 5 e 9 X 4 e 1 67984.8978 67984.8590 0.0388 .036600 0.0000 : sR31(4) : Excited dv=7 5 6 F3e - X Xv=0 4 4 F1e :1597:<Linelist>
Excited 6 e 9 X 5 e 1 67980.9838 67981.0370 -0.0532 .037300 0.0000 : sR31(5) : Excited dv=7 6 7 F3e - X Xv=0 5 5 F1e :1598:<Linelist>
Excited 7 e 10 X 6 e 1 67976.0239 67976.0013 0.0226 .033900 0.0000 : sR31(6) : Excited dv=7 7 8 F3e - X Xv=0 6 6 F1e :1599:<Linelist>
Excited 8 e 10 X 7 e 1 67969.7192 67969.7061 0.0131 .034600 0.0000 : sR31(7) : Excited dv=7 8 9 F3e - X Xv=0 7 7 F1e :1600:<Linelist>
Excited 9 e 10 X 8 e 1 67962.2366 67962.1699 0.0667 .033700 0.0000 : sR31(8) : Excited dv=7 9 10 F3e - X Xv=0 8 8 F1e :1601:<Linelist>
Excited 10 e 10 X 9 e 1 67953.4055 67953.3831 0.0224 .033900 0.0000 : sR31(9) : Excited dv=7 10 11 F3e - X Xv=0 9 9 F1e :1602:<Linelist>
Excited 11 e 10 X 10 e 1 67943.3567 67943.3393 0.0174 .032200 0.0000 : sR31(10) : Excited dv=7 11 12 F3e - X Xv=0 10 10 F1e :1603:<Linelist>
Excited 12 e 10 X 11 e 1 67932.0327 67932.0325 0.0002 .033200 0.0000 : sR31(11) : Excited dv=7 12 13 F3e - X Xv=0 11 11 F1e :1604:<Linelist>
Excited 13 e 10 X 12 e 1 67919.4637 67919.4570 0.0067 .030700 0.0000 : sR31(12) : Excited dv=7 13 14 F3e - X Xv=0 12 12 F1e :1605:<Linelist>
Excited 14 e 10 X 13 e 1 67905.6725 67905.6075 0.0650 .030200 0.0000 : sR31(13) : Excited dv=7 14 15 F3e - X Xv=0 13 13 F1e :1606:<Linelist>
Excited 16 e 10 X 15 e 1 67874.0440 67874.0668 -0.0228 .034800 0.0000 : sR31(15) : Excited dv=7 16 17 F3e - X Xv=0 15 15 F1e :1608:<Linelist>
Excited 17 e 10 X 16 e 1 67856.3627 67856.3667 -0.0040 .029200 0.0000 : sR31(16) : Excited dv=7 17 18 F3e - X Xv=0 16 16 F1e :1609:<Linelist>
Excited 18 e 10 X 17 e 1 67837.3834 67837.3746 0.0088 .028400 0.0000 : sR31(17) : Excited dv=7 18 19 F3e - X Xv=0 17 17 F1e :1610:<Linelist>
Excited 19 e 10 X 18 e 1 67817.0428 67817.0871 -0.0443 .030200 0.0000 : sR31(18) : Excited dv=7 19 20 F3e - X Xv=0 18 18 F1e :1611:<Linelist>
Excited 20 e 10 X 19 e 1 67795.5304 67795.5008 0.0296 .035500 0.0000 : sR31(19) : Excited dv=7 20 21 F3e - X Xv=0 19 19 F1e :1612:<Linelist>
Excited 21 e 10 X 20 e 1 67772.5972 67772.6132 -0.0160 .028900 0.0000 : sR31(20) : Excited dv=7 21 22 F3e - X Xv=0 20 20 F1e :1613:<Linelist>
Excited 22 e 10 X 21 e 1 67748.4032 67748.4218 -0.0186 .025800 0.0000 : sR31(21) : Excited dv=7 22 23 F3e - X Xv=0 21 21 F1e :1614:<Linelist>
Excited 23 e 10 X 22 e 1 67722.8916 67722.9247 -0.0331 .028300 0.0000 : sR31(22) : Excited dv=7 23 24 F3e - X Xv=0 22 22 F1e :1615:<Linelist>
Excited 24 e 10 X 23 e 1 67696.0801 67696.1208 -0.0407 .029100 0.0000 : sR31(23) : Excited dv=7 24 25 F3e - X Xv=0 23 23 F1e :1616:<Linelist>
Excited 25 e 10 X 24 e 1 67667.9651 67668.0096 -0.0445 .030000 0.0000 : sR31(24) : Excited dv=7 25 26 F3e - X Xv=0 24 24 F1e :1617:<Linelist>
Excited 26 e 10 X 25 e 1 67638.5177 67638.5918 -0.0741 .027600 0.0000 : sR31(25) : Excited dv=7 26 27 F3e - X Xv=0 25 25 F1e :1618:<Linelist>
Excited 1 e 7 X 2 e 1 67982.1694 67982.1651 0.0043 .042500 3.0000 : pP21(2) : Excited a'v=12 1 1 F2e - X Xv=0 2 2 F1e :1643:<Linelist>
Excited 2 e 9 X 3 e 1 67975.8705 67975.8376 0.0329 .033100 3.0000 : pP21(3) : Excited a'v=12 2 2 F2e - X Xv=0 3 3 F1e :1644:<Linelist>
Excited 3 e 10 X 4 e 1 67968.0673 67968.0930 -0.0257 .073600 3.0000 : pP21(4) : Excited a'v=12 3 3 F2e - X Xv=0 4 4 F1e :1645:<Linelist>
Excited 4 e 10 X 5 e 1 67958.9253 67958.9315 -0.0062 .049300 3.0000 : pP21(5) : Excited a'v=12 4 4 F2e - X Xv=0 5 5 F1e :1646:<Linelist>
Excited 5 e 10 X 6 e 1 67948.3636 67948.3535 0.0101 .037900 3.0000 : pP21(6) : Excited a'v=12 5 5 F2e - X Xv=0 6 6 F1e :1647:<Linelist>
Excited 6 e 10 X 7 e 1 67936.3819 67936.3624 0.0195 .043600 3.0000 : pP21(7) : Excited a'v=12 6 6 F2e - X Xv=0 7 7 F1e :1648:<Linelist>
Excited 8 e 9 X 9 e 1 67908.1249 67908.1108 0.0141 .051200 3.0000 : pP21(9) : Excited a'v=12 8 8 F2e - X Xv=0 9 9 F1e :1650:<Linelist>
Excited 9 e 9 X 10 e 1 67891.8791 67891.8660 0.0131 .046100 3.0000 : pP21(10) : Excited a'v=12 9 9 F2e - X Xv=0 10 10 F1e :1651:<Linelist>
Excited 10 e 9 X 11 e 1 67874.2747 67874.2041 0.0706 .033000 3.0000 : pP21(11) : Excited a'v=12 10 10 F2e - X Xv=0 11 11 F1e :1652:<Linelist>
Excited 11 e 9 X 12 e 1 67855.1179 67855.1256 -0.0077 .051900 3.0000 : pP21(12) : Excited a'v=12 11 11 F2e - X Xv=0 12 12 F1e :1653:<Linelist>
Excited 12 e 9 X 13 e 1 67834.6176 67834.6307 -0.0131 .046200 3.0000 : pP21(13) : Excited a'v=12 12 12 F2e - X Xv=0 13 13 F1e :1654:<Linelist>
Excited 13 e 9 X 14 e 1 67812.7158 67812.7197 -0.0039 .030100 3.0000 : pP21(14) : Excited a'v=12 13 13 F2e - X Xv=0 14 14 F1e :1655:<Linelist>
Excited 14 e 9 X 15 e 1 67789.4157 67789.3924 0.0233 .070300 3.0000 : pP21(15) : Excited a'v=12 14 14 F2e - X Xv=0 15 15 F1e :1656:<Linelist>
Excited 15 e 9 X 16 e 1 67764.5850 67764.6492 -0.0642 .049300 3.0000 : pP21(16) : Excited a'v=12 15 15 F2e - X Xv=0 16 16 F1e :1657:<Linelist>
Excited 16 e 9 X 17 e 1 67738.5042 67738.4900 0.0142 .048100 3.0000 : pP21(17) : Excited a'v=12 16 16 F2e - X Xv=0 17 17 F1e :1658:<Linelist>
Excited 17 e 9 X 18 e 1 67710.9121 67710.9150 -0.0029 .033600 3.0000 : pP21(18) : Excited a'v=12 17 17 F2e - X Xv=0 18 18 F1e :1659:<Linelist>
Excited 19 e 9 X 20 e 1 67651.5326 67651.5180 0.0146 .026500 3.0000 : pP21(20) : Excited a'v=12 19 19 F2e - X Xv=0 20 20 F1e :1661:<Linelist>
Excited 30 e 8 X 31 e 1 67223.6258 67223.6928 -0.0670 .032200 3.0000 : pP21(31) : Excited a'v=12 30 30 F2e - X Xv=0 31 31 F1e :1672:<Linelist>
Excited 31 e 8 X 32 e 1 67176.4155 67176.4213 -0.0058 .055500 3.0000 : pP21(32) : Excited a'v=12 31 31 F2e - X Xv=0 32 32 F1e :1673:<Linelist>
Excited 32 e 7 X 33 e 1 67127.6464 67127.5897 0.0567 .028400 3.0000 : pP21(33) : Excited a'v=12 32 32 F2e - X Xv=0 33 33 F1e :1674:<Linelist>
Excited 33 e 7 X 34 e 1 67078.0818 67078.0889 -0.0071 .040000 3.0000 : pP21(34) : Excited a'v=12 33 33 F2e - X Xv=0 34 34 F1e :1675:<Linelist>
Excited 37 e 5 X 38 e 1 66861.8870 66861.9182 -0.0312 .032400 3.0000 : pP21(38) : Excited a'v=12 37 37 F2e - X Xv=0 38 38 F1e :1679:<Linelist>
Excited 2 f 9 X 2 e 1 67982.9331 67982.9521 -0.0190 .038700 3.0000 : pQ1(2) : Excited a'v=12 2 1 F1f - X Xv=0 2 2 F1e :1693:<Linelist>
Excited 3 f 9 X 3 e 1 67976.6773 67976.7064 -0.0291 .060300 3.0000 : pQ1(3) : Excited a'v=12 3 2 F1f - X Xv=0 3 3 F1e :1694:<Linelist>
Excited 4 f 9 X 4 e 1 67968.9997 67969.0113 -0.0116 .060800 3.0000 : pQ1(4) : Excited a'v=12 4 3 F1f - X Xv=0 4 4 F1e :1695:<Linelist>
Excited 5 f 9 X 5 e 1 67959.8716 67959.8791 -0.0075 .049300 3.0000 : pQ1(5) : Excited a'v=12 5 4 F1f - X Xv=0 5 5 F1e :1696:<Linelist>
Excited 6 f 9 X 6 e 1 67949.2949 67949.3192 -0.0243 .053700 3.0000 : pQ1(6) : Excited a'v=12 6 5 F1f - X Xv=0 6 6 F1e :1697:<Linelist>
Excited 7 f 9 X 7 e 1 67937.3159 67937.3361 -0.0202 .029200 3.0000 : pQ1(7) : Excited a'v=12 7 6 F1f - X Xv=0 7 7 F1e :1698:<Linelist>
Excited 8 f 9 X 8 e 1 67923.8983 67923.9320 -0.0337 .040100 3.0000 : pQ1(8) : Excited a'v=12 8 7 F1f - X Xv=0 8 8 F1e :1699:<Linelist>
Excited 9 f 9 X 9 e 1 67909.1056 67909.1083 -0.0027 .035500 3.0000 : pQ1(9) : Excited a'v=12 9 8 F1f - X Xv=0 9 9 F1e :1700:<Linelist>
Excited 10 f 9 X 10 e 1 67892.8213 67892.8660 -0.0447 .040300 3.0000 : pQ1(10) : Excited a'v=12 10 9 F1f - X Xv=0 10 10 F1e :1701:<Linelist>
Excited 11 f 9 X 11 e 1 67875.1597 67875.2057 -0.0460 .032400 3.0000 : pQ1(11) : Excited a'v=12 11 10 F1f - X Xv=0 11 11 F1e :1702:<Linelist>
Excited 12 f 9 X 12 e 1 67856.1317 67856.1279 0.0038 .043300 3.0000 : pQ1(12) : Excited a'v=12 12 11 F1f - X Xv=0 12 12 F1e :1703:<Linelist>
Excited 13 f 9 X 13 e 1 67835.6079 67835.6329 -0.0250 .033100 3.0000 : pQ1(13) : Excited a'v=12 13 12 F1f - X Xv=0 13 13 F1e :1704:<Linelist>
Excited 14 f 9 X 14 e 1 67813.6744 67813.7210 -0.0466 .034900 3.0000 : pQ1(14) : Excited a'v=12 14 13 F1f - X Xv=0 14 14 F1e :1705:<Linelist>
Excited 16 f 8 X 16 e 1 67765.6319 67765.6333 -0.0014 .035600 3.0000 : pQ1(16) : Excited a'v=12 16 15 F1f - X Xv=0 16 16 F1e :1707:<Linelist>
Excited 17 f 8 X 17 e 1 67739.4645 67739.4850 -0.0205 .032700 3.0000 : pQ1(17) : Excited a'v=12 17 16 F1f - X Xv=0 17 17 F1e :1708:<Linelist>
Excited 18 f 8 X 18 e 1 67711.8956 67711.9078 -0.0122 .034800 3.0000 : pQ1(18) : Excited a'v=12 18 17 F1f - X Xv=0 18 18 F1e :1709:<Linelist>
Excited 19 f 8 X 19 e 1 67682.9447 67682.9144 0.0303 .033800 3.0000 : pQ1(19) : Excited a'v=12 19 18 F1f - X Xv=0 19 19 F1e :1710:<Linelist>
Excited 31 f 5 X 31 e 1 67223.7987 67223.7141 0.0846 .027900 3.0000 : pQ1(31) : Excited a'v=12 31 30 F1f - X Xv=0 31 31 F1e :1722:<Linelist>
Excited 32 f 5 X 32 e 1 67176.9914 67176.9535 0.0379 .028600 3.0000 : pQ1(32) : Excited a'v=12 32 31 F1f - X Xv=0 32 32 F1e :1723:<Linelist>
Excited 33 f 5 X 33 e 1 67128.2350 67128.2400 -0.0050 .071300 3.0000 : pQ1(33) : Excited a'v=12 33 32 F1f - X Xv=0 33 33 F1e :1724:<Linelist>
Excited 34 f 5 X 34 e 1 67078.1739 67078.0983 0.0756 .031800 3.0000 : pQ1(34) : Excited a'v=12 34 33 F1f - X Xv=0 34 34 F1e :1725:<Linelist>
Excited 36 f 4 X 36 e 1 66973.2180 66973.2115 0.0065 .027400 3.0000 : pQ1(36) : Excited a'v=12 36 35 F1f - X Xv=0 36 36 F1e :1727:<Linelist>
Excited 37 f 3 X 37 e 1 66918.5280 66918.6252 -0.0972 .025500 3.0000 : pQ1(37) : Excited a'v=12 37 36 F1f - X Xv=0 37 37 F1e :1728:<Linelist>
Excited 38 f 3 X 38 e 1 66862.7941 66862.8704 -0.0763 .057900 3.0000 : pQ1(38) : Excited a'v=12 38 37 F1f - X Xv=0 38 38 F1e :1729:<Linelist>
Excited 1 f 8 X 1 e 1 67994.9301 67994.9809 -0.0508 .258700 3.0000 : rQ31(1) : Excited a'v=12 1 2 F3f - X Xv=0 1 1 F1e :1742:<Linelist>
Excited 2 f 10 X 2 e 1 67993.9811 67994.0273 -0.0462 .050800 3.0000 : rQ31(2) : Excited a'v=12 2 3 F3f - X Xv=0 2 2 F1e :1743:<Linelist>
Excited 3 f 11 X 3 e 1 67991.7072 67991.7609 -0.0537 .077500 3.0000 : rQ31(3) : Excited a'v=12 3 4 F3f - X Xv=0 3 3 F1e :1744:<Linelist>
Excited 4 f 11 X 4 e 1 67988.1083 67988.1161 -0.0078 .056600 3.0000 : rQ31(4) : Excited a'v=12 4 5 F3f - X Xv=0 4 4 F1e :1745:<Linelist>
Excited 5 f 11 X 5 e 1 67983.0217 67983.0719 -0.0502 .069700 3.0000 : rQ31(5) : Excited a'v=12 5 6 F3f - X Xv=0 5 5 F1e :1746:<Linelist>
Excited 6 f 11 X 6 e 1 67976.5289 67976.6198 -0.0909 .070900 3.0000 : rQ31(6) : Excited a'v=12 6 7 F3f - X Xv=0 6 6 F1e :1747:<Linelist>
Excited 7 f 11 X 7 e 1 67968.7380 67968.7555 -0.0175 .056500 3.0000 : rQ31(7) : Excited a'v=12 7 8 F3f - X Xv=0 7 7 F1e :1748:<Linelist>
Excited 8 f 11 X 8 e 1 67959.4586 67959.4766 -0.0180 .062900 3.0000 : rQ31(8) : Excited a'v=12 8 9 F3f - X Xv=0 8 8 F1e :1749:<Linelist>
Excited 9 f 11 X 9 e 1 67948.7756 67948.7817 -0.0061 .056600 3.0000 : rQ31(9) : Excited a'v=12 9 10 F3f - X Xv=0 9 9 F1e :1750:<Linelist>
Excited 10 f 11 X 10 e 1 67936.6039 67936.6699 -0.0660 .090600 3.0000 : rQ31(10) : Excited a'v=12 10 11 F3f - X Xv=0 10 10 F1e :1751:<Linelist>
Excited 11 f 11 X 11 e 1 67923.1082 67923.1404 -0.0322 .079900 3.0000 : rQ31(11) : Excited a'v=12 11 12 F3f - X Xv=0 11 11 F1e :1752:<Linelist>
Excited 12 f 11 X 12 e 1 67908.1648 67908.1928 -0.0280 .039900 3.0000 : rQ31(12) : Excited a'v=12 12 13 F3f - X Xv=0 12 12 F1e :1753:<Linelist>
Excited 13 f 11 X 13 e 1 67891.7958 67891.8266 -0.0308 .045700 3.0000 : rQ31(13) : Excited a'v=12 13 14 F3f - X Xv=0 13 13 F1e :1754:<Linelist>
Excited 14 f 11 X 14 e 1 67873.9383 67874.0415 -0.1032 .035200 3.0000 : rQ31(14) : Excited a'v=12 14 15 F3f - X Xv=0 14 14 F1e :1755:<Linelist>

Excited 15 f 11 X 15 e 1 67854.8377 67854.8372 0.0005 .049200 3.0000 : rQ31(15) : Excited a'v=12 15 16 F3f - X Xv=0 15 15 F1e :1756:<Linelist>
Excited 16 f 11 X 16 e 1 67834.2353 67834.2134 0.0219 .052200 3.0000 : rQ31(16) : Excited a'v=12 16 17 F3f - X Xv=0 16 16 F1e :1757:<Linelist>
Excited 17 f 11 X 17 e 1 67812.1456 67812.1698 -0.0242 .068000 3.0000 : rQ31(17) : Excited a'v=12 17 18 F3f - X Xv=0 17 17 F1e :1758:<Linelist>
Excited 18 f 11 X 18 e 1 67788.7282 67788.7063 0.0219 .083400 3.0000 : rQ31(18) : Excited a'v=12 18 19 F3f - X Xv=0 18 18 F1e :1759:<Linelist>
Excited 19 f 11 X 19 e 1 67763.8562 67763.8225 0.0337 .025200 3.0000 : rQ31(19) : Excited a'v=12 19 20 F3f - X Xv=0 19 19 F1e :1760:<Linelist>
Excited 20 f 11 X 20 e 1 67737.5371 67737.5184 0.0187 .044100 3.0000 : rQ31(20) : Excited a'v=12 20 21 F3f - X Xv=0 20 20 F1e :1761:<Linelist>
Excited 21 f 11 X 21 e 1 67709.8122 67709.7936 0.0186 .077000 3.0000 : rQ31(21) : Excited a'v=12 21 22 F3f - X Xv=0 21 21 F1e :1762:<Linelist>
Excited 23 f 11 X 23 e 1 67650.0979 67650.0814 0.0165 .055800 3.0000 : rQ31(23) : Excited a'v=12 23 24 F3f - X Xv=0 23 23 F1e :1764:<Linelist>
Excited 37 f 8 X 37 e 1 67072.5414 67072.4983 0.0431 .033300 3.0000 : rQ31(37) : Excited a'v=12 37 38 F3f - X Xv=0 37 37 F1e :1778:<Linelist>
Excited 49 f 6 X 49 e 1 66355.6510 66355.6287 0.0223 .027200 3.0000 : rQ31(49) : Excited a'v=12 49 50 F3f - X Xv=0 49 49 F1e :1790:<Linelist>
Excited 1 e 7 X 0 e 1 67992.5873 67992.6434 -0.0561 .045700 3.0000 : rR21(0) : Excited a'v=12 1 1 F2e - X Xv=0 0 0 F1e :1792:<Linelist>
Excited 2 e 9 X 1 e 1 67993.2788 67993.3010 -0.0222 .035500 3.0000 : rR21(1) : Excited a'v=12 2 2 F2e - X Xv=0 1 1 F1e :1793:<Linelist>
Excited 3 e 10 X 2 e 1 67992.4807 67992.5409 -0.0602 .045800 3.0000 : rR21(2) : Excited a'v=12 3 3 F2e - X Xv=0 2 2 F1e :1794:<Linelist>
Excited 4 e 10 X 3 e 1 67990.3521 67990.3630 -0.0109 .029600 3.0000 : rR21(3) : Excited a'v=12 4 4 F2e - X Xv=0 3 3 F1e :1795:<Linelist>
Excited 5 e 10 X 4 e 1 67986.7584 67986.7676 -0.0092 .035300 3.0000 : rR21(4) : Excited a'v=12 5 5 F2e - X Xv=0 4 4 F1e :1796:<Linelist>
Excited 6 e 10 X 5 e 1 67981.8196 67981.7577 0.0619 .040200 3.0000 : rR21(5) : Excited a'v=12 6 6 F2e - X Xv=0 5 5 F1e :1797:<Linelist>
Excited 7 e 9 X 6 e 1 67975.2097 67975.3097 -0.1000 .099100 3.0000 : rR21(6) : Excited a'v=12 7 7 F2e - X Xv=0 6 6 F1e :1798:<Linelist>
Excited 8 e 9 X 7 e 1 67967.4802 67967.4636 0.0166 .065400 3.0000 : rR21(7) : Excited a'v=12 8 8 F2e - X Xv=0 7 7 F1e :1799:<Linelist>
Excited 9 e 9 X 8 e 1 67958.1909 67958.1946 -0.0037 .039600 3.0000 : rR21(8) : Excited a'v=12 9 9 F2e - X Xv=0 8 8 F1e :1800:<Linelist>
Excited 10 e 9 X 9 e 1 67947.4825 67947.5062 -0.0237 .038800 3.0000 : rR21(9) : Excited a'v=12 10 10 F2e - X Xv=0 9 9 F1e :1801:<Linelist>
Excited 11 e 9 X 10 e 1 67935.3138 67935.3986 -0.0848 .042300 3.0000 : rR21(10) : Excited a'v=12 11 11 F2e - X Xv=0 10 10 F1e :1802:<Linelist>
Excited 12 e 9 X 11 e 1 67921.8689 67921.8719 -0.0030 .028400 3.0000 : rR21(11) : Excited a'v=12 12 12 F2e - X Xv=0 11 11 F1e :1803:<Linelist>
Excited 13 e 9 X 12 e 1 67906.9223 67906.9259 -0.0036 .036800 3.0000 : rR21(12) : Excited a'v=12 13 13 F2e - X Xv=0 12 12 F1e :1804:<Linelist>
Excited 15 e 9 X 14 e 1 67872.7752 67872.7757 -0.0005 .028300 3.0000 : rR21(14) : Excited a'v=12 15 15 F2e - X Xv=0 14 14 F1e :1806:<Linelist>
Excited 16 e 9 X 15 e 1 67853.5736 67853.5711 0.0025 .027800 3.0000 : rR21(15) : Excited a'v=12 16 16 F2e - X Xv=0 15 15 F1e :1807:<Linelist>
Excited 17 e 9 X 16 e 1 67832.9537 67832.9467 0.0070 .027800 3.0000 : rR21(16) : Excited a'v=12 17 17 F2e - X Xv=0 16 16 F1e :1808:<Linelist>
Excited 18 e 9 X 17 e 1 67810.9080 67810.9024 0.0056 .028800 3.0000 : rR21(17) : Excited a'v=12 18 18 F2e - X Xv=0 17 17 F1e :1809:<Linelist>
Excited 19 e 9 X 18 e 1 67787.4608 67787.4379 0.0229 .037700 3.0000 : rR21(18) : Excited a'v=12 19 19 F2e - X Xv=0 18 18 F1e :1810:<Linelist>
Excited 20 e 9 X 19 e 1 67762.5699 67762.5533 0.0166 .040200 3.0000 : rR21(19) : Excited a'v=12 20 20 F2e - X Xv=0 19 19 F1e :1811:<Linelist>
Excited 21 e 9 X 20 e 1 67736.2996 67736.2484 0.0512 .052400 3.0000 : rR21(20) : Excited a'v=12 21 21 F2e - X Xv=0 20 20 F1e :1812:<Linelist>
Excited 22 e 9 X 21 e 1 67708.4881 67708.5232 -0.0351 .031300 3.0000 : rR21(21) : Excited a'v=12 22 22 F2e - X Xv=0 21 21 F1e :1813:<Linelist>
Excited 23 e 9 X 22 e 1 67679.3933 67679.3776 0.0157 .053900 3.0000 : rR21(22) : Excited a'v=12 23 23 F2e - X Xv=0 22 22 F1e :1814:<Linelist>
Excited 40 e 4 X 39 e 1 66965.9428 66966.0274 -0.0846 .025500 3.0000 : rR21(39) : Excited a'v=12 40 40 F2e - X Xv=0 39 39 F1e :1831:<Linelist>
Excited 42 e 3 X 41 e 1 66855.1260 66855.0911 0.0349 .027100 3.0000 : rR21(41) : Excited a'v=12 42 42 F2e - X Xv=0 41 41 F1e :1833:<Linelist>
Excited 10 e 11 X 11 e 1 68267.5062 68267.4868 0.0194 .309900 2.0000 : pP(11) : Excited Dsv=3 10 10 F1e - X Xv=0 11 11 F1e :1850:<Linelist>
Excited 11 e 11 X 12 e 1 68249.5354 68249.5417 -0.0063 .238400 2.0000 : pP(12) : Excited Dsv=3 11 11 F1e - X Xv=0 12 12 F1e :1851:<Linelist>
Excited 12 e 11 X 13 e 1 68230.3075 68230.2825 0.0250 .145000 2.0000 : pP(13) : Excited Dsv=3 12 12 F1e - X Xv=0 13 13 F1e :1852:<Linelist>
Excited 13 e 11 X 14 e 1 68209.7105 68209.7093 0.0012 .333700 2.0000 : pP(14) : Excited Dsv=3 13 13 F1e - X Xv=0 14 14 F1e :1853:<Linelist>
Excited 15 e 11 X 16 e 1 68164.6139 68164.6200 -0.0061 .047900 2.0000 : pP(16) : Excited Dsv=3 15 15 F1e - X Xv=0 16 16 F1e :1855:<Linelist>
Excited 16 e 11 X 17 e 1 68140.0845 68140.1038 -0.0193 .030300 2.0000 : pP(17) : Excited Dsv=3 16 16 F1e - X Xv=0 17 17 F1e :1856:<Linelist>
Excited 17 e 11 X 18 e 1 68114.2695 68114.2728 -0.0033 .036800 2.0000 : pP(18) : Excited Dsv=3 17 17 F1e - X Xv=0 18 18 F1e :1857:<Linelist>
Excited 19 e 11 X 20 e 1 68058.7388 68058.6661 0.0727 .106300 2.0000 : pP(20) : Excited Dsv=3 19 19 F1e - X Xv=0 20 20 F1e :1859:<Linelist>
Excited 20 e 11 X 21 e 1 68028.9701 68028.8900 0.0801 .035400 2.0000 : pP(21) : Excited Dsv=3 20 20 F1e - X Xv=0 21 21 F1e :1860:<Linelist>
Excited 21 e 11 X 22 e 1 67997.7147 67997.7983 -0.0836 .150000 2.0000 : pP(22) : Excited Dsv=3 21 21 F1e - X Xv=0 22 22 F1e :1861:<Linelist>
Excited 45 e 11 X 46 e 1 66856.2466 66856.2469 -0.0003 .044000 2.0000 : pP(46) : Excited Dsv=3 45 45 F1e - X Xv=0 46 46 F1e :1885:<Linelist>
Excited 5 f 12 X 5 e 1 68358.4864 68358.4578 0.0286 .261500 2.0000 : qQ(5) : Excited Dsv=3 5 5 F1f - X Xv=0 5 5 F1e :1893:<Linelist>
Excited 6 f 12 X 6 e 1 68350.6657 68350.5719 0.0938 .436300 2.0000 : qQ(6) : Excited Dsv=3 6 6 F1f - X Xv=0 6 6 F1e :1894:<Linelist>
Excited 7 f 12 X 7 e 1 68341.3933 68341.3714 0.0219 .139600 2.0000 : qQ(7) : Excited Dsv=3 7 7 F1f - X Xv=0 7 7 F1e :1895:<Linelist>
Excited 8 f 12 X 8 e 1 68330.8919 68330.8562 0.0357 .239300 2.0000 : qQ(8) : Excited Dsv=3 8 8 F1f - X Xv=0 8 8 F1e :1896:<Linelist>
Excited 9 f 12 X 9 e 1 68319.0432 68319.0261 0.0171 .170600 2.0000 : qQ(9) : Excited Dsv=3 9 9 F1f - X Xv=0 9 9 F1e :1897:<Linelist>
Excited 10 f 12 X 10 e 1 68305.8946 68305.8809 0.0137 .136700 2.0000 : qQ(10) : Excited Dsv=3 10 10 F1f - X Xv=0 10 10 F1e :1898:<Linelist>
Excited 11 f 12 X 11 e 1 68291.4270 68291.4206 0.0064 .132100 2.0000 : qQ(11) : Excited Dsv=3 11 11 F1f - X Xv=0 11 11 F1e :1899:<Linelist>
Excited 12 f 12 X 12 e 1 68275.6267 68275.6448 -0.0181 .175600 2.0000 : qQ(12) : Excited Dsv=3 12 12 F1f - X Xv=0 12 12 F1e :1900:<Linelist>
Excited 13 f 12 X 13 e 1 68258.5289 68258.5533 -0.0244 .063300 2.0000 : qQ(13) : Excited Dsv=3 13 13 F1f - X Xv=0 13 13 F1e :1901:<Linelist>
Excited 14 f 12 X 14 e 1 68240.1413 68240.1459 -0.0046 .118400 2.0000 : qQ(14) : Excited Dsv=3 14 14 F1f - X Xv=0 14 14 F1e :1902:<Linelist>
Excited 15 f 12 X 15 e 1 68220.4159 68220.4224 -0.0065 .146800 2.0000 : qQ(15) : Excited Dsv=3 15 15 F1f - X Xv=0 15 15 F1e :1903:<Linelist>
Excited 16 f 12 X 16 e 1 68199.3676 68199.3825 -0.0149 .136400 2.0000 : qQ(16) : Excited Dsv=3 16 16 F1f - X Xv=0 16 16 F1e :1904:<Linelist>
Excited 17 f 12 X 17 e 1 68177.0687 68177.0258 0.0429 .088900 2.0000 : qQ(17) : Excited Dsv=3 17 17 F1f - X Xv=0 17 17 F1e :1905:<Linelist>
Excited 18 f 12 X 18 e 1 68153.2796 68153.3521 -0.0725 .082700 2.0000 : qQ(18) : Excited Dsv=3 18 18 F1f - X Xv=0 18 18 F1e :1906:<Linelist>
Excited 19 f 12 X 19 e 1 68128.3268 68128.3610 -0.0342 .166000 2.0000 : qQ(19) : Excited Dsv=3 19 19 F1f - X Xv=0 19 19 F1e :1907:<Linelist>
Excited 20 f 12 X 20 e 1 68102.0195 68102.0522 -0.0327 .126400 2.0000 : qQ(20) : Excited Dsv=3 20 20 F1f - X Xv=0 20 20 F1e :1908:<Linelist>
Excited 21 f 12 X 21 e 1 68074.3660 68074.4253 -0.0593 .245800 2.0000 : qQ(21) : Excited Dsv=3 21 21 F1f - X Xv=0 21 21 F1e :1909:<Linelist>
Excited 22 f 12 X 22 e 1 68045.4641 68045.4800 -0.0159 .219300 2.0000 : qQ(22) : Excited Dsv=3 22 22 F1f - X Xv=0 22 22 F1e :1910:<Linelist>
Excited 23 f 12 X 23 e 1 68015.1547 68015.2158 -0.0611 .298400 2.0000 : qQ(23) : Excited Dsv=3 23 23 F1f - X Xv=0 23 23 F1e :1911:<Linelist>
Excited 50 f 11 X 50 e 1 66700.7564 66700.7503 0.0061 .183400 2.0000 : qQ(50) : Excited Dsv=3 50 50 F1f - X Xv=0 50 50 F1e :1938:<Linelist>
Excited 9 e 11 X 8 e 1 68350.4962 68350.4467 0.0495 .197700 2.0000 : rR(8) : Excited Dsv=3 9 9 F1e - X Xv=0 8 8 F1e :1946:<Linelist>
Excited 10 e 11 X 9 e 1 68340.8150 68340.7889 0.0261 .180000 2.0000 : rR(9) : Excited Dsv=3 10 10 F1e - X Xv=0 9 9 F1e :1947:<Linelist>
Excited 11 e 11 X 10 e 1 68329.8066 68329.8147 -0.0081 .229200 2.0000 : rR(10) : Excited Dsv=3 11 11 F1e - X Xv=0 10 10 F1e :1948:<Linelist>
Excited 12 e 11 X 11 e 1 68317.5398 68317.5237 0.0161 .141500 2.0000 : rR(11) : Excited Dsv=3 12 12 F1e - X Xv=0 11 11 F1e :1949:<Linelist>
Excited 13 e 11 X 12 e 1 68303.9391 68303.9155 0.0236 .049700 2.0000 : rR(12) : Excited Dsv=3 13 13 F1e - X Xv=0 12 12 F1e :1950:<Linelist>
Excited 14 e 11 X 13 e 1 68288.9767 68288.9900 -0.0133 .180900 2.0000 : rR(13) : Excited Dsv=3 14 14 F1e - X Xv=0 13 13 F1e :1951:<Linelist>
Excited 15 e 11 X 14 e 1 68272.7305 68272.7465 -0.0160 .151900 2.0000 : rR(14) : Excited Dsv=3 15 15 F1e - X Xv=0 14 14 F1e :1952:<Linelist>
Excited 16 e 11 X 15 e 1 68255.1279 68255.1848 -0.0569 .143000 2.0000 : rR(15) : Excited Dsv=3 16 16 F1e - X Xv=0 15 15 F1e :1953:<Linelist>
Excited 17 e 11 X 16 e 1 68236.3244 68236.3045 0.0199 .109500 2.0000 : rR(16) : Excited Dsv=3 17 17 F1e - X Xv=0 16 16 F1e :1954:<Linelist>
Excited 18 e 11 X 17 e 1 68216.0785 68216.1050 -0.0265 .051600 2.0000 : rR(17) : Excited Dsv=3 18 18 F1e - X Xv=0 17 17 F1e :1955:<Linelist>
Excited 19 e 11 X 18 e 1 68194.5055 68194.5860 -0.0805 .247000 2.0000 : rR(18) : Excited Dsv=3 19 19 F1e - X Xv=0 18 18 F1e :1956:<Linelist>
Excited 20 e 11 X 19 e 1 68171.6902 68171.7470 -0.0568 .067300 2.0000 : rR(19) : Excited Dsv=3 20 20 F1e - X Xv=0 19 19 F1e :1957:<Linelist>
Excited 21 e 11 X 20 e 1 68147.4920 68147.5875 -0.0955 .207500 2.0000 : rR(20) : Excited Dsv=3 21 21 F1e - X Xv=0 20 20 F1e :1958:<Linelist>
Excited 22 e 11 X 21 e 1 68122.1026 68122.1071 -0.0045 .204000 2.0000 : rR(21) : Excited Dsv=3 22 22 F1e - X Xv=0 21 21 F1e :1959:<Linelist>
Excited 23 e 11 X 22 e 1 68095.2117 68095.3051 -0.0934 .090400 2.0000 : rR(22) : Excited Dsv=3 23 23 F1e - X Xv=0 22 22 F1e :1960:<Linelist>
B 0 e 1 X 1 e 1 86913.8618 86913.8645 -0.0027 .010300 1.0000 : pP(1) : B Bv=0 0 0 F1e - X Xv=0 1 1 F1e :1988:<Linelist>
B 1 e 1 X 2 e 1 86910.4175 86910.4208 -0.0033 .010100 1.0000 : pP(2) : B Bv=0 1 1 F1e - X Xv=0 2 2 F1e :1989:<Linelist>
B 2 e 1 X 3 e 1 86907.0197 86907.0241 -0.0044 .010000 1.0000 : pP(3) : B Bv=0 2 2 F1e - X Xv=0 3 3 F1e :1990:<Linelist>
B 3 e 1 X 4 e 1 86903.6685 86903.6715 -0.0030 .010000 1.0000 : pP(4) : B Bv=0 3 3 F1e - X Xv=0 4 4 F1e :1991:<Linelist>
B 4 e 1 X 5 e 1 86900.3628 86900.3658 -0.0030 .010000 1.0000 : pP(5) : B Bv=0 4 4 F1e - X Xv=0 5 5 F1e :1992:<Linelist>
B 5 e 1 X 6 e 1 86897.1071 86897.1084 -0.0013 .010000 1.0000 : pP(6) : B Bv=0 5 5 F1e - X Xv=0 6 6 F1e :1993:<Linelist>
B 6 e 1 X 7 e 1 86893.8966 86893.8978 -0.0012 .010000 1.0000 : pP(7) : B Bv=0 6 6 F1e - X Xv=0 7 7 F1e :1994:<Linelist>
B 7 e 1 X 8 e 1 86890.7332 86890.7349 -0.0017 .010000 1.0000 : pP(8) : B Bv=0 7 7 F1e - X Xv=0 8 8 F1e :1995:<Linelist>
B 8 e 1 X 9 e 1 86887.6174 86887.6192 -0.0018 .010000 1.0000 : pP(9) : B Bv=0 8 8 F1e - X Xv=0 9 9 F1e :1996:<Linelist>
B 9 e 1 X 10 e 1 86884.5487 86884.5507 -0.0020 .010000 1.0000 : pP(10) : B Bv=0 9 9 F1e - X Xv=0 10 10 F1e :1997:<Linelist>
B 10 e 1 X 11 e 1 86881.5279 86881.5294 -0.0015 .010000 1.0000 : pP(11) : B Bv=0 10 10 F1e - X Xv=0 11 11 F1e :1998:<Linelist>
B 11 e 1 X 12 e 1 86878.5527 86878.5554 -0.0027 .010000 1.0000 : pP(12) : B Bv=0 11 11 F1e - X Xv=0 12 12 F1e :1999:<Linelist>

B 12 e 1 X 13 e 1 86875.6290 86875.6313 -0.0023 .010000 1.0000 : pP(13) : B Bv=0 12 12 F1e - X Xv=0 13 13 F1e :2000:<Linelist>
B 13 e 1 X 14 e 1 86872.7481 86872.7507 -0.0026 .010000 1.0000 : pP(14) : B Bv=0 13 13 F1e - X Xv=0 14 14 F1e :2001:<Linelist>
B 14 e 1 X 15 e 1 86869.9176 86869.9187 -0.0011 .010000 1.0000 : pP(15) : B Bv=0 14 14 F1e - X Xv=0 15 15 F1e :2002:<Linelist>
B 15 e 1 X 16 e 1 86867.1331 86867.1343 -0.0012 .010000 1.0000 : pP(16) : B Bv=0 15 15 F1e - X Xv=0 16 16 F1e :2003:<Linelist>
B 16 e 1 X 17 e 1 86864.3941 86864.3957 -0.0016 .010000 1.0000 : pP(17) : B Bv=0 16 16 F1e - X Xv=0 17 17 F1e :2004:<Linelist>
B 17 e 1 X 18 e 1 86861.6624 86861.6645 -0.0021 .010000 1.0000 : pP(18) : B Bv=0 17 17 F1e - X Xv=0 18 18 F1e :2005:<Linelist>
B 18 e 1 X 19 e 1 86859.0571 86859.0606 -0.0035 .010000 1.0000 : pP(19) : B Bv=0 18 18 F1e - X Xv=0 19 19 F1e :2006:<Linelist>
B 19 e 1 X 20 e 1 86856.4697 86856.4718 -0.0021 .010100 1.0000 : pP(20) : B Bv=0 19 19 F1e - X Xv=0 20 20 F1e :2007:<Linelist>
B 20 e 1 X 21 e 1 86853.9190 86853.9215 -0.0025 .010100 1.0000 : pP(21) : B Bv=0 20 20 F1e - X Xv=0 21 21 F1e :2008:<Linelist>
B 21 e 1 X 22 e 1 86851.4162 86851.4189 -0.0027 .010100 1.0000 : pP(22) : B Bv=0 21 21 F1e - X Xv=0 22 22 F1e :2009:<Linelist>
B 22 e 1 X 23 e 1 86848.9608 86848.9638 -0.0030 .010100 1.0000 : pP(23) : B Bv=0 22 22 F1e - X Xv=0 23 23 F1e :2010:<Linelist>
B 23 e 1 X 24 e 1 86846.5508 86846.5551 -0.0043 .010100 1.0000 : pP(24) : B Bv=0 23 23 F1e - X Xv=0 24 24 F1e :2011:<Linelist>
B 24 e 1 X 25 e 1 86844.1902 86844.1920 -0.0018 .010100 1.0000 : pP(25) : B Bv=0 24 24 F1e - X Xv=0 25 25 F1e :2012:<Linelist>
B 25 e 1 X 26 e 1 86841.8719 86841.8736 -0.0017 .010100 1.0000 : pP(26) : B Bv=0 25 25 F1e - X Xv=0 26 26 F1e :2013:<Linelist>
B 26 e 1 X 27 e 1 86839.6059 86839.6066 -0.0007 .010100 1.0000 : pP(27) : B Bv=0 26 26 F1e - X Xv=0 27 27 F1e :2014:<Linelist>
B 27 e 1 X 28 e 1 86837.3860 86837.3860 -0.0000 .010200 1.0000 : pP(28) : B Bv=0 27 27 F1e - X Xv=0 28 28 F1e :2015:<Linelist>
B 28 e 1 X 29 e 1 86835.2037 86835.2057 -0.0020 .010200 1.0000 : pP(29) : B Bv=0 28 28 F1e - X Xv=0 29 29 F1e :2016:<Linelist>
B 29 e 1 X 30 e 1 86833.0668 86833.0751 -0.0083 .010200 1.0000 : pP(30) : B Bv=0 29 29 F1e - X Xv=0 30 30 F1e :2017:<Linelist>
B 30 e 1 X 31 e 1 86830.9825 86830.9846 -0.0021 .010200 1.0000 : pP(31) : B Bv=0 30 30 F1e - X Xv=0 31 31 F1e :2018:<Linelist>
B 31 e 1 X 32 e 1 86828.9279 86828.9350 -0.0071 .010200 1.0000 : pP(32) : B Bv=0 31 31 F1e - X Xv=0 32 32 F1e :2019:<Linelist>
B 32 e 1 X 33 e 1 86826.9288 86826.9328 -0.0040 .010300 1.0000 : pP(33) : B Bv=0 32 32 F1e - X Xv=0 33 33 F1e :2020:<Linelist>
B 33 e 1 X 34 e 1 86824.9841 86824.9899 -0.0058 .010400 1.0000 : pP(34) : B Bv=0 33 33 F1e - X Xv=0 34 34 F1e :2021:<Linelist>
B 34 e 1 X 35 e 1 86823.0779 86823.0804 -0.0025 .010400 1.0000 : pP(35) : B Bv=0 34 34 F1e - X Xv=0 35 35 F1e :2022:<Linelist>
B 35 e 1 X 36 e 1 86821.2268 86821.2326 -0.0058 .010500 1.0000 : pP(36) : B Bv=0 35 35 F1e - X Xv=0 36 36 F1e :2023:<Linelist>
B 36 e 1 X 37 e 1 86819.3852 86819.3859 -0.0007 .010700 1.0000 : pP(37) : B Bv=0 36 36 F1e - X Xv=0 37 37 F1e :2024:<Linelist>
B 37 e 1 X 38 e 1 86817.6024 86817.6104 -0.0080 .010800 1.0000 : pP(38) : B Bv=0 37 37 F1e - X Xv=0 38 38 F1e :2025:<Linelist>
B 38 e 1 X 39 e 1 86815.8717 86815.8770 -0.0053 .011100 1.0000 : pP(39) : B Bv=0 38 38 F1e - X Xv=0 39 39 F1e :2026:<Linelist>
B 39 e 1 X 40 e 1 86814.1784 86814.1935 -0.0151 .011400 1.0000 : pP(40) : B Bv=0 39 39 F1e - X Xv=0 40 40 F1e :2027:<Linelist>
B 40 e 1 X 41 e 1 86812.5204 86812.5075 0.0129 .011900 1.0000 : pP(41) : B Bv=0 40 40 F1e - X Xv=0 41 41 F1e :2028:<Linelist>
B 41 e 1 X 42 e 1 86810.9109 86810.9115 -0.0006 .012400 1.0000 : pP(42) : B Bv=0 41 41 F1e - X Xv=0 42 42 F1e :2029:<Linelist>
B 42 e 1 X 43 e 1 86809.3213 86809.3214 -0.0001 .013400 1.0000 : pP(43) : B Bv=0 42 42 F1e - X Xv=0 43 43 F1e :2030:<Linelist>
B 43 e 1 X 44 e 1 86807.7995 86807.8006 -0.0011 .014400 1.0000 : pP(44) : B Bv=0 43 43 F1e - X Xv=0 44 44 F1e :2031:<Linelist>
B 44 e 1 X 45 e 1 86806.3191 86806.3251 -0.0060 .016000 1.0000 : pP(45) : B Bv=0 44 44 F1e - X Xv=0 45 45 F1e :2032:<Linelist>
B 45 e 1 X 46 e 1 86804.8694 86804.8699 -0.0005 .017900 1.0000 : pP(46) : B Bv=0 45 45 F1e - X Xv=0 46 46 F1e :2033:<Linelist>
B 46 e 1 X 47 e 1 86803.4940 86803.4970 -0.0030 .020900 1.0000 : pP(47) : B Bv=0 46 46 F1e - X Xv=0 47 47 F1e :2034:<Linelist>
B 47 e 1 X 48 e 1 86802.0789 86802.0788 0.0001 .024100 1.0000 : pP(48) : B Bv=0 47 47 F1e - X Xv=0 48 48 F1e :2035:<Linelist>
B 48 e 1 X 49 e 1 86800.7304 86800.7198 0.0106 .029100 1.0000 : pP(49) : B Bv=0 48 48 F1e - X Xv=0 49 49 F1e :2036:<Linelist>
B 49 e 1 X 50 e 1 86799.4713 86799.4712 0.0001 .034600 1.0000 : pP(50) : B Bv=0 49 49 F1e - X Xv=0 50 50 F1e :2037:<Linelist>
B 1 e 1 X 0 e 1 86920.8958 86920.8991 -0.0033 .010100 1.0000 : rR(0) : B Bv=0 1 1 F1e - X Xv=0 0 0 F1e :2048:<Linelist>
B 2 e 1 X 1 e 1 86924.4831 86924.4874 -0.0043 .010000 1.0000 : rR(1) : B Bv=0 2 2 F1e - X Xv=0 1 1 F1e :2049:<Linelist>
B 3 e 1 X 2 e 1 86928.1164 86928.1193 -0.0029 .010000 1.0000 : rR(2) : B Bv=0 3 3 F1e - X Xv=0 2 2 F1e :2050:<Linelist>
B 4 e 1 X 3 e 1 86931.7944 86931.7973 -0.0029 .010000 1.0000 : rR(3) : B Bv=0 4 4 F1e - X Xv=0 3 3 F1e :2051:<Linelist>
B 5 e 1 X 4 e 1 86935.5212 86935.5224 -0.0012 .010000 1.0000 : rR(4) : B Bv=0 5 5 F1e - X Xv=0 4 4 F1e :2052:<Linelist>
B 6 e 1 X 5 e 1 86939.2919 86939.2932 -0.0013 .010000 1.0000 : rR(5) : B Bv=0 6 6 F1e - X Xv=0 5 5 F1e :2053:<Linelist>
B 7 e 1 X 6 e 1 86943.1082 86943.1099 -0.0017 .010000 1.0000 : rR(6) : B Bv=0 7 7 F1e - X Xv=0 6 6 F1e :2054:<Linelist>
B 8 e 1 X 7 e 1 86946.9702 86946.9720 -0.0018 .010000 1.0000 : rR(7) : B Bv=0 8 8 F1e - X Xv=0 7 7 F1e :2055:<Linelist>
B 9 e 1 X 8 e 1 86950.8773 86950.8793 -0.0020 .010000 1.0000 : rR(8) : B Bv=0 9 9 F1e - X Xv=0 8 8 F1e :2056:<Linelist>
B 10 e 1 X 9 e 1 86954.8299 86954.8315 -0.0016 .010000 1.0000 : rR(9) : B Bv=0 10 10 F1e - X Xv=0 9 9 F1e :2057:<Linelist>
B 11 e 1 X 10 e 1 86958.8257 86958.8284 -0.0027 .010000 1.0000 : rR(10) : B Bv=0 11 11 F1e - X Xv=0 10 10 F1e :2058:<Linelist>
B 12 e 1 X 11 e 1 86962.8701 86962.8724 -0.0023 .010000 1.0000 : rR(11) : B Bv=0 12 12 F1e - X Xv=0 11 11 F1e :2059:<Linelist>
B 13 e 1 X 12 e 1 86966.9544 86966.9570 -0.0026 .010000 1.0000 : rR(12) : B Bv=0 13 13 F1e - X Xv=0 12 12 F1e :2060:<Linelist>
B 14 e 1 X 13 e 1 86971.0857 86971.0869 -0.0012 .010000 1.0000 : rR(13) : B Bv=0 14 14 F1e - X Xv=0 13 13 F1e :2061:<Linelist>
B 15 e 1 X 14 e 1 86975.2595 86975.2607 -0.0012 .010000 1.0000 : rR(14) : B Bv=0 15 15 F1e - X Xv=0 14 14 F1e :2062:<Linelist>
B 16 e 1 X 15 e 1 86979.4751 86979.4768 -0.0017 .010000 1.0000 : rR(15) : B Bv=0 16 16 F1e - X Xv=0 15 15 F1e :2063:<Linelist>
B 17 e 1 X 16 e 1 86983.6940 86983.6962 -0.0022 .010000 1.0000 : rR(16) : B Bv=0 17 17 F1e - X Xv=0 16 16 F1e :2064:<Linelist>
B 18 e 1 X 17 e 1 86988.0351 86988.0386 -0.0035 .010000 1.0000 : rR(17) : B Bv=0 18 18 F1e - X Xv=0 17 17 F1e :2065:<Linelist>
B 19 e 1 X 18 e 1 86992.3895 86992.3917 -0.0022 .010100 1.0000 : rR(18) : B Bv=0 19 19 F1e - X Xv=0 18 18 F1e :2066:<Linelist>
B 20 e 1 X 19 e 1 86996.7760 86996.7785 -0.0025 .010100 1.0000 : rR(19) : B Bv=0 20 20 F1e - X Xv=0 19 19 F1e :2067:<Linelist>
B 21 e 1 X 20 e 1 87001.2054 87001.2082 -0.0028 .010100 1.0000 : rR(20) : B Bv=0 21 21 F1e - X Xv=0 20 20 F1e :2068:<Linelist>
B 22 e 1 X 21 e 1 87005.6770 87005.6801 -0.0031 .010100 1.0000 : rR(21) : B Bv=0 22 22 F1e - X Xv=0 21 21 F1e :2069:<Linelist>
B 23 e 1 X 22 e 1 87010.1886 87010.1929 -0.0043 .010100 1.0000 : rR(22) : B Bv=0 23 23 F1e - X Xv=0 22 22 F1e :2070:<Linelist>
B 24 e 1 X 23 e 1 87014.7439 87014.7458 -0.0019 .010100 1.0000 : rR(23) : B Bv=0 24 24 F1e - X Xv=0 23 23 F1e :2071:<Linelist>
B 25 e 1 X 24 e 1 87019.3356 87019.3373 -0.0017 .010100 1.0000 : rR(24) : B Bv=0 25 25 F1e - X Xv=0 24 24 F1e :2072:<Linelist>
B 26 e 1 X 25 e 1 87023.9733 87023.9741 -0.0008 .010100 1.0000 : rR(25) : B Bv=0 26 26 F1e - X Xv=0 25 25 F1e :2073:<Linelist>
B 27 e 1 X 26 e 1 87028.6508 87028.6509 -0.0001 .010200 1.0000 : rR(26) : B Bv=0 27 27 F1e - X Xv=0 26 26 F1e :2074:<Linelist>
B 28 e 1 X 27 e 1 87033.3592 87033.3613 -0.0021 .010200 1.0000 : rR(27) : B Bv=0 28 28 F1e - X Xv=0 27 27 F1e :2075:<Linelist>
B 29 e 1 X 28 e 1 87038.1063 87038.1146 -0.0083 .010200 1.0000 : rR(28) : B Bv=0 29 29 F1e - X Xv=0 28 28 F1e :2076:<Linelist>
B 30 e 1 X 29 e 1 87042.8987 87042.9008 -0.0021 .010200 1.0000 : rR(29) : B Bv=0 30 30 F1e - X Xv=0 29 29 F1e :2077:<Linelist>
B 31 e 1 X 30 e 1 87047.7135 87047.7207 -0.0072 .010200 1.0000 : rR(30) : B Bv=0 31 31 F1e - X Xv=0 30 30 F1e :2078:<Linelist>
B 32 e 1 X 31 e 1 87052.5762 87052.5802 -0.0040 .010300 1.0000 : rR(31) : B Bv=0 32 32 F1e - X Xv=0 31 31 F1e :2079:<Linelist>
B 33 e 1 X 32 e 1 87057.4854 87057.4912 -0.0058 .010400 1.0000 : rR(32) : B Bv=0 33 33 F1e - X Xv=0 32 32 F1e :2080:<Linelist>
B 34 e 1 X 33 e 1 87062.4251 87062.4277 -0.0026 .010400 1.0000 : rR(33) : B Bv=0 34 34 F1e - X Xv=0 33 33 F1e :2081:<Linelist>
B 35 e 1 X 34 e 1 87067.4115 87067.4174 -0.0059 .010500 1.0000 : rR(34) : B Bv=0 35 35 F1e - X Xv=0 34 34 F1e :2082:<Linelist>
B 36 e 1 X 35 e 1 87072.3989 87072.3997 -0.0008 .010700 1.0000 : rR(35) : B Bv=0 36 36 F1e - X Xv=0 35 35 F1e :2083:<Linelist>
B 37 e 1 X 36 e 1 87077.4362 87077.4444 -0.0082 .010800 1.0000 : rR(36) : B Bv=0 37 37 F1e - X Xv=0 36 36 F1e :2084:<Linelist>
B 38 e 1 X 37 e 1 87082.5167 87082.5221 -0.0054 .011100 1.0000 : rR(37) : B Bv=0 38 38 F1e - X Xv=0 37 37 F1e :2085:<Linelist>
B 39 e 1 X 38 e 1 87087.6253 87087.6405 -0.0152 .011400 1.0000 : rR(38) : B Bv=0 39 39 F1e - X Xv=0 38 38 F1e :2086:<Linelist>
B 40 e 1 X 39 e 1 87092.7596 87092.7469 0.0127 .011900 1.0000 : rR(39) : B Bv=0 40 40 F1e - X Xv=0 39 39 F1e :2087:<Linelist>
B 41 e 1 X 40 e 1 87097.9328 87097.9335 -0.0007 .012400 1.0000 : rR(40) : B Bv=0 41 41 F1e - X Xv=0 40 40 F1e :2088:<Linelist>
B 42 e 1 X 41 e 1 87103.1158 87103.1160 -0.0002 .013400 1.0000 : rR(41) : B Bv=0 42 42 F1e - X Xv=0 41 41 F1e :2089:<Linelist>
B 43 e 1 X 42 e 1 87108.3564 87108.3577 -0.0013 .014400 1.0000 : rR(42) : B Bv=0 43 43 F1e - X Xv=0 42 42 F1e :2090:<Linelist>
B 44 e 1 X 43 e 1 87113.6280 87113.6342 -0.0062 .016000 1.0000 : rR(43) : B Bv=0 44 44 F1e - X Xv=0 43 43 F1e :2091:<Linelist>
B 45 e 1 X 44 e 1 87118.9196 87118.9202 -0.0006 .017900 1.0000 : rR(44) : B Bv=0 45 45 F1e - X Xv=0 44 44 F1e :2092:<Linelist>
B 46 e 1 X 45 e 1 87124.2746 87124.2777 -0.0031 .020900 1.0000 : rR(45) : B Bv=0 46 46 F1e - X Xv=0 45 45 F1e :2093:<Linelist>
B 47 e 1 X 46 e 1 87129.5786 87129.5787 -0.0001 .024100 1.0000 : rR(46) : B Bv=0 47 47 F1e - X Xv=0 46 46 F1e :2094:<Linelist>
B 48 e 1 X 47 e 1 87134.9379 87134.9274 0.0105 .029100 1.0000 : rR(47) : B Bv=0 48 48 F1e - X Xv=0 47 47 F1e :2095:<Linelist>
B 49 e 1 X 48 e 1 87140.3749 87140.3750 -0.0001 .034600 1.0000 : rR(48) : B Bv=0 49 49 F1e - X Xv=0 48 48 F1e :2096:<Linelist>
B 50 e 1 X 49 e 1 87145.7391 87145.7392 -0.0001 .042800 1.0000 : rR(49) : B Bv=0 50 50 F1e - X Xv=0 49 49 F1e :2097:<Linelist>
C 0 e 1 X 1 e 1 91915.3222 91915.3218 0.0004 .010284 0.0000 : pP(1) : C Cv=0 0 0 F1e - X Xv=0 1 1 F1e :2108:<Linelist>
C 1 e 1 X 2 e 1 91911.8699 91911.8719 -0.0020 .010127 0.0000 : pP(2) : C Cv=0 1 1 F1e - X Xv=0 2 2 F1e :2109:<Linelist>
C 2 e 1 X 3 e 1 91908.4572 91908.4584 -0.0012 .010098 0.0000 : pP(3) : C Cv=0 2 2 F1e - X Xv=0 3 3 F1e :2110:<Linelist>
C 3 e 1 X 4 e 1 91905.0838 91905.0826 0.0012 .010084 0.0000 : pP(4) : C Cv=0 3 3 F1e - X Xv=0 4 4 F1e :2111:<Linelist>

C 4 e 1 X 5 e 1 91901.7513 91901.7515 -0.0002 .010084 0.0000 : pP(5) : C Cv=0 4 4 F1e - X Xv=0 5 5 F1e :2112:<Linelist>
C 5 e 1 X 6 e 1 91898.4618 91898.4645 -0.0027 .010084 0.0000 : pP(6) : C Cv=0 5 5 F1e - X Xv=0 6 6 F1e :2113:<Linelist>
C 6 e 1 X 7 e 1 91895.2120 91895.2133 -0.0013 .010072 0.0000 : pP(7) : C Cv=0 6 6 F1e - X Xv=0 7 7 F1e :2114:<Linelist>
C 7 e 1 X 8 e 1 91892.0040 91892.0058 -0.0018 .010072 0.0000 : pP(8) : C Cv=0 7 7 F1e - X Xv=0 8 8 F1e :2115:<Linelist>
C 8 e 1 X 9 e 1 91888.8331 91888.8371 -0.0040 .010072 0.0000 : pP(9) : C Cv=0 8 8 F1e - X Xv=0 9 9 F1e :2116:<Linelist>
C 9 e 1 X 10 e 1 91885.7061 91885.7111 -0.0050 .010084 0.0000 : pP(10) : C Cv=0 9 9 F1e - X Xv=0 10 10 F1e :2117:<Linelist>
C 10 e 1 X 11 e 1 91882.6177 91882.6208 -0.0031 .010084 0.0000 : pP(11) : C Cv=0 10 10 F1e - X Xv=0 11 11 F1e :2118:<Linelist>
C 11 e 1 X 12 e 1 91879.5741 91879.5799 -0.0058 .010084 0.0000 : pP(12) : C Cv=0 11 11 F1e - X Xv=0 12 12 F1e :2119:<Linelist>
C 12 e 1 X 13 e 1 91876.5745 91876.5786 -0.0041 .010084 0.0000 : pP(13) : C Cv=0 12 12 F1e - X Xv=0 13 13 F1e :2120:<Linelist>
C 13 e 1 X 14 e 1 91873.6078 91873.6097 -0.0019 .010098 0.0000 : pP(14) : C Cv=0 13 13 F1e - X Xv=0 14 14 F1e :2121:<Linelist>
C 14 e 1 X 15 e 1 91870.6874 91870.6902 -0.0028 .010098 0.0000 : pP(15) : C Cv=0 14 14 F1e - X Xv=0 15 15 F1e :2122:<Linelist>
C 15 e 1 X 16 e 1 91867.8065 91867.8120 -0.0055 .010112 0.0000 : pP(16) : C Cv=0 15 15 F1e - X Xv=0 16 16 F1e :2123:<Linelist>
C 16 e 1 X 17 e 1 91864.9699 91864.9742 -0.0043 .010127 0.0000 : pP(17) : C Cv=0 16 16 F1e - X Xv=0 17 17 F1e :2124:<Linelist>
C 17 e 1 X 18 e 1 91862.1742 91862.1775 -0.0033 .010143 0.0000 : pP(18) : C Cv=0 17 17 F1e - X Xv=0 18 18 F1e :2125:<Linelist>
C 18 e 1 X 19 e 1 91859.4184 91859.4228 -0.0044 .010161 0.0000 : pP(19) : C Cv=0 18 18 F1e - X Xv=0 19 19 F1e :2126:<Linelist>
C 19 e 1 X 20 e 1 91856.7060 91856.7115 -0.0055 .010198 0.0000 : pP(20) : C Cv=0 19 19 F1e - X Xv=0 20 20 F1e :2127:<Linelist>
C 20 e 1 X 21 e 1 91854.0369 91854.0405 -0.0036 .010239 0.0000 : pP(21) : C Cv=0 20 20 F1e - X Xv=0 21 21 F1e :2128:<Linelist>
C 21 e 1 X 22 e 1 91851.4092 91851.4102 -0.0010 .010308 0.0000 : pP(22) : C Cv=0 21 21 F1e - X Xv=0 22 22 F1e :2129:<Linelist>
C 22 e 1 X 23 e 1 91848.8223 91848.8235 -0.0012 .010385 0.0000 : pP(23) : C Cv=0 22 22 F1e - X Xv=0 23 23 F1e :2130:<Linelist>
C 23 e 1 X 24 e 1 91846.2799 91846.2797 0.0002 .010530 0.0000 : pP(24) : C Cv=0 23 23 F1e - X Xv=0 24 24 F1e :2131:<Linelist>
C 24 e 1 X 25 e 1 91843.7876 91843.7889 -0.0013 .010663 0.0000 : pP(25) : C Cv=0 24 24 F1e - X Xv=0 25 25 F1e :2132:<Linelist>
C 25 e 1 X 26 e 1 91841.3318 91841.3312 0.0006 .010846 0.0000 : pP(26) : C Cv=0 25 25 F1e - X Xv=0 26 26 F1e :2133:<Linelist>
C 26 e 1 X 27 e 1 91838.9172 91838.9167 0.0005 .010966 0.0000 : pP(27) : C Cv=0 26 26 F1e - X Xv=0 27 27 F1e :2134:<Linelist>
C 27 e 1 X 28 e 1 91836.5415 91836.5420 -0.0005 .011092 0.0000 : pP(28) : C Cv=0 27 27 F1e - X Xv=0 28 28 F1e :2135:<Linelist>
C 28 e 1 X 29 e 1 91834.2099 91834.2102 -0.0003 .011136 0.0000 : pP(29) : C Cv=0 28 28 F1e - X Xv=0 29 29 F1e :2136:<Linelist>
C 29 e 1 X 30 e 1 91831.9225 91831.9226 -0.0001 .011136 0.0000 : pP(30) : C Cv=0 29 29 F1e - X Xv=0 30 30 F1e :2137:<Linelist>
C 30 e 1 X 31 e 1 91829.6755 91829.6755 0.0000 .011180 0.0000 : pP(31) : C Cv=0 30 30 F1e - X Xv=0 31 31 F1e :2138:<Linelist>
C 31 e 1 X 32 e 1 91827.4729 91827.4729 0.0000 .011180 0.0000 : pP(32) : C Cv=0 31 31 F1e - X Xv=0 32 32 F1e :2139:<Linelist>
C 32 e 1 X 33 e 1 91825.3141 91825.3141 0.0000 .011180 0.0000 : pP(33) : C Cv=0 32 32 F1e - X Xv=0 33 33 F1e :2140:<Linelist>
C 33 e 1 X 34 e 1 91823.1981 91823.1981 -0.0000 .011180 0.0000 : pP(34) : C Cv=0 33 33 F1e - X Xv=0 34 34 F1e :2141:<Linelist>
C 34 e 1 X 35 e 1 91821.1243 91821.1243 0.0000 .011180 0.0000 : pP(35) : C Cv=0 34 34 F1e - X Xv=0 35 35 F1e :2142:<Linelist>
C 35 e 1 X 36 e 1 91819.0887 91819.0887 -0.0000 .011180 0.0000 : pP(36) : C Cv=0 35 35 F1e - X Xv=0 36 36 F1e :2143:<Linelist>
C 36 e 1 X 37 e 1 91817.0997 91817.0997 -0.0000 .011180 0.0000 : pP(37) : C Cv=0 36 36 F1e - X Xv=0 37 37 F1e :2144:<Linelist>
C 37 e 1 X 38 e 1 91815.1515 91815.1515 -0.0000 .011180 0.0000 : pP(38) : C Cv=0 37 37 F1e - X Xv=0 38 38 F1e :2145:<Linelist>
C 38 e 1 X 39 e 1 91813.2494 91813.2494 0.0000 .011180 0.0000 : pP(39) : C Cv=0 38 38 F1e - X Xv=0 39 39 F1e :2146:<Linelist>
C 39 e 1 X 40 e 1 91811.3875 91811.3875 0.0000 .011180 0.0000 : pP(40) : C Cv=0 39 39 F1e - X Xv=0 40 40 F1e :2147:<Linelist>
C 40 e 1 X 41 e 1 91809.5721 91809.5721 -0.0000 .010308 0.0000 : pP(41) : C Cv=0 40 40 F1e - X Xv=0 41 41 F1e :2148:<Linelist>
C 41 e 1 X 42 e 1 91807.8005 91807.8004 0.0001 .010332 0.0000 : pP(42) : C Cv=0 41 41 F1e - X Xv=0 42 42 F1e :2149:<Linelist>
C 42 e 1 X 43 e 1 91806.0719 91806.0718 0.0001 .010358 0.0000 : pP(43) : C Cv=0 42 42 F1e - X Xv=0 43 43 F1e :2150:<Linelist>
C 43 e 1 X 44 e 1 91804.3794 91804.3794 0.0000 .010412 0.0000 : pP(44) : C Cv=0 43 43 F1e - X Xv=0 44 44 F1e :2151:<Linelist>
C 44 e 1 X 45 e 1 91802.7365 91802.7364 0.0001 .010500 0.0000 : pP(45) : C Cv=0 44 44 F1e - X Xv=0 45 45 F1e :2152:<Linelist>
C 45 e 1 X 46 e 1 91801.1362 91801.1361 0.0001 .010628 0.0000 : pP(46) : C Cv=0 45 45 F1e - X Xv=0 46 46 F1e :2153:<Linelist>
C 46 e 1 X 47 e 1 91799.5889 91799.5888 0.0001 .010808 0.0000 : pP(47) : C Cv=0 46 46 F1e - X Xv=0 47 47 F1e :2154:<Linelist>
C 47 e 1 X 48 e 1 91798.0647 91798.0646 0.0001 .011049 0.0000 : pP(48) : C Cv=0 47 47 F1e - X Xv=0 48 48 F1e :2155:<Linelist>
C 48 e 1 X 49 e 1 91796.5959 91796.5959 0.0000 .011413 0.0000 : pP(49) : C Cv=0 48 48 F1e - X Xv=0 49 49 F1e :2156:<Linelist>
C 49 e 1 X 50 e 1 91795.1658 91795.1657 0.0001 .011927 0.0000 : pP(50) : C Cv=0 49 49 F1e - X Xv=0 50 50 F1e :2157:<Linelist>
C 50 e 1 X 51 e 1 91793.7865 91793.7865 0.0000 .012744 0.0000 : pP(51) : C Cv=0 50 50 F1e - X Xv=0 51 51 F1e :2158:<Linelist>
C 51 e 1 X 52 e 1 91792.4344 91792.4343 0.0001 .013862 0.0000 : pP(52) : C Cv=0 51 51 F1e - X Xv=0 52 52 F1e :2159:<Linelist>
C 52 e 1 X 53 e 1 91791.1226 91791.1225 0.0001 .015544 0.0000 : pP(53) : C Cv=0 52 52 F1e - X Xv=0 53 53 F1e :2160:<Linelist>
C 53 e 1 X 54 e 1 91789.8744 91789.8743 0.0001 .017945 0.0000 : pP(54) : C Cv=0 53 53 F1e - X Xv=0 54 54 F1e :2161:<Linelist>
C 54 e 1 X 55 e 1 91788.6700 91788.6699 0.0001 .021294 0.0000 : pP(55) : C Cv=0 54 54 F1e - X Xv=0 55 55 F1e :2162:<Linelist>
C 55 e 1 X 56 e 1 91787.4886 91787.4885 0.0001 .025908 0.0000 : pP(56) : C Cv=0 55 55 F1e - X Xv=0 56 56 F1e :2163:<Linelist>
C 56 e 1 X 57 e 1 91786.3325 91786.3324 0.0001 .032288 0.0000 : pP(57) : C Cv=0 56 56 F1e - X Xv=0 57 57 F1e :2164:<Linelist>
C 57 e 1 X 58 e 1 91785.2689 91785.2689 0.0000 .041037 0.0000 : pP(58) : C Cv=0 57 57 F1e - X Xv=0 58 58 F1e :2165:<Linelist>
C 58 e 1 X 59 e 1 91784.2351 91784.2350 0.0001 .052756 0.0000 : pP(59) : C Cv=0 58 58 F1e - X Xv=0 59 59 F1e :2166:<Linelist>
C 59 e 1 X 60 e 1 91783.1933 91783.1932 0.0001 .068731 0.0000 : pP(60) : C Cv=0 59 59 F1e - X Xv=0 60 60 F1e :2167:<Linelist>
C 60 e 1 X 61 e 1 91782.2686 91782.2686 0.0000 .090355 0.0000 : pP(61) : C Cv=0 60 60 F1e - X Xv=0 61 61 F1e :2168:<Linelist>
C 1 e 1 X 0 e 1 91922.3481 91922.3502 -0.0021 .010127 0.0000 : rR(0) : C Cv=0 1 1 F1e - X Xv=0 0 0 F1e :2171:<Linelist>
C 2 e 1 X 1 e 1 91925.9206 91925.9218 -0.0012 .010098 0.0000 : rR(1) : C Cv=0 2 2 F1e - X Xv=0 1 1 F1e :2172:<Linelist>
C 3 e 1 X 2 e 1 91929.5317 91929.5305 0.0012 .010084 0.0000 : rR(2) : C Cv=0 3 3 F1e - X Xv=0 2 2 F1e :2173:<Linelist>
C 4 e 1 X 3 e 1 91933.1828 91933.1830 -0.0002 .010084 0.0000 : rR(3) : C Cv=0 4 4 F1e - X Xv=0 3 3 F1e :2174:<Linelist>
C 5 e 1 X 4 e 1 91936.8758 91936.8786 -0.0028 .010084 0.0000 : rR(4) : C Cv=0 5 5 F1e - X Xv=0 4 4 F1e :2175:<Linelist>
C 6 e 1 X 5 e 1 91940.6073 91940.6086 -0.0013 .010072 0.0000 : rR(5) : C Cv=0 6 6 F1e - X Xv=0 5 5 F1e :2176:<Linelist>
C 7 e 1 X 6 e 1 91944.3789 91944.3808 -0.0019 .010072 0.0000 : rR(6) : C Cv=0 7 7 F1e - X Xv=0 6 6 F1e :2177:<Linelist>
C 8 e 1 X 7 e 1 91948.1859 91948.1899 -0.0040 .010072 0.0000 : rR(7) : C Cv=0 8 8 F1e - X Xv=0 7 7 F1e :2178:<Linelist>
C 9 e 1 X 8 e 1 91952.0346 91952.0397 -0.0051 .010084 0.0000 : rR(8) : C Cv=0 9 9 F1e - X Xv=0 8 8 F1e :2179:<Linelist>
C 10 e 1 X 9 e 1 91955.9197 91955.9229 -0.0032 .010084 0.0000 : rR(9) : C Cv=0 10 10 F1e - X Xv=0 9 9 F1e :2180:<Linelist>
C 11 e 1 X 10 e 1 91959.8470 91959.8529 -0.0059 .010084 0.0000 : rR(10) : C Cv=0 11 11 F1e - X Xv=0 10 10 F1e :2181:<Linelist>
C 12 e 1 X 11 e 1 91963.8156 91963.8197 -0.0041 .010084 0.0000 : rR(11) : C Cv=0 12 12 F1e - X Xv=0 11 11 F1e :2182:<Linelist>
C 13 e 1 X 12 e 1 91967.8140 91967.8160 -0.0020 .010098 0.0000 : rR(12) : C Cv=0 13 13 F1e - X Xv=0 12 12 F1e :2183:<Linelist>
C 14 e 1 X 13 e 1 91971.8555 91971.8583 -0.0028 .010098 0.0000 : rR(13) : C Cv=0 14 14 F1e - X Xv=0 13 13 F1e :2184:<Linelist>
C 15 e 1 X 14 e 1 91975.9329 91975.9384 -0.0055 .010112 0.0000 : rR(14) : C Cv=0 15 15 F1e - X Xv=0 14 14 F1e :2185:<Linelist>
C 16 e 1 X 15 e 1 91980.0508 91980.0553 -0.0045 .010127 0.0000 : rR(15) : C Cv=0 16 16 F1e - X Xv=0 15 15 F1e :2186:<Linelist>
C 17 e 1 X 16 e 1 91984.2058 91984.2091 -0.0033 .010143 0.0000 : rR(16) : C Cv=0 17 17 F1e - X Xv=0 16 16 F1e :2187:<Linelist>
C 18 e 1 X 17 e 1 91988.3964 91988.4008 -0.0044 .010161 0.0000 : rR(17) : C Cv=0 18 18 F1e - X Xv=0 17 17 F1e :2188:<Linelist>
C 19 e 1 X 18 e 1 91992.6258 91992.6315 -0.0057 .010198 0.0000 : rR(18) : C Cv=0 19 19 F1e - X Xv=0 18 18 F1e :2189:<Linelist>
C 20 e 1 X 19 e 1 91996.8938 91996.8976 -0.0038 .010239 0.0000 : rR(19) : C Cv=0 20 20 F1e - X Xv=0 19 19 F1e :2190:<Linelist>
C 21 e 1 X 20 e 1 92001.1983 92001.1995 -0.0012 .010308 0.0000 : rR(20) : C Cv=0 21 21 F1e - X Xv=0 20 20 F1e :2191:<Linelist>
C 22 e 1 X 21 e 1 92005.5385 92005.5398 -0.0013 .010385 0.0000 : rR(21) : C Cv=0 22 22 F1e - X Xv=0 21 21 F1e :2192:<Linelist>
C 23 e 1 X 22 e 1 92009.9177 92009.9176 0.0001 .010530 0.0000 : rR(22) : C Cv=0 23 23 F1e - X Xv=0 22 22 F1e :2193:<Linelist>
C 24 e 1 X 23 e 1 92014.3413 92014.3427 -0.0014 .010663 0.0000 : rR(23) : C Cv=0 24 24 F1e - X Xv=0 23 23 F1e :2194:<Linelist>
C 25 e 1 X 24 e 1 92018.7950 92018.7949 0.0001 .010846 0.0000 : rR(24) : C Cv=0 25 25 F1e - X Xv=0 24 24 F1e :2195:<Linelist>
C 26 e 1 X 25 e 1 92023.2845 92023.2842 0.0003 .010966 0.0000 : rR(25) : C Cv=0 26 26 F1e - X Xv=0 25 25 F1e :2196:<Linelist>
C 27 e 1 X 26 e 1 92027.8061 92027.8069 -0.0008 .011049 0.0000 : rR(26) : C Cv=0 27 27 F1e - X Xv=0 26 26 F1e :2197:<Linelist>
C 28 e 1 X 27 e 1 92032.3655 92032.3659 -0.0004 .011136 0.0000 : rR(27) : C Cv=0 28 28 F1e - X Xv=0 27 27 F1e :2198:<Linelist>
C 29 e 1 X 28 e 1 92036.9620 92036.9621 -0.0001 .011136 0.0000 : rR(28) : C Cv=0 29 29 F1e - X Xv=0 28 28 F1e :2199:<Linelist>
C 30 e 1 X 29 e 1 92041.5917 92041.5917 -0.0000 .011180 0.0000 : rR(29) : C Cv=0 30 30 F1e - X Xv=0 29 29 F1e :2200:<Linelist>
C 31 e 1 X 30 e 1 92046.2585 92046.2585 -0.0000 .011180 0.0000 : rR(30) : C Cv=0 31 31 F1e - X Xv=0 30 30 F1e :2201:<Linelist>
C 32 e 1 X 31 e 1 92050.9615 92050.9615 -0.0000 .011180 0.0000 : rR(31) : C Cv=0 32 32 F1e - X Xv=0 31 31 F1e :2202:<Linelist>
C 33 e 1 X 32 e 1 92055.6995 92055.6995 0.0000 .011180 0.0000 : rR(32) : C Cv=0 33 33 F1e - X Xv=0 32 32 F1e :2203:<Linelist>
C 34 e 1 X 33 e 1 92060.4715 92060.4715 -0.0000 .011180 0.0000 : rR(33) : C Cv=0 34 34 F1e - X Xv=0 33 33 F1e :2204:<Linelist>
C 35 e 1 X 34 e 1 92065.2735 92065.2735 0.0000 .011180 0.0000 : rR(34) : C Cv=0 35 35 F1e - X Xv=0 34 34 F1e :2205:<Linelist>

```
C   36 e  1     X   35 e  1   92070.1135    92070.1135      0.0000 .011180 0.0000 : rR(35) : C Cv=0 36 36 F1e - X Xv=0 35 35 F1e            :2206:<Linelist>
C   37 e  1     X   36 e  1   92074.9855    92074.9855      0.0000 .011180 0.0000 : rR(36) : C Cv=0 37 37 F1e - X Xv=0 36 36 F1e            :2207:<Linelist>
C   38 e  1     X   37 e  1   92079.8945    92079.8945     -0.0000 .011180 0.0000 : rR(37) : C Cv=0 38 38 F1e - X Xv=0 37 37 F1e            :2208:<Linelist>
C   39 e  1     X   38 e  1   92084.8345    92084.8345     -0.0000 .011180 0.0000 : rR(38) : C Cv=0 39 39 F1e - X Xv=0 38 38 F1e            :2209:<Linelist>
C   40 e  1     X   39 e  1   92089.8115    92089.8115      0.0000 .010308 0.0000 : rR(39) : C Cv=0 40 40 F1e - X Xv=0 39 39 F1e            :2210:<Linelist>
C   41 e  1     X   40 e  1   92094.8224    92094.8225     -0.0001 .010332 0.0000 : rR(40) : C Cv=0 41 41 F1e - X Xv=0 40 40 F1e            :2211:<Linelist>
C   42 e  1     X   41 e  1   92099.8664    92099.8665     -0.0001 .010358 0.0000 : rR(41) : C Cv=0 42 42 F1e - X Xv=0 41 41 F1e            :2212:<Linelist>
C   43 e  1     X   42 e  1   92104.9364    92104.9364     -0.0000 .010412 0.0000 : rR(42) : C Cv=0 43 43 F1e - X Xv=0 42 42 F1e            :2213:<Linelist>
C   44 e  1     X   43 e  1   92110.0454    92110.0455     -0.0001 .010500 0.0000 : rR(43) : C Cv=0 44 44 F1e - X Xv=0 43 43 F1e            :2214:<Linelist>
C   45 e  1     X   44 e  1   92115.1864    92115.1865     -0.0001 .010628 0.0000 : rR(44) : C Cv=0 45 45 F1e - X Xv=0 44 44 F1e            :2215:<Linelist>
C   46 e  1     X   45 e  1   92120.3694    92120.3695     -0.0001 .010808 0.0000 : rR(45) : C Cv=0 46 46 F1e - X Xv=0 45 45 F1e            :2216:<Linelist>
C   47 e  1     X   46 e  1   92125.5644    92125.5645     -0.0001 .011049 0.0000 : rR(46) : C Cv=0 47 47 F1e - X Xv=0 46 46 F1e            :2217:<Linelist>
C   48 e  1     X   47 e  1   92130.8034    92130.8034     -0.0000 .011413 0.0000 : rR(47) : C Cv=0 48 48 F1e - X Xv=0 47 47 F1e            :2218:<Linelist>
C   49 e  1     X   48 e  1   92136.0694    92136.0695     -0.0001 .011927 0.0000 : rR(48) : C Cv=0 49 49 F1e - X Xv=0 48 48 F1e            :2219:<Linelist>
C   50 e  1     X   49 e  1   92141.3744    92141.3744     -0.0000 .012744 0.0000 : rR(49) : C Cv=0 50 50 F1e - X Xv=0 49 49 F1e            :2220:<Linelist>
C   51 e  1     X   50 e  1   92146.6944    92146.6945     -0.0001 .013862 0.0000 : rR(50) : C Cv=0 51 51 F1e - X Xv=0 50 50 F1e            :2221:<Linelist>
C   52 e  1     X   51 e  1   92152.0424    92152.0425     -0.0001 .015544 0.0000 : rR(51) : C Cv=0 52 52 F1e - X Xv=0 51 51 F1e            :2222:<Linelist>
C   53 e  1     X   52 e  1   92157.4414    92157.4415     -0.0001 .017945 0.0000 : rR(52) : C Cv=0 53 53 F1e - X Xv=0 52 52 F1e            :2223:<Linelist>
C   54 e  1     X   53 e  1   92162.8714    92162.8715     -0.0001 .021294 0.0000 : rR(53) : C Cv=0 54 54 F1e - X Xv=0 53 53 F1e            :2224:<Linelist>
C   55 e  1     X   54 e  1   92168.3114    92168.3115     -0.0001 .025908 0.0000 : rR(54) : C Cv=0 55 55 F1e - X Xv=0 54 54 F1e            :2225:<Linelist>
C   56 e  1     X   55 e  1   92173.7634    92173.7635     -0.0001 .032288 0.0000 : rR(55) : C Cv=0 56 56 F1e - X Xv=0 55 55 F1e            :2226:<Linelist>
C   57 e  1     X   56 e  1   92179.2944    92179.2944     -0.0000 .041037 0.0000 : rR(56) : C Cv=0 57 57 F1e - X Xv=0 56 56 F1e            :2227:<Linelist>
C   58 e  1     X   57 e  1   92184.8414    92184.8415     -0.0001 .052756 0.0000 : rR(57) : C Cv=0 58 58 F1e - X Xv=0 57 57 F1e            :2228:<Linelist>
C   59 e  1     X   58 e  1   92190.3664    92190.3665     -0.0001 .068731 0.0000 : rR(58) : C Cv=0 59 59 F1e - X Xv=0 58 58 F1e            :2229:<Linelist>
C   60 e  1     X   59 e  1   92195.9944    92195.9944     -0.0000 .090355 0.0000 : rR(59) : C Cv=0 60 60 F1e - X Xv=0 59 59 F1e            :2230:<Linelist>
B   29 e  1 Excited   30 e  3    19586.9298    19586.9288      0.0010 .060000 0.0000 : qP1(30) : B Bv=0 29 29 F1e - Excited dv=7 30 29 F1e      :2234:<Linelist>
B   31 e  1 Excited   30 e  3    19801.5764    19801.5744      0.0020 .014500 0.0000 : sR1(30) : B Bv=0 31 31 F1e - Excited dv=7 30 29 F1e      :2235:<Linelist>
B   30 e  1 Excited   31 e  4    19610.4472    19610.4149      0.0323 .020800 0.0000 : oP13(31) : B Bv=0 30 30 F1e - Excited ev=4 31 32 F3e      :2238:<Linelist>
B   31 e  1 Excited   32 e  6    19617.3395    19617.3567     -0.0172 .007300 0.0000 : pP12(32) : B Bv=0 31 31 F1e - Excited a3Pv=13 32 32 F2e   :2239:<Linelist>
B   31 e  1 Excited   31 f  5    19716.3195    19716.3301     -0.0106 .013900 0.0000 : rQ1(31) : B Bv=0 31 31 F1e - Excited a'v=12 31 30 F1f    :2240:<Linelist>
B   32 e  1 Excited   31 e  4    19832.0333    19832.0105      0.0228 .032800 0.0000 : qR13(31) : B Bv=0 32 32 F1e - Excited ev=4 31 32 F3e      :2241:<Linelist>

SVD fit: 1008 Observations,  38 Parameters (scaled)
Initial Average Error: 1.1502025653437  (Unweighted: .0295955485827221)
Predicted New Error:   1.14717009167763  (Unweighted: .029465037504618)
Parameters:
 # Old                New                Std Dev   Change/Std Sens    Summary              Name
 1 67557.9263807482    67557.9263810701    .00087862    0.0004  3.44e-5 67557.92638(88)      Excited Av=2 Origin
 2 1.41770507032897    1.41770506363778    3.5132e-6   -0.0019  6.56e-8 1.4177051(35)        Excited Av=2 B
 3 6.24399485577261e-6 6.24399108608897e-6 2.5362e-9   -0.0015  5.7e-11 6.2440(25)e-6        Excited Av=2 D
 4 67579.2383904013    67579.2384055729    .01547960    0.0010  .000530 67579.238(15)        Excited Iv=3 Origin
 5 1.09918701526773    1.09918688539230    .00022665   -0.0006  2.57e-6 1.09919(23)          Excited Iv=3 B
 6 5.77973239037241e-6 5.77953594122483e-6 4.0437e-7   -0.0005  5.41e-9 5.78(40)e-6          Excited Iv=3 D
 7 -.0504224240425148  -.0504228416460905  .00047719   -0.0009  3.77e-5 -0.05042(48)         Excited <Av=2|J+L-|Iv=3> Value
 8 67802.7849165451    67802.7848875454    .00556060   -0.0052  .000236 67802.7849(56)       Excited ev=4 Origin
 9 1.09715336874161    1.09715353997320    3.0073e-5    0.0057  4.52e-7 1.097154(30)         Excited ev=4 B
10 .424603205929654    .424579055599668    .01151692   -0.0021  .000607 0.425(12)            Excited ev=4 LambdaSS
11 5.34314012526997e-6 5.34329820601173e-6 2.5163e-8    0.0063  4.8e-10 5.343(25)e-6         Excited ev=4 D
12 -13.1135785220975   -13.1135722759008   .00455887    0.0014  .000284 -13.1136(46)         Excited <Av=2|LS|ev=4> Value
13 67946.2486612564    67946.2486639027    .00364353    0.0007  .000196 67946.2487(36)       Excited dv=7 Origin
14 1.07973878060191    1.07973873112873    1.3904e-5   -0.0036  3.34e-7 1.079739(14)         Excited dv=7 B
15 -16.9899246280871   -16.9899214894736   .00331410    0.0009  .000186 -16.9899(33)         Excited dv=7 A
16 1.23163175363889    1.23163200196255    .00538419    0.0000  .000413 1.2316(54)           Excited dv=7 LambdaSS
17 -.00290463321309688 -.00290540266173026 .00025755   -0.0030  1.37e-5 -2.91(26)e-3         Excited dv=7 gamma
18 5.12786503880413e-6 5.12783957207506e-6 8.0424e-9   -0.0032  2.2e-10 5.1278(80)e-6        Excited dv=7 D
19 -9.51850056737858   -9.51852615949637   .01047844   -0.0024  .000673 -9.519(10)           Excited <Av=2|LS|dv=7> Value
20 67991.2956897487    67991.2956970936    .00589195    0.0012  .000312 67991.2957(59)       Excited a'v=12 Origin
21 1.03759347733243    1.03759347486748    2.2493e-5   -0.0001  4.73e-7 1.037593(22)         Excited a'v=12 B
22 -1.14571380522651   -1.14571729964806   .00908994   -0.0004  .000700 -1.1457(91)          Excited a'v=12 LambdaSS
23 -.00527072942354211 -.00527095098307920 .00031172   -0.0007  2.10e-5 -5.27(31)e-3         Excited a'v=12 gamma
24 5.18562812216526e-6 5.18562893250139e-6 1.2228e-8    0.0001  2.9e-10 5.186(12)e-6         Excited a'v=12 D
25 -5.90222571840932   -5.90222290190178   .01148474    0.0002  .000759 -5.902(11)           Excited <Av=2|LS|a'v=12> Value
26 68377.9123376615    68377.9123516970    .05036832    0.0003  .001195 68377.912(50)        Excited Dsv=3 Origin
27 1.08921917955507    1.08921912165493    .00018753   -0.0003  1.78e-6 1.089219(188)        Excited Dsv=3 B
28 5.85540454591073e-6 5.85537632665456e-6 7.9672e-8   -0.0004  9.4e-10 5.855(80)e-6         Excited Dsv=3 D
29 67670.6782977856    67670.6779227765    .00506836   -0.0740  .000248 67670.6779(51)       Excited a3Pv=13 Origin
30 1.30813665955102    1.30813765113497    5.5672e-7    1.7811  3.60e-8 1.30813765(56)       Excited a3Pv=13 B
31 38.1154482644111    38.1153846346168    .01295243   -0.0049  .000670 38.115(13)           Excited a3Pv=13 A
32 6.06110427245085e-6 6.06155910230364e-6 4.6454e-9    0.0979  3.2e-10 6.0616(46)e-6        Excited a3Pv=13 D
33 24.7288685468983    24.7287011256680    .01972215   -0.0085  .000958 24.729(20)           Excited <a3Pv=13|LS|ev=4> Value
34 -.0598221863162786  -.0598230826106589  .00025564   -0.0035  1.37e-5 -0.05982(26)         Excited <a3Pv=13|J+L-|ev=4> Value
35 23.2188654276278    23.2189357988461    .04412555    0.0016  .003062 23.219(44)           Excited <a3Pv=13|LS|dv=7> Value
36 -.0627951499543510  -.0627945906974129  .00045347    0.0012  3.19e-5 -0.06279(45)         Excited <a3Pv=13|J+L-|dv=7> Value
37 2.39113881014602    2.39157392255636    .20400913    0.0021  .010788 2.39(20)             Excited <a3Pv=13|LS|a'v=12> Value
38 -.0112468031812311  -.0112360932879575  .00125602    0.0085  6.81e-5 -0.0112(13)          Excited <a3Pv=13|J+L-|a'v=12> Value

Correlation Matrix
Largest off-diagonal element = 0.988 at 28,27 = Excited Dsv=3 D, Excited Dsv=3 B
      1      2      3      4      5      6      7      8      9     10     11     12     13     14     15     16     17     18     19     20     21     22     23     24     25     26     27     28     29     30     31     32
33     34     35     36     37     38
 1   1.000
 2  -0.611  1.000
 3  -0.395  0.931  1.000
 4   0.033 -0.038 -0.027  1.000
 5  -0.043  0.040  0.026 -0.876  1.000
 6  -0.042  0.039  0.026 -0.823  0.984  1.000
 7   0.087  0.106  0.100 -0.133  0.103  0.097  1.000
 8  -0.003  0.018  0.012 -0.005  0.003  0.002  0.032  1.000
 9   0.039 -0.109 -0.095  0.007 -0.002 -0.002 -0.032 -0.672  1.000
10   0.009  0.035  0.040 -0.011  0.007  0.006  0.043  0.529 -0.514  1.000
11   0.032 -0.092 -0.087  0.005 -0.001  0.000 -0.023 -0.454  0.939 -0.365  1.000
12  -0.538  0.096 -0.058 -0.006  0.011  0.011 -0.076  0.006  0.019 -0.134  0.020  1.000
```

13 0.005 0.004 -0.005 -0.001 0.000 0.000 -0.006 -0.004 -0.018 -0.017 -0.026 -0.052 1.000
14 0.006 0.005 0.035 0.003 -0.002 -0.002 0.001 0.003 0.017 0.034 0.032 0.030 -0.612 1.000
15 -0.008 0.025 0.032 -0.002 0.001 0.001 0.002 -0.005 0.000 -0.014 -0.005 0.000 0.267 -0.061 1.000
16 0.001 -0.007 -0.006 0.000 0.000 0.000 0.000 -0.005 -0.002 -0.005 -0.004 0.000 0.212 -0.066 0.357 1.000
17 0.021 -0.072 -0.088 0.003 -0.004 -0.004 -0.007 -0.007 -0.008 0.026 0.003 -0.001 -0.066 -0.025 -0.626 -0.168 1.000
18 0.008 -0.007 0.011 0.003 -0.002 -0.002 -0.001 0.009 0.012 0.030 0.025 0.019 -0.449 0.918 0.025 -0.053 -0.171 1.000
19 -0.178 -0.200 -0.268 0.000 -0.001 -0.001 -0.030 0.044 -0.084 0.031 -0.080 0.170 -0.030 -0.035 -0.065 0.009 0.170 -0.013 1.000
20 0.021 0.053 0.025 0.005 -0.004 -0.004 -0.025 -0.015 0.020 -0.019 0.019 -0.084 -0.050 0.027 -0.021 -0.005 0.033 0.024 0.020 1.000
21 -0.013 -0.001 0.040 -0.005 0.005 0.005 0.021 0.001 0.003 0.012 0.003 0.066 0.015 -0.008 0.003 0.001 -0.009 -0.008 -0.006 -0.569 1.000
22 0.004 -0.035 -0.040 0.001 -0.002 -0.002 -0.004 0.006 -0.005 0.004 -0.003 0.010 0.021 -0.017 0.011 -0.002 -0.019 -0.013 -0.020 -0.254 -0.085 1.000
23 0.008 -0.050 -0.075 0.002 -0.002 -0.002 -0.006 0.005 -0.009 -0.002 -0.008 -0.006 0.017 -0.011 0.012 0.003 -0.016 -0.006 -0.001 0.073 -0.506 0.204 1.000
24 -0.011 0.004 0.037 -0.005 0.004 0.004 0.016 -0.001 0.003 0.012 0.004 0.053 0.004 -0.002 0.000 0.000 -0.001 -0.004 0.007 -0.424 0.948 -0.125 -0.545 1.000
25 -0.015 -0.103 -0.126 0.002 -0.002 -0.002 0.003 -0.004 0.024 -0.012 0.031 -0.015 0.063 -0.042 0.054 0.006 -0.092 -0.040 -0.283 -0.101 -0.112 0.114 0.167 -0.119 1.000
26 0.000 0.000 0.000 0.000 0.000 0.000 0.000 0.000 -0.001 0.000 -0.002 0.000 0.000 0.001 0.000 0.000 0.001 0.000 0.001 0.000 0.000 0.000 0.000 0.000 0.000 1.000
27 0.000 0.000 0.000 0.000 0.000 0.000 0.000 0.000 0.001 0.000 0.002 0.000 0.000 -0.001 0.000 0.000 -0.001 0.000 -0.001 0.000 0.000 0.000 0.000 0.000 0.000 -
0.953 1.000
28 0.000 0.000 0.000 0.000 0.000 0.000 0.000 -0.001 0.002 0.000 0.002 0.000 0.000 -0.001 0.000 0.000 -0.001 0.000 -0.001 0.000 0.000 0.000 0.000 0.000 0.000 -
0.925 0.988 1.000
29 -0.004 0.064 0.075 0.008 -0.004 -0.004 0.031 0.068 -0.091 0.035 -0.070 0.116 -0.261 0.188 -0.050 -0.045 0.029 0.123 -0.009 -0.020 0.017 -0.005 -0.010 0.015 -0.070
0.002 -0.002 -0.002 1.000
30 -0.033 0.012 0.006 -0.541 0.474 0.445 0.074 0.017 -0.031 -0.039 -0.019 0.016 -0.050 0.025 -0.018 -0.013 0.013 0.013 0.013 -0.186 0.101 0.048 -0.010 0.074 0.030
0.000 0.000 -0.001 -0.149 1.000
31 -0.012 0.099 0.102 0.000 0.009 0.009 0.036 -0.094 0.224 -0.204 0.221 0.135 -0.183 0.120 -0.030 -0.040 -0.043 0.092 -0.092 0.013 -0.003 -0.006 -0.011 -0.003 -0.012
0.000 0.000 0.001 0.636 -0.097 1.000
32 -0.010 0.053 0.048 -0.039 0.033 0.031 0.031 0.054 -0.130 0.029 -0.143 0.012 -0.104 0.093 -0.027 -0.002 0.100 0.013 0.049 -0.003 -0.008 -0.001 0.014 -0.007 -0.015
0.006 -0.006 -0.008 0.454 0.055 0.137 1.000
33 -0.055 0.047 0.037 -0.004 0.000 0.000 0.015 0.259 -0.559 0.209 -0.550 0.009 0.062 -0.052 0.011 0.016 0.031 -0.050 0.147 -0.038 0.008 0.005 0.010 0.004 -0.060
0.002 -0.003 -0.003 0.125 0.102 -0.256 0.231 1.000
34 -0.044 0.031 0.016 0.001 0.000 0.000 0.024 -0.043 -0.083 -0.417 -0.126 0.142 0.043 -0.044 0.012 0.010 0.010 -0.044 0.067 0.004 -0.016 -0.002 0.006 -0.017 -0.032
0.001 -0.001 -0.001 0.137 0.100 0.125 0.136 0.500 1.000
35 -0.026 -0.039 -0.055 -0.001 0.000 -0.001 -0.001 -0.041 -0.002 -0.018 -0.008 0.053 -0.145 0.039 0.063 0.066 0.232 -0.085 0.179 0.031 -0.009 -0.017 -0.005 0.004 -0.038
0.001 -0.001 -0.001 0.036 0.033 -0.090 0.146 0.065 0.038 1.000
36 -0.010 -0.058 -0.074 -0.003 0.003 0.004 0.018 0.073 -0.092 -0.028 -0.105 0.067 -0.131 -0.065 0.135 0.034 -0.285 -0.041 0.115 -0.049 0.009 0.024 -0.005 0.004 0.041 -
0.001 0.001 0.001 0.179 -0.019 0.164 -0.015 0.126 0.127 0.118 1.000
37 0.056 -0.016 -0.034 0.005 -0.003 -0.003 -0.004 -0.011 0.055 -0.053 0.054 -0.108 0.049 -0.031 0.020 -0.003 -0.056 -0.013 -0.268 0.284 -0.281 -0.009 0.098 -0.263 0.436
0.000 0.000 0.001 -0.088 -0.031 0.037 -0.075 -0.105 -0.004 -0.140 0.063 1.000
38 0.110 -0.037 -0.056 0.000 0.001 0.001 0.013 -0.065 0.162 -0.030 0.176 -0.126 -0.032 0.006 0.023 -0.007 -0.046 0.007 -0.271 0.216 -0.171 -0.011 0.029 -0.147 0.334 -
0.002 0.002 0.002 -0.183 -0.035 0.015 -0.190 -0.302 -0.172 0.019 0.047 0.679 1.000
PGOPHER 11.0.142 1 Apr 2021 8:30 64 bit Unicode (Delphi 33/33)
pgopher 44b8c50a90149a6ce53e669f4a0e44afc86f1abd
lutils 65f951dfa553c7f9bc501632d3360a85cd1a0165
utils ec0cac7863199e65355ae4c6ab78dbe3037fae3b
lforms d4dfa0031586b0cf5ffd76ff4d8f7aee2a9f8932
wutils 9bb7abdf4436dd4d7dfed61ce7fb7c6bc914386c
GraphicEx bfb0df9494191c3a10d6089c21fe439fcd763090
Mixture: 13C18O Deperturbation of the A(v=2) level.pgo
Units cm1
PrintLevel LineList
Precision 4
BasisOrder boDefault
QuantumNumberFormat J
FitCycles 1
FitDetails False
AutoReplot True
AllowComplex True
NoLineList False
ShowEstUnc False
ShowDerivatives False
SVDlimit 5
ShowSubBasis False
AllowZeroI False
ScaledParameters True
FixTransitionPhase False
OldHyperfineOrder False
OldColourOrder False
SingleResidualPlot False
UpperStateColour False
SVDThresh 1E-6 Discard Singular Value Decomposition eigenvalues below this fraction of the largest eigenvalue; <0 to use QR factorisation.
ScaleChanges 1 Scale changes produced by fitting by this - normally 1
SmallE 2E-18 Matrix elements smaller than this are discarded
Species: CO
Jmax 63
Concentration 1 Relative concentration of species
LinearMolecule: 13C18O
JAdjustSym True
BlockMatrix True
AbsoluteE False
Symmetric False
SymWt 1
AsymWt 1
RSquaredH False
ShowJ True
ShowOmega False
ShowN True
ShowFNumber True
Showef True
Abundance 1 Abundance of this isotope; default is 1
AssumedOrigin 1E99 Energy origin for Boltzmann calculations; 1e99 implies automatic estimate. Currently 0
LinearManifold: X

Initial True
EigenSearch True

```
  LimitSearch      True
  AutoQConverge    True
  UsePopParams     False
  OverrideEnergies True
  Linear: X Xv=0
    Status     Active
    Lambda     Sigma+
    OmegaOrder Auto => Regular
    B  1.746408199                  Rotational constant.
    D  5.0488146E-6                  Quartic centrifugal distortion.
    H  4.35471E-12                  Sextic centrifugal distortion.
    L  -2.4821E-17                 J^8  centrifugal distortion.
    M  2.99E-23                    J^10  centrifugal distortion.
    PP -4.5E-28                    J^12 centrifugal distortion. (Not P to avoid confusion with p.)
    <J,Lambda=0,Omega=0|B|J,Lambda=0,Omega=0> = J*(J+1)
      Symmetrized:
    <J,Lambda=0,Omega=0|B|J,Lambda=0,Omega=0> = J*(J+1)

    <J,Lambda=0,Omega=0|D|J,Lambda=0,Omega=0> = -J^2*(J*(J+2)+1)
      Symmetrized:
    <J,Lambda=0,Omega=0|D|J,Lambda=0,Omega=0> = -J^2*(J*(J+2)+1)

    <J,Lambda=0,Omega=0|H|J,Lambda=0,Omega=0> = J^3*(J*(J*(J+3)+3)+1)
      Symmetrized:
    <J,Lambda=0,Omega=0|H|J,Lambda=0,Omega=0> = J^3*(J*(J*(J+3)+3)+1)

    <J,Lambda=0,Omega=0|L|J,Lambda=0,Omega=0> = J^4*(J*(J*(J*(J+4)+6)+4)+1)
      Symmetrized:
    <J,Lambda=0,Omega=0|L|J,Lambda=0,Omega=0> = J^4*(J*(J*(J*(J+4)+6)+4)+1)

    <J,Lambda=0,Omega=0|M|J,Lambda=0,Omega=0> = J^5*(J*(J*(J*(J*(J+5)+10)+10)+5)+1)
      Symmetrized:
    <J,Lambda=0,Omega=0|M|J,Lambda=0,Omega=0> = J^5*(J*(J*(J*(J*(J+5)+10)+10)+5)+1)

    <J,Lambda=0,Omega=0|PP|J,Lambda=0,Omega=0> = J^6*(J*(J*(J*(J*(J*(J+6)+15)+20)+15)+6)+1)
      Symmetrized:
    <J,Lambda=0,Omega=0|PP|J,Lambda=0,Omega=0> = J^6*(J*(J*(J*(J*(J*(J+6)+15)+20)+15)+6)+1)

  LinearPerturbation: X <Xv=0|N^14|Xv=0>
    O (rotational constant)
    Value -2.2E-33                   Size of perturbation.
  <J,Lambda=0,Omega=0|N^14|J,Lambda=0,Omega=0> = J^7*(J*(J*(J*(J*(J*(J*(J+7)+21)+35)+35)+21)+7)+1)

  Symmetrized Matrix Elements of <Xv=0|N^14|Xv=0>
  <J,Lambda=0,Omega=0|N^14|J,Lambda=0,Omega=0> = J^7*(J*(J*(J*(J*(J*(J*(J+7)+21)+35)+35)+21)+7)+1)

LinearManifold: C
  Initial          True
  EigenSearch      True
  LimitSearch      False
  AutoQConverge    True
  UsePopParams     False
  OverrideEnergies True
  E 0 e 1  91918.81461769            Energy for J=E 0 e 1 level
  E 1 e 1  91922.3501979315           Energy for J=E 1 e 1 level
  E 2 e 1  91929.41460931            Energy for J=E 2 e 1 level
  E 3 e 1  91940.0087911758           Energy for J=E 3 e 1 level
  E 4 e 1  91954.1391754158           Energy for J=E 4 e 1 level
  E 5 e 1  91971.8047483651           Energy for J=E 5 e 1 level
  E 6 e 1  91992.9963415625           Energy for J=E 6 e 1 level
  E 7 e 1  92017.7210158785           Energy for J=E 7 e 1 level
  E 8 e 1  92045.9729147246           Energy for J=E 8 e 1 level
  E 9 e 1  92077.7548931101           Energy for J=E 9 e 1 level
  E 10 e 1 92113.0587701987           Energy for J=E 10 e 1 level
  E 11 e 1 92151.8966952636           Energy for J=E 11 e 1 level
  E 12 e 1 92194.2576663392           Energy for J=E 12 e 1 level
  E 13 e 1 92240.1328568783           Energy for J=E 13 e 1 level
  E 14 e 1 92289.5374084555           Energy for J=E 14 e 1 level
  E 15 e 1 92342.4615531966           Energy for J=E 15 e 1 level
  E 16 e 1 92398.9025159008           Energy for J=E 16 e 1 level
  E 17 e 1 92458.8587353746           Energy for J=E 17 e 1 level
  E 18 e 1 92522.3290731028           Energy for J=E 18 e 1 level
  E 19 e 1 92589.3126987891           Energy for J=E 19 e 1 level
  E 20 e 1 92659.8038563447           Energy for J=E 20 e 1 level
  E 21 e 1 92733.8006230299           Energy for J=E 21 e 1 level
  E 22 e 1 92811.3031429715           Energy for J=E 22 e 1 level
  E 23 e 1 92892.3079897717           Energy for J=E 23 e 1 level
  E 24 e 1 92976.822314948           Energy for J=E 24 e 1 level
  E 25 e 1 93064.8231465903           Energy for J=E 25 e 1 level
  E 26 e 1 93156.3176170672           Energy for J=E 26 e 1 level
  E 27 e 1 93251.2988822767           Energy for J=E 27 e 1 level
  E 28 e 1 93349.7667497678           Energy for J=E 28 e 1 level
  E 29 e 1 93451.7189577959           Energy for J=E 29 e 1 level
  E 30 e 1 93557.148271242           Energy for J=E 30 e 1 level
  E 31 e 1 93666.0549149757           Energy for J=E 31 e 1 level
  E 32 e 1 93778.4343213518           Energy for J=E 32 e 1 level
  E 33 e 1 93894.2815297818           Energy for J=E 33 e 1 level
  E 34 e 1 94013.591760117           Energy for J=E 34 e 1 level
  E 35 e 1 94136.3569127458           Energy for J=E 35 e 1 level
  E 36 e 1 94262.5809686937           Energy for J=E 36 e 1 level
  E 37 e 1 94392.253689726           Energy for J=E 37 e 1 level
  E 38 e 1 94525.3757684531           Energy for J=E 38 e 1 level
```

```
  E 39 e 1 94661.9366784378            Energy for J=E 39 e 1 level
  E 40 e 1 94801.9378743053            Energy for J=E 40 e 1 level
  E 41 e 1 94945.3716418562            Energy for J=E 41 e 1 level
  E 42 e 1 95092.2322481804            Energy for J=E 42 e 1 level
  E 43 e 1 95242.507641775            Energy for J=E 43 e 1 level
  E 44 e 1 95396.2059026634            Energy for J=E 44 e 1 level
  E 45 e 1 95553.3147425169            Energy for J=E 45 e 1 level
  E 46 e 1 95713.8389547786            Energy for J=E 46 e 1 level
  E 47 e 1 95877.7430647896            Energy for J=E 47 e 1 level
  E 48 e 1 96045.0535799167            Energy for J=E 48 e 1 level
  E 49 e 1 96215.7478896832            Energy for J=E 49 e 1 level
  E 50 e 1 96389.8321659005            Energy for J=E 50 e 1 level
  E 51 e 1 96567.2766128028            Energy for J=E 51 e 1 level
  E 52 e 1 96748.0881671828            Energy for J=E 52 e 1 level
  E 53 e 1 96932.2837485302            Energy for J=E 53 e 1 level
  E 54 e 1 97119.8371091712            Energy for J=E 54 e 1 level
  E 55 e 1 97310.7208844101            Energy for J=E 55 e 1 level
  E 56 e 1 97504.930642673            Energy for J=E 56 e 1 level
  E 57 e 1 97702.5267856525            Energy for J=E 57 e 1 level
  E 58 e 1 97903.4396484547            Energy for J=E 58 e 1 level
  E 59 e 1 98107.6243997471            Energy for J=E 59 e 1 level
  E 60 e 1 98315.1990419085            Energy for J=E 60 e 1 level
  E 61 e 1 98526.0359657192            Energy for J=E 61 e 1 level
  E 62 e 1 98740.1821395715            Energy for J=E 62 e 1 level
  E 63 e 1 98957.6207414145            Energy for J=E 63 e 1 level
  E 64 e 1 99178.3440172057            Energy for J=E 64 e 1 level
  E 65 e 1 99402.3440898222            Energy for J=E 65 e 1 level
  Linear: C Cv=0
   Status     Active
   Lambda     Sigma+
   OmegaOrder Auto => Regular
   Origin 91918.79                    State Origin. Estimated lowest level = 91918.79
   B      1.76641191315359            Rotational constant.
   D      5.12833508732762E-6         Quartic centrifugal distortion.
   <J,Lambda=0,Omega=0|B|J,Lambda=0,Omega=0> = J*(J+1)
     Symmetrized:
   <J,Lambda=0,Omega=0|B|J,Lambda=0,Omega=0> = J*(J+1)

   <J,Lambda=0,Omega=0|D|J,Lambda=0,Omega=0> = -J^2*(J*(J+2)+1)
     Symmetrized:
   <J,Lambda=0,Omega=0|D|J,Lambda=0,Omega=0> = -J^2*(J*(J+2)+1)

LinearManifold: B
  Initial          True
  EigenSearch      True
  LimitSearch      False
  AutoQConverge    True
  UsePopParams     False
  OverrideEnergies True
  E 0 e 1  86917.3573438145           Energy for J=E 0 e 1 level
  E 1 e 1  86920.8990798371           Energy for J=E 1 e 1 level
  E 2 e 1  86927.9802287834           Energy for J=E 2 e 1 level
  E 3 e 1  86938.5975955474           Energy for J=E 3 e 1 level
  E 4 e 1  86952.7535165513           Energy for J=E 4 e 1 level
  E 5 e 1  86970.4485900942           Energy for J=E 5 e 1 level
  E 6 e 1  86991.6808705965           Energy for J=E 6 e 1 level
  E 7 e 1  87016.4501212347           Energy for J=E 7 e 1 level
  E 8 e 1  87044.7550139408           Energy for J=E 8 e 1 level
  E 9 e 1  87076.5945102899           Energy for J=E 9 e 1 level
  E 10 e 1 87111.9673665656            Energy for J=E 10 e 1 level
  E 11 e 1 87150.8722462898            Energy for J=E 11 e 1 level
  E 12 e 1 87193.3103335359            Energy for J=E 12 e 1 level
  E 13 e 1 87239.2738059414            Energy for J=E 13 e 1 level
  E 14 e 1 87288.7659451284            Energy for J=E 14 e 1 level
  E 15 e 1 87341.7838423505            Energy for J=E 15 e 1 level
  E 16 e 1 87398.3239892327            Energy for J=E 16 e 1 level
  E 17 e 1 87458.3457386672            Energy for J=E 17 e 1 level
  E 18 e 1 87521.9669176787            Energy for J=E 18 e 1 level
  E 19 e 1 87589.0729980635            Energy for J=E 19 e 1 level
  E 20 e 1 87659.6848539994            Energy for J=E 20 e 1 level
  E 21 e 1 87733.8093501407            Energy for J=E 21 e 1 level
  E 22 e 1 87811.4434750138            Energy for J=E 22 e 1 level
  E 23 e 1 87892.5833524621            Energy for J=E 23 e 1 level
  E 24 e 1 87977.2254359612            Energy for J=E 24 e 1 level
  E 25 e 1 88065.3655819688            Energy for J=E 25 e 1 level
  E 26 e 1 88157.0074638094            Energy for J=E 26 e 1 level
  E 27 e 1 88252.1428727514            Energy for J=E 27 e 1 level
  E 28 e 1 88350.7622038786            Energy for J=E 28 e 1 level
  E 29 e 1 88452.8714734024            Energy for J=E 29 e 1 level
  E 30 e 1 88558.4573652084            Energy for J=E 30 e 1 level
  E 31 e 1 88667.5170675484            Energy for J=E 31 e 1 level
  E 32 e 1 88780.0530157549            Energy for J=E 32 e 1 level
  E 33 e 1 88896.0732697678            Energy for J=E 33 e 1 level
  E 34 e 1 89015.5479117331            Energy for J=E 34 e 1 level
  E 35 e 1 89138.5007753021            Energy for J=E 35 e 1 level
  E 36 e 1 89264.8671434128            Energy for J=E 36 e 1 level
  E 37 e 1 89394.712613204            Energy for J=E 37 e 1 level
  E 38 e 1 89528.0033885644            Energy for J=E 38 e 1 level
  E 39 e 1 89664.7427100465            Energy for J=E 39 e 1 level
  E 40 e 1 89804.8732821819            Energy for J=E 40 e 1 level
  E 41 e 1 89948.4826520769            Energy for J=E 41 e 1 level
  E 42 e 1 90095.4817872327            Energy for J=E 42 e 1 level
```

```
 E 43 e 1 90245.9288501256      Energy for J=E 43 e 1 level
 E 44 e 1 90399.7945833577      Energy for J=E 44 e 1 level
 E 45 e 1 90557.048472636       Energy for J=E 45 e 1 level
 E 46 e 1 90717.7471747185      Energy for J=E 46 e 1 level
 E 47 e 1 90881.7572647896      Energy for J=E 47 e 1 level
 E 48 e 1 91049.1774976252      Energy for J=E 48 e 1 level
 E 49 e 1 91220.0533896832      Energy for J=E 49 e 1 level
 E 50 e 1 91394.1969159005      Energy for J=E 50 e 1 level
 E 51 e 1 91571.8211856604      Energy for J=E 51 e 1 level
 E 52 e 1 91752.7615313151      Energy for J=E 52 e 1 level
 E 53 e 1 91937.0582069911      Energy for J=E 53 e 1 level
 E 54 e 1 92124.7041674308      Energy for J=E 54 e 1 level
 E 55 e 1 92315.6922344469      Energy for J=E 55 e 1 level
 Linear: B Bv=0
  Status     Active
  Lambda     Sigma+
  OmegaOrder Auto => Regular
  Origin 86917.360320496        State Origin. Estimated lowest level = 86917.360320496
  B      1.76976443875216       Rotational constant.
  D      5.53872463797865E-6    Quartic centrifugal distortion.
  Matrix elements as for C Cv=0
LinearManifold: Excited
 Initial         False
 EigenSearch     True
 LimitSearch     True
 AutoQConverge   True
 UsePopParams    False
 OverrideEnergies True
 Linear: Excited Av=2
  Status     Active
  Lambda     Pi
  OmegaOrder Auto => Regular
  Origin 67557.9263810701       Float, Std Dev = 0.000878618160689914  State Origin. Estimated lowest level = 67557.9263810701
  B      1.41770506363778       Float, Std Dev = 3.51316122242332E-6  Rotational constant.
  q      -1.187706E-5           Lambda doubling constant.
  D      6.24399108608897E-6    Float, Std Dev = 2.5362343397168E-9  Quartic centrifugal distortion.
  H      -1.280902E-11          Sextic centrifugal distortion.
  <J,Lambda=-1,Omega=-1|B|J,Lambda=-1,Omega=-1> = J*(J+1)
  <J,Lambda=1,Omega=1|B|J,Lambda=1,Omega=1> = J*(J+1)
   Symmetrized:
  <J,Lambda=1,Omega=1|B|J,Lambda=1,Omega=1> = J*(J+1)
  <J,Lambda=1,Omega=1|B|J,Lambda=1,Omega=1> = J*(J+1)

  <J,Lambda=-1,Omega=-1|q|J,Lambda=1,Omega=1> = (J*(J+1))/2
  <J,Lambda=1,Omega=1|q|J,Lambda=-1,Omega=-1> = (J*(J+1))/2
   Symmetrized:
  <J,Lambda=1,Omega=1|q|J,Lambda=1,Omega=1> = (J*(J+1))/2
  <J,Lambda=1,Omega=1|q|J,Lambda=1,Omega=1> = (J*(J+1))/2

  <J,Lambda=-1,Omega=-1|D|J,Lambda=-1,Omega=-1> = -J^2*(J*(J+2)+1)
  <J,Lambda=1,Omega=1|D|J,Lambda=1,Omega=1> = -J^2*(J*(J+2)+1)
   Symmetrized:
  <J,Lambda=1,Omega=1|D|J,Lambda=1,Omega=1> = -J^2*(J*(J+2)+1)
  <J,Lambda=1,Omega=1|D|J,Lambda=1,Omega=1> = -J^2*(J*(J+2)+1)

  <J,Lambda=-1,Omega=-1|H|J,Lambda=-1,Omega=-1> = J^3*(J*(J*(J+3)+3)+1)
  <J,Lambda=1,Omega=1|H|J,Lambda=1,Omega=1> = J^3*(J*(J*(J+3)+3)+1)
   Symmetrized:
  <J,Lambda=1,Omega=1|H|J,Lambda=1,Omega=1> = J^3*(J*(J*(J+3)+3)+1)
  <J,Lambda=1,Omega=1|H|J,Lambda=1,Omega=1> = J^3*(J*(J*(J+3)+3)+1)

 Linear: Excited Iv=3
  Status     Active
  Colour     Fuchsia
  Lambda     Sigma-
  OmegaOrder Auto => Regular
  Origin 67579.2384055729       Float, Std Dev = 0.0154796027418882  State Origin. Estimated lowest level = 67579.2384055729
  B      1.0991868853923        Float, Std Dev = 0.000226647963134712  Rotational constant.
  D      5.77953594122483E-6    Float, Std Dev = 4.04368762739283E-7  Quartic centrifugal distortion.
  H      2.247197E-12           Sextic centrifugal distortion.
  <J,Lambda=0,Omega=0|B|J,Lambda=0,Omega=0> = J*(J+1)
   Symmetrized:
  <J,Lambda=0,Omega=0|B|J,Lambda=0,Omega=0> = J*(J+1)

  <J,Lambda=0,Omega=0|D|J,Lambda=0,Omega=0> = -J^2*(J*(J+2)+1)
   Symmetrized:
  <J,Lambda=0,Omega=0|D|J,Lambda=0,Omega=0> = -J^2*(J*(J+2)+1)

  <J,Lambda=0,Omega=0|H|J,Lambda=0,Omega=0> = J^3*(J*(J*(J+3)+3)+1)
   Symmetrized:
  <J,Lambda=0,Omega=0|H|J,Lambda=0,Omega=0> = J^3*(J*(J*(J+3)+3)+1)

 Linear: Excited Iv=4
  Status     Active
  Colour     Fuchsia
  Lambda     Sigma-
  OmegaOrder Auto => Regular
  Origin 68544.2939028          State Origin. Estimated lowest level = 68544.2939028
  B      1.0840499034           Rotational constant.
  D      5.712347E-6            Quartic centrifugal distortion.
  H      2.247197E-12           Sextic centrifugal distortion.
  <J,Lambda=0,Omega=0|B|J,Lambda=0,Omega=0> = J*(J+1)
```

Symmetrized:
<J,Lambda=0,Omega=0|B|J,Lambda=0,Omega=0> = J*(J+1)

<J,Lambda=0,Omega=0|D|J,Lambda=0,Omega=0> = -J^2*(J*(J+2)+1)
Symmetrized:
<J,Lambda=0,Omega=0|D|J,Lambda=0,Omega=0> = -J^2*(J*(J+2)+1)

<J,Lambda=0,Omega=0|H|J,Lambda=0,Omega=0> = J^3*(J*(J*(J+3)+3)+1)
Symmetrized:
<J,Lambda=0,Omega=0|H|J,Lambda=0,Omega=0> = J^3*(J*(J*(J+3)+3)+1)

LinearPerturbation: Excited <Av=2|J+L-|Iv=3>
Value -0.0504228416460905 Float, Std Dev = 0.00047718670991937 Size of perturbation.
<J,Lambda=-1,Omega=-1|J+L-|J,Lambda=0,Omega=0> = -sqrt(J*(J+1))
<J,Lambda=1,Omega=1|J+L-|J,Lambda=0,Omega=0> = sqrt(J*(J+1))

Symmetrized Matrix Elements of <Av=2|J+L-|Iv=3>
<J,Lambda=1,Omega=1|J+L-|J,Lambda=0,Omega=0> = -sqrt(J*(J+1))

LinearPerturbation: Excited <Av=2|J+L-|Iv=4>
Value 0.06495 Size of perturbation.
<J,Lambda=-1,Omega=-1|J+L-|J,Lambda=0,Omega=0> = -sqrt(J*(J+1))
<J,Lambda=1,Omega=1|J+L-|J,Lambda=0,Omega=0> = sqrt(J*(J+1))

Symmetrized Matrix Elements of <Av=2|J+L-|Iv=4>
<J,Lambda=1,Omega=1|J+L-|J,Lambda=0,Omega=0> = -sqrt(J*(J+1))

Linear: Excited ev=4
Status Active
Colour Lime
S 1
Lambda Sigma-
OmegaOrder Auto => Inverted
Origin 67802.7848875454 Float, Std Dev = 0.00556060271589252 State Origin. Estimated lowest level = 67802.218782138
B 1.0971535399732 Float, Std Dev = 3.00733412851795E-5 Rotational constant.
LambdaSS 0.424579055599668 Float, Std Dev = 0.0115169208698887 Spin - spin coupling constant (=1.5*epsilon).
D 5.34329820601173E-6 Float, Std Dev = 2.51634003019302E-8 Quartic centrifugal distortion.
H -1.498131E-12 Sextic centrifugal distortion.
<J,Lambda=0,Omega=-1|B|J,Lambda=0,Omega=-1> = J*(J+1)
<J,Lambda=0,Omega=-1|B|J,Lambda=0,Omega=0> = -sqrt(2*J*(J+1))
<J,Lambda=0,Omega=0|B|J,Lambda=0,Omega=-1> = -sqrt(2*J*(J+1))
<J,Lambda=0,Omega=0|B|J,Lambda=0,Omega=0> = J*(J+1)+2
<J,Lambda=0,Omega=0|B|J,Lambda=0,Omega=1> = -sqrt(2*J*(J+1))
<J,Lambda=0,Omega=1|B|J,Lambda=0,Omega=0> = -sqrt(2*J*(J+1))
<J,Lambda=0,Omega=1|B|J,Lambda=0,Omega=1> = J*(J+1)
Symmetrized:
<J,Lambda=0,Omega=1|B|J,Lambda=0,Omega=1> = J*(J+1)
<J,Lambda=0,Omega=1|B|J,Lambda=0,Omega=1> = -sqrt(2*J*(J+1))
<J,Lambda=0,Omega=1|B|J,Lambda=0,Omega=0> = -sqrt(2*J*(J+1))
<J,Lambda=0,Omega=0|B|J,Lambda=0,Omega=1> = J*(J+1)+2
<J,Lambda=0,Omega=0|B|J,Lambda=0,Omega=0> = -sqrt(2*J*(J+1))

<J,Lambda=0,Omega=-1|LambdaSS|J,Lambda=0,Omega=-1> = 2/3
<J,Lambda=0,Omega=0|LambdaSS|J,Lambda=0,Omega=0> = -4/3
<J,Lambda=0,Omega=1|LambdaSS|J,Lambda=0,Omega=1> = 2/3
Symmetrized:
<J,Lambda=0,Omega=1|LambdaSS|J,Lambda=0,Omega=1> = 2/3
<J,Lambda=0,Omega=1|LambdaSS|J,Lambda=0,Omega=1> = -4/3
<J,Lambda=0,Omega=0|LambdaSS|J,Lambda=0,Omega=0> = 2/3

<J,Lambda=0,Omega=-1|D|J,Lambda=0,Omega=-1> = -J*(J*(J*(J+2)+3)+2)
<J,Lambda=0,Omega=-1|D|J,Lambda=0,Omega=0> = 2*(J*(J+1)+1)*sqrt(2*J*(J+1))
<J,Lambda=0,Omega=-1|D|J,Lambda=0,Omega=1> = -2*J*(J+1)
<J,Lambda=0,Omega=0|D|J,Lambda=0,Omega=-1> = 2*(J*(J+1)+1)*sqrt(2*J*(J+1))
<J,Lambda=0,Omega=0|D|J,Lambda=0,Omega=0> = -J*(J*(J*(J+2)+9)+8)-4
<J,Lambda=0,Omega=0|D|J,Lambda=0,Omega=1> = 2*(J*(J+1)+1)*sqrt(2*J*(J+1))
<J,Lambda=0,Omega=1|D|J,Lambda=0,Omega=-1> = -2*J*(J+1)
<J,Lambda=0,Omega=1|D|J,Lambda=0,Omega=0> = 2*(J*(J+1)+1)*sqrt(2*J*(J+1))
<J,Lambda=0,Omega=1|D|J,Lambda=0,Omega=1> = -J*(J*(J*(J+2)+3)+2)
Symmetrized:
<J,Lambda=0,Omega=1|D|J,Lambda=0,Omega=1> = -J*(J*(J*(J+2)+3)+2)
<J,Lambda=0,Omega=1|D|J,Lambda=0,Omega=1> = 2*(J*(J+1)+1)*sqrt(2*J*(J+1))
<J,Lambda=0,Omega=1|D|J,Lambda=0,Omega=0> = -2*J*(J+1)
<J,Lambda=0,Omega=0|D|J,Lambda=0,Omega=1> = 2*(J*(J+1)+1)*sqrt(2*J*(J+1))
<J,Lambda=0,Omega=0|D|J,Lambda=0,Omega=0> = -J*(J*(J*(J+2)+9)+8)-4

<J,Lambda=0,Omega=-1|H|J,Lambda=0,Omega=-1> = J*(J*(J*(J*(J*(J+3)+9)+13)+10)+4)
<J,Lambda=0,Omega=-1|H|J,Lambda=0,Omega=0> = (J*(J*(-3*J*(J+2)-13)-10)-4)*sqrt(2*J*(J+1))
<J,Lambda=0,Omega=-1|H|J,Lambda=0,Omega=1> = 2*J*(J*(3*J*(J+2)+5)+2)
<J,Lambda=0,Omega=0|H|J,Lambda=0,Omega=-1> = (J*(J*(-3*J*(J+2)-13)-10)-4)*sqrt(2*J*(J+1))
<J,Lambda=0,Omega=0|H|J,Lambda=0,Omega=0> = J*(J*(J*(J*(J*(J+3)+21)+37)+46)+28)+8
<J,Lambda=0,Omega=0|H|J,Lambda=0,Omega=1> = (J*(J*(-3*J*(J+2)-13)-10)-4)*sqrt(2*J*(J+1))
<J,Lambda=0,Omega=1|H|J,Lambda=0,Omega=-1> = 2*J*(J*(3*J*(J+2)+5)+2)
<J,Lambda=0,Omega=1|H|J,Lambda=0,Omega=0> = (J*(J*(-3*J*(J+2)-13)-10)-4)*sqrt(2*J*(J+1))
<J,Lambda=0,Omega=1|H|J,Lambda=0,Omega=1> = J*(J*(J*(J*(J*(J+3)+9)+13)+10)+4)
Symmetrized:
<J,Lambda=0,Omega=1|H|J,Lambda=0,Omega=1> = J*(J*(J*(J*(J*(J+3)+9)+13)+10)+4)
<J,Lambda=0,Omega=1|H|J,Lambda=0,Omega=1> = (J*(J*(-3*J*(J+2)-13)-10)-4)*sqrt(2*J*(J+1))
<J,Lambda=0,Omega=1|H|J,Lambda=0,Omega=0> = 2*J*(J*(3*J*(J+2)+5)+2)
<J,Lambda=0,Omega=0|H|J,Lambda=0,Omega=1> = (J*(J*(-3*J*(J+2)-13)-10)-4)*sqrt(2*J*(J+1))
<J,Lambda=0,Omega=0|H|J,Lambda=0,Omega=0> = J*(J*(J*(J*(J*(J+3)+21)+37)+46)+28)+8

```
LinearPerturbation: Excited <Av=2|LS|ev=4>
  Value -13.1135722759008          Float, Std Dev = 0.00455886954697559  Size of perturbation.
<J,Lambda=-1,Omega=-1|LS|J,Lambda=0,Omega=-1> = -1/sqrt(3)
<J,Lambda=1,Omega=1|LS|J,Lambda=0,Omega=1> = -1/sqrt(3)

Symmetrized Matrix Elements of <Av=2|LS|ev=4>
<J,Lambda=1,Omega=1|LS|J,Lambda=0,Omega=1> = -1/sqrt(3)
<J,Lambda=1,Omega=1|LS|J,Lambda=0,Omega=1> = -1/sqrt(3)

Linear: Excited dv=7
  Status     Active
  Colour     Blue
  S          1
  Lambda     Delta
  OmegaOrder Auto => Inverted
  Origin   67946.2486639027           Float, Std Dev = 0.00364353428464716  State Origin. Estimated lowest level = 67913.0899089251
  B        1.07973873112873           Float, Std Dev = 1.39040855133839E-5  Rotational constant.
  A        -16.9899214894736          Float, Std Dev = 0.00331409986625827  Spin-orbit coupling constant.
  LambdaSS 1.23163200196255           Float, Std Dev = 0.00538419115269944  Spin - spin coupling constant (=1.5*epsilon).
  gamma    -0.00290540266173026       Float, Std Dev = 0.000257551998475258  Spin rotation coupling constant.
  D        5.12783957207506E-6        Float, Std Dev = 8.04240611537847E-9  Quartic centrifugal distortion.
  H        -5.992526E-13              Sextic centrifugal distortion.
  AD       -9.172647E-5               Centrifugal distortion of A.
  <J,Lambda=-2,Omega=-3|B|J,Lambda=-2,Omega=-3> = J*(J+1)-4
  <J,Lambda=-2,Omega=-3|B|J,Lambda=-2,Omega=-2> = -sqrt(2*(J*(J+1)-6))
  <J,Lambda=-2,Omega=-2|B|J,Lambda=-2,Omega=-3> = -sqrt(2*(J*(J+1)-6))
  <J,Lambda=-2,Omega=-2|B|J,Lambda=-2,Omega=-2> = J*(J+1)+2
  <J,Lambda=-2,Omega=-2|B|J,Lambda=-2,Omega=-1> = -sqrt(2*(J*(J+1)-2))
  <J,Lambda=-2,Omega=-1|B|J,Lambda=-2,Omega=-2> = -sqrt(2*(J*(J+1)-2))
  <J,Lambda=-2,Omega=-1|B|J,Lambda=-2,Omega=-1> = J*(J+1)+4
  <J,Lambda=2,Omega=1|B|J,Lambda=2,Omega=1> = J*(J+1)+4
  <J,Lambda=2,Omega=1|B|J,Lambda=2,Omega=2> = -sqrt(2*(J*(J+1)-2))
  <J,Lambda=2,Omega=2|B|J,Lambda=2,Omega=1> = -sqrt(2*(J*(J+1)-2))
  <J,Lambda=2,Omega=2|B|J,Lambda=2,Omega=2> = J*(J+1)+2
  <J,Lambda=2,Omega=2|B|J,Lambda=2,Omega=3> = -sqrt(2*(J*(J+1)-6))
  <J,Lambda=2,Omega=3|B|J,Lambda=2,Omega=2> = -sqrt(2*(J*(J+1)-6))
  <J,Lambda=2,Omega=3|B|J,Lambda=2,Omega=3> = J*(J+1)-4
    Symmetrized:
  <J,Lambda=2,Omega=3|B|J,Lambda=2,Omega=3> = J*(J+1)-4
  <J,Lambda=2,Omega=3|B|J,Lambda=2,Omega=3> = -sqrt(2*(J*(J+1)-6))
  <J,Lambda=2,Omega=3|B|J,Lambda=2,Omega=2> = -sqrt(2*(J*(J+1)-6))
  <J,Lambda=2,Omega=3|B|J,Lambda=2,Omega=2> = J*(J+1)+2
  <J,Lambda=2,Omega=2|B|J,Lambda=2,Omega=3> = -sqrt(2*(J*(J+1)-2))
  <J,Lambda=2,Omega=2|B|J,Lambda=2,Omega=3> = -sqrt(2*(J*(J+1)-2))
  <J,Lambda=2,Omega=2|B|J,Lambda=2,Omega=2> = J*(J+1)+4
  <J,Lambda=2,Omega=2|B|J,Lambda=2,Omega=2> = J*(J+1)+4
  <J,Lambda=2,Omega=2|B|J,Lambda=2,Omega=1> = -sqrt(2*(J*(J+1)-2))
  <J,Lambda=2,Omega=2|B|J,Lambda=2,Omega=1> = -sqrt(2*(J*(J+1)-2))
  <J,Lambda=2,Omega=1|B|J,Lambda=2,Omega=2> = J*(J+1)+2
  <J,Lambda=2,Omega=1|B|J,Lambda=2,Omega=2> = -sqrt(2*(J*(J+1)-6))
  <J,Lambda=2,Omega=1|B|J,Lambda=2,Omega=1> = -sqrt(2*(J*(J+1)-6))
  <J,Lambda=2,Omega=1|B|J,Lambda=2,Omega=1> = J*(J+1)-4

  <J,Lambda=-2,Omega=-3|A|J,Lambda=-2,Omega=-3> = 2
  <J,Lambda=-2,Omega=-1|A|J,Lambda=-2,Omega=-1> = -2
  <J,Lambda=2,Omega=1|A|J,Lambda=2,Omega=1> = -2
  <J,Lambda=2,Omega=3|A|J,Lambda=2,Omega=3> = 2
    Symmetrized:
  <J,Lambda=2,Omega=3|A|J,Lambda=2,Omega=3> = 2
  <J,Lambda=2,Omega=3|A|J,Lambda=2,Omega=3> = -2
  <J,Lambda=2,Omega=1|A|J,Lambda=2,Omega=1> = -2
  <J,Lambda=2,Omega=1|A|J,Lambda=2,Omega=1> = 2

  <J,Lambda=-2,Omega=-3|LambdaSS|J,Lambda=-2,Omega=-3> = 2/3
  <J,Lambda=-2,Omega=-2|LambdaSS|J,Lambda=-2,Omega=-2> = -4/3
  <J,Lambda=-2,Omega=-1|LambdaSS|J,Lambda=-2,Omega=-1> = 2/3
  <J,Lambda=2,Omega=1|LambdaSS|J,Lambda=2,Omega=1> = 2/3
  <J,Lambda=2,Omega=2|LambdaSS|J,Lambda=2,Omega=2> = -4/3
  <J,Lambda=2,Omega=3|LambdaSS|J,Lambda=2,Omega=3> = 2/3
    Symmetrized:
  <J,Lambda=2,Omega=3|LambdaSS|J,Lambda=2,Omega=3> = 2/3
  <J,Lambda=2,Omega=3|LambdaSS|J,Lambda=2,Omega=3> = -4/3
  <J,Lambda=2,Omega=2|LambdaSS|J,Lambda=2,Omega=2> = 2/3
  <J,Lambda=2,Omega=2|LambdaSS|J,Lambda=2,Omega=2> = 2/3
  <J,Lambda=2,Omega=1|LambdaSS|J,Lambda=2,Omega=1> = -4/3
  <J,Lambda=2,Omega=1|LambdaSS|J,Lambda=2,Omega=1> = 2/3

  <J,Lambda=-2,Omega=-3|gamma|J,Lambda=-2,Omega=-3> = 1
  <J,Lambda=-2,Omega=-3|gamma|J,Lambda=-2,Omega=-2> = sqrt(2*(J*(J+1)-6))/2
  <J,Lambda=-2,Omega=-2|gamma|J,Lambda=-2,Omega=-3> = sqrt(2*(J*(J+1)-6))/2
  <J,Lambda=-2,Omega=-2|gamma|J,Lambda=-2,Omega=-2> = -2
  <J,Lambda=-2,Omega=-2|gamma|J,Lambda=-2,Omega=-1> = sqrt(2*(J*(J+1)-2))/2
  <J,Lambda=-2,Omega=-1|gamma|J,Lambda=-2,Omega=-2> = sqrt(2*(J*(J+1)-2))/2
  <J,Lambda=-2,Omega=-1|gamma|J,Lambda=-2,Omega=-1> = -3
  <J,Lambda=2,Omega=1|gamma|J,Lambda=2,Omega=1> = -3
  <J,Lambda=2,Omega=1|gamma|J,Lambda=2,Omega=2> = sqrt(2*(J*(J+1)-2))/2
  <J,Lambda=2,Omega=2|gamma|J,Lambda=2,Omega=1> = sqrt(2*(J*(J+1)-2))/2
  <J,Lambda=2,Omega=2|gamma|J,Lambda=2,Omega=2> = -2
  <J,Lambda=2,Omega=2|gamma|J,Lambda=2,Omega=3> = sqrt(2*(J*(J+1)-6))/2
  <J,Lambda=2,Omega=3|gamma|J,Lambda=2,Omega=2> = sqrt(2*(J*(J+1)-6))/2
  <J,Lambda=2,Omega=3|gamma|J,Lambda=2,Omega=3> = 1
    Symmetrized:
```

<J,Lambda=2,Omega=3|gamma|J,Lambda=2,Omega=3> = 1
<J,Lambda=2,Omega=3|gamma|J,Lambda=2,Omega=3> = sqrt(2*(J*(J+1)-6))/2
<J,Lambda=2,Omega=3|gamma|J,Lambda=2,Omega=2> = sqrt(2*(J*(J+1)-6))/2
<J,Lambda=2,Omega=3|gamma|J,Lambda=2,Omega=2> = -2
<J,Lambda=2,Omega=2|gamma|J,Lambda=2,Omega=3> = sqrt(2*(J*(J+1)-2))/2
<J,Lambda=2,Omega=2|gamma|J,Lambda=2,Omega=3> = sqrt(2*(J*(J+1)-2))/2
<J,Lambda=2,Omega=2|gamma|J,Lambda=2,Omega=2> = -3
<J,Lambda=2,Omega=2|gamma|J,Lambda=2,Omega=2> = -3
<J,Lambda=2,Omega=2|gamma|J,Lambda=2,Omega=1> = sqrt(2*(J*(J+1)-2))/2
<J,Lambda=2,Omega=2|gamma|J,Lambda=2,Omega=1> = sqrt(2*(J*(J+1)-2))/2
<J,Lambda=2,Omega=1|gamma|J,Lambda=2,Omega=2> = -2
<J,Lambda=2,Omega=1|gamma|J,Lambda=2,Omega=2> = sqrt(2*(J*(J+1)-6))/2
<J,Lambda=2,Omega=1|gamma|J,Lambda=2,Omega=1> = sqrt(2*(J*(J+1)-6))/2
<J,Lambda=2,Omega=1|gamma|J,Lambda=2,Omega=1> = 1

<J,Lambda=-2,Omega=-3|D|J,Lambda=-2,Omega=-3> = -J*(J*(J*(J+2)-5)-6)-4
<J,Lambda=-2,Omega=-3|D|J,Lambda=-2,Omega=-2> = -2*(-J*(J+1)+1)*sqrt(2*(J*(J+1)-6))
<J,Lambda=-2,Omega=-3|D|J,Lambda=-2,Omega=-1> = -2*sqrt(J*(J*(J*(J+2)-7)-8)+12)
<J,Lambda=-2,Omega=-2|D|J,Lambda=-2,Omega=-3> = -2*(-J*(J+1)+1)*sqrt(2*(J*(J+1)-6))
<J,Lambda=-2,Omega=-2|D|J,Lambda=-2,Omega=-2> = -J*(J*(J*(J+2)+9)+8)+12
<J,Lambda=-2,Omega=-2|D|J,Lambda=-2,Omega=-1> = 2*(J*(J+1)+3)*sqrt(2*(J*(J+1)-2))
<J,Lambda=-2,Omega=-1|D|J,Lambda=-2,Omega=-3> = -2*sqrt(J*(J*(J*(J+2)-7)-8)+12)
<J,Lambda=-2,Omega=-1|D|J,Lambda=-2,Omega=-2> = 2*(J*(J+1)+3)*sqrt(2*(J*(J+1)-2))
<J,Lambda=-2,Omega=-1|D|J,Lambda=-2,Omega=-1> = -J*(J*(J*(J+2)+11)+10)-12
<J,Lambda=2,Omega=1|D|J,Lambda=2,Omega=1> = -J*(J*(J*(J+2)+11)+10)-12
<J,Lambda=2,Omega=1|D|J,Lambda=2,Omega=2> = 2*(J*(J+1)+3)*sqrt(2*(J*(J+1)-2))
<J,Lambda=2,Omega=1|D|J,Lambda=2,Omega=3> = -2*sqrt(J*(J*(J*(J+2)-7)-8)+12)
<J,Lambda=2,Omega=2|D|J,Lambda=2,Omega=1> = 2*(J*(J+1)+3)*sqrt(2*(J*(J+1)-2))
<J,Lambda=2,Omega=2|D|J,Lambda=2,Omega=2> = -J*(J*(J*(J+2)+9)+8)+12
<J,Lambda=2,Omega=2|D|J,Lambda=2,Omega=3> = -2*(-J*(J+1)+1)*sqrt(2*(J*(J+1)-6))
<J,Lambda=2,Omega=3|D|J,Lambda=2,Omega=1> = -2*sqrt(J*(J*(J*(J+2)-7)-8)+12)
<J,Lambda=2,Omega=3|D|J,Lambda=2,Omega=2> = -2*(-J*(J+1)+1)*sqrt(2*(J*(J+1)-6))
<J,Lambda=2,Omega=3|D|J,Lambda=2,Omega=3> = -J*(J*(J*(J+2)-5)-6)-4
Symmetrized:
<J,Lambda=2,Omega=3|D|J,Lambda=2,Omega=3> = -J*(J*(J*(J+2)-5)-6)-4
<J,Lambda=2,Omega=3|D|J,Lambda=2,Omega=3> = -2*(-J*(J+1)+1)*sqrt(2*(J*(J+1)-6))
<J,Lambda=2,Omega=3|D|J,Lambda=2,Omega=2> = -2*sqrt(J*(J*(J*(J+2)-7)-8)+12)
<J,Lambda=2,Omega=3|D|J,Lambda=2,Omega=2> = -2*(-J*(J+1)+1)*sqrt(2*(J*(J+1)-6))
<J,Lambda=2,Omega=3|D|J,Lambda=2,Omega=1> = -J*(J*(J*(J+2)+9)+8)+12
<J,Lambda=2,Omega=3|D|J,Lambda=2,Omega=1> = 2*(J*(J+1)+3)*sqrt(2*(J*(J+1)-2))
<J,Lambda=2,Omega=2|D|J,Lambda=2,Omega=3> = -2*sqrt(J*(J*(J*(J+2)-7)-8)+12)
<J,Lambda=2,Omega=2|D|J,Lambda=2,Omega=3> = 2*(J*(J+1)+3)*sqrt(2*(J*(J+1)-2))
<J,Lambda=2,Omega=2|D|J,Lambda=2,Omega=2> = -J*(J*(J*(J+2)+11)+10)-12
<J,Lambda=2,Omega=2|D|J,Lambda=2,Omega=2> = -J*(J*(J*(J+2)+11)+10)-12
<J,Lambda=2,Omega=2|D|J,Lambda=2,Omega=1> = 2*(J*(J+1)+3)*sqrt(2*(J*(J+1)-2))
<J,Lambda=2,Omega=2|D|J,Lambda=2,Omega=1> = -2*sqrt(J*(J*(J*(J+2)-7)-8)+12)
<J,Lambda=2,Omega=1|D|J,Lambda=2,Omega=3> = 2*(J*(J+1)+3)*sqrt(2*(J*(J+1)-2))
<J,Lambda=2,Omega=1|D|J,Lambda=2,Omega=3> = -J*(J*(J*(J+2)+9)+8)+12
<J,Lambda=2,Omega=1|D|J,Lambda=2,Omega=2> = -2*(-J*(J+1)+1)*sqrt(2*(J*(J+1)-6))
<J,Lambda=2,Omega=1|D|J,Lambda=2,Omega=2> = -2*sqrt(J*(J*(J*(J+2)-7)-8)+12)
<J,Lambda=2,Omega=1|D|J,Lambda=2,Omega=1> = -2*(-J*(J+1)+1)*sqrt(2*(J*(J+1)-6))
<J,Lambda=2,Omega=1|D|J,Lambda=2,Omega=1> = -J*(J*(J*(J+2)-5)-6)-4

<J,Lambda=-2,Omega=-3|H|J,Lambda=-2,Omega=-3> = J^2*(J*(J*(J*(J+3)-3)-11)-6)+8
<J,Lambda=-2,Omega=-3|H|J,Lambda=-2,Omega=-2> = (J*(J*(-3*J*(J+2)-1)+2)+4)*sqrt(2*(J*(J+1)-6))
<J,Lambda=-2,Omega=-3|H|J,Lambda=-2,Omega=-1> = 2*(3*J*(J+1)+2)*sqrt(J*(J*(J*(J+2)-7)-8)+12)
<J,Lambda=-2,Omega=-2|H|J,Lambda=-2,Omega=-3> = J*(J*(-3*J*(J+2)+1)+4)*sqrt(2*(J*(J+1)-6))-2*sqrt(2*(J*(J*(J*(J*(J*(J+3)-7)-19)+18)+28)-24))
<J,Lambda=-2,Omega=-2|H|J,Lambda=-2,Omega=-2> = J*(J*(J*(J*(J*(J+3)+21)+37)-2)-20)-24
<J,Lambda=-2,Omega=-2|H|J,Lambda=-2,Omega=-1> = J*(J*(-3*J*(J+2)-23)-20)*sqrt(2*(J*(J+1)-2))-2*sqrt(2*(J*(J*(J*(J*(J*(J+3)-11)-27)+46)+60)-72))-24*sqrt(2*(J*(J+1)-2))
<J,Lambda=-2,Omega=-1|H|J,Lambda=-2,Omega=-3> = 2*(3*J*(J+1)+2)*sqrt(J*(J*(J*(J+2)-7)-8)+12)
<J,Lambda=-2,Omega=-1|H|J,Lambda=-2,Omega=-2> = (J*(J*(-3*J*(J+2)-25)-22)-12)*sqrt(2*(J*(J+1)-2))
<J,Lambda=-2,Omega=-1|H|J,Lambda=-2,Omega=-1> = J*(J*(J*(J*(J*(J+3)+21)+37)+74)+56)+24
<J,Lambda=2,Omega=1|H|J,Lambda=2,Omega=1> = J*(J*(J*(J*(J*(J+3)+21)+37)+74)+56)+24
<J,Lambda=2,Omega=1|H|J,Lambda=2,Omega=2> = (J*(J*(-3*J*(J+2)-25)-22)-12)*sqrt(2*(J*(J+1)-2))
<J,Lambda=2,Omega=1|H|J,Lambda=2,Omega=3> = 2*(3*J*(J+1)+2)*sqrt(J*(J*(J*(J+2)-7)-8)+12)
<J,Lambda=2,Omega=2|H|J,Lambda=2,Omega=1> = J*(J*(-3*J*(J+2)-23)-20)*sqrt(2*(J*(J+1)-2))-2*sqrt(2*(J*(J*(J*(J*(J*(J+3)-11)-27)+46)+60)-72))-24*sqrt(2*(J*(J+1)-2))
<J,Lambda=2,Omega=2|H|J,Lambda=2,Omega=2> = J*(J*(J*(J*(J*(J+3)+21)+37)-2)-20)-24
<J,Lambda=2,Omega=2|H|J,Lambda=2,Omega=3> = J*(J*(-3*J*(J+2)+1)+4)*sqrt(2*(J*(J+1)-6))-2*sqrt(2*(J*(J*(J*(J*(J*(J+3)-7)-19)+18)+28)-24))
<J,Lambda=2,Omega=3|H|J,Lambda=2,Omega=1> = 2*(3*J*(J+1)+2)*sqrt(J*(J*(J*(J+2)-7)-8)+12)
<J,Lambda=2,Omega=3|H|J,Lambda=2,Omega=2> = (J*(J*(-3*J*(J+2)-1)+2)+4)*sqrt(2*(J*(J+1)-6))
<J,Lambda=2,Omega=3|H|J,Lambda=2,Omega=3> = J^2*(J*(J*(J*(J+3)-3)-11)-6)+8
Symmetrized:
<J,Lambda=2,Omega=3|H|J,Lambda=2,Omega=3> = J^2*(J*(J*(J*(J+3)-3)-11)-6)+8
<J,Lambda=2,Omega=3|H|J,Lambda=2,Omega=3> = (J*(J*(-3*J*(J+2)-1)+2)+4)*sqrt(2*(J*(J+1)-6))
<J,Lambda=2,Omega=3|H|J,Lambda=2,Omega=2> = 2*(3*J*(J+1)+2)*sqrt(J*(J*(J*(J+2)-7)-8)+12)
<J,Lambda=2,Omega=3|H|J,Lambda=2,Omega=2> = J*(J*(-3*J*(J+2)+1)+4)*sqrt(2*(J*(J+1)-6))-2*sqrt(2*(J*(J*(J*(J*(J*(J+3)-7)-19)+18)+28)-24))
<J,Lambda=2,Omega=3|H|J,Lambda=2,Omega=1> = J*(J*(J*(J*(J*(J+3)+21)+37)-2)-20)-24
<J,Lambda=2,Omega=3|H|J,Lambda=2,Omega=1> = J*(J*(-3*J*(J+2)-23)-20)*sqrt(2*(J*(J+1)-2))-2*sqrt(2*(J*(J*(J*(J*(J*(J+3)-11)-27)+46)+60)-72))-24*sqrt(2*(J*(J+1)-2))
<J,Lambda=2,Omega=2|H|J,Lambda=2,Omega=3> = 2*(3*J*(J+1)+2)*sqrt(J*(J*(J*(J+2)-7)-8)+12)
<J,Lambda=2,Omega=2|H|J,Lambda=2,Omega=3> = (J*(J*(-3*J*(J+2)-25)-22)-12)*sqrt(2*(J*(J+1)-2))
<J,Lambda=2,Omega=2|H|J,Lambda=2,Omega=2> = J*(J*(J*(J*(J*(J+3)+21)+37)+74)+56)+24
<J,Lambda=2,Omega=2|H|J,Lambda=2,Omega=2> = J*(J*(J*(J*(J*(J+3)+21)+37)+74)+56)+24
<J,Lambda=2,Omega=2|H|J,Lambda=2,Omega=1> = (J*(J*(-3*J*(J+2)-25)-22)-12)*sqrt(2*(J*(J+1)-2))
<J,Lambda=2,Omega=2|H|J,Lambda=2,Omega=1> = 2*(3*J*(J+1)+2)*sqrt(J*(J*(J*(J+2)-7)-8)+12)
<J,Lambda=2,Omega=1|H|J,Lambda=2,Omega=3> = J*(J*(-3*J*(J+2)-23)-20)*sqrt(2*(J*(J+1)-2))-2*sqrt(2*(J*(J*(J*(J*(J*(J+3)-11)-27)+46)+60)-72))-24*sqrt(2*(J*(J+1)-2))
<J,Lambda=2,Omega=1|H|J,Lambda=2,Omega=3> = J*(J*(J*(J*(J*(J+3)+21)+37)-2)-20)-24
<J,Lambda=2,Omega=1|H|J,Lambda=2,Omega=2> = J*(J*(-3*J*(J+2)+1)+4)*sqrt(2*(J*(J+1)-6))-2*sqrt(2*(J*(J*(J*(J*(J*(J+3)-7)-19)+18)+28)-24))
<J,Lambda=2,Omega=1|H|J,Lambda=2,Omega=2> = 2*(3*J*(J+1)+2)*sqrt(J*(J*(J*(J+2)-7)-8)+12)
<J,Lambda=2,Omega=1|H|J,Lambda=2,Omega=1> = (J*(J*(-3*J*(J+2)-1)+2)+4)*sqrt(2*(J*(J+1)-6))
<J,Lambda=2,Omega=1|H|J,Lambda=2,Omega=1> = J^2*(J*(J*(J*(J+3)-3)-11)-6)+8

<J,Lambda=-2,Omega=-3|AD|J,Lambda=-2,Omega=-3> = 2*(J*(J+1)-4)

```
<J,Lambda=-2,Omega=-3|AD|J,Lambda=-2,Omega=-2> = -sqrt(2*(J*(J+1)-6))
<J,Lambda=-2,Omega=-2|AD|J,Lambda=-2,Omega=-3> = -sqrt(2*(J*(J+1)-6))
<J,Lambda=-2,Omega=-2|AD|J,Lambda=-2,Omega=-1> = sqrt(2*(J*(J+1)-2))
<J,Lambda=-2,Omega=-1|AD|J,Lambda=-2,Omega=-2> = sqrt(2*(J*(J+1)-2))
<J,Lambda=-2,Omega=-1|AD|J,Lambda=-2,Omega=-1> = 2*(-J*(J+1)-4)
<J,Lambda=2,Omega=1|AD|J,Lambda=2,Omega=1> = 2*(-J*(J+1)-4)
<J,Lambda=2,Omega=1|AD|J,Lambda=2,Omega=2> = sqrt(2*(J*(J+1)-2))
<J,Lambda=2,Omega=2|AD|J,Lambda=2,Omega=1> = sqrt(2*(J*(J+1)-2))
<J,Lambda=2,Omega=2|AD|J,Lambda=2,Omega=3> = -sqrt(2*(J*(J+1)-6))
<J,Lambda=2,Omega=3|AD|J,Lambda=2,Omega=2> = -sqrt(2*(J*(J+1)-6))
<J,Lambda=2,Omega=3|AD|J,Lambda=2,Omega=3> = 2*(J*(J+1)-4)
  Symmetrized:
<J,Lambda=2,Omega=3|AD|J,Lambda=2,Omega=3> = 2*(J*(J+1)-4)
<J,Lambda=2,Omega=3|AD|J,Lambda=2,Omega=3> = -sqrt(2*(J*(J+1)-6))
<J,Lambda=2,Omega=3|AD|J,Lambda=2,Omega=2> = -sqrt(2*(J*(J+1)-6))
<J,Lambda=2,Omega=3|AD|J,Lambda=2,Omega=2> = sqrt(2*(J*(J+1)-2))
<J,Lambda=2,Omega=2|AD|J,Lambda=2,Omega=3> = sqrt(2*(J*(J+1)-2))
<J,Lambda=2,Omega=2|AD|J,Lambda=2,Omega=3> = 2*(-J*(J+1)-4)
<J,Lambda=2,Omega=2|AD|J,Lambda=2,Omega=1> = 2*(-J*(J+1)-4)
<J,Lambda=2,Omega=2|AD|J,Lambda=2,Omega=1> = sqrt(2*(J*(J+1)-2))
<J,Lambda=2,Omega=1|AD|J,Lambda=2,Omega=2> = sqrt(2*(J*(J+1)-2))
<J,Lambda=2,Omega=1|AD|J,Lambda=2,Omega=2> = -sqrt(2*(J*(J+1)-6))
<J,Lambda=2,Omega=1|AD|J,Lambda=2,Omega=1> = -sqrt(2*(J*(J+1)-6))
<J,Lambda=2,Omega=1|AD|J,Lambda=2,Omega=1> = 2*(J*(J+1)-4)

LinearPerturbation: Excited <Av=2|LS|dv=7>
  Value -9.51852615949637          Float, Std Dev = 0.0104784401443708  Size of perturbation.
<J,Lambda=-1,Omega=-1|LS|J,Lambda=-2,Omega=-1> = 1/sqrt(3)
<J,Lambda=1,Omega=1|LS|J,Lambda=2,Omega=1> = -1/sqrt(3)

Symmetrized Matrix Elements of <Av=2|LS|dv=7>
<J,Lambda=1,Omega=1|LS|J,Lambda=2,Omega=1> = 1/sqrt(3)
<J,Lambda=1,Omega=1|LS|J,Lambda=2,Omega=1> = -1/sqrt(3)

Linear: Excited a'v=12
  Status     Active
  Colour     Red
  S          1
  Lambda     Sigma+
  OmegaOrder Auto => Inverted
  Origin   67991.2956970935           Float, Std Dev = 0.00589195259730653  State Origin. Estimated lowest level = 67990.5318855605
  B        1.03759347486748           Float, Std Dev = 2.24934982641142E-5  Rotational constant.
  LambdaSS -1.14571729964806          Float, Std Dev = 0.0090899360677818  Spin - spin coupling constant (=1.5*epsilon).
  gamma    -0.0052709509830792        Float, Std Dev = 0.000311718263193545  Spin rotation coupling constant.
  D        5.18562893250139E-6        Float, Std Dev = 1.22277506611042E-8  Quartic centrifugal distortion.
  H        -2.996263E-13              Sextic centrifugal distortion.
<J,Lambda=0,Omega=-1|B|J,Lambda=0,Omega=-1> = J*(J+1)
<J,Lambda=0,Omega=-1|B|J,Lambda=0,Omega=0> = -sqrt(2*J*(J+1))
<J,Lambda=0,Omega=0|B|J,Lambda=0,Omega=-1> = -sqrt(2*J*(J+1))
<J,Lambda=0,Omega=0|B|J,Lambda=0,Omega=0> = J*(J+1)+2
<J,Lambda=0,Omega=0|B|J,Lambda=0,Omega=1> = -sqrt(2*J*(J+1))
<J,Lambda=0,Omega=1|B|J,Lambda=0,Omega=0> = -sqrt(2*J*(J+1))
<J,Lambda=0,Omega=1|B|J,Lambda=0,Omega=1> = J*(J+1)
  Symmetrized:
<J,Lambda=0,Omega=1|B|J,Lambda=0,Omega=1> = J*(J+1)
<J,Lambda=0,Omega=1|B|J,Lambda=0,Omega=1> = -sqrt(2*J*(J+1))
<J,Lambda=0,Omega=1|B|J,Lambda=0,Omega=0> = -sqrt(2*J*(J+1))
<J,Lambda=0,Omega=0|B|J,Lambda=0,Omega=1> = J*(J+1)+2
<J,Lambda=0,Omega=0|B|J,Lambda=0,Omega=0> = -sqrt(2*J*(J+1))

<J,Lambda=0,Omega=-1|LambdaSS|J,Lambda=0,Omega=-1> = 2/3
<J,Lambda=0,Omega=0|LambdaSS|J,Lambda=0,Omega=0> = -4/3
<J,Lambda=0,Omega=1|LambdaSS|J,Lambda=0,Omega=1> = 2/3
  Symmetrized:
<J,Lambda=0,Omega=1|LambdaSS|J,Lambda=0,Omega=1> = 2/3
<J,Lambda=0,Omega=1|LambdaSS|J,Lambda=0,Omega=1> = -4/3
<J,Lambda=0,Omega=0|LambdaSS|J,Lambda=0,Omega=0> = 2/3

<J,Lambda=0,Omega=-1|gamma|J,Lambda=0,Omega=-1> = -1
<J,Lambda=0,Omega=-1|gamma|J,Lambda=0,Omega=0> = sqrt(2*J*(J+1))/2
<J,Lambda=0,Omega=0|gamma|J,Lambda=0,Omega=-1> = sqrt(2*J*(J+1))/2
<J,Lambda=0,Omega=0|gamma|J,Lambda=0,Omega=0> = -2
<J,Lambda=0,Omega=0|gamma|J,Lambda=0,Omega=1> = sqrt(2*J*(J+1))/2
<J,Lambda=0,Omega=1|gamma|J,Lambda=0,Omega=0> = sqrt(2*J*(J+1))/2
<J,Lambda=0,Omega=1|gamma|J,Lambda=0,Omega=1> = -1
  Symmetrized:
<J,Lambda=0,Omega=1|gamma|J,Lambda=0,Omega=1> = -1
<J,Lambda=0,Omega=1|gamma|J,Lambda=0,Omega=1> = sqrt(2*J*(J+1))/2
<J,Lambda=0,Omega=1|gamma|J,Lambda=0,Omega=0> = sqrt(2*J*(J+1))/2
<J,Lambda=0,Omega=0|gamma|J,Lambda=0,Omega=1> = -2
<J,Lambda=0,Omega=0|gamma|J,Lambda=0,Omega=0> = sqrt(2*J*(J+1))/2

<J,Lambda=0,Omega=-1|D|J,Lambda=0,Omega=-1> = -J*(J*(J*(J+2)+3)+2)
<J,Lambda=0,Omega=-1|D|J,Lambda=0,Omega=0> = 2*(J*(J+1)+1)*sqrt(2*J*(J+1))
<J,Lambda=0,Omega=-1|D|J,Lambda=0,Omega=1> = -2*J*(J+1)
<J,Lambda=0,Omega=0|D|J,Lambda=0,Omega=-1> = 2*(J*(J+1)+1)*sqrt(2*J*(J+1))
<J,Lambda=0,Omega=0|D|J,Lambda=0,Omega=0> = -J*(J*(J*(J+2)+9)+8)-4
<J,Lambda=0,Omega=0|D|J,Lambda=0,Omega=1> = 2*(J*(J+1)+1)*sqrt(2*J*(J+1))
<J,Lambda=0,Omega=1|D|J,Lambda=0,Omega=-1> = -2*J*(J+1)
<J,Lambda=0,Omega=1|D|J,Lambda=0,Omega=0> = 2*(J*(J+1)+1)*sqrt(2*J*(J+1))
<J,Lambda=0,Omega=1|D|J,Lambda=0,Omega=1> = -J*(J*(J*(J+2)+3)+2)
  Symmetrized:
```

```
  <J,Lambda=0,Omega=1|D|J,Lambda=0,Omega=1> = -J*(J*(J*(J+2)+3)+2)
  <J,Lambda=0,Omega=1|D|J,Lambda=0,Omega=1> = 2*(J*(J+1)+1)*sqrt(2*J*(J+1))
  <J,Lambda=0,Omega=1|D|J,Lambda=0,Omega=0> = -2*J*(J+1)
  <J,Lambda=0,Omega=0|D|J,Lambda=0,Omega=1> = 2*(J*(J+1)+1)*sqrt(2*J*(J+1))
  <J,Lambda=0,Omega=0|D|J,Lambda=0,Omega=0> = -J*(J*(J*(J+2)+9)+8)-4

  <J,Lambda=0,Omega=-1|H|J,Lambda=0,Omega=-1> = J*(J*(J*(J*(J*(J+3)+9)+13)+10)+4)
  <J,Lambda=0,Omega=-1|H|J,Lambda=0,Omega=0> = (J*(J*(-3*J*(J+2)-13)-10)-4)*sqrt(2*J*(J+1))
  <J,Lambda=0,Omega=-1|H|J,Lambda=0,Omega=1> = 2*J*(J*(3*J*(J+2)+5)+2)
  <J,Lambda=0,Omega=0|H|J,Lambda=0,Omega=-1> = (J*(J*(-3*J*(J+2)-13)-10)-4)*sqrt(2*J*(J+1))
  <J,Lambda=0,Omega=0|H|J,Lambda=0,Omega=0> = J*(J*(J*(J*(J*(J+3)+21)+37)+46)+28)+8
  <J,Lambda=0,Omega=0|H|J,Lambda=0,Omega=1> = (J*(J*(-3*J*(J+2)-13)-10)-4)*sqrt(2*J*(J+1))
  <J,Lambda=0,Omega=1|H|J,Lambda=0,Omega=-1> = 2*J*(J*(3*J*(J+2)+5)+2)
  <J,Lambda=0,Omega=1|H|J,Lambda=0,Omega=0> = (J*(J*(-3*J*(J+2)-13)-10)-4)*sqrt(2*J*(J+1))
  <J,Lambda=0,Omega=1|H|J,Lambda=0,Omega=1> = J*(J*(J*(J*(J*(J+3)+9)+13)+10)+4)
   Symmetrized:
  <J,Lambda=0,Omega=1|H|J,Lambda=0,Omega=1> = J*(J*(J*(J*(J*(J+3)+9)+13)+10)+4)
  <J,Lambda=0,Omega=1|H|J,Lambda=0,Omega=1> = (J*(J*(-3*J*(J+2)-13)-10)-4)*sqrt(2*J*(J+1))
  <J,Lambda=0,Omega=1|H|J,Lambda=0,Omega=0> = 2*J*(J*(3*J*(J+2)+5)+2)
  <J,Lambda=0,Omega=0|H|J,Lambda=0,Omega=1> = (J*(J*(-3*J*(J+2)-13)-10)-4)*sqrt(2*J*(J+1))
  <J,Lambda=0,Omega=0|H|J,Lambda=0,Omega=0> = J*(J*(J*(J*(J*(J+3)+21)+37)+46)+28)+8

LinearPerturbation: Excited <Av=2|LS|a'v=12>
  Value -5.90222290190178          Float, Std Dev = 0.0114847379051065  Size of perturbation.
<J,Lambda=-1,Omega=-1|LS|J,Lambda=0,Omega=-1> = 1/sqrt(3)
<J,Lambda=1,Omega=1|LS|J,Lambda=0,Omega=1> = -1/sqrt(3)

Symmetrized Matrix Elements of <Av=2|LS|a'v=12>
<J,Lambda=1,Omega=1|LS|J,Lambda=0,Omega=1> = 1/sqrt(3)
<J,Lambda=1,Omega=1|LS|J,Lambda=0,Omega=1> = -1/sqrt(3)

Linear: Excited Dsv=3
  Status     Active
  Colour     Orange
  Lambda     Delta
  OmegaOrder Auto => Regular
  Origin 68377.912351697            Float, Std Dev = 0.0503683180380362  State Origin. Estimated lowest level = 68377.912351697
  B      1.08921912165493           Float, Std Dev = 0.000187528245029578  Rotational constant.
  D      5.85537632665456E-6        Float, Std Dev = 7.96715100187949E-8  Quartic centrifugal distortion.
  H      -2.247197E-13              Sextic centrifugal distortion.
  <J,Lambda=-2,Omega=-2|B|J,Lambda=-2,Omega=-2> = J*(J+1)
  <J,Lambda=2,Omega=2|B|J,Lambda=2,Omega=2> = J*(J+1)
   Symmetrized:
  <J,Lambda=2,Omega=2|B|J,Lambda=2,Omega=2> = J*(J+1)
  <J,Lambda=2,Omega=2|B|J,Lambda=2,Omega=2> = J*(J+1)

  <J,Lambda=-2,Omega=-2|D|J,Lambda=-2,Omega=-2> = -J^2*(J*(J+2)+1)
  <J,Lambda=2,Omega=2|D|J,Lambda=2,Omega=2> = -J^2*(J*(J+2)+1)
   Symmetrized:
  <J,Lambda=2,Omega=2|D|J,Lambda=2,Omega=2> = -J^2*(J*(J+2)+1)
  <J,Lambda=2,Omega=2|D|J,Lambda=2,Omega=2> = -J^2*(J*(J+2)+1)

  <J,Lambda=-2,Omega=-2|H|J,Lambda=-2,Omega=-2> = J^3*(J*(J*(J+3)+3)+1)
  <J,Lambda=2,Omega=2|H|J,Lambda=2,Omega=2> = J^3*(J*(J*(J+3)+3)+1)
   Symmetrized:
  <J,Lambda=2,Omega=2|H|J,Lambda=2,Omega=2> = J^3*(J*(J*(J+3)+3)+1)
  <J,Lambda=2,Omega=2|H|J,Lambda=2,Omega=2> = J^3*(J*(J*(J+3)+3)+1)

LinearPerturbation: Excited <Av=2|J+L-|Dsv=3>
  Value 0.02597                     Size of perturbation.
<J,Lambda=-1,Omega=-1|J+L-|J,Lambda=-2,Omega=-2> = sqrt(J*(J+1)-2)
<J,Lambda=1,Omega=1|J+L-|J,Lambda=2,Omega=2> = sqrt(J*(J+1)-2)

Symmetrized Matrix Elements of <Av=2|J+L-|Dsv=3>
<J,Lambda=1,Omega=1|J+L-|J,Lambda=2,Omega=2> = sqrt(J*(J+1)-2)
<J,Lambda=1,Omega=1|J+L-|J,Lambda=2,Omega=2> = sqrt(J*(J+1)-2)

Linear: Excited a3Pv=13
  Status     Active
  Colour     Silver
  S          1
  Lambda     Pi
  OmegaOrder Auto => Regular
  Origin   67670.6779227765         Float, Std Dev = 0.00506836342284951  State Origin. Estimated lowest level = 67632.5503916685
  B        1.30813765113497         Float, Std Dev = 5.56718911902302E-7  Rotational constant.
  A        38.1153846346168         Float, Std Dev = 0.0129524282362973  Spin-orbit coupling constant.
  LambdaSS -0.01821971              Spin - spin coupling constant (=1.5*epsilon).
  o        0.6497647                Spin-spin coupling constant giving lambda doubling for Omega = 0 states.
  gamma    0.003153713              Spin rotation coupling constant.
  p        0.008831129              Lambda doubling constant.
  q        2.786443E-5              Lambda doubling constant.
  D        6.06155910230364E-6      Float, Std Dev = 4.64536445607895E-9  Quartic centrifugal distortion.
  AD       -0.0001940953            Centrifugal distortion of A.
  <J,Lambda=-1,Omega=-2|B|J,Lambda=-1,Omega=-2> = J*(J+1)-2
  <J,Lambda=-1,Omega=-2|B|J,Lambda=-1,Omega=-1> = -sqrt(2*(J*(J+1)-2))
  <J,Lambda=-1,Omega=-1|B|J,Lambda=-1,Omega=-2> = -sqrt(2*(J*(J+1)-2))
  <J,Lambda=-1,Omega=-1|B|J,Lambda=-1,Omega=-1> = J*(J+1)+2
  <J,Lambda=-1,Omega=-1|B|J,Lambda=-1,Omega=0> = -sqrt(2*J*(J+1))
  <J,Lambda=-1,Omega=0|B|J,Lambda=-1,Omega=-1> = -sqrt(2*J*(J+1))
  <J,Lambda=-1,Omega=0|B|J,Lambda=-1,Omega=0> = J*(J+1)+2
  <J,Lambda=1,Omega=0|B|J,Lambda=1,Omega=0> = J*(J+1)+2
  <J,Lambda=1,Omega=0|B|J,Lambda=1,Omega=1> = -sqrt(2*J*(J+1))
```

<J,Lambda=1,Omega=1|B|J,Lambda=1,Omega=0> = -sqrt(2*J*(J+1))
<J,Lambda=1,Omega=1|B|J,Lambda=1,Omega=1> = J*(J+1)+2
<J,Lambda=1,Omega=1|B|J,Lambda=1,Omega=2> = -sqrt(2*(J*(J+1)-2))
<J,Lambda=1,Omega=2|B|J,Lambda=1,Omega=1> = -sqrt(2*(J*(J+1)-2))
<J,Lambda=1,Omega=2|B|J,Lambda=1,Omega=2> = J*(J+1)-2
Symmetrized:
<J,Lambda=1,Omega=2|B|J,Lambda=1,Omega=2> = J*(J+1)-2
<J,Lambda=1,Omega=2|B|J,Lambda=1,Omega=2> = -sqrt(2*(J*(J+1)-2))
<J,Lambda=1,Omega=2|B|J,Lambda=1,Omega=1> = -sqrt(2*(J*(J+1)-2))
<J,Lambda=1,Omega=2|B|J,Lambda=1,Omega=1> = J*(J+1)+2
<J,Lambda=1,Omega=1|B|J,Lambda=1,Omega=2> = -sqrt(2*J*(J+1))
<J,Lambda=1,Omega=1|B|J,Lambda=1,Omega=2> = -sqrt(2*J*(J+1))
<J,Lambda=1,Omega=1|B|J,Lambda=1,Omega=1> = J*(J+1)+2
<J,Lambda=1,Omega=1|B|J,Lambda=1,Omega=1> = J*(J+1)+2
<J,Lambda=1,Omega=1|B|J,Lambda=1,Omega=0> = -sqrt(2*J*(J+1))
<J,Lambda=1,Omega=1|B|J,Lambda=1,Omega=0> = -sqrt(2*J*(J+1))
<J,Lambda=1,Omega=0|B|J,Lambda=1,Omega=1> = J*(J+1)+2
<J,Lambda=1,Omega=0|B|J,Lambda=1,Omega=1> = -sqrt(2*(J*(J+1)-2))
<J,Lambda=1,Omega=0|B|J,Lambda=1,Omega=0> = -sqrt(2*(J*(J+1)-2))
<J,Lambda=1,Omega=0|B|J,Lambda=1,Omega=0> = J*(J+1)-2

<J,Lambda=-1,Omega=-2|A|J,Lambda=-1,Omega=-2> = 1
<J,Lambda=-1,Omega=0|A|J,Lambda=-1,Omega=0> = -1
<J,Lambda=1,Omega=0|A|J,Lambda=1,Omega=0> = -1
<J,Lambda=1,Omega=2|A|J,Lambda=1,Omega=2> = 1
Symmetrized:
<J,Lambda=1,Omega=2|A|J,Lambda=1,Omega=2> = 1
<J,Lambda=1,Omega=2|A|J,Lambda=1,Omega=2> = -1
<J,Lambda=1,Omega=0|A|J,Lambda=1,Omega=0> = -1
<J,Lambda=1,Omega=0|A|J,Lambda=1,Omega=0> = 1

<J,Lambda=-1,Omega=-2|LambdaSS|J,Lambda=-1,Omega=-2> = 2/3
<J,Lambda=-1,Omega=-1|LambdaSS|J,Lambda=-1,Omega=-1> = -4/3
<J,Lambda=-1,Omega=0|LambdaSS|J,Lambda=-1,Omega=0> = 2/3
<J,Lambda=1,Omega=0|LambdaSS|J,Lambda=1,Omega=0> = 2/3
<J,Lambda=1,Omega=1|LambdaSS|J,Lambda=1,Omega=1> = -4/3
<J,Lambda=1,Omega=2|LambdaSS|J,Lambda=1,Omega=2> = 2/3
Symmetrized:
<J,Lambda=1,Omega=2|LambdaSS|J,Lambda=1,Omega=2> = 2/3
<J,Lambda=1,Omega=2|LambdaSS|J,Lambda=1,Omega=2> = -4/3
<J,Lambda=1,Omega=1|LambdaSS|J,Lambda=1,Omega=1> = 2/3
<J,Lambda=1,Omega=1|LambdaSS|J,Lambda=1,Omega=1> = 2/3
<J,Lambda=1,Omega=0|LambdaSS|J,Lambda=1,Omega=0> = -4/3
<J,Lambda=1,Omega=0|LambdaSS|J,Lambda=1,Omega=0> = 2/3

<J,Lambda=-1,Omega=0|o|J,Lambda=1,Omega=0> = 1
<J,Lambda=1,Omega=0|o|J,Lambda=-1,Omega=0> = 1
Symmetrized:
<J,Lambda=1,Omega=0|o|J,Lambda=1,Omega=0> = 1
<J,Lambda=1,Omega=0|o|J,Lambda=1,Omega=0> = 1

<J,Lambda=-1,Omega=-2|gamma|J,Lambda=-1,Omega=-1> = sqrt(2*(J*(J+1)-2))/2
<J,Lambda=-1,Omega=-1|gamma|J,Lambda=-1,Omega=-2> = sqrt(2*(J*(J+1)-2))/2
<J,Lambda=-1,Omega=-1|gamma|J,Lambda=-1,Omega=-1> = -2
<J,Lambda=-1,Omega=-1|gamma|J,Lambda=-1,Omega=0> = sqrt(2*J*(J+1))/2
<J,Lambda=-1,Omega=0|gamma|J,Lambda=-1,Omega=-1> = sqrt(2*J*(J+1))/2
<J,Lambda=-1,Omega=0|gamma|J,Lambda=-1,Omega=0> = -2
<J,Lambda=1,Omega=0|gamma|J,Lambda=1,Omega=0> = -2
<J,Lambda=1,Omega=0|gamma|J,Lambda=1,Omega=1> = sqrt(2*J*(J+1))/2
<J,Lambda=1,Omega=1|gamma|J,Lambda=1,Omega=0> = sqrt(2*J*(J+1))/2
<J,Lambda=1,Omega=1|gamma|J,Lambda=1,Omega=1> = -2
<J,Lambda=1,Omega=1|gamma|J,Lambda=1,Omega=2> = sqrt(2*(J*(J+1)-2))/2
<J,Lambda=1,Omega=2|gamma|J,Lambda=1,Omega=1> = sqrt(2*(J*(J+1)-2))/2
Symmetrized:
<J,Lambda=1,Omega=2|gamma|J,Lambda=1,Omega=1> = sqrt(2*(J*(J+1)-2))/2
<J,Lambda=1,Omega=2|gamma|J,Lambda=1,Omega=1> = sqrt(2*(J*(J+1)-2))/2
<J,Lambda=1,Omega=1|gamma|J,Lambda=1,Omega=2> = -2
<J,Lambda=1,Omega=1|gamma|J,Lambda=1,Omega=2> = sqrt(2*J*(J+1))/2
<J,Lambda=1,Omega=1|gamma|J,Lambda=1,Omega=1> = sqrt(2*J*(J+1))/2
<J,Lambda=1,Omega=1|gamma|J,Lambda=1,Omega=1> = -2
<J,Lambda=1,Omega=1|gamma|J,Lambda=1,Omega=0> = -2
<J,Lambda=1,Omega=1|gamma|J,Lambda=1,Omega=0> = sqrt(2*J*(J+1))/2
<J,Lambda=1,Omega=0|gamma|J,Lambda=1,Omega=1> = sqrt(2*J*(J+1))/2
<J,Lambda=1,Omega=0|gamma|J,Lambda=1,Omega=1> = -2
<J,Lambda=1,Omega=0|gamma|J,Lambda=1,Omega=0> = sqrt(2*(J*(J+1)-2))/2
<J,Lambda=1,Omega=0|gamma|J,Lambda=1,Omega=0> = sqrt(2*(J*(J+1)-2))/2

<J,Lambda=-1,Omega=-1|p|J,Lambda=1,Omega=0> = (-sqrt(2*J*(J+1)))/2
<J,Lambda=-1,Omega=0|p|J,Lambda=1,Omega=0> = 1
<J,Lambda=-1,Omega=0|p|J,Lambda=1,Omega=1> = (-sqrt(2*J*(J+1)))/2
<J,Lambda=1,Omega=0|p|J,Lambda=-1,Omega=-1> = (-sqrt(2*J*(J+1)))/2
<J,Lambda=1,Omega=0|p|J,Lambda=-1,Omega=0> = 1
<J,Lambda=1,Omega=1|p|J,Lambda=-1,Omega=0> = (-sqrt(2*J*(J+1)))/2
Symmetrized:
<J,Lambda=1,Omega=1|p|J,Lambda=1,Omega=0> = (-sqrt(2*J*(J+1)))/2
<J,Lambda=1,Omega=1|p|J,Lambda=1,Omega=0> = 1
<J,Lambda=1,Omega=0|p|J,Lambda=1,Omega=0> = (-sqrt(2*J*(J+1)))/2
<J,Lambda=1,Omega=0|p|J,Lambda=1,Omega=0> = (-sqrt(2*J*(J+1)))/2
<J,Lambda=1,Omega=0|p|J,Lambda=1,Omega=1> = 1
<J,Lambda=1,Omega=0|p|J,Lambda=1,Omega=1> = (-sqrt(2*J*(J+1)))/2

<J,Lambda=-1,Omega=-2|q|J,Lambda=1,Omega=0> = sqrt(J*(J*(J+1)-2)*(J+1))/2

<J,Lambda=-1,Omega=-1|q|J,Lambda=1,Omega=0> = -sqrt(2*J*(J+1))
<J,Lambda=-1,Omega=-1|q|J,Lambda=1,Omega=1> = (J*(J+1))/2
<J,Lambda=-1,Omega=0|q|J,Lambda=1,Omega=0> = 1
<J,Lambda=-1,Omega=0|q|J,Lambda=1,Omega=1> = -sqrt(2*J*(J+1))
<J,Lambda=-1,Omega=0|q|J,Lambda=1,Omega=2> = sqrt(J*(J*(J+1)-2)*(J+1))/2
<J,Lambda=1,Omega=0|q|J,Lambda=-1,Omega=-2> = sqrt(J*(J*(J+1)-2)*(J+1))/2
<J,Lambda=1,Omega=0|q|J,Lambda=-1,Omega=-1> = -sqrt(2*J*(J+1))
<J,Lambda=1,Omega=0|q|J,Lambda=-1,Omega=0> = 1
<J,Lambda=1,Omega=1|q|J,Lambda=-1,Omega=-1> = (J*(J+1))/2
<J,Lambda=1,Omega=1|q|J,Lambda=-1,Omega=0> = -sqrt(2*J*(J+1))
<J,Lambda=1,Omega=2|q|J,Lambda=-1,Omega=0> = sqrt(J*(J*(J+1)-2)*(J+1))/2
Symmetrized:
<J,Lambda=1,Omega=2|q|J,Lambda=1,Omega=0> = sqrt(J*(J*(J+1)-2)*(J+1))/2
<J,Lambda=1,Omega=2|q|J,Lambda=1,Omega=0> = -sqrt(2*J*(J+1))
<J,Lambda=1,Omega=1|q|J,Lambda=1,Omega=0> = (J*(J+1))/2
<J,Lambda=1,Omega=1|q|J,Lambda=1,Omega=0> = 1
<J,Lambda=1,Omega=1|q|J,Lambda=1,Omega=1> = -sqrt(2*J*(J+1))
<J,Lambda=1,Omega=1|q|J,Lambda=1,Omega=1> = sqrt(J*(J*(J+1)-2)*(J+1))/2
<J,Lambda=1,Omega=0|q|J,Lambda=1,Omega=0> = sqrt(J*(J*(J+1)-2)*(J+1))/2
<J,Lambda=1,Omega=0|q|J,Lambda=1,Omega=0> = -sqrt(2*J*(J+1))
<J,Lambda=1,Omega=0|q|J,Lambda=1,Omega=1> = 1
<J,Lambda=1,Omega=0|q|J,Lambda=1,Omega=1> = (J*(J+1))/2
<J,Lambda=1,Omega=0|q|J,Lambda=1,Omega=2> = -sqrt(2*J*(J+1))
<J,Lambda=1,Omega=0|q|J,Lambda=1,Omega=2> = sqrt(J*(J*(J+1)-2)*(J+1))/2

<J,Lambda=-1,Omega=-2|D|J,Lambda=-1,Omega=-2> = -J*(J*(J*(J+2)-1)-2)
<J,Lambda=-1,Omega=-2|D|J,Lambda=-1,Omega=-1> = 2*J*(J+1)*sqrt(2*(J*(J+1)-2))
<J,Lambda=-1,Omega=-2|D|J,Lambda=-1,Omega=0> = -2*sqrt(J*(J*(J*(J+2)-1)-2))
<J,Lambda=-1,Omega=-1|D|J,Lambda=-1,Omega=-2> = 2*J*(J+1)*sqrt(2*(J*(J+1)-2))
<J,Lambda=-1,Omega=-1|D|J,Lambda=-1,Omega=-1> = -J*(J*(J*(J+2)+9)+8)
<J,Lambda=-1,Omega=-1|D|J,Lambda=-1,Omega=0> = 2*(J*(J+1)+2)*sqrt(2*J*(J+1))
<J,Lambda=-1,Omega=0|D|J,Lambda=-1,Omega=-2> = -2*sqrt(J*(J*(J*(J+2)-1)-2))
<J,Lambda=-1,Omega=0|D|J,Lambda=-1,Omega=-1> = 2*(J*(J+1)+2)*sqrt(2*J*(J+1))
<J,Lambda=-1,Omega=0|D|J,Lambda=-1,Omega=0> = -J*(J*(J*(J+2)+7)+6)-4
<J,Lambda=1,Omega=0|D|J,Lambda=1,Omega=0> = -J*(J*(J*(J+2)+7)+6)-4
<J,Lambda=1,Omega=0|D|J,Lambda=1,Omega=1> = 2*(J*(J+1)+2)*sqrt(2*J*(J+1))
<J,Lambda=1,Omega=0|D|J,Lambda=1,Omega=2> = -2*sqrt(J*(J*(J*(J+2)-1)-2))
<J,Lambda=1,Omega=1|D|J,Lambda=1,Omega=0> = 2*(J*(J+1)+2)*sqrt(2*J*(J+1))
<J,Lambda=1,Omega=1|D|J,Lambda=1,Omega=1> = -J*(J*(J*(J+2)+9)+8)
<J,Lambda=1,Omega=1|D|J,Lambda=1,Omega=2> = 2*J*(J+1)*sqrt(2*(J*(J+1)-2))
<J,Lambda=1,Omega=2|D|J,Lambda=1,Omega=0> = -2*sqrt(J*(J*(J*(J+2)-1)-2))
<J,Lambda=1,Omega=2|D|J,Lambda=1,Omega=1> = 2*J*(J+1)*sqrt(2*(J*(J+1)-2))
<J,Lambda=1,Omega=2|D|J,Lambda=1,Omega=2> = -J*(J*(J*(J+2)-1)-2)
Symmetrized:
<J,Lambda=1,Omega=2|D|J,Lambda=1,Omega=2> = -J*(J*(J*(J+2)-1)-2)
<J,Lambda=1,Omega=2|D|J,Lambda=1,Omega=2> = 2*J*(J+1)*sqrt(2*(J*(J+1)-2))
<J,Lambda=1,Omega=2|D|J,Lambda=1,Omega=1> = -2*sqrt(J*(J*(J*(J+2)-1)-2))
<J,Lambda=1,Omega=2|D|J,Lambda=1,Omega=1> = 2*J*(J+1)*sqrt(2*(J*(J+1)-2))
<J,Lambda=1,Omega=2|D|J,Lambda=1,Omega=0> = -J*(J*(J*(J+2)+9)+8)
<J,Lambda=1,Omega=2|D|J,Lambda=1,Omega=0> = 2*(J*(J+1)+2)*sqrt(2*J*(J+1))
<J,Lambda=1,Omega=1|D|J,Lambda=1,Omega=2> = -2*sqrt(J*(J*(J*(J+2)-1)-2))
<J,Lambda=1,Omega=1|D|J,Lambda=1,Omega=2> = 2*(J*(J+1)+2)*sqrt(2*J*(J+1))
<J,Lambda=1,Omega=1|D|J,Lambda=1,Omega=1> = -J*(J*(J*(J+2)+7)+6)-4
<J,Lambda=1,Omega=1|D|J,Lambda=1,Omega=1> = -J*(J*(J*(J+2)+7)+6)-4
<J,Lambda=1,Omega=1|D|J,Lambda=1,Omega=0> = 2*(J*(J+1)+2)*sqrt(2*J*(J+1))
<J,Lambda=1,Omega=1|D|J,Lambda=1,Omega=0> = -2*sqrt(J*(J*(J*(J+2)-1)-2))
<J,Lambda=1,Omega=0|D|J,Lambda=1,Omega=2> = 2*(J*(J+1)+2)*sqrt(2*J*(J+1))
<J,Lambda=1,Omega=0|D|J,Lambda=1,Omega=2> = -J*(J*(J*(J+2)+9)+8)
<J,Lambda=1,Omega=0|D|J,Lambda=1,Omega=1> = 2*J*(J+1)*sqrt(2*(J*(J+1)-2))
<J,Lambda=1,Omega=0|D|J,Lambda=1,Omega=1> = -2*sqrt(J*(J*(J*(J+2)-1)-2))
<J,Lambda=1,Omega=0|D|J,Lambda=1,Omega=0> = 2*J*(J+1)*sqrt(2*(J*(J+1)-2))
<J,Lambda=1,Omega=0|D|J,Lambda=1,Omega=0> = -J*(J*(J*(J+2)-1)-2)

<J,Lambda=-1,Omega=-2|AD|J,Lambda=-1,Omega=-2> = J*(J+1)-2
<J,Lambda=-1,Omega=-2|AD|J,Lambda=-1,Omega=-1> = (-sqrt(2*(J*(J+1)-2)))/2
<J,Lambda=-1,Omega=-1|AD|J,Lambda=-1,Omega=-2> = (-sqrt(2*(J*(J+1)-2)))/2
<J,Lambda=-1,Omega=-1|AD|J,Lambda=-1,Omega=0> = sqrt(2*J*(J+1))/2
<J,Lambda=-1,Omega=0|AD|J,Lambda=-1,Omega=-1> = sqrt(2*J*(J+1))/2
<J,Lambda=-1,Omega=0|AD|J,Lambda=-1,Omega=0> = -J*(J+1)-2
<J,Lambda=1,Omega=0|AD|J,Lambda=1,Omega=0> = -J*(J+1)-2
<J,Lambda=1,Omega=0|AD|J,Lambda=1,Omega=1> = sqrt(2*J*(J+1))/2
<J,Lambda=1,Omega=1|AD|J,Lambda=1,Omega=0> = sqrt(2*J*(J+1))/2
<J,Lambda=1,Omega=1|AD|J,Lambda=1,Omega=2> = (-sqrt(2*(J*(J+1)-2)))/2
<J,Lambda=1,Omega=2|AD|J,Lambda=1,Omega=1> = (-sqrt(2*(J*(J+1)-2)))/2
<J,Lambda=1,Omega=2|AD|J,Lambda=1,Omega=2> = J*(J+1)-2
Symmetrized:
<J,Lambda=1,Omega=2|AD|J,Lambda=1,Omega=2> = J*(J+1)-2
<J,Lambda=1,Omega=2|AD|J,Lambda=1,Omega=2> = (-sqrt(2*(J*(J+1)-2)))/2
<J,Lambda=1,Omega=2|AD|J,Lambda=1,Omega=1> = (-sqrt(2*(J*(J+1)-2)))/2
<J,Lambda=1,Omega=2|AD|J,Lambda=1,Omega=1> = sqrt(2*J*(J+1))/2
<J,Lambda=1,Omega=1|AD|J,Lambda=1,Omega=2> = sqrt(2*J*(J+1))/2
<J,Lambda=1,Omega=1|AD|J,Lambda=1,Omega=2> = -J*(J+1)-2
<J,Lambda=1,Omega=1|AD|J,Lambda=1,Omega=0> = -J*(J+1)-2
<J,Lambda=1,Omega=1|AD|J,Lambda=1,Omega=0> = sqrt(2*J*(J+1))/2
<J,Lambda=1,Omega=0|AD|J,Lambda=1,Omega=1> = sqrt(2*J*(J+1))/2
<J,Lambda=1,Omega=0|AD|J,Lambda=1,Omega=1> = (-sqrt(2*(J*(J+1)-2)))/2
<J,Lambda=1,Omega=0|AD|J,Lambda=1,Omega=0> = (-sqrt(2*(J*(J+1)-2)))/2
<J,Lambda=1,Omega=0|AD|J,Lambda=1,Omega=0> = J*(J+1)-2

LinearPerturbation: Excited <a3Pv=13|LS|Iv=3>
Value -11.5278 Size of perturbation.
<J,Lambda=-1,Omega=0|LS|J,Lambda=0,Omega=0> = -1/sqrt(3)

<J,Lambda=1,Omega=0|LS|J,Lambda=0,Omega=0> = -1/sqrt(3)

Symmetrized Matrix Elements of <a3Pv=13|LS|Iv=3>
<J,Lambda=1,Omega=0|LS|J,Lambda=0,Omega=0> = -1/sqrt(3)

LinearPerturbation: Excited <a3Pv=13|LS|Iv=4>
Value -16.442 Size of perturbation.
<J,Lambda=-1,Omega=0|LS|J,Lambda=0,Omega=0> = -1/sqrt(3)
<J,Lambda=1,Omega=0|LS|J,Lambda=0,Omega=0> = -1/sqrt(3)

Symmetrized Matrix Elements of <a3Pv=13|LS|Iv=4>
<J,Lambda=1,Omega=0|LS|J,Lambda=0,Omega=0> = -1/sqrt(3)

LinearPerturbation: Excited <a3Pv=13|LS|ev=4>
Value 24.728701125668 Float, Std Dev = 0.0197221506962047 Size of perturbation.
<J,Lambda=-1,Omega=-1|LS|J,Lambda=0,Omega=-1> = 1/sqrt(6)
<J,Lambda=-1,Omega=0|LS|J,Lambda=0,Omega=0> = 1/sqrt(6)
<J,Lambda=1,Omega=0|LS|J,Lambda=0,Omega=0> = -1/sqrt(6)
<J,Lambda=1,Omega=1|LS|J,Lambda=0,Omega=1> = -1/sqrt(6)

Symmetrized Matrix Elements of <a3Pv=13|LS|ev=4>
<J,Lambda=1,Omega=1|LS|J,Lambda=0,Omega=1> = 1/sqrt(6)
<J,Lambda=1,Omega=1|LS|J,Lambda=0,Omega=1> = 1/sqrt(6)
<J,Lambda=1,Omega=0|LS|J,Lambda=0,Omega=0> = -1/sqrt(6)

LinearPerturbation: Excited <a3Pv=13|J+L-|ev=4>
Value -0.0598230826106589 Float, Std Dev = 0.000255639894261763 Size of perturbation.
<J,Lambda=-1,Omega=-2|J+L-|J,Lambda=0,Omega=-1> = -sqrt(J*(J+1)-2)
<J,Lambda=-1,Omega=-1|J+L-|J,Lambda=0,Omega=0> = -sqrt(J*(J+1))
<J,Lambda=-1,Omega=0|J+L-|J,Lambda=0,Omega=1> = -sqrt(J*(J+1))
<J,Lambda=1,Omega=0|J+L-|J,Lambda=0,Omega=-1> = sqrt(J*(J+1))
<J,Lambda=1,Omega=1|J+L-|J,Lambda=0,Omega=0> = sqrt(J*(J+1))
<J,Lambda=1,Omega=2|J+L-|J,Lambda=0,Omega=1> = sqrt(J*(J+1)-2)

Symmetrized Matrix Elements of <a3Pv=13|J+L-|ev=4>
<J,Lambda=1,Omega=2|J+L-|J,Lambda=0,Omega=1> = -sqrt(J*(J+1)-2)
<J,Lambda=1,Omega=2|J+L-|J,Lambda=0,Omega=1> = -sqrt(J*(J+1))
<J,Lambda=1,Omega=1|J+L-|J,Lambda=0,Omega=0> = -sqrt(J*(J+1))
<J,Lambda=1,Omega=0|J+L-|J,Lambda=0,Omega=1> = sqrt(J*(J+1))
<J,Lambda=1,Omega=0|J+L-|J,Lambda=0,Omega=1> = sqrt(J*(J+1))

LinearPerturbation: Excited <a3Pv=13|LS|dv=7>
Value 23.2189357988461 Float, Std Dev = 0.0441255453837693 Size of perturbation.
<J,Lambda=-1,Omega=-2|LS|J,Lambda=-2,Omega=-2> = -1/sqrt(6)
<J,Lambda=-1,Omega=-1|LS|J,Lambda=-2,Omega=-1> = -1/sqrt(6)
<J,Lambda=1,Omega=1|LS|J,Lambda=2,Omega=1> = -1/sqrt(6)
<J,Lambda=1,Omega=2|LS|J,Lambda=2,Omega=2> = -1/sqrt(6)

Symmetrized Matrix Elements of <a3Pv=13|LS|dv=7>
<J,Lambda=1,Omega=2|LS|J,Lambda=2,Omega=2> = -1/sqrt(6)
<J,Lambda=1,Omega=2|LS|J,Lambda=2,Omega=2> = -1/sqrt(6)
<J,Lambda=1,Omega=1|LS|J,Lambda=2,Omega=1> = -1/sqrt(6)
<J,Lambda=1,Omega=1|LS|J,Lambda=2,Omega=1> = -1/sqrt(6)

LinearPerturbation: Excited <a3Pv=13|J+L-|dv=7>
Value -0.0627945906974129 Float, Std Dev = 0.000453469558882703 Size of perturbation.
<J,Lambda=-1,Omega=-2|J+L-|J,Lambda=-2,Omega=-3> = sqrt(J*(J+1)-6)
<J,Lambda=-1,Omega=-1|J+L-|J,Lambda=-2,Omega=-2> = sqrt(J*(J+1)-2)
<J,Lambda=-1,Omega=0|J+L-|J,Lambda=-2,Omega=-1> = sqrt(J*(J+1))
<J,Lambda=1,Omega=0|J+L-|J,Lambda=2,Omega=1> = sqrt(J*(J+1))
<J,Lambda=1,Omega=1|J+L-|J,Lambda=2,Omega=2> = sqrt(J*(J+1)-2)
<J,Lambda=1,Omega=2|J+L-|J,Lambda=2,Omega=3> = sqrt(J*(J+1)-6)

Symmetrized Matrix Elements of <a3Pv=13|J+L-|dv=7>
<J,Lambda=1,Omega=2|J+L-|J,Lambda=2,Omega=3> = sqrt(J*(J+1)-6)
<J,Lambda=1,Omega=2|J+L-|J,Lambda=2,Omega=3> = sqrt(J*(J+1)-2)
<J,Lambda=1,Omega=1|J+L-|J,Lambda=2,Omega=2> = sqrt(J*(J+1))
<J,Lambda=1,Omega=1|J+L-|J,Lambda=2,Omega=2> = sqrt(J*(J+1))
<J,Lambda=1,Omega=0|J+L-|J,Lambda=2,Omega=1> = sqrt(J*(J+1)-2)
<J,Lambda=1,Omega=0|J+L-|J,Lambda=2,Omega=1> = sqrt(J*(J+1)-6)

LinearPerturbation: Excited <a3Pv=13|LS|a'v=12>
Value 2.39157392255636 Float, Std Dev = 0.204009133001652 Size of perturbation.
<J,Lambda=-1,Omega=-1|LS|J,Lambda=0,Omega=-1> = -1/sqrt(6)
<J,Lambda=-1,Omega=0|LS|J,Lambda=0,Omega=0> = -1/sqrt(6)
<J,Lambda=1,Omega=0|LS|J,Lambda=0,Omega=0> = -1/sqrt(6)
<J,Lambda=1,Omega=1|LS|J,Lambda=0,Omega=1> = -1/sqrt(6)

Symmetrized Matrix Elements of <a3Pv=13|LS|a'v=12>
<J,Lambda=1,Omega=1|LS|J,Lambda=0,Omega=1> = -1/sqrt(6)
<J,Lambda=1,Omega=1|LS|J,Lambda=0,Omega=1> = -1/sqrt(6)
<J,Lambda=1,Omega=0|LS|J,Lambda=0,Omega=0> = -1/sqrt(6)

LinearPerturbation: Excited <a3Pv=13|J+L-|a'v=12>
Value -0.0112360932879575 Float, Std Dev = 0.00125602151495791 Size of perturbation.
<J,Lambda=-1,Omega=-2|J+L-|J,Lambda=0,Omega=-1> = sqrt(J*(J+1)-2)
<J,Lambda=-1,Omega=-1|J+L-|J,Lambda=0,Omega=0> = sqrt(J*(J+1))
<J,Lambda=-1,Omega=0|J+L-|J,Lambda=0,Omega=1> = sqrt(J*(J+1))
<J,Lambda=1,Omega=0|J+L-|J,Lambda=0,Omega=-1> = sqrt(J*(J+1))
<J,Lambda=1,Omega=1|J+L-|J,Lambda=0,Omega=0> = sqrt(J*(J+1))
<J,Lambda=1,Omega=2|J+L-|J,Lambda=0,Omega=1> = sqrt(J*(J+1)-2)

Symmetrized Matrix Elements of <a3Pv=13|J+L-|a'v=12>
<J,Lambda=1,Omega=2|J+L-|J,Lambda=0,Omega=1> = sqrt(J*(J+1)-2)
<J,Lambda=1,Omega=2|J+L-|J,Lambda=0,Omega=1> = sqrt(J*(J+1))
<J,Lambda=1,Omega=1|J+L-|J,Lambda=0,Omega=0> = sqrt(J*(J+1))
<J,Lambda=1,Omega=0|J+L-|J,Lambda=0,Omega=1> = sqrt(J*(J+1))
<J,Lambda=1,Omega=0|J+L-|J,Lambda=0,Omega=1> = sqrt(J*(J+1))

LinearPerturbation: Excited <a3Pv=13|LS|Dsv=3>
Value -22.8568 Size of perturbation.
<J,Lambda=-1,Omega=-2|LS|J,Lambda=-2,Omega=-2> = 1/sqrt(3)
<J,Lambda=1,Omega=2|LS|J,Lambda=2,Omega=2> = -1/sqrt(3)

Symmetrized Matrix Elements of <a3Pv=13|LS|Dsv=3>
<J,Lambda=1,Omega=2|LS|J,Lambda=2,Omega=2> = 1/sqrt(3)
<J,Lambda=1,Omega=2|LS|J,Lambda=2,Omega=2> = -1/sqrt(3)

TransitionMoments: <Excited|mu|X>
SphericalTransitionMoment: <Excited|mu|X> <Av=2|T(1)|Xv=0>
Rank 1
Component Auto => 1
Strength 1 Transition (dipole) moment / Debye - note that relative intensities are proportional to the square of this value.
Reduced Matrix Elements of <Av=2|T(1)|Xv=0>
<J-1,Lambda=-1,Omega=-1|T(1)|J,Lambda=0,Omega=0> = sqrt(2*(J-1))/2
<J-1,Lambda=1,Omega=1|T(1)|J,Lambda=0,Omega=0> = sqrt(2*(J-1))/2
<J,Lambda=-1,Omega=-1|T(1)|J,Lambda=0,Omega=0> = (2*J+1)/sqrt(2*(2*J+1))
<J,Lambda=1,Omega=1|T(1)|J,Lambda=0,Omega=0> = (-2*J-1)/sqrt(2*(2*J+1))
<J+1,Lambda=-1,Omega=-1|T(1)|J,Lambda=0,Omega=0> = sqrt(2*(J+2))/2
<J+1,Lambda=1,Omega=1|T(1)|J,Lambda=0,Omega=0> = sqrt(2*(J+2))/2
Line Strengths from <Av=2|T(1)|Xv=0>
S<J-1,Lambda=-1,Omega=-1|T(1)|J,Lambda=0,Omega=0> = (J-1)/2
S<J-1,Lambda=1,Omega=1|T(1)|J,Lambda=0,Omega=0> = (J-1)/2
S<J,Lambda=-1,Omega=-1|T(1)|J,Lambda=0,Omega=0> = (2*J+1)/2
S<J,Lambda=1,Omega=1|T(1)|J,Lambda=0,Omega=0> = (-2*J-1)/-2
S<J+1,Lambda=-1,Omega=-1|T(1)|J,Lambda=0,Omega=0> = (J+2)/2
S<J+1,Lambda=1,Omega=1|T(1)|J,Lambda=0,Omega=0> = (J+2)/2
Symmetrized Reduced Matrix Elements of <Av=2|T(1)|Xv=0>
<J-1,Lambda=1,Omega=1 + |T(1)|J,Lambda=0,Omega=0> = sqrt(J-1)
<J,Lambda=1,Omega=1 - |T(1)|J,Lambda=0,Omega=0> = (-2*J-1)/sqrt(2*J+1)
<J+1,Lambda=1,Omega=1 + |T(1)|J,Lambda=0,Omega=0> = sqrt(J+2)
Symmetrized Line Strengths from <Av=2|T(1)|Xv=0>
S<J-1,Lambda=1,Omega=1 + |T(1)|J,Lambda=0,Omega=0> = J-1
S<J,Lambda=1,Omega=1 - |T(1)|J,Lambda=0,Omega=0> = (-2*J-1)/-1
S<J+1,Lambda=1,Omega=1 + |T(1)|J,Lambda=0,Omega=0> = J+2

ForbiddenTransitionMoment: <Excited|mu|X> <a3Pv=13|fT(1,1)|Xv=0>
Rank 1
Component 1
Magnetic False
Strength 1 Transition (dipole) moment / Debye - note that relative intensities are proportional to the square of this value.
Reduced Matrix Elements of <a3Pv=13|fT(1,1)|Xv=0>
<J-1,Lambda=-1,Omega=-1|fT(1,1)|J,Lambda=0,Omega=0> = (-sqrt(2*(J-1)))/2
<J-1,Lambda=1,Omega=1|fT(1,1)|J,Lambda=0,Omega=0> = sqrt(2*(J-1))/2
<J,Lambda=-1,Omega=-1|fT(1,1)|J,Lambda=0,Omega=0> = (-2*J-1)/sqrt(2*(2*J+1))
<J,Lambda=1,Omega=1|fT(1,1)|J,Lambda=0,Omega=0> = (-2*J-1)/sqrt(2*(2*J+1))
<J+1,Lambda=-1,Omega=-1|fT(1,1)|J,Lambda=0,Omega=0> = (-sqrt(2*(J+2)))/2
<J+1,Lambda=1,Omega=1|fT(1,1)|J,Lambda=0,Omega=0> = sqrt(2*(J+2))/2
Line Strengths from <a3Pv=13|fT(1,1)|Xv=0>
S<J-1,Lambda=-1,Omega=-1|fT(1,1)|J,Lambda=0,Omega=0> = (J-1)/2
S<J-1,Lambda=1,Omega=1|fT(1,1)|J,Lambda=0,Omega=0> = (J-1)/2
S<J,Lambda=-1,Omega=-1|fT(1,1)|J,Lambda=0,Omega=0> = (-2*J-1)/-2
S<J,Lambda=1,Omega=1|fT(1,1)|J,Lambda=0,Omega=0> = (-2*J-1)/-2
S<J+1,Lambda=-1,Omega=-1|fT(1,1)|J,Lambda=0,Omega=0> = (J+2)/2
S<J+1,Lambda=1,Omega=1|fT(1,1)|J,Lambda=0,Omega=0> = (J+2)/2
Symmetrized Reduced Matrix Elements of <a3Pv=13|fT(1,1)|Xv=0>
<J-1,Lambda=1,Omega=1 - |fT(1,1)|J,Lambda=0,Omega=0> = sqrt(J-1)
<J,Lambda=1,Omega=1 + |fT(1,1)|J,Lambda=0,Omega=0> = (-2*J-1)/sqrt(2*J+1)
<J+1,Lambda=1,Omega=1 - |fT(1,1)|J,Lambda=0,Omega=0> = sqrt(J+2)
Symmetrized Line Strengths from <a3Pv=13|fT(1,1)|Xv=0>
S<J-1,Lambda=1,Omega=1 - |fT(1,1)|J,Lambda=0,Omega=0> = J-1
S<J,Lambda=1,Omega=1 + |fT(1,1)|J,Lambda=0,Omega=0> = (-2*J-1)/-1
S<J+1,Lambda=1,Omega=1 - |fT(1,1)|J,Lambda=0,Omega=0> = J+2

ForbiddenTransitionMoment: <Excited|mu|X> <dv=7|fT(1,1)|Xv=0>
Rank 1
Component 1
Magnetic False
Strength 1 Transition (dipole) moment / Debye - note that relative intensities are proportional to the square of this value.
Reduced Matrix Elements of <dv=7|fT(1,1)|Xv=0>
<J-1,Lambda=-2,Omega=-1|fT(1,1)|J,Lambda=0,Omega=0> = (-sqrt(2*(J-1)))/2
<J-1,Lambda=2,Omega=1|fT(1,1)|J,Lambda=0,Omega=0> = sqrt(2*(J-1))/2
<J,Lambda=-2,Omega=-1|fT(1,1)|J,Lambda=0,Omega=0> = (-2*J-1)/sqrt(2*(2*J+1))
<J,Lambda=2,Omega=1|fT(1,1)|J,Lambda=0,Omega=0> = (-2*J-1)/sqrt(2*(2*J+1))
<J+1,Lambda=-2,Omega=-1|fT(1,1)|J,Lambda=0,Omega=0> = (-sqrt(2*(J+2)))/2
<J+1,Lambda=2,Omega=1|fT(1,1)|J,Lambda=0,Omega=0> = sqrt(2*(J+2))/2
Line Strengths from <dv=7|fT(1,1)|Xv=0>
S<J-1,Lambda=-2,Omega=-1|fT(1,1)|J,Lambda=0,Omega=0> = (J-1)/2
S<J-1,Lambda=2,Omega=1|fT(1,1)|J,Lambda=0,Omega=0> = (J-1)/2
S<J,Lambda=-2,Omega=-1|fT(1,1)|J,Lambda=0,Omega=0> = (-2*J-1)/-2
S<J,Lambda=2,Omega=1|fT(1,1)|J,Lambda=0,Omega=0> = (-2*J-1)/-2
S<J+1,Lambda=-2,Omega=-1|fT(1,1)|J,Lambda=0,Omega=0> = (J+2)/2
S<J+1,Lambda=2,Omega=1|fT(1,1)|J,Lambda=0,Omega=0> = (J+2)/2
Symmetrized Reduced Matrix Elements of <dv=7|fT(1,1)|Xv=0>

<J-1,Lambda=2,Omega=1 - |fT(1,1)|J,Lambda=0,Omega=0> = sqrt(J-1)
<J,Lambda=2,Omega=1 + |fT(1,1)|J,Lambda=0,Omega=0> = (-2*J-1)/sqrt(2*J+1)
<J+1,Lambda=2,Omega=1 - |fT(1,1)|J,Lambda=0,Omega=0> = sqrt(J+2)
Symmetrized Line Strengths from <dv=7|fT(1,1)|Xv=0>
S<J-1,Lambda=2,Omega=1 - |fT(1,1)|J,Lambda=0,Omega=0> = J-1
S<J,Lambda=2,Omega=1 + |fT(1,1)|J,Lambda=0,Omega=0> = (-2*J-1)/-1
S<J+1,Lambda=2,Omega=1 - |fT(1,1)|J,Lambda=0,Omega=0> = J+2

ForbiddenTransitionMoment: <Excited|mu|X> <a'v=12|fT(1,1)|Xv=0>
Rank 1
Component 1
Magnetic False
Strength 1 Transition (dipole) moment / Debye - note that relative intensities are proportional to the square of this value.
Reduced Matrix Elements of <a'v=12|fT(1,1)|Xv=0>
<J-1,Lambda=0,Omega=-1|fT(1,1)|J,Lambda=0,Omega=0> = (-sqrt(2*(J-1)))/2
<J-1,Lambda=0,Omega=1|fT(1,1)|J,Lambda=0,Omega=0> = sqrt(2*(J-1))/2
<J,Lambda=0,Omega=-1|fT(1,1)|J,Lambda=0,Omega=0> = (-2*J-1)/sqrt(2*(2*J+1))
<J,Lambda=0,Omega=1|fT(1,1)|J,Lambda=0,Omega=0> = (-2*J-1)/sqrt(2*(2*J+1))
<J+1,Lambda=0,Omega=-1|fT(1,1)|J,Lambda=0,Omega=0> = (-sqrt(2*(J+2)))/2
<J+1,Lambda=0,Omega=1|fT(1,1)|J,Lambda=0,Omega=0> = sqrt(2*(J+2))/2
Line Strengths from <a'v=12|fT(1,1)|Xv=0>
S<J-1,Lambda=0,Omega=-1|fT(1,1)|J,Lambda=0,Omega=0> = (J-1)/2
S<J-1,Lambda=0,Omega=1|fT(1,1)|J,Lambda=0,Omega=0> = (J-1)/2
S<J,Lambda=0,Omega=-1|fT(1,1)|J,Lambda=0,Omega=0> = (-2*J-1)/-2
S<J,Lambda=0,Omega=1|fT(1,1)|J,Lambda=0,Omega=0> = (-2*J-1)/-2
S<J+1,Lambda=0,Omega=-1|fT(1,1)|J,Lambda=0,Omega=0> = (J+2)/2
S<J+1,Lambda=0,Omega=1|fT(1,1)|J,Lambda=0,Omega=0> = (J+2)/2
Symmetrized Reduced Matrix Elements of <a'v=12|fT(1,1)|Xv=0>
<J-1,Lambda=0,Omega=1 - |fT(1,1)|J,Lambda=0,Omega=0> = sqrt(J-1)
<J,Lambda=0,Omega=1 + |fT(1,1)|J,Lambda=0,Omega=0> = (-2*J-1)/sqrt(2*J+1)
<J+1,Lambda=0,Omega=1 - |fT(1,1)|J,Lambda=0,Omega=0> = sqrt(J+2)
Symmetrized Line Strengths from <a'v=12|fT(1,1)|Xv=0>
S<J-1,Lambda=0,Omega=1 - |fT(1,1)|J,Lambda=0,Omega=0> = J-1
S<J,Lambda=0,Omega=1 + |fT(1,1)|J,Lambda=0,Omega=0> = (-2*J-1)/-1
S<J+1,Lambda=0,Omega=1 - |fT(1,1)|J,Lambda=0,Omega=0> = J+2

ForbiddenTransitionMoment: <Excited|mu|X> <ev=4|fT(1,1)|Xv=0>
Rank 1
Component 1
Magnetic False
Strength 1 Transition (dipole) moment / Debye - note that relative intensities are proportional to the square of this value.
Reduced Matrix Elements of <ev=4|fT(1,1)|Xv=0>
<J-1,Lambda=0,Omega=-1|fT(1,1)|J,Lambda=0,Omega=0> = sqrt(2*(J-1))/2
<J-1,Lambda=0,Omega=1|fT(1,1)|J,Lambda=0,Omega=0> = sqrt(2*(J-1))/2
<J,Lambda=0,Omega=-1|fT(1,1)|J,Lambda=0,Omega=0> = (2*J+1)/sqrt(2*(2*J+1))
<J,Lambda=0,Omega=1|fT(1,1)|J,Lambda=0,Omega=0> = (-2*J-1)/sqrt(2*(2*J+1))
<J+1,Lambda=0,Omega=-1|fT(1,1)|J,Lambda=0,Omega=0> = sqrt(2*(J+2))/2
<J+1,Lambda=0,Omega=1|fT(1,1)|J,Lambda=0,Omega=0> = sqrt(2*(J+2))/2
Line Strengths from <ev=4|fT(1,1)|Xv=0>
S<J-1,Lambda=0,Omega=-1|fT(1,1)|J,Lambda=0,Omega=0> = (J-1)/2
S<J-1,Lambda=0,Omega=1|fT(1,1)|J,Lambda=0,Omega=0> = (J-1)/2
S<J,Lambda=0,Omega=-1|fT(1,1)|J,Lambda=0,Omega=0> = (2*J+1)/2
S<J,Lambda=0,Omega=1|fT(1,1)|J,Lambda=0,Omega=0> = (-2*J-1)/-2
S<J+1,Lambda=0,Omega=-1|fT(1,1)|J,Lambda=0,Omega=0> = (J+2)/2
S<J+1,Lambda=0,Omega=1|fT(1,1)|J,Lambda=0,Omega=0> = (J+2)/2
Symmetrized Reduced Matrix Elements of <ev=4|fT(1,1)|Xv=0>
<J-1,Lambda=0,Omega=1 + |fT(1,1)|J,Lambda=0,Omega=0> = sqrt(J-1)
<J,Lambda=0,Omega=1 - |fT(1,1)|J,Lambda=0,Omega=0> = (-2*J-1)/sqrt(2*J+1)
<J+1,Lambda=0,Omega=1 + |fT(1,1)|J,Lambda=0,Omega=0> = sqrt(J+2)
Symmetrized Line Strengths from <ev=4|fT(1,1)|Xv=0>
S<J-1,Lambda=0,Omega=1 + |fT(1,1)|J,Lambda=0,Omega=0> = J-1
S<J,Lambda=0,Omega=1 - |fT(1,1)|J,Lambda=0,Omega=0> = (-2*J-1)/-1
S<J+1,Lambda=0,Omega=1 + |fT(1,1)|J,Lambda=0,Omega=0> = J+2

ForbiddenTransitionMoment: <Excited|mu|X> <Iv=3|fT(1,1)|Xv=0>
Rank 1
Component 1
Magnetic False
Strength 1 Transition (dipole) moment / Debye - note that relative intensities are proportional to the square of this value.
ForbiddenTransitionMoment: <Excited|mu|X> <Dsv=3|fT(1,1)|Xv=0>
Rank 1
Component 1
Magnetic False
Strength 1 Transition (dipole) moment / Debye - note that relative intensities are proportional to the square of this value.
TransitionMoments: <B|mu|X>
SphericalTransitionMoment: <B|mu|X> <Bv=0|T(1)|Xv=0>
Rank 1
Component Auto => 0
Strength 1 Transition (dipole) moment / Debye - note that relative intensities are proportional to the square of this value.
Reduced Matrix Elements of <Bv=0|T(1)|Xv=0>
<J-1,Lambda=0,Omega=0|T(1)|J,Lambda=0,Omega=0> = -sqrt(J)
<J+1,Lambda=0,Omega=0|T(1)|J,Lambda=0,Omega=0> = (J+1)/sqrt(J+1)
Line Strengths from <Bv=0|T(1)|Xv=0>
S<J-1,Lambda=0,Omega=0|T(1)|J,Lambda=0,Omega=0> = J
S<J+1,Lambda=0,Omega=0|T(1)|J,Lambda=0,Omega=0> = J+1
Symmetrized Reduced Matrix Elements of <Bv=0|T(1)|Xv=0>
<J-1,Lambda=0,Omega=0|T(1)|J,Lambda=0,Omega=0> = -sqrt(J)
<J+1,Lambda=0,Omega=0|T(1)|J,Lambda=0,Omega=0> = (J+1)/sqrt(J+1)
Symmetrized Line Strengths from <Bv=0|T(1)|Xv=0>
S<J-1,Lambda=0,Omega=0|T(1)|J,Lambda=0,Omega=0> = J
S<J+1,Lambda=0,Omega=0|T(1)|J,Lambda=0,Omega=0> = J+1

TransitionMoments: <B|mu|Excited>
SphericalTransitionMoment: <B|mu|Excited> <Bv=0|T(1)|Av=2>
Rank 1
Component Auto => 1
Strength 1 Transition (dipole) moment / Debye - note that relative intensities are proportional to the square of this value.
Reduced Matrix Elements of <Bv=0|T(1)|Av=2>
<J-1,Lambda=0,Omega=0|T(1)|J,Lambda=-1,Omega=-1> = sqrt(2*(J+1))/2
<J-1,Lambda=0,Omega=0|T(1)|J,Lambda=1,Omega=1> = sqrt(2*(J+1))/2
<J,Lambda=0,Omega=0|T(1)|J,Lambda=-1,Omega=-1> = (-2*J-1)/sqrt(2*(2*J+1))
<J,Lambda=0,Omega=0|T(1)|J,Lambda=1,Omega=1> = (2*J+1)/sqrt(2*(2*J+1))
<J+1,Lambda=0,Omega=0|T(1)|J,Lambda=-1,Omega=-1> = sqrt(2*J)/2
<J+1,Lambda=0,Omega=0|T(1)|J,Lambda=1,Omega=1> = sqrt(2*J)/2
Line Strengths from <Bv=0|T(1)|Av=2>
S<J-1,Lambda=0,Omega=0|T(1)|J,Lambda=-1,Omega=-1> = (J+1)/2
S<J-1,Lambda=0,Omega=0|T(1)|J,Lambda=1,Omega=1> = (J+1)/2
S<J,Lambda=0,Omega=0|T(1)|J,Lambda=-1,Omega=-1> = (-2*J-1)/-2
S<J,Lambda=0,Omega=0|T(1)|J,Lambda=1,Omega=1> = (2*J+1)/2
S<J+1,Lambda=0,Omega=0|T(1)|J,Lambda=-1,Omega=-1> = J/2
S<J+1,Lambda=0,Omega=0|T(1)|J,Lambda=1,Omega=1> = J/2
Symmetrized Reduced Matrix Elements of <Bv=0|T(1)|Av=2>
<J-1,Lambda=0,Omega=0|T(1)| + J,Lambda=1,Omega=1> = sqrt(J+1)
<J,Lambda=0,Omega=0|T(1)| - J,Lambda=1,Omega=1> = (2*J+1)/sqrt(2*J+1)
<J+1,Lambda=0,Omega=0|T(1)| + J,Lambda=1,Omega=1> = sqrt(J)
Symmetrized Line Strengths from <Bv=0|T(1)|Av=2>
S<J-1,Lambda=0,Omega=0|T(1)| + J,Lambda=1,Omega=1> = J+1
S<J,Lambda=0,Omega=0|T(1)| - J,Lambda=1,Omega=1> = 2*J+1
S<J+1,Lambda=0,Omega=0|T(1)| + J,Lambda=1,Omega=1> = J

TransitionMoments: <C|mu|X>
SphericalTransitionMoment: <C|mu|X> <Cv=0|T(1)|Xv=0>
Rank 1
Component Auto => 0
Strength 1 Transition (dipole) moment / Debye - note that relative intensities are proportional to the square of this value.
Matrix elements as for <B|mu|X> <Bv=0|T(1)|Xv=0>
TransitionMoments: <C|mu|Excited>
SphericalTransitionMoment: <C|mu|Excited> <Cv=0|T(1)|Av=2>
Rank 1
Component Auto => 1
Strength 1 Transition (dipole) moment / Debye - note that relative intensities are proportional to the square of this value.
Matrix elements as for <B|mu|Excited> <Bv=0|T(1)|Av=2>
Correlations: Correlations
CO.13C18O.Excited.Av=2.Origin CO.13C18O.Excited.Av=2.B CO.13C18O.Excited.Av=2.D CO.13C18O.Excited.Iv=3.Origin CO.13C18O.Excited.Iv=3.B
CO.13C18O.Excited.Iv=3.D CO.13C18O.Excited.<Av=2|J+L-|Iv=3>.Value CO.13C18O.Excited.ev=4.Origin CO.13C18O.Excited.ev=4.B CO.13C18O.Excited.ev=4.LambdaSS
CO.13C18O.Excited.ev=4.D CO.13C18O.Excited.<Av=2|LS|ev=4>.Value CO.13C18O.Excited.dv=7.Origin CO.13C18O.Excited.dv=7.B CO.13C18O.Excited.dv=7.A
CO.13C18O.Excited.dv=7.LambdaSS CO.13C18O.Excited.dv=7.gamma CO.13C18O.Excited.dv=7.D CO.13C18O.Excited.<Av=2|LS|dv=7>.Value CO.13C18O.Excited.a'v=12.Origin
CO.13C18O.Excited.a'v=12.B CO.13C18O.Excited.a'v=12.LambdaSS CO.13C18O.Excited.a'v=12.gamma CO.13C18O.Excited.a'v=12.D CO.13C18O.Excited.<Av=2|LS|a'v=12>.Value
CO.13C18O.Excited.Dsv=3.Origin CO.13C18O.Excited.Dsv=3.B CO.13C18O.Excited.Dsv=3.D CO.13C18O.Excited.a3Pv=13.Origin CO.13C18O.Excited.a3Pv=13.B
CO.13C18O.Excited.a3Pv=13.A CO.13C18O.Excited.a3Pv=13.D CO.13C18O.Excited.<a3Pv=13|LS|ev=4>.Value CO.13C18O.Excited.<a3Pv=13|J+L-|ev=4>.Value
CO.13C18O.Excited.<a3Pv=13|LS|dv=7>.Value CO.13C18O.Excited.<a3Pv=13|J+L-|dv=7>.Value CO.13C18O.Excited.<a3Pv=13|LS|a'v=12>.Value CO.13C18O.Excited.<a3Pv=13|J+L-
|a'v=12>.Value
CO.13C18O.Excited.Av=2.Origin 1.000000
CO.13C18O.Excited.Av=2.B -0.611097 1.000000
CO.13C18O.Excited.Av=2.D -0.394999 0.931474 1.000000
CO.13C18O.Excited.Iv=3.Origin 0.032699 -0.038056 -0.026507 1.000000
CO.13C18O.Excited.Iv=3.B -0.042506 0.039516 0.026229 -0.876261 1.000000
CO.13C18O.Excited.Iv=3.D -0.041999 0.038630 0.025513 -0.822677 0.983707 1.000000
CO.13C18O.Excited.<Av=2|J+L-|Iv=3>.Value 0.086756 0.106406 0.100399 -0.132875 0.102512 0.097275 1.000000
CO.13C18O.Excited.ev=4.Origin -0.003057 0.018384 0.011713 -0.005347 0.002599 0.002316 0.032342
1.000000
CO.13C18O.Excited.ev=4.B 0.039228 -0.108622 -0.095386 0.007055 -0.002469 -0.001935 -0.032034
-0.671857 1.000000
CO.13C18O.Excited.ev=4.LambdaSS 0.008759 0.034764 0.040224 -0.011285 0.007018 0.006486 0.042605
0.528819 -0.513685 1.000000
CO.13C18O.Excited.ev=4.D 0.031625 -0.092400 -0.086637 0.004892 -0.000841 -0.000404 -0.023026
-0.453502 0.938560 -0.365005 1.000000
CO.13C18O.Excited.<Av=2|LS|ev=4>.Value -0.538280 0.096021 -0.057563 -0.005926 0.011475 0.011400 -
0.075553 0.006246 0.018663 -0.133845 0.019953 1.000000
CO.13C18O.Excited.dv=7.Origin 0.004827 0.003795 -0.004605 -0.001072 0.000105 -0.000023 -
0.006026 -0.004311 -0.018079 -0.016744 -0.025702 -0.051928 1.000000
CO.13C18O.Excited.dv=7.B 0.005560 0.004537 0.035495 0.002746 -0.001806 -0.001603 0.000674
0.003442 0.017375 0.033633 0.031833 0.030412 -0.611571 1.000000
CO.13C18O.Excited.dv=7.A -0.007808 0.024737 0.032140 -0.001568 0.001397 0.001337 0.002364
-0.005373 0.000233 -0.013746 -0.005216 0.000034 0.267315 -0.060700 1.000000
CO.13C18O.Excited.dv=7.LambdaSS 0.000983 -0.007159 -0.006359 -0.000398 0.000050 0.000010 0.000411
-0.004915 -0.001761 -0.005370 -0.003972 -0.000060 0.211893 -0.066184 0.357139 1.000000
CO.13C18O.Excited.dv=7.gamma 0.021028 -0.072130 -0.087712 0.003399 -0.004035 -0.003986 -
0.006941 -0.006735 -0.008331 0.025878 0.003011 -0.001349 -0.066485 -0.025139 -0.625889
-0.167714 1.000000
CO.13C18O.Excited.dv=7.D 0.007570 -0.007281 0.011171 0.002637 -0.001739 -0.001542 -0.000557
0.009037 0.012332 0.030057 0.024712 0.019486 -0.448784 0.918072 0.025408 -0.052520
-0.171101 1.000000
CO.13C18O.Excited.<Av=2|LS|dv=7>.Value -0.177600 -0.200233 -0.267842 0.000459 -0.000717 -0.000812 -
0.030321 0.044336 -0.084127 0.030574 -0.080210 0.169721 -0.029569 -0.035230 -0.064983
0.009009 0.169763 -0.012971 1.000000
CO.13C18O.Excited.a'v=12.Origin 0.021190 0.052536 0.024845 0.005224 -0.004262 -0.004118 -
0.024521 -0.014699 0.019616 -0.018687 0.018715 -0.084042 -0.050199 0.026841 -0.021118
-0.005458 0.033318 0.024332 0.019816 1.000000
CO.13C18O.Excited.a'v=12.B -0.012897 -0.001229 0.039821 -0.005454 0.004798 0.004663 0.020823
0.001268 0.002780 0.011861 0.002753 0.065898 0.014897 -0.008091 0.003147 0.000614
-0.008950 -0.008241 -0.005872 -0.568855 1.000000

CO.13C18O.Excited.a'v=12.LambdaSS 0.004052 -0.034970 -0.040242 0.001472 -0.001547 -0.001528 -
0.004469 0.005757 -0.005372 0.004264 -0.003444 0.010050 0.021276 -0.016871 0.011453
-0.002302 -0.018642 -0.012890 -0.020312 -0.254081 -0.084794 1.000000
CO.13C18O.Excited.a'v=12.gamma 0.007713 -0.050414 -0.075382 0.002371 -0.002370 -0.002313 -
0.005713 0.004892 -0.008964 -0.001536 -0.008263 -0.006183 0.017333 -0.010752 0.012392
0.002908 -0.016030 -0.006100 -0.001231 0.073295 -0.506491 0.204477 1.000000
CO.13C18O.Excited.a'v=12.D -0.010930 0.003768 0.036819 -0.004527 0.003957 0.003839 0.016397
-0.000515 0.003399 0.011593 0.003759 0.052698 0.003861 -0.002060 -0.000470 -0.000128
-0.001109 -0.004378 0.006758 -0.424419 0.948354 -0.124615 -0.545285 1.000000
CO.13C18O.Excited.<Av=2|LS|a'v=12>.Value -0.015475 -0.102886 -0.126416 0.001821 -0.001663 -0.001594 0.003214
-0.004453 0.023560 -0.012215 0.030556 -0.015292 0.062857 -0.041558 0.054128 0.005980
-0.092007 -0.040226 -0.283280 -0.100598 -0.112127 0.113784 0.166559 -0.118548 1.000000
CO.13C18O.Excited.Dsv=3.Origin -0.000059 0.000361 0.000251 -0.000161 0.000122 0.000112 0.000212
0.000449 -0.001314 0.000254 -0.001505 -0.000339 -0.000241 0.000519 -0.000128 0.000108
0.000898 0.000017 0.000575 -0.000009 -0.000161 0.000025 0.000222 -0.000157 0.000140
1.000000
CO.13C18O.Excited.Dsv=3.B 0.000068 -0.000422 -0.000302 0.000181 -0.000137 -0.000126 -
0.000243 -0.000496 0.001451 -0.000268 0.001665 0.000361 0.000303 -0.000606 0.000152
-0.000113 -0.001010 -0.000035 -0.000634 0.000008 0.000180 -0.000026 -0.000248 0.000175
-0.000150 -0.952676 1.000000
CO.13C18O.Excited.Dsv=3.D 0.000078 -0.000480 -0.000356 0.000209 -0.000155 -0.000142 -
0.000270 -0.000607 0.001740 -0.000407 0.001984 0.000457 0.000299 -0.000654 0.000177
-0.000138 -0.001198 0.000022 -0.000738 -0.000011 0.000209 -0.000022 -0.000280 0.000198
-0.000150 -0.925213 0.988436 1.000000
CO.13C18O.Excited.a3Pv=13.Origin -0.003583 0.064382 0.074685 0.007821 -0.004451 -0.003743 0.030535
0.067955 -0.091302 0.035415 -0.069691 0.115703 -0.261368 0.187859 -0.049504 -0.044980
0.029182 0.122578 -0.008556 -0.019600 0.016916 -0.004998 -0.010161 0.014820 -
0.069582 0.001950 -0.002303 -0.002484 1.000000
CO.13C18O.Excited.a3Pv=13.B -0.033294 0.011821 0.005798 -0.540810 0.473944 0.445100 0.073837
0.016802 -0.030734 -0.038832 -0.018636 0.016494 -0.049915 0.024934 -0.018425 -0.012655
0.012782 0.013061 0.012889 -0.186050 0.101170 0.047968 -0.010231 0.073862 0.030304
0.000427 -0.000461 -0.000569 -0.148687 1.000000
CO.13C18O.Excited.a3Pv=13.A -0.011819 0.099048 0.101995 -0.000061 0.008650 0.009412 0.036412
-0.093786 0.224425 -0.204001 0.221317 0.134912 -0.183310 0.119637 -0.030004 -0.040058
-0.042929 0.091542 -0.091600 0.012681 -0.003385 -0.005855 -0.011038 -0.003359 -
0.011846 -0.000473 0.000408 0.000830 0.636282 -0.096731 1.000000
CO.13C18O.Excited.a3Pv=13.D -0.009759 0.053017 0.047579 -0.039242 0.033013 0.030773 0.030935
0.054404 -0.130493 0.028705 -0.143333 0.011839 -0.104188 0.092714 -0.026930 -0.002369
0.100049 0.012761 0.049320 -0.003074 -0.008109 -0.000917 0.013895 -0.006767 -
0.014959 0.005692 -0.006468 -0.008184 0.453807 0.054904 0.136522 1.000000
CO.13C18O.Excited.<a3Pv=13|LS|ev=4>.Value -0.055405 0.047328 0.036882 0.009342 -0.003994 -0.000014 -0.000407
0.015152 0.259061 -0.558728 0.209430 -0.549803 0.009342 0.062405 -0.052361 0.011457
0.015924 0.030654 -0.050176 0.146875 -0.038118 0.007621 0.005064 0.009696 0.003753
-0.060195 0.002303 -0.002558 -0.003012 0.124553 0.101987 -0.256276 0.231312 1.000000
CO.13C18O.Excited.<a3Pv=13|J+L-|ev=4>.Value -0.043812 0.030884 0.015548 0.000825 0.000000 0.000187
0.023557 -0.042817 -0.082843 -0.417042 -0.126262 0.141820 0.042940 -0.043739 0.012050
0.010298 0.010054 -0.043658 0.066713 0.004447 -0.016271 -0.002352 0.006082 -0.016685
-0.031609 0.001204 -0.001363 -0.001469 0.137388 0.099891 0.125289 0.136177 0.500347
1.000000
CO.13C18O.Excited.<a3Pv=13|LS|dv=7>.Value -0.026411 -0.038548 -0.054881 -0.001469 -0.000342 -0.000613 -
0.000891 -0.040824 -0.001752 -0.018028 -0.008215 0.052732 -0.145139 0.038658 0.063301
0.066438 0.231665 -0.085357 0.178636 0.030516 -0.009486 -0.017459 -0.004600 0.003997
-0.038382 0.001083 -0.001210 -0.001438 0.036230 0.033402 -0.089783 0.145611 0.064526
0.038445 1.000000
CO.13C18O.Excited.<a3Pv=13|J+L-|dv=7>.Value -0.010466 -0.058041 -0.074032 -0.002516 0.003430 0.003555
0.018261 0.073316 -0.091577 -0.028447 -0.105439 0.066683 -0.131114 -0.065253 0.134675
0.034185 -0.285067 -0.040743 0.114609 -0.049484 0.009143 0.023765 -0.005236 0.003874
0.041015 -0.000631 0.000693 0.000911 0.179351 -0.018751 0.164184 -0.015460 0.125545
0.127174 0.117835 1.000000
CO.13C18O.Excited.<a3Pv=13|LS|a'v=12>.Value 0.055603 -0.015861 -0.033832 0.004524 -0.003008 -0.002784 -
0.003867 -0.011265 0.055408 -0.053383 0.053839 -0.107621 0.049129 -0.030596 0.019892
-0.002537 -0.056303 -0.013328 -0.268162 0.283978 -0.280938 -0.008792 0.098043 -0.263458
0.436119 -0.000349 0.000396 0.000503 -0.087966 -0.031377 0.036832 -0.074711 -0.104746
-0.003578 -0.140228 0.062745 1.000000
CO.13C18O.Excited.<a3Pv=13|J+L-|a'v=12>.Value 0.109949 -0.036578 -0.056443 -0.000427 0.000740 0.000779
0.013024 -0.064613 0.162359 -0.029603 0.175950 -0.126007 -0.032354 0.005686 0.022524
-0.007383 -0.045900 0.007189 -0.271075 0.215613 -0.171464 -0.011225 0.029132 -0.147314
0.334120 -0.001733 0.001951 0.002217 -0.183206 -0.034947 0.014691 -0.189676 -0.302143
-0.172013 0.018689 0.046982 0.678533 1.000000

Simulation: Simulation
Intensity factor:0.333333333333333 (1 from Spol)
IntensityUnits Arbitrary
LifeModel lmNone
PlotUnits cm1
Fluorescence False
nDF 50000
WidthMult 6
AutoMin True
AutoMax True
ShowSum False
ShowParts True
ShowFortrat True
UseUpper False
ShowSymmetry False
ShowDeltaJ True
ScaleMarkSize True
UseSymmetry False
UseStateNumber False
FortratQno J
PreserveArea False
SeparateScales False

LogScale False
Show2DPlot False
Auto2DY True
Fmin 19713.4264274956 Left edge of plot range.
Fmax 19725.7352925044 Right edge of plot range.
Ymax 7.90431110974814E-183 Top of plot range; updated automatically if AutoMax set.
Temperature 300 Rotational temeperature (Kelvin).
Gaussian 0.08 Gaussian contribution to linewidth (full width half maximum). If both this and a Lorentzian width (below) is set the result is a convolution of the two, a Voigt profile.
Lorentzian 0.02 Lorentzian contribution to linewidth (full width half maximum). This and the Gaussian width can also be set from the main window.
SMargin -1 Extra points to calculate off each end of required spectrum.
OThreshold 0.0001 Ignore peaks smaller than this fraction of the maximum peak intensity.
Tvib -1 Vibrational Temperature (Kelvin); set to -1 (default) to use rotational temperature for all Boltzmann factors.
MinI 1E-15 Weakest transitions to plot in Fortrat plots, as a fraction of strongest.
Tspin -1 Spin Temperature; set to -1 (default) to equilibrate different spinstates
IScale 1 Scale factor applied to all line intensities
PlotSplit 0.5 Fraction of plot area to use for simulation where plots separated (-1..1)
Fundamental Constants from CODATA 2018: h=6.62607015E-34 k=1.380649E-23 NA=6.02214076E23 e=1.602176634E-19 me=9.1093837015E-31

| Molecu | Manifol' | J' | S' | #' | Manifol" | J" | S" | #" | Position | I | Eupper | Elower | Spol | A | Width | |
|---|---|---|---|---|---|---|---|---|---|---|---|---|---|---|---|---|
| 13C18O | C | 1 | e | 1 | X | 0 | e | 1 | 91922.3502 | .3333333 | 91922.3502 | 0.0000 | .3333333 | 81197804 | 0.000000 : | rR(0) : C Cv=0 1 1 F1e - X Xv=0 0 0 F1e |
| 13C18O | C | 0 | e | 1 | X | 1 | e | 1 | 91915.3218 | .3277961 | 91918.8146 | 3.4928 | .3333333 | 2.4354e8 | 0.000000 : | pP(1) : C Cv=0 0 0 F1e - X Xv=0 1 1 F1e |
| 13C18O | C | 2 | e | 1 | X | 1 | e | 1 | 91925.9218 | .6555922 | 91929.4146 | 3.4928 | .6666667 | 97448723 | 0.000000 : | rR(1) : C Cv=0 2 2 F1e - X Xv=0 1 1 F1e |
| 13C18O | C | 1 | e | 1 | X | 2 | e | 1 | 91911.8719 | .6339926 | 91922.3502 | 10.4783 | .6666667 | 1.6234e8 | 0.000000 : | pP(2) : C Cv=0 1 1 F1e - X Xv=0 2 2 F1e |
| 13C18O | C | 3 | e | 1 | X | 2 | e | 1 | 91929.5305 | .9509888 | 91940.0088 | 10.4783 | 1.000000 | 1.0442e8 | 0.000000 : | rR(2) : C Cv=0 3 3 F1e - X Xv=0 2 2 F1e |
| 13C18O | C | 2 | e | 1 | X | 3 | e | 1 | 91908.4584 | .9043813 | 91929.4146 | 20.9562 | 1.000000 | 1.4609e8 | 0.000000 : | pP(3) : C Cv=0 2 2 F1e - X Xv=0 3 3 F1e |
| 13C18O | C | 4 | e | 1 | X | 3 | e | 1 | 91933.1830 | 1.205842 | 91954.1392 | 20.9562 | 1.333333 | 1.0830e8 | 0.000000 : | rR(3) : C Cv=0 4 4 F1e - X Xv=0 3 3 F1e |
| 13C18O | C | 3 | e | 1 | X | 4 | e | 1 | 91905.0826 | 1.127699 | 91940.0088 | 34.9261 | 1.333333 | 1.3912e8 | 0.000000 : | pP(4) : C Cv=0 3 3 F1e - X Xv=0 4 4 F1e |
| 13C18O | C | 5 | e | 1 | X | 4 | e | 1 | 91936.8786 | 1.409623 | 91971.8047 | 34.9261 | 1.666667 | 1.1078e8 | 0.000000 : | rR(4) : C Cv=0 5 5 F1e - X Xv=0 4 4 F1e |
| 13C18O | C | 4 | e | 1 | X | 5 | e | 1 | 91901.7515 | 1.296383 | 91954.1392 | 52.3877 | 1.666667 | 1.3524e8 | 0.000000 : | pP(5) : C Cv=0 4 4 F1e - X Xv=0 5 5 F1e |
| 13C18O | C | 6 | e | 1 | X | 5 | e | 1 | 91940.6086 | 1.555660 | 91992.9963 | 52.3877 | 2.000000 | 1.1249e8 | 0.000000 : | rR(5) : C Cv=0 6 6 F1e - X Xv=0 5 5 F1e |
| 13C18O | C | 5 | e | 1 | X | 6 | e | 1 | 91898.4645 | 1.406934 | 91971.8047 | 73.3402 | 2.000000 | 1.3277e8 | 0.000000 : | pP(6) : C Cv=0 5 5 F1e - X Xv=0 6 6 F1e |
| 13C18O | C | 7 | e | 1 | X | 6 | e | 1 | 91944.3808 | 1.641423 | 92017.7210 | 73.3402 | 2.333333 | 1.1376e8 | 0.000000 : | rR(6) : C Cv=0 7 7 F1e - X Xv=0 6 6 F1e |
| 13C18O | C | 6 | e | 1 | X | 7 | e | 1 | 91895.2133 | 1.459856 | 91992.9963 | 97.7830 | 2.333333 | 1.3105e8 | 0.000000 : | pP(7) : C Cv=0 6 6 F1e - X Xv=0 7 7 F1e |
| 13C18O | C | 8 | e | 1 | X | 7 | e | 1 | 91948.1899 | 1.668407 | 92045.9729 | 97.7830 | 2.666667 | 1.1473e8 | 0.000000 : | rR(7) : C Cv=0 8 8 F1e - X Xv=0 7 7 F1e |
| 13C18O | C | 7 | e | 1 | X | 8 | e | 1 | 91892.0058 | 1.459230 | 92017.7210 | 125.7152 | 2.666667 | 1.2979e8 | 0.000000 : | pP(8) : C Cv=0 7 7 F1e - X Xv=0 8 8 F1e |
| 13C18O | C | 9 | e | 1 | X | 8 | e | 1 | 91952.0397 | 1.641634 | 92077.7549 | 125.7152 | 3.000000 | 1.1550e8 | 0.000000 : | rR(8) : C Cv=0 9 9 F1e - X Xv=0 8 8 F1e |
| 13C18O | C | 8 | e | 1 | X | 9 | e | 1 | 91888.8371 | 1.411991 | 92045.9729 | 157.1358 | 3.000000 | 1.2882e8 | 0.000000 : | pP(9) : C Cv=0 8 8 F1e - X Xv=0 9 9 F1e |
| 13C18O | C | 10 | e | 1 | X | 9 | e | 1 | 91955.9229 | 1.568879 | 92113.0588 | 157.1358 | 3.333333 | 1.1612e8 | 0.000000 : | rR(9) : C Cv=0 10 10 F1e - X Xv=0 9 9 F1e |
| 13C18O | C | 9 | e | 1 | X | 10 | e | 1 | 91885.7111 | 1.327033 | 92077.7549 | 192.0438 | 3.333333 | 1.2805e8 | 0.000000 : | pP(10) : C Cv=0 9 9 F1e - X Xv=0 10 10 F1e |
| 13C18O | C | 11 | e | 1 | X | 10 | e | 1 | 91959.8529 | 1.459736 | 92151.8967 | 192.0438 | 3.666667 | 1.1664e8 | 0.000000 : | rR(10) : C Cv=0 11 11 F1e - X Xv=0 10 10 F1e |
| 13C18O | C | 10 | e | 1 | X | 11 | e | 1 | 91882.6208 | 1.214243 | 92113.0588 | 230.4379 | 3.666667 | 1.2743e8 | 0.000000 : | pP(11) : C Cv=0 10 10 F1e - X Xv=0 11 11 F1e |
| 13C18O | C | 12 | e | 1 | X | 11 | e | 1 | 91963.8197 | 1.324628 | 92194.2577 | 230.4379 | 4.000000 | 1.1708e8 | 0.000000 : | rR(11) : C Cv=0 12 12 F1e - X Xv=0 11 11 F1e |
| 13C18O | C | 11 | e | 1 | X | 12 | e | 1 | 91879.5799 | 1.083595 | 92151.8967 | 272.3168 | 4.000000 | 1.2691e8 | 0.000000 : | pP(12) : C Cv=0 11 11 F1e - X Xv=0 12 12 F1e |
| 13C18O | C | 13 | e | 1 | X | 12 | e | 1 | 91967.8160 | 1.173894 | 92240.1329 | 272.3168 | 4.333333 | 1.1746e8 | 0.000000 : | rR(12) : C Cv=0 13 13 F1e - X Xv=0 12 12 F1e |
| 13C18O | C | 12 | e | 1 | X | 13 | e | 1 | 91876.5786 | .9443794 | 92194.2577 | 317.6791 | 4.333333 | 1.2648e8 | 0.000000 : | pP(13) : C Cv=0 12 12 F1e - X Xv=0 13 13 F1e |
| 13C18O | C | 14 | e | 1 | X | 13 | e | 1 | 91971.8583 | 1.017024 | 92289.5374 | 317.6791 | 4.666667 | 1.1779e8 | 0.000000 : | rR(13) : C Cv=0 14 14 F1e - X Xv=0 13 13 F1e |
| 13C18O | C | 13 | e | 1 | X | 14 | e | 1 | 91873.6097 | .8046310 | 92240.1329 | 366.5231 | 4.666667 | 1.2611e8 | 0.000000 : | pP(14) : C Cv=0 13 13 F1e - X Xv=0 14 14 F1e |
| 13C18O | C | 15 | e | 1 | X | 14 | e | 1 | 91975.9384 | .8621046 | 92342.4616 | 366.5231 | 5.000000 | 1.1807e8 | 0.000000 : | rR(14) : C Cv=0 15 15 F1e - X Xv=0 14 14 F1e |
| 13C18O | C | 14 | e | 1 | X | 15 | e | 1 | 91870.6902 | .6707753 | 92289.5374 | 418.8472 | 5.000000 | 1.2578e8 | 0.000000 : | pP(15) : C Cv=0 14 14 F1e - X Xv=0 15 15 F1e |
| 13C18O | C | 16 | e | 1 | X | 15 | e | 1 | 91980.0553 | .7154937 | 92398.9025 | 418.8472 | 5.333333 | 1.1833e8 | 0.000000 : | rR(15) : C Cv=0 16 16 F1e - X Xv=0 15 15 F1e |
| 13C18O | C | 15 | e | 1 | X | 16 | e | 1 | 91867.8120 | .5474926 | 92342.4616 | 474.6496 | 5.333333 | 1.2550e8 | 0.000000 : | pP(16) : C Cv=0 15 15 F1e - X Xv=0 16 16 F1e |
| 13C18O | C | 17 | e | 1 | X | 16 | e | 1 | 91984.2091 | .5817108 | 92458.8587 | 474.6496 | 5.666667 | 1.1856e8 | 0.000000 : | rR(16) : C Cv=0 17 17 F1e - X Xv=0 16 16 F1e |
| 13C18O | C | 16 | e | 1 | X | 17 | e | 1 | 91864.9742 | .4377629 | 92398.9025 | 533.9283 | 5.666667 | 1.2525e8 | 0.000000 : | pP(17) : C Cv=0 16 16 F1e - X Xv=0 17 17 F1e |
| 13C18O | C | 18 | e | 1 | X | 17 | e | 1 | 91988.4008 | .4635137 | 92522.3291 | 533.9283 | 6.000000 | 1.1876e8 | 0.000000 : | rR(17) : C Cv=0 18 18 F1e - X Xv=0 17 17 F1e |
| 13C18O | C | 17 | e | 1 | X | 18 | e | 1 | 91862.1775 | .3430504 | 92458.8587 | 596.6812 | 6.000000 | 1.2503e8 | 0.000000 : | pP(18) : C Cv=0 17 17 F1e - X Xv=0 18 18 F1e |
| 13C18O | C | 19 | e | 1 | X | 18 | e | 1 | 91992.6315 | .3621088 | 92589.3127 | 596.6812 | 6.333333 | 1.1895e8 | 0.000000 : | rR(18) : C Cv=0 19 19 F1e - X Xv=0 18 18 F1e |
| 13C18O | C | 18 | e | 1 | X | 19 | e | 1 | 91859.4228 | .2635741 | 92522.3291 | 662.9063 | 6.333333 | 1.2483e8 | 0.000000 : | pP(19) : C Cv=0 18 18 F1e - X Xv=0 19 19 F1e |
| 13C18O | C | 20 | e | 1 | X | 19 | e | 1 | 91996.8976 | .2774464 | 92659.8039 | 662.9063 | 6.666667 | 1.1912e8 | 0.000000 : | rR(19) : C Cv=0 20 20 F1e - X Xv=0 19 19 F1e |
| 13C18O | C | 19 | e | 1 | X | 20 | e | 1 | 91856.7115 | .1986167 | 92589.3127 | 732.6012 | 6.666667 | 1.2465e8 | 0.000000 : | pP(20) : C Cv=0 19 19 F1e - X Xv=0 20 20 F1e |
| 13C18O | C | 21 | e | 1 | X | 20 | e | 1 | 92001.1995 | .2085475 | 92733.8006 | 732.6012 | 7.000000 | 1.1927e8 | 0.000000 : | rR(20) : C Cv=0 21 21 F1e - X Xv=0 20 20 F1e |
| 13C18O | C | 20 | e | 1 | X | 21 | e | 1 | 91854.0405 | .1468316 | 92659.8039 | 805.7634 | 7.000000 | 1.2449e8 | 0.000000 : | pP(21) : C Cv=0 20 20 F1e - X Xv=0 21 21 F1e |
| 13C18O | C | 22 | e | 1 | X | 21 | e | 1 | 92005.5398 | .1538236 | 92811.3031 | 805.7634 | 7.333333 | 1.1941e8 | 0.000000 : | rR(21) : C Cv=0 22 22 F1e - X Xv=0 21 21 F1e |
| 13C18O | C | 21 | e | 1 | X | 22 | e | 1 | 91851.4102 | .1065175 | 92733.8006 | 882.3904 | 7.333333 | 1.2434e8 | 0.000000 : | pP(22) : C Cv=0 21 21 F1e - X Xv=0 22 22 F1e |
| 13C18O | C | 23 | e | 1 | X | 22 | e | 1 | 92009.9176 | .1113592 | 92892.3080 | 882.3904 | 7.666667 | 1.1955e8 | 0.000000 : | rR(22) : C Cv=0 23 23 F1e - X Xv=0 22 22 F1e |
| 13C18O | C | 22 | e | 1 | X | 23 | e | 1 | 91848.8235 | .0758425 | 92811.3031 | 962.4797 | 7.666667 | 1.2420e8 | 0.000000 : | pP(23) : C Cv=0 22 22 F1e - X Xv=0 23 23 F1e |
| 13C18O | C | 24 | e | 1 | X | 23 | e | 1 | 92014.3427 | .0791400 | 92976.8223 | 962.4797 | 8.000000 | 1.1967e8 | 0.000000 : | rR(23) : C Cv=0 24 24 F1e - X Xv=0 23 23 F1e |
| 13C18O | C | 23 | e | 1 | X | 24 | e | 1 | 91846.2797 | .0530124 | 92892.3080 | 1046.0283 | 8.000000 | 1.2408e8 | 0.000000 : | pP(24) : C Cv=0 23 23 F1e - X Xv=0 24 24 F1e |
| 13C18O | C | 25 | e | 1 | X | 24 | e | 1 | 92018.7949 | .0552212 | 93064.8231 | 1046.0283 | 8.333333 | 1.1978e8 | 0.000000 : | rR(24) : C Cv=0 25 25 F1e - X Xv=0 24 24 F1e |
| 13C18O | C | 24 | e | 1 | X | 25 | e | 1 | 91843.7889 | .0363821 | 92976.8223 | 1133.0334 | 8.333333 | 1.2396e8 | 0.000000 : | pP(25) : C Cv=0 24 24 F1e - X Xv=0 25 25 F1e |
| 13C18O | C | 26 | e | 1 | X | 25 | e | 1 | 92023.2842 | .0378374 | 93156.3176 | 1133.0334 | 8.666667 | 1.1989e8 | 0.000000 : | rR(25) : C Cv=0 26 26 F1e - X Xv=0 25 25 F1e |
| 13C18O | C | 25 | e | 1 | X | 26 | e | 1 | 91841.3312 | .0245194 | 93064.8231 | 1223.4920 | 8.666667 | 1.2386e8 | 0.000000 : | pP(26) : C Cv=0 25 25 F1e - X Xv=0 26 26 F1e |
| 13C18O | C | 27 | e | 1 | X | 26 | e | 1 | 92027.8069 | .0254625 | 93251.2989 | 1223.4920 | 9.000000 | 1.1999e8 | 0.000000 : | rR(26) : C Cv=0 27 27 F1e - X Xv=0 26 26 F1e |
| 13C18O | C | 26 | e | 1 | X | 27 | e | 1 | 91838.9167 | .0162294 | 93156.3176 | 1317.4009 | 9.000000 | 1.2376e8 | 0.000000 : | pP(27) : C Cv=0 26 26 F1e - X Xv=0 27 27 F1e |
| 13C18O | C | 28 | e | 1 | X | 27 | e | 1 | 92032.3659 | .0168305 | 93349.7667 | 1317.4009 | 9.333333 | 1.2009e8 | 0.000000 : | rR(27) : C Cv=0 28 28 F1e - X Xv=0 27 27 F1e |
| 13C18O | C | 27 | e | 1 | X | 28 | e | 1 | 91836.5420 | .0105517 | 93251.2989 | 1414.7569 | 9.333333 | 1.2366e8 | 0.000000 : | pP(28) : C Cv=0 27 27 F1e - X Xv=0 28 28 F1e |
| 13C18O | C | 29 | e | 1 | X | 28 | e | 1 | 92036.9621 | .0109285 | 93451.7190 | 1414.7569 | 9.666667 | 1.2018e8 | 0.000000 : | rR(28) : C Cv=0 29 29 F1e - X Xv=0 28 28 F1e |
| 13C18O | C | 28 | e | 1 | X | 29 | e | 1 | 91834.2102 | .0067392 | 93349.7667 | 1515.5565 | 9.666667 | 1.2358e8 | 0.000000 : | pP(29) : C Cv=0 28 28 F1e - X Xv=0 29 29 F1e |
| 13C18O | C | 30 | e | 1 | X | 29 | e | 1 | 92041.5917 | .0069716 | 93557.1483 | 1515.5565 | 10.00000 | 1.2027e8 | 0.000000 : | rR(29) : C Cv=0 30 30 F1e - X Xv=0 29 29 F1e |
| 13C18O | C | 29 | e | 1 | X | 30 | e | 1 | 91831.9226 | .0042288 | 93451.7190 | 1619.7964 | 10.00000 | 1.2350e8 | 0.000000 : | pP(30) : C Cv=0 29 29 F1e - X Xv=0 30 30 F1e |
| 13C18O | C | 31 | e | 1 | X | 30 | e | 1 | 92046.2585 | .0043698 | 93666.0549 | 1619.7964 | 10.33333 | 1.2035e8 | 0.000000 : | rR(30) : C Cv=0 31 31 F1e - X Xv=0 30 30 F1e |
| 13C18O | C | 30 | e | 1 | X | 31 | e | 1 | 91829.6755 | .0026073 | 93557.1483 | 1727.4728 | 10.33333 | 1.2342e8 | 0.000000 : | pP(31) : C Cv=0 30 30 F1e - X Xv=0 31 31 F1e |
| 13C18O | C | 32 | e | 1 | X | 31 | e | 1 | 92050.9615 | .0026914 | 93778.4343 | 1727.4728 | 10.66667 | 1.2043e8 | 0.000000 : | rR(31) : C Cv=0 32 32 F1e - X Xv=0 31 31 F1e |
| 13C18O | C | 31 | e | 1 | X | 32 | e | 1 | 91827.4729 | .0015796 | 93666.0549 | 1838.5820 | 10.66667 | 1.2335e8 | 0.000000 : | pP(32) : C Cv=0 31 31 F1e - X Xv=0 32 32 F1e |
| 13C18O | C | 33 | e | 1 | X | 32 | e | 1 | 92055.6995 | .0016290 | 93894.2815 | 1838.5820 | 11.00000 | 1.2050e8 | 0.000000 : | rR(32) : C Cv=0 33 33 F1e - X Xv=0 32 32 F1e |
| 13C18O | C | 32 | e | 1 | X | 33 | e | 1 | 91825.3141 | .0009405 | 93778.4343 | 1953.1202 | 11.00000 | 1.2328e8 | 0.000000 : | pP(33) : C Cv=0 32 32 F1e - X Xv=0 33 33 F1e |
| 13C18O | C | 34 | e | 1 | X | 33 | e | 1 | 92060.4715 | .0009690 | 94013.5918 | 1953.1202 | 11.33333 | 1.2057e8 | 0.000000 : | rR(33) : C Cv=0 34 34 F1e - X Xv=0 33 33 F1e |
| 13C18O | C | 33 | e | 1 | X | 34 | e | 1 | 91823.1981 | .0005503 | 93894.2815 | 2071.0834 | 11.33333 | 1.2321e8 | 0.000000 : | pP(34) : C Cv=0 33 33 F1e - X Xv=0 34 34 F1e |
| 13C18O | C | 35 | e | 1 | X | 34 | e | 1 | 92065.2735 | .0005665 | 94136.3569 | 2071.0834 | 11.66667 | 1.2064e8 | 0.000000 : | rR(34) : C Cv=0 35 35 F1e - X Xv=0 34 34 F1e |
| 13C18O | C | 34 | e | 1 | X | 35 | e | 1 | 91821.1243 | .0003165 | 94013.5918 | 2192.4675 | 11.66667 | 1.2315e8 | 0.000000 : | pP(35) : C Cv=0 34 34 F1e - X Xv=0 35 35 F1e |
| 13C18O | C | 36 | e | 1 | X | 35 | e | 1 | 92070.1135 | .0003255 | 94262.5810 | 2192.4675 | 12.00000 | 1.2071e8 | 0.000000 : | rR(35) : C Cv=0 36 36 F1e - X Xv=0 35 35 F1e |
| 13C18O | C | 35 | e | 1 | X | 36 | e | 1 | 91819.0887 | .0001789 | 94136.3569 | 2317.2682 | 12.00000 | 1.2310e8 | 0.000000 : | pP(36) : C Cv=0 35 35 F1e - X Xv=0 36 36 F1e |

| | | | | | | | | | | |
|---|---|---|---|---|---|---|---|---|---|---|
| 13C18O | C 37 e 1 | X 36 e 1 | 92074.9855 | .0001839 | 94392.2537 | 2317.2682 | 12.33333 | 1.2077e8 | 0.000000 | : rR(36) : C Cv=0 37 37 F1e - X Xv=0 36 36 F1e |
| 13C18O | C 38 e 1 | X 37 e 1 | 92079.8945 | .0001021 | 94525.3758 | 2445.4813 | 12.66667 | 1.2083e8 | 0.000000 | : rR(37) : C Cv=0 38 38 F1e - X Xv=0 37 37 F1e |
| 13C18O | B 1 e 1 | X 0 e 1 | 86920.8991 | .3333333 | 86920.8991 | 0.0000 | .3333333 | 68652057 | 0.000000 | : rR(0) : B Bv=0 1 1 F1e - X Xv=0 0 0 F1e |
| 13C18O | B 0 e 1 | X 1 e 1 | 86913.8645 | .3277961 | 86917.3573 | 3.4928 | .3333333 | 2.0591e8 | 0.000000 | : pP(1) : B Bv=0 0 0 F1e - X Xv=0 1 1 F1e |
| 13C18O | B 2 e 1 | X 1 e 1 | 86924.4874 | .6555922 | 86927.9802 | 3.4928 | .6666667 | 82392672 | 0.000000 | : rR(1) : B Bv=0 2 2 F1e - X Xv=0 1 1 F1e |
| 13C18O | B 1 e 1 | X 2 e 1 | 86910.4208 | .6339926 | 86920.8991 | 10.4783 | .6666667 | 1.3725e8 | 0.000000 | : pP(2) : B Bv=0 1 1 F1e - X Xv=0 2 2 F1e |
| 13C18O | B 3 e 1 | X 2 e 1 | 86928.1193 | .9509888 | 86938.5976 | 10.4783 | 1.000000 | 88288929 | 0.000000 | : rR(2) : B Bv=0 3 3 F1e - X Xv=0 2 2 F1e |
| 13C18O | B 2 e 1 | X 3 e 1 | 86907.0241 | .9043813 | 86927.9802 | 20.9562 | 1.000000 | 1.2351e8 | 0.000000 | : pP(3) : B Bv=0 2 2 F1e - X Xv=0 3 3 F1e |
| 13C18O | B 4 e 1 | X 3 e 1 | 86931.7973 | 1.205842 | 86952.7535 | 20.9562 | 1.333333 | 91570511 | 0.000000 | : rR(3) : B Bv=0 4 4 F1e - X Xv=0 3 3 F1e |
| 13C18O | B 3 e 1 | X 4 e 1 | 86903.6715 | 1.127699 | 86938.5976 | 34.9261 | 1.333333 | 1.1762e8 | 0.000000 | : pP(4) : B Bv=0 3 3 F1e - X Xv=0 4 4 F1e |
| 13C18O | B 5 e 1 | X 4 e 1 | 86935.5224 | 1.409623 | 86970.4486 | 34.9261 | 1.666667 | 93663699 | 0.000000 | : rR(4) : B Bv=0 5 5 F1e - X Xv=0 4 4 F1e |
| 13C18O | B 4 e 1 | X 5 e 1 | 86900.3658 | 1.296383 | 86952.7535 | 52.3877 | 1.666667 | 1.1434e8 | 0.000000 | : pP(5) : B Bv=0 4 4 F1e - X Xv=0 5 5 F1e |
| 13C18O | B 6 e 1 | X 5 e 1 | 86939.2932 | 1.555660 | 86991.6809 | 52.3877 | 2.000000 | 95117055 | 0.000000 | : rR(5) : B Bv=0 6 6 F1e - X Xv=0 5 5 F1e |
| 13C18O | B 5 e 1 | X 6 e 1 | 86897.1084 | 1.406934 | 86970.4486 | 73.3402 | 2.000000 | 1.1225e8 | 0.000000 | : pP(6) : B Bv=0 5 5 F1e - X Xv=0 6 6 F1e |
| 13C18O | B 7 e 1 | X 6 e 1 | 86943.1099 | 1.641423 | 87016.4501 | 73.3402 | 2.333333 | 96186578 | 0.000000 | : rR(6) : B Bv=0 7 7 F1e - X Xv=0 6 6 F1e |
| 13C18O | B 6 e 1 | X 7 e 1 | 86893.8978 | 1.459856 | 86991.6809 | 97.7830 | 2.333333 | 1.1080e8 | 0.000000 | : pP(7) : B Bv=0 6 6 F1e - X Xv=0 7 7 F1e |
| 13C18O | B 8 e 1 | X 7 e 1 | 86946.9720 | 1.668407 | 87044.7550 | 97.7830 | 2.666667 | 97007795 | 0.000000 | : rR(7) : B Bv=0 8 8 F1e - X Xv=0 7 7 F1e |
| 13C18O | B 7 e 1 | X 8 e 1 | 86890.7349 | 1.459230 | 87016.4501 | 125.7152 | 2.666667 | 1.0973e8 | 0.000000 | : pP(8) : B Bv=0 7 7 F1e - X Xv=0 8 8 F1e |
| 13C18O | B 9 e 1 | X 8 e 1 | 86950.8793 | 1.641634 | 87076.5945 | 125.7152 | 3.000000 | 97659169 | 0.000000 | : rR(8) : B Bv=0 9 9 F1e - X Xv=0 8 8 F1e |
| 13C18O | B 8 e 1 | X 9 e 1 | 86887.6192 | 1.411991 | 87044.7550 | 157.1358 | 3.000000 | 1.0891e8 | 0.000000 | : pP(9) : B Bv=0 8 8 F1e - X Xv=0 9 9 F1e |
| 13C18O | B 10 e 1 | X 9 e 1 | 86954.8315 | 1.568879 | 87111.9674 | 157.1358 | 3.333333 | 98189272 | 0.000000 | : rR(9) : B Bv=0 10 10 F1e - X Xv=0 9 9 F1e |
| 13C18O | B 9 e 1 | X 10 e 1 | 86884.5507 | 1.327033 | 87076.5945 | 192.0438 | 3.333333 | 1.0826e8 | 0.000000 | : pP(10) : B Bv=0 9 9 F1e - X Xv=0 10 10 F1e |
| 13C18O | B 11 e 1 | X 10 e 1 | 86958.8284 | 1.459736 | 87150.8722 | 192.0438 | 3.666667 | 98629781 | 0.000000 | : rR(10) : B Bv=0 11 11 F1e - X Xv=0 10 10 F1e |
| 13C18O | B 10 e 1 | X 11 e 1 | 86881.5294 | 1.214243 | 87111.9674 | 230.4379 | 3.666667 | 1.0774e8 | 0.000000 | : pP(11) : B Bv=0 10 10 F1e - X Xv=0 11 11 F1e |
| 13C18O | B 12 e 1 | X 11 e 1 | 86962.8724 | 1.324628 | 87193.3103 | 230.4379 | 4.000000 | 99002246 | 0.000000 | : rR(11) : B Bv=0 12 12 F1e - X Xv=0 11 11 F1e |
| 13C18O | B 11 e 1 | X 12 e 1 | 86878.5554 | 1.083595 | 87150.8722 | 272.3168 | 4.000000 | 1.0730e8 | 0.000000 | : pP(12) : B Bv=0 11 11 F1e - X Xv=0 12 12 F1e |
| 13C18O | B 13 e 1 | X 12 e 1 | 86966.9570 | 1.173894 | 87239.2738 | 272.3168 | 4.333333 | 99321802 | 0.000000 | : rR(12) : B Bv=0 13 13 F1e - X Xv=0 12 12 F1e |
| 13C18O | B 12 e 1 | X 13 e 1 | 86875.6313 | .9443794 | 87193.3103 | 317.6791 | 4.333333 | 1.0693e8 | 0.000000 | : pP(13) : B Bv=0 12 12 F1e - X Xv=0 13 13 F1e |
| 13C18O | B 14 e 1 | X 13 e 1 | 86971.0869 | 1.017024 | 87288.7659 | 317.6791 | 4.666667 | 99599443 | 0.000000 | : rR(13) : B Bv=0 14 14 F1e - X Xv=0 13 13 F1e |
| 13C18O | B 13 e 1 | X 14 e 1 | 86872.7507 | .8046310 | 87239.2738 | 366.5231 | 4.666667 | 1.0661e8 | 0.000000 | : pP(14) : B Bv=0 13 13 F1e - X Xv=0 14 14 F1e |
| 13C18O | B 15 e 1 | X 14 e 1 | 86975.2607 | .8621046 | 87341.7838 | 366.5231 | 5.000000 | 99843309 | 0.000000 | : rR(14) : B Bv=0 15 15 F1e - X Xv=0 14 14 F1e |
| 13C18O | B 14 e 1 | X 15 e 1 | 86869.9187 | .6707753 | 87288.7659 | 418.8472 | 5.000000 | 1.0634e8 | 0.000000 | : pP(15) : B Bv=0 14 14 F1e - X Xv=0 15 15 F1e |
| 13C18O | B 16 e 1 | X 15 e 1 | 86979.4768 | .7154937 | 87398.3240 | 418.8472 | 5.333333 | 1.0006e8 | 0.000000 | : rR(15) : B Bv=0 16 16 F1e - X Xv=0 15 15 F1e |
| 13C18O | B 15 e 1 | X 16 e 1 | 86867.1343 | .5474926 | 87341.7838 | 474.6496 | 5.333333 | 1.0610e8 | 0.000000 | : pP(16) : B Bv=0 15 15 F1e - X Xv=0 16 16 F1e |
| 13C18O | B 17 e 1 | X 16 e 1 | 86983.6962 | .5817108 | 87458.3457 | 474.6496 | 5.666667 | 1.0025e8 | 0.000000 | : rR(16) : B Bv=0 17 17 F1e - X Xv=0 16 16 F1e |
| 13C18O | B 16 e 1 | X 17 e 1 | 86864.3957 | .4377629 | 87398.3240 | 533.9283 | 5.666667 | 1.0589e8 | 0.000000 | : pP(17) : B Bv=0 16 16 F1e - X Xv=0 17 17 F1e |
| 13C18O | B 18 e 1 | X 17 e 1 | 86988.0386 | .4635137 | 87521.9669 | 533.9283 | 6.000000 | 1.0043e8 | 0.000000 | : rR(17) : B Bv=0 18 18 F1e - X Xv=0 17 17 F1e |
| 13C18O | B 17 e 1 | X 18 e 1 | 86861.6645 | .3430504 | 87458.3457 | 596.6812 | 6.000000 | 1.0570e8 | 0.000000 | : pP(18) : B Bv=0 17 17 F1e - X Xv=0 18 18 F1e |
| 13C18O | B 19 e 1 | X 18 e 1 | 86992.3917 | .3621088 | 87589.0730 | 596.6812 | 6.333333 | 1.0059e8 | 0.000000 | : rR(18) : B Bv=0 19 19 F1e - X Xv=0 18 18 F1e |
| 13C18O | B 18 e 1 | X 19 e 1 | 86859.0606 | .2635741 | 87521.9669 | 662.9063 | 6.333333 | 1.0554e8 | 0.000000 | : pP(19) : B Bv=0 18 18 F1e - X Xv=0 19 19 F1e |
| 13C18O | B 20 e 1 | X 19 e 1 | 86996.7785 | .2774464 | 87659.6849 | 662.9063 | 6.666667 | 1.0073e8 | 0.000000 | : rR(19) : B Bv=0 20 20 F1e - X Xv=0 19 19 F1e |
| 13C18O | B 19 e 1 | X 20 e 1 | 86856.4718 | .1986167 | 87589.0730 | 732.6012 | 6.666667 | 1.0538e8 | 0.000000 | : pP(20) : B Bv=0 19 19 F1e - X Xv=0 20 20 F1e |
| 13C18O | B 21 e 1 | X 20 e 1 | 87001.2082 | .2085475 | 87733.8094 | 732.6012 | 7.000000 | 1.0086e8 | 0.000000 | : rR(20) : B Bv=0 21 21 F1e - X Xv=0 20 20 F1e |
| 13C18O | B 20 e 1 | X 21 e 1 | 86853.9215 | .1468316 | 87659.6849 | 805.7634 | 7.000000 | 1.0525e8 | 0.000000 | : pP(21) : B Bv=0 20 20 F1e - X Xv=0 21 21 F1e |
| 13C18O | B 22 e 1 | X 21 e 1 | 87005.6801 | .1538236 | 87811.4435 | 805.7634 | 7.333333 | 1.0098e8 | 0.000000 | : rR(21) : B Bv=0 22 22 F1e - X Xv=0 21 21 F1e |
| 13C18O | B 21 e 1 | X 22 e 1 | 86851.4189 | .1065175 | 87733.8094 | 882.3904 | 7.333333 | 1.0512e8 | 0.000000 | : pP(22) : B Bv=0 21 21 F1e - X Xv=0 22 22 F1e |
| 13C18O | B 23 e 1 | X 22 e 1 | 87010.1929 | .1113592 | 87892.5834 | 882.3904 | 7.666667 | 1.0110e8 | 0.000000 | : rR(22) : B Bv=0 23 23 F1e - X Xv=0 22 22 F1e |
| 13C18O | B 22 e 1 | X 23 e 1 | 86848.9638 | .0758425 | 87811.4435 | 962.4797 | 7.666667 | 1.0501e8 | 0.000000 | : pP(23) : B Bv=0 22 22 F1e - X Xv=0 23 23 F1e |
| 13C18O | B 24 e 1 | X 23 e 1 | 87014.7458 | .0791400 | 87977.2254 | 962.4797 | 8.000000 | 1.0120e8 | 0.000000 | : rR(23) : B Bv=0 24 24 F1e - X Xv=0 23 23 F1e |
| 13C18O | B 23 e 1 | X 24 e 1 | 86846.5551 | .0530124 | 87892.5834 | 1046.0283 | 8.000000 | 1.0490e8 | 0.000000 | : pP(24) : B Bv=0 23 23 F1e - X Xv=0 24 24 F1e |
| 13C18O | B 25 e 1 | X 24 e 1 | 87019.3373 | .0552212 | 88065.3656 | 1046.0283 | 8.333333 | 1.0130e8 | 0.000000 | : rR(24) : B Bv=0 25 25 F1e - X Xv=0 24 24 F1e |
| 13C18O | B 24 e 1 | X 25 e 1 | 86844.1920 | .0363821 | 87977.2254 | 1133.0334 | 8.333333 | 1.0480e8 | 0.000000 | : pP(25) : B Bv=0 24 24 F1e - X Xv=0 25 25 F1e |
| 13C18O | B 26 e 1 | X 25 e 1 | 87023.9741 | .0378374 | 88157.0075 | 1133.0334 | 8.666667 | 1.0139e8 | 0.000000 | : rR(25) : B Bv=0 26 26 F1e - X Xv=0 25 25 F1e |
| 13C18O | B 25 e 1 | X 26 e 1 | 86841.8736 | .0245194 | 88065.3656 | 1223.4920 | 8.666667 | 1.0471e8 | 0.000000 | : pP(26) : B Bv=0 25 25 F1e - X Xv=0 26 26 F1e |
| 13C18O | B 27 e 1 | X 26 e 1 | 87028.6509 | .0254625 | 88252.1429 | 1223.4920 | 9.000000 | 1.0148e8 | 0.000000 | : rR(26) : B Bv=0 27 27 F1e - X Xv=0 26 26 F1e |
| 13C18O | B 26 e 1 | X 27 e 1 | 86839.6066 | .0162294 | 88157.0075 | 1317.4009 | 9.000000 | 1.0463e8 | 0.000000 | : pP(27) : B Bv=0 26 26 F1e - X Xv=0 27 27 F1e |
| 13C18O | B 28 e 1 | X 27 e 1 | 87033.3613 | .0168305 | 88350.7622 | 1317.4009 | 9.333333 | 1.0156e8 | 0.000000 | : rR(27) : B Bv=0 28 28 F1e - X Xv=0 27 27 F1e |
| 13C18O | B 27 e 1 | X 28 e 1 | 86837.3860 | .0105517 | 88252.1429 | 1414.7569 | 9.333333 | 1.0455e8 | 0.000000 | : pP(28) : B Bv=0 27 27 F1e - X Xv=0 28 28 F1e |
| 13C18O | B 29 e 1 | X 28 e 1 | 87038.1146 | .0109285 | 88452.8715 | 1414.7569 | 9.666667 | 1.0164e8 | 0.000000 | : rR(28) : B Bv=0 29 29 F1e - X Xv=0 28 28 F1e |
| 13C18O | B 28 e 1 | X 29 e 1 | 86835.2057 | .0067392 | 88350.7622 | 1515.5565 | 9.666667 | 1.0448e8 | 0.000000 | : pP(29) : B Bv=0 28 28 F1e - X Xv=0 29 29 F1e |
| 13C18O | B 30 e 1 | X 29 e 1 | 87042.9008 | .0069716 | 88558.4574 | 1515.5565 | 10.00000 | 1.0172e8 | 0.000000 | : rR(29) : B Bv=0 30 30 F1e - X Xv=0 29 29 F1e |
| 13C18O | B 29 e 1 | X 30 e 1 | 86833.0751 | .0042288 | 88452.8715 | 1619.7964 | 10.00000 | 1.0441e8 | 0.000000 | : pP(30) : B Bv=0 29 29 F1e - X Xv=0 30 30 F1e |
| 13C18O | B 31 e 1 | X 30 e 1 | 87047.7207 | .0043698 | 88667.5171 | 1619.7964 | 10.33333 | 1.0179e8 | 0.000000 | : rR(30) : B Bv=0 31 31 F1e - X Xv=0 30 30 F1e |
| 13C18O | B 30 e 1 | X 31 e 1 | 86830.9846 | .0026073 | 88558.4574 | 1727.4728 | 10.33333 | 1.0434e8 | 0.000000 | : pP(31) : B Bv=0 30 30 F1e - X Xv=0 31 31 F1e |
| 13C18O | B 32 e 1 | X 31 e 1 | 87052.5802 | .0026914 | 88780.0530 | 1727.4728 | 10.66667 | 1.0186e8 | 0.000000 | : rR(31) : B Bv=0 32 32 F1e - X Xv=0 31 31 F1e |
| 13C18O | B 31 e 1 | X 32 e 1 | 86828.9350 | .0015796 | 88667.5171 | 1838.5820 | 10.66667 | 1.0428e8 | 0.000000 | : pP(32) : B Bv=0 31 31 F1e - X Xv=0 32 32 F1e |
| 13C18O | B 33 e 1 | X 32 e 1 | 87057.4912 | .0016290 | 88896.0733 | 1838.5820 | 11.00000 | 1.0192e8 | 0.000000 | : rR(32) : B Bv=0 33 33 F1e - X Xv=0 32 32 F1e |
| 13C18O | B 32 e 1 | X 33 e 1 | 86826.9328 | .0009405 | 88780.0530 | 1953.1202 | 11.00000 | 1.0422e8 | 0.000000 | : pP(33) : B Bv=0 32 32 F1e - X Xv=0 33 33 F1e |
| 13C18O | B 34 e 1 | X 33 e 1 | 87062.4277 | .0009690 | 89015.5479 | 1953.1202 | 11.33333 | 1.0198e8 | 0.000000 | : rR(33) : B Bv=0 34 34 F1e - X Xv=0 33 33 F1e |
| 13C18O | B 33 e 1 | X 34 e 1 | 86824.9899 | .0005503 | 88896.0733 | 2071.0834 | 11.33333 | 1.0417e8 | 0.000000 | : pP(34) : B Bv=0 33 33 F1e - X Xv=0 34 34 F1e |
| 13C18O | B 35 e 1 | X 34 e 1 | 87067.4174 | .0005665 | 89138.5008 | 2071.0834 | 11.66667 | 1.0204e8 | 0.000000 | : rR(34) : B Bv=0 35 35 F1e - X Xv=0 34 34 F1e |
| 13C18O | B 34 e 1 | X 35 e 1 | 86823.0804 | .0003165 | 89015.5479 | 2192.4675 | 11.66667 | 1.0412e8 | 0.000000 | : pP(35) : B Bv=0 34 34 F1e - X Xv=0 35 35 F1e |
| 13C18O | B 36 e 1 | X 35 e 1 | 87072.3997 | .0003255 | 89264.8671 | 2192.4675 | 12.00000 | 1.0210e8 | 0.000000 | : rR(35) : B Bv=0 36 36 F1e - X Xv=0 35 35 F1e |
| 13C18O | B 35 e 1 | X 36 e 1 | 86821.2326 | .0001789 | 89138.5008 | 2317.2682 | 12.00000 | 1.0407e8 | 0.000000 | : pP(36) : B Bv=0 35 35 F1e - X Xv=0 36 36 F1e |
| 13C18O | B 37 e 1 | X 36 e 1 | 87077.4444 | .0001839 | 89394.7126 | 2317.2682 | 12.33333 | 1.0216e8 | 0.000000 | : rR(36) : B Bv=0 37 37 F1e - X Xv=0 36 36 F1e |
| 13C18O | B 38 e 1 | X 37 e 1 | 87082.5221 | .0001021 | 89528.0034 | 2445.4813 | 12.66667 | 1.0221e8 | 0.000000 | : rR(37) : B Bv=0 38 38 F1e - X Xv=0 37 37 F1e |
| 13C18O | Excited 1 e 1 | X 0 e 1 | 67560.4283 | .5936401 | 67560.4283 | 0.0000 | .5936401 | 57411986 | 0.000000 | : rR(0) : Excited Av=2 1 1 F1e - X Xv=0 0 0 F1e |
| 13C18O | Excited 1 e 2 | X 0 e 1 | 67635.7481 | .0038056 | 67635.7481 | 0.0000 | .0038056 | 369274.0 | 0.000000 | : rR1(0) : Excited a3Pv=13 1 1 F1e - X Xv=0 0 0 F1e |
| 13C18O | Excited 1 e 3 | X 0 e 1 | 67675.0375 | .8221443 | 67675.0375 | 0.0000 | .8221443 | 79916370 | 0.000000 | : sR21(0) : Excited a3Pv=13 1 2 F2e - X Xv=0 0 0 F1e |
| 13C18O | Excited 1 e 4 | X 0 e 1 | 67803.8618 | .3684130 | 67803.8618 | 0.0000 | .3684130 | 36016407 | 0.000000 | : qR1(0) : Excited ev=4 1 0 F1e - X Xv=0 0 0 F1e |
| 13C18O | Excited 1 e 5 | X 0 e 1 | 67810.2129 | .2262309 | 67810.2129 | 0.0000 | .2262309 | 22122763 | 0.000000 | : sR31(0) : Excited ev=4 1 2 F3e - X Xv=0 0 0 F1e |
| 13C18O | Excited 1 e 6 | X 0 e 1 | 67987.8961 | .6252107 | 67987.8961 | 0.0000 | .6252107 | 61620236 | 0.000000 | : sR31(0) : Excited dv=7 1 2 F3e - X Xv=0 0 0 F1e |
| 13C18O | Excited 1 e 7 | X 0 e 1 | 67992.6434 | .6938889 | 67992.6434 | 0.0000 | .6938889 | 68403433 | 0.000000 | : rR21(0) : Excited a'v=12 1 1 F2e - X Xv=0 0 0 F1e |
| 13C18O | Excited 1 f 1 | X 1 e 1 | 67556.9350 | .8756357 | 67560.4278 | 3.4928 | .8904272 | 86101441 | 0.000000 | : qQ(1) : Excited Av=2 1 1 F1f - X Xv=0 1 1 F1e |
| 13C18O | Excited 1 f 3 | X 1 e 1 | 67636.0698 | .0061963 | 67639.5626 | 3.4928 | .0063009 | 611422.4 | 0.000000 | : qQ1(1) : Excited a3Pv=13 1 1 F1f - X Xv=0 1 1 F1e |
| 13C18O | Excited 1 f 4 | X 1 e 1 | 67671.5585 | 1.212153 | 67675.0513 | 3.4928 | 1.232629 | 1.1980e8 | 0.000000 | : rQ21(1) : Excited a3Pv=13 1 2 F2f - X Xv=0 1 1 F1e |
| 13C18O | Excited 1 f 5 | X 1 e 1 | 67802.7807 | .8770851 | 67806.2734 | 3.4928 | .8919011 | 87188935 | 0.000000 | : qQ21(1) : Excited ev=4 1 1 F2f - X Xv=0 1 1 F1e |
| 13C18O | Excited 1 f 6 | X 1 e 1 | 67984.4030 | .9145211 | 67987.8958 | 3.4928 | .9299694 | 91642879 | 0.000000 | : rQ31(1) : Excited dv=7 1 2 F3f - X Xv=0 1 1 F1e |
| 13C18O | Excited 1 f 7 | X 1 e 1 | 67987.6666 | .8269986 | 67991.1594 | 3.4928 | .8409685 | 82884305 | 0.000000 | : pQ1(1) : Excited a'v=12 1 0 F1f - X Xv=0 1 1 F1e |
| 13C18O | Excited 1 f 8 | X 1 e 1 | 67994.9809 | .2042618 | 67998.4737 | 3.4928 | .2077123 | 20478349 | 0.000000 | : rQ31(1) : Excited a'v=12 1 2 F3f - X Xv=0 1 1 F1e |
| 13C18O | Excited 2 e 1 | X 1 e 1 | 67562.6044 | .8751035 | 67566.0972 | 3.4928 | .8898860 | 51642462 | 0.000000 | : rR(1) : Excited Av=2 2 2 F1e - X Xv=0 1 1 F1e |
| 13C18O | Excited 2 e 2 | X 1 e 1 | 67637.1319 | .0165778 | 67640.6247 | 3.4928 | .0168579 | 981547.7 | 0.000000 | : qR1(1) : Excited a3Pv=13 2 1 F1e - X Xv=0 1 1 F1e |

| | | | | | | | | | | | | |
|---|---|---|---|---|---|---|---|---|---|---|---|---|
| 13C18O Excited | 2 e 3 | X | 1 e 1 | 67676.7120 | 1.188186 | 67680.2048 | 3.4928 | 1.208257 | 70474259 | 0.000000 | : | rR21(1) : Excited a3Pv=13 2 2 F2e - X Xv=0 1 1 F1e |
| 13C18O Excited | 2 e 4 | X | 1 e 1 | 67710.2945 | .0150916 | 67713.7873 | 3.4928 | .0153465 | 896452.2 | 0.000000 | : | sR31(1) : Excited a3Pv=13 2 3 F3e - X Xv=0 1 1 F1e |
| 13C18O Excited | 2 e 5 | X | 1 e 1 | 67802.5469 | .4989555 | 67806.0397 | 3.4928 | .5073840 | 29759675 | 0.000000 | : | qR1(1) : Excited ev=4 2 1 F1e - X Xv=0 1 1 F1e |
| 13C18O Excited | 2 e 6 | X | 1 e 1 | 67813.3319 | .3772948 | 67816.8247 | 3.4928 | .3836681 | 22514090 | 0.000000 | : | sR31(1) : Excited ev=4 2 3 F3e - X Xv=0 1 1 F1e |
| 13C18O Excited | 2 e 7 | X | 1 e 1 | 67949.8912 | .0058195 | 67953.3840 | 3.4928 | .0059178 | 349367.8 | 0.000000 | : | rR21(1) : Excited dv=7 2 2 F2e - X Xv=0 1 1 F1e |
| 13C18O Excited | 2 e 8 | X | 1 e 1 | 67988.9639 | .9136111 | 67992.4567 | 3.4928 | .9290440 | 54942067 | 0.000000 | : | sR31(1) : Excited dv=7 2 3 F3e - X Xv=0 1 1 F1e |
| 13C18O Excited | 2 e 9 | X | 1 e 1 | 67993.3010 | 1.026302 | 67996.7938 | 3.4928 | 1.043639 | 61730809 | 0.000000 | : | rR21(1) : Excited a'v=12 2 2 F2e - X Xv=0 1 1 F1e |
| 13C18O Excited | 1 e 1 | X | 2 e 1 | 67549.9500 | .2822725 | 67560.4283 | 10.4783 | .2968200 | 28692639 | 0.000000 | : | pP(2) : Excited Av=2 1 1 F1e - X Xv=0 2 2 F1e |
| 13C18O Excited | 1 e 2 | X | 2 e 1 | 67625.2698 | .0018095 | 67635.7481 | 10.4783 | .0019028 | 184551.2 | 0.000000 | : | pP1(2) : Excited a3Pv=13 1 1 F1e - X Xv=0 2 2 F1e |
| 13C18O Excited | 1 e 3 | X | 2 e 1 | 67664.5592 | .3909250 | 67675.0375 | 10.4783 | .4110722 | 39939628 | 0.000000 | : | qP21(2) : Excited a3Pv=13 1 2 F2e - X Xv=0 2 2 F1e |
| 13C18O Excited | 1 e 4 | X | 2 e 1 | 67793.3835 | .1751783 | 67803.8618 | 10.4783 | .1842065 | 17999856 | 0.000000 | : | oP1(2) : Excited ev=4 1 0 F1e - X Xv=0 2 2 F1e |
| 13C18O Excited | 1 e 5 | X | 2 e 1 | 67799.7347 | .1075715 | 67810.2129 | 10.4783 | .1131154 | 11056255 | 0.000000 | : | qP31(2) : Excited ev=4 1 2 F3e - X Xv=0 2 2 F1e |
| 13C18O Excited | 1 e 6 | X | 2 e 1 | 67977.4179 | .2972842 | 67987.8961 | 10.4783 | .3126053 | 30795875 | 0.000000 | : | qP31(2) : Excited dv=7 1 2 F3e - X Xv=0 2 2 F1e |
| 13C18O Excited | 1 e 7 | X | 2 e 1 | 67982.1651 | .3299403 | 67992.6434 | 10.4783 | .3469444 | 34185907 | 0.000000 | : | pP21(2) : Excited a'v=12 1 1 F2e - X Xv=0 2 2 F1e |
| 13C18O Excited | 2 f 1 | X | 2 e 1 | 67555.6174 | 1.410271 | 67566.0957 | 10.4783 | 1.482952 | 86032986 | 0.000000 | : | qQ(2) : Excited Av=2 2 2 F1f - X Xv=0 2 2 F1e |
| 13C18O Excited | 2 f 2 | X | 2 e 1 | 67573.8534 | .0004560 | 67584.3317 | 10.4783 | .0004795 | 27839.03 | 0.000000 | : | qQ(2) : Excited Iv=3 2 2 F1f - X Xv=0 2 2 F1e |
| 13C18O Excited | 2 f 3 | X | 2 e 1 | 67633.9221 | .0293555 | 67644.4004 | 10.4783 | .0308684 | 1797057 | 0.000000 | : | pQ1(2) : Excited a3Pv=13 2 1 F1f - X Xv=0 2 2 F1e |
| 13C18O Excited | 2 f 4 | X | 2 e 1 | 67669.7669 | 1.912436 | 67680.2452 | 10.4783 | 2.010997 | 1.1726e8 | 0.000000 | : | qQ21(2) : Excited a3Pv=13 2 2 F2f - X Xv=0 2 2 F1e |
| 13C18O Excited | 2 f 5 | X | 2 e 1 | 67703.3092 | .0243488 | 67713.7874 | 10.4783 | .0256037 | 1495153 | 0.000000 | : | rQ31(2) : Excited a3Pv=13 2 3 F3f - X Xv=0 2 2 F1e |
| 13C18O Excited | 2 f 6 | X | 2 e 1 | 67800.1911 | 1.411990 | 67810.6694 | 10.4783 | 1.484760 | 87076816 | 0.000000 | : | qQ21(2) : Excited ev=4 2 2 F2f - X Xv=0 2 2 F1e |
| 13C18O Excited | 2 f 7 | X | 2 e 1 | 67942.9057 | .0093790 | 67953.3840 | 10.4783 | .0098624 | 582060.1 | 0.000000 | : | qQ21(2) : Excited dv=7 2 2 F2f - X Xv=0 2 2 F1e |
| 13C18O Excited | 2 f 8 | X | 2 e 1 | 67981.9756 | 1.352766 | 67992.4539 | 10.4783 | 1.422484 | 84097326 | 0.000000 | : | rQ31(2) : Excited dv=7 2 3 F3f - X Xv=0 2 2 F1e |
| 13C18O Excited | 2 f 9 | X | 2 e 1 | 67982.9521 | 1.281743 | 67993.4304 | 10.4783 | 1.347801 | 79685464 | 0.000000 | : | pQ1(2) : Excited a'v=12 2 1 F1f - X Xv=0 2 2 F1e |
| 13C18O Excited | 2 f 10 | X | 2 e 1 | 67994.0274 | .4921605 | 68004.5056 | 10.4783 | .5175250 | 30612377 | 0.000000 | : | rQ31(2) : Excited a'v=12 2 3 F3f - X Xv=0 2 2 F1e |
| 13C18O Excited | 3 e 1 | X | 2 e 1 | 67564.1219 | 1.127253 | 67574.6002 | 10.4783 | 1.185348 | 49138260 | 0.000000 | : | rR(2) : Excited Av=2 3 3 F1e - X Xv=0 2 2 F1e |
| 13C18O Excited | 3 e 2 | X | 2 e 1 | 67637.4662 | .0417763 | 67647.9445 | 10.4783 | .0439293 | 1827012 | 0.000000 | : | qR1(2) : Excited a3Pv=13 3 2 F1e - X Xv=0 2 2 F1e |
| 13C18O Excited | 3 e 3 | X | 2 e 1 | 67677.4818 | 1.487100 | 67687.9601 | 10.4783 | 1.563740 | 65151228 | 0.000000 | : | rR21(2) : Excited a3Pv=13 3 3 F2e - X Xv=0 2 2 F1e |
| 13C18O Excited | 3 e 4 | X | 2 e 1 | 67711.7558 | .0469988 | 67722.2341 | 10.4783 | .0494210 | 2062192 | 0.000000 | : | sR31(2) : Excited a3Pv=13 3 4 F3e - X Xv=0 2 2 F1e |
| 13C18O Excited | 3 e 5 | X | 2 e 1 | 67799.9593 | .6184577 | 67810.4376 | 10.4783 | .6503312 | 27242584 | 0.000000 | : | qR1(2) : Excited ev=4 3 2 F1e - X Xv=0 2 2 F1e |
| 13C18O Excited | 3 e 6 | X | 2 e 1 | 67815.1327 | .5095974 | 67825.6110 | 10.4783 | .5358606 | 22462446 | 0.000000 | : | sR31(2) : Excited ev=4 3 4 F3e - X Xv=0 2 2 F1e |
| 13C18O Excited | 3 e 8 | X | 2 e 1 | 67949.3909 | .0184130 | 67959.8692 | 10.4783 | .0193620 | 816453.9 | 0.000000 | : | rR21(2) : Excited dv=7 3 3 F2e - X Xv=0 2 2 F1e |
| 13C18O Excited | 3 e 9 | X | 2 e 1 | 67988.8161 | 1.160220 | 67999.2944 | 10.4783 | 1.220015 | 51535078 | 0.000000 | : | sR31(2) : Excited dv=7 3 4 F3e - X Xv=0 2 2 F1e |
| 13C18O Excited | 3 e 10 | X | 2 e 1 | 67992.5409 | 1.330064 | 68003.0192 | 10.4783 | 1.398612 | 59088968 | 0.000000 | : | rR21(2) : Excited a'v=12 3 3 F2e - X Xv=0 2 2 F1e |
| 13C18O Excited | 2 e 1 | X | 3 e 1 | 67545.1411 | .5365308 | 67566.0972 | 20.9562 | .5932573 | 34401618 | 0.000000 | : | pP(3) : Excited Av=2 2 2 F1e - X Xv=0 3 3 F1e |
| 13C18O Excited | 2 e 2 | X | 3 e 1 | 67619.6686 | .0101640 | 67640.6247 | 20.9562 | .0112386 | 653858.4 | 0.000000 | : | oP1(3) : Excited a3Pv=13 2 1 F1e - X Xv=0 3 3 F1e |
| 13C18O Excited | 2 e 3 | X | 3 e 1 | 67659.2486 | .7284834 | 67680.2048 | 20.9562 | .8055047 | 46946478 | 0.000000 | : | pP21(3) : Excited a3Pv=13 2 2 F2e - X Xv=0 3 3 F1e |
| 13C18O Excited | 2 e 4 | X | 3 e 1 | 67692.8312 | .0092527 | 67713.7873 | 20.9562 | .0102310 | 597172.5 | 0.000000 | : | qP31(3) : Excited a3Pv=13 2 3 F3e - X Xv=0 3 3 F1e |
| 13C18O Excited | 2 e 5 | X | 3 e 1 | 67785.0835 | .3059124 | 67806.0397 | 20.9562 | .3382560 | 19824458 | 0.000000 | : | oP1(3) : Excited ev=4 2 1 F1e - X Xv=0 3 3 F1e |
| 13C18O Excited | 2 e 6 | X | 3 e 1 | 67795.8685 | .2313215 | 67816.8247 | 20.9562 | .2557787 | 14997801 | 0.000000 | : | qP31(3) : Excited ev=4 2 3 F3e - X Xv=0 3 3 F1e |
| 13C18O Excited | 2 e 7 | X | 3 e 1 | 67932.4278 | .0035680 | 67953.3840 | 20.9562 | .0039452 | 232732.4 | 0.000000 | : | pP21(3) : Excited dv=7 2 2 F2e - X Xv=0 3 3 F1e |
| 13C18O Excited | 2 e 8 | X | 3 e 1 | 67971.5005 | .5601400 | 67992.4567 | 20.9562 | .6193627 | 36599828 | 0.000000 | : | qP31(3) : Excited dv=7 2 3 F3e - X Xv=0 3 3 F1e |
| 13C18O Excited | 2 e 9 | X | 3 e 1 | 67975.8376 | .6292315 | 67996.7938 | 20.9562 | .6957591 | 41122171 | 0.000000 | : | pP21(3) : Excited a'v=12 2 2 F2e - X Xv=0 3 3 F1e |
| 13C18O Excited | 3 f 1 | X | 3 e 1 | 67553.6406 | 1.875399 | 67574.5968 | 20.9562 | 2.073682 | 85923876 | 0.000000 | : | qQ(3) : Excited Av=2 3 3 F1f - X Xv=0 3 3 F1e |
| 13C18O Excited | 3 f 2 | X | 3 e 1 | 67569.9977 | .0013339 | 67590.9538 | 20.9562 | .0014749 | 61156.64 | 0.000000 | : | qQ(3) : Excited Iv=3 3 3 F1f - X Xv=0 3 3 F1e |
| 13C18O Excited | 3 f 3 | X | 3 e 1 | 67630.7083 | .0758728 | 67651.6645 | 20.9562 | .0838947 | 3488123 | 0.000000 | : | pQ1(3) : Excited a3Pv=13 3 2 F1f - X Xv=0 3 3 F1e |
| 13C18O Excited | 3 f 4 | X | 3 e 1 | 67667.0820 | 2.468486 | 67688.0382 | 20.9562 | 2.729474 | 1.1367e8 | 0.000000 | : | qQ21(3) : Excited a3Pv=13 3 3 F2f - X Xv=0 3 3 F1e |
| 13C18O Excited | 3 f 5 | X | 3 e 1 | 67701.2784 | .0783695 | 67722.2345 | 20.9562 | .0866554 | 3614194 | 0.000000 | : | rQ31(3) : Excited a3Pv=13 3 4 F3f - X Xv=0 3 3 F1e |
| 13C18O Excited | 3 f 6 | X | 3 e 1 | 67796.3071 | 1.876509 | 67817.2632 | 20.9562 | 2.074910 | 86904583 | 0.000000 | : | qQ21(3) : Excited ev=4 3 3 F2f - X Xv=0 3 3 F1e |
| 13C18O Excited | 3 f 8 | X | 3 e 1 | 67938.9130 | .0306392 | 67959.8692 | 20.9562 | .0338786 | 1427930 | 0.000000 | : | qQ21(3) : Excited dv=7 3 3 F2f - X Xv=0 3 3 F1e |
| 13C18O Excited | 3 f 9 | X | 3 e 1 | 67976.7064 | 1.252935 | 67997.6626 | 20.9562 | 1.385406 | 58490165 | 0.000000 | : | pQ1(3) : Excited a'v=12 3 2 F1f - X Xv=0 3 3 F1e |
| 13C18O Excited | 3 f 10 | X | 3 e 1 | 67978.3417 | 2.125495 | 67999.2979 | 20.9562 | 2.350220 | 99230580 | 0.000000 | : | rQ31(3) : Excited dv=7 3 4 F3f - X Xv=0 3 3 F1e |
| 13C18O Excited | 3 f 11 | X | 3 e 1 | 67991.7609 | .7660003 | 68012.7171 | 20.9562 | .8469882 | 35782577 | 0.000000 | : | rQ31(3) : Excited a'v=12 3 4 F3f - X Xv=0 3 3 F1e |
| 13C18O Excited | 4 e 1 | X | 3 e 1 | 67564.9806 | 1.338211 | 67585.9368 | 20.9562 | 1.479698 | 47711040 | 0.000000 | : | rR(3) : Excited Av=2 4 4 F1e - X Xv=0 3 3 F1e |
| 13C18O Excited | 4 e 2 | X | 3 e 1 | 67636.7566 | .0803364 | 67657.7128 | 20.9562 | .0888302 | 2873360 | 0.000000 | : | qR1(3) : Excited a3Pv=13 4 3 F1e - X Xv=0 3 3 F1e |
| 13C18O Excited | 4 e 3 | X | 3 e 1 | 67677.3515 | 1.701583 | 67698.3077 | 20.9562 | 1.881488 | 60969474 | 0.000000 | : | rR21(3) : Excited a3Pv=13 4 4 F2e - X Xv=0 3 3 F1e |
| 13C18O Excited | 4 e 4 | X | 3 e 1 | 67712.5216 | .0962729 | 67733.4778 | 20.9562 | .1064516 | 3454937 | 0.000000 | : | sR31(3) : Excited a3Pv=13 4 5 F3e - X Xv=0 3 3 F1e |
| 13C18O Excited | 4 e 5 | X | 3 e 1 | 67796.0845 | .7175200 | 67817.0407 | 20.9562 | .7933821 | 25845037 | 0.000000 | : | qR1(3) : Excited ev=4 4 3 F1e - X Xv=0 3 3 F1e |
| 13C18O Excited | 4 e 6 | X | 3 e 1 | 67815.6299 | .6204987 | 67836.5861 | 20.9562 | .6861030 | 22369673 | 0.000000 | : | sR31(3) : Excited ev=4 4 5 F3e - X Xv=0 3 3 F1e |
| 13C18O Excited | 4 e 7 | X | 3 e 1 | 67908.5618 | .0002169 | 67929.5180 | 20.9562 | .0002399 | 7852.734 | 0.000000 | : | qR1(3) : Excited dv=7 4 3 F1e - X Xv=0 3 3 F1e |
| 13C18O Excited | 4 e 8 | X | 3 e 1 | 67947.5591 | .0384519 | 67968.5152 | 20.9562 | .0425173 | 1394340 | 0.000000 | : | rR21(3) : Excited dv=7 4 4 F2e - X Xv=0 3 3 F1e |
| 13C18O Excited | 4 e 9 | X | 3 e 1 | 67987.4493 | 1.346579 | 68008.4055 | 20.9562 | 1.488950 | 48915599 | 0.000000 | : | sR31(3) : Excited dv=7 4 5 F3e - X Xv=0 3 3 F1e |
| 13C18O Excited | 4 e 10 | X | 3 e 1 | 67990.3630 | 1.596842 | 68011.3192 | 20.9562 | 1.765673 | 58014070 | 0.000000 | : | rR21(3) : Excited a'v=12 4 4 F2e - X Xv=0 3 3 F1e |
| 13C18O Excited | 3 e 1 | X | 4 e 1 | 67539.6741 | .7519026 | 67574.6002 | 34.9261 | .8890112 | 36813703 | 0.000000 | : | pP(4) : Excited Av=2 3 3 F1e - X Xv=0 4 4 F1e |
| 13C18O Excited | 3 e 2 | X | 4 e 1 | 67613.0183 | .0278657 | 67647.9445 | 34.9261 | .0329470 | 1368774 | 0.000000 | : | oP1(4) : Excited a3Pv=13 3 2 F1e - X Xv=0 4 4 F1e |
| 13C18O Excited | 3 e 3 | X | 4 e 1 | 67653.0339 | .9919282 | 67687.9601 | 34.9261 | 1.172805 | 48810485 | 0.000000 | : | pP21(4) : Excited a3Pv=13 3 3 F2e - X Xv=0 4 4 F1e |
| 13C18O Excited | 3 e 4 | X | 4 e 1 | 67687.3079 | .0313492 | 67722.2341 | 34.9261 | .0370657 | 1544969 | 0.000000 | : | qP31(4) : Excited a3Pv=13 3 4 F3e - X Xv=0 4 4 F1e |
| 13C18O Excited | 3 e 5 | X | 4 e 1 | 67775.5114 | .4125249 | 67810.4376 | 34.9261 | .4877484 | 20409844 | 0.000000 | : | oP1(4) : Excited ev=4 3 2 F1e - X Xv=0 4 4 F1e |
| 13C18O Excited | 3 e 6 | X | 4 e 1 | 67790.6848 | .3399127 | 67825.6110 | 34.9261 | .4018954 | 16828621 | 0.000000 | : | qP31(4) : Excited ev=4 3 4 F3e - X Xv=0 4 4 F1e |
| 13C18O Excited | 3 e 8 | X | 4 e 1 | 67924.9430 | .0122819 | 67959.8692 | 34.9261 | .0145215 | 611679.7 | 0.000000 | : | pP21(4) : Excited dv=7 3 3 F2e - X Xv=0 4 4 F1e |
| 13C18O Excited | 3 e 9 | X | 4 e 1 | 67964.3682 | .7738926 | 67999.2944 | 34.9261 | .9150111 | 38609628 | 0.000000 | : | qP31(4) : Excited dv=7 3 4 F3e - X Xv=0 4 4 F1e |
| 13C18O Excited | 3 e 10 | X | 4 e 1 | 67968.0930 | .8871821 | 68003.0192 | 34.9261 | 1.048959 | 44268938 | 0.000000 | : | pP21(4) : Excited a'v=12 3 3 F2e - X Xv=0 4 4 F1e |
| 13C18O Excited | 4 f 1 | X | 4 e 1 | 67551.0038 | 2.250952 | 67585.9299 | 34.9261 | 2.661410 | 85760674 | 0.000000 | : | qQ(4) : Excited Av=2 4 4 F1f - X Xv=0 4 4 F1e |
| 13C18O Excited | 4 f 2 | X | 4 e 1 | 67564.8563 | .0031554 | 67599.7824 | 34.9261 | .0037308 | 120294.9 | 0.000000 | : | qQ(4) : Excited Iv=3 4 4 F1f - X Xv=0 4 4 F1e |
| 13C18O Excited | 4 f 3 | X | 4 e 1 | 67626.4371 | .1463355 | 67661.3633 | 34.9261 | .1730196 | 5594044 | 0.000000 | : | pQ1(4) : Excited a3Pv=13 4 3 F1f - X Xv=0 4 4 F1e |
| 13C18O Excited | 4 f 4 | X | 4 e 1 | 67663.5063 | 2.853026 | 67698.4324 | 34.9261 | 3.373272 | 1.0924e8 | 0.000000 | : | qQ21(4) : Excited a3Pv=13 4 4 F2f - X Xv=0 4 4 F1e |
| 13C18O Excited | 4 f 5 | X | 4 e 1 | 67698.5529 | .1625560 | 67733.4791 | 34.9261 | .1921979 | 6234014 | 0.000000 | : | rQ31(4) : Excited a3Pv=13 4 5 F3f - X Xv=0 4 4 F1e |
| 13C18O Excited | 4 f 6 | X | 4 e 1 | 67791.1287 | 2.250653 | 67826.0549 | 34.9261 | 2.661057 | 86666983 | 0.000000 | : | qQ21(4) : Excited ev=4 4 4 F2f - X Xv=0 4 4 F1e |
| 13C18O Excited | 4 f 7 | X | 4 e 1 | 67894.5919 | .0003651 | 67929.5180 | 34.9261 | .0004317 | 14124.40 | 0.000000 | : | pQ1(4) : Excited dv=7 4 3 F1f - X Xv=0 4 4 F1e |
| 13C18O Excited | 4 f 8 | X | 4 e 1 | 67933.5891 | .0647121 | 67968.5152 | 34.9261 | .0765123 | 2507643 | 0.000000 | : | qQ21(4) : Excited dv=7 4 4 F2f - X Xv=0 4 4 F1e |
| 13C18O Excited | 4 f 9 | X | 4 e 1 | 67969.0113 | 1.522798 | 68003.9375 | 34.9261 | 1.800478 | 59101928 | 0.000000 | : | pQ1(4) : Excited a'v=12 4 3 F1f - X Xv=0 4 4 F1e |
| 13C18O Excited | 4 f 10 | X | 4 e 1 | 67973.4818 | 2.432569 | 68008.4080 | 34.9261 | 2.876146 | 94430079 | 0.000000 | : | rQ31(4) : Excited dv=7 4 5 F3f - X Xv=0 4 4 F1e |
| 13C18O Excited | 4 f 11 | X | 4 e 1 | 67988.1161 | .9994887 | 68023.0422 | 34.9261 | 1.181744 | 38824290 | 0.000000 | : | rQ31(4) : Excited a'v=12 4 5 F3f - X Xv=0 4 4 F1e |
| 13C18O Excited | 5 e 1 | X | 4 e 1 | 67565.1800 | 1.499191 | 67600.1062 | 34.9261 | 1.772567 | 46762998 | 0.000000 | : | rR(4) : Excited Av=2 5 5 F1e - X Xv=0 4 4 F1e |
| 13C18O Excited | 5 e 2 | X | 4 e 1 | 67635.0103 | .1304718 | 67669.9365 | 34.9261 | .1542632 | 4082326 | 0.000000 | : | qR1(4) : Excited a3Pv=13 5 4 F1e - X Xv=0 4 4 F1e |
| 13C18O Excited | 5 e 3 | X | 4 e 1 | 67676.3264 | 1.824760 | 67711.2525 | 34.9261 | 2.157503 | 57199520 | 0.000000 | : | rR21(4) : Excited a3Pv=13 5 5 F2e - X Xv=0 4 4 F1e |
| 13C18O Excited | 5 e 4 | X | 4 e 1 | 67712.5783 | .1597010 | 67747.5045 | 34.9261 | .1888223 | 5014090 | 0.000000 | : | sR31(4) : Excited a3Pv=13 5 6 F3e - X Xv=0 4 4 F1e |
| 13C18O Excited | 5 e 5 | X | 4 e 1 | 67790.9192 | .7909836 | 67825.8454 | 34.9261 | .9352186 | 24920598 | 0.000000 | : | qR1(4) : Excited ev=4 5 4 F1e - X Xv=0 4 4 F1e |
| 13C18O Excited | 5 e 6 | X | 4 e 1 | 67814.8272 | .7062359 | 67849.7534 | 34.9261 | .8350172 | 22274100 | 0.000000 | : | sR31(4) : Excited ev=4 5 6 F3e - X Xv=0 4 4 F1e |
| 13C18O Excited | 5 e 7 | X | 4 e 1 | 67904.7964 | .0006204 | 67939.7225 | 34.9261 | .0007336 | 19645.54 | 0.000000 | : | qR1(4) : Excited dv=7 5 4 F1e - X Xv=0 4 4 F1e |
| 13C18O Excited | 5 e 8 | X | 4 e 1 | 67944.3952 | .0651662 | 67979.3214 | 34.9261 | .0770492 | 2067091 | 0.000000 | : | rR21(4) : Excited dv=7 5 5 F2e - X Xv=0 4 4 F1e |
| 13C18O Excited | 5 e 9 | X | 4 e 1 | 67984.8590 | 1.446067 | 68019.7851 | 34.9261 | 1.709756 | 45951696 | 0.000000 | : | sR31(4) : Excited dv=7 5 6 F3e - X Xv=0 4 4 F1e |
| 13C18O Excited | 5 e 10 | X | 4 e 1 | 67986.7676 | 1.834543 | 68021.6938 | 34.9261 | 2.169070 | 58301193 | 0.000000 | : | rR21(4) : Excited a'v=12 5 5 F2e - X Xv=0 4 4 F1e |

| | | | | | | | | | | | |
|---|---|---|---|---|---|---|---|---|---|---|---|
| 13C18O Excited | 4 e 1 | X | 5 e 1 | 67533.5491 | .9207625 | 67585.9368 | 52.3877 | 1.183758 | 38115588 | 0.000000 : | pP(5) : Excited Av=2 4 4 F1e - X Xv=0 5 5 F1e |
| 13C18O Excited | 4 e 2 | X | 5 e 1 | 67605.3251 | .0552758 | 67657.7128 | 52.3877 | .0710642 | 2295485 | 0.000000 : | oP1(5) : Excited a3Pv=13 4 3 F1e - X Xv=0 5 5 F1e |
| 13C18O Excited | 4 e 3 | X | 5 e 1 | 67645.9200 | 1.170782 | 67698.3077 | 52.3877 | 1.505190 | 48707652 | 0.000000 : | pP21(5) : Excited a3Pv=13 4 4 F2e - X Xv=0 5 5 F1e |
| 13C18O Excited | 4 e 4 | X | 5 e 1 | 67681.0901 | .0662410 | 67733.4778 | 52.3877 | .0851613 | 2760102 | 0.000000 : | qP31(5) : Excited a3Pv=13 4 5 F3e - X Xv=0 5 5 F1e |
| 13C18O Excited | 4 e 5 | X | 5 e 1 | 67764.6530 | .4936931 | 67817.0407 | 52.3877 | .6347057 | 20647286 | 0.000000 : | oP1(5) : Excited ev=4 4 3 F1e - X Xv=0 5 5 F1e |
| 13C18O Excited | 4 e 6 | X | 5 e 1 | 67784.1984 | .4269371 | 67836.5861 | 52.3877 | .5488824 | 17870867 | 0.000000 : | qP31(5) : Excited ev=4 4 5 F3e - X Xv=0 5 5 F1e |
| 13C18O Excited | 4 e 7 | X | 5 e 1 | 67877.1303 | .0001493 | 67929.5180 | 52.3877 | .0001919 | 6273.468 | 0.000000 : | oP1(5) : Excited dv=7 4 3 F1e - X Xv=0 5 5 F1e |
| 13C18O Excited | 4 e 8 | X | 5 e 1 | 67916.1275 | .0264570 | 67968.5152 | 52.3877 | .0340139 | 1113924 | 0.000000 : | pP21(5) : Excited dv=7 4 4 F2e - X Xv=0 5 5 F1e |
| 13C18O Excited | 4 e 9 | X | 5 e 1 | 67956.0178 | .9265201 | 68008.4055 | 52.3877 | 1.191160 | 39078230 | 0.000000 : | qP31(5) : Excited dv=7 4 5 F3e - X Xv=0 5 5 F1e |
| 13C18O Excited | 4 e 10 | X | 5 e 1 | 67958.9315 | 1.098715 | 68011.3192 | 52.3877 | 1.412539 | 46346919 | 0.000000 : | pP21(5) : Excited a'v=12 4 4 F2e - X Xv=0 5 5 F1e |
| 13C18O Excited | 5 f 1 | X | 5 e 1 | 67547.7051 | 2.522905 | 67600.0928 | 52.3877 | 3.243518 | 85502504 | 0.000000 : | qQ(5) : Excited Av=2 5 5 F1f - X Xv=0 5 5 F1e |
| 13C18O Excited | 5 f 2 | X | 5 e 1 | 67558.4301 | .0071857 | 67610.8178 | 52.3877 | .0092381 | 243641.6 | 0.000000 : | qQ(5) : Excited Iv=3 5 5 F1f - X Xv=0 5 5 F1e |
| 13C18O Excited | 5 f 3 | X | 5 e 1 | 67621.1193 | .2356971 | 67673.5070 | 52.3877 | .3030188 | 8013964 | 0.000000 : | pQ1(5) : Excited a3Pv=13 5 4 F1f - X Xv=0 5 5 F1e |
| 13C18O Excited | 5 f 4 | X | 5 e 1 | 67659.0424 | 3.060302 | 67711.4301 | 52.3877 | 3.934410 | 1.0423e8 | 0.000000 : | qQ21(5) : Excited a3Pv=13 5 5 F2f - X Xv=0 5 5 F1e |
| 13C18O Excited | 5 f 5 | X | 5 e 1 | 67695.1195 | .2704075 | 67747.5072 | 52.3877 | .3476434 | 9224373 | 0.000000 : | rQ31(5) : Excited a3Pv=13 5 6 F3f - X Xv=0 5 5 F1e |
| 13C18O Excited | 5 f 6 | X | 5 e 1 | 67784.6566 | 2.521469 | 67837.0443 | 52.3877 | 3.241672 | 86356290 | 0.000000 : | qQ21(5) : Excited ev=4 5 5 F2f - X Xv=0 5 5 F1e |
| 13C18O Excited | 5 f 7 | X | 5 e 1 | 67887.3348 | .0010459 | 67939.7225 | 52.3877 | .0013446 | 35981.98 | 0.000000 : | pQ1(5) : Excited dv=7 5 4 F1f - X Xv=0 5 5 F1e |
| 13C18O Excited | 5 f 8 | X | 5 e 1 | 67926.9336 | .1098306 | 67979.3214 | 52.3877 | .1412014 | 3785259 | 0.000000 : | qQ21(5) : Excited dv=7 5 5 F2f - X Xv=0 5 5 F1e |
| 13C18O Excited | 5 f 9 | X | 5 e 1 | 67959.8791 | 1.678515 | 68012.2668 | 52.3877 | 2.157945 | 57933398 | 0.000000 : | pQ1(5) : Excited a'v=12 5 4 F1f - X Xv=0 5 5 F1e |
| 13C18O Excited | 5 f 10 | X | 5 e 1 | 67967.4005 | 2.671773 | 68019.7882 | 52.3877 | 3.434906 | 92246015 | 0.000000 : | rQ31(5) : Excited dv=7 5 6 F3f - X Xv=0 5 5 F1e |
| 13C18O Excited | 5 f 11 | X | 5 e 1 | 67983.0719 | 1.181084 | 68035.4596 | 52.3877 | 1.518435 | 40806502 | 0.000000 : | rQ31(5) : Excited a'v=12 5 6 F3f - X Xv=0 5 5 F1e |
| 13C18O Excited | 6 e 1 | X | 5 e 1 | 67564.7200 | 1.605082 | 67617.1077 | 52.3877 | 2.063538 | 46063028 | 0.000000 : | rR(5) : Excited Av=2 6 6 F1e - X Xv=0 5 5 F1e |
| 13C18O Excited | 6 e 2 | X | 5 e 1 | 67632.2355 | .1881021 | 67684.6232 | 52.3877 | .2418293 | 5414399 | 0.000000 : | qR1(5) : Excited a3Pv=13 6 5 F1e - X Xv=0 5 5 F1e |
| 13C18O Excited | 6 e 3 | X | 5 e 1 | 67674.4113 | 1.859531 | 67726.7990 | 52.3877 | 2.390666 | 53625632 | 0.000000 : | rR21(5) : Excited a3Pv=13 6 6 F2e - X Xv=0 5 5 F1e |
| 13C18O Excited | 6 e 4 | X | 5 e 1 | 67711.9113 | .2313107 | 67764.2990 | 52.3877 | .2973795 | 6681690 | 0.000000 : | sR31(5) : Excited a3Pv=13 6 7 F3e - X Xv=0 5 5 F1e |
| 13C18O Excited | 6 e 5 | X | 5 e 1 | 67784.4625 | .8360993 | 67836.8502 | 52.3877 | 1.074913 | 24229462 | 0.000000 : | qR1(5) : Excited ev=4 6 5 F1e - X Xv=0 5 5 F1e |
| 13C18O Excited | 6 e 6 | X | 5 e 1 | 67812.7260 | .7643333 | 67865.1137 | 52.3877 | .9826484 | 22177461 | 0.000000 : | sR31(5) : Excited ev=4 6 7 F3e - X Xv=0 5 5 F1e |
| 13C18O Excited | 6 e 7 | X | 5 e 1 | 67899.5909 | .0013532 | 67951.9786 | 52.3877 | .0017398 | 39415.87 | 0.000000 : | qR1(5) : Excited dv=7 6 5 F1e - X Xv=0 5 5 F1e |
| 13C18O Excited | 6 e 8 | X | 5 e 1 | 67939.8989 | .0965022 | 67992.2866 | 52.3877 | .1240660 | 2815837 | 0.000000 : | rR21(5) : Excited dv=7 6 6 F2e - X Xv=0 5 5 F1e |
| 13C18O Excited | 6 e 9 | X | 5 e 1 | 67981.0370 | 1.307724 | 68033.4247 | 52.3877 | 1.681247 | 38227416 | 0.000000 : | sR31(5) : Excited dv=7 6 7 F3e - X Xv=0 5 5 F1e |
| 13C18O Excited | 6 e 10 | X | 5 e 1 | 67981.7577 | 2.184644 | 68034.1454 | 52.3877 | 2.808640 | 63863596 | 0.000000 : | rR21(5) : Excited a'v=12 6 6 F2e - X Xv=0 5 5 F1e |
| 13C18O Excited | 5 e 1 | X | 6 e 1 | 67526.7660 | 1.039119 | 67600.1062 | 73.3402 | 1.477140 | 38902735 | 0.000000 : | pP(6) : Excited Av=2 5 5 F1e - X Xv=0 6 6 F1e |
| 13C18O Excited | 5 e 2 | X | 6 e 1 | 67596.5962 | .0904325 | 67669.9365 | 73.3402 | .1285526 | 3396145 | 0.000000 : | oP1(6) : Excited a3Pv=13 5 4 F1e - X Xv=0 6 6 F1e |
| 13C18O Excited | 5 e 3 | X | 6 e 1 | 67637.9123 | 1.264777 | 67711.2525 | 73.3402 | 1.797919 | 47585145 | 0.000000 : | pP21(6) : Excited a3Pv=13 5 5 F2e - X Xv=0 6 6 F1e |
| 13C18O Excited | 5 e 4 | X | 6 e 1 | 67674.1643 | .1106919 | 67747.5045 | 73.3402 | .1573519 | 4171301 | 0.000000 : | qP31(6) : Excited a3Pv=13 5 6 F3e - X Xv=0 6 6 F1e |
| 13C18O Excited | 5 e 5 | X | 6 e 1 | 67752.5051 | .5482463 | 67825.8454 | 73.3402 | .7793489 | 20731882 | 0.000000 : | oP1(6) : Excited ev=4 5 4 F1e - X Xv=0 6 6 F1e |
| 13C18O Excited | 5 e 6 | X | 6 e 1 | 67776.4131 | .4895060 | 67849.7534 | 73.3402 | .6958477 | 18530225 | 0.000000 : | qP31(6) : Excited ev=4 5 6 F3e - X Xv=0 6 6 F1e |
| 13C18O Excited | 5 e 7 | X | 6 e 1 | 67866.3823 | .0004300 | 67939.7225 | 73.3402 | .0006113 | 16343.52 | 0.000000 : | oP1(6) : Excited dv=7 5 4 F1e - X Xv=0 6 6 F1e |
| 13C18O Excited | 5 e 8 | X | 6 e 1 | 67905.9811 | .0451680 | 67979.3214 | 73.3402 | .0642077 | 1719656 | 0.000000 : | pP21(6) : Excited dv=7 5 5 F2e - X Xv=0 6 6 F1e |
| 13C18O Excited | 5 e 9 | X | 6 e 1 | 67946.4449 | 1.002298 | 68019.7851 | 73.3402 | 1.424797 | 38228205 | 0.000000 : | qP31(6) : Excited dv=7 5 6 F3e - X Xv=0 6 6 F1e |
| 13C18O Excited | 5 e 10 | X | 6 e 1 | 67948.3536 | 1.271558 | 68021.6938 | 73.3402 | 1.807558 | 48502020 | 0.000000 : | pP21(6) : Excited a'v=12 5 5 F2e - X Xv=0 6 6 F1e |
| 13C18O Excited | 6 f 1 | X | 6 e 1 | 67543.7379 | 2.679802 | 67617.0781 | 73.3402 | 3.809420 | 84956037 | 0.000000 : | qQ(6) : Excited Av=2 6 6 F1f - X Xv=0 6 6 F1e |
| 13C18O Excited | 6 f 2 | X | 6 e 1 | 67550.7243 | .0195973 | 67624.0645 | 73.3402 | .0278581 | 621472.3 | 0.000000 : | qQ(6) : Excited Iv=3 6 6 F1f - X Xv=0 6 6 F1e |
| 13C18O Excited | 6 f 3 | X | 6 e 1 | 67614.7667 | .3348036 | 67688.1070 | 73.3402 | .4759336 | 10647585 | 0.000000 : | pQ1(6) : Excited a3Pv=13 6 5 F1f - X Xv=0 6 6 F1e |
| 13C18O Excited | 6 f 4 | X | 6 e 1 | 67653.6931 | 3.103157 | 67727.0333 | 73.3402 | 4.411233 | 98858635 | 0.000000 : | qQ21(6) : Excited a3Pv=13 6 6 F2f - X Xv=0 6 6 F1e |
| 13C18O Excited | 6 f 5 | X | 6 e 1 | 67690.9635 | .3906145 | 67764.3037 | 73.3402 | .5552705 | 12464554 | 0.000000 : | rQ31(6) : Excited a3Pv=13 6 7 F3f - X Xv=0 6 6 F1e |
| 13C18O Excited | 6 f 6 | X | 6 e 1 | 67776.8911 | 2.683632 | 67850.2314 | 73.3402 | 3.814864 | 85961526 | 0.000000 : | qQ21(6) : Excited ev=4 6 6 F2f - X Xv=0 6 6 F1e |
| 13C18O Excited | 6 f 7 | X | 6 e 1 | 67878.6384 | .0022722 | 67951.9786 | 73.3402 | .0032301 | 73112.34 | 0.000000 : | pQ1(6) : Excited dv=7 6 5 F1f - X Xv=0 6 6 F1e |
| 13C18O Excited | 6 f 8 | X | 6 e 1 | 67918.9464 | .1619891 | 67992.2866 | 73.3402 | .2302724 | 5221495 | 0.000000 : | qQ21(6) : Excited dv=7 6 6 F2f - X Xv=0 6 6 F1e |
| 13C18O Excited | 6 f 9 | X | 6 e 1 | 67949.3192 | 1.762260 | 68022.6595 | 73.3402 | 2.505107 | 56880276 | 0.000000 : | pQ1(6) : Excited a'v=12 6 5 F1f - X Xv=0 6 6 F1e |
| 13C18O Excited | 6 f 10 | X | 6 e 1 | 67960.0920 | 2.797999 | 68033.4323 | 73.3402 | 3.977441 | 90353646 | 0.000000 : | rQ31(6) : Excited dv=7 6 7 F3f - X Xv=0 6 6 F1e |
| 13C18O Excited | 6 f 11 | X | 6 e 1 | 67976.6198 | 1.305660 | 68049.9600 | 73.3402 | 1.856036 | 42193457 | 0.000000 : | rQ31(6) : Excited a'v=12 6 7 F3f - X Xv=0 6 6 F1e |
| 13C18O Excited | 7 e 1 | X | 6 e 1 | 67563.6000 | 1.654636 | 67636.9402 | 73.3402 | 2.352115 | 45501850 | 0.000000 : | rR(6) : Excited Av=2 7 7 F1e - X Xv=0 6 6 F1e |
| 13C18O Excited | 7 e 2 | X | 6 e 1 | 67628.4411 | .2477368 | 67701.7813 | 73.3402 | .3521654 | 6832300 | 0.000000 : | qR1(6) : Excited a3Pv=13 7 6 F1e - X Xv=0 6 6 F1e |
| 13C18O Excited | 7 e 3 | X | 6 e 1 | 67671.6110 | 1.816496 | 67744.9512 | 73.3402 | 2.582205 | 50192910 | 0.000000 : | rR21(6) : Excited a3Pv=13 7 7 F2e - X Xv=0 6 6 F1e |
| 13C18O Excited | 7 e 4 | X | 6 e 1 | 67710.5050 | .3038734 | 67783.8453 | 73.3402 | .4319654 | 8411032 | 0.000000 : | sR31(6) : Excited a3Pv=13 7 8 F3e - X Xv=0 6 6 F1e |
| 13C18O Excited | 7 e 5 | X | 6 e 1 | 67776.7146 | .8522179 | 67850.0548 | 73.3402 | 1.211454 | 23658137 | 0.000000 : | qR1(6) : Excited ev=4 7 6 F1e - X Xv=0 6 6 F1e |
| 13C18O Excited | 7 e 6 | X | 6 e 1 | 67809.3268 | .7940902 | 67882.6671 | 73.3402 | 1.128823 | 22076311 | 0.000000 : | sR31(6) : Excited ev=4 7 8 F3e - X Xv=0 6 6 F1e |
| 13C18O Excited | 7 e 7 | X | 6 e 1 | 67892.9510 | .0024717 | 67966.2912 | 73.3402 | .0035135 | 68968.34 | 0.000000 : | qR1(6) : Excited dv=7 7 6 F1e - X Xv=0 6 6 F1e |
| 13C18O Excited | 7 e 8 | X | 6 e 1 | 67934.0698 | .1295311 | 68007.4101 | 73.3402 | .1841324 | 3620972 | 0.000000 : | rR21(6) : Excited dv=7 7 7 F2e - X Xv=0 6 6 F1e |
| 13C18O Excited | 7 e 9 | X | 6 e 1 | 67975.3097 | 1.519675 | 68048.6499 | 73.3402 | 2.160264 | 42559119 | 0.000000 : | rR21(6) : Excited a'v=12 7 7 F2e - X Xv=0 6 6 F1e |
| 13C18O Excited | 7 e 10 | X | 6 e 1 | 67976.0013 | 2.058834 | 68049.3416 | 73.3402 | 2.926696 | 57660260 | 0.000000 : | sR31(6) : Excited dv=7 7 8 F3e - X Xv=0 6 6 F1e |
| 13C18O Excited | 6 e 1 | X | 7 e 1 | 67519.3247 | 1.106621 | 67617.1077 | 97.7830 | 1.768747 | 39403066 | 0.000000 : | pP(7) : Excited Av=2 6 6 F1e - X Xv=0 7 7 F1e |
| 13C18O Excited | 6 e 2 | X | 7 e 1 | 67586.8402 | .1296867 | 67684.6232 | 97.7830 | .2072823 | 4631575 | 0.000000 : | oP1(7) : Excited a3Pv=13 6 5 F1e - X Xv=0 7 7 F1e |
| 13C18O Excited | 6 e 3 | X | 7 e 1 | 67629.0160 | 1.282051 | 67726.7990 | 97.7830 | 2.049142 | 45872391 | 0.000000 : | pP21(7) : Excited a3Pv=13 6 6 F2e - X Xv=0 7 7 F1e |
| 13C18O Excited | 6 e 4 | X | 7 e 1 | 67666.5160 | .1594768 | 67764.2990 | 97.7830 | .2548967 | 5715652 | 0.000000 : | qP31(7) : Excited a3Pv=13 6 7 F3e - X Xv=0 7 7 F1e |
| 13C18O Excited | 6 e 5 | X | 7 e 1 | 67739.0672 | .5764475 | 67836.8502 | 97.7830 | .9213538 | 20726413 | 0.000000 : | oP1(7) : Excited ev=4 6 5 F1e - X Xv=0 7 7 F1e |
| 13C18O Excited | 6 e 6 | X | 7 e 1 | 67767.3307 | .5269685 | 67865.1137 | 97.7830 | .8422700 | 18971102 | 0.000000 : | qP31(7) : Excited ev=4 6 7 F3e - X Xv=0 7 7 F1e |
| 13C18O Excited | 6 e 7 | X | 7 e 1 | 67854.1956 | .0009330 | 67951.9786 | 97.7830 | .0014912 | 33717.31 | 0.000000 : | oP1(7) : Excited dv=7 6 5 F1e - X Xv=0 7 7 F1e |
| 13C18O Excited | 6 e 8 | X | 7 e 1 | 67894.5036 | .0665333 | 67992.2866 | 97.7830 | .1063423 | 2408740 | 0.000000 : | pP21(7) : Excited dv=7 6 6 F2e - X Xv=0 7 7 F1e |
| 13C18O Excited | 6 e 9 | X | 7 e 1 | 67935.6416 | .9016086 | 68033.4247 | 97.7830 | 1.441069 | 32700760 | 0.000000 : | qP31(7) : Excited dv=7 6 7 F3e - X Xv=0 7 7 F1e |
| 13C18O Excited | 6 e 10 | X | 7 e 1 | 67936.3624 | 1.506200 | 68034.1454 | 97.7830 | 2.407406 | 54630638 | 0.000000 : | pP21(7) : Excited a'v=12 6 6 F2e - X Xv=0 7 7 F1e |
| 13C18O Excited | 7 f 1 | X | 7 e 1 | 67539.0504 | 2.631311 | 67636.8334 | 97.7830 | 4.205706 | 81271068 | 0.000000 : | qQ(7) : Excited Av=2 7 7 F1f - X Xv=0 7 7 F1e |
| 13C18O Excited | 7 f 2 | X | 7 e 1 | 67541.7894 | .1327399 | 67639.5724 | 97.7830 | .2121623 | 4100323 | 0.000000 : | qQ(7) : Excited Iv=3 7 7 F1f - X Xv=0 7 7 F1e |
| 13C18O Excited | 7 f 3 | X | 7 e 1 | 67607.3922 | .4326540 | 67705.1752 | 97.7830 | .6915242 | 13403620 | 0.000000 : | pQ1(7) : Excited a3Pv=13 7 6 F1f - X Xv=0 7 7 F1e |
| 13C18O Excited | 7 f 4 | X | 7 e 1 | 67647.4606 | 3.007646 | 67745.2436 | 97.7830 | 4.807214 | 93342650 | 0.000000 : | qQ21(7) : Excited a3Pv=13 7 7 F2f - X Xv=0 7 7 F1e |
| 13C18O Excited | 7 f 5 | X | 7 e 1 | 67686.0696 | .5100080 | 67783.8526 | 97.7830 | .8151615 | 15855274 | 0.000000 : | rQ31(7) : Excited a3Pv=13 7 8 F3f - X Xv=0 7 7 F1e |
| 13C18O Excited | 7 f 6 | X | 7 e 1 | 67767.8332 | 2.739239 | 67865.6163 | 97.7830 | 4.378211 | 85467232 | 0.000000 : | qQ21(7) : Excited ev=4 7 7 F2f - X Xv=0 7 7 F1e |
| 13C18O Excited | 7 f 7 | X | 7 e 1 | 67868.5082 | .0041201 | 67966.2912 | 97.7830 | .0065853 | 129125.0 | 0.000000 : | pQ1(7) : Excited dv=7 7 6 F1f - X Xv=0 7 7 F1e |
| 13C18O Excited | 7 f 8 | X | 7 e 1 | 67909.6270 | .2158206 | 68007.4101 | 97.7830 | .3449527 | 6776193 | 0.000000 : | qQ21(7) : Excited dv=7 7 7 F2f - X Xv=0 7 7 F1e |
| 13C18O Excited | 7 f 9 | X | 7 e 1 | 67937.3361 | 1.782313 | 68035.1191 | 97.7830 | 2.848726 | 56028417 | 0.000000 : | pQ1(7) : Excited a'v=12 7 6 F1f - X Xv=0 7 7 F1e |
| 13C18O Excited | 7 f 10 | X | 7 e 1 | 67951.5510 | 2.812666 | 68049.3340 | 97.7830 | 4.495570 | 88473886 | 0.000000 : | rQ31(7) : Excited dv=7 7 8 F3f - X Xv=0 7 7 F1e |
| 13C18O Excited | 7 f 11 | X | 7 e 1 | 67968.7555 | 1.372798 | 68066.5385 | 97.7830 | 2.194186 | 43214898 | 0.000000 : | rQ31(7) : Excited a'v=12 7 8 F3f - X Xv=0 7 7 F1e |
| 13C18O Excited | 8 e 1 | X | 7 e 1 | 67561.8196 | 1.650285 | 67659.6027 | 97.7830 | 2.637701 | 45019861 | 0.000000 : | rR(7) : Excited Av=2 8 8 F1e - X Xv=0 7 7 F1e |
| 13C18O Excited | 8 e 2 | X | 7 e 1 | 67623.6363 | .3035484 | 67721.4193 | 97.7830 | .4851708 | 8303568 | 0.000000 : | qR1(7) : Excited a3Pv=13 8 7 F1e - X Xv=0 7 7 F1e |
| 13C18O Excited | 8 e 3 | X | 7 e 1 | 67667.9290 | 1.711178 | 67765.7121 | 97.7830 | 2.735028 | 46901310 | 0.000000 : | rR21(7) : Excited a3Pv=13 8 8 F2e - X Xv=0 7 7 F1e |
| 13C18O Excited | 8 e 4 | X | 7 e 1 | 67708.3442 | .3704399 | 67806.1273 | 97.7830 | .5920856 | 10171513 | 0.000000 : | sR31(7) : Excited a3Pv=13 8 9 F3e - X Xv=0 7 7 F1e |
| 13C18O Excited | 8 e 5 | X | 7 e 1 | 67767.6763 | .8405385 | 67865.4594 | 97.7830 | 1.343459 | 23140172 | 0.000000 : | qR1(7) : Excited ev=4 8 7 F1e - X Xv=0 7 7 F1e |
| 13C18O Excited | 8 e 6 | X | 7 e 1 | 67804.6300 | .7966041 | 67902.4130 | 97.7830 | 1.273237 | 21966545 | 0.000000 : | sR31(7) : Excited ev=4 8 9 F3e - X Xv=0 7 7 F1e |
| 13C18O Excited | 8 e 7 | X | 7 e 1 | 67884.8825 | .0039652 | 67982.6656 | 97.7830 | .0063378 | 109730.9 | 0.000000 : | qR1(7) : Excited dv=7 8 7 F1e - X Xv=0 7 7 F1e |
| 13C18O Excited | 8 e 8 | X | 7 e 1 | 67926.9075 | .1609937 | 68024.6905 | 97.7830 | .2573213 | 4463502 | 0.000000 : | rR21(7) : Excited dv=7 8 8 F2e - X Xv=0 7 7 F1e |
| 13C18O Excited | 8 e 9 | X | 7 e 1 | 67967.4636 | 1.783703 | 68065.2466 | 97.7830 | 2.850948 | 49541253 | 0.000000 : | rR21(7) : Excited a'v=12 8 8 F2e - X Xv=0 7 7 F1e |

13C18O Excited 8 e 10 X 7 e 1 67969.7061 1.763535 68067.4891 97.7830 2.818712 48985930 0.000000 : sR31(7) : Excited dv=7 8 9 F3e - X Xv=0 7 7 F1e
13C18O Excited 7 e 1 X 8 e 1 67511.2250 1.126216 67636.9402 125.7152 2.058101 39721600 0.000000 : pP(8) : Excited Av=2 7 7 F1e - X Xv=0 8 8 F1e
13C18O Excited 7 e 2 X 8 e 1 67576.0661 .1686202 67701.7813 125.7152 .3081447 5964384 0.000000 : oP1(8) : Excited a3Pv=13 7 6 F1e - X Xv=0 8 8 F1e
13C18O Excited 7 e 3 X 8 e 1 67619.2360 1.236385 67744.9512 125.7152 2.259429 43816901 0.000000 : pP21(8) : Excited a3Pv=13 7 7 F2e - X Xv=0 8 8 F1e
13C18O Excited 7 e 4 X 8 e 1 67658.1300 .2068292 67783.8453 125.7152 .3779697 7342588 0.000000 : qP31(8) : Excited a3Pv=13 7 8 F3e - X Xv=0 8 8 F1e
13C18O Excited 7 e 5 X 8 e 1 67724.3396 .5800559 67850.0548 125.7152 1.060022 20652917 0.000000 : oP1(8) : Excited ev=4 7 6 F1e - X Xv=0 8 8 F1e
13C18O Excited 7 e 6 X 8 e 1 67756.9518 .5404917 67882.6671 125.7152 .9877205 19272047 0.000000 : qP31(8) : Excited ev=4 7 8 F3e - X Xv=0 8 8 F1e
13C18O Excited 7 e 7 X 8 e 1 67840.5760 .0016823 67966.2912 125.7152 .0030743 60207.75 0.000000 : oP1(8) : Excited dv=7 7 6 F1e - X Xv=0 8 8 F1e
13C18O Excited 7 e 8 X 8 e 1 67881.6949 .0881644 68007.4101 125.7152 .1611158 3161028 0.000000 : pP21(8) : Excited dv=7 7 7 F2e - X Xv=0 8 8 F1e
13C18O Excited 7 e 9 X 8 e 1 67922.9347 1.034356 68048.6499 125.7152 1.890231 37153217 0.000000 : pP21(8) : Excited a'v=12 7 7 F2e - X Xv=0 8 8 F1e
13C18O Excited 7 e 10 X 8 e 1 67923.6264 1.401331 68049.3416 125.7152 2.560859 50336197 0.000000 : qP31(8) : Excited dv=7 7 8 F3e - X Xv=0 8 8 F1e
13C18O Excited 8 f 1 X 8 e 1 67531.2381 .1109356 67656.9533 125.7152 .2027290 3455449 0.000000 : qQ(8) : Excited Iv=3 8 8 F1f - X Xv=0 8 8 F1e
13C18O Excited 8 f 2 X 8 e 1 67534.0283 2.621586 67659.7435 125.7152 4.790812 81667936 0.000000 : qQ(8) : Excited Av=2 8 8 F1f - X Xv=0 8 8 F1e
13C18O Excited 8 f 3 X 8 e 1 67599.0086 .5186955 67724.7238 125.7152 .9478892 16205150 0.000000 : pQ1(8) : Excited a3Pv=13 8 7 F1f - X Xv=0 8 8 F1e
13C18O Excited 8 f 4 X 8 e 1 67640.3467 2.806914 67766.0620 125.7152 5.129490 87854933 0.000000 : qQ21(8) : Excited a3Pv=13 8 8 F2f - X Xv=0 8 8 F1e
13C18O Excited 8 f 5 X 8 e 1 67680.4225 .6163513 67806.1377 125.7152 1.126350 19325781 0.000000 : rQ31(8) : Excited a3Pv=13 8 9 F3f - X Xv=0 8 8 F1e
13C18O Excited 8 f 6 X 8 e 1 67757.4841 2.696920 67883.1993 125.7152 4.928482 84851494 0.000000 : qQ21(8) : Excited ev=4 8 8 F2f - X Xv=0 8 8 F1e
13C18O Excited 8 f 7 X 8 e 1 67856.9503 .0065473 67982.6656 125.7152 .0119649 206903.5 0.000000 : pQ1(8) : Excited dv=7 8 7 F1f - X Xv=0 8 8 F1e
13C18O Excited 8 f 8 X 8 e 1 67898.9753 .2656483 68024.6905 125.7152 .4854584 8410392 0.000000 : qQ21(8) : Excited dv=7 8 8 F2f - X Xv=0 8 8 F1e
13C18O Excited 8 f 9 X 8 e 1 67923.9320 1.746107 68049.6472 125.7152 3.190921 55342524 0.000000 : pQ1(8) : Excited a'v=12 8 7 F1f - X Xv=0 8 8 F1e
13C18O Excited 8 f 10 X 8 e 1 67941.7716 2.728657 68067.4868 125.7152 4.986478 86552389 0.000000 : rQ31(8) : Excited dv=7 8 9 F3f - X Xv=0 8 8 F1e
13C18O Excited 8 f 11 X 8 e 1 67959.4766 1.385953 68085.1918 125.7152 2.532758 43996511 0.000000 : rQ31(8) : Excited a'v=12 8 9 F3f - X Xv=0 8 8 F1e
13C18O Excited 9 e 1 X 8 e 1 67559.3783 1.597613 67685.0936 125.7152 2.919555 44580343 0.000000 : rR(8) : Excited Av=2 9 9 F1e - X Xv=0 8 8 F1e
13C18O Excited 9 e 2 X 8 e 1 67617.8302 .3503712 67743.5454 125.7152 .6402853 9802278 0.000000 : qR1(8) : Excited a3Pv=13 9 8 F1e - X Xv=0 8 8 F1e
13C18O Excited 9 e 3 X 8 e 1 67663.3683 1.561238 67789.0835 125.7152 2.853081 43766792 0.000000 : rR21(8) : Excited a3Pv=13 9 9 F2e - X Xv=0 8 8 F1e
13C18O Excited 9 e 4 X 8 e 1 67705.4138 .4255794 67831.1290 125.7152 .7777242 11952690 0.000000 : sR31(8) : Excited a3Pv=13 9 10 F3e - X Xv=0 8 8 F1e
13C18O Excited 9 e 5 X 8 e 1 67757.3493 .8037208 67883.0645 125.7152 1.468758 22625038 0.000000 : qR1(8) : Excited ev=4 9 8 F1e - X Xv=0 8 8 F1e
13C18O Excited 9 e 6 X 8 e 1 67798.6359 .7745563 67924.3511 125.7152 1.415461 21843929 0.000000 : sR31(8) : Excited ev=4 9 10 F3e - X Xv=0 8 8 F1e
13C18O Excited 9 e 7 X 8 e 1 67875.3917 .0057490 68001.1070 125.7152 .0105060 162683.7 0.000000 : qR1(8) : Excited dv=7 9 8 F1e - X Xv=0 8 8 F1e
13C18O Excited 9 e 8 X 8 e 1 67918.4116 .1878532 68044.1268 125.7152 .3432919 5325937 0.000000 : rR21(8) : Excited dv=7 9 9 F2e - X Xv=0 8 8 F1e
13C18O Excited 9 e 9 X 8 e 1 67958.1946 1.783235 68083.9098 125.7152 3.258769 50646440 0.000000 : rR21(8) : Excited a'v=12 9 9 F2e - X Xv=0 8 8 F1e
13C18O Excited 9 e 10 X 8 e 1 67962.1699 1.630271 68087.8851 125.7152 2.979235 46310164 0.000000 : sR31(8) : Excited dv=7 9 10 F3e - X Xv=0 8 8 F1e
13C18O Excited 8 e 1 X 9 e 1 67502.4668 1.103529 67659.6027 157.1358 2.344623 39912280 0.000000 : pP(9) : Excited Av=2 8 8 F1e - X Xv=0 9 9 F1e
13C18O Excited 8 e 2 X 9 e 1 67564.2835 .2029798 67721.4193 157.1358 .4312629 7361531 0.000000 : oP1(9) : Excited a3Pv=13 8 7 F1e - X Xv=0 9 9 F1e
13C18O Excited 8 e 3 X 9 e 1 67608.5762 1.144248 67765.7121 157.1358 2.431136 41580448 0.000000 : pP21(9) : Excited a3Pv=13 8 8 F2e - X Xv=0 9 9 F1e
13C18O Excited 8 e 4 X 9 e 1 67648.9914 .2477095 67806.1273 157.1358 .5262983 9017589 0.000000 : qP31(9) : Excited a3Pv=13 8 9 F3e - X Xv=0 9 9 F1e
13C18O Excited 8 e 5 X 9 e 1 67708.3235 .5620597 67865.4594 157.1358 1.194185 20515044 0.000000 : oP1(9) : Excited ev=4 8 7 F1e - X Xv=0 9 9 F1e
13C18O Excited 8 e 6 X 9 e 1 67745.2772 .5326812 67902.4130 157.1358 1.131766 19474587 0.000000 : qP31(9) : Excited ev=4 8 9 F3e - X Xv=0 9 9 F1e
13C18O Excited 8 e 7 X 9 e 1 67825.5297 .0026515 67982.6656 157.1358 .0056336 97282.97 0.000000 : oP1(9) : Excited dv=7 8 7 F1e - X Xv=0 9 9 F1e
13C18O Excited 8 e 8 X 9 e 1 67867.5547 .1076549 68024.6905 157.1358 .2287300 3957166 0.000000 : pP21(9) : Excited dv=7 8 8 F2e - X Xv=0 9 9 F1e
13C18O Excited 8 e 9 X 9 e 1 67908.1108 1.192745 68065.2466 157.1358 2.534176 43921404 0.000000 : pP21(9) : Excited a'v=12 8 8 F2e - X Xv=0 9 9 F1e
13C18O Excited 8 e 10 X 9 e 1 67910.3532 1.179258 68067.4891 157.1358 2.505522 43429080 0.000000 : qP31(9) : Excited dv=7 8 9 F3e - X Xv=0 9 9 F1e
13C18O Excited 9 f 1 X 9 e 1 67519.7238 .0092463 67676.8596 157.1358 .0196453 299446.7 0.000000 : qQ(9) : Excited Iv=3 9 9 F1f - X Xv=0 9 9 F1e
13C18O Excited 9 f 2 X 9 e 1 67528.0167 2.609114 67685.1525 157.1358 5.543477 84528676 0.000000 : qQ(9) : Excited Av=2 9 9 F1f - X Xv=0 9 9 F1e
13C18O Excited 9 f 3 X 9 e 1 67589.6284 .5846175 67746.7643 157.1358 1.242113 18992019 0.000000 : pQ1(9) : Excited a3Pv=13 9 8 F1f - X Xv=0 9 9 F1e
13C18O Excited 9 f 4 X 9 e 1 67632.3529 2.535682 67789.4887 157.1358 5.387460 82531049 0.000000 : qQ21(9) : Excited a3Pv=13 9 9 F2f - X Xv=0 9 9 F1e
13C18O Excited 9 f 5 X 9 e 1 67674.0069 .7003699 67831.1428 157.1358 1.488047 22837694 0.000000 : rQ31(9) : Excited a3Pv=13 9 10 F3f - X Xv=0 9 9 F1e
13C18O Excited 9 f 6 X 9 e 1 67745.8452 2.570393 67902.9810 157.1358 5.461209 84082691 0.000000 : qQ21(9) : Excited ev=4 9 9 F2f - X Xv=0 9 9 F1e
13C18O Excited 9 f 7 X 9 e 1 67843.9711 .0093885 68001.1070 157.1358 .0199474 308453.2 0.000000 : pQ1(9) : Excited dv=7 9 8 F1f - X Xv=0 9 9 F1e
13C18O Excited 9 f 8 X 9 e 1 67886.9910 .3064627 68044.1268 157.1358 .6511287 10087797 0.000000 : qQ21(9) : Excited dv=7 9 9 F2f - X Xv=0 9 9 F1e
13C18O Excited 9 f 9 X 9 e 1 67909.1083 1.662699 68066.2442 157.1358 3.532670 54784404 0.000000 : pQ1(9) : Excited a'v=12 9 8 F1f - X Xv=0 9 9 F1e
13C18O Excited 9 f 10 X 9 e 1 67930.7480 2.564762 68087.8839 157.1358 5.449245 84587338 0.000000 : rQ31(9) : Excited dv=7 9 10 F3f - X Xv=0 9 9 F1e
13C18O Excited 9 f 11 X 9 e 1 67948.7817 1.351615 68105.9176 157.1358 2.871722 44612575 0.000000 : rQ31(9) : Excited a'v=12 9 10 F3f - X Xv=0 9 9 F1e
13C18O Excited 10 e 1 X 9 e 1 67556.2755 1.504583 67713.4114 157.1358 3.196725 44157708 0.000000 : rR(9) : Excited Av=2 10 10 F1e - X Xv=0 9 9 F1e
13C18O Excited 10 e 2 X 9 e 1 67611.0316 .3844310 67768.1675 157.1358 .8167849 11310050 0.000000 : qR1(9) : Excited a3Pv=13 10 9 F1e - X Xv=0 9 9 F1e
13C18O Excited 10 e 3 X 9 e 1 67657.9306 1.384122 67815.0665 157.1358 2.940788 40805990 0.000000 : rR21(9) : Excited a3Pv=13 10 10 F2e - X Xv=0 9 9 F1e
13C18O Excited 10 e 4 X 9 e 1 67701.6985 .4662162 67858.8344 157.1358 .9905504 13771442 0.000000 : sR31(9) : Excited a3Pv=13 10 11 F3e - X Xv=0 9 9 F1e
13C18O Excited 10 e 5 X 9 e 1 67745.7361 .7453104 67902.8720 157.1358 1.583531 22058526 0.000000 : qR1(9) : Excited ev=4 10 9 F1e - X Xv=0 9 9 F1e
13C18O Excited 10 e 6 X 9 e 1 67791.3447 .7318482 67948.4806 157.1358 1.554928 21703871 0.000000 : sR31(9) : Excited ev=4 10 11 F3e - X Xv=0 9 9 F1e
13C18O Excited 10 e 7 X 9 e 1 67864.4848 .0076752 68021.6207 157.1358 .0163071 228353.7 0.000000 : qR1(9) : Excited dv=7 10 9 F1e - X Xv=0 9 9 F1e
13C18O Excited 10 e 8 X 9 e 1 67908.5818 .2077414 68065.7176 157.1358 .4413797 6192848 0.000000 : rR21(9) : Excited dv=7 10 10 F2e - X Xv=0 9 9 F1e
13C18O Excited 10 e 9 X 9 e 1 67947.5062 1.707169 68104.6420 157.1358 3.627152 50978905 0.000000 : rR21(9) : Excited a'v=12 10 10 F2e - X Xv=0 9 9 F1e
13C18O Excited 10 e 10 X 9 e 1 67953.3831 1.489739 68110.5190 157.1358 3.165188 44497626 0.000000 : sR31(9) : Excited dv=7 10 11 F3e - X Xv=0 9 9 F1e
13C18O Excited 9 e 1 X 10 e 1 67493.0498 1.046073 67685.0936 192.0438 2.627599 40004251 0.000000 : pP(10) : Excited Av=2 9 9 F1e - X Xv=0 10 10 F1e
13C18O Excited 9 e 2 X 10 e 1 67551.5016 .2294135 67743.5454 192.0438 .5762567 8796114 0.000000 : oP1(10) : Excited a3Pv=13 9 8 F1e - X Xv=0 10 10 F1e
13C18O Excited 9 e 3 X 10 e 1 67597.0397 1.022256 67789.0835 192.0438 2.567773 39274387 0.000000 : pP21(10) : Excited a3Pv=13 9 9 F2e - X Xv=0 10 10 F1e
13C18O Excited 9 e 4 X 10 e 1 67639.0852 .2786577 67831.1290 192.0438 .6999518 10725836 0.000000 : qP31(10) : Excited a3Pv=13 9 10 F3e - X Xv=0 10 10 F1e
13C18O Excited 9 e 5 X 10 e 1 67691.0207 .5262543 67883.0645 192.0438 1.321882 20302793 0.000000 : oP1(10) : Excited ev=4 9 8 F1e - X Xv=0 10 10 F1e
13C18O Excited 9 e 6 X 10 e 1 67732.3073 .5071582 67924.3511 192.0438 1.273915 19601892 0.000000 : qP31(10) : Excited ev=4 9 10 F3e - X Xv=0 10 10 F1e
13C18O Excited 9 e 7 X 10 e 1 67809.0631 .0037643 68001.1070 192.0438 .0094554 145986.5 0.000000 : oP1(10) : Excited dv=7 9 8 F1e - X Xv=0 10 10 F1e
13C18O Excited 9 e 8 X 10 e 1 67852.0830 .1230011 68044.1268 192.0438 .3089627 4779313 0.000000 : pP21(10) : Excited dv=7 9 9 F2e - X Xv=0 10 10 F1e
13C18O Excited 9 e 9 X 10 e 1 67891.8660 1.167613 68083.9098 192.0438 2.932892 45448460 0.000000 : pP21(10) : Excited a'v=12 9 9 F2e - X Xv=0 10 10 F1e
13C18O Excited 9 e 10 X 10 e 1 67895.8413 1.067457 68087.8851 192.0438 2.681312 41557234 0.000000 : qP31(10) : Excited dv=7 9 10 F3e - X Xv=0 10 10 F1e
13C18O Excited 10 f 1 X 10 e 1 67506.8499 .0019735 67698.8937 192.0438 .0049571 68324.06 0.000000 : qQ(10) : Excited Iv=3 10 10 F1f - X Xv=0 10 10 F1e
13C18O Excited 10 f 2 X 10 e 1 67521.4105 2.436527 67713.4543 192.0438 6.120237 84410582 0.000000 : qQ(10) : Excited Av=2 10 10 F1f - X Xv=0 10 10 F1e
13C18O Excited 10 f 3 X 10 e 1 67579.2638 .6253710 67771.3076 192.0438 1.570851 21720978 0.000000 : pQ1(10) : Excited a3Pv=13 10 9 F1f - X Xv=0 10 10 F1e
13C18O Excited 10 f 4 X 10 e 1 67623.4797 2.226094 67815.5235 192.0438 5.591659 77470677 0.000000 : qQ21(10) : Excited a3Pv=13 10 10 F2f - X Xv=0 10 10 F1e
13C18O Excited 10 f 5 X 10 e 1 67666.8081 .7568005 67858.8519 192.0438 1.900984 26388196 0.000000 : rQ31(10) : Excited a3Pv=13 10 11 F3f - X Xv=0 10 10 F1e
13C18O Excited 10 f 6 X 10 e 1 67732.9190 2.376694 67924.9628 192.0438 5.969944 83113928 0.000000 : qQ21(10) : Excited ev=4 10 10 F2f - X Xv=0 10 10 F1e
13C18O Excited 10 f 7 X 10 e 1 67829.5768 .0123825 68021.6207 192.0438 .0311033 434879.1 0.000000 : pQ1(10) : Excited dv=7 10 9 F1f - X Xv=0 10 10 F1e
13C18O Excited 10 f 8 X 10 e 1 67873.6738 .3346380 68065.7176 192.0438 .8405671 11775534 0.000000 : qQ21(10) : Excited dv=7 10 10 F2f - X Xv=0 10 10 F1e
13C18O Excited 10 f 9 X 10 e 1 67892.8660 1.542522 68084.9098 192.0438 3.874613 54325646 0.000000 : pQ1(10) : Excited a'v=12 10 9 F1f - X Xv=0 10 10 F1e
13C18O Excited 10 f 10 X 10 e 1 67918.4743 2.342472 68110.5181 192.0438 5.883983 82592258 0.000000 : rQ31(10) : Excited dv=7 10 11 F3f - X Xv=0 10 10 F1e
13C18O Excited 10 f 11 X 10 e 1 67936.6699 1.278370 68128.7137 192.0438 3.211098 45109749 0.000000 : rQ31(10) : Excited a'v=12 10 11 F3f - X Xv=0 10 10 F1e
13C18O Excited 11 e 1 X 10 e 1 67552.5104 1.380627 67744.5542 192.0438 3.467954 43731402 0.000000 : rR(10) : Excited Av=2 11 11 F1e - X Xv=0 10 10 F1e
13C18O Excited 11 e 2 X 10 e 1 67603.2488 .4037178 67795.2926 192.0438 1.014087 12816615 0.000000 : qR1(10) : Excited a3Pv=13 11 10 F1e - X Xv=0 10 10 F1e
13C18O Excited 11 e 3 X 10 e 1 67651.6171 1.195375 67843.6609 192.0438 3.002625 38030429 0.000000 : rR21(10) : Excited a3Pv=13 11 11 F2e - X Xv=0 10 10 F1e
13C18O Excited 11 e 4 X 10 e 1 67697.1826 .4922223 67889.2264 192.0438 1.236398 15691545 0.000000 : sR31(10) : Excited a3Pv=13 11 12 F3e - X Xv=0 10 10 F1e
13C18O Excited 11 e 5 X 10 e 1 67732.8418 .6688616 67924.8856 192.0438 1.680093 21356340 0.000000 : qR1(10) : Excited ev=4 11 10 F1e - X Xv=0 10 10 F1e
13C18O Excited 11 e 6 X 10 e 1 67782.7570 .6731557 67974.8008 192.0438 1.690879 21541004 0.000000 : sR31(10) : Excited ev=4 11 12 F3e - X Xv=0 10 10 F1e
13C18O Excited 11 e 7 X 10 e 1 67852.1679 .0095601 68044.2117 192.0438 .0240137 306863.9 0.000000 : qR1(10) : Excited dv=7 11 10 F1e - X Xv=0 10 10 F1e
13C18O Excited 11 e 8 X 10 e 1 67897.4177 .2192334 68089.4615 192.0438 .5506857 7051137 0.000000 : rR21(10) : Excited dv=7 11 11 F2e - X Xv=0 10 10 F1e

13C18O Excited 11 e 9 X 10 e 1 67935.3986 1.585559 68127.4424 192.0438 3.982717 51081478 0.000000 : rR21(10) : Excited a'v=12 11 11 F2e - X Xv=0 10 10 F1e
13C18O Excited 11 e 10 X 10 e 1 67943.3393 1.333886 68135.3831 192.0438 3.350547 42988470 0.000000 : sR31(10) : Excited dv=7 11 12 F3e - X Xv=0 10 10 F1e
13C18O Excited 10 e 1 X 11 e 1 67482.9734 .9623801 67713.4114 230.4379 2.906114 40012839 0.000000 : pP(11) : Excited Av=2 10 10 F1e - X Xv=0 11 11 F1e
13C18O Excited 10 e 2 X 11 e 1 67537.7296 .2458946 67768.1675 230.4379 .7425317 10248458 0.000000 : oP1(11) : Excited a3Pv=13 10 9 F1e - X Xv=0 11 11 F1e
13C18O Excited 10 e 3 X 11 e 1 67584.6286 .8853296 67815.0665 230.4379 2.673443 36975912 0.000000 : pP21(11) : Excited a3Pv=13 10 10 F2e - X Xv=0 11 11 F1e
13C18O Excited 10 e 4 X 11 e 1 67628.3965 .2982071 67858.8344 230.4379 .9005004 12478872 0.000000 : qP31(11) : Excited a3Pv=13 10 11 F3e - X Xv=0 11 11 F1e
13C18O Excited 10 e 5 X 11 e 1 67672.4341 .4767248 67902.8720 230.4379 1.439573 19988182 0.000000 : oP1(11) : Excited ev=4 10 9 F1e - X Xv=0 11 11 F1e
13C18O Excited 10 e 6 X 11 e 1 67718.0426 .4681140 67948.4806 230.4379 1.413571 19666857 0.000000 : qP31(11) : Excited ev=4 10 11 F3e - X Xv=0 11 11 F1e
13C18O Excited 10 e 7 X 11 e 1 67791.1827 .0049093 68021.6207 230.4379 .0148246 206922.3 0.000000 : oP1(11) : Excited dv=7 10 9 F1e - X Xv=0 11 11 F1e
13C18O Excited 10 e 8 X 11 e 1 67835.2797 .1328782 68065.7176 230.4379 .4012542 5611651 0.000000 : pP21(11) : Excited dv=7 10 10 F2e - X Xv=0 11 11 F1e
13C18O Excited 10 e 9 X 11 e 1 67874.2041 1.091961 68104.6420 230.4379 3.297411 46194631 0.000000 : pP21(11) : Excited a'v=12 10 10 F2e - X Xv=0 11 11 F1e
13C18O Excited 10 e 10 X 11 e 1 67880.0811 .9528857 68110.5190 230.4379 2.877443 40321619 0.000000 : qP31(11) : Excited dv=7 10 11 F3e - X Xv=0 11 11 F1e
13C18O Excited 11 f 1 X 11 e 1 67492.6711 .0004973 67723.1091 230.4379 .0015016 18885.10 0.000000 : qQ(11) : Excited Iv=3 11 11 F1f - X Xv=0 11 11 F1e
13C18O Excited 11 f 2 X 11 e 1 67514.1534 2.210969 67744.5914 230.4379 6.676497 84048261 0.000000 : qQ(11) : Excited Av=2 11 11 F1f - X Xv=0 11 11 F1e
13C18O Excited 11 f 3 X 11 e 1 67567.9257 .6394007 67798.3636 230.4379 1.930808 24364440 0.000000 : pQ1(11) : Excited a3Pv=13 11 10 F1f - X Xv=0 11 11 F1e
13C18O Excited 11 f 4 X 11 e 1 67613.7274 1.905140 67844.1653 230.4379 5.752981 72743323 0.000000 : qQ21(11) : Excited a3Pv=13 11 11 F2f - X Xv=0 11 11 F1e
13C18O Excited 11 f 5 X 11 e 1 67658.8111 .7845491 67889.2490 230.4379 2.369115 30016138 0.000000 : rQ31(11) : Excited a3Pv=13 11 12 F3f - X Xv=0 11 11 F1e
13C18O Excited 11 f 6 X 11 e 1 67718.7089 2.134288 67949.1468 230.4379 6.444943 81872997 0.000000 : qQ21(11) : Excited ev=4 11 11 F2f - X Xv=0 11 11 F1e
13C18O Excited 11 f 7 X 11 e 1 67813.7738 .0152241 68044.2117 230.4379 .0459725 586472.7 0.000000 : pQ1(11) : Excited dv=7 11 10 F1f - X Xv=0 11 11 F1e
13C18O Excited 11 f 8 X 11 e 1 67859.0235 .3482924 68089.4615 230.4379 1.051744 13443996 0.000000 : qQ21(11) : Excited dv=7 11 11 F2f - X Xv=0 11 11 F1e
13C18O Excited 11 f 9 X 11 e 1 67875.2057 1.396613 68105.6436 230.4379 4.217373 53947467 0.000000 : pQ1(11) : Excited a'v=12 11 10 F1f - X Xv=0 11 11 F1e
13C18O Excited 11 f 10 X 11 e 1 67904.9446 2.083462 68135.3825 230.4379 6.291463 80584482 0.000000 : rQ31(11) : Excited dv=7 11 12 F3f - X Xv=0 11 11 F1e
13C18O Excited 11 f 11 X 11 e 1 67923.1404 1.175916 68153.5783 230.4379 3.550932 45518843 0.000000 : rQ31(11) : Excited a'v=12 11 12 F3f - X Xv=0 11 11 F1e
13C18O Excited 12 e 1 X 11 e 1 67548.0821 1.235717 67778.5200 230.4379 3.731512 43282002 0.000000 : rR(11) : Excited Av=2 12 12 F1e - X Xv=0 11 11 F1e
13C18O Excited 12 e 2 X 11 e 1 67594.4891 .4080106 67824.9270 230.4379 1.232076 14320386 0.000000 : qR1(11) : Excited a3Pv=13 12 11 F1e - X Xv=0 11 11 F1e
13C18O Excited 12 e 3 X 11 e 1 67644.4280 1.007645 67874.8659 230.4379 3.042800 35444835 0.000000 : rR21(11) : Excited a3Pv=13 12 12 F2e - X Xv=0 11 11 F1e
13C18O Excited 12 e 4 X 11 e 1 67691.8455 .5074748 67922.2834 230.4379 1.532429 17888459 0.000000 : sR31(11) : Excited a3Pv=13 12 13 F3e - X Xv=0 11 11 F1e
13C18O Excited 12 e 5 X 11 e 1 67718.6767 .5762028 67949.1147 230.4379 1.739968 20335279 0.000000 : qR1(11) : Excited ev=4 12 11 F1e - X Xv=0 11 11 F1e
13C18O Excited 12 e 6 X 11 e 1 67772.8733 .6034653 68003.3112 230.4379 1.822293 21348601 0.000000 : sR31(11) : Excited ev=4 12 13 F3e - X Xv=0 11 11 F1e
13C18O Excited 12 e 7 X 11 e 1 67838.4471 .0112185 68068.8850 230.4379 .0338766 398025.6 0.000000 : qR1(11) : Excited dv=7 12 11 F1e - X Xv=0 11 11 F1e
13C18O Excited 12 e 8 X 11 e 1 67884.9191 .2219279 68115.3570 230.4379 .6701590 7890077 0.000000 : rR21(11) : Excited dv=7 12 12 F2e - X Xv=0 11 11 F1e
13C18O Excited 12 e 9 X 11 e 1 67921.8719 1.434807 68152.3098 230.4379 4.332707 51094210 0.000000 : rR21(11) : Excited a'v=12 12 12 F2e - X Xv=0 11 11 F1e
13C18O Excited 12 e 10 X 11 e 1 67932.0325 1.168602 68162.4704 230.4379 3.528845 41633201 0.000000 : sR31(11) : Excited dv=7 12 13 F3e - X Xv=0 11 11 F1e
13C18O Excited 11 e 1 X 12 e 1 67472.2374 .8611754 67744.5542 272.3168 3.178958 39944381 0.000000 : pP(12) : Excited Av=2 11 11 F1e - X Xv=0 12 12 F1e
13C18O Excited 11 e 2 X 12 e 1 67522.9758 .2518218 67795.2926 272.3168 .9295794 11706762 0.000000 : oP1(12) : Excited a3Pv=13 11 10 F1e - X Xv=0 12 12 F1e
13C18O Excited 11 e 3 X 12 e 1 67571.3441 .7456231 67843.6609 272.3168 2.752406 34737279 0.000000 : pP21(12) : Excited a3Pv=13 11 11 F2e - X Xv=0 12 12 F1e
13C18O Excited 11 e 4 X 12 e 1 67616.9095 .3070270 67889.2264 272.3168 1.133365 14332809 0.000000 : qP31(12) : Excited a3Pv=13 11 12 F3e - X Xv=0 12 12 F1e
13C18O Excited 11 e 5 X 12 e 1 67652.5688 .4172070 67924.8856 272.3168 1.540085 19507124 0.000000 : oP1(12) : Excited ev=4 11 10 F1e - X Xv=0 12 12 F1e
13C18O Excited 11 e 6 X 12 e 1 67702.4839 .4198855 67974.8008 272.3168 1.549973 19675849 0.000000 : qP31(12) : Excited ev=4 11 12 F3e - X Xv=0 12 12 F1e
13C18O Excited 11 e 7 X 12 e 1 67771.8949 .0059632 68044.2117 272.3168 .0220126 280294.7 0.000000 : oP1(12) : Excited dv=7 11 10 F1e - X Xv=0 12 12 F1e
13C18O Excited 11 e 8 X 12 e 1 67817.1447 .1367483 68089.4615 272.3168 .5047952 6440645 0.000000 : pP21(12) : Excited dv=7 11 11 F2e - X Xv=0 12 12 F1e
13C18O Excited 11 e 9 X 12 e 1 67855.1256 .9890033 68127.4424 272.3168 3.650824 46658899 0.000000 : pP21(12) : Excited a'v=12 11 11 F2e - X Xv=0 12 12 F1e
13C18O Excited 11 e 10 X 12 e 1 67863.0663 .8320205 68135.3831 272.3168 3.071335 39266591 0.000000 : qP31(12) : Excited dv=7 11 12 F3e - X Xv=0 12 12 F1e
13C18O Excited 12 f 1 X 12 e 1 67477.1931 .0001134 67749.5100 272.3168 .0004186 4840.576 0.000000 : qQ(12) : Excited Iv=3 12 12 F1f - X Xv=0 12 12 F1e
13C18O Excited 12 f 2 X 12 e 1 67506.2383 1.955585 67778.5552 272.3168 7.218880 83576647 0.000000 : qQ(12) : Excited Av=2 12 12 F1f - X Xv=0 12 12 F1e
13C18O Excited 12 f 3 X 12 e 1 67555.6244 .6282489 67827.9412 272.3168 2.319129 26908709 0.000000 : pQ1(12) : Excited a3Pv=13 12 11 F1f - X Xv=0 12 12 F1e
13C18O Excited 12 f 4 X 12 e 1 67603.0958 1.593495 67875.4126 272.3168 5.882256 68395432 0.000000 : qQ21(12) : Excited a3Pv=13 12 12 F2f - X Xv=0 12 12 F1e
13C18O Excited 12 f 5 X 12 e 1 67650.0002 .7862360 67922.3170 272.3168 2.902325 33816831 0.000000 : rQ31(12) : Excited a3Pv=13 12 13 F3f - X Xv=0 12 12 F1e
13C18O Excited 12 f 6 X 12 e 1 67703.2206 1.861243 67975.5375 272.3168 6.870625 80243098 0.000000 : qQ21(12) : Excited ev=4 12 12 F2f - X Xv=0 12 12 F1e
13C18O Excited 12 f 7 X 12 e 1 67796.5682 .0176222 68068.8850 272.3168 .0650508 762885.7 0.000000 : pQ1(12) : Excited dv=7 12 11 F1f - X Xv=0 12 12 F1e
13C18O Excited 12 f 8 X 12 e 1 67843.0402 .3472828 68115.3570 272.3168 1.281966 15065229 0.000000 : qQ21(12) : Excited dv=7 12 12 F2f - X Xv=0 12 12 F1e
13C18O Excited 12 f 9 X 12 e 1 67856.1279 1.235797 68128.4447 272.3168 4.561842 53640266 0.000000 : pQ1(12) : Excited a'v=12 12 11 F1f - X Xv=0 12 12 F1e
13C18O Excited 12 f 10 X 12 e 1 67890.1531 1.807672 68162.4699 272.3168 6.672871 78580816 0.000000 : rQ31(12) : Excited dv=7 12 13 F3f - X Xv=0 12 12 F1e
13C18O Excited 12 f 11 X 12 e 1 67908.1928 1.054147 68180.5096 272.3168 3.891296 45861078 0.000000 : rQ31(12) : Excited a'v=12 12 13 F3f - X Xv=0 12 12 F1e
13C18O Excited 13 e 1 X 12 e 1 67542.9896 1.079506 67815.3064 272.3168 3.984908 42787691 0.000000 : rR(12) : Excited Av=2 13 13 F1e - X Xv=0 12 12 F1e
13C18O Excited 13 e 2 X 12 e 1 67584.7588 .3986314 67857.0756 272.3168 1.471515 15829624 0.000000 : qR1(12) : Excited a3Pv=13 13 12 F1e - X Xv=0 12 12 F1e
13C18O Excited 13 e 3 X 12 e 1 67636.3628 .8303059 67908.6796 272.3168 3.065006 33046975 0.000000 : rR21(12) : Excited a3Pv=13 13 13 F2e - X Xv=0 12 12 F1e
13C18O Excited 13 e 4 X 12 e 1 67685.6479 .5247109 67957.9648 272.3168 1.936927 20929684 0.000000 : sR31(12) : Excited a3Pv=13 13 14 F3e - X Xv=0 12 12 F1e
13C18O Excited 13 e 5 X 12 e 1 67703.2719 .4617035 67975.5888 272.3168 1.704340 18430832 0.000000 : qR1(12) : Excited ev=4 13 12 F1e - X Xv=0 12 12 F1e
13C18O Excited 13 e 6 X 12 e 1 67761.6946 .5276450 68034.0115 272.3168 1.947758 21117741 0.000000 : sR31(12) : Excited ev=4 13 14 F3e - X Xv=0 12 12 F1e
13C18O Excited 13 e 7 X 12 e 1 67823.3282 .0124954 68095.6450 272.3168 .0461256 501463.5 0.000000 : qR1(12) : Excited dv=7 13 12 F1e - X Xv=0 12 12 F1e
13C18O Excited 13 e 8 X 12 e 1 67871.0859 .2163572 68143.4027 272.3168 .7986646 8701187 0.000000 : rR21(12) : Excited dv=7 13 13 F2e - X Xv=0 12 12 F1e
13C18O Excited 13 e 9 X 12 e 1 67906.9259 1.267791 68179.2428 272.3168 4.679947 51067296 0.000000 : rR21(12) : Excited a'v=12 13 13 F2e - X Xv=0 12 12 F1e
13C18O Excited 13 e 10 X 12 e 1 67919.4570 1.001821 68191.7738 272.3168 3.698141 40376230 0.000000 : sR31(12) : Excited dv=7 13 14 F3e - X Xv=0 12 12 F1e
13C18O Excited 12 e 1 X 13 e 1 67460.8409 .7506667 67778.5200 317.6791 3.444473 39798015 0.000000 : pP(13) : Excited Av=2 12 12 F1e - X Xv=0 13 13 F1e
13C18O Excited 12 e 2 X 13 e 1 67507.2479 .2478562 67824.9270 317.6791 1.137301 13167701 0.000000 : oP1(13) : Excited a3Pv=13 12 11 F1e - X Xv=0 13 13 F1e
13C18O Excited 12 e 3 X 13 e 1 67557.1868 .6121188 67874.8659 317.6791 2.808739 32591882 0.000000 : pP21(13) : Excited a3Pv=13 12 12 F2e - X Xv=0 13 13 F1e
13C18O Excited 12 e 4 X 13 e 1 67604.6043 .3082781 67922.2834 317.6791 1.414550 16448663 0.000000 : qP31(13) : Excited a3Pv=13 12 13 F3e - X Xv=0 13 13 F1e
13C18O Excited 12 e 5 X 13 e 1 67631.4356 .3500287 67949.1147 317.6791 1.606125 18698573 0.000000 : oP1(13) : Excited ev=4 12 11 F1e - X Xv=0 13 13 F1e
13C18O Excited 12 e 6 X 13 e 1 67685.6321 .3665900 68003.3112 317.6791 1.682117 19630397 0.000000 : qP31(13) : Excited ev=4 12 13 F3e - X Xv=0 13 13 F1e
13C18O Excited 12 e 7 X 13 e 1 67751.2059 .0068149 68068.8850 317.6791 .0312707 365992.6 0.000000 : oP1(13) : Excited dv=7 12 11 F1e - X Xv=0 13 13 F1e
13C18O Excited 12 e 8 X 13 e 1 67797.6780 .1348156 68115.3570 317.6791 .6186083 7255104 0.000000 : pP21(13) : Excited dv=7 12 12 F2e - X Xv=0 13 13 F1e
13C18O Excited 12 e 9 X 13 e 1 67834.6308 .8716088 68152.3098 317.6791 3.999422 46982382 0.000000 : pP21(13) : Excited a'v=12 12 12 F2e - X Xv=0 13 13 F1e
13C18O Excited 12 e 10 X 13 e 1 67844.7913 .7098962 68162.4704 317.6791 3.257395 38282775 0.000000 : qP31(13) : Excited dv=7 12 13 F3e - X Xv=0 13 13 F1e
13C18O Excited 13 f 2 X 13 e 1 67497.6626 1.688130 67815.3417 317.6791 7.746072 83005601 0.000000 : qQ(13) : Excited Av=2 13 13 F1f - X Xv=0 13 13 F1e
13C18O Excited 13 f 3 X 13 e 1 67542.3690 .5957714 67860.0481 317.6791 2.733728 29352412 0.000000 : pQ1(13) : Excited a3Pv=13 13 12 F1f - X Xv=0 13 13 F1e
13C18O Excited 13 f 4 X 13 e 1 67591.5839 1.305457 67909.2630 317.6791 5.990156 64457827 0.000000 : qQ21(13) : Excited a3Pv=13 13 13 F2f - X Xv=0 13 13 F1e
13C18O Excited 13 f 5 X 13 e 1 67640.3574 .7674929 67958.0365 317.6791 3.521681 37977582 0.000000 : rQ31(13) : Excited a3Pv=13 13 14 F3f - X Xv=0 13 13 F1e
13C18O Excited 13 f 6 X 13 e 1 67686.4636 1.573550 68004.1426 317.6791 7.220316 78022762 0.000000 : qQ21(13) : Excited ev=4 13 13 F2f - X Xv=0 13 13 F1e
13C18O Excited 13 f 7 X 13 e 1 67777.9659 .0193505 68095.6450 317.6791 .0887907 963369.0 0.000000 : pQ1(13) : Excited dv=7 13 12 F1f - X Xv=0 13 13 F1e
13C18O Excited 13 f 8 X 13 e 1 67825.7235 .3328781 68143.4026 317.6791 1.527428 16607481 0.000000 : qQ21(13) : Excited dv=7 13 13 F2f - X Xv=0 13 13 F1e
13C18O Excited 13 f 9 X 13 e 1 67835.6329 1.070015 68153.3120 317.6791 4.909820 53407085 0.000000 : pQ1(13) : Excited a'v=12 13 12 F1f - X Xv=0 13 13 F1e
13C18O Excited 13 f 10 X 13 e 1 67874.0943 1.531995 68191.7734 317.6791 7.029640 76595791 0.000000 : rQ31(13) : Excited dv=7 13 14 F3f - X Xv=0 13 13 F1e
13C18O Excited 13 f 11 X 13 e 1 67891.8266 .9223578 68209.5057 317.6791 4.232286 46151644 0.000000 : rQ31(13) : Excited a'v=12 13 14 F3f - X Xv=0 13 13 F1e
13C18O Excited 14 e 1 X 13 e 1 67537.2315 .9206266 67854.9105 317.6791 4.224342 42219618 0.000000 : rR(13) : Excited Av=2 14 14 F1e - X Xv=0 13 13 F1e
13C18O Excited 14 e 2 X 13 e 1 67574.0631 .3780429 67891.7422 317.6791 1.734669 17365296 0.000000 : qR1(13) : Excited a3Pv=13 14 13 F1e - X Xv=0 13 13 F1e
13C18O Excited 14 e 3 X 13 e 1 67627.4207 .6695298 67945.0998 317.6791 3.072172 30827578 0.000000 : rR21(13) : Excited a3Pv=13 14 14 F2e - X Xv=0 13 13 F1e
13C18O Excited 14 e 4 X 13 e 1 67678.4124 .6024867 67996.0915 317.6791 2.764541 27803470 0.000000 : sR31(13) : Excited a3Pv=13 14 15 F3e - X Xv=0 13 13 F1e
13C18O Excited 14 e 5 X 13 e 1 67686.7977 .2744168 68004.4768 317.6791 1.259176 12668457 0.000000 : qR1(13) : Excited ev=4 14 13 F1e - X Xv=0 13 13 F1e
13C18O Excited 14 e 6 X 13 e 1 67749.2225 .4500912 68066.9016 317.6791 2.065267 20836008 0.000000 : sR31(13) : Excited ev=4 14 15 F3e - X Xv=0 13 13 F1e
13C18O Excited 14 e 7 X 13 e 1 67806.8168 .0132893 68124.4959 317.6791 .0609788 616771.8 0.000000 : qR1(13) : Excited dv=7 14 13 F1e - X Xv=0 13 13 F1e
13C18O Excited 14 e 8 X 13 e 1 67855.9178 .2037758 68173.5968 317.6791 .9350358 9478004 0.000000 : rR21(13) : Excited dv=7 14 14 F2e - X Xv=0 13 13 F1e

13C18O Excited 14 e 9 X 13 e 1 67890.5606 1.095294 68208.2397 317.6791 5.025811 51022270 0.000000 : rR21(13) : Excited a'v=12 14 14 F2e - X Xv=0 13 13 F1e
13C18O Excited 14 e 10 X 13 e 1 67905.6075 .8407897 68223.2866 317.6791 3.858007 39192715 0.000000 : sR31(13) : Excited dv=7 14 15 F3e - X Xv=0 13 13 F1e
13C18O Excited 13 e 1 X 14 e 1 67448.7833 .6380043 67815.3064 366.5231 3.700272 39565412 0.000000 : pP(14) : Excited Av=2 13 13 F1e - X Xv=0 14 14 F1e
13C18O Excited 13 e 2 X 14 e 1 67490.5525 .2355971 67857.0756 366.5231 1.366407 14637556 0.000000 : oP1(14) : Excited a3Pv=13 13 12 F1e - X Xv=0 14 14 F1e
13C18O Excited 13 e 3 X 14 e 1 67542.1565 .4907232 67908.6796 366.5231 2.846077 30558432 0.000000 : pP21(14) : Excited a3Pv=13 13 13 F2e - X Xv=0 14 14 F1e
13C18O Excited 13 e 4 X 14 e 1 67591.4417 .3101120 67957.9648 366.5231 1.798575 19353671 0.000000 : qP31(14) : Excited a3Pv=13 13 14 F3e - X Xv=0 14 14 F1e
13C18O Excited 13 e 5 X 14 e 1 67609.0656 .2728736 67975.5888 366.5231 1.582602 17043001 0.000000 : oP1(14) : Excited ev=4 13 12 F1e - X Xv=0 14 14 F1e
13C18O Excited 13 e 6 X 14 e 1 67667.4883 .3118461 68034.0115 366.5231 1.808633 19527659 0.000000 : qP31(14) : Excited ev=4 13 14 F3e - X Xv=0 14 14 F1e
13C18O Excited 13 e 7 X 14 e 1 67729.1219 .0073850 68095.6450 366.5231 .0428309 463707.1 0.000000 : oP1(14) : Excited dv=7 13 12 F1e - X Xv=0 14 14 F1e
13C18O Excited 13 e 8 X 14 e 1 67776.8796 .1278703 68143.4027 366.5231 .7416171 8046076 0.000000 : pP21(14) : Excited dv=7 13 13 F2e - X Xv=0 14 14 F1e
13C18O Excited 13 e 9 X 14 e 1 67812.7197 .7492836 68179.2428 366.5231 4.345665 47222552 0.000000 : pP21(14) : Excited a'v=12 13 13 F2e - X Xv=0 14 14 F1e
13C18O Excited 13 e 10 X 14 e 1 67825.2507 .5920914 68191.7738 366.5231 3.433988 37336422 0.000000 : qP31(14) : Excited dv=7 13 14 F3e - X Xv=0 14 14 F1e
13C18O Excited 14 f 2 X 14 e 1 67488.4248 1.423249 67854.9479 366.5231 8.254503 82319790 0.000000 : qQ(14) : Excited Av=2 14 14 F1f - X Xv=0 14 14 F1e
13C18O Excited 14 f 3 X 14 e 1 67528.1677 .5472049 67894.6908 366.5231 3.173657 31705919 0.000000 : pQ1(14) : Excited a3Pv=13 14 13 F1f - X Xv=0 14 14 F1e
13C18O Excited 14 f 4 X 14 e 1 67579.1905 1.049595 67945.7136 366.5231 6.087397 60953126 0.000000 : qQ21(14) : Excited a3Pv=13 14 14 F2f - X Xv=0 14 14 F1e
13C18O Excited 14 f 5 X 14 e 1 67629.8578 .7364666 67996.3809 366.5231 4.271330 42865113 0.000000 : rQ31(14) : Excited a3Pv=13 14 15 F3f - X Xv=0 14 14 F1e
13C18O Excited 14 f 6 X 14 e 1 67668.4550 1.283525 68034.9781 366.5231 7.444136 74833920 0.000000 : qQ21(14) : Excited ev=4 14 14 F2f - X Xv=0 14 14 F1e
13C18O Excited 14 f 7 X 14 e 1 67757.9728 .0202797 68124.4959 366.5231 .1176174 1187075 0.000000 : pQ1(14) : Excited dv=7 14 13 F1f - X Xv=0 14 14 F1e
13C18O Excited 14 f 8 X 14 e 1 67807.0735 .3070724 68173.5967 366.5231 1.780946 18013632 0.000000 : qQ21(14) : Excited dv=7 14 14 F2f - X Xv=0 14 14 F1e
13C18O Excited 14 f 9 X 14 e 1 67813.7210 .9080392 68180.2441 366.5231 5.266410 53283510 0.000000 : pQ1(14) : Excited a'v=12 14 13 F1f - X Xv=0 14 14 F1e
13C18O Excited 14 f 10 X 14 e 1 67856.7631 1.269587 68223.2862 366.5231 7.363303 74641012 0.000000 : rQ31(14) : Excited dv=7 14 15 F3f - X Xv=0 14 14 F1e
13C18O Excited 14 f 11 X 14 e 1 67874.0415 .7886573 68240.5646 366.5231 4.574023 46401814 0.000000 : rQ31(14) : Excited a'v=12 14 15 F3f - X Xv=0 14 14 F1e
13C18O Excited 15 e 1 X 14 e 1 67530.8059 .7661623 67897.3290 366.5231 4.443558 41533482 0.000000 : rR(14) : Excited Av=2 15 15 F1e - X Xv=0 14 14 F1e
13C18O Excited 15 e 2 X 14 e 1 67562.4058 .3494099 67928.9289 366.5231 2.026494 18968034 0.000000 : qR1(14) : Excited a3Pv=13 15 14 F1e - X Xv=0 14 14 F1e
13C18O Excited 15 e 3 X 14 e 1 67617.6000 .5286565 67984.1232 366.5231 3.066081 28768988 0.000000 : rR21(14) : Excited a3Pv=13 15 15 F2e - X Xv=0 14 14 F1e
13C18O Excited 15 e 4 X 14 e 1 67667.1477 .7098740 68033.6708 366.5231 4.117099 38715661 0.000000 : qR1(14) : Excited ev=4 15 14 F1e - X Xv=0 14 14 F1e
13C18O Excited 15 e 5 X 14 e 1 67672.2396 .0526059 68038.7627 366.5231 .3051015 2869708 0.000000 : sR31(14) : Excited a3Pv=13 15 16 F3e - X Xv=0 14 14 F1e
13C18O Excited 15 e 6 X 14 e 1 67735.4590 .3744748 68101.9822 366.5231 2.171864 20485311 0.000000 : sR31(14) : Excited ev=4 15 16 F3e - X Xv=0 14 14 F1e
13C18O Excited 15 e 7 X 14 e 1 67788.9185 .0135631 68155.4417 366.5231 .0786625 743713.4 0.000000 : qR1(14) : Excited dv=7 15 14 F1e - X Xv=0 14 14 F1e
13C18O Excited 15 e 8 X 14 e 1 67839.4146 .1858889 68205.9378 366.5231 1.078111 10215777 0.000000 : rR21(14) : Excited dv=7 15 15 F2e - X Xv=0 14 14 F1e
13C18O Excited 15 e 9 X 14 e 1 67872.7757 .9260878 68239.2988 366.5231 5.371087 50969546 0.000000 : rR21(14) : Excited a'v=12 15 15 F2e - X Xv=0 14 14 F1e
13C18O Excited 15 e 10 X 14 e 1 67890.4790 .6911681 68257.0021 366.5231 4.008609 38069930 0.000000 : sR31(14) : Excited dv=7 15 16 F3e - X Xv=0 14 14 F1e
13C18O Excited 14 e 1 X 15 e 1 67436.0633 .5289358 67854.9105 418.8472 3.942720 39228161 0.000000 : pP(15) : Excited Av=2 14 14 F1e - X Xv=0 15 15 F1e
13C18O Excited 14 e 2 X 15 e 1 67472.8949 .2172003 67891.7422 418.8472 1.619025 16134924 0.000000 : oP1(15) : Excited a3Pv=13 14 13 F1e - X Xv=0 15 15 F1e
13C18O Excited 14 e 3 X 15 e 1 67526.2526 .3846709 67945.0998 418.8472 2.867360 28643472 0.000000 : pP21(15) : Excited a3Pv=13 14 14 F2e - X Xv=0 15 15 F1e
13C18O Excited 14 e 4 X 15 e 1 67577.2443 .3461521 67996.0915 418.8472 2.580238 25833706 0.000000 : qP31(15) : Excited a3Pv=13 14 15 F3e - X Xv=0 15 15 F1e
13C18O Excited 14 e 5 X 15 e 1 67585.6296 .1576632 68004.4768 418.8472 1.175231 11770954 0.000000 : oP1(15) : Excited ev=4 14 13 F1e - X Xv=0 15 15 F1e
13C18O Excited 14 e 6 X 15 e 1 67648.0543 .2585949 68066.9016 418.8472 1.927582 19359952 0.000000 : qP31(15) : Excited ev=4 14 15 F3e - X Xv=0 15 15 F1e
13C18O Excited 14 e 7 X 15 e 1 67705.6487 .0076352 68124.4959 418.8472 .0569136 573080.8 0.000000 : oP1(15) : Excited dv=7 14 13 F1e - X Xv=0 15 15 F1e
13C18O Excited 14 e 8 X 15 e 1 67754.7496 .1170771 68173.5968 418.8472 .8727001 8806629 0.000000 : pP21(15) : Excited dv=7 14 14 F2e - X Xv=0 15 15 F1e
13C18O Excited 14 e 9 X 15 e 1 67789.3924 .6292888 68208.2397 418.8472 4.690757 47408214 0.000000 : pP21(15) : Excited a'v=12 14 14 F2e - X Xv=0 15 15 F1e
13C18O Excited 14 e 10 X 15 e 1 67804.4394 .4830664 68223.2866 418.8472 3.600806 36416616 0.000000 : qP31(15) : Excited dv=7 14 15 F3e - X Xv=0 15 15 F1e
13C18O Excited 15 f 2 X 15 e 1 67478.5235 1.172313 67897.3707 418.8472 8.738493 81488251 0.000000 : qQ(15) : Excited Av=2 15 15 F1f - X Xv=0 15 15 F1e
13C18O Excited 15 f 3 X 15 e 1 67513.0275 .4882820 67931.8747 418.8472 3.639684 33992895 0.000000 : pQ1(15) : Excited a3Pv=13 15 14 F1f - X Xv=0 15 15 F1e
13C18O Excited 15 f 4 X 15 e 1 67565.9134 .8297860 67984.7606 418.8472 6.185275 57903353 0.000000 : qQ21(15) : Excited a3Pv=13 15 15 F2f - X Xv=0 15 15 F1e
13C18O Excited 15 f 5 X 15 e 1 67618.4592 .7042670 68037.3064 418.8472 5.249649 49259251 0.000000 : rQ31(15) : Excited a3Pv=13 15 16 F3f - X Xv=0 15 15 F1e
13C18O Excited 15 f 6 X 15 e 1 67649.2311 .9977702 68068.0784 418.8472 7.437440 69883363 0.000000 : qQ21(15) : Excited ev=4 15 15 F2f - X Xv=0 15 15 F1e
13C18O Excited 15 f 7 X 15 e 1 67736.5944 .0203872 68155.4416 418.8472 .1519671 1433447 0.000000 : pQ1(15) : Excited dv=7 15 14 F1f - X Xv=0 15 15 F1e
13C18O Excited 15 f 8 X 15 e 1 67787.0901 .2703156 68205.9373 418.8472 2.014949 19048763 0.000000 : qQ21(15) : Excited dv=7 15 15 F2f - X Xv=0 15 15 F1e
13C18O Excited 15 f 9 X 15 e 1 67790.3927 .7589321 68209.2399 418.8472 5.657126 53488704 0.000000 : pQ1(15) : Excited a'v=12 15 14 F1f - X Xv=0 15 15 F1e
13C18O Excited 15 f 10 X 15 e 1 67838.1546 1.029691 68257.0018 418.8472 7.675381 72725006 0.000000 : rQ31(15) : Excited dv=7 15 16 F3f - X Xv=0 15 15 F1e
13C18O Excited 15 f 11 X 15 e 1 67854.8372 .6595957 68273.6844 418.8472 4.916667 46620287 0.000000 : rQ31(15) : Excited a'v=12 15 16 F3f - X Xv=0 15 15 F1e
13C18O Excited 16 e 1 X 15 e 1 67523.7099 .6212894 67942.5572 418.8472 4.631129 40650443 0.000000 : rR(15) : Excited Av=2 16 16 F1e - X Xv=0 15 15 F1e
13C18O Excited 16 e 2 X 15 e 1 67549.7893 .3162497 67968.6366 418.8472 2.357344 20715934 0.000000 : qR1(15) : Excited a3Pv=13 16 15 F1e - X Xv=0 15 15 F1e
13C18O Excited 16 e 3 X 15 e 1 67606.8985 .4087021 68025.7457 418.8472 3.046490 26839989 0.000000 : rR21(15) : Excited a3Pv=13 16 16 F2e - X Xv=0 15 15 F1e
13C18O Excited 16 e 4 X 15 e 1 67648.5771 .4365777 68067.4243 418.8472 3.254276 28723671 0.000000 : qR1(15) : Excited ev=4 16 15 F1e - X Xv=0 15 15 F1e
13C18O Excited 16 e 5 X 15 e 1 67662.8697 .2130680 68081.7169 418.8472 1.588222 14027229 0.000000 : sR31(15) : Excited a3Pv=13 16 17 F3e - X Xv=0 15 15 F1e
13C18O Excited 16 e 6 X 15 e 1 67720.4079 .3035939 68139.2551 418.8472 2.263008 20037993 0.000000 : sR31(15) : Excited ev=4 16 17 F3e - X Xv=0 15 15 F1e
13C18O Excited 16 e 7 X 15 e 1 67769.6387 .0133415 68188.4859 418.8472 .0994486 882498.0 0.000000 : qR1(15) : Excited dv=7 16 15 F1e - X Xv=0 15 15 F1e
13C18O Excited 16 e 8 X 15 e 1 67821.5765 .1645751 68240.4237 418.8472 1.226753 10911144 0.000000 : rR21(15) : Excited dv=7 16 16 F2e - X Xv=0 15 15 F1e
13C18O Excited 16 e 9 X 15 e 1 67853.5711 .7668713 68272.4183 418.8472 5.716305 50914681 0.000000 : rR21(15) : Excited a'v=12 16 16 F2e - X Xv=0 15 15 F1e
13C18O Excited 16 e 10 X 15 e 1 67874.0668 .5567913 68292.9141 418.8472 4.150356 37000403 0.000000 : sR31(15) : Excited dv=7 16 17 F3e - X Xv=0 15 15 F1e
13C18O Excited 15 e 1 X 16 e 1 67422.6794 .4276432 67897.3290 474.6496 4.165835 38750905 0.000000 : pP(16) : Excited Av=2 15 15 F1e - X Xv=0 16 16 F1e
13C18O Excited 15 e 2 X 16 e 1 67454.2793 .1950276 67928.9289 474.6496 1.899838 17697292 0.000000 : oP1(16) : Excited a3Pv=13 15 14 F1e - X Xv=0 16 16 F1e
13C18O Excited 15 e 3 X 16 e 1 67509.4736 .2950763 67984.1232 474.6496 2.874451 26841746 0.000000 : pP21(16) : Excited a3Pv=13 15 15 F2e - X Xv=0 16 16 F1e
13C18O Excited 15 e 4 X 16 e 1 67559.0212 .3962252 68033.6708 474.6496 3.859780 36122217 0.000000 : oP1(16) : Excited ev=4 15 14 F1e - X Xv=0 16 16 F1e
13C18O Excited 15 e 5 X 16 e 1 67564.1131 .0293626 68038.7627 474.6496 .2860326 2677476 0.000000 : qP31(16) : Excited a3Pv=13 15 16 F3e - X Xv=0 16 16 F1e
13C18O Excited 15 e 6 X 16 e 1 67627.3326 .2090178 68101.9822 474.6496 2.036122 19113155 0.000000 : qP31(16) : Excited ev=4 15 16 F3e - X Xv=0 16 16 F1e
13C18O Excited 15 e 7 X 16 e 1 67680.7921 .0075704 68155.4417 474.6496 .0737461 693900.2 0.000000 : oP1(16) : Excited dv=7 15 14 F1e - X Xv=0 16 16 F1e
13C18O Excited 15 e 8 X 16 e 1 67731.2882 .1037562 68205.9378 474.6496 1.010729 9531570 0.000000 : pP21(16) : Excited dv=7 15 15 F2e - X Xv=0 16 16 F1e
13C18O Excited 15 e 9 X 16 e 1 67764.6492 .5169077 68239.2988 474.6496 5.035394 47555943 0.000000 : pP21(16) : Excited a'v=12 15 15 F2e - X Xv=0 16 16 F1e
13C18O Excited 15 e 10 X 16 e 1 67782.3525 .3857842 68257.0021 474.6496 3.758071 35520302 0.000000 : qP31(16) : Excited dv=7 15 16 F3e - X Xv=0 16 16 F1e
13C18O Excited 16 f 2 X 16 e 1 67467.9570 .9432934 67942.6066 474.6496 9.188980 80458045 0.000000 : qQ(16) : Excited Av=2 16 16 F1f - X Xv=0 16 16 F1e
13C18O Excited 16 f 3 X 16 e 1 67496.9547 .4245278 67971.6043 474.6496 4.135487 36256734 0.000000 : pQ1(16) : Excited a3Pv=13 16 15 F1f - X Xv=0 16 16 F1e
13C18O Excited 16 f 4 X 16 e 1 67551.7498 .6463817 68026.3994 474.6496 6.296650 55338684 0.000000 : qQ21(16) : Excited a3Pv=13 16 16 F2f - X Xv=0 16 16 F1e
13C18O Excited 16 f 5 X 16 e 1 67606.0674 .6881780 68080.7169 474.6496 6.703804 59059233 0.000000 : rQ31(16) : Excited a3Pv=13 16 17 F3f - X Xv=0 16 16 F1e
13C18O Excited 16 f 6 X 16 e 1 67628.8820 .7129879 68103.5316 474.6496 6.945487 61250382 0.000000 : qQ21(16) : Excited ev=4 16 16 F2f - X Xv=0 16 16 F1e
13C18O Excited 16 f 7 X 16 e 1 67713.8363 .0197467 68188.4859 474.6496 .1923597 1702769 0.000000 : pQ1(16) : Excited dv=7 16 15 F1f - X Xv=0 16 16 F1e
13C18O Excited 16 f 8 X 16 e 1 67765.6333 .3549501 68240.2829 474.6496 3.457704 30677912 0.000000 : pQ1(16) : Excited a'v=12 16 15 F1f - X Xv=0 16 16 F1e
13C18O Excited 16 f 9 X 16 e 1 67765.7879 .4980536 68240.4375 474.6496 4.851729 43046470 0.000000 : qQ21(16) : Excited dv=7 16 16 F2f - X Xv=0 16 16 F1e
13C18O Excited 16 f 10 X 16 e 1 67818.2642 .8178808 68292.9137 474.6496 7.967288 70853293 0.000000 : rQ31(16) : Excited dv=7 16 17 F3f - X Xv=0 16 16 F1e
13C18O Excited 16 f 11 X 16 e 1 67834.2134 .5400080 68308.8630 474.6496 5.260423 46814088 0.000000 : rQ31(16) : Excited a'v=12 16 17 F3f - X Xv=0 16 16 F1e
13C18O Excited 17 e 1 X 16 e 1 67515.9389 .4889251 67990.5885 474.6496 4.762805 39403715 0.000000 : rR(16) : Excited Av=2 17 17 F1e - X Xv=0 16 16 F1e
13C18O Excited 17 e 2 X 16 e 1 67536.2150 .2823597 68010.8646 474.6496 2.750573 22776591 0.000000 : qR1(16) : Excited a3Pv=13 17 16 F1e - X Xv=0 16 16 F1e
13C18O Excited 17 e 3 X 16 e 1 67595.3127 .3088507 68069.9623 474.6496 3.008632 24978953 0.000000 : rR21(16) : Excited a3Pv=13 17 17 F2e - X Xv=0 16 16 F1e
13C18O Excited 17 e 4 X 16 e 1 67628.1266 .3232776 68102.7761 474.6496 3.149170 26183856 0.000000 : qR1(16) : Excited ev=4 17 16 F1e - X Xv=0 16 16 F1e
13C18O Excited 17 e 5 X 16 e 1 67653.2592 .2205326 68127.9087 474.6496 2.148291 17881945 0.000000 : sR31(16) : Excited a3Pv=13 17 18 F3e - X Xv=0 16 16 F1e
13C18O Excited 17 e 6 X 16 e 1 67704.0748 .2393234 68178.7244 474.6496 2.331340 19449368 0.000000 : sR31(16) : Excited ev=4 17 18 F3e - X Xv=0 16 16 F1e
13C18O Excited 17 e 7 X 16 e 1 67748.9826 .0127008 68223.6322 474.6496 .1237231 1034224 0.000000 : qR1(16) : Excited dv=7 17 16 F1e - X Xv=0 16 16 F1e
13C18O Excited 17 e 8 X 16 e 1 67802.4031 .1416490 68277.0527 474.6496 1.379856 11561775 0.000000 : rR21(16) : Excited dv=7 17 17 F2e - X Xv=0 16 16 F1e
13C18O Excited 17 e 9 X 16 e 1 67832.9467 .6222813 68307.5963 474.6496 6.061879 50860978 0.000000 : rR21(16) : Excited a'v=12 17 17 F2e - X Xv=0 16 16 F1e
13C18O Excited 17 e 10 X 16 e 1 67856.3667 .4397455 68331.0163 474.6496 4.283728 35978997 0.000000 : sR31(16) : Excited dv=7 17 18 F3e - X Xv=0 16 16 F1e

| | | | | | | | | | | |
|---|---|---|---|---|---|---|---|---|---|---|
| 13C18O Excited | 16 e 1 | X 17 e 1 | 67408.6289 | .3367203 | 67942.5572 | 533.9283 | 4.358710 | 38063957 | 0.000000 : | pP(17) : Excited Av=2 16 16 F1e - X Xv=0 17 17 F1e |
| 13C18O Excited | 16 e 2 | X 17 e 1 | 67434.7083 | .1713979 | 67968.6366 | 533.9283 | 2.218677 | 19397869 | 0.000000 : | oP1(17) : Excited a3Pv=13 16 15 F1e - X Xv=0 17 17 F1e |
| 13C18O Excited | 16 e 3 | X 17 e 1 | 67491.8175 | .2215043 | 68025.7457 | 533.9283 | 2.867285 | 25132386 | 0.000000 : | pP21(17) : Excited a3Pv=13 16 16 F2e - X Xv=0 17 17 F1e |
| 13C18O Excited | 16 e 4 | X 17 e 1 | 67533.4960 | .2366120 | 68067.4243 | 533.9283 | 3.062848 | 26896310 | 0.000000 : | oP1(17) : Excited ev=4 16 15 F1e - X Xv=0 17 17 F1e |
| 13C18O Excited | 16 e 5 | X 17 e 1 | 67547.7887 | .1154765 | 68081.7169 | 533.9283 | 1.494797 | 13134850 | 0.000000 : | qP31(17) : Excited a3Pv=13 16 17 F3e - X Xv=0 17 17 F1e |
| 13C18O Excited | 16 e 6 | X 17 e 1 | 67605.3268 | .1645388 | 68139.2551 | 533.9283 | 2.129890 | 18763305 | 0.000000 : | qP31(17) : Excited ev=4 16 17 F3e - X Xv=0 17 17 F1e |
| 13C18O Excited | 16 e 7 | X 17 e 1 | 67654.5576 | .0072307 | 68188.4859 | 533.9283 | .0935987 | 826362.3 | 0.000000 : | oP1(17) : Excited dv=7 16 15 F1e - X Xv=0 17 17 F1e |
| 13C18O Excited | 16 e 8 | X 17 e 1 | 67706.4954 | .0891948 | 68240.4237 | 533.9283 | 1.154591 | 10217125 | 0.000000 : | pP21(17) : Excited dv=7 16 16 F2e - X Xv=0 17 17 F1e |
| 13C18O Excited | 16 e 9 | X 17 e 1 | 67738.4900 | .4156213 | 68272.4183 | 533.9283 | 5.380052 | 47676295 | 0.000000 : | pP21(17) : Excited a'v=12 16 16 F2e - X Xv=0 17 17 F1e |
| 13C18O Excited | 16 e 10 | X 17 e 1 | 67758.9858 | .3017642 | 68292.9141 | 533.9283 | 3.906217 | 34647076 | 0.000000 : | qP31(17) : Excited dv=7 16 17 F3e - X Xv=0 17 17 F1e |
| 13C18O Excited | 17 f 2 | X 17 e 1 | 67456.7236 | .7408527 | 67990.6518 | 533.9283 | 9.590044 | 79131921 | 0.000000 : | qQ(17) : Excited Av=2 17 17 F1f - X Xv=0 17 17 F1e |
| 13C18O Excited | 17 f 3 | X 17 e 1 | 67479.9548 | .3608260 | 68013.8831 | 533.9283 | 4.670749 | 38580361 | 0.000000 : | pQ1(17) : Excited a3Pv=13 17 16 F1f - X Xv=0 17 17 F1e |
| 13C18O Excited | 17 f 4 | X 17 e 1 | 67536.6954 | .4973243 | 68070.6237 | 533.9283 | 6.437666 | 53309332 | 0.000000 : | qQ21(17) : Excited a3Pv=13 17 17 F2f - X Xv=0 17 17 F1e |
| 13C18O Excited | 17 f 5 | X 17 e 1 | 67592.3899 | .7210742 | 68126.3182 | 533.9283 | 9.334019 | 77484970 | 0.000000 : | rQ31(17) : Excited a3Pv=13 17 18 F3f - X Xv=0 17 17 F1e |
| 13C18O Excited | 17 f 6 | X 17 e 1 | 67607.6973 | .4061760 | 68141.6256 | 533.9283 | 5.257787 | 43676394 | 0.000000 : | qQ21(17) : Excited ev=4 17 17 F2f - X Xv=0 17 17 F1e |
| 13C18O Excited | 17 f 7 | X 17 e 1 | 67689.7039 | .0185046 | 68223.6322 | 533.9283 | .2395352 | 1997067 | 0.000000 : | pQ1(17) : Excited dv=7 17 16 F1f - X Xv=0 17 17 F1e |
| 13C18O Excited | 17 f 8 | X 17 e 1 | 67739.4850 | .4751872 | 68273.4133 | 533.9283 | 6.151109 | 51396613 | 0.000000 : | pQ1(17) : Excited a'v=12 17 16 F1f - X Xv=0 17 17 F1e |
| 13C18O Excited | 17 f 9 | X 17 e 1 | 67743.1251 | .2168062 | 68277.0534 | 533.9283 | 2.806470 | 23453708 | 0.000000 : | qQ21(17) : Excited dv=7 17 17 F2f - X Xv=0 17 17 F1e |
| 13C18O Excited | 17 f 10 | X 17 e 1 | 67797.0877 | .6365774 | 68331.0160 | 533.9283 | 8.240242 | 69028504 | 0.000000 : | rQ31(17) : Excited dv=7 17 18 F3f - X Xv=0 17 17 F1e |
| 13C18O Excited | 17 f 11 | X 17 e 1 | 67812.1698 | .4330432 | 68346.0981 | 533.9283 | 5.605572 | 46989222 | 0.000000 : | rQ31(17) : Excited a'v=12 17 18 F3f - X Xv=0 17 17 F1e |
| 13C18O Excited | 18 e 1 | X 17 e 1 | 67507.4823 | .3687468 | 68041.4106 | 533.9283 | 4.773281 | 37341729 | 0.000000 : | rR(17) : Excited Av=2 18 18 F1e - X Xv=0 17 17 F1e |
| 13C18O Excited | 18 e 2 | X 17 e 1 | 67521.6853 | .2526573 | 68055.6136 | 533.9283 | 3.270548 | 25601890 | 0.000000 : | qR1(17) : Excited a3Pv=13 18 17 F1e - X Xv=0 17 17 F1e |
| 13C18O Excited | 18 e 3 | X 17 e 1 | 67582.8361 | .2266739 | 68116.7644 | 533.9283 | 2.934203 | 23031441 | 0.000000 : | rR21(17) : Excited a3Pv=13 18 18 F2e - X Xv=0 17 17 F1e |
| 13C18O Excited | 18 e 4 | X 17 e 1 | 67606.3488 | .2491568 | 68140.2771 | 533.9283 | 3.225236 | 25342277 | 0.000000 : | qR1(17) : Excited ev=4 18 17 F1e - X Xv=0 17 17 F1e |
| 13C18O Excited | 18 e 5 | X 17 e 1 | 67642.8544 | .2003500 | 68176.7827 | 533.9283 | 2.593451 | 20411061 | 0.000000 : | sR31(17) : Excited a3Pv=13 18 19 F3e - X Xv=0 17 17 F1e |
| 13C18O Excited | 18 e 6 | X 17 e 1 | 67686.4700 | .1826340 | 68220.3983 | 533.9283 | 2.364125 | 18642225 | 0.000000 : | sR31(17) : Excited ev=4 18 19 F3e - X Xv=0 17 17 F1e |
| 13C18O Excited | 18 e 7 | X 17 e 1 | 67726.9561 | .0117515 | 68260.8844 | 533.9283 | .1521189 | 1201682 | 0.000000 : | qR1(17) : Excited dv=7 18 17 F1e - X Xv=0 17 17 F1e |
| 13C18O Excited | 18 e 8 | X 17 e 1 | 67781.8947 | .1186854 | 68315.8230 | 533.9283 | 1.536335 | 12166024 | 0.000000 : | rR21(17) : Excited dv=7 18 18 F2e - X Xv=0 17 17 F1e |
| 13C18O Excited | 18 e 9 | X 17 e 1 | 67810.9024 | .4950467 | 68344.8306 | 533.9283 | 6.408183 | 50810698 | 0.000000 : | rR21(17) : Excited a'v=12 18 18 F2e - X Xv=0 17 17 F1e |
| 13C18O Excited | 18 e 10 | X 17 e 1 | 67837.3746 | .3406196 | 68371.3029 | 533.9283 | 4.409185 | 35001541 | 0.000000 : | sR31(17) : Excited dv=7 18 19 F3e - X Xv=0 17 17 F1e |
| 13C18O Excited | 17 e 1 | X 18 e 1 | 67393.9073 | .2571852 | 67990.5885 | 596.6812 | 4.498205 | 37013193 | 0.000000 : | pP(18) : Excited Av=2 17 17 F1e - X Xv=0 18 18 F1e |
| 13C18O Excited | 17 e 2 | X 18 e 1 | 67414.1834 | .1485273 | 68010.8646 | 596.6812 | 2.597763 | 21394829 | 0.000000 : | oP1(18) : Excited a3Pv=13 17 16 F1e - X Xv=0 18 18 F1e |
| 13C18O Excited | 17 e 3 | X 18 e 1 | 67473.2810 | .1624622 | 68069.9623 | 596.6812 | 2.841486 | 23463694 | 0.000000 : | pP21(18) : Excited a3Pv=13 17 17 F2e - X Xv=0 18 18 F1e |
| 13C18O Excited | 17 e 4 | X 18 e 1 | 67506.0949 | .1700510 | 68102.7761 | 596.6812 | 2.974216 | 24595571 | 0.000000 : | oP1(18) : Excited ev=4 17 16 F1e - X Xv=0 18 18 F1e |
| 13C18O Excited | 17 e 5 | X 18 e 1 | 67531.2275 | .1160049 | 68127.9087 | 596.6812 | 2.028942 | 16797279 | 0.000000 : | qP31(18) : Excited a3Pv=13 17 18 F3e - X Xv=0 18 18 F1e |
| 13C18O Excited | 17 e 6 | X 18 e 1 | 67582.0431 | .1258893 | 68178.7244 | 596.6812 | 2.201821 | 18269701 | 0.000000 : | qP31(18) : Excited ev=4 17 18 F3e - X Xv=0 18 18 F1e |
| 13C18O Excited | 17 e 7 | X 18 e 1 | 67626.9510 | .0066809 | 68223.6322 | 596.6812 | .1168496 | 971498.3 | 0.000000 : | oP1(18) : Excited dv=7 17 16 F1e - X Xv=0 18 18 F1e |
| 13C18O Excited | 17 e 8 | X 18 e 1 | 67680.3715 | .0745104 | 68277.0527 | 596.6812 | 1.303198 | 10860602 | 0.000000 : | pP21(18) : Excited dv=7 17 17 F2e - X Xv=0 18 18 F1e |
| 13C18O Excited | 17 e 9 | X 18 e 1 | 67710.9150 | .3273335 | 68307.5963 | 596.6812 | 5.725108 | 47776587 | 0.000000 : | pP21(18) : Excited a'v=12 17 17 F2e - X Xv=0 18 18 F1e |
| 13C18O Excited | 17 e 10 | X 18 e 1 | 67734.3350 | .2313156 | 68331.0163 | 596.6812 | 4.045743 | 33797166 | 0.000000 : | qP31(18) : Excited dv=7 17 18 F3e - X Xv=0 18 18 F1e |
| 13C18O Excited | 18 f 2 | X 18 e 1 | 67444.8202 | .5665018 | 68041.5015 | 596.6812 | 9.908195 | 77296899 | 0.000000 : | qQ(18) : Excited Av=2 18 18 F1f - X Xv=0 18 18 F1e |
| 13C18O Excited | 18 f 3 | X 18 e 1 | 67462.0328 | .3013791 | 68058.7140 | 596.6812 | 5.271162 | 41153460 | 0.000000 : | pQ1(18) : Excited a3Pv=13 18 17 F1f - X Xv=0 18 18 F1e |
| 13C18O Excited | 18 f 4 | X 18 e 1 | 67520.7438 | .3791236 | 68117.4251 | 596.6812 | 6.630925 | 51904795 | 0.000000 : | qQ21(18) : Excited a3Pv=13 18 18 F2f - X Xv=0 18 18 F1e |
| 13C18O Excited | 18 f 5 | X 18 e 1 | 67576.2494 | .8083246 | 68172.9307 | 596.6812 | 14.13771 | 1.1094e8 | 0.000000 : | rQ31(18) : Excited a3Pv=13 18 19 F3f - X Xv=0 18 18 F1e |
| 13C18O Excited | 18 f 6 | X 18 e 1 | 67586.8546 | .0778323 | 68183.5358 | 596.6812 | 1.361298 | 10687145 | 0.000000 : | qQ21(18) : Excited ev=4 18 18 F2f - X Xv=0 18 18 F1e |
| 13C18O Excited | 18 f 7 | X 18 e 1 | 67664.2031 | .0168512 | 68260.8843 | 596.6812 | .2947305 | 2321794 | 0.000000 : | pQ1(18) : Excited dv=7 18 17 F1f - X Xv=0 18 18 F1e |
| 13C18O Excited | 18 f 8 | X 18 e 1 | 67711.9078 | .3739523 | 68308.5890 | 596.6812 | 6.540478 | 51632879 | 0.000000 : | pQ1(18) : Excited a'v=12 18 17 F1f - X Xv=0 18 18 F1e |
| 13C18O Excited | 18 f 9 | X 18 e 1 | 67719.1422 | .1757743 | 68315.8234 | 596.6812 | 3.074318 | 24277546 | 0.000000 : | qQ21(18) : Excited dv=7 18 18 F2f - X Xv=0 18 18 F1e |
| 13C18O Excited | 18 f 10 | X 18 e 1 | 67774.6214 | .4857120 | 68371.3026 | 596.6812 | 8.495171 | 67250438 | 0.000000 : | rQ31(18) : Excited dv=7 18 19 F3f - X Xv=0 18 18 F1e |
| 13C18O Excited | 18 f 11 | X 18 e 1 | 67788.7063 | .3403346 | 68385.3875 | 596.6812 | 5.952500 | 47151243 | 0.000000 : | rQ31(18) : Excited a'v=12 18 19 F3f - X Xv=0 18 18 F1e |
| 13C18O Excited | 19 e 1 | X 18 e 1 | 67498.3030 | .2510096 | 68094.9843 | 596.6812 | 4.390192 | 32570239 | 0.000000 : | rR(18) : Excited Av=2 19 19 F1e - X Xv=0 18 18 F1e |
| 13C18O Excited | 19 e 2 | X 18 e 1 | 67506.2223 | .2394526 | 68102.9036 | 596.6812 | 4.188060 | 31081586 | 0.000000 : | qR1(18) : Excited a3Pv=13 19 18 F1e - X Xv=0 18 18 F1e |
| 13C18O Excited | 19 e 3 | X 18 e 1 | 67569.4509 | .1568482 | 68166.1322 | 596.6812 | 2.743296 | 20416568 | 0.000000 : | rR21(18) : Excited a3Pv=13 19 19 F2e - X Xv=0 18 18 F1e |
| 13C18O Excited | 19 e 4 | X 18 e 1 | 67583.3318 | .1980062 | 68180.0130 | 596.6812 | 3.463155 | 25789897 | 0.000000 : | qR1(18) : Excited ev=4 19 18 F1e - X Xv=0 18 18 F1e |
| 13C18O Excited | 19 e 5 | X 18 e 1 | 67631.5744 | .1742502 | 68228.2557 | 596.6812 | 3.047661 | 22744374 | 0.000000 : | sR31(18) : Excited a3Pv=13 19 20 F3e - X Xv=0 18 18 F1e |
| 13C18O Excited | 19 e 6 | X 18 e 1 | 67667.6132 | .1336390 | 68264.2945 | 596.6812 | 2.337365 | 17471413 | 0.000000 : | sR31(18) : Excited ev=4 19 20 F3e - X Xv=0 18 18 F1e |
| 13C18O Excited | 19 e 7 | X 18 e 1 | 67703.5655 | .0106231 | 68300.2468 | 596.6812 | .1858002 | 1391040 | 0.000000 : | qR1(18) : Excited dv=7 19 18 F1e - X Xv=0 18 18 F1e |
| 13C18O Excited | 19 e 8 | X 18 e 1 | 67760.0512 | .0969170 | 68356.7324 | 596.6812 | 1.695092 | 12722527 | 0.000000 : | rR21(18) : Excited dv=7 19 19 F2e - X Xv=0 18 18 F1e |
| 13C18O Excited | 19 e 9 | X 18 e 1 | 67787.4379 | .3862518 | 68384.1192 | 596.6812 | 6.755598 | 50765690 | 0.000000 : | rR21(18) : Excited a'v=12 19 19 F2e - X Xv=0 18 18 F1e |
| 13C18O Excited | 19 e 10 | X 18 e 1 | 67817.0871 | .2588378 | 68413.7683 | 596.6812 | 4.527110 | 34064125 | 0.000000 : | sR31(18) : Excited dv=7 19 20 F3e - X Xv=0 18 18 F1e |
| 13C18O Excited | 18 e 1 | X 19 e 1 | 67378.5043 | .1881942 | 68041.4106 | 662.9063 | 4.522055 | 35173995 | 0.000000 : | pP(19) : Excited Av=2 18 18 F1e - X Xv=0 19 19 F1e |
| 13C18O Excited | 18 e 2 | X 19 e 1 | 67392.7073 | .1289466 | 68055.6136 | 662.9063 | 3.098414 | 24115697 | 0.000000 : | oP1(19) : Excited a3Pv=13 18 17 F1e - X Xv=0 19 19 F1e |
| 13C18O Excited | 18 e 3 | X 19 e 1 | 67453.8581 | .1156856 | 68116.7644 | 662.9063 | 2.779771 | 21694576 | 0.000000 : | pP21(19) : Excited a3Pv=13 18 18 F2e - X Xv=0 19 19 F1e |
| 13C18O Excited | 18 e 4 | X 19 e 1 | 67477.3708 | .1271601 | 68140.2771 | 662.9063 | 3.055487 | 23871327 | 0.000000 : | oP1(19) : Excited ev=4 18 17 F1e - X Xv=0 19 19 F1e |
| 13C18O Excited | 18 e 5 | X 19 e 1 | 67513.8764 | .1022509 | 68176.7827 | 662.9063 | 2.456954 | 19226394 | 0.000000 : | qP31(19) : Excited a3Pv=13 18 19 F3e - X Xv=0 19 19 F1e |
| 13C18O Excited | 18 e 6 | X 19 e 1 | 67557.4920 | .0932094 | 68220.3983 | 662.9063 | 2.239697 | 17560287 | 0.000000 : | qP31(19) : Excited ev=4 18 19 F3e - X Xv=0 19 19 F1e |
| 13C18O Excited | 18 e 7 | X 19 e 1 | 67597.9781 | .0059975 | 68260.8844 | 662.9063 | .1441127 | 1131944 | 0.000000 : | oP1(19) : Excited dv=7 18 17 F1e - X Xv=0 19 19 F1e |
| 13C18O Excited | 18 e 8 | X 19 e 1 | 67652.9167 | .0605724 | 68315.8230 | 662.9063 | 1.455475 | 11460037 | 0.000000 : | pP21(19) : Excited dv=7 18 18 F2e - X Xv=0 19 19 F1e |
| 13C18O Excited | 18 e 9 | X 19 e 1 | 67681.9243 | .2526528 | 68344.8306 | 662.9063 | 6.070910 | 47862302 | 0.000000 : | pP21(19) : Excited a'v=12 18 18 F2e - X Xv=0 19 19 F1e |
| 13C18O Excited | 18 e 10 | X 19 e 1 | 67708.3966 | .1738392 | 68371.3029 | 662.9063 | 4.177123 | 32970578 | 0.000000 : | qP31(19) : Excited dv=7 18 19 F3e - X Xv=0 19 19 F1e |
| 13C18O Excited | 19 f 2 | X 19 e 1 | 67432.2419 | .4180454 | 68095.1482 | 662.9063 | 10.04507 | 74304429 | 0.000000 : | qQ(19) : Excited Av=2 19 19 F1f - X Xv=0 19 19 F1e |
| 13C18O Excited | 19 f 3 | X 19 e 1 | 67443.1944 | .2507217 | 68106.1007 | 662.9063 | 6.024508 | 44585616 | 0.000000 : | pQ1(19) : Excited a3Pv=13 19 18 F1f - X Xv=0 19 19 F1e |
| 13C18O Excited | 19 f 4 | X 19 e 1 | 67503.8849 | .2876460 | 68166.7912 | 662.9063 | 6.911750 | 51290055 | 0.000000 : | qQ21(19) : Excited a3Pv=13 19 19 F2f - X Xv=0 19 19 F1e |
| 13C18O Excited | 19 f 5 | X 19 e 1 | 67555.3109 | .6700120 | 68218.2172 | 662.9063 | 16.09949 | 1.1974e8 | 0.000000 : | qQ21(19) : Excited ev=4 19 19 F2f - X Xv=0 19 19 F1e |
| 13C18O Excited | 19 f 6 | X 19 e 1 | 67568.6927 | .0101894 | 68231.5990 | 662.9063 | .2448378 | 1822107 | 0.000000 : | rQ31(19) : Excited a3Pv=13 19 20 F3f - X Xv=0 19 19 F1e |
| 13C18O Excited | 19 f 7 | X 19 e 1 | 67637.3404 | .0149942 | 68300.2467 | 662.9063 | .3602912 | 2689503 | 0.000000 : | pQ1(19) : Excited dv=7 19 18 F1f - X Xv=0 19 19 F1e |
| 13C18O Excited | 19 f 8 | X 19 e 1 | 67682.9144 | .2872197 | 68345.8207 | 662.9063 | 6.901505 | 51622592 | 0.000000 : | pQ1(19) : Excited a'v=12 19 18 F1f - X Xv=0 19 19 F1e |
| 13C18O Excited | 19 f 9 | X 19 e 1 | 67693.8265 | .1405809 | 68356.7328 | 662.9063 | 3.377972 | 25279127 | 0.000000 : | qQ21(19) : Excited dv=7 19 19 F2f - X Xv=0 19 19 F1e |
| 13C18O Excited | 19 f 10 | X 19 e 1 | 67750.8617 | .3634242 | 68413.7680 | 662.9063 | 8.732600 | 65515923 | 0.000000 : | rQ31(19) : Excited dv=7 19 20 F3f - X Xv=0 19 19 F1e |
| 13C18O Excited | 19 f 11 | X 19 e 1 | 67763.8225 | .2622603 | 68426.7289 | 662.9063 | 6.301764 | 47305829 | 0.000000 : | rQ31(19) : Excited a'v=12 19 20 F3f - X Xv=0 19 19 F1e |
| 13C18O Excited | 20 e 1 | X 19 e 1 | 67487.9125 | .0375327 | 68150.8188 | 662.9063 | .9018609 | 6361464 | 0.000000 : | rR(19) : Excited Av=2 20 20 F1e - X Xv=0 19 19 F1e |
| 13C18O Excited | 20 e 2 | X 19 e 1 | 67490.2877 | .3417084 | 68153.1940 | 662.9063 | 8.210796 | 57922685 | 0.000000 : | qR1(19) : Excited a3Pv=13 20 19 F1e - X Xv=0 19 19 F1e |
| 13C18O Excited | 20 e 3 | X 19 e 1 | 67554.9933 | .0678010 | 68217.8996 | 662.9063 | 1.629168 | 11525980 | 0.000000 : | rR21(19) : Excited a3Pv=13 20 20 F2e - X Xv=0 19 19 F1e |
| 13C18O Excited | 20 e 4 | X 19 e 1 | 67559.2784 | .1934055 | 68222.1847 | 662.9063 | 4.647275 | 32884631 | 0.000000 : | qR1(19) : Excited ev=4 20 19 F1e - X Xv=0 19 19 F1e |
| 13C18O Excited | 20 e 5 | X 19 e 1 | 67619.3548 | .1492284 | 68282.2611 | 662.9063 | 3.585758 | 25440970 | 0.000000 : | sR31(19) : Excited a3Pv=13 20 21 F3e - X Xv=0 19 19 F1e |
| 13C18O Excited | 20 e 6 | X 19 e 1 | 67647.5475 | .0915870 | 68310.4538 | 662.9063 | 2.200712 | 15633601 | 0.000000 : | sR31(19) : Excited ev=4 20 21 F3e - X Xv=0 19 19 F1e |
| 13C18O Excited | 20 e 7 | X 19 e 1 | 67678.8196 | .0094533 | 68341.7259 | 662.9063 | .2271508 | 1615891 | 0.000000 : | qR1(19) : Excited dv=7 20 19 F1e - X Xv=0 19 19 F1e |
| 13C18O Excited | 20 e 8 | X 19 e 1 | 67736.8727 | .0771981 | 68399.7790 | 662.9063 | 1.854968 | 13229743 | 0.000000 : | rR21(19) : Excited dv=7 20 20 F2e - X Xv=0 19 19 F1e |
| 13C18O Excited | 20 e 9 | X 19 e 1 | 67762.5533 | .2956700 | 68425.4596 | 662.9063 | 7.104557 | 50727780 | 0.000000 : | rR21(19) : Excited a'v=12 20 20 F2e - X Xv=0 19 19 F1e |
| 13C18O Excited | 20 e 10 | X 19 e 1 | 67795.5008 | .1930091 | 68458.4071 | 662.9063 | 4.637751 | 33162682 | 0.000000 : | sR31(19) : Excited dv=7 20 21 F3e - X Xv=0 19 19 F1e |
| 13C18O Excited | 19 e 1 | X 20 e 1 | 67362.3831 | .1242551 | 68094.9843 | 732.6012 | 4.170683 | 30755183 | 0.000000 : | pP(20) : Excited Av=2 19 19 F1e - X Xv=0 20 20 F1e |
| 13C18O Excited | 19 e 2 | X 20 e 1 | 67370.3024 | .1185341 | 68102.9036 | 732.6012 | 3.978657 | 29349509 | 0.000000 : | oP1(20) : Excited a3Pv=13 19 18 F1e - X Xv=0 20 20 F1e |
| 13C18O Excited | 19 e 3 | X 20 e 1 | 67433.5310 | .0776432 | 68166.1322 | 732.6012 | 2.606132 | 19278928 | 0.000000 | : pP21(20) : Excited a3Pv=13 19 19 F2e - X Xv=0 20 20 F1e |
| 13C18O Excited | 19 e 4 | X 20 e 1 | 67447.4119 | .0980172 | 68180.0130 | 732.6012 | 3.289998 | 24352878 | 0.000000 | : oP1(20) : Excited ev=4 19 18 F1e - X Xv=0 20 20 F1e |
| 13C18O Excited | 19 e 5 | X 20 e 1 | 67495.6545 | .0862576 | 68228.2557 | 732.6012 | 2.895278 | 21477145 | 0.000000 | : qP31(20) : Excited a3Pv=13 19 20 F3e - X Xv=0 20 20 F1e |
| 13C18O Excited | 19 e 6 | X 20 e 1 | 67531.6933 | .0661542 | 68264.2945 | 732.6012 | 2.220497 | 16498026 | 0.000000 | : qP31(20) : Excited ev=4 19 20 F3e - X Xv=0 20 20 F1e |
| 13C18O Excited | 19 e 7 | X 20 e 1 | 67567.6456 | .0052587 | 68300.2468 | 732.6012 | .1765102 | 1313545 | 0.000000 | : oP1(20) : Excited dv=7 19 18 F1e - X Xv=0 20 20 F1e |
| 13C18O Excited | 19 e 8 | X 20 e 1 | 67624.1313 | .0479760 | 68356.7324 | 732.6012 | 1.610338 | 12013814 | 0.000000 | : pP21(20) : Excited dv=7 19 19 F2e - X Xv=0 20 20 F1e |
| 13C18O Excited | 19 e 9 | X 20 e 1 | 67651.5180 | .1912028 | 68384.1192 | 732.6012 | 6.417818 | 47937886 | 0.000000 | : pP21(20) : Excited a'v=12 19 19 F2e - X Xv=0 20 20 F1e |
| 13C18O Excited | 19 e 10 | X 20 e 1 | 67681.1671 | .1281302 | 68413.7683 | 732.6012 | 4.300754 | 32166734 | 0.000000 | : qP31(20) : Excited dv=7 19 20 F3e - X Xv=0 20 20 F1e |
| 13C18O Excited | 20 f 2 | X 20 e 1 | 67418.9693 | .2792962 | 68151.5704 | 732.6012 | 9.374716 | 65924070 | 0.000000 | : qQ(20) : Excited Av=2 20 20 F1f - X Xv=0 20 20 F1e |
| 13C18O Excited | 20 f 3 | X 20 e 1 | 67423.4576 | .2246707 | 68156.0588 | 732.6012 | 7.541185 | 53041067 | 0.000000 | : pQ1(20) : Excited a3Pv=13 20 19 F1f - X Xv=0 20 20 F1e |
| 13C18O Excited | 20 f 4 | X 20 e 1 | 67486.1012 | .2187307 | 68218.7023 | 732.6012 | 7.341805 | 51782790 | 0.000000 | : qQ21(20) : Excited a3Pv=13 20 20 F2f - X Xv=0 20 20 F1e |
| 13C18O Excited | 20 f 5 | X 20 e 1 | 67530.7047 | .4558742 | 68263.3059 | 732.6012 | 15.30164 | 1.0814e8 | 0.000000 | : qQ21(20) : Excited ev=4 20 20 F2f - X Xv=0 20 20 F1e |
| 13C18O Excited | 20 f 6 | X 20 e 1 | 67552.0920 | .0529049 | 68284.6931 | 732.6012 | 1.775778 | 12561593 | 0.000000 | : rQ31(20) : Excited a3Pv=13 20 21 F3f - X Xv=0 20 20 F1e |
| 13C18O Excited | 20 f 7 | X 20 e 1 | 67609.1247 | .0131453 | 68341.7258 | 732.6012 | .4412276 | 3129091 | 0.000000 | : pQ1(20) : Excited dv=7 20 19 F1f - X Xv=0 20 20 F1e |
| 13C18O Excited | 20 f 8 | X 20 e 1 | 67652.5052 | .2162019 | 68385.1063 | 732.6012 | 7.256923 | 51563661 | 0.000000 | : pQ1(20) : Excited a'v=12 20 19 F1f - X Xv=0 20 20 F1e |
| 13C18O Excited | 20 f 9 | X 20 e 1 | 67667.1782 | .1100282 | 68399.7793 | 732.6012 | 3.693152 | 26258562 | 0.000000 | : qQ21(20) : Excited dv=7 20 20 F2f - X Xv=0 20 20 F1e |
| 13C18O Excited | 20 f 10 | X 20 e 1 | 67725.8057 | .2667166 | 68458.4069 | 732.6012 | 8.952475 | 63818304 | 0.000000 | : rQ31(20) : Excited dv=7 20 21 F3f - X Xv=0 20 20 F1e |
| 13C18O Excited | 20 f 11 | X 20 e 1 | 67737.5184 | .1982455 | 68470.1195 | 732.6012 | 6.654210 | 47459583 | 0.000000 | : rQ31(20) : Excited a'v=12 20 21 F3f - X Xv=0 20 20 F1e |
| 13C18O Excited | 21 e 1 | X 20 e 1 | 67471.7850 | .0672121 | 68204.3862 | 732.6012 | 2.256006 | 15162186 | 0.000000 | : qR1(20) : Excited a3Pv=13 21 20 F1e - X Xv=0 20 20 F1e |
| 13C18O Excited | 21 e 2 | X 20 e 1 | 67478.3580 | .2198298 | 68210.9591 | 732.6012 | 7.378695 | 49605297 | 0.000000 | : rR(20) : Excited Av=2 21 21 F1e - X Xv=0 20 20 F1e |
| 13C18O Excited | 21 e 3 | X 20 e 1 | 67533.3597 | .0617004 | 68265.9608 | 732.6012 | 2.071005 | 13956975 | 0.000000 | : qR1(20) : Excited ev=4 21 20 F1e - X Xv=0 20 20 F1e |
| 13C18O Excited | 21 e 4 | X 20 e 1 | 67540.3427 | .1281592 | 68272.9438 | 732.6012 | 4.301726 | 28999302 | 0.000000 | : rR21(20) : Excited a3Pv=13 21 21 F2e - X Xv=0 20 20 F1e |
| 13C18O Excited | 21 e 5 | X 20 e 1 | 67606.0647 | .1290315 | 68338.6659 | 732.6012 | 4.331007 | 29282006 | 0.000000 | : sR31(20) : Excited a3Pv=13 21 22 F3e - X Xv=0 20 20 F1e |
| 13C18O Excited | 21 e 6 | X 20 e 1 | 67626.3872 | .0546712 | 68358.9883 | 732.6012 | 1.835067 | 12418111 | 0.000000 | : sR31(20) : Excited ev=4 21 22 F3e - X Xv=0 20 20 F1e |
| 13C18O Excited | 21 e 7 | X 20 e 1 | 67652.7323 | .0083946 | 68385.3334 | 732.6012 | .2817695 | 1908996 | 0.000000 | : qR1(20) : Excited dv=7 21 20 F1e - X Xv=0 20 20 F1e |
| 13C18O Excited | 21 e 8 | X 20 e 1 | 67712.3594 | .0600209 | 68444.9606 | 732.6012 | 2.014631 | 13685302 | 0.000000 | : rR21(20) : Excited dv=7 21 21 F2e - X Xv=0 20 20 F1e |
| 13C18O Excited | 21 e 9 | X 20 e 1 | 67736.2484 | .2221203 | 68468.8496 | 732.6012 | 7.455578 | 50699041 | 0.000000 | : rR21(20) : Excited a'v=12 21 21 F2e - X Xv=0 20 20 F1e |
| 13C18O Excited | 21 e 10 | X 20 e 1 | 67772.6132 | .1412515 | 68505.2143 | 732.6012 | 4.741178 | 32292667 | 0.000000 | : sR31(20) : Excited dv=7 21 22 F3e - X Xv=0 20 20 F1e |
| 13C18O Excited | 20 e 1 | X 21 e 1 | 67345.0554 | .0180166 | 68150.8188 | 805.7634 | .8589151 | 6020145 | 0.000000 | : pP(21) : Excited Av=2 20 20 F1e - X Xv=0 21 21 F1e |
| 13C18O Excited | 20 e 2 | X 21 e 1 | 67347.4307 | .1640278 | 68153.1940 | 805.7634 | 7.819806 | 54814902 | 0.000000 | : oP1(21) : Excited a3Pv=13 20 19 F1e - X Xv=0 21 21 F1e |
| 13C18O Excited | 20 e 3 | X 21 e 1 | 67412.1363 | .0325460 | 68217.8996 | 805.7634 | 1.551589 | 10907632 | 0.000000 | : pP21(21) : Excited a3Pv=13 20 20 F2e - X Xv=0 21 21 F1e |
| 13C18O Excited | 20 e 4 | X 21 e 1 | 67416.4213 | .0928390 | 68222.1847 | 805.7634 | 4.425976 | 31120442 | 0.000000 | : oP1(21) : Excited ev=4 20 19 F1e - X Xv=0 21 21 F1e |
| 13C18O Excited | 20 e 5 | X 21 e 1 | 67476.4977 | .0716330 | 68282.2611 | 805.7634 | 3.415008 | 24076253 | 0.000000 | : qP31(21) : Excited a3Pv=13 20 21 F3e - X Xv=0 21 21 F1e |
| 13C18O Excited | 20 e 6 | X 21 e 1 | 67504.6904 | .0439638 | 68310.4538 | 805.7634 | 2.095917 | 14795015 | 0.000000 | : qP31(21) : Excited ev=4 20 21 F3e - X Xv=0 21 21 F1e |
| 13C18O Excited | 20 e 7 | X 21 e 1 | 67535.9625 | .0045378 | 68341.7259 | 805.7634 | .2163341 | 1529219 | 0.000000 | : oP1(21) : Excited dv=7 20 19 F1e - X Xv=0 21 21 F1e |
| 13C18O Excited | 20 e 8 | X 21 e 1 | 67594.0156 | .0370569 | 68399.7790 | 805.7634 | 1.766636 | 12520204 | 0.000000 | : pP21(21) : Excited dv=7 20 20 F2e - X Xv=0 21 21 F1e |
| 13C18O Excited | 20 e 9 | X 21 e 1 | 67619.6963 | .1419284 | 68425.4596 | 805.7634 | 6.766245 | 48007260 | 0.000000 | : pP21(21) : Excited a'v=12 20 20 F2e - X Xv=0 21 21 F1e |
| 13C18O Excited | 20 e 10 | X 21 e 1 | 67652.6438 | .0926488 | 68458.4071 | 805.7634 | 4.416906 | 31384271 | 0.000000 | : qP31(21) : Excited dv=7 20 21 F3e - X Xv=0 21 21 F1e |
| 13C18O Excited | 21 f 2 | X 21 e 1 | 67402.6636 | .0777403 | 68208.4270 | 805.7634 | 3.706166 | 24831959 | 0.000000 | : pQ1(21) : Excited a3Pv=13 21 20 F1f - X Xv=0 21 21 F1e |
| 13C18O Excited | 21 f 3 | X 21 e 1 | 67405.1567 | .2933987 | 68210.9201 | 805.7634 | 13.98739 | 93728375 | 0.000000 | : qQ(21) : Excited Av=2 21 21 F1f - X Xv=0 21 21 F1e |
| 13C18O Excited | 21 f 4 | X 21 e 1 | 67467.3594 | .1686918 | 68273.1228 | 805.7634 | 8.042154 | 54039147 | 0.000000 | : qQ21(21) : Excited a3Pv=13 21 21 F2f - X Xv=0 21 21 F1e |
| 13C18O Excited | 21 f 5 | X 21 e 1 | 67504.1970 | .3059317 | 68309.9604 | 805.7634 | 14.58488 | 98163547 | 0.000000 | : qQ21(21) : Excited ev=4 21 21 F2f - X Xv=0 21 21 F1e |
| 13C18O Excited | 21 f 6 | X 21 e 1 | 67535.3094 | .0630514 | 68341.0727 | 805.7634 | 3.005893 | 20259149 | 0.000000 | : rQ31(21) : Excited a3Pv=13 21 22 F3f - X Xv=0 21 21 F1e |
| 13C18O Excited | 21 f 7 | X 21 e 1 | 67579.5701 | .0115360 | 68385.3335 | 805.7634 | .5499636 | 3713943 | 0.000000 | : pQ1(21) : Excited dv=7 21 20 F1f - X Xv=0 21 21 F1e |
| 13C18O Excited | 21 f 8 | X 21 e 1 | 67620.6804 | .1596650 | 68426.4437 | 805.7634 | 7.611815 | 51496985 | 0.000000 | : pQ1(21) : Excited a'v=12 21 20 F1f - X Xv=0 21 21 F1e |
| 13C18O Excited | 21 f 9 | X 21 e 1 | 67639.1976 | .0841791 | 68444.9610 | 805.7634 | 4.013127 | 27172724 | 0.000000 | : qQ21(21) : Excited dv=7 21 21 F2f - X Xv=0 21 21 F1e |
| 13C18O Excited | 21 f 10 | X 21 e 1 | 67699.4507 | .1920107 | 68505.2140 | 805.7634 | 9.153853 | 62146166 | 0.000000 | : rQ31(21) : Excited dv=7 21 22 F3f - X Xv=0 21 21 F1e |
| 13C18O Excited | 21 f 11 | X 21 e 1 | 67709.7936 | .1470666 | 68515.5570 | 805.7634 | 7.011204 | 47621388 | 0.000000 | : rQ31(21) : Excited a'v=12 21 22 F3f - X Xv=0 21 21 F1e |
| 13C18O Excited | 22 e 1 | X 21 e 1 | 67453.2433 | .0727834 | 68259.0067 | 805.7634 | 3.469850 | 22265381 | 0.000000 | : qR1(21) : Excited a3Pv=13 22 21 F1e - X Xv=0 21 21 F1e |
| 13C18O Excited | 22 e 2 | X 21 e 1 | 67467.0114 | .1390852 | 68272.7747 | 805.7634 | 6.630702 | 42574038 | 0.000000 | : rR(21) : Excited Av=2 22 22 F1e - X Xv=0 21 21 F1e |
| 13C18O Excited | 22 e 3 | X 21 e 1 | 67506.9737 | .0578731 | 68312.7371 | 805.7634 | 2.759021 | 17746467 | 0.000000 | : qR1(21) : Excited ev=4 22 21 F1e - X Xv=0 21 21 F1e |
| 13C18O Excited | 22 e 4 | X 21 e 1 | 67524.1955 | .0792312 | 68329.9588 | 805.7634 | 3.777243 | 24314429 | 0.000000 | : rR21(21) : Excited a3Pv=13 22 22 F2e - X Xv=0 21 21 F1e |
| 13C18O Excited | 22 e 5 | X 21 e 1 | 67591.2758 | .1161974 | 68397.0392 | 805.7634 | 5.539553 | 35764948 | 0.000000 | : sR31(21) : Excited a3Pv=13 22 23 F3e - X Xv=0 21 21 F1e |
| 13C18O Excited | 22 e 6 | X 21 e 1 | 67604.5307 | .0203877 | 68410.2941 | 805.7634 | .9719567 | 6278925 | 0.000000 | : sR31(21) : Excited ev=4 22 23 F3e - X Xv=0 21 21 F1e |
| 13C18O Excited | 22 e 7 | X 21 e 1 | 67625.3361 | .0076790 | 68431.0995 | 805.7634 | .3660866 | 2367135 | 0.000000 | : qR1(21) : Excited dv=7 22 21 F1e - X Xv=0 21 21 F1e |
| 13C18O Excited | 22 e 8 | X 21 e 1 | 67686.5118 | .0455680 | 68492.2752 | 805.7634 | 2.172394 | 14084968 | 0.000000 | : rR21(21) : Excited dv=7 22 22 F2e - X Xv=0 21 21 F1e |
| 13C18O Excited | 22 e 9 | X 21 e 1 | 67708.5232 | .1638078 | 68514.2866 | 805.7634 | 7.809317 | 50682031 | 0.000000 | : rR21(21) : Excited a'v=12 22 22 F2e - X Xv=0 21 21 F1e |
| 13C18O Excited | 22 e 10 | X 21 e 1 | 67748.4218 | .1014650 | 68554.1851 | 805.7634 | 4.837206 | 31448727 | 0.000000 | : sR31(21) : Excited dv=7 22 23 F3e - X Xv=0 21 21 F1e |
| 13C18O Excited | 21 e 1 | X 22 e 1 | 67321.9957 | .0312792 | 68204.3862 | 882.3904 | 2.153460 | 14376818 | 0.000000 | : oP1(22) : Excited a3Pv=13 21 20 F1e - X Xv=0 22 22 F1e |
| 13C18O Excited | 21 e 2 | X 22 e 1 | 67328.5687 | .1023047 | 68210.9591 | 882.3904 | 7.043300 | 47035882 | 0.000000 | : pP(22) : Excited Av=2 21 21 F1e - X Xv=0 22 22 F1e |
| 13C18O Excited | 21 e 3 | X 22 e 1 | 67383.5704 | .0287142 | 68265.9608 | 882.3904 | 1.976869 | 13234115 | 0.000000 | : oP1(22) : Excited ev=4 21 20 F1e - X Xv=0 22 22 F1e |
| 13C18O Excited | 21 e 4 | X 22 e 1 | 67390.5534 | .0596429 | 68272.9438 | 882.3904 | 4.106193 | 27497389 | 0.000000 | : pP21(22) : Excited a3Pv=13 21 21 F2e - X Xv=0 22 22 F1e |
| 13C18O Excited | 21 e 5 | X 22 e 1 | 67456.2754 | .0600489 | 68338.6659 | 882.3904 | 4.134143 | 27765631 | 0.000000 | : pP31(22) : Excited a3Pv=13 21 22 F3e - X Xv=0 22 22 F1e |
| 13C18O Excited | 21 e 6 | X 22 e 1 | 67476.5979 | .0254430 | 68358.9883 | 882.3904 | 1.751655 | 11775060 | 0.000000 | : qP31(22) : Excited ev=4 21 22 F3e - X Xv=0 22 22 F1e |
| 13C18O Excited | 21 e 7 | X 22 e 1 | 67502.9430 | .0039067 | 68385.3334 | 882.3904 | .2689618 | 1810146 | 0.000000 | : oP1(22) : Excited dv=7 21 20 F1e - X Xv=0 22 22 F1e |
| 13C18O Excited | 21 e 8 | X 22 e 1 | 67562.5701 | .0279326 | 68444.9606 | 882.3904 | 1.923057 | 12976741 | 0.000000 | : pP21(22) : Excited dv=7 21 21 F2e - X Xv=0 22 22 F1e |
| 13C18O Excited | 21 e 9 | X 22 e 1 | 67586.4591 | .1033707 | 68468.8496 | 882.3904 | 7.116689 | 48074195 | 0.000000 | : pP21(22) : Excited a'v=12 21 21 F2e - X Xv=0 22 22 F1e |
| 13C18O Excited | 21 e 10 | X 22 e 1 | 67622.8239 | .0657358 | 68505.2143 | 882.3904 | 4.525670 | 30620885 | 0.000000 | : qP31(22) : Excited dv=7 21 22 F3e - X Xv=0 22 22 F1e |
| 13C18O Excited | 22 f 2 | X 22 e 1 | 67381.1304 | .0911147 | 68263.5208 | 882.3904 | 6.272907 | 40123114 | 0.000000 | : pQ1(22) : Excited a3Pv=13 22 21 F1f - X Xv=0 22 22 F1e |
| 13C18O Excited | 22 f 3 | X 22 e 1 | 67390.4764 | .1755630 | 68272.8668 | 882.3904 | 12.08686 | 77342846 | 0.000000 | : qQ(22) : Excited Av=2 22 22 F1f - X Xv=0 22 22 F1e |
| 13C18O Excited | 22 f 4 | X 22 e 1 | 67447.5871 | .1348551 | 68329.9775 | 882.3904 | 9.284278 | 59560495 | 0.000000 | : qQ21(22) : Excited a3Pv=13 22 22 F2f - X Xv=0 22 22 F1e |
| 13C18O Excited | 22 f 5 | X 22 e 1 | 67476.3382 | .1996280 | 68358.7287 | 882.3904 | 13.74365 | 88281054 | 0.000000 | : qQ21(22) : Excited ev=4 22 22 F2f - X Xv=0 22 22 F1e |
| 13C18O Excited | 22 f 6 | X 22 e 1 | 67517.8434 | .0562879 | 68400.2338 | 882.3904 | 3.875214 | 24938038 | 0.000000 | : rQ31(22) : Excited a3Pv=13 22 23 F3f - X Xv=0 22 22 F1e |
| 13C18O Excited | 22 f 7 | X 22 e 1 | 67548.7091 | .0105381 | 68431.0995 | 882.3904 | .7255066 | 4675235 | 0.000000 | : pQ1(22) : Excited dv=7 22 21 F1f - X Xv=0 22 22 F1e |
| 13C18O Excited | 22 f 8 | X 22 e 1 | 67587.4402 | .1157391 | 68469.8306 | 882.3904 | 7.968208 | 51436283 | 0.000000 | : pQ1(22) : Excited a'v=12 22 21 F1f - X Xv=0 22 22 F1e |
| 13C18O Excited | 22 f 9 | X 22 e 1 | 67609.8852 | .0629481 | 68492.2756 | 882.3904 | 4.333745 | 28003019 | 0.000000 | : qQ21(22) : Excited dv=7 22 22 F2f - X Xv=0 22 22 F1e |
| 13C18O Excited | 22 f 10 | X 22 e 1 | 67671.7944 | .1355812 | 68554.1848 | 882.3904 | 9.334264 | 60480320 | 0.000000 | : rQ31(22) : Excited dv=7 22 23 F3f - X Xv=0 22 22 F1e |
| 13C18O Excited | 22 f 11 | X 22 e 1 | 67680.6480 | .1071249 | 68563.0384 | 882.3904 | 7.375152 | 47805230 | 0.000000 | : rQ31(22) : Excited a'v=12 22 23 F3f - X Xv=0 22 22 F1e |
| 13C18O Excited | 23 e 1 | X 22 e 1 | 67433.4418 | .0625315 | 68315.8322 | 882.3904 | 4.305059 | 26425954 | 0.000000 | : qR1(22) : Excited a3Pv=13 23 22 F1e - X Xv=0 22 22 F1e |
| 13C18O Excited | 23 e 2 | X 22 e 1 | 67454.8458 | .0872866 | 68337.2362 | 882.3904 | 6.009361 | 36922690 | 0.000000 | : rR(22) : Excited Av=2 23 23 F1e - X Xv=0 22 22 F1e |
| 13C18O Excited | 23 e 3 | X 22 e 1 | 67479.6941 | .0470950 | 68362.0846 | 882.3904 | 3.242320 | 19943470 | 0.000000 | : qR1(22) : Excited ev=4 23 22 F1e - X Xv=0 22 22 F1e |
| 13C18O Excited | 23 e 4 | X 22 e 1 | 67507.2161 | .0540470 | 68389.6065 | 882.3904 | 3.720935 | 22915442 | 0.000000 | : rR21(22) : Excited a3Pv=13 23 23 F2e - X Xv=0 22 22 F1e |
| 13C18O Excited | 23 e 5 | X 22 e 1 | 67573.3539 | .0980029 | 68455.7443 | 882.3904 | 6.747140 | 41674627 | 0.000000 | : sR31(22) : Excited ev=4 23 24 F3e - X Xv=0 22 22 F1e |
| 13C18O Excited | 23 e 7 | X 22 e 1 | 67596.7607 | .0080695 | 68479.1511 | 882.3904 | .5555524 | 3435012 | 0.000000 | : qR1(22) : Excited dv=7 23 22 F1e - X Xv=0 22 22 F1e |
| 13C18O Excited | 23 e 8 | X 22 e 1 | 67659.3303 | .0337828 | 68541.7207 | 882.3904 | 2.325823 | 14420666 | 0.000000 | : rR21(22) : Excited dv=7 23 23 F2e - X Xv=0 22 22 F1e |
| 13C18O Excited | 23 e 9 | X 22 e 1 | 67679.3776 | .1186209 | 68561.7680 | 882.3904 | 8.166612 | 50679998 | 0.000000 | : rR21(22) : Excited a'v=12 23 23 F2e - X Xv=0 22 22 F1e |
| 13C18O Excited | 23 e 10 | X 22 e 1 | 67722.9247 | .0715405 | 68605.3152 | 882.3904 | 4.925299 | 30624238 | 0.000000 | : sR31(22) : Excited dv=7 23 24 F3e - X Xv=0 22 22 F1e |
| 13C18O Excited | 22 e 1 | X 23 e 1 | 67296.5271 | .0328331 | 68259.0067 | 962.4797 | 3.318987 | 21149223 | 0.000000 | : oP1(23) : Excited a3Pv=13 22 21 F1e - X Xv=0 23 23 F1e |
| 13C18O Excited | 22 e 2 | X 23 e 1 | 67310.2951 | .0627423 | 68272.7747 | 962.4797 | 6.342411 | 40439871 | 0.000000 | : pP(23) : Excited Av=2 22 22 F1e - X Xv=0 23 23 F1e |
| 13C18O Excited | 22 e 3 | X 23 e 1 | 67350.2575 | .0261069 | 68312.7371 | 962.4797 | 2.639064 | 16856935 | 0.000000 | : oP1(23) : Excited ev=4 22 21 F1e - X Xv=0 23 23 F1e |
| 13C18O Excited | 22 e 4 | X 23 e 1 | 67367.4792 | .0357418 | 68329.9588 | 962.4797 | 3.613015 | 23095723 | 0.000000 | : pP21(23) : Excited a3Pv=13 22 22 F2e - X Xv=0 23 23 F1e |
| 13C18O Excited | 22 e 5 | X 23 e 1 | 67434.5595 | .0524174 | 68397.0392 | 962.4797 | 5.298703 | 33972545 | 0.000000 | : qP31(23) : Excited a3Pv=13 22 23 F3e - X Xv=0 23 23 F1e |
| 13C18O Excited | 22 e 6 | X 23 e 1 | 67447.8144 | .0091970 | 68410.2941 | 962.4797 | .9296978 | 5964257 | 0.000000 | : qP31(23) : Excited ev=4 22 23 F3e - X Xv=0 23 23 F1e |
| 13C18O Excited | 22 e 7 | X 23 e 1 | 67468.6198 | .0034641 | 68431.0995 | 962.4797 | .3501698 | 2248511 | 0.000000 | : oP1(23) : Excited dv=7 22 21 F1e - X Xv=0 23 23 F1e |
| 13C18O Excited | 22 e 8 | X 23 e 1 | 67529.7955 | .0205560 | 68492.2752 | 962.4797 | 2.077942 | 13379214 | 0.000000 | : pP21(23) : Excited dv=7 22 22 F2e - X Xv=0 23 23 F1e |
| 13C18O Excited | 22 e 9 | X 23 e 1 | 67551.8069 | .0738948 | 68514.2866 | 962.4797 | 7.469782 | 48142622 | 0.000000 | : pP21(23) : Excited a'v=12 22 22 F2e - X Xv=0 23 23 F1e |
| 13C18O Excited | 22 e 10 | X 23 e 1 | 67591.7055 | .0457715 | 68554.1851 | 962.4797 | 4.626893 | 29873120 | 0.000000 | : qP31(23) : Excited dv=7 22 23 F3e - X Xv=0 23 23 F1e |
| 13C18O Excited | 23 f 2 | X 23 e 1 | 67358.6359 | .0715025 | 68321.1155 | 962.4797 | 7.227954 | 44220209 | 0.000000 | : pQ1(23) : Excited a3Pv=13 23 22 F1f - X Xv=0 23 23 F1e |
| 13C18O Excited | 23 f 3 | X 23 e 1 | 67375.1303 | .1148074 | 68337.6100 | 962.4797 | 11.60549 | 71053919 | 0.000000 | : qQ(23) : Excited Av=2 23 23 F1f - X Xv=0 23 23 F1e |
| 13C18O Excited | 23 f 4 | X 23 e 1 | 67426.5936 | .1164050 | 68389.0733 | 962.4797 | 11.76699 | 72207910 | 0.000000 | : qQ21(23) : Excited a3Pv=13 23 23 F2f - X Xv=0 23 23 F1e |
| 13C18O Excited | 23 f 5 | X 23 e 1 | 67447.4643 | .1189016 | 68409.9440 | 962.4797 | 12.01937 | 73825099 | 0.000000 | : qQ21(23) : Excited ev=4 23 23 F2f - X Xv=0 23 23 F1e |
| 13C18O Excited | 23 f 6 | X 23 e 1 | 67499.4421 | .0428940 | 68461.9218 | 962.4797 | 4.336007 | 26694151 | 0.000000 | : rQ31(23) : Excited a3Pv=13 23 24 F3f - X Xv=0 23 23 F1e |
| 13C18O Excited | 23 f 7 | X 23 e 1 | 67516.6660 | .0114842 | 68479.1457 | 962.4797 | 1.160899 | 7152420 | 0.000000 | : pQ1(23) : Excited dv=7 23 22 F1f - X Xv=0 23 23 F1e |
| 13C18O Excited | 23 f 8 | X 23 e 1 | 67552.7850 | .0823717 | 68515.2647 | 962.4797 | 8.326684 | 51383947 | 0.000000 | : pQ1(23) : Excited a'v=12 23 22 F1f - X Xv=0 23 23 F1e |
| 13C18O Excited | 23 f 9 | X 23 e 1 | 67579.2416 | .0460071 | 68541.7212 | 962.4797 | 4.650701 | 28733198 | 0.000000 | : qQ21(23) : Excited dv=7 23 23 F2f - X Xv=0 23 23 F1e |
| 13C18O Excited | 23 f 10 | X 23 e 1 | 67642.8351 | .0938622 | 68605.3148 | 962.4797 | 9.488214 | 58786209 | 0.000000 | : rQ31(23) : Excited dv=7 23 24 F3f - X Xv=0 23 23 F1e |
| 13C18O Excited | 23 f 11 | X 23 e 1 | 67650.0814 | .0766750 | 68612.5611 | 962.4797 | 7.750816 | 48037235 | 0.000000 | : rQ31(23) : Excited a'v=12 23 24 F3f - X Xv=0 23 23 F1e |
| 13C18O Excited | 24 e 1 | X 23 e 1 | 67412.2344 | .0509456 | 68374.7140 | 962.4797 | 5.149921 | 30293126 | 0.000000 | : qR1(23) : Excited a3Pv=13 24 23 F1e - X Xv=0 23 23 F1e |
| 13C18O Excited | 24 e 2 | X 23 e 1 | 67440.9076 | .0378965 | 68403.3873 | 962.4797 | 3.830834 | 22562690 | 0.000000 | : rR(23) : Excited Av=2 24 24 F1e - X Xv=0 23 23 F1e |
| 13C18O Excited | 24 e 3 | X 23 e 1 | 67452.6850 | .0509945 | 68415.1647 | 962.4797 | 5.154860 | 30376798 | 0.000000 | : qR1(23) : Excited ev=4 24 23 F1e - X Xv=0 23 23 F1e |
| 13C18O Excited | 24 e 4 | X 23 e 1 | 67489.3228 | .0375707 | 68451.8024 | 962.4797 | 3.797896 | 22416903 | 0.000000 | : rR21(23) : Excited a3Pv=13 24 24 F2e - X Xv=0 23 23 F1e |
| 13C18O Excited | 24 e 5 | X 23 e 1 | 67550.5200 | .0598208 | 68512.9997 | 962.4797 | 6.047084 | 35789813 | 0.000000 | : sR31(23) : Excited ev=4 24 25 F3e - X Xv=0 23 23 F1e |
| 13C18O Excited | 24 e 6 | X 23 e 1 | 67563.2998 | .0006354 | 68525.7795 | 962.4797 | .0642333 | 380382.3 | 0.000000 | : sR31(23) : Excited a3Pv=13 24 25 F3e - X Xv=0 23 23 F1e |
| 13C18O Excited | 24 e 7 | X 23 e 1 | 67568.7627 | .0160063 | 68531.2424 | 962.4797 | 1.618028 | 9584097 | 0.000000 | : qR1(23) : Excited dv=7 24 23 F1e - X Xv=0 23 23 F1e |
| 13C18O Excited | 24 e 8 | X 23 e 1 | 67630.8159 | .0244433 | 68593.2956 | 962.4797 | 2.470887 | 14676218 | 0.000000 | : rR21(23) : Excited dv=7 24 24 F2e - X Xv=0 23 23 F1e |
| 13C18O Excited | 24 e 9 | X 23 e 1 | 67648.8117 | .0843684 | 68611.2914 | 962.4797 | 8.528522 | 50696925 | 0.000000 | : rR21(23) : Excited a'v=12 24 24 F2e - X Xv=0 23 23 F1e |
| 13C18O Excited | 24 e 10 | X 23 e 1 | 67696.1208 | .0495060 | 68658.6005 | 962.4797 | 5.004396 | 29810579 | 0.000000 | : sR31(23) : Excited dv=7 24 25 F3e - X Xv=0 23 23 F1e |
| 13C18O Excited | 23 e 1 | X 24 e 1 | 67269.8040 | .0273390 | 68315.8322 | 1046.0283 | 4.125682 | 25140955 | 0.000000 | : oP1(24) : Excited a3Pv=13 23 22 F1e - X Xv=0 24 24 F1e |
| 13C18O Excited | 23 e 2 | X 24 e 1 | 67291.2080 | .0381621 | 68337.2362 | 1046.0283 | 5.758971 | 35127354 | 0.000000 | : pP(24) : Excited Av=2 23 23 F1e - X Xv=0 24 24 F1e |
| 13C18O Excited | 23 e 3 | X 24 e 1 | 67316.0563 | .0205902 | 68362.0846 | 1046.0283 | 3.107223 | 18973786 | 0.000000 | : oP1(24) : Excited ev=4 23 22 F1e - X Xv=0 24 24 F1e |
| 13C18O Excited | 23 e 4 | X 24 e 1 | 67343.5783 | .0236296 | 68389.6065 | 1046.0283 | 3.565896 | 21801321 | 0.000000 | : pP21(24) : Excited a3Pv=13 23 23 F2e - X Xv=0 24 24 F1e |
| 13C18O Excited | 23 e 5 | X 24 e 1 | 67409.7160 | .0428473 | 68455.7443 | 1046.0283 | 6.466009 | 39648740 | 0.000000 | : qP31(24) : Excited ev=4 23 24 F3e - X Xv=0 24 24 F1e |
| 13C18O Excited | 23 e 7 | X 24 e 1 | 67433.1229 | .0035280 | 68479.1511 | 1046.0283 | .5324044 | 3268037 | 0.000000 | : oP1(24) : Excited dv=7 23 22 F1e - X Xv=0 24 24 F1e |
| 13C18O Excited | 23 e 8 | X 24 e 1 | 67495.6924 | .0147700 | 68541.7207 | 1046.0283 | 2.228914 | 13719775 | 0.000000 | : pP21(24) : Excited dv=7 23 23 F2e - X Xv=0 24 24 F1e |
| 13C18O Excited | 23 e 9 | X 24 e 1 | 67515.7397 | .0518616 | 68561.7680 | 1046.0283 | 7.826337 | 48216892 | 0.000000 | : pP21(24) : Excited a'v=12 23 23 F2e - X Xv=0 24 24 F1e |
| 13C18O Excited | 23 e 10 | X 24 e 1 | 67559.2869 | .0312778 | 68605.3152 | 1046.0283 | 4.720078 | 29136001 | 0.000000 | : qP31(24) : Excited dv=7 23 24 F3e - X Xv=0 24 24 F1e |
| 13C18O Excited | 24 f 2 | X 24 e 1 | 67335.2219 | .0532921 | 68381.2502 | 1046.0283 | 8.042211 | 47144353 | 0.000000 | : pQ1(24) : Excited a3Pv=13 24 23 F1f - X Xv=0 24 24 F1e |
| 13C18O Excited | 24 f 3 | X 24 e 1 | 67359.0321 | .0722299 | 68405.0604 | 1046.0283 | 10.90007 | 63965274 | 0.000000 | : qQ(24) : Excited Av=2 24 24 F1f - X Xv=0 24 24 F1e |
| 13C18O Excited | 24 f 4 | X 24 e 1 | 67403.7726 | .1141176 | 68449.8009 | 1046.0283 | 17.22127 | 1.0126e8 | 0.000000 | : qQ21(24) : Excited a3Pv=13 24 24 F2f - X Xv=0 24 24 F1e |
| 13C18O Excited | 24 f 5 | X 24 e 1 | 67418.2650 | .0515218 | 68464.2933 | 1046.0283 | 7.775052 | 45747081 | 0.000000 | : qQ21(24) : Excited ev=4 24 24 F2f - X Xv=0 24 24 F1e |
| 13C18O Excited | 24 f 6 | X 24 e 1 | 67478.5930 | .0138757 | 68524.6213 | 1046.0283 | 2.093951 | 12353557 | 0.000000 | : rQ31(24) : Excited a3Pv=13 24 25 F3f - X Xv=0 24 24 F1e |
| 13C18O Excited | 24 f 7 | X 24 e 1 | 67484.9117 | .0280970 | 68530.9400 | 1046.0283 | 4.240072 | 25021922 | 0.000000 | : pQ1(24) : Excited dv=7 24 23 F1f - X Xv=0 24 24 F1e |
| 13C18O Excited | 24 f 8 | X 24 e 1 | 67516.7153 | .0575450 | 68562.7436 | 1046.0283 | 8.684008 | 51319392 | 0.000000 | : pQ1(24) : Excited a'v=12 24 23 F1f - X Xv=0 24 24 F1e |
| 13C18O Excited | 24 f 9 | X 24 e 1 | 67547.2679 | .0328521 | 68593.2962 | 1046.0283 | 4.957651 | 29337741 | 0.000000 | : qQ21(24) : Excited dv=7 24 24 F2f - X Xv=0 24 24 F1e |
| 13C18O Excited | 24 f 10 | X 24 e 1 | 67612.5717 | .0636333 | 68658.6000 | 1046.0283 | 9.602774 | 56991021 | 0.000000 | : rQ31(24) : Excited dv=7 24 25 F3f - X Xv=0 24 24 F1e |
| 13C18O Excited | 24 f 11 | X 24 e 1 | 67618.0937 | .0540025 | 68664.1220 | 1046.0283 | 8.149409 | 48377367 | 0.000000 | : rQ31(24) : Excited a'v=12 24 25 F3f - X Xv=0 24 24 F1e |
| 13C18O Excited | 25 e 1 | X 24 e 1 | 67389.2494 | .0402627 | 68435.2777 | 1046.0283 | 6.075968 | 34303670 | 0.000000 | : qR1(24) : Excited a3Pv=13 25 24 F1e - X Xv=0 24 24 F1e |
| 13C18O Excited | 25 e 2 | X 24 e 1 | 67419.2386 | .0007528 | 68465.2669 | 1046.0283 | .1136046 | 642244.9 | 0.000000 | : qR1(24) : Excited ev=4 25 24 F1e - X Xv=0 24 24 F1e |
| 13C18O Excited | 25 e 3 | X 24 e 1 | 67432.2809 | .0559221 | 68478.3092 | 1046.0283 | 8.439104 | 47736782 | 0.000000 | : rR(24) : Excited Av=2 25 25 F1e - X Xv=0 24 24 F1e |
| 13C18O Excited | 25 e 4 | X 24 e 1 | 67470.4803 | .0263982 | 68516.5086 | 1046.0283 | 3.983709 | 22572633 | 0.000000 | : rR21(24) : Excited a3Pv=13 25 25 F2e - X Xv=0 24 24 F1e |
| 13C18O Excited | 25 e 5 | X 24 e 1 | 67524.6438 | .0347277 | 68570.6721 | 1046.0283 | 5.240697 | 29766598 | 0.000000 | : sR31(24) : Excited ev=4 25 26 F3e - X Xv=0 24 24 F1e |
| 13C18O Excited | 25 e 6 | X 24 e 1 | 67534.1084 | .0017492 | 68580.1367 | 1046.0283 | .2639634 | 1499915 | 0.000000 | : qR1(24) : Excited dv=7 25 24 F1e - X Xv=0 24 24 F1e |
| 13C18O Excited | 25 e 7 | X 24 e 1 | 67549.3498 | .0175450 | 68595.3781 | 1046.0283 | 2.647689 | 15055101 | 0.000000 | : sR31(24) : Excited a3Pv=13 25 26 F3e - X Xv=0 24 24 F1e |
| 13C18O Excited | 25 e 8 | X 24 e 1 | 67600.9703 | .0172274 | 68646.9986 | 1046.0283 | 2.599754 | 14816452 | 0.000000 | : rR21(24) : Excited dv=7 25 25 F2e - X Xv=0 24 24 F1e |
| 13C18O Excited | 25 e 9 | X 24 e 1 | 67616.8258 | .0589512 | 68662.8541 | 1046.0283 | 8.896213 | 50736766 | 0.000000 | : rR21(24) : Excited a'v=12 25 25 F2e - X Xv=0 24 24 F1e |
| 13C18O Excited | 25 e 10 | X 24 e 1 | 67668.0096 | .0336140 | 68714.0379 | 1046.0283 | 5.072626 | 28995880 | 0.000000 | : sR31(24) : Excited dv=7 25 26 F3e - X Xv=0 24 24 F1e |
| 13C18O Excited | 24 e 1 | X 25 e 1 | 67241.6806 | .0215845 | 68374.7140 | 1133.0334 | 4.943924 | 28861230 | 0.000000 | : oP1(25) : Excited a3Pv=13 24 23 F1e - X Xv=0 25 25 F1e |
| 13C18O Excited | 24 e 2 | X 25 e 1 | 67270.3539 | .0160559 | 68403.3873 | 1133.0334 | 3.677600 | 21496266 | 0.000000 | : pP(25) : Excited Av=2 24 24 F1e - X Xv=0 25 25 F1e |
| 13C18O Excited | 24 e 3 | X 25 e 1 | 67282.1313 | .0216052 | 68415.1647 | 1133.0334 | 4.948666 | 28941079 | 0.000000 | : oP1(25) : Excited ev=4 24 23 F1e - X Xv=0 25 25 F1e |
| 13C18O Excited | 24 e 4 | X 25 e 1 | 67318.7690 | .0159178 | 68451.8024 | 1133.0334 | 3.645980 | 21357486 | 0.000000 | : pP21(25) : Excited a3Pv=13 24 24 F2e - X Xv=0 25 25 F1e |
| 13C18O Excited | 24 e 5 | X 25 e 1 | 67379.9663 | .0253447 | 68512.9997 | 1133.0334 | 5.805200 | 34098631 | 0.000000 | : qP31(25) : Excited ev=4 24 25 F3e - X Xv=0 25 25 F1e |
| 13C18O Excited | 24 e 6 | X 25 e 1 | 67392.7461 | .0002692 | 68525.7795 | 1133.0334 | .0616640 | 362408.6 | 0.000000 | : qP31(25) : Excited a3Pv=13 24 25 F3e - X Xv=0 25 25 F1e |
| 13C18O Excited | 24 e 7 | X 25 e 1 | 67398.2090 | .0067815 | 68531.2424 | 1133.0334 | 1.553307 | 9131237 | 0.000000 | : oP1(25) : Excited dv=7 24 23 F1e - X Xv=0 25 25 F1e |
| 13C18O Excited | 24 e 8 | X 25 e 1 | 67460.2622 | .0103560 | 68593.2956 | 1133.0334 | 2.372052 | 13982846 | 0.000000 | : pP21(25) : Excited dv=7 24 24 F2e - X Xv=0 25 25 F1e |
| 13C18O Excited | 24 e 9 | X 25 e 1 | 67478.2580 | .0357449 | 68611.2914 | 1133.0334 | 8.187381 | 48301868 | 0.000000 | : pP21(25) : Excited a'v=12 24 24 F2e - X Xv=0 25 25 F1e |
| 13C18O Excited | 24 e 10 | X 25 e 1 | 67525.5671 | .0209745 | 68658.6005 | 1133.0334 | 4.804221 | 28402399 | 0.000000 | : qP31(25) : Excited dv=7 24 25 F3e - X Xv=0 25 25 F1e |
| 13C18O Excited | 25 f 2 | X 25 e 1 | 67310.8894 | .0387246 | 68443.9228 | 1133.0334 | 8.869882 | 49903052 | 0.000000 | : pQ1(25) : Excited a3Pv=13 25 24 F1f - X Xv=0 25 25 F1e |
| 13C18O Excited | 25 f 3 | X 25 e 1 | 67342.0644 | .0410139 | 68475.0978 | 1133.0334 | 9.394233 | 52926585 | 0.000000 | : qQ(25) : Excited Av=2 25 25 F1f - X Xv=0 25 25 F1e |
| 13C18O Excited | 25 f 4 | X 25 e 1 | 67377.4995 | .1112784 | 68510.5329 | 1133.0334 | 25.48833 | 1.4383e8 | 0.000000 | : qQ21(25) : Excited ev=4 25 25 F2f - X Xv=0 25 25 F1e |
| 13C18O Excited | 25 f 5 | X 25 e 1 | 67390.4863 | .0067548 | 68523.5197 | 1133.0334 | 1.547178 | 8735531 | 0.000000 | : qQ21(25) : Excited a3Pv=13 25 25 F2f - X Xv=0 25 25 F1e |
| 13C18O Excited | 25 f 6 | X 25 e 1 | 67446.8410 | .0001338 | 68579.8743 | 1133.0334 | .0306576 | 173530.7 | 0.000000 | : pQ1(25) : Excited dv=7 25 24 F1f - X Xv=0 25 25 F1e |
| 13C18O Excited | 25 f 7 | X 25 e 1 | 67461.8906 | .0313139 | 68594.9240 | 1133.0334 | 7.172462 | 40625340 | 0.000000 | : rQ31(25) : Excited a3Pv=13 25 26 F3f - X Xv=0 25 25 F1e |
| 13C18O Excited | 25 f 8 | X 25 e 1 | 67479.2320 | .0393322 | 68612.2654 | 1133.0334 | 9.009055 | 51067300 | 0.000000 | : pQ1(25) : Excited a'v=12 25 24 F1f - X Xv=0 25 25 F1e |
| 13C18O Excited | 25 f 9 | X 25 e 1 | 67513.9660 | .0228876 | 68646.9994 | 1133.0334 | 5.242409 | 29762197 | 0.000000 | : qQ21(25) : Excited dv=7 25 25 F2f - X Xv=0 25 25 F1e |
| 13C18O Excited | 25 f 10 | X 25 e 1 | 67581.0037 | .0420850 | 68714.0371 | 1133.0334 | 9.639577 | 54888981 | 0.000000 | : rQ31(25) : Excited dv=7 25 26 F3f - X Xv=0 25 25 F1e |
| 13C18O Excited | 25 f 11 | X 25 e 1 | 67584.6848 | .0375727 | 68717.7181 | 1133.0334 | 8.606035 | 49011865 | 0.000000 | : rQ31(25) : Excited a'v=12 25 26 F3f - X Xv=0 25 25 F1e |
| 13C18O Excited | 26 e 1 | X 25 e 1 | 67363.8467 | .0303020 | 68496.8801 | 1133.0334 | 6.940673 | 37664292 | 0.000000 | : qR1(25) : Excited a3Pv=13 26 25 F1e - X Xv=0 25 25 F1e |
| 13C18O Excited | 26 e 3 | X 25 e 1 | 67417.2963 | .0351662 | 68550.3297 | 1133.0334 | 8.054815 | 43814426 | 0.000000 | : rR(25) : Excited Av=2 26 26 F1e - X Xv=0 25 25 F1e |
| 13C18O Excited | 26 e 4 | X 25 e 1 | 67450.6398 | .0188469 | 68583.6732 | 1133.0334 | 4.316882 | 23516680 | 0.000000 | : rR21(25) : Excited a3Pv=13 26 26 F2e - X Xv=0 25 25 F1e |
| 13C18O Excited | 26 e 5 | X 25 e 1 | 67496.8704 | .0188317 | 68629.9038 | 1133.0334 | 4.313394 | 23546025 | 0.000000 | : sR31(25) : Excited ev=4 26 27 F3e - X Xv=0 25 25 F1e |
| 13C18O Excited | 26 e 6 | X 25 e 1 | 67501.8146 | .0040085 | 68634.8480 | 1133.0334 | .9181486 | 5013106 | 0.000000 | : qR1(25) : Excited dv=7 26 25 F1e - X Xv=0 25 25 F1e |
| 13C18O Excited | 26 e 7 | X 25 e 1 | 67531.5243 | .0146121 | 68664.5577 | 1133.0334 | 3.346909 | 18298314 | 0.000000 | : sR31(25) : Excited a3Pv=13 26 27 F3e - X Xv=0 25 25 F1e |
| 13C18O Excited | 26 e 8 | X 25 e 1 | 67569.7971 | .0117609 | 68702.8305 | 1133.0334 | 2.693841 | 14752901 | 0.000000 | : rR21(25) : Excited dv=7 26 26 F2e - X Xv=0 25 25 F1e |
| 13C18O Excited | 26 e 9 | X 25 e 1 | 67583.4205 | .0404720 | 68716.4539 | 1133.0334 | 9.270105 | 50798716 | 0.000000 | : rR21(25) : Excited a'v=12 26 26 F2e - X Xv=0 25 25 F1e |
| 13C18O Excited | 26 e 10 | X 25 e 1 | 67638.5918 | .0223827 | 68771.6252 | 1133.0334 | 5.126770 | 28162749 | 0.000000 | : sR31(25) : Excited dv=7 26 27 F3e - X Xv=0 25 25 F1e |
| 13C18O Excited | 25 e 1 | X 26 e 1 | 67211.7857 | .0165288 | 68435.2777 | 1223.4920 | 5.842277 | 32724400 | 0.000000 | : oP1(26) : Excited a3Pv=13 25 24 F1e - X Xv=0 26 26 F1e |
| 13C18O Excited | 25 e 2 | X 26 e 1 | 67241.7749 | .0003090 | 68465.2669 | 1223.4920 | .1092352 | 612679.4 | 0.000000 | : oP1(26) : Excited ev=4 25 24 F1e - X Xv=0 26 26 F1e |
| 13C18O Excited | 25 e 3 | X 26 e 1 | 67254.8172 | .0229573 | 68478.3092 | 1223.4920 | 8.114523 | 45539309 | 0.000000 | : pP(26) : Excited Av=2 25 25 F1e - X Xv=0 26 26 F1e |
| 13C18O Excited | 25 e 4 | X 26 e 1 | 67293.0166 | .0108371 | 68516.5086 | 1223.4920 | 3.830489 | 21533641 | 0.000000 | : pP21(26) : Excited a3Pv=13 25 25 F2e - X Xv=0 26 26 F1e |
| 13C18O Excited | 25 e 5 | X 26 e 1 | 67347.1801 | .0142565 | 68570.6721 | 1223.4920 | 5.039132 | 28396656 | 0.000000 | : qP31(26) : Excited ev=4 25 26 F3e - X Xv=0 26 26 F1e |
| 13C18O Excited | 25 e 6 | X 26 e 1 | 67356.6447 | .0007181 | 68580.1367 | 1223.4920 | .2538110 | 1430886 | 0.000000 | : oP1(26) : Excited dv=7 25 24 F1e - X Xv=0 26 26 F1e |
| 13C18O Excited | 25 e 7 | X 26 e 1 | 67371.8861 | .0072026 | 68595.3781 | 1223.4920 | 2.545854 | 14362265 | 0.000000 | : qP31(26) : Excited a3Pv=13 25 26 F3e - X Xv=0 26 26 F1e |
| 13C18O Excited | 25 e 8 | X 26 e 1 | 67423.5066 | .0070722 | 68646.9986 | 1223.4920 | 2.499763 | 14134684 | 0.000000 | : pP21(26) : Excited dv=7 25 25 F2e - X Xv=0 26 26 F1e |
| 13C18O Excited | 25 e 9 | X 26 e 1 | 67439.3622 | .0242008 | 68662.8541 | 1223.4920 | 8.554051 | 48402241 | 0.000000 | : pP21(26) : Excited a'v=12 25 25 F2e - X Xv=0 26 26 F1e |

| | | | | | | | | | | | |
|---|---|---|---|---|---|---|---|---|---|---|---|
| 13C18O Excited | 25 e 10 | X 26 e 1 | 67490.5459 | .0137993 | 68714.0379 | 1223.4920 | 4.877525 | 27661872 | 0.000000 | qP31(26) | Excited dv=7 25 26 F3e - X Xv=0 26 26 F1e |
| 13C18O Excited | 26 f 2 | X 26 e 1 | 67285.6335 | .0276673 | 68509.1255 | 1223.4920 | 9.779328 | 52883917 | 0.000000 | pQ1(26) | Excited a3Pv=13 26 25 F1f - X Xv=0 26 26 F1e |
| 13C18O Excited | 26 f 3 | X 26 e 1 | 67323.8216 | .0165860 | 68547.3136 | 1223.4920 | 5.862487 | 31756728 | 0.000000 | qQ(26) | Excited Av=2 26 26 F1f - X Xv=0 26 26 F1e |
| 13C18O Excited | 26 f 4 | X 26 e 1 | 67347.3457 | .0877365 | 68570.8376 | 1223.4920 | 31.01144 | 1.6816e8 | 0.000000 | qQ21(26) | Excited ev=4 26 26 F2f - X Xv=0 26 26 F1e |
| 13C18O Excited | 26 f 6 | X 26 e 1 | 67410.9328 | .0005797 | 68634.4248 | 1223.4920 | .2049170 | 1114337 | 0.000000 | pQ1(26) | Excited dv=7 26 25 F1f - X Xv=0 26 26 F1e |
| 13C18O Excited | 26 f 7 | X 26 e 1 | 67440.2937 | .0387110 | 68663.7857 | 1223.4920 | 13.68284 | 74504446 | 0.000000 | pQ1(26) | Excited a'v=12 26 25 F1f - X Xv=0 26 26 F1e |
| 13C18O Excited | 26 f 8 | X 26 e 1 | 67440.9116 | .0101421 | 68664.4036 | 1223.4920 | 3.584846 | 19520386 | 0.000000 | rQ31(26) | Excited a3Pv=13 26 27 F3f - X Xv=0 26 26 F1e |
| 13C18O Excited | 26 f 9 | X 26 e 1 | 67479.3397 | .0154892 | 68702.8317 | 1223.4920 | 5.474828 | 29862798 | 0.000000 | qQ21(26) | Excited dv=7 26 26 F2f - X Xv=0 26 26 F1e |
| 13C18O Excited | 26 f 10 | X 26 e 1 | 67548.1314 | .0265684 | 68771.6234 | 1223.4920 | 9.390903 | 51380099 | 0.000000 | rQ31(26) | Excited dv=7 26 27 F3f - X Xv=0 26 26 F1e |
| 13C18O Excited | 26 f 11 | X 26 e 1 | 67549.8551 | .0263751 | 68773.3471 | 1223.4920 | 9.322567 | 51010119 | 0.000000 | rQ31(26) | Excited a'v=12 26 27 F3f - X Xv=0 26 26 F1e |
| 13C18O Excited | 27 e 1 | X 26 e 1 | 67335.3499 | .0209350 | 68558.8418 | 1223.4920 | 7.399719 | 38646072 | 0.000000 | qR1(26) | Excited ev=4 27 26 F1e - X Xv=0 26 26 F1e |
| 13C18O Excited | 27 e 2 | X 26 e 1 | 67364.1174 | .0006906 | 68587.6094 | 1223.4920 | .2440831 | 1276393 | 0.000000 | qR1(26) | Excited a3Pv=13 27 26 F1e - X Xv=0 26 26 F1e |
| 13C18O Excited | 27 e 3 | X 26 e 1 | 67402.4329 | .0214302 | 68625.9249 | 1223.4920 | 7.574741 | 39678501 | 0.000000 | rR(26) | Excited Av=2 27 27 F1e - X Xv=0 26 26 F1e |
| 13C18O Excited | 27 e 4 | X 26 e 1 | 67429.7073 | .0138711 | 68653.1993 | 1223.4920 | 4.902898 | 25713871 | 0.000000 | rR21(26) | Excited a3Pv=13 27 27 F2e - X Xv=0 26 26 F1e |
| 13C18O Excited | 27 e 5 | X 26 e 1 | 67466.1991 | .0032285 | 68689.6911 | 1223.4920 | 1.141162 | 5994693 | 0.000000 | qR1(26) | Excited dv=7 27 26 F1e - X Xv=0 26 26 F1e |
| 13C18O Excited | 27 e 6 | X 26 e 1 | 67469.6664 | .0108633 | 68693.1584 | 1223.4920 | 3.839761 | 20173936 | 0.000000 | sR31(26) | Excited ev=4 27 28 F3e - X Xv=0 26 26 F1e |
| 13C18O Excited | 27 e 7 | X 26 e 1 | 67513.1350 | .0115025 | 68736.6270 | 1223.4920 | 4.065683 | 21402233 | 0.000000 | sR31(26) | Excited a3Pv=13 27 28 F3e - X Xv=0 26 26 F1e |
| 13C18O Excited | 27 e 8 | X 26 e 1 | 67537.3067 | .0076189 | 68760.7986 | 1223.4920 | 2.692967 | 14191328 | 0.000000 | rR21(26) | Excited dv=7 27 27 F2e - X Xv=0 26 26 F1e |
| 13C18O Excited | 27 e 9 | X 26 e 1 | 67548.5973 | .0272856 | 68772.0893 | 1223.4920 | 9.644406 | 50849328 | 0.000000 | rR21(26) | Excited a'v=12 27 27 F2e - X Xv=0 26 26 F1e |
| 13C18O Excited | 27 e 10 | X 26 e 1 | 67607.8701 | .0146020 | 68831.3620 | 1223.4920 | 5.161234 | 27283876 | 0.000000 | sR31(26) | Excited dv=7 27 28 F3e - X Xv=0 26 26 F1e |
| 13C18O Excited | 26 e 1 | X 27 e 1 | 67179.4792 | .0120524 | 68496.8801 | 1317.4009 | 6.683611 | 35972337 | 0.000000 | oP1(27) | Excited a3Pv=13 26 25 F1e - X Xv=0 27 27 F1e |
| 13C18O Excited | 26 e 3 | X 27 e 1 | 67232.9288 | .0139870 | 68550.3297 | 1317.4009 | 7.756489 | 41846468 | 0.000000 | pP(27) | Excited Av=2 26 26 F1e - X Xv=0 27 27 F1e |
| 13C18O Excited | 26 e 4 | X 27 e 1 | 67266.2723 | .0074962 | 68583.6732 | 1317.4009 | 4.156998 | 22460502 | 0.000000 | pP21(27) | Excited a3Pv=13 26 26 F2e - X Xv=0 27 27 F1e |
| 13C18O Excited | 26 e 5 | X 27 e 1 | 67312.5029 | .0074901 | 68629.9038 | 1317.4009 | 4.153639 | 22488656 | 0.000000 | qP31(27) | Excited ev=4 26 27 F3e - X Xv=0 27 27 F1e |
| 13C18O Excited | 26 e 6 | X 27 e 1 | 67317.4471 | .0015943 | 68634.8480 | 1317.4009 | .8841431 | 4787988 | 0.000000 | oP1(27) | Excited dv=7 26 25 F1e - X Xv=0 27 27 F1e |
| 13C18O Excited | 26 e 7 | X 27 e 1 | 67347.1569 | .0058118 | 68664.5577 | 1317.4009 | 3.222949 | 17476675 | 0.000000 | qP31(27) | Excited a3Pv=13 26 27 F3e - X Xv=0 27 27 F1e |
| 13C18O Excited | 26 e 8 | X 27 e 1 | 67385.4296 | .0046778 | 68702.8305 | 1317.4009 | 2.594070 | 14090525 | 0.000000 | pP21(27) | Excited dv=7 26 26 F2e - X Xv=0 27 27 F1e |
| 13C18O Excited | 26 e 9 | X 27 e 1 | 67399.0531 | .0160974 | 68716.4539 | 1317.4009 | 8.926768 | 48518034 | 0.000000 | pP21(27) | Excited a'v=12 26 26 F2e - X Xv=0 27 27 F1e |
| 13C18O Excited | 26 e 10 | X 27 e 1 | 67454.2243 | .0089025 | 68771.6252 | 1317.4009 | 4.936889 | 26898522 | 0.000000 | qP31(27) | Excited dv=7 26 27 F3e - X Xv=0 27 27 F1e |
| 13C18O Excited | 27 f 2 | X 27 e 1 | 67259.4437 | .0195526 | 68576.8446 | 1317.4009 | 10.84284 | 56436970 | 0.000000 | pQ1(27) | Excited a3Pv=13 27 26 F1f - X Xv=0 27 27 F1e |
| 13C18O Excited | 27 f 3 | X 27 e 1 | 67302.1995 | .0001728 | 68619.6004 | 1317.4009 | .0958064 | 499623.5 | 0.000000 | qQ(27) | Excited Av=2 27 27 F1f - X Xv=0 27 27 F1e |
| 13C18O Excited | 27 f 4 | X 27 e 1 | 67317.1076 | .0656349 | 68634.5085 | 1317.4009 | 36.39770 | 1.8994e8 | 0.000000 | qQ21(27) | Excited ev=4 27 27 F2f - X Xv=0 27 27 F1e |
| 13C18O Excited | 27 f 5 | X 27 e 1 | 67339.8821 | .0012495 | 68657.2830 | 1317.4009 | .6929182 | 3619593 | 0.000000 | qQ21(27) | Excited a3Pv=13 27 27 F2f - X Xv=0 27 27 F1e |
| 13C18O Excited | 27 f 6 | X 27 e 1 | 67373.5879 | .0004942 | 68690.9888 | 1317.4009 | .2740389 | 1433646 | 0.000000 | pQ1(27) | Excited dv=7 27 26 F1f - X Xv=0 27 27 F1e |
| 13C18O Excited | 27 f 7 | X 27 e 1 | 67400.0250 | .0184142 | 68717.4259 | 1317.4009 | 10.21157 | 53485170 | 0.000000 | pQ1(27) | Excited a'v=12 27 26 F1f - X Xv=0 27 27 F1e |
| 13C18O Excited | 27 f 8 | X 27 e 1 | 67419.1909 | .0152490 | 68736.5918 | 1317.4009 | 8.456307 | 44329423 | 0.000000 | rQ31(27) | Excited a3Pv=13 27 28 F3f - X Xv=0 27 27 F1e |
| 13C18O Excited | 27 f 9 | X 27 e 1 | 67443.3998 | .0100082 | 68760.8007 | 1317.4009 | 5.550050 | 29125674 | 0.000000 | qQ21(27) | Excited dv=7 27 27 F2f - X Xv=0 27 27 F1e |
| 13C18O Excited | 27 f 10 | X 27 e 1 | 67513.5902 | .0107206 | 68830.9911 | 1317.4009 | 5.945088 | 31296275 | 0.000000 | rQ31(27) | Excited a'v=12 27 28 F3f - X Xv=0 27 27 F1e |
| 13C18O Excited | 27 f 11 | X 27 e 1 | 67513.9719 | .0238035 | 68831.3728 | 1317.4009 | 13.20017 | 69489826 | 0.000000 | rQ31(27) | Excited dv=7 27 28 F3f - X Xv=0 27 27 F1e |
| 13C18O Excited | 28 e 1 | X 27 e 1 | 67303.6839 | .0133308 | 68621.0848 | 1317.4009 | 7.392566 | 37201485 | 0.000000 | qR1(27) | Excited ev=4 28 27 F1e - X Xv=0 27 27 F1e |
| 13C18O Excited | 28 e 2 | X 27 e 1 | 67337.3500 | .0014798 | 68654.7509 | 1317.4009 | .8206346 | 4135866 | 0.000000 | qR1(27) | Excited a3Pv=13 28 27 F1e - X Xv=0 27 27 F1e |
| 13C18O Excited | 28 e 3 | X 27 e 1 | 67386.9785 | .0123126 | 68704.3794 | 1317.4009 | 6.827924 | 34487780 | 0.000000 | rR(27) | Excited Av=2 28 28 F1e - X Xv=0 27 27 F1e |
| 13C18O Excited | 28 e 4 | X 27 e 1 | 67407.4607 | .0107993 | 68724.8616 | 1317.4009 | 5.988707 | 30276493 | 0.000000 | rR21(27) | Excited a3Pv=13 28 28 F2e - X Xv=0 27 27 F1e |
| 13C18O Excited | 28 e 5 | X 27 e 1 | 67431.6230 | .0004109 | 68749.0239 | 1317.4009 | .2278393 | 1153103 | 0.000000 | qR1(27) | Excited dv=7 28 27 F1e - X Xv=0 27 27 F1e |
| 13C18O Excited | 28 e 6 | X 27 e 1 | 67439.1627 | .0079650 | 68756.5636 | 1317.4009 | 4.416992 | 22362054 | 0.000000 | sR31(27) | Excited ev=4 28 29 F3e - X Xv=0 27 27 F1e |
| 13C18O Excited | 28 e 7 | X 27 e 1 | 67493.8872 | .0096099 | 68811.2881 | 1317.4009 | 5.329167 | 27045887 | 0.000000 | sR31(27) | Excited a3Pv=13 28 29 F3e - X Xv=0 27 27 F1e |
| 13C18O Excited | 28 e 8 | X 27 e 1 | 67503.5638 | .0039782 | 68820.9647 | 1317.4009 | 2.206092 | 11200884 | 0.000000 | rR21(27) | Excited dv=7 28 28 F2e - X Xv=0 27 27 F1e |
| 13C18O Excited | 28 e 9 | X 27 e 1 | 67512.3613 | .0179561 | 68829.7622 | 1317.4009 | 9.957508 | 50576543 | 0.000000 | rR21(27) | Excited a'v=12 28 28 F2e - X Xv=0 27 27 F1e |
| 13C18O Excited | 28 e 10 | X 27 e 1 | 67575.8503 | .0093155 | 68893.2512 | 1317.4009 | 5.165888 | 26312866 | 0.000000 | sR31(27) | Excited dv=7 28 29 F3e - X Xv=0 27 27 F1e |
| 13C18O Excited | 27 e 1 | X 28 e 1 | 67144.0850 | .0080669 | 68558.8418 | 1414.7569 | 7.135444 | 36949197 | 0.000000 | oP1(28) | Excited ev=4 27 26 F1e - X Xv=0 28 28 F1e |
| 13C18O Excited | 27 e 2 | X 28 e 1 | 67172.8525 | .0002661 | 68587.6094 | 1414.7569 | .2353659 | 1220353 | 0.000000 | oP1(28) | Excited a3Pv=13 27 26 F1e - X Xv=0 28 28 F1e |
| 13C18O Excited | 27 e 3 | X 28 e 1 | 67211.1680 | .0082577 | 68625.9249 | 1414.7569 | 7.304215 | 37936617 | 0.000000 | pP(28) | Excited Av=2 27 27 F1e - X Xv=0 28 28 F1e |
| 13C18O Excited | 27 e 4 | X 28 e 1 | 67238.4424 | .0053449 | 68653.1993 | 1414.7569 | 4.727795 | 24585118 | 0.000000 | pP21(28) | Excited a3Pv=13 27 27 F2e - X Xv=0 28 28 F1e |
| 13C18O Excited | 27 e 5 | X 28 e 1 | 67274.9342 | .0012440 | 68689.6911 | 1414.7569 | 1.100407 | 5731573 | 0.000000 | oP1(28) | Excited dv=7 27 26 F1e - X Xv=0 28 28 F1e |
| 13C18O Excited | 27 e 6 | X 28 e 1 | 67278.4016 | .0041859 | 68693.1584 | 1414.7569 | 3.702627 | 19288465 | 0.000000 | qP31(28) | Excited ev=4 27 28 F3e - X Xv=0 28 28 F1e |
| 13C18O Excited | 27 e 7 | X 28 e 1 | 67321.8701 | .0044322 | 68736.6270 | 1414.7569 | 3.920480 | 20462962 | 0.000000 | qP31(28) | Excited a3Pv=13 27 28 F3e - X Xv=0 28 28 F1e |
| 13C18O Excited | 27 e 8 | X 28 e 1 | 67346.0418 | .0029358 | 68760.7986 | 1414.7569 | 2.596790 | 13568561 | 0.000000 | pP21(28) | Excited dv=7 27 27 F2e - X Xv=0 28 28 F1e |
| 13C18O Excited | 27 e 9 | X 28 e 1 | 67357.3324 | .0105139 | 68772.0893 | 1414.7569 | 9.299963 | 48617943 | 0.000000 | pP21(28) | Excited a'v=12 27 27 F2e - X Xv=0 28 28 F1e |
| 13C18O Excited | 27 e 10 | X 28 e 1 | 67416.6052 | .0056266 | 68831.3620 | 1414.7569 | 4.976904 | 26086792 | 0.000000 | qP31(28) | Excited dv=7 27 28 F3e - X Xv=0 28 28 F1e |
| 13C18O Excited | 28 f 2 | X 28 e 1 | 67232.2986 | .0137691 | 68647.0555 | 1414.7569 | 12.17924 | 61094584 | 0.000000 | pQ1(28) | Excited a3Pv=13 28 27 F1f - X Xv=0 28 28 F1e |
| 13C18O Excited | 28 f 3 | X 28 e 1 | 67272.0381 | .0063376 | 68686.7950 | 1414.7569 | 5.605808 | 28170243 | 0.000000 | qQ21(28) | Excited ev=4 28 28 F2f - X Xv=0 28 28 F1e |
| 13C18O Excited | 28 f 4 | X 28 e 1 | 67292.9542 | .0339443 | 68707.7111 | 1414.7569 | 30.02500 | 1.5102e8 | 0.000000 | qQ(28) | Excited Av=2 28 28 F1f - X Xv=0 28 28 F1e |
| 13C18O Excited | 28 f 5 | X 28 e 1 | 67314.0269 | .0017307 | 68728.7838 | 1414.7569 | 1.530838 | 7707164 | 0.000000 | qQ21(28) | Excited a3Pv=13 28 28 F2f - X Xv=0 28 28 F1e |
| 13C18O Excited | 28 f 6 | X 28 e 1 | 67335.2557 | .0002441 | 68750.0126 | 1414.7569 | .2159114 | 1088057 | 0.000000 | pQ1(28) | Excited dv=7 28 27 F1f - X Xv=0 28 28 F1e |
| 13C18O Excited | 28 f 7 | X 28 e 1 | 67358.3126 | .0121572 | 68773.0695 | 1414.7569 | 10.75352 | 54246651 | 0.000000 | pQ1(28) | Excited a'v=12 28 27 F1f - X Xv=0 28 28 F1e |
| 13C18O Excited | 28 f 8 | X 28 e 1 | 67396.5986 | .0118115 | 68811.3555 | 1414.7569 | 10.44772 | 52793957 | 0.000000 | rQ31(28) | Excited a3Pv=13 28 29 F3f - X Xv=0 28 28 F1e |
| 13C18O Excited | 28 f 9 | X 28 e 1 | 67406.2137 | .0053365 | 68820.9706 | 1414.7569 | 4.720353 | 23862882 | 0.000000 | qQ21(28) | Excited dv=7 28 28 F2f - X Xv=0 28 28 F1e |
| 13C18O Excited | 28 f 10 | X 28 e 1 | 67475.9248 | .0100578 | 68890.6817 | 1414.7569 | 8.896503 | 45114322 | 0.000000 | rQ31(28) | Excited a'v=12 28 29 F3f - X Xv=0 28 28 F1e |
| 13C18O Excited | 28 f 11 | X 28 e 1 | 67478.4966 | .0120118 | 68893.2535 | 1414.7569 | 10.62492 | 53885308 | 0.000000 | rQ31(28) | Excited dv=7 28 29 F3f - X Xv=0 28 28 F1e |
| 13C18O Excited | 29 e 1 | X 28 e 1 | 67269.3876 | .0081558 | 68684.1444 | 1414.7569 | 7.214133 | 35019344 | 0.000000 | qR1(28) | Excited ev=4 29 28 F1e - X Xv=0 28 28 F1e |
| 13C18O Excited | 29 e 2 | X 28 e 1 | 67310.9278 | .0017156 | 68725.6847 | 1414.7569 | 1.517485 | 7379936 | 0.000000 | qR1(28) | Excited a3Pv=13 29 28 F1e - X Xv=0 28 28 F1e |
| 13C18O Excited | 29 e 3 | X 28 e 1 | 67370.6013 | .0056583 | 68785.3582 | 1414.7569 | 5.005009 | 24405491 | 0.000000 | rR(28) | Excited Av=2 29 29 F1e - X Xv=0 28 28 F1e |
| 13C18O Excited | 29 e 4 | X 28 e 1 | 67383.2025 | .0090232 | 68797.9593 | 1414.7569 | 7.981384 | 38940774 | 0.000000 | rR21(28) | Excited a3Pv=13 29 29 F2e - X Xv=0 28 28 F1e |
| 13C18O Excited | 29 e 5 | X 28 e 1 | 67395.8910 | .0008832 | 68810.6479 | 1414.7569 | .7812467 | 3813817 | 0.000000 | qR1(28) | Excited dv=7 29 28 F1e - X Xv=0 28 28 F1e |
| 13C18O Excited | 29 e 6 | X 28 e 1 | 67408.5256 | .0042113 | 68823.2825 | 1414.7569 | 3.725073 | 18194942 | 0.000000 | sR31(28) | Excited ev=4 29 30 F3e - X Xv=0 28 28 F1e |
| 13C18O Excited | 29 e 7 | X 28 e 1 | 67467.9757 | .0057311 | 68882.7326 | 1414.7569 | 5.069370 | 24826668 | 0.000000 | rR21(28) | Excited dv=7 29 29 F2e - X Xv=0 28 28 F1e |
| 13C18O Excited | 29 e 8 | X 28 e 1 | 67473.9577 | .0121923 | 68888.7146 | 1414.7569 | 10.78458 | 52830339 | 0.000000 | sR31(28) | Excited a3Pv=13 29 30 F3e - X Xv=0 28 28 F1e |
| 13C18O Excited | 29 e 9 | X 28 e 1 | 67475.0577 | .0031662 | 68889.8146 | 1414.7569 | 2.800588 | 13719882 | 0.000000 | rR21(28) | Excited a'v=12 29 29 F2e - X Xv=0 28 28 F1e |
| 13C18O Excited | 29 e 10 | X 28 e 1 | 67542.5451 | .0057896 | 68957.3019 | 1414.7569 | 5.121108 | 25163303 | 0.000000 | sR31(28) | Excited dv=7 29 30 F3e - X Xv=0 28 28 F1e |
| 13C18O Excited | 28 e 1 | X 29 e 1 | 67105.5283 | .0049761 | 68621.0848 | 1515.5565 | 7.137650 | 35602353 | 0.000000 | oP1(29) | Excited ev=4 28 27 F1e - X Xv=0 29 29 F1e |
| 13C18O Excited | 28 e 2 | X 29 e 1 | 67139.1943 | .0005524 | 68654.7509 | 1515.5565 | .7923369 | 3958100 | 0.000000 | oP1(29) | Excited a3Pv=13 28 27 F1e - X Xv=0 29 29 F1e |
| 13C18O Excited | 28 e 3 | X 29 e 1 | 67188.8228 | .0045960 | 68704.3794 | 1515.5565 | 6.592479 | 33005659 | 0.000000 | pP(29) | Excited Av=2 28 28 F1e - X Xv=0 29 29 F1e |
| 13C18O Excited | 28 e 4 | X 29 e 1 | 67209.3051 | .0040311 | 68724.8616 | 1515.5565 | 5.782200 | 28975431 | 0.000000 | pP21(29) | Excited a3Pv=13 28 28 F2e - X Xv=0 29 29 F1e |
| 13C18O Excited | 28 e 5 | X 29 e 1 | 67233.4674 | .0001534 | 68749.0239 | 1515.5565 | .2199828 | 1103555 | 0.000000 | oP1(29) | Excited dv=7 28 27 F1e - X Xv=0 29 29 F1e |
| 13C18O Excited | 28 e 6 | X 29 e 1 | 67241.0071 | .0029732 | 68756.5636 | 1515.5565 | 4.264682 | 21401186 | 0.000000 | qP31(29) | Excited ev=4 28 29 F3e - X Xv=0 29 29 F1e |
| 13C18O Excited | 28 e 7 | X 29 e 1 | 67295.7315 | .0035872 | 68811.2881 | 1515.5565 | 5.145403 | 25883947 | 0.000000 | qP31(29) | Excited a3Pv=13 28 29 F3e - X Xv=0 29 29 F1e |
| 13C18O Excited | 28 e 8 | X 29 e 1 | 67305.4081 | .0014850 | 68820.9647 | 1515.5565 | 2.130020 | 10719687 | 0.000000 | pP21(29) | Excited dv=7 28 28 F2e - X Xv=0 29 29 F1e |
| 13C18O Excited | 28 e 9 | X 29 e 1 | 67314.2056 | .0067026 | 68829.7622 | 1515.5565 | 9.614146 | 48403799 | 0.000000 | pP21(29) | Excited a'v=12 28 28 F2e - X Xv=0 29 29 F1e |
| 13C18O Excited | 28 e 10 | X 29 e 1 | 67377.6946 | .0034773 | 68893.2512 | 1515.5565 | 4.987754 | 25182688 | 0.000000 | qP31(29) | Excited dv=7 28 29 F3e - X Xv=0 29 29 F1e |
| 13C18O Excited | 29 f 2 | X 29 e 1 | 67204.1512 | .0097823 | 68719.7078 | 1515.5565 | 14.03161 | 67915256 | 0.000000 | pQ1(29) | Excited a3Pv=13 29 28 F1f - X Xv=0 29 29 F1e |
| 13C18O Excited | 29 f 3 | X 29 e 1 | 67236.3401 | .0069684 | 68751.8967 | 1515.5565 | 9.995329 | 48448544 | 0.000000 | qQ21(29) | Excited ev=4 29 29 F2f - X Xv=0 29 29 F1e |
| 13C18O Excited | 29 f 4 | X 29 e 1 | 67272.1964 | .0165891 | 68787.7530 | 1515.5565 | 23.79516 | 1.1552e8 | 0.000000 | qQ(29) | Excited Av=2 29 29 F1f - X Xv=0 29 29 F1e |

| | | | | | | | | | | | | | | | | | |
|---|---|---|---|---|---|---|---|---|---|---|---|---|---|---|---|---|---|
| 13C18O | Excited | 29 | f | 5 | X | 29 | e | 1 | 67285.5009 | .0018033 | 68801.0574 | 1515.5565 | 2.586664 | 12565387 | 0.000000 | qQ21(29) | Excited a3Pv=13 29 29 F2f - X Xv=0 29 29 F1e |
| 13C18O | Excited | 29 | f | 7 | X | 29 | e | 1 | 67315.2112 | .0081455 | 68830.7677 | 1515.5565 | 11.68375 | 56832041 | 0.000000 | pQ1(29) | Excited a'v=12 29 28 F1f - X Xv=0 29 29 F1e |
| 13C18O | Excited | 29 | f | 8 | X | 29 | e | 1 | 67367.1819 | .0072118 | 68882.7384 | 1515.5565 | 10.34446 | 50434119 | 0.000000 | qQ21(29) | Excited dv=7 29 29 F2f - X Xv=0 29 29 F1e |
| 13C18O | Excited | 29 | f | 9 | X | 29 | e | 1 | 67373.6496 | .0042477 | 68889.2061 | 1515.5565 | 6.092853 | 29714082 | 0.000000 | rQ31(29) | Excited a3Pv=13 29 30 F3f - X Xv=0 29 29 F1e |
| 13C18O | Excited | 29 | f | 10 | X | 29 | e | 1 | 67436.8297 | .0066072 | 68952.3862 | 1515.5565 | 9.477299 | 46349756 | 0.000000 | rQ31(29) | Excited a'v=12 29 30 F3f - X Xv=0 29 29 F1e |
| 13C18O | Excited | 29 | f | 11 | X | 29 | e | 1 | 67441.7475 | .0071987 | 68957.3040 | 1515.5565 | 10.32571 | 50510040 | 0.000000 | rQ31(29) | Excited dv=7 29 30 F3f - X Xv=0 29 29 F1e |
| 13C18O | Excited | 30 | e | 1 | X | 29 | e | 1 | 67233.0110 | .0049203 | 68748.5676 | 1515.5565 | 7.057593 | 33082466 | 0.000000 | qR1(29) | Excited ev=4 30 29 F1e - X Xv=0 29 29 F1e |
| 13C18O | Excited | 30 | e | 2 | X | 29 | e | 1 | 67284.3288 | .0015148 | 68799.8854 | 1515.5565 | 2.172781 | 10208250 | 0.000000 | qR1(29) | Excited a3Pv=13 30 29 F1e - X Xv=0 29 29 F1e |
| 13C18O | Excited | 30 | e | 4 | X | 29 | e | 1 | 67355.4122 | .0044065 | 68870.9688 | 1515.5565 | 6.320553 | 29789704 | 0.000000 | rR(29) | Excited Av=2 30 30 F1e - X Xv=0 29 29 F1e |
| 13C18O | Excited | 30 | e | 5 | X | 29 | e | 1 | 67362.9644 | .0061076 | 68878.5209 | 1515.5565 | 8.760707 | 41304401 | 0.000000 | rR21(29) | Excited a3Pv=13 30 30 F2e - X Xv=0 29 29 F1e |
| 13C18O | Excited | 30 | e | 6 | X | 29 | e | 1 | 67378.2078 | .0016436 | 68893.7644 | 1515.5565 | 2.357590 | 11122957 | 0.000000 | sR31(29) | Excited ev=4 30 31 F3e - X Xv=0 29 29 F1e |
| 13C18O | Excited | 30 | e | 7 | X | 29 | e | 1 | 67431.6803 | .0025784 | 68947.2368 | 1515.5565 | 3.698435 | 17490553 | 0.000000 | rR21(29) | Excited dv=7 30 30 F2e - X Xv=0 29 29 F1e |
| 13C18O | Excited | 30 | e | 8 | X | 29 | e | 1 | 67435.6091 | .0084727 | 68951.1656 | 1515.5565 | 12.15309 | 57484154 | 0.000000 | rR21(29) | Excited a'v=12 30 30 F2e - X Xv=0 29 29 F1e |
| 13C18O | Excited | 30 | e | 9 | X | 29 | e | 1 | 67453.2470 | .0028938 | 68968.8035 | 1515.5565 | 4.150835 | 19648872 | 0.000000 | sR31(29) | Excited a3Pv=13 30 31 F3e - X Xv=0 29 29 F1e |
| 13C18O | Excited | 30 | e | 10 | X | 29 | e | 1 | 67507.9821 | .0034753 | 69023.5387 | 1515.5565 | 4.984926 | 23654716 | 0.000000 | sR31(29) | Excited dv=7 30 31 F3e - X Xv=0 29 29 F1e |
| 13C18O | Excited | 29 | e | 1 | X | 30 | e | 1 | 67064.3480 | .0029490 | 68684.1444 | 1619.7964 | 6.973662 | 33543429 | 0.000000 | oP1(30) | Excited ev=4 29 28 F1e - X Xv=0 30 30 F1e |
| 13C18O | Excited | 29 | e | 2 | X | 30 | e | 1 | 67105.8883 | .0006203 | 68725.6847 | 1619.7964 | 1.466902 | 7068943 | 0.000000 | oP1(30) | Excited a3Pv=13 29 28 F1e - X Xv=0 30 30 F1e |
| 13C18O | Excited | 29 | e | 3 | X | 30 | e | 1 | 67165.5618 | .0020460 | 68785.3582 | 1619.7964 | 4.838175 | 23377226 | 0.000000 | pP(30) | Excited Av=2 29 29 F1e - X Xv=0 30 30 F1e |
| 13C18O | Excited | 29 | e | 4 | X | 30 | e | 1 | 67178.1630 | .0032627 | 68797.9593 | 1619.7964 | 7.715338 | 37300164 | 0.000000 | pP21(30) | Excited a3Pv=13 29 29 F2e - X Xv=0 30 30 F1e |
| 13C18O | Excited | 29 | e | 5 | X | 30 | e | 1 | 67190.8515 | .0003194 | 68810.6479 | 1619.7964 | .7552052 | 3653144 | 0.000000 | oP1(30) | Excited dv=7 29 28 F1e - X Xv=0 30 30 F1e |
| 13C18O | Excited | 29 | e | 6 | X | 30 | e | 1 | 67203.4861 | .0015228 | 68823.2825 | 1619.7964 | 3.600904 | 17428433 | 0.000000 | qP31(30) | Excited ev=4 29 30 F3e - X Xv=0 30 30 F1e |
| 13C18O | Excited | 29 | e | 7 | X | 30 | e | 1 | 67262.9362 | .0020723 | 68882.7326 | 1619.7964 | 4.900391 | 23780972 | 0.000000 | pP21(30) | Excited dv=7 29 29 F2e - X Xv=0 30 30 F1e |
| 13C18O | Excited | 29 | e | 8 | X | 30 | e | 1 | 67268.9182 | .0044086 | 68888.7146 | 1619.7964 | 10.42510 | 50605173 | 0.000000 | qP31(30) | Excited a3Pv=13 29 30 F3e - X Xv=0 30 30 F1e |
| 13C18O | Excited | 29 | e | 9 | X | 30 | e | 1 | 67270.0182 | .0011448 | 68889.8146 | 1619.7964 | 2.707235 | 13142015 | 0.000000 | pP21(30) | Excited a'v=12 29 29 F2e - X Xv=0 30 30 F1e |
| 13C18O | Excited | 29 | e | 10 | X | 30 | e | 1 | 67337.5055 | .0020934 | 68957.3019 | 1619.7964 | 4.950405 | 24103672 | 0.000000 | qP31(30) | Excited dv=7 29 30 F3e - X Xv=0 30 30 F1e |
| 13C18O | Excited | 30 | f | 2 | X | 30 | e | 1 | 67174.8792 | .0071842 | 68794.6756 | 1619.7964 | 16.98879 | 79428580 | 0.000000 | pQ1(30) | Excited a3Pv=13 30 29 F1f - X Xv=0 30 30 F1e |
| 13C18O | Excited | 30 | f | 3 | X | 30 | e | 1 | 67198.5347 | .0043917 | 68818.3311 | 1619.7964 | 10.38509 | 48605282 | 0.000000 | qQ21(30) | Excited ev=4 30 30 F2f - X Xv=0 30 30 F1e |
| 13C18O | Excited | 30 | f | 5 | X | 30 | e | 1 | 67251.9040 | .0090526 | 68871.7004 | 1619.7964 | 21.40697 | 1.0043e8 | 0.000000 | qQ(30) | Excited Av=2 30 30 F1f - X Xv=0 30 30 F1e |
| 13C18O | Excited | 30 | f | 6 | X | 30 | e | 1 | 67265.6477 | .0002502 | 68885.4441 | 1619.7964 | .5915482 | 2776922 | 0.000000 | qQ21(30) | Excited a3Pv=13 30 30 F2f - X Xv=0 30 30 F1e |
| 13C18O | Excited | 30 | f | 7 | X | 30 | e | 1 | 67270.8246 | .0060717 | 68890.6210 | 1619.7964 | 14.35801 | 67416781 | 0.000000 | pQ1(30) | Excited a'v=12 30 29 F1f - X Xv=0 30 30 F1e |
| 13C18O | Excited | 30 | f | 8 | X | 30 | e | 1 | 67327.4496 | .0036325 | 68947.2460 | 1619.7964 | 8.589952 | 40435316 | 0.000000 | qQ21(30) | Excited dv=7 30 30 F2f - X Xv=0 30 30 F1e |
| 13C18O | Excited | 30 | f | 9 | X | 30 | e | 1 | 67349.1656 | .0039682 | 68968.9620 | 1619.7964 | 9.383783 | 44214854 | 0.000000 | rQ31(30) | Excited a3Pv=13 30 31 F3f - X Xv=0 30 30 F1e |
| 13C18O | Excited | 30 | f | 10 | X | 30 | e | 1 | 67396.3126 | .0041964 | 69016.1090 | 1619.7964 | 9.923254 | 46855018 | 0.000000 | rQ31(30) | Excited a'v=12 30 31 F3f - X Xv=0 30 30 F1e |
| 13C18O | Excited | 30 | f | 11 | X | 30 | e | 1 | 67403.7451 | .0042207 | 69023.5415 | 1619.7964 | 9.980783 | 47142247 | 0.000000 | rQ31(30) | Excited dv=7 30 31 F3f - X Xv=0 30 30 F1e |
| 13C18O | Excited | 31 | e | 1 | X | 30 | e | 1 | 67194.8908 | .0029461 | 68814.6872 | 1619.7964 | 6.966627 | 31565610 | 0.000000 | qR1(30) | Excited ev=4 31 30 F1e - X Xv=0 30 30 F1e |
| 13C18O | Excited | 31 | e | 2 | X | 30 | e | 1 | 67257.2167 | .0011686 | 68877.0131 | 1619.7964 | 2.763435 | 12555928 | 0.000000 | qR1(30) | Excited a3Pv=13 31 30 F1e - X Xv=0 30 30 F1e |
| 13C18O | Excited | 31 | e | 3 | X | 30 | e | 1 | 67314.4096 | .0003937 | 68934.2060 | 1619.7964 | .9310528 | 4241127 | 0.000000 | qR1(30) | Excited dv=7 31 30 F1e - X Xv=0 30 30 F1e |
| 13C18O | Excited | 31 | e | 4 | X | 30 | e | 1 | 67328.2462 | .0005767 | 68948.0426 | 1619.7964 | 1.363655 | 6215545 | 0.000000 | sR31(30) | Excited ev=4 31 32 F3e - X Xv=0 30 30 F1e |
| 13C18O | Excited | 31 | e | 5 | X | 30 | e | 1 | 67339.3907 | .0058303 | 68959.1871 | 1619.7964 | 13.78719 | 62873311 | 0.000000 | rR(30) | Excited Av=2 31 31 F1e - X Xv=0 30 30 F1e |
| 13C18O | Excited | 31 | e | 6 | X | 30 | e | 1 | 67349.3961 | .0004534 | 68969.1925 | 1619.7964 | 1.072280 | 4892064 | 0.000000 | rR21(30) | Excited a3Pv=13 31 31 F2e - X Xv=0 30 30 F1e |
| 13C18O | Excited | 31 | e | 7 | X | 30 | e | 1 | 67393.8988 | .0007261 | 69013.6952 | 1619.7964 | 1.716925 | 7848667 | 0.000000 | rR21(30) | Excited dv=7 31 31 F2e - X Xv=0 30 30 F1e |
| 13C18O | Excited | 31 | e | 8 | X | 30 | e | 1 | 67395.2069 | .0062519 | 69015.0033 | 1619.7964 | 14.78394 | 67586543 | 0.000000 | rR21(30) | Excited a'v=12 31 31 F2e - X Xv=0 30 30 F1e |
| 13C18O | Excited | 31 | e | 9 | X | 30 | e | 1 | 67431.2999 | .0022388 | 69051.0963 | 1619.7964 | 5.294107 | 24241539 | 0.000000 | sR31(30) | Excited a3Pv=13 31 32 F3e - X Xv=0 30 30 F1e |
| 13C18O | Excited | 31 | e | 10 | X | 30 | e | 1 | 67472.2309 | .0019681 | 69092.0273 | 1619.7964 | 4.654083 | 21349723 | 0.000000 | sR31(30) | Excited dv=7 31 32 F3e - X Xv=0 30 30 F1e |
| 13C18O | Excited | 30 | e | 1 | X | 31 | e | 1 | 67021.0947 | .0017233 | 68748.5676 | 1727.4728 | 6.829928 | 31713509 | 0.000000 | oP1(31) | Excited ev=4 30 29 F1e - X Xv=0 31 31 F1e |
| 13C18O | Excited | 30 | e | 2 | X | 31 | e | 1 | 67072.4126 | .0005305 | 68799.8854 | 1727.4728 | 2.102691 | 9785902 | 0.000000 | oP1(31) | Excited a3Pv=13 30 29 F1e - X Xv=0 31 31 F1e |
| 13C18O | Excited | 30 | e | 4 | X | 31 | e | 1 | 67143.4960 | .0015433 | 68870.9688 | 1727.4728 | 6.116664 | 28557495 | 0.000000 | pP(31) | Excited Av=2 30 30 F1e - X Xv=0 31 31 F1e |
| 13C18O | Excited | 30 | e | 5 | X | 31 | e | 1 | 67151.0481 | .0021392 | 68878.5209 | 1727.4728 | 8.478104 | 39595945 | 0.000000 | pP21(31) | Excited a3Pv=13 30 30 F2e - X Xv=0 31 31 F1e |
| 13C18O | Excited | 30 | e | 6 | X | 31 | e | 1 | 67166.2916 | .0005757 | 68893.7644 | 1727.4728 | 2.281539 | 10662905 | 0.000000 | qP31(31) | Excited ev=4 30 31 F3e - X Xv=0 31 31 F1e |
| 13C18O | Excited | 30 | e | 7 | X | 31 | e | 1 | 67219.7640 | .0009031 | 68947.2368 | 1727.4728 | 3.579131 | 16767261 | 0.000000 | pP21(31) | Excited dv=7 30 30 F2e - X Xv=0 31 31 F1e |
| 13C18O | Excited | 30 | e | 8 | X | 31 | e | 1 | 67223.6928 | .0029675 | 68951.1656 | 1727.4728 | 11.76105 | 55107023 | 0.000000 | pP21(31) | Excited a'v=12 30 30 F2e - X Xv=0 31 31 F1e |
| 13C18O | Excited | 30 | e | 9 | X | 31 | e | 1 | 67241.3307 | .0010135 | 68968.8035 | 1727.4728 | 4.016937 | 18836382 | 0.000000 | qP31(31) | Excited a3Pv=13 30 31 F3e - X Xv=0 31 31 F1e |
| 13C18O | Excited | 30 | e | 10 | X | 31 | e | 1 | 67296.0659 | .0012172 | 69023.5387 | 1727.4728 | 4.824122 | 22676757 | 0.000000 | qP31(31) | Excited dv=7 30 31 F3e - X Xv=0 31 31 F1e |
| 13C18O | Excited | 31 | f | 2 | X | 31 | e | 1 | 67144.0706 | .0057010 | 68871.5435 | 1727.4728 | 22.59479 | 1.0214e8 | 0.000000 | pQ1(31) | Excited a3Pv=13 31 30 F1f - X Xv=0 31 31 F1e |
| 13C18O | Excited | 31 | f | 3 | X | 31 | e | 1 | 67159.7063 | .0017960 | 68887.1791 | 1727.4728 | 7.117954 | 32200633 | 0.000000 | qQ21(31) | Excited ev=4 31 31 F2f - X Xv=0 31 31 F1e |
| 13C18O | Excited | 31 | f | 4 | X | 31 | e | 1 | 67207.0598 | .0001809 | 68934.5326 | 1727.4728 | .7170823 | 3250847 | 0.000000 | pQ1(31) | Excited dv=7 31 30 F1f - X Xv=0 31 31 F1e |
| 13C18O | Excited | 31 | f | 5 | X | 31 | e | 1 | 67223.7141 | .0005821 | 68951.1869 | 1727.4728 | 2.306873 | 10465838 | 0.000000 | pQ1(31) | Excited a'v=12 31 30 F1f - X Xv=0 31 31 F1e |
| 13C18O | Excited | 31 | f | 6 | X | 31 | e | 1 | 67231.8857 | .0080156 | 68959.3585 | 1727.4728 | 31.76815 | 1.4418e8 | 0.000000 | qQ(31) | Excited Av=2 31 31 F1f - X Xv=0 31 31 F1e |
| 13C18O | Excited | 31 | f | 7 | X | 31 | e | 1 | 67236.9593 | .0001964 | 68964.4321 | 1727.4728 | .7785815 | 3534363 | 0.000000 | qQ21(31) | Excited a3Pv=13 31 31 F2f - X Xv=0 31 31 F1e |
| 13C18O | Excited | 31 | f | 8 | X | 31 | e | 1 | 67286.2835 | .0022258 | 69013.7563 | 1727.4728 | 8.821285 | 40132327 | 0.000000 | qQ21(31) | Excited dv=7 31 31 F2f - X Xv=0 31 31 F1e |
| 13C18O | Excited | 31 | f | 9 | X | 31 | e | 1 | 67323.8269 | .0028491 | 69051.2997 | 1727.4728 | 11.29161 | 51457058 | 0.000000 | rQ31(31) | Excited a3Pv=13 31 32 F3f - X Xv=0 31 31 F1e |
| 13C18O | Excited | 31 | f | 10 | X | 31 | e | 1 | 67354.3746 | .0026040 | 69081.8474 | 1727.4728 | 10.32021 | 47094351 | 0.000000 | rQ31(31) | Excited a'v=12 31 32 F3f - X Xv=0 31 31 F1e |
| 13C18O | Excited | 31 | f | 11 | X | 31 | e | 1 | 67364.5598 | .0023423 | 69092.0326 | 1727.4728 | 9.283236 | 42381539 | 0.000000 | rQ31(31) | Excited dv=7 31 32 F3f - X Xv=0 31 31 F1e |
| 13C18O | Excited | 32 | e | 1 | X | 31 | e | 1 | 67155.2080 | .0017497 | 68882.6808 | 1727.4728 | 6.934649 | 30400005 | 0.000000 | qR1(31) | Excited ev=4 32 31 F1e - X Xv=0 31 31 F1e |
| 13C18O | Excited | 32 | e | 2 | X | 31 | e | 1 | 67229.3867 | .0008373 | 68956.8595 | 1727.4728 | 3.318366 | 14595260 | 0.000000 | qR1(31) | Excited a3Pv=13 32 31 F1e - X Xv=0 31 31 F1e |
| 13C18O | Excited | 32 | e | 3 | X | 31 | e | 1 | 67274.1489 | .0001691 | 69001.6217 | 1727.4728 | .6702383 | 2953820 | 0.000000 | qR1(31) | Excited dv=7 32 31 F1e - X Xv=0 31 31 F1e |
| 13C18O | Excited | 32 | e | 4 | X | 31 | e | 1 | 67295.1393 | .0011516 | 69022.6121 | 1727.4728 | 4.564215 | 20133869 | 0.000000 | sR31(31) | Excited ev=4 32 33 F3e - X Xv=0 31 31 F1e |
| 13C18O | Excited | 32 | e | 5 | X | 31 | e | 1 | 67319.5423 | .0019666 | 69047.0151 | 1727.4728 | 7.794279 | 34419892 | 0.000000 | rR(31) | Excited Av=2 32 32 F1e - X Xv=0 31 31 F1e |
| 13C18O | Excited | 32 | e | 6 | X | 31 | e | 1 | 67322.6877 | .0007629 | 69050.1605 | 1727.4728 | 3.023627 | 13354347 | 0.000000 | rR21(31) | Excited a3Pv=13 32 32 F2e - X Xv=0 31 31 F1e |
| 13C18O | Excited | 32 | e | 7 | X | 31 | e | 1 | 67353.2371 | .0022135 | 69080.7099 | 1727.4728 | 8.772604 | 38798418 | 0.000000 | rR21(31) | Excited a'v=12 32 32 F2e - X Xv=0 31 31 F1e |
| 13C18O | Excited | 32 | e | 8 | X | 31 | e | 1 | 67355.1270 | .0023666 | 69082.5998 | 1727.4728 | 9.379434 | 41485724 | 0.000000 | rR21(31) | Excited dv=7 32 32 F2e - X Xv=0 31 31 F1e |
| 13C18O | Excited | 32 | e | 9 | X | 31 | e | 1 | 67408.1523 | .0016955 | 69135.6251 | 1727.4728 | 6.719680 | 29791745 | 0.000000 | sR31(31) | Excited a3Pv=13 32 33 F3e - X Xv=0 31 31 F1e |
| 13C18O | Excited | 32 | e | 10 | X | 31 | e | 1 | 67435.5109 | .0009646 | 69162.9837 | 1727.4728 | 3.822868 | 16969353 | 0.000000 | sR31(31) | Excited dv=7 32 33 F3e - X Xv=0 31 31 F1e |
| 13C18O | Excited | 31 | e | 1 | X | 32 | e | 1 | 66976.1052 | .0009994 | 68814.6872 | 1838.5820 | 6.748920 | 30281460 | 0.000000 | oP1(32) | Excited ev=4 31 30 F1e - X Xv=0 32 32 F1e |
| 13C18O | Excited | 31 | e | 2 | X | 32 | e | 1 | 67038.4311 | .0003964 | 68877.0131 | 1838.5820 | 2.677078 | 12045238 | 0.000000 | oP1(32) | Excited a3Pv=13 31 30 F1e - X Xv=0 32 32 F1e |
| 13C18O | Excited | 31 | e | 3 | X | 32 | e | 1 | 67095.6239 | .0001336 | 68934.2060 | 1838.5820 | .9019574 | 4068661 | 0.000000 | oP1(32) | Excited dv=7 31 30 F1e - X Xv=0 32 32 F1e |
| 13C18O | Excited | 31 | e | 4 | X | 32 | e | 1 | 67109.4605 | .0001956 | 68948.0426 | 1838.5820 | 1.321040 | 5962800 | 0.000000 | qP31(32) | Excited ev=4 31 32 F3e - X Xv=0 32 32 F1e |
| 13C18O | Excited | 31 | e | 5 | X | 32 | e | 1 | 67120.6051 | .0019779 | 68959.1871 | 1838.5820 | 13.35634 | 60316771 | 0.000000 | pP(32) | Excited Av=2 31 31 F1e - X Xv=0 32 32 F1e |
| 13C18O | Excited | 31 | e | 6 | X | 32 | e | 1 | 67130.6105 | .0001538 | 68969.1925 | 1838.5820 | 1.038771 | 4693151 | 0.000000 | pP21(32) | Excited a3Pv=13 31 31 F2e - X Xv=0 32 32 F1e |
| 13C18O | Excited | 31 | e | 7 | X | 32 | e | 1 | 67175.1132 | .0002463 | 69013.6952 | 1838.5820 | 1.663271 | 7529585 | 0.000000 | pP21(32) | Excited dv=7 31 31 F2e - X Xv=0 32 32 F1e |
| 13C18O | Excited | 31 | e | 8 | X | 32 | e | 1 | 67176.4212 | .0021209 | 69015.0033 | 1838.5820 | 14.32195 | 64838880 | 0.000000 | pP21(32) | Excited a'v=12 31 31 F2e - X Xv=0 32 32 F1e |
| 13C18O | Excited | 31 | e | 9 | X | 32 | e | 1 | 67212.5143 | .0007595 | 69051.0963 | 1838.5820 | 5.128666 | 23256145 | 0.000000 | qP31(32) | Excited a3Pv=13 31 32 F3e - X Xv=0 32 32 F1e |
| 13C18O | Excited | 31 | e | 10 | X | 32 | e | 1 | 67253.4453 | .0006677 | 69092.0273 | 1838.5820 | 4.508643 | 20482000 | 0.000000 | qP31(32) | Excited dv=7 31 32 F3e - X Xv=0 32 32 F1e |
| 13C18O | Excited | 32 | f | 2 | X | 32 | e | 1 | 67110.0911 | .0046108 | 68948.6731 | 1838.5820 | 31.13513 | 1.3621e8 | 0.000000 | pQ1(32) | Excited a3Pv=13 32 31 F1f - X Xv=0 32 32 F1e |
| 13C18O | Excited | 32 | f | 4 | X | 32 | e | 1 | 67163.1356 | .0001013 | 69001.7177 | 1838.5820 | .6843096 | 3000928 | 0.000000 | pQ1(32) | Excited dv=7 32 31 F1f - X Xv=0 32 32 F1e |
| 13C18O | Excited | 32 | f | 5 | X | 32 | e | 1 | 67176.9535 | .0012352 | 69015.5355 | 1838.5820 | 8.340711 | 36599408 | 0.000000 | pQ1(32) | Excited a'v=12 32 31 F1f - X Xv=0 32 32 F1e |
| 13C18O | Excited | 32 | f | 6 | X | 32 | e | 1 | 67208.0271 | .0008439 | 69046.6091 | 1838.5820 | 5.698710 | 25040908 | 0.000000 | qQ21(32) | Excited a3Pv=13 32 32 F2f - X Xv=0 32 32 F1e |
| 13C18O | Excited | 32 | f | 7 | X | 32 | e | 1 | 67209.8059 | .0029254 | 69048.3880 | 1838.5820 | 19.75427 | 86809862 | 0.000000 | qQ(32) | Excited Av=2 32 32 F1f - X Xv=0 32 32 F1e |
| 13C18O | Excited | 32 | f | 8 | X | 32 | e | 1 | 67243.8956 | .0014449 | 69082.4776 | 1838.5820 | 9.756829 | 42941515 | 0.000000 | qQ21(32) | Excited dv=7 32 32 F2f - X Xv=0 32 32 F1e |
| 13C18O | Excited | 32 | f | 9 | X | 32 | e | 1 | 67297.2715 | .0021030 | 69135.8535 | 1838.5820 | 14.20125 | 62651160 | 0.000000 | rQ31(32) | Excited a3Pv=13 32 33 F3f - X Xv=0 32 32 F1e |
| 13C18O | Excited | 32 | f | 10 | X | 32 | e | 1 | 67311.0177 | .0015674 | 69149.5997 | 1838.5820 | 10.58396 | 46721511 | 0.000000 | rQ31(32) | Excited a'v=12 32 33 F3f - X Xv=0 32 32 F1e |
| 13C18O | Excited | 32 | f | 11 | X | 32 | e | 1 | 67324.4157 | .0011240 | 69162.9978 | 1838.5820 | 7.589808 | 33524219 | 0.000000 | rQ31(32) | Excited dv=7 32 33 F3f - X Xv=0 32 32 F1e |

13C18O Excited 33 e 1 X 32 e 1 67114.0605 .0010288 68952.6425 1838.5820 6.947395 29492471 0.000000 : qR1(32) : Excited ev=4 33 32 F1e - X Xv=0 32 32 F1e
13C18O Excited 33 e 2 X 32 e 1 67200.6607 .0005762 69039.2428 1838.5820 3.890962 16581592 0.000000 : qR1(32) : Excited a3Pv=13 33 32 F1e - X Xv=0 32 32 F1e
13C18O Excited 33 e 4 X 32 e 1 67259.1811 .0008552 69097.7631 1838.5820 5.774601 24673193 0.000000 : sR31(32) : Excited ev=4 33 34 F3e - X Xv=0 32 32 F1e
13C18O Excited 33 e 5 X 32 e 1 67294.0246 .0002352 69132.6066 1838.5820 1.588234 6796616 0.000000 : rR21(32) : Excited a3Pv=13 33 33 F2e - X Xv=0 32 32 F1e
13C18O Excited 33 e 6 X 32 e 1 67300.6823 .0004217 69139.2644 1838.5820 2.847286 12188158 0.000000 : rR(32) : Excited Av=2 33 33 F1e - X Xv=0 32 32 F1e
13C18O Excited 33 e 7 X 32 e 1 67310.5903 .0017689 69149.1723 1838.5820 11.94500 51154639 0.000000 : rR21(32) : Excited a'v=12 33 33 F2e - X Xv=0 32 32 F1e
13C18O Excited 33 e 8 X 32 e 1 67315.5901 .0017928 69154.1721 1838.5820 12.10616 51856350 0.000000 : rR21(32) : Excited dv=7 33 33 F2e - X Xv=0 32 32 F1e
13C18O Excited 33 e 9 X 32 e 1 67382.7738 .0014109 69221.3558 1838.5820 9.527183 40931700 0.000000 : sR31(32) : Excited a3Pv=13 33 34 F3e - X Xv=0 32 32 F1e
13C18O Excited 33 e 10 X 32 e 1 67398.8333 .0002355 69237.4153 1838.5820 1.590333 6837444 0.000000 : sR31(32) : Excited dv=7 33 34 F3e - X Xv=0 32 32 F1e
13C18O Excited 32 e 1 X 33 e 1 66929.5605 .0005749 68882.6808 1953.1202 6.724508 29182636 0.000000 : oP1(33) : Excited ev=4 32 31 F1e - X Xv=0 33 33 F1e
13C18O Excited 32 e 2 X 33 e 1 67003.7393 .0002751 68956.8595 1953.1202 3.217810 14010948 0.000000 : oP1(33) : Excited a3Pv=13 32 31 F1e - X Xv=0 33 33 F1e
13C18O Excited 32 e 4 X 33 e 1 67069.4918 .0003784 69022.6121 1953.1202 4.425905 19328014 0.000000 : qP31(33) : Excited ev=4 32 33 F3e - X Xv=0 33 33 F1e
13C18O Excited 32 e 5 X 33 e 1 67093.8948 .0006462 69047.0151 1953.1202 7.558089 33042363 0.000000 : pP(33) : Excited Av=2 32 32 F1e - X Xv=0 33 33 F1e
13C18O Excited 32 e 6 X 33 e 1 67097.0402 .0002507 69050.1605 1953.1202 2.932002 12819894 0.000000 : pP21(33) : Excited a3Pv=13 32 32 F2e - X Xv=0 33 33 F1e
13C18O Excited 32 e 7 X 33 e 1 67127.5897 .0007273 69080.7099 1953.1202 8.506767 37245842 0.000000 : pP21(33) : Excited a'v=12 32 32 F2e - X Xv=0 33 33 F1e
13C18O Excited 32 e 8 X 33 e 1 67129.4795 .0007776 69082.5998 1953.1202 9.095208 39825623 0.000000 : pP21(33) : Excited dv=7 32 32 F2e - X Xv=0 33 33 F1e
13C18O Excited 32 e 9 X 33 e 1 67182.5049 .0005571 69135.6251 1953.1202 6.516053 28599819 0.000000 : qP31(33) : Excited a3Pv=13 32 33 F3e - X Xv=0 33 33 F1e
13C18O Excited 32 e 10 X 33 e 1 67209.8635 .0003169 69162.9837 1953.1202 3.707023 16290500 0.000000 : qP31(33) : Excited dv=7 32 33 F3e - X Xv=0 33 33 F1e
13C18O Excited 33 f 2 X 33 e 1 67070.6644 .0027050 69023.7847 1953.1202 31.63775 1.3405e8 0.000000 : qQ21(33) : Excited ev=4 33 33 F2f - X Xv=0 33 33 F1e
13C18O Excited 33 f 3 X 33 e 1 67086.5940 .0001810 69039.7143 1953.1202 2.117173 8976608 0.000000 : pQ1(33) : Excited a3Pv=13 33 32 F1f - X Xv=0 33 33 F1e
13C18O Excited 33 f 5 X 33 e 1 67128.2400 .0008112 69081.3603 1953.1202 9.488234 40304153 0.000000 : pQ1(33) : Excited a'v=12 33 32 F1f - X Xv=0 33 33 F1e
13C18O Excited 33 f 6 X 33 e 1 67178.4414 .0002734 69131.5616 1953.1202 3.197986 13614917 0.000000 : qQ21(33) : Excited a3Pv=13 33 33 F2f - X Xv=0 33 33 F1e
13C18O Excited 33 f 7 X 33 e 1 67187.0463 .0014392 69140.1665 1953.1202 16.83315 71692002 0.000000 : qQ(33) : Excited Av=2 33 33 F1f - X Xv=0 33 33 F1e
13C18O Excited 33 f 8 X 33 e 1 67200.7745 .0011919 69153.8948 1953.1202 13.94040 59408254 0.000000 : qQ21(33) : Excited dv=7 33 33 F2f - X Xv=0 33 33 F1e
13C18O Excited 33 f 9 X 33 e 1 67266.2238 .0011977 69219.3440 1953.1202 14.00879 59874298 0.000000 : rQ31(33) : Excited a'v=12 33 34 F3f - X Xv=0 33 33 F1e
13C18O Excited 33 f 10 X 33 e 1 67268.4659 .0014449 69221.5861 1953.1202 16.90003 72238808 0.000000 : rQ31(33) : Excited a3Pv=13 33 34 F3f - X Xv=0 33 33 F1e
13C18O Excited 33 f 11 X 33 e 1 67284.3549 .0002625 69237.4751 1953.1202 3.069742 13130850 0.000000 : rQ31(33) : Excited dv=7 33 34 F3f - X Xv=0 33 33 F1e
13C18O Excited 34 e 1 X 33 e 1 67071.5040 .0005979 69024.6243 1953.1202 6.992857 28770217 0.000000 : qR1(33) : Excited ev=4 34 33 F1e - X Xv=0 33 33 F1e
13C18O Excited 34 e 2 X 33 e 1 67170.6535 .0003905 69123.7737 1953.1202 4.567217 18874031 0.000000 : qR1(33) : Excited a3Pv=13 34 33 F1e - X Xv=0 33 33 F1e
13C18O Excited 34 e 4 X 33 e 1 67221.3017 .0005456 69174.4219 1953.1202 6.381718 26432165 0.000000 : sR31(33) : Excited ev=4 34 35 F3e - X Xv=0 33 33 F1e
13C18O Excited 34 e 5 X 33 e 1 67263.8146 .0008132 69216.9348 1953.1202 9.511410 39469693 0.000000 : rR21(33) : Excited a'v=12 34 34 F2e - X Xv=0 33 33 F1e
13C18O Excited 34 e 8 X 33 e 1 67285.0860 .0016211 69238.2062 1953.1202 18.96096 78757353 0.000000 : rR(33) : Excited Av=2 34 34 F1e - X Xv=0 33 33 F1e
13C18O Excited 34 e 9 X 33 e 1 67350.8680 .0009879 69303.9882 1953.1202 11.55480 48135580 0.000000 : sR31(33) : Excited dv=7 34 35 F3e - X Xv=0 33 33 F1e
13C18O Excited 33 e 1 X 34 e 1 66881.5591 .0003274 68952.6425 2071.0834 6.743060 28328580 0.000000 : oP1(34) : Excited ev=4 33 32 F1e - X Xv=0 34 34 F1e
13C18O Excited 33 e 2 X 34 e 1 66968.1593 .0001834 69039.2428 2071.0834 3.776522 15927430 0.000000 : oP1(34) : Excited a3Pv=13 33 32 F1e - X Xv=0 34 34 F1e
13C18O Excited 33 e 4 X 34 e 1 67026.6797 .0002722 69097.7631 2071.0834 5.604760 23700023 0.000000 : qP31(34) : Excited ev=4 33 34 F3e - X Xv=0 34 34 F1e
13C18O Excited 33 e 6 X 34 e 1 67068.1810 .0001342 69139.2644 2071.0834 2.763542 11707503 0.000000 : pP(34) : Excited Av=2 33 33 F1e - X Xv=0 34 34 F1e
13C18O Excited 33 e 7 X 34 e 1 67078.0889 .0005630 69149.1723 2071.0834 11.59368 49137368 0.000000 : pP21(34) : Excited a'v=12 33 33 F2e - X Xv=0 34 34 F1e
13C18O Excited 33 e 8 X 34 e 1 67083.0887 .0005706 69154.1721 2071.0834 11.75009 49811446 0.000000 : pP21(34) : Excited dv=7 33 33 F2e - X Xv=0 34 34 F1e
13C18O Excited 33 e 9 X 34 e 1 67150.2724 .0004490 69221.3558 2071.0834 9.246971 39318006 0.000000 : qP31(34) : Excited a3Pv=13 33 34 F3e - X Xv=0 34 34 F1e
13C18O Excited 34 f 2 X 34 e 1 67027.6740 .0014376 69098.7574 2071.0834 29.60663 1.2157e8 0.000000 : qQ21(34) : Excited ev=4 34 34 F2f - X Xv=0 34 34 F1e
13C18O Excited 34 f 3 X 34 e 1 67052.3974 .0003207 69123.4808 2071.0834 6.603981 27147076 0.000000 : pQ1(34) : Excited a3Pv=13 34 33 F1f - X Xv=0 34 34 F1e
13C18O Excited 34 f 5 X 34 e 1 67078.0982 .0004730 69149.1816 2071.0834 9.741377 40090079 0.000000 : pQ1(34) : Excited a'v=12 34 33 F1f - X Xv=0 34 34 F1e
13C18O Excited 34 f 6 X 34 e 1 67146.9246 .0002427 69218.0080 2071.0834 4.999103 20636916 0.000000 : qQ21(34) : Excited a3Pv=13 34 34 F2f - X Xv=0 34 34 F1e
13C18O Excited 34 f 8 X 34 e 1 67166.3243 .0013572 69237.4077 2071.0834 27.94998 1.1548e8 0.000000 : qQ(34) : Excited Av=2 34 34 F1f - X Xv=0 34 34 F1e
13C18O Excited 34 f 9 X 34 e 1 67220.0532 .0006144 69291.1367 2071.0834 12.65278 52403081 0.000000 : rQ31(34) : Excited a'v=12 34 35 F3f - X Xv=0 34 34 F1e
13C18O Excited 34 f 10 X 34 e 1 67232.9949 .0011091 69304.0784 2071.0834 22.84082 94652775 0.000000 : rQ31(34) : Excited dv=7 34 35 F3f - X Xv=0 34 34 F1e
13C18O Excited 35 e 1 X 34 e 1 67027.5717 .0003429 69098.6551 2071.0834 7.062312 28182043 0.000000 : qR1(34) : Excited ev=4 35 34 F1e - X Xv=0 34 34 F1e
13C18O Excited 35 e 2 X 34 e 1 67137.0764 .0002242 69208.1598 2071.0834 4.617081 18514834 0.000000 : qR1(34) : Excited a3Pv=13 35 34 F1e - X Xv=0 34 34 F1e
13C18O Excited 35 e 4 X 34 e 1 67181.8286 .0003317 69252.9120 2071.0834 6.832053 27451860 0.000000 : sR31(34) : Excited ev=4 35 36 F3e - X Xv=0 34 34 F1e
13C18O Excited 35 e 5 X 34 e 1 67218.2972 .0005117 69289.3806 2071.0834 10.53704 42407842 0.000000 : rR21(34) : Excited a'v=12 35 35 F2e - X Xv=0 34 34 F1e
13C18O Excited 35 e 6 X 34 e 1 67227.4253 .0001342 69298.5087 2071.0834 2.762934 11124361 0.000000 : rR21(34) : Excited dv=7 35 35 F2e - X Xv=0 34 34 F1e
13C18O Excited 35 e 8 X 34 e 1 67264.0243 .0007247 69335.1077 2071.0834 14.92419 60187224 0.000000 : rR(34) : Excited Av=2 35 35 F1e - X Xv=0 34 34 F1e
13C18O Excited 35 e 9 X 34 e 1 67311.8189 .0005202 69382.9024 2071.0834 10.71226 43293252 0.000000 : sR31(34) : Excited dv=7 35 36 F3e - X Xv=0 34 34 F1e
13C18O Excited 34 e 1 X 35 e 1 66832.1568 .0001843 69024.6243 2192.4675 6.793061 27650074 0.000000 : oP1(35) : Excited ev=4 34 33 F1e - X Xv=0 35 35 F1e
13C18O Excited 34 e 2 X 35 e 1 66931.3062 .0001204 69123.7737 2192.4675 4.436725 18139475 0.000000 : oP1(35) : Excited a3Pv=13 34 33 F1e - X Xv=0 35 35 F1e
13C18O Excited 34 e 4 X 35 e 1 66981.9545 .0001682 69174.4219 2192.4675 6.199384 25403661 0.000000 : qP31(35) : Excited ev=4 34 35 F3e - X Xv=0 35 35 F1e
13C18O Excited 34 e 5 X 35 e 1 67024.4674 .0002507 69216.9348 2192.4675 9.239656 37934141 0.000000 : pP21(35) : Excited a'v=12 34 34 F2e - X Xv=0 35 35 F1e
13C18O Excited 34 e 8 X 35 e 1 67045.7387 .0004997 69238.2062 2192.4675 18.41922 75693587 0.000000 : pP(35) : Excited Av=2 34 34 F1e - X Xv=0 35 35 F1e
13C18O Excited 34 e 9 X 35 e 1 67111.5208 .0003045 69303.9882 2192.4675 11.22466 46263526 0.000000 : qP31(35) : Excited dv=7 34 35 F3e - X Xv=0 35 35 F1e
13C18O Excited 35 f 2 X 35 e 1 66982.7328 .0007780 69175.2003 2192.4675 28.67857 1.1421e8 0.000000 : qQ21(35) : Excited ev=4 35 35 F2f - X Xv=0 35 35 F1e
13C18O Excited 35 f 3 X 35 e 1 67015.3574 .0002598 69207.8248 2192.4675 9.575790 38191147 0.000000 : pQ1(35) : Excited dv=7 35 34 F1f - X Xv=0 35 35 F1e
13C18O Excited 35 f 4 X 35 e 1 67025.2790 .0001904 69217.7465 2192.4675 7.019373 28007821 0.000000 : pQ1(35) : Excited a3Pv=13 35 34 F1f - X Xv=0 35 35 F1e
13C18O Excited 35 f 6 X 35 e 1 67105.9123 .0001977 69298.3798 2192.4675 7.287422 29182426 0.000000 : qQ21(35) : Excited dv=7 35 35 F2f - X Xv=0 35 35 F1e
13C18O Excited 35 f 8 X 35 e 1 67142.1297 .0006275 69334.5972 2192.4675 23.12930 92771129 0.000000 : qQ(35) : Excited Av=2 35 35 F1f - X Xv=0 35 35 F1e
13C18O Excited 35 f 9 X 35 e 1 67172.4976 .0003787 69364.9651 2192.4675 13.95870 56064050 0.000000 : rQ31(35) : Excited a'v=12 35 36 F3f - X Xv=0 35 35 F1e
13C18O Excited 35 f 10 X 35 e 1 67190.4674 .0005773 69382.9348 2192.4675 21.28050 85540127 0.000000 : rQ31(35) : Excited dv=7 35 36 F3f - X Xv=0 35 35 F1e
13C18O Excited 35 f 11 X 35 e 1 67218.0415 .0001075 69410.5090 2192.4675 3.962415 15947133 0.000000 : rQ31(35) : Excited a3Pv=13 35 36 F3f - X Xv=0 35 35 F1e
13C18O Excited 36 e 1 X 35 e 1 66982.2841 .0001940 69174.7516 2192.4675 7.149625 27692597 0.000000 : qR1(35) : Excited ev=4 36 35 F1e - X Xv=0 35 35 F1e
13C18O Excited 36 e 4 X 35 e 1 67140.9033 .0001983 69333.3708 2192.4675 7.309118 28511959 0.000000 : sR31(35) : Excited ev=4 36 37 F3e - X Xv=0 35 35 F1e
13C18O Excited 36 e 5 X 35 e 1 67170.7687 .0003033 69363.2362 2192.4675 11.17819 43662948 0.000000 : rR21(35) : Excited a'v=12 36 36 F2e - X Xv=0 35 35 F1e
13C18O Excited 36 e 8 X 35 e 1 67242.9566 .0003385 69435.4241 2192.4675 12.47860 48899761 0.000000 : rR(35) : Excited Av=2 36 36 F1e - X Xv=0 35 35 F1e
13C18O Excited 36 e 9 X 35 e 1 67270.0769 .0003099 69462.5444 2192.4675 11.42272 44816276 0.000000 : sR31(35) : Excited dv=7 36 37 F3e - X Xv=0 35 35 F1e
13C18O Excited 35 e 1 X 36 e 1 66781.3869 .0001024 69098.6551 2317.2682 6.866137 27098413 0.000000 : oP1(36) : Excited ev=4 35 34 F1e - X Xv=0 36 36 F1e
13C18O Excited 35 e 5 X 36 e 1 66972.1124 .0001527 69289.3806 2317.2682 10.24434 40778494 0.000000 : pP21(36) : Excited a'v=12 35 35 F2e - X Xv=0 36 36 F1e
13C18O Excited 35 e 8 X 36 e 1 67017.8395 .0002163 69335.1077 2317.2682 14.50963 57875211 0.000000 : pP(36) : Excited Av=2 35 35 F1e - X Xv=0 36 36 F1e
13C18O Excited 35 e 9 X 36 e 1 67065.6341 .0001553 69382.9024 2317.2682 10.41470 41630524 0.000000 : qP31(36) : Excited dv=7 35 36 F3e - X Xv=0 36 36 F1e
13C18O Excited 36 f 2 X 36 e 1 66936.2564 .0004238 69253.5246 2317.2682 28.42532 1.0987e8 0.000000 : qQ21(36) : Excited ev=4 36 36 F2f - X Xv=0 36 36 F1e
13C18O Excited 36 f 3 X 36 e 1 66969.6517 .0001773 69286.9199 2317.2682 6.852784 26527830 0.000000 : pQ1(36) : Excited dv=7 36 35 F1f - X Xv=0 36 36 F1e
13C18O Excited 36 f 4 X 36 e 1 66973.2116 .0001606 69290.4798 2317.2682 10.76894 41694331 0.000000 : pQ1(36) : Excited a'v=12 36 35 F1f - X Xv=0 36 36 F1e
13C18O Excited 36 f 6 X 36 e 1 67058.7190 .0001125 69375.9872 2317.2682 7.542089 29312845 0.000000 : qQ21(36) : Excited dv=7 36 36 F2f - X Xv=0 36 36 F1e
13C18O Excited 36 f 8 X 36 e 1 67117.0591 .0001560 69434.3273 2317.2682 10.46558 40781444 0.000000 : qQ(36) : Excited Av=2 36 36 F1f - X Xv=0 36 36 F1e
13C18O Excited 36 f 9 X 36 e 1 67124.1501 .0003479 69441.4183 2317.2682 23.33260 90949470 0.000000 : rQ31(36) : Excited a'v=12 36 37 F3f - X Xv=0 36 36 F1e
13C18O Excited 36 f 10 X 36 e 1 67145.2986 .0003406 69462.5668 2317.2682 22.84501 89133060 0.000000 : rQ31(36) : Excited dv=7 36 37 F3f - X Xv=0 36 36 F1e
13C18O Excited 37 e 1 X 36 e 1 66935.6550 .0001081 69252.9232 2317.2682 7.250471 27277273 0.000000 : qR1(36) : Excited ev=4 37 36 F1e - X Xv=0 36 36 F1e
13C18O Excited 37 e 4 X 36 e 1 67098.6130 .0001198 69415.8812 2317.2682 8.035592 30452340 0.000000 : sR31(36) : Excited ev=4 37 38 F3e - X Xv=0 36 36 F1e
13C18O Excited 37 e 5 X 36 e 1 67121.7522 .0001729 69439.0204 2317.2682 11.59517 43987480 0.000000 : rR21(36) : Excited a'v=12 37 37 F2e - X Xv=0 36 36 F1e
13C18O Excited 37 e 9 X 36 e 1 67228.2648 .0002966 69545.5330 2317.2682 19.89330 75827135 0.000000 : sR31(36) : Excited dv=7 37 38 F3e - X Xv=0 36 36 F1e
13C18O Excited 37 f 2 X 37 e 1 66888.3682 .0002300 69333.8495 2445.4813 28.52923 1.0710e8 0.000000 : qQ21(37) : Excited ev=4 37 37 F2f - X Xv=0 37 37 F1e
13C18O Excited 37 f 3 X 37 e 1 66918.6252 .0001299 69364.1065 2445.4813 16.11036 60563153 0.000000 : pQ1(37) : Excited a'v=12 37 36 F1f - X Xv=0 37 37 F1e
13C18O Excited 37 f 10 X 37 e 1 67100.1015 .0003257 69545.5828 2445.4813 40.40118 1.5312e8 0.000000 : rQ31(37) : Excited dv=7 37 38 F3f - X Xv=0 37 37 F1e
13C18O Excited 38 e 9 X 37 e 1 67200.9529 .0001538 69646.4342 2445.4813 19.07569 70735805 0.000000 : rR(37) : Excited Av=2 38 38 F1e - X Xv=0 37 37 F1e

13C18O Excited 38 f 2 X 38 e 1 66839.1162 .0001237 69416.2183 2577.1022 28.84754 1.0525e8 0.000000 : qQ21(38) : Excited ev=4 38 38 F2f - X Xv=0 38 38 F1e
13C18O Excited 38 f 10 X 38 e 1 67069.3632 .0001642 69646.4654 2577.1022 38.29437 1.4117e8 0.000000 : qQ(38) : Excited Av=2 38 38 F1f - X Xv=0 38 38 F1e
1178 Lines above OThreshold, Total intensity = 479.2413324369